\input epsf.tex
\epsfclipon
 \documentclass[11pt,a4paper,amsart]{article}
 \usepackage{jheppub}
\usepackage[toc,page]{appendix}
\usepackage[compat=1.1.0]{tikz-feynman}

\usepackage{amsmath,amssymb,mathtools}
\usetikzlibrary{decorations.pathreplacing}
\usetikzlibrary{hobby}

\newcommand{\sgn}{\operatorname{sgn}}

\usepackage{hyperref}
\usepackage{multirow,tabularx}

 \usepackage{color}
\definecolor{myblue}{rgb}{.8, .8, 1}

\usepackage{amsmath}
\usepackage{empheq}
\usepackage{slashed}
\newlength\mytemplen
\newsavebox\mytempbox

\makeatletter
\newcommand\mybluebox{%
    \@ifnextchar[
       {\@mybluebox}%
       {\@mybluebox[0pt]}}

\def\@mybluebox[#1]{%
    \@ifnextchar[
       {\@@mybluebox[#1]}%
       {\@@mybluebox[#1][0pt]}}

\def\@@mybluebox[#1][#2]#3{
    \sbox\mytempbox{#3}%
    \mytemplen\ht\mytempbox
    \advance\mytemplen #1\relax
    \ht\mytempbox\mytemplen
    \mytemplen\dp\mytempbox
    \advance\mytemplen #2\relax
    \dp\mytempbox\mytemplen
    \colorbox{myblue}{\hspace{1em}\usebox{\mytempbox}\hspace{1em}}}

\makeatother

\usepackage{float}
\usepackage{mathtools}
\usepackage{color}

\usepackage{epsfig}
\usepackage{epstopdf}
\usepackage{epsf}
\input{epsf.sty}
\usepackage{graphicx,amsmath,amssymb}
\usepackage{graphicx}
\usepackage{epsfig,multicol}
\usepackage{multirow}
\allowdisplaybreaks

\usepackage{tikz} 

\usepackage{bbm,bm,amsmath,amssymb}
 
\newcommand{\Su}{\color{blue}}
\newcommand{\red}{\color{red}}
\usepackage[normalem]{ulem}

\newcommand{\Hc}{{\cal H}}
\usepackage{comment}
\def\be{\begin{equation}}
\def\ee{\end{equation}}
\def\figs/B{B}
\def\bea{\begin{eqnarray}}
\def\eea{\end{eqnarray}}
\def\bg{\begin{eqnarray}}
\def\nd{\end{eqnarray}}
\def\sin{{\rm sin}}
\def\cos{{\rm cos}}

\def\log{{\rm log}}
\def\ln{{\rm log}}
\def\xxy{\left({g_s\over {\rm H}(y){\rm H}_o({\bf x})}\right)}
\def\xoxo{{\mathbb{T}^2\over {\cal G}}}
\renewcommand{\d}{{\mathrm{d}}}
\renewcommand{\[}{\left[}

\def\be{\begin{equation}}
\def\ee{\end{equation}}

\def\doi{http://doi.org}

\def\d{\mathrm{d}}
\def\g{\mathrm{g}}

\usepackage{physics}
\usepackage{tikz}
\usetikzlibrary{arrows.meta}
\usetikzlibrary{angles,quotes} 
\usetikzlibrary{decorations.markings,arrows.meta}
\tikzset{>=latex} 

\tikzset{
  midarr/.style={decoration={markings,mark=at position #1 with {\arrow{stealth}}},postaction={decorate}},
  midarr/.default=0.5
}
\colorlet{xcol}{blue!70!black}

\title{Transient de Sitter and Quasi de Sitter States in ${\rm SO}(32)$ and ${\rm E}_8 \times {\rm E}_8$ Heterotic String Theories}
\author{{\Su Keshav Dasgupta}$^{1}$, {\Su Suddhasattwa Brahma}$^{2}$, {\Su Bohdan Kulinich}$^{1}$, {\Su Archana Maji}$^3$, {\Su Pichai Ramadevi}$^3$ and
{\Su Radu Tatar}$^{4}$\\
	\vskip.03in
        ${}^1$ Department of Physics, McGill University, Montr\'{e}al, Qu\'{e}bec, H3A 2T8, Canada \\    
        ${}^2$ Higgs Centre for Theoretical Physics, School of Physics \& 
	Astronomy, University of Edinburgh, Edinburgh, EH9 3FD, United 
        Kingdom\\ 
	${}^3$ Department of Physics, Indian Institute of Technology Bombay, Mumbai 400076, India\\
       ${}^4$ Department of Mathematical Sciences,
        University of Liverpool,  Liverpool, L69 7ZL, United Kingdom \\	
        {\tt ramadevi@iitb.ac.in, Radu.Tatar@Liverpool.ac.uk}  
        {\tt bohdan.kulinich@mail.mcgill.ca, archana${}_-$phy@iitb.ac.in}  {\tt suddhasattwa.brahma@gmail.com, keshav@hep.physics.mcgill.ca}}

\date{\today}

\abstract{One of the long-standing puzzles in string theory has been on the existence of a four-dimensional de Sitter and quasi de Sitter configurations, the latter being defined with a temporally varying dark energy, in ${\rm E}_8 \times {\rm E}_8$ and ${\rm SO}(32)$ heterotic theories. In this work, novel dynamical duality-sequences are devised that provide natural constructions of de Sitter and quasi de Sitter excited states in the aforementioned theories allowing no late-time singularities. The emergent positive dark energies $-$ including the intriguing possibility of their slow temporal variations $-$ appear from Borel resumming Gevrey series from the path-integral representations of such states. Additionally, precise ways to handle the equations of motion, Bianchi identities, flux quantizations and anomaly cancellations $-$ consistent with the underlying axionic cosmology, with the temporal dependence, and with the probability of forming wormholes that connect baby universes $-$ are presented for the $SO(32)$ and the ${\rm E}_8 \times {\rm E}_8$ theories 
within a framework that systematically incorporates perturbative and non-perturbative corrections in the far infrared. The temporally varying dark energy, which is much more natural in our set-up because of its emergent nature, surprisingly simplifies many of the aforementioned computations.
Interestingly, 
our analysis also provides, probably for the first time, a set-up to consistently embed four-dimensional standard model degrees of freedom at late time in a realistic gravitational background with positive dark energy from string theory.} 

\begin{document}

\maketitle

\newpage

\hskip3.0in {\Su {\it The devil, as always, is in the details}.}

\section{Introduction and conceptual framework}	
\label{sec:intro}

Accelerating universes, four-dimensional de Sitter space-time being one prime example, challenge many of the long-held beliefs that we usually take in for granted. For example the existence of a global notion of time, existence of a well-defined Hamiltonian driving the dynamics of the system, the concept of the conservation of energy, mode expansion of the fields in terms of Fourier components {\it et cetera} are generically considered to be the cornerstones for building an effective field theory. Unfortunately none of the aforementioned list of accepted facts remain true in accelerating backgrounds: there is generically no true notion of a global time\footnote{Let us clarify two things that are often mixed up: existence of a global timelike killing vector and the existence of a global time coordinate. For a de Sitter space
there is \emph{no globally timelike Killing vector} on all of $\mathrm{dS}_4$.
In other words, there is no symmetry that looks like ``time translation everywhere.''
(Only in a static patch---inside one observer's horizon---does $\partial_t$ stay timelike
and generate a time-translation symmetry.) On the other hand, 
de Sitter does admit a global time function and a global synchronous slicing.
In global coordinates, with the Hubble constant ${\rm H}$:
\bg\label{peribindu00}
ds^2 \;=\; -\,dt^2 \;+\; {1\over {\rm H}^{2}}\,\cosh^2({\rm H}t)\, d\Omega_3^2,
\nd
the constant-$t$ slices are ${\bf S}^3$ and $t\in(-\infty,\infty)$. This is synchronous
(lapse $=1$, shift $=0$) and covers all of de Sitter. Thus 
we \emph{can} foliate all of dS with spacelike slices labeled by a global time $t$
(a global time direction in the foliation sense). However
we \emph{cannot} find a single vector field that is timelike everywhere \emph{and} Killing
(a global time-translation symmetry). Each static patch has its own timelike Killing vector,
but it becomes spacelike outside that patch. Thus when 
we say ``de Sitter has no global time direction,'' we mean there is
no \emph{global timelike Killing vector}, not that there is no global time coordinate. Of course caveats that could spoil a truly global slicing aren’t specific to string theory: e.g., metastability (bubble nucleation), domain walls, or other time-dependent ingredients that end de Sitter evolution can terminate the foliation globally. Absence of such events, the de Sitter geometry itself has a perfectly good global synchronous slicing.}, the Hamiltonian is simply a constraint (and therefore usually vanishes), there is no energy conservation per se\footnote{Locally there does exist energy conservation because 
$\nabla^{\rm M} {\bf T}_{\rm MN} = 0$ coming from the symmetry of the Einstein tensor. Globally however the story is very different.}, and simple mode expansions typically lead to the so-called backreaction issues. All these problems have an immediate bearing on another fundamental concept of high energy physics, namely the Wilsonian effective action. Due to the non-conservation of energy, any UV mode quickly gets red-shifted to an IR mode and therefore constructing a simple effective-field theory by {\it integrating out} the UV modes does not\footnote{One might expect that working in the static patch would alleviate the difficulty. Unfortunately, this merely obscures the issue rather than resolving it, and only with limited effectiveness \cite{joydeep}.} make sense! 

The aforementioned issues have been raised earlier in \cite{joydeep, wdwpaper}, and a simple way out of these conundrums is to view the accelerating backgrounds as {\it excited states} over a given supersymmetric Minkowski (or solitonic) background in the low energy effective action of string theory: the so-called Glauber-Sudarshan (GS) states \cite{coherbeta, coherbeta2}. Being supersymmetric {\it and} Minkowski (with possibly warped but static internal dimensions), none of the aforementioned problems would be there\footnote{Backreactions of the modes could in principle still exist, but if we are in the vacuum state such backreactions can be easily quantified. In fact in our construction of the Glauber-Sudarshan states the backreactions in the Minkowski spacetime play a crucial role \cite{coherbeta, coherbeta2}.}! The only other alternative to this, as far as we know, is to look for an {\it open field theory} description in string theory. The latter has many issues related to its actual implementation in string theory, for example, at this stage it is not clear how to separate degrees of freedom into {\it relevant} and {\it environmental} to perform efficient transfer of energies between them. Additionally, due to its reliance on Markovian processes and due to the absence of an underlying action governing the dynamics, it becomes harder to realize this in the low energy stringy set-up.

Many questions related to the consistency of the above picture arise, and in \cite{joydeep, coherbeta, coherbeta2, wdwpaper} we have tried to answer some of them. Here however we want to raise new questions that we haven't asked yet. For example, the question of realizing the Glauber-Sudarshan states from a world-sheet description in string theory is a particularly intriguing one. The reason is twofold. First, since our aim in this paper is to realize transient de Sitter excited states in $SO(32)$ and ${\rm E}_8 \times {\rm E}_8$ heterotic theories, and second, the fact that certain no-go theorems have been proposed 
precisely for such theories in \cite{KMMS} from world-sheet point of view, answering questions related to this become necessary. 

Realizing transient de Sitter excited states in heterotic theories bring forth more urgent questions like: how do we know such backrounds are consistent? Do they satisfy EOMs, and if yes, what do the EOMs mean when we have transient states? What about consistency related to the flux quantizations, anomaly cancellations, Bianchi identities {\it et cetera} as we are dealing with excited states that are inherently time-dependent? All these and more have been answered in details in sections \ref{sec6}, \ref{sec7s}, \ref{sec4.5}, \ref{secanomaly} and \ref{secmetric}.

There are also  more {\it exotic} questions that could appear in our framework because the consistency of our picture relies heavily on non-perturbative effects coming from the five-brane instantons (in M-theory). In fact both {\it real} and {\it complex} instantons contribute leading to a resurgent trans-series structure of the far IR action (see for example \eqref{kimkarol}). Since both the non-perturbative instanton effects and resurgence are absolutely essential, questions arise whether such structures would change if our Minkowski background spits out baby universes which are in turn connected to the Minkowski universe via wormholes. 

In the following we will give brief qualitative introductions to the various issues that we raised above starting with the world-sheet representation of the Glauber-Sudarshan states, and then provide more quantitative analysis in the main text.

\subsection{World-sheet representation of the Glauber-Sudarshan states? \label{ethantag} }

In M-theory, the answer is simple: there isn’t a single, simple ``world-sheet"–style CFT for a de Sitter-as-Glauber–Sudarshan (GS) state in M-theory, because M-theory has no perturbative world-sheet. On the other hand, in ten-dimensional string theory
this is a very interesting and subtle question. We are essentially asking whether the Glauber--Sudarshan (GS) representation of a de Sitter-like (transient) state in quantum gravity or string theory could be realized in terms of a world-sheet description, the way certain backgrounds (flat space, black holes, AdS, orbifolds, etc.) are. Let us parse the layers carefully.

\section*{(a) Glauber--Sudarshan states and de Sitter}

\noindent $\bullet$ A GS state (in quantum optics/field theory) is a representation of a mixed or coherent-like state in terms of an ensemble of coherent states, i.e.~as a diagonal representation of the density matrix in the overcomplete coherent-state basis.
\vskip.1in
\noindent $\bullet$
In recent proposals \cite{coherbeta, coherbeta2}, de Sitter is not a true \emph{vacuum} of string/M-theory, but rather a \emph{transient state} that can be represented as a Glauber--Sudarshan--type statistical mixture over more fundamental coherent states.
\vskip.1in
\noindent $\bullet$
The key feature is that such a state captures the quasi-classical properties of de Sitter while being fundamentally non-stationary (so consistent with the ``no-go'' theorems against stable de Sitter vacua).
\vskip.1in

\section*{(b) World-sheet representations in string theory}

\noindent $\bullet$
World-sheet CFTs describe backgrounds that solve the full string equations of motion (beta functions vanish). This usually corresponds to exact vacua: Minkowski, AdS, Calabi--Yau compactifications, WZW models, etc.
\vskip.1in
\noindent $\bullet$
A Glauber--Sudarshan--like de Sitter state is not an exact vacuum. It is, by construction, a statistical mixture or ensemble over microscopic states solving Schwinger-Dyson equations.
\vskip.1in
\noindent $\bullet$
Therefore, on the world-sheet one would not expect a \emph{single local 2D CFT} to describe it. Instead, the correct description would probably look like a weighted ensemble of world-sheet theories, or maybe as a path integral with an additional ``density matrix weight'' rather than a pure-state amplitude.

\section*{(c) Possible approaches to a world-sheet picture}

\noindent $\bullet$
\textbf{Mixed-state path integrals:} In open string field theory and in string thermodynamics, mixed states can be represented via thermal traces (e.g.~torus partition function with Euclidean time compactification). Analogously, a GS ensemble could be represented as a sum/integral over boundary conditions or coherent-state deformations of the world-sheet action.

\vskip.1in
\noindent $\bullet$
\textbf{Liouville/time-dependent backgrounds:} World-sheet CFTs like Liouville theory, or rolling tachyon setups, already give transient, non-equilibrium states. These may be the closest analogs to a GS representation on the world-sheet.
\vskip.1in
\noindent $\bullet$
\textbf{String field theory representation:} In string field theory (second quantized), coherent states correspond to classical field configurations. A GS representation would then correspond to a density matrix over such coherent configurations. Translating that back to a world-sheet language would likely mean a non-local insertion in the {\it 2D} theory, encoding the statistical mixture.
\vskip.1in
\noindent $\bullet$
\textbf{Matrix cosmology/dual pictures:} In certain dual formulations (e.g.~matrix models for {\it 2D} strings), mixed states have clearer statistical interpretations. One could imagine embedding the GS state in such a dual, then mapping back to a ``smeared'' world-sheet description.

\section*{(d) Glauber--Sudarshan states in M-Theory}

\noindent $\bullet$
As mentioned earlier, in M-theory there is no perturbative world-sheet CFT, so one cannot expect a ``simple'' {\it 2D} description of a transient de Sitter-like state.

\vskip.1in
\noindent $\bullet$
Instead, the Glauber--Sudarshan (GS) framework suggests representing such a state as an \emph{ensemble} over more fundamental configurations, rather than a single vacuum.
\vskip.1in
\noindent $\bullet$
Three natural viewpoints arise in eleven dimensions:
\vskip.1in
\noindent (i)~
\textbf{Bulk field ensemble:} We treat the 11D metric and 3-form as coherent-state variables, and build a GS mixture over classical solutions. This matches the idea that de Sitter is a coarse-grained, quasi-classical transient state.
\vskip.1in\noindent (ii)~
\textbf{Flux-sector ensemble:} Compactifications involve quantized ${\bf G}_4$ fluxes and discrete choices of internal geometry. The GS representation can be understood as a statistical mixture over these flux/geometry sectors, with brane instantons mediating transitions between them.
\vskip.1in\noindent (iii)~
\textbf{Brane-worldvolume ensemble:} Instead of a world-sheet, M-theory has M2 and M5 brane worldvolumes. A GS picture corresponds to an ensemble over such brane embeddings together with the bulk fields, capturing the transient and unstable character of the state.

\vskip.1in
\noindent $\bullet$ In all cases, the key point is that de Sitter is represented not by a single stable vacuum but by a \emph{statistical mixture} of underlying coherent configurations.
\vskip.1in
\noindent $\bullet$
This makes the construction consistent with the well-known no-go theorems against static de Sitter vacua in M-theory, while still allowing a quasi-de Sitter phase as an emergent, metastable state.

\section*{(e) Bottom line}

\noindent $\bullet$ A single world-sheet CFT for de Sitter as a Glauber--Sudarshan state does not exist, because the GS state is not a stationary vacuum.  
\vskip.1in
\noindent $\bullet$ However, a world-sheet representation as an \emph{ensemble of world-sheet theories} (each corresponding to a coherent-like configuration) is possible in principle. This would mean:
  \begin{equation}
    Z_{\rm GS} \;=\; \int d\mu(\sigma)\; P(\sigma)\; Z_{\rm ws}[\sigma],
  \end{equation}
  where $Z_{\rm ws}[\sigma]$ is the world-sheet partition function in background controlled by a state $\vert\sigma\rangle$, and $P(\sigma)$ is the GS weight functional.
\vskip.1in
\noindent $\bullet$  
This is very reminiscent of thermal string theory or time-dependent Liouville backgrounds, where the ``state'' is not captured by a single exact CFT, but by a deformation or ensemble.

\bigskip
\noindent
To conclude, therefore there cannot be a simple single world-sheet CFT for de Sitter as a Glauber--Sudarshan state, but one can probably represent it as an ensemble over world-sheet theories, much like how mixed or transient string backgrounds (thermal, Liouville, rolling tachyon) are handled --- so the ``world-sheet'' is statistical rather than pure. An explicit construction of such a world-sheet is technically challenging so a simple take-away of the above discussion is to study such transient states from the bulk (or the target space) point of view using displacement operators -- much like how we have been doing so far \cite{coherbeta, coherbeta2} -- and avoid any world-sheet representations. Interestingly, our above discussion also has a direct bearing on another front: it prepares us to address other questions related to the world-sheet representation of such transient states, namely the KMMS \cite{KMMS} no-go conditions. This is what we turn to next.

\subsection{The KMMS no-go theorem for classical de-Sitter}

In quantum field theory, a Wick rotation refers to the analytic continuation of real time 
$t$ into imaginary time $\tau = i t$. This maps a Lorentzian spacetime with signature 
$(-+++)$ into a Euclidean one with all positive signs. The advantage is that many 
technical tools, such as path integrals and conformal field theory constructions, are best 
defined or better controlled in Euclidean signature. In string theory, however, a background (as long as it is not transient)
is only consistent if it corresponds to a worldsheet conformal field theory. For Lorentzian 
spacetimes with a given symmetry group, consistency requires that the Wick-rotated 
Euclidean background also define a valid, compact, unitary CFT that realizes the symmetry 
group in the Euclidean signature. This is the playground where the Kutasov-Maxfield-Melnikov-Sethi (KMMS) \cite{KMMS} no go condition arises.

In essence, KMMS \cite{KMMS} argue that if de Sitter space were Wick-rotatable, the resulting 
Euclidean background would require a compact, unitary CFT with an affine 
$SO(n+1)$ symmetry. Such algebras quickly exhaust the allowed central charge, and 
for $n \geq 4$ no consistent worldsheet theory can be built. This leads them to 
conclude that classical heterotic string theory, and type II without RR fluxes, 
cannot realize higher-dimensional de Sitter vacua.

The interesting part is that Wick rotatability may be too strong an assumption for 
cosmological applications. Realistic de Sitter-like universes are typically only 
approximate, metastable, or explicitly time-dependent. They need not have the full 
$SO(1,4)$ isometry of exact de Sitter space, and often they do not Wick-rotate to a compact 
Euclidean sphere. For example, our ``de Sitter as a Glauber--Sudarshan state'' program \cite{coherbeta, coherbeta2}
constructs de Sitter as an excited coherent state rather than a vacuum geometry, and such 
states do not admit a compact Euclidean continuation with $SO(5)$ symmetry. In such 
scenarios, the KMMS current algebra argument no longer applies. Thus, Wick rotatability is 
both the engine that drives their proof and the assumption that limits its reach. In the following let us elaborate the story in some more detail.

\subsubsection{The KMMS no-go condition}

The work by KMMS \cite{KMMS} presents a refined no-go theorem for de Sitter vacua within certain regimes of string 
theory. Its central claim is that heterotic string theory, and similarly type II string theory 
without RR fluxes, does not admit de Sitter solutions in four or higher dimensions. This conclusion 
holds even after including all perturbative $\alpha'$ corrections and worldsheet instantons, though 
it does not incorporate quantum loop corrections in the string coupling $g_s$. In this way, the 
paper extends older no-go theorems (see the first three papers in \cite{GMN}), which had only been established in the low-energy supergravity 
limit, to more stringy, high-curvature regimes.

\begin{figure}[h]
\centering
\begin{tabular}{c}
\includegraphics[width=5in]{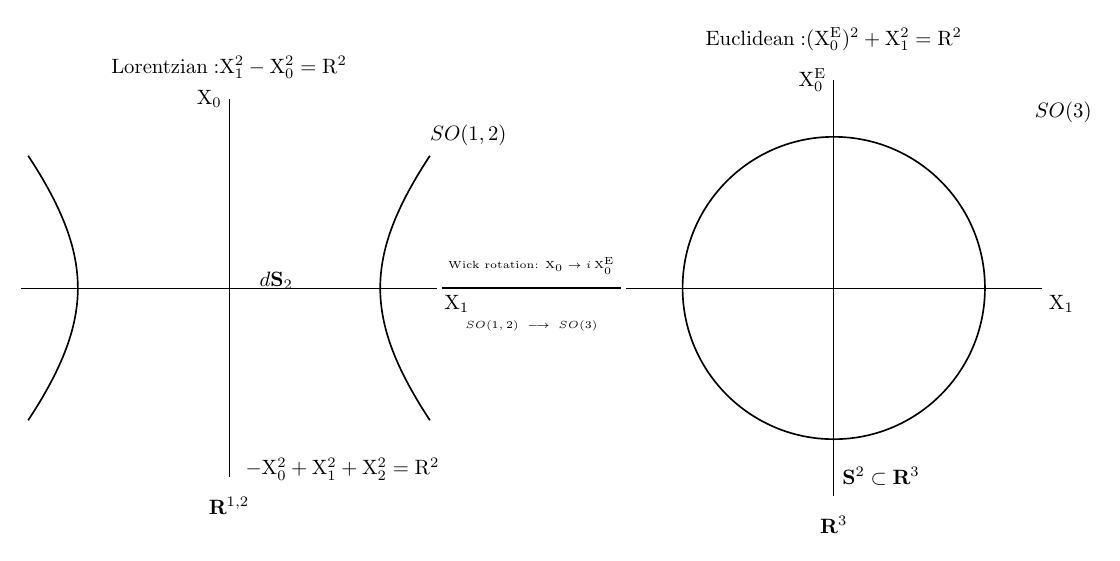}
\end{tabular}
\caption[]{Wick rotation from $d{\bf S}_2 \subset {\bf R}^{1,2}$ with isometry group $SO(1,2)$
to ${\bf S}^2 \subset {\bf R}^3$ with compact isometry group $SO(3)$.}
\label{kmmsfig}
\end{figure}

The argument relies on several key assumptions. First, the authors assume the existence of a 
consistent worldsheet conformal field theory description of the background. In the heterotic case, 
this CFT carries $(1,0)$ worldsheet supersymmetry, and in Lorentzian signature it must realize the 
full $SO(1,n)$ isometry of de Sitter space. After Wick rotation, this symmetry is promoted to the 
compact group $SO(n+1)$, which the Euclidean CFT must also represent consistently. The assumption 
of Wick rotatability is central: the Lorentzian de Sitter background is expected to map to a 
compact, unitary Euclidean theory with the enlarged symmetry group. In such a setup, the symmetries 
of spacetime appear on the worldsheet as conserved currents. These currents split into left- and 
right-moving pieces, generating affine Kac--Moody algebras. This construction imposes stringent 
constraints from unitarity and from central charge accounting, since the affine symmetry algebras 
contribute heavily to the central charge budget\footnote{This is easy to see from an illustrative example with a Lorentzian $d{\bf S}_2$ (see {\bf figure \ref{kmmsfig}}). For a Lorentzian $d{\bf S}_2$, realized as a one-sheeted hyperboloid embedded in ${\bf R}^{1,2}$, Wick rotation gives a
two-sphere ${\bf S}^2$ with an isometry group $SO(3)$.
In string theory, a consistent background corresponds to a {\it 2D} world-sheet CFT. For ${\bf S}^2$, the sigma-model action leads to Noether currents $J_a(z,\bar z)$ and $\bar J_a(z,\bar z)$
for the $SO(3)$ Killing vectors. At a conformal fixed point, exact conservation implies $\bar\partial J_a=0$ and $\partial \bar J_a=0$, so the currents are chiral and generate an affine Kac--Moody algebra,
 $ J_a(z)J_b(0) \sim \frac{k\,\delta_{ab}}{z^2} + \frac{i f_{abc} J_c(0)}{z} + \cdots$ with positive integer level $k$. Such affine currents contribute to the stress tensor through the Sugawara construction. For $G=SO(3)\simeq SU(2)$ one finds
 $c(G,k) = \frac{3k}{k+2}$
per chiral sector. This central charge must fit into the total budget of the world-sheet theory (e.g.\ $c_L=26$, $c_R=15$ for heterotic strings). For higher $d{\bf S}_n \to {\bf S}^n$, the group $SO(n+1)$ has dimension $\tfrac{1}{2}n(n+1)$, so the affine symmetry rapidly consumes central charge, making consistent worldsheet theories impossible for $n \geq 4$ under the KMMS assumptions.}.

The technical strategy of the paper is to analyze these constraints systematically. Because the 
affine symmetry must be realized at positive level, the possible values of the central charge are 
bounded. For large $n$, the required $SO(n+1)$ symmetry consumes so much of the central charge 
that it cannot be accommodated within the heterotic or type II worldsheet theory. The authors show 
that for $n \geq 8$ the constraints categorically rule out any such worldsheet realization\footnote{The KMMS paper \cite{KMMS} makes two slightly different points that must be considered together. 
Their overall conclusion is that heterotic string theory (and type II without RR fluxes) 
cannot support de Sitter vacua in dimensions $n \geq 4$. This follows from combining 
assumptions such as Wick rotatability and the requirement that spacetime isometries 
appear as affine currents in a compact, unitary worldsheet CFT. However, a sharper 
technical statement emerges from their current algebra analysis: for $n \geq 8$ the 
algebraic and central charge constraints are so restrictive that no affine $SO(n+1)$ 
symmetry can be realized at all. In the intermediate range $4 \leq n \leq 7$, the algebra 
may fit within the bounds, but no consistent or physically meaningful de Sitter background 
arises. Thus, $n \geq 8$ marks the point of absolute algebraic obstruction, while 
$n \geq 4$ expresses the broader physical exclusion.}. For 
smaller dimensions, up to $n=7$, there are a few special CFT constructions that technically fit the 
bounds, but these do not correspond to genuine classical de Sitter spacetimes in string theory. As 
a result, the general conclusion is that heterotic string theory cannot produce classical de Sitter 
vacua in four or higher dimensions. The same reasoning applies to type II theories without RR 
fluxes, since the argument is worldsheet-based and does not involve RR backgrounds.

It is important to note what the paper does not claim. The analysis does not include loop 
corrections in $g_s$, so quantum effects at the string coupling level may alter the picture. 
Similarly, the results do not apply to type II compactifications with RR fluxes, orientifold planes, 
or other non-perturbative ingredients, which could change the structure of the worldsheet or bypass 
the assumptions of Wick rotatability. The theorem also does not rule out metastable or approximate 
de Sitter states; it only excludes exact vacua under the specified conditions.

The significance of this work is that it provides a stronger barrier against de Sitter vacua than 
earlier supergravity-based theorems. By including all $\alpha'$ corrections and worldsheet 
instantons, it shows that the obstruction persists deep into the stringy regime. This tightens the 
requirements on any attempt to realize de Sitter in string theory: one must appeal to ingredients 
like RR fluxes, orientifolds, or genuinely quantum corrections. The result highlights the need to 
revisit some landscape arguments, since many na\"ive compactification scenarios may fail once 
worldsheet consistency and central charge constraints are taken into account.

\subsubsection{Glauber-Sudarshan states and the KMMS theorem}

Here we present an expanded comparison between the assumptions of 
Kutasov--Maxfield--Melnikov--Sethi (KMMS) and the framework of de Sitter 
as a Glauber--Sudarshan (GS) state. The discussion highlights the points 
of divergence between the two approaches.

\vskip.1in

\noindent $\bullet$ {\bf Exact vs.\ approximate de Sitter}:
KMMS assume the existence of an \emph{exact} de Sitter vacuum with full 
$SO(1,4)$ symmetry in Lorentzian signature, which Wick-rotates to a compact, 
unitary Euclidean CFT with $SO(5)$ affine currents. This exactness is the 
cornerstone of their current-algebra reasoning. By contrast, the GS approach 
describes de Sitter as a \emph{time-dependent excited state} rather than a 
vacuum. The full isometry need not be preserved, and no compact Euclidean 
worldsheet CFT with affine $SO(5)$ currents is required\footnote{By this we mean single {\it 2D} world-sheet CFT and not an ensemble over world-sheet theories. For details on this, see discussion in section \ref{ethantag}.}. Consequently, the 
central-charge and current-algebra bounds invoked by KMMS are not relevant.

\vskip.1in

\noindent $\bullet$ {\bf Worldsheet perturbation vs.\ resummation and quantum effects}:
KMMS restrict their analysis to the classical worldsheet regime: all 
$\alpha'$ corrections are included, but string-loop ($g_s$) effects are 
excluded. Their no-go result is valid only in this tree-level domain. 
GS constructions, however, go beyond perturbation theory. They employ 
resurgence and Borel resummation to control infinite towers of corrections, 
and in some formulations incorporate loop or strong-coupling physics via 
Schwinger--Dyson equations or duality arguments. In this way, the GS 
framework lies outside the perturbative assumptions of KMMS.

\vskip.1in

\noindent $\bullet$ {\bf Euclidean CFT vs.\ state-based description}:
A central KMMS requirement is that the background admit a compact, unitary 
Euclidean CFT with separately conserved left- and right-moving $SO(n+1)$ 
currents at positive level. The GS state, in contrast, is defined directly 
in the Hilbert space of spacetime quantum fields, such as in minisuperspace 
or Wheeler--DeWitt formulations. It need not correspond to any Euclidean 
CFT at all. Moreover, it can break isometries softly and persist only over 
a finite time interval, rendering worldsheet current-algebra constraints 
irrelevant.

\vskip.1in

\noindent $\bullet$ {\bf Treatment of stability}:
KMMS analyze \emph{vacua}, meaning stationary, stable backgrounds with 
preserved symmetry. The GS picture explicitly abandons this requirement, 
treating de Sitter as a \emph{metastable, transient state} realized by a 
coherent excitation. Such a state may decay or evolve after a finite time, 
which avoids the exact stability conditions assumed by KMMS.

\vskip.1in

\noindent $\bullet$ {\bf Role of background ingredients}:
The KMMS analysis is confined to heterotic and type II theories without 
RR fluxes or orientifold planes, and does not incorporate exotic 
non-perturbative elements. GS models often invoke dual frames, 
strong-coupling effects, and emergent nonlocal phenomena, all of which 
lie beyond the scope of the KMMS framework. These additional ingredients 
further separate the GS construction from the assumptions of the no-go theorem.

\vskip.1in

\noindent $\bullet$ {\bf Role of symmetry}:
In KMMS, spacetime 
symmetries must be realized \emph{exactly} on the worldsheet as conserved 
currents. In the GS scenario, symmetry can be realized only 
\emph{approximately or dynamically}, with expectation values reproducing 
quasi-de Sitter evolution rather than exact invariance. Finally, while KMMS 
aim to rule out \emph{vacua}, the GS approach is geared toward constructing 
\emph{states} that satisfy consistency conditions such as Schwinger--Dyson 
equations, positivity bounds, and resurgent expansions.

\vskip.1in

\vskip.1in

\noindent $\bullet$ {\bf Infrared vs.\ ultraviolet control}:
KMMS rely on a fully consistent worldsheet CFT, which is a UV-complete description with exact modular invariance and affine symmetry. 
The GS picture instead emphasizes infrared consistency, requiring Schwinger--Dyson equations, positivity and causality bounds, and controlled backreaction over a finite time window. 
GS construction does not insist that the state arise from a compact Euclidean CFT. 
Moreover, KMMS assume exact worldsheet supersymmetry and current conservation, which force strict Kac--Moody realizations, while GS works with approximate Ward identities for expectation values, allowing soft breaking of symmetry in time.

\vskip.1in

\noindent $\bullet$ {\bf Dilaton and coupling evolution}:
KMMS keep to a regime where the background can be treated at tree level in $g_s$, even while resumming all $\alpha'$ corrections. 
This ensures a bounded coupling and a stationary vacuum. 
By contrast, GS constructions often allow a time-dependent dilaton or moduli evolution, as long as couplings remain under control during the quasi-de Sitter epoch. 
This time dependence helps realize transient expansion but places the solution outside the stationary, Wick-rotatable class considered by KMMS.

\vskip.1in

\noindent $\bullet$ {\bf Boundary conditions and initial data}:
KMMS look for vacua: solutions that require no special initial state. 
GS begins with explicit coherent initial data in the Hilbert space, so the cosmology is an excited configuration whose lifetime, Hubble scale, and decay channels are tied to the chosen state and the resummation scheme. 
In this sense, KMMS test the existence of backgrounds, while GS tests the viability of states.

\vskip.1in

\noindent $\bullet$ {\bf Energy conditions and stress tensors}:
KMMS show that, even with higher-curvature corrections that can violate classical energy conditions, the symmetry and central-charge budget forbid exact de Sitter vacua. 
GS instead allows expectation values in coherent states to source an effective stress tensor that transiently mimics positive vacuum energy, while tracking backreaction and eventual exit from the quasi-de Sitter phase. 
The emphasis is on controlled metastability rather than exact balance.

\vskip.1in

\noindent $\bullet$ {\bf Extra ingredients}:
KMMS deliberately exclude RR fluxes, orientifolds, and brane sources that can change the worldsheet realization or spoil unitarity upon Euclidean continuation. 
GS constructions are more permissive, importing dual descriptions or nonlocal effects that can effectively encode such ingredients, even when their direct worldsheet incarnations are unclear. 
This broadens the mechanism space but lies beyond KMMS’s clean worldsheet framework.

\vskip.1in

\noindent $\bullet$ {\bf Criteria for success}:
For KMMS, success would mean constructing a compact, unitary Euclidean CFT with affine $SO(n+1)$ symmetry and the correct central charge---something argued to be impossible for $n \geq 4$ and algebraically untenable for $n \geq 8$. 
For GS, success is defined more loosely: producing a finite-duration quasi-de Sitter epoch with calculable observables such as an emergent Hubble parameter, controlled corrections, and a consistent exit. 
Thus KMMS target exact vacua with stringent worldsheet realizations, while GS targets metastable, physically reasonable states verified by IR consistency and resummation.

\vskip.1in 
\noindent $\bullet$ {\bf Summary}: The KMMS theorem shows that a consistent, unitary, modular-invariant world-sheet conformal field theory cannot describe a space-time with a strictly positive cosmological constant. In other words, perturbative string theory admits no exact world-sheet CFT that corresponds to a de Sitter vacuum. This result is often taken as a strong indication of the ``no-go'' for stable de Sitter vacua in string theory, in close analogy to the supergravity no-go theorems in higher dimensions.

By contrast, the Glauber--Sudarshan (GS) proposal interprets de Sitter not as a true vacuum, but as a transient, statistical mixture of coherent states. In this framework, the world-sheet description is not a single exact CFT but rather an ensemble or deformation of allowed backgrounds. The GS state is inherently non-stationary and metastable, so the assumptions of the KMMS theorem do not apply. Thus, there is no contradiction: KMMS forbids exact de Sitter vacua, while the GS construction allows de Sitter as an emergent, coarse-grained phase built from an ensemble of consistent string/M-theory configurations\footnote{A similar logic applies in M-theory. The Maldacena--Nuñez class of no-go theorems \cite{GMN} shows that smooth, flux-supported compactifications cannot yield a stable de Sitter solution in eleven dimensions. The GS picture, however, does not claim the existence of such a vacuum. Instead, it frames de Sitter as a metastable statistical state, arising from mixtures of flux sectors, brane instantons, and geometric configurations. In this sense, the GS construction remains fully consistent with the M-theory no-go theorems: it provides a mechanism for transient de Sitter phases without contradicting the prohibition of exact, static vacua.}.

In summary, the GS program avoids the KMMS no-go by relaxing nearly every 
assumption: it does not posit an exact vacuum, it is not confined to 
tree-level worldsheet perturbation theory, it does not rely on a Wick-rotated 
Euclidean CFT with affine currents, it embraces metastability rather than 
stability, and it allows ingredients excluded from the KMMS setup. 
Accordingly, the two frameworks address different physical regimes: 
KMMS constrain exact, perturbative vacua, while GS seeks to realize 
excited, metastable states with resummed or quantum-corrected dynamics. In {\bf Table \ref{chatrchange}} we summarize the key points that show how the GS constructions avoid the KMMS no-go conditions.

\begin{table}[h]
\centering
\renewcommand{\arraystretch}{1}
\begin{tabular}{|p{6cm}|p{9cm}|}
\hline
\textbf{KMMS Assumption (worldsheet no-go)} & \textbf{How GS de Sitter Evades It} \\
\hline
Exact de Sitter vacuum with full $SO(1,4)$ isometries, Wick-rotatable to a compact, unitary Euclidean CFT realizing $SO(5)$ as affine currents
& GS describes a time-dependent \emph{excited state}, not an exact vacuum. No requirement exists for a compact Euclidean CFT with affine $SO(5)$ currents, so the central-charge and current-algebra constraints of KMMS do not apply. \\
\hline
Work restricted to the classical worldsheet regime: include all $\alpha'$ corrections, but neglect $g_s$ (string-loop) effects
& GS constructions rely on \emph{resurgence and Borel resummation} to control infinite corrections, and in some cases explicitly use loop or strong-coupling effects captured by dualities or Schwinger--Dyson equations. These go beyond the perturbative assumptions of KMMS. \\
\hline
Background must admit a unitary Euclidean CFT with separately conserved $SO(n+1)$ currents at positive level
& A GS coherent state is defined in the Hilbert space of spacetime EFT or in a Wheeler--DeWitt/minisuperspace setup, not as a Euclidean CFT. Such states can \emph{softly break isometries} and exist only for finite time intervals, bypassing the need for affine currents. \\
\hline
Vacua must be stable, stationary backgrounds with full symmetry preserved
& GS explicitly treats de Sitter as a \emph{metastable, transient state}, not a vacuum. Its finite lifetime avoids the strict stability conditions underlying the KMMS no-go. \\
\hline
Focus on heterotic and type II setups without RR fluxes or orientifolds
& GS proposals often appeal to \emph{dual frames, strong-coupling effects, and emergent nonlocal phenomena}, which are outside the KMMS heterotic/type II tree-level setup. \\
\hline
Symmetry must be realized exactly on the worldsheet
& In GS, symmetry can be realized \emph{approximately or dynamically}, with expectation values reproducing quasi-de Sitter behavior without enforcing strict worldsheet current conservation. \\
\hline
\end{tabular}
\caption{\Su Comparison of KMMS assumptions \cite{KMMS} and how the GS de Sitter program \cite{coherbeta, coherbeta2} avoids them.}
\label{chatrchange}
\end{table}


\subsection{Gevrey growths of the diagrams in QFTs and string theories
\label{papindigo}}

Our above discussion suggests the importance of non-perturbative effects $-$ like five-brane instantons as we shall see later $-$ in constructing the GS states. In fact
the asymptotic nature of the Feynman diagrams in usual QFT \cite{dyson} is one of the main reason why we have non-perturbative effects like the instantons. In simple cases such asymptotic growths $-$ which typically show factorial behaviors $-$ are Borel summable, but in more complicated cases they may not even be Borel summable \cite{unsal, dorigoni}. In string theory, the diagrams show more non-trivial growths and they are in general {\it not} Borel summable\footnote{Borel {\it not summable} implies the presence of poles in the Borel plane. This typically leads to certain amount of ambiguities but the resolution is simple as pointed out by \'Ecalle \cite{ecalle}. Thus in a more modern language, all these are Borel {\it summable} but with due considerations to the poles and the Stokes lines in the Borel plane. See also \cite{dorigoni} for an excellent review on the subject.} \cite{grossper}. These growths are typically called the Gevrey growths \cite{gevrey} and they are summed over using the Borel-\'Ecalle resummation \cite{borelborel, ecalle} that takes into account the ambiguities associated with the poles in the Borel 
plane. In string theory, such a procedure leads to the D-brane instantons \cite{shenker}, so the question is what happens when we have excited states like the Glauber-Sudarshan states in M-theory. A study using a simpler set-up in M-theory revealed similar Gevrey growths \cite{borel2}, so the natural question is whether we expect complicated Gevrey behavior when we study a more generic set-up. Such a analysis will have direct bearings on the trans-series form of the action itself as advocated in \cite{joydeep}, so a careful study is needed here. In the following subsection we will first argue that the Gevrey behavior that we see in string theory may continue to persist even with a generic set-up with fields other than scalar fields (taken in \cite{borel2}) in M-theory with the GS states.

\subsubsection{Factorial growths in simple quantum field theories \label{bibi}}

Let us start by asking why we get factorial growths in QFT, {\it i.e.} why do the Feynman diagrams show factorial growths? The answer is quite straightforward and may be easily exemplified by a simple $\lambda\phi^4$ theory. The kinetic term for such a theory can be expressed in terms of the Fourier components as:
\bg\label{natyrob}
\int d^4 x ~\partial_\mu \phi \partial^\mu \phi = {1\over {\rm V}}\sum_k k^2 \varphi(k) \varphi^\ast(k) =  
\frac{1}{{\rm V}} \sum\limits_{k} k^2 \quad
    \begin{tikzpicture}[baseline={(0, 0)}, scale=0.8, thick,
        main/.style = {draw, circle, fill=black, minimum size=2pt},
        dot/.style={inner sep=0pt,fill=black, circle, minimum size=3pt},
        dots/.style={inner sep=0pt,fill=black, circle, minimum size=1pt}, 
        ball/.style={ball  color=white, circle,  minimum size=15pt}]

         \node[ball] (b) at (0,0) {} ;
         \node[main, label=0:$\varphi(k)$] (m1) at (0, 1.1) {};
         \node[main, label=0:$\varphi^*(k)$] (m2) at (0, -1.1) {};

         \draw (b) -- node {} (m1);
         \draw (b) [dashed] -- node {} (m2);

    \end{tikzpicture}  
\nd
where ${\rm V}$ is the four-dimensional volume and we used $\varphi(k)$ to denote the Fourier components. On the RHS of \eqref{natyrob} the series is denoted using the so-called {\it nodal diagrams} that were initiated in \cite{borel2}. The dotted line signify momentum conservation as $\varphi(-k) = \varphi^\ast(k)$. In a similar vein the interaction term takes the following form:

{\footnotesize
\bg\label{amilenonu}
{\lambda\over 4!}\int d^4x~\phi^4(x) =    \frac{\lambda}{4!{\rm V}^3} \sum\limits_{\{k_j\}}\prod_{i = 1}^3 \varphi(k_i) \varphi^*(k_1 + k_2 + k_3)  = \frac{\lambda}{4!{\rm V}^3} \sum\limits_{\{k_i\}}   
 
\nd}
where again the dotted line connecting to a field denotes the momentum conservation at a given vertex.  Note the sum over all the set of three momenta $\{k_j\}$ which would collect the series of terms coming from the interaction vertex. (For $k_1\ne k_2\ne k_3$, the dominant contribution will kill the $4!$ factor in the denominator.) At ${\rm N}$-th order the dominant contribution takes the following form:

{\footnotesize
\bg\label{amilmey}
    {1\over {\rm N}!} \left({\lambda \over 4!{\rm V}^3}\right)^{\rm N} {\rm N}! \sum\limits_{\{k_i\}}    
    %
   \nd}
and we do not show the other sub-dominant contributions. All the dominant contributions have an overall scalings of ${\rm N}!$ (as well as $(4!)^{\rm N}$, although this factor is hidden inside every chain of the nodal diagram). This is canceled by the ${1\over (4!)^{\rm N}{\rm N}!}$ suppression coming from expanding the exponential. What remains are the ${\rm N}$ copies of 
%
 which would have ${\rm N}!$ permutations between them. In the presence of external legs these are the permutations that {\it contribute to the ${N}!$ growth of the Feynman diagrams.} In fact the ${\rm N}!$ permutation of the first set of ${\rm N}$ objects is enough to justify the factorial growth because the other set of terms would simply convert the summation to an integral appropriately in the presence of external inputs. 

There is however one puzzling feature that would require more clarifications. The factorial growth argued above depends on the {\it permutation} of the ${\rm N}$ vertices. However this seems like over-counting because we have already removed the ${\rm N}!$ permutations by the ${1\over {\rm N}!}$ factors from expanding the exponential. So why do this again now? The answer is subtle and may be clarified by first counting the total number of ways four legs at each of the N vertices join together. The answer is:
\bg\label{shonalola}
\left(2m{\rm N} - 1\right)!! = {\left(2m{\rm N}\right)!\over 2^{m{\rm N}} \left(m{\rm N}\right)!} \approx (m{\rm N})!, \nd
for $m \ge 2$. The severely over-counts the total number of possible Feynman diagrams at ${\cal O}({\rm N})$. In fact most of the diagrams 
from \eqref{shonalola} are removed because they do not satisfy the momentum conservation\footnote{Recall that in QFT two lines can be joined to form a propagator if and only if they have momenta with opposite signs.}. This means that \eqref{shonalola} cannot be the answer. To find the actual answer we will go in two steps: find the lower and the upper bounds of the number of diagrams. The lower bound is easy to find: choose a possible {diagram} by first joining the vertices in {\it one} specific way. In the absence of external legs $-$ we are looking at the vacuum diagrams only $-$ this creates a chain with N vertices. We can now ask, how many possible ways can we permute the vertices keeping the topology unchanged. The answer is clearly ${\rm N}!$. Note that these ${\rm N}!$ permutations may provide the {\it same} amplitudes but they not only count as distinct contributions, but are also physically {\it different} from the ${\rm N}!$ that we got from the binomial expansion. This is shown in the following figure for say $\lambda \varphi^4$ theory where the permutations of the labels lead to different diagrams but with similar amplitudes. 

\tikzset{
  wavy/.style={
    decorate,
    decoration={snake, amplitude=1.2pt, segment length=6pt},
    thick
  },
  vertex/.style={circle, fill=black, inner sep=1.8pt}
}

\begin{tikzpicture}[scale=1]

\begin{scope}[xshift=0cm]

\node[vertex,label=above:{1}] (A) at (90:1.2) {};
\node[vertex,label=below left:{2}] (B) at (210:1.2) {};
\node[vertex,label=below right:{3}] (C) at (330:1.2) {};

\draw[wavy] (A) to[bend left=15] (B);
\draw[wavy] (A) to[bend right=15] (B);

\draw[wavy] (B) to[bend left=15] (C);
\draw[wavy] (B) to[bend right=15] (C);

\draw[wavy] (C) to[bend left=15] (A);
\draw[wavy] (C) to[bend right=15] (A);

\end{scope}

\begin{scope}[xshift=4.2cm]

\node[vertex,label=above:{2}] (A) at (90:1.2) {};
\node[vertex,label=below left:{3}] (B) at (210:1.2) {};
\node[vertex,label=below right:{1}] (C) at (330:1.2) {};

\draw[wavy] (A) to[bend left=15] (B);
\draw[wavy] (A) to[bend right=15] (B);

\draw[wavy] (B) to[bend left=15] (C);
\draw[wavy] (B) to[bend right=15] (C);

\draw[wavy] (C) to[bend left=15] (A);
\draw[wavy] (C) to[bend right=15] (A);

\end{scope}

\begin{scope}[xshift=8.4cm]

\node[vertex,label=above:{3}] (A) at (90:1.2) {};
\node[vertex,label=below left:{1}] (B) at (210:1.2) {};
\node[vertex,label=below right:{2}] (C) at (330:1.2) {};

\draw[wavy] (A) to[bend left=15] (B);
\draw[wavy] (A) to[bend right=15] (B);

\draw[wavy] (B) to[bend left=15] (C);
\draw[wavy] (B) to[bend right=15] (C);

\draw[wavy] (C) to[bend left=15] (A);
\draw[wavy] (C) to[bend right=15] (A);

\end{scope}

\end{tikzpicture}

\noindent The lower bound is now determined in the following way. With the specific choice of the leg contractions, {\it i.e.} a specific choice of the topology, we can look at any one of the ${\rm N}$ vertices and determine the {\it smallest} number of possible variations that one may perform to generate new topologies. Assume that there are $c_1$ possible distinct choices, {\it i.e.} $c_1$ allowed distinct topologies of Feynman diagrams at the designated vertex\footnote{The answer doesn't change even if multiple vertices are allowed.}. Then the total number of diagrams ${\cal N}_{\rm diag}$ is clearly bounded from below by:
\bg\label{tequlazoe}
{\cal N}_{\rm diag} ~\ge ~c_1^{\rm N}~ {\rm N}!
\nd
capturing both the factorial and the polynomial growths. In fact $c_1$ may be fixed quantitatively by demanding that all $c_1^{\rm N}$ should in principle lead to actual momentum conserving diagrams.  For the upper bound, we can again look at all the ${\rm N}$ vertices and ask what is the {\it largest} allowed variations that one may perform to generate distinct topologies. Let the answer be $c_2$ at a given vertex ${\rm V}_k$ with $1 \le k \le {\rm N}$ for a fixed $k$. Then ${\cal N}_{\rm diag}$ for the total number of diagrams is bounded from top and bottom by:
\bg\label{tequlazoe2}
c_1^{\rm N} ~{\rm N}! ~ \le ~ {\cal N}_{\rm diag} ~\le ~c_2^{\rm N}~ {\rm N}!
\nd
where we combined \eqref{tequlazoe}. In terms of practical construction, not all $c_2^{\rm N}$ could in principle lead to actual momentum conserving diagrams, but that's besides the point because we are only aiming for an upper bound here. 

The analysis in \eqref{tequlazoe2} not only shows that the number of Feynman diagrams grows factorially $-$ up to an overall polynomial factor $-$ but also that the complications at the individual vertices 
do not influence the factorial nature. For a more complicated interaction of the form:
\bg\label{pobitradud}
\int d^4x \sum_{\{n_i, p_i\}}{\lambda_{n_1..n_lp_1..p_l}\over p_1!p_2!...p_l!} \prod_{j = 1}^l\partial^{n_j}\otimes\phi_j^{p_j}, \nd
where $\otimes$ simply implies all possible ways one could take the derivative action on $p_j$ scalar fields $\varphi_j$ for $j = l$ sets of different scalar fields\footnote{See also \cite{joydeep} and \cite{borel2} for notations.}, similar story unfolds. For illustrative purpose here we can take a slightly simpler model of the following form by replacing $\otimes$ by a simple product:

{\scriptsize
\bg\label{gulabx}
&& \int d^4x \sum_{\{n_i, p_i\}}{\lambda_{n_1..n_lp_1..p_l}\over p_1!p_2!...p_l!} \prod_{j = 1}^l\partial^{n_j}\phi_j^{p_j} =  \sum_{\{n_s, p_s\}} \lambda(\{n_s, p_s\}; {\rm V})\sum_{\{k_j^{(i)}\}}\prod_{i = 1}^{l}\Big(\sum_{j = 1}^{p_i} k_j^{(i)}\Big)^{n_i} \prod_{r=1}^{p_i} \varphi_i(k_r^{(i)})\delta\Big(\sum\limits_{i = 1}^l \sum\limits_{j = 1}^{p_i}k_j^{(i)}\Big)\nonumber\\
& = &\sum_{\{n_s, p_s\}}{\lambda_{n_1..n_lp_1..p_l}\over p_1!...p_l!{\rm V}^{p_1+..+p_l-1}}\sum_{\{k_j^{(i)}\}}\prod_{i = 1}^{l}\Big(\sum_{j = 1}^{p_i} k_j^{(i)}\Big)^{n_i}\delta\Big(\sum\limits_{i, j}k_j^{(i)}\Big)
   \nonumber
\nd}
where $\lambda(\{n_s, p_s\}; {\rm V}) \equiv{\lambda_{n_1..n_lp_1..p_l}\over p_1!...p_l!{\rm V}^{p_1+..+p_l-1}}$, and the delta function imposes momentum conservation at the vertices similar to what we saw for the $\lambda \phi^4$ case earlier. The story here is more involved because of multiple scalar interactions at a given vertex. To study the ${\rm N}$-th order let us, for simplicity, fix the values of $(n_s, p_s)$ to $(0, p)$ so that the subtlety with the momenta do not enter the diagrams. The {\it simplicity} alluded to above is to avoid cluttering of terms because  the generic case (although not important for the present discussion) is equally easy to do. With this in mind, at N-th order, the dominant contribution takes the following form:
\begin{equation}
      \frac{\lambda^{\rm N}(0, p; {\rm V})}{{\rm N!}} \sum\limits_{\{k_i\}} {\rm N}!\quad
   \nonumber
\end{equation}
\noindent where we see that the ${\rm N}!$ term cancels out. What remains are the ${\rm N}$ copies of the vertices which, in the presence of external legs, will lead to ${\rm N}!$ Feynman diagrams exactly in the way illustrated earlier in \eqref{tequlazoe2} by sampling the upper and the lower bounds. This is precisely the ${\rm N}!$ growth of the dominant diagrams, which may also be made clear once we express it in the following {\it split} form at ${\cal O}({\rm N})$:
\bg\label{lolashonaek}
&&{1\over {\rm N}!}\left({\lambda_1\over 4!}\phi_1^4 + {\lambda_2\over 6!}\phi_2^6 + {\lambda_3\over 8!} \phi_3^8 + .... + {\lambda_n\over (2n+2)!} \phi_n^{2n+2}\right)^{\rm N} \nonumber\\
& = & \sum_{\{k_i\}} {1\over {\rm N}!} \cdot {{\rm N}!\over k_1! k_2! k_3!... k_n!} {\lambda_1^{k_1} \lambda_2^{k_2} \lambda_3^{k_3}... \lambda_n^{k_n}}~\underbrace{\phi_1^{4k_1} \phi_2^{6k_2} \phi_3^{8k_3}....\phi_n^{(2n+2)k_n}}_{\sum\limits_{i = 1}^n k_i = {\rm N}}, \nd
where we have ignored the external momenta factors and the cross-terms from \eqref{pobitradud} just to avoid clutter. Looking at the $\phi_l^{2l+2}$ piece, and using our earlier analysis, we see that it may be expressed as a tower of momentum modes $\varphi_l(k)$ with discrete $k$. Thus raising this to $k_l$ power in \eqref{lolashonaek} leads to two immediate outcomes: {\Su one}, the dominant terms in the momentum tower all grow as $k_l!$, and {\Su two}, there are exactly $k_l$ copies of the momentum pieces coupled together. Therefore in \eqref{lolashonaek}, the ${\rm N}!$ piece cancels out trivially, and the $k_l!$ pieces in the denominator are now cancelled by the individual momentum factors for all $1 \le l \le n$. What remains are the $k_l$ coupled pieces $\forall l$ which sum up together as 
$\sum\limits_{l = 1}^n k_l = {\rm N}$. In the presence of external lines, there are ${\rm N}!$ permutations leading again to the ${\rm N}!$ growth of the Feynman diagrams. The story continues in a similar way for the generic set-up of \eqref{pobitradud} with momenta factors. 

\subsubsection{Gevrey growths in closed string perturbation theories \label{agabibilada}}

Our above analysis tells us that, in usual QFTs, it is in general not possible to have growths of the Feynman diagrams bigger than ${\rm N}!$ at ${\cal O}({\rm N})$. So how can we understand the Gevrey growths in string theory and in the nodal diagrams from \cite{borel2}? Let us start with string theory, and we can simply take the bosonic string into account to illustrate the idea. From 26-dimensional {\it supergravity} point of view $-$ where we concentrate only on the NS-NS sector $-$ the story cannot be very different from what we had above, implying that the growth of the supergravity loops cannot be more complicated than ${\rm N}!$. This argument should extend to other string theories like type II and heterotic. 

However, there must be a flaw in our arguments regarding the 
${\rm N}!$ growth in string theory. If the reasoning presented above were correct, D-branes would not exist in, for example, type II string theories. The resolution lies in the fact that string theory is not supergravity $-$ nor even an approximate realization of it.

The key idea that distinguishes string theory from say supergravity is the world-sheet realization of the Feynman diagrams. This is not just replacing supergravity loops by Riemann surfaces with non-trivial world-sheet CFTs, but also the realization that theories on the Riemann surfaces may be split into right and left moving degrees of freedom. The latter holds the key that distinguishes usual factorial growths of the Feynman diagrams in QFTs to more complicated Gevrey growths of the Riemann surfaces in string theories.

There is yet another factor that distinguishes our analysis from the usual QFTs and plays an important role here: a non-trivial Riemann surface may be split into pair of pants in the following suggestive way:



\noindent where ${\rm V}_i$ are the vertex operators that insert in 
external states to any given Riemann surface. In the above figure, there are four vertex operators attaching themselves to a genus $g = 2$ Riemann surface. We can quantify the above statement in the following way. Let $\Sigma_g$ be a closed, oriented Riemann surface of genus $g \ge 2$.
A \emph{pair of pants} is a three-holed sphere, {\it i.e.}\ a genus-zero surface
with three boundary components. The Euler characteristic of $\Sigma_g$ is $\chi(\Sigma_g) = 2 - 2g$ and
each pair of pants has Euler characteristic of
$\chi(\text{pants}) = 2 - 3 = -1$. Therefore
if $\Sigma_g$ is decomposed into ${\rm N}$ pairs of pants, additivity of Euler
characteristic gives ${\rm N} = 2g - 2$. For $g = 2$ we have ${\rm N} = 2$ shown in the figure above. For $g = 3$ we get:



\noindent implying that there are four pair of pants as depicted in the above figure. Generically, 
each pair of pants has three boundary components, so the total number of
boundary components before gluing is
$3(2g - 2) = 6g - 6$.
Since $\Sigma_g$ is closed (no boundary), these $6g-6$ cuffs must be glued
pairwise, producing
$\frac{6g - 6}{2} = 3g - 3$ internal gluing curves.
Each internal gluing curve carries one length parameter and one twist
parameter, yielding:
\begin{equation}
\dim \mathcal{M}_g = 2(3g - 3) = 6g - 6,
\end{equation}
which is the real dimension of the moduli space of genus-$g$ Riemann
surfaces. In fact this has some analogy with ${\rm N}$ vertices with cubic interactions at each of the vertices. The difference is that the cubic interactions are replaced by pair of pants. 
The factorial growth in Weil--Petersson volumes arises from the
number of distinct ways to glue the $6g-6$ cuffs pairwise and to arrange
the $2g-2$ pants, modulo automorphisms. The naive answer for the 
number of distinct gluings of cuffs into pairs is:
\begin{equation}\label{agabibi}
{\cal N}_{\text{gluings}} =
\frac{(6g - 6)!}{2^{\,3g-3}(3g - 3)!}
\; = \;
(6g - 7)!! \, ,
\end{equation}
where the denominator accounts for pairwise identification of cuffs and
indistinguishability of the $3g-3$ pairs. Unfortunately, \eqref{agabibi} severely over-counts the number of  possibilities. Despite this, the factorial growth is the combinatorial origin of
the large--genus growth of Weil-Petersson volumes proved by Mirzakhani \cite{mirzakhanipaper}, the $(2g)!$ behavior of closed-string perturbation theory, and the subsequent non--Borel summability of the genus expansion
\begin{figure}[h]
\centering
\begin{tabular}{c}
\includegraphics[width=5in]{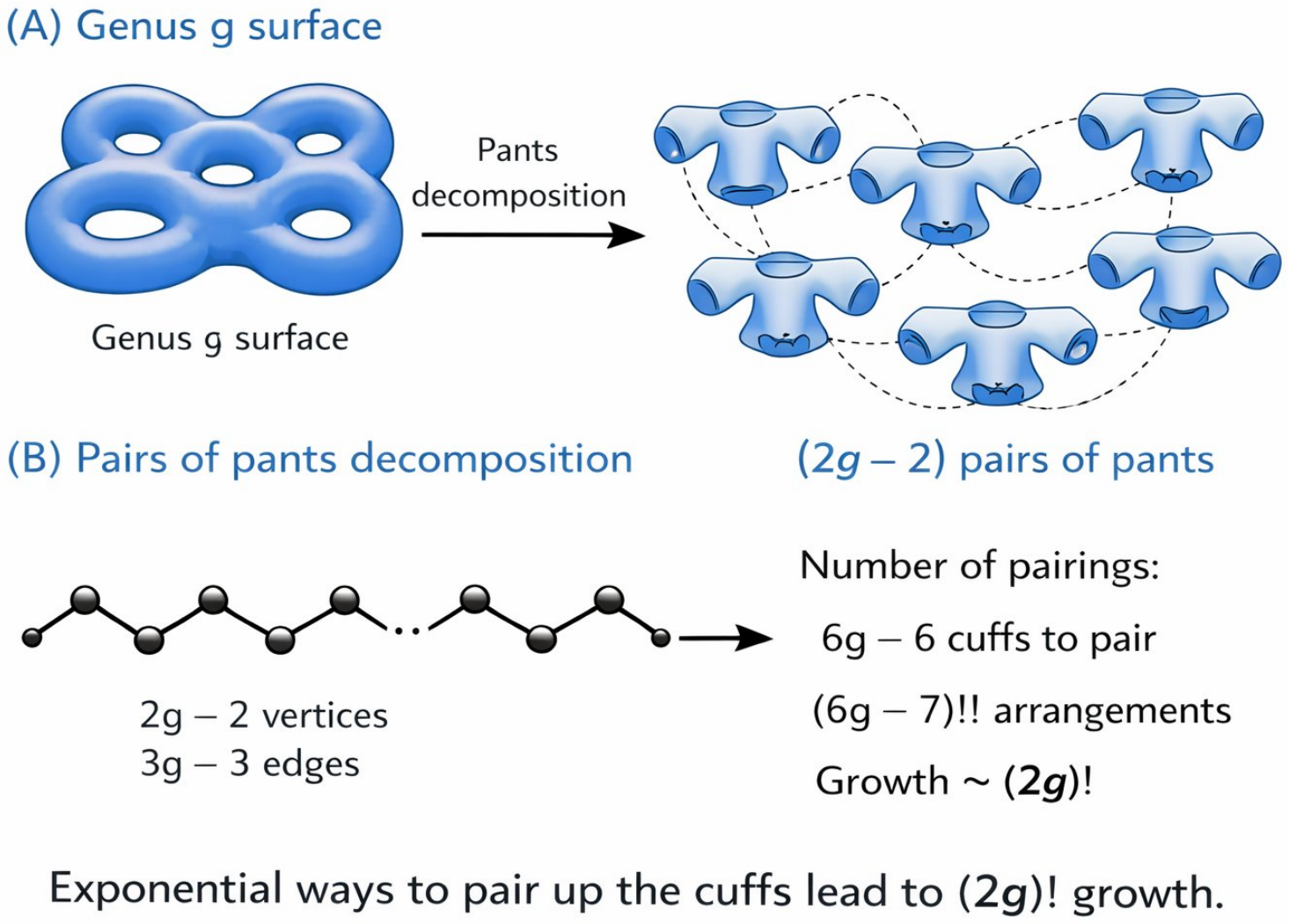}
\end{tabular}
\vskip-.2in
\caption[]{A closed genus-$g$ surface can be cut along $3g-3$ disjoint simple closed curves into $2g-2$
three-holed spheres (“pairs of pants”). Each pant contributes three boundary circles (cuffs), so there are
$6g-6$ cuffs total, which are glued back in pairs to reconstruct the surface.
The dual graph has one trivalent vertex per pant and one edge per glued cuff-pair, hence
$V=2g-2$ vertices and $E=3g-3$ edges. The large number of possible cuff-pairings gives factorial
combinatorics.}
\label{mirzakhani}
\end{figure} 
\noindent as shown in {\bf figure \ref{mirzakhani}}. In other words, in closed-string perturbation theory, the genus-$g$ contribution integrates the worldsheet CFT correlator
over the moduli space $\mathcal{M}_g$.
A pants decomposition corresponds to sewing $2g-2$ three-point building blocks (three-holed spheres)
with $3g-3$ propagator tubes. Summing over inequivalent sewings (i.e.\ pairings of the $6g-6$ cuffs,
together with twist parameters) is a combinatorial skeleton behind the growth of the number of
degenerations/channels at large $g$, and is one way to see why genus-$g$ amplitudes can exhibit
factorial growth (often summarized as $\sim (2g)!$ up to subfactorial factors), which underlies the
expected non-Borel-summability of the closed-string genus expansion.

We can quantify the above paragraph in the following way. At genus $g$, the closed-string vacuum amplitude takes the schematic form:
\begin{equation}
\mathcal{A}_g \;\sim\; g_s^{2g-2} 
\int_{\mathcal{M}_g} d\mu_{\rm WP}\; \mathcal{I}_g ,
\end{equation}
where $\mathcal{M}_g$ is the moduli space of closed genus-$g$ Riemann surfaces, $d\mu_{\rm WP}$ is the Weil--Petersson measure, and $\mathcal{I}_g$ is the worldsheet CFT correlator. A fundamental result due to Mirzakhani \cite{mirzakhanipaper} shows that the Weil--Petersson volume of $\mathcal{M}_g$ grows factorially at large genus in the following way:
\begin{equation}
\mathrm{Vol}(\mathcal{M}_g) \;\sim\; (2g)! \qquad (g \to \infty).
\end{equation}
This geometric growth of moduli space volume is the primary origin of the factorial divergence of perturbative string theory.
Using Stirling’s approximation, one finds:
\begin{equation}
(2g)! \;\sim\; (g!)^2 \, 4^g \times \big(\text{subleading factors}\big).
\end{equation}
Thus the frequently quoted $({\rm N}!)^2$ behavior is simply a repackaging of the fundamental $(2g)!$ growth. No additional combinatorial enhancement is implied by this rewriting.
In degeneration limits of moduli space, closed-string amplitudes factorize schematically as a product of left-moving and right-moving sectors, thus
reflecting the Hilbert space decomposition:
\begin{equation}
\mathcal{H}_{\rm closed} \;=\; \mathcal{H}_{\rm left} \otimes \mathcal{H}_{\rm right}.
\end{equation}
This factorized structure makes it natural to view closed-string amplitudes as a ``square'' of simpler building blocks. However, it is crucial to emphasize that each open-string sector already exhibits factorial growth due to its own moduli space geometry implying that
the closed-string factorial growth does \emph{not} arise from multiplying two quantum-field-theoretic perturbation series. In fact
both open and closed strings probe moduli spaces whose volumes grow factorially. Thus open--closed factorization explains the \emph{structure} of the amplitudes, but the \emph{origin} of the factorial growth is intrinsically geometric.

\subsubsection{Gevrey growths of the nodal diagrams in \cite{joydeep} and \cite{borel2} \label{vanmeyaagu}}

Our analysis in the above subsections now leads us to the question as to why would there be Gevrey growths of the {\it nodal diagrams} $-$ first studied in \cite{borel2} and then elaborated further in \cite{joydeep}? The answer is instructive and lies at the heart of the problem where the intersection of string/M-theory and QFT matters. 

Let us start by looking at the lower bound on the nodal diagrams. These diagrams are similar to the ones, for say $\lambda \phi^m$ with $m \ge 4$ theories, that we studied in section \ref{bibi} and in \cite{borel2}. Therefore we would naively expect the lower bound \eqref{tequlazoe} to hold again. This is naive because we are ignoring one aspect of the nodal diagrams for the present case that was not there for the QFT results from section \ref{bibi}, namely that the nodal diagrams are now computed over a path-integral whose action is shifted by the displacement operators (see for example eq. (3.4) in \cite{borel2}). How does this change the lower bound from \eqref{tequlazoe}? The answer is that now we are allowed to have many more diagrams including 1-point functions that would typically vanish for the QFT cases. Therefore, for ${\rm N}!$ possible permutations of the ${\rm N}$ vertices considered earlier, and taking one possible configuration we can change $c_1$ from say \eqref{tequlazoe} to the following:
\bg\label{agabibi2}
c_1 ~\to~ \sum_{r= 0}^m \alpha_{rm} c_{10}^{(r)}~ c_{11}^{(m-r)}, \nd
where $\alpha_{rm}$ are constants and we can define $c_{10}^{(0)} = c_{11}^{(0)} \equiv 1$ without loss of generality. The quantity $c^{(r)}_{10}$ is related to the {\it lowest} possible number of variations by {\it joining the legs} that one may perform to generate new topologies in $\lambda \phi^m$ theory. This is clearly related to $c_1$ discussed earlier in say \eqref{tequlazoe}, however now we also have new possibilities to study 1-point functions. The superscript $r$, with $r \le m$, is related to number of legs that are joined together with opposite momenta as required for the usual Feynman diagrams. However this leaves $m-r$ possibilities for whom we find the smallest number of ``contractions" with the displacement operators\footnote{See for example the computations done in section 3 of \cite{borel2}.}. They are denoted by $c_{11}^{(m-r)}$. \eqref{agabibi2} is the quantitative statement that the total number of possibilities may be captured by summing over all of the aforementioned choices.  Putting everything together then suggests that the total number of diagrams should at least be {\it bigger} than the smallest number of possibilities appearing from imposing \eqref{agabibi2}. In other words:
\bg\label{agabibi3}
{\cal N}_{\rm diag} \ge \Big( \sum_{r= 0}^m \alpha_{rm} c_{10}^{(r)}~ c_{11}^{(m-r)}\Big)^{\rm N} ~{\rm N}! \nd
which clearly reduces to \eqref{tequlazoe} for $\alpha_{rm} = \delta_{rm}$ and $c_{10}^{(m)} \equiv c_1$. The string/M-theory input to \eqref{agabibi3} is the observation that at far IR there would {\it always} be $m$ such that $m \gg {\rm N}$ for any choice of ${\rm N}$. This would imply that the most dominant contribution comes from:
\bg\label{agabibi4}
\Big( \sum_{r= 0}^m \alpha_{rm} c_{10}^{(r)}~ c_{11}^{(m-r)}\Big)^{\rm N} = {\rm N}!~\underbrace{\sum_l \prod_{r = l}^{l + {\rm N}} \alpha_{rm} c_{10}^{(r)}~ c_{11}^{(m-r)}}_{\hat{c}^{\rm N}_1} + {\cal O}\left({{\rm N}!\over {\rm N}^q}\right),  \nd
where we expect $(q, l) \in (\mathbb{Z}_+, \mathbb{Z}_+)$. The series in $l \in \mathbb{Z}_+$ does not contribute additional growths in much the same fashion as the series in the discrete momenta do not add to new contributions. Rather they help to convert the sums to integrals. Combining \eqref{agabibi4} and \eqref{agabibi3} then gives us the following lower bound:
\bg\label{agabibi5}
{\cal N}_{\rm diag} ~\ge ~ \hat{c}^{\rm N}_1 ~ \left({\rm N}!\right)^2, \nd
which provides the first indication that there may be more non-trivial growths for the nodal diagrams as predicted in \cite{borel2}. To see that the growth is bounded from below by Gevrey-2 we have to go through the discussion presented in section 4 of \cite{joydeep}. After the dust settles, the lower bound on the growth of the nodal diagrams may be presented succinctly by:
\bg\label{agabibi6}
{\cal N}_{\rm diag} ~ \ge ~ \hat{c}_1^{\rm N} ~\left(2{\rm N}\right)! \nd
where we have kept the factor $\hat{c}_1$ unchanged. The result \eqref{agabibi6} is in concordance with what one expects from the underlying string/M-theory picture but the derivation of \eqref{agabibi6} is different from the stringy computation presented in section \ref{agabibilada}. Nevertheless the similarity to Mirzakhani's result \cite{mirzakhanipaper} is a useful guide to study the non-perturbative contributions as emphasized earlier in \cite{joydeep, borel2}.

What about the upper bound? Here one may be tempted to replace $\hat{c}_1$ in \eqref{agabibi6} by say $\hat{c}_2$ much like \eqref{tequlazoe2} keeping in mind the appropriate modification as in \eqref{agabibi2} but replacing $c^{(r)}_{1j}$ with $c^{(r)}_{2j}$ respectively. Such a choice would suggest that the nodal diagrams show a Gevrey-2 growth. However, as pointed out in \cite{borel2, joydeep}, the story could in-principle be more non-trivial. To see this consider for example a path-integral of the form given by eq. (3.4) in \cite{borel2}. The growth of the nodal diagrams, as computed in eq. (4.10) therein is given by:
\bg\label{odalisa}
{\cal N}_{\rm diag} = {1\over {\rm N}!} \cdot \left({\rm N}!\right)^3 \left({\rm N}q+1\right)! \left({\rm N}r\right)!\left({\rm N}(s-1)\right)! \nd
where the first ${1\over {\rm N}!}$ comes from expanding the exponential, the second ${\rm N}!$ comes from choosing the dominant contribution, the third ${\rm N}!$ comes from the permutation of the ${\rm N}$ vertices, the fourth ${\rm N}!$ comes from similar argument that lead to \eqref{agabibi4}, and the remaining three comes from the actual proliferations of the nodal diagrams. If \eqref{odalisa} were true, then the nodal diagrams would show a Gevrey-$(q+r+s+1)$ growth leading to non-perturbative corrections of the form:
\bg\label{vanandgorom}
{\rm exp}\left[-{1\over g^{1/(q+r+s+1)}}\right], \nd
where $g \equiv c_{nmpqrs}$ in eq. (3.4) of \cite{borel2}, but is not related to the string coupling $g_s$, instead it is related to powers of inverse ${\rm M}_p$ (recall that we are in M-theory) {\it i.e.} $g = {1\over {\rm M}_p^{m+n+p}}$. Thus we are not predicting new inverse $g_s$ states. This is good, but there is something seriously off about \eqref{vanandgorom}: since $g \to 0$, the non-perturbative effects increase as $(q, r, s)$ increase! Thus \eqref{odalisa} cannot be the correct growth of the nodal diagrams. 

The resolution, as pointed out in \cite{borel2} lies on the fact that the last three factorial growths $-$ that capture the proliferations of the nodal diagrams $-$ are in fact {\it spurious}. The actual growth is only $({\rm N}!)^2$, or more appropriately Gevrey-2. However the result in \cite{borel2} was for three scalar degrees of freedom. Do we expect higher Gevrey growth if we have, say 256 degrees of freedom in M-theory? The answer is tricky because, as we shall discuss here, many of the 256 massless degrees of freedom are integrated away to allow for non-local interactions so that we only have a small number of on-shell degrees of freedom (especially in the gravitational sector). 

Even if we ignore the integrating out procedure, the story is interesting in its own right. Let us consider the scenario as depicted in \eqref{pobitradud} with $\sigma$ degrees of freedom $-$ all represented by dimensionless scalar fields $\phi_j$. In this language 
the coupling constant $\lambda_{n_1..n_l p_1..p_l} \propto {1\over {\rm M}_p^{n_1 + ..+n_l}}$ and therefore vanishes for ${\rm M}_p \to \infty$. The difference with the QFT results from section \ref{bibi} now becomes apparent when we study the growth of the diagrams. For $l = \sigma$ in \eqref{pobitradud} we can split $c_2^{\rm N}$ in \eqref{tequlazoe2} to $\prod\limits_{j = 1}^\sigma c^{\rm N}_{1+j}$ to take care of the $\sigma$ fields but that will not alter the factorial growth of the Feynman diagrams. The polynomial growth changes but the non-perturbative behavior $-$ that relies on the factorial growth $-$ remains unchanged. This also tells us that in QFT it is in general difficult to get behavior that is beyond Gevrey-1.

For the nodal diagrams, the story is slightly different. We are now looking at the placement of ${\rm N}$ vertices with $\sigma$ nodes such that at every $j$-th node we have $p_j$ fields as evident from \eqref{pobitradud}. We are also ignoring the $\otimes$ operation and the nodal diagram takes the form as shown above \eqref{lolashonaek}. We can now impose:
\bg\label{agabibi22}
c_{1+j} ~\to~ \sum_{r= 0}^{p_j} \alpha^{(j)}_{rp_j} c_{1+j,0}^{(r)}~ c_{1+j,1}^{(p_j-r)}, \nd
where the subscripts 0 and 1 denote the two-point and one-point functions respectively and the superscripts $(r, p_j-r)$ denote the number of such possible contractions. It should also be clear that the splitting of $c_{1+j}$ in \eqref{agabibi22} captures the {\it largest} possible variations that one may perform to generate distinct topologies. Thus for the nodal diagrams, we want:
\bg\label{vanandraat}
\prod\limits_{j = 1}^\sigma c^{\rm N}_{1+j} {\rm N}! ~\to ~ 
\prod_{j = 1}^\sigma \left(\sum_{r= 0}^{p_j} \alpha^{(j)}_{rp_j} c_{1+j,0}^{(r)}~ c_{1+j,1}^{(p_j-r)}\right)^{\rm N} {\rm N}! \nd
capturing both the polynomial and the factorial growths. In fact the above decomposition \eqref{vanandraat} shows that the dominant contribution from each of the $j$ terms is ${\rm N}!$ implying that \eqref{vanandraat} allows a dominant growth of atmost $({\rm N}!)^{1+\sigma}$. This is the upper bound of growth exactly as anticipated in \cite{borel2}, although we expect the actual growth of the nodal diagrams to be significantly lower because many of the gravitational degrees are integrated out to generate non-local interactions. Even for the case where the degrees of freedom are not integrated out one may find a better upper bound by asking for the effective number of {\it decoupled} degrees of freedom. These are some linear combinations of the $\sigma$ degrees of freedom that allow contractions between the same species of fields. We can denote them by $\hat\sigma$ and clearly $1 \le \hat\sigma << \sigma$. In principle this would be the right way to go, but generically and following the manipulations presented in \cite{joydeep}, we get the following bounds:
\bg\label{vanand2mey}
\hat{c}_1^{\rm N}(2{\rm N})! ~\le ~ {\cal N}_{\rm diag} ~ < ~ \hat{c}_2^{\rm N} \big((1+\sigma){\rm N}\big)!\nd
which is bounded from below by Gevrey-2 and from above by Gevrey-$(1+\sigma)$. The other parameter, $\hat{c}_2^{\rm N}$, may be easily extracted from \eqref{vanandraat}.  In the language of $\hat\sigma$ we typically expect a Gevrey-2 behavior for simple cases. For more complicated case with $\sigma = 256$, one expects ${\cal N}_{\rm diag}$ to still lie closer to the lower bound in \eqref{vanand2mey}. In fact, in the language of $\hat\sigma$, we can interpret the lower and the upper bounds in \eqref{vanand2mey} roughly as though we have taken $\hat\sigma = 1$ and $\hat\sigma = \sigma$ respectively. Interestingly, we can also see how \eqref{vanandgorom} is modified. Taking the upper limit in \eqref{vanand2mey}, the non-perturbative corrections now go as:
\bg\label{vanandgorom2}
{\rm exp}\left(-{\rm M}_p^{n_1+...+n_\sigma\over 1+\sigma}\right), \nd
which when compared to \eqref{vanandgorom} shows that, with dimensionless field contents and as we go to higher derivatives, the terms are both exponentially and polynomially suppressed by powers of ${\rm M}_p$, unlike the situation in \eqref{vanandgorom}. One may replace $\sigma$ in the denominator of the ${\rm M}_p$ power in \eqref{vanandgorom2} to $\hat\sigma$, but the physics will not change. Moreover our analysis shows that the $g_s$ behavior does not change either, {\it i.e.} it continues to show a Gevrey-2 behavior.

\subsection{Trans-series actions, path-integrals and wormhole dressing \label{trans00}}

\begin{figure}[h]
\centering
\begin{tabular}{c}
\includegraphics[width=5in]{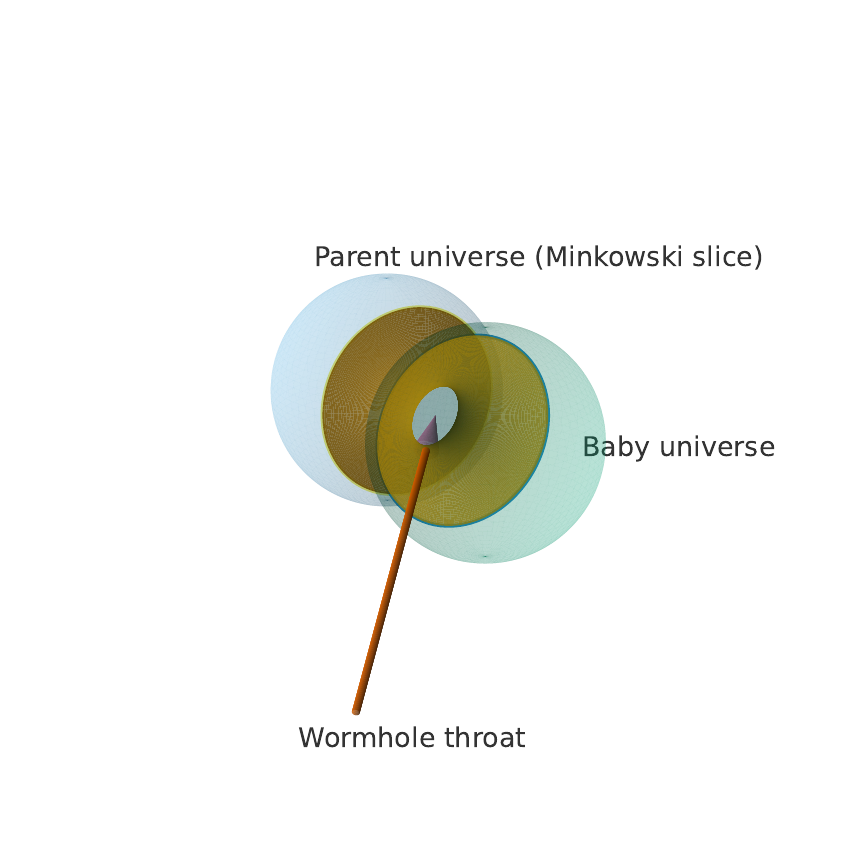}
\end{tabular}
\vskip-.2in
\caption[]{A representation of a baby universe appearing from a Minkowski spacetime with a wormhole throat connecting them.}
\label{wormholefigure}
\end{figure}

The existence of de Sitter as an excited state depends crucially on how we express the action at the Minkowski minima. As elaborated in \cite{joydeep}, string theory (or M-theory) doesn't appear to harbor any minima other than Minkowski (and possibly AdS) ones. For our case it is enough to consider Minkowski minima, and one can study the case where the dynamics is governed by one such Minkowski minimum. For such a choice, the action can be expressed as a trans-series where we consider all possible instanton saddles and the fluctuations around them. In M-theory, these instanton saddles could not only come from M5 and M2-brane instantons, but also from other non-perturbative instantons (both real and complex kinds). A detailed study of this appears in section \ref{trans1}, and in particular in \eqref{kimkarol}, which will form the key structure on which we will base our understanding of the existence of a de Sitter excited state in heterotic string theory. The trans-series form of the action is important because it is the only way to control the asymptotic nature of perturbative corrections, so it is important to ask whether this form could be effected by any means. One such influence comes from the formation of {\it baby universes} which are {\it external} effects that could in principle change the trans-series structure of the action governing our universe (see {\bf figure \ref{wormholefigure}}). Question is to what extent a baby universe would influence the physics of the excited state. 
This will be studied in details in section \ref{trans2}, but here we want to provide a simple construction that illustrates the process. Before that however let us start by justifying the trans-series structure of the action itself.

\subsubsection{Justification of the trans-series action structure}

In our study of the Schwinger-Dyson equations and the Glauber-Sudarshan states, an important step is to express the action itself in a trans-series form. This was first proposed in \cite{joydeep} as a useful way to incorporate the Gevrey growths of the diagrams and still allow for a controlled computation of the expectation values. A precise application of this formalism to the Schwinger-Dyson equations will be undertaken here using a 
trans-series form of the source action given by ~\eqref{kimkarol}. This typically takes the form:
\begin{equation}\label{vonstag}
\hat{\bf S}_{\rm tot}({\bf \Xi}) \;=\; \sum_{\rm saddles}
\mathcal{G}[{\bf Q}_{\rm pert},\mathbb{F}]\,
\exp\!\big[\mathcal{H}[{\bf Q}_{\rm pert},\mathbb{F}]\big],
\end{equation}
where $\mathcal{G}[{\bf Q}_{\rm pert},\mathbb{F}]$ is the perturbative fluctuations around a given saddle expressed in terms of a perturbative series ${\bf Q}_{\rm pert}({\bf \Xi})$ and possible non-locality function $\mathbb{F}(x -y)$ (or more generically $\mathbb{F}(x, y)$) for the class of on-shell states ${\bf \Xi}$. \eqref{vonstag}
is also a generic outcome of integrating out auxiliary or off-shell fields whose propagators generate non-local couplings among perturbative operators ${\bf Q}_{\rm pert}$.  
To justify the form \eqref{vonstag} concretely, it suffices to examine a few low-dimensional analogues that illustrate the same mechanism.

\subsubsection*{Example 1: Gaussian elimination and exponential resummation}

We can start with a simple real scalar field $\phi$ with an auxiliary field $\chi$ in $d$-dimensions. Consider also that the scalar field $\phi$ and the auxiliary field $\chi$ couples in the following way:
\begin{equation}
S[\phi,\chi] \;=\;
\int d^d x\;
\Big[\frac{1}{2}\chi^2 + g_s\,\chi\,Q(\phi)\Big],
\end{equation}
where $Q(\phi)$ represents any local functional built from the physical field $\phi$ (e.g. $Q=\phi^2$ or a curvature polynomial in the gravitational case).  
Integrating out the auxiliary field $\chi$ gives:
\begin{equation}
\int D\chi\,e^{-S[\phi,\chi]}
\;\propto\;
\exp\!\Big[-\frac{g_s^2}{2}\int d^d x\,Q^2(\phi)\Big].
\end{equation}
The resulting effective action is \emph{non-local} when written in terms of the Green's function $G(x-y)$ of the eliminated field:
\begin{equation}
S_{\rm eff}[\phi]
\;=\;
\frac{g_s^2}{2}\int d^d x\,d^d y\;Q(\phi(x))\,G(x-y)\,Q(\phi(y)).
\end{equation}
This structure matches the schematic trans-series kernel
$\mathbb{F}(x,y)\!\equiv\!G(x-y)$ in $\mathcal{G}[{\bf Q}_{\rm pert},\mathbb{F}]$.
If the path integral also includes a determinant prefactor $\det(\partial^2+m^2)^{-1/2}$, its logarithm expands as an infinite sum of local curvature terms, giving rise to the $\mathcal{H}[{\bf Q}_{\rm pert},\mathbb{F}]$ part.  
Hence the combined structure:
\begin{equation}
\mathcal{G}[{\bf Q}_{\rm pert},\mathbb{F}]\,e^{\mathcal{H}[{\bf Q}_{\rm pert},\mathbb{F}]}
\end{equation}
emerges directly from integrating out $\chi$. An alternative way, which may be a bit more transparent, would to view $\chi$ as an off-shell field with a quadratic kernel $\mathbb{K}$ that may depend on the background configuration ${\bf \Xi}$, and whose inverse $\mathbb{K}^{-1} = {\rm G}$ is the propagator (or non-local kernel $\mathbb{F}$). The interaction with the on-shell fields $\phi$ is encoded through a source term $J[\phi]$ and possibly higher-order couplings ${\rm V}_{\rm int}[\chi;\phi]$ in the following way:
\begin{equation}
{\bf S}[\chi;\phi] \;=\; \frac{1}{2}\int \chi\,\mathbb{K}\,\chi 
\;-\; \int {\rm J}[\phi]\,\chi 
\;+\; {\rm V}_{\rm int}[\chi;\phi].
\end{equation}
We start by defining the effective action obtained after integrating out an off-shell field $\chi$ in the usual way as 
$e^{-{\bf S}_{\rm eff}[\phi]} 
= 
\int D\chi\, e^{-{\bf S}[\chi;\phi]}$.
Taking the logarithm and then performing the Gaussian integral gives:
\bg\label{etolengmey}
{\bf S}_{\rm eff}[\phi] & = &
{\bf S}_{\rm cl}[\phi]
-\log \!\int D\chi\, e^{-{\bf S}_{\rm quad}[\chi;\phi] - {\rm V}_{\rm int}[\chi;\phi]} \nonumber\\
& = &
{\bf S}_{\rm cl}[\phi]
+\frac{1}{2}\log\det \mathbb{K}[{\bf \Xi}]
-\frac{1}{2} {\rm J} \mathbb{K}^{-1} {\rm J}
-\log\Big\langle e^{-{\rm V}_{\rm int}[\chi+\mathbb{K}^{-1}{\rm J};\phi]}\Big\rangle_{\mathbb{K}}, \nd
where $\langle \mathcal O \rangle_{\mathbb{K}}$ is the expectation value with respect to the Gaussian measure whose quadratic kernel is $\mathbb{K}$. Each term in \eqref{etolengmey} carries its own hierarchy in the coupling $g_s$:
\vskip.1in \noindent $\bullet$ 
The term $\tfrac{1}{2}\log\det \mathbb{K}$ expands through the heat-kernel (Seeley--DeWitt) expansion as an infinite series of local curvature terms, representing perturbative loop corrections.
\vskip.1in \noindent $\bullet$
The bilocal term $\tfrac{1}{2}{\rm J} \mathbb{K}^{-1}{\rm J}$ represents connected tree-level and one-instanton effects.
\vskip.1in \noindent $\bullet$
The Gaussian average $\langle e^{-{\rm V}_{\rm int}}\rangle_{\mathbb{K}}$ generates higher-order multi-instanton and non-local corrections.
\vskip.12in

\noindent Each contribution contains non-analytic pieces in the coupling, typically of the form $e^{-{\rm A}/g_s^2}$, multiplied by a power-series in $g_s$.  
Consequently, the effective action becomes a trans-series:
\begin{equation}
{\bf S}_{\rm eff}[\phi]
=
{\bf S}_{\rm pert}[\phi]
+\sum_{n\ge1} e^{-n{\rm A}/g_s^2}
\sum_{k\ge0} c_{n,k}(\phi)\, g_s^{\,k} \equiv \sum_{\rm saddles}
\mathcal{G}[{\bf Q}_{\rm pert},\mathbb{F}]\,
\exp\!\big[\mathcal{H}[{\bf Q}_{\rm pert},\mathbb{F}]\big],
\label{eq:trans_series_action}
\end{equation}
where each exponential term corresponds to an $n$-instanton sector, and the coefficients $c_{n,k}(\phi)$ are local functionals built from curvature and flux invariants. Thus the effective source action inherits the same exponential hierarchy already visible at the level of the path integral.

\subsubsection*{Example 2: Instanton saddle and trans-series expansion}

Another illustrative example comes from viewing the action as an Airy function (this will be useful when we study the wormhole effects).
Take a one-dimensional toy path integral:
\begin{equation}
{\bf Z}({\rm J}) \;=\; \int dz\,\exp\!\big[-{\bf S}(z;{\rm J})\big],
\qquad
{\bf S}(z;{\rm J})=\frac{1}{3}z^3 - {\rm J}z.
\end{equation}
The saddle points satisfy $z^2={\rm J}$, giving $z_\pm=\pm\sqrt{{\rm J}}$.  
The full integral receives contributions from each saddle in the following way:
\begin{equation}
{\bf Z}({\rm J})
\;\sim\;
\sum_{\sigma=\pm}
n_\sigma({\rm J})\,
\exp[-{\bf S}(z_\sigma;{\rm J})]\,
\big(1+\text{fluctuations around }z_\sigma\big),
\end{equation}
which is the simplest example of a \emph{trans-series}:
a discrete sum of exponentials multiplied by asymptotic series in ${\rm J}^{-1/2}$. But that's not the full story, because with the Glauber-Sudarshan states there would still be the Gevrey growth\footnote{For a discussion of Gevrey growth, see section \ref{papindigo}. A brief summary is the following. Gevrey-$k$ growth \cite{gevrey} is a generalization of the usual factorial growth. For the standard Feynman diagrams in field theory and gravity, the typical growth is factorial {\it i.e.} ${\rm N}!$ at ${\rm N}$-th order, so generically Gevrey-1. In string theory it is known that the growth can be $({\rm N}!)^2 \approx (2{\rm N})!$ {\it i.e.} Gevrey-2 from world-sheet proliferation as shown in section \eqref{agabibilada} (see also \cite{mirzakhanipaper, shenker}). For nodal diagrams studied in \cite{borel2}, the growth could in principle be Gevrey-$k$ with $k \ge 2$ as shown in \eqref{vanand2mey}.} when we take expectation values. Such asymptotic expansion may be tamed by converting the effective action itself to a trans-series over the non-perturbative saddles.  
In field theory, the instantons play the role of these saddles, and the fluctuation determinants around them correspond to the prefactors $\mathcal{G}$ and $\mathcal{H}$ in equation~\eqref{kimkarol}.  

\subsubsection*{Example 3: Integrating out an off-shell metric component}

To connect more directly with the gravitational setup, let ${\bf g}_{im}$ denote a cross-term metric component (off-shell) coupled linearly to a perturbative curvature operator $Q_1({\bf \Xi})$ and quadratically to itself through its propagator $G_{im,jm'}$.  
Schematically,
\begin{equation}\label{omasara}
S[{\bf g}_{im}]
=\frac{1}{2}\int {\bf g}_{im}\,{\cal O}^{im,jm'}\,{\bf g}_{jm'}
+\int {\bf g}_{im}\,Q_1^{im}({\bf \Xi}),
\end{equation}
where we will not worry too much about the consistency of such a representation in the absence of ghosts {\it et cetera}. A more detailed and careful study will be performed in sections \ref{sec6} and \ref{grace}. Meanwhile we will take \eqref{omasara} for illustrative purpose only.
Eliminating ${\bf g}_{im}$ gives:
\begin{equation}
S_{\rm eff}[{\bf \Xi}]
=\frac{1}{2}\int Q_1^{im}({\bf \Xi})
\,{\cal O}^{-1}_{im,jm'}\,
Q_1^{jm'}({\bf \Xi}),
\end{equation}
where ${\cal O}^{-1}\equiv G$ acts as the non-local kernel $\mathbb{F}({\rm Y,Y_Z})$, where ${\rm Y, Y_Z}$ are two different points.
If the background also contains instanton-induced exponentials
$e^{-{\rm S}_d^{(1)}{\rm M}_p^d I_d^{(1)}({\rm Y},x_{\rm Z})}$,
these multiply the Gaussian result to yield an effective action:
\begin{equation}
\hat{\bf S}_{\rm eff}({\bf \Xi})
\;=\; \sum_{\rm saddles}
\mathcal{G}[{\bf Q}_{\rm pert},\mathbb{F}]
\exp\!\Big[\mathcal{H}[{\bf Q}_{\rm pert},\mathbb{F}]\Big],
\end{equation}
where $\mathcal{G}$ collects the polynomial combinations of $Q_1$ and its derivatives,
and $\mathcal{H}$ exponentiates the determinant and instanton factors. We have also summed over all the allowed saddles.
This is precisely the structure of equation~\eqref{kimkarol}.

\subsubsection*{Example 4: Linked-cluster expansion}

Finally, when the coupling between the off-shell and on-shell sectors is non-linear,
\begin{equation}
S_{\rm int}=\sum_{n\ge1}\frac{a_n}{n!}\int d^d x\;
\phi^n(x)\,Q_1(x),
\end{equation}
integrating out $\phi$ produces a \emph{linked-cluster expansion} of connected diagrams:
\begin{equation}
\ln Z[Q_1]
=\sum_{r\ge1}\frac{1}{r!}
\int \!Q_1(x_1)\cdots Q_1(x_r)\,
G(x_1-x_2)\cdots G(x_{r-1}-x_r),
\end{equation}
which resums naturally into the exponential of a functional $\mathcal{H}[Q_1,G]$.  
Hence the logarithm of the partition function generates the $\mathcal{H}$ term, while the prefactor polynomials in $Q_1$ define $\mathcal{G}$. This is similar to the trans-series structure of the effective action.

\subsubsection*{Summary of the justification}

Let us summarize the justification of using a trans-series form for the action.
Across all these examples, we see that the same pattern repeats:
\vskip.1in \noindent $\bullet$
Integrating out an auxiliary or off-shell degree of freedom produces a kernel $G(x,y)$ that mediates non-local couplings among perturbative operators.
\vskip.1in \noindent $\bullet$
The resulting effective action contains both polynomial prefactors (the $\mathcal{G}$ function) and exponentials of determinant or instanton contributions (the $\mathcal{H}$ function).
\vskip.1in \noindent $\bullet$ 
 When multiple instanton saddles contribute, their weighted sum yields a full {trans-series} action of the following form:
  \begin{equation}
  \hat{\bf S}_{\rm eff}({\bf \Xi})
  =\sum_{d,S_d^{(i)}}\!
  \mathcal{G}_d({\bf Q}_{\rm pert},\mathbb{F})
  \exp\!\big[-S_d^{(i)}{\rm M}_p^d I_d^{(i)}+\mathcal{H}_d({\bf Q}_{\rm pert},\mathbb{F})\big],
  \end{equation}
  matching the structure used in \eqref{kimkarol} and in sections~\ref{sec6} to \ref{secmetric}. The finiteness emanating from the above choice amidst Gevrey growths of the diagrams, further justifies the trans-series form.
  
\vskip.12in
\noindent The three points above essentially provide the main reason, but one may also see it from the following example. Consider an action of the form ${\bf S}_{\rm kin} + {\bf S}_{\rm int}$. In the presence of the displacement operator $\mathbb{D}(\sigma)$, one-point functions are non-zero and they appear in the path-integral accompanied by $\sum\limits_n (n!)^{-1}\left({\bf S}_{\rm int}\right)^n$. They typically show Gevrey growths, and only after Borel resummations finite results could be extracted. On the other hand, with an action of the form 
${\bf S}_{\rm kin} + \sum\limits_{\rm saddles} \widetilde{\bf S}_{\rm int} ~{\rm exp}\left(-{\bf S}_{\rm inst}\right)$, the one-point functions are now typically accompanied by $\sum\limits_{n,{\rm saddles}} (n!)^{-1} \left(\widetilde{\bf S}_{\rm int}\right)^n {\rm exp}\left(-n{\bf S}_{\rm inst}\right)$, which make them finite! Thus there are two equivalent viewpoints. {\Su One}, we consider the Gevrey growths and the then extract finite answers {\it after} Borel resumming, or {\Su two}, consider the action in the trans-series form right from the start and simply extract finite answers by Taylor expanding the interaction terms to any order in the expansion parameters.

\vskip.1in

\noindent Therefore to summarize: the trans-series action in equation~\eqref{kimkarol} is not an assumption but the inevitable result of (i) integrating out off-shell fields with propagators $G(x,y)$, and (ii) summing over non-perturbative saddles weighted by their fluctuation determinants.  
In the full eleven-dimensional theory, the cross-term metric components like ${\bf g}_{im},{\bf g}_{i\rho},{\bf g}_{m\rho}$ {\it et cetera} play the role of $\chi$ or $\phi$ in the toy models above, and the functions $\mathcal{G}$ and $\mathcal{H}$ encode the resulting non-local, instanton-dressed couplings among the physical curvature and flux operators ${\bf Q}_{\rm pert}$ (see \eqref{kimkarol}). In the following section we will see how such a trans-series form remains intact even in the presence of wormholes and baby universes.

\subsubsection{Action in trans-series form and wormhole effects}

When the gravitational path integral admits multiple disconnected asymptotic regions --- {\it i.e.}\ when ``baby universes'' appear --- the correct way to think about their quantum dynamics is that \emph{each universe has its own path integral}, but the actions entering these path integrals admit trans-series expansions built around the chosen vacuum. Wormhole effects then provide correlations between the otherwise independent universes by \emph{dressing} the trans-series structure\footnote{As far as we are aware of, the dressing mechanism studied here and in section \ref{trans2}, have not been addressed in the literature. Earlier studies by \cite{coleman} have shown that the effective couplings become $\alpha$-dependent because of wormholes, and one must integrate over $\alpha$ in the gravitational path integral. But this is always described at the level of couplings or partition functions, not at the level of a trans-series expansion of the action itself. On the other hand, the trans-series structure in field theory has been studied more recently in \cite{dyson, unsal} (see also \cite{gevrey, borelborel}) but not in the context of the wormholes. Thus our proposal here is to combine these two apparently disparate facts. As we shall show here, and in section \ref{trans2} in more quantitative details, a simple and a self-consistent picture emerges from these considerations.}. We now explain this step by step.

\paragraph{(a)~ Independent path integrals for each universe.}
Suppose the total Euclidean path integral of M-theory (or any gravitational theory) admits two disconnected boundaries, $\mathcal U_1$ and $\mathcal U_2$. From the perspective of the low-energy effective theory, these correspond to two distinct universes, each with its own Hilbert space, set of fields, and path integral. At this level the full partition function factorizes as:
\begin{equation}\label{ruccaP}
\mathcal Z_{\rm tot} = \mathcal Z_1 \times \mathcal Z_2,
\end{equation}
where $\mathcal Z_i$ denotes the partition function of the universe $\mathcal U_i$. In principle $i > 1$ and could take any positive integer values but, as mentioned above, we will take $i = 1, 2$. 
The individual contributions are:
\begin{equation}
\mathcal Z_i = \int \mathcal D{\bf \Xi}_i \, e^{-\hat{\bf S}^{(i)}_{\rm tot}[{\bf \Xi}_i]} , \qquad i = 1,2.
\end{equation}
Here ${\bf \Xi}_i$ denotes the collection of on-shell fields (metric, $p$-forms, moduli, matter fields, etc.) living in universe $i$, and $\hat{\bf S}^{(i)}_{\rm tot}({\bf \Xi}_i)$ is the effective action functional describing its dynamics\footnote{The appearance of only the on-shell fields will become clear from section \ref{sec6}.}.

\paragraph{(b)~ Choice of vacuum and trans-series expansion of the action.}
Within each universe, one first chooses a vacuum configuration around which the semiclassical description is constructed. For instance, this could be a Minkowski minimum of the scalar potential, or some other solitonic solution. Once the vacuum ${\bf \Xi}_{\rm vac}^{(i)}$ is chosen, the effective action admits a trans-series expansion of the form
\begin{equation}
\hat{\bf S}^{(i)}_{\rm tot}[{\bf \Xi}_i] 
= {\bf S}^{(i)}_{\rm pert}[{\bf \Xi}_i] 
+ \sum_{k=1}^\infty e^{-k {\bf S}_{\rm inst}^{(i)}[{\bf \Xi}_i]} {\bf S}^{(i)}_{k\text{-inst}}[{\bf \Xi}_i] 
+ \cdots ,
\end{equation}
which, as mentioned earlier, is the only possible way to control the underlying resurgent structure associated with the asymptotic growth of the Feynman diagrams governing the system.
The terms in this expansion have the following meaning:
\vskip.1in \noindent $\bullet$  ${\bf S}^{(i)}_{\rm pert}({\bf \Xi}_i)$ is the perturbative effective action obtained by expanding around the chosen minimum.
\vskip.1in \noindent $\bullet$ Each term proportional to $e^{-k {\bf S}_{\rm inst}^{(i)}({\bf \Xi}_i)}$ arises from a $k$-instanton sector --- a finite-action Euclidean solution interpolating between adjacent minima in the same universe’s potential landscape.
\vskip.1in \noindent $\bullet$ ${\bf S}^{(i)}_{k\text{-inst}}({\bf \Xi}_i)$ contains the one-loop determinants and higher-order fluctuation corrections associated with the $k$-instanton background.
\vskip.12in \noindent 
This trans-series is a property of the \emph{action functional itself}, not of the partition function. It reflects the complete semiclassical structure of the theory around the chosen vacuum.

\paragraph{(c)~ Path integrals built from trans-series actions.}
The partition function in each universe is then computed by inserting the trans-series-expanded action into the path integral:
\begin{equation}
\mathcal Z_i 
= \int \mathcal D{\bf \Xi}_i \, 
\exp\!\left[ 
- {\bf S}^{(i)}_{\rm pert}[{\bf \Xi}_i]
- \sum_{k=1}^\infty e^{-k {\bf S}_{\rm inst}^{(i)}({\bf \Xi}_i)} {\bf S}^{(i)}_{k\text{-inst}}[{\bf \Xi}_i] 
- \cdots 
\right] .
\end{equation}
All the standard non-perturbative physics --- vacuum tunneling, multi-instanton contributions, and resurgence structure --- is contained within this integral. Importantly, all of these effects are ``horizontal'' in the sense that they describe processes \emph{within a single universe} and require no reference to other asymptotic regions\footnote{To keep the structure simple, we will ignore the ghosts for the time being. They will be inserted in when we allow for the full quantitative analysis from section \ref{sec6} onwards.}.

\paragraph{(d)~ Wormholes and the breakdown of factorization.}
The situation changes fundamentally once we include wormhole configurations in the gravitational path integral. A Euclidean wormhole connecting $\mathcal U_1$ and $\mathcal U_2$ is a classical solution of the full gravitational theory that glues together two otherwise disconnected universes. Because such configurations contribute to the saddle-point expansion of the total path integral, the factorization 
\eqref{ruccaP} is no longer exact. This is important and is related to the concern that we raised earlier. Question is how will the presence of a wormhole effect the dynamics of our universe. To see this, note that the correct expression for the total partition function becomes:
\begin{equation}
\mathcal Z_{\rm tot} 
= \int \mathcal D\alpha \, \mathcal P[\alpha] 
\int \mathcal D{\bf \Xi}_1 \mathcal D{\bf \Xi}_2 \,
\exp\!\left[
- \hat{\bf S}^{(1)}_{\rm tot}[{\bf \Xi}_1; \alpha]
- \hat{\bf S}^{(2)}_{\rm tot}[{\bf \Xi}_2; \alpha]
\right] ,
\end{equation}
where the measure $\mathcal P[\alpha]$ encodes the distribution of $\alpha$-parameters --- effective couplings that emerge from integrating over the wormhole sector.

\paragraph{(e)~ Physical meaning of wormhole dressing.}
The dependence on $\alpha$ represents the \emph{wormhole dressing} of the trans-series actions. This dressing modifies the original structure in several ways:
\vskip.1in \noindent $\bullet$  Wormholes effectively introduce a shared set of hidden variables that shift the couplings in each universe. After averaging over these variables, the apparent “constants” become stochastic—and, crucially, correlated across sectors—according to a single kernel fixed by the wormhole physics.\vskip.1in \noindent $\bullet$ Because instanton actions and prefactors depend on the couplings, the same hidden variables that shift the couplings also shift instanton data. Instanton amplitudes therefore become correlated with perturbative coefficients and with each other across universes; once the wormhole kernel is fixed, these cannot be tuned independently.\vskip.1in \noindent $\bullet$ Eliminating the hidden variables and working in the averaged theory reproduces their influence as bilocal terms that couple operators between universes. These mixings are not an extra freedom; they are the macroscopic imprint of the same coupling shifts and are determined by the same kernel.

\vskip.12in
\noindent All three—stochastic couplings, instanton shifts, and bilocal mixings—are different faces of one wormhole–dressing mechanism; fixing the underlying kernel fixes all three together.
Importantly, none of these effects corresponds to new instantons in the trans-series of a single universe. Instead, they represent ``vertical'' correlations that link the trans-series structures of different universes.

\paragraph{(f)~ Conceptual picture.}
The resulting picture is conceptually clean:
\vskip.1in \noindent $\bullet$
Each universe has its own path integral and its own trans-series expansion of the effective action built around a chosen vacuum.
\vskip.1in \noindent $\bullet$ The non-perturbative terms in this trans-series --- the instantons --- describe tunneling between minima within the same universe and exist independently of any other universe.
\vskip.1in \noindent $\bullet$ The presence of other universes is encoded entirely in the wormhole-induced $\alpha$-dependence of the action. This dressing correlates the couplings and modifies the trans-series coefficients, thereby allowing ``foreign'' information to imprint itself on the local physics without introducing new instantons. 
\vskip.12in

\begin{figure}[h]
\centering
\begin{tabular}{c}
\includegraphics[width=5in]{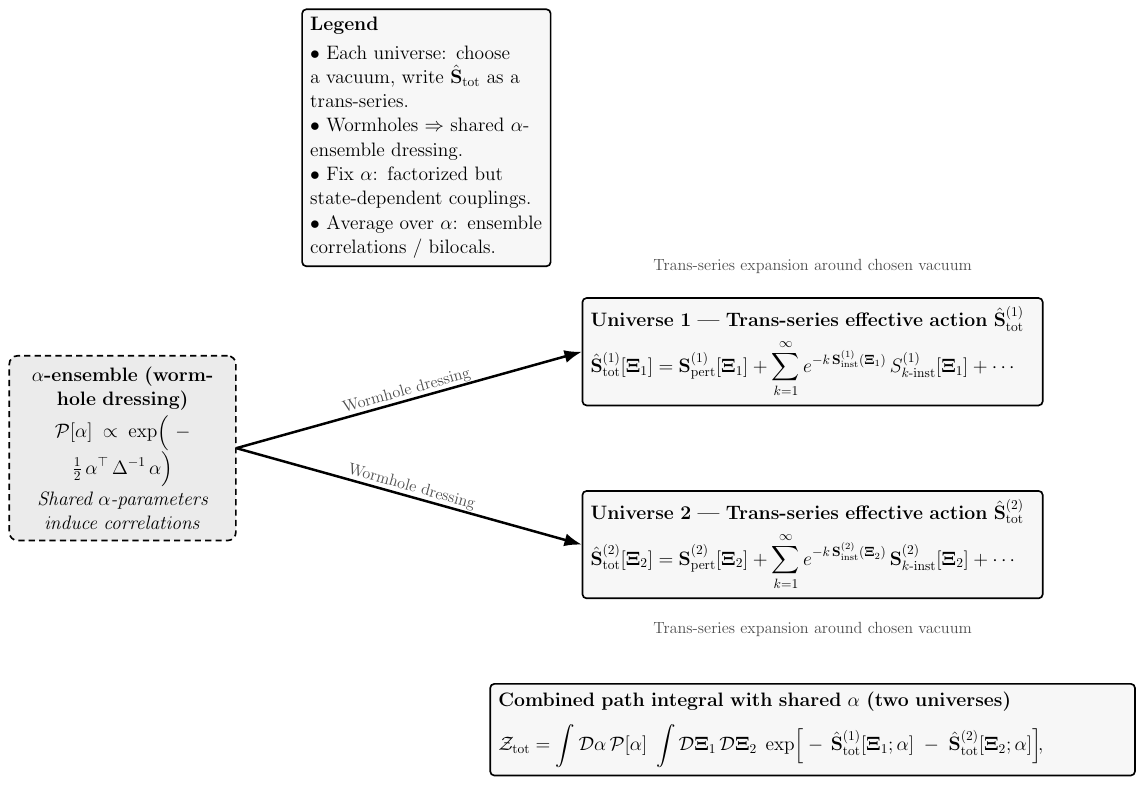}
\end{tabular}
\vskip-3.3in
\caption[]{Simple diagram to illustrate how the trans-series action in a given universe is effected by the presence of a baby universe. Note that it is the wormhole that is responsible for the dressing of the total action.}
\label{wormhole}
\end{figure}

\paragraph{(g)~ Final interpretation.}
From the standpoint of an observer in $\mathcal U_1$, the path integral continues to look like a trans-series expansion around a chosen minimum. However, the numerical values of the couplings and coefficients in that expansion are no longer fundamental constants: they fluctuate, acquire statistical distributions, or are correlated with nonlocal data because of the existence of $\mathcal U_2$. In this way, the physics of one universe carries an imprint of the other --- not through tunneling between minima, but through wormhole dressing of the action that underlies its path integral. In {\bf figure \ref{wormhole}} we have provided a summary of the underlying physics.

\subsubsection{A double-well toy model, wormholes and $\alpha$-dressing}

The discussion that we presented in the above subsection was qualitative and therefore could raise the question whether one can provide quantitative details of the above process. In the following let us present a simple model with a double-well potential that illustrates the idea clearly.
Consider quantum mechanics with the following Euclidean action:
\begin{equation}
S[x] \;=\; \int d\tau \left[ \frac{1}{2}\,\dot x^2 \;+\; V(x) \right],
\qquad
V(x) \;=\; \frac{\lambda}{4}\,\big(x^2 - a^2\big)^2 ,
\end{equation}
which has two degenerate minima at $x=\pm a$. Let us pick the vacuum at $x=+a$ and expand semiclassically about it.

\paragraph{(a)~Trans-series structure of the \emph{action}.}
The effective action around the chosen minimum admits the standard instanton trans-series:
\begin{equation}
{\bf S}_{\rm eff}[x]
\;=\;
{\bf S}_{\rm pert}[x]
\;+\; \sum_{k=1}^{\infty} e^{-k\,{\bf S}_{\rm inst}[x]}\, {\bf S}_{k\text{-inst}}[x]
\;+\; \cdots ,
\label{eq:DW-transseries}
\end{equation}
where ${\bf S}_{\rm inst}[x]$ is the single-instanton action connecting $+a \to -a$ (its explicit value is not needed for what follows) and ${\bf S}_{k\text{-inst}}[x]$ contains fluctuations/determinants in the $k$-instanton sector. Inserting \eqref{eq:DW-transseries} into the path integral produces the usual multi-sector expansion, but the \emph{trans-series belongs to the action}.

\paragraph{(b)~A simple $\alpha$-dressing model.}
Let the wormhole sector induce a single Gaussian $\alpha$-parameter that \emph{linearly shifts} the instanton action (using for example \cite{hubbard}):
\begin{equation}
\mathcal{P}(\alpha)\;=\;\frac{1}{\sqrt{2\pi\sigma^2}}\,
\exp\!\Big(-\frac{\alpha^2}{2\sigma^2}\Big),
\qquad
{\bf S}_{\rm inst}[x]\;\longrightarrow\; {\bf S}_{\rm inst}[x;\alpha]
\;=\; {\bf S}_{\rm inst}[x] \;-\; \mu\,\alpha ,
\label{eq:DW-alpha-shift}
\end{equation}
with constants $\mu$ (the sensitivity of the instanton action to the baby-universe sector) and $\sigma^2$ (the wormhole-induced variance).
Physically, \eqref{eq:DW-alpha-shift} means the $\alpha$-state slightly \emph{renormalizes} the cost of an instanton event. Equivalently, it dresses the $k$-instanton sector by a factor:
\begin{equation}
e^{-k\,{\bf S}_{\rm inst}[x;\alpha]}
\;=\;
e^{-k\,{\bf S}_{\rm inst}[x]}\; e^{\,k\,\mu\,\alpha}\, .
\end{equation}

\paragraph{(c)~ Effect on the $k$-instanton coefficients.}
Let the undressed $k$-instanton coefficient (including determinants/fluctuations) be $\mathbb{C}_k$.
After averaging over the $\alpha$-ensemble one finds the following dressed structure:
\bg \label{eq:DW-dressed-coeff}
\big\langle \mathbb{C}_k\, e^{-k\,S_{\rm inst}[x;\alpha]} \big\rangle_{\alpha}
& \equiv &
\mathbb{C}_k\, e^{-k\,{\bf S}_{\rm inst}[x]}\;
\int d\alpha\,\mathcal{P}(\alpha)\, e^{\,k\,\mu\,\alpha}
\;=\;
{~\mathbb{C}_k^{\rm (dressed)}\, e^{-k\,{\bf S}_{\rm inst}[x]}~,} \nonumber\\
&& 
\mathbb{C}_k^{\rm (dressed)} \;=\; \mathbb{C}_k\, \exp\!\Big(\frac{1}{2}k^2\mu^2\sigma^2\Big).
\nd
Thus the wormhole dressing \emph{does not create new instanton sectors}; rather, it \emph{rescales} the weight of each existing sector by an $\alpha$-average. The modification grows with $k^2$ (multi-instanton sectors are enhanced more strongly).

\paragraph{(d)~ Explicit leading sectors.}
Keeping only $k=0,1$ sectors for illustration (and suppressing fluctuation functionals for brevity), we get:
\bg
{\bf S}_{\rm eff}[x;\alpha]
& \approx &
{\bf S}_{\rm pert}[x]
\;+\; e^{-{\bf S}_{\rm inst}[x]+\mu\,\alpha}\, {\bf S}_{1\text{-inst}}[x] \;+\; \cdots\\
&\Longrightarrow & 
\big\langle e^{-{\bf S}_{\rm eff}[x;\alpha]}\big\rangle_{\alpha}
\;\supset\;
e^{-{\bf S}_{\rm pert}[x]}\;
\Big[\, 1 \;+\; e^{-{\bf S}_{\rm inst}[x]}\,e^{\tfrac{1}{2}\mu^2\sigma^2}\, {\bf S}_{1\text{-inst}}[x] \;+\; \cdots \Big],\nonumber
\nd
which when compared to the undressed case shows the single-instanton amplitude multiplied by $e^{\tfrac{1}{2}\mu^2\sigma^2}$, in agreement with \eqref{eq:DW-dressed-coeff}.

\paragraph{(e)~ Bilocal (kernel) viewpoint.}
The same result follows by integrating out $\alpha$ at the level of the action. If $\alpha$ couples linearly to the (topological) instanton number operator $\mathbb{N}_{\rm inst}[x]$ via $\delta {\bf S}_{\rm inst} = -\,\mu\,\alpha\,\mathbb{N}_{\rm inst}[x]$, then it is easy to infer:
\begin{equation}
\int d\alpha\, \mathcal{P}(\alpha)\;
\exp\!\big(+\mu\,\alpha\,\mathbb{N}_{\rm inst}[x]\big)
\;=\;
\exp\!\Big(\frac{1}{2}\mu^2\sigma^2\, \mathbb{N}^2_{\rm inst}[x]\Big),
\end{equation}
{\it i.e.}\ $\alpha$-averaging generates a \emph{nonlocal, quadratic} interaction in the topological charge, which precisely reproduces the sector-wise enhancement $e^{\frac{1}{2}k^2\mu^2\sigma^2}$ upon projecting onto fixed $k$.

\paragraph{(f)~Takeaway.}
In this double-well toy model, wormhole dressing is captured by a simple Gaussian average over $\alpha$ that \emph{rescales} the $k$-instanton coefficients as in \eqref{eq:DW-dressed-coeff}. This cleanly illustrates our general claim: wormholes do not produce new “horizontal” instanton sectors; they \emph{dress} the trans-series coefficients of the existing sectors through $\alpha$-dependent couplings. This immediately implies that in the M-theory set-up with a trans-series action of the form \eqref{kimkarol}, the wormholes' contributions would only {\it dress} the action in the aforementioned ways. We will have more to say on this  in section \ref{trans2}.

\begin{figure}[h]
\centering
\begin{tabular}{c}
\includegraphics[width=3in]{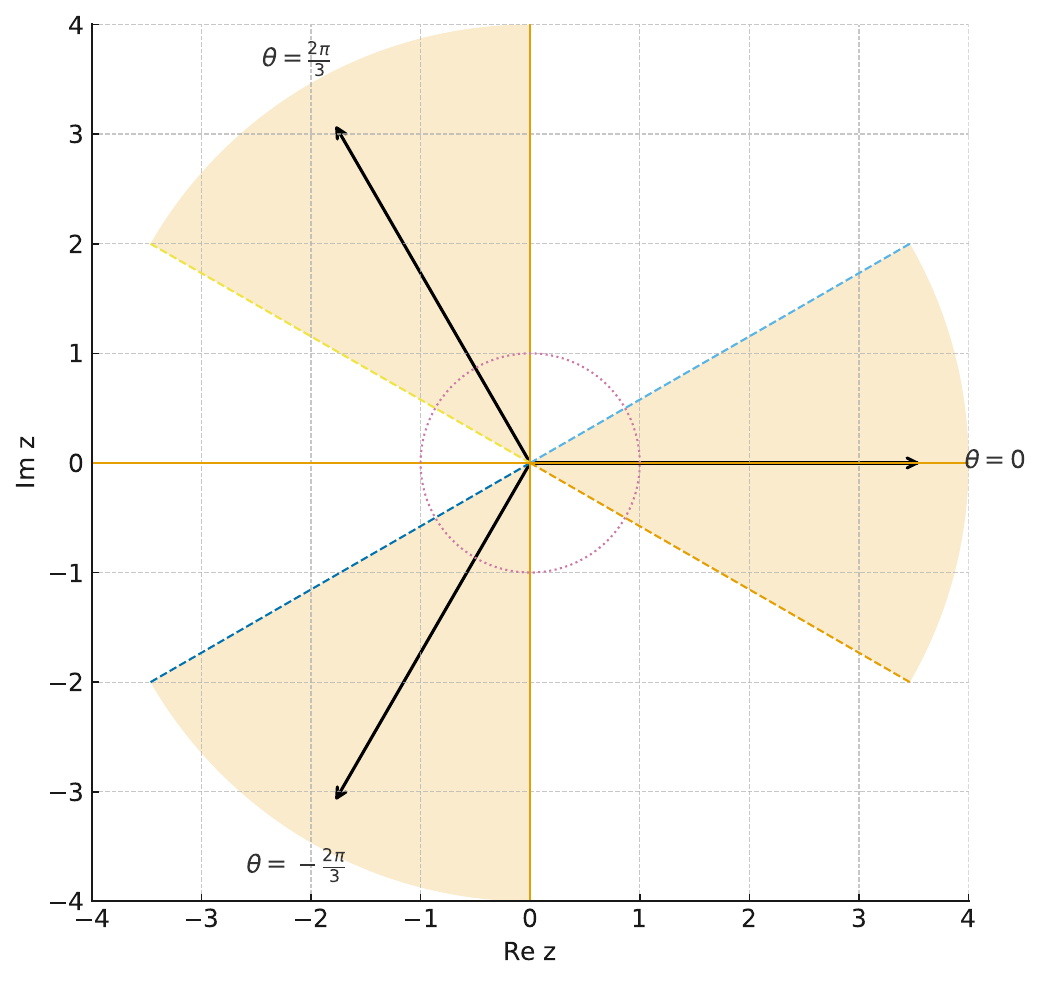}
\end{tabular}
\caption[]{The three decay wedges (where ${\rm Re}~{\rm S} > 0$) for the Airy cubic ${\rm S}(z)={z^3\over 3}$. The shaded $60^\circ$ sectors are centered at $\theta=0$ and $\theta = \pm {2\pi\over 3}$; any admissible steepest–descent contour must approach infinity inside two such wedges.}
\label{decaywedge}
\end{figure}

\subsubsection{On contour integrations and Lefschetz thimbles \label{picontour}}

The $\mathcal P(\alpha)$ function that we defined above creates no issues as long as it is a Gaussian function with $\alpha^2$ remaining strictly positive. Unfortunately, once we go away from Gaussian functions, new subtleties show up that suggest that the simple integration contour (on the real axis) may no longer be the right one. Question is why would one go for a more complicated contour and is there a logical way to derive the precise shape of the contour. This is where Lefschetz thimbles become very useful\footnote{A beautiful and self-contained exposition of Lefschetz thimbles and the Picard-Lefschetz theory is given by Witten in \cite{wittenlefschetz} which we strongly recommend the readers unfamiliar with the subject to look into.}.

We have opened up many new ideas, like Lefschetz thimbles, non-trivial integration contour et cetera, so our aim here will be to provide a simple way to understand the importance of these concepts. More quantitative details will be left for section \ref{trans2}. First, let us clarify the appearance of the non-trivial contour of integration. The Gaussian $\mathcal P(\alpha)$ that we took in \eqref{eq:DW-alpha-shift}, suggests that a more general integral with a non-Gaussian $\mathcal P(\alpha)$ could be entertained. The generic wormhole effect is to take the new $\mathcal P(\alpha)$ and shift the instanton term linearly, as in \eqref{eq:DW-alpha-shift}. If we take $\mathcal P(\alpha) = \mathcal A(\zeta){\rm exp}\left(-{\alpha^3\over 3\zeta^3}\right)$ with a parameter $\zeta$ and a normalization constant $\mathcal A(\zeta)$, then redefining $z\equiv {\alpha\over \zeta}$ we are basically looking at an 
Airy-type integral of the form:
\bg\label{skittiesmey}
{\rm I}({\rm J})=\int_{\mathcal C} e^{-{\bf S}(z;{\rm J})}\,dz,\qquad {\bf S}(z;{\rm J})=\frac{1}{3}z^{3}-{\rm J}z,
\nd
where the integrand $e^{-{\bf S}}$ is entire (no poles anywhere) and ${\rm J} \equiv \mu\zeta$ in \eqref{eq:DW-alpha-shift}. The original integration contour is the real axis but, as one may easily see, for negative $z$ the integration blows up for large $z$. Thus the real-axis contour cannot provide a finite value of this integral, and we need to deform the contour. This {deformation} cannot be a {\it small one} like we are used to deforming slightly around a pole for usual complex integration. Question however is: how are we allowed to deform drastically away from the original integration contour? The answer is that by Cauchy’s theorem one may deform $\mathcal C$ to any other contour $\mathcal C'$ provided (i) no singularities are crossed (there are none here), and (ii) $\mathcal C$ and $\mathcal C'$ lie in the same relative homology class with respect to the admissible ``ends at infinity'' where the integral converges. This is why one can move far from the real axis: there are no obstructions in the plane\footnote{In residue calculus for rational functions one keeps to small detours because poles obstruct large motions. For Airy integrals there are no poles; the only constraints are at infinity. The analogue of Jordan’s lemma is that arcs at infinity vanish when they stay inside a decay wedge. Hence one may slide the contour onto ``thimbles'' that look nothing like the real line. We will explain what thimbles are momentarily.}.

Question now is what decides where the new contour should lie. In other words, along the new contour we expect convergent results. How do we determine this? It turns out that the convergence is set by ``decay wedges'' at infinity, not by staying near the real line. This is depicted in {\bf figure \ref{decaywedge}} which we shall explain in the following.
As $|z|\to\infty$, ${\bf S}(z; {\rm J})\sim {\bf S}(z; 0) = z^{3}/3$. Writing $z=re^{i\theta}$, the real and the imaginary parts of ${\bf S}(z; 0)$ become:
\begin{equation}
{\rm Re}~ {\bf S}(z; 0) \;=\; \frac{r^{3}}{3}\,\cos(3\theta), ~~~~~ {\rm Im}~ {\bf S}(z; 0) \;=\; \frac{r^{3}}{3}\,\sin(3\theta).\end{equation}
A non-compact contour converges only if it approaches infinity inside wedges where ${\rm Re}~ {\bf S}\to +\infty$, i.e.\ where $\cos(3\theta)>0$. These are three wedges of opening angle $\pi/3$, centered at:
\begin{equation}
\theta=0,\qquad \theta=\frac{2\pi}{3},\qquad \theta=-\frac{2\pi}{3},
\end{equation}
as shown in {\bf figure \ref{decaywedge}}.
Any admissible contour must start and end in (a choice of) these decay wedges. Lefschetz thimbles are precisely middle-dimensional cycles that (i) flow out of a saddle $z_\sigma$ along steepest descent so that ${\rm Im}~{\bf S}(z; {\rm J})$ is constant and ${\rm Re}~{\bf S}(z; {\rm J})\to+\infty$ at the ends, and (ii) asymptote to the allowed decay wedges. Replacing the real axis by such a thimble (even if it arcs far into the complex plane) is therefore legal: both contours anchor to the same decay sectors at infinity and no poles are crossed.

Our above discussion provides us a quantitative way to determine the thimbles. First, the saddles are at $z = \pm 1$ and we will express $z$ as $z = x + iy$. Plugging this in the definition of ${\bf S}(z; {\rm J})$
from \eqref{skittiesmey} gives us:
\bg\label{mumbmeyaxion}
{\rm Re}~{\bf S}(z; {\rm J}) = {x\over 3}\left(x^2 - 3y^2 - 3{\rm J}\right), ~~~~~~ {\rm Im}~{\bf S}(z; {\rm J}) = {y\over 3}\left(3x^2 - y^2 - 3{\rm J}\right), \nd
where, along the thimble, we expect ${\rm Re}~{\bf S}(z; {\rm J}) > 0$ and to increase monotonically, and ${\rm Im}~{\bf S}(z; {\rm J})$, which will determine the phase, to remain a constant. For simplicity we can take this constant to be zero, and keep ${\rm J} = 1$. This provides the following curves:
\bg\label{mumbmeykitag}
y = 0, ~~~~~~ y^2 = 3(x^2 - 1), \nd
near {\it both} the saddles. One could question here whether we should instead take $z = 1 + x + iy$ and $z = -1 + x + iy$ respectively near each of the saddles and then expand ${\bf S}(z; 1)$.  Both viewpoints are the same. On a thimble through a saddle $z_\sigma$ of a holomorphic ${\bf S}(z; 1)$ we
must have
\begin{equation}
\mathrm{Im}\,{\bf S}(z; 1)\;=\;\mathrm{Im}\,{\bf S}(z_\sigma; 1).
\end{equation}
Since ${\bf S}(\pm1; 1)\in\mathbb{R}$, this means $\mathrm{Im}\,{\bf S}(z; 1)=0$ near \emph{both}
$z=+1$ and $z=-1$. We can enforce it globally by writing $z=x+iy$ and solving $\mathrm{Im}\,{\bf S}(x,y; 1)=0$,
or locally by expanding $z=z_\sigma + u$ with $u=x+iy$ small and setting
$\mathrm{Im}\,[{\bf S}(z_\sigma+u; 1)-{\bf S}(z_\sigma; 1)]=0$.

The rest of the analysis can be made more precise using the Picard-Lefschetz theory which will require a bit more machinery than we can develop here. We will elaborate further on the story in section \ref{thimbles}. Meanwhile we can define 
 downward flows by
$\dot z \;=\; -\,\overline{\partial {\bf S}}$ with the dot defined with respect to a parameter $\tau$, such that 
along a flow line, we can satisfy equations like \eqref{stepsilver4} which would further justify keeping ${\rm Im}~{\bf S}(z; {\rm J}) = 0$.
The thimble $\mathcal J_\sigma$ through a saddle $z_\sigma$ is the union of its downward flows and is a convergent integration cycle as shown in {\bf figures \ref{thimbleL}} and {\bf \ref{greencontour}}. The original contour $\mathcal C$ is homologous to a finite integer combination of thimbles:
\begin{equation}
\int_{\mathcal C} e^{-{\bf S}(z; {\rm J})}\,dz \;=\; \sum_{\sigma} n_\sigma(\mathcal C)\,\int_{\mathcal J_\sigma} e^{-{\bf S}(z; {\rm J})}\,dz,
\end{equation}
where the integers $n_\sigma(C)$ are intersection numbers with the dual (upward) cycles as shown in {\bf figures \ref{contourinteg}} and {\bf \ref{airycontour}}. These coefficients change only when parameters cross Stokes lines where two saddles have equal ${\rm Im}~ {\bf S}(z; {\rm J})$. 

Thus here is the bottom line. For an integral of the form \eqref{skittiesmey}, 
 the integrand is entire, and convergence is controlled by asymptotics at infinity. Lefschetz thimbles are globally deformed contours lying entirely in decay directions as shown in {\bf figures \ref{contourinteg}} and {\bf \ref{airycontour}}. In fact they are in the same homology class (relative to the decay sectors) as the original contour, so the integral is unchanged and therefore there is no need to restrict to small deformations. With this powerful machinery we can now tackle any ${\bf S}(z; {\rm J})$ and study the influence of baby universes on the effective action. Our analysis here was with a toy quantum mechanical example. More details on how this extends to an actual setting with eleven-dimensional M-theory and its influence on effective action like \eqref{kimkarol} will be discussed in section \ref{trans2} and specifically in sub-section \ref{thimbles}.

\section{Summaries of the various sections of the paper}

In this section we provide concise summaries of the principal sections and subsections of the manuscript. Our
aim is to clarify the main line of reasoning and highlight the logical progression of the arguments, thereby allowing readers to locate the material most relevant to their interests without needing to follow the full sequential development.

\subsection{Summaries of the sections \ref{sec2} and \ref{sec3}}

In sections \ref{sec2} and \ref{sec3} our aim is to show that 
late-time de Sitter arises $-$ albeit as an excited state $-$ as a stable endpoint (or near-endpoint) of the temporal evolution under well-posed constraints on couplings, fluxes, and corrections; the construction is duality-compatible and EFT-consistent. Moreover the quantum corrections, packaged in a trans-series, naturally points to an emergent dynamical dark-energy sector. In the following we will provide a slightly more detailed summaries of section \ref{sec2} and \ref{sec3} by pointing out the key features of each of the two sections.

\subsubsection{Summary of section \ref{sec2}}

In this section we construct late-time de Sitter states in string theory by starting from M-theory backgrounds defined through expectation values over Glauber–Sudarshan states. In the type IIB case, temporal evolution of the M-theory torus leads to a flat-sliced de Sitter background, while in the heterotic SO(32) case the process involves orbifolds, duality chains, and a more intricate warp-factor scaling, ultimately yielding a weakly coupled configuration with gauge group broken to $\left(SO(8)\right)^4$.
A parallel construction also reaches the heterotic ${\rm E}_8 \times {\rm E}_8$ theory, indicating a unifying endpoint across dual descriptions. Central to the analysis is the ambiguity of Newton’s constant in different frames: the string frame allows a static internal volume and fixed coupling, while the Einstein frame often implies shrinking cycles or singularities. Higher-order $\alpha'$
 corrections are shown not to destabilize the construction, as they can be absorbed into the solitonic backgrounds without altering the Glauber–Sudarshan expectation values. The framework yields a naturally small positive cosmological constant, with a controlled possibility of time dependence consistent with observations, and suggests that while the string frame is most natural from a duality perspective, careful warp-factor choices can render both string and Einstein frames viable in the late-time regime. In the following let us provide, in slightly more detail, capsule summaries of the various subsections of section \ref{sec2}.

In section \ref{secc2.1} we construct a late-time de Sitter state in type IIB string theory using an M-theory uplift and expectation values of metric and flux operators over Glauber–Sudarshan states. Starting from an M-theory configuration with warped products of compact manifolds, one traces the duality chain through type IIA and finally to type IIB, where the flat-slicing de Sitter metric emerges naturally. The warp factors ${\rm F}_1(t)$ and ${\rm F}_2(t)$ are expanded in powers of the type IIA coupling $g_s$, and their behavior is constrained to allow the type IIB background to remain de Sitter at late times. Einstein tensor computations confirm consistency with a cosmological constant $\Lambda$ when $a(t)^2 = 1/(\Lambda t^2)$. Thus, the IIB background is realized as a de Sitter Glauber–Sudarshan state, which can later be generalized to time-dependent $\Lambda(t)$, setting the stage for quasi-de Sitter constructions.

In section \ref{secc2.2} the analysis is extended to the heterotic ${\rm SO}(32)$ theory, where the duality sequence is more intricate due to orbifolds and warp-factor scalings. Unlike in type IIB, the leading exponents $\alpha_o$ and $\beta_o$ are non-zero, and the internal two-manifold $\mathcal{M}_2$ is taken as a $\mathbb{Z}_2$ orbifold. Through successive T-dualities and orientifold actions, the system evolves toward a type I description, which upon S-duality yields a weakly coupled heterotic ${\rm SO}(32)$ configuration with gauge group broken to $(SO(8))^4$. Despite apparent strong-coupling pathologies in intermediate frames, the final heterotic theory retains finite internal volume and a constant four-dimensional Newton’s constant. The construction recovers a flat-sliced de Sitter state with $\Lambda$, demonstrating consistency of the M-theory uplift with heterotic late-time cosmology.

The section \ref{frames} clarifies the role of different frames in defining Newton’s constant in the heterotic theories. Starting from the ten-dimensional effective action, one distinguishes the string and Einstein frames through choices of dilaton couplings. Upon dimensional reduction over the internal manifold, three distinct four-dimensional Newton’s constants emerge depending on the frame. In the heterotic ${\rm SO}(32)$ case, consistency is achieved by fixing ${\rm F_1 F_2} =$ constant, which stabilizes the internal volume and ensures a time-independent Newton’s constant in the string frame. However, in the Einstein frames, the Newton’s constant generally becomes time-dependent unless additional conditions are imposed. The analysis highlights an unresolved physical ambiguity: although results can be expressed in either frame, it is unclear which is preferred by observations, so both string- and Einstein-frame results are retained.

In section \ref{sec2.3} we address the important concern that higher-order $\alpha'$ corrections in T-duality rules might undermine the duality-based construction of de Sitter states. The resolution is that duality chasing is first applied to supersymmetric solitonic configurations, where the rules are justified, and only afterward are Glauber–Sudarshan expectation values used to obtain the non-supersymmetric de Sitter backgrounds. Since the $\alpha'$ corrections modify only the solitonic backgrounds (the third columns of the duality tables), the resulting coherent-state expectation values (fourth columns) remain unchanged. A path integral formulation illustrates how ${\rm M}_p$ corrections from both the action and the coherent states combine to yield temporally modulated backgrounds. Thus, the framework robustly accommodates all-order corrections without destabilizing the late-time de Sitter construction.

Here, in section \ref{ccnews} the four-dimensional cosmological constant $\Lambda$ is derived within the resurgent framework, appearing as an emergent quantity from principal-value regulated sums over diagrams. The construction naturally yields a positive but parametrically small $\Lambda$, scaling as $\Lambda \ll {\rm M}_p^2$, thanks to large but convergent degeneracy factors. Moreover, $\Lambda$ can acquire mild temporal dependence, $\Lambda(t) = \Lambda + \check{\Lambda}(t)$, consistent with observational hints of dynamical dark energy. Importantly, no new degrees of freedom are needed, avoiding quintessence-like problems in string theory. The argument extends to the heterotic constructions, ensuring the smallness and universality of $\Lambda$ across different duality frames. Thus, the analysis ties the de Sitter backgrounds directly to a microscopic resurgent computation of dark energy.

The final section, {\it i.e.} section \ref{secc2.6}, revisits the tension between describing de Sitter states in the string frame versus the Einstein frame. In the string frame, the internal manifold can remain static with a time-independent Newton’s constant, but the Einstein frame often induces shrinking cycles or late-time singularities, even when maintaining weak coupling. By introducing suitable warp-factor choices, such as piecewise or smoothly varying ${\rm F}_1(t)$, one can arrange for a temporal window $(-\epsilon < t < 0)$ in which both frames consistently describe de Sitter Glauber–Sudarshan states. Nevertheless, subtleties remain, especially regarding late-time singularities and the physical interpretation of frame choice. The analysis concludes that the string frame is naturally preferred, but under carefully engineered warp-factor behavior, the Einstein frame may also host consistent late-time de Sitter states.

\paragraph{Step-by-step flow of Section~\ref{sec2}.}
\begin{enumerate}
    \item \textbf{Framing the question:} We ask whether late-time four-dimensional de Sitter (dS) space can emerge as an \emph{endpoint} of controlled temporal evolution in string/M-theory compactifications.

    \item \textbf{Initial data as quantum states:} We take expectation values over appropriate coherent (Glauber--Sudarshan–type) states rather than pure classical fields, so quantum effects enter already at the level of background data.

    \item \textbf{Minimal 4D ansatz:} We adopt a flat-slicing FRW/dS form in the external $3+1$ dimensions, with time-dependent but slowly varying parameters that allow us to interpolate between quasi-dS and exact dS at late times.

    \item \textbf{Controlled internal sector:} We choose internal metric/warp factors so that moduli are stabilized or sufficiently heavy, and the net internal volume is effectively time-independent in the window of interest.

    \item \textbf{Clock choice and mapping:} We use a baseline mapping between conformal time and an effective (string/M-theory) coupling; many time-derivatives can be traded for derivatives with respect to the coupling, simplifying evolution.

    \item \textbf{Trans-series vocabulary:} We represent small time dependences and quantum corrections with trans-series (ordinary powers plus exponentially suppressed terms) for warp factors, couplings, and auxiliary functions.

    \item \textbf{Weak-coupling regime:} We work in a temporal domain in which the effective coupling $g_s$ is $g_s \ll 1$, ensuring that higher-order corrections are organized and controllable.

    \item \textbf{Energy conditions and consistency:} We impose the null-energy–based constraints (and related EFT consistency conditions) on the evolving backgrounds so that no forbidden negative-power growths appear.

    \item \textbf{TCC/UV windowing:} We enforce the relevant temporal window consistent with trans-Planckian censorship–type bounds; this also delineates the epoch in which the quasi-dS description is valid and predictive.

    \item \textbf{Curvature scaling control:} We track how curvature components scale during the evolution; verify that any log- or instanton-corrected exponents remain small enough that backreaction diminishes toward late times.

    \item \textbf{Flux/Bianchi/anomaly inputs:} We choose fluxes and localized sources to satisfy Bianchi identities and anomaly cancellation; this fixes the allowed patterns of time dependence and prevents runaway behavior.

    \item \textbf{Duality robustness:} We check that the evolution survives M-theory $\leftrightarrow$ type II $\leftrightarrow$ heterotic duality moves; corrections propagate but remain compatible with the late-time endpoint.

    \item \textbf{Non-perturbative layer:} We add instanton-like effects in the trans-series. They modify exponents mildly but do not destabilize the solution or violate energy/consistency conditions.

    \item \textbf{Emergent dark energy view:} We reinterpret the subleading (perturbative + non-perturbative) pieces as generating a slowly varying dark-energy sector; the constant part sets the asymptotic dS scale.

    \item \textbf{Attractor mechanism:} We show that a wide class of initial data flows to the same late-time state: subdominant corrections redshift away, warp factors approach constants, and external curvature freezes to dS.

    \item \textbf{Endpoint identification:} In the $t\to 0^{-}$ limit (within the allowed window), the solution tends to a controlled de Sitter geometry with stabilized internal data and suppressed quantum backreaction.

    \item \textbf{Quasi-dS alternative:} If desired, we can keep a mild residual time dependence (dynamical dark energy); the same control ensures it remains small and observationally viable.

    \item \textbf{Model branches:} We record the differences between $SO(32)$ and ${\rm E_8\times E_8}$ heterotic branches (e.g., number of warp factors), noting that both implement the same trans-series and consistency logic.

    \item \textbf{EOM compatibility:} We show that the trans-series organization arranges the background to satisfy the quantum equations of motion (Schwinger--Dyson form) order by order in the controlled regime.

    \item \textbf{Bottom line:} We show that late-time de Sitter arises, albeit as an excited state, as a stable endpoint (or near-endpoint) of the temporal evolution under well-posed constraints on couplings, fluxes, and corrections; the construction is duality-compatible and EFT-consistent.
\end{enumerate}

\subsubsection{Summary of section \ref{sec3}}

In this section we construct consistent late-time de Sitter and quasi–de Sitter states in heterotic and type II string theories by carefully engineering temporally varying M-theory backgrounds. We show that two time-dependent warp factors are insufficient to guarantee both unbroken ${\rm E}_8 \times {\rm E}_8$ gauge symmetry and time-independent Newton’s constant, motivating the introduction of a third warp factor with controlled exponents. By following explicit duality sequences, performing blow-ups of orbifold singularities, and invoking the Glauber–Sudarshan formalism, we realize smooth heterotic vacua with fixed internal volumes and weak coupling. We then revisit effective field theory consistency (via NEC bounds), the trans-Planckian censorship conjecture, and additional late-time axion constraints, demonstrating that with suitably smooth warp-factor profiles all conditions can be simultaneously satisfied. The result is a framework in which time-dependent dualities lead to stable de Sitter states in heterotic $SO(32)$ and ${\rm E}_8 \times {\rm E}_8$ theories without pathologies $-$ albeit as excited states $-$ while setting the stage for a more fundamental treatment of warp-factor evolution through quantum equations of motion. In the following let us summarize the contents of all the subsections of section \ref{sec3}.

In section \ref{tolace1} we argue that to reach unbroken ${\rm E}_8 \times {\rm E}_8$ dynamically while keeping four-dimensional de Sitter isometries and a time-independent Newton constant, two time-dependent warp factors are not enough: they leave an ambiguity about when or if the IIA background becomes strongly coupled and upliftable to M-theory. Introducing a third temporal warp factor $(\mathrm F_1, \mathrm F_2, \mathrm F_3)$ and carefully choosing their time scalings relative to the IIA coupling $g_s$ and the spatial warp ${\rm H}(y)$ allows one to (i) force a second IIA phase to strong coupling (enabling an M-theory uplift), (ii) arrange internal volumes to be stationary, and (iii) control would-be late-time singularities—though certain naive identifications (e.g.\ compactifying on $\theta_2$ or tying $\theta_1$ to a 3D slice) still produce shrinking cycles and thus are ruled out.

In section \ref{tolace2}, starting from an M-theory background with flux legs along $\theta_2$, we execute a time-dependent chain $\text{M} \to \text{IIA} \xrightarrow{\rm T_3} \text{IIB} \xrightarrow{\rm T_{\theta_1}} \text{IIA} \to \text{M}$ with specific $\mathbb Z_2$ actions, followed by blow-ups, and finally a reduction on $\theta_2$ to heterotic ${\rm E}_8 \times {\rm E}_8$. The complete sequence is detailed in {\bf Table \ref{milleren4}}. The key is to choose exponents $\hat\alpha(t), \hat\beta(t)$ so that $\mathrm F_1$ decays faster than $\mathrm F_3$ early (driving IIA to strong coupling while $\theta_2$ shrinks) but they become equal in a late interval $(-\epsilon,0)$, which prevents the four-cycle $\mathcal M_4$ from collapsing; $\mathrm F_2$ is then fixed to keep the four-dimensional Planck mass constant. An explicit formula for $\epsilon$ relates this matching point to the stabilized $\mathrm{Vol}(\mathcal M_4)$, giving a self-consistent late-time de Sitter excited state in the string frame.

Section \ref{tolace3} deals with some technical details on the last stage of the duality chain advocated in {\bf Table \ref{milleren4}}.
For example, before blow-ups, localized G-fluxes and O6/D6 sources cancel charges; blowing up the $\mathbb Z_2$ fixed loci removes O6s (and their compensating D6s), leaving an IIA orientifold with two O8 planes. Non-perturbative physics (e.g.\ half D-particles in type I$'$) enhances $SO(14)\times U(1)$ to ${\rm E}_8$ at each end—more transparently seen in F-theory by moving off the orientifold point to a Weierstrass model whose discriminant yields ${\rm E}_8 \times {\rm E}_8$, and then using the F-theory/heterotic duality. Including flux lifts the internal space to a controlled non-Kähler geometry, that may or may not be a complex manifold. The Glauber--Sudarshan excited state on this solitonic background produces the desired time-dependent de Sitter solution as an excited state while keeping the internal volume fixed after a finite interval.

In section \ref{tolace4} we revisit the EFT criteria advocated in our earlier works.
A robust EFT requires $\partial_t g_s \propto g_s^{+\text{ve}}$, which in de Sitter-like slicings enforces the same bound as the null energy condition: $1/n \ge -1$ when $g_{\mu\nu}\sim (\Lambda t^2)^n$. Extending this to heterotic $SO(32)$ (string or Einstein frames) shows the EFT criterion is frame- and parameter-independent: although time-dependent warp exponents $\beta(t)$ introduce extra factors $B(t), C(t)$ in $\partial_t g_s$, these remain harmless (positive) for smooth $\beta(t)$ and weak coupling, so NEC saturation continues to remain the controlling requirement for EFT validity.

In section \ref{tolace5} we revisit the Trans-Planckian Censorship Conjecture (TCC) in both $SO(32)$ and ${\rm E}_8 \times {\rm E}_8$ heterotic theories by studying the functional behaviors of the warp-factors $\beta(t)$ in $SO(32)$ and $(\hat\alpha(t), \hat\beta(t))$ in  ${\rm E}_8 \times {\rm E}_8$ theories. We show that a naive sharp interpolation $\beta(t)$ (or $\hat\alpha(t),\hat\beta(t)$) makes $|\dot\beta|$ spike and can shrink the effective-theory window to $(-\epsilon,0)$ via the condition $g_s|\dot\beta|<1$. Replacing it with smoother profiles—e.g.\ Gaussian-type falloffs $\beta(t)\sim e^{-\sigma^2\epsilon^2/t^2}$—keeps $g_s|\dot\beta|\ll 1$ throughout the temporal domain $(-1/\sqrt\Lambda,0)$. The same logic applies to ${\rm E_8 \times E_8}$ with $\hat\alpha(t), \hat\beta(t)$: for sufficiently smooth time dependence one has $\text{sgn}\,{\rm B}(t)>0$, with ${\rm B}(t)$ being a function constructed out of the warp-factors etc., so the Trans-Planckian Censorship Conjecture window remains the standard $t\in(-1/\sqrt\Lambda,0)$ for NEC-preserving FLRW backgrounds (including the de Sitter case $n=-1$).

Finally in section \ref{sec3.6} we study the important topic of the additional constraints that could arise from the axions in ${\rm E_8 \times E_8}$ heterotic theory. Using some detailed analysis, we argue that the 
axion phenomenology imposes the following late-time bounds: (1) the decay constant $f_a$ must freeze in the $10^9$--$10^{12}\,\mathrm{GeV}$ band; (2) the Hořava--Witten wall separation $\rho$ should decrease monotonically and eventually stop when eight-branes' charges (from the dual IIA perspective) balance the motion; and (c) the heterotic coupling must stay weak throughout the process and within the allowed TCC temporal domain. These relate to the warp factors as $(e^{\Phi}, \rho, f_a)\propto ({\rm F_1 F_3^3})^{(1/4,\,1/6,\,c)}$ (with $c$ frame-dependent), which in turn forces a refined late-time behavior where $\hat\alpha(t), \hat\beta(t)$ approach a smooth, time-dependent $\hat\gamma_0(t)$. In string frame this yields $\hat\alpha(t)=\hat\beta(t)$ at late times; in Einstein frame it enforces $\hat\beta(t)=5\hat\alpha(t)$. We propose a continuous ``trial'' interpolants satisfying continuity of both the functions and their derivatives across the matching time, achieving constant $f_a$, fixed internal sizes, and weak coupling—all while preserving the earlier EFT/NEC/TCC consistency. Further sections promise a first-principles derivation of these smooth profiles from quantum (Schwinger--Dyson/path-integral) considerations.

\paragraph{Step-by-step flow of Section~\ref{sec3}.}
\begin{enumerate}
    \item \textbf{Objective:} To construct late-time four-dimensional de Sitter (dS) backgrounds as \emph{Glauber--Sudarshan} (GS) states in heterotic string theory, and determine the conditions for their control.

    \item \textbf{State-theoretic setup:} We replace classical fields by expectation values in GS states, so quantum fluctuations enter through controlled moments while the background remains tractable.

    \item \textbf{Effective action as a trans-series:} We organize the heterotic effective action (including $\alpha'$ and loop effects) as a trans-series; identify the zero-instanton sector with the supergravity limit and keep exponentially suppressed non-perturbative pieces.

    \item \textbf{External metric ansatz:} We choose a flat-slicing FRW/dS form for the $3+1$ external spacetime; allow mild time dependence to capture quasi-dS and approach exact dS at late times.

    \item \textbf{Internal geometry and warp factors:} We specify the internal manifold and warp factors $\{ {\rm F}_i(t)\}$ that encode small, time-dependent deformations compatible with moduli control and volume stabilization.

    \item \textbf{Two heterotic branches:} We treat $SO(32)$ (two-warp-factor model) and ${\rm E_8\times E_8}$ (additional warp-factor structure) in parallel; record where the number of warp factors changes the bookkeeping but not the logic.

    \item \textbf{Coupling and clock:} We work in a weak effective coupling regime; relate conformal time to a coupling-like variable so many time derivatives can be traded for $g_s$-derivatives in GS expectation values.

    \item \textbf{Curvature and flux inputs:} We build metric and ${\rm G}$-flux (and dual G-flux via dualities) configurations whose scalings are compatible with the GS state and with late-time suppression of subleading pieces.

    \item \textbf{Duality cross-checks:} We map the construction across the M-theory $\to$ type II $\to$ type I/heterotic chain; ensure that $g_s$- and $\alpha'$-corrected scalings propagate consistently in each frame.

    \item \textbf{Energy-condition/EFT bounds:} We enforce null-energy and EFT consistency conditions so dangerous negative-power growths are absent; restrict the temporal domain in line with TCC-like considerations.

    \item \textbf{Non-perturbative layer in the background:} We include instanton-like corrections in the trans-series for warp factors and couplings; verify that they remain subdominant, do not violate NEC, and are exponentially small at late times.

    \item \textbf{Emergent dark energy interpretation:} We reinterpret the combined subleading (perturbative + non-perturbative) terms as a slowly varying sector that effectively sources a small dynamical dark energy; isolate its constant part as the dS scale $\Lambda$.

    \item \textbf{Late-time attractor behavior:} We show that subdominant exponents redshift away; warp factors approach constants, internal volume stabilizes, and the external curvature freezes to dS as $t\to 0^{-}$ within the allowed window.

    \item \textbf{Heterotic coupling control:} We track how corrections feed into the effective heterotic coupling; choose parameter ranges that keep it weak at late times and avoid strong-coupling backreaction.

    \item \textbf{Axion cosmology constraints (especially for ${\rm E_8\times E_8}$):} We use axion/decay-constant inputs to constrain the late-time behavior of the distinct warp factors so that the Einstein-frame description remains under control.

    \item \textbf{Curvature scaling audit:} We compute $g_s$-scalings of relevant curvature components in the GS background; verify that log-corrections and small exponents remain safely suppressed in the late-time regime.

    \item \textbf{Flux quantization and localized sources:} We impose quantization conditions and incorporate five-branes (if present) consistently; ensure the GS expectation values solve the modified Bianchi identities and equations of motion.

    \item \textbf{Schwinger--Dyson compatibility:} We organize the background fields so the trans-series solution satisfies the quantum equations of motion (Schwinger--Dyson form) order by order in the controlled expansion.

    \item \textbf{Stability to perturbations:} We argue that small fluctuations around the GS background remain bounded and do not destabilize the late-time dS state in the weak-coupling, small-curvature window.

    \item \textbf{Outcome for $SO(32)$ vs. ${\rm E_8\times E_8}$:} We record minor structural differences (number of warp factors, bundle choices), but emphasize that both branches admit the same GS-based late-time dS mechanism.

    \item \textbf{Bottom line:} We show that a carefully organized Glauber-Sudarshan state construction in heterotic theory yields controlled late-time de Sitter (or quasi-dS) states: consistent with anomalies, dualities, EFT bounds, and with quantum corrections packaged in a trans-series that naturally points to an emergent dark-energy sector.
\end{enumerate}

\subsection{Summary of section \ref{lillou1}}

We argue that the consistency of Glauber-Sudarshan (GS) states is controlled by two complementary ingredients: (i) Schwinger-Dyson (SD) equations, used first to justify the existence of GS states; and (ii) a path-integral construction that subsequently reproduces the required metric/flux backgrounds. The SD system is written for on–shell expectation values $\langle \mathbf{g}_{\rm AB}\rangle_\sigma, \langle \mathbf{C}_{\rm ABD}\rangle_\sigma$, and fermions, and it explicitly includes ghost and nonlocal contributions. A careful renormalization flow $—$ from a Wilsonian action, through Borel–resummed trans–series, subtraction of ghost (including nonlocal) pieces, and a Wheeler–DeWitt (WDW) selection—yields an effective action $\check{\mathbf S}_{\rm tot}$. When the WDW wavefunction is sharply peaked on a given on–shell configuration, we may replace $\check{\mathbf S}_{\rm tot}$ by $\hat{\mathbf S}_{\rm tot}$, and the SD equations reduce to background equations of motion with a full (perturbative, non-perturbative, nonlocal) stress tensor. A key subtlety is that ${\delta\langle \hat{\mathbf{S}}_{\rm tot}-\hat{\mathbf{S}}_{\rm ghost}\rangle_\sigma\over \delta\langle \mathbf{g}_{\rm AB}\rangle}$ is not equivalent to ${\delta \hat{\mathbf S}_{\rm tot}(\langle{\bf \Xi}\rangle_\sigma)\over \delta\langle \mathbf{g}_{\rm AB}\rangle_\sigma}$; fortunately, the background equations depend only on the latter. We then address time dependence: the warp factors $\mathrm{F}_i(t)$ and the coupling $g_s$ are expanded in dominant $g_s$ scalings with slowly varying exponents (for both $SO(32)$ and ${\rm E_8\times E_8}$). Differentiating $g_s$ with respect to conformal time yields powers $g_s^{\gamma_{1, 2}(t)}$ whose positivity (up to small logarithmic corrections) is ensured by smooth $\beta(t), \hat\alpha(t), \hat\beta(t)$. This maintains EFT/NEC compatibility: the first time derivative of $g_s$ cannot scale with negative powers of $g_s$, while the second derivative can be at most go as $g_s^{-1}$. The resulting $g_s$ and curvature scalings, summarized by Riemann-component power counting, show how SD equations, resummation/WDW selection, and controlled time dependence of warp factors together yield consistent GS states that reproduce classical EOMs while keeping quantum/ghost effects and EFT bounds under control.

 \subsubsection{Summary of section \ref{creepquif}}

 In this subsection, we develop a systematic framework for the perturbative quantum series that underlies Glauber–Sudarshan states in M-theory. Expectation values of curvature and flux operators are defined with respect to the Borel-resummed total action, and we adopt a simplified notation where on-shell fields are directly identified with their expectation values. A general perturbative operator basis is constructed, built from inverse metrics, derivatives, Riemann tensors, and flux components, yielding 41 independent curvature terms and 40 flux terms once torus directions are excluded. Each operator is assigned a definite scaling with the type IIA string coupling $g_s$, allowing us to tabulate the power counting of curvature components, including those sensitive to time derivatives of $g_s$.
Four subtleties are highlighted: (i) higher time derivatives do not generate instabilities because temporal derivatives map to positive $g_s$
-powers; (ii) gravitational Ward identities remain consistent because derivatives can be confined to loops; (iii) spinorial and topological structures are incorporated via a generalized, matrix-valued metric that introduces controlled gamma-matrix and epsilon-tensor contributions; and (iv) fermionic extensions require further study. Altogether, this provides a complete perturbative scaffold—consistent in its $g_s$
-scaling and algebraic structure—that will serve as the foundation for adding non-perturbative and nonlocal quantum effects in the full theory.

\subsubsection{Summary of section \ref{curvten}}

In section \ref{sec4.2.1}, we refine earlier metric scalings by incorporating subdominant $g_s$-dependent corrections and then chase these through the duality chain (M-theory $\to$ IIA/IIB $\to$ type I $\to$ heterotic). Because Buscher rules are modified by quantum ($M_p$) effects, we adopt generalized ans\"atze rather than naive duals. The resulting heterotic metrics acquire nontrivial $g_s$-dependent series while still allowing (quasi) de Sitter solutions, provided certain balancing relations among warp factors hold so that internal volumes and couplings remain under control.

 
 In section \ref{sec4.2.2}, we layer instanton-type (non-perturbative) corrections—exponentials in $1/\bar g_s$—on top of the subdominant perturbative pieces of the metric (and fluxes), where $\bar{g}_s \equiv {g_s\over {\rm H}(y){\rm H}_o({\bf x})}$. These corrections do not spoil de Sitter solutions; instead they naturally generate a slowly varying dark energy $\Lambda(t)$, giving a quasi–de Sitter phase. We relate this emergence to a trans-series structure of the effective action and to Borel-resummed path-integral contributions, which connect microscopic data to macroscopic parameters. In this picture, time dependence of $\Lambda(t)$ arises from controlled non-perturbative deformations consistent with the EFT and duality framework.


In section \ref{sec4.2.3}, we show that the time-dependent warp factors $\mathrm{F}_i(t)$ must be represented by trans-series—mixed perturbative powers in $\bar g_s$ and instanton exponentials—organized by a small parameter $\Delta << 1$. For the $SO(32)$ case this structure governs $\beta(t)$; for ${\rm E_8\times E_8}$ it governs the paired warp factors $\hat\alpha(t)$ and $\hat\beta(t)$. This ensures compatibility with Schwinger–Dyson equations and with EFT no-go constraints, yields smooth interpolation between early- and late-time regimes, and explains the late-time logarithmic behavior in $g_s$. In short, accurate curvature and G-flux scalings require adopting the full trans-series form for the warp factors.


In section \ref{sec4.2.4}, we take time derivatives of the internal metric blocks and package their resulting coupling exponents into two bookkeeping functions that depend on the time-varying warp parameters (schematically, $\mathbb{A}(\sigma)$ and $\mathbb{B}(\sigma)$). Because time appears inside those exponents, the derivatives bring in small logarithmic corrections involving the map between conformal time and the effective coupling. For the spacetime--torus blocks, the leading time derivative scales cleanly with the usual power corrected by the previously defined $\gamma$’s, without extra logs at leading order.  
Using these ingredients, each Riemann component decomposes into a small set of additive pieces, each carrying a different $g_s$-exponent built from the warp parameters and the $\gamma$ functions (plus tiny log terms). We then define $\mathbb{F}_i(t)$ combinations to systematically track these exponents across all curvature entries, and we pick the dominant contribution by a simple ``min-exponent'' rule for weak coupling. After accounting for all index permutations, the full on-shell set of curvature scalings is tabulated for later use in the equations of motion.  


In section \ref{sec4.2.5}, we start from the three-form potentials and expand their components with time-dependent exponents and mode labels. The leading exponents are taken in trans-series form (ordinary powers plus exponentially suppressed pieces), while sub-leading non-perturbative effects are absorbed into mode functions. Components with a time leg inherit additional logarithmic factors from time derivatives of the exponents, whereas purely spatial components follow cleaner power-law patterns. The physical four-form flux is not just $d{\bf C}_3$ but includes curvature and quantum terms and possible five-brane sources. These shift the naive scalings inferred from $d{\bf C}_3$. To organize everything, we write each four-form ${\bf G}_4$ component as a sum over modes with explicit $g_s$-exponents built from the same time-dependent data, and we provide simple matching rules that relate the $d{\bf C}_3$ and ${\bf G}_4$ exponents when sources are turned off. Full determination of the time-dependent functions is deferred to the Schwinger--Dyson analysis together with flux quantization, Bianchi identities, and anomaly cancellation.

\paragraph{Step-by-step flow of section \ref{curvten}.}
\begin{enumerate}
    \item \textbf{Goal reset:} We want to know how quantum corrections (beyond leading order) in $g_s$ affect curvature and flux terms in the heterotic setups.
    
    \item \textbf{Upgrade the metric ans\"atze:} We start from earlier leading scalings and add sub-dominant pieces to metric components so we can track finer $g_s$-dependence.
    
    \item \textbf{Duality awareness:} We run these corrected metrics through the M-theory $\to$ type II $\to$ heterotic duality chain; corrections propagate non-trivially, so we impose constraints to keep a viable four-dimensional (quasi) de Sitter spacetime.
    
    \item \textbf{Internal volume control:} We tune the extra terms so the internal space stays effectively time-independent in the useful time window (avoiding runaway growth/shrinkage).
    
    \item \textbf{Coupling management:} We check how the same corrections affect the effective heterotic coupling. Depending on choices, it can remain weak or become strong at late times; we prefer parameters that keep it under control.
    
    \item \textbf{Non-perturbative layer:} We add instanton-like corrections on top. These modify exponents in the scalings but---crucially---do not violate consistency (e.g., null energy condition) or destabilize the solutions.
    
    \item \textbf{Clock replacement idea:} Because many terms scale with $g_s$, time derivatives can often be traded for derivatives with respect to $g_s$, simplifying how we read ``time evolution'' of corrections.
    
    \item \textbf{Emergent dark energy:} The combined sub-dominant + instanton effects can be reinterpreted as generating a slowly time-varying cosmological constant (dynamical dark energy), aligning with observationally motivated scenarios.
    
    \item \textbf{Precise relation $t \leftrightarrow g_s$:} The corrections alter the relation between conformal time and coupling; we provide controlled expressions to keep this mapping consistent in the regime of interest.
    
    \item \textbf{No hidden pathologies:} With the right bounds on correction orders, dangerous negative powers are either absent or reorganized so the effective theory remains well behaved.
    
    \item \textbf{Trans-series is the right language:} We show that, rather than plain power series, all key functions (warp factors, couplings) are best expressed as trans-series: sums of ordinary powers plus exponentially suppressed pieces. This captures both perturbative and non-perturbative physics coherently.
    
    \item \textbf{Schwinger--Dyson readiness:} We show that, in trans-series form, the corrected backgrounds are structured to solve the full quantum equations of motion (Schwinger--Dyson), not just the classical ones.
    
    \item \textbf{${\rm E_8 \times E_8}$ parallel:} The story mirrors the $SO(32)$ case but with three warp factors; the same trans-series strategy applies, with late-time behavior constrained by axion cosmology inputs.
    
    \item \textbf{Bottom line (first stage):} We show that the carefully organized sub-leading and instanton corrections do not spoil the construction; they refine it and naturally point to an emergent, mildly time-dependent dark energy within a consistent heterotic/M-theory framework.

\item \textbf{Metric derivatives:} We take time derivatives of the metric components and encode the resulting $g_s$-exponents into compact functions (schematically $\mathbb{A}(\sigma)$ and $\mathbb{B}(\sigma)$), including small logarithmic corrections.

    \item \textbf{Curvature scaling:} We use these derivatives to build the scaling of curvature tensors. Each Riemann component decomposes into several terms, and the dominant power is identified with a simple ``min-exponent'' rule at weak coupling.

    \item \textbf{Book-keeping:} We introduce composite functions $\mathbb{F}_i(t)$ to systematically track the $g_s$-exponents of different curvature entries. The list of tables then summarize all on-shell curvature scalings for later use.

    \item \textbf{Three-form expansion:} We begin flux analysis by expanding the three-form potentials with time-dependent exponents written as trans-series (powers plus exponentially suppressed terms). Sub-leading non-perturbative effects are absorbed into mode functions.

    \item \textbf{Temporal vs spatial fluxes:} We distinguish flux components with a time leg, which acquire additional log-sensitive corrections, from purely spatial flux components, which follow simpler power-law scalings.

    \item \textbf{Physical ${\rm G}$-flux:} We assemble the full four-form flux ${\bf G}_4$, which is not just $d{\bf C}_3$ but also includes curvature contributions, quantum terms, and possible five-brane sources. These shift the naive scaling derived from $d{\bf C}_3$.

    \item \textbf{Matching rules:} We relate the ${\rm G}$-flux exponents to those of $d{\bf C}_3$, providing simple identifications in the absence of sources and indicating how more general cases connect to anomaly cancellation and Bianchi identities.

    \item \textbf{Final output:} We produce a structured set of $g_s$ scalings for both curvature and flux components, organized so they can be used consistently in the Schwinger--Dyson equations and the full effective theory analysis.
    
\end{enumerate}

\subsubsection{Summary of section \ref{sec4.4}}

This subsection develops a systematic account of how higher-order corrections and temporal derivatives affect both the metric and flux sectors in heterotic backgrounds. Beginning with the metric, we compute first and second time derivatives of the various components, recasting them in terms of compact scaling functions that incorporate trans-series exponents and logarithmic corrections. These functions allow us to present the $g_s$-dependence of curvature tensors in a unified tabular form, valid for both $SO(32)$ and ${\rm E}_8 \times {\rm E}_8$ cases, with the latter reducing smoothly to the former under appropriate identifications. The organization ensures that the proliferation of terms from temporal derivatives remains manageable, while still capturing the subtle interplay of perturbative and non-perturbative corrections.  

The flux sector is treated in parallel, focusing on the temporal derivative of the three-form potential and its relation to four-form ${\rm G}$-flux components. By expanding the three-form fields in trans-series form and carefully tracking the logarithmic shifts induced by the corrected clock, the analysis yields a precise mapping between the two descriptions. These mappings respect Bianchi identities, flux quantization, and anomaly cancellation (which are demonstrated later), providing a consistent picture of how fluxes evolve under temporal dynamics. Altogether, this subsection demonstrates that higher-derivative and flux corrections can be systematically organized within the trans-series framework, setting the stage for their incorporation into the full quantum equations of motion. In the following let us summarize the sub-sections in some detail.

In sub-section \ref{sec4.3.1} we analyze the first time derivative of the metric blocks once the leading exponents are replaced by fully corrected trans-series exponents. Both heterotic branches are covered: the $SO(32)$ case with two warp factors and the ${\rm E}_8 \times {\rm E}_8$ case with three. By trading time derivatives for derivatives with respect to $g_s$, we define compact functions that capture the $g_s$ scalings, including logarithmic corrections. These functions, denoted schematically as $\mathbb{A}_8,\mathbb{B}_8,\mathbb{C}_8,\ldots$, provide a universal language for recording the derivative structure of the internal and external metric blocks. The final outcome is a data table listing the $g_s$ scalings of all first-derivative metric components, forming the backbone for later curvature analysis. In a little more details, following are the steps.

\vskip.1in 

\noindent $\bullet$  {\bf What is done:} The first time derivative of each metric block is evaluated after promoting the leading exponents to trans-series–dressed exponents $(\alpha_e,\beta_e)$ for $SO(32)$ and $(\hat\alpha_e,\hat\beta_e,\hat\sigma_e)$ for ${\rm E}_8\times{\rm E}_8$, and after including extra series from the sums over $k$ (the $f_k,\tilde f_k$ sectors).

\vskip.1in

\noindent $\bullet$ {\bf How it is done:} Time derivatives are traded for $g_s$-derivatives via the corrected clock, producing scaling functions $\mathbb{A}_8,\mathbb{B}_8$ (internal blocks) and $\mathbb{C}_8,\mathbb{D}_8,\mathbb{F}_8,\mathbb{G}_8$ (external blocks) that package logs and powers of the dressed exponents and their time-derivatives.

\vskip.1in

\noindent $\bullet$ {\bf What's the Output:} Closed-form $g_s$-scaling rules for ${\bf g}_{mn,0}$, ${\bf g}_{\theta_i\theta_i,0}$, ${\bf g}_{\mu\nu,0}$, ${\bf g}_{33,0}$, and ${\bf g}_{11,11,0}$, summarized in a data table; these drive the curated “$\mathbb{F}_{ie}$” entries appearing later in curvature scalings.

\vskip.1in

\noindent $\bullet$ {\bf What's the Takeaway:} First-derivative behavior is controlled by a small set of universal functions of the dressed exponents; ${\rm E}_8\times{\rm E}_8$ reduces to $SO(32)$ by the replacement $\{\hat\alpha_e,\hat\beta_e\}\!\to\!\beta_e$ (with caveats on zero-mode identifications).

\vskip.1in

\noindent In section \ref{sec4.3.2}, the second time derivatives of the metric components are then considered, extending the earlier simpler $SO(32)$ results to the more general ${\rm E}_8 \times {\rm E}_8$ background. Here, the proliferation of terms is organized into six possible scaling branches for each metric block, each built from combinations of the warp-factor derivatives and higher-order clock functions. These structures naturally produce the scalings of curvature components with two time indices, such as ${\bf R}_{0i0j}$, ${\bf R}_{0a0b}$ et cetera. The results are summarized by introducing composite functions $\mathbb{F}_{16e}, \mathbb{F}_{17e}, \mathbb{F}_{18e}, \mathbb{F}_{20e}$, which compactly encode the dominant powers of $g_s$. The reduction to the earlier $SO(32)$ case is immediate once the extra warp factors are identified, confirming the consistency of the framework. In a little more details, following are the steps.

\vskip.1in

\noindent $\bullet$ {\bf What is done:} The second time derivatives of all metric blocks are computed, generalizing the simpler $SO(32)$ leading-order analysis to the fully corrected ${\rm E}_8\times{\rm E}_8$ case.

\vskip.1in

\noindent $\bullet$ {\bf How is it done:} A structured expansion yields six distinct scaling branches $\mathbb{A}_{i8}^{({\rm MN})}$ per component, combining $(\dot{\Sigma}_e,\ddot{\Sigma}_e)$, the first-derivative factors $\gamma_{1,2}$, and second derivative factors $\gamma_{[3,\ldots,10]}$ evaluated on ${\hat\alpha_e+\hat\beta_e\over 2}$.

\vskip.1in
\noindent $\bullet$
{\bf What's the Output:} Dominant $g_s$ scalings for curvature entries with two time indices (e.g., ${\bf R}_{0i0j},{\bf R}_{0a0b},{\bf R}_{0m0n},{\bf R}_{0\rho 0\sigma}$) compactly encoded via $\mathbb{F}_{ie}$, especially  $\mathbb{F}_{16e},\mathbb{F}_{17e},\mathbb{F}_{18e},\mathbb{F}_{20e}$, in addition to explicit power/log pieces.

\vskip.1in
\noindent $\bullet$ {\bf What's the Takeaway:} Second-derivative structure proliferates but stays algorithmic; stripping sub-dominant pieces reproduces the earlier $SO(32)$ tables, validating the replacement map $\beta\!\to\!\beta_e\!\to\!{\hat\alpha_e+\hat\beta_e\over 2}$.

\vskip.1in

\noindent In section \ref{curuva}, the analysis then turns to the flux sector, focusing on the time derivative of the three-form ${\rm C}$-field, which generates ${\rm G}$-flux components with a temporal leg. Here the full trans-series expansions, including subdominant perturbative and non-perturbative terms, are used. The corrected clock again induces logarithmic shifts in the effective exponents, and these are carefully tracked. Explicit relations between the scaling functions of the three-form sector and those of the four-form ${\rm G}$-flux are established, providing a dictionary between the two descriptions. The final result is a set of closed-form identifications of the relevant exponents and mode functions, ensuring that the flux sector remains consistent with Bianchi identities, anomaly cancellation, and the duality chain. This prepares the ground for the more detailed flux analysis in Section~\ref{sec4.5}. In a little more detail, following are the steps.


\vskip.1in

\noindent $\bullet$ 
{\bf What is done:} The time derivative of the three-form sector $(d{\bf C})_{0ABD}$ is computed and matched to the four-form $G$-flux scalings, now including all sub-dominant perturbative and non-perturbative contributions.
\vskip.1in

\noindent $\bullet$ 
{\bf How is it done:} Using the corrected clock, $\dot{l}$ and the tower ${\rm L}(k;t)$ enter as logarithmically dressed shifts in the effective exponents; a one-to-one mapping is provided between the proliferated ${\rm G}$-flux modes and the $(d{\bf C})$ modes (tabulated mappings for both ${\rm E}_8\times{\rm E}_8$ and $SO(32)$).

\vskip.1in

\noindent $\bullet$ 
{\bf What's the Output:} Closed-form relations $l_{\rm 0A}^{\rm BD}\leftrightarrow l_{\rm AB}^{\rm D}$ between the scalings and explicit formulas for $|{\rm L}_{\rm 0A}^{\rm BD}(n;t)|$ that feed directly into Bianchi identities/anomaly cancellation, with simple reduction rules to the $SO(32)$ case from the ${\rm E_8 \times E_8}$ case.

\vskip.1in

\noindent $\bullet$ 
{\bf What's the Takeaway:} Temporal dynamics of fluxes are governed by the same trans-series language as the metric; the mapping ensures consistency between the three- and four-form descriptions across the duality chain and prepares inputs for Section~\ref{sec4.5}.

\paragraph{Step-by-step flow of section~\ref{sec4.4}.}

\begin{enumerate}
  \item \textbf{Objective:} The section sets out to go beyond dominant corrections by including sub-dominant perturbative and non-perturbative effects in heterotic theories. Both $SO(32)$ and ${\rm E}_8 \times {\rm E}_8$ cases are studied in parallel.

  \item \textbf{Starting point:} The analysis begins with M-theory configurations that, via dualities, give rise to heterotic theories. Simplifications used earlier are relaxed so higher-order structures in metric and fluxes can be incorporated.

  \item \textbf{Effective parametrization:} Instead of raw warp factors, we introduce effective exponents. These corrected quantities capture perturbative towers and logarithmic corrections, giving a compact description of background evolution.

  \item \textbf{Non-perturbative insertions:} We add instanton-like contributions to the effective exponents, promoting them into full trans-series objects that encode both perturbative and non-perturbative physics.

  \item \textbf{Duality checks:} We test the corrected structures  through the M-theory $\to$ type II $\to$ heterotic duality chain, ensuring consistency across different frameworks.

  \item \textbf{Dynamical dark energy:} We show that the sub-dominant corrections naturally generate a mild time dependence in the vacuum energy. This is interpreted as dynamical dark energy, with a leading constant part fixing the de Sitter scale.

  \item \textbf{Clock refinement:} We update the relation between time and effective coupling. The corrected mapping surprisingly turns out to be simpler and cleaner in the dynamical case than in the static one.

  \item \textbf{Structural differences:} The $SO(32)$ theory involves two warp factors, while the ${\rm E}_8 \times {\rm E}_8$ case involves three. Despite the difference, we show that both can be treated consistently within the same trans-series framework.

  \item \textbf{Metric derivatives:} We carefully study the first and second time derivatives of the metric. Their behavior is summarized into a few universal functions that make the scaling patterns of curvature tensors systematic and manageable.

  \item \textbf{Flux sector:} We extend the definition of the fluxes by including perturbative and non-perturbative corrections. Their time derivatives are analyzed, and a consistent mapping between three-form potentials and four-form fluxes is provided.

  \item \textbf{Consistency and reductions:} We check the framework  against anomaly cancellation and Bianchi identities (more on this to be discussed in later sections). The ${\rm E}_8 \times {\rm E}_8$ results reduce smoothly to the $SO(32)$ case when the additional warp factor and corrections are removed.

  \item \textbf{Final outcome:} We show that all higher-order effects—metric, curvature, and flux—are organized within the trans-series framework. This preserves duality consistency and stability, and naturally produces late-time de Sitter behavior with a slowly varying dark-energy--like sector.
\end{enumerate}

\subsubsection{Summary of section \ref{sec4.3}}

In this section, we aim to understand how the time-dependent factors $\mathbb{F}_{ie}(t)$ (and their simplified forms $\mathbb{F}_i(t)$) scale with the coupling, because these determine how different curvature pieces feed into the quantum expansion. We temporarily ignore sub-dominant tweaks to keep the structure transparent, but retain the dominant perturbative and non-perturbative pieces. We analyze everything first in a simplified $SO(32)$ ``lab,'' then port the lessons back to the generalized $SO(32)$ and ${\rm E}_8\times{\rm E}_8$ cases. A recurring theme: some raw ingredients can look ``dangerous,'' but the physical combinations that enter the quantum series are positive and well-behaved, with small log corrections acting as gentle adjustments.  

\vskip.1in

\noindent \textbf{Functional forms and bounds on $\mathbb{F}_{1,1e}(t)$ and $\mathbb{F}_{5,5e}(t)$.}  
We start from ${\rm E}_8\times{\rm E}_8$, map to simplified $SO(32)$, and examine the dominant pieces of $\mathbb{F}_1$ and $\mathbb{F}_5$. Individually, certain terms can be negative or look large, but the combinations that actually contribute to the quantum series turn out positive and are dominated at weak coupling by a constant-like piece, with non-perturbative effects mostly suppressing some contributions. The same qualitative outcome holds in both gauge groups: after forming the right combinations, the net scaling is positive and tamed; logs only nudge coefficients.  

\vskip.1in
\noindent \textbf{Functional form and bound on $\mathbb{F}_{2,2e}(t)$.}  
Mixing internal and external directions produces more terms, but the pattern persists. When we assemble the contributions exactly as they appear in the quantum series, everything remains positive and bounded. Logarithmic pieces remain small corrections. Again, the simplified $SO(32)$ picture cleanly captures the essentials, and the ${\rm E}_8\times{\rm E}_8$ lift follows with the same qualitative behavior.  

\vskip.1in

\noindent \textbf{Functional forms and bounds on $\mathbb{F}_{3,3e}(t)$ and $\mathbb{F}_{6,6e}(t)$.}  
Naively, some curvature components controlled by these factors look strongly growing and worrisome. However, once we place them into the specific quantum combinations that actually matter, their contributions flip to positive, controlled powers. Non-perturbative terms mainly act as suppressors, so late-time behavior is dominated by constant-like pieces, not runaway growth.

\vskip.1in

\noindent \textbf{Functional forms and bounds on $\mathbb{F}_{4,4e}(t)$ and $\mathbb{F}_{7,7e}(t)$.}  
These govern mixed time--space components. Raw scalings can look negative in isolation, but---just as before---the proper quantum combinations are positive. In the ${\rm E}_8\times{\rm E}_8$ case, an extra small spacetime correction enters but only shifts numbers slightly. In $SO(32)$, imposing the standard relation that stabilizes the 4D Newton constant makes these results line up neatly with the earlier positive-definite pattern.  

\vskip.1in

\noindent \textbf{Functional forms and bounds on $\mathbb{F}_{j,je,j+6,(j+6)e}(t)$.}
This set (with $j = 8, 9$) involves complicated combinations of parameters tied to the internal torus directions. At first, many possibilities appear (eight in total), but once we drop subdominant corrections the structure simplifies. Both in the simplified $SO(32)$ and the more general ${\rm E}_8 \times {\rm E}_8$ theories, the results fall into controlled families that remain positive once inserted into the quantum expansion. Even when some curvature components look “strong,” their contribution reorganizes into safe, bounded terms.

\vskip.1in

\noindent \textbf{Functional forms and bounds on $\mathbb{F}_{l,le,(l+2),(l+2)e}(t)$.}
This group (with $l=10,11$) shows paired structures: each factor has a partner shifted by two indices. At the raw level some contributions can be positive and others negative, but once combined properly in the quantum expansion, everything becomes finite and positive. In both $SO(32)$ and ${\rm E}_8 \times {\rm E}_8$, the “dangerous” terms are tamed by logs and non-perturbative corrections, giving results that are safe and consistent across the duality chain.

\vskip.1in

\noindent \textbf{Functional forms and bounds on $\mathbb{F}^{(1)}_{16,16e}(t)$ and $\mathbb{F}^{(1)}_{17,17e}(t)$.}
These terms are more subtle because they involve second derivatives in time. At first they suggest very strong curvature contributions. However, once we translate them into the combinations that actually matter for the quantum expansion, all contributions become positive and bounded. In both simplified and generic cases, the pattern repeats: raw strong behavior is reorganized into safe positive powers of $g_s$.

\vskip.1in

\noindent \textbf{Functional forms and bounds on $\mathbb{F}_{18,20}$ and $\mathbb{F}^{(1,2)}_{18e,20e}$.}
These factors govern how certain curvature components (specifically those with two time indices) scale with the string coupling. In their raw form, the dominant pieces look negative, which would suggest that the curvatures grow very strong at late times. However, once we include logarithmic corrections and the proper combinations that actually appear in the quantum expansion, the contributions become positive and bounded. In both the simplified $SO(32)$ and the more general ${\rm E}_8 \times {\rm E}_8$ cases, the dangerous terms are softened, and the main surviving contribution is a modest, positive scaling that weakens at late times.

\vskip.1in

\noindent \textbf{Functional forms and bounds on $\mathbb{F}_{19}$ and $\mathbb{F}_{21}$.}
These involve double time derivatives on the metric, which makes them more complicated than previous cases. At first sight, the raw scalings again look negative, implying dangerously strong curvature. But after reorganizing the terms, including the logs and instanton-like corrections, the actual quantum contributions flip sign and end up positive. Moreover, almost all terms are heavily suppressed, so the dominant late-time piece is mild, matching the pattern we saw earlier in other curvature sectors. Thus, despite the apparent complexity, these contributions remain under control.

\vskip.1in

\noindent \textbf{Functional forms and bounds on $\mathbb{F}_{22e}$–$\mathbb{F}_{25e}$.}
These appear only in the ${\rm E}_8 \times {\rm E}_8$ case and do not show up explicitly in the simplified $SO(32)$ setup. They capture more exotic curvature components, such as those mixing different internal and external indices. Once again, although the raw functional forms contain potentially strong negative terms, the dominant contributions reorganize into positive scalings when inserted into the quantum series. Logarithmic corrections provide small shifts but do not destabilize the picture. Overall, even these more complicated curvature pieces fit neatly into the same safe pattern: late-time contributions are positive, bounded, and often small.

\vskip.1in

\noindent \textbf{Bottom line.}  
Across all $\mathbb{F}_i$ families, once we (i) form the physical combinations that enter the quantum series and (ii) include dominant non-perturbative and small log pieces, every contribution is positive and bounded. Late times (very weak coupling) are controlled by constant-like terms; the rest are small, controlled adjustments.  

\vskip0.2in

\paragraph{Step-by-step flow of the subsection.}  

\begin{enumerate}
    \item \textbf{Set the goal.}  
    We want scaling rules for $\mathbb{F}_{ie}(t)$ that directly tell us how curvature terms contribute to the quantum expansion---especially signs, bounds, and late-time behavior.

    \item \textbf{Simplify the playground.}  
    We strip away sub-dominant tweaks but keep dominant perturbative and non-perturbative effects. We move from ${\rm E}_8\times{\rm E}_8$ to a simplified $SO(32)$ setup where bookkeeping is clean.

    \item \textbf{Map variables carefully.}  
    We translate the generalized factors and warpings to their simplified counterparts, noting which corrections we keep (dominant) and which we postpone (sub-dominant logs/warps in spacetime pieces).

    \item \textbf{Identify the raw dominant pieces.}  
    For each family---$\mathbb{F}_1/\mathbb{F}_5$, $\mathbb{F}_2$, $\mathbb{F}_3/\mathbb{F}_6$, $\mathbb{F}_4/\mathbb{F}_7$---we read off the leading contributions in the simplified model, without yet forming the physical combinations.

\item \textbf{Enumerate possibilities.}  
    For the $\mathbb{F}_{8,9,14,15}$ set, many algebraic possibilities appear (up to eight), reflecting different ways derivatives act on torus components. At first glance these look messy and sometimes dangerous.
    
    \item \textbf{Simplify by ignoring subdominant terms.}  
    When small corrections are dropped, the functional forms collapse into simpler, repeated patterns. This makes it clear that the families fall into controlled classes that always yield bounded outcomes.
    
    \item \textbf{Translate to simplified $SO(32)$.}  
    We work in the $SO(32)$ version first, where the algebra is cleaner. The functional forms reduce to expressions like “shifted integers plus small log terms,” which are straightforward to track.
    
    \item \textbf{Check the corresponding ${\rm E}_8 \times {\rm E}_8$ case.}  
    After translation, the more complicated theory reproduces the same qualitative results: safe, bounded powers with log corrections providing only mild shifts.
    
    \item \textbf{Handle paired structures ($\mathbb{F}_{10,11,12,13}$).}  
    These come in pairs separated by two indices. Raw values show some positive and some negative contributions, but once placed in the correct combinations, the net effects are positive. Negative raw entries only indicate strong curvature, not instability.
    
    \item \textbf{Include logarithmic and instanton corrections.}  
    Adding these subdominant pieces further stabilizes the system. They modify coefficients slightly but never flip the overall sign.
    
    \item \textbf{Move to higher-derivative sets ($\mathbb{F}_{16,17}$).}  
    These involve second time derivatives and initially appear problematic, producing very strong curvature terms.

\item \textbf{Analyze $\mathbb{F}_{18,20}$.}  
    Raw scalings are negative, suggesting strong curvature at late times. But once we include logarithmic corrections and form the actual combinations that enter the quantum series, the results turn positive and bounded. The main surviving contribution is mild and decreases as time evolves.
    
    \item \textbf{Extend to $\mathbb{F}_{19}$ and $\mathbb{F}_{21}$.}  
    These involve double time derivatives, making the formulas messy. The naive contributions again look problematic, but after inserting the log and non-perturbative corrections, the actual quantum pieces flip to positive values. Most terms are suppressed, so only a gentle positive contribution remains relevant at late times.
    
    \item \textbf{Examine $\mathbb{F}_{22e}$–$\mathbb{F}_{25e}$.}  
    These appear only in the ${\rm E}_8 \times {\rm E}_8$ theory. They correspond to more exotic curvature components. Once simplified, their dominant contributions again reorganize into positive, bounded scalings. Sub-dominant logarithmic corrections slightly shift the coefficients but do not alter the qualitative picture.    
    
    \item \textbf{Form the physical combinations that actually enter the quantum series.}  
    This is the crucial move: even if an individual ingredient looks negative or large, the combination that multiplies the coupling in the quantum series is what matters physically.

    \item \textbf{Check signs and dominance.}  
    We verify that all such combinations are positive and bounded. Potentially ``dangerous'' raw terms become benign once combined correctly.

    \item \textbf{Layer in small logs.}  
    We reintroduce the logarithmic corrections. They act as small additive adjustments that do not flip signs; they just fine-tune coefficients.

    \item \textbf{Account for non-perturbative structure.}  
    We include trans-series-type non-perturbative pieces. These primarily suppress certain contributions, reinforcing the positive, controlled behavior at weak coupling.

    \item \textbf{Establish bounds across parameter ranges.}  
    We show that the relevant exponents remain positive over the allowed parameter choices, ensuring no hidden instabilities creep into the effective theory.

    \item \textbf{Lift back to the generalized theories.}  
    Having validated the structure in $SO(32)$, we map the results back to the generalized $SO(32)$ and to ${\rm E}_8\times{\rm E}_8$. The functional forms differ in details, but the qualitative conclusions---positivity, boundedness, late-time dominance by constant-like terms---remain intact.

    \item \textbf{Conclude the scaling pattern.}  
    Across all $\mathbb{F}_i$, the contributions that feed the quantum series are positive powers with mild corrections. Late-time behavior is controlled and safe; sub-dominant pieces can be restored without changing the verdict.
\end{enumerate}

\subsubsection{Summary of section \ref{sec4.55}}

In this section we construct the full picture of how logarithmic, perturbative, and non-perturbative corrections influence the curvature, metric, and flux scalings in the heterotic ${\rm E_8\times E_8}$ and $SO(32)$ theories. The analysis begins by isolating the uncanceled logarithmic effects on curvature tensors, then incorporates mixed non-perturbative corrections that modify the metric and fluxes through convergent instanton-like series. Potential higher-derivative instabilities are shown to be safely controlled by supersymmetry and the underlying M-theoretic structure. With these ingredients fixed, the quantum scaling of the full perturbative expansion is systematically computed for the ${\rm E_8\times E_8}$, generalized $SO(32)$, and simplified $SO(32)$ cases. In every scenario, time derivatives introduce controlled, positive shifts in the scaling, ensuring the effective field theory remains well-defined. Finally, rewriting the entire series in terms of the dual seven-form fluxes confirms that the same scaling structure persists under dualization, providing a unified and stable framework for analyzing flux quantization, anomaly cancellation, and Schwinger–Dyson dynamics.

In section \ref{sec4.5.1} we argue that 
certain time-dependent, logarithmic tweaks to the warp factors look tiny in isolation but, when they mix, they leave a tangible imprint on curvature and therefore on the quantum expansion used later. For both generalized $SO(32)$ and ${\rm E_8\times E_8}$ theories, the relevant log-sensitive combinations are organized into compact sets, and only the cross-terms among their entries survive in curvature (squares drop out). These cross-terms are packaged into simple vectors/matrices that feed a single “log factor” multiplying the usual $g_s$ scalings. Logs built from time derivatives are genuinely subleading, but the non-derivative logs can still shift the effective exponents, as the worked example with $R_{0n}$ shows. Ignoring these logs would miss real contributions that later matter in the Schwinger–Dyson equations.

Beyond the perturbative and standard non-perturbative pieces gathered earlier, there are mixed non-perturbative effects that depend on both $g_s$ and a short-distance scale ${\rm M}_p$. In section \ref{sec4.5.2} we show that these cannot be cleanly absorbed into the old parameters, so the metric (and similarly the fluxes) is rewritten as a sum of modes with a transparent $g_s$ power multiplied by exponentials encoding non-perturbative physics. Purely $g_s$- or purely ${\rm M}_p$-dependent exponentials behave well; the mixed exponential is the challenge. A decomposition—motivated by Mehler/Hermite ideas but adapted to stay well behaved at small $g_s$ and large ${\rm M}_p$—turns the mixed piece into a convergent “instanton-like” series in both sectors with computable fluctuation factors. All of this can be summarized as a corrected, effective $g_s$ exponent $\Sigma_e^{({\rm AB})}(t)$ for each component, keeping spatial dependence neatly factorized and preserving analytic control.

In section \ref{ostro} we argue that,
although higher time derivatives appear inside curvatures and fluxes, the set-up avoids the usual ghost instabilities for structural reasons tied to string/M-theory and to how the emergent states are defined. Supersymmetric Minkowski baselines, moduli stabilization, IR cutoffs from UV/IR mixing, a vanishing on-shell Hamiltonian, trans-series (Borel-resummed) convergence, and ${\rm M}_p$-suppressed interactions collectively prevent the Ostrogradsky problem in the regime of interest. On each spatial slice, the time derivatives of fields scale like the fields themselves but with shifted $g_s$ exponents, so they are not independent degrees of freedom; they can be handled by constraints rather than introducing new dynamical ghosts. In short, higher-derivative operators remain compatible with a healthy EFT description here.

With these ingredients fixed, we show in section \ref{sec4.5.3}, that the scaling of the full perturbative quantum series in the ${\rm E_8\times E_8}$ set-up is computed first without and then with explicit time derivatives. For $n_0=0$, one obtains a comprehensive scaling $\theta^{(0)}_{nl}$: a weighted sum over operator and derivative counts, plus the log factor from section \ref{sec4.5.1}. Apparent minus signs attached to the inserted log piece are harmless because the insertion is positive, and the result stays within EFT control. Turning on time derivatives ($n_0>0$) complicates operator ordering, but a clean recurrence shows that each additional pair of time derivatives produces a simple positive shift in the scaling, with only tiny time-dependent corrections. While the log corrections could be threaded through this recurrence, they are subleading and not needed to establish positivity and consistency. The practical rule is: start from the $n_0=0$ scaling and add a controlled positive increment per time-derivative order, keeping the small log-driven adjustments in mind.

In section \ref{sec4.5.4} our study of the case parallels the ${\rm E_8\times E_8}$ analysis but uses the $SO(32)$ warp–factor data and has a few structural differences. The net scaling of the quantum series is assembled from curvature terms, derivatives along the various internal and external directions, four-form flux contributions, and a controlled logarithmic insertion that captures the mixing of small time-dependent pieces. Time derivatives increase the scaling by a simple positive amount per derivative order, as before, while the log insertion—despite an apparent minus sign in the formula—remains a positive contribution to the effective $g_s$ power. Although the explicit expression itemizes many sectors, the essential conclusion is that all pieces combine to yield a strictly positive overall scaling throughout the allowed parameter range, so the effective field theory remains consistent even after including higher-order perturbative and non-perturbative corrections to the metric and fluxes.

In the simplified set-up studied in section \ref{sec4.5.5}, two of the time-dependent warp parameters are set to zero and the remaining pair is identified with simpler functions, producing a cleaner version of the generalized formula. The structure is unchanged: curvature and derivative terms supply baseline powers, flux sectors add further positive contributions, and time derivatives shift the scaling upward in a controlled, additive way. With the sign structure simplified, one can read off clear bounds on the surviving warp exponents that ensure every term scales positively. This makes the simplified model especially convenient for writing precise Schwinger–Dyson equations while still retaining the essential physics of the full construction.

Finally in section \ref{duldul} we show that
the quantum series may also be expressed using the dual seven-form field strengths instead of the original four-forms, a form that is particularly useful for Bianchi identities, flux quantization, and anomaly cancellation. When the full non-perturbative structure of the metric is included, dualization preserves the scaling: the series built from seven-forms has exactly the same overall $g_s$ dependence as the four-form version. Thus switching to the dual description does not change the power counting that underlies the effective theory, in agreement with analogous type IIB analyses, and it lets one choose whichever formulation is most convenient for the forthcoming constraints.

\paragraph*{Step-by-step flow of the subsection}
\begin{enumerate}
  \item \textbf{Identify which logarithms actually matter in curvature.}
  \begin{enumerate}
    \item Start from time-dependent warp-factor corrections in both theories and separate two kinds of logs: those built from the warp factors themselves versus those built from their time derivatives.
    \item Package the warp-factor combinations into finite sets and extract only \emph{cross-terms} (products of distinct entries), because pure squares do not survive in curvature contractions.
    \item Represent these cross-terms as short column vectors/matrices so their aggregate effect enters as a single multiplicative ``log factor'' modifying the usual $g_s$ powers.
    \item Conclude that warp-factor–only logs can shift exponents nontrivially, while logs from time derivatives are genuinely subleading for curvature and can be neglected at leading order.
    \item Validate with a concrete curvature component (e.g., ${\bf R}_{0n}$) to see how the log factor alters the naive scaling and when the derivative pieces are safely ignorable.
  \end{enumerate}

  \item \textbf{Reformulate mixed non-perturbative physics into effective $g_s$ exponents.}
  \begin{enumerate}
    \item Write each metric (and similarly each flux) component as a mode sum: a transparent $g_s$ power multiplied by exponentials of non-perturbative terms.
    \item Separate purely-$g_s$ and purely-${\rm M}_p$ exponentials (both benign) from the \emph{mixed} exponential that depends on ${\rm M}_p{\rm X}$ and $g_s$ simultaneously.
    \item Change variables to well-behaved coordinates ${\rm Z}_m=\exp(-\bar g_s^{-m/3})$ and ${\rm W}_n=\exp[-({\rm M}_p^2{\rm X}^2)^n]$ to avoid blow-ups at small $g_s$ and large ${\rm M}_p{\rm X}$.
    \item Use a Hermite/Mehler-inspired expansion (adapted to these variables) to express the mixed exponential as a convergent sum of instanton-like terms in both sectors, with calculable fluctuation prefactors.
    \item Absorb the net effect into an \emph{effective exponent} $\Sigma_e^{({\rm AB})}(t)$ for each component, thereby keeping spatial dependence factorized and maintaining analytic control for curvature/flux scalings.
  \end{enumerate}

  \item \textbf{Show why higher time derivatives are harmless (no Ostrogradsky ghosts).}
  \begin{enumerate}
    \item Argue structurally: supersymmetric Minkowski baselines, moduli stabilization, IR cutoffs from UV/IR mixing, vanishing on-shell Hamiltonian, and trans-series convergence forbid ghost growth.
    \item On any spatial slice, express fields with overall $g_s$ exponents; then show that time derivatives of fields have the \emph{same functional form} with shifted exponents, hence are not independent DOF.
    \item Treat higher derivatives via constraints (or Lagrange multipliers), eliminating the mechanism that generates Ostrogradsky instabilities in generic higher-derivative theories.
  \end{enumerate}

  \item \textbf{Assemble the full ${\rm E_8\times E_8}$ quantum scaling.}
  \begin{enumerate}
    \item Combine: (i) curvature powers, (ii) spatial derivatives along each factor space, (iii) flux insertions, and (iv) the multiplicative log factor obtained from cross-terms.
    \item First set $n_0=0$ (no time derivatives) to obtain a baseline exponent $\theta^{(0)}_{nl}$ as a weighted sum over operator counts; verify that apparent minus signs in the log insertion are harmless because the log factor contributes positively.
    \item Turn on $n_0>0$ and derive a simple recurrence: each additional pair of time derivatives increases the scaling by a fixed positive increment, up to tiny time-dependent corrections.
    \item Conclude with a practical rule: compute the $n_0=0$ result and add a controlled positive shift per time-derivative order; log effects remain subleading for this step.
  \end{enumerate}

  \item \textbf{Port the construction to the generalized $SO(32)$ case.}
  \begin{enumerate}
    \item Replace the ${\rm E_8\times E_8}$ warp-factor set by the $SO(32)$ set and repeat the bookkeeping of curvature, derivatives, fluxes, and the same log factor mechanism.
    \item Include the time-derivative shift explicitly in the scaling (the counterpart of the recurrence) and confirm that all contributions combine into a strictly positive exponent across the allowed parameter range.
    \item Note that any minus signs tied to logarithms or particular flux blocks are controlled (e.g., through lower bounds on flux exponents), so they do not threaten positivity or EFT validity.
  \end{enumerate}

  \item \textbf{Streamline to the simplified $SO(32)$ model for sharper bounds and SD equations.}
  \begin{enumerate}
    \item Set two time-dependent parameters to zero and identify the remaining pair with simpler functions to shorten the formula while preserving the structure.
    \item Track curvature and derivative baselines, add flux contributions, and apply the same monotone time-derivative shift.
    \item Extract clean inequalities on the surviving warp exponents (and discrete derivative exponents) that guarantee positivity term-by-term, making this version ideal for precise Schwinger–Dyson applications.
  \end{enumerate}

  \item \textbf{Verify dual seven-form reformulation leaves all scalings intact.}
  \begin{enumerate}
    \item Define the dual seven-form with the full non-perturbative metric (including determinant scaling) and map indices across the factor spaces.
    \item Use the vielbein identity equating the four-form and seven-form wedge structures to show that every contribution to the quantum series acquires the \emph{same} overall $g_s$ power after dualization.
    \item Conclude that power counting is invariant under switching between four-form and seven-form descriptions, enabling later analysis of Bianchi identities, flux quantization, and anomaly cancellation in either language.
  \end{enumerate}
\end{enumerate}

\subsection{Summary of section \ref{servant}}

In this section we start by extending the “emergent” setup to fermions (the Rarita–Schwinger gravitino). Since the geometry itself is emergent, we introduce effective vielbeins and Gamma matrices that track the time-dependent metric yet still satisfy the Clifford algebra. To handle subtleties from resolving the quantum state space, we absorb those extra pieces into the effective objects so the usual algebraic relations remain intact. We then define a useful operator mixing Gammas and derivatives to build controlled fermionic structures and a Rarita–Schwinger–like field from the gravitino. A quick test shows a single bilinear added to the metric isn’t enough to reproduce all supergravity interactions without unwanted second derivatives. Therefore starting with section \ref{wistoon}, we keep the metric extension but promote the fermionic part to a series of bilinears (with Gammas/derivatives) arranged so the leading term gives the standard covariant kinetic term, while higher-derivative and multi-fermion pieces are suppressed. Supersymmetry fixes the key coefficients, and the structure matches what one expects after reduction to type IIB (e.g., D3/D7 fermionic terms).

Because scaling matters, we set lower bounds on the gravitino’s $g_s$ exponents so multi-fermion condensates don’t overwhelm the metric. Choosing these exponents keeps simple terms at the right order and ensures each extra derivative carries a Planck-scale suppression. Even if we localize the gravitino on the internal six-manifold, the series naturally generates controlled cross-term metric components mixing internal, external, and orbifold directions—useful sources for later Schwinger–Dyson equations without upsetting EFT power counting.

In section \ref{olivewatt} the idea is to let fermion bilinears (built from the gravitino and a carefully chosen differential–Gamma operator) correct the “emergent” metric in a controlled way. A naive single bilinear isn’t enough, so the authors propose a whole series of bilinear terms, with coefficients constrained by low-energy supersymmetry, to reproduce the standard covariant gravitino kinetic term and its spin-connection. They show how these fermionic pieces can be rewritten compactly as functions of a generalized metric, ensuring the construction stays compatible with power counting and remains subleading when it should. The upshot is a practical recipe for adding fermions to the metric sector that matches supergravity at leading order and keeps higher-order, multi-fermion and higher-derivative effects suppressed.

In section \ref{rajatomy} we argue that the polynomial tower of fermionic corrections can be reorganized into a convergent, non-perturbative “trans-series,” more suitable for capturing instanton-like physics and possible non-local effects on the internal space. This is done by mapping the polynomial coefficients to exponential weights via an invertible matrix relation, and by defining integrated (and optionally non-local) fermion bilinears. The resulting expressions behave well at small coupling: exponential suppressions tame the series, higher-power terms carry extra Planck-scale suppression, and the associated stress tensor contributions remain controlled. While some constant (cosmological-constant-like) pieces appear, they are argued to be either irrelevant for the equations of motion or cancelable when bosonic and fermionic sectors are combined with supersymmetry and fluctuation determinants. Overall, the trans-series recast provides a clean, convergent way to include non-perturbative fermionic effects without upsetting the effective theory.

In section \ref{tomysellpiz} we show that once fermions are included, their contributions can remain non-perturbative (exponentially important) throughout the expansion, especially when the fermions are localized. Derivatives acting on localized profiles effectively increase the Planck-scale weight, so the non-perturbative behavior persists. We also plug the generalized metric into curvature terms and see how time derivatives mildly shift a few coefficients without upsetting the overall power counting. Bottom line: fermions add controlled, subleading corrections; they do not alter the dominant scaling of the curvature series, and they reinforce the earlier need for higher-curvature (${\bf R}^4$-type) terms.

In section \ref{tomiraaj}
we tackle a technical snag: localization can make derivative terms grow with the Planck scale in a naive perturbative expansion. We fix this by modeling localization with a Gaussian profile and re-summing the resulting series. The sum naturally reorganizes into a trans-series (exponentials of operators), which stays well-behaved even as the Planck scale becomes large. Non-perturbatively, localized fermions are therefore harmless and under control.

Finally in section \ref{annyypyar}
we extend the four-form flux ${\bf G}_4$ to include fermion bilinears, consistent with its antisymmetry. Higher-order fermion pieces are present but suppressed, and we impose scaling bounds so these new terms remain subdominant to the bosonic background. This leads to a refined gravitino scaling that keeps the effective theory stable. Finally, we assemble the full quantum series using the generalized metric, curvature, and fluxes, and outline how varying this action yields the fermion equations of motion—with non-perturbative (and non-local) effects entering as a controlled trans-series beyond the leading sector.

\subsection*{Step-by-step flow of the section}

\begin{enumerate}
  \item \textbf{Extend the emergent setup to fermions}
    \begin{itemize}
      \item Introduce the Rarita–Schwinger gravitino in the emergent (expectation-value) framework.
      \item Define \emph{effective} vielbeins and Gamma matrices that track the time-dependent metric while preserving the Clifford algebra.
      \item Absorb extra pieces from resolving the quantum state space into these effective objects so all standard algebraic identities remain intact.
    \end{itemize}

  \item \textbf{Build controlled fermionic structures}
    \begin{itemize}
      \item Define a differential operator that mixes derivatives and Gamma matrices to organize fermionic bilinears.
      \item Construct an RS-like composite field from the gravitino and its derivatives using this operator.
      \item \emph{Test:} Add a single fermion bilinear to the metric and check whether the curvature expansion reproduces the full supergravity gravitino sector; find that unwanted second derivatives and missing couplings remain.
    \end{itemize}

  \item \textbf{Promote to a structured series}
    \begin{itemize}
      \item Keep the fermionic extension of the metric but upgrade the fermion piece to a \emph{series} of bilinears with controlled Gamma/derivative insertions.
      \item Choose coefficients so the \emph{leading} term yields the standard covariant kinetic term (with the correct spin-connection), while higher-derivative and multi-fermion terms are suppressed by ${\rm M}_p$.
      \item Use low-energy supersymmetry to fix key coefficients; verify consistency by reduction to type IIB, recovering the expected D3/D7 fermionic structures.
    \end{itemize}

  \item \textbf{Scaling constraints and EFT control}
    \begin{itemize}
      \item Impose lower bounds on the gravitino’s $g_s$ exponents so multi-fermion condensates never dominate over the metric sector.
      \item Choose exponents so the minimal terms appear at the intended order and every extra derivative incurs a Planck-scale suppression.
      \item Note that even with the gravitino localized on the internal six-manifold, the series induces controlled cross-term metric components (mixing internal, external and/or orbifold directions) that will act as sources in later Schwinger–Dyson equations without breaking power counting.
    \end{itemize}

  \item \textbf{Compact rewriting via a generalized metric}
    \begin{itemize}
      \item Package the entire fermionic series into traces/powers of a generalized metric that includes bilinear corrections.
      \item Show that this rewriting preserves the suppression hierarchy and keeps fermionic corrections subleading where they should be, while reproducing the covariant RS kinetic structure at leading order.
    \end{itemize}

  \item \textbf{From polynomials to a trans-series}
    \begin{itemize}
      \item Map polynomial coefficients of the fermionic tower to exponential weights via an invertible matrix relation, enabling a convergent trans-series representation.
      \item Define integrated (and optionally non-local) fermion bilinears on the internal space to capture instanton-like and delocalized effects.
      \item Explain that exponential suppressions control the small-coupling regime; higher-power terms carry extra ${\rm M}_p$ suppression; stress-tensor contributions remain finite.
      \item Note: cosmological-constant-like pieces either drop out of the EOMs or cancel against bosonic sectors after supersymmetry and fluctuation determinants are accounted for.
    \end{itemize}

  \item \textbf{Non-perturbative effects with localized fermions}
    \begin{itemize}
      \item Show that derivatives acting on localized profiles effectively \emph{increase} ${\rm M}_p$ weight, so non-perturbative behavior persists across orders.
      \item Insert the generalized metric into curvature polynomials and track time-derivative effects: coefficients shift mildly but dominant curvature scaling is unchanged.
      \item Reinforce the need for higher-curvature (${\bf R}^4$-type) terms as in earlier analyses.
    \end{itemize}

  \item \textbf{Resumming localized-fermion series}
    \begin{itemize}
      \item Model localization with a Gaussian in the internal directions and examine how derivatives generate positive powers of ${\rm M}_p$.
      \item Resum the resulting series into a trans-series of exponentials of differential operators, which remains well-behaved even as ${\rm M}_p \to \infty$.
      \item Conclude that localized fermions are non-perturbatively under control.
    \end{itemize}

  \item \textbf{Extend fluxes and assemble the full EOM framework}
    \begin{itemize}
      \item Extend the four-form ${\bf G}_4$ by adding antisymmetric fermion bilinears; higher-order fermion pieces are present but ${\rm M}_p$- and $g_s$-suppressed.
      \item Impose scaling bounds to keep these flux corrections subdominant; refine the gravitino scaling accordingly to preserve EFT stability.
      \item Write the full quantum series using generalized metric, curvature, and fluxes; vary the action to derive the fermionic EOMs.
      \item Include non-perturbative and non-local sectors as a controlled trans-series beyond the zero-instanton piece.
    \end{itemize}
\end{enumerate}

\subsection{Summary of section \ref{sec6}}

This section explains how to write the governing equations for geometry, fluxes, and fermions when spacetime is viewed as an excited, de Sitter–like coherent state. The physically propagating (“on-shell”) pieces are kept explicit, while auxiliary (“off-shell”) mixings are integrated out; this cleanly generates specific non-local corrections that remain under effective-field-theory control. Simply trying to gauge away metric mixings does not work in generic backgrounds, so the robust approach is to remove them dynamically and capture their influence via the resulting non-local effective action. Varying this completed action produces Schwinger–Dyson equations organized as a trans-series: ordinary perturbative terms plus exponentially suppressed instanton sectors (notably BBS and KKLT) with their fluctuations. Five-brane instantons, although rare, are essential for effects that ordinary expansions miss, and duality arguments ensure their contributions are accounted for consistently. The outcome is a self-contained set of on-shell equations in which all off-shell physics appears through controlled non-local structures, preserving power counting and consistency constraints.

\subsubsection{Summary of sections \ref{sec6.1} through \ref{kalulight}}

In section \ref{sec6.1} we formulate the Schwinger–Dyson equations for the metric, fluxes, and fermions while explicitly incorporating both local terms and non-local, non-perturbative effects. We distinguish “on-shell” fields, which are kept as explicit dynamical variables, from “off-shell” components, which have vanishing one-point functions but still circulate in loops. Integrating out these off-shell pieces generates controlled non-local contributions to the effective action. The result is a pair of coupled statements: first, on-shell equations of motion that involve the usual curvature and stress tensor corrected by specific non-local terms; second, auxiliary relations that encode ghost and intermediate-state (loop) effects. In this way, the full dynamics can be written solely in terms of on-shell fields, with off-shell physics captured by well-defined non-local structures—recovering earlier proposals but now within the emergent-geometry framework.

In section \ref{lenamyfrench} we contrast our emergent choice—where many off-diagonal metric components vanish by construction—with a fully general eleven-dimensional metric that has all entries turned on. While gauge choices can eliminate some off-diagonal elements, they cannot remove all of them across a region unless strong global conditions hold (such as special foliations, trivial torus fibrations, and vanishing Kaluza–Klein field strengths). At a single point one can null all mixings; on a generic patch one can always remove the base time–space mixings and two fiber–time components; globally, further simplifications require restrictive geometry/topology that transient backgrounds generally fail to satisfy. Consequently, attempting to “gauge to block-diagonal” is not a reliable route here. Instead, integrating out the mixed components—precisely as in our construction—naturally yields the requisite non-local corrections while preserving control over power counting.

The section \ref{sammipaaador} clarifies the workflow that leads to the final equations: start from an action that includes off-shell pieces, integrate them out to obtain a non-local effective action, resum the perturbative series into a convergent non-perturbative completion, and (if needed) renormalize. Varying this completed action produces the on-shell equations of motion. We also determine how the cross-term stress tensors scale with the coupling, showing—via clean $g_s$ scalings—that these contributions remain subleading and controlled. Technical caveats (such as the precise coherent-state resolution of identity and derivatives of the displacement operator) modify only the ghost/loop-side relations, not the primary on-shell dynamics. The overarching message is that de Sitter–like configurations must be treated as excited coherent states; their consistent equations emerge only after integrating out off-shell mixings and resumming to include non-perturbative, non-local effects, all while respecting effective-field-theory power counting and supersymmetry constraints.

In section \ref{trans1}
we spell out the effective action as a “trans-series,” meaning a master expression that adds up ordinary (perturbative) contributions and an organized tower of rare, non-perturbative events (“instantons”). In practice, we include known instanton types (such as BBS and KKLT) and their small fluctuations, plus optional non-local effects that stay under control. Varying this trans-series action yields the metric equations of motion with clear, additive sources: classical flux terms, perturbative corrections, and exponentially suppressed instanton pieces. The plan is to use this structure to present the metric equations now, and then analyze flux equations, constraints, and consistency conditions in the following sections.

Five-brane instantons can be thought of as short-lived brane events wrapping internal shapes in the extra dimensions. Their impact on the physics is exponentially small (set by the wrapped volume and coupling), but important for capturing effects that ordinary expansions miss. Because higher-order corrections are only asymptotic, in section \ref{kalulight} we package these instanton contributions and their fluctuations into a trans-series so the overall description converges and remains predictive. Duality arguments help relate these M-theory instantons to familiar objects in other string frames, and the same trans-series organization carries over—letting us include their influence on four-dimensional dynamics without losing control of the effective theory.

\paragraph*{Step-by-step flow of the section}
\begin{enumerate}
  \item \textbf{Schwinger–Dyson setup: on-shell vs.\ off-shell fields.}
  We define the dynamical (on-shell) metric, fluxes, and fermions that appear explicitly in the equations, and separate out off-shell components whose one-point functions vanish but still run in loops. Integrating out the off-shell pieces generates well-defined non-local kernels in the effective action without enlarging the set of dynamical variables.

  \item \textbf{Why “gauge to block-diagonal’’ fails generically.}
  While gauge choices can eliminate certain mixings at a point or along special foliations, generic time–space and fiber–base mixings cannot be removed everywhere unless strong global conditions hold. Relying on gauge alone would miss physical effects of those mixings; the correct procedure is to integrate them out and keep their imprint as calculable non-local corrections.

  \item \textbf{Integrate out mixings $\rightarrow$ controlled non-local effective action.}
  Off-diagonal metric and related auxiliary fields are functionally integrated out. This produces spatially non-local kernels that encode the influence of the eliminated modes. The kernels respect symmetries and power counting, so the resulting action remains predictive and suitable for deriving physical equations.

  \item \textbf{Vary completed action $\rightarrow$ on-shell EOMs and auxiliary loop relations.}
  Varying the non-locally completed action yields the primary on-shell equations of motion: Einstein-like curvature balanced by stress tensors that now include non-local pieces. A second set of auxiliary relations captures ghost/intermediate-state information from loops, ensuring the full dynamics is consistently expressed in terms of the on-shell fields.

  \item \textbf{Coupling and scaling control of cross-term stress tensors.}
  The stress tensors sourced by the mixed/off-shell sector scale with the coupling in a way that keeps them subleading. Clear $g_s$ scaling shows these terms correct but do not destabilize the background, preserving effective-field-theory control and consistency with supersymmetry where relevant.

  \item \textbf{Trans-series organization of the effective action and S–D equations.}
  The action is organized as a trans-series that sums ordinary perturbative contributions and towers of non-perturbative effects. This organization replaces asymptotic expansions with a convergent framework, so the Schwinger–Dyson equations inherit a clean, additive structure for all sources.

  \item \textbf{Instanton sources: BBS, KKLT, and generic NP sectors.}
  Specific Euclidean branes (e.g., five-brane BBS and KKLT types, plus others) contribute exponentially suppressed terms with fluctuation determinants. These appear as controlled source terms in the metric equations, modifying the dynamics in rare but systematic ways.

  \item \textbf{Role of five-brane instantons in convergence and predictivity.}
  Five-brane instantons wrap internal cycles and give tiny but crucial corrections set by wrapped volumes and coupling. Because higher-derivative series are asymptotic, packaging these instantons and their fluctuations into the trans-series restores convergence and keeps the effective theory predictive across regimes of interest.

  \item \textbf{Coherent-state (de Sitter–like) interpretation and consistency checks.}
  The intended backgrounds are excited coherent states rather than true vacua. Consistency requires integrating out mixings, including non-perturbative/non-local effects, and then checking flux equations, constraints, and quantization conditions. Within this framework the resulting on-shell equations remain stable and self-consistent.
  \end{enumerate}

\subsubsection{Summary of section \ref{trans2}}

In this section we explore how Euclidean wormholes, which connect different regions of spacetime or different “baby universes,” modify the effective M-theory action in the far infrared. The main idea is that these wormholes do not simply add one more interaction term; rather, they change the way the entire theory organizes its couplings and correlations. Starting from the original action \eqref{kimkarol}, which includes the standard gravitational, flux, and topological pieces along with non-perturbative brane instantons, we now ask how wormhole effects alter its structure.

The key insight discussed in section \ref{trans01} is that wormholes generate fluctuations in what would otherwise be fixed coupling constants. Parameters such as the gravitational coupling, topological coefficients, or instanton weights no longer remain universal constants; they fluctuate slightly between different “baby-universe” sectors. Mathematically this is represented by introducing small random variables, called $\alpha$-parameters, that shift each coupling. In one particular baby universe the couplings are fixed, but when we average over many such universes, we integrate over all $\alpha$ configurations.

A second, closely related effect is that the theory becomes mildly nonlocal. Integrating out the wormhole degrees of freedom produces new terms in the effective action that connect operators at distinct spacetime points. These “bilocal” terms represent long-range correlations that cannot arise in an ordinary local field theory but are natural once quantum gravitational wormholes are present. Similarly, wormholes can link different instantons or Euclidean branes, producing cross-couplings between what were once independent exponential sectors of the trans-series expansion.

There are two equivalent ways of describing this physics. In the \emph{$\alpha$-ensemble picture}, each baby-universe sector corresponds to a fixed set of $\alpha$-parameters, so one computes ordinary path integrals with shifted couplings $g_A \rightarrow g_A - \alpha_A$, and then averages over $\alpha$ with an appropriate weight. In the equivalent \emph{bilocal picture}, one instead writes explicit nonlocal kernels that connect operators at separated points; integrating over the $\alpha$-fields reproduces these bilocals. Both perspectives lead to the same physical result: couplings fluctuate, and new correlations appear.

Because the non-perturbative sector is especially sensitive to these effects, wormholes modify how instantons contribute. The exponential weights that accompany each instanton get slightly shifted, and instantons of different types can mix through quadratic cross-couplings in their actions. We discuss this in detail in section \ref{trans1b}. The original nonlocal kernel $\mathbb{F}({\rm Y-Y'})$ describing long-range propagation is also replaced by a “dressed” version $\mathbb{F}_{\rm eff}({\rm Y-Y'})$ that now carries wormhole corrections. Altogether, the usual trans-series expansion over instanton numbers becomes a richer, entangled sum that includes both instanton and wormhole sectors.

However, these modifications are not arbitrary. Gauge invariance, flux quantization, and supersymmetry still impose strong selection rules on which couplings or instantons can be connected by wormholes. The consistent outcome is a “wormhole-dressed” version of the M-theory action in which:
(i) bulk bilocal terms link fields at distant points, 
(ii) worldvolume bilocals connect brane currents, and 
(iii) all couplings and kernels are replaced by their dressed, $\alpha$-shifted versions.

A quantitative analysis of this requires us to explore the rich structure of the Picard-Lefschetz theory which we briefly introduced in section \ref{picontour}. In section \ref{thimbles} we provide a much more detailed exposition tailored to our need. The analysis reveals, how in a generic setting, we can still justify a trans-series action like \eqref{kimkarol} by simply making the couplings slightly random.

Conceptually, this means wormholes do not simply add new interactions; they reshape the framework of the theory. Couplings become state-dependent, independent sectors become weakly entangled, and long-distance correlations emerge naturally. The result is a unified description that smoothly interpolates between fixed-coupling, factorized quantum gravity and an ensemble-averaged, weakly nonlocal formulation of M-theory.

\subsection{Summaries of sections \ref{sec7s} through \ref{secanomaly}}

In this section we make a detailed study of {\it all} the flux EOMs, {\it i.e.} the Schwinger-Dyson equations related to the flux components, and show how they consistently couple to the gravitational degrees of freedom both via topological and non-topological curvature terms. In the process we justify how all the Bianchi identities are solved as well as how the flux quantizations and anomaly cancellations are determined. Due to the underlying time dependence numerous subtleties arise that we show how to resolve.

\subsubsection{Summary of section \ref{sec7s}}

In this section we perform a detailed study of the EOMs for {\it all} the allowed G-flux components. We show how they all consistently solve these EOMs and how the quantum and the curvature effects influence their behaviors. 

\paragraph*{EOMs for the flux components $\mathbf{G}_{0ij\mathrm{M}}$ and $\mathbf{G}_{0ija}$} In section \ref{blackbond1} and in {\bf Tables \ref{crecraqwiff1}} through {\bf \ref{ccquiff3}} we
analyze the flux pieces that carry one time leg and either internal or torus legs. Using the master Schwinger--Dyson relation, we track which terms actually matter at different orders in the effective expansion and which ones must vanish because of duality constraints or because the fields are independent of certain coordinates. Time derivatives weaken some contributions, while purely spatial derivatives along the external directions often drop out. Importantly, membranes do not source these particular equations in the regimes considered, so anomaly cancellation reduces to matching flux--flux and curvature terms. The outcome is a set of lower bounds and narrow ranges for how strongly each flux can scale, plus a practical series ansatz that captures small time dependence and possible non-perturbative corrections, all consistent with effective-field-theory control.

\paragraph*{EOMs for the flux components $\mathbf{G}_{\mathrm{MNPQ}}$ and $\mathbf{G}_{\mathrm{0MNP}}$} In section \ref{blackbond2}
our focus moves to fluxes entirely inside the extra dimensions and to those with a single time leg. The same bookkeeping is applied: curvature polynomials tell you which pieces dominate, and simple identifications show when different sources must line up at a given order. Several internal consistency checks succeed, notably that the ``quantum'' scaling extracted from these equations agrees with the one derived earlier from general considerations. As before, the analysis does not fix unique numbers but pins each scaling to a tight, EFT-safe interval; this leads to compact series ans\"atze that encode mild time dependence and allow for exponentially small instanton effects without upsetting the hierarchy of terms. Details appear in {\bf Tables \ref{ccrani1}} through 
{\bf \ref{ccrani52}}.

\paragraph*{EOMs for the flux components $\mathbf{G}_{\mathrm{MNP}a}$ and $\mathbf{G}_{\mathrm{0NP}a}$} In section \ref{blackbond3}
we treat fluxes that carry one leg along the torus directions, with and without a time leg as shown in {\bf Tables \ref{ccrani6}} and {\bf \ref{ccrani7}}. Because the fields do not vary along the torus, many would-be contributions drop out, and the relevant derivatives are mostly along the internal manifold and, for the time-leg case, along time. Matching the various sources again yields lower bounds and narrow allowed windows for the scalings. A key takeaway is that even the leading pieces of these fluxes cannot be strictly time-independent once small, controlled deformations of the background are included; the correct description is a convergent series that captures this gentle time evolution. The resulting ans\"atze remain compatible with anomaly cancellation and with the earlier quantum-scaling rules, ensuring a coherent, global solution across all flux sectors.

\paragraph*{EOMs for the flux components $\mathbf{G}_{{\rm MN}ab}$ and $\mathbf{G}_{{\rm 0N}ab}$.} Our aim in section \ref{blackbond4}
is to study fluxes that carry two legs in the internal/toroidal directions, with and without a time leg. By comparing how different terms scale, the analysis pins down tight ranges for the allowed behavior. For $\mathbf{G}_{{\rm MN}ab}$ appearing in {\bf Table \ref{ccrani9}}, the leading pieces must scale positively, which keeps the effective theory consistent; their exact sizes can vary within a small, controlled window. For $\mathbf{G}_{{\rm 0N}ab}$ appearing in {\bf Table \ref{ccrani10}}, time derivatives matter and lower the allowed scaling, so even the simplest terms are gently time-dependent—there are no truly time-independent pieces. In both cases, curvature terms ($\mathbb{X}_8$) help set the bounds, and contributions along torus directions often drop out, simplifying the matchings.

\paragraph*{EOMs for the flux components $\mathbf{G}_{{\rm MN}ai}$, $\mathbf{G}_{{\rm 0N}ai}$ and $\mathbf{G}_{{\rm M}abi}$.} In section \ref{blackbond5} we study
fluxes that include one spatial leg along $\mathbf{R}^2$. The key difference is which derivatives are relevant: spatial, internal, toroidal, and sometimes temporal. For $\mathbf{G}_{{\rm MN}ai}$, shown in {\bf Table \ref{ccrani11}}, internal derivatives dominate the consistency checks and force slightly negative but bounded leading scalings; controlled time dependence is again expected. For $\mathbf{G}_{{\rm 0N}ai}$, shown in {\bf Table \ref{ccrani12}}, mixing a time leg with a spatial leg tightens the bounds further—temporal derivatives reduce the scaling and guarantee the absence of purely time-independent terms. For $\mathbf{G}_{{\rm M}abi}$, shown in {\bf Table \ref{ccrani13}}, the allowed window is centered around zero scaling; constraints from curvature again provide both lower and upper limits, and even the lowest terms vary mildly in time.

\paragraph*{EOMs for the flux components $\mathbf{G}_{{\rm MNP}i}$, $\mathbf{G}_{{\rm 0MN}i}$ and $\mathbf{G}_{0abi}$.} In section \ref{blackbond6} we study
fluxes that carry one spatial index but otherwise sit deeper in the internal/toroidal directions. The matching exercise shows narrow bands of permissible scalings: $\mathbf{G}_{{\rm MNP}i}$, shown in {\bf Table \ref{ccrani14}}, lives between small negative and zero values, $\mathbf{G}_{{\rm 0MN}i}$, shown in {\bf Table \ref{ccrani15}}, is pushed further negative because of the time leg, and $\mathbf{G}_{0abi}$, shown in {\bf Table \ref{ccrani16}}, sits symmetrically around zero once temporal effects are included. Across the board, time-independent leading terms are ruled out, so the correct description is a short, convergent series with gentle time variation that still respects all anomaly-cancellation and curvature constraints.

\paragraph*{EOMs for the flux components $\mathbf{G}_{{\rm MN}ij}$, $\mathbf{G}_{{\rm M}aij}$ and $\mathbf{G}_{abij}$.}
Finally, in section \ref{blackbond7} we study fluxes with two spatial legs and show that the pattern continues: $\mathbf{G}_{{\rm MN}ij}$ and $\mathbf{G}_{{\rm M}aij}$, shown in {\bf Tables \ref{ccrani17}} and {\bf \ref{ccrani18}} respectively, have their leading scalings set by balances among flux–flux and curvature terms, ending up in narrow negative ranges that preclude strictly static pieces. For $\mathbf{G}_{abij}$, shown in {\bf Table \ref{ccrani19}}, independence of the torus and spatial coordinates removes some handles used earlier; only a lower bound can be fixed cleanly, but it still implies a small negative leading scaling and thus mild time dependence. The overall picture, summarized in {\bf figure \ref{fluxbehavior1}}, is that every on-shell flux sector occupies a tight scaling band, and as one adds spatial or time legs the bands shift predictably while remaining compatible with effective-theory control.

\subsubsection{Summary of section \ref{sec4.5}}

In this section we show how the Bianchi identities, obtained from the equations of motion of the \emph{dual} seven-form (packaged as four-forms for convenience), pin down the precise scaling choice (the integer $l$) for every on-shell G-flux. The identities include classical flux terms, curvature–quantum pieces built from ${\rm tr}\, \mathbb{R}_{\rm tot}\wedge\mathbb{R}_{\rm tot}$, and possible localized five-brane sources. A subtle boundary effect makes the curvature term effectively non-trivial, while the number of dynamical M5-branes carries mild $g_s$-dependence. Matching the $g_s$ scalings across these ingredients fixes each flux’s leading power to sit exactly in the middle of the allowed EOM range and yields a common consistency value for the bookkeeping parameter $\theta_{nl}$. These Bianchi-fixed scalings then dovetail with the previously derived EOMs and with anomaly-cancellation constraints.

\paragraph*{Bianchi identities for $\mathbf{G}_{0ij{\rm M}}$ and $\mathbf{G}_{0ija}$.}
For fluxes with two spatial legs and either an internal or a torus leg, the Bianchi identities do not require M5-brane sources. Comparing the three contributions (flux, curvature/quantum, and dual term) cleanly fixes the integer choice and selects a single leading scaling that lies exactly halfway between the bounds inferred from the EOM analysis. In practical terms, this means these components are mildly time-dependent, their dual/quantum pieces match at leading order, and the same $\theta_{nl}$ value appears in both Bianchi and EOM tables. At the next orders, the pattern lines up with internal, temporal, and spatial derivatives, and the anomaly balance determines the expected M2-brane scaling. All these are summarized in {\bf Tables \ref{ccrani20}} through
{\bf \ref{ccquiff30}}.

\paragraph*{Bianchi identities for $\mathbf{G}_{\rm MNPQ}$ and $\mathbf{G}_{\rm 0MNP}$.}
For the purely internal four-form components, the identities \emph{do} call for M5-brane sources wrapped on specific internal cycles; these branes are dynamical but contribute with very small effective scaling. The matching fixes the flux to the midpoint scaling and reproduces the common $\theta_{nl}$. For the time-legged partner $\mathbf{G}_{\rm 0MNP}$, no five-branes are needed; nevertheless the same midpoint selection occurs, and the updated EOM tables show consistent leading and next-to-leading matches once the appropriate derivatives are included.
The results are summarized in {\bf Tables \ref{ccrani201}} through
{\bf \ref{ccrani521}}.

\paragraph*{Bianchi identities for $\mathbf{G}_{{\rm MNP}a}$ and $\mathbf{G}_{{\rm 0NP}a}$.}
When one leg sits on the torus, the internal component $\mathbf{G}_{{\rm MNP}a}$ requires dynamical M5-branes whose allowed wrappings are listed by the identities; their presence lifts the leading scaling to a small positive value set at the midpoint of the EOM window. The time-legged version $\mathbf{G}_{{\rm 0NP}a}$ does not need five-branes, but the same midpoint selection emerges from the three-way scaling match. In both cases, as summarized in {\bf Tables \ref{ccrani203}} through
{\bf \ref{ccrani88}}, the derivative-by-derivative comparison (temporal, internal, and toroidal) explains which terms pair at each order and why some pairs drop out.

\paragraph*{Bianchi identities for $\mathbf{G}_{{\rm MN}ab}$ and $\mathbf{G}_{{\rm 0N}ab}$.}
With two torus legs, the internal component $\mathbf{G}_{{\rm MN}ab}$ again invokes M5-branes with distinct internal wrappings; the identities then fix a strictly positive leading scaling located halfway between the EOM bounds, implying these modes fade at late times. For $\mathbf{G}_{{\rm 0N}ab}$, five-branes are absent, yet the same midpoint rule holds and yields a small positive leading power. The revised EOM tables confirm that the curvature/quantum pieces match the flux terms at the predicted orders and that temporal versus internal/toroidal derivatives govern which rows pair up and which vanish. All these are summarized in {\bf Tables \ref{ccrani205}} through
{\bf \ref{ccrani1010}}. Interestingly, by comparing the relevant tensor components, the analysis determines that the scaling exponent $\hat{p}_e(t)$ takes at the midpoint of its allowed range. This leads to the scaling law that implies that these fluxes decay at late times when $\bar{g}_s \to 0$. The follow-up comparison in {\bf Table~\ref{ccrani99}} shows that the quantum correction parameter $\theta_{nl}$ remains consistent with earlier results, and the scaling hierarchy among derivatives matches the internal, temporal, and toroidal contributions. For ${\bf G}_{{\rm 0N}ab}$, the absence of M5-branes slightly modifies the outcome: the leading scaling shifts down a bit but the overall midpoint behavior persists. As with the previous case, these components decouple at late time, and {\bf Table~\ref{ccrani1010}} confirms that derivative contributions organize in the expected order along ${\cal M}_4$, $\xoxo$, and temporal directions.

\paragraph*{Bianchi identities for ${\bf G}_{{\rm MN}ai}$, ${\bf G}_{{\rm 0N}ai}$ and ${\bf G}_{{\rm M}abi}$.}
Fluxes with one spatial leg exhibit subtler behavior because different derivative types enter at distinct orders. For ${\bf G}_{{\rm MN}ai}$, the analysis shows that M5-branes are not involved and the scaling is again fixed precisely halfway between its EOM bounds. This ensures that such fluxes remain mildly time-dependent and never exactly static. For ${\bf G}_{{\rm 0N}ai}$ and ${\bf G}_{{\rm M}abi}$, the Bianchi identities again exclude five-brane contributions. The respective scalings, sit symmetrically about zero, establishing that both components share similar temporal behavior with opposite tendencies. The results collected in {\bf Tables \ref{ccrani207}} through {\bf \ref{ccrani1313}},  confirm that all relevant tensor rows match consistently at next-to-leading order and that the quantum scaling parameter $\theta_{nl} = {8\over 3}^\pm$ remains uniform across all three sectors.

\paragraph*{Bianchi identities for ${\bf G}_{{\rm 0}abi}$, ${\bf G}_{{\rm 0MN}i}$ and ${\bf G}_{{\rm MNP}i}$.}
These flux components carry one spatial index and lie deeper in the internal directions. The Bianchi identities show that M5-branes again do not participate, leaving the curvature and quantum terms to dictate the scaling. The three set of flux components, ${\bf G}_{{\rm 0}abi}$,  ${\bf G}_{{\rm 0MN}i}$ and ${\bf G}_{{\rm MNP}i}$  sit exactly halfway between their upper and lower bounds, ensuring consistent matching among the tensor terms listed in {\bf Tables~\ref{ccrani212}} through {\bf \ref{ccrani1414}}. These results confirm that all such fluxes remain dynamically evolving in time, with their $g_s$ dependence tightly constrained. The overall scaling pattern continues the trend established earlier: components with additional time or spatial legs are systematically more suppressed. As before, $\theta_{nl}$ stays fixed at ${8\over 3}^\pm$.

\paragraph*{Bianchi identities for ${\bf G}_{abij}$, ${\bf G}_{{\rm M}aij}$ and ${\bf G}_{{\rm MN}ij}$.}
The final set of fluxes, carrying two spatial indices, confirms the universal midpoint rule. Here all fluxes are free from five-brane effects, and their exact scalings are determined by comparing the flux, curvature, and quantum pieces. The corresponding exponents for the three flux components ${\bf G}_{abij}$, ${\bf G}_{{\rm M}aij}$, and  ${\bf G}_{{\rm MN}ij}$ may be easily read from {\bf Tables \ref{ccrani215}} through {\bf \ref{ccrani1717}}. These values precisely bisect the ranges obtained from the EOMs, ensuring that all derivative contributions  align between leading and next-to-leading orders. As illustrated in {\bf figure~\ref{fluxbehavior2}}, every flux component occupies its half-way scaling point, and positive exponents signal late-time decoupling as $\bar{g}_s \to 0$. The consistency of $\theta_{nl} = {8\over 3}^\pm$ throughout all tables reinforces that the quantum sector remains uniformly controlled across all flux configurations.

\paragraph*{Conclusion and synthesis.}
Collectively, these subsections establish that the Bianchi identities, together with the Schwinger–Dyson equations, determine the precise scaling hierarchy of all G-flux components. Each flux sits at a half-way scaling point within its allowed range (as summarized in {\bf figure \ref{fluxbehavior2}}), a result that guarantees smooth matching between curvature, quantum, and flux terms across all tensor structures. The positive exponents indicate late-time suppression, and both M2- and M5-brane sources decouple dynamically in this limit. The residual $\pm$ superscripts encode small curvature-induced corrections but do not affect the overall consistency, as confirmed by the curvature tables and the trace relations involving ${\rm tr}~\mathbb{R}_{\rm tot}\wedge\mathbb{R}_{\rm tot}$. Ultimately, the combined EOM and Bianchi analyses show that all fluxes evolve coherently and that the system remains anomaly-free and well-behaved both perturbatively and non-perturbatively at late times.

\subsubsection{Summary of section \ref{secanomaly}}

In this section we explain how to keep track of fluxes and anomalies in a time-dependent string/M-theory setup without breaking consistency. First, it shows that although a certain curvature quantity $ \mathbb{X}_8 $ can vary point-by-point in time, its total over a smooth compact space does not change—except when there are orbifold boundaries, which localize the time-variation to those boundaries. With that clarified, the section then lays out how to quantize the \emph{global} pieces of all relevant 4-form flux components even when everything is time-dependent and quantum corrections are present: the quantum effects are absorbed by \emph{localized} flux pieces so that the global pieces still obey the familiar Witten quantization rules, with possible shifts counted by dynamical five-branes. Finally, it shows how anomalies cancel: for M2-branes, a specific combination of flux and curvature terms balances; for M5-branes, splitting the 4-form flux into global and local parts lets the local piece soak up quantum corrections, while the global piece reduces (through dualities) to the standard heterotic Bianchi identity with torsion, reproducing the usual anomaly-cancellation formula.

\subsection*{Revisiting time-dependent $\mathbb{X}_8$ and the Chern–Weil transgression}
In section \ref{jesskibhaluu} we ask whether the \emph{total} of the curvature 8-form $ \mathbb{X}_8 $ changes in time when geometry is time-dependent. Using Chern–Weil transgression, the time-derivative satisfies $ \partial_t \mathbb{X}_8 = d\Theta_7 $, {\it i.e.}, it is a pure boundary term. Hence on a smooth closed 8-manifold, $ \int \mathbb{X}_8 $ is time-independent. In our setups there are effective boundaries from orbifold singularities (e.g., Horava–Witten walls or O7-planes), so any time-variation is localized there. Possible singularities from moving M2-branes do not add new contributions: after excising small spheres around them, the associated Chern–Simons integrals vanish. Conclusion: only orbifold-type boundaries matter for the time-variation of $ \int \mathbb{X}_8 $, thus fitting consistently within out set-up.

\subsection*{Flux quantizations for the global fluxes over internal manifolds}
In section \ref{kolamey} we show that to define flux quantization with time-dependent backgrounds and quantum corrections, we need to carefully split the 4-form flux into \emph{global} (bulk) and \emph{localized} parts. Quantum corrections deform local equations, but the localized pieces can be chosen to absorb these corrections so that the global fluxes still satisfy time-independent Witten-type quantization, possibly shifted by dynamical five-branes. This is worked out for three classes:

\vskip.1in

\noindent \textbf{Flux quantization for $ {\bf G}_{\rm MNPQ} $}: In section \ref{sutton1} we argue that integrating the Bianchi identity over a 5D region shows quantum terms can be traded for localized flux. The global integral equals a curvature term plus five-brane contributions, giving Witten-like quantization with brane shifts.

\vskip.1in
  
\noindent \textbf{Flux quantization for $ {\bf G}_{{\rm MNP}a} $}: With one leg on the torus fiber, wrapped 1-cycles are globally sensible because the torus is fibered in a non-K\"ahler space. In section \ref{sutton2} we show that the same cancellation logic applies, yielding Witten-like quantization with appropriate five-brane counts.

\vskip.1in

\noindent \textbf{Flux quantization for $ {\bf G}_{{\rm MN}ab} $}: Time-dependence is required for EFT consistency, but leading scalings of fluxes and five-brane sources match so localized pieces again absorb quantum terms. The global flux remains quantized with five-brane shifts as we demonstrate in section \ref{sutton3}.

\vskip.1in

\noindent In all cases, localized two-form fluxes themselves are \emph{not} generically quantized (they typically map to abelian combinations on the heterotic side).

\subsection*{Anomaly cancellations with global fluxes, branes and curvatures}
In a time-dependent background of the kind that we study here, it is important to show how the anomalies get canceled. This is important for the underlying consistency of our whole construction. In section \ref{suttonjen} the 
global consistency is checked by showing anomalies cancel in the following two ways.
\vskip.1in
\noindent \textbf{With dynamical M2-branes and fluxes:} For equations involving “electric” flux components, we show in section \ref{suttonjen1} that uniform scalings make time-dependence drop out once integrated over the internal space. One obtains a time-independent balance between a flux bilinear and a specific curvature polynomial, mirroring standard anomaly-cancellation relations.
\vskip.1in
\noindent \textbf{With dynamical M5-branes and curvatures:} Starting from the 4-form Bianchi identity and splitting $ {\bf G}_4 $ into global/local parts, we demonstrate in section \ref{suttonjen2} that the local piece matches quantum corrections while the global piece couples to curvature and five-brane sources. Through dualities to the heterotic theory, the global part becomes the heterotic three-form $ \mathcal H $ and the curvature uses a torsional connection $ \omega_+ $, yielding the familiar identity $ d\mathcal H = \tfrac{\alpha'}{4}\big(\mathrm{tr}\,{\bf R}_+\wedge {\bf R}_+ - \tfrac{1}{30}\mathrm{tr}\,{\bf F}\wedge {\bf F}\big) $, solved iteratively in $ \alpha' $.

\section*{Step-by-step flow of the full section}

\begin{enumerate}
  \item \textbf{Scope:} Work in the zero-instanton sector to analyze flux quantization and anomaly cancellation using prior curvature scalings and Bianchi identities.
  \item \textbf{Question:} Does the total $ \int \mathbb{X}_8 $ change in time?
  \item \textbf{Key fact:} Chern–Weil transgression gives $ \partial_t \mathbb{X}_8 = d\Theta_7 $.
  \item \textbf{Consequence:} On smooth compact 8-manifolds without boundary, $ \int \mathbb{X}_8 $ is time-independent.
  \item \textbf{Localization:} In our setups, orbifold singularities act as boundaries; time-variation lives on those loci.
  \item \textbf{M2 subtlety:} Excision shows moving M2-branes do not contribute to $ \partial_t \!\int \mathbb{X}_8 $.
  \item \textbf{Flux split:} Decompose $ {\bf G}_4 = {\bf G}_4^{(\mathrm{global})} + {\bf G}_4^{(\mathrm{local})} $.
  \item \textbf{Quantization goal:} Impose time-independent, Witten-type quantization on \emph{global} fluxes.
  \item \textbf{Mechanism:} Let \emph{localized} fluxes absorb quantum-correction series so global quantization survives.
  \item \textbf{Case A ($ {\bf G}_{\rm MNPQ} $):} Integrate the Bianchi identity; localized terms cancel quantum pieces; global integral equals curvature plus five-brane numbers $\Rightarrow$ Witten-like quantization with shifts.
  \item \textbf{Case B ($ {\bf G}_{{\rm MNP}a} $):} Repeat with one fiber leg; fibered 1-cycles are allowed; same conclusion with appropriate five-brane counts.
  \item \textbf{Case C ($ {\bf G}_{{\rm MN}ab} $):} Time-dependence required but leading scalings match; localized terms remove quantum pieces; global flux quantized with five-brane shifts.
  \item \textbf{Note:} Localized two-form fluxes are generally not quantized (abelian heterotic combinations).
  \item \textbf{M2 anomalies:} Integrate the EOMs; uniform scalings remove time-dependence; obtain a curvature–flux balance like standard anomaly cancellation.
  \item \textbf{M5 anomalies:} Start from the Bianchi identity; split into global/local; local matches quantum corrections; global couples to curvature and five-branes.
  \item \textbf{Duality to heterotic:} Map global piece to $ \mathcal H $; use torsional connection $ \omega_+ $.
  \item \textbf{Final identity:} Recover $ d\mathcal H = \tfrac{\alpha'}{4}\big(\mathrm{tr}\,{\bf R}_+\wedge {\bf R}_+ - \tfrac{1}{30}\mathrm{tr}\,{\bf F}\wedge {\bf F}\big) $ (to be solved order-by-order in $ \alpha' $).
  \item \textbf{Takeaway:} Despite time-dependence and quantum corrections, the global quantization and anomaly-cancellation structures persist once fluxes are split into global and localized parts and orbifold-localized effects are handled carefully.
\end{enumerate}

\subsection{Summary of section \ref{secmetric}}

This is a crucial section which not only ties up all the loose ends, but also provides answers to the questions like: how are all the on-shell and cross-term metric EOMs satisfied? What causes our excited states to {\it inflate}? How is the dynamical nature of dark energy manifested? How is the equation of state (EoS) satisfied? And other related topics. 

\subsubsection{Summary of section \ref{eomgmn}}

In this section we test what really drives the ``Einstein equation'' for the internal metric $g_{mn}$ in our emergent, off-shell framework. Ordinary ingredients---fluxes, branes, orientifolds, and all perturbative quantum corrections---cannot supply the leading-order stress-energy needed (this echoes classic no-go results). The first effects that \emph{do} match the required scalings are specific non-perturbative M5-brane instantons (BBS type \cite{bbs}) wrapping the full internal six-manifold; they induce quartic-in-curvature terms that naturally support accelerated expansion. Non-local versions of these instantons generate even higher-derivative terms but are strongly suppressed. By contrast, KKLT-type five-brane instantons \cite{kklt} on four-cycles struggle to match the needed scalings unless one allows time-growing non-localities, which is physically unappealing. Net: quartic curvature terms from BBS instantons dominate and remain consistent with a finite, controllable EFT.

Therefore the main goal is to determine which ingredients can balance the Einstein tensor for the internal metric sector in the Schwinger--Dyson (SD) equations. All fields (metric, fluxes, fermions) are ``emergent'' in the sense that they are path-integral expectation values, so ``on-shell/off-shell'' refers to satisfying the SD equations rather than classical equations of motion. A careful bookkeeping of how every contribution scales with the effective coupling $\bar g_s$ shows:

\begin{itemize}
\item \textbf{Fluxes/branes/O-planes at leading order:} Their stress-energy scales too differently to appear on the right-hand side at the same order as the Einstein tensor; this mirrors standard no-go theorems that such ingredients alone cannot sustain positive energy needed for de Sitter--like behavior.
\item \textbf{Perturbative quantum series (local and non-local counterterms):} Matching the Einstein tensor would require a specific exponent $\theta_{nl}$ from \eqref{botsuga} and \eqref{brittbaba007}; comparing with the allowed menu of scalings forces a particular selection that actually kills all nontrivial perturbative candidates at leading order. Even when non-local kernels are included, staying in the perturbative regime leaves no solution.
\item \textbf{Non-perturbative M5 (BBS) instantons on the six-manifold:} These \emph{do} shift the scaling just right. They generate {quartic curvature} terms ${\bf R}^4$ (and certain derivative-dressings) that land precisely where the Einstein tensor lives in the hierarchy. The set of allowed corrections is {finite} (countable) because the integer labels that control them are restricted. Non-local BBS contributions can also appear, producing higher-order curvature structures (e.g., ${\bf R}^5, {\bf R}^7$ and mixed ${\bf R}$--flux--derivative terms), but these are Planck-suppressed and subleading relative to the quartic piece.
\item \textbf{Instanton ``charges'' and consistency:} The Chern--Simons (charge) sector of BBS instantons receives small, $\bar g_s$--dependent renormalizations (tiny deviations from the Born--Infeld scaling). These do {not} enter stress-energy (being topological) but do feed correctly into Bianchi identities; localized fluxes absorb the resulting charges, so consistency is maintained.
\item \textbf{KKLT-type five-brane instantons on $\Sigma_4\times \mathbb{T}^2$:} Their volume scalings bring in a ``torus'' exponent $\hat\eta_e$ that does not appear in the curvature sector at the needed order, spoiling the match unless one invokes non-localities that {grow with time}---a behavior we want to avoid. Hence they are not the leading drivers here.
\item \textbf{Other/exotic instantons:} Space-filling M2 configurations or eight-/seven-brane--like exotic objects typically mis-match the required Lorentz structure or come in at much higher effective order; they cannot lead the dynamics.
\item \textbf{Including exponential suppressions:} Treating the instanton exponent systematically (via a Lambert-$\mathbb{W}$ solution) shows the dominant net scaling remains the BBS-driven value; the exponential only seeds a controlled perturbative tower around it and is therefore sub-dominant.
\end{itemize}

\noindent\textbf{Bottom line:} For $g_{mn}$, the only robust leading-order source compatible with the SD equations is the {BBS instanton--induced quartic curvature sector}. Everything else is either too weak, mismatched, or safely subleading.

\subsection*{Step-by-step flow for the section}
\begin{enumerate}
\item \textbf{Set the rule of the game:} ``On/off-shell'' means ``satisfies SD equations.'' Identify the dominant $\bar g_s$ scaling of the Einstein tensor for $g_{mn}$.
\item \textbf{Test classical sources:} Insert flux/brane/O-plane stress-energy; compare their scalings to the Einstein tensor. {Result:} too suppressed $\Rightarrow$ cannot balance (aligned with no-go theorems).
\item \textbf{Scan perturbative quantum terms:} Evaluate both local and non-local counterterms. Matching the needed exponent forces a specific selection that {kills all nontrivial perturbative contributions} at leading order.
\item \textbf{Turn on BBS instantons (local):} Their worldvolume action shifts the exponent exactly as needed. The induced {${\bf R}^4$} (plus allowed derivative dressings) is the \emph{first viable} right-hand side.
\item \textbf{Account for exponentials carefully:} Solve the instanton exponential with a Lambert-$\mathbb{W}$ approach; the dominant piece remains the BBS-matching power, with a controlled expansion around it.
\item \textbf{Consider non-local BBS corrections:} These introduce higher-order curvature (e.g., ${\bf R}^5, {\bf R}^7$) and mixed terms, but all are {Planck-suppressed} and {subleading} to the quartic term.
\item \textbf{Check KKLT-type instantons:} Their four-cycle volumes inject the torus exponent $\hat\eta_e$, which {doesn’t appear} in the curvature sector at the required order. Only by allowing {time-growing non-localities} (undesirable) could they contribute; we discard as leading drivers.
\item \textbf{Survey exotic instantons:} Space-filling M2 or 7/8-brane--type effects mis-match Lorentz structure or enter at too high order; they don’t lead.
\item \textbf{Consistency and charges:} Small renormalizations of instanton ``charges'' are topological and routed through Bianchi identities; localized fluxes absorb them without spoiling the EFT.
\item \textbf{Conclusion:} {BBS-induced quartic curvature terms are the dominant, controlled source} that balances the SD equation for $g_{mn}$; everything else is subleading or inconsistent.
\end{enumerate}

\subsubsection{Summary of section \ref{eomgbeta}}

Repeating the analysis for the remaining metric blocks (two-cycle, torus, and spacetime pieces) leads to the same qualitative picture with different numerical exponents: fluxes and perturbative effects still fail at leading order; BBS instantons again provide the first viable, controlled source via quartic curvature terms; non-local/higher-order pieces are subleading; KKLT-type instantons face the same mismatch issues. The Einstein tensors for these blocks scale differently, but the matching pattern---and the conclusion that BBS-driven quartic curvature dominates---is unchanged. The same SD matching is repeated for the remaining metric blocks, each with its own leading Einstein-tensor scaling:

\begin{itemize}
\item \textbf{$g_{\alpha\beta}$ (two-cycle block):} Einstein equation scales like $0^{\pm}$. Flux stress-energy scales too high; perturbative terms reduce again to the trivial selection that doesn’t contribute at leading order; {BBS instantons} supply the right shift, again pointing to {quartic curvature} dominance.
\item \textbf{$g_{ab}$ (torus block):} Einstein scales like $2^{\pm}$. Flux terms start at $4^{\pm}$ and thus cannot balance. The perturbative menu still collapses to the trivial choice; BBS instantons restore the match, again through quartic curvature structures.
\item \textbf{$g_{\mu\nu}$ (spacetime block):} Einstein scales like $-2^{\pm}$. Fluxes sit at $0^{\pm}$, so they are again too high to balance the leading left-hand side. The perturbative sector fails for the same reason as before. Non-perturbative BBS effects remain the first consistent source, with higher-order non-local pieces subleading.
\end{itemize}

\noindent Across all blocks: KKLT-type instantons run into the same $\hat\eta_e$--driven mismatch; only BBS instantons reliably supply the leading, consistent stress-energy via quartic curvature terms. The EFT remains controlled (finite tower, Planck suppressions), and the leading driver of acceleration is universal.


\subsection*{Step-by-step flow for the section}
\begin{enumerate}
\item \textbf{Map the targets:} Read off the leading scalings of the Einstein tensors for each block: $g_{\alpha\beta}\sim 0^{\pm}$, $g_{ab}\sim 2^{\pm}$, $g_{\mu\nu}\sim -2^{\pm}$.
\item \textbf{Flux sanity check:} Compare flux stress-energy scalings to each block’s Einstein scaling. {Outcome:} fluxes are too high to appear at leading order in all cases.
\item \textbf{Perturbative scan:} As with $g_{mn}$, the only consistent selection annihilates nontrivial perturbative contributions at leading order, even with non-local kernels.
\item \textbf{Switch on BBS instantons:} For each block, the same exponent shift appears; the {first consistent right-hand side} is {quartic curvature} on the instanton worldvolume.
\item \textbf{Include exponentials/non-localities:} Exponentials leave the dominant power intact; non-local BBS pieces add higher-order, {subleading} corrections.
\item \textbf{Re-check KKLT instantons:} The $\hat\eta_e$ dependence again prevents a clean match at leading order unless one admits time-growing non-localities; we exclude them as primary drivers.
\item \textbf{Close the loop:} Across all blocks, the {unified leading mechanism} is the {BBS-induced ${\bf R}^4$} sector; higher-order terms are finite and suppressed, keeping the EFT controlled and predictive.
\end{enumerate}

\noindent\textbf{One-line takeaway:} Throughout the metric SD equations, \emph{only} BBS M5-brane instantons reliably generate the leading, properly scaled {quartic curvature} stress-energy required to balance the Einstein tensors; all perturbative pieces are red herrings, KKLT-type instantons are mismatched at leading order, and any higher-order/non-local effects are safely subleading.

\subsubsection{Summary of section \ref{counting}}

\noindent
This subsection investigates how many consistent combinations of terms can appear when balancing the contributions from curvature and derivative sectors against those from quantum corrections and instanton effects. At first, the analysis assumed a specific simplification ($l_{14} = 1$ in \eqref{brittbaba007}), which gave a one-to-one match between the curvature terms and the quantum series. The present goal, however, is to check whether other possible consistent combinations exist beyond this simple case. For this, a general balance equation, eq.~\eqref{melisrain}, is introduced. It links seventeen integer variables $(x_1, x_2, \dots, x_{17})$—each representing a curvature or derivative contribution—to several possible target expressions on the right-hand side. These targets depend on small parameters 
$(\hat{\alpha}_e(t), \hat{\beta}_e(t), \hat{\sigma}_e(t), \hat{\zeta}_e(t))$ that slightly shift the scaling of different sectors. The task is then to find all sets of integer tuples that satisfy the equation for every possible right-hand side. Each valid tuple represents one consistent “embedding” of instanton contributions into the overall structure. To make the counting manageable, related groups of $x_i$’s and flux components $l_p$ are combined into fewer effective quantities:
\begin{align}
{\rm B} &= \sum_{i=2}^4 x_i + x_{17}, &
{\rm C} &= \sum_{j=5}^9 x_j + \sum_{l=11}^{16} x_l, &
{\rm L} &= \sum_{p=42}^{81} l_p,
\end{align}
so that the master equation simplifies to
\begin{equation}
4x_1 + 5{\rm B} + 2{\rm C} + 8x_{10} + {\rm L} = {\rm N},
\end{equation}
where ${\rm N}=8$ or ${\rm N}=14$ for the relevant instanton configurations. 
Solving for non-negative integers gives ${\bf 12}$ possible solutions for ${\rm N}=8$ and ${\bf 35}$ for ${\rm N}=14$, establishing that the number of allowed combinations is finite. Interestingly, when the small correction parameters vanish, the total number of allowed $17$-tuples is large: $1069$ for ${\rm N}=8$ and $23715$ for ${\rm N}=14$. 
However, once the small parameters 
$(\hat{\zeta}_e(t), \hat{\sigma}_e(t), \hat{\alpha}_e(t), \hat{\beta}_e(t), \hat{\eta}_e(t))$
are restored, almost all of these are ruled out, leaving only a handful of consistent possibilities. In fact, after analyzing all the possible right-hand-side values of eq.~\eqref{melisrain}, the surviving solutions show a repeating pattern:
\begin{align}
x_5 = x_6 = 1, \qquad x_{11}=2, \qquad x_i = 0 \text{ for } i\ne(5,6,11),
\end{align}
and small variants thereof, depending on which row of \eqref{melisrain} is considered. 
These nearly unique $17$-tuples match the earlier results of eqs.~\eqref{isaferr} and \eqref{isaferr2}. Thus, what begins as a large combinatorial problem—with potentially tens of thousands of possibilities—collapses to almost a single, physically consistent configuration once all the scaling constraints and small corrections are applied.

\subsubsection*{Step-by-step flow for this part}

\begin{enumerate}
\item \textbf{Motivation:}  
Earlier analyses assumed $l_{14}=1$, ensuring a simple one-to-one mapping.  
Here, the goal is to test if other consistent solutions may exist.

\item \textbf{Setting up the master equation:}  
Equation~\eqref{melisrain} connects all curvature and derivative scalings ($x_i$) to possible target values, each modified by small time-dependent corrections.

\item \textbf{Definition of the $x_i$:}  
Each $x_i$ corresponds to a grouped term from the quantum expansion.  
The challenge is to find which $17$-tuples of integers make the equality hold.

\item \textbf{Special case — BBS instantons:}  
For $m_1=1, m_2=0$, the system reduces to eq.~\eqref{falooda}, giving simpler numerical constraints that depend only on two possible totals ($N=8,14$).

\item \textbf{Grouping the variables:}  
The $17$ variables are reduced to a $5$-tuple $(x_1, {\rm B}, {\rm C}, x_{10}, {\rm L})$ through suitable combinations, yielding a compact equation:
$$
4x_1 + 5{\rm B} + 2{\rm C} + 8x_{10} + {\rm L} = {\rm N}.
$$

\item \textbf{Counting the integer solutions:}  
There are ${\bf 12}$ solutions for ${\rm N}=8$ and ${\bf 35}$ for ${\rm N}=14$, ensuring only finitely many options.

\item \textbf{Including small parameters:}  
Without corrections, thousands of possible tuples exist.  
With small parameters switched on, nearly all are eliminated, leaving very few consistent patterns.

\item \textbf{Checking each right-hand side row:}  
Each choice of the RHS in eq.~\eqref{melisrain} gives one or two possible $17$-tuples.  
Most are discarded if inconsistent with previous constraints, leaving one dominant pattern.

\item \textbf{Consistency check:}  
The surviving $17$-tuples perfectly match the earlier forms in eqs.~\eqref{isaferr} and \eqref{isaferr2}, confirming internal consistency.

\item \textbf{Final outcome:}  
The intricate algebraic network with initially many solutions collapses to an almost unique configuration due to the small correction parameters, ensuring that only physically meaningful instanton embeddings remain.
\end{enumerate}

\noindent
In our next stage, we study the BBS instanton case \emph{with} non-local effects by setting $m_1=m_2=2$ and writing the non-locality as
$a_{\rm F}=-|a_{\rm F}|=-{x_{18}\over 3}$. We begin with the top row of the master balance equation and ask how many non-negative
$18$-tuples $(x_1,\dots,x_{18})$ solve it. Interestingly, when the small time-dependent parameters are turned off
$(\hat\zeta_e,\hat\alpha_e,\hat\beta_e,\hat\sigma_e)=(0,0,0,0)$, the minus sign in front of $x_{18}$ permits
\emph{infinitely many} solutions. Turning the small parameters on introduces coefficient-matching constraints that bound many
$x_i$'s from above and collapses the infinite family to exactly \textbf{24} explicit solutions. In these, $x_{18}=3|a_{\rm F}|\in\{0,1,2\}$,
so the non-locality scales at most like $\bar g_s^{-2/3}$. More importantly, the pertinent question is the {\it source} of the bounds, {\it i.e.} where do the bounds come from? Following intuitive discussion helps us to understand this.
\begin{itemize}
  \item Matching the constant term yields a Diophantine relation with $-2x_{18}$; this is the source of the infinite branch when the small parameters vanish. If we hypothetically flip the sign to $+2x_{18}$, all terms become non-negative and each grouped variable can be bounded, yielding finitely many options (for instance, $71$ for the compressed $5$-tuple in that thought experiment).
  \item Matching $\hat\zeta_e$ produces a linear relation that caps several variables at $1$ or $2$ (specifically $x_1,x_3,x_4,x_7,x_8,x_9,x_{17}\le 2$ and $x_2,x_{10},x_{14}\le 1$).
  \item Matching $\hat\sigma_e$ further constrains the system (for example $x_{11}\le 4$) and isolates a small block to determine.
  \item Matching $\hat\alpha_e$ and $\hat\beta_e$ yields two linear relations whose sum/difference imply $(x_{12},x_{13}$, $x_{15},x_{16})\le 2$. Substituting these bounds back into the constant-term relation also bounds $x_{18}$.
\end{itemize}
These tiny parameters therefore \emph{regularize} the counting and restore EFT control despite the presence of a non-local term.
{We can then look at the other rows of the master equation} and repeat the analysis for the remaining six RHS rows:
\begin{itemize}
  \item When $\hat\zeta_e$ is absent on the LHS (two successive rows), the $\hat\zeta_e$-matching forces a block of $x_i$'s to vanish, simplifying the system. We find $5$ solutions in one case and $11$ in the next; swapping $\hat\alpha_e\leftrightarrow\hat\beta_e$ yields another $11$. In all of these, $x_{18}=0$.
  \item For the middle row, $\hat\zeta_e$ \emph{is} present on the LHS; we obtain $25$ solutions and allow $x_{18}\in\{0,2,3,4\}$, so the dominant non-local scaling can be as strong as $\bar g_s^{-4/3}$. This indicates a pattern: when $\hat\zeta_e$ appears on both sides of the comparison, non-zero $x_{18}$ can occur.
  \item For the two rows without $\hat\zeta_e$, we find $9$ solutions each, again with $x_{18}=0$.
\end{itemize}

\noindent
Putting everything together now shows an interesting pattern which we depict in {\red red} across all rows. From each row’s list, exactly one highlighted solution satisfies \emph{all} equations simultaneously. Collectively, these have
$x_5=x_6=2$ \text{ (always)}, 
$x_{11}=2$ \text{ (in all but one case)},
$x_{18}=0$ \text{ (always)},
and exactly one of $x_{10},x_{11},x_{12},x_{13},x_{14},x_{15},x_{16}$ equals $1$ (a single ``1'' distributed among those slots depending on the row). The single ``1'' implies $l_{14}=1$. The repeated ``2''s in $x_5=x_6=x_{11}$ imply $\sum\limits_{q=29}^{33} l_q=4$ and the remaining 
realized either with
$\sum\limits_{i=1}^{13}l_i=2$ or $n_2=4$.
Hence the non-local BBS instanton contributions that accelerate the universe come with:
\begin{equation}
a_{\rm F}=0, \quad
\big(l_{14}=1,\ \sum_{q=29}^{33}l_q=4,\ \sum_{i=1}^{13}l_i=2\big)
\ \text{or}\
\big(l_{14}=1,\ \sum_{q=29}^{33}l_q=4,\ n_2=4\big).
\end{equation}
This confirms earlier counting and shows the non-locality factor is $\bar g_s$-independent in all consistent solutions. One may also do {a sign flip test.} Re-running the middle row with a \emph{plus} sign in front of $x_{18}$ yields $19$ solutions, all with $x_{18}=0$. Repeating for the other rows gives the same result. Thus, even with the plus sign, consistent solutions have $x_{18}=0$. The actual contributors are septic-order curvature terms on the instanton world-volume.

\subsubsection*{Step-by-step flow for this part}

\begin{enumerate}
  \item \textbf{Set the case.} We choose BBS instantons with non-localities: $m_1=m_2=2$, $a_{\rm F}=-x_{18}/3$, and start from the top row of the master equation.
  \item \textbf{Switch off small parameters.} With $(\hat\zeta_e,\hat\alpha_e,\hat\beta_e,\hat\sigma_e)=(0,0,0,0)$, the $-\tfrac{2}{3}x_{18}$ term produces \emph{infinitely many} solutions.
  \item \textbf{Switch on small parameters.} Matching the coefficients of $\hat\zeta_e,\hat\sigma_e,\hat\alpha_e,\hat\beta_e$ and the constant term imposes upper bounds on many $x_i$'s and makes the solution set finite.
  \item \textbf{Extract concrete bounds.} From $\hat\zeta_e$-matching we bound $x_1,x_3,x_4,x_7,x_8,x_9,x_{17}\le 2$ and $x_2,x_{10},x_{14}\le 1$; from $\hat\sigma_e$ we get $x_{11}\le 4$; from $\hat\alpha_e,\hat\beta_e$ we find $(x_{12},x_{13},x_{15},x_{16})\le 2$; together these also bound $x_{18}$.
  \item \textbf{Top-row outcome.} The infinite set collapses to \textbf{24} explicit solutions with $x_{18}\in\{0,1,2\}$; non-locality scales at most as $\bar g_s^{-2/3}$.
  \item \textbf{Rows without $\hat\zeta_e$ on the LHS.} The $\hat\zeta_e$-match forces a block of variables to vanish. We find $5$ solutions in one case and $11$ in the next; swapping $\hat\alpha_e\leftrightarrow\hat\beta_e$ gives another $11$. In all these, $x_{18}=0$.
  \item \textbf{Middle row with $\hat\zeta_e$ present.} We obtain \textbf{25} solutions with $x_{18}\in\{0,2,3,4\}$, allowing non-local scaling up to $\bar g_s^{-4/3}$. This suggests that $\hat\zeta_e$ on both sides correlates with non-zero $x_{18}$.
  \item \textbf{Remaining rows without $\hat\zeta_e$.} We find $9$ solutions each, again with $x_{18}=0$.
  \item \textbf{Select global solutions.} From each row, exactly one highlighted solution satisfies \emph{all} equations. Collectively we have
  $x_5=x_6=2$ always, $x_{11}=2$ all but one, $x_{18}=0$ always
    and exactly one of $x_{10}\text{--}x_{16}$ equals $1$.
  \item \textbf{Map to series data.} The single ``1'' implies $l_{14}=1$; the $2$'s imply $\sum\limits_{q=29}^{33}l_q=4$, realized either via $\sum\limits_{i=1}^{13}l_i=2$ or $n_2=4$. Thus the contributing non-local BBS terms that accelerate the universe satisfy
  \begin{equation}
  a_{\rm F}=0,\quad
  (l_{14}=1,\ \sum_{q=29}^{33}l_q=4,\ \sum_{i=1}^{13}l_i=2)\ \text{or}\ (l_{14}=1,\ \sum_{q=29}^{33}l_q=4,\ n_2=4).
  \end{equation}
  \item \textbf{Flip the $x_{18}$ sign (check).} Replacing $-\tfrac{2}{3}x_{18}$ by $+\tfrac{2}{3}x_{18}$ in the middle row yields $19$ solutions, all still with $x_{18}=0$; the same happens for the other rows.
  \item \textbf{Conclusion.} The small parameters enforce EFT-friendly finiteness and drive us to unique, consistent patterns with \emph{no} time-dependent non-local factor ($a_{\rm F}=0$). The actual contributions arise from septic-order curvature terms on the instanton world-volume.
\end{enumerate}

\noindent
In our final stage we first examine whether KKLT-type five-brane instantons---wrapping the circle $\xoxo$ and suitable four-cycles---can satisfy the master scaling balance (cf.\ \eqref{melisrain3}) and contribute acceptable quantum terms in the curvature/derivative sector that support late-time acceleration. We immediately face an initial obstruction. A parameter $\hat\eta_e(t)$ appears on the RHS but not on the LHS of the balance. With $\hat\eta_e(t)>0$ and/or $a_{\rm F}>0$, we find no solutions. Even the choice $\hat\eta_e(t)=a_{\rm F}=0$ does not yield solutions in the curvature/derivative sector. Solutions arise only if we set:
\begin{equation}
\hat\eta_e(t)=0, 
\qquad 
a_{\rm F}=-\vert a_{\rm F}\vert \equiv -\frac{x_{18}}{3}<0,
\end{equation}
i.e.\ we allow a \emph{negative} non-locality exponent so that the non-local factor grows with conformal time. This is a bit unsatisfactory, but we continue with the analysis to see what's the possible outcome once we perform
{row-by-row enumeration to look for consistent subsets.} In fact working through the rows of the balance relations and scanning non-negative 18-tuples $(x_1,\dots,x_{18})$, we obtain finite sets of candidates. From each set, we identify a small subset of globally consistent tuples that simultaneously satisfy all constraints. Two embeddings are relevant:

\begin{itemize}
  \item \textbf{Embedding A:} Five-branes on $\xoxo\times\mathcal{M}_4$. The consistent tuples imply
  \begin{equation}
  a_{\rm F}=-\frac{4}{3},
  \qquad
  \big(l_{14}=1,\ \sum_{i=1}^{13} l_i=4\big)
  \ \ \text{or}\ \ 
  \big(l_{14}=1,\ n_2=8\big),
  \end{equation}
  corresponding schematically to quintic/septic structures such as $\mathbf{R}^5$ and $\square^4\mathbf{R}$.

  \item \textbf{Embedding B:} Five-branes on $\xoxo\times\mathcal{M}_2\times\boldsymbol{\Sigma}_2\subset\mathcal{M}_4$. The consistent tuples give
  \begin{equation}
  a_{\rm F}=-\frac{4}{3},
  \qquad
  l_{14}=1,
  \qquad
  \sum_{q=29}^{33} l_q=4,
  \end{equation}
  again indicating $\mathbf{R}^5$-type contributions.
\end{itemize}

\medskip
\noindent{There are however two caveats for the KKLT five-branes for both the aforementioned embeddings.}
\begin{enumerate}
  \item We must take $a_{\rm F}<0$ to make the counting work. Physically this means non-local effects grow in conformal time, unlike the BBS case where $a_{\rm F}=0$ (no time dependence).
  \item To get solutions we set $\hat\eta_e(t)=0$. However, duality chasing suggested a near-cancellation $\hat\zeta_e(t)+\hat\eta_e(t)\approx 0$. If we impose $\hat\eta_e(t)=-\hat\zeta_e(t)$ strictly in \eqref{melisrain3}, we generate a linear relation among the $x_i$'s that has no non-negative integer solution, collapsing the solution set. A softer interpretation is that the cancellation holds only in an intermediate IIB frame \emph{up to} small corrections, so working in M-theory with $\hat\eta_e(t)=0$ is not immediately inconsistent—but it constrains $\hat\zeta_e(t)$.
\end{enumerate}

\medskip
\noindent
{We can also study other instanton species that could potentially enter our set-up as viable candidates to balance the SD equations.}
\begin{itemize}
  \item \textbf{Space-time filling two-branes:} even with non-local terms, solutions force $x_1$ to be odd $(1,3)$, yielding Lorentz-violating operators; we discard them in the curvature/derivative sector.
  \item \textbf{Eight-branes on ${\rm Euc}(\mathbb{R}^3)\times\mathcal{M}_4\times\mathcal{M}_2$:} They are non-BPS states in the theory but have similar obstruction as above; acceptable (Lorentz-invariant) terms do not arise.
  \item \textbf{Seven-branes filling the internal eight-manifold:} These are also non-BPS states in the theory but with non-localities and $\hat\eta_e(t)=0$, highlighted tuples exist and point to quartic structures such as $\mathbf{R}^4$ and $\square\mathbf{R}^3$, summarized by
  \begin{equation}
  a_{\rm F}=-\frac{2}{3},
  \qquad
  \Big(\sum_{q=29}^{33} l_q=2,\ l_{14}=1,\ \sum_{i=1}^{13} l_i=1\Big)
  \ \ \text{or}\ \ 
  \Big(\sum_{q=29}^{33} l_q=2,\ l_{14}=1,\ n_2=2\Big),
  \end{equation}
  but these share the same two concerns: $a_{\rm F}<0$ and $\hat\eta_e(t)=0$.
\end{itemize}

\medskip
\subsection*{Step-by-step flow for the last part}

\begin{enumerate}
  \item \textbf{Question.} We ask whether KKLT five-brane instantons (on $\xoxo$ and appropriate four-cycles) can satisfy the balance equations and supply useful quantum terms.
  \item \textbf{Asymmetry.} Because $\hat\eta_e(t)$ appears only on one side, choices with $\hat\eta_e(t)>0$ and/or $a_{\rm F}>0$ admit no solutions. Even $\hat\eta_e(t)=a_{\rm F}=0$ fails.
  \item \textbf{Working window.} We set $\hat\eta_e(t)=0$ and allow $a_{\rm F}<0$. This opens up finite sets of solutions.
  \item \textbf{Enumeration.} We scan each row of the balance and list all non-negative 18-tuples $(x_1,\dots,x_{18})$ that solve it for two embeddings: $\xoxo\times\mathcal{M}_4$ and $\xoxo\times\mathcal{M}_2\times\boldsymbol{\Sigma}_2$.
  \item \textbf{Consistency filter.} From each row, we retain the highlighted tuples that also solve the other rows; these show a simple diagonal ``one-unit'' pattern across variables.
  \item \textbf{Parameter read-off.} Two possibilities arise. {\Su One},
$\xoxo\times\mathcal{M}_4$ with $a_{\rm F}=-\frac{4}{3}$ and either $(l_{14}=1,\ \sum\limits_{i=1}^{13}l_i=4)$ or $(l_{14}=1,\ n_2=8)$, implying $\mathbf{R}^5$, $\square^4\mathbf{R}$ type interactions. And {\Su two}, 
    $\xoxo\times\mathcal{M}_2\times\boldsymbol{\Sigma}_2$ with $a_{\rm F}=-\frac{4}{3}$,  and $l_{14}=1$, $\sum\limits_{q=29}^{33}l_q=4$,  implying $\mathbf{R}^5$ type interactions.
    \item \textbf{Caveats.} We must accept $a_{\rm F}<0$ (growing non-locality) and $\hat\eta_e(t)=0$ (tension with duality-motivated $\hat\zeta_e+\hat\eta_e\approx 0$ if enforced strictly).
  \item \textbf{Cross-check other instantons.} Two-branes and non-BPS eight-branes fail (Lorentz-violating or no acceptable operators). Non-BPS Seven-branes allow quartic structures but still require $a_{\rm F}<0$ and $\hat\eta_e(t)=0$.
  \item \textbf{Synthesis.} KKLT-type contributions exist but are less attractive than BBS: BBS instantons remain the dominant, clean source in our setting.
\end{enumerate}

\medskip
\noindent
{Our conclusion then is the following.} Across all objects examined, BBS instantons \cite{bbs} still provide the cleanest, dominant, EFT-friendly contributions (with $a_{\rm F}=0$ and consistent counting). KKLT five-brane instantons \cite{kklt} can contribute, but only with $\hat\eta_e(t)=0$ and $a_{\rm F}<0$, which are less appealing phenomenologically and under dualities. Other BPS or non-BPS instantons either do not contribute, or contribute sub-dominantly with similar unsatisfactory with $a_{\rm F}<0$.

\subsubsection{Summary of section \ref{grace}}

After working out the “on-shell” (physical) parts of the metric, in this section we turn towards the cross-term metric components. These are “off-shell” helpers: they don’t survive in the final description because we integrate them out, but they still obey equations of motion (EOMs) and they leave fingerprints in the effective theory. The goal for this section is basically to 
understand how these off-shell cross-terms contribute to the emergent energy-momentum tensor once non-perturbative effects and non-localities are included. To quantify this, 
we write the path integral with both the physical fields and the off-shell cross-terms, plus ghosts and a spacetime-dependent determinant factor (from the metric’s volume element). The energy-momentum tensor that showed up earlier covered only perturbative pieces. Here we upgrade to include non-perturbative (instanton-like) pieces and the non-local structures they generate. It turns out that there are 
two equivalent routes to the same effective action.
\begin{enumerate}
\item Start perturbatively, integrate out off-shell fields, then resum (Borel + non-local resummations).
\item New route: first promote the action to its full non-perturbative form (including off-shell fields), then integrate out off-shell fields.
\end{enumerate}
Both routes land on the same effective “on-shell” action, but the second route keeps non-perturbative structure visible from the start, which makes the bookkeeping cleaner here. The question then is what the non-perturbative source action looks like.
There are three perturbative series that sit around each non-perturbative saddle:
\begin{itemize}
    \item $Q^{(1)}$: depends only on on-shell fields.
    \item $Q^{(2)}$: on-shell fields contracted with cross-term metric pieces.
    \item $Q^{(3)}$: explicitly depends on at least two off-shell components.
\end{itemize}

\noindent Each series is multiplied by non-perturbative exponentials whose integrands include world-volume determinants (volumes of wrapped submanifolds) and absolute values of the $Q$’s. (Technical observation: 
a special “$\otimes$” marker reminds us that fluctuation determinants (the 1-loop corrections around each instanton) are distributed properly among single- and multi-instanton sectors.) What we now perform in detail is to vary the action w.r.t. a cross-term metric component. 
The variation pulls contributions from:
\begin{itemize}
    \item the 11D volume factor (standard $-\tfrac{1}{2}\,g\,\delta g$ structure),
    \item the $Q^{(2)}$ and $Q^{(3)}$ families (the only ones directly sensitive to off-shell metrics),
    \item and the non-perturbative integrals through their inner integrands (because the off-shell pieces also appear under those integrals).
\end{itemize}

\noindent The result naturally splits into parts that do or do not contain off-shell fields in the perturbative series. Importantly, the world-volume determinants for the wrapped instantons may themselves depend on cross-terms, so they can also feed the variation. Question is:
What does this buy us?
Inserting that variation back into the path integral and then integrating out the cross-terms produces an emergent non-perturbative energy-momentum tensor for the remaining (on-shell) fields. The 
key upshot is the following. As soon as the instanton sector tied to $Q^{(2)}$ is active, the emergent tensor is inherently non-local (because the relevant exponentials depend on integrated quantities that link separated points). In contrast, a purely $Q^{(1)}$-driven non-perturbative sector (with the other two turned off) can still generate local contributions for on-shell fields. This results 
to a clean mapping between “big” action pieces. 
The total non-perturbative source piece splits into: a part depending only on on-shell fields and a part that still involves off-shell fields.
The kinetic and ghost sectors split similarly. After integrating out off-shell fields, these pieces match the earlier effective non-local actions (including ghost non-localities), confirming internal consistency. Interestingly, even if we ignore extra determinant dressings, the instanton integral tied to $Q^{(2)}$ remains, and that integral couples distant points. Thus non-locality is unavoidable here. Therefore the emergent non-perturbative energy-momentum tensor generically carries non-local kernels. To illustrate the above story, we use a toy model to see the mechanism transparently. The model consists of 
replacing one cross-term metric component by a massless scalar $\phi$ (off-shell proxy), and on-shell metric bits by scalars $\psi, \chi$.
The path integral contains: a bulk coupling $f(\phi)\,Q_1(\psi)$ and a point insertion $e^{-\phi(y)\,Q_2(\chi(y))}$ (this mimics the instanton-induced factor evaluated at a point).

\paragraph{Case 1 (linear coupling) $f(\phi) = g\phi$:}
The integral is Gaussian. Integrating out $\phi$ produces the following interactions:
\begin{itemize}
    \item a bilocal kernel $\int Q_1(x)\,G(x-x')\,Q_1(x')$,
    \item a mixed non-local kernel $Q_2(y)\int G(y-x)\,Q_1(x)$,
    \item a local (removable) contact self-contraction at $y$.
\end{itemize}
This already shows how non-localities arise from eliminating $\phi$.

\paragraph{Case 2 (general polynomial) $f(\phi)=\sum a_n\,\phi^n/n!$:}
Linked-cluster (connected) expansion builds multi-point non-local kernels out of propagators $G$. UV self-contractions at the insertion are removed by normal ordering/renormalization, leaving the clean non-local kernels among the $Q$’s. The schematic Feynman-like diagrams in {\bf figure \ref{feynmanNL}} with squares and a marked point joined by lines, simply visualize these non-local couplings.

The bottom line is the following.
Cross-term (off-shell) metric components don’t disappear harmlessly: when we integrate them out with non-perturbative sectors turned on, they generate controlled non-local structures in the emergent energy-momentum tensor. Purely on-shell non-perturbative pieces can remain local, but once the $Q^{(2)}$ channel participates, non-locality is built-in. The two derivation routes (resummation-then-integrate vs. integrate-within-NP-completion) are equivalent and cross-check each other.

\section*{Step-by-step flow for this part}

\begin{enumerate}
\item \textbf{Identify the actors:} On-shell fields (physical), off-shell cross-term metric components (to be integrated out), ghosts, and determinant factors from the metric volume.

\item \textbf{State the task:} Write the full path integral including off-shell cross-terms and derive how they feed into the emergent energy-momentum tensor once non-perturbative effects are included.

\item \textbf{Choose the route:} Adopt the non-perturbative-first action (keeps instanton structure explicit), then integrate out off-shell fields. Note: this is equivalent to the previous perturbative-first route.

\item \textbf{Organize contributions:} Three perturbative series around instantons:
    \begin{itemize}
        \item $Q^{(1)}$: on-shell only,
        \item $Q^{(2)}$: on-shell times off-shell metric,
        \item $Q^{(3)}$: depends on multiple off-shell pieces.
    \end{itemize}
    Each comes with non-perturbative exponentials built from world-volume determinants and $|Q|$’s. The “$\otimes$” symbol enforces correct fluctuation-determinant assignments.

\item \textbf{Vary with respect to a cross-term:} Differentiate the action with respect to an off-shell metric component. Collect terms from the volume factor, from $Q^{(2)}$ and $Q^{(3)}$, and from the inner non-perturbative integrals. Note where dependence on cross-terms in sub-determinants appears.

\item \textbf{Insert back and integrate out:} Put the variation into the master path integral. Integrate over off-shell metrics to obtain the emergent non-perturbative energy-momentum tensor. Observe: if the $Q^{(2)}$ instanton channel is active, the result is non-local.

\item \textbf{Match to earlier split:} Split non-perturbative sources/kinetic/ghosts into “on-shell-only” and “still-off-shell” parts. After integration, these reproduce the earlier effective non-local actions, ensuring consistency.

\item \textbf{Key inference about locality:} Local non-perturbative terms exist for purely $Q^{(1)}$ sectors. Non-local terms are unavoidable once $Q^{(2)}$ (and often $Q^{(3)}$) contributes.

\item \textbf{Toy model confirmation:} Replace one cross-term by a massless scalar $\phi$, couple it linearly or polynomially to on-shell data $Q_1(\psi)$ and a point insertion $Q_2(\chi(y))$. The linear case (Gaussian) produces explicit bilocal and mixed non-local kernels via $G$, while the polynomial case builds multi-point non-local kernels after normal ordering.

\item \textbf{Conclusion:} Integrating out cross-term metrics in the presence of non-perturbative sectors naturally generates non-localities in the emergent energy-momentum tensor. The framework cleanly tracks where these non-local pieces come from (determinants, instanton integrals, and $Q^{(2)}/Q^{(3)}$ channels) and shows equivalence of the two derivation routes.
\end{enumerate}

\noindent In section \ref{pabondi}, we
connect the earlier toy-model lesson about non-locality to the actual cross-term Einstein equations for mixed metric components (the ones that couple different subspaces): ${\bf R}_{m\rho}$, ${\bf R}_{im}$, and ${\bf R}_{i\rho}$. The key idea is that when we eliminate (integrate out) those off-shell cross-terms in the presence of instantons, they generate non-local contributions to the effective energy-momentum tensor that sources these Ricci components. In the toy model discussed in section \ref{nolock}, the non-local kernel looked like a Green’s function $G(x-y)$. In the full theory, that kernel is more general: we write it as $\mathbb F(x,y)$ (not just a function of the distance $x-y$). This richer kernel is plugged into the non-perturbative structures used earlier. The complicated path-integral over cross-terms collapses into a structured object:
\begin{equation}
\mathbb A(\cdots)=\mathcal G[\text{perturbative series},\,\mathbb F]\times
\exp\{\mathcal H[\text{perturbative series},\,\mathbb F]\},
\end{equation}
where $\mathcal G$ and $\mathcal H$ are polynomial functions of the known perturbative series and the non-local kernel $\mathbb F$. Using this, the emergent non-perturbative energy-momentum tensor ${\bf T}^{\rm NP}_{\rm C'D'}$ gets a compact, trans-series-like form that is generically non-local. The equations of motion for the cross-terms take the clean form:
\bg\label{minerca}
{{\bf R}_{\rm C'D'}(\langle {\bf \Xi}\rangle_\sigma)=
{\bf T}^{\rm NP}_{\rm C'D'}(\langle {\bf \Xi}\rangle_\sigma)}.
\nd
The left-hand side (Ricci) is evaluated on the averaged on-shell background, and the right-hand side is the non-perturbative, non-local tensor coming from integrating out the cross-terms. The question then is how scalings and 18-tuples are fixed.
The non-local tensor’s scaling in $\bar g_s$ comes from:
\begin{enumerate}
  \item a purely on-shell series $\mathcal Q_1$ (positive powers only),
  \item the non-local polynomial/exponential dressing $\mathcal G,\mathcal H$ (positive powers only if expanded normally),
  \item the cross-term series $\mathcal Q^{(2)}_{\rm C'D'}$ (fixed by the perturbative scaling table),
  \item and the wrapped-cycle determinant $\sqrt{|{\bf g}_d|}$ (which can introduce the required negative powers).
\end{enumerate}
Matching the net scaling on both sides gives allowed 18-tuples (integer labels encoding which curvature/derivative monomials are active). In all solutions here, $x_{18}=0$ (no extra time-dependence from the non-local kernel). To see what happens for each cross-term Ricci component now, we need to look at the scalings carefully to see how the 18-tuples behave. The results are listed as follows.

\vskip.1in

\noindent \textbf{(1) ${\bf R}_{im}$ (mixing ${\bf R}^2$ with ${\cal M}_4$):}  
${\bf R}_{im}$ scales to zero on the left-hand side, so the right-hand side must do the same.  
We obtain three admissible 18-tuples (explicitly listed in the text), all with $x_{18}=0$.  
They correspond to quartic-in-curvature quantum corrections, the same pattern as in the on-shell sector.

\vskip.1in

\noindent \textbf{(2) ${\bf R}_{i\rho}$ (mixing ${\bf R}^2$ with ${\cal M}_2$):}  
For $\rho=\alpha$ and $\rho=\beta$, the Ricci tensor again scales to zero.  
Each case gives a small set of allowed 18-tuples, again with $x_{18}=0$, pointing to quartic curvature terms.

\vskip.1in

\noindent \textbf{(3) ${\bf R}_{m\rho}$ (mixing ${\cal M}_4$ with ${\cal M}_2$):}  
Both $(m,\alpha)$ and $(m,\beta)$ components have left-hand side scaling zero.  
Each yields two allowed 18-tuples, once more quartic-curvature dominated and with $x_{18}=0$.

\vskip.1in

\noindent One may also perform the zero-instanton sector baseline check.
For ${\bf R}_{im}$, ${\bf R}_{i\rho}$, and ${\bf R}_{m\rho}$, the zero-instanton choices force 18-tuples that do not switch on any useful quantum terms, so they cannot satisfy the EOM by themselves.  
Therefore, at least one BBS instanton must be included for a non-trivial, consistent solution.

\noindent Therefore the big picture takeaway is the following.
The cross-term EOMs are sourced by a non-local, non-perturbative energy-momentum tensor built from instanton sectors and perturbative series, organized by $\mathcal G$ and $\mathcal H$, and coupled through a general kernel $\mathbb F(x,y)$.  
The dominant consistent contributions across all three families of cross-term equations arise from quartic curvature structures, mirroring the on-shell analysis.  
No extra time-dependence from the non-local kernel is needed here ($x_{18}=0$ throughout these solutions).

\subsection*{Step-by-step flow for this part}

\begin{enumerate}
\item \textbf{Promote the kernel:} Replace the toy $G(x-y)$ by a general non-local kernel $\mathbb F(x,y)$ and plug it into the non-perturbative framework.

\item \textbf{Package the cross-term integral:} Summarize the path-integral over off-shell cross-terms in the following generic way:
\begin{equation}
\mathbb A=\mathcal G[\text{perturbative series},\mathbb F]\times e^{\mathcal H[\text{perturbative series},\mathbb F]}.
\end{equation}

\item \textbf{Write the effective source:} Use $\mathbb A$ to build the emergent, non-local ${\bf T}^{\rm NP}_{\rm C'D'}$ that will source the cross-term Einstein equations.

\item \textbf{Impose the EOM:} Set ${\bf R}_{\rm C'D'}(\langle{\bf \Xi}\rangle_\sigma)={\bf T}^{\rm NP}_{\rm C'D'}(\langle{\bf \Xi}\rangle_\sigma)$ for each mixed index pair $(i,m)$, $(i,\rho)$, $(m,\rho)$.

\item \textbf{Track $\bar g_s$ scalings:} Combine contributions from $\mathcal Q_1$, $\mathcal G/\mathcal H$, $\mathcal Q^{(2)}_{\rm C'D'}$, and $\sqrt{|{\bf g}_d|}$ to match the left-hand side scaling (zero in these cases).

\item \textbf{Solve for allowed 18-tuples:} Pick the integer tuples that satisfy the scaling balance.  
Result: specific small sets of 18-tuples for each Ricci component, all with $x_{18}=0$.

\item \textbf{Interpret the structures:} The allowed tuples turn on quartic curvature operators as the dominant quantum corrections.

\item \textbf{Check the zero-instanton limit:} With no instantons, the tuples that solve the scaling do not activate useful quantum terms—so the EOMs cannot be satisfied.  
Therefore, include at least one BBS instanton.

\item \textbf{Conclude on locality:} Because $\mathbb F(x,y)$ enters ${\bf T}^{\rm NP}_{\rm C'D'}$, the source is non-local.  
The working solutions (with instantons) are non-local and dominated by quartic-curvature terms, consistently across all three cross-term families.

\item \textbf{Consistency with earlier sections:} The pattern (quartic dominance, need for instantons, $x_{18}=0$) aligns with the on-shell counting done earlier, confirming internal consistency.
\end{enumerate}

\noindent Finally, in section \ref{aidapril} we probe the mixed time–space Einstein equations for three kinds of cross-term Ricci components:
${\bf R}_{0n}$ (time–$\mathcal M_4$), ${\bf R}_{0\rho}$ (time–$\mathcal M_2$ with $\rho=\alpha,\beta$), and ${\bf R}_{0i}$ (time–$\mathbb R^2$).
The goal is twofold: (i) confirm that non-perturbative (BBS) instantons with quartic-curvature terms can consistently source these EOMs; and
(ii) explore whether this sector can relate the small time-dependent parameters $p_e(t)\equiv(\hat\zeta_e,\hat\sigma_e,\hat\alpha_e,\hat\beta_e,\hat\eta_e)$.

\noindent We can also justify again why instantons are needed.
From the earlier tables (for ${\bf R}_{0n}$, ${\bf R}_{0i}$, ${\bf R}_{0\alpha}$), all these Ricci components scale as
$-1 + \tfrac{\hat\alpha_e+\hat\beta_e}{4}$.
At the \emph{zero}-instanton level, the corresponding source terms (energy–momentum tensors) cannot match these EOMs.
Therefore, as before, one must include non-perturbative sectors, and the equations are of the form \eqref{minerca}. 
With instantons on, the required $\bar g_s$-scalings of the non-perturbative sources for each index choice are summarized by four targets ${\rm L}_k$:
\begin{align}
{\rm L}_1:&\;\; {11\over 3} - \hat\zeta_e - 2\hat\sigma_e - {\hat\alpha_e+\hat\beta_e\over 4}\quad\;\;({\bf T}^{\rm NP}_{0i}),\\
{\rm L}_2:&\;\; {8\over 3} - {\hat\zeta_e\over 2} - {5\hat\sigma_e\over 2} - {\hat\alpha_e+\hat\beta_e\over 4}\quad~\;\;({\bf T}^{\rm NP}_{0m}),\\
{\rm L}_3:&\;\; {8\over 3} - {\hat\zeta_e\over 2} - 2\hat\sigma_e - {3\hat\alpha_e+\hat\beta_e\over 4}\;\;\;\;\;\;({\bf T}^{\rm NP}_{0\alpha}),\\
{\rm L}_4:&\;\; {8\over 3} - {\hat\zeta_e\over 2} - 2\hat\sigma_e - {\hat\alpha_e+3\hat\beta_e\over 4}\;\;\;\;\;\;({\bf T}^{\rm NP}_{0\beta}).
\end{align}

\noindent The matching of the 18-tuples becomes a combinatorial problem once more.
Each target ${\rm L}_k$ must be reproduced by a non-negative integer 18-tuple $(x_1,\dots,x_{18})$ weighted by a fixed menu of basis scalings ${\rm B}_i$ (the curvature/derivative “building blocks”), exactly as encoded in:
\begin{equation}
\sum_{i=1}^{17} x_i\,{\rm B}_i(\hat\sigma_e,\hat\zeta_e,\hat\alpha_e,\hat\beta_e)\;-\;\frac{2}{3}x_{18} \;=\; {\rm L}_k,
\end{equation}
with the ${\rm B}_i$ given in {\bf Table \ref{millerkhul}}.
Working in the gravitational sector and keeping $x_{18}=0$ (time-independent non-locality), one finds:
\begin{itemize}
  \item For ${\rm L}_1$ (${\bf R}_{0i}$): eight admissible 18-tuples, each turning on quartic-curvature corrections.
  \item For ${\rm L}_2$ (${\bf R}_{0n}$): three admissible 18-tuples, again quartic-curvature dominated.
  \item For ${\rm L}_3$ (${\bf R}_{0\alpha}$): five admissible 18-tuples with quartic curvature.
  \item For ${\rm L}_4$ (${\bf R}_{0\beta}$): five admissible 18-tuples with quartic curvature.
\end{itemize}
These reproduce the required constants and the coefficients of $(\hat\sigma_e,\hat\alpha_e,\hat\beta_e)$ while keeping all entries non-negative and satisfying the additional “constant-budget” constraint (the sum of $5/3$ and $2/3$ units adds up exactly to each ${\rm L}_k$).

\paragraph{Flux sector does not help here.}
In the flux part of the theory the dominant scalings cluster around $4/3^\pm$,
and the only integer patterns available do not generate non-trivial interactions compatible with the ${\rm L}_k$ targets.
Hence the solutions are furnished entirely by the gravitational (curvature) sector, consistent with quartic-order dominance.

\paragraph{Keeping the small parameters small.}
So far, nothing \emph{forces} relations among the five small functions $p_e(t)=(\hat\zeta_e,\hat\sigma_e,\hat\alpha_e,\hat\beta_e,\hat\eta_e)$.
For internal consistency (e.g.\ table checks) one imposes $|p_e(t)|\ll 1$.
This motivates seeking a relation where $\hat\zeta_e$ depends on $(\hat\sigma_e,\hat\alpha_e,\hat\beta_e)$.
A general polynomial ansatz is proposed and then truncated to its dominant linear part:
\begin{equation}\label{camilsara}
{~\hat\zeta_e(t) \;=\; a\,\hat\sigma_e(t) \;+\; \frac{p\,\hat\alpha_e(t) + q\,\hat\beta_e(t)}{2}\,,\quad a,p,q\in\mathbb Z~.}
\end{equation}

\paragraph{What the linear relation buys us (lattice picture).}
Substituting the linear ans\"atze simply shifts the coefficients of $\hat\sigma_e,\hat\alpha_e,\hat\beta_e$ in ${\rm L}_k$ by
multiples of $\tfrac{a}{2},\tfrac{p}{4},\tfrac{q}{4}$, i.e.\ all targets live on the same rational lattice
$\Lambda_{\hat\sigma_e,\hat\alpha_e,\hat\beta_e}=\mathbb Z/2\times\mathbb Z/4\times\mathbb Z/4$.
On the left-hand side, the ${\rm B}_i$ generators span the \emph{same} lattice after the same substitution.
Thus, for any fixed $(a,p,q)$, one can still find the same sets of 18-tuples that solve each ${\rm L}_k$ and pass the constant-budget constraint.
This confirms consistency \emph{without} yet fixing $(a,p,q)$.

\paragraph{Uniqueness (up to controlled swaps).}
A constant-budget identity forces one $5/3$-term plus three $2/3$-terms for ${\rm L}_1$,
and analogous fixed budgets for the other ${\rm L}_k$.
Moreover, certain “pair identities” among coefficient vectors $v_i$ allow swapping specific pairs (e.g.\ $v_2{+}v_{11}=v_9{+}v_{17}$) without changing totals,
which explains \emph{why} there are \emph{sets} (not a single tuple) but also \emph{why} there are no more solutions beyond those listed.
In short: the admissible families are unique up to these controlled swaps.

\paragraph{Bottom line.}
All three cross-term sectors (${\bf R}_{0n},{\bf R}_{0\rho},{\bf R}_{0i}$) are consistently solved by non-perturbative BBS instantons with quartic-curvature corrections and $x_{18}=0$.
This remains true even after imposing the linear relation for $\hat\zeta_e$.
The values of $(a,p,q)$ are \emph{not} fixed here; their determination is deferred to the next section, which connects to dynamical dark-energy considerations.

\subsection*{Step-by-step flow for this part}

\begin{enumerate}
\item \textbf{Identify the targets.}
Work with the three mixed time–space Ricci components: ${\bf R}_{0i}$, ${\bf R}_{0n}$, ${\bf R}_{0\rho}$ ($\rho=\alpha,\beta$).
Read off their $\bar g_s$-scalings from the tables.

\item \textbf{Turn on instantons.}
Zero-instanton sources cannot satisfy the EOMs. Use the non-perturbative equation \eqref{minerca}
with BBS instantons included (local and non-local pieces), and define four target scalings ${\rm L}_1,\dots,{\rm L}_4$ for the sources.

\item \textbf{Set up the matching.}
Express the source as a non-negative integer combination of basis scalings ${\rm B}_i$:
\begin{equation}
\sum_{i=1}^{17} x_i\,{\rm B}_i\;-\;\frac{2}{3}\,x_{18} \;=\; {\rm L}_k,\qquad x_i\in\mathbb Z_{\ge0}.
\end{equation}
Impose $x_{18}=0$ to keep the non-locality time-independent.

\item \textbf{Solve for each sector.}
Find all admissible 18-tuples for each target:
8 for ${\rm L}_1$ (${\bf R}_{0i}$), 3 for ${\rm L}_2$ (${\bf R}_{0n}$), 5 for ${\rm L}_3$ (${\bf R}_{0\alpha}$), and 5 for ${\rm L}_4$ (${\bf R}_{0\beta}$).
All activate quartic-curvature terms and satisfy the constant-budget constraint.

\item \textbf{Note what does \emph{not} help.}
Check the flux sector: its dominant $4/3^\pm$ scalings cannot deliver the required non-trivial interactions compatible with ${\rm L}_k$.
Thus, the working solutions arise from the curvature sector.

\item \textbf{Keep parameters small.}
For internal consistency one requires $|p_e(t)|\ll 1$.
Motivated by this, posit a relation that makes $\hat\zeta_e$ depend on $(\hat\sigma_e,\hat\alpha_e,\hat\beta_e)$.
Use the dominant linear truncation:
\begin{equation}
\hat\zeta_e \;=\; a\,\hat\sigma_e \;+\; \frac{p\,\hat\alpha_e + q\,\hat\beta_e}{2}.
\end{equation}

\item \textbf{Check lattice compatibility.}
Show that substituting the linear relation shifts coefficients by rational steps so that both sides (targets and generators) lie on the same lattice
$\mathbb Z/2\times\mathbb Z/4\times\mathbb Z/4$. Hence, for any fixed $(a,p,q)$, the same families of 18-tuples solve all four ${\rm L}_k$.

\item \textbf{Explain the families and uniqueness.}
Use the fixed “constant budget” and pair identities among coefficient vectors to justify why there are specific \emph{sets} of solutions and no others (up to allowed swaps).

\item \textbf{Conclude for this subsection.}
All three cross-term EOM classes are solved consistently by BBS instantons with quartic-curvature corrections and $x_{18}=0$.
The linear relation keeps parameters small but does not yet determine $(a,p,q)$.

\item \textbf{Preview.}
The subsequent section investigates how far one can fix $(a,p,q)$, linking the construction to a dynamical dark-energy picture and its equation of state.
\end{enumerate}

\subsubsection{Summary of section \ref{dynamical}}

\noindent In our last section, {\it i.e.} section \ref{dynamical}
we make an attempt to answer the following questions:
How dark energy can be \emph{dynamical} (slowly changing), what actually drives the current acceleration, how the axion decay constant $f_a$ behaves when dark energy runs, and what equation of state $w$ the $4$D universe exhibits (and how it departs from $-1$). The key ingredients used here are the following.
\begin{itemize}
  \item Revisit the duality chains (M/II/heterotic) while tracking two extra warp factors $\mathrm F_4,\mathrm F_5$ that encode time dependence tied to dark energy.
  \item Keep sub-dominant ${\rm M}_p$ and $g_s$ corrections to the duality sequence in mind; ignore them when they do not alter qualitative conclusions.
  \item Fix the $4$D Newton constant by suitable relations among warp factors so that ${\rm G}_N$ does not drift in time.
\end{itemize}

\paragraph{Dynamical dark energy via warp factors.}
\begin{itemize}
  \item In the ${\rm SO}(32)$ chain, relations among warp factors link the time variation of $\mathrm F_4(t)$ to those of $\mathrm F_1(t),\mathrm F_2(t)$. With $\mathrm F_i(t)\sim \bar g_s^{\hat\Delta_e(t)}$, where $\hat\Delta_e(t) =$ small time-dependent exponent, a slowly running $\Lambda(t)$ is geometrized.
  \item In the ${\rm E}_8\times{\rm E}_8$ chain, internal-volume preservation gives linear relations among the small exponent functions. Comparing with the ans\"atze \eqref{camilsara}
  fixes the integers \emph{uniquely} to
  $(a,p,q)=(2,1,3)$.
\end{itemize}

\paragraph{What powers the acceleration?}
\begin{itemize}
  \item BBS instantons dominate the emergent energy-momentum tensor; their time-scaling matches the Einstein-tensor components, so as $\bar g_s\to 0$ (late time in conformal slicing) they remain the leading driver of acceleration.
  \item Other non-perturbative/higher-curvature effects are present but sub-dominant in the established scaling hierarchy.
\end{itemize}

\paragraph{Axion decay constant with running dark energy.}
\begin{itemize}
  \item The same warp factor that tracks $\Lambda(t)$ enters the axion kinetic term, giving $f_a(t)$ a mild, controlled time dependence.
  \item If $\check\Lambda(t)\to 0$ at late times, $f_a(t)$ smoothly returns to the constant-$\Lambda$ behavior derived earlier.
\end{itemize}

\paragraph{Equation of state (EoS).}
\begin{itemize}
  \item After compactification, in conformal time $\eta$ (which we defined previously as $t$) the scale factor behaves as:
  $$
  a^{2}(\eta)\sim \frac{1}{\Lambda(\eta)\,\eta^{2}},
  $$
  with $\Lambda(\eta)$ defined as in \eqref{marapaug}. Such a definition encodes the ``running" of the dark energy $\Lambda(\eta)$.
  \item Up to small internal backreaction terms, the EoS, as seen from the SD equations and the duality sequence from {\bf Tables \ref{milleren222}} and {\bf \ref{milleren444}}, is:
  \begin{equation}
  w(\eta)\;=\;-\frac{1}{3}\;-\;\frac{2}{3}\,\frac{\mathcal H'(\eta)}{\mathcal H^{2}(\eta)},
  \qquad \mathcal H(\eta)\equiv \frac{a'(\eta)}{a(\eta)},
  \end{equation}
  where the backreaction terms appearing from the internal dimensions only have sub-dominant corrections to $w(\eta)$.
  \item Writing $\mathbb X(\eta)\equiv {\Lambda'(\eta)\over \Lambda(\eta)}$ and using $a^{2}(\eta)={1\over \Lambda(\eta)\eta^{2}}$, one convenient closed form is:
  \begin{equation}\label{claudouble}
  w(\eta)\;=\;-\frac{1}{3}\;-\;\frac{4}{3}\,
  \left[\frac{2-\eta^{2}\,\mathbb X'(\eta)}{\big(2+\eta\,\mathbb X(\eta)\big)^{2}}\right],
  \end{equation}
  which is a generic result for our case and, depending on how we parametrize $\Lambda(\eta)$, we can easily work out the corresponding $w(\eta)$.
  \item As a concrete example, for small departures from de Sitter which would keep $\mathbb X(\eta)$ and  $\mathbb X'(\eta)$ small, we can expand \eqref{claudouble} to get:
  \begin{equation}
  w(\eta)\;\approx\;-1\;+\;\frac{2}{3}\,\eta\,\mathbb X(\eta)\;+\;\frac{1}{3}\,\eta^{2}\,\mathbb X'(\eta)\;+\;\mathcal O\!\big(\mathbb X^{2}\big),
  \end{equation}
  which would imply that the deviations from $w(\eta)=-1$ are governed by the first two derivatives of $\ln\Lambda(\eta)$. In our case since $\mathbb{X}(\eta)$ depends on the warp-factors ${\rm F}_i(t)$, and specifically on ${\rm F}_4(t)$ (as seen from {\bf Tables \ref{milleren222}} and {\bf \ref{milleren444}}), we naturally have dynamical dark energy.
\end{itemize}

\paragraph{Bottom line.}
\begin{itemize}
  \item Extra warp factors tie late-time acceleration and the mild running of dark energy directly to higher-dimensional geometry.
  \item The duality-consistent bookkeeping \emph{fixes} $(a,p,q)=(2,1,3)$ in \eqref{camilsara}.
  \item BBS instantons are the leading engine of acceleration; both $f_a(t)$ and $w(\eta)$ inherit gentle time dependence controlled by the same small parameters, while ${\rm G}_N$ remains fixed.
\end{itemize}

\subsection*{Step-by-step flow of this section}

\begin{enumerate}
  \item \textbf{Set goals:} explain dynamical dark energy, its driver, the behavior of $f_a(t)$, and the resulting EoS $w(\eta)$.
  \item \textbf{Turn on extra warp factors:} include $\mathrm F_4,\mathrm F_5$ in the duality chains; impose relations to keep ${\rm G}_N$ constant.
  \item \textbf{Map to small exponents:} write $\mathrm F_i(t)\sim \bar g_s^{\hat\Delta_e(t)}$ where $\hat\Delta_e(t)=$ small time-dependent exponent, and relate them via internal-volume constraints.
  \item \textbf{Fix remaining integers:} compare with $\hat\zeta_e(t)=a\,\hat\sigma_e(t)+\frac{p\,\hat\alpha_e(t)+q\,\hat\beta_e(t)}{2}$; deduce $(a,p,q)=(2,1,3)$.
  \item \textbf{Identify the accelerator:} BBS instantons dominate the emergent stress tensor and source late-time acceleration.
  \item \textbf{Track $\Lambda(\eta)$ in $a(\eta)$:} use $a^{2}(\eta)={1\over \Lambda(\eta)\eta^{2}}$ to parametrize departures from de Sitter.
  \item \textbf{Predict $f_a(t)$:} the same warp-factor dependence induces mild running of $f_a(t)$; constant $\Lambda$ recovers the earlier result.
  \item \textbf{Compute EoS:} from the expressions above, obtain $w(\eta)$; for slow running use the expansion for small $\mathbb X$.
  \item \textbf{Consistency checks:} ${\rm G}_N$ fixed; backreaction corrections small; $w(\eta)\to -1$ as $\Lambda(\eta)\to \Lambda$.
  \item \textbf{Conclude:} controlled, stringy realization of dynamical dark energy with gentle time variation in both $f_a(t)$ and $w(\eta)$, and with $(a,p,q)$ uniquely fixed.
\end{enumerate}

\noindent Therefore our analysis reveals, probably for the first time, a completely stringy set-up to realize the dynamical nature of dark energy
as \eqref{marapaug}.


\section{Late-time de Sitter states as end-points of temporal evolutions \label{sec2}}

One of the useful viewpoint to study temporally varying backgrounds, such as the de Sitter states, in various string theories is to (a) study them from a well-defined M-theory set-up and (b) implement the duality chasings via temporal evolutions only. The latter point is particularly useful when we want to ascertain the behaviors of the late time de Sitter or quasi de Sitter states in type II and  heterotic theories. (A quasi de Sitter state is defined with a temporally varying dark energy, which we shall elaborate soon.) In the following let us give a few examples from our earlier works to justify the aforementioned criteria.

\subsection{Late-time de Sitter state in type IIB string theory \label{secc2.1}}

Our first example is the oft-studied case of de Sitter state in type IIB string theory\footnote{Our approach here will be to start with a de Sitter space, defined using a cosmological constant $\Lambda$, and then go to the more intriguing case with a temporally varying dark energy scenario namely, the quasi de Sitter case. The latter will be a non-trivial extension of the constant $\Lambda$ scenario, as will become clearer soon from section \ref{ccnews} onwards (see \eqref{marapaug} therein). \label{4chum}}. The details have appeared in \cite{desitter2, coherbeta, coherbeta2, borel2}, so here we will avoid elaborating on them and directly address the consequences. The M-theory configuration studied in \cite{coherbeta, coherbeta2, borel2} appears from taking the expectation values of the metric and the flux operators over the Glauber-Sudarshan state\footnote{How the Glauber-Sudarshan states differ from the usual notion of {\it localized} coherent states, and how they overcome the Hamiltonian constraints placed by the Wheeler-De Witt equations, have been discussed in details in \cite{wdwpaper}. We will not repeat any of these details here, and the interested readers may look up \cite{wdwpaper}.} $\vert\sigma\rangle$. The metric, for example, takes the form:

{\footnotesize
\bg\label{giuraman1}
ds^2 & = & \left\langle {\bf g}_{\rm AB} \right\rangle_\sigma d{\rm Y}^{\rm A} d{\rm Y}^{\rm B} \nonumber\\
& = & g_s^{-8/3}\left(-dt^2 + \delta_{ij} dx^i dx^j\right) + 
g_s^{-2/3} {\rm H}^2(y)\left[{\rm F}_1(g_s/{\rm H})g_{\alpha\beta} dy^\alpha dy^\beta +{\rm F}_2(g_s/{\rm H})g_{mn} dy^m dy^n\right] \nonumber\\ 
& + & g_s^{4/3} \delta_{ab} dw^a dw^b, \nd}
where ${\rm Y}^{\rm A} \in (x^\mu, y^\alpha, y^m, w^a)$ with $x^\mu \in {\bf R}^{2, 1}, y^\alpha \in {\cal M}_2, y^m \in {\cal M}_4$, and $w^a \in {\mathbb{T}^2\over {\cal G}}$. The two and four manifolds, ${\cal M}_2$ and ${\cal M}_4$ respectively, are compact and non-K\"ahler and ${\cal G}$ is a group action without fixed points (we will discuss the case with orbifold fixed points soon). The type IIA coupling $g_s \to 0$ at late time because of its temporal dependence is as $g_s/ {\rm H}(y) = 1/a(t)$ with $t$ $-$ appearing as the conformal time in the type IIB side $-$ that lies in the domain $-\infty < t \le 0$ and $a(t) \to \infty$ as $t \to 0$. The remaining variables are defined in the following way. ${\rm H}(y)$ is the warp-factor, $\Lambda$ is the cosmological constant, and ${\rm F}_i(t)$ are the additional warp-factors that are arranged in a way that as $t \to 0$, they approach identities respectively. One may easily see that as $g_s \to 0$, {\rm i.e.} as $t \to 0$ at late time, the M-theory torus 
${{\mathbb T}^2\over {\cal G}}$ shrinks to zero size. From type IIA point of view, this shrinking would imply that at late time the radius of the IIA circle, parametrized by say $x^3$, crosses the self-dual radius. Such a consideration means that the description is succinctly captured by the {\it non-compact} ten-dimensional type IIB theory with the following metric configuration:

\begin{table}[tb]  
 \begin{center}
\renewcommand{\arraystretch}{2.6}
\resizebox{\textwidth}{!}{\begin{tabular}{|c|c|c|c|c|c|}\hline
Theory & duality & Solitonic Configuration & $\langle {\bf g}_{\rm AB}\rangle_{\sigma}$ & coupling \\ \hline
M &  & $\mathbb{R}^{2, 1} \times {\cal M}_4 \times {\cal M}_2 \times {\mathbb{T}^2_{3, 11}\over {\cal G}}$
& $\left(g_s^{-8/3}, g_s^{-2/3}{\rm F}_2, g_s^{-2/3}{\rm F}_1, g_s^{4/3}\right)$ & \\ \hline
IIA &  dim. red. along $x^{11}$ & $\mathbb{R}^{2, 1} \times {\cal M}_4 \times {\cal M}_2 \times {\bf S}^1_3 $  & $\left(g_s^{-2},~ {\rm F}_2,~ {\rm F}_1, ~g_s^2\right)$ & $g_s$ \\ \hline
IIB &T duality along $x^3$ &$\mathbb{R}^{3, 1} \times {\cal M}_4 \times 
{\cal M}_2$
&  $\left(g_s^{-2}, ~ {\rm F}_2, ~{\rm F}_1\right)$ & 1 \\ \hline
\end{tabular}}
\renewcommand{\arraystretch}{1}
\end{center}
 \caption[]{\Su The dynamical duality sequence that takes a M-theory configuration \eqref{giuraman1} to type IIB theory with a de Sitter Glauber-Sudarshan state.}
 \label{milleren1}
 \end{table}

{\footnotesize
\bg\label{viomyer1}
ds^2 & = & \left\langle {\bf g}_{\rm MN}\right\rangle_\rho d{\rm X}^{\rm M} d{\rm X}^{\rm N} \nonumber\\
& = & {a^2(t)\over {\rm H}^2(y)}\left[-dt^2 + \delta_{ij}dx^i dx^j + (dx^3)^2\right] + 
{\rm H}^2(y)\left[{\rm F}_2(t) g_{mn} dy^m dy^n + {\rm F}_1(t) g_{\alpha\beta} dy^\alpha dy^\beta\right], 
\nd}
where $\vert \rho\rangle$ should be viewed as the corresponding Glauber-Sudarshan state in the type IIB side. It is now clear that if we choose 
$a^2(t) = {1\over \Lambda t^2}$, we do reproduce the de Sitter state in a {\it flat-slicing}. Other equivalent slicings could also be entertained, but we will not do so here\footnote{We will not include de Sitter state in a {\it static patch}, or other equivalent patches related to the staic patch, because of the issues mentioned in \cite{coherbeta2}. In other words, despite the apparent {\it static} nature of the space-time, the trans-Planckian problems still exist as shown in \cite{joydeep, lalu}.} and the readers may look up the details in \cite{desitter2, coherbeta, coherbeta2}. 

There is however one point that we would like to emphasize at this stage. The type IIB metric \eqref{viomyer1} is expressed using four functions $(a(t), {\rm F}_1(t), {\rm F}_2(t), {\rm H}(y))$. The M-theory uplift is interestingly {\it always} of the form \eqref{giuraman1} with 
the dual IIA coupling given by $g_s = {{\rm H}(y)\over a(t)}$ irrespective of the choice of $a(t)$. Clearly weak coupling appears once we allow for an expanding universe, {\it i.e.} when $a(t)$ increases with the increase of the conformal time $t$. Therefore to allow for a temporally varying background, we will use ${g_s\over {\rm H}(y)}$ (or, as we shall see later, ${g_s\over {\rm H}(y) {\rm H}_o({\bf x})} \equiv \bar{g}_s$ with ${\bf x} \in {\bf R}^2$) as the expansion parameter. For time-independent background, $a(t) = 1$, and the expansion parameter becomes identity. 

The second set of warp-factors, namely ${\rm F}_1(t)$ and ${\rm F}_2(t)$, are important because their behavior with respect to $g_s$ will set up the stage for the duality transformations. Let us then propose the following behavior of the ${\rm F}_i$ warp-factors (which will be modified soon):
\bg\label{sezoe71}
{\rm F}_1(t) \equiv \sum_{k = 0}^\infty {\rm A}_k\left({g_s\over {\rm H}(y)}\right)^{\beta_o + 2k/3}, ~~~~~~
{\rm F}_2(t) \equiv \sum_{k = 0}^\infty {\rm B}_k\left({g_s\over {\rm H}(y)}\right)^{\alpha_o + 2k/3},  \nd
where ${\rm A}_k$ and ${\rm B}_k$ are integers (they can be positive or negative) with $\alpha_o$ and $\beta_o$ being the dominant scalings
and $k \in \mathbb{Z}_+$. Both $\beta_o$ and $\alpha_o$ should at least be bounded from above by $\beta_o < {2\over 3}$ and $\alpha_o < {2\over 3}$ so as not to offset the duality to the IIB side from M-theory. In fact for the present case, and as discussed in \cite{coherbeta, coherbeta2}, we require $\beta_o = \alpha_o = 0$. This way the later time behavior will always be given by the 
type IIB de Sitter state \eqref{viomyer1}. One may equivalently find the corresponding flux configuration that would support such a state in the IIB side. In {\bf Table \ref{milleren1}} we elucidate the detailed behavior of the de Sitter state and the corresponding duality sequence leading to it. Taking\footnote{The choice ${\rm H}(y) = 1$ simplifies the expressions but is not essential because we can always measure the conformal temporal coordinate using the ratio ${g_s\over {\rm H}(y)}$. In fact, as mentioned earlier and also shown in \cite{desitter2, coherbeta} we could even incorporate spatial warp-factor via ${\rm H}_o({\bf x})$ and express the temporal coordinate as ${g_s\over {\rm H}(y) {\rm H}_o({\bf x})}$. We will avoid these subtle nuances for the time being, but will get back to them very soon.} ${\rm H}(y) = 1$ and choosing $g_s^{-2} = {1\over \Lambda t^2}$ will provide the required Glauber-Sudarshan state that reproduces a de Sitter state in type IIB with a flat-slicing.

Before ending this section, let us point out another interesting aspect of the story related to the metric choices \eqref{giuraman1} and \eqref{viomyer1}. Computing the Einstein tensors along ${\bf R}^{2, 1}$ using the M-theory metric \eqref{giuraman1}, the $ij$ component gives us (see also \cite{desitter2}):
\bg\label{mameybike}
{\bf G}_{ij} & = & -a^2(t)\eta_{ij}\left[\frac{{\rm R}(y, t)}{2{\rm H}^4(y)}+\frac{4{g}^{\alpha\beta}\partial_\alpha {\rm H}(y)\partial_\beta {\rm H}(y)}{{\rm H}^6(y){\rm F}_1(t)}
+
\frac{4{g}^{mn}\partial_m {\rm H}(y)\partial_n {\rm H}(y)}{{\rm H}^6(y){\rm F}_2(t)}\right] \\
&& + a^2(t)\eta_{ij}\left[\frac{\Box_\alpha {\rm H}^4(y)}{2{\rm H}^8{\rm F}_1(t)}
+
\frac{\Box_m {\rm H}^4(y)}{2{\rm H}^8(y){\rm F}_2(t)} +
\frac{\dot a^2(t)}{a^4(t)}- \frac{2\ddot a(t)}{a^3(t)}\right]
\nonumber\\
&& + \eta_{ij}\left[ \frac{\dot{\rm F}_1^2(t)}{4{\rm F}_1^2(t)}
-\frac{\dot a(t)}{a(t)}
\frac{\dot{\rm F}_1(t)}{{\rm F}_1(t)}
- \frac{\ddot{\rm F}_1(t)}{{\rm F}_1(t)}
-
\frac{\dot{\rm F}_2^2(t)}{2{\rm F}_2^2(t)}
-\frac{\dot a(t)}{a(t)}\frac{2\dot{\rm F}_2(t)}{{\rm F}_2(t)}
- \frac{2\ddot{\rm F}_2(t)}{{\rm F}_2(t)}
- \frac{2\dot{\rm F}_1(t)\dot{\rm F}_2(t)}{{\rm F}_1(t) {\rm F}_2(t)}
\right], \nonumber
\nd
where all the metric components $g_{\rm MN}(y)$ with $({\rm M, N}) \in {\cal M}_4 \times {\cal M}_2$, and the Ricci curvature ${\rm R}(y, t) \equiv {{\rm R}_1(y)\over {\rm F}_1(t)} +{{\rm R}_2(y)\over {\rm F}_2(t)}$ are related to the unwarped ones once we view ${\rm R}_1(y)$ and ${\rm R}_2(y)$ as the Ricci scalars for the unwarped metric components $g_{\alpha\beta}(y)$ and $g_{mn}(y)$ respectively. In a similar vein, the $00$ component of Einstein tensor is:
\bg\label{mameybike2}
{\bf G}_{00} & = & -a^2(t)\eta_{00}\left[\frac{{\rm R}(y, t)}{2{\rm H}^4(y)}+\frac{4{g}^{\alpha\beta}\partial_\alpha {\rm H}(y)\partial_\beta {\rm H}(y)}{{\rm H}^6(y){\rm F}_1(t)}
+
\frac{4{g}^{mn}\partial_m {\rm H}(y)\partial_n {\rm H}(y)}{{\rm H}^6(y){\rm F}_2(t)}\right] \\
&& + a^2(t)\eta_{00}\left[\frac{\Box_\alpha {\rm H}^4(y)}{2{\rm H}^8{\rm F}_1(t)}
+
\frac{\Box_m {\rm H}^4(y)}{2{\rm H}^8(y){\rm F}_2(t)} - \frac{3\dot{a}^2(t)}{a^4(t)}\right]
\nonumber\\
&&-\eta_{00}\left[ \frac{\dot{\rm F}_1^2(t)}{4{\rm F}_1^2(t)}
+\frac{3\dot a(t)}{a(t)}
\frac{\dot{\rm F}_1(t)}{{\rm F}_1(t)}
+
\frac{3\dot{\rm F}_2^2(t)}{2{\rm F}_2^2(t)}
+\frac{\dot a(t)}{a(t)}\frac{6\dot{\rm F}_2(t)}{{\rm F}_2(t)}
+\frac{2\dot{\rm F}_1(t)\dot{\rm F}_2(t)}{{\rm F}_1(t) {\rm F}_2(t)}
\right], \nonumber \nd
where comparing to \eqref{mameybike} we see that the first line match up exactly, whereas in the second line there is a mismatch with the terms containing $a(t)$. The third line with ${\rm F}_i(t)$ and $a(t)$ is quite different. We can ask whether there exists a choice of $a(t)$ which can equate the first two lines in \eqref{mameybike} and \eqref{mameybike2}. Imposing this by hand gives us:
\bg\label{2indig2metro}
\frac{\dot a^2(t)}{a^4(t)}- \frac{2\ddot a(t)}{a^3(t)} = - \frac{3\dot{a}^2(t)}{a^4(t)}, ~~~~ \implies ~~~ a^2(t) = {1\over \Lambda t^2}, \nd
which is surprisingly the flat-slicing of a de Sitter space that we took earlier. Thus in a flat-slicing, and in the limit where ${\rm F}_i(t) = 1$, the two Einstein tensor would be identical up to an overall sign. However since ${\rm F}_i(t)$ are generically non-trivial, this identification does not extend beyond the first two lines in \eqref{mameybike} and \eqref{mameybike2}. The consequence of the generic choice of $a(t)$ and ${\rm F}_i(t)$, for example on the equation of state, will be discussed in section \ref{dynamical}. In this paper we will start with the flat-slicing, {\it i.e.} with a cosmological constant $\Lambda$, and then make $\Lambda$ dynamical as $\Lambda(t)$ in \eqref{marapaug}. As we shall see, our choice of the flat-slicing $-$ and it's later generalization to include running $\Lambda(t)$ $-$ will not only be useful to quantify the dark energy, but will also give us a handle to study the equations of motion, Bianchi identities et cetera when we go to the $SO(32)$ and the ${\rm E}_8 \times {\rm E}_8$ heterotic theories. 

\begin{table}[tb]  
 \begin{center}
\renewcommand{\arraystretch}{2.6}
\resizebox{\textwidth}{!}{
\begin{tabular}{|c|c|c|c|c|c|}\hline
Theory & duality & Solitonic Configuration & $\langle {\bf g}_{\rm AB}\rangle_{\sigma}$ & coupling & vector bundle \\ \hline
M &  & $\mathbb{R}^{2, 1} \times {\cal M}_4 \times {\mathbb{T}^2_{45} \times \mathbb{T}^2_{3, 11}\over {\cal I}_4 {\cal I}_5 {\cal I}_3 {\cal I}_{11}}$
& $\left(g_s^{-8/3}, g_s^{-2/3}{\rm F}_2, g_s^{-2/3}{\rm F}_1, g_s^{4/3}\right)$ & & \\ \hline
IIA &  dim. red. along $x^{11}$ & $\mathbb{R}^{2, 1} \times {\cal M}_4 \times {\mathbb{T}^2_{45} \times {\bf S}^1_3\over \Omega {\cal I}_4 {\cal I}_5 {\cal I}_3}$  & $\left(g_s^{-2},~ {\rm F}_2,~ {\rm F}_1, ~g_s^2\right)$ & $g_s$ & 
${\rm O6} \oplus {\rm D6}$ \\ \hline
IIB &T duality along $x^3$ &$\mathbb{R}^{3, 1} \times {\cal M}_4 \times {{\bf S}^1_{4} \times {\bf S}^1_{5}\over \Omega (-1)^{{\rm F}_{\rm L}}{\cal I}_4 {\cal I}_5}$
&  $\left(g_s^{-2}, ~ {\rm F}_2, ~{\rm F}_1, ~{\rm F}_1\right)$ & 1 & ${\rm O7} \oplus {\rm D7}$ \\ \hline
IIA & T duality along $y^4$ & $\mathbb{R}^{3, 1} \times \big({\cal M}_4 \rtimes \widetilde{\bf S}^1_{4}\big) \times {{\bf S}^1_{5} \over \Omega {\cal I}_5}$ & 
$\left(g_s^{-2}, ~ {\rm F}_2, ~ {1\over {\rm F}_1}, ~ {{\rm F}_1}\right)$ &
${1\over \sqrt{{\rm F}_1}}$ & ${\rm O8} \oplus {\rm D8}$ \\ \hline
Type I & T duality along $y^5$ & $\mathbb{R}^{3, 1} \times {\cal M}_4 \rtimes \big(\widetilde{\bf S}^1_{5} \times \widetilde{\bf S}^1_{4}\big)$ & 
$\left(g_s^{-2}, ~ {\rm F}_2, ~ {1\over {\rm F}_1}, ~ {1\over {\rm F}_1}\right)$ & ${1\over {\rm F}_1}$ & ${\rm O9} \oplus {\rm D9}$ \\ \hline
Het ${\rm SO(32)}$ & S duality & $\mathbb{R}^{3, 1} \times {\cal M}_4 \rtimes \big(\widetilde{\bf S}^1_{5} \times \widetilde{\bf S}^1_{4}\big)$ & $\left(g_s^{-2} {\rm F}_1, ~ {\rm F_1 F_2}, ~ 1, ~ 1\right)$ & ${\rm F}_1$ &
$\left({\rm SO(8)}\right)^4$ \\
\hline
\end{tabular}}
\renewcommand{\arraystretch}{1}
\end{center}
 \caption[]{\Su The dynamical duality sequence that takes a M-theory configuration \eqref{giuraman1} to heterotic SO(32) theory with the gauge group broken to $\left({\rm SO(8)}\right)^4$. The symbol ${\cal M}_4 \rtimes {\cal M}$ denotes non-trivial fibration of the manifold ${\cal M}$ over the base manifold ${\cal M}_4$. We have not specified the precise numbers of ${\rm O}p$'s and ${\rm D}p$'s in various theories for global charge cancellations.}
 \label{milleren2}
 \end{table}

\subsection{Late-time de Sitter state in heterotic ${\rm SO}(32)$ 
theory \label{secc2.2}}

The story in the heterotic ${\rm SO}(32)$ side differs from the type IIB example studied above in at least {\it three} possible ways. {\Su One}, the choices of $\beta_o$ and $\alpha_o$ are no longer vanishing, but $\beta_o$ lies in the range $0 < \beta_o < {2\over 3}$ 
and $\alpha_o  = -\beta_o$. {\Su Two}, 
the internal manifold ${\cal M}_2$ in the M-theory side is now a toroidal $\mathbb{Z}_2$ orbifold instead of a smooth space taken earlier. In fact now ${\cal G}$ is also an orbifold action, {\it i.e.} ${\cal G} = \mathbb{Z}_2$, so that the eleven-dimensional torus ${\mathbb{T}^2\over {\cal G}}$ is now an orbifold. And {\Su three}, the functional form of $g_s$ is more complicated despite the fact that 
the M-theory metric continues to have the form given in \eqref{giuraman1}. All these have been explained in \cite{hetborel}, but here we want to provide a slightly different perspective to the story studied therein with the aim of unifying this discussion with the heterotic ${\rm E}_8 \times {\rm E}_8$ story in the next section. The duality sequence that we are interested in here may be presented as:
\bg\label{momkune}
{\mathbb{T}^4\over {\cal I}_4 {\cal I}_5 {\cal I}_3 {\cal I}_{11}} 
~~ \xrightarrow{{\bf S}^1_{11} \to 0} ~~ {\mathbb{T}^3\over \Omega {\cal I}_4 {\cal I}_5 {\cal I}_3}~~ \xrightarrow{{\bf T}_3}~~ {\mathbb{T}^2\over \Omega (-1)^{{\rm F}_{\rm L}}{\cal I}_4 {\cal I}_5}~~ \xrightarrow{{\bf T}_4} ~~{{\bf S}_5^1\over \Omega {\cal I}_5} ~~\xrightarrow{{\bf T}_5} ~~ {1\over \Omega}, \nd
where the coordinate choices are $y^\alpha \equiv (y^4, y^5) \in {\cal M}_2,~ y^m \equiv (y^6, y^7, y^8, y^9) \in {\cal M}_4,~ w^a \equiv (x^3, x^{11}) \in {\mathbb{T}^2\over {\cal G}}$ with ${\cal I}_j$ being the coordinate reversal along the $j$ direction, $(-1)^{{\rm F}_{\rm L}}$ being the reversal of the left-moving world-sheet fermion in type IIB theory, $\Omega$ being the orientifold action in either type IIA or IIB theories, and ${\bf T}_j$ being the T-duality operation along the compact $j$ direction. 

The duality sequence in \eqref{momkune} may again be implemented via temporal evolution of $g_s$ and the warp-factors ${\rm F}_i$. This may be explained in the following way. Let us start with the M-theory configuration as given in \eqref{giuraman1}. The internal sub-manifold ${\cal M}_2 \times {\mathbb{T}^2\over {\cal G}}$ is now an orbifold of the form ${\mathbb{T}^4\over{\cal I}_4 {\cal I}_5 {\cal I}_3 {\cal I}_{11}}$, which forms the starting point of the duality sequence in \eqref{momkune}. Since both $\alpha_o < {2\over 3}$ and $\beta_o < {2\over 3}$ and $g_s \to 0$ at late time, the late-time behavior is given by a type IIB orientifold with four set of orientifold seven-planes at the four fixed points of ${\cal M}_2$. One may now cancel the global and local charges by adding an appropriate number of seven-branes (see \cite{hetborel}) but the duality sequence does not end here as the manifold ${\cal M}_2$ shrinks further because of the ${\rm F}_1$ warp-factor (since $0 < \beta_o < {2\over 3}$). This leads us to a type I configuration $-$ as shown by the 
${1\over \Omega}$ action in \eqref{momkune} $-$ at {\it strong} coupling, which implies the existence of a weakly coupled late-time de Sitter state in heterotic ${\rm SO}(32)$ theory with the gauge group broken to $\left({\rm SO}(8)\right)^4$. The temporal independence of the four-dimensional Newton's constant, as well as the aforementioned weak-coupling behavior, tells us that this is the {\it final} stage of the duality sequence \eqref{momkune}. In {\bf Table \ref{milleren2}} we summarize this duality sequence and the final de Sitter state in heterotic ${\rm SO}(32)$ theory broken to $\left({\rm SO}(8)\right)^4$. The content of this table may be elaborated in the following way. As in {\bf Table \ref{milleren1}}, the expectation values of the metric operators are taken over the respective Glauber-Sudarshan states $\vert \sigma \rangle$ in the corresponding theories. The various branes and O-planes are parallel to each other and are located at the fixed points of the orbifolds. For example the O8/D8 system are stretched along 
 ${\mathbb R}^{3, 1} \times {\cal M}_4 \times \widetilde{\bf S}^1_4$ and are located at the fixed points of ${{\bf S}^1_5\over \Omega {\cal I}_5}$. Since ${\rm F}_1 \to 0$ at late time, and $\alpha_o = -\beta_o$ (see \cite{hetborel}), the internal manifold in the type I side would appear to simultaneously decompactify and reach strong coupling. However the heterotic configuration, after a S-duality, shows none of these pathologies: it is perfectly weakly coupled and has a finite-sized internal manifold because ${\rm F}_1 {\rm F}_2 =$ constant. The latter would also keep the four-dimensional Newton's constant time-independent. If we now claim that $g_s^{-2} {\rm F}_1 \equiv {1\over \Lambda t^2}$, with $\Lambda$ being the four-dimensional cosmological constant and $t$ the conformal time, we not only recover a de Sitter state in heterotic SO(32) theory in the flat-slicing but also fix the form of $g_s$ in \eqref{giuraman1}. Note that we have taken ${\rm H}(y) = 1$ and have  avoided showing the fibration structure in the metric for simplification. Again, all these missing details could be inferred 
 from \cite{hetborel}.

There is however a subtlety that arises when we carefully look at the duality sequence from {\bf Table \ref{milleren2}}. When we reach the type IIB orientifold side (with the orientifold seven-planes and the charge cancelling seven-branes), a subsequent T-duality, shown by ${\bf T}_4$ in \eqref{momkune}, leads to type IIA theory with a coupling constant 
${1\over \sqrt{{\rm F}_1}}$ (see {\bf Table \ref{milleren2}}). Since ${\rm F}_1 \to 0$ at late time, the theory appears to become strongly coupled at late time. On the other hand, since the $y^5 \in {\cal M}_2$ circle comes with the metric warp-factor ${\rm F}_1$, it appears to simultaneously shrink to zero size at late time. It seems we now have an ambiguity: we can either go to the dual type I theory by T-dualizing as in \eqref{momkune}, or lift the configuration to M-theory\footnote{One might argue that, since both the cycles of ${\cal M}_2$ are simultaneously shrinking to zero sizes 
 $-$ because of their metric warp-factor ${\rm F}_1$ $-$ it makes more sense to simultaneously T-dualize both the cycles and go to the type I configuration as in \eqref{momkune}. While this may be true, a slightly different choices of warp-factors, namely ${\rm F}_1$ and ${\rm F}_3$ for the two cycles with different positive powers of $g_s$, could in principle have led to the aforementioned ambiguity. Therefore it would make sense to entertain such a scenario here. In section \ref{sec3} we will discuss such a case for the heterotic ${\rm E}_8 \times {\rm E}_8$ theory.}.  If we do the latter, then the duality sequence becomes:

{\footnotesize
\bg\label{momkune2}
{\mathbb{T}^4\over {\cal I}_4 {\cal I}_5 {\cal I}_3 {\cal I}_{11}} 
~~ \xrightarrow{{\bf S}^1_{11} \to 0} ~~ {\mathbb{T}^3\over \Omega {\cal I}_4 {\cal I}_5 {\cal I}_3}~~ \xrightarrow{{\bf T}_3}~~ {\mathbb{T}^2\over \Omega (-1)^{{\rm F}_{\rm L}}{\cal I}_4 {\cal I}_5}~~ \xrightarrow{{\bf T}_4} ~~{{\bf S}_5^1\over \Omega {\cal I}_5} ~~\xrightarrow{\widetilde{\bf S}^1_{11} \to \infty} ~~ {{\bf S}_5^1 \times \widetilde{\bf S}^1_{11}\over 
{\cal I}_5} ~~ \xrightarrow{{\bf S}^1_5 \to 0} ~~ \widetilde{\bf S}^1_{11}, \nonumber\\ \nd}
leading us to two Horava-Witten walls \cite{horava} at the two fixed points of ${\bf S}^1_5$. A further dimensional reduction then gives us the ${\rm E}_8 \times {\rm E}_8$ heterotic string theory with the gauge group broken down to exactly $({\rm SO}(8))^4$ what we had earlier. This happens because of the Wilson lines that break each ${\rm E}_8$ to $\left({\rm SO}(8)\right)^2$ on each of the two Horava-Witten walls. In fact, as one may infer from {\bf Table \ref{milleren3}} that both the coupling constant and the metric structure match to what we had derived for the SO(32) heterotic case. This means the ambiguity that we were worrying about earlier, does not produce different results.

\begin{table}[tb]  
 \begin{center}
\renewcommand{\arraystretch}{2.6}
\resizebox{\textwidth}{!}{
\begin{tabular}{|c|c|c|c|c|c|}\hline
Theory & duality & Solitonic Configuration & $\langle {\bf g}_{\rm AB}\rangle_{\sigma}$ & coupling & vector bundle \\ \hline
M &  & $\mathbb{R}^{2, 1} \times {\cal M}_4 \times {\mathbb{T}^2_{45} \times \mathbb{T}^2_{3, 11}\over {\cal I}_4 {\cal I}_5 {\cal I}_3 {\cal I}_{11}}$
& $\left(g_s^{-8/3}, g_s^{-2/3}{\rm F}_2, g_s^{-2/3}{\rm F}_1, g_s^{4/3}\right)$ & & \\ \hline
IIA &  dim. red. along $x^{11}$ & $\mathbb{R}^{2, 1} \times {\cal M}_4 \times {\mathbb{T}^2_{45} \times {\bf S}^1_3\over \Omega {\cal I}_4 {\cal I}_5 {\cal I}_3}$  & $\left(g_s^{-2},~ {\rm F}_2,~ {\rm F}_1, ~g_s^2\right)$ & $g_s$ & 
${\rm O6} \oplus {\rm D6}$ \\ \hline
IIB &T duality along $x^3$ &$\mathbb{R}^{3, 1} \times {\cal M}_4 \times {{\bf S}^1_{4} \times {\bf S}^1_{5}\over \Omega (-1)^{{\rm F}_{\rm L}}{\cal I}_4 {\cal I}_5}$
&  $\left(g_s^{-2}, ~ {\rm F}_2, ~{\rm F}_1, ~{\rm F}_1\right)$ & 1 & ${\rm O7} \oplus {\rm D7}$ \\ \hline
IIA & T duality along $y^4$ & $\mathbb{R}^{3, 1} \times \big({\cal M}_4 \rtimes \widetilde{\bf S}^1_{4}\big) \times {{\bf S}^1_{5} \over \Omega {\cal I}_5}$ & 
$\left(g_s^{-2}, ~ {\rm F}_2, ~ {1\over {\rm F}_1}, ~ {{\rm F}_1}\right)$ &
${1\over \sqrt{{\rm F}_1}}$ & ${\rm O8} \oplus {\rm D8}$ \\ \hline
M & uplift along $\widetilde{x}^{11}$ & $\mathbb{R}^{3, 1} \times {\cal M}_4 \rtimes \big (\widetilde{\bf S}^1_{4}\times \widetilde{\bf S}^1_{11}\big) \times {{\bf S}^1_{5} \over {\cal I}_5} $ & $\left({\rm F}_1^{1/3} g_s^{-2},  {\rm F}_1^{1/3} {\rm F}_2,  {\rm F}_1^{-2/3}, {\rm F}_1^{-2/3}, {\rm F}_1^{4/3}\right)$ & & Horava-Witten walls \\ \hline
Het ${\rm E}_8 \times {\rm E}_8$ & dim. red. along $y^5$ & $\mathbb{R}^{3, 1} \times {\cal M}_4 \rtimes \big(\widetilde{\bf S}^1_{4} \times \widetilde{\bf S}^1_{11}\big)$ & $\left({\rm F}_1 g_s^{-2}, ~ {\rm F}_1 {\rm F}_2, ~ 1, ~ 1\right)$ & ${\rm F}_1$ & $\left({\rm SO(8)}\right)^4$\\
\hline
\end{tabular}}
\renewcommand{\arraystretch}{1}
\end{center}
 \caption[]{\Su The dynamical duality sequence that takes a M-theory configuration \eqref{giuraman1} to heterotic ${\rm E}_8 \times {\rm E}_8$ theory with the gauge group broken to $\left({\rm SO(8)}\right)^4$. As before, we have not specified the precise numbers of ${\rm O}p$'s and ${\rm D}p$'s in various theories for global charge cancellations.}
 \label{milleren3}
 \end{table}

\subsection{String and Einstein frames and Newton's constants \label{frames}}

In the previous section we briefly discussed Newton's constant and the possibility of it remaining time-independent. However now that there is a chance of having multiple definitions of the Newton's constant in the heterotic theories, let us clarify what we mean by it here. For this let us first express the zero instanton sector of the action (expressed as a trans-series but now in ten-dimensions) in the following way:

{\footnotesize
\bg\label{doclogan}
{\bf S}_{10} = {1\over 2\kappa^2}\int d^{10}x \sqrt{-{\bf g}_{10}}\left[ 
e^{-\alpha_1{\bf \Phi}} {\bf R}_{10}({\bf g}_{\rm AB}) - {e^{-\alpha_2 {\bf \Phi}}\over 2} \left(\nabla {\bf \Phi}\right)^2 - e^{-\alpha_3{\bf \Phi}} {\bf H}_{10}^2 - {\alpha' e^{-\alpha_4{\bf \Phi}}\over 120} {\bf Tr}~{\bf F}^2 + ...
\right], \nd}
where the trace is in the adjoint representation of either $SO(32)$ or 
${\rm E}_8 \times {\rm E}_8$ gauge group, ${\bf \Phi} = \langle {\bf \Phi}\rangle_\sigma$ and ${\bf g}_{\rm PQ} = \langle {\bf g}_{\rm PQ}\rangle_\sigma$ with $\vert\sigma\rangle$ being the Glauber-Sudarshan state in the heterotic theories and $({\rm P, Q}) \in {\bf R}^{3, 1} \times {\cal M}_6$. The string and the Einstein frames in ten-dimensions are given by the following choices of $\alpha_i$ in \eqref{doclogan}:
\bg\label{taggilley}
(\alpha_1, \alpha_2, \alpha_3, \alpha_4) = (2, 2, 2, 2), ~~~~~ 
(\alpha_1, \alpha_2, \alpha_3, \alpha_4) = \Big(0, 0, 1, {1\over 2}\Big), \nd
with the former related to the string frame and the latter to the Einstein frame. (To avoid clutter, we do not provide distinguishing marks over the fields in the string and the Einstein frames\footnote{For example all field contractions that involve metric components would look different in each of the two frames. For example ${\bf H}^2$ in the string and the Einstein frames would be: 
\bg
{\bf H}^2 = \begin{cases} {\bf H}_{\rm ABC} {\bf H}_{\rm A'B'C'} {\bf g}^{\rm (s)AA'}{\bf g}^{\rm (s)BB'}{\bf g}^{\rm (s)CC'}\\
~~~ \\
{\bf H}_{\rm ABC} {\bf H}_{\rm A'B'C'} {\bf g}^{\rm (e)AA'}{\bf g}^{\rm (e)BB'}{\bf g}^{\rm (e)CC'}\end{cases} \nonumber \nd
where ${\bf H}_{\rm DEF} = \partial_{[{\rm D}}\langle {\bf B}_{\rm EF]}\rangle_\sigma + ...$ with the dotted terms coming from the Chern-Simons corrections and ${\bf g}^{\rm (s, e) AB} \equiv \langle {\bf g}^{\rm (s, e)AB}\rangle_\sigma$ being the metric in the string or the Einstein frames.}.) Dimensionally reducing the theory over a six-dimensional compact non-K\"ahler manifold ${\cal M}_6$, we can easily see that there are at least three possible Newton's constant appearing here. They are as follows:

{\footnotesize
 \bg\label{kashori}
 {\bf S}_4 = \int d^4x {{e^{-2{\bf \Phi}}}\over {\rm G}^{(\rm s)}_{\rm N}}\sqrt{-{\bf g}_4^{({\rm s})}} {\bf R}_4({\bf g}^{({\rm s})}_{\mu\nu}) =  \int d^4x {1\over {\rm G}^{(\rm e)}_{\rm N}}\sqrt{-{\bf g}_4^{({\rm e})}} {\bf R}_4({\bf g}^{({\rm e})}_{\mu\nu}) = \int d^4x {1\over {\rm G}^{(\rm e')}_{\rm N}}\sqrt{-{\bf g}^{({\rm e}')}} {\bf R}({\bf g}^{({\rm e')})}_{\mu\nu}), \nd}
 where $(\mu, \nu) \in {\bf R}^{3, 1}$; and the superscript s, e and ${\rm e}'$ denote the four-dimensional string frame, Einstein frame and the second Einstein frame. The second Einstein frame will be another realization of the Einstein frame that we shall explain below. The three choices of the Newton's constants, at least for the $SO(32)$ case, are\footnote{For the ${\rm E}_8 \times {\rm E}_8$ case one would get similar results albeit slightly more complicated because of the usage of {\it three} warp-factors ${\rm F}_1, {\rm F}_2$ and ${\rm F}_3$ as we shall see in section \ref{sec3}.}:

 {\scriptsize
 \bg\label{aspenbru}
\left({\rm G}^{({\rm s})}_{\rm N}\right)^{-1} = \int d^6 y \sqrt{{\bf g}^{({\rm s})}} = {\rm F}^2_1{\rm F}_2^2, ~~~\left({\rm G}^{({\rm e})}_{\rm N}\right)^{-1} = \int d^6 y \sqrt{{\bf g}^{({\rm e})}} = \sqrt{ {\rm F}_1}{\rm F}_2^2, ~~~\left({\rm G}^{({\rm e'})}_{\rm N}\right)^{-1} = \int d^6 y e^{-2{\bf \Phi}}\sqrt{{\bf g}^{({\rm s})}} = {\rm F}_2^2, \nonumber\\ \nd}
which clearly have different behaviors (we have taken ${\rm H}(y)= 1$ for simplicity). Interestingly if we keep 
${\rm G}^{({\rm e'})}_{\rm N}$ constant, then ${\rm F}_2$ = constant and for $0 < \beta_o < {2\over 3}$ we can still study the dynamical duality sequence from {\bf Table \ref{milleren2}}. However this can become very constrained, so we will not consider this and instead we will define the Newton's constant using ${\rm G}^{({\rm s})}_{\rm N}$. This would imply ${\rm F}_1 {\rm F}_2 =$ constant, which is what we take here. One advantage of this choice is that the internal manifold in the heterotic $SO(32)$ side will have no time-dependence\footnote{And also in the heterotic ${\rm E}_8 \times {\rm E}_8$ side at late time as we shall explain soon.}, thus allowing us to easily compute the flux quantizations, anomaly cancellations et cetera. Unfortunately this makes the other two definitions of the Newton's constant time-dependent. However if $\beta(t)$ takes the form as in \eqref{afleis}, then at late time all three Newton's constant will become time-independent and identical.

There is a small subtlety that we did not explain above. The appearance of the dilaton factor in the four-dimensional string-frame action $-$ sometime also called as the Jordan frame $-$ which allows us to express the string and Einstein frames' Newton's constant without invoking the dilaton has been much debated in the literature (see for example \cite{jordan}). However the Einstein frame that enters the discussion of \cite{jordan} isn't exactly the one we discussed in \eqref{aspenbru}, rather it is the one we get from the rescaling the four-dimensional metric in the string (or the Jordan) frame in the following way:
\bg\label{imperialwalk}
{\bf g}^{\rm (s)}_{\mu\nu} \equiv \langle{\bf g}^{\rm (s)}_{\mu\nu}\rangle_\sigma = e^{2{\bf \Phi}}{\bf g}^{\rm (e'')}_{\mu\nu} \equiv e^{2\langle{\bf \Phi}\rangle_\sigma}\langle{\bf g}^{\rm (e'')}_{\mu\nu}\rangle_\sigma, \nd
where ${\rm e}''$ is the third Einstein frame with $(\mu, \nu) \in {\bf R}^{3, 1}$. This mapping clearly differs from the mapping of ${\bf g}^{\rm (s)}_{\rm AB}$ to either ${\bf g}^{\rm (e)}_{\rm AB}$ or 
${\bf g}^{\rm (e')}_{\rm AB}$. The question now is two-fold: {\Su one}, which frame is preferred, string or Einstein? And {\Su two}, in Einstein frame, which Einstein metric is preferred, ${\bf g}_{\mu\nu}^{\rm (e)}, {\bf g}_{\mu\nu}^{\rm (e')}$ or ${\bf g}_{\mu\nu}^{\rm (e'')}$? From computational point of view, one can choose any frame and then conformally transform to other frames. The question however is from a {\it physical} point of view: in four-dimensions is the string-frame or the Einstein-frame preferred by experiments? If it is the latter, which of the three metrics is preferred? As far as we are aware of, the answer is not yet settled (see for example \cite{corda}), so we will present our results in terms of ${\bf g}_{\mu\nu}^{\rm(s)}$ and ${\bf g}_{\mu\nu}^{\rm(e)}$ in this paper, and leave the other two possibilities, with metric components ${\bf g}_{\mu\nu}^{\rm(e')}$ and ${\bf g}_{\mu\nu}^{\rm(e'')}$, for the diligent readers to work them out.

\subsection{$\alpha'$ corrections, Buscher's rules and duality sequences \label{sec2.3}}

There is one important question that could be raised regarding the duality chasings that we performed in {\bf Tables \ref{milleren1}, \ref{milleren2}}
and {\bf \ref{milleren3}} to get to the required de Sitter states: the duality transformations, in particular the T-duality transformations, have $\alpha'$ corrections, so shouldn't these corrections change the outcome of our computations? Additionally, we are dealing with non-supersymmetric backgrounds where these corrections are relevant, so how are the contents of the aforementioned tables justified? The answer lies rather remarkably on two aspects of the transformations: {\Su one}, the original duality transformations are done over {\it supersymmetric} solitonic configurations where the duality chasings could be justified, and {\Su two}, the corresponding {\it non-supersymmetric} backgrounds are determined using expectation values of the graviton, flux and the fermionic operators over the corresponding Glauber-Sudarshan states (which in turn are states defined over supersymmetric Minkowski background). The $\alpha'$ corrections, for example in the T-duality rules, could in-principle be incorporated in determining the supersymmetric solitonic background, {\it i.e.} the third columns of {\bf Tables \ref{milleren1}, \ref{milleren2}} and {\bf \ref{milleren3}}, but this will not effect the outcomes (coming from the expectation values) shown in the fourth columns of the three tables. One way to justify this is to note that all the configurations in the fourth columns of 
{\bf Tables \ref{milleren1}, \ref{milleren2}} and {\bf \ref{milleren3}} have to satisfy Schwinger-Dyson equations. To analyze this will however require some more ingredients than what we have developed so far, so we will come back to it in section \ref{sec6}.  Meanwhile a slightly simpler, duality motivated argument may be presented in the following way. 

Let us start with {\bf Table \ref{milleren1}}. Our aim would be to keep the de Sitter configuration in the desired theory $-$ here it will be type IIB $-$ simple and introduce $\alpha'$ corrections to the dual configurations. This means we will fix the third row by taking the metric of the solitonic configuration to be of the form:

{\footnotesize
\bg\label{ingrpitt}
g^{(0)}_{\mu\nu} = \eta_{\mu\nu}, ~~ g^{(0)}_{\rm MN}(y) = \sum\limits_{i, j = 0}^\infty {(g_s^{(0, b)})^i\over {\rm M}_p^j}\left[1 
+ {\cal O}\left(e^{-{\rm M}_p/g_s^{(0,b)}}\right)\right]
{g_{\rm MN}^{(0ij)}(y)}, ~~({\rm M, N}) \in {\cal M}_4 \times {\cal M}_2, \nd}
where for simplicity we have taken ${\rm H}(y) \equiv 1$ and have used a series in ${\rm M}_p$ instead of $\alpha'$ to keep in tune with the underlying M-theory configuration\footnote{The string scale $\alpha'$ can be defined over time-independent type IIA solitonic background as 
$\alpha' = l_s^2 = {(g_s^{(0,a)})^{-2/3}\over {\rm M}_p^2} \propto {1\over {\rm M}_p^2}$ where $g_s^{(0,a)}$ is the constant type IIA string coupling ($g_s^{(0,b)}$ denotes the corresponding constant IIB coupling). Thus even for a temporally varying background where the type IIA coupling $g_s$ has both spatial and temporal dependences, the correct way to relate $\alpha'$ to ${\rm M}_p$ will be to use the aforementioned definition.}. Additionally, $g^{(0ij)}_{\rm MN}$ for $(i, j) \ge 1$ are dimensionful such that both $g^{(0)}_{\rm MN}$ and $g^{(000)}_{\rm MN}$ are kept dimensionless. The above metric configuration would solve the background EOMs in the presence of higher order curvature and flux corrections to the action. For weak string coupling $g_s^{(0)} << 1$, and ${\rm M}_p >> 1$ the non-perturbative corrections are negligible, so \eqref{ingrpitt} in the presence of appropriate fluxes would solve the all-order perturbative EOMs.

Let us now T-dualize along $x^3$, by compactifying the $x^3$ direction, and then uplift to M-theory. The M-theory metric components will receive ${\rm M}_p$ corrections from the T-duality rules themselves, but we will maintain the topology as depicted in the first-row third-column of {\bf Table \ref{milleren1}}. This is possible by choosing appropriate choice of G-fluxes and consequently add in higher order perturbative corrections over the Minkowski minimum (the instanton corrections are negligible, so will ignore those saddles for simplicity, although could be taken for completeness). The M-theory metric components may be expressed as 
$\widetilde{g}^{(0)}_{\rm AB} \equiv \sum\limits_{i,j = 0}^\infty{(g_s^{(0, a)})^i\widetilde{g}^{(0ij)}_{\rm AB}\over {\rm M}_p^j} + {\cal O}(e^{-{\rm M}_p/g_s^{(0,a)}})$, or as 
$(\widetilde{g}^{(0)}_{\mu\nu}, \widetilde{g}^{(0)}_{mn}, \widetilde{g}^{(0)}_{\alpha\beta}, \widetilde{g}^{(0)}_{ab})$ where 
$(\mu, \nu) \in {\mathbb R}^{2, 1}, (m, n) \in {\cal M}_4, (\alpha, \beta) \in {\cal M}_2$ and $(a, b) \in {\mathbb{T}^2\over {\cal G}}$. Note that, due to ${\rm M}_p$ corrections, $\widetilde{g}^{(0)}_{\rm AB}$ components are in general different from the $(g^{(0)}_{\mu\nu}, g^{(0)}_{\rm MN})$ components from \eqref{ingrpitt}, although the lowest order Buscher's rule would relate $(g^{(000)}_{\mu\nu}, g^{(000)}_{\rm MN})$ with $\widetilde{g}^{(000)}_{\rm AB}$ in the standard way \cite{berg1}. (To keep the analysis simple, we will henceforth ignore the instanton corrections to the solitonic configurations.)

Once the solitonic configuration is fixed in M-theory, the first-row and fourth-column of {\bf Table \ref{milleren1}} provides the required metric 
after taking the expectation value over the Glauber-Sudarshan state as
$\langle {\bf g}_{\rm AB}\rangle_\sigma \equiv (g_s^{-8/3}, g_s^{-2/3}{\rm F}_2, g_s^{-2/3}{\rm F}_1, g_s^{4/3})$ respectively. Since 
$\langle {\bf g}_{\rm AB}\rangle_\sigma = \widetilde{g}^{(0)}_{\rm AB} + \langle \delta {\bf g}_{\rm AB}\rangle_\sigma$, where $\widetilde{g}^{(0)}_{\rm AB}$ is the supersymmetric Minkowski background with the aforementioned ${\rm M}_p$ corrections, and $\langle \delta {\bf g}_{\rm AB}\rangle_\sigma$ appears from resurgent sum as elaborated in \cite{borel2}, one may incorporate additional ${\rm M}_p$ corrections through latter. This may be seen in the following way. Demanding:

{\footnotesize
\bg\label{ebrowning}
\langle {\bf g}_{\rm AB} \rangle_{\sigma} &\equiv & {\int [{\cal D} g_{\rm MN}] [{\cal D}{\rm C}_{\rm MNP}] [{\cal D}\overline\Psi_{\rm M}] [{\cal D}
\Psi_{\rm N}][{\cal D}\Upsilon_g]~e^{-{\bf S}_{\rm tot}}~ \mathbb{D}^\dagger({\alpha}, {\beta}, {\gamma}; \Xi)~ g_{\rm AB}(x, y, z)~
\mathbb{D}({\alpha}, {\beta}, {\gamma}; \Xi) \over 
\int [{\cal D} g_{\rm MN}] [{\cal D}{\rm C}_{\rm MNP}] [{\cal D}\overline\Psi_{\rm M}] [{\cal D}
\Psi_{\rm N}][{\cal D}\Upsilon_g]
~e^{-{\bf S}_{\rm tot}} ~\mathbb{D}^\dagger({\alpha}, {\beta}, {\gamma}; \Xi) 
\mathbb{D}({\alpha}, {\beta}, {\gamma}; \Xi)} \\
& = &  \sum_{\{{\bf s}\}}\left[{\mathbb{F}_{({\bf s}, {\rm A}, {\rm B})}\over g_{({\bf s}, {\rm A}, {\rm B})}^{1/l}}
\int_0^\infty d{\rm S} ~{{\rm exp}\left(-{{\rm S}/ g_{({\bf s}, {\rm A}, {\rm B})}^{1/l}}\right) \over 1 - {\cal A}_{({\bf s}, {\rm A}, {\rm B})}{\rm S}^l}\right]_{\Su {\rm P. V}}
\int_{k_{\rm IR}}^\mu d^{11}k ~{\overline\alpha_{\rm AB}(k)\over a(k)}~
{\bf Re}\left(\psi_k({\rm X})~e^{-i(k_0 - \overline\kappa_{\rm IR})t}\right)\nonumber\\
& \equiv & (g_s^{-8/3}, ~g_s^{-2/3}{\rm F}_2, ~g_s^{-2/3}{\rm F}_1, ~g_s^{4/3}) ~ = ~ \big(g_s^{-8/3} \widetilde{\rm g}^{(0)}_{\mu\nu}, ~g_s^{-2/3}{\rm F}_2 ~\widetilde{\rm g}^{(0)}_{mn}, ~g_s^{-2/3}{\rm F}_1~ \widetilde{\rm g}^{(0)}_{\alpha\beta}, ~ g_s^{4/3} \widetilde{\rm g}^{(0)}_{ab}\big), \nonumber \nd}
where $g_{\rm AB} = \widetilde{g}^{(0)}_{\rm AB} + \delta g_{\rm AB}$ and $\Upsilon_g$ is the ghost sector.
The ${\rm M}_p$ corrections come from two possible sources\footnote{$g_{({\bf s})}$ should not be confused with $g_s$. The former is proportional to inverse powers of ${\rm M}_p$ for configurations labelled by ${\bf s}$ \cite{borel3}, whereas the latter is the type IIA coupling which is generically a temporally dependent function \cite{desitter2}. See also section \ref{vanmeyaagu} and \eqref{vanandgorom2} where $1+\sigma$ (or $1+\hat\sigma$) is related to $l$ in \eqref{ebrowning}.}, {\Su one}, from the action itself, which is now written as 
${\bf S}_{\rm tot} \equiv {\bf S}_{\rm kin} + {\bf S}_{\rm int} + 
{\bf S}_{\rm ghost} + {\bf S}_{\rm gf}$, and {\Su two}, from the form of the Glauber-Sudarshan state $\overline\alpha_{\rm AB} \equiv \sum\limits_{i,j = 0}^\infty {(g_s^{(0,a)})^i\overline\alpha^{(0ij)}_{\rm AB}\over {\rm M}_p^j}$ related to the metric components (similar decompositions for $\overline\beta_{\rm ABCD}$, for the G-flux components, and $\overline\gamma_{\rm A}$, for the Rarita-Schwinger fermions with the set of on-shell fields given by $\Xi$ such that 
$\Xi \equiv (g_{\rm AB}, {\rm C}_{\rm ABD}, \Psi_{\rm A})$) defined over the corresponding solitonic configurations. Note two things. {\Su One}, ${\bf S}_{\rm tot}$ as it appears above is not the full result: we have kept the off-shell DOFs massive and haven't expressed the result in a trans-series form \cite{joydeep}. This will be rectified later. {\Su Two}, the ${\rm M}_p$ dependence from 
${\bf S}_{\rm tot}$ shows the usual Gevrey growth of $l$ \cite{gevrey, borelborel}, which can then be summed over as in \cite{borel2}. The ${\rm M}_p$ dependence in $\overline\alpha_{\rm AB}$ (as well as in $\overline\beta_{\rm ABCD}$ and $\overline\gamma_{\rm A}$) also effects ${\cal A}_{({\bf s}, {\rm A}, {\rm B})}$ as may be inferred from \cite{borel2, borel3}, where the subscript ${\bf s}$ ranges over the types of allowed interactions (in \cite{borel2} only scalar degrees of freedom were taken, with coupling $g_{({\bf s}, {\rm A}, {\rm B})}$, where ${\bf s}$ represents configurations with all allowed powers of the scalar fields and their derivatives). $\mathbb{F}_{({\bf s}, {\rm A}, {\rm B})}$ represents the number of degeneracies for a given choice of ${\bf s}, {\rm A}$ and ${\rm B}$
\cite{borel3}, but the details are not important for this work. The other parameters appearing in \eqref{ebrowning} are defined in the following way. The coordinates $(x, y, z) \in \big({\bf R}^{2, 1}, {\cal M}_4 \times {\cal M}_2, {\mathbb{T}^2\over {\cal G}}\big)$;  $\vert\sigma\rangle \equiv \vert\alpha, \beta, \gamma\rangle = \vert\alpha\rangle \otimes \vert\beta\rangle \otimes \vert \gamma\rangle$ is the required Glauber-Sudarshan state corresponding to the metric, flux and the Rarita-Schwinger fermions; $\mathbb{D}(\sigma) \equiv \mathbb{D}(\alpha, \beta, \gamma; \Xi)$ is a non-unitary displacement operator with $\mathbb{D}^\dagger \mathbb{D} \ne 1$; and $l$ is the Gevrey growth \cite{gevrey} of the underlying nodal diagrams \cite{borel2}. The energy scale is chosen such that:
\bg\label{greeneve}
\Lambda_{\rm UV} ~>~ {\rm M}_p ~>~ {\rm M}_s ~>~ \hat{\mu} ~>~ \mu ~>~ \overline\kappa_{\rm IR} ~>~ k_{\rm IR}, \nd 
with $k_{\rm IR} \le k \le \mu$, where $\hat\mu$ is the scale associated with the size of the internal compact manifold ${\cal M}_4 \times {\cal M}_2$, $\Lambda_{\rm UV}$ is the UV cutoff, $\overline\kappa_{\rm IR}$ is an IR scale and $k_{\rm IR}$ is the IR cutoff. $a(k)$ is the massless graviton propagator.

In writing $\overline\alpha_{\rm AB}$ in \eqref{ebrowning}, we have hidden a subtlety related to the off-shell states coming from the solitonic configurations. This may be seen by comparing the first and the second lines of \eqref{ebrowning}. In the first line we have $g_{\rm AB} = \widetilde{g}^{(0)}_{\rm AB} + \delta g_{\rm AB}$ that contains both the solitonic configuration as well as the fluctuations over it. This doesn't influence the path integral because, for example, ${\cal D}g_{\rm AB} = {\cal D}\delta g_{\rm AB}$ and therefore the integration in \eqref{ebrowning} is over the fluctuations. To see how this works in the presence of the displacement operator, let us define it in the following way (see also \cite{coherbeta, borel2, wdwpaper, joydeep}): 

{\footnotesize
\bg\label{maneely}
\mathbb{D}(\sigma) \equiv \mathbb{D}(\alpha, \beta, \gamma; \Xi)
= {\rm exp}\left(\prod_{i = 1}^4\int d^{11}k_i \sum_{n, m, p} z_{n_1..n_4mp}~ k_i^{2n_i} \odot \sigma^{(p, m)} \otimes \left(\Xi\right)^m + {\rm c.c}\right) \nd}
where $\odot$ attaches the four sector of momenta with the four sectors\footnote{These four sectors come from the metric fluctuations along ${\bf R}^{2, 1}$, the metric fluctuations along the internal eight-manifold, the fluctuations of the G-flux components, and the fluctuations of the Rarita-Schwinger fermions. In other words $\sigma \equiv (\alpha_{\mu\nu}, \alpha_{\rm MN}, \beta_{\rm ABCD}, \gamma_{\rm A})$ 
for $({\rm A, B}) \in {\bf R}^{2, 1} \times {\cal M}_4 \times {\cal M}_2 \times {\mathbb{T}^2\over {\cal G}};  ({\mu, \nu}) \in {\bf R}^{2, 1}$; and $({\rm M, N}) \in {\cal M}_4 \times {\cal M}_2 \times {\mathbb{T}^2\over {\cal G}}$ where, with $\sigma \equiv \widehat\sigma + \overline\sigma$,  we reserve $\widehat\sigma$ to take care of the solitonic components and $\overline\sigma$ for the actual positive cosmological constant background. For more details see \cite{joydeep}.} in the definition of $\sigma$; $\otimes$ takes care of the appropriate contractions of the tensor indices from the set $\sigma$ with the on-shell fields from the set $\Xi$; $z_{n_1..n_4mp}$ are dimensionful coefficients; and $(n_i, m, p) \in (\mathbb{Z}_+, \mathbb{Z}_+, \mathbb{Z}_+)$ where $m, p$ are integer set distributed over the four set of fields in $\sigma$ and $\Xi$ respectively. It is now easy to see that in the linear part of $\mathbb{D}(\sigma)$, {\it i.e.} for $m = 1$ in \eqref{maneely},  
the solitonic pieces of the fields do not effect the path integral analysis and are typically eliminated by the denominator of the path integral. The subtlety alluded to above appears exactly from the solitonic parts of the fields. For example, going from the first line of \eqref{ebrowning} to the second line, we see only the presence of $\overline{\alpha}_{\rm AB}$. The reason is that the solitonic part is eliminated by another piece from $\alpha_{\rm AB} \in \sigma$, which we can write as $\alpha_{\rm AB} = \overline{\alpha}_{\rm AB} + \widehat\alpha_{\rm AB}$. The second part, {\it i.e.} $\widehat\alpha_{\rm AB}$, is the off-shell piece that takes the form as given in eq. (5.4) in \cite{joydeep}, whereas the first part, {\it i.e.} $\overline\alpha_{\rm AB}$, takes the form as in eq. (5.2) of \cite{joydeep}. As explicitly demonstrated in \cite{joydeep}, this off-shell piece is responsible in removing the solitonic background, and therefore does not appear in the second line of \eqref{ebrowning}\footnote{There is yet another subtlety here that may be explained in the following way. Since the off-shell state has a wave-function, the cancellation with the solitonic background only happens for the most dominant amplitude implying that the $g_s$ scaling shown in the last line of \eqref{ebrowning} is related to the most dominant amplitudes for the two wave-functions associated with $\overline{\alpha}_{\rm AB}$ and 
$\widehat{\alpha}_{\rm AB}$.}. 

There is still one more interesting point that needs some discussion when we look at the final expression for the expectation values in \eqref{ebrowning}, namely that the Glauber-Sudarshan state in M-theory simply produces a temporally modulated background over the solitonic configuration. In other words it appears to convert:
\bg\label{gulabs}
\big(\widetilde{g}^{(0)}_{\mu\nu},  ~\widetilde{g}^{(0)}_{mn}, ~ \widetilde{g}^{(0)}_{\alpha\beta}, ~ \widetilde{g}^{(0)}_{ab}\big)
~ \to ~ \big(g_s^{-8/3} \widetilde{\rm g}^{(0)}_{\mu\nu}, ~g_s^{-2/3}{\rm F}_2 ~\widetilde{\rm g}^{(0)}_{mn}, ~g_s^{-2/3}{\rm F}_1~ \widetilde{\rm g}^{(0)}_{\alpha\beta}, ~ g_s^{4/3} \widetilde{\rm g}^{(0)}_{ab}\big), \nd
where conservatively we expect $\widetilde{\rm g}^{(0)}_{\rm AB} = \widetilde{g}^{(0)}_{\rm AB}$, with $\widetilde{g}^{(0)}_{\rm AB}$ being the metric component at the solitonic level. The powers of $g_s$ are the dominant contributions and they cannot change, so if there are $g_s$ corrections they have to be sub-dominant ones. Thus generically we can expect the following mapping between the solitonic and the coherent state variables:
\bg\label{ramyakri}
\widetilde{\rm g}^{(0)}_{\rm AB}({\bf x}, y, w; g_s) = \widetilde{g}^{(0)}_{\rm AB}({\bf x}, y, w) + 
\sum_{j = 0}^\infty\sum_{k = 1}^\infty \left({g_s\over {\rm HH}_o}\right)^{2k/3} {\hat{g}^{(0jk)}_{\rm AB}\over {\rm M}_p^{j}}, \nd
where $\hat{g}^{(0jk)}_{\rm AB} \equiv \hat{g}^{(0jk)}_{\rm AB}({\bf x}, y, w)$ are dimensionful metric components with $(j, k) \in \mathbb{Z}$, ${\rm H} \equiv {\rm H}(y)$ and ${\rm H}_o \equiv {\rm H}_o({\bf x})$. (In this paper we have kept ${\rm H}(y) = {\rm H}_o({\bf x}) = 1$ for simplicity. See \cite{desitter2} for the inclusion of these parameters.) 
If $\hat{g}_{\rm AB}^{(0jk)} = 0$, then
it is clearly the simplest scenario with temporal modulations. Our earlier analysis with the Schwinger-Dyson's equations in \cite{desitter2, coherbeta2} however provided some indication that in fact \eqref{ramyakri}, with $\hat{g}_{\rm AB}^{(0jk)} = c_{jk}\widetilde{g}^{(0)}_{\rm AB}({\bf x}, y, w)$, could be the case for all metric components except the space-times ones where $\hat{g}_{\mu\nu}^{(0jk)} = 0$ for $(\mu, \nu) \in {\bf R}^{1, 2}$. 
Both of these could be generalized which, in the language of \eqref{maneely}, would appear from switching on $\sigma^{(p, 1)}$ with 
$p \in \mathbb{Z}_+$. For every choice of $p \in \mathbb{Z}_+$, we can split the Fourier coefficients $\sigma^{(p, 1)}$ into two parts, one to take care of the $\alpha'$ corrected solitonic background, and the other to insert in the $\alpha'$ (or more appropriately ${\rm M}_p$) corrections to the de Sitter space-time (both along internal eight-manifold and the space-time ${\bf R}^{2, 1}$ directions). This way all order ${\rm M}_p$ corrections may be incorporated in all the duality frames, both at the solitonic and at the de Sitter levels. For the flux degrees of freedom we expect similar consideration too $-$ see section 4 in the second reference of \cite{coherbeta2} $-$ although a more detailed study is required to pin-point the exact functional forms of the $g_s$ independent pieces in the flux components. We will discuss a bit more on the story in sections \ref{sec4.2.1} and \ref{sec4.2.2} where slightly more complicated choices of the metric, compared to \eqref{ramyakri}, and fluxes, compared to \cite{desitter2, coherbeta, coherbeta2}, will be considered.

\subsection{Revisiting the four-dimensional cosmological constant \label{ccnews}}

The above analysis provides a precise way to see how the original supersymmetric background is now modified in the presence of ${\rm M}_p$ corrections to the duality rules. In fact the two sources of ${\rm M}_p$, one from the action and one from the Glauber-Sudarshan state, intersect only in the analysis of the principal value integral that fixes the {\it four-dimensional} cosmological constant to (see \cite{borel2, joydeep}):
\bg\label{montehan2}
{1\over \Lambda^{\kappa}} \equiv \sum_{\{{\bf s}\}}\left[{\mathbb{F}_{({\bf s}, \mu, \nu)}\over g_{({\bf s}, \mu, \nu)}^{1/l} {\rm M}_p^{2\kappa}}
\int_0^\infty d{\rm S} ~{\rm exp}\Bigg(-{{\rm S}\over g_{({\bf s}, \mu, \nu)}^{1/l}}\Bigg) {1\over 1 - {\cal A}_{({\bf s}, \mu, \nu)}{\rm S}^l}\right]_{\Su {\rm P. V}}, \nd
where $\kappa = {4\over 3}$, $(\mu, \nu) \in {\bf R}^{2, 1}$ with the ${\rm M}_p$ dependence of ${\cal A}_{({\bf s}, \mu, \nu)}$ coming from the corresponding ${\rm M}_p$ dependence in 
$(\overline\alpha_{\rm AB}, \overline\beta_{\rm ABCD}, \overline\gamma_{\rm A})$ as mentioned earlier; and $l$ is the same as $1+\sigma$ (or $1+\hat\sigma$) from \eqref{vanandgorom2}. Our analysis in the previous section provided a compelling reason for the ${\rm M}_p$ dependence in $\overline\alpha_{\rm AB}$ (as also in $\overline\beta_{\rm ABCD}$ and $\overline\gamma_{\rm A}$). They appear from the definition of $\mathbb{D}(\sigma)$ itself in \eqref{maneely}: the non-zero coefficients $z_{n_1..n_4m1}$ in \eqref{maneely} precisely indicates the presence of the inverse powers of ${\rm M}_p^{2n}$ (once we put back the dimensions in $k^{2n}$) in $\overline\alpha_{\rm AB}$! Similar arguments can be given for the presence of ${\rm M}_p$ dependence in $\overline\beta_{\rm ABCD}$ and $\overline\gamma_{\rm A}$. This hopefully resolves the last loose end in our chain of arguments.

There is one more point that we should quickly mention before moving ahead. The expression for $\Lambda$ in \eqref{montehan2} is a {\it constant}, {\it i.e.} not varying with respect to time. However this is not the most generic case, which, in fact, will require us to make $\Lambda$ itself a temporally varying quantity. Since the $\Lambda$ that appears in our set-up is an {\it emergent} quantity, it is not too hard to argue its temporal dependence. In fact we can express the temporally varying dark energy as $\Lambda(t)$, with $\Lambda(t)$ given by \cite{joydeep}:
\bg\label{marapaug}
\Lambda(t) = \Lambda + \check{\Lambda}(t), \nd
where $\Lambda$ is the constant {\it bare} part of the cosmological constant whose value is determined in \eqref{montehan2}. However the actual determination of $\check{\Lambda}(t)$ will require us to first develop some groundwork to address this line of computations\footnote{See also section 5.3 in \cite{joydeep}.}. We will come back to a detailed analysis of this in section \ref{curvten}. Meanwhile, since we expect ${\check{\Lambda}(t) \over \Lambda} << 1$, we will continue with $\Lambda(t) \approx \Lambda$.

The three parameters appearing in \eqref{montehan2} or in the second line of \eqref{ebrowning}, namely ${\cal A}_{({\bf s}, {\rm A}, {\rm B})}$, $g_{({\bf s}, {\rm A}, {\rm B})}$ and $\mathbb{F}_{({\bf s}, {\rm A}, {\rm B})}$, may be bounded either from above or below by ${\cal A}_{({\bf s})}$, $g_{({\bf s})}$ and $\mathbb{F}_{({\bf s})}$ respectively via ${\cal A}_{({\bf s}, {\rm A}, {\rm B})} \le {\cal A}_{({\bf s})}$, ${g}_{({\bf s}, {\rm A}, {\rm B})} \ge g_{({\bf s})}$ and $\mathbb{F}_{({\bf s}, {\rm A}, {\rm B})} \le 
\mathbb{F}_{({\bf s})}$ $\forall ~({\rm A, B}) \in {\bf R}^{2, 1} \times {\cal M}_4 \times {\cal M}_2 \times {\mathbb{T}^2\over {\cal G}}$ \cite{borel2, borel3}. Under this bound, and defining $u_{({\bf s})} \equiv {\cal A}^{1/l}_{({\bf s})} {\rm S}$ and $c^l_{({\bf s})} \equiv {\cal A}_{({\bf s})} g_{({\bf s})}$, in the limit ${\rm M}_p \to \infty$ or $g_{({\bf s})} \to 0$ the four-dimensional {\it positive} cosmological constant \eqref{montehan2} takes the following form \cite{borel2, joydeep, borel3}:
\bg\label{laumoon}
\Lambda = \lim_{c_{({\bf s})}\to 0} {{\rm M}_p^2 \over \Big[\sum\limits_{\{{\bf s}\}}
{\mathbb{F}_{({\bf s})}\over c_{({\bf s})}}
\int_0^\infty du_{({\bf s})}~{{\rm exp}\left(-u_{({\bf s})}/c_{({\bf s})}\right) \over 1 - 
u_{({\bf s})}^l}\Big]^{3/4}_{\Su {\rm P. V}}} = {{\rm M}_p^2\over\Big(\sum\limits_{\{{\bf s}\}} {\mathbb{F}_{({\bf s})}}\Big)^{3/4}} ~ << ~{\rm M}_p^2, \nd
justifying its smallness {\it with respect to} ${\rm M}_p^2$. (How \eqref{laumoon} appears from using new technical ingredients like {\it Borel Boxes} have been discussed in \cite{joydeep} so we will not repeat them here and the readers may look up the details in \cite{joydeep}.) Note that the result in this limit is completely independent of ${\cal A}_{({\bf s}, {\rm A}, {\rm B})}$ but depends only the fact that the sum of the degeneracy $\mathbb{F}_{({\bf s})}$ is very large but {\it convergent} and ${\rm M}_p \to \infty$. Thus in principle there is no {\it classical} limit of ${\rm M}_p \to \infty$ as such. This will be elaborated further in \cite{borel3}. Also under the aforementioned bound, the second line of \eqref{ebrowning}, may be expressed completely in terms of ${\cal A}_{({\bf s})}$, $g_{({\bf s})}$ and $\mathbb{F}_{({\bf s})}$ instead of ${\cal A}_{({\bf s}, {\rm A}, {\rm B})}$, $g_{({\bf s}, {\rm A}, {\rm B})}$ and $\mathbb{F}_{({\bf s}, {\rm A}, {\rm B})}$ $-$ which is exactly how it appeared in \cite{hetborel}. This simplification however raises the following question. In the second line of \eqref{ebrowning}, the factor in front of the Fourier integral is the {\it same} for all components of the metric tensor. Now that we have identified that factor with $({\rm M}_p^2/\Lambda)^{-4/3}$ giving us the $g_s^{-8/3} \equiv (\sqrt{\Lambda} \vert t\vert)^{-8/3}$ coefficient for the metric along ${\bf R}^{2, 1}$, how are we to justify the other powers of $g_s$ along the internal directions? Of course while the {\it dimensionless}\footnote{Recall that at the beginning of section 3 in \cite{borel2} we expressed all parameters using dimensionless variables by measuring everything with respect to ${\rm M}_p$ (or with respect to $\hat\mu$ from \eqref{greeneve} in \cite{borel3}).} conformal time $t$ could be easily recovered from the Fourier factor $\overline\alpha_{\rm AB}$, the powers of the cosmological constant would appear to {\it not} match with the expected powers of $g_s$. The resolution is simple: we can absorb the extra powers of $\Lambda$ in the definitions of $\overline{\alpha}_{mn}, \overline{\alpha}_{\alpha\beta}$ and $\overline{\alpha}_{ab}$ respectively (or in the definitions of the respective wave-functions). This seems like a viable possibility because any changes to ${\cal A}_{({\bf s}, \mu, \nu)}$
{\it do not} effect the value of the cosmological constant $\Lambda$ in \eqref{laumoon}. However this raises a different question: if we are allowed to change $\overline\alpha_{\mu\nu}$ by absorbing arbitrary factors of ${\rm M}_p$, shouldn't this then give us {\it any} value of the cosmological constant $\Lambda$? The answer is no because of the reason given earlier (see also \cite{joydeep}): the $\widehat\alpha_{\mu\nu}$ piece that {\it cancels} the solitonic configuration also fixes the boundary conditions appropriately to give us \eqref{laumoon}. This fits well with the fact that both the conformal time $t$ and the wave-function normalization are fixed: the former is fixed by the Hubble constant and the latter by the IR cut-off $k_{\rm IR}$, implying that any arbitrary changes to $\overline\alpha_{\mu\nu}$ or the wave-function are {\it not} allowed for a given value of $p \in \mathbb{Z}_+$ in $\sigma^{(p, 1)}$. After the dust settles, the form of the four-dimensional cosmological constant is as given in \eqref{laumoon}  without any ambiguous factors.

The cosmological constants appearing in {\bf Tables \ref{milleren2}} and 
{\bf \ref{milleren3}} follow similar arguments as above. We will fix the metric of the solitonic configurations in the sixth rows and third columns of both the tables as in \eqref{ingrpitt} with $({\rm M, N})$ now scanning the coordinates of ${\cal M}_4$ as well as ${\cal M}_2 = \widetilde{\bf S}^1_\alpha \times \widetilde{\bf S}^1_\beta$ with $(\alpha, \beta)$ respectively taking values as in the sixth rows and third columns of {\bf Tables \ref{milleren2}} and {\bf \ref{milleren3}}. Once this is fixed, we can access other solitonic theories by going up the rows in the two tables. At every stage we can add ${\rm M}_p$ corrections to the metric and flux configurations of the corresponding solitonic configurations. The difference from {\bf Table \ref{milleren1}} is that, at every rows of the two tables, the metric configurations would differ such that only the zeroth order configurations would be related by the usual Buscher's dualities of \cite{berg1}. The ${\rm M}_p$ corrections at every stage could either be estimated from the higher order ${\rm M}_p$ corrections in 
${\bf S}_{\rm tot}$ or from the higher order ${\rm M}_p$ corrections to the Buscher's rules (although the former is preferable here because going up the rows in the two tables involve generic U-dualities and not just T-dualities).

The expectation values of the metric operators, which appear in the fourth columns of the two tables, could now be computed using the path-integral approach as in \eqref{ebrowning}. The cosmological constant may be determined at every stages of the duality sequences, but we will prefer to do this right at the top rows where we study the M-theory uplifts by taking advantage of our earlier computational machinery of \cite{desitter2, coherbeta, coherbeta2, borel2, joydeep}. Interestingly, while the four-dimensional cosmological constants take the same forms as in  \eqref{montehan2} for both cases $-$ with values of ${\cal A}_{({\bf s}, \mu, \nu)}$ changing due to the differences in the functional forms of $\overline\alpha_{\rm AB}$ $-$ the actual values of the cosmological constants
would only depend on the degeneracy factors exactly as in \eqref{laumoon} in the limit ${\rm M}_p \to \infty$. This should justify both the {\it similarities} and the {\it smallness} of the cosmological constants for the models in {\bf Tables \ref{milleren1}, \ref{milleren2}} and {\bf \ref{milleren3}}. More details will be presented in \cite{borel3}.

There is however one more point related to \eqref{marapaug} that needs some discussion. The deviation from a constant $\Lambda$, given by $\check{\Lambda}(t)$, is small but a temporally varying number. Such a choice, while consistent with the recent result from DESI \cite{desibao}, nevertheless will depend whether the statistical significance of the DESI result increases or decreases in the five-year run. However our analysis suggests that such a small temporal variation may be easily accommodated in our set-up by slightly changing the Glauber-Sudarshan state. Note that we do not have to introduce any new degrees of freedom to accommodate such a result, and therefore our construction does not suffer from the usual problems associated with quintessence in string theory. The question however is whether such temporally varying cosmological constant scenario solves the Schwinger-Dyson equations or is consistent with the duality sequence. In sections \ref{curvten} and \ref{sec6} we will give further elaboration on these and other related scenarios.

\subsection{The curious case of the string versus the Einstein 
frames \label{secc2.6}}

Before ending this section let us bring forth yet another subtlety that we briefly alluded to in section \ref{frames}. This appears when we try to describe the configurations, both for the SO(32) and ${\rm E_8} \times {\rm E_8}$ theories, in the Einstein frames as shown in the last rows of {\bf Table \ref{milleren2}} and {\bf Table \ref{milleren3}} respectively. The physics isn't exactly similar: 
we cannot get four-dimensional de Sitter configurations in both the frames
simultaneously unless ${\rm F}_i \to 1$ (or a constant) at late times. As an example, if we demand de Sitter configurations in the Einstein frames
for both the heterotic theories, we will have to identify $g_s^{-2} \sqrt{\rm F_1} \equiv {1\over \Lambda t^2}$ with $\alpha_o = -{\beta_o\over 4}$ and $\Lambda$ from \eqref{marapaug}, while still maintaining $\beta_o < {2\over 3}$ as to not off-set the duality sequence and keep the four-dimensional Newton's constant time-independent. Unfortunately this makes the internal four-manifold ${\cal M}_4$ shrink to zero size and the toroidal fibre  infinite, although the total volume remains time-independent. The string coupling however continues to be weak. Additionally keeping ${\rm F}_i \to 1$ (or a constant) at late times {\rm i.e.} $t \to 0$, although appears to provide de Sitter configurations at late times in both the frames, cannot simultaneously keep time-independent Newton's constant in the string frame for the range $-{1\over \sqrt{\Lambda}} < t < 0$. In type IIB, as shown in {\bf Table \ref{milleren1}}, this issue doesn't appear because the string coupling is stabilized to 1 (or a constant).

\begin{figure}[h]
\centering
\begin{tabular}{c}
\includegraphics[width=3in]{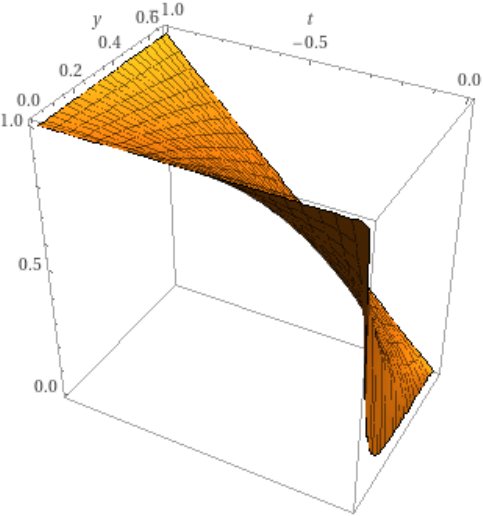}
\end{tabular}
\caption[]{The plot of ${\rm F}_1(t) = \left(\Lambda t^2\right)^{\beta_o\over 2 - \beta_o}$ for $-1 \le t \le 0$, $\Lambda \equiv 1$ and $0 < \beta_o \equiv y \le {2\over 3}$ in the string frame. For small values of $t$, for example $\epsilon < t < 0$ with $\epsilon$ as in \eqref{tegtrex}, one may control the behavior of ${\rm F}_1(t)$ using \eqref{evalike} thus avoiding the late-time singularity of ${\cal M}_4$ in the Einstein frame.}
\label{f1plot}
\end{figure} 

We could however try a different line of thought by switching on a non-trivial dilaton in the type IIB side. The scenario is depicted in {\bf Table \ref{millerenoo}}. Keeping $0 < \beta_o < {2}$ may still guarantees safe transit to the heterotic side, but now there are subtleties. First, the string couplings in both the Type I and the heterotic sides are now identities, so they are all strongly coupled. Secondly, there are late time singularities because $\alpha_o = {\beta_o\over 2}$. This already means that both ${\cal M}_4$ and $\mathbb{T}^2_{45}$ are shrinking in the Type IIB side, although the former is shrinking at a {\it slower} rate than the latter at late time. This leads to the late-time singularities in both the string and the Einstein frames, which were absent in the string frames for the configuration studies in {\bf Tables \ref{milleren2}} and {\bf \ref{milleren3}}. Interestingly however, although we do seem to get de Sitter configurations in all the theories starting from Type IIB and moving down the rows in {\bf Table \ref{millerenoo}}, they all suffer from late-time singularities (and the possibilities of having time-dependent Newton's constants\footnote{For example, demanding time-independent Newton's constants in Type I and the heterotic theories would  necessitate having time-dependent Newton's constants in the Type II theories; and so on.}). Additionally the seven-branes in the IIB side become dynamical to account for the time-dependent dilaton which is now reflected in the complicated metric components in the M-theory uplift.
For example:
\bg\label{rindrag}
\langle {\bf g}_{\rm AB}\rangle_\sigma & = &
\Big(\langle {\bf g}_{\mu\nu}\rangle_\sigma, ~\langle {\bf g}_{mn}\rangle_\sigma, ~\langle {\bf g}_{\alpha\beta}\rangle_\sigma, ~
\langle {\bf g}_{33}\rangle_\sigma, ~\langle {\bf g}_{11, 11}\rangle_\sigma\Big)\\
& = & \left(g_s^{-8/3 - 2\beta_o/3}, ~ g_s^{-2/3 - \beta_o/6}\widetilde{g}^{(0)}_{mn},  ~ g_s^{-2/3 + \beta_o/3}\widetilde{g}^{(0)}_{\alpha\beta}, ~g_s^{4/3 - 2\beta_o/3}, ~g_s^{4/3 + 4\beta_o/3}\right), \nonumber \nd
becomes the metric components in the M-theory uplift as shown in the first row of {\bf Table \ref{millerenoo}}, where $g_s = \sqrt{\Lambda} \vert t \vert$, $\widetilde{g}^{(0)}_{\rm AB}$ is given in \eqref{ramyakri}; and $0 < \beta_o < 2$.
Thus demanding de Sitter configurations in both the string and the Einstein frames appear to make the analysis more involved.

\begin{table}[tb]  
 \begin{center}
\renewcommand{\arraystretch}{2.6}
\resizebox{\textwidth}{!}{
\begin{tabular}{|c|c|c|c|c|c|}\hline
Theory & duality & Solitonic Configuration & $\langle {\bf g}_{\rm AB}\rangle_{\sigma}$ & coupling & vector bundle \\ \hline
M &  & $\mathbb{R}^{2, 1} \times {\cal M}_4 \times {\mathbb{T}^2_{45} \times {\bf S}^1_{3}\times {\bf S}^1_{11}\over {\cal I}_4 {\cal I}_5 {\cal I}_3 {\cal I}_{11}}$
& $\left(g_s^{-8/3} {\rm F}_1^{-2/3}, (g_s {\rm F}_1)^{-2/3}{\rm F}_2, g_s^{-2/3}{\rm F}^{1/3}_1, g_s^{4/3}{\rm F}_1^{-2/3}, (g_s{\rm F}_1)^{4/3}\right)$ & & \\ \hline
IIA &  dim. red. along $x^{11}$ & $\mathbb{R}^{2, 1} \times {\cal M}_4 \times {\mathbb{T}^2_{45} \times {\bf S}^1_3\over \Omega {\cal I}_4 {\cal I}_5 {\cal I}_3}$  & $\left(g_s^{-2},~ {\rm F}_2,~ {\rm F}_1, ~g_s^2\right)$ & $g_s {\rm F}_1$ & 
${\rm O6} \oplus {\rm D6}$ \\ \hline
IIB &T duality along $x^3$ &$\mathbb{R}^{3, 1} \times {\cal M}_4 \times {{\bf S}^1_{4} \times {\bf S}^1_{5}\over \Omega (-1)^{{\rm F}_{\rm L}}{\cal I}_4 {\cal I}_5}$
&  $\left(g_s^{-2}, ~ {\rm F}_2, ~{\rm F}_1, ~{\rm F}_1\right)$ & ${\rm F}_1$ & ${\rm O7} \oplus {\rm D7}$ \\ \hline
IIA & T duality along $y^4$ & $\mathbb{R}^{3, 1} \times \big({\cal M}_4 \rtimes \widetilde{\bf S}^1_{4}\big) \times {{\bf S}^1_{5} \over \Omega {\cal I}_5}$ & 
$\left(g_s^{-2}, ~ {\rm F}_2, ~ {1\over {\rm F}_1}, ~ {{\rm F}_1}\right)$ &
${\sqrt{{\rm F}_1}}$ & ${\rm O8} \oplus {\rm D8}$ \\ \hline
Type I & T duality along $y^5$ & $\mathbb{R}^{3, 1} \times {\cal M}_4 \rtimes \big(\widetilde{\bf S}^1_{5} \times \widetilde{\bf S}^1_{4}\big)$ & 
$\left(g_s^{-2}, ~ {\rm F}_2, ~ {1\over {\rm F}_1}, ~ {1\over {\rm F}_1}\right)$ & $1$ & ${\rm O9} \oplus {\rm D9}$ \\ \hline
Het ${\rm SO(32)}$ & S duality & $\mathbb{R}^{3, 1} \times {\cal M}_4 \rtimes \big(\widetilde{\bf S}^1_{5} \times \widetilde{\bf S}^1_{4}\big)$ & $\left(g_s^{-2}, ~ {\rm F_2}, ~{1\over {\rm F}_1}, ~{1\over {\rm F}_1}\right)$ & $1$ &
$\left({\rm SO(8)}\right)^4$ \\
\hline
\end{tabular}}
\renewcommand{\arraystretch}{1}
\end{center}
 \caption[]{\Su The dynamical duality sequence that takes a M-theory configuration \eqref{giuraman1} to heterotic SO(32) theory with the gauge group broken to $\left({\rm SO(8)}\right)^4$. The difference from {\bf Table \ref{milleren2}} is that now we have non-trivial IIB coupling switched on. As before, symbol ${\cal M}_4 \rtimes {\cal M}$ denotes non-trivial fibration of the manifold ${\cal M}$ over the base manifold ${\cal M}_4$. Again, we have not specified the precise numbers of ${\rm O}p$'s and ${\rm D}p$'s in various theories for global charge cancellations.}
 \label{millerenoo}
 \end{table}

We seem to have a conundrum now. Should we go with the de Sitter Glauber-Sudarshan state in the string frame or in the Einstein frame? Cosmologically the latter is preferred \cite{jordan, corda}, although most of the computations leading to the de Sitter Glauber-Sudarshan state in the Einstein frame appears directly from the string frame analysis because the duality sequence is mostly designed in the string frame. On the other hand, the Einstein frame configuration shows the pathology that the internal space typically remains time-dependent with the possibility of late-time singularities. Taking these into account, there appears to be two possibilities:

\vskip.1in

\noindent $\bullet$ de Sitter Glauber-Sudarshan state in the string frame with time-{\it independent} internal space and the Newton's constant dual to a non de Sitter Glauber-Sudarshan state in the Einstein frame with time-{\it dependent} internal space and the Newton's constant. The latter also shows late-time decompactification of the internal six-manifold. 

\vskip.1in

\noindent $\bullet$ de Sitter Glauber-Sudarshan state in the Einstein frame with time-{\it dependent} internal space but time-{\it independent} Newton's constant dual to a non de Sitter Glauber-Sudarshan state in the string frame with time-{\it dependent} internal space and the Newton's constant. Both the configurations show late-time singularities of the internal four-manifold ${\cal M}_4$ but the former also show decompactification of the internal toroidal direction while keeping the overall volume of the six-manifold temporally invariant. 

\vskip.1in

\noindent It seems, from our above discussion, the Glauber-Sudarshan state prefers the string frame over the Einstein frame. On the other hand, if we demand the following behavior of ${\rm F}_1$:
\bg\label{evalike}
{\rm F}_1(t) = \begin{cases} \left(\Lambda t^2\right)^{v\beta_o \over 2v -\beta_o}~~~~~~ -{1\over \sqrt{\Lambda}} < t < -\epsilon \\
~~~ \\
~~ 1~~~~~~~~~~~~~~~~~ -\epsilon < t < 0
\end{cases}
\nd
with $\epsilon$ being a temporal point where ${\rm F}_1$ (and hence ${\rm F}_2$) becomes a constant (see {\bf figure \ref{f1plot}}), and $v = 1, 2$ for the string and the Einstein frames respectively, then in the regime $-\epsilon < t < 0$, either frame would provide a consistent Glauber-Sudarshan state. The warp-factor choice \eqref{evalike} can be implemented by first defining ${\rm F}_1(t) = (\Lambda t^2)^{v\beta(t)\over 2v - \beta(t)}$ and then taking $\beta(t)$ as a function of the conformal time $t$ in the following way\footnote{There are two $\beta$ parameters here. One is the coordinate $y^\beta \in {\cal M}_2$ whereas the other is a function of time $\beta(t)$ as in \eqref{afleis}. Similarly there are three $\alpha$ parameters, $\alpha'$ is the string scale, $y^\alpha \in {\cal M}_2$ and is a constant, and $\alpha(t) = -\beta(t)$ which appears as an exponent of $g_s$. Unfortunately this has become standard in the literature so, hopefully this will not be confusing. \label{confusing}}:
\bg\label{afleis}
\beta(t) = \begin{cases} ~ \beta_o ~~~~~~ -{1\over \sqrt{\Lambda}} < t < -\epsilon \\
~~~ \\
~ 0 ~~~~~~~ -\epsilon < t < 0
\end{cases}
\nd
where $0 < \beta_o < {2\over 3}$ as before, and there is a range around the neighborhood of $ t = \epsilon$ where $\beta(t)$ varies from $\beta_o$ to $0$. This is a consistent choice because the string coupling ${g_s\over {\rm H}(y)} = (\Lambda t^2)^{v\over 2v - \beta(t)}$, with $\beta(t)$ from \eqref{afleis} and ${\rm H}(y) \ne 1$, provides the necessary temporal behavior of $g_s$ to realize de Sitter states in either frame. Interestingly, we can represent \eqref{afleis} {\it exactly} using a continuous function but with a discontinuous derivative (for example \eqref{aflies3}), or {\it approximately} using a continuous function with a well-defined derivative (for example \eqref{fermige}). The former will have issues that we will discuss soon (for example {\bf figures \ref{comparison}, \ref{F1behavior33}}), so opting for the second choice will be better.
Question however is whether we can determine an unambiguous value of $\epsilon$. It turns out there is a simple way to determine $\epsilon$. It can be estimated when the volume of the internal four-manifold ${\cal M}_4$ attains a size ${\rm V}_1$ with ${\rm V}_1 >> \alpha'^2$. Using \eqref{gulabs}, and keeping 
$\hat{g}_{\rm AB}^{(0j_k)} = 0$ in \eqref{ramyakri}, this may be reached at:
\bg\label{tegtrex}
\epsilon = {1\over \sqrt{\Lambda}}\left[{{\rm V}_1 \over {\rm V}_o(-1/\sqrt{\Lambda})}\right]^{4-\beta_o\over 2\beta_o}, \nd
where ${\rm V}_o(-1/\sqrt{\Lambda}) >> {\rm V}_1$ is the stabilized volume of the four-manifold ${\cal M}_4$ at $t = -{1\over \sqrt{\Lambda}}$. (Around this small neighborhood of $t = \epsilon$, we expect $\beta(t)$ to vary in the aforementioned way.)
The {\it dynamical moduli stabilization}, discussed in \cite{coherbeta, coherbeta2}, provides the recipe to fix the moduli at a certain fixed interval so as to avoid the Dine-Seiberg runaway. (We will make an estimate of this a little later.) After which the size of ${\cal M}_4$ will be governed by the temporal behavior of the metric. Interestingly, since ${\rm V}_o(-1/\sqrt{\Lambda}) >> {\rm V}_1 >> \alpha'^2$, the temporal interval $\epsilon << {1\over \sqrt{\Lambda}}$ and is therefore expectedly close to the late time.

Our simple analysis above then provides the following resolution of the aforementioned conundrum. While the de Sitter Glauber-Sudarshan state is better expressed in the string frame, a choice of ${\rm F}_1$ as in \eqref{evalike} at late-times shows that in the temporal domain $-\epsilon < t < 0$, with $\epsilon$ as in \eqref{tegtrex}, a de Sitter Glauber-Sudarshan state may be described unequivocally in either the string or the Einstein frame. Despite this, there is more subtleties associated with the last stage of temporal evolution in the interval $-\epsilon < t < 0$ as studied recently in \cite{axion}. We will come back to this a little later.

\section{Late-time de Sitter Glauber-Sudarshan states in heterotic theories \label{sec3}}

Our aim now would be to study in detail the properties of de Sitter Glauber-Sudarshan states in heterotic theories with special emphasis on the ${\rm E}_8 \times {\rm E}_8$ theory. This will turn out to be  bit more non-trivial from what we have done so far. The reason is that the dynamical duality sequence in {\bf Table \ref{milleren3}}, while provides us with a hint as how to reach the ${\rm E}_8$ point, does not tell us how to switch-off the Wilson lines to avoid breaking the ${\rm E}_8$ groups. Additionally, as we shall see, subtleties due to late-time singularities could arise for this case even in the string frame (which were absent for the ${\rm SO(32)}$ case). 

\subsection{Necessity of three temporal warp-factors \label{tolace1}}

Our plan would be to avoid taking the path shown in {\bf Table \ref{milleren2}}, and follow the path taken in {\bf Table \ref{milleren3}}. For this to happen, we have to make sure that after the second IIA duality sequence, the theory should automatically become strongly coupled, so that a M-theory uplift could be realized. In other words, the {\it ambiguity} that we discussed earlier should not occur here. This turns out to be easier said than done because, with the choice of two warp-factors ${\rm F}_1$ and ${\rm F}_2$, there is no simple way to avoid the ambiguity discussed earlier. It appears then we might have to consider at least three warp-factors. In other words, we consider the following variation of \eqref{giuraman1}\footnote{Henceforth we avoid tildes and zeroes from \eqref{ramyakri}.}:

{\footnotesize
\bg\label{giuraman2}
ds^2 & = & \left\langle {\bf g}_{\rm AB} \right\rangle_\sigma d{\rm Y}^{\rm A} d{\rm Y}^{\rm B}\\
& = & g_s^{-8/3}\left(-dt^2 + \delta_{ij} dx^i dx^j\right) + 
g_s^{-2/3} {\rm H}^2(y)\Big[{\rm F}_1(g_s/{\rm H})g_{\theta_1\theta_1} d\theta_1^2 +{\rm F}_3(g_s/{\rm H})g_{\theta_2\theta_2} d\theta_2^2\Big] \nonumber\\ 
& + & g_s^{-2/3} {\rm H}^2(y)~{\rm F}_2(g_s/{\rm H})g_{mn} dy^m dy^n + g_s^{4/3} \delta_{ab} dw^a dw^b, \nonumber \nd}
over a supersymmetric solitonic configuration with a topology of 
${\bf R}^{2, 1} \times {\bf S}^1_{\theta_1} \times {{\bf S}^1_{\theta_2}\over {\cal I}_{\theta_2}} \times {\cal M}_4 \times {\mathbb{T}^2\over {\cal G}}$, 
where ${\rm Y}^{\rm A} \in (x^\mu, \theta_1, \theta_2, y^m, w^a)$ with $x^\mu \in {\bf R}^{2, 1}, y^\alpha = (\theta_1, \theta_2) \in {\cal M}_2 \equiv {\bf S}^1_{\theta_1} \times {{\bf S}^1_{\theta_2}\over {\cal I}_{\theta_2}}, y^m \in {\cal M}_4$, and $w^a \in {\mathbb{T}^2\over {\cal G}}$. The four-manifold, ${\cal M}_4$, is again compact and non-K\"ahler, but the two-manifold ${\cal M}_2$ is now locally a product manifold of the form ${\bf S}^1_{\theta_1} \times {{\bf S}^1_{\theta_2}\over {\cal I}_{\theta_2}}$. ${\cal G}$ remains a group action without fixed points.

Few questions immediately arise from the metric choice \eqref{giuraman2}. The local ${\bf S}^1$ orbifold nature of the compact manifold should suggest that we could compactify directly along $\theta_2$ direction and get the heterotic ${\rm E}_8 \times {\rm E}_8$ theory with two Horava-Witten walls \cite{horava}. Of course, dynamically this is only possible if the $\theta_2$ direction shrinks {\it faster} than the toroidal direction. Assuming this to be the case $-$ which could be arranged by allowing $\hat{b} > 2$ in ${\rm F}_3 \equiv \left({g_s\over {\rm H}}\right)^{\hat{b}}$ $-$ this still doesn't work because of couple of reasons. {\Su One}, the third direction $x^3$ remains compact compared to the other three directions spanning ${\bf R}^{2, 1}$ with the former evolving as $g_s \sqrt{{\rm F}_3}$ and the latter evolving as $g_s^{-3} \sqrt{{\rm F}_3}$. This ends up ruining the four-dimensional de Sitter isometries.  {\Su Two}, the volume of the internal six-manifold becomes proportional to $g_s^{-4}{\rm F}_3^3 {\rm F}_2^4 {\rm F}_1$, and since with the aforementioned choice of $g_s$ there is no simple way to fix the space-time metric to be de Sitter, the internal volume (and hence the four-dimensional Newton's constant) generically remains time-dependent. 

One might entertain yet another possible way by identifying the $\theta_1$ direction with the third spatial direction without worring too much about the compactness of $\theta_1$. We will also keep ${\rm H}(y) = g_{\theta_1\theta_1} \equiv 1$ for simplicity. It is easy to see that, if we want to keep the internal six-dimensional volume time-independent, we require $2 < \hat{b} < 6$ and the warp-factors and $g_s$ to take the following form:
\bg\label{samthai}
{\rm F}_1 = g_s^{-2}, ~~~~~ {\rm F_2} = g_s^{2-3\hat{b}\over 4}, ~~~~~ 
{\rm F}_3 = g_s^{\hat{b}}, ~~~~~ g_s = \left(\Lambda t^2\right)^{2\over 6 - \hat{b}}, \nd
where the internal six-manifold has a topology of ${\cal M}_4 \times {\mathbb{T}^2\over {\cal G}}$. The metric on the heterotic side now has the right four-dimensional isometries, and takes the following form:

{\footnotesize
\bg\label{annasta}
ds^2 = {1\over \Lambda\vert t\vert^2}(-dt^2 + \delta_{ij} dx^i dx^j + d\theta_1^2) + \left(\Lambda\vert t\vert^2\right)^{-{\hat{b}+2\over 2(6 - \hat{b})}} g_{mn} dy^m dy^n + \left(\Lambda\vert t\vert^2\right)^{{\hat{b}+2\over 6 - \hat{b}}} \delta_{ab}dw^a dw^b, \nd}
from where one may easily see that there is a late-time singularity when the toroidal manifold shrinks to zero size while the volume of the six-manifold remains unchanged. The string coupling however continues to remain weak. Choosing a different direction, say one of the isometry direction of ${\cal M}_4$ by expressing it as ${\cal M}_3 \times {\rm S}^1$ to replace $\theta_1$ in \eqref{annasta}, now changes the definitions of ${\rm F}_1$ and ${\rm F}_2$ to the following:
\bg\label{samthai2}
{\rm F}_1 = g_s^{8 - 3\hat{b}}, ~~~~~ {\rm F_2} = g_s^{-2}, \nd
with $g_s$ and ${\rm F}_3$ remaining the same as in \eqref{samthai}. Due to the change in the form of ${\rm F}_1$, we now require ${8\over 3} < b < 6$ otherwise the aforementioned duality to heterotic theory may not go through smoothly. Assuming this to be the case, the four-dimensional metric now takes the following form:

{\scriptsize
\bg\label{annasta2}
ds^2 = {1\over \Lambda\vert t\vert^2}(-dt^2 + \delta_{ij} dx^i dx^j + dy_1^2) + {g_{mn} \over \Lambda\vert t\vert^2}~dy^m dy^n + 
\left(\Lambda \vert t\vert^2\right)^{14 - 5\hat{b}\over 6 - \hat{b}} g_{\theta_1\theta_1} d\theta_1^2 + \left(\Lambda\vert t\vert^2\right)^{{\hat{b}+2\over 6 - \hat{b}}} \delta_{ab}dw^a dw^b, \nd}
where the topology of the internal manifold is now ${\cal M}_3 \times {\rm S}^1_{\theta_1} \times {\mathbb{T}^2\over {\cal G}}$ and we have chosen $g_{y_1y_1} = {\rm H}(y) \equiv 1$ and $\Lambda$ from \eqref{marapaug} but $g_{\theta_1\theta_1} \ne 1$. It is easy to check that the volume of the internal six-manifold remains time-independent, but now there is a more severe late-time singularities coming from both the toroidal manifold and the circle ${\rm S}^1_{\theta_1}$ shrinking to zero sizes for ${8\over 3} < b < {14\over 5}$. For ${14\over 5} < b < 6$, the late-time singularity comes from the shrinking of the toroidal manifold ${\mathbb{T}^2\over {\cal G}}$. The choice of the warp-factors in \eqref{samthai} and \eqref{samthai2} should also satisfy the Schwinger-Dyson equations, but since in both cases the metric \eqref{annasta} and \eqref{annasta2} show late-time singularities, the dualities do not lead to consistent de Sitter backgrounds at late times.

Our aforementioned failure to directly follow the route to ${\rm E}_8$ heterotic theory via dimensional reduction along $\theta_2$ direction dynamically might suggest the choice of ${\bf R}^{3, 1} \times {\bf S}^1_{\theta_1} \times {{\bf S}^1_{\theta_2}\over {\cal I}_{\theta_2}} \times {\cal M}_4 \times {\bf S}^1_{11}$ to be a more suitable solitonic configuration. One may then construct the Glauber-Sudarshan $\vert\rho\rangle$ state over this vacuum and determine the four-dimensional de Sitter metric via 
$\langle {\bf g}_{\mu\nu}\rangle_\rho$. Again this is easier said than done, because the construction of the state $\vert\rho\rangle$ $-$ which requires consistency from the Schwinger-Dyson type equations for the metric and the flux expectation values $-$ is even harder to work out as may be evident from \cite{desitter2}. How should we then proceed? This is where the dynamical duality sequence from the metric choice \eqref{giuraman2} ({\Su constructed using the Glauber-Sudarshan state $\vert \sigma \rangle$ via the expectation value $\langle {\bf g}^{\rm AB}\rangle_\sigma$}) to the required metric ({\Su constructed using the Glauber-Sudarshan state $\vert\rho\rangle$ using the expectation value $\langle {\bf g}^{\rm AB}\rangle_\rho$}) becomes immensely useful. This means our earlier works in \cite{desitter2, coherbeta, coherbeta2, borel2} $-$ which justified the metric choice \eqref{giuraman2} both from the path-integral computations and the Schwinger-Dyson's equations $-$ suggest that starting with the Glauber-Sudarshan state $\vert\sigma\rangle$ and following the duality sequence {\it dynamically} may be the simplest way to reach our goal. The duality sequence now becomes:

{\footnotesize
\bg\label{momkune4}
&& {{\bf S}^1_{\theta_1} \times {\bf S}^1_{\theta_2} \times \mathbb{T}^2_{3, 11}\over {\cal I}_{\theta_2}} 
~~ \xrightarrow{{\bf S}^1_{11} \to 0} ~~ {{\bf S}^1_{\theta_1} \times {\bf S}^1_{\theta_2} \times {\bf S}^1_3 \over \Omega {\cal I}_{\theta_2}}~~ \xrightarrow{{\bf T}_3}~~ {{\bf S}^1_{\theta_1} \times {\bf S}^1_{\theta_2} \times {\bf R}_3 \over \Omega (-1)^{{\rm F}_{\rm L}}{\cal I}_{\theta_2} {\cal I}_3}~~ \xrightarrow{{\bf T}_{\theta_1}} ~~{\widetilde{\bf S}^1_{\theta_1} \times {\bf S}^1_{\theta_2} \times {\bf R}_3\over \Omega {\cal I}_{\theta_1} {\cal I}_{\theta_2} {\cal I}_3} \\
&& \nonumber\\
 && \xrightarrow{{\bf S}^1_{11} \to \infty} ~~ {\widetilde{\bf S}^1_{\theta_1} \times {\bf S}^1_{\theta_2} \times {\bf R}_3 \times {\bf S}^1_{11}\over 
{\cal I}_{\theta_1} {\cal I}_{\theta_2} {\cal I}_3} ~~ 
\xrightarrow[{\rm blow-ups}]{{{\bf R}_3\over {\cal I}_3} \to \hat{\bf R}_3,~~~ {\widetilde{\bf S}^1_{\theta_1}\over {\cal I}_{\theta_1}} \to \hat{\bf S}^1_{\theta_1}} ~~ {\hat{\bf R}_3 \times \hat{\bf S}^1_{\theta_1} \times {\bf S}^1_{\theta_2} \times {\bf S}_{11}\over {\cal I}_{\theta_2}} ~~ \xrightarrow{{\bf S}^1_{\theta_2} \to 0} ~~ \hat{\bf S}^1_{\theta_2} \times {\bf S}^1_{11}, \nonumber \nd}

\begin{table}[tb]  
 \begin{center}
\renewcommand{\arraystretch}{2.6}
\resizebox{\textwidth}{!}    
{
\begin{tabular}{|c|c|c|c|c|c|}\hline
Theory & duality & Solitonic Configuration & $\langle {\bf g}_{\rm AB}\rangle_{\sigma}$ & coupling & vector bundle \\ \hline
\multirow{2}{*}{M} & \multirow{2}{*}{} & \multirow{2}{*}{$\mathbb{R}^{2, 1} \times {\bf S}^1_{\theta_1} \times {{\bf S}^1_{\theta_2}\over {\cal I}_{\theta_2}}\times {\cal M}_4 \times {\mathbb{T}^2_{3, 11}\over 
{\cal G}}$}
& \multirow{2}{*}{$\left(g_s^{-8/3}, g_s^{-2/3}{\rm F}_1, g_s^{-2/3} {\rm F}_3, g_s^{-2/3}{\rm F}_2, g_s^{4/3}\right)$} & \multirow{2}{*}{} & \multirow{2}{*}{Horava-Witten walls} \\ 
& & & & & $\left({\red{01234_{~-~}6789, 11}}\right)$ \\ \hline
\multirow{2}{*}{IIA} & \multirow{2}{*}{dim. red. along $x^{11}$} & \multirow{2}{*}{$\mathbb{R}^{2, 1} \times {\bf S}^1_{\theta_1} \times {{\bf S}^1_{\theta_2}\over \Omega ~{\cal I}_{\theta_2}}\times {\cal M}_4 \times {\bf S}^1_3$} & \multirow{2}{*}{$\left(g_s^{-2},~ {\rm F}_1, ~ {\rm F}_3,~ {\rm F}_2, ~g_s^2\right)$} & \multirow{2}{*}{$g_s$} & \multirow{2}{*}{${\rm O8} \oplus {\rm D8}$} \\
& & & & & $\left({\red{01234_{~-~}6789}}\right)$ \\ \hline 
\multirow{2}{*}{IIB} & \multirow{2}{*}{T duality along $x^3$} & \multirow{2}{*}{$\mathbb{R}^{2, 1} \times {{\bf R}_3 \times {\bf S}^1_{\theta_1} \times {\bf S}^1_{\theta_2} \over \Omega ~{\cal I}_{\theta_2}~ {\cal I}_3 ~(-1)^{{\rm F}_{\rm L}}} \times 
{\cal M}_4$}
& \multirow{2}{*}{$\Big(g_s^{-2}, ~g_s^{-2}, ~{\rm F}_1, ~ {\rm F}_3, ~{\rm F}_2\Big)$} & \multirow{2}{*}{1} & \multirow{2}{*}{${\rm O7} \oplus {\rm D7}$} \\
& & & & & $\left({\red{012_{~-~}4_{~-~}6789}}\right)$\\
\hline
\end{tabular}}
\renewcommand{\arraystretch}{1}
\end{center}
 \caption[]{\Su The dynamical duality sequence that takes a M-theory configuration \eqref{giuraman2} to the unbroken ${\rm E}_8 \times {\rm E}_8$ heterotic theory. For convenience we have taken $(\theta_1, \theta_2) = (y^4, y^5)$ with the four-manifold ${\cal M}_4$ spanned by the coordinates $(y^6, y^7, y^8, y^9)$. Again, we have not specified the precise numbers of ${\rm O}p$'s and ${\rm D}p$'s in various theories for global charge cancellations. The dashed lines in the last column denote the points where the branes, O-planes or other defects are located. See also {\bf figure \ref{tab4row4}} for the configuration in the fourth row. The equality between the string and the Einstein frames happen at late time for different relations between ${\rm F}_1$ and ${\rm F}_3$. In fact in the Einstein frame, the system has to go thorough an intermediate ${\rm F}_1 = {\rm F}_3$ state to finally reach the state where ${\rm F}_3 = {\rm F}_1^5$. We will discuss the consequence of this in section \ref{sec3.6}.}
 \label{milleren4}
 \end{table}
\noindent where, as before, we have not shown the sub-spaces on which we do not have any $\mathbb{Z}_2$ actions (for example ${\cal M}_4$). The 
detailed application of the duality sequence, starting with the metric configuration \eqref{giuraman2}, appears in {\bf Table \ref{milleren4}} which we explain in the following.

\subsection{The dynamical duality sequence and unbroken ${\rm E}_8 \times {\rm E}_8$ \label{tolace2}}

Our starting point then remains \eqref{giuraman2} with the flux state 
$\langle {\bf C}_{\rm ABC}\rangle_\sigma$ arranged such that it has one leg along $\theta_2$, so as not to be projected out by the orbifold action \cite{horava}, and the other two legs along ${\bf R}^{2, 1} \times {\bf S}^1_3 \times {\bf S}^1_{\theta_1} \times {\cal M}_4 \times {\mathbb{T}^2\over {\cal G}}$. Due to the duality sequence \eqref{momkune4} we expect the functional dependence to be only on $y^m \in {\cal M}_4$ and on possible temporal direction \cite{desitter2}\footnote{In other words the G-flux components are $\langle {\bf G}_{{\rm MNP}\theta_2}\rangle_\sigma$ where $({\rm M, N, P}) \in {\bf R}^{2, 1} \times {\bf S}^1_3 \times {\bf S}^1_{\theta_1} \times {\cal M}_4 \times{\mathbb{T}^2\over {\cal G}}$ and $\langle {\bf G}_{\rm ABDE}\rangle_\sigma = \partial_{[{\rm A}}\langle {\bf C}_{{\rm BDE}]}\rangle_\sigma$. (The latter could be verified directly from the path-integral computation of $\langle {\bf C}_{\rm BDE}\rangle_\sigma$.) There would also be localized G-flux components which would lead to gauge fluxes on the D8-branes.}. As to why such components are not projected out by intermediate orientifold actions, or by demanding four-dimensional de Sitter isometries will become clearer as we go along.

In {\bf Table \ref{milleren4}} our aim is to get the ${\rm E}_8 \times {\rm E}_8$ theory as the final goal with time-independent Newton's constant. We will also require ${\rm F}_1$ warp-factor to go to zero {\it faster} than the ${\rm F}_3$ warp-factor (compared to the example studied earlier). To keep the volume of the internal six-manifold in the final heterotic side time-independent in the string frame\footnote{Unless mentioned otherwise, the emphasis here is to get heterotic results in the string frame with ${\rm H}(y) \equiv 1$. What happens in the Einstein frame will be discussed a little later.}, the ${\rm F}_2$ warp-factor needs to grow. Taking two parameters $\hat\alpha_o$ and $\hat\beta_o$ with ${\hat\alpha_o\over 9} < \hat\beta_o < \hat\alpha_o < 1$, we may quantify the first of the two decreasing warp-factors in the following suggestive way:
\bg\label{tigbenso}
{\rm F}_1(t) = \begin{cases} \left(\Lambda t^2\right)^{2\hat\alpha_o \over 6 -\hat\alpha_o - \hat\beta_o}~~~~~~ -{1\over \sqrt{\Lambda}} < t <  -\epsilon \\
~~~ \\
\left(\Lambda t^2\right)^{\hat\gamma_o \over 3 -\hat\gamma_o}~~~~~~~~~~ -\epsilon < t < 0 \end{cases} \nd
which may be compared to \eqref{evalike} for the earlier cases. (Here $\Lambda$ is the constant piece from \eqref{marapaug}, and henceforth we will use this unless mentioned otherwise.) The difference is in the choice of the exponents: the two exponents at the two temporal domains are different from what we had in \eqref{evalike}. This is also visible in the second of the two decreasing warp-factor:
\bg\label{tigbenso2}
{\rm F}_3(t) = \begin{cases} \left(\Lambda t^2\right)^{2\hat\beta_o \over 6 -\hat\alpha_o - \hat\beta_o}~~~~~~ -{1\over \sqrt{\Lambda}} < t < -\epsilon \\
~~~ \\
\left(\Lambda t^2\right)^{\hat\gamma_o \over 3 -\hat\gamma_o}~~~~~~~~~~ -\epsilon < t < 0\end{cases}
\nd
 The condition $\hat\alpha_o > \hat\beta_o$ makes ${\rm F}_1$ to decay {\it faster} than ${\rm F}_3$ in the temporal domain $-{1\over \sqrt{\Lambda}} < t < -\epsilon$. As in \eqref{afleis}, the behaviors of the two warp-factors in \eqref{tigbenso} and \eqref{tigbenso2} implies 
 that we may express ${\rm F}_1(t) \equiv (\Lambda t^2)^{2\hat\alpha(t) \over 6 - \hat\alpha(t) - \hat\beta(t)}$ and ${\rm F}_3(t) \equiv (\Lambda t^2)^{2\hat\beta(t) \over 6 - \hat\alpha(t) - \hat\beta(t)}$ with  $\hat\alpha(t)$ and $\hat\beta(t)$ defined in the following 
 way\footnote{This will be modified to incorporate a more finer splitting in \eqref{afleis4}.}:
\bg\label{afleis2}
\hat\alpha(t) = \begin{cases} ~ \hat\alpha_o ~~~~~~ -{1\over \sqrt{\Lambda}} < t < -\epsilon \\
~~~ \\
~ \hat\gamma_o ~~~~~~ -\epsilon < t < 0
\end{cases}, ~~~~~ \hat\beta(t) = \begin{cases} ~ \hat\beta_o ~~~~~~ -{1\over \sqrt{\Lambda}} < t < -\epsilon \\
~~~ \\
~ \hat\gamma_o ~~~~~~ -\epsilon < t < 0
\end{cases}
\nd 
in the respective temporal domain, where we expect both $\hat\alpha(t)$ and $\hat\beta(t)$ to vary in the small neighborhood of $t = \epsilon$ from $\hat\alpha_o$ to $\hat\gamma_o$ and from $\hat\beta_o$ to $\hat\gamma_o$ respectively. Additionally the respective transitions from $\hat\alpha_o$ and $\hat\beta_o$ to $\hat\gamma_o$ in the string frame should be accomplished by smooth functions with well-defined derivatives. (We will discuss more on this below.) This in-turn implies that the type IIA configuration would approach 
strong-coupling as $g_{\rm IIA} = \left({g_s\over {\rm H}}\right)^{-\hat\alpha(t)/3}$, where $g_s$ is the original type IIA coupling after the first duality sequence in {\bf Table \ref{milleren4}}. This growth is {\it faster} than the rate at which the ${\bf S}^1_{\theta_2}$ is shrinking, namely $\left({g_s\over {\rm H}}\right)^{2\hat\beta(t)/3}$, resulting in an uplift to M-theory. In M-theory, as may be inferred from {\bf Table \ref{milleren4}}, most of the internal directions are growing except ${\bf S}^1_{\theta_2}$ and ${\cal M}_4$ as long as $\hat\alpha_o > 9\hat\beta_o$. However if we demand that only the circle ${\bf S}^1_{\theta_2}$ shrinks, but not ${\cal M}_4$, then $\hat\beta_o > {\hat\alpha_o\over 9}$. This provides the range of choices for $\hat\alpha_o$ and $\hat\beta_o$ mentioned above \eqref{tigbenso}. Unfortunately despite this, the duality sequence leading to the four-cycle in the final heterotic configuration, namely ${\cal M}_4$, still shrinks to zero size as $\left(\sqrt{\Lambda}t\right)^{\hat\alpha_o - \hat\beta_o \over 6 - \hat\alpha_o -\hat\beta_o}$ leading to late-time singularities. This can only be avoided if at a certain interval $-\epsilon < t < 0$ we demand that $\hat\alpha(t) = \hat\beta(t) \equiv \hat\gamma_o$ with ${\rm F}_1$ and ${\rm F}_3$ taking the late-time values as in \eqref{tigbenso} and \eqref{tigbenso2} respectively. The value of $\epsilon$ can again be easily determined using similar considerations that lead to \eqref{tegtrex}, as:
\bg\label{tegtrex2}
\epsilon \equiv {1\over \sqrt{\Lambda}}\left[{{\rm V}_1\over {\rm V}_o(-1/\sqrt{\Lambda})}\right]^{6-\hat\alpha_o - \hat\beta_o\over 2\hat\alpha_o - 2\hat\beta_o}, \nd
where ${\rm V}_o(-1/\sqrt{\Lambda}) >> {\rm V}_1 >> \alpha'^2$ is the stabilized volume of the four-manifold ${\cal M}_4$ at $t = -{1\over \sqrt{\Lambda}}$. In fact in this temporal domain the form of the metric is almost similar to what we had in {\bf Tables \ref{milleren2}} and {\bf \ref{milleren3}}. We may also express the ${\rm F}_2$ warp-factor, that keeps the Newton's constant time-independent at any given instant of time, in the following way:
\bg\label{tigbenso3}
{\rm F}_2(t) = \left(\sqrt{\Lambda} t\right)^{-\left({\hat\alpha(t) + 3\hat\beta(t)\over 6 - \hat\alpha(t) - \hat\beta(t)}\right)}
= \begin{cases} \left(\sqrt{\Lambda} t\right)^{-\left({\hat\alpha_o + 3\hat\beta_o\over 6 -\hat\alpha_o - \hat\beta_o}\right)}~~~~~~ -{1\over \sqrt{\Lambda}} < t < -\epsilon \\
~~~ \\
\left(\Lambda t^2\right)^{-{\hat\gamma_o \over 3 -\hat\gamma_o}}~~~~~~~~~~~~~~~ -\epsilon < t < 0 \end{cases} 
\nd
forming a self-consistent framework in the string frame. One may also easily verify that, in the temporal domain $-{1\over \sqrt{\Lambda}} < t < 0$, the aforementioned forms of ${\rm F}_i$ may be expressed using $g_s$ as ${\rm F}_1 = \left({g_s\over {\rm H}}\right)^{2\hat\alpha(t)\over 3}, 
{\rm F}_3 = \left({g_s\over {\rm H}}\right)^{2\hat\beta(t)\over 3}$
and ${\rm F}_2 \equiv \left({g_s\over {\rm H}}\right)^{2\hat\sigma\over 3} = \left({g_s\over {\rm H}}\right)^{-{1\over 6}(\hat\alpha(t) + 3\hat\beta(t))}$ with ${\rm H}(y)$ being the spatial warp-factor appearing in \eqref{giuraman2}. Additionally, as is clear from our above computations, the following choice of $g_s$ in the original type IIA side:
\bg\label{harmonkal}
{g_s\over {\rm H}(y)} = \left(\Lambda t^2\right)^{3\over 6 - \hat\alpha(t) - \hat\beta(t)}, \nd
would consistently reproduce the de Sitter Glauber-Sudarshan state in a flat-slicing in the string frame of the ${\rm E}_8 \times {\rm E}_8$ heterotic theory for any values of $\hat\alpha_o$ and $\hat\beta_o$ satisfying the original criterion ${\hat\alpha_o\over 9} < \hat\beta_o < \hat\alpha_o < 1$. 

\vskip-1in
\begin{figure}[h]
\centering
\begin{tabular}{c}
\includegraphics[width=7in]{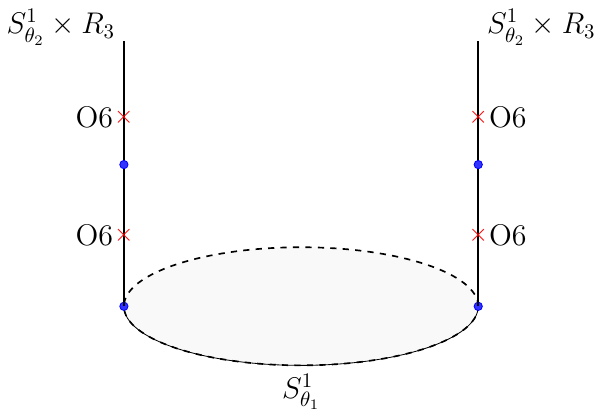}
\end{tabular}
\vskip-6.0in
\caption[]{Distribution of the O6-planes for the configuration in the fourth row of {\bf Table \ref{milleren4}}. At the two fixed points of ${\widetilde{\bf S}_{\theta_1}^1\over {\cal I}_{\theta_1}}$, the $\mathbb{Z}_2$ action now acts on ${\bf R}_3 \times {\bf S}^1_{\theta_2}$ creating four fixed points (because of the non-compact ${\bf R}_3$) which form the locations of O6-planes.}
\label{tab4row4}
\end{figure}

There are however a few subtleties that we kept under the rug in describing the {\it unbroken} ${\rm E}_8 \times {\rm E}_8$ gauge group in the heterotic side, namely blow-ups, NEC and EFT criteria. Additionally we haven't specified the precise form of the Glauber-Sudarshan state in the original M-theory leading to \eqref{giuraman2} that is responsible for the heterotic de Sitter state. In the following section we will clarify these details. 

\subsection{Blow-ups and the M-theory Glauber-Sudarshan state \label{tolace3}}

To fully appreciate the considerations leading to the aforementioned blow-ups, we will need to first understand how the gauge groups are distributed in the $SO(32)$ case.  For this let us start with the second row in {\bf Table \ref{milleren2}} which, at the solitonic level, is type IIA compactified on an orbifold (the orientifold nature will become important soon). The fixed point counting herein is as follows. The $\mathbb{Z}_2$ orbifold of ${\bf S}^1_3$ creates two fixed points. At the fixed points the $\mathbb{Z}_2$ operation now extends to the toroidal manifold $\mathbb{T}^2_{45}$ creating four more fixed points. Together they lead to $2^3 = 8$ fixed points. In the presence of the orientifold action $\Omega$, at each fixed points we expect a gauge group of $SO(4)$, so that the total gauge group on the ${\rm O6} \oplus {\rm D6}$ system becomes $(SO(4))^8$. This is puzzling in light of what is known about the type IIB dual of the above compactification \cite{senkd}: the gauge group of the dual ${\rm O7} \oplus {\rm D7}$ is $(SO(8))^4$. Question is, what happens to the Wilson lines that are expected to break each $SO(8)$ group to $SO(4) \times SO(4)$ in type IIB?

The answer is actually simple. To go to type IIB, we are in principle shrinking the ${\bf S}^1_3$ cycle to a very small size. This should also be clear from the $g_s^2$ factor for $\langle {\bf g}_{33}\rangle_\sigma$
from the second row and fourth column of {\bf Table \ref{milleren2}}. The two fixed points of ${\bf S}^1_3/{\mathbb Z}_2$ now practically 
overlap\footnote{In fact this is where our previous consideration of the temporally modulated solitonic metric factors from \eqref{gulabs} becomes useful. In the solitonic side of {\bf Table \ref{milleren2}}, if we want to go to the type IIB side, we will have to shrink the ${\bf S}^1_3$ circle to a very small size. The mapping \eqref{gulabs} suggests that a similar temporal modulation by $g_s^2$ in the type IIA side makes the ${\bf S}^1_3$ circle further small, thus bringing the fixed points arbitrarily close.}, so in the type IIB dual the Wilson lines vanish. This enhances:
\bg\label{amergods}
SO(4) \times SO(4) ~ \to ~ SO(8), \nd
leading to the expected gauge group in the type IIB side. The Wilson lines now only appear in the type I side breaking $SO(32) \to (SO(8))^4$ and therefore fits consistently with the type IIB configuration \cite{senkd}. The rest of the story, both here and in {\bf Table \ref{milleren3}}, follows in the usual fashion with no surprises.

Let us now clarify the {\it blow-ups} that convert the Atiyah-Hitchin and the Taub-NUT configurations to the Horava-Witten walls from {\bf Table \ref{milleren4}}. The precise blow-ups that we are looking at appear in the sixth row of {\bf Table \ref{milleren4}} as:
\bg\label{dija}
{{\bf R}_3\over {\cal I}_3} \rightarrow \hat{\bf{R}}_3, ~~~~~~~~ {\widetilde{\bf S}^1_{\theta_1}\over {\cal I}_{\theta_1}} \rightarrow \hat{\bf S}^1_{\theta_1}, \nd
where ${\cal I}_j: \zeta^j \to -\zeta^j$ is the orbifold action in M-theory. Before the blow-ups, the localized G-fluxes at the fixed points of ${\bf R}_3$ and $\widetilde{\bf S}^1_{\theta_1}$ (now appearing as localized NS two-form ${\bf B}_2$ fluxes in type IIA) would provide the necessary number of D6-branes to cancel the O6-planes' charges locally (see {\bf figure \ref{tab4row4}}). Once we blow-up the orbifold points as in \eqref{dija}, the O6-planes would be gone and so would be the D6-branes (because of the vanishing of the localized fluxes) with only the O8-planes remaining. In other words, dimensionally reducing this back on ${\bf S}^1_{11}$, now produces a type IIA orientifold of the form ${{\bf S}^1_{\theta_2}\over \Omega {\cal I}_{\theta_2}}$ with two O8-planes of charges $-16$ each (in the unit of D8-brane charge). One might naively put 8 D8-branes at the two boundaries to cancel the charges locally, but this {\it does not} lead to the expected ${\rm E}_8 \times {\rm E}_8$ gauge-group enhancement\footnote{For example in the decomposition ${\bf 248} = {\bf 128} + {\bf 120}$, the spinor representation ${\bf 120}$ is much harder to see. Plus the group $SO(16) \times SO(16)$ may lead to perturbative susy breaking at the solitonic level.}. A better way would be to separate one D8-brane from each end of the boundaries to create type I$'$ {\it half} D-particles, as shown in \cite{Gorbatov}. These half D-particles remain stuck at each of the two orientifold points, thus producing extra generators that would enhance the gauge group as:
\bg\label{laurphil}
SO(14) ~ \times ~ U(1) ~ \to ~ {\rm E}_8, \nd
at each end of the interval ${{\bf S}^1_{\theta_2}\over {\cal I}_{\theta_2}}$ \cite{Gorbatov}. This is however {\it not} a perturbative process, due to the strong coupling effects, so the enhancement may only be seen non-perturbatively. In other words most of the generators, other than 
${\bf 1}$ and ${\bf 91}$, will have to appear non-perturbatively, namely:
\bg\label{mari_c}
{\bf 248} ~=~ {\bf 1}_0 ~\oplus~ {\bf 91}_0 ~\oplus~ {\bf 14}_{+1} ~\oplus~ 
{\bf 14}_{-1} ~\oplus~ {\bf 64}_{+1/2} ~\oplus~  {\bf 64}_{-1/2}, \nd
where the subscripts show the necessary $U(1)$ charges. The above analysis is for the solitonic configuration $-$ appearing on the sixth row and third column of {\bf Table \ref{milleren4}} $-$ but the question is whether we can quantify the non-perturbative effects more clearly even at this level. In other words, when we go away from the orientifold points, can we represent the dynamics at the solitonic level? Once this is known, the temporal considerations from the Glauber-Sudarshan states may be easily inserted-in following \eqref{gulabs}.

\begin{figure}[h]
\centering
\begin{tabular}{c}
\includegraphics[width=5in]{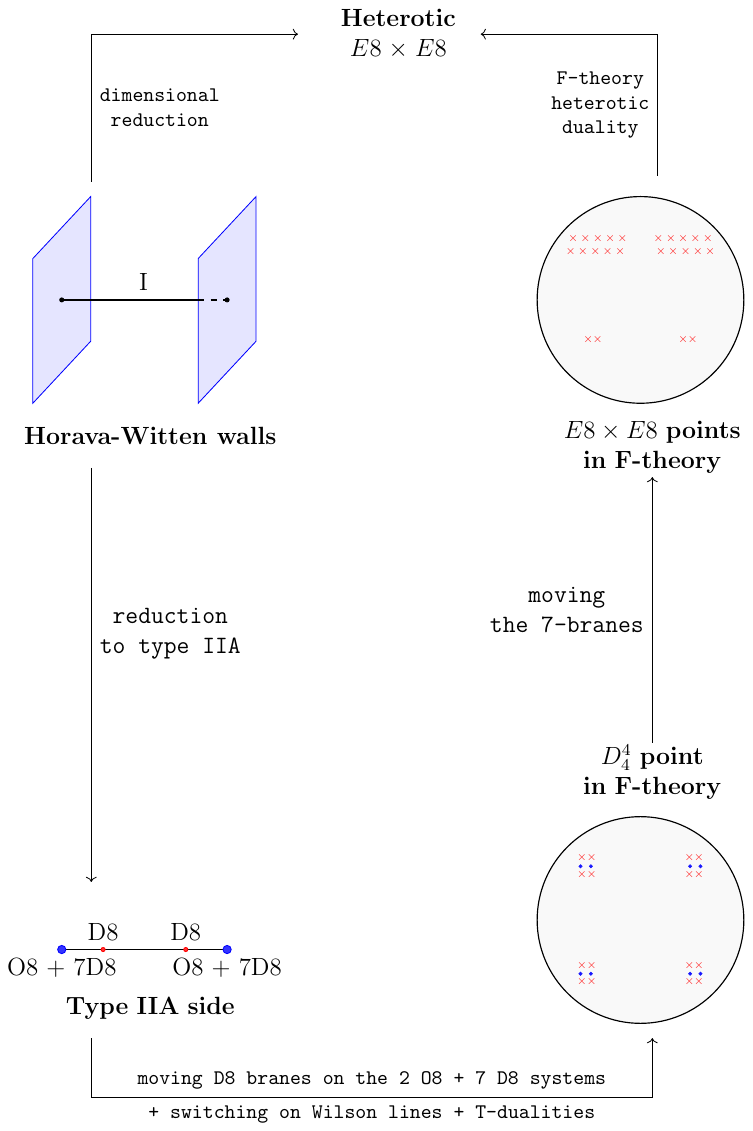}
\end{tabular}
\vskip-1in
\caption[]{Reaching the heterotic ${\rm E}_8 \times {\rm E}_8$ point from both M-theory and F-theory. For the latter certain series of duality transformations from IIA, along with switching on Wilson lines, lead to the configuration in \eqref{jailer}. This is shown in the bottom right circle, signifying a sphere, with the blue dots being the $[p, q]$ seven-branes and the crosses are the D7-branes. One may derive the F-theory configuration \eqref{remki} by moving away from the original O-planes' configuration and distributing the 24 branes seven branes as a set of $10 + 10 + 2 + 2$ as shown in the top right circle/sphere. Under a subsequent F-theory/heterotic duality gives us the required ${\rm E}_8 \times {\rm E}_8$ theory.}
\label{hwe8}
\end{figure}

The best way to capture the non-perturbative effects is to use F-theory \cite{VafaF}. We can use the following trick to go to the F-theory side (and avoid taking the type I$'$ route). We bring the individual D8-branes respectively on top of the two ${\rm O8 + 7D8}$ systems at the two ends of the interval and then switch on Wilson lines. Dimensionally reducing the solitonic configuration along ${\bf S}^1_{11}$ and then T-dualizing along $\theta_1$ direction produces type IIB theory compactified on a geometry that is locally given by: 
\bg\label{jailer}
\mathbb{R}^{3, 1} ~ \times ~ {\cal M}_4 ~\times~ {\hat{\bf S}^1_{\theta_1} \times {\bf S}^1_{\theta_2}\over \Omega \cdot{\cal I}_{\theta_1} {\cal I}_{\theta_2} \cdot(-1)^{\rm F_L}}, \nd
which should now be compared to the solitonic configuration we had earlier in the third row of {\bf Table \ref{milleren4}} before the blow-ups \eqref{dija}. The internal manifold in \eqref{jailer} is a familiar one and clearly shows the $u$-plane, with coordinate $z \equiv \theta_1 + i \theta_2$, once we map this to the Seiberg-Witten theory \cite{seibergw, senkd}. Expectedly, due to the Wilson lines in the type IIA (or the type I$'$) side, the above is a perturbative picture and therefore can only display the $(SO(8))^4$ gauge group. To see the full gauge-group enhancement we have to go away from the orientifold point.
The F-theory curve for the scenario when we are away from the orientifold point may be easily constructed by going to branch II in the second reference of \cite{senkd}, and then expressing the Weierstrass equation as:
\bg\label{remki}
y^2 = x^3 + (z - z_1)^5 (z - z_2)^5 (z- z_3) (z - z_4), \nd
where $z$ is the same coordinate on the $u$-plane, and here it parametrizes the toroidal base $(\theta_1, \theta_2)$ of the F-theory compactification. An analysis of the discriminant, and mapping it to the Tate's algorithm \cite{bershadsky}, clearly shows an 
${\rm E}_8 \times {\rm E}_8$ enhancement of gauge symmetry. Finally, using the F-theory/Heterotic duality \cite{VafaF}, we get the ${\rm E_8 \times E_8}$ heterotic theory compactified on a six-dimensional internal space whose local geometry is ${\cal M}_4 \times \mathbb{T}^2$. The whole process is depicted in {\bf figure \ref{hwe8}}.

The aforementioned solitonic geometry is still not the full picture. We have avoided the inclusion of fluxes in our duality chain. Once we do that and follow the F-theory route, the internal space becomes a non-K\"ahler space. The analysis is now slightly more complicated than the ones presented in \cite{DRS} because we are at the ${\rm E_8 \times E_8}$ point, but the non-K\"ahlerity of the internal space can still be quantified. From F-theory point of view, we are dealing with a non-trivial fibration of the form
${\cal M}_4 \rtimes K3$, with the assumption that (a) the $K3$ is at a special point, and (b) the metrics on both ${\cal M}_4$ and $K3$ are non-K\"ahler.

The Glauber-Sudarshan state now creates the necessary de Sitter state that appears in the seventh and the eighth rows of {\bf Table \ref{milleren4}}. For the mapping to the solitonic configuration, we can use the simplest identification from \eqref{gulabs}, {\it i.e.} keeping $\hat{g}_{\rm AB}^{(0j_k)} = 0$ in \eqref{ramyakri}, and ignoring the non-trivial fibration. The size of the toroidal fibre grows and the ${\cal M}_4$ base shrinks temporally, keeping the total volume constant, till the interval \eqref{tegtrex2}. After which the internal manifold remains time-independent.

\subsection{Revisiting EFT constraints and NEC in $SO(32)$ theory \label{tolace4}}

The computations that we presented above rely heavily on the existence of an EFT both at the solitonic, {\it i.e.} at the supersymmetric Minkowski, level and at the excited state, {\it i.e.} at the Glauber-Sudarshan state, level. Looking at the first rows of {\bf Tables \ref{milleren1}} to 
{\bf \ref{milleren4}}, wherein we used the type IIA coupling $g_s$, the EFT condition boils down to \cite{coherbeta2}:
\bg\label{katudi}
{\partial g_s\over \partial t} ~ \propto ~ g_s^{+ve}, \nd
where $t$ is the conformal time in four-dimensions (and not the {\it coordinate} time). The condition \eqref{katudi} is crucial and should work even if $g_s$ is a non-trivial function of the conformal time $t$. For example if $g_s = f(t)$, then:
\bg\label{sarshonali}
{\partial g_s\over \partial t} \equiv \dot{g}_s = \dot{f}(t) = 
\dot{f}(f^{-1}\odot g_s) ~ \propto ~ g_s^{+ve}, \nd
where the $\odot$ operation inverts the equation to express the conformal time $t$ as a function of $g_s$. For simple cases, studied in \cite{coherbeta2}, one may show that the condition \eqref{sarshonali} remains valid. In our case, if we look at the third row of {\bf Table \ref{milleren1}} and demand that the space-time part of the metric grows as $(\Lambda t^2)^n$, {\it i.e.} $\langle {\bf g}_{\mu\nu}\rangle_\sigma = 
(\Lambda t^2)^n$, then the condition \eqref{katudi} or \eqref{sarshonali} tells us that ${1\over n} \ge -1$ for EFT to be valid at the level of the Glauber-Sudarshan state. This is exactly the NEC condition \cite{coherbeta2, bernmir}, and therefore for the simple case of  de Sitter in type IIB theory, the existence of an effective field theory {\it i.e.} the condition \eqref{katudi}, boils down to the non-violation of the null energy condition. 

\begin{figure}[h]
\centering
\begin{tabular}{c}
\includegraphics[width=3in]{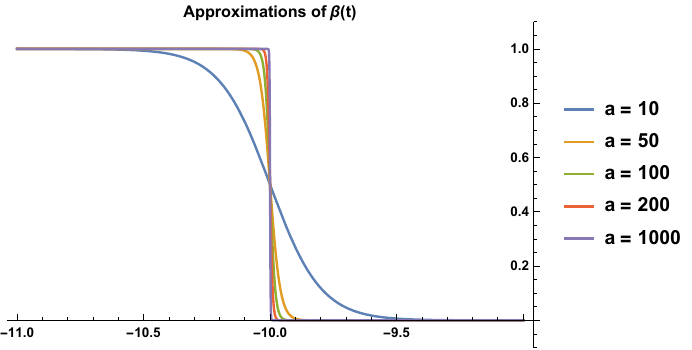}
\end{tabular}
\caption[]{Plot of $\beta(t)$ from \eqref{aflies3} for $\beta_o = 1, \Lambda = a^{-2}$ and $\epsilon = -0.1$.}
\label{plot-heaviside-o}
\end{figure} 

What happens in the heterotic theories? Let us start with the heterotic $SO(32)$ theory in the string frame and, as a warm-up exercise, we will assume that $\beta(t) \equiv \beta_o$ and ignore the temporal dependence from \eqref{afleis}. (We shall elaborate the full story a bit later.) If we demand that the four-dimensional spacetime metric in the fourth columns and the sixth and the seventh rows of {\bf Table \ref{milleren2}} to take the form $\langle {\bf g}_{\mu\nu}\rangle_\sigma \equiv (\Lambda t^2)^n$, then with $g_s = (\Lambda t^2)^{nv\over \beta_0 - 2v}$, where $v = (1, 2)$ for the string and the Einstein frames respectively, the EFT condition implies:
\bg\label{kelryan}
{1\over n} ~ \ge ~ -{1\over 1 - {\beta_0\over 2v}}, \nd
with $0 < \beta_0 < {2\over 3}$. Naively, the condition \eqref{kelryan} appears to improve upon the EFT criterion proposed in \cite{coherbeta2}, but a quick thought will tell us that this cannot be the case. First, the bound cannot depend on which frame we are in, and secondly, the bound cannot depend on an arbitrary parameter $\beta_0$ in the theory\footnote{For example, once we replace $\beta_o$ by $\beta(t)$ from \eqref{afleis} in \eqref{kelryan}, the temporal independence of $n$ will tell us that the bound cannot depend on a temporally varying function $\beta(t)$. \label{karkapr}}. Thus the EFT condition should be independent of both the frames and the specific values of $\beta_0$. This means ${1\over n} \ge -1$ should still be the criterion for EFT. Note that in $\dot{g}_s \propto g_s^{1 + {1\over n} - {\beta_0\over 2nv}}$, for $-1\le {1\over n} \le 0$, the third term in the exponent, namely ${\beta_0\over 2nv}$, is always negative definite and therefore keeps the EFT criterion intact. Similar argument can be give for ${1\over n} > 0$ because $0 < {\beta_0 \over 2v} < {1\over 3}$, and thus there is again no violation of the EFT criterion of \cite{coherbeta2} $\forall \beta_0 \in [0, {2\over 3}]$. For the ${\rm E}_8 \times {\rm E}_8$ case appearing in the seventh row of {\bf Table \ref{milleren4}}, a similar analysis, again ignoring the temporal dependence of $\alpha(t)$ and $\beta(t)$ from \eqref{afleis2} for the time being, now leads to:
\bg\label{korwitch}
{1\over n} ~ \ge ~ - {1\over 1 - {(5-v)\alpha_o + (7-3v)\beta_o\over 24}}, \nd
where $0 < {\alpha_o\over 9} < \beta_o < \alpha_o < 1$; and $v = (1, 2)$ for the string and the Einstein frames respectively. As before, demanding the EFT criterion to be independent of either frames or arbitrary parameters $(\alpha_o, \beta_o)$, immediately tells us that 
${1\over n} \ge -1$ to be the correct EFT criterion. 

\begin{figure}[h]
\centering
\begin{tabular}{c}
\includegraphics[width=2.5in]{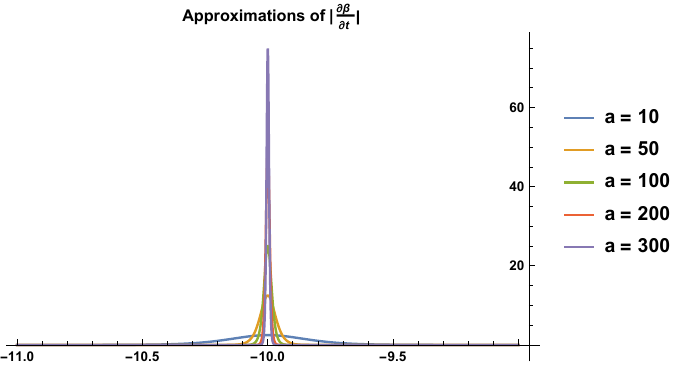}
\end{tabular}
\caption[]{Plot of $\vert\dot\beta(t)\vert$ from \eqref{aflies3} for $\beta_o = 1, \Lambda = {1\over a^{2}}$ and $\epsilon = -0.1$.}
\label{plot-delta-o}
\end{figure} 

The actual story with temporal dependence is however far more involved. This is because the temporal derivative in the EFT condition \eqref{katudi} would now have to extend to the parameters in \eqref{afleis} and \eqref{afleis2}. For the heterotic $SO(32)$ case, both in the string and the Einstein frames, the EFT condition in \eqref{katudi} now implies:
\bg\label{tamana}
{\partial g_s\over \partial t} = g_s^{1 + {\rm B}(t)} - 
2nv\sqrt{\Lambda}~ g_s^{1 + {1\over n} - {\beta(t)\over 2nv} + {\rm C}(t)}, \nd
where $\beta(t)$ is given by \eqref{afleis}. The power of $g_s$ in the second term is a familiar one that we encountered above, up-to the factor ${\rm C}(t)$, but the first term is new. There is also a subtlety here that needs some explanation. For the choice of $g_s = (\sqrt{\Lambda} \vert t\vert)^{2v\over 2v - \beta(t)}$, one will have to invert this equation to express the conformal time $t$ as $t \equiv t(g_s)$ so that the RHS of \eqref{tamana} may be expressed only as functions of $g_s$. This is unfortunately a much harder exercise because generically there may not be a simple analytical form for $t(g_s)$. In view of that the RHS of \eqref{tamana} will be expressed as $g_s$ raised to some temporally dependent function $-$ ${\rm B}(t)$ and ${\rm C}(t)$ here $-$ which we shall demand to be positive for the validity of the underlying EFT. The functional forms of the parameters ${\rm B}(t)$ and ${\rm C}(t)$ appearing respectively in both the terms may be expressed in the following form:
\bg\label{fl2mey}
{\rm B}(t)  = {{\rm log}\left[\left({\dot\beta(t)\over 2v - \beta(t)}\right){\rm log}~g_s\right]\over {\rm log}~g_s},~~~~~~ 
{\rm C}(t) = -{{\rm log}~(2v - \beta(t))\over {\rm log}~g_s}, \nd
where $\dot\beta(t) \equiv {\partial\beta(t)\over \partial t}$. Since $g_s << 1$ and $2v - \beta(t) > 1$ for $v = (1, 2)$ $-$ the latter may be inferred from \eqref{afleis} $-$ the sign of ${\rm C}(t)$ is clearly positive. In other words,
\bg\label{sekdrem18}
{\bf sgn}~{\rm C}(t) = {\bf sgn}\left({{\rm log}(2v-\beta(t))\over \vert{\rm log}~g_s\vert}\right) ~ > ~ 0. \nd 
On the other hand, 
the term inside the square bracket for ${\rm B}(t)$ in \eqref{fl2mey} would only make sense if $\dot\beta(t) < 0$, implying that $\beta(t)$ can only be a decreasing function. This is at least consistent with the functional form in \eqref{afleis}. The sign of ${\rm B}(t)$ would however depend on the precise value of 
$\vert\dot\beta(t)\vert$. For example, if:
\bg\label{bosema}
{\vert\dot\beta(t)\vert\over 2v - \beta(t)} ~ < ~ 
{1\over \vert{\rm log}~g_s\vert}, \nd
then with $g_s << 1$, ${\bf sgn}~{\rm B}(t) > 0$. In this case the EFT condition, from the first term in \eqref{tamana}, doesn't depend on the precise magnitude of ${\rm B}(t)$. On the other hand, if $\vert\dot\beta(t)\vert >> 1$, then ${\bf sgn}~{\rm B}(t) < 0$. In fact this happens because:
\bg\label{saramac}
{\bf sgn}~{\rm B}(t) = {\bf sgn}\left({{\rm log}(2v -\beta(t)) - {\rm log}\vert {\rm log}~g_s\vert - {\rm log}\vert \dot\beta(t)\vert \over 
\vert {\rm log}~g_s\vert}\right) ~ < ~ 0, \nd
as $1 < 2v - \beta(t) < 4$ and therefore  can never dominate over the last two terms in \eqref{saramac}. On the other hand, since 
$\vert {\rm log}~g_s\vert$ dominates over both the ${\rm log}(2v - \beta(t))$ and the ${\rm log}\vert{\rm log}~g_s\vert$ terms, the magnitude of ${\rm B}(t)$ will solely depend upon the rate $\dot\beta(t)$ for $\vert\dot\beta(t)\vert >> 1$. However at late time, when $g_s \to 0$, we expect 
$\vert{\rm log}~g_s\vert >> {\rm log}\vert \dot\beta(t)\vert$ for any smoothly varying function, {\it i.e.} for any continuous and differentiable
function, $\beta(t)$. Generically, as long as $\vert\dot\beta\vert < {1\over g_s}$, the EFT criterion \eqref{katudi} is always satisfied. The story in the ${\rm E}_8 \times {\rm E}_8$ is slightly more involved, and will be discussed in the next section.

Since the aforementioned criterion was identified with the non-violation of the null-energy condition (NEC) in type IIB \cite{coherbeta2}, our analysis then extends it to the heterotic $SO(32)$ also. The second term in \eqref{tamana} fixes the NEC constraint ${1\over n} \ge -1$. Again this is because NEC cannot depend on any temporally varying parameters. (For example as inferred in footnote \ref{karkapr}, any dependence on $\beta(t)$ will imply that $-{3v\over 3v - 1} < {1\over n} < -1$ ceases to have an EFT description at late time.) On the other hand the first term would fix the TCC constraint. This is what we turn to next.

\begin{figure}[h]
\centering
\begin{tabular}{c}
\includegraphics[width=3in]{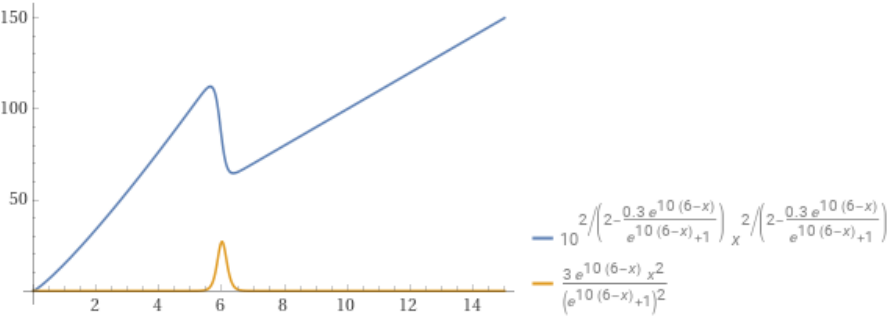}
\end{tabular}
\caption[]{Plots of $g_s^{-1}(1/x)$ (in blue) and $\vert\dot\beta(1/x)\vert$ (in orange) from \eqref{fuhrmann} for $x \equiv {1\over \vert t\vert}, \beta_o = 0.3, \Lambda = 0.01$ and $\vert\epsilon\vert = {1\over 6}$. Note that there are no non-trivial intersection points.}
\label{so324}
\end{figure}

\subsection{Revisiting TCC constraints in $SO(32)$ and ${\rm E}_8 \times {\rm E}_8$ theories \label{tolace5}}

The generic criterion $\vert\dot\beta(t)\vert < {1\over g_s(t)}$ unfortunately leads to certain insurmountable issues. The trouble starts from early on. Since both $\beta(t)$ and $g_s(t)$ are functions of conformal time $t$, the above inequality  should lead to a finer determination of the TCC temporal domain. Recall that we used $g_s < 1$ to determine the TCC domain in type IIB theory, {\it i.e.} the one from {\bf Table \ref{milleren1}}. For the heterotic $SO(32)$ case with ${\rm H}(y) = 1$, $g_s = 
(\sqrt{\Lambda}t)^{2v\over 2v - \beta(t)}$ and therefore $g_s < 1$ would still appear to retain the same temporal domain for the conformal time $t$
as in the type IIB case, namely $-{1\over \sqrt{\Lambda}} < t < 0$. However now the first term in \eqref{tamana} appears to reduce the temporal domain because $g_s \vert\dot\beta(t)\vert < 1$. This is intriguing because it tells us that there is a temporal region where $g_s < 1$ but the theory {\it does not} have an EFT description. The conformal time then spans a shorter domain where both $g_s < 1$ and $g_s \vert\dot\beta(t)\vert < 1$. Can we determine the new domain? To analyze this we will have to make an ans\"atze for a functional form for $\beta(t)$ that would reproduce \eqref{afleis}. One such functional form may be expressed in the following way:
\bg\label{aflies3}
\beta(t) = {\beta_o~ {\rm exp}\left[{1\over \sqrt{\Lambda}}\left({1\over \vert\epsilon\vert} - {1\over \vert t \vert}\right)\right] \over 
1 +{\rm exp}\left[{1\over \sqrt{\Lambda}}\left({1\over \vert\epsilon\vert} - {1\over \vert t \vert}\right)\right]}, \nd
where $\epsilon$ is given by \eqref{tegtrex}. A plot of this function appears in {\bf figure \ref{plot-heaviside-o}} wherein it should be evident that the behavior in \eqref{afleis} is reproduced precisely\footnote{We can take all parameters {\it dimensionless} by measuring with respect to ${\rm M}_p$. An example would be from \eqref{laumoon} where the dimensionless $\Lambda$, measured with respect to ${\rm M}_p^2$, is given completely in terms of the degeneracy factors only. \label{hisable}}. The derivative of this 
function, shown in {\bf figure \ref{plot-delta-o}}, is a sharply peaked one at $\vert t \vert = \vert\epsilon\vert$ implying that $\dot\beta(\vert\epsilon\vert) = -{\beta_o\over 4\epsilon^2 \sqrt{\Lambda}}$ and zero elsewhere. This is actually the root cause of the problem as we shall see in the following.

\begin{figure}[h]
\centering
\begin{tabular}{c}
\includegraphics[width=3in]{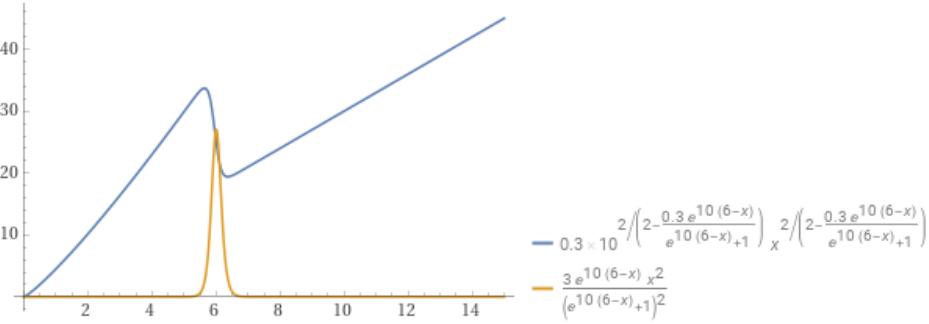}
\end{tabular}
\caption[]{Plots of $g_s^{-1}(1/x)$ (in blue) and $\vert\dot\beta(1/x)\vert$ (in orange) from \eqref{fuhrmann} for the same values of the parameters as in {\bf figure \ref{so324}} but with $g_s(1/x) \vert\dot\beta(1/x)\vert = 0.3$. There are now two non-trivial intersection points very close to $ x \equiv {1\over \vert t\vert} = {1\over 6}$.}
\label{so323}
\end{figure} 

\begin{figure}[h]
\centering
\begin{tabular}{c}
\includegraphics[width=3in]{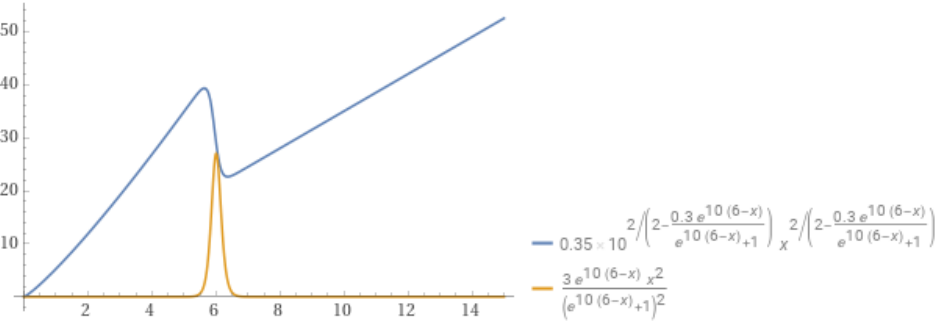}
\end{tabular}
\caption[]{Similar plots of $g_s^{-1}(1/x)$ (in blue) and $\vert\dot\beta(1/x)\vert$ (in orange) from \eqref{fuhrmann} for the same values of the parameters as in {\bf figure \ref{so324}} but now with $g_s(1/x) \vert\dot\beta(1/x)\vert = 0.35$. The two non-trivial intersecting points in 
{\bf figure \ref{so323}} are now closer together.}
\label{so322}
\end{figure}

\begin{figure}[h]
\centering
\begin{tabular}{c}
\includegraphics[width=3in]{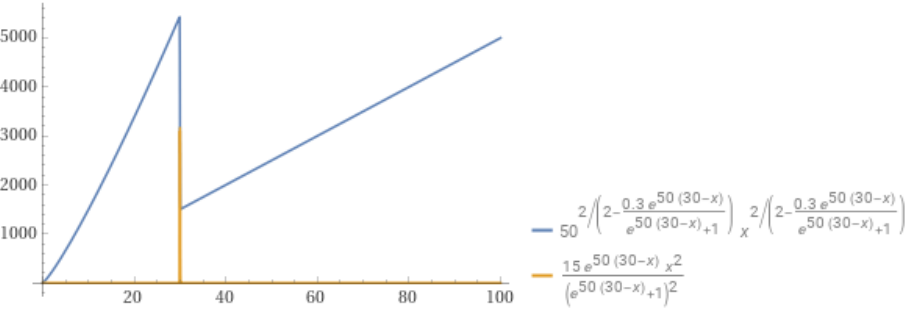}
\end{tabular}
\caption[]{The two graphs in the above figures become more sharply peaked once we decrease the value of $\Lambda$ to $\Lambda = 0.0004$. Here $\vert\epsilon\vert = {1\over 30}$ and $\beta_o = 0.3$. The graphs intersect exactly at $x \equiv {1\over \vert t\vert} = {1\over 30}$.}
\label{so321}
\end{figure}

For large values of $\Lambda$, as shown in {\bf figure \ref{so324}}, the condition $g_s \vert\dot\beta(t)\vert = 1$ has no solution. Solutions start appearing once we equate $g_s \vert\dot\beta(t)\vert$ to a quantity much smaller than 1, or when we make the cosmological constant small. This is because:
\bg\label{fuhrmann}
g_s = \left(\sqrt{\Lambda}\vert t \vert\right)^{2v\over 2v - \beta(t)}, ~~~
\dot\beta(t) = -{\beta_o\over \sqrt{\Lambda}\vert t\vert^2}
{{\rm exp}\left({1\over \sqrt{\Lambda}\vert\epsilon\vert} - {1\over \sqrt{\Lambda}\vert t \vert}\right) \over\left[
1 +{\rm exp}\left({1\over \sqrt{\Lambda}\vert\epsilon\vert} - {1\over \sqrt{\Lambda}\vert t \vert}\right)\right]^2}, \nd
and therefore the two graphs would intersect at two points which are close neighborhood of $\vert t\vert = \vert\epsilon\vert$ for $g_s \vert\dot\beta(t)\vert < 1$. This is shown in {\bf figures \ref{so323}} and {\bf \ref{so322}}.  
As $\Lambda$ decreases, the derivative in \eqref{fuhrmann} becomes more sharply peaked and the two graphs intersect exactly at $\vert t\vert = \vert\epsilon\vert$ as shown in {\bf figure \ref{so321}}. Unfortunately this would mean that the temporal domain in which EFT is valid is now much smaller and is given by $-\epsilon < t < 0$. This in turn would imply that the temporal domain 
$-{1\over \sqrt{\Lambda}} < t \le -\epsilon$, where much of the interesting dynamics happen, {\it does not} have an EFT description. 

\begin{figure}%
        \centering
        {\includegraphics[scale=0.40]{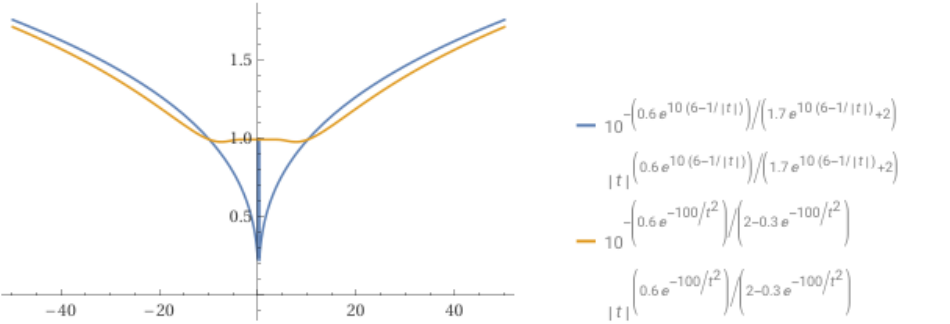}}%
        \qquad
        {\includegraphics[scale=0.40]{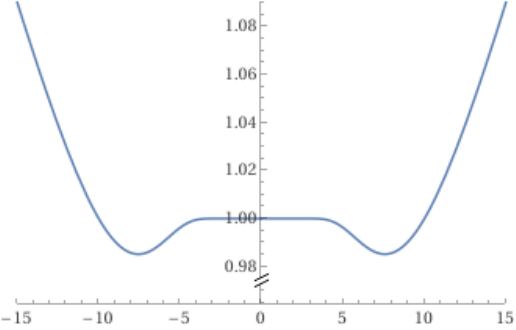}}%
        \caption{The behavior of ${\rm F}_1(t) = (\Lambda t^2)^{\beta(t)\over 2- \beta(t)}$ with $\beta(t)$ as defined in 
        \eqref{aflies3} and \eqref{fermige} with $\Lambda = {1\over 100}$. For the choice \eqref{aflies3}, ${\rm F}_1(t)$ shows a sharp rise at late time (shown in blue in the first figure), whereas for the choice \eqref{fermige}, the behavior is smooth as shown in orange in the first figure and in blue in the second figure. As before, only the negative axes are important.}%
        \label{F1behavior33}%
    \end{figure}

\begin{figure}[h]
\centering
\begin{tabular}{c}
\includegraphics[width=3in]{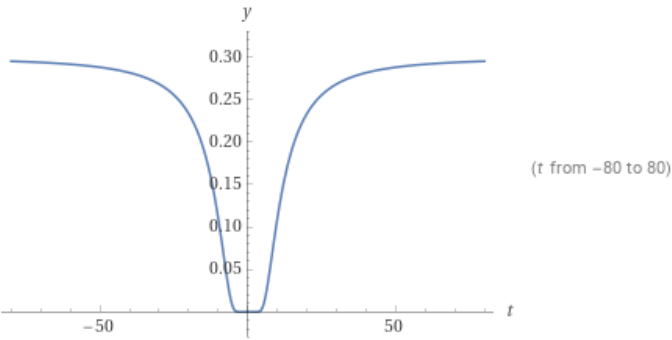}
\end{tabular}
\caption[]{Plot of $y(t) \equiv \beta(t)$ from \eqref{fermige} with $\beta_o = 0.3$ and $\sigma\vert\epsilon\vert = 10$. (We should only consider the negative time interval.) Note that $\beta(t)$ approaches $0.3$ for $t \approx -80$ and almost vanishes for $t > -5$ implying that $\Lambda \approx {1\over 6400}$ and $\epsilon \approx -5$. This further implies that $5 < \sigma << 16$ for consistency.}
\label{so325}
\end{figure}

\begin{figure}[h]
\centering
\begin{tabular}{c}
\includegraphics[width=3in]{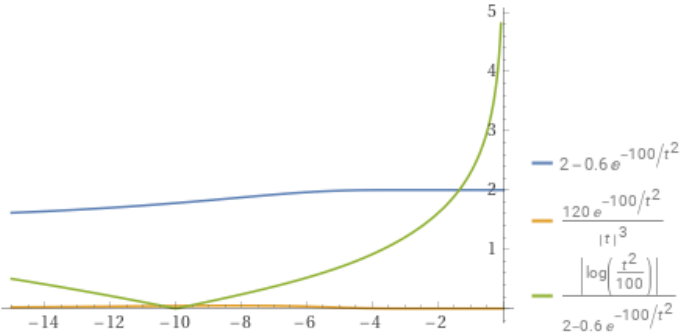}
\end{tabular}
\caption[]{Plot of three functions, $y_1(t) = 2 - \beta(t)$, (in blue) $y_2(t) = \vert\dot\beta(t)\vert$ (in orange) and $y_3(t) = \vert\log~g_s\vert$ (in green) from \eqref{fermige} with $\Lambda = 0.01$, $\beta_o = 0.6$ and $v = 1$ for the string frame. Note that, while the orange and the blue lines never intersect, the green line intersects both the blue and the the orange lines at two non-coincident points.}
\label{so326}
\end{figure}

\begin{figure}[h]
\centering
\begin{tabular}{c}
\includegraphics[width=3in]{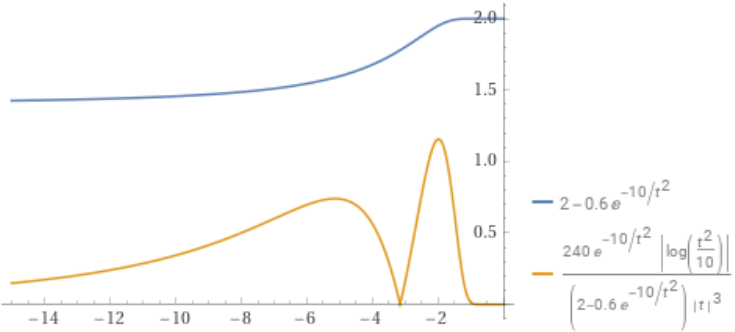}
\end{tabular}
\caption[]{Plot of two functions, $y_1(t) = 2 - \beta(t)$, (in blue) $y_2(t) = 40\vert\dot\beta(t)\vert \vert\log~g_s\vert$ (in orange) from \eqref{fermige} with $\Lambda = 0.1$, $\beta_o = 0.6$ and $v = 1$ for the string frame. Note that the orange and the blue lines never intersect despite increasing the second function $y_2(t)$ by 40 times the original. It is clear that $y_1(t) > 40 y_2(t)$ and therefore \eqref{bosema} is satisfied implying ${\bf sgn}~{\rm B}(t) > 0$ from \eqref{fl2mey}.}
\label{so327}
\end{figure}

\begin{figure}[h]
\centering
\begin{tabular}{c}
\includegraphics[width=3in]{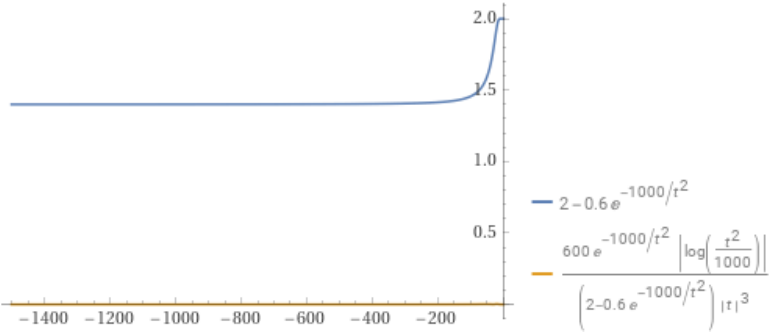}
\end{tabular}
\caption[]{Flattening of the orange curve for large value of $\sigma^2\vert\epsilon\vert^2 = 1000$. The remaining parameters remain the same as before. We now see no intersection between the blue and the orange curves.}
\label{so328}
\end{figure}

Our failure to get a consistent dynamics with the functional choice \eqref{aflies3} suggests that we should go for a smoother function that interpolates between two points in \eqref{afleis}. This is because the original condition $g_s = (\sqrt{\Lambda}\vert t\vert)^{2v\over 2v-\beta(t)} < {1\over \vert\dot\beta(t)\vert}$ reduces the temporal domain 
to $-\epsilon < t < 0$ as $\vert\dot\beta(t)\vert$ is sharply peaked at $\vert t\vert = \vert\epsilon\vert$. Once we take a smoother function, {\it i.e.} a function whose derivative is not sharply peaked near any specific value of $\vert t\vert$ and in particular near $\vert t\vert = \vert\epsilon\vert$, we might be able to increase the temporal domain of the validity of the EFT so as to lie in the region $-{1\over \sqrt{\Lambda}} < t \le -\epsilon$. 
One possibility would be the following choice:
\bg\label{fermige}
\beta(t) = \beta_o ~{\rm exp}\left(-{\sigma^2\vert\epsilon\vert^2\over \vert t \vert^2}\right), ~~~~~~ \dot\beta(t) = -{2\beta_o\sigma^2\vert\epsilon\vert^2\over \vert t\vert^3}~ {\rm exp}\left(-{\sigma^2\vert\epsilon\vert^2\over \vert t \vert^2}\right), \nd
for the temporal domain $-{1\over \sqrt{\Lambda}} < t < 0$ with the dimensionless factor $\sigma >> 1$. It is clear that for the temporal domain $-\epsilon < t < 0$, $\beta(t) \approx 0$ and for $t \to -{1\over \sqrt{\Lambda}}$, $\beta(t) \to \beta_o$ provided $\vert\epsilon\vert << {1\over \sigma\sqrt{\Lambda}}$ for a given value of $\Lambda$ from say \eqref{laumoon}. This matches the interpolating regimes of \eqref{afleis}. (A comparison between the $g_s$ $-$ from \eqref{fuhrmann} $-$ and 
${\rm F}_1$ $-$ from \eqref{evalike} $-$ for the two choices of $\beta(t)$, namely \eqref{aflies3} and \eqref{fermige}, are shown in {\bf figure \ref{comparison}}  and {\bf figure \ref{F1behavior33}} respectively. One can easily see the {\it jumps} in the values of $g_s$ and ${\rm F}_1$ for the choice \eqref{aflies3}.) For some choice of the parameters in \eqref{fermige}, the function $\beta(t)$ is plotted in {\bf figure \ref{so325}}. Note that now the derivative peaks at 
$\vert t\vert =  \sqrt{2\sigma^2\vert\epsilon\vert^2\over 3} << \sqrt{2\over 3\Lambda^2}$. The plot of the three functions, $y_1(t) \equiv 2v - \beta(t), y_2(t) \equiv \vert\dot\beta(t)\vert$ and $y_3(t) \equiv \vert{\rm log}~g_s\vert$ is given in {\bf figure \ref{so326}}. It is easy to see that the condition \eqref{bosema} is satisfied and therefore ${\bf sgn}~{\rm B}(t) > 0$ from \eqref{fl2mey}. This is depicted in {\bf figure \ref{so327}}. As the cosmological constant, or the inverse of the quantity $\sigma^2\vert\epsilon\vert^2$, is taken to be very small, the orange curve in {\bf figure \ref{so327}} practically flattens out in the temporal domain of interest, implying zero intersection with the blue curve as shown in {\bf figure \ref{so328}}. This means the temporal domain in which EFT is valid remains $-{1\over \sqrt{\Lambda}} < t < 0$.

The story in the ${\rm E}_8 \times {\rm E}_8$ heterotic is almost similar to the $SO(32)$ case with some minor but interesting deviations. The EFT condition should again be derived from the derivative constraint as in \eqref{tamana}, {\it i.e.} now in the following way:
\bg\label{tamana2}
{\partial g_s\over \partial t} = g_s^{1 + {\rm B}(t)} - 
6n\sqrt{\Lambda}~ g_s^{1 + {1\over n} - {\hat\alpha(t) + \hat\beta(t)\over 6n} + {\rm C}(t)}, \nd
which basically implies that ${\beta(t)\over 2}$ from \eqref{tamana} changes to ${\hat\alpha(t) + \hat\beta(t)\over 6}$ from \eqref{afleis2} in the string frame ($v = 1$); and the other parameters $g_s, {\rm B}(t)$ and ${\rm C}(t)$ from the $SO(32)$ case change to the take the following functional form:

\begin{figure}%
        \centering
        {\includegraphics[scale=0.40]{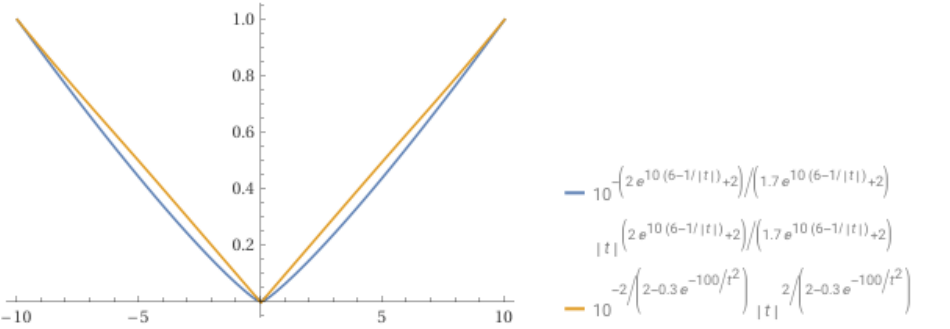}}%
        \qquad
        {\includegraphics[scale=0.40]{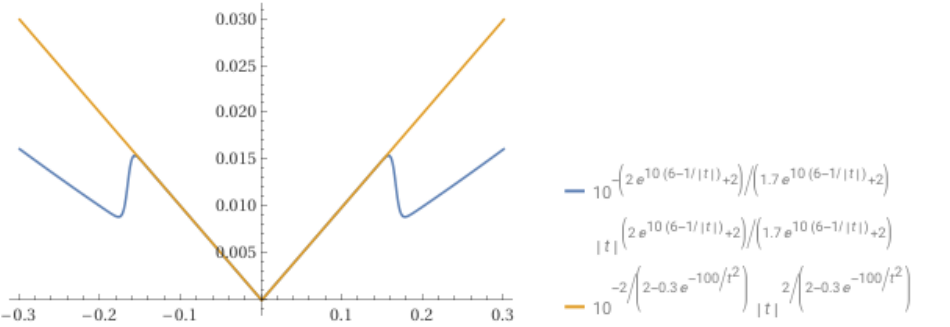}}%
        \caption{The behavior of $g_s$ from \eqref{fuhrmann} for the two choices of $\beta(t)$ in \eqref{aflies3} (in {blue}) and \eqref{fermige} (in {orange}). Both behaves in almost similar way except the choice \eqref{aflies3} shows a jump at late time. Only the negative axes are relevant here, and the positive axes are shown to quantify the behavior near $t \to \pm 0$.}%
        \label{comparison}%
    \end{figure}

{\footnotesize
\bg\label{nobodi}
g_s = \left(\sqrt{\Lambda}\vert t\vert\right)^{6n\over \hat\alpha(t)+\hat\beta(t)-6}, ~~~ {\rm B}(t) = {\log\left[\left({\dot{\hat\alpha}(t) + \dot{\hat\beta}(t)\over 6 - \hat\alpha(t) - \hat\beta(t)}\right)\log~g_s\right]\over \log~g_s}, ~~~ {\rm C}(t) = 
-{\log(6 - \hat\alpha(t) - \hat\beta(t))\over \log~g_s}, \nd}
which may be compared to \eqref{fl2mey}. Since $\hat\beta(t) < \hat\alpha(t) < 1$ and $g_s << 1$, ${\bf sgn}~{\rm C}(t) > 0$. It should also be clear that $\dot{\hat\alpha}(t) + \dot{\hat\beta}(t) \le 0$ for 
${\rm B}(t)$ in \eqref{nobodi} to make sense. Before studying the implications of the temporal derivatives, let us ask for the behavior of 
$\hat\alpha(t)$ and $\hat\beta(t)$ in the whole range of temporal domain 
$-{1\over \sqrt{\Lambda}} < t < 0$ for the system to be consistent in both the string and the Einstein frames. In a string frame one would expect:
\bg\label{afleis4}
\hat\alpha(t) = \begin{cases} ~ \hat\alpha_o ~~~~~~ -{1\over \sqrt{\Lambda}} < t < -\epsilon_1 \\
~~~ \\
~ \hat\gamma_o ~~~~~~ -\epsilon_1 < t < -\epsilon_2\\
~~~\\
~0 ~~~~~~~ -\epsilon_2 < t < 0
\end{cases}, ~~~~~ 
\hat\beta(t) = \begin{cases} ~ \hat\beta_o ~~~~~~ -{1\over \sqrt{\Lambda}} < t < -\epsilon_1 \\
~~~ \\
~ \hat\gamma_o ~~~~~~ -\epsilon_1 < t < -\epsilon_2\\
~~~\\
~0 ~~~~~~~ -\epsilon_2 < t < 0
\end{cases}
\nd 
which is basically a more finer distribution of the parameters than the ones in \eqref{afleis2} with $\epsilon_1 > \epsilon_2$. As before, we expect that near the local neighborhoods of $t = \epsilon_1$ and $t = \epsilon_2$ the functions $\hat\alpha(t)$ and $\hat\beta(t)$ are allowed to vary smoothly (with well-defined derivatives) according to \eqref{afleis4} to avoid jumps shown in {\bf figures \ref{comparison}, \ref{F1behavior33}}. (It is assumed that $\epsilon_i$ are positive definite, {\it i.e.} $\epsilon_i \equiv \vert\epsilon_i\vert$.) The derivative condition $\dot{\hat\alpha}(t) + \dot{\hat\beta}(t) \le 0$ now imposes, at least in the first go, the following six interesting scenarios\footnote{These will have to be modified further once we impose constraints from axionic cosmology as we shall see in section \ref{sec3.6}.} (see {\bf figure \ref{6cases}}):
\begin{itemize}
    \item {\Su $\dot{\hat\alpha}(t) < 0$ and $\dot{\hat\beta}(t) < 0$ in the small interval in the neighborhood of $t = -\epsilon_1$ till they both hit $\hat\gamma_o$. After which they become equal and go to zero 
    for $-\epsilon_ 2 < t < 0$.}
\item{\Su $\dot{\hat\alpha}(t) < 0$ and $\dot{\hat\beta}(t) > 0$ but 
$\dot{\hat\alpha}(t) + \dot{\hat\beta}(t) < 0$ near the small neighborhood of 
$t = -\epsilon_1$ till they both hit $\hat\gamma_o$. After which they become equal and go to zero 
    for $-\epsilon_ 2 < t < 0$.}
\item{\Su $\dot{\hat\alpha}(t) = 0$ and $\dot{\hat\beta}(t) < 0$ in the small neighborhood of $t = -\epsilon_1$, but both $\dot{\hat\alpha}(t) < 0$ and $\dot{\hat\beta}(t) < 0$ in the small interval in the neighborhood of $t = -\epsilon_2$. This means $\hat\beta(t)$ goes through the point $\hat\gamma_o$ but $\hat\alpha(t)$ does not.}
\item{\Su $\dot{\hat\alpha}(t) < 0$ and $\dot{\hat\beta}(t) = 0$ in the small neighborhood of $t = -\epsilon_1$, but both $\dot{\hat\alpha}(t) < 0$ and $\dot{\hat\beta}(t) < 0$ in the small interval in the neighborhood of $t = -\epsilon_2$. This means $\hat\alpha(t)$ goes through the point $\hat\gamma_o$ but $\hat\beta(t)$ does not.}
\item{\Su $\dot{\hat\alpha}(t) = \dot{\hat\beta}(t) = 0$ in the temporal domain $-{1\over \sqrt{\Lambda}} < t < -\epsilon_2$ but shows the behavior $\dot{\hat\alpha}(t) < 0$ and $\dot{\hat\beta}(t) < 0$ in the small interval in the neighborhood of $t = -\epsilon_2$. This means both the functions never go through the intermediate point $\hat\gamma_o$.}
\item{\Su$\dot{\hat\alpha}(t) = \dot{\hat\beta}(t) = 0$ in the temporal domain $-{1\over \sqrt{\lambda}} < t < 0$, implying that $\hat\alpha(t)$ and $\hat\beta(t)$ are constants {\it i.e.} $\hat\alpha(t) = \hat\alpha_o$ and $\hat\beta(t) = \hat\beta_o$.}
\end{itemize}
Clearly for the first two cases, the equality ${\rm F}_1 = {\rm F}_3$ happens much earlier, {\it i.e.} for the interval $-\epsilon_1 < t < 0$, whereas for the next two cases the equality happens much later, {\it i.e.} for the interval $-\epsilon_2 < t < 0$. For the latter, the temporal domain $-\epsilon_2 < t < 0$ is also the domain where the Einstein frame physics starts displaying de Sitter like behavior. The fifth case is interesting. This is simply \eqref{afleis2} with $\hat\gamma_o = 0$, {\it i.e.} just a one-step process. Whereas for the last case we have constant $\hat\alpha(t)$ and $\hat\beta(t)$. Such a scenario would work well in the string frame but will have late-time singularities. All these cases are shown in {\bf figure \ref{6cases}}, but once we consider the constraints coming from axionic cosmology, as discussed in section \ref{sec3.6}, the last stage within the temporal domain $-\epsilon_1 < t < 0$ gets {\it smoothened out} (see {\bf figure \ref{gamma00}}). We will motivate this in section \ref{sec3.6}. Meanwhile, for the second case, realizing the following possibility:
\bg\label{popeexor}
\dot{\hat\alpha}(t) + \dot{\hat\beta}(t) = 0, ~~~~ -{1\over \sqrt{\Lambda}} < t < -\epsilon_1, \nd
which would imply no contributions to ${\rm B}(t)$ in \eqref{nobodi} in the aforementioned temporal domain $-$ with the contribution starting only when both the functions (now equal) go from $\hat\gamma_o$ to zero in the temporal domain $-\epsilon_2 < t < 0$ $-$ may not be easy to realize for all $t$ in the aforementioned temporal domain. This may be easily inferred from {\bf figure \ref{e81}} for the following choice of $\hat\alpha(t)$ and $\hat\beta(t)$:
\bg\label{hanasmit}
\hat\alpha(t) = (\hat\alpha_o - \hat\gamma_o) {\rm exp}\left(-{\sigma^2\vert\epsilon\vert^2\over \vert t \vert^2}\right) + \hat\gamma_o, ~~~~ \hat\beta(t) = \hat\beta_o~{\rm exp}\left(-{\sigma'^2\vert t\vert^2\over \vert\epsilon\vert^2}\right) + \hat\gamma_o - \hat\beta_o, \nd
corresponding to \eqref{afleis2} $-$ which we take for simplicity (and for illustrative purpose) here $-$ rather than to the finer splitting of \eqref{afleis4}. ($\sigma$ and $\sigma'$ are dimensionless constants with $\sigma >> \sigma'$. See {\bf figure \ref{e81}}.) The derivative condition \eqref{popeexor} clearly does not happen for all values of $t$ in the required domain. 
For the fourth case where $\dot{\hat\beta}(t) = 0$, the value of $\hat\beta(t)$ cannot be bigger than $\hat\gamma_o$, otherwise in the temporal domain $-\epsilon_1 < t < \epsilon_2$ there is a possibility that $\hat\beta(t) > \hat\alpha(t)$ thus ruining the duality cycle of {\bf Table \ref{milleren4}}.

Comparing to the $SO(32)$ case, once we take {\it smooth} functional choice for 
$\hat\alpha(t)$ and $\hat\beta(t)$, much like the ones depicted in 
{\bf figure \ref{6cases}}, we would expect:
\bg\label{cardavs}
6 - \hat\alpha(t) - \hat\beta(t) ~ > ~ \vert \dot{\hat\alpha}(t) + 
\dot{\hat\beta}(t)\vert \vert {\rm log}~g_s\vert, \nd
implying that ${\bf sgn}~{\rm B}(t) > 0$ for ${\rm B}(t)$ as in \eqref{nobodi}. (Recall that $g_s << 1$.) In fact this happens for all the six cases studied above, implying further that the temporal domain where EFT is valid, in both $SO(32)$ and ${\rm E}_8 \times {\rm E}_8$ theories, is $-{1\over \sqrt{\Lambda}} < t < 0$.  
With a little effort, one could even extend this to the type IIA de Sitter case too although we shall not do so here. Thus it appears that the non-violation of NEC is an important criterion to follow for an EFT description in any string theories
that allow four-dimensional FLRW universes (in conformal coordinates $t$).

Our conclusion from the aforementioned discussion is then the following. The TCC criterion, namely $-{1\over \sqrt{\Lambda}} < t < 0$, appears to remain unchanged in all string theories once we restrict ourselves to $n = -1$, {\it i.e.} with four-dimensional de Sitter isometries in the flat-slicings. One might even extend the TCC criterion to all NEC preserving FLRW universes with $-1 \le {1\over n} < 0$ in any string theories.

\subsection{Additional constraints from the axions in the ${\rm E}_8 \times {\rm E}_8$ theory \label{sec3.6}}

The behavior of $\hat\alpha(t)$ and $\hat\beta(t)$ that we proposed in \eqref{afleis4} and \eqref{hanasmit} still needs improvement to be consistent with the constraints coming from the experimental results. These experimental results appear from studying the axionic cosmology that has recently been discussed briefly in \cite{axion}. (For a previous study, in the context of heterotic theories but not in the context of temporally varying heterotic de Sitter state, the readers may refer to \cite{svrcek, hassan}.) For definiteness, we will discuss the heterotic ${\rm E}_8 \times {\rm E}_8$ theory in both the Einstein and the string frames. 

In {\bf Table \ref{milleren4}} we have shown how the metric in the ${\rm E}_8 \times {\rm E}_8$ theory may be expressed in the Einstein frame. Clearly we cannot get de Sitter excited states in both string and Einstein frames {\it simultaneously}, but we can demand de Sitter excited states in each of the two frames separately. Since both the results appear from M-theory, it would be  good to compare the results in both the frames carefully. The M-theory metric is always given by:

{\footnotesize
\bg\label{aztecho}
ds^2_{11} = g_s^{-8/3} ds^2_{\mathbb{R}^{2, 1}} + g_s^{-2/3} {\rm H}^2(y) \left[{\rm F}_1(g_s/{\rm H}) ds^2_{\theta_1} + {\rm F}_3(g_s/{\rm H}) ds^2_{\theta_2} + {\rm F}_2(g_s/{\rm H}) ds^2_{{\cal M}_4}\right] + 
g_s^{-4/3} ds^2_{\mathbb{T}^2_{3, 11}}, \nd}
where the internal eight-manifold has a topology of ${\bf S}^1_{\theta_1} \times {{\bf S}^1_{\theta_2}\over {\cal I}_{\theta_2}} \times {\cal M}_4 \times {\mathbb{T}^2\over {\cal G}} \equiv {\cal M}_7 \times {{\bf S}^1_{\theta_2}\over {\cal I}_{\theta_2}}$ with three-form flux oriented along ${\bf C}_{\theta_2 {\rm AB}}$ where $({\rm A, B}) \in {\cal M}_7 \times \mathbb{R}^{2,1}$ so that it is not projected by the orbifold action. The flux configuration on the heterotic ${\rm E}_8 \times {\rm E}_8$ side is affected by the intermediate orientifold projections, but because of the blow-ups that we perform in \eqref{dija}, many of the flux components that are projected out by the orientifold projections are regained back (albeit with appropriately changed magnitude). This means that, in the language of axions, we can have both model-dependent and model-independent axions (see \cite{svrcek, hassan, axion} for details). Ignoring the fluxes for the time being, demanding a de Sitter excited state in the Einstein frame, the ten-dimensional metric takes the following form \cite{axion}:
\bg\label{aztecmey}
ds^2_{10} & = & {\rm H}(y) g_s^{-2} {\rm F}_1^{3/8} {\rm F}_3^{1/8} ds^2_{\mathbb{R}^{3, 1}} + {\rm H}^3(y) {\rm F}_2 {\rm F}_1^{3/8} {\rm F}_3^{1/8} ds^2_{{\cal M}_4} + {\rm H}^{-1}(y){\rm F}_3^{1/8}{\rm F}_1^{-5/8} ds^2_{\mathbb{T}^2}\\
& = & {1\over \Lambda t^2 {\rm H}(y)} ds^2_{\mathbb{R}^{3, 1}}
+ {\rm H}^3(y) \left(\Lambda t^2 \right)^{5\hat\alpha(t) - \hat\beta(t)\over 48 - 6\hat\alpha(t) - 2\hat\beta(t)} ds^2_{{\cal M}_4}
+ {\rm H}^{-1}(y)
\left(\Lambda t^2 \right)^{\hat\beta(t) - 5\hat\alpha(t)\over 24 - 3\hat\alpha(t) - \hat\beta(t)} ds^2_{\mathbb{T}^2},\nonumber
\nd
where $\mathbb{T}^2 \equiv \hat{\bf S}^1_{\theta_1} \times {\bf S}^1_{11}$
as depicted in {\bf  Table \ref{milleren4}} and we have inserted back the warp-factor ${\rm H}(y)$. Note that not only the powers of the warp-factor are different from what one would have expected in the string frame, but also are the factors accompanying $\hat\alpha(t)$ and $\hat\beta(t)$. The latter implies that at late time $\hat\beta(t) = 5\hat\alpha(t)$ in the Einstein frame compared to $\hat\alpha(t) = \hat\beta(t)$ in the string frame. In other words, we now expect:
\bg\label{afleis44}
\hat\alpha(t) = \begin{cases} ~ \hat\alpha_o ~~~~~~ -{1\over \sqrt{\Lambda}} < t < -\epsilon_1 \\
~~~ \\
~ \hat\gamma_o ~~~~~~ -\epsilon_1 < t < -\epsilon_2\\
~~~\\
~0 ~~~~~~~ -\epsilon_2 < t < 0
\end{cases}, ~~~~~ 
\hat\beta(t) = \begin{cases} ~ \hat\beta_o ~~~~~~ -{1\over \sqrt{\Lambda}} < t < -\epsilon_1 \\
~~~ \\
~ 5\hat\gamma_o ~~~~~~ -\epsilon_1 < t < -\epsilon_2\\
~~~\\
~0 ~~~~~~~ -\epsilon_2 < t < 0
\end{cases}
\nd 
compared to $(\hat\alpha(t), \hat\beta(t))$ defined in the string frame in \eqref{afleis4}. Such a choice would essentially keep the internal six-manifold ${\cal M}_4 \times \mathbb{T}^2$ time-independent at late time (not just the volume). But the question regarding which of the two choices, \eqref{afleis4} and \eqref{afleis44}, is consistent with cosmological observation remains to be seen. Keeping this in mind, the type IIA string coupling and the heterotic ${\rm E}_8 \times {\rm E}_8$ coupling take the following form:
\begin{figure}[h]
\centering
\begin{tabular}{c}
\includegraphics[width=5in]{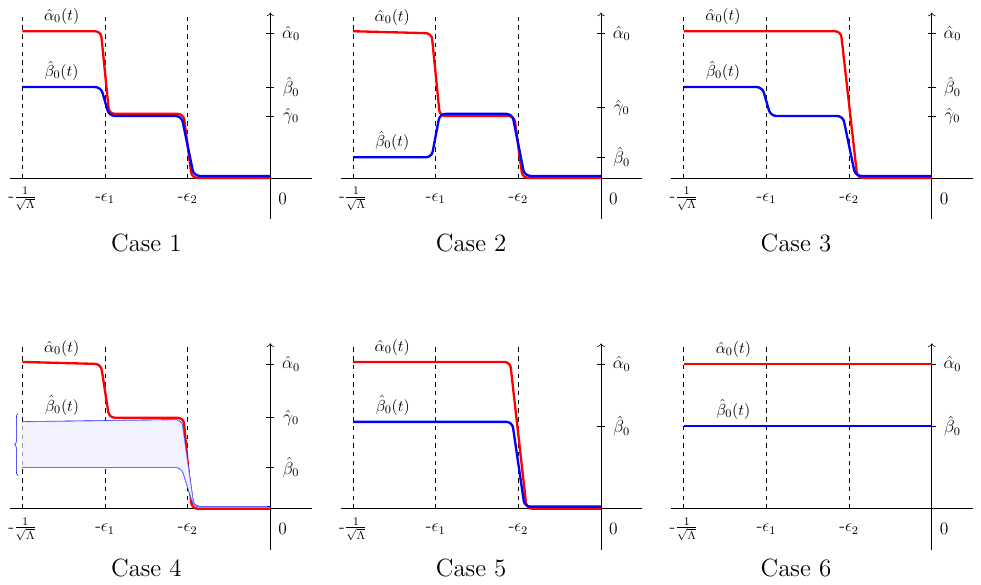}
\end{tabular}
\vskip-4.5in
\caption[]{Plots of all the six cases discussed after \eqref{afleis4}. Note that the last stages, between the temporal domain $-\epsilon_1 < t < 0$ would get significantly modified from the constraints coming from axionic cosmology (see {\bf figure \ref{gamma00}}).}
\label{6cases}
\end{figure} 
\bg\label{fountainbleau}
{g_s\over {\rm H}(y)} = \left(\Lambda t^2\right)^{12\over 24 - 3\hat\alpha(t) - \hat\beta(t)}, ~~~~~ g_{\rm het} = e^{\bf\Phi} = {\rm H}^2(y) \left(\Lambda t^2\right)^{2\hat\alpha(t) + 6\hat\beta(t)\over 24 - 3\hat\alpha(t) - \hat\beta(t)}, \nd
suggesting that both the couplings remain weak at late time as $\hat\alpha(t) < 1$ and $\hat\beta(t) < 1$ even though $\hat\beta(t) < \hat\alpha(t) < 1$ for $-{1\over \sqrt{\Lambda}} < t < -\epsilon_1$ and 
$\hat\beta(t) = 5 \hat\alpha(t)$ for $-\epsilon_1 < t < -\epsilon_2$. On the other hand, the string frame metric takes the following form:
\bg\label{aztecmey2}
ds^2_{10} & = & {\rm H}^2(y) g_s^{-2} \sqrt{{\rm F}_1{\rm F}_3} ds^2_{\mathbb{R}^{3, 1}} + {\rm H}^4(y) {\rm F}_2 \sqrt{{\rm F}_1 {\rm F}_3} ds^2_{{\cal M}_4} + \sqrt{{\rm F}_3\over {\rm F}_1} ds^2_{\mathbb{T}^2}\\
& = & {1\over \Lambda t^2} ds^2_{\mathbb{R}^{3, 1}}
+ {\rm H}^4(y) \left(\Lambda t^2 \right)^{\hat\alpha(t) - \hat\beta(t)\over 12 - 2\hat\alpha(t) - 2\hat\beta(t)} ds^2_{{\cal M}_4}
+ \left(\Lambda t^2 \right)^{\hat\beta(t) - \hat\alpha(t)\over 6 - \hat\alpha(t) - \hat\beta(t)} ds^2_{\mathbb{T}^2},\nonumber
\nd
where note the placements of the warp-factor ${\rm H}(y)$ as well as $(\hat\alpha(t), \hat\beta(t))$ defined in \eqref{afleis4} differ from the Einstein frame metric \eqref{aztecmey}. Expectedly, the type IIA string coupling and the heterotic ${\rm E}_8 \times {\rm E}_8$ coupling also differ from \eqref{fountainbleau} in the following way:
\bg\label{fountainbleaugu}
{g_s\over {\rm H}(y)} = \left(\Lambda t^2\right)^{3 \over 6 - \hat\alpha(t) - \hat\beta(t)}, ~~~~~ g_{\rm het} = e^{\bf\Phi} = {\rm H}^2(y) \left(\Lambda t^2\right)^{\hat\alpha(t) + 3\hat\beta(t)\over 12 - 2\hat\alpha(t) - 2\hat\beta(t)}, \nd
although both remains weakly coupled at late time as in the Einstein frame. The two sets of configurations: (\eqref{aztecmey}, \eqref{afleis44}, \eqref{fountainbleau}) and (\eqref{aztecmey2}, \eqref{afleis4}, \eqref{fountainbleaugu}) can now be used to determine the behavior of the axionic coupling $f_a$ following the strategy laid out in \cite{axion}. All we need are the two parameters\footnote{ $\beta(t)$, which only appears in this section and is unfortunately a standard notation, should not be confused with the $SO(32)$ warp-factor $\beta(t)$ which appears everywhere else.}: $\beta(t)$ and 
${\rm M}_p^2$. They are defined as:
\bg\label{stratmey}
{\rm M}_p^2 = 4\pi {\rm M}_s^2 \int d^6y~{\rm H}^4(y) \sqrt{{\rm det}~g_{{\cal M}_4}\cdot {\rm det}~g_{\mathbb{T}^2}}, ~~~
\beta(t) = \left(\Lambda t^2\right)^{v\hat\alpha(t) + 3v \hat\beta(t)\over (2v-1)\hat\alpha(t) + \hat\beta(t) -6v^2}, \nd
where $v = (1, 2)$ represent the string and the Einstein frames respectively (with $\hat\alpha(t)$ and $\hat\beta(t)$ taking their values as in \eqref{afleis44} and \eqref{afleis4} respectively). Note that the definition of ${\rm M}_p$, in terms of the string scale ${\rm M}_s$, warp-factor ${\rm H}(y)$ and the unwarped volume
of the internal six-manifold, remains the same in either frames. Putting everything together, the axionic coupling can be expressed as (see also \cite{hassan, axion}):
\begin{figure}[h]
\centering
\begin{tabular}{c}
\includegraphics[width=3in]{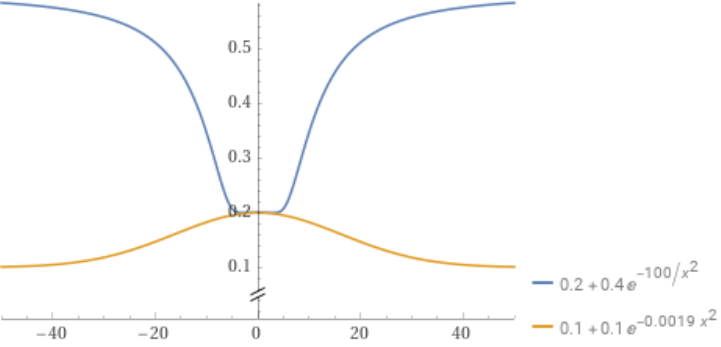}
\end{tabular}
\caption[]{Plot of two functions $\hat\alpha(t)$ (in blue) and $\hat\beta(t)$ (in orange) from \eqref{hanasmit} for $t \equiv x, \hat\alpha_o = 0.6, \hat\beta_o = 0.1, \hat\gamma_o = 0.2, \sigma^2\vert\epsilon\vert^2 = 100$ and ${\sigma'^2\over \vert\epsilon\vert^2} = 0.0019$. As expected the minima and maxima of the two functions are respectively $\hat\gamma_o = 0.2$. However the derivative condition \eqref{popeexor} is generically not satisfied at all points of $x \equiv t$ in the required domain. Note that only the negative axis of $x$ should be considered. The positive axis is just shown here for symmetry. Further modification of the flat region at late time appears in {\bf figure \ref{gamma00}}.}
\label{e81}
\end{figure}

{\scriptsize
\bg\label{TItostrat}
f_a(t) = \sqrt{2\over \beta(t)}~{{\rm M}_s^2\over {\rm M}_p}  = {1\over 2\pi}
\left(\sqrt{\Lambda}\vert t\vert\right)^{v\hat\alpha(t) + 3v \hat\beta(t)\over 6v^2 - (2v-1)\hat\alpha(t) - \hat\beta(t)}
 \Bigg({2\pi {\rm M}_s^{-4} \over \int d^6 y ~{\rm H}^4(y) \sqrt{{\rm det}~g_{{\cal M}_4}(y)\cdot {\rm det}~g_{\mathbb{T}^2}(y)}}\Bigg)^{1/2}, \nd}
which clearly {\it decreases} at late time and in fact vanishes as $t \to 0$. This, as also noticed in \cite{axion}, is problematic and the only way out is to change the behavior of $\hat\alpha(t)$ and $\hat\beta(t)$ in \eqref{afleis4} and \eqref{afleis44}. To do this we have to first make a few observations from the M-theory metric. From {\bf Table \ref{milleren4}}, the M-theory metric after the blow-ups may be written as:

{\footnotesize
\bg\label{kingtut}
ds^2_{11} = {\rm H}^{2/3}{\rm F}_1^{1/3} g_s^{-2} ds^2_{{\mathbb{R}^{3, 1}}} + {\rm H}^{8/3}
{\rm F}_1^{1/3}{\rm F}_2 ds^2_{{\cal M}_4} + {\rm H}^{-4/3}{\rm F}_1^{-2/3} ds^2_{\mathbb{T}^2} + {\rm H}^{8/3}{\rm F}_1^{1/3} {\rm F}_3 ds^2_{\mathbb{I}}, \nd}
where $\mathbb{T}^2 = {\rm S}^1_{\theta_1} \times {\rm S}_{11}$ and 
$\mathbb{I} = {{\rm S}^1_{\theta_2}\over {\cal I}_{\theta_2}}$. 
From \eqref{kingtut} a few things may be easily inferred: the heterotic dilaton $e^{\bf \Phi}$, the distance between the two Horava-Witten walls $\rho$, and the axionic decay constant  $f_a$ are all related to each other via the following functional choice:
\bg\label{shamitagra}
\left(e^{\bf \Phi}, \rho, f_a\right) \propto \left(({\rm F}_1 {\rm F}_3^3)^a, ({\rm F}_1 {\rm F}_3^3)^b, ({\rm F}_1 {\rm F}_3^3)^c\right), \nd
where the functional form for ${\rm F}_1$ and ${\rm F}_3$ would be frame-dependent if we demand de Sitter space in both frames (but not simultaneously). The exponents $(a, b)$ are frame-independent and take values $a = {1\over 4}$ and $b = {1\over 6}$, while $c = {1\over 2}$ in the Einstein frame and $c = 1$ in the string frame. This remarkable connection between three hitherto unrelated quantities now imposes some constraints that would affect the functional form for ${\rm F}_1$ and ${\rm F}_3$. These constraints are:
\begin{figure}%
        \centering
        {\includegraphics[scale=0.40]{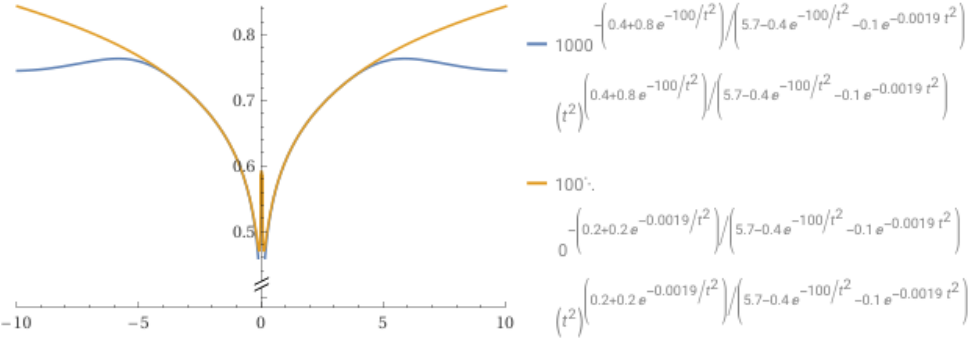}}%
        \qquad
        {\includegraphics[scale=0.40]{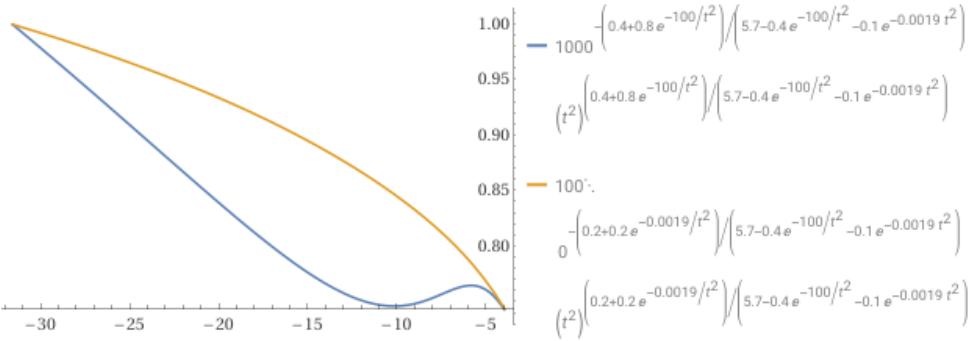}}%
        \caption{The behavior of ${\rm F}_1(t) = (\Lambda t^2)^{2\hat\alpha(t)\over 6 - \hat\alpha(t)-\hat\beta(t)}$ and 
        ${\rm F}_3(t) = (\Lambda t^2)^{2\hat\beta(t)\over 6 - \hat\alpha(t)-\hat\beta(t)}$ with $\hat\alpha(t)$ and $\hat\beta(t)$ as defined in \eqref{hanasmit}. Since $\hat\alpha_o > \hat\beta_o$ in \eqref{afleis4}, ${\rm F}_1(t)$ decays faster than ${\rm F}_3(t)$ as shown in the second figure. At later time, they become equal as shown in the first figure. However \eqref{hanasmit} doesn't quite represent correctly the behavior at the very late time as ${\rm F}_3(t)$ shows a jump to higher value before going to zero. The parameters chosen to plot the various functions are the same as in {\bf figure \ref{e81}}. See also {\bf figure \ref{f1f32}} below.}%
        \label{F1F3}%
    \end{figure}

\vskip.1in

\noindent $\bullet$ The axionic decay constant $f_a(t)$ cannot become zero because cosmological observation restricts its value to be bounded as $10^9~ {\rm GeV} < f_a(t) < 10^{12}~ {\rm GeV}$.

\vskip.1in

\noindent $\bullet$ The distance between the two Horava-Witten walls can decrease monotonically, but should not show any sudden jump at late time that violates the monotonic decrease.

\vskip.1in

\noindent $\bullet$ The heterotic coupling should remain weak throughout the dynamical process within the allowed temporal domain of $-{1\over \sqrt{\Lambda}} < t < 0$.

\vskip.1in

\noindent The first two constraints tell us that the monotonic decrease in $f_a(t)$ should {\it terminate} at late-time, implying that the motion of the two Horava-Witten walls should consequently stop. Thus at late time the Horava-Witten walls would come close to each other but should {\it never coincide}. The second constraint further implies that there should be no bounce, thus we should not see any Ekpyrotic \cite{ekpyrotic} like behavior. In fact the absence of an Ekpyrotic cosmology here is related to the non-violation of the NEC that we studied earlier. Any violation of NEC would imply the breakdown of an EFT \cite{coherbeta2, bernmir}.

\begin{figure}%
        \centering
        {\includegraphics[scale=0.85]{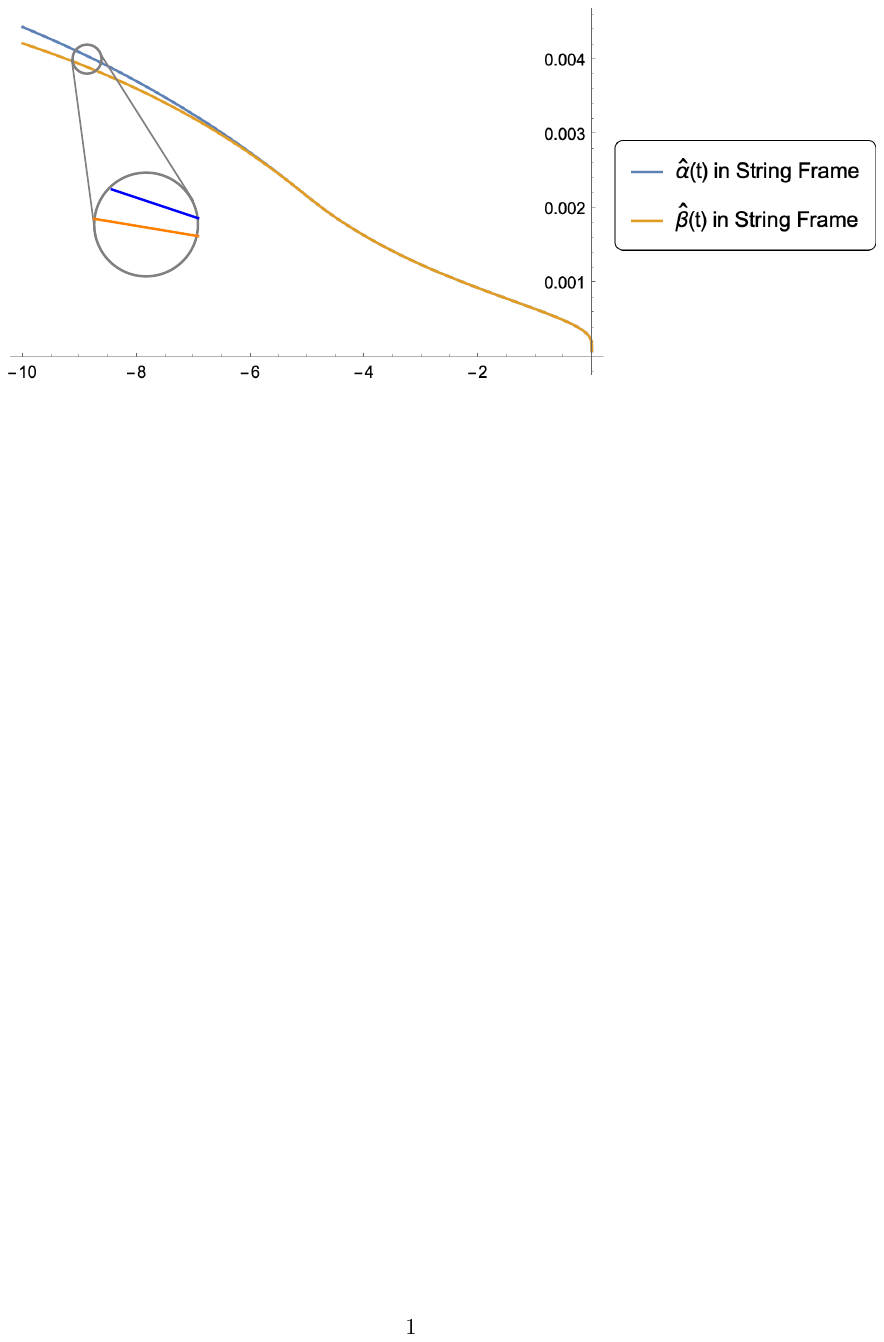}}%
        \vskip-6.5in
        \caption{Plots of $\hat\alpha(t)$ and $\hat\beta(t))$   in the string frame with $v = 1, \alpha_o = 0.3$,  $\beta_o = 0.015$, $\epsilon_1 = -5, \Lambda = {1\over 100}, c_o = 0.001$. With the choice of $\alpha_o, \beta_o$ in \eqref{TIhostsmey}, it is clear that $\hat\alpha(t) > \hat\beta(t)$ in the temporal domain $-10 < t < -5$, whereas they expectedly coincide in the domain $-5 \le t < 0$.}%
        \label{alphabetahats}%
    \end{figure}

\begin{figure}%
        \centering
        {\includegraphics[scale=0.80]{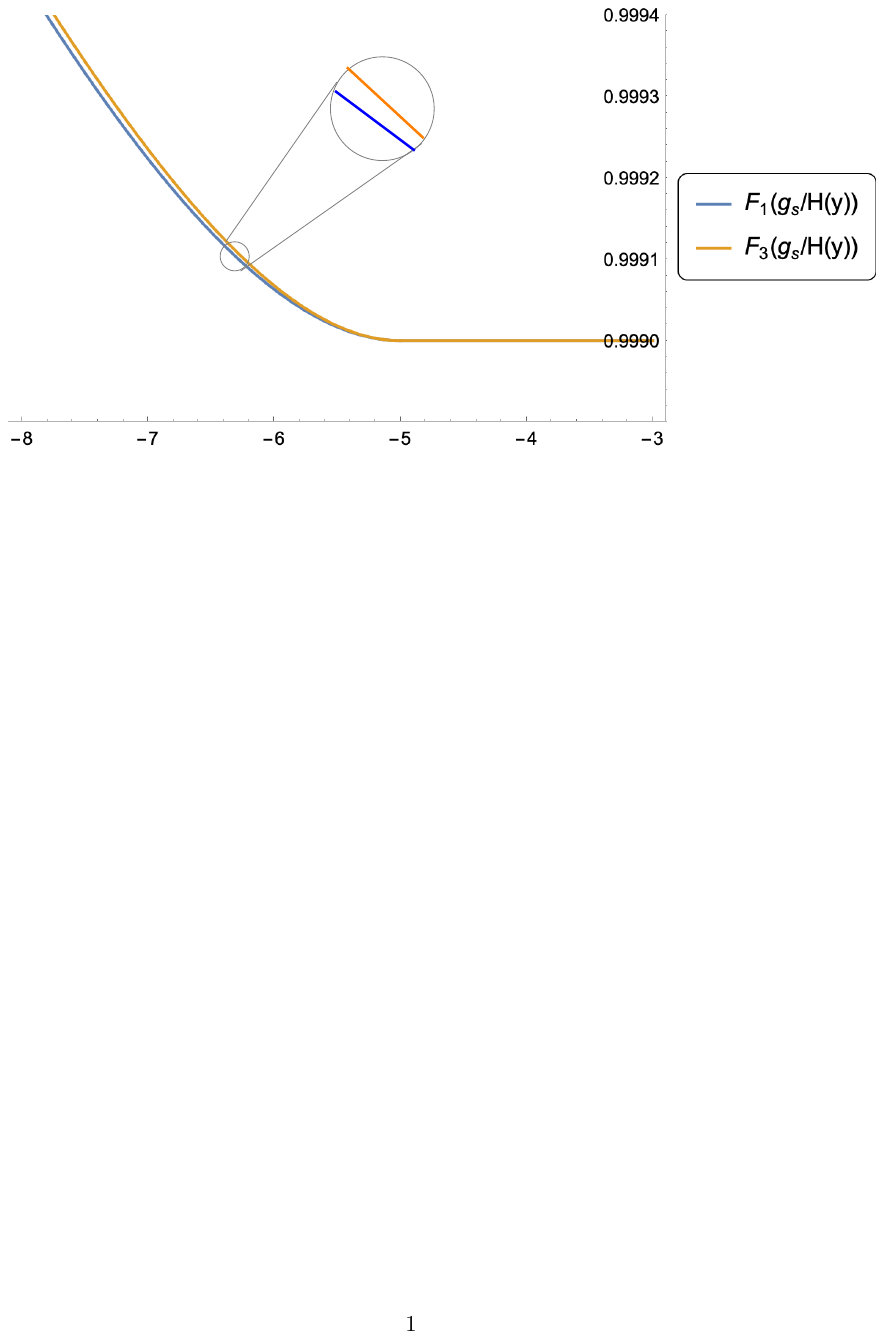}}%
        \vskip-6in
        \caption{Plots of ${\rm F}_1(t(g_s))$ and ${\rm F}_3(t(g_s))$ with ${\rm H}_o({\bf x}) \equiv 1$ in the string frame with the same choice of the parameters as above.  With the choice of $\alpha_o, \beta_o$ in \eqref{TIhostsmey}, it is clear that  ${\rm F}_3(t) > {\rm F}_1(t)$ in the temporal domain $-10 < t < -5$, whereas they expectedly coincide in the domain $-5 \le t < 0$.}%
        \label{f11f33}%
    \end{figure}

\begin{figure}%
        \centering
        {\includegraphics[scale=0.30]{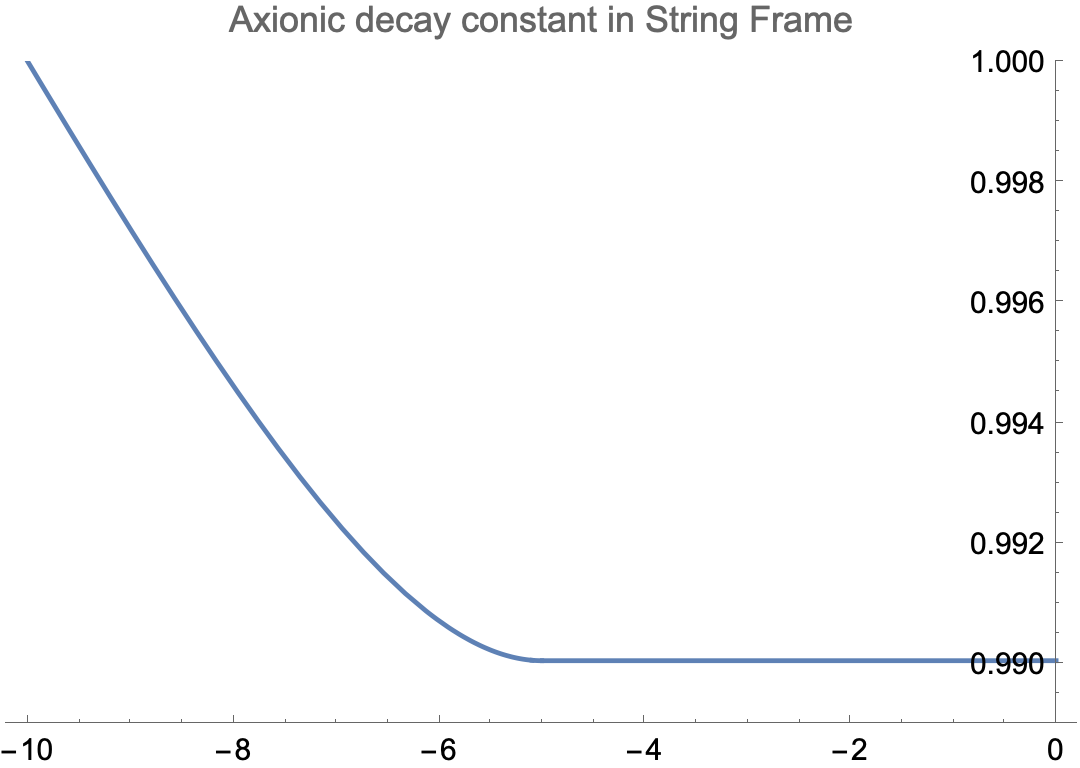}}%
        \qquad \qquad
        {\includegraphics[scale=0.30]{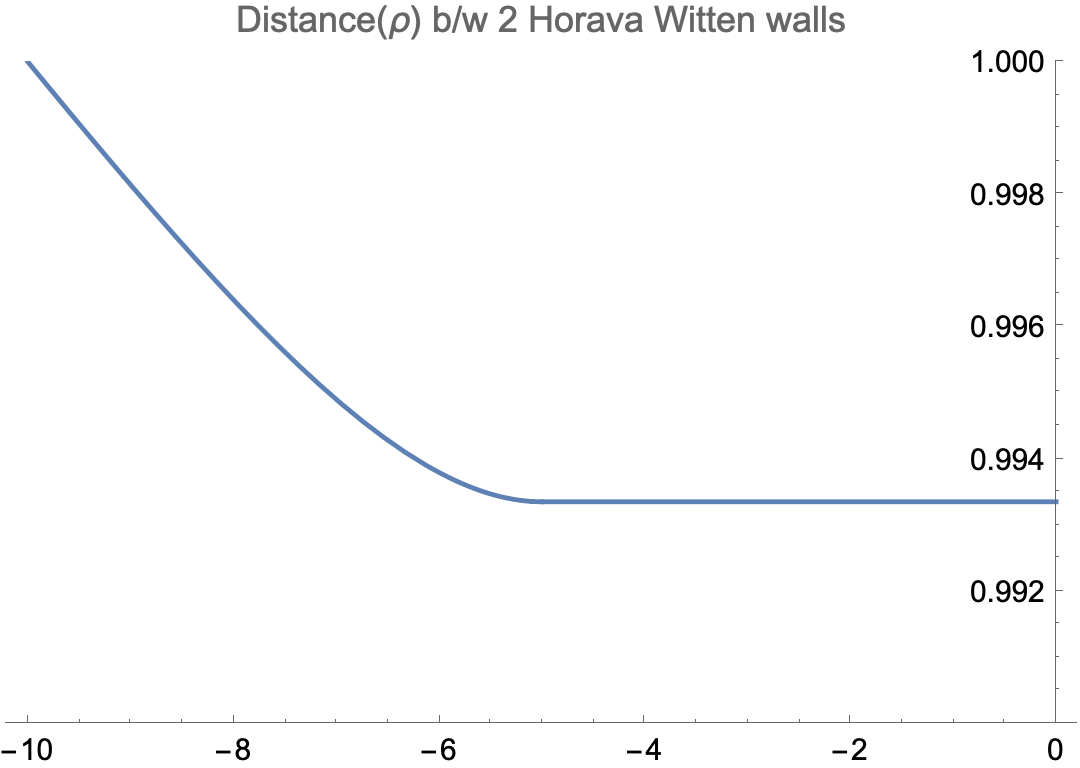}}%
        \qquad \qquad \qquad
        {\includegraphics[scale=0.30]{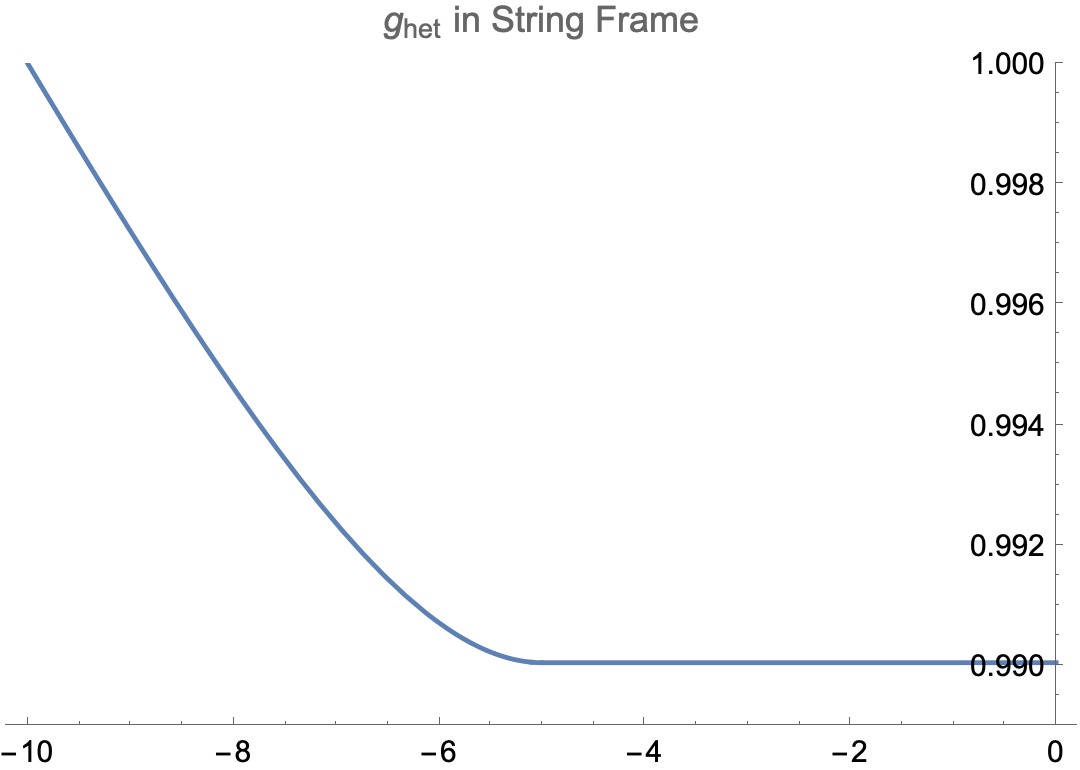}}%
        
        \caption{Plots of the axion decay constant $f_a(t)$, the distance $\rho(t)$ between the two Horava-Witten walls and the heterotic coupling $e^{\Phi}$ from \eqref{shamitagra} in the string frame for the following choice of the parameters: $\alpha_o = 0.3$,  $\beta_o = 0.15$, $\epsilon_1 = -5, \Lambda = {1\over 100}, c_o = 0.01, a_1 = 0.77143, a_2 = 0.835915, \sigma_1 = -9.81128$ and $\sigma_2 = -85.5891$. Note that at late time in the temporal domain $-5 < t < 0$ all the three plots show no temporal dependence as expected from the discussion presented here.}%
        \label{axdecay}%
    \end{figure}

The third constraint is interesting, as it tells us that the heterotic coupling should never be bigger than 1 in the temporal domain $-{1\over \sqrt{\Lambda}} < t < 0$. The warp-factor ${\rm H}(y)$ enters the heterotic coupling in both the frames as seen from \eqref{fountainbleau} and \eqref{fountainbleaugu}, so for $t \approx -{1\over \sqrt{\Lambda}}$, we expect ${\rm H}(y) < 1$ to keep the heterotic couplings small in both frames. At late time the temporal factors in \eqref{fountainbleau} and \eqref{fountainbleaugu} decrease significantly, so we can allow ${\rm H}(y) > 1$ keeping the couplings small. For the case with ${\rm H}(y) = 1$, it is known that $f_a \approx 10^{16}~{\rm GeV}$, a value too large to be consistent with experimental observations \cite{svrcek}. If we switch on a constant warp-factor, then we need to go to a slightly later time to keep $f_a(t) < 10^{12}~{\rm GeV}$ with ${\rm H}(y) = {\rm H} \approx 100$. However if we make ${\rm H}(y)$ a function of $y^m$ with $m \in {\cal M}_4 \times \mathbb{T}^2$, then the integrated function can make ${\rm M}_p$ small in \eqref{stratmey} with ${\rm H}(y) < 1$ everywhere in the internal space while still keeping the KK modes heavy enough to be integrated away in the energy scale of interest. Taking all the aforementioned points into account then tell us that $\hat\gamma_o$ in both \eqref{afleis44} and \eqref{afleis4} should develop time dependence
as:
\begin{figure}[h]
\centering
\begin{tabular}{c}
\includegraphics[width=3in]{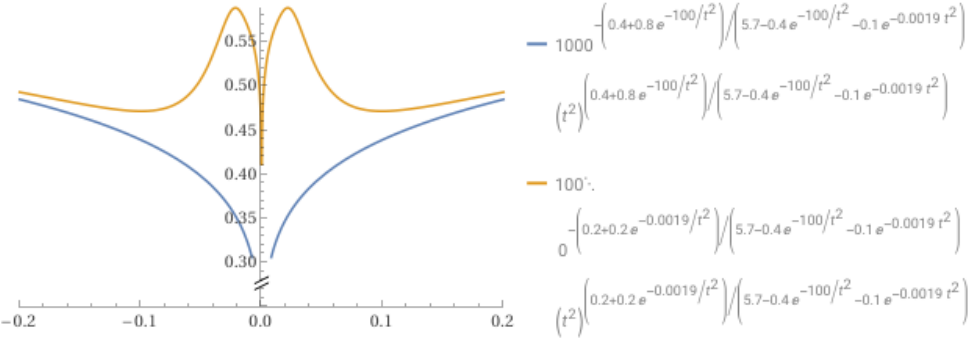}
\end{tabular}
\caption[]{At very late time, or when $t \to \pm 0$, \eqref{hanasmit} doesn't quite represent the behavior of ${\rm F}_1(t)$ and ${\rm F}_3(t)$ correctly. In fact axionic cosmology gives certain constraints at late time which would modify the behavior from the naive ans\"atze \eqref{hanasmit} as we shall discuss in section \ref{sec3.6}. A more accurate representation will be presented later in \eqref{horseman}.}
\label{f1f32}
\end{figure} 
\bg\label{kitagTI}
\hat\gamma_o(t) = {3v^2\over 3v-2}\left({c_o\over c_o + v\vert{\rm log}(\Lambda t^2)\vert}\right), ~~~~ -\epsilon_1 < t < 0\nd 
where $v = (1, 2)$ for the string and the Einstein frames respectively; and $c_o$ is a constant given by $c_o = {v\hat\alpha(-\epsilon_1) + 3v\hat\beta(-\epsilon_1)\over 6v^2 - (2v-1)\hat\alpha(-\epsilon_1) -\hat\beta(-\epsilon_1)} \big\vert {\rm log}(\sqrt{\Lambda} \vert\epsilon_1\vert)\big\vert$, where $(\hat\alpha(-\epsilon_1), \hat\beta(-\epsilon_1))$ are the values appearing in the definitions of \eqref{afleis44} and \eqref{afleis4} at $t = -\epsilon_1$. Note that we no longer need to specify the intermediate temporal domain $-\epsilon_1 < t < -\epsilon_2$ in \eqref{afleis4} and \eqref{afleis44}. One can also see that the late time behavior, in the temporal domain $-\epsilon_1 < t < 0$, is significantly modified from what appears in cases 1 and 2 in {\bf figure \ref{6cases}} and in {\bf figure \ref{gamma00}}. The smooth function $\hat\gamma_o(t)$ replaces the step functions in \eqref{afleis4} and \eqref{afleis44}. In other words, \eqref{afleis4} and \eqref{afleis44} get replaced by:
\bg\label{afleis45}
\hat\alpha(t) = \begin{cases} ~ \hat\alpha_o ~~~~~~ -{1\over \sqrt{\Lambda}} < t < -\epsilon_1 \\
~~~ \\
~ \hat\gamma_o(t) ~~ -\epsilon_1 < t < 0
\end{cases}, ~~
\hat\beta(t) = \begin{cases} ~ \hat\beta_o ~~~~~~~~~~~~~~~ -{1\over \sqrt{\Lambda}} < t < -\epsilon_1 \\
~~~ \\
~ (4v-3)\hat\gamma_o(t) ~ -\epsilon_1 < t < 0
\end{cases}
\nd 
for $v = (1, 2)$ in the string and the Einstein frames. While this is a significant modification from \eqref{afleis4} and \eqref{afleis44}, one would still need to find interpolating functions between $\hat\alpha_o$ and 
$\hat\gamma_o(t)$ as well as between $\hat\beta_o(t)$ and $(4v-3)\hat\gamma_o(t)$ that smoothly connects them. This is a much harder exercise because it will require us to work out the Schwinger-Dyson equations carefully to ascertain that such a choice remains consistent with the full dynamics of the system. We will discuss this approach later, but meanwhile let us try the following trial functions:
\bg\label{TIhostsmey}
&&\hat\alpha(t) = \alpha_o\Theta(-t - \vert\epsilon_1\vert) 
{{\rm exp}\left[{1\over \sqrt{\Lambda}}\left({1\over \sigma_1} - {a_1\over \vert t \vert}\right)\right] \over 
1 +{\rm exp}\left[{1\over \sqrt{\Lambda}}\left({1\over \sigma_1} - {a_1\over \vert t \vert}\right)\right]} + 
\Theta(t + \vert\epsilon_1\vert) {\alpha_v c_o \over  c_o + v \vert {\rm log}(\Lambda t^2)\vert} \nonumber\\
&& \hat\beta(t) = \beta_o\Theta(-t - \vert\epsilon_1\vert) 
{{\rm exp}\left[{1\over \sqrt{\Lambda}}\left({1\over \sigma_2} - {a_2\over \vert t \vert}\right)\right] \over 
1 +{\rm exp}\left[{1\over \sqrt{\Lambda}}\left({1\over \sigma_2} - {a_2\over \vert t \vert}\right)\right]} + 
\Theta(t + \vert\epsilon_1\vert) {\alpha_v (4v-3) c_o \over  c_o + v \vert {\rm log}(\Lambda t^2)\vert}, \nd
where $\alpha_o$ and $\beta_o$ are two constants that will be related to 
$\hat\alpha_o$ and $\hat\beta_o$ below, $v = (1, 2)$ respectively for the string and the Einstein frames, $\Theta(x)$ is the Heaviside function defined in the standard way, $\alpha_v = {3v^2\over 3v -2}$, $c_o$ is defined just below \eqref{kitagTI}, and $\sigma_i$ and $a_i$ are additional parameters that may be fixed by demanding the continuity of the two functions as well as the continuity of the derivatives of the two functions near $t = -\epsilon_1$. Imposing them gives us the following relations:
\begin{figure}%
        \centering
        {\includegraphics[scale=0.45]{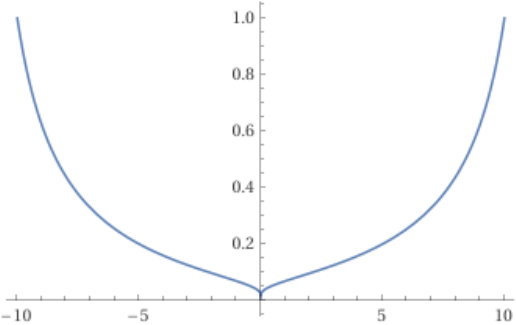}}%
        \qquad \qquad
        {\includegraphics[scale=0.45]{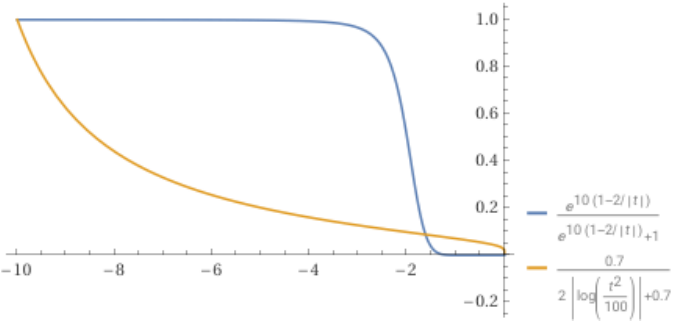}}%
        \caption{The behavior of $\hat\gamma_0(t)$ from \eqref{kitagTI} for $c_0 = 0.7, \Lambda = {1\over 100}$ and $\epsilon_1 = -5$ in the Einstein frame. In the second figure we show that the last stages between the temporal domain $-\epsilon_1 < t < 0$ in {\bf figure \ref{6cases}} from cases 1 and 2 (shown in blue here) are replaced by a smooth function (shown in orange here). The latter appears from the constraints imposed by axionic cosmology.}%
        \label{gamma00}%
    \end{figure}
\bg\label{sirimey}
&&c_o = {v\hat\alpha(-\epsilon_1)+ 3v\hat\beta(-\epsilon_1)\over 6v^2 - (2v-1)\hat\alpha(-\epsilon_1)-\hat\beta(-\epsilon_1)} \big\vert {\rm log}(\sqrt{\Lambda} \vert\epsilon_1\vert)\big\vert\\
&& {1\over \sigma_1} = {a_1\over \vert\epsilon_1\vert} - 
\sqrt{\Lambda} ~{\rm log}\left[{\alpha_o\over \alpha_v}\left(1 + {v\over c_o}\left\vert{\rm log}(\Lambda \epsilon_1^2)\right\vert \right) - 1\right]\nonumber\\
&& {1\over \sigma_2} = {a_2\over \vert\epsilon_1\vert} - 
\sqrt{\Lambda} ~{\rm log}\left[{\beta_o\over \alpha_v(4v-3)}\left(1 + {v\over c_o}\left\vert{\rm log}(\Lambda \epsilon_1^2)\right\vert \right) - 1\right]\nonumber\\
&& \hat\alpha_o =  {\alpha_o~{\rm exp}\left[{1\over \sqrt{\Lambda}}\left({1\over \sigma_1} - {a_1\sqrt{\Lambda}}\right)\right] \over 
1 +{\rm exp}\left[{1\over \sqrt{\Lambda}}\left({1\over \sigma_1} - {a_1\sqrt{\Lambda}}\right)\right]}, ~~
\hat\beta_o =  {\beta_o~{\rm exp}\left[{1\over \sqrt{\Lambda}}\left({1\over \sigma_2} - {a_2\sqrt{\Lambda}}\right)\right] \over 
1 +{\rm exp}\left[{1\over \sqrt{\Lambda}}\left({1\over \sigma_2} - {a_2\sqrt{\Lambda}}\right)\right]}\nonumber\\
&& a_1 = {2\alpha_o v \vert\epsilon_1\vert \sqrt{\Lambda} \over \alpha_o\left(c_o + v\vert{\rm log}(\Lambda \epsilon_1^2)\vert\right) - \alpha_v c_o},~~a_2 = {2\beta_o v \vert\epsilon_1\vert \sqrt{\Lambda} \over \beta_o\left(c_o + v\vert{\rm log}(\Lambda \epsilon_1^2)\vert\right) - (4v-3) \alpha_v c_o} \nonumber \nd
where, without loss of generalities, we can absorb $\alpha_v \equiv {3v^2\over 3v -2}$ in the definition for $\alpha_o$ and $\beta_o$ (still keeping $\hat\alpha_o > \hat\beta_o$). Note that $(\alpha_o, \beta_o)$ could be greater than 1 as long as ${\hat\alpha_o\over 9} < \hat\beta_o <  {\hat\alpha_o} < 1$. In this way the trial functions in \eqref{TIhostsmey} would be able to reproduce \eqref{afleis4} and \eqref{afleis44} as smooth functions interpolating between the values therein. The plots of the behavior of 
$(\hat\alpha(t), \hat\beta(t))$ and $({\rm F}_1(t), {\rm F}_3(t))$ in the string frame are shown in {\bf figure \ref{alphabetahats}} and {\bf figure \ref{f11f33}} respectively.  The choice of $\epsilon_1$ is further constrained due to our choice of the functions in \eqref{afleis45}. It is clear that:
\bg\label{linqplaza}
\big\vert{\rm log}(\Lambda \epsilon_1^2)\big\vert > {c_o\over v}\left({\alpha_v(4v-3)\over \beta_o} - 1\right), ~~~~ {\rm or}~~~
\big\vert{\rm log}(\Lambda \epsilon_1^2)\big\vert > {c_o\over v}\left({\alpha_v\over \alpha_o} - 1\right), \nd
whichever is bigger. Clearly in the string frame, it is the former that imposes the constraint because $\alpha_o > \beta_o$ (consequently making $\hat\alpha_o > \hat\beta_o$). In the Einstein frame, if ${\beta_o\over 5} < \alpha_o$, it is again the former that constrains the choice of $\epsilon_1$. Otherwise it is the latter. In {\bf figure \ref{axdecay}}, we plot the behaviors of the axion decay constant, the distance between the Horava-Witten walls and the heterotic dilaton in the string frame. The expected constancy with respect to time should be clear from the plots.

\begin{figure}%
        \centering
        {\includegraphics[scale=0.80]{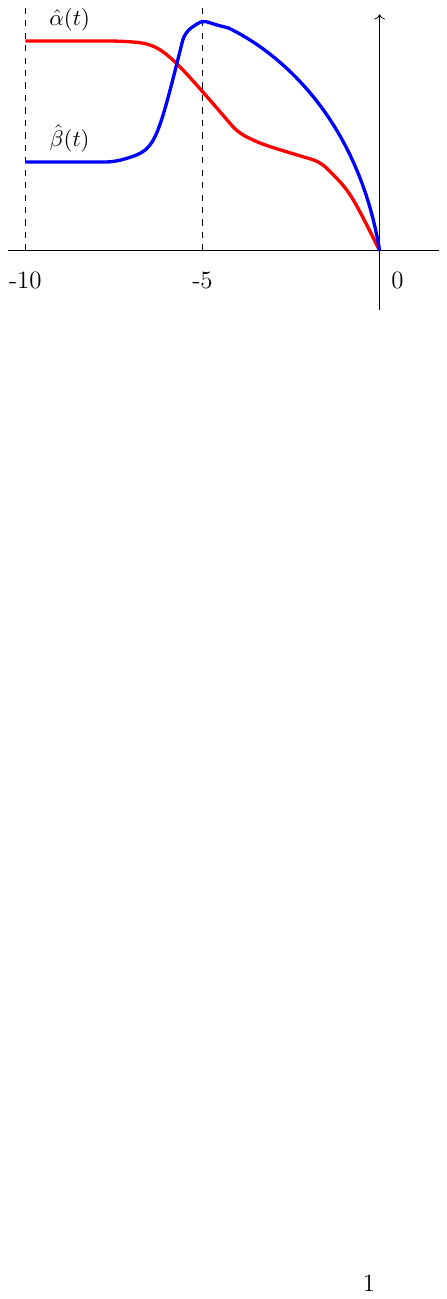}}%
        \vskip-6.2in
        \caption{Plots of $\hat\alpha(t)$ and $\hat\beta(t)$ in the Einstein frame. This is the expected behavior but getting it from a functional choice like \eqref{TIhostsmey} is more delicate. Despite this many of the features can be easily explained. For example, within the domain $-10 < t < -6$, we expect $\hat\alpha(t) > \hat\beta(t)$. At $t \equiv \epsilon_a = -6$, we expect $\hat\alpha(t) = \hat\beta(t)$. Finally in the temporal domain $-5 < t < 0$, we expect $\hat\beta(t) = 5 \hat\alpha(t)$. The last phase is clearly controlled by the function $\hat\gamma_o(t)$ from \eqref{kitagTI}. (Note that the figure is not drawn to scale unlike {\bf figures \ref{alphabetahats}}, {\bf \ref{f11f33}} and {\bf \ref{axdecay}}.)}%
        \label{abeinstein}%
    \end{figure}

The story in the Einstein frame is slightly more involved with the choice of the step function \eqref{TIhostsmey}. The reason is easy to see. We need $\hat\alpha(t) > \hat\beta(t)$ for $-{1\over\sqrt{\Lambda}} < t < -\epsilon_a$, then at $t = \epsilon_a$, $\hat\alpha(t) = \hat\beta(t)$, and finally for $-\epsilon_1 < t < 0$, $\hat\beta(t) = 5 \hat\alpha(t)$. The tentative behavior is plotted in {\bf figure \ref{abeinstein}}. In the language of ${\rm F}_1(t)$ and ${\rm F}_3(t)$, the constraints on them are as follows: (a) ${\rm F}_3(t) > {\rm F}_1(t)$ for $-{1\over\sqrt{\Lambda}} < t < -\epsilon_a$, (b) ${\rm F}_1(t) = {\rm F}_3(t)$ at $t = - \epsilon_a$ and (c) ${\rm F}_3(t) = {\rm F}_1^5$ for $-\epsilon_1 < t < 0$, as shown in {\bf Table \ref{milleren4}}. On the other hand, if we want to keep $\hat\alpha(t) > \hat\beta(t)$, or alternatively, $\hat\alpha_o > \hat\beta_o$ in the temporal domain $-{1\over \sqrt{\Lambda}} < t < \epsilon_a$, the condition becomes:
\bg\label{mehagu}
{{\cal A}_1({\cal A}_2 - 1)\over \beta_o} - {{\cal A}_2({\cal A}_1 - 1)\over \alpha_o} ~ > ~ {(4v-3){\cal A}_2 - {\cal A}_1\over (4v-3)\alpha_v}\left(1 + {v\over c_o}\vert\log(\Lambda \epsilon^2_1)\vert\right), \nd
where ${\cal A}_i = {\rm exp}\left[a_i\left({1\over \sqrt{\Lambda} \vert\epsilon_1\vert} -1\right)\right]$, with the parameters defined in \eqref{sirimey}. For $v = 1$, the results are shown in {\bf figures \ref{alphabetahats}} and {\bf \ref{f11f33}}. For $v = 2$, it will be easier to take specific parameters to see if \eqref{mehagu} could be implemented. For example, we can take $\Lambda = {1\over 100}$ and $\epsilon_1 = -5$. Also, taking $c_o$ as a free parameter fixes $(a_i, \sigma)$ from \eqref{sirimey} in the following way:
\bg\label{vendeta}
&& {1\over a_1} = c_o + 2\log~4 - {2c_o\over \alpha_o} \approx 2\log~4, ~~~
{1\over a_2} = c_o + 2\log~4 - {2c_o\over \beta_o} \approx 2\log~4 \\
&& {\rm exp}\left({1\over \sqrt{\Lambda}\sigma_1}\right) = {3~{\rm exp}(2a_1)\over \alpha_o\left(1 + {2\log~4\over c_o}\right) - 3} \approx 
{3~{\rm exp}\left({1\over \log~4}\right)\over \alpha_o\left(1 + {2\log~4\over c_o}\right) - 3} \le {3c_o~{\rm exp}\left({1\over \log~4}\right) \over 2\alpha_o \log~4 - 3c_o}\nonumber\\
&& {\rm exp}\left({1\over \sqrt{\Lambda}\sigma_2}\right) = {15~{\rm exp}(2a_2)\over \beta_o\left(1 + {2\log~4\over c_o}\right) - 15} \approx 
{15~{\rm exp}\left({1\over \log~4}\right)\over \beta_o\left(1 + {2\log~4\over c_o}\right) - 15}\le {15c_o~{\rm exp}\left({1\over \log~4}\right) \over 2\beta_o \log~4 - 15c_o}, \nonumber \nd
where we have taken $c_o << (\alpha_o, \beta_o, 2\log~4)$ so that \eqref{linqplaza} remains satisfied. Plugging \eqref{vendeta} into \eqref{mehagu} then provides the following bound on a certain combination of $\alpha_o$ and $\beta_o$:
\bg\label{1461lya}
{1\over \beta_o} - {1\over \alpha_o} ~ > ~ {8\log~4\over 15 c_o\left(e^{1\over 2\log~4} - 1\right)} \approx {1.70243 \over c_o}, \nd
which would keep $\hat\alpha_o > \hat\beta_o$ in the temporal domain 
$-10 < t < -\epsilon_a$. On the other hand, we should also keep in mind that $\hat\alpha_o$ and $\hat\beta_o$ are bounded as ${\hat\alpha_o \over 9} < \hat\beta_o < \hat\alpha_o < 1$. Unfortunately with all these constraints, the  condition \eqref{1461lya} {\it cannot} be satisfied. However this does not immediately rule out \eqref{mehagu} for $v = 2$ because we are not obligated to impose $c_o << (\alpha_o, \beta_o, 2\log~4)$, but finding a range of parameters satisfying \eqref{1461lya} as well as the bounds on $\hat\alpha_o$ and $\hat\beta_o$ is a delicate affair in the Einstein frame. (The string frame behavior works out well as we saw earlier.) Moreover we have to make sure that, in the temporal domain $-\epsilon_1 < t < 0$ for $\vert\epsilon_1\vert = 5$, $\hat\alpha(t)$ and $\hat\beta(t)$ decay as $\hat\gamma_o(t)$ and 
$5\hat\gamma_o(t)$ respectively with $\hat\gamma_o(t)$ defined as in \eqref{kitagTI}. 

Our analysis above has revealed that, imposing constraints from axionic cosmology, the late time behavior of $\hat\alpha(t)$ and $\hat\beta(t)$ in \eqref{afleis45} is severely restricted by the function $\hat\gamma_o(t)$ 
from \eqref{kitagTI}. The early time behavior in the temporal domain $-{1\over \sqrt{\Lambda}} < t \le -\epsilon_1$, however remains to be fixed appropriately because, as we saw for the $v = 2$ case, a function like \eqref{TIhostsmey} doesn't suffice. This therefore raises multiple questions. For example, is there a way to justify that the choices in \eqref{afleis} and \eqref{afleis45} respectively for the two heterotic theories solve the EOMs with smooth functions? Also,
how do we know that the conditions \eqref{bosema} and \eqref{cardavs} for the $SO(32)$ and the ${\rm E}_8 \times {\rm E}_8$ theories respectively, are the {\it only} conditions for EFT? To answer them we will have to dig deeper into the quantum effects to see how they scale when the parameters are themselves time-dependent. For example in section \ref{sec4.2.3} we will find smooth functions that provide the behavior of the warp-factors, for example  \eqref{ryanfan} for $\beta(t)$ in the $SO(32)$ theory and \eqref{horseman} for $(\hat\alpha(t), \hat\beta(t))$ in the ${\rm E}_8 \times {\rm E}_8$ theory. However to arrive at these results require us to carefully study the quantum behavior of the system. This is what we turn to next.

\section{Detailed analysis of the quantum terms and EFT from M-theory \label{lillou1}}

Consistency of the Glauber-Sudarshan states comes not only from the path-integral analysis that we performed in \cite{borel2, joydeep} but also from the Schwinger-Dyson equations discussed in \cite{coherbeta} and \cite{coherbeta2}. Both these steps are essential, and in fact the correct procedure would be to first justify the existence of the Glauber-Sudarshan states using Schwinger-Dyson's equations and then reproduce the metric configuration using the path-integral formulation as in \cite{borel2}. The typical Schwinger-Dyson's equations for the Glauber-Sudarshan states may be expressed as:
\bg\label{mariesole}
\left\langle {\delta\left(\hat{\bf S}_{\rm tot} - \hat{\bf S}_{\rm ghost}\right)\over \delta {\bf g}^{\rm AB}}\right\rangle_\sigma &= &
 {\delta\hat{\bf S}_{\rm tot}(\langle{\bf \Xi}\rangle_\sigma)\over \delta \left\langle{\bf g}^{\rm AB}\right\rangle_\sigma}   +\sum_{{\bf Q}, \sigma'}  \left\langle {\delta\left(\hat{\bf S}_{\rm tot} - \hat{\bf S}_{\rm ghost}\right)\over \delta {\bf g}^{\rm AB}}\right\rangle_{(\sigma\vert {\bf Q}, \sigma')} \nonumber\\ 
 &=& 
-\left\langle {\delta\hat{\bf S}_{\rm ghost}\over \delta {\bf g}^{\rm AB}}\right\rangle_\sigma - \left\langle {\delta\over \delta {\bf g}^{\rm AB}}~{\rm log}\left(\mathbb{D}^\dagger(\sigma)\mathbb{D}(\sigma)\right)\right\rangle_\sigma\nonumber\\
\left\langle {\delta\left(\hat{\bf S}_{\rm tot} - \hat{\bf S}_{\rm ghost}\right)\over \delta {\bf C}^{\rm ABD}}\right\rangle_\sigma & = & 
{\delta\hat{\bf S}_{\rm tot}(\langle{\bf \Xi}\rangle_\sigma)\over \delta \left\langle{\bf C}^{\rm ABD}\right\rangle_\sigma} +\sum_{{\bf Q}, \sigma'}  \left\langle {\delta\left(\hat{\bf S}_{\rm tot} - \hat{\bf S}_{\rm ghost}\right)\over \delta {\bf C}^{\rm ABD}}\right\rangle_{(\sigma\vert {\bf Q}, \sigma')} \nonumber\\ 
 &=& 
-\left\langle {\delta\hat{\bf S}_{\rm ghost}\over \delta {\bf C}^{\rm ABD}}\right\rangle_\sigma - \left\langle {\delta\over \delta {\bf C}^{\rm ABD}}~{\rm log}\left(\mathbb{D}^\dagger(\sigma)\mathbb{D}(\sigma)\right)\right\rangle_\sigma, \nd
where $\langle{\bf \Xi}\rangle_\sigma =\left(\langle {\bf g}_{\rm AB}\rangle_\sigma, \langle {\bf C}_{\rm ABD}\rangle_\sigma, \langle {\bf \Psi}_{\rm A}\rangle_\sigma, \langle \overline{\bf \Psi}_{\rm A}\rangle_\sigma\right)$ is the collection of the on-shell fields;    $\hat{\bf S}_{\rm tot}$ is the total action that differs from ${\bf S}_{\rm tot}$ that appeared earlier in \eqref{ebrowning} (including the topological terms\footnote{Wherein we had assumed the off-shell states to be massive. As discussed briefly in \cite{joydeep}, and to be elaborated here later in section \ref{sec6.1}, once we include the massless off-shell states and sum over the Gevrey growths appropriately the non-perturbative and non-local parts appear naturally. For the time being, with massive off-shell states, ${\bf S}_{\rm tot}({\bf \Xi, \Upsilon})= {\bf S}_{\rm kin}({\bf \Xi})+ {\bf S}_{\rm pert}({\bf \Xi})+ {\bf S}_{\rm ghost}({\bf \Xi, \Upsilon})$ and $\hat{\bf S}_{\rm tot}({\bf \Xi, \Upsilon})={\bf S}_{\rm kin}({\bf \Xi})+ {\bf S}_{\rm NP}({\bf \Xi})+ \hat{\bf S}_{\rm ghost}({\bf \Xi, \Upsilon})$ with no non-local contributions. As mentioned above, this will be modified to include the non-local interactions in section \ref{sec6}. See also \cite{joydeep}. \label{loicry}}) with $\vert\sigma\rangle = \vert\alpha,\beta,\gamma\rangle$ as defined in \eqref{ebrowning}, and $({\bf Q}, \vert\sigma'\rangle)$ are intermediate operators and states respectively (for details the readers may see section 6.1 in \cite{joydeep}). The action in \eqref{ebrowning} was simpler because we restricted ourselves to only scalar degrees of freedom with perturbative interactions, but now we will have to consider a more elaborate action that involves actual metric and three-form flux components as well as the Rarita-Schwinger fermions with full non-perturbative, non-local and other contributions (to be discussed soon). The action that appears from these considerations is $\check{\bf S}_{\rm tot}$ {\it and not exactly} $\hat{\bf S}_{\rm tot}$. The difference between $\check{\bf S}_{\rm tot}$ and $\hat{\bf S}_{\rm tot}$ is much more non-trivial and is discussed recently in \cite{wdwpaper}. This may be represented in the following way:

{\footnotesize
\bg\label{monfluent}
&& {\bf S}_{\rm tot}({\bf \Xi, \Upsilon})~ \xrightarrow{\rm resum}~ \hat{\bf S}_{\rm tot}({\bf \Xi, \Upsilon}) \longrightarrow \hat{\bf S}_{\rm tot}(\langle{\bf \Xi}\rangle_\sigma, \langle{\bf \Upsilon}\rangle_\sigma) \nonumber\\  
&& \xrightarrow{\rm transf}~ \hat{\bf S}_{\rm tot}(\langle{\bf \Xi}\rangle_\sigma, \langle{\bf \Upsilon}\rangle_\sigma) - \hat{\bf S}^{(g)}(\langle{\bf \Xi}\rangle_\sigma, \langle{\bf \Upsilon}\rangle_\sigma) -\hat{\bf S}^{(g)}_{\rm nloc}(\langle{\bf \Xi}\rangle_\sigma, \langle{\bf \Upsilon}\rangle_\sigma)~\xrightarrow{\rm renorm}~ \check{\bf S}_{\rm tot}(\langle{\bf \Xi}\rangle_\sigma), \nd}
where notice that the final action is $\check{\bf S}_{\rm tot}(\langle{\bf \Xi}\rangle_\sigma)$ and not exactly $\hat{\bf S}_{\rm tot}(\langle{\bf \Xi}\rangle_\sigma)$. The former is got from the renormalization procedure that we outlined in \cite{wdwpaper} which appears from the Wheeler-De Witt wave-function. This is an {\it envelope} wave-function covering all the on- and off-shell Glauber-Sudarshan states and if the wave-function {\it sharply} peaks over
a specific on-shell configuration (leading to the four-dimensional de Sitter or quasi de Sitter space-time), then we can replace $\check{\bf S}_{\rm tot}(\langle{\bf \Xi}\rangle_\sigma)$ by $\hat{\bf S}_{\rm tot}(\langle{\bf \Xi}\rangle_\sigma)$. Assuming this to be the case\footnote{Henceforth, unless mentioned otherwise, we will only consider this case in the paper. In fact, even taking a dynamical dark energy of the form \eqref{marapaug}, we will assume that the wave-functional sharply peaks over such a configuration.} \cite{wdwpaper}, the Schwinger-Dyson equations take the form \eqref{mariesole} above, where 
$\hat{\bf S}_{\rm ghost}({\bf \Xi, \Upsilon}) = \hat{\bf S}^{(g)}({\bf \Xi, \Upsilon}) + \hat{\bf S}^{(g)}_{\rm nloc}({\bf \Xi, \Upsilon})$. It is also understood that, in the aforementioned case\footnote{Alternatively, $\hat{\bf S}_{\rm tot}(\langle {\bf \Xi}\rangle_\sigma, 0) \equiv\hat{\bf S}_{\rm tot}(\langle {\bf \Xi}\rangle_\sigma)$ because $\hat{\bf S}_{\rm ghost}(\langle{\bf \Xi}\rangle_\sigma, 0) = 0$. \label{desclaud}}:

{\footnotesize
\bg\label{vkrieps}
\hat{\bf S}_{\rm tot}(\langle {\bf \Xi}\rangle_\sigma) \equiv  \hat{\bf S}_{\rm tot}(\langle {\bf \Xi}\rangle_\sigma, \langle {\bf \Upsilon}\rangle_\sigma) - \hat{\bf S}_{\rm ghost}(\langle{\bf \Xi}\rangle_\sigma, \langle{\bf \Upsilon}\rangle_\sigma) =
\hat{\bf S}_{\rm tot}\left(\langle {\bf g}_{\rm AB}\rangle_\sigma, \langle {\bf C}_{\rm ABD}\rangle_\sigma, \langle {\bf \Psi}_{\rm A}\rangle_\sigma, \langle \overline{\bf \Psi}_{\rm A}\rangle_\sigma\right), \nd} 
with the set of {on-shell fields} $\langle {\bf \Xi}\rangle_\sigma \equiv \big\{\langle {\bf g}_{\rm AB}\rangle_\sigma, \langle {\bf C}_{\rm ABD}\rangle_\sigma, \langle {\bf \Psi}_{\rm A}\rangle_\sigma,$ $\langle \overline{\bf \Psi}_{\rm A}\rangle_\sigma\big\}$. ({\Su Previously the bold-faced letters were mostly reserved for operators and light-faced ones for functions. We will henceforth avoid all these nuances and stick to only bold-faced notations to avoid unnecessary clutter. It should be clear from the context which is which.}) The detailed derivation of this appears in section 3.3 of \cite{coherbeta}, and in section \ref{sec6.1} of \cite{joydeep}. To avoid any confusion at this stage, let us repeat that the action \eqref{vkrieps} is expressed completely in terms of the {\it on-shell} degrees of freedom, which means that it will have non-local counter-terms that appear from integrating out massless off-shell degrees of freedom for components that do not appear in the {\it classical} EOMs. (These may in turn be expressed by exponentiating interaction terms involving on-shell degrees of freedom \cite{coherbeta} that are suppressed by ${\rm exp}\left(-{1\over g_s^p}\right)$ where values of $p$ are determined in section 3.2 of \cite{coherbeta}.) There would also be instanton corrections from wrapped five- and two-branes on six and three cycles respectively. These details have all been derived in \cite{desitter2, coherbeta, coherbeta2} and \cite{joydeep}, which the readers may look up. We will however revisit some of these computations and especially the ones involving the fermionic interactions, that were not addressed in details in \cite{coherbeta, joydeep}, soon.

The first terms on the RHS of the equalities in the first and the third lines of \eqref{mariesole} would resemble the background EOMs if we put them to zero. In fact this is what we did in \cite{coherbeta, coherbeta2, joydeep}, and the result we got therein was:
\bg\label{fulashut}
{\delta\hat{\bf S}_{\rm tot}(\langle{\bf \Xi}\rangle_\sigma) \over \delta \left\langle{\bf g}^{\rm AB}\right\rangle_\sigma}= 0 
 =  {\bf R}_{\rm AB}(\langle {\bf \Xi}\rangle_\sigma) - {1\over 2} \langle {\bf g}_{\rm AB}\rangle_\sigma {\bf R}(\langle {\bf \Xi}\rangle_\sigma) - {\bf T}_{\rm AB}(\langle {\bf \Xi}\rangle_\sigma), \nd
with the remaining equation from \eqref{mariesole} associated to ghosts (similar story is for the flux and the fermionic EOMs)\footnote{The expression ${\bf R}_{\rm AB}(\langle {\bf \Xi}\rangle_\sigma)$ works for a more generalized description of the curvature (that we shall study soon). For the present case of \eqref{fulashut}, one may restrict $\langle {\bf \Xi}\rangle_\sigma$ to $\langle {\bf g}_{\rm AB}\rangle_\sigma$.}. ${\bf T}_{\rm AB}$ denotes the full energy-momentum tensor, meaning that it contains perturbative, non-perturbative, non-local and other possible contributions (see section 6 in \cite{joydeep} for a simpler derivation). The above EOM is interesting because (a) it is precisely the EOM for the metric degree of freedom $\langle {\bf g}_{\rm AB}\rangle_\sigma$, and (b) it cleanly decouples the ghost EOMs from the other ones. In retrospect this was expected because if the Glauber-Sudarshan states were to reproduce {\it classical} physics they have to satisfy some EOMs. Note that the Glauber-Sudarshan states by themselves are definitely {\it not} classical, but they conspire to reproduce the ``classical" physics using purely quantum effects $-$ with the main difference being, not as backgrounds but as excited states over a supersymmetric Minkowski vacuum.

In fact this is all there is to it if we consider (a) the EOM \eqref{fulashut} and (b) the nodal diagram ways to construct $\langle {\bf g}_{\rm AB}\rangle_\sigma$ (with equivalent ones for the flux and the fermionic degrees of freedom), but we avoided a potential pitfall in the process. This appears if we consider ${\delta\left\langle\hat{\bf S}_{\rm tot} - \hat{\bf S}_{\rm ghost}\right\rangle_\sigma\over \delta \left\langle{\bf g}^{\rm AB}\right\rangle_\sigma}$ instead of ${\delta\hat{\bf S}_{\rm tot}(\langle{\bf \Xi}\rangle_\sigma)\over \delta \left\langle{\bf g}^{\rm AB}\right\rangle_\sigma}$. For the former 
there is a deeper level of subtlety that we did not elaborate in \cite{coherbeta, coherbeta2, joydeep}. The subtlety appears when we consider the following expression, again with massive off-shell states:

{\footnotesize
\bg\label{bdreamred}
\left\langle {\bf S}_{\rm tot} - {\bf S}_{\rm ghost}\right\rangle_\sigma = 
{\int [{\cal D}{\bf g}_{\rm AB}] [{\cal D}{\bf C}_{\rm ABD}] [{\cal D}{\bf \Psi}_{\rm A}] [{\cal D}\overline{\bf \Psi}_{\rm A}][{\cal D}{\bf \Upsilon}_{g}]~{\rm exp}\left(-{\bf S}_{\rm tot}\right)~\left({\bf S}_{\rm tot} - {\bf S}_{\rm ghost}\right)\mathbb{D}^\dagger(\sigma) \mathbb{D}(\sigma) \over 
{\int [{\cal D} {\bf g}_{\rm AB}] [{\cal D}{\bf C}_{\rm ABD}] [{\cal D}{\bf \Psi}_{\rm A}] [{\cal D}\overline{\bf \Psi}_{\rm A}][{\cal D}{\bf \Upsilon}_{g}]~{\rm exp}\left(-{\bf S}_{\rm tot}\right)~\mathbb{D}^\dagger(\sigma) \mathbb{D}(\sigma)}}, \nd}
expressed as a path-integral over the Glauber-Sudarshan state $\vert \sigma \rangle \equiv \mathbb{D}(\sigma)\vert\Omega\rangle$, where $\vert\Omega\rangle$ is the stable interacting vacuum, and ${\bf \Upsilon}_g$ is now a set of all possible ghosts: Faddeev-Popov, Batalin-Vilkovisky, non-local et cetera. In \eqref{bdreamred} since every piece of the action has to go through Borel resummation of certain Gevrey series (we took a linear source in \cite{borel2}, but here all higher orders terms have to taken as {\it sources}). We expect:
\bg\label{ckcigaret}
\left\langle{\bf S}_{\rm tot} - {\bf S}_{\rm ghost}\right\rangle_\sigma ~\xrightarrow{\rm resum}~ \langle\hat{\bf S}_{\rm tot} - \hat{\bf S}_{\rm ghost}\rangle_\sigma, \nd
with the hatted terms expressed as trans-series in a way described earlier and in \cite{joydeep}.
This clearly raises the question of the compatibility of the Wilson effective action (over a static Minkowski vacuum) with Borel resummations\footnote{A way to compare would be the following. Compute the two sides of \eqref{ckcigaret} at two different scales and compare the coefficients of each terms with the ones we get from \eqref{bdreamred}. Compatibility of Wilsonian effective action (assuming we are doing exact RG) at a low-energy scale would imply the similarity of the coefficients up to some overall scaling.}. The results in the literature are not conclusive enough to go one way or other (see however \cite{compatibility} and \cite{joydeep}), but we are fortunate that the so-called background EOMs are controlled by 
$\hat{\bf S}_{\rm tot}(\langle{\bf \Xi}\rangle_\sigma)$ and not by 
$\langle \hat{\bf S}_{\rm tot} - \hat{\bf S}_{\rm ghost}\rangle_\sigma$. This, as alluded to above, is completely expected.

We are not out of the water yet: there is a different level of subtlety that becomes important now, namely the temporal dependence of the exponents of $g_s$ for the warp-factors ${\rm F}_1(t)$ and ${\rm F}_2(t)$ for the $SO(32)$ theory and ${\rm F}_1(t), {\rm F}_3(t)$ and ${\rm F}_2(t)$ for the ${\rm E}_8 \times {\rm E}_8$ theory. We can then propose the following dominant scalings for the $SO(32)$ theory:
\bg\label{sezoe70}
&& {\partial\over \partial t}\left({g_s\over {\rm HH}_o}\right) \equiv \sum_{i = 1}^n\sum_{k_i = 0}^\infty {\rm C}_{k_i}\left({g_s\over {\rm HH}_o}\right)^{\gamma_i(t) + {2k_i\vert\gamma_{k_i}(t)\vert\over 3}}\nonumber\\
&& {\rm F}_1 \equiv \sum_{k = 0}^\infty {\rm A}_k\left({g_s\over {\rm HH}_o}\right)^{\beta(t) + {2k\vert\beta_k(t)\vert\over 3}}, ~~~~
{\rm F}_2 \equiv \sum_{k = 0}^\infty {\rm B}_k\left({g_s\over {\rm HH}_o}\right)^{\alpha(t) + {2k\vert\alpha_k(t)\vert\over 3}},\nd
where $({\rm A}_k, {\rm B}_k, {\rm C}_{k_i})$ are all integers, positive or negative with $(\alpha(t), \beta(t), \gamma_i(t))$ being the dominant scalings with $\beta(t)$ being given by \eqref{afleis}, $\alpha(t) = -\beta(t)$, but $\gamma_i(t)$ are non-trivial and so are $(\alpha_k(t), \beta_k(t), \gamma_{k_i}(t))$. (We will determine the value of $n$ soon, but a more detailed analysis using $(\alpha_k(t), \beta_k(t), \gamma_{k_i}(t))$ will have to wait till we use the Schwinger-Dyson's equations \eqref{mariesole}. For the time being we will proceed with $k = k_i = 0$.) 
(Note that we have expressed the $g_s$ scalings using ${g_s\over {\rm H}(y) {\rm H}_o({\bf x})}$, where ${\bf x} \in {\bf R}^2$, instead of just ${g_s\over {\rm H}(y)}$. The former choice is much more complete because it takes into account the various possible slicings of de Sitter space, instead of just the flat-slicing.) For the ${\rm E}_8 \times {\rm E}_8$ theory the ${\rm F}_i(t)$ parameters are distributed as:
\bg\label{sezoe71}
&& {\rm F}_1 \equiv \sum_{k = 0}^\infty {\rm A}_k\left({g_s\over {\rm HH}_o}\right)^{{2\hat\alpha(t)\over 3} + {2k\vert\hat\alpha_k(t)\vert\over 3}}, ~~
{\rm F}_3 \equiv \sum_{k = 0}^\infty {\rm B}_k\left({g_s\over {\rm HH}_o}\right)^{{2\hat\beta(t)\over 3} + {2k\vert\hat\beta_k(t)\vert\over 3}}\\
&& {\rm F}_2 \equiv \sum_{k = 0}^\infty {\rm C}_k\left({g_s\over {\rm HH}_o}\right)^{{2\hat\sigma(t)\over 3} + {2k\vert\hat\sigma_k(t)\vert\over 3}}, ~~
{\partial\over \partial t}\left({g_s\over {\rm HH}_o}\right) \equiv \sum_{i = 1}^n\sum_{k_i = 0}^\infty {\rm C}_{k_i}\left({g_s\over {\rm HH}_o}\right)^{\hat\gamma_i(t) + {2k_i\vert\hat\gamma_{k_i}(t)\vert\over 3}}, \nonumber \nd
where $\hat\alpha(t)$ and $\hat\beta(t)$ satisfy \eqref{afleis45} with 
$\hat\sigma(t) = -{1\over 4}(\hat\alpha(t) + 3\hat\beta(t))$, but $\hat\gamma_i(t)$ are again non-trivial, including the functions $(\hat\alpha_k(t), \hat\beta_k(t), \hat\sigma_k(t), \hat\gamma_{k_i}(t))$. As before, we will suffice with $k = k_i = 0$ in the following unless mentioned otherwise\footnote{We will come back to non-zero sub-dominant corrections with $(k, k_i) \ne 0$ in section \ref{sec4.4}.}, and use the complete form when we apply the Schwinger-Dyson's equations \eqref{mariesole}. The full behavior of $\hat\alpha(t)$ and $\hat\beta(t)$ follows the plots in {\bf figure \ref{6cases}}, or more correctly as in {\bf figure \ref{gamma00}} in the string frame. The string coupling for the $SO(32)$ and the ${\rm E}_8 \times {\rm E}_8$ theories respectively in the string frames are\footnote{A more careful study in section \ref{curvten} will show that there could be sub-leading corrections to \eqref{covda} coming from perturbative and non-perturbative terms as well as taking the dynamical dark energy \eqref{marapaug}. However, this will not change any of the main conclusions of this section.}:

{\footnotesize
\bg\label{covda}
{g_s\over {\rm H}(y) {\rm H}_o({\bf x})} \equiv \bar{g}_s[\beta(t)] = \left(\sqrt{\Lambda} \vert t\vert\right)^{{2\over 2 - \beta(t)}}, ~~~~~~ 
{g_s\over {\rm H}(y) {\rm H}_o({\bf x})} \equiv \bar{g}_s[\hat\alpha(t), \hat\beta(t)] = \left(\sqrt{\Lambda} \vert t\vert\right)^{{6\over 6 - \hat\alpha(t) - \hat\beta(t)}}, \nd}
where $(y^{m}, y^n) \in {\cal M}_4$, $(y^\alpha, y^\beta) \in {\cal M}_2$ and ${\bf x} = x^i \in {\bf R}^2$ (see footnote \ref{confusing} on notations. Compared to \eqref{giuraman1} and \eqref{viomyer1} we have now made non-trivial dependence on all coordinates except $(x^3, w^{11}) \in {\mathbb{T}^2\over {\cal G}}$.). There is however the following immediate problem: when $\beta(t) = \beta_0$, $\hat\alpha(t) = \hat\alpha_o$ and $\hat\beta(t) = \hat\beta_o$, we can easily invert the two expressions and express the conformal time $\vert t\vert$ as functions of $g_s$ for the two heterotic theories, but now, due to the complicated nature of the functions 
$\beta(t)$, $\hat\alpha(t)$ and $\hat\beta(t)$, we no longer have the privilege of doing this. So how should we proceed?

One way would be to simply take the derivative with respect to the conformal time $t$ of the two expressions in \eqref{covda}. Choosing the first one from \eqref{covda}, {\it i.e.} for the $SO(32)$ case, the derivative yields:
\bg\label{ishena}
{\partial\over \partial t}\left({g_s\over {\rm HH}_o}\right)
& = & \sum_{k = 1}^{2} e_k \left({g_s\over {\rm HH}_o}\right)^{\gamma_k} \equiv e_{1, 2}\left({g_s\over {\rm HH}_o}\right)^{\gamma_{1, 2}(t)}\nonumber\\
& = &  \left({g_s\over {\rm HH}_o}\right)^{1 - {\log(\vert\dot\beta\vert\vert\log~\bar g_s\vert)\over \vert\log~\bar g_s\vert} + {\log(2-\beta)\over \vert\log~\bar g_s\vert}} + 2\sqrt{\Lambda}\left({g_s\over {\rm HH}_o}\right)^{{\beta\over 2} + {\log(2-\beta)\over \vert\log~\bar g_s\vert}}, \nd
where all terms are kept dimensionless following footnote \ref{hisable}, $\bar g_s \equiv {g_s \over {\rm H}(y){\rm H}_o({\bf x})}$; and $e_1 = 1, e_2 = 2\sqrt{\Lambda}$. If we view the $g_s$ exponents to be simply functions of the conformal time $t$ (dimensionless with respect to ${\rm M}_p$), then $n = 2$ in \eqref{sezoe70}, and they could respectively be identified with the dominant scalings $\gamma_1(t)$ and $\gamma_2(t)$. Since $0<\beta(t) < {2/3}$, $\log(2-\beta) > 0$ and therefore $\gamma_2(t) > 0$. On the other hand, $\gamma_1(t) > 0$ if at least 
$2-\beta > \vert\dot\beta\vert\vert\log~g_s\vert$. As shown earlier in {\bf figures \ref{so327}} and {\bf \ref{so328}} with a smooth choice of 
$\beta(t)$, as in say {\bf figure \ref{so325}}, this condition is easily satisfied without constraining the system any further. For the second derivative of $g_s$, the result is a bit more complicated than \eqref{ishena} and is given by the following expression:
\bg\label{ishena2}
{\partial^2\over \partial t^2} \left({g_s\over {\rm HH}_o}\right) 
& = &
\sum_{k = 3}^{10} e_k \left({g_s\over {\rm HH}_o}\right)^{\gamma_k}
 \equiv  e_{[3,.., 10]}\left({g_s\over {\rm HH}_o}\right)^{\gamma_{[3,..,10]}}\\
& = & 2\sqrt{\Lambda} a_\Lambda\left({g_s\over {\rm HH}_o}\right)^{{\beta\over 2} - 1 + 
\gamma_{1, 2} + {\log(2-\beta)\over \vert\log~\bar g_s\vert}} - 2\Lambda 
\left({g_s\over {\rm HH}_o}\right)^{\beta - 1 + {\log(2-\beta)\over \vert\log~\bar g_s\vert}}\nonumber\\
& \pm & \left({g_s\over {\rm HH}_o}\right)^{1 -{\log(\mp\ddot{\beta}\vert\log~\bar g_s\vert)\over \vert\log~\bar g_s\vert} + {\log(2-\beta)\over \vert\log~\bar g_s\vert}} + 2\left({g_s\over {\rm HH}_o}\right)^{1 - {\log({\dot\beta}^2\vert\log~\bar g_s\vert)\over \vert\log~\bar g_s\vert} + {2\log(2-\beta)\over \vert\log~\bar g_s\vert}}\nonumber\\
& + & a_\Lambda\left({g_s\over {\rm HH}_o}\right)^{\gamma_{1, 2} - {\log(\vert\dot{\beta}\vert\vert\log~\bar g_s\vert)\over \vert\log~\bar g_s\vert} + {\log(2-\beta)\over \vert\log~\bar g_s\vert}} + 4\sqrt{\Lambda} 
\left({g_s\over {\rm HH}_o}\right)^{{\beta\over 2} -{\log(\vert\dot{\beta}\vert)\over \vert\log~\bar g_s\vert} + {2\log(2-\beta)\over \vert\log~\bar g_s\vert}}, \nonumber \nd
where $a_\Lambda \equiv (1, 2\sqrt{\Lambda})$ is the set associated with $\gamma_1$ and $\gamma_2$ respectively. As before, $e_k$ may be read from above, and since $g_s < 1$, all the logarithmic corrections are small and hence contribute (see {\bf Table \ref{rindrag}}). The sign ambiguity of the first exponent on the first line is to take care of the sign for $\ddot\beta(t)$ (we have already assumed ${\bf sgn}~\dot\beta(t) < 0$ from earlier considerations). Note that most $\gamma_i > 0$ except for
$\gamma_{4}$ and $\gamma_{5}$. Ignoring the logarithmic corrections and plugging in the values of $\gamma_{1, 2}$ from \eqref{ishena}, we see that 
$\gamma_{4}$ and $\gamma_{5}$ are both proportional to $\beta - 1$. Since $0 < \beta < {2\over 3}$, this implies $\gamma_{4} < 0$ and 
$\gamma_{5} < 0$. This is not immediately worrisome because the positivity of the exponent is connected to the perturbative quantum series first, and from there constraints on first and the second temporal derivatives of $g_s$ are determined. ($\gamma_3 = \gamma_9$ as can be seen from {\bf Table \ref{rindrag}}.) For example let us consider the Riemann tensor ${\bf R}_{0i0j}$. This takes the following 
form\footnote{Riemann tensors are defined more carefully in \eqref{lilsect} and \eqref{katusigel}. See also details in {\bf Tables \ref{firzacut}, \ref{firzacut5}} and {\bf \ref{firzacut2}}.}:

{\footnotesize
\bg\label{mca5rose}
{\bf R}_{0i0j}({\bf x}, y; g_s) & = & {\bf R}_{0i0j}^{(1)}({\bf x}, y)\left({g_s\over {\rm HH}_o}\right)^{-{8\over 3}} + 
{\bf R}_{0i0j}^{(2)}({\bf x}, y)\left({g_s\over {\rm HH}_o}\right)^{-{14\over 3} - \alpha(t)} + 
{\bf R}_{0i0j}^{(3)}({\bf x}, y)\left({g_s\over {\rm HH}_o}\right)^{-{14\over 3} - \beta(t)}\nonumber\\
& + &  
{\bf R}_{0i0j}^{(4)}({\bf x}, y)\left({g_s\over {\rm HH}_o}\right)^{-{14\over 3}}\left[{\partial\over \partial t}\left({g_s\over {\rm HH}_o}\right)\right]^2 + 
{\bf R}_{0i0j}^{(5)}({\bf x}, y)\left({g_s\over {\rm HH}_o}\right)^{-{11\over 3}}{\partial^2\over \partial t^2}\left({g_s\over {\rm HH}_o}\right), \nd}
where in the first line, since $0 < \beta(t) < {2\over 3}$ and $\alpha(t) = -\beta(t)$, there is no inconsistency. (In fact one may turn the argument around and say that the non-violating NEC implies that both $\alpha(t)$ and $\beta(t)$ have to be smaller than ${2\over 3}$.) In the second line we see that, since the scaling jumps by $\pm {\mathbb{Z}\over 3}$, the first derivative\footnote{Comparing the third and the fourth terms in \eqref{mca5rose}, one might think that ${\partial\over \partial t}\left({g_s\over {\rm HH}_o}\right) \propto \left({g_s\over {\rm HH}_o}\right)^{-{\beta\over 2}}$. However this would clash with \eqref{ishena}, thus avoiding a potential conflict with the non-violating NEC condition from \cite{coherbeta2}.} of $g_s$ cannot be a negative power of $g_s$ whereas the second derivative could scale as $\left({g_s\over {\rm HH}_o}\right)^a$ with $a \ge -{1}$. This, while justifying the no go condition proposed in \cite{coherbeta2}, allows some leeway for the second derivative.

\begin{table}[tb]  
 \begin{center}
\renewcommand{\arraystretch}{1.5}
\begin{tabular}{|c||c||c||c|}\hline ${\cal F}_i$  & ${\rm value}({\cal F}_i)$ & $\bar{g}_s^{{\cal F}_i}$  \\ \hline\hline
$\gamma_1$ & $1 - {\log(\vert\dot\beta\vert\vert\log~\bar g_s\vert)\over \vert\log~\bar g_s\vert} + {\log(2-\beta)\over \vert\log~\bar g_s\vert}$ & $\bar{g}_s \cdot {\vert\dot\beta\vert\vert\log~\bar g_s\vert \over 2-\beta}$\\ \hline
$\gamma_2$ & ${\beta\over 2} + {\log(2-\beta)\over \vert\log~\bar g_s\vert}$ & $\bar{g}_s^{\beta/2} \cdot {1\over 2-\beta}$\\ \hline
$\gamma_3$ & ${\beta\over 2} - 1 + 
\gamma_{1} + {\log(2-\beta)\over \vert\log~\bar g_s\vert}$ & $\bar{g}^{\beta/2}_s \cdot {\vert\dot\beta\vert\vert\log~\bar g_s\vert \over (2-\beta)^2}$\\ \hline
$\gamma_4$ &   ${\beta\over 2} - 1 + 
\gamma_{2} + {\log(2-\beta)\over \vert\log~\bar g_s\vert}$ & $\bar{g}^{\beta - 1}_s \cdot {1 \over (2-\beta)^2}$\\ \hline
$\gamma_5$ & $\beta - 1 + {\log(2-\beta)\over \vert\log~\bar g_s\vert}$ & $\bar{g}^{\beta - 1}_s \cdot {1 \over 2-\beta}$\\ \hline
$\gamma_6$ & $1 -{\log(\mp\ddot{\beta}\vert\log~\bar g_s\vert)\over \vert\log~\bar g_s\vert} + {\log(2-\beta)\over \vert\log~\bar g_s\vert}$ &$\bar{g}_s \cdot {\vert\ddot\beta\vert\vert\log~\bar g_s\vert \over 2-\beta}$\\ \hline
$\gamma_7$ & $1 - {\log({\dot\beta}^2\vert\log~\bar g_s\vert)\over \vert\log~\bar g_s\vert} + {2\log(2-\beta)\over \vert\log~\bar g_s\vert}$ &$\bar{g}_s \cdot {\vert\dot\beta\vert^2\vert\log~\bar g_s\vert \over (2-\beta)^2}$\\
\hline
\end{tabular}
\renewcommand{\arraystretch}{1}
\end{center}
 \caption[]{\Su The functional form of the various parameters appearing in \eqref{ishena}, \eqref{ishena2} and \eqref{corsage2}. We have defined $\bar{g}_s \equiv {g_s\over {\rm H}(y) {\rm H}_o({\bf x})}$. Note that $\gamma_3 = \gamma_9$, and one could combine $\gamma_4$ and $\gamma_5$ to further simplify \eqref{ishena2}. Note also that while the log corrections, appearing in the second column, are small, they do contribute non-trivially as seen in the third column and therefore cannot be ignored.} 
\label{rindrag}
 \end{table}

For the ${\rm E}_8 \times {\rm E}_8$ case, the derivative yields the same answer as in \eqref{ishena} and \eqref{ishena2} except with the replacement: $\beta(t) \to {\hat\alpha(t) + \hat\beta(t)\over 3}$. Since ${\hat\alpha(t)\over 9} < \hat\beta(t) < \hat\alpha(t) < 1$, the condition remains similar: $\hat\alpha(t)$ and 
$\hat\beta(t)$ should remain smooth functions, much like \eqref{TIhostsmey} or the ones shown in {\bf figure \ref{alphabetahats}}. However the issues mentioned above appear here too. This means a careful study of the quantum terms is now required before we can make a definitive statement for the $SO(32)$ and the ${\rm E}_8 \times {\rm E}_8$ cases.
These and other details will be investigated in the following sections.

\begin{table}[tb]  
 \begin{center}
\renewcommand{\arraystretch}{1.5}
\begin{tabular}{|c||c||c|}\hline Riemann tensors  & $g_s$ scalings  \\ \hline\hline
${\bf R}_{mnpq}$ &  ${\rm dom}\left(-{2\over 3} + \alpha(t), ~\mathbb{F}_1(t),~ {4\over 3} + 2\alpha(t),~ -{2\over 3} + 2\alpha(t) - \beta(t)\right)$ \\ \hline 
${\bf R}_{mn\rho\sigma}$ &  ${\rm dom}\left(-{2\over 3} + \alpha(t),~ -{2\over 3} + \beta(t), ~\mathbb{F}_2(t), ~{4\over 3} + \alpha(t) + \beta(t)\right)$ \\ \hline 
${\bf R}_{\rho\sigma i0}$ &  ${\rm dom}\left(-{5\over 3} + \beta(t) + \gamma_{1,2}(t), ~ \mathbb{F}_3(t)\right)$ \\ \hline 
${\bf R}_{m\rho\sigma 0}$ &  ${\rm dom}\left(\mathbb{F}_3(t), ~ \mathbb{F}_4(t)\right)$ \\ \hline 
${\bf R}_{0\sigma\sigma\rho}$ &  ${\rm dom}~\mathbb{F}_3(t)$ \\ \hline 
${\bf R}_{\rho\sigma\rho\sigma}$ &  ${\rm dom}\left(-{2\over 3} + \beta(t), ~\mathbb{F}_5(t), ~-{2\over 3} + 2\beta(t) - \alpha(t), ~{4\over 3} + 2\beta(t)\right)$ \\ \hline
${\bf R}_{mn i0}$ &  ${\rm dom}\left(-{5\over 3} + \alpha(t) + \gamma_{1,2}(t), ~ \mathbb{F}_6(t)\right)$ \\
\hline
\end{tabular}
\renewcommand{\arraystretch}{1}
\end{center}
 \caption[]{\Su The ${g_s\over {\rm H H}_o}$ scalings of the curvature tensors assuming no dependence on ${\mathbb{T}^2\over {\cal G}}$ directions. Here $(m, n) \in {\cal M}_4, (\rho, \sigma) \in {\cal M}_2, (i, j) \in {\bf R}^2$, $(a, b) \in {\mathbb{T}^2\over {\cal G}}$, $\gamma_{1, 2}$ are defined in \eqref{ishena}, and $\mathbb{F}_i(t)$ are 
 defined in {\bf Table \ref{firzacut3}}. In computing the scalings all permutations of the curvature indices are taken into account.} 
\label{firzacut}
 \end{table}

\subsection{An $\acute{\bf e}$tude on perturbative quantum series from M-theory \label{creepquif}}

To study the quantum series, let us first get some of the notations right. 
So far we have denoted bold-faced alphabets as operators and the light-faced ones as functions (with certain exceptions). Within this terminology we can, for example, define\footnote{As mentioned earlier, we will continue to assume the off-shell states to be massive. This will be rectified soon.}:

{\scriptsize
\bg\label{lilsect}
\langle {\bf R}_{\rm ABCD}\rangle_\sigma  &= & 
{\int [{\cal D}{\bf g}_{\rm AB}] [{\cal D}{\bf C}_{\rm ABD}] [{\cal D}{\bf \Psi}_{\rm A}] [{\cal D}\overline{\bf \Psi}_{\rm A}][{\cal D}{\bf \Upsilon}_{g}]~{\rm exp}\left(-\hat{\bf S}_{\rm tot}\right)~{\bf R}_{\rm ABCD}~\mathbb{D}^\dagger(\sigma; {\bf \Xi}) \mathbb{D}(\sigma; {\bf \Xi}) \over 
{\int [{\cal D}{\bf g}_{\rm AB}] [{\cal D}{\bf C}_{\rm ABD}] [{\cal D}{\bf \Psi}_{\rm A}] [{\cal D}\overline{\bf \Psi}_{\rm A}][{\cal D}{\bf \Upsilon}_{g}]~{\rm exp}\left(-\hat{\bf S}_{\rm tot}\right)~\mathbb{D}^\dagger(\sigma; {\bf \Xi}) \mathbb{D}(\sigma; {\bf \Xi})}} = {\bf R}_{\rm ABCD}(\langle {\bf g}_{\rm EF}\rangle_\sigma) + ....,
\nonumber\\ \nd}
where $\hat{\bf S}_{\rm tot} \approx {\bf S}_{\rm tot}$ if we ignore the instanton saddles \cite{joydeep}; and the first term in the last equality in \eqref{lilsect} is the standard curvature term that goes in say the EOM \eqref{fulashut}, and the dotted terms, which come from intermediate Glauber-Sudarshan states, contribute to the ghost equations in \eqref{mariesole}. Similar construction works for $\langle{\bf G}_{\rm ABDE}\rangle_\sigma$, with ${\bf G}_{\rm ABDE}(\langle {\bf C}_{\rm ABD}\rangle_\sigma)$ contributing to the flux EOMs similar to \eqref{fulashut}. Regarding notations, and as mentioned earlier, we have made the following judicious choice:
\bg\label{katusigel}
&&\langle{\bf g}_{\rm AB}\rangle_\sigma \equiv {\bf g}_{\rm AB} = {\bf g}_{\rm AB}({\bf x}, y; g_s(t)), ~~~~
\langle{\bf C}_{\rm ABD}\rangle_\sigma \equiv {\bf C}_{\rm ABD} 
= {\bf C}_{\rm ABD}({\bf x}, y; g_s(t))\nonumber\\
&&{\bf R}_{\rm ABCD}(\langle{\bf g}_{\rm EF}\rangle_\sigma) \equiv {\bf R}_{\rm ABCD}({\bf g}_{\rm EF}), ~~~~~
{\bf G}_{\rm ABED}(\langle{\bf C}_{\rm FGH}\rangle_\sigma) \equiv {\bf G}_{\rm ABED}({\bf C}_{\rm FGH}), \nd
which would not only avoid clutter, but would also provide a direct connection to certain known facts like {\it background} EOMs, anomaly cancellations, flux quantizations {\it et cetera} that we will discuss soon. With these at hand, our starting point will be simple: we will first analyze the perturbative quantum series, despite the fact that the existence of de Sitter Glauber-Sudarshan state relies heavily on the non-perturbative and non-local quantum terms. The reason is that the perturbative series form the foundational structure of both the standard non-perturbative terms as well as the non-local counter-terms as shown in \cite{desitter2, coherbeta, coherbeta2} and \cite{joydeep}.  The perturbative series in terms of the on-shell fields is given by:
\bg\label{botsuga}
\mathbb{Q}_{\rm T}^{(\{l_i\}, n_i)} &= & \left[{\bf g}^{-1}\right] \prod_{i = 0}^3 \left[\partial\right]^{n_i} 
\prod_{{\rm k} = 1}^{41} \left({\bf R}_{\rm A_k B_k C_k D_k}\right)^{l_{\rm k}} \prod_{{\rm r} = 42}^{81} 
\left({\bf G}_{\rm A_r B_r C_r D_r}\right)^{l_{\rm r}}\\
& = & {\bf g}^{m_i m'_i}.... {\bf g}^{j_k j'_k} 
\{\partial_m^{n_1}\} \{\partial_\alpha^{n_2}\} \{\partial_i^{n_3}\}\{\partial_0^{n_0}\}
\left({\bf R}_{mnpq}\right)^{l_1} \left({\bf R}_{mnab}\right)^{l_2}\left({\bf R}_{\alpha\beta ab}\right)^{l_3}\left({\bf R}_{abab}\right)^{l_4} \nonumber\\
&\times& \left({\bf R}_{mnij}\right)^{l_5}\left({\bf R}_{\alpha\beta ij}\right)^{l_6}
\left({\bf R}_{0m0n}\right)^{l_7}\left({\bf R}_{0\alpha 0\beta}\right)^{l_8}\left({\bf R}_{ijij}\right)^{l_9}
\left({\bf R}_{0i0j}\right)^{l_{10}}\left({\bf R}_{abij}\right)^{l_{11}}
\nonumber\\
& \times & \left({\bf R}_{0a0b}\right)^{l_{12}}\left({\bf R}_{\alpha\beta\alpha\beta}\right)^{l_{13}}\left({\bf R}_{mn\alpha\beta}\right)^{l_{14}}\left({\bf R}_{ijk0}\right)^{l_{15}}\left({\bf R}_{abi0}\right)^{l_{16}}\left({\bf R}_{\alpha\beta i0}\right)^{l_{17}}
\left({\bf R}_{mni0}\right)^{l_{18}}
\nonumber\\
& \times & \left({\bf R}_{m\alpha\beta i}\right)^{l_{19}}\left({\bf R}_{mnpi}\right)^{l_{20}}
\left({\bf R}_{m0i0}\right)^{l_{21}}\left({\bf R}_{mijk}\right)^{l_{22}}
\left({\bf R}_{mabi}\right)^{l_{23}}\left({\bf R}_{i\alpha\alpha\beta}\right)^{l_{24}}
\left({\bf R}_{mn\alpha i}\right)^{l_{25}}
\nonumber\\
&\times& \left({\bf R}_{\alpha 0i0}\right)^{l_{26}}\left({\bf R}_{\alpha ijk}\right)^{l_{27}}
\left({\bf R}_{\alpha abi}\right)^{l_{28}}\left({\bf R}_{mnp\alpha}\right)^{l_{29}}
\left({\bf R}_{m\alpha\alpha\beta}\right)^{l_{30}}\left({\bf R}_{m\alpha ab}\right)^{l_{31}}
\left({\bf R}_{m\alpha ij}\right)^{l_{32}}\nonumber\\
&\times& \left({\bf R}_{0m0\alpha}\right)^{l_{33}} 
\left({\bf R}_{mn\alpha 0}\right)^{l_{34}}
\left({\bf R}_{0\alpha\alpha\beta}\right)^{l_{35}}
\left({\bf R}_{\alpha 0ij}\right)^{l_{36}}\left({\bf R}_{0\alpha ab}\right)^{l_{37}}\left({\bf R}_{mnp0}\right)^{l_{38}}
\left({\bf R}_{m\alpha\beta 0}\right)^{l_{39}}\nonumber\\
&\times&\left({\bf R}_{m0ij}\right)^{l_{40}}
\left({\bf R}_{0mab}\right)^{l_{41}}\left({\bf G}_{mnpq}\right)^{l_{42}}\left({\bf G}_{mnp\alpha}\right)^{l_{43}}
\left({\bf G}_{mnpa}\right)^{l_{44}}\left({\bf G}_{mn\alpha\beta}\right)^{l_{45}}
\left({\bf G}_{mn\alpha a}\right)^{l_{46}}\nonumber\\
&\times&\left({\bf G}_{m\alpha\beta a}\right)^{l_{47}}\left({\bf G}_{0ijm}\right)^{l_{48}} 
\left({\bf G}_{0ij\alpha}\right)^{l_{49}}
\left({\bf G}_{mnab}\right)^{l_{50}}\left({\bf G}_{ab\alpha\beta}\right)^{l_{51}}
\left({\bf G}_{m\alpha ab}\right)^{l_{52}} \left({\bf G}_{mnpi}\right)^{l_{53}} \nonumber\\
&\times&\left({\bf G}_{m\alpha\beta i}\right)^{l_{54}}\left({\bf G}_{mn\alpha i}\right)^{l_{55}} 
\left({\bf G}_{mnai}\right)^{l_{56}}
\left({\bf G}_{mabi}\right)^{l_{57}}\left({\bf G}_{a\alpha\beta i}\right)^{l_{58}}
\left({\bf G}_{\alpha ab i}\right)^{l_{59}} \left({\bf G}_{ma\alpha i}\right)^{60} \nonumber\\
&\times&\left({\bf G}_{mn ij}\right)^{l_{61}}\left({\bf G}_{m\alpha ij}\right)^{l_{62}} 
\left({\bf G}_{\alpha\beta ij}\right)^{l_{63}}
\left({\bf G}_{maij}\right)^{l_{64}}\left({\bf G}_{\alpha a ij}\right)^{l_{65}}
\left({\bf G}_{ab ij}\right)^{l_{66}} \left({\bf G}_{0ija}\right)^{l_{67}} \nonumber\\
&\times& \left({\bf G}_{0mnp}\right)^{l_{68}}
\left({\bf G}_{0mn\alpha}\right)^{l_{69}} 
\left({\bf G}_{0m\alpha\beta}\right)^{l_{70}}\left({\bf G}_{0mab}\right)^{l_{71}}
\left({\bf G}_{0\alpha ab}\right)^{l_{72}}\left({\bf G}_{0mna}\right)^{l_{73}}
\left({\bf G}_{0m\alpha a}\right)^{l_{74}}\nonumber\\
&\times&\left({\bf G}_{0\alpha\beta a}\right)^{l_{75}}
\left({\bf G}_{0mni}\right)^{l_{76}} \left({\bf G}_{0m\alpha i}\right)^{l_{77}} 
\left({\bf G}_{0\alpha\beta i}\right)^{l_{78}} \left({\bf G}_{0ab i}\right)^{l_{79}}\left({\bf G}_{0mia}\right)^{l_{80}}
\left({\bf G}_{0\alpha ia}\right)^{l_{81}},\nonumber \nd
for any choices of $(l_i, n_i) \in \left(\mathbb{Z}, \mathbb{Z}\right)$. Note that we have used partial derivatives instead of covariant derivatives. The reason was made clear earlier in \cite{desitter2, coherbeta, coherbeta2}: the covariant derivatives involve additional interactions and these interactions are already contained within the framework of the perturbative expansion, thus removing any extra constraints on \eqref{botsuga}. Compared to \cite{desitter2, coherbeta, coherbeta2}, there are at now 41 independent choices of the Riemann tensors and 40 possible choices of G-flux components (modulo their permutations). In \cite{coherbeta2} we had 60 choices of the Riemann tensors and 40 choices for the G-flux components. This was because we allowed non-trivial dependences on the toroidal directions $w^a \equiv (x^3, w^{11})$, which here is not necessary. Additionally, the perturbative expansion \eqref{botsuga} is a clear improvement over the simpler perturbative series we took in \cite{borel2} and in \eqref{ebrowning} to compute $\langle {\bf g}_{\rm AB}\rangle_\sigma$. However since \eqref{botsuga} is expressed using expectation values \eqref{katusigel} it is not amenable to be used in path-integrals, but only in the Schwinger-Dyson's equations \eqref{mariesole}.

\begin{table}[tb]  
 \begin{center}
\renewcommand{\arraystretch}{1.5}
\begin{tabular}{|c||c||c|}\hline Riemann tensors  & $g_s$ scalings  \\ \hline\hline
${\bf R}_{m\rho\sigma i}, {\bf R}_{i\sigma \sigma\rho}, {\bf R}_{m\sigma\sigma\rho}$ &  $-{2 \over 3} + \beta(t)$ \\ \hline 
${\bf R}_{mnpi}, {\bf R}_{mnp\sigma}, {\bf R}_{mn\sigma i}$ &  $-{2\over 3} + \alpha(t)$ \\ \hline 
${\bf R}_{mabi}, {\bf R}_{\sigma ab i}, {\bf R}_{m\sigma a b}$ &  $~~{4\over 3}$ \\ \hline 
${\bf R}_{m0i0}, {\bf R}_{mijk}, {\bf R}_{0\sigma i0}$ & $ -{8\over 3}$ \\ \hline
${\bf R}_{\sigma ijk}, {\bf R}_{m\sigma ij}, {\bf R}_{0m 0\sigma}$ & $ -{8\over 3}$ \\ \hline
\end{tabular}
\renewcommand{\arraystretch}{1}
\end{center}
 \caption[]{\Su The ${g_s\over {\rm H H}_o}$ scalings of the curvature tensors that involve no logarithmic corrections and no dependence on M-theory toroidal direction.} 
\label{firzacut5}
 \end{table}

Let us clarify a few subtleties that may appear in using \eqref{botsuga}. {\Su One}, is the possibility of Ostrogradsky instability due to the presence of higher-order time derivatives {\it i.e.} $n_0 > 0$. In section \ref{ostro} we will discuss this in details and show that such instabilities may not occur in our case because of the underlying IR cut-off we impose on the nodal diagrams (see \cite{borel2}). For the quantum scaling of \eqref{botsuga}, this is again {\it not} an issue here because the quantum terms have definite scalings with respect to the type IIA string coupling $g_s$, and consequently the temporal derivatives simply become equivalent to another rescalings of these terms with respect to $g_s$. Saying differently, $g_s$ can be viewed as a {\it clock} here, and the temporal derivatives get replaced by derivatives with respect to $g_s$ that surprisingly only induce {\it positive} powers of $g_s$ as 
$\left({g_s\over {\rm H}{\rm H}_o}\right)^{{n_0\over 3} + n_0\gamma_{1, 2}}$ for $n_0 > 0$ and $\gamma_{1, 2}$ defined in \eqref{rindrag} (the first term ${n_0\over 3}$ was shown in \cite{desitter2, coherbeta2}). More detailed derivation appears in \eqref{cortanaeve}.

{\Su Two},  
is the possibility of implementing the gravitational Ward-identities in loops constructed from the above series at any given energy scales. Due to the presence of the inverse metric components {\it i.e.} $\left[{\bf g}^{-1}\right]$, the external legs of possible Feynman diagrams from \eqref{botsuga}, attaching to any currents, may {\it not} involve any derivatives. All derivatives could in principle occur in the loops thus avoiding any trivialization emanating from  possible Ward-identities. {\Su Three}, it appears that the terms like:
\bg\label{ozark}
\prod_k \Gamma^{[{\rm A}_k {\rm B}_k]}\Gamma^{[{\rm C}_k {\rm D}_k]} {\bf R}_{{\rm A}_k{\rm B}_k {\rm C}_k{\rm D}_k}, \nd
are absent in the series \eqref{botsuga}. This is actually not the case, and \eqref{ozark} may be easily accommodated by noting that $\Gamma^{{\rm A}_k}$ is a curved space Gamma-matrix and therefore 
one may, as a first attempt, take $\Gamma^{{\rm A}_k} \equiv \Gamma^a {\bf e}^{{\rm A}_k}_a$ by using eleven-dimensional vielbeins ${\bf e}^{{\rm A}_k}_a \equiv \langle{\bf e}^{{\rm A}_k}_a\rangle_\sigma$. This definition suggests that the Gamma matrices should themselves be viewed as expectation values much like the on-shell degrees of freedom from \eqref{katusigel}. Thus one expects $\Gamma^{{\rm A}_k} \equiv \langle \Gamma^{{\rm A}_k}\rangle_\sigma = \langle {\bf e}^{{\rm A}_k}_a\rangle_\sigma \Gamma_a$. Unfortunately such a choice collides with some compatibility issues that we will discuss in section \ref{servant}. For the time being we will simply call the Gamma matrices as $\Gamma_{{\rm A}_k}$ without worrying about their origins. Using this, one could replace the metric ${\bf g}_{\rm CD}$, which is symmetric in ${\rm C}$ and ${\rm D}$, by the following:
\bg\label{charlot}
{\bf g}_{({\rm CD})} ~ \to ~ \hat{\bf g}_{\rm CD} \equiv {\bf e}_{\rm C}^a {\bf e}_{\rm D}^b\Big[\eta_{ab} \mathbb{I} + c_1\Gamma_{(ab)} + c_2 \Gamma_{[ab]} + ...\Big], \nd
for some constants $(c_1, c_2)$; the dotted terms are additional contributions coming from the fermionic sector and from the resolution of identity that we will elaborate in section \ref{servant}; and $\mathbb{I}$ is the identity matrix in the same representations as the Gamma-matrices ($(a, b)$ are the {\it internal} coordinates and should not be confused with the toroidal coordinates $(w^a, w^b)$. Also since $\Gamma_{(ab)} \equiv 2\eta_{ab}$, the term proportional to $c_1$ can be absorbed in the definition of the first term, so it's only the $c_2$ coefficient that provides a non-trivial extension of the usual metric tensor\footnote{The Gamma matrices presented here would typically be in the Majorana representations $-$ because of the presence of Majorana gravitinos $-$ and therefore they are in purely imaginary representations. As such this leads to an intriguing possibility of a {\it complex} metric of the form $\hat{\bf g}_{\rm CD} \equiv {\bf g}_{({\rm CD})}\mathbb{I} + c_2 {\Gamma}_{[{\rm CD}]}$ with a real symmetric part and an imaginary anti-symmetric piece because $\Gamma^\dagger_{[{\rm CD}]} = -\Gamma_{[{\rm CD}]}$ and $c_2^\ast = c_2$. Introducing complex metrics in GR is not new, see for example the recent work \cite{wittencomplex}, although here it would appear to make the system more involved (notwithstanding the addition constraints that complex metrics need to satisfy to be ``good" metrics). It is interesting that with a relative factor of $ic_2 \equiv c_2\sqrt{-1}$, $\hat{\bf g}^\dagger_{\rm CD} = \hat{\bf g}_{\rm CD}$ and is therefore {\it real} in the usual sense but the terms in the matrix $\hat{\bf g}_{\rm CD}$ are imaginary. To avoid such un-necessary complications, we will use the definition \eqref{charlot} with $c_1 = 0$ (and without any factors of $i \equiv \sqrt{-1}$) to simply compute additional terms in the quantum series \eqref{botsuga} without recoursing to its real or complex nature. For more details see \cite{wittencomplex}}). Thus replacing $\left[{\bf g}^{-1}\right]$ by $\left[\hat{\bf g}^{-1}\right]$, the terms proportional to $c_2$ (not involving the G-flux components and derivatives) would indeed reproduce \eqref{ozark}.
Interestingly, from the presence of Gamma-matrices in \eqref{charlot} and from the product of the generalized metric components in \eqref{botsuga} (see \eqref{fahingsha10} also), we can construct epsilon tensor\footnote{By this we will always mean the Levi-Civita {\it symbol} and not the Levi-Civita {\it tensor} $\varepsilon_{{\rm M_1...M}_d}$ unless indicated otherwise. They are related by $\varepsilon_{{\rm M_1..M}_d} = \sqrt{|{\bf g}_d|} ~\epsilon_{{\rm M_1...M}_d}$.} as:
\bg\label{indig2thi}
\Gamma^{{\rm M_1M_2...M_{11}}} = {\pm i \over \sqrt{|{\bf g}_{11}|}} \epsilon_{11}^{{\rm M_1M_2...M_{11}}} \mathbb{I}, \nd
where the sign ambiguity comes from the choice of the signatures. Presence of multiple such epsilon tensors instead of the metric components can help us construct curvature and flux forms, thus extending \eqref{botsuga} and \eqref{fahingsha10} to incorporate topological interactions. Moreover,
including the G-flux components, we could even extract a more generic quantum series of the form\footnote{See \cite{gatpurali} for a discussion on the higher order terms in M-theory.}:

{\footnotesize
\bg\label{rooth}
\mathbb{Q}_8 \equiv \sum_{k = 1}^\infty c_{k + 1} \Big[b_1 {\rm M}_p^{-2} t_8 {\bf G}^2 + b_2 {\rm M}_p^{-8} t_8 t_8 \left({\bf R}^4 + b_2' {\bf G}^2 {\bf R}^3\right)+ b_3{\rm M}_p^{-8} \epsilon_{11} \epsilon_{11} \left({\bf R}^4 + b_3'{\bf G}^2 {\bf R}^3\right) + ... \Big]^k, \nd}
from \eqref{botsuga} and \eqref{fahingsha10}, where $t_8$ is a tensor constructed from the Gamma-matrices and involve the Levi-Civita tensor density and inverse metric components; the dotted terms are of the form $\square {\bf R}^3$ et cetera; and $(b_i, b'_i)$ are constant coefficients. The curvature ${\bf R}_{{\rm A}_k{\rm B}_k {\rm C}_k{\rm D}_k}$ should now be defined with respect to the generalized metric components \eqref{charlot}, and therefore will involve appropriate traces to deal with the matrix valued form of $\hat{\bf g}_{\rm CD}$. The coefficients $c_{k + 1}$ for $k > 1$ are functions of $c_2$ from \eqref{charlot} and the coefficients governing the original quantum series 
\eqref{botsuga}. 
Note that the scalings of the $\hat{\bf g}_{\rm CD}$ with $\left({\rm C}, {\rm D}\right) \in {\bf R}^{2, 1} \times {\cal M}_4 \times {\cal M}_2 \times
{\mathbb{T}^2\over {\cal G}}$ remain similar to the scalings of the original metric components ${\bf g}_{\rm CD}$. And finally, subtlety number {\Su four}: the possibility of fermionic extensions of the quantum series \eqref{botsuga}. This will actually be much more non-trivial and therefore will require detailed study. We will come back to this in section \ref{servant} after we discuss the more urgent matter of the $g_s$ scaling of \eqref{botsuga}.

\begin{table}[tb]  
 \begin{center}
\renewcommand{\arraystretch}{1.5}
\begin{tabular}{|c||c||c|}\hline Riemann tensors  & $g_s$ scalings \\ \hline\hline
${\bf R}_{0i0j}$ & ${\rm dom}\left(-{8\over 3}, ~ \mathbb{F}_{16}(t), ~-{14\over 3} - \alpha(t), 
~ -{14\over 3} - \beta(t)\right)$ \\ \hline
${\bf R}_{0a0b}$ & ${\rm dom}\left({4\over 3}, ~\mathbb{F}_{17}(t), ~-{2\over 3} - \alpha(t), 
~ -{2\over 3} - \beta(t)\right)$ \\ \hline
${\bf R}_{0m0n}$ & ${\rm dom}\left(-{8\over 3}, ~-{2\over 3} + \alpha(t),~\mathbb{F}_{14}(t), ~ \mathbb{F}_{18}(t),~\mathbb{F}_{19}(t),~ 
-{8\over 3} +\alpha(t) - \beta(t)\right)$ \\ \hline
${\bf R}_{0\rho 0\sigma}$ & ${\rm dom}\left(-{8\over 3}, ~-{2\over 3} + \beta(t),~\mathbb{F}_{15}(t), ~ \mathbb{F}_{20}(t), ~\mathbb{F}_{21}(t), ~-{8\over 3} +\beta(t) - \alpha(t)\right)$ \\ \hline
\end{tabular} 
\renewcommand{\arraystretch}{1}
\end{center}
 \caption[]{\Su The ${g_s\over {\rm H H}_o}$ scalings of the curvature tensors that involve second derivative $\ddot{g}_s$. Again no dependence on the toroidal direction is assumed; and $\mathbb{F}_i(t)$ are defined in {\bf Table \ref{firzacut3}}.} 
  \label{firzacut2}
 \end{table}

\subsection{$g_s$ scalings of the curvature tensors and the G-flux components \label{curvten}}

Our aim for this section would be to see how the quantum series in \eqref{botsuga} influences the effective field theory discussion that we had earlier.  For this, the precise behavior of the 
${g_s\over {\rm HH}_o}$ scaling of the quantum series \eqref{botsuga} is now called for, in the light of the fact that the ${\rm F}_i(t)$ warp-factors for both the heterotic theories now involve $g_s$ scalings that are now no longer constants. The ${\rm E}_8 \times {\rm E}_8$ theory has an additional complication from the presence of {\it three} warp-factors. This unfortunately will now make the subsequent analysis 
much more involved compared to \cite{desitter2, coherbeta, coherbeta2}. We can simplify slightly by studying the two warp-factor case, {\it i.e.} the case associated with $SO(32)$ heterotic theory and in fact, as the result below will show, extending from two warp-factors to three will not be too hard. We will however not attempt the latter, but will provide enough hints for the diligent reader to complete the story for herself or himself. 

\subsubsection{$g_s$ scalings of the metric tensors and dualities \label{sec4.2.1}}

Few more necessary details are required before we compute the $g_s$ scaling of the quantum series in \eqref{botsuga}, and they have to do with the $g_s$ scalings of the flux, metric and the curvature components.
So far we have been using only the dominant $g_s$ scalings of the metric components, with a brief mention of the sub-dominant contributions in \eqref{sezoe70} for the $SO(32)$ and in \eqref{sezoe71} for the ${\rm E}_8 \times {\rm E}_8$ cases. (See also the discussion in section \ref{sec2.3}.) We can add the sub-dominant contributions in the following way\footnote{From \eqref{ebrowning}, \eqref{gulabs} and \eqref{ramyakri}, we see that a non-trivial temporal modulation of the Minkowski metric is needed to solve the corresponding Schwinger-Dyson's equations associated to type IIB de Sitter state \cite{desitter2, coherbeta, coherbeta2}. On the other hand, \eqref{gwenphone} tells us that even more non-trivial metric ans\"atze is needed to solve the corresponding Schwinger-Dyson's equations \eqref{mariesole} associated to a $SO(32)$ heterotic de Sitter state. (For the ${\rm E}_8 \times {\rm E}_8$ case, a slightly more involved ans\"atze is needed.) In section \ref{sec4.2.2} we will argue that \eqref{gwenphone} is not enough: we have to go beyond that by incorporating the non-perturbative corrections if we want the metric components to solve the EOMs coming from the trans-series form of the action \cite{joydeep}. For the time being we will proceed in small steps. \label{lenmune}}:
\bg\label{gwenphone}
&& \langle{\bf g}_{\kappa\rho}\rangle_\sigma \equiv {\bf g}_{\kappa\rho}({\bf x}, y; g_s) = 
\sum_{k = 0}^\infty {\bf g}^{(k)}_{\kappa\rho}({\bf x}, y) \left({g_s\over {\rm H}(y){\rm H}_o({\bf x})}\right)^{-{2\over 3} + \beta(t) + {2k\vert\beta_k(t)\vert\over 3}}\nonumber\\
&& \langle{\bf g}_{mn}\rangle_\sigma \equiv {\bf g}_{mn}({\bf x}, y; g_s) = 
\sum_{k = 0}^\infty {\bf g}^{(k)}_{mn}({\bf x}, y) \left({g_s\over {\rm H}(y){\rm H}_o({\bf x})}\right)^{-{2\over 3} + \alpha(t) + {2k\vert\alpha_k(t)\vert\over 3}}, \nd
where the functional forms for $(\alpha_k(t), \beta_k(t))$ can be determined from the Schwinger-Dyson's equations \eqref{mariesole}; and $k \in {\mathbb{Z}\over 2}$. The other set of metric components $\langle {\bf g}_{\mu\nu}\rangle_\sigma \equiv {\bf g}_{\mu\nu}({\bf x}, y; g_s)$ and  $\langle {\bf g}_{ab}\rangle_\sigma \equiv {\bf g}_{ab}({\bf x}, y; g_s)$ scale as 
$\left({g_s\over {\rm HH}_o}\right)^{-{8\over 3}}$ and $\left({g_s\over {\rm HH}_o}\right)^{{4\over 3}}$ respectively without any other sub-dominant contributions. This is because, any sub-dominant contributions to the aforementioned metric components will naively take us away from a simple realization of a four-dimensional de Sitter configuration that we seek in the heterotic side. This is interesting because it provides us a way to study {\it quasi} de Sitter configuration with a dynamical dark energy of the form \eqref{marapaug}. One could then relax the aforementioned condition, and  consider the case where the four-dimensional cosmological constant is slowly varying with time \cite{desibao}. However including these changes will make the analysis a bit more involved. To see this let us assign sub-dominant contributions to the space-time metric 
 components and one of the toroidal direction of ${\mathbb{T}^2\over {\cal G}}$ in M-theory. In other words, let us consider the following metric configurations in addition to the ones in \eqref{gwenphone}:
 \bg\label{infinitypool}
&& \langle {\bf g}_{33}\rangle_\sigma \equiv {\bf g}_{33}({\bf x}, y; g_s) = \sum_{k = 0}^\infty {\bf g}^{(k)}_{33}({\bf x}, y) \left({g_s\over {\rm H}(y){\rm H}_o({\bf x})}\right)^{{4\over 3} + {2k\vert \widetilde{f}_k(t)\vert\over 3}}\nonumber\\
&&\langle {\bf g}_{\mu\nu}\rangle_\sigma \equiv {\bf g}_{\mu\nu}({\bf x}, y; g_s) = \sum_{k = 0}^\infty {\bf g}^{(k)}_{\mu\nu}({\bf x}, y) \left({g_s\over {\rm H}(y){\rm H}_o({\bf x})}\right)^{-{8\over 3} + {2k\vert f_k(t)\vert\over 3}}\nonumber\\
&& \langle {\bf g}_{11, 11}\rangle_\sigma \equiv {\bf g}_{11, 11}({\bf x}, y; g_s) = \sum_{k=0}^\infty {\bf \widetilde{g}}_{11, 11}({\bf x}, y) \left({g_s\over {\rm H}(y){\rm H}_o({\bf x})}\right)^{{4\over 3}+ {2k\vert {g}_k(t)\vert\over 3}},
 \nd
 where we have made a special choice for the modes for ${\bf g}_{33}({\bf x}, y; g_s)$  and ${\bf g}_{\mu\nu}({\bf x}, y; g_s)$ with ${\bf g}^{(k)}_{33}({\bf x}, y) = \widetilde{h}_k$ and ${\bf g}_{\mu\nu}^{(k)}({\bf x}, y) = h_k \eta_{\mu\nu}$ with a constant $h_k$ and $k \in {\mathbb{Z}\over 2}$ so as to maintain a de Sitter configuration in the heterotic side once we impose the following conditions on the functions $f_k(t)$ and $\widetilde{f}_k(t)$:
 \bg\label{englishgulab}
\sum_{k = 0}^\infty \sum_{l = 0}^\infty \widetilde{h}'_k h'_l \left({g_s\over {\rm H}(y) {\rm H}_o({\bf x})}\right)^{{2\over 3}\left(k\vert\widetilde{f}'_k(t)\vert + l\vert f'_l(t)\vert\right)} = 1, \nd 
where ${f'}_k(t) = f_k(t) + {\cal O}({\rm M}_p, g_s)$, $\widetilde{f}_k(t) = \widetilde{f}_k(t) + {\cal O}({\rm M}_p, g_s)$ with $h_k \to h'_k$ and $\widetilde{h}_k \to \widetilde{h}'_k$, see \eqref{ramyakri} (note that $f'_k(t)$ is {\it not} a derivative of $f_k(t)$).
The LHS in \eqref{englishgulab} can be split into product of two separate sums over $k$ and $l$ for $\widetilde{f}'_k$ and $f'_l$, which we shall use below to express the metric in the type IIB side. In \eqref{infinitypool}, we have also chosen $g_k(t)$ 
in ${\bf g}_{11, 11}({\bf x}, y; g_s)$ in such a way so as to allow the type IIA coupling to remain $g_s$ with no additional corrections. Shrinking the toroidal direction to zero size makes the type IIB coupling $\bar{g}_s^{(b)}$ time-dependent as:
 \bg\label{influencer}
 \left(\bar{g}_s^{(b)}\right)^2 = \sum_{k = 0}^\infty h'_k \left({g_s\over {\rm H}(y){\rm H}_o({\bf x})}\right)^{{2k\vert {\Su{f'}_k(t)}\vert\over 3}} ~ > ~ 1, \nd
 compared to what we had in {\bf Table \ref{milleren2}} for the $SO(32)$ case. The strength of the type IIB coupling now depends on the functional form for $f'_0(t)$: if $\lim\limits_{k = 0} k\vert {f'}_k(t)\vert = {\rm constant}$, then the IIB coupling can be weak in the allowed temporal domain, otherwise for finite ${f'}_0(t)$, the IIB coupling continues to be strong. In fact in this language one may show that the above duality chasings, using the simplifying conditions mentioned after \eqref{infinitypool} and the constraint \eqref{englishgulab}, produce the following metric configuration in the type IIB side:
 
 {\footnotesize
 \bg\label{engrose}
 ds^2 &= & \sum_{k=0}^\infty h'_k \left({g_s\over {\rm H}(y){\rm H}_o({\bf x})}\right)^{-2 + {2k\vert {\Su{f'}_k(t)}\vert\over 3}}\left(-dt^2 + \sum_{i = 1}^3 dx_i^2\right) \\
 &+ & \sum_{k=0}^\infty {\Su {\bf g}'^{(k)}_{mn}({\bf x}, y)} 
\left({g_s\over {\rm H}(y){\rm H}_o({\bf x})}\right)^{{\Su\alpha'(t)} + {2k\vert{\Su\alpha'_k(t)}\vert\over 3}} + \sum_{k=0}^\infty {\Su {\bf g}'^{(k)}_{\rho\sigma}({\bf x}, y)}
\left({g_s\over {\rm H}(y){\rm H}_o({\bf x})}\right)^{{\Su\beta'(t)} + {2k\vert{\Su\beta'_k(t)}\vert\over 3}}, \nonumber \nd}
which may be compared to \eqref{viomyer1} and we have take $\widetilde{\bf g}_{11, 11}({\bf x}, y) \equiv 1$ in \eqref{infinitypool}. (Note also the parameters defining the IIB backgrounds are $({\Su{f'}_k(t)}, {\Su\alpha'_k(t)}, {\Su\beta'_k(t)}, {\Su {\bf g}'^{(k)}_{mn}}, {\Su {\bf g}'^{(k)}_{\alpha\beta}}, {\Su \alpha'(t)}, {\Su \beta'(t)})$ compared to $(f_k(t), \alpha_k(t), \beta_k(t), {\bf g}^{(k)}_{mn}, {\bf g}^{(k)}_{\alpha\beta}, \alpha(t), \beta(t))$ in M-theory, following \eqref{ramyakri}\footnote{More generic metric with the coefficient of $dx_3^2$ different from $(dt^2, dx_1^2, dx_2^2)$ is allowed but we will stick with \eqref{engrose} and adjust the fluxes accordingly. A concrete example will be shown later.}. Additionally, any cross-term in the metric is integrated out to produce {\it non-local} quantum terms, which would appear in the corresponding Schwinger-Dyson's equations, much like what we had in \cite{desitter2, coherbeta, coherbeta2, joydeep}\footnote{Such a scenario is expected in all theories that follow from the duality chasing a specific configuration, like \eqref{gwenphone} and \eqref{infinitypool}, in M-theory. Thus henceforth we will refrain from mentioning it, but the readers should keep in mind that this is at play at every stages of the duality sequences.}.) The metric configuration in \eqref{engrose} differs significantly from a simple Buscher's dual of the M-theory configuration due to the ${\rm M}_p$ corrections as shown in \eqref{ramyakri}. However we could restore some of the M-theory components in the metric by expressing \eqref{engrose} as:

{\footnotesize
 \bg\label{engrose2}
 ds^2 &= & \sum_{k=0}^\infty h_k \left({g_s\over {\rm H}(y){\rm H}_o({\bf x})}\right)^{-2 + {2k\vert {\Su{f'}_k(t)}\vert\over 3}}\left(-dt^2 + \sum_{i = 1}^3 dx_i^2\right) \\
 &+ & \sum_{k=0}^\infty {{\bf g}^{(k)}_{mn}({\bf x}, y)} 
\left({g_s\over {\rm H}(y){\rm H}_o({\bf x})}\right)^{{\alpha(t)} + {2k\vert{\Su\alpha'_k(t)}\vert\over 3}} + \sum_{k=0}^\infty {{\bf g}^{(k)}_{\rho\sigma}({\bf x}, y)}
\left({g_s\over {\rm H}(y){\rm H}_o({\bf x})}\right)^{{\beta(t)} + {2k\vert{\Su\beta'_k(t)}\vert\over 3}}, \nonumber \nd}
where the only changes are in $(f_k(t), \alpha_k(t), \beta_k(t))$. One might however wonder how we can justify the transition from \eqref{engrose} to \eqref{engrose2}. As discussed in section \ref{sec2.3}, the string dualities work well over the supersymmetric solitonic background. For the expectation values over the Glauber-Sudarshan states, where the supersymmetry is spontaneously broken, we expect some level of duality sequences to work because the effective action $-$ as shown in \cite{joydeep} and as we shall discuss later $-$ takes a trans-series form whose zero instanton sector mimics the perturbative (but beyond) supergravity action. This means, for a complete picture in the presence of the non-zero instanton sectors, the correct way would be to resort to Schwinger-Dyson equations that take the form as in eq. (7.57) in \cite{joydeep}. The metric configuration \eqref{engrose2} should then solve exactly the aforementioned EOM to all orders in $g_s$ and ${\rm M}_p$. While this leads to a metric structure
simpler than \eqref{engrose}, the compromise now is a more involved flux configuration from what we had with the choice \eqref{engrose}. 
Henceforth we will stick with \eqref{engrose2} and work out the fluxes using the Schwinger-Dyson's equations\footnote{One might envision an even simpler metric structure with $(f'_k(t), \alpha'_k(t), \beta'_k(t)) \to 
(f_k(t), \alpha_k(t), \beta_k(t))$ but, since it is not clear at this stage whether the duality sequence allow this, we will continue with \eqref{engrose2}.}.
Once we take ${\cal M}_2 = {\mathbb{T}^2\over \Omega (-1)^{\rm F_L} {\cal I}_{\rho\sigma}}$, {\it i.e.} a toroidal orientifold, two T-dualities will lead to type I metric configuration. 
 (The ${\rm E}_8 \times {\rm E}_8$ story is similar and may easily be worked out from our earlier discussions.) Following the aforementioned duality sequence as depicted in {\bf Table \ref{milleren2}}, the metric components in the heterotic $SO(32)$ side take the following form:

{\footnotesize
\bg\label{gethmia}
&&\langle {\bf g}^{({\rm het})}_{\rho\sigma}\rangle_{\sigma'} \equiv 
{\bf g}^{({\rm het})}_{\rho\sigma}({\bf x}, y; g_s) = {\delta{\rho\sigma} \over \left[
\sum\limits_{k = 0}^\infty h_k
\left({g_s\over {\rm H}(y){\rm H}_o({\bf x})}\right)^{{2k\vert {\red\check{f}_k(t)}\vert \over 3}}\right]^{1\over 2}}\\
&&\langle {\bf g}^{({\rm het})}_{mn}\rangle_{\sigma'} \equiv 
{\bf g}^{({\rm het})}_{mn}({\bf x}, y; g_s) = \sum\limits_{k = 0}^\infty\sum\limits_{l = 0}^\infty
{\bf g}^{(k, l; {\rm het})}_{mn}({\bf x}, y) c_k b_l
{\left({g_s\over {\rm H}(y){\rm H}_o({\bf x})}\right)^{\alpha(t) + \beta(t) + {2k\vert{\red\check\alpha_k(t)}\vert \over 3} + {2l\vert{\red\check\beta_l(t)}\vert\over 3}}
\over \left[\sum\limits_{p = 0}^\infty h_p \left({g_s\over {\rm H}(y){\rm H}_o({\bf x})}\right)^{2p\vert {\red\check{f}_p(t)}\vert \over 3}\right]^{1\over 2}}\nonumber\\
&&\langle {\bf g}^{({\rm het})}_{\mu\nu}\rangle_{\sigma'} \equiv 
{\bf g}^{({\rm het})}_{\mu\nu}({\bf x}, y; g_s) =\sum_{l = 0}^\infty\eta_{\mu\nu} b_l
\left({g_s\over {\rm H}(y){\rm H}({\bf x})_o}\right)^{-2 + \beta(t) +{2l\vert{\red\check\beta_l(t)}\vert\over 3}}\left[
\sum_{k = 0}^\infty h_k
\left({g_s\over {\rm H}(y){\rm H}({\bf x})_o}\right)^{2k\vert {\red\check{f}_k(t)}\vert \over 3}\right]^{1\over 2}, \nonumber
\nd}
where $(\mu, \nu) \in {\bf R}^{1, 3}, (m, n) \in {\cal M}_4$, $(\rho, \sigma) \in {\cal M}_2$ and we have taken $h'_k \to h_k$ in \eqref{influencer}, ${\bf g}^{(k)}_{mn}({\bf x}, y) = c_k \widetilde{\bf g}_{mn}({\bf x}, y)$ and ${\bf g}_{k\rho}^{(k)}({\bf x}, y) = b_k\delta_{\rho\sigma}$ to help with the duality sequence\footnote{We can keep a non-trivial metric on ${\cal M}_2$ as ${\bf g}_{\rho\sigma}^{(k)}({\bf x}, y) \equiv b_k\bar{\bf g}_{\rho\sigma}({\bf x}, y^m)$, where $y^m \in {\cal M}_4$ and constant $b_k$, but then we have to carefully keep track of ${\rm det}~{\bf g}_{\rho\sigma}({\bf x}, y^m)$ as well as ${\bf g}^{\rho\sigma}({\bf x}, y^m)$ appearing in the intermediate type I coupling and elsewhere. To avoid these unnecessary complications we make ${\bf g}_{\rho\sigma}^{(k)}({\bf x}, y) \equiv b_k \delta_{\rho\sigma}$.}; and $\vert\sigma'\rangle$ is the Glauber-Sudarshan state over a supersymmetric Minkowski background in the heterotic $SO(32)$ side with the parameters now being $({\red\check{f}_k(t)}, {\red\check\alpha_k(t)}, {\red\check\beta_k(t)})$ following \eqref{ramyakri} and the discussion after \eqref{engrose2}. Looking at \eqref{gethmia}, we see the problem now. Since we haven't provided any bounds on $\check{f}_k(t)$, and if $\check{f}_0(t)$ blows up when $g_s \to 0$ keeping $k\check{f}_k(t)$ fixed, then the toroidal manifold ${\cal M}_2$ increases in size as $g_s \to 0$, and the four-manifold ${\cal M}_4$ has to shrink to zero size to keep the volume small in the allowed temporal domain where $g_s < 1$. Unfortunately this doesn't generically mean that the volume remains {\it time independent} unless we make $c_k b_l {\bf g}_{mn}^{(k, l; {\rm het})}({\bf x}, y) = c_{kl} \widetilde{\bf g}_{mn}^{({\rm het})}({\bf x}, y)$ with constant $c_{kl}$. This fits perfectly well with the ${\rm F}_1 {\rm F}_2$ factor appearing in the sixth row and fourth column of {\bf Table \ref{milleren2}}. Putting everything together amounts to the following condition:
\bg\label{didigoth}
{\left[\sum\limits_{p=0}^\infty h_p \left({g_s\over {\rm H}(y){\rm H}_o({\bf x})}\right)^{{2p\vert {\red\check{f}_p(t)}\vert \over 3}}\right]^{3\over 2} \over 
\left[\sum\limits_{k= 0}^\infty \sum\limits_{l = 0}^\infty~c_{kl}~\left({g_s\over {\rm H}(y){\rm H}_o({\bf x})}\right)^{{2k\vert{\red\check\alpha_k(t)}\vert \over 3} + {2l\vert{\red\check\beta_l(t)}\vert\over 3}}\right]^2} ~ = ~ \left({g_s\over {\rm H}(y){\rm H}_o({\bf x})}\right)^{2(\alpha(t) + \beta(t))} \nd
which would relate $(\check\alpha_k(t), \check\beta_k(t), \check{f}_k(t))$ with $(\alpha(t), \beta(t))$. Note that we cannot simply compare the $g_s$ scalings on both sides because of the temporal dependence of the exponents as the conformal time $t \equiv t(g_s)$. The dependence of $t$ on $g_s$, or the other way around, comes from solving the following equation:
\bg\label{ccoleman}
\sum_{l = 0}^\infty b_l
\left({g_s\over {\rm H}(y){\rm H}({\bf x})_o}\right)^{-2 + \beta(t) +{2l\vert{\red\check\beta_l(t)}\vert\over 3}}\left[
\sum_{k = 0}^\infty h_k
\left({g_s\over {\rm H}(y){\rm H}({\bf x})_o}\right)^{2k\vert {\red\check{f}_k(t)}\vert \over 3}\right]^{1\over 2} 
= \begin{cases} ~ {1\over \Lambda \vert t\vert^2}\\
~~~~\\
~{1\over \Lambda(t) \vert t \vert^2},
\end{cases}
\nd
where the first case would give us a de Sitter configuration in the $SO(32)$ heterotic theory with a flat-slicing, while the second case will be the one with a dynamical dark energy \eqref{marapaug}. Comparing this to \eqref{covda}, we see that \eqref{ccoleman} {\it modifies} the relation between $g_s$ and $t$. For example if $\check\beta_1(t)$ and $\check{f}_1(t)$ are respectively the smallest values in the set 
$\{\check\beta_k(t), \check{f}_k(t)\}$, then the relation between $g_s$ and $t$ becomes:

{\scriptsize
\bg\label{cassnaud}
\left({\Lambda(t)} \vert t \vert^2\right)^{r(t)} & = & 
{{g_s\over {\rm H}(y){\rm H}_o({\bf x})} \over 
\left[\sum\limits_{l = 0}^\infty b_l
\left({g_s\over {\rm H}(y){\rm H}({\bf x})_o}\right)^{{2l\vert{\red\check\beta_l(t)}\vert\over 3}}\right]^{1\over {2-\beta(t)}}\left[
\sum\limits_{k = 0}^\infty h_k
\left({g_s\over {\rm H}(y){\rm H}({\bf x})_o}\right)^{2k\vert {\red\check{f}_k(t)}\vert \over 3}\right]^{1\over 4 - 2\beta(t)}}\\ 
& = &
{{g_s\over {\rm H}(y){\rm H}_o({\bf x})} \over \left[ 1 + {h_1\over 2} \left({g_s\over {\rm H}(y){\rm H}_o({\bf x})}\right)^{2\vert {\red\check{f}_1(t)}\vert\over 3} + b_1\left({g_s\over {\rm H}(y){\rm H}_o({\bf x})}\right)^{2\vert {\red\check\beta_1(t)}\vert\over 3} + {h_1b_1\over 2} \left({g_s\over {\rm H}(y){\rm H}_o({\bf x})}\right)^{{2\vert{\red\check\beta_1(t)}\vert \over 3} + {2\vert {\red\check{f}_1(t)}\vert\over 3}} + ..\right]^{r(t)}}, \nonumber\nd}
where $r(t) \equiv {1\over 2-\beta(t)}$ and we have taken only the case \eqref{marapaug}. \eqref{cassnaud}
may be compared to \eqref{covda}, and indeed \eqref{covda} remains reassuringly the dominant contribution, and as long as the following condition is maintained:
\bg\label{annypyar}
{\rm min}\{k\vert \check{f}_k(t)\vert\} ~ < ~ 4 - 2\beta(t), \nd
the TCC condition: $-{1\over \sqrt\Lambda} < t < 0$ does not change.
A similar analysis may be done for the heterotic ${\rm E}_8 \times {\rm E}_8$ theory but we will not do so here and leave the exercise for the diligent reader. In fact, looking at the above analysis one might get the hope that the M-theory metric configuration \eqref{gwenphone} with the modification \eqref{infinitypool} could in principle lead to a consistent quasi de Sitter configuration in the heterotic $SO(32)$ side. This would appear possible if the choice of the parameters do not lead to late-time singularity from the shrinking of the internal four-manifold ${\cal M}_4$. However there can be another issue: the heterotic coupling, given by the following series:
\bg\label{CWsara}
\bar{g}_s^{({\rm het})}~ = ~ {\sum\limits_{k = 0}^\infty b_k\left({g_s\over {\rm H}(y){\rm H}_o({\bf x})}\right)^{\beta(t) + {2k\vert{\red\check\beta_k(t)}\vert \over 3}}
\over \left[\sum\limits_{l = 0}^\infty h_l\left({g_s\over {\rm H}(y){\rm H}_o({\bf x})}\right)^{2l\vert {\red\check{f}_l(t)}\vert \over 3}\right]^{1\over 2}}, \nd
{can} become {\it strong} if ~ $\lim\limits_{k = 0} k\vert \check{f}_k(t)\vert = $
constant, thus making $\bar{g}_s^{({\rm het})} >> 1$ as $g_s \to 0$ at late time. One way out would be to keep $\check{f}_k(t) = 0$, then the heterotic $SO(32)$ becomes weakly coupled at late time. In fact keeping $f_k(t) = \widetilde{f}_k(t) = {f'}_k(t) = \check{f}_k(t) = 0$ also implies that the metric components in \eqref{infinitypool} will have no subdominant contributions, although the metric components in \eqref{gwenphone} continue to have the subdominant contributions. This is of course consistent with the point that we made just below \eqref{gwenphone} but our little exercise from \eqref{gwenphone} till \eqref{CWsara} is instructive because it tells us precisely how the subdominant contributions in \eqref{gwenphone} continue to keep the internal manifold time-{\it independent} provided the requirement \eqref{didigoth} $-$ but now with $(f_k(t), \widetilde{f}_k(t), {f'}_k(t), \check{f}_k(t))$ contributions removed $-$ is satisfied. Unfortunately the actual demonstration of time-independence of the internal manifold does not simplify substantially even if we make $f_k(t) = \widetilde{f}_k(t) = {f'}_k(t) = \check{f}_k(t)= 0$ because the relation between $g_s$ and $t$ is still nontrivial (see \eqref{cassnaud}) so an order by order analysis may be a bit harder to do. Additionally the precise determination of 
the sub-dominant contributions in \eqref{gwenphone} is quite non-trivial because the analysis involves invoking \eqref{mariesole} as well as other constraints like flux quantizations and anomaly cancellations. On the other hand, if we demand that $\check{f}_k(t)$ and $\check\beta_k(t)$ are related by:
\bg\label{brebeach}
\sum\limits_{l = 0}^\infty\sum\limits_{p = 0}^\infty 
\sum\limits_{k = 0}^\infty b_l b_p h_k
\left({g_s\over {\rm H}(y){\rm H}({\bf x})_o}\right)^{{2l\vert{\red\check\beta_l(t)}\vert\over 3} + {2p \vert {\red\check\beta_p(t)}\vert \over 3} + {2k\vert {\red\check{f}_k(t)}\vert \over 3}} = 1, \nd
then $g_s$ and $t$ are related by ${g_s\over {\rm H}(y) {\rm H}_o({\bf x})} = (\sqrt{\Lambda(t)} \vert t\vert)^{2\over 2 - \beta(t)}$, almost like \eqref{covda} but now with a dynamical dark energy $\Lambda(t)$ from \eqref{marapaug}! This simple relation between $\check{f}_k(t)$ and $\check\beta_k(t)$ in the heterotic $SO(32)$ theory
 however does not guarantee a simple relation between $f_k(t)$ and $\beta_k(t)$ in M-theory because of the $\alpha'$ corrections to the T-duality rules connecting these two theories. Nevertheless if \eqref{brebeach} is true then clearly the relations in \eqref{covda} are exact (the ${\rm E}_8 \times {\rm E}_8$ analysis is straightforward). For the time being, to avoid such complications at this stage,
we can keep $k = k_i = 0$ which would eliminate these contributions in \eqref{gwenphone}, \eqref{sezoe70} and \eqref{sezoe71}, as well as all the subsequent sub-dominant contributions studied above. We shall come back to the full analysis, and in particular
the validity of \eqref{brebeach}, when we study the Schwinger-Dyson's equations.

\subsubsection{$g_s$ scalings with non-perturbative corrections \label{sec4.2.2}}

Before incorporating the aforementioned simplifications, there are couple more points that need to be discussed at this stage. The {\Su first} point is the instanton corrections to the metric components themselves. We expect the metric components in all the theories (connected by the aforementioned duality sequences) to get further corrections to the already-existing subdominant corrections that we added above starting with \eqref{gwenphone} and \eqref{infinitypool} in M-theory. The surprising thing is that, even with such modifications, we can still get de Sitter (or a quasi de Sitter) states in heterotic theories. To see this let us change the metric components, say along the space-time directions, in the following way:

{\footnotesize
\bg\label{tranpart}
\langle {\bf g}_{\mu\nu}\rangle_\sigma \equiv {\bf g}_{\mu\nu}({\bf x}, y; g_s) = \sum_{l = 0}^\infty {\bf g}^{(l)}_{\mu\nu}({\bf x}, y) \left({g_s\over {\rm H}(y){\rm H}_o({\bf x})}\right)^{-{8\over 3} + {2l\vert f_l(t)\vert\over 3}}~{\rm exp}\left[-\sum_{a\in\mathbb{Z}} {n_a(l)\over \bar{g}_s^{a/3}}\right], \nd}
with similar modifications to all other components in \eqref{gwenphone} and \eqref{infinitypool}. Here $n_a(l) \equiv n_a(l; {\rm M}_p, k_{\rm IR}, \mu)$ with $n_a(0) \equiv 0$ so that the corrections are sub-dominant (which is what we expect from the trans-series form of the EFT in \cite{joydeep}). (The simpler case of $a = 1$ was studied in \cite{desitter2} and here we generalize the story even further.) Unfortunately \eqref{tranpart} is still {\it not} the most generic ans\"atze for the metric that we can allow here. Once we consider the ${\rm M}_p$ scalings $-$ and not just the $g_s$ scalings $-$ the story gets a bit more involved. We will elaborate this later, but for the time being we shall assume the validity of the ans\"atze \eqref{tranpart} and 
similar modifications to all other components in \eqref{gwenphone} and \eqref{infinitypool} as mentioned above. This means, in the presence of the non-perturbative corrections to the metric components, it is easy to see that sub-dominant corrections given by $\vert f_l(t)\vert$ changes as:
\bg\label{ivbres}
\vert f_l(t(\bar{g}_s))\vert ~ \to ~ \vert{\rm F}_l(t(\bar{g}_s))\vert = \vert f_l(t(\bar{g}_s))\vert + {3\over 2l}
\sum_{a \in \mathbb{Z}} {n_a(l) \over \bar{g}_s^{a/3} \vert {\rm log}~\bar{g}_s\vert}, \nd
where $\bar{g}_s \equiv{g_s\over {\rm H}(y){\rm H}_o({\bf x})}$ and $l > 0$. For $l = 0$, $n_a(0) = 0$ and so does $l\vert f_l(t)\vert = 0\vert f_0(t)\vert = 0$. In fact \eqref{ivbres} suggests that the addition of non-perturbative corrections cannot change the NEC criterion of \cite{coherbeta2}. This may also be seen by first defining:
\bg\label{manucox}
{\cal H}(\bar{g}_s, l) \equiv \left({g_s\over {\rm H}(y){\rm H}_o({\bf x})}\right)^{-{\check{b}\over 3} + {2l\over 3} {\rm F}_l(t(\bar{g}_s))}, \nd
where ${\rm F}_l(t(\bar{g}_s))$ is as in \eqref{ivbres}, and $\check{b} = \big(8, 2 - {3} \beta(t), 2 - {3}\alpha(t), -4\big)$. Such a term could be part of the metric components, much like \eqref{tranpart}, and therefore we should ask how the derivative with respect to $\bar{g}_s$ behaves. One can easily show that:
\bg\label{lorocket}
{\partial{\cal H}(\bar{g}_s, l; a)\over \partial\bar{g}_s} \propto \left({g_s\over {\rm H}(y) {\rm H}_o({\bf x})}\right)^{-{\check{b}+3\over 3} + {2l\over 3}\vert{\rm F}_l(t(\bar{g}_s))\vert - {a\over 3} + \Delta + {\rm positive}}, \nd
where $\Delta \to 0$, $a > 0$ is the exponent in \eqref{ivbres}, and the additional positive exponent comes from $\vert f_l(t(\bar{g}_s))\vert$ which is expressed as a series in {\it positive} powers of $\bar{g}_s$. (We will discuss a generalization of the latter idea soon.) The most dominant term $-$ coming from say the Riemann curvature ${\bf R}_{ijij}$ for $(i, j) \in {\bf R}^2$ $-$ where we have used the notation from \eqref{katusigel}, would imply a $g_s$ scaling of:
\bg\label{nightwatch}
\left({g_s\over {\rm H}(y) {\rm H}_o({\bf x})}\right)^{-{14\over 3} + {2l\over 3}\vert{\rm F}_l(t(\bar{g}_s))\vert +{2p\over 3}\vert{\rm F}_p(t(\bar{g}_s))\vert +{2{\rm N}q\over 3}\vert{\rm F}_q(t(\bar{g}_s))\vert- {2a\over 3} + 2\Delta + {\rm positive}}\left({\partial \bar{g}_s\over \partial t}\right)^2, \nd
where ${\rm N} \in \mathbb{Z}$, and we will eventually sum over it as well as over $(l, p, q) \in (\mathbb{Z}, \mathbb{Z}, \mathbb{Z})$. Note that we have taken $(l, p, q) > 0$ because for $l = p = q = 0$ the non-perturbative correction vanishes (recall $n_a(0) = 0$ in \eqref{tranpart}). The above scaling differs from an equivalent term in \eqref{mca5rose} due to the presence of $-{2a\over 3}$ as well as other positive exponents, and therefore one might worry that the derivative of $\bar{g}_s$ with respect to the conformal time $t$ can also pick up an arbitrary {\it negative} power of $\bar{g}_s$. This is not true because of three reasons. {\Su One}, the $l = p= q = 0$ term would still contribute 
$\bar{g}_s^{+2/3}$ to the quantum series \eqref{botsuga}, so an arbitrary negative power of ${\partial \bar{g}_s\over \partial t}$ is not allowed. {\Su Two}, the form of $\bar{g}_s$ is actually fixed as:

{\scriptsize
\bg\label{cassnaud2}
\left(\sqrt{\Lambda(t)} \vert t \vert\right)^{2\over 2 - \beta(t)}  =  
{{g_s\over {\rm H}(y){\rm H}_o({\bf x})} \over 
\left[\sum\limits_{l = 0}^\infty b_l
\left({g_s\over {\rm H}(y){\rm H}({\bf x})_o}\right)^{{2l\vert{\red\check{\rm B}_l(t)}\vert\over 3}}\right]^{1\over {2-\beta(t)}}\left[
\sum\limits_{k = 0}^\infty h_k
\left({g_s\over {\rm H}(y){\rm H}({\bf x})_o}\right)^{2k\vert {\red\check{\rm F}_k(t)}\vert \over 3}\right]^{1\over 4 - 2\beta(t)}},\nd}
using \eqref{marapaug}, and therefore the derivative with respect to the conformal time $t$ cannot be arbitrary. And {\Su three}, even though ${\partial \bar{g}_s\over \partial t}$ is fixed, the possibility of a relative minus sign coming solely from large values of the factor $-{2a\over 3}$ is now remote because of the relative suppression of these terms by the exponential factor ${\rm exp}\Big[-\sum\limits_{a\in\mathbb{Z}} {n_a(l)\over \bar{g}_s^{a/3}}\Big]$ in \eqref{tranpart} and because of the presence of positive exponent in \eqref{lorocket}. (We will discuss more on this, and in particular on the ans\"atze \eqref{tranpart}, in the following sub-section.)

In writing \eqref{cassnaud2}, we have defined 
$\vert{\red\check{\rm B}_l(t)}\vert$ and $\vert {\red\check{\rm F}_k(t)}\vert$ exactly as in \eqref{ivbres} with the exception that now
we allow $\vert{\red\check{\beta}_l(t)}\vert$ and $\vert{\red\check{f}_l(t)}\vert$ from \eqref{cassnaud} to respectively change by the corresponding non-perturbative corrections. This justifies the non-violation of the NEC condition proposed in \cite{coherbeta2}. Note however that in the rest of the section we will not explicitly involve the subdominant terms including the non-perturbative effects, unless mentioned otherwise. 
 Since this does not change the aforementioned conclusion, we will leave a more detailed analysis for the diligent readers. 

There is however one worrisome feature \eqref{cassnaud2} related to the behavior of $\Lambda(t)$ from \eqref{marapaug}: the temporal derivative ${\partial \bar{g}_s\over \partial t}$ will also involve a term proportional to ${\partial \check{\Lambda}(t)\over \partial t}$. How should we deal with such a term? This brings us 
to the {\Su second} point, namely, we can reinterpret the sub-dominant corrections in a slightly different way by emphasizing that these corrections make the cosmological {constant} to slowly {\it vary} with respect to time. This is the point of view taken in section 5.3 of \cite{joydeep} and here we will elaborate on this in the context of the heterotic theories. The metric components along the $3+1$ dimensional spacetime in the heterotic $SO(32)$ case, involve two pieces:
\bg\label{steelepata}
\sum_{l = 0}^\infty b_l
\left({g_s\over {\rm H}(y){\rm H}({\bf x})_o}\right)^{-2 + \beta(t) +{2l\vert{\red\check{\rm B}_l(t)}\vert\over 3}} ~~{\rm and}~~ \left[
\sum_{k = 0}^\infty h_k
\left({g_s\over {\rm H}(y){\rm H}({\bf x})_o}\right)^{2k\vert {\red\check{\rm F}_k(t)}\vert \over 3}\right]^{1\over 2}, \nd
with the $({\red\check{\rm B}_l(t), \red\check{\rm F}_l(t)})$ factors coming from the equivalent $({{\rm B}_l(t), {\rm F}_l(t)})$ factors in M-theory via the duality chasing and incorporating the $(g_s, {\rm M}_p)$ corrections. In general we expect: ${\red\check{\rm B}_l(t)} = {\rm B}_l(t) + {\cal O}(g_s, {\rm M}_p)$ and ${\red\check{\rm F}_l(t)} = {\rm F}_l(t) + {\cal O}(g_s, {\rm M}_p)$. We can however resort to a slightly simplified case where ${\red\check{\rm B}_l(t)} = {\rm B}_l(t)$, but keep  ${\red\check{\rm F}_l(t)}$ and ${\rm F}_l(t)$ unequal. We can then rewrite \eqref{ccoleman} as:
\bg\label{leylajon}
\left({g_s\over {\rm H}(y){\rm H}({\bf x})_o}\right)^{-2 + \beta_e(t)}\left[
\sum_{k = 0}^\infty h_k
\left({g_s\over {\rm H}(y){\rm H}({\bf x})_o}\right)^{2k\vert {\red\check{\rm F}_k(t)}\vert \over 3}\right]^{1\over 2} 
= \begin{cases} ~ {1\over \Lambda \vert t\vert^2}\\
~~~~\\
~{1\over \Lambda(t) \vert t \vert^2},
\end{cases}
\nd
where, as before, we kept the option for both cosmological constant $\Lambda$ and dynamical dark energy $\Lambda(t)$ from \eqref{marapaug}. The other quantity $\beta_e(t)$ is now defined as:
\bg\label{paigest}
\beta_e(t) = \beta(t) - {1\over \vert\log~\bar{g}_s\vert}~\log\left[1 + 
\sum_{l = 1}^\infty b_l\left({g_s\over {\rm H}(y) {\rm H}_o({\bf x})}\right)^{2l\vert{\red{\rm B}_l(t)}\vert\over 3}\right], \nd
where we have taken $b_0 = 1$ without loss of generalities, We can also  define ${\red\check{\rm B}_l(t)} = {\red{\rm B}_l(t)} + {\red{\rm C}_{l}(t)}$ to accommodate the additional $(g_s, {\rm M}_p)$ corrections from the duality chasing, but here, to avoid too much complications, we will assume ${\red\check{\rm B}_l(t)} \approx {\red{\rm B}_l(t)}$ as alluded to earlier\footnote{We will address a more generic construction in section \ref{sec4.4}.}. The relation \eqref{leylajon} now leads to the following remarkable simplification:
\bg\label{johnsonsteel}
{g_s\over {\rm H}(y) {\rm H}_o({\bf x})} = \left(\sqrt{\Lambda}\vert t\vert\right)^{2\over 2 - \beta_e(t)}, \nd
with $\Lambda$ being the constant bare part of the dynamical dark energy \eqref{marapaug}, and $\beta_e(t)$ is as in \eqref{paigest} defined with ${\rm B}_l(t)$. One might ask whether we can simplify the relation between $\bar{g}_s$ and $t$ even further by expressing the relation simply as $\bar{g}_s = \sqrt{\Lambda}\vert t \vert$. If we do so, then \eqref{leylajon} leads to the following identification:

{\footnotesize
\bg\label{daisisaint}
1 + {\check{\Lambda}(t)\over \Lambda} = \left(\sqrt{\Lambda}\vert t \vert\right)^{-\beta(t)} \left[ 1 + \sum_{l = 1}^\infty b_l\left(\sqrt{\Lambda}\vert t \vert\right)^{2l\vert {\red\check{\rm B}_l(t)}\vert\over 3}\right]^{-1}  \left[ 1 + \sum_{k = 1}^\infty h_k\left(\sqrt{\Lambda}\vert t \vert\right)^{2k\vert {\red\check{\rm F}_l(t)}\vert\over 3}\right]^{-{1\over 2}}, \nd}
which is not only inconsistent with \cite{desibao}, but also the RHS blows up as we approach $t = 0$. One could then ask whether we can simplify the identification from \eqref{johnsonsteel} to the one 
in \eqref{covda}. The answer appears to be {\it no} because of the perturbative and the non-perturbative $\bar{g}_s$ corrections to the other metric and flux components. This will hopefully become clearer when we compute the $\bar{g}_s$ scalings of all the fluxes, two-forms and the curvature components. 

The remaining part of $\Lambda(t)$ from \eqref{marapaug}, namely $\check{\Lambda}(t)$, can now be fixed by plugging in \eqref{johnsonsteel} in \eqref{leylajon}. This reproduces the following explicit form for $\check{\Lambda}(t)$ in terms of $\check{\rm F}_l(t)$:
\bg\label{mmilan}
\check{\Lambda}(t) & = & \Lambda \sum_{n = 1}^\infty {1\over n!}\prod_{p = 0}^{n-1} \left(-{1\over 2} - p\right)\prod_{i=1}^n \sum_{l_i=0}^\infty h_{l_i} \left(\sqrt{\Lambda}\vert t\vert\right)^{ {4l_i\vert\check{\rm F}_{l_i}(t)\vert\over 3(2 - \beta_e(t)}} \nonumber\\
&= & -{\Lambda\over 2}\sum_{l = 1}^\infty h_l\left(\sqrt{\Lambda}\vert t\vert\right)^{4l\vert \check{\rm F}_l(t)\vert\over 3(2 - \beta_e(t))} + {3\Lambda\over 8}\sum_{l, m = 1}^\infty h_{lm}\left(\sqrt{\Lambda}\vert t\vert\right)^{4(l\vert \check{\rm F}_l(t)\vert + m\vert \check{\rm F}_m(t)\vert)\over 3(2 - \beta_e(t))} \nonumber\\ 
&& -
{5\Lambda\over 16}\sum_{l..n = 1}^\infty h_{lmn}\left(\sqrt{\Lambda}\vert t\vert\right)^{4(l\vert \check{\rm F}_l(t)\vert + m\vert \check{\rm F}_m(t)\vert + n\vert \check{\rm F}_n(t)\vert)\over 3(2 - \beta_e(t))}
\nonumber\\
&& + {35\Lambda\over 128}\sum_{l..p = 1}^\infty h_{lmnp}\left(\sqrt{\Lambda}\vert t\vert\right)^{4(l\vert \check{\rm F}_l(t)\vert + m\vert \check{\rm F}_m(t)\vert + n\vert \check{\rm F}_n(t)\vert + p\vert \check{\rm F}_p(t)\vert)\over 3(2 - \beta_e(t))} +
..., 
\nd
where $h_{kl..m} \equiv h_kh_l..h_m$, and $\check{\rm F}_k(t)$ are related to the coefficients appearing in \eqref{cassnaud2} and \eqref{leylajon}. Following \cite{desibao}, we expect $h_1 = -\vert h_1\vert$ and $h_0 \equiv 0$. It should also be clear from \eqref{mmilan} that $\check{\Lambda}(t) << \Lambda$ in the temporal domain $-{1\over \sqrt{\Lambda}} < t < 0$ because of the sub-dominant nature of the terms on the RHS of \eqref{mmilan}. The analysis following \eqref{mmilan} is a bit more complicated now compared to what we had in \cite{joydeep} because the exponents appearing on the LHS of \eqref{mmilan} are non-trivial functions of the conformal time $t$. Nevertheless the smallness of the sub-dominant contributions controlled by $\bar{g}_s$ and $\check{\rm F}_k(t)$, justifies the smallness of $\check{\Lambda}(t)$. 

While the aforementioned rewriting of \eqref{cassnaud2} as \eqref{mmilan} could in principle be connected to the temporal variation of the cosmological constant as recently discussed in \cite{desibao}, an alternative description with a {\it constant} cosmological constant exists. For the latter, we would simply have to replace $\Lambda(t)$ by $\Lambda$ in \eqref{cassnaud2}. Notice that this would make the relation between $\bar{g}_s$ and the conformal time $t$ much more complicated than
\eqref{johnsonsteel}. However,
it is too early to say which description fits nature because the results from \cite{desibao} are not conclusive enough for us to decide which ones to choose\footnote{In fact there are two possibilities here. It could be that in the third run in 2027, the statistical significance of the analysis of DESI-BAO \cite{desibao} increases beyond what is predicted now. In that case we can check whether the functional form we get for $\check{\Lambda}(t)$ in \eqref{mmilan} matches with the result from \cite{desibao}. However there is also a possibility that the statistical significance decreases over subsequent runs of DESI-BAO (see for example \cite{jabbari} for a prediction on this). In that case we can continue with the {\it constant} cosmological constant scenario that we get from \eqref{leylajon}, and use \eqref{leylajon} to determine the form for $g_s$.}. Thus the result from \eqref{mmilan} should be viewed with some caution.  

Nevertheless, the success of predicting the dynamical nature of the dark energy from our construction suggest that we can continue with \eqref{johnsonsteel} to express $\bar{g}_s$ in terms of the conformal time $t$. The reason why this worked out in a way that matches the analysis of \cite{desibao} is because of the {\it emergent} nature of the dark energy in our model. For example, the emergent nature dictates the following identification of the Borel resummed factor from the path-integral, {\it i.e.} from \cite{borel2, joydeep} (see also {\bf Table \ref{scarstan25}}):

{\scriptsize
\bg\label{mercifit}
\sum_{l = 0}^\infty b_l \left({g_s\over {\rm H}(y) {\rm H}_o({\bf x})}\right)^{-{8\over 3} + {2l\vert {\rm F}_l(t)\vert\over 3}}\eta_{\mu\nu} &= & 
\int_{k_{\rm IR}}^\mu d^{11} k~ {\sigma_{\mu\nu}(k)\over a^2(k)}~\psi_k(x, y, w)\nonumber\\
& \times& \sum_{\{{\bf s}\}}\left[{\mathbb{F}_{({\bf s}, \mu, \nu)}\over g_{({\bf s}, \mu, \nu)}^{1/l} {\rm M}_p^{2\kappa}}
\int_0^\infty d{\rm S} ~{\rm exp}\Bigg(-{{\rm S}\over g_{({\bf s}, \mu, \nu)}^{1/l}}\Bigg) {1\over 1 - {\cal A}_{({\bf s}, \mu, \nu)}{\rm S}^l}\right]_{\Su {\rm P. V}}, \nd}
with $(\mu, \nu) \in {\bf R}^{2, 1}$ and $\bar{g}_s$ is given by \eqref{johnsonsteel}. The aforementioned identification
consistently reproduces \eqref{montehan2} relating the constant part of $\Lambda(t)$ from \eqref{marapaug}, {\it i.e.} $\Lambda$, with the Borel-resummed principal value integral from \eqref{mercifit}. All the parameters appearing in 
\eqref{mercifit} are defined earlier after \eqref{montehan2}, and the remaining parameters are defined as follows: $\sigma_{\mu\nu}(k)$ is related to the Glauber-Sudarshan state $\vert\sigma\rangle$, $a^2(k)$ is the gravitational propagator and $\psi_k(x, y, w)$ is the $k$-th mode of the bulk wave-function. One can now make the following identification to determine the explicit form for $\sigma_{\mu\nu}(k)$:

{\footnotesize
\bg\label{privsocmooth}
\int_{k_{\rm IR}}^\mu d^{11} k ~{\sigma_{\mu\nu}(k)\over a^2(k)}~\psi_k(x, y, w) = 
{\bf FT}_{\rm RL}\left[({\rm M}_pt)^{-{8\over 3}} \sum_{l = 0}^\infty \left(\Lambda \vert t\vert^2\right)^{-{2\beta_e(t)\over 3} + {\cal O}(\beta_e^2(t)) + {2l\vert {\rm F}_l(t)\vert\over 3}}\right]\eta_{\mu\nu}, \nd}
where ${\bf FT}_{\rm RL}$ is the Fourier-transform with the usage of the Riemann-Lebesgue theorem to express it within the range $k_{\rm IR} < k < \mu$. (See \cite{borel2} for some illustrative examples.) Such a procedure reproduces the precise form of the Glauber-Sudarshan state $\vert\sigma\rangle$ along the ${\bf R}^{2, 1}$ directions needed for constructing the quasi de Sitter metric. 

Before ending this section, let us make a few observations. {\Su One}, knowing the functional form for $\sigma_{\mu\nu}(k)$ from \eqref{privsocmooth} does not fix the complete functional form for 
$\check{\Lambda}(t)$ because of the latter's dependence on $\check{\rm F}_l(t)$ (see \eqref{mmilan}) compared to the former's dependence on ${\rm F}_l(t)$. Since $\check{\rm F}_l(t) = {\rm F}_l(t) + {\cal O}({\rm M}_p, g_s)$, its value depends on the ${\rm M}_p$ (or $\alpha'$) and $g_s$ corrections encountered during the duality chasing from M-theory to the heterotic theories. {\Su Two}, we haven't discussed how the identifications are done in the ${\rm E}_8 \times {\rm E}_8$ side. Additionally, we haven't specified the internal manifold using similar identifications as above. Answering both of these would require us to first develop a few technical details. We will come back to this in section \ref{sec4.4}.

\begin{figure}[h]
\centering
\begin{tabular}{c}
\includegraphics[width=5in]{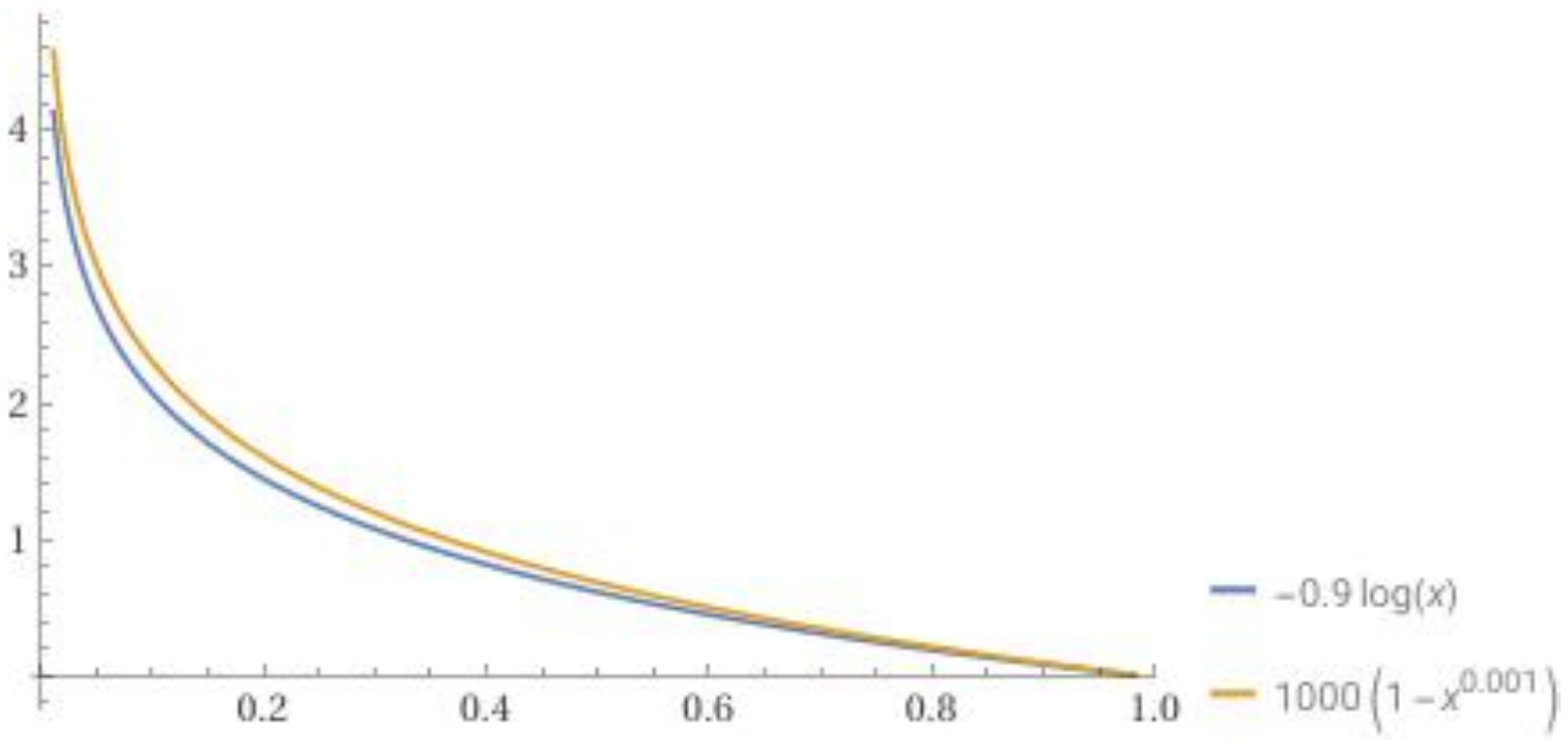}
\end{tabular}
\vskip-.5in
\caption[]{Plots of $y_1 = -\vert a\vert~\log~x$ (in blue) and $y_2 = {1-x^\Delta\over \Delta}$ (in orange), for $x = \bar{g}_s, a = 0.9$ and $\Delta = 0.001$. The two graphs exactly coincide when $a \to 1$, showing that one may replace $\vert\log~\bar{g}_s\vert$ by the function ${1 - \bar{g}_s^\Delta \over \Delta}$ for $\Delta << 1$ and $\bar{g}_s < 1$. In the limit $\bar{g}_s$ goes to zero as $\bar{g}_s \to \epsilon$, and $\Delta = \epsilon^{|b|}$ with $b > 1$, the two functions continue to coincide.}
\label{logdelta}
\end{figure}

\subsubsection{$g_s$ scalings of ${\rm F}_i(t)$ in the two heterotic theories \label{sec4.2.3}}

\begin{figure}[h]
\centering
\begin{tabular}{c}
\includegraphics[width=5in]{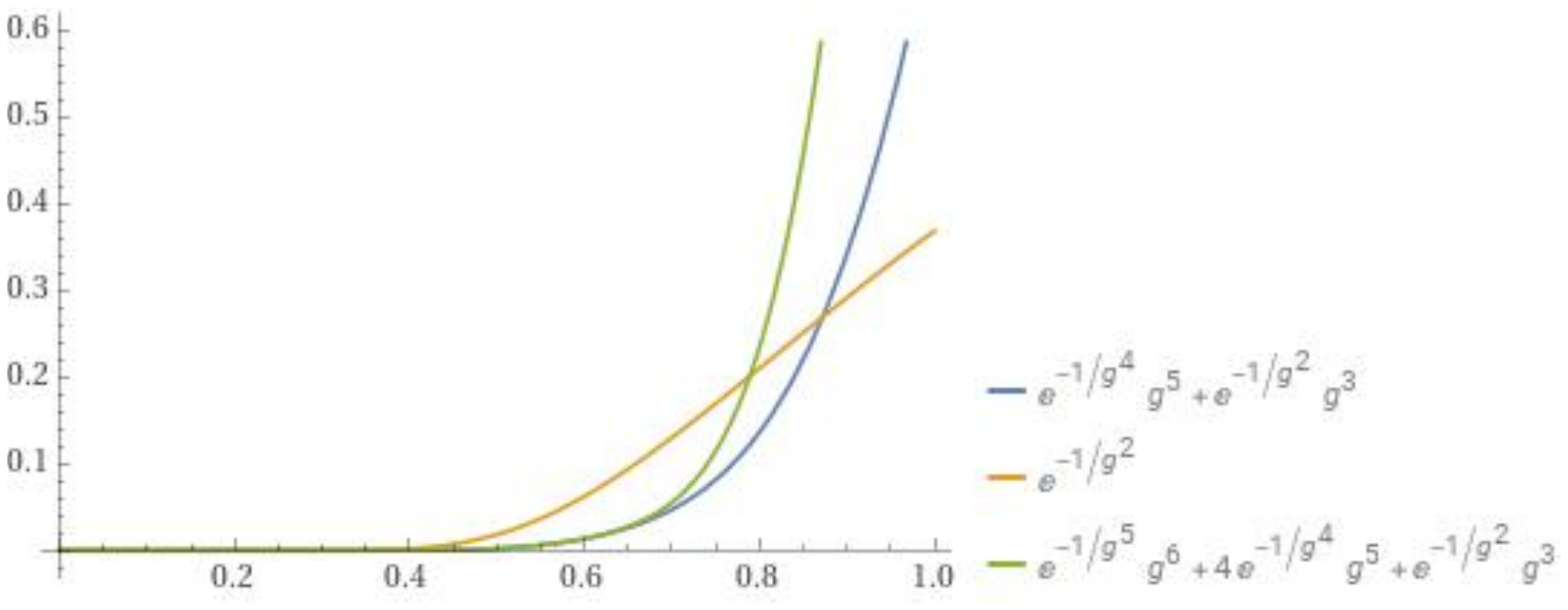}
\end{tabular}
\vskip-.8in
\caption[]{Plot of three functional forms for $\beta(t(g_s))$ from \eqref{tokio} with $\beta(t(g)) = {\rm exp}\left(-{1\over g^2}\right)$ (in orange), $\beta(t(g)) = g^3 {\rm exp}\left(-{1\over g^2}\right) + 
g^5 {\rm exp}\left(-{1\over g^4}\right)$ (in blue) and 
$\beta(t(g)) =g^3 {\rm exp}\left(-{1\over g^2}\right) +4 g^5 {\rm exp}\left(-{1\over g^4}\right) +g^6 {\rm exp}\left(-{1\over g^5}\right)$ (in green),
where $g \equiv \bar{g}_s = {\g_s\over {\rm H}(y) {\rm H}_o({\bf x})}$ for $0 \le g \le 1$. Note that as we increase the number of terms in the trans-series description of $\beta(t(g))$, it starts resembling the required condition from \eqref{afleis} more clearly.}
\label{boudi}
\end{figure}

In computing the $g_s$ scaling of the derivative of ${\cal H}$ with respect to $g_s$ in \eqref{lorocket}, we did not explain the origin of $\Delta$ appearing in the exponent. Of course it is not too hard to see that ${\rm log}~\bar{g}_s = \lim\limits_{\Delta \to 0} {1\over \Delta} \left(\bar{g}_s^\Delta - 1\right)$, which would explain the origin of this factor in \eqref{lorocket} and \eqref{nightwatch}. However this raises the question of how to deal with the ${1\over \Delta}$ factor. We also notice something else from {\bf Table \ref{rindrag}} (and from other subsequent tables): the ${\rm log}~\bar{g}_s$ factor always comes accompanied with either $\dot{\beta}(t)$ or $\ddot{\beta}(t)$ (recall that $\alpha(t) = -\beta(t)$ so a similar story with $\alpha(t)$ too).
Since $0 < \beta(t) < {2\over 3}$, we can define $\beta(t)$ as a series in $\bar{g}_s \equiv {g_s\over {\rm H}(y) {\rm H}({\bf x})}$ in the following way:
\bg\label{tokio}
\beta(t(g_s)) = \Delta \sum_{b}^\infty c_{b} \left( {g_s\over {\rm H}(y) {\rm H}({\bf x})}\right)^{b\over 3}~{\rm exp}\left(-\sum_d {n_{d}(b)\over \bar{g}_s^{d/3}}\right), \nd
with $0 < \Delta << 1$,  and constants $c_{b}$ and $n_{bd}$. (The difference is that, now $\Delta << 1$, but {\it not} zero. See {\bf figure \ref{logdelta}}.) Comparing \eqref{tokio} with \eqref{fermige}, we see interesting similarities. In fact \eqref{tokio} is a better representation of $\beta(t)$ than \eqref{fermige} because of it's trans-series form (see {\bf figure \ref{boudi}}). This also shows that $d \ge 0$ and $b \ge 0$, but the precise connection between $b$ and $d$ (and $a$ in \eqref{tranpart}) will be ascertained soon. Moreover since expressing the conformal time $t$ in terms of $g_s$ is much harder because the relation between them, as in \eqref{cassnaud2}, is non-trivial, the trans-series form of the expression suits better here. However some immediate comparison between \eqref{tokio} and \eqref{fermige} may be made. The coefficient $\beta_o$ in \eqref{fermige} may be identified with $\Delta$ in \eqref{tokio}. It is also clear that 
$\sigma^2|\epsilon|^2$ from \eqref{fermige} is connected to $n_{bd}$ here (for some appropriate choice of $d$) in \eqref{tokio}. Additionally, the trans-series form 
of expressing $\beta(t)$ simplifies many of the subsequent expressions. For example, once we define ${\rm log}~\bar{g}_s = \lim\limits_{\Delta << 1} {1\over \Delta} \left(\bar{g}_s^\Delta - 1\right)$, it is easy to see that:

{\footnotesize
\bg\label{bellachao}
\left( {g_s\over {\rm H}(y) {\rm H}({\bf x})}\right)^{\beta(t)} = 1 + \sum_{b}^\infty c_{b}\left[\left({g_s\over {\rm H}(y) {\rm H}({\bf x})}\right)^{\Delta + {b\over 3}} -\left({g_s\over {\rm H}(y) {\rm H}({\bf x})}\right)^{{b\over 3}}\right]{\rm exp}\left(-\sum_d {n_{d}(b)\over \bar{g}_s^{d/3}}\right) + {\cal O}(\beta^2), \nd}
which does not vanish precisely because $0 < \Delta << 1$. The RHS of \eqref{bellachao} is a series in powers of $\bar{g}_s$, so can become useful once we use the Schwinger-Dyson equation. Whether this would consistently fix the value of $\Delta$ within $0 < \Delta << 1$ will be discussed later. 
Because of the smallness of $\Delta$, we can also ignore the ${\cal O}(\beta^2)$ corrections in \eqref{bellachao}.

\begin{figure}[h]
\centering
\begin{tabular}{c}
\includegraphics[width=5in]{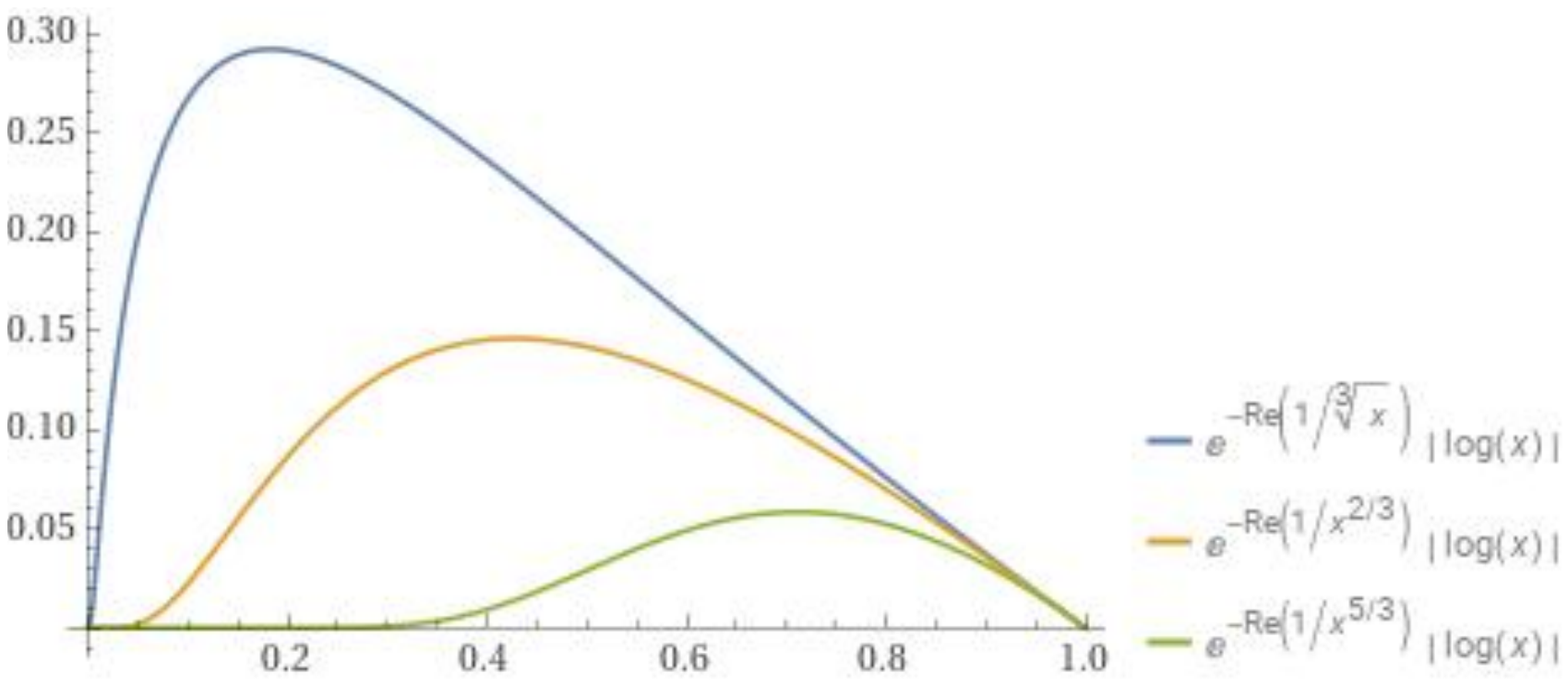}
\end{tabular}
\vskip-.8in
\caption[]{Plot of three functional forms for ${1\over \Delta}\vert \beta(t(x))~\log~x\vert$ with $x \equiv \bar{g}_s$ and $d = (1, 2, 5)$ from \eqref{tokio}. All the three functions remain smaller than 1 in the region $0 \le x \le 1$, and continues to do so $\forall d\in \mathbb{Z}$. This justifies the form in \eqref{stan22}.}
\label{whitelotus}
\end{figure}

\begin{figure}[h]
\centering
\begin{tabular}{c}
\includegraphics[width=5in]{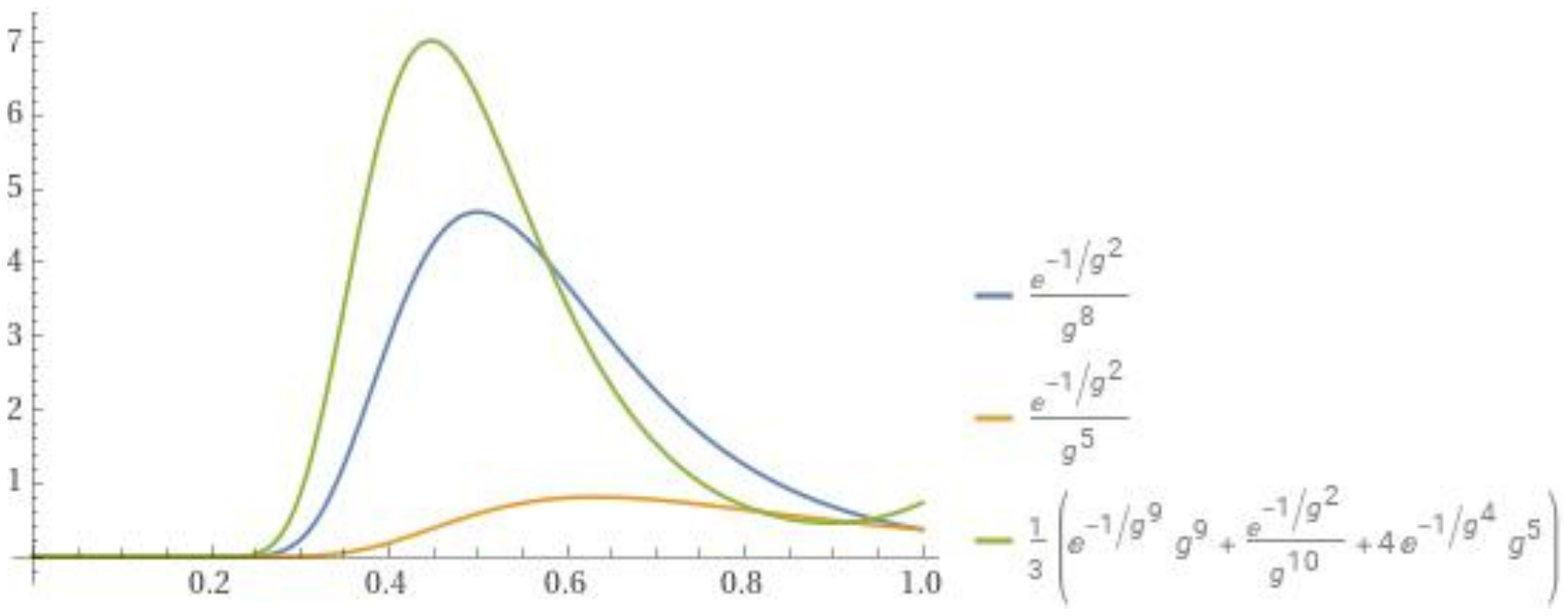}
\end{tabular}
\vskip-.8in
\caption[]{Plot of three functions $g^{-8} {\rm exp}\left(-{1\over g^2}\right)$ (in blue), $g^{-5} {\rm exp}\left(-{1\over g^2}\right)$ (in orange) and 
${1\over 3} g^{-10} {\rm exp}\left(-{1\over g^2}\right) +{4\over 3} g^5 {\rm exp}\left(-{1\over g^4}\right) +{1\over 3} g^9 {\rm exp}\left(-{1\over g^9}\right)$ (in green),
where $g \equiv \bar{g}_s = {\g_s\over {\rm H}(y) {\rm H}_o({\bf x})}$ for $0 \le g \le 1$. It should be clear that as $g \to 0$ all the three functions vanish implying that the denominator of \eqref{stan22} approaches 1 as $g \equiv \bar{g}_s \to 0$ as shown in {\bf figure \ref{whitelotus}}.}
\label{enmill}
\end{figure}

Before moving further with others terms involving $\dot{\beta}(t)$ and $\ddot{\beta}(t)$, we should note that the non-perturbative corrections to the metric components (say from \eqref{tranpart}) and the non-perturbative corrections to $\beta(t)$ (from \eqref{tokio}), are related by \eqref{cassnaud2} as:

{\scriptsize
\bg\label{cassnaud3}
 \left[\sum\limits_{l = 0}^\infty b_l
\left({g_s\over {\rm H}(y){\rm H}({\bf x})_o}\right)^{{2l\vert{\red\check{\rm B}_l(t)}\vert\over 3}}\right]^{1\over 2}\left[
\sum\limits_{k = 0}^\infty h_k
\left({g_s\over {\rm H}(y){\rm H}({\bf x})_o}\right)^{2k\vert {\red\check{\rm F}_k(t)}\vert \over 3}\right]^{1\over 4}
\begin{cases}  \sqrt{\Lambda} \vert t \vert\\
~~~~ \\
\sqrt{\Lambda(t)} \vert t \vert \end{cases}
=  
\left({g_s\over {\rm H}(y){\rm H}_o({\bf x})}\right)^{1 - {\beta(t)\over 2}},\nd}
with either a constant $\Lambda$, or a dynamical $\Lambda(t)$. Turning \eqref{cassnaud3} around should relate $g_s$ with the conformal time $t$ but, as mentioned earlier, this is easier said than done when we have a {\it constant} $\Lambda$. Complications come from $\beta_e(t)$ in \eqref{paigest}, as well as from $\check{\rm F}_l(t)$ in \eqref{leylajon} which is now a function of $t(\bar{g}_s)$. Simplifications happen when we consider \eqref{johnsonsteel} and relate $\bar{g}_s$ to $\beta_e(t)$. For such a case, we expect:
\bg\label{claudhostess}
{\partial \bar{g}_s\over \partial t} = {\sqrt{\Lambda}\over \left(1 - {\beta_e(t(g_s))\over 2} - {1\over 2}{\partial \beta_e(t(g_s))\over \partial \bar{g}_s}~\bar{g}_s \log~\bar{g}_s\right) \bar{g}_s^{\beta_e(t(g_s))/2}}, \nd
with $\beta_e(t(g_s))$ is from \eqref{paigest}. This is an exact answer, but is a bit harder to work with because $\check{\rm B}_l(t(g_s))$ involves additional non-perturbative corrections. If we ignore the log corrections in \eqref{paigest}, then $\beta_e(t(g_s))  \approx \beta(t(g_s))$. Alternatively, if we take $k = l = 0$ in \eqref{cassnaud3}), it would resemble \eqref{covda}, for which we have already worked out the derivative in \eqref{ishena} in terms of $\gamma_1$ and $\gamma_2$. However now that we have a more explicit form for $\beta(t(g_s))$, we can re-evaluate ${\partial \bar{g}_s\over \partial t}$ to give us:

{\scriptsize
\bg\label{stan22}
{\partial \bar{g}_s\over \partial t} & = & {\sqrt{\Lambda}\over 1 - {\beta(t(g_s))\over 2}(1 + \log~\bar{g}_s) - {1\over 2}{\partial \beta(t(g_s))\over \partial \bar{g}_s}~\bar{g}_s \log~\bar{g}_s}\\
& = & {\sqrt{\Lambda} \over 
1 - {1\over 2}\sum\limits_{b} c_{b}\big(\bar{g}_s^{b/3 + \Delta} -  \bar{g}_s^{b/3}\big){\rm exp}\big(-\sum\limits_d{n_{d}(b)\over \bar{g}_s^{d/3}}\big)- {1\over 6}\sum\limits_{b, d} c_{b}\big(\bar{g}_s^\Delta - 1\big)\big(b \bar{g}_s^{b/3} + n_{d}(b) d \bar{g}_s^{(b-d)/3}\big)~{\rm exp}\big(-\sum\limits_{d'}{n_{d'}(b)\over \bar{g}_s^{d'/3}}\big)}, \nonumber \nd}
which would come from first computing $\gamma_1$ and $\gamma_2$ factors from {\bf Table \ref{rindrag}} and then inserting them in \eqref{ishena} taking into account small $\beta$ from \eqref{tokio}. (Note that $\vert \beta(t(g_s))~\log~\bar{g}_s\vert < 1$ $\forall d\in\mathbb{Z}$ and $0 \le \bar{g}_s \le 1$ as shown in {\bf figure \ref{whitelotus}}.) The expression in \eqref{stan22} is a series in $\bar{g}_s^{b/3}, \bar{g}_s^{(b-d)/3}$ and their $\Delta$ extensions, accompanied by the non-perturbative factor
${\rm exp}\big(-{n_d(b)\over \bar{g}_s^{d/3}}\big)$. It is easy to infer that the denominator approaches 1 as $\bar{g}_s \to 0$ irrespective of the polynomial powers of $\bar{g}_s$ (see {\bf figure \ref{enmill}}). 
The worrisome feature however is the part of the series that goes as $\bar{g}_s^{(b-d)/3}$: for $d > b$, this will go as negative powers of $\bar{g}_s$ thus appearing to {\it violate} the no-go conditions from \cite{coherbeta2, joydeep}. However this is not exactly the case because of the following reason. Let us choose a specific value of $d$, say $d_o$, which is bigger than some chosen value of $b$, say $b_o$. To facilitate the computation, we can rewrite \eqref{stan22} as:

{\footnotesize
\bg\label{tundezotka}
&&{\partial \bar{g}_s\over \partial t} = {\sqrt{\Lambda} \over 
1 - \sum\limits_b f_1(b) - \sum\limits_{b, d} f_2(b, d)} = 
{\sqrt{\Lambda} \over 
1 - f_1(b_o)-f_2(b_o, d_o) - \sum\limits_{b\ne b_o} f_1(b) - \sum\limits_{z\ne z_o} f_2(b, d)}\nonumber\\
&& f_1(b) \equiv {1\over 2}c_{b}\big(\bar{g}_s^{b/3 + \Delta} -  \bar{g}_s^{b/3}\big){\rm exp}\Big(-\sum\limits_{d\ge 0}{n_{d}(b)\over \bar{g}_s^{d/3}}\Big)\\
&& f_2(b, d) = {1\over 6}c_{b}\big(\bar{g}_s^\Delta - 1\big)\big(b \bar{g}_s^{b/3} + n_{d}(b) d \bar{g}_s^{(b-d)/3}\big)~{\rm exp}\Big(-\sum\limits_{d'\ge 0}{n_{d'}(b)\over \bar{g}_s^{d'/3}}\Big), \nonumber
\nd}
where $z = (b, d)$ and $z_o = (b_o, d_o)$. For $d_o > b_o$, we can multiply the numerator and the denominator of ${\partial \bar{g}_s\over \partial t}$ from \eqref{tundezotka} by $\bar{g}_s^{(d_o - b_o)/3}$. This will convert $f_1(b_o)$ and $f_2(b_o, d_o)$ to:
\bg\label{tundezotka2}
&& f_1(b_o) ~ \to ~ {1\over 2} c_{b_o}\big(\bar{g}_s^{d_o/3 + \Delta} -  \bar{g}_s^{d_o/3}\big){\rm exp}\Big(-\sum\limits_{d\ge 0}{n_{d}(b_o)\over \bar{g}_s^{d/3}}\Big)\nonumber\\
&& f_2(b_o, d_o) = {1\over 6}c_{b_o}\big(\bar{g}_s^\Delta - 1\big)\big(b_o \bar{g}_s^{d_o/3} + n_{d_o}(b_o) d_o\big)\left[1 + {\rm exp}\Big(-\sum\limits_{d'> 0}{n_{d'}(b_o)\over \bar{g}_s^{d'/3}}\Big)\right], \nonumber \nd
where we have assumed, without any loss of generalities, the zero instanton sector to satisfy $n_0(b) = 0$. Clearly now ${\partial \bar{g}_s\over \partial t}$ has a Taylor expansion with positive powers of $\bar{g}_s$ so no potential conflicts result. Once we bound $d$ as $d \le d_{\rm max} \in \mathbb{Z}$ by imposing $n_{d > d_{\rm max}(b)} = 0$, 
$\forall b \in \mathbb{Z}$, then it is easy to see that ${\partial \bar{g}_s\over \partial t}$ remains positive definite by multiplying the numerator and the denominator of \eqref{stan22} by:
\bg\label{hatpathanda}
\left({g_s\over {\rm H}(y){\rm H}_o({\bf x})}\right)^{d_{\rm max} - b_{\rm min}}, \nd
where $b_{\rm min}$ is the minimum value of $b$, again leading to no conflict with the EFT. Since $b_{\rm min} = 0$, the multiplication factor simply becomes $\bar{g}_s^{d_{\rm max}}$.
However question can be raised for $b = d = 0$, because $\beta(t(g_s)) \sim \Delta > 0$ so would appear to {\it not} vanish as $g_s \to 0$. Fortunately this poses no problem as we shall soon see. Thus one may impose $b \ge 0$ and $d \ge 0$ or alternatively for the latter $d < d_{\rm max}$. One may also ask at this stage whether the bound that we discussed above is consistent with axion cosmology as well as the scalings of the quantum terms. In fact the former, {\it i.e.} the axion cosmology, provides stronger constraints on the choice of $b$ and $d$ in \eqref{stan22} such that $d$ can be negative without violating any of the EFT criteria, although $b$ has to be strictly greater than or equal to zero for consistency with EFT from \cite{coherbeta2, joydeep}. We will however take $(b, d) \ge 0$ as negative values of $d$ will just be another perturbative series. With this in mind, we can even generalize \eqref{tokio} to take the complete trans-series form as:
\bg\label{ryanfan}
\beta(t(g_s)) =\Delta \sum_{b, d} \sum_{l = 0}^\infty c^{(l)}_{b} \left( {g_s\over {\rm H}(y) {\rm H}({\bf x})}\right)^{{b\over 3} + {2l\over 3}}~{\rm exp}\left(-\sum_d{n_{d}(b, l)\over \bar{g}_s^{d/3}}\right), \nd
where $c_{d}^{(0)} \equiv c_{d}$ from \eqref{tokio} and $n_d \ge 0$ with $n_0 = 0$.
 The series in \eqref{ryanfan} may be understood in the following way. For a given value of $b \in \mathbb{Z}$, we choose all $d \le d_{\rm max}$ with $d_{\rm max} \in \mathbb{Z}$. Thus with $d_{\rm max}$ controlling the non-perturbative instanton-like effects and $b$ determining the perturbative fluctuations around such saddles, will  
 render \eqref{ryanfan} consistent with the EFT criteria from \cite{coherbeta2, joydeep}.

 The above conclusion is still not the full story, because the temporal derivative of $\bar{g}_s$ in \eqref{stan22} would involve additional terms coming from ${\partial\check{\rm B}_l(t(\bar{g}_s)) \over \partial \bar{g}_s}$ once we remove the approximation $\beta_e(t(g_s)) \approx \beta(t(g_s))$. Using similar definition as in \eqref{ivbres} for the two variables would imply couple of things. {\Su One}, the $g_s$ powers of the perturbative terms could always be bigger or smaller than the corresponding powers in the non-perturbative terms. In particular this means, for example in \eqref{tranpart}, $a$ could be bounded as $a \le a_{\rm max} \in \mathbb{Z}$ and as such should have no relation to the powers of $\bar{g}_s$ in the perturbative series for $f_l(t(g_s))$ for any value of $l > 0$. And {\Su two}, all the sub-dominant pieces in the metric (and also the flux) components $-$ as in \eqref{gwenphone} and \eqref{infinitypool} including their dual companions $-$ should be expressed in the trans-series form as \eqref{ryanfan} implying that they would all be controlled by the small parameter $\Delta$. This will be consistent with the expectation from \eqref{mmilan} once we consider the concept of a temporally varying dark energy \eqref{marapaug} more seriously \cite{desibao}.

\begin{figure}[h]
\centering
\begin{tabular}{c}
\includegraphics[width=5in]{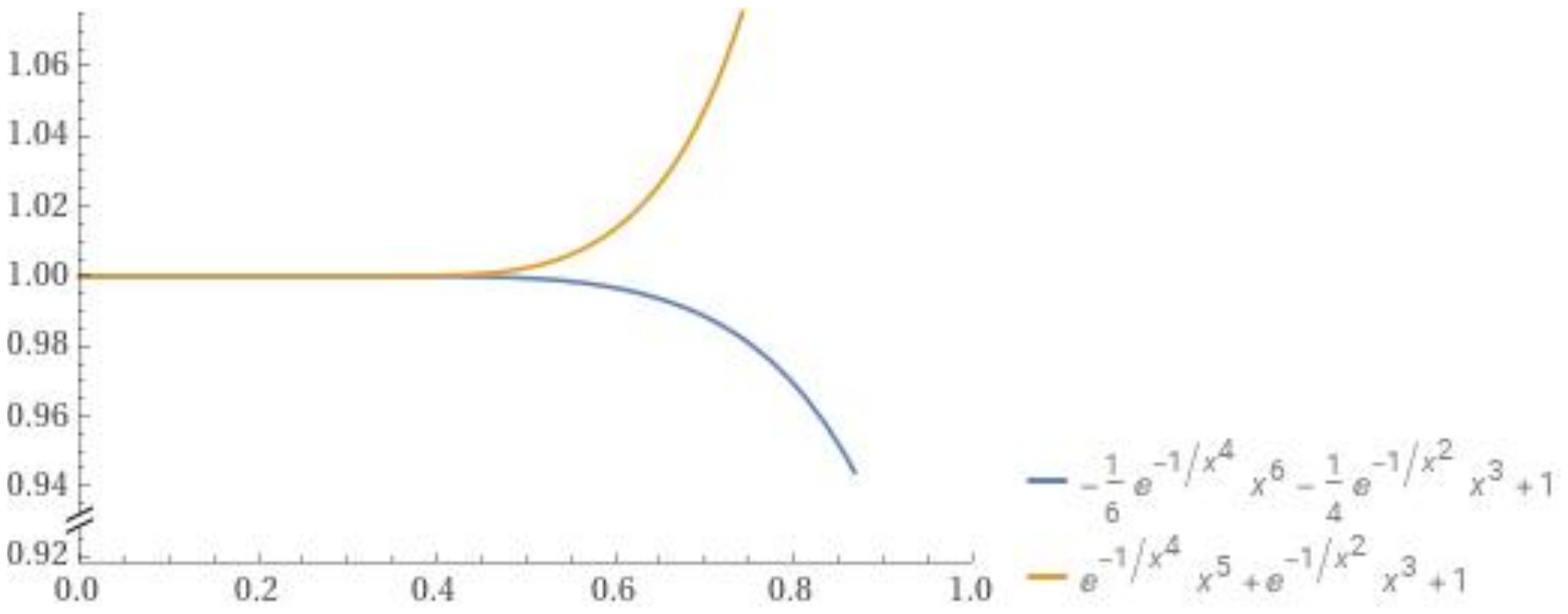}
\end{tabular}
\vskip-.8in
\caption[]{Plot of the two warp-factors $\hat{\alpha}(x) = 1 + x^3 {\rm exp}\left(-{1\over x^2}\right) + x^5 {\rm exp}\left(-{1\over x^4}\right)$ (in orange), and 
$\hat\beta(x) = 1 - {1\over 4} x^{3} {\rm exp}\left(-{1\over x^2}\right) - {1\over 6} {\rm exp}\left(-{1\over x^4}\right)$ (in blue),
where $x \equiv \bar{g}_s = {\g_s\over {\rm H}(y) {\rm H}_o({\bf x})}$ for $0 \le x \le 1$. It should be clear that as $x \to 0$, $\hat\beta(x) = \hat\alpha(x) \to 1$, which would make the two warp-factors ${\rm F}_1$ and ${\rm F}_3$ equal to each other for the ${\rm E}_8 \times {\rm E}_8$ case. However, as we discuss here, this doesn't provide the full answer because the correct choice for $\hat\alpha(t)$ and $\hat\beta(t)$ satisfying the constraint coming from the axionic decay appears from \eqref{horseman}. Nevertheless the plot will be very useful in determining the precise behavior of $\hat\alpha(x)$ and $\hat\beta(x)$.}
\label{kelsika}
\end{figure} 

 The story in heterotic ${\rm E}_8 \times {\rm E}_8$ theory is similar, which we shall elaborate further in section \ref{sec4.4}. A hint of how the two warp-factors $\hat\alpha(t)$ and $\hat\beta(t)$ should behave was originally pointed out in \eqref{hanasmit}. The plot of their behavior appears in {\bf figure \ref{e81}}.  Unfortunately they do not represent ${\rm F}_1(t)$ and ${\rm F}_3(t)$ accurately at very late time as shown in {\bf figures \ref{F1F3}} and {\bf \ref{f1f32}}. 
 Once we take the constraints from the axionic cosmology, a possible modification of \eqref{hanasmit} appears in \eqref{TIhostsmey} and \eqref{sirimey}. While the late-time behavior, from \eqref{kitagTI}, is well under control, the early-time behavior in \eqref{TIhostsmey} is still an ans\"atze despite the plots in {\bf figure \ref{alphabetahats}} and {\bf figure \ref{f11f33}} being well motivated. 
 Additionally, the expressions in \eqref{TIhostsmey} are expressed using conformal time $t$, but inverting it and writing in terms of $g_s$ will be harder because the relation between\footnote{A more correct way would be to express $t \equiv t(\bar{g}_s)$. But we will use $t \equiv t(g_s)$ to avoid unnecessary clutter.} $t$ and $g_s$ is more involved now (see the second relation in \eqref{covda} expressed without incorporating the sub-dominant corrections). However, motivated from the above arguments, we can use trans-series forms to express $\hat\alpha(t({g}_s))$ and $\hat\beta(t({g}_s))$ in the following way:
\bg\label{horseman}
&&  \hat\alpha(t(g_s)) = \Delta\sum_{b} \sum_{l = 0}^\infty  e^{(l)}_{b} \left( {g_s\over {\rm H}(y) {\rm H}({\bf x})}\right)^{{b\over 3} + {2l\over 3}}~{\rm exp}\left(-\sum_d{n_{d}(b, l)\over \bar{g}_s^{d/3}}\right)\nonumber\\
&&  \hat\beta(t(g_s)) = \Delta \sum_{b'} \sum_{l = 0}^\infty f^{(l)}_{b'}\left( {g_s\over {\rm H}(y) {\rm H}({\bf x})}\right)^{{b'\over 3} + {2l\over 3}}~{\rm exp}\left(-\sum_{d'}{m_{d'}(b', l)\over \bar{g}_s^{d'/3}}\right), \nd
where both $\hat\alpha(t)$ and $\hat\beta(t)$ go to zero as ${g}_s \to 0$, and we take $e^{(l)}_{b} > f^{(l)}_{b'}$ so that $\hat\alpha(t) > \hat\beta(t)$ for $t \to -{1\over \sqrt{\Lambda}}$. We have also chosen $(b, d) \ge 0$ and $(b', d') \ge 0$ because the late time behavior is now controlled by the function $\hat\gamma_o(t)$ from \eqref{kitagTI}. (This will be demonstrated soon.) To see if the series representation in \eqref{horseman} makes sense, let us consider the special case where $\hat\gamma_o(t) \equiv \hat\gamma_o$, {\it i.e.} a constant. This will be close to the choices \eqref{afleis4} and \eqref{afleis44}, albeit inconsistent with the axionic cosmology. If we ignore the latter constraint, then we can rewrite \eqref{horseman} in the following suggestive way:
 \bg\label{horseman1}
&&  \hat\alpha(t(g_s)) = \bar{\Delta}(g_s) \left[1 +  \sum_{b } \sum_{l = 0}^\infty \vert \bar{e}^{(l)}_{b}\vert \left( {g_s\over {\rm H}(y) {\rm H}({\bf x})}\right)^{{b\over 3} + {2l\over 3}}~{\rm exp}\left(-\sum_d{n_{d}(b, l)\over \bar{g}_s^{d/3}}\right)\right]\\
&&  \hat\beta(t(g_s)) = \bar{\Delta}(g_s)\left[1 -  \sum_{b'} \sum_{l = 0}^\infty \vert \bar{f}^{(l)}_{b'}\vert \left( {g_s\over {\rm H}(y) {\rm H}({\bf x})}\right)^{{b'\over 3} + {2l\over 3}}~{\rm exp}\left(-\sum_{d'}{m_{d'}(b', l)\over \bar{g}_s^{d'/3}}\right)\right], \nonumber \nd
where we can keep $\vert \bar{e}^{(l)}_{b}\vert >\vert \bar{f}^{(l)}_{b}\vert$, but is not essential. The relative sign difference will help us to maintain the required condition for ${\rm F}_1(t)$ and ${\rm F}_3(t)$. We have also defined $\bar{\Delta}(g_s)$ to be another trans-series in $\bar{g}_s \equiv {g_s\over {\rm H}(y) {\rm H}_o({\bf x})}$, but here we can keep it just a constant, {\it i.e.} $\bar{\Delta}(g_s) \approx \Delta$. An example is shown in {\bf figure \ref{kelsika}} with some generic choice of the relative coefficients. The ans\"atze works very well in the string-frame, 
because both $\hat\alpha(t(g_s))$ and $\hat\beta(t(g_s))$ quickly 
approach $\Delta$. However to resolve issues with the Einstein-frame, one needs to further modify \eqref{horseman1}, so that at late time both
$\hat\beta(t(g_s))$ approach $5\hat\gamma_o$. To go from \eqref{horseman1} to either \eqref{horseman} or \eqref{TIhostsmey}, we resort back to the functional dependence $\bar{\Delta}(g_s)$ so that for $\vert t\vert > \vert \epsilon_1\vert$ where $\epsilon_1$ is temporal domain defined in \eqref{TIhostsmey}. This is basically similar to what we discussed earlier in \eqref{afleis4} and \eqref{afleis44}, and in the subsequent plots in {\bf figures \ref{alphabetahats}}, {\bf \ref{f11f33}} and {\bf \ref{axdecay}} in the string frame. (One should also compare them to case 2 in {\bf figure \ref{6cases}} which gets modified at late time.) The point is that the trans-series form for $\beta(t(g_s))$ in \eqref{ryanfan} and the trans-series form for $(\hat\alpha(t(g_s)), \hat\beta(t(g_s))$ in \eqref{horseman} are the correct way to represent the parameters in the two heterotic theories. We expect, only in these trans-series forms, they would solve the corresponding Schwinger-Dyson equations. 

One may also verify whether the late-time behavior, like the one shown in \eqref{kitagTI} for $\hat\gamma_o(t)$, can appear naturally in our choice of $(\hat\alpha(t), \hat\beta(t))$. For this, let us express $\Lambda t^2$ in terms of the type IIA coupling $g_s$ in the following way:
\bg\label{tailordais}
\Lambda t^2 = \left({g_s\over {\rm H}(y) {\rm H}_o({\bf x})}\right)^{2 - \left({2v-1\over 3v^2}\right)\hat\alpha(t) - {\hat\beta(t)\over 3v^2}}, \nd
which is typically complicated to invert because $t$ appears on both sides of the mapping. This means substituting \eqref{tailordais} in \eqref{kitagTI} can in principle make it harder to express in terms of 
$g_s$\footnote{For example taking \eqref{kitagTI} at face value and plugging in the expressions for $\hat\alpha(t(g_s))$ and $\hat\beta(t(g_s))$ from \eqref{horseman} gives us the following relation for $\hat\gamma_o(t(g_s))$ in terms of $\bar{g}_s \equiv {g_s\over {\rm H}(y){\rm H}_o({\bf x})}$:

{\scriptsize
\bg\label{deppmey}
\hat\gamma_o(t(g_s)) &= & {9v^3\over 3v-2} \Bigg({c_o\Delta\over 
3vc_o \Delta + \Big(6v^2 - \sum\limits_{b \ge pd}\prod\limits_d\left((2v-1)\Delta e^{(l)}_{b} + \Delta f^{(l)}_{b}\right)\bar{g}_s^{b/3+2l/3} e^{-n_{d}(b)/\bar{g}_s^{d/3}}\Big)(1-\bar{g}_s^\Delta)}\Bigg)\nonumber\\
& = & {3c_o v\Delta\over 6v-4} \sum\limits_{{\rm N} = 0}^\infty 
\left(\bar{g}_s^\Delta + {1\over 6v^2} \sum\limits_{b \ge pd}\prod\limits_d\left((2v-1)\Delta e^{(l)}_{b} + \Delta f^{(l)}_{b}\right)\bar{g}_s^{b/3+2l/3} e^{-n_{d}(b)/\bar{g}_s^{d/3}}\Big)(1-\bar{g}_s^\Delta) -{3c_o\Delta \over 6v}\right)^{\rm N}, \nonumber
\nd}
where we have utilized the trans-series representation of $\hat\alpha(t(g_s))$ and $\hat\beta(t(g_s))$ from \eqref{horseman} but kept $n_{d}(b) \approx m_{d}(b)$ for simplicity (generalization is possible but we will not attempt it here). In this way of writing $\hat\gamma_o(t(g_s))$, it is slightly hard to see how it goes to zero  when $g_s \to 0$. An easier way will be discussed below wherein we will see that, using the {\it correct choice} of the functions $\hat\alpha(t)$ and $\hat\beta(t)$, the relation for $\hat\gamma_o(t(g_s))$ can be simplified even further.}, the simplification occurs because in the temporal domain $-\epsilon_1 < t < 0$, both $\hat\alpha(t)$ and $\hat\beta(t)$ behave as $\hat\gamma_o(t)$ and $(4v-3) \hat\gamma_o(t)$ respectively as seen from \eqref{afleis45}. Therefore in the temporal domain $-\epsilon_1 < t < 0$,
$\Lambda t^2 = \left({g_s\over {\rm H}(y){\rm H}_o({\bf x})}\right)^{2-\left({6v-4\over 3v^2}\right)\hat\gamma_o(t)}$, and plugging this in \eqref{kitagTI} gives us the following remarkably simple form for $\hat\gamma_o(t(g_s))$:
\bg\label{maryjane}
\hat\gamma_o(t(g_s)) = {3v c_o\over (6v-4) \vert\log~\bar{g}_s\vert}, \nd
which is expressed as an inverse logarithm of $\bar{g}_s = {g_s\over {\rm H}(y){\rm H}_o({\bf x})}$. Interestingly, there is another solution for $\hat\gamma_o(t(g_s))$ given by $\hat\gamma_o(t(g_s)) = {3v^2\over 3v -2}$ which is just a constant. A plot of it appears in {\bf figure \ref{kelsika}} but, as mentioned earlier, this does not satisfy the boundary condition so we will only take \eqref{maryjane}. Moreover, from \eqref{horseman}, one possibility becomes:

{\footnotesize
\bg\label{isabferr}
n_{d}(b, l) = m_{d'}(b', l) = 0, ~~~b = b' = 3\Delta {\rm N}, ~~~
e_b^{(l)} = {3vc_o \over 6v - 4}~\delta_{l0}, ~~~ f_{b'}^{(l)} = 
{3(4v-3)v c_o\over 6v - 4}~\delta_{l0}, \nd}
in \eqref{horseman}, showing that in the temporal domain $-\epsilon_1 < t < 0$, $\hat\alpha(t(g_s))$ and $\hat\beta(t(g_s))$ are only represented by the perturbative series with zero contributions from the non-perturbative terms. Assuming \eqref{isabferr} is possible, it would appears that we can express \eqref{maryjane} as:
\bg\label{maryjane2}
\hat\gamma_o(t(g_s)) =  {3vc_o\Delta \over 6v-4} \sum_{b \ge 0}^\infty \left({g_s\over {\rm H}(y){\rm H}_o({\bf x})}\right)^{b\over 3}, \nd
with $0 < \Delta << 1$. Unfortunately, 
this could not be the right behavior for $\hat\alpha(t(g_s))$ and $\hat\beta(t(g_s))$ because it doesn't quite reproduce both the inverse log behavior in \eqref{maryjane} for $-\epsilon_1 < t < 0$ {\it and} the expected early time behavior for $-{1\over \sqrt{\Lambda}} < t \le -\epsilon_1$. Additionally, since $\bar{g}_s < 1$, the limiting process that leads to \eqref{isabferr} from \eqref{maryjane} can be subtle.
Could one use the non-perturbative pieces that go as ${\rm exp}\left(-{n_d(b)\over \bar{g}_s^{d/3}}\right)$, along with the perturbative ones in \eqref{maryjane2}, to reproduce the behavior of $\hat\gamma_o(t)$ from \eqref{maryjane}? This would then exactly reproduce the form that we anticipated in 
\eqref{horseman} for the ${\rm E}_8 \times {\rm E}_8$ case, and the form 
\eqref{ryanfan} for  the $SO(32)$ case.

The aforementioned discussion may be quantified a bit more precisely, which would also help us to answer the question raised. Our aim would be to reproduce the behavior for $\hat\alpha(t)$ and $\hat\beta(t)$ from \eqref{afleis45} in say the string frame keeping in mind the late-time behavior \eqref{maryjane}. We first note that \eqref{maryjane} may be rewritten as:
\bg\label{snowmilena}
{1\over \vert\log~\bar{g}_s(\epsilon_1)\vert} = {\Delta\over 1 - \bar{g}_s^\Delta(\epsilon_1)} = 
\Delta\left(1 + \bar{g}_s^\Delta + \bar{g}_s^{2\Delta} + ... + 
\bar{g}_s^{{\rm N}\Delta} + ..\right)\big\vert_{|t| =\epsilon_1}, \nd
where $\Delta << 1$ and $\bar{g}_s(\epsilon_1) < 1$ with $\Delta\vert\log~\bar{g}_s(\epsilon_1)\vert << 1$, which makes the series well-defined\footnote{Although we already showed in {\bf figure \ref{logdelta}} the validity of an expression like \eqref{snowmilena} for all $\bar{g}_s$, we would still like to continue with this line of thought to dispel any doubts on the functional forms in \eqref{horseman} and \eqref{ryanfan} for the ${\rm E}_8 \times {\rm E}_8$ and the $SO(32)$ cases respectively.}. Question is, how do we construct the behavior for $\hat\alpha(t)$ in the full temporal domain 
$-{1\over\sqrt{\Lambda}} < t < 0$ now? A hint comes from  the graphical representation in {\bf figure \ref{kelsika}}. We can make the following ans\"atze for the functional form for $\hat\alpha(g_s)$ and $\hat\beta(g_s)$:

{\footnotesize
\bg\label{areseriesmey}
\hat\alpha(g_s(\epsilon_1)) & = & \Delta\left({c_0(\bar{g}_s)\over \vert c_0\vert} + 
{c_1(\bar{g}_s) \bar{g}_s^\Delta \over \vert c_1\vert} + {c_2(\bar{g}_s) \bar{g}_s^{2\Delta} \over \vert c_2\vert} + ... \right)\bigg\vert_{|t| = \epsilon_1} = \Delta\sum_{{\rm N} = 0}^\infty{c_{\rm N}(\bar{g}_s) \bar{g}_s^{{\rm N}\Delta} \over \vert c_{\rm N}\vert}\bigg\vert_{|t| = \epsilon_1}\nonumber\\
\hat\beta(g_s(\epsilon_1)) & = & \Delta\left({b_0(\bar{g}_s)\over \vert b_0\vert} + 
{b_1(\bar{g}_s) \bar{g}_s^\Delta \over \vert b_1\vert} + {b_2(\bar{g}_s) \bar{g}_s^{2\Delta} \over \vert b_2\vert} + ... \right)\bigg\vert_{|t| = \epsilon_1}= \Delta \sum_{{\rm N} = 0}^\infty{b_{\rm N}(\bar{g}_s) \bar{g}_s^{{\rm N}\Delta} \over \vert b_{\rm N}\vert}\bigg\vert_{|t| = \epsilon_1},
\nd}
where each term of \eqref{snowmilena} is modulated by the corresponding 
$c_{\rm N}(\bar{g}_s)$ and $b_{\rm N}(\bar{g}_s)$ functions such that they behave with respect to $\bar{g}_s$ in exactly the way depicted in {\bf figure \ref{kelsika}}, namely, they reach constant values of $\vert c_{\rm N}\vert$ and $\vert b_{\rm N}\vert$ respectively in the temporal domain $-\epsilon_1 < t < 0$.  We can then make the following ans\"atze for the functional form for $c_{\rm N}(\bar{g}_s)$ and $b_{\rm N}(\bar{g}_s)$:

{\footnotesize
\bg\label{uvuru}
c_{\rm N}(\bar{g}_s(\epsilon_1)) & = & \vert c_{\rm N}\vert + \sum_{j > 0} \sum_{l = 1}^\infty c^{(j)}_{\rm N}(\epsilon_1) \left({g_s\over {\rm H}(y){\rm H}_o({\bf x})}\right)^{{j\Delta\over 3} + {2l\over 3}} ~{\rm exp}\left(-\sum_{d > 0} {n_d(j, l; {\rm N})\over \bar{g}_s^{d/3}}\right)\bigg\vert_{|t| = \epsilon_1}\nonumber\\
b_{\rm N}(\bar{g}_s(\epsilon_1)) & = & \vert b_{\rm N}\vert + \sum_{j > 0} \sum_{l = 1}^\infty b^{(j)}_{\rm N}(\epsilon_1) \left({g_s\over {\rm H}(y){\rm H}_o({\bf x})}\right)^{{j\Delta\over 3} + {2l\over 3}} ~{\rm exp}\left(-\sum_{d > 0} {m_d(j, l; {\rm N})\over \bar{g}_s^{d/3}}\right) 
\bigg\vert_{|t| = \epsilon_1}, \nd}
where the coefficients $c_{\rm N}^{(j)}(\epsilon_1)$ and $c_{\rm N}^{(j)}(\epsilon_1)$ are chosen in such a way that the trans-series part of \eqref{uvuru} dies off very fast at $ t \ge -\epsilon_1$, much like what we see from {\bf figure \ref{kelsika}}. This means we are no longer required to restrict $\bar{g}_s$ at the value $\bar{g}_s(\epsilon_1)$ because each of the series in \eqref{areseriesmey} would quickly approach the inverse log behavior. Plugging \eqref{uvuru} in \eqref{areseriesmey} then provides the following $\bar{g}_s$ expansion\footnote{We will use $\hat\alpha(t(g_s)), \hat\beta(t(g_s))$ (or sometime $\hat\alpha(g_s), \hat\beta(g_s)$ to avoid clutter) henceforth.} for $\hat\alpha(t(g_s))$ and $\hat\beta(t(g_s))$:

{\scriptsize
\bg\label{mercyfitnes}
\hat\alpha(t(g_s)) & = & \Delta \sum_{{\rm M}= 0}^\infty\left({g_s\over {\rm H}(y){\rm H}_o({\bf x})}\right)^{{\rm M}\Delta} + \Delta\sum_{{\rm M} = 0}^\infty \sum_{j > 0} \sum_{l = 1}^\infty {c^{(j)}_{\rm M}(\epsilon_1)\over \vert c_{\rm M}\vert} \left({g_s\over {\rm H}(y){\rm H}_o({\bf x})}\right)^{{j\Delta\over 3} + {\rm M}\Delta + {2l\over 3}} ~{\rm exp}\left(-\sum_{d > 0}^{d_{\rm max}} {n_d(j, l; {\rm M})\over \bar{g}_s^{d/3}}\right) \nonumber\\
\hat\beta(t(g_s)) & = & \Delta \sum_{{\rm M}= 0}^\infty\left({g_s\over {\rm H}(y){\rm H}_o({\bf x})}\right)^{{\rm M}\Delta} + \Delta\sum_{{\rm M} = 0}^\infty\sum_{j > 0} \sum_{l = 1}^\infty {b^{(j)}_{\rm M}(\epsilon_1)\over \vert b_{\rm M}\vert} \left({g_s\over {\rm H}(y){\rm H}_o({\bf x})}\right)^{{j\Delta\over 3} + {\rm M}\Delta + {2l\over 3}} ~{\rm exp}\left(-\sum_{d > 0}^{d_{\rm max}} {m_d(j, l; {\rm M})\over \bar{g}_s^{d/3}}\right), \nonumber\\ \nd}
where as before we have assumed an upper limit $d_{\rm max}$ for the instanton series and the only remnants of $\vert t \vert = \epsilon_1$ are in the coefficients $c^{(j)}_{\rm M}(\epsilon_1)$ and $b^{(j)}_{\rm M}(\epsilon_1)$. To see how the above expressions resemble the trans-series forms advocated in \eqref{horseman}, we can rewrite for example 
$\hat\alpha(g_s)$ in the following suggestive way:

{\scriptsize
\bg\label{marceahbby}
\hat\alpha(t(g_s)) & = & \Delta \sum_{{\rm N}= 0}^\infty \bar{g}_s^{{\rm N}\Delta} + \Delta \sum_{{\rm N} = 0}^\infty\left[\sum_{{\rm M}} \sum_{j > 0} \sum_{l = 1}^\infty {c^{(j)}_{\rm M}(\epsilon_1)\over \vert c_{\rm M}\vert} \bar{g}_s^{({j\over 3} + {\rm M})\Delta + {2l\over 3}}~\delta_{{\rm M} + {j\over 3}, {\rm N}} ~{\rm exp}\left(-\sum_{d > 0}^{d_{\rm max}} {n_d(j, l; {\rm M})\over \bar{g}_s^{d/3}}\right)\right], \nd}
with a similar expression for $\hat\beta(t(g_s))$ by making the replacements $c^{(j)}_{\rm M}(\epsilon_1) \to b^{(j)}_{\rm M}(\epsilon_1), c_{\rm M} \to b_{\rm M}$ and $n_d \to m_d$ in \eqref{marceahbby}. The Kronecker delta identifies all combinations of $(j, {\rm M})$ with ${\rm N}$, {\it i.e.} imposes ${\rm M} + {j\over 3} = {\rm N}$. Such a rearrangement now brings \eqref{mercyfitnes} to precisely resemble \eqref{horseman}! Therefore 
\eqref{areseriesmey}, or \eqref{mercyfitnes}, is a rewriting of \eqref{horseman} but with a key difference: we can now see that in the temporal domain $-\epsilon_1 < t < 0$ we do reproduce the functional form for $\hat\gamma_o(t(g_s))$ from \eqref{maryjane}. Moreover, \eqref{areseriesmey} is a smooth function that interpolates between the two ranges from \eqref{afleis44}, thus eliminating the tension we had earlier with the functional choice \eqref{TIhostsmey}. Similar story goes through for the $SO(32)$ case, and a detailed derivation of which is left for the diligent reader.

Returning to the $SO(32)$ case and looking at {\bf Table \ref{rindrag}}, we note that the terms that occur frequently are 
$\vert\dot\beta\vert \vert {\rm log}~\bar{g}_s\vert$ and 
$\vert\ddot\beta\vert \vert {\rm log}~\bar{g}_s\vert$ (including the sole appearances of $\vert\dot\beta\vert$). It is also interesting that, while in the metric components \eqref{tranpart} the $\bar{g}_s^{|f_l(t)|}$ and the ${\rm exp}\left(-{n_d\over \bar{g}_s^{d/3}}\right)$ pieces are {\it sub-dominant} contributions; in the definition of $\beta(t(g_s))$ from \eqref{ryanfan}, they appear as the {\it dominant} contributions\footnote{Exactly similar criteria are for the metric components \eqref{tranpart}, including the trans-series form for the other components, and $(\hat\alpha(t(g_s)), \hat\beta(t(g_s)))$ in \eqref{horseman} for the ${\rm E}_8 \times {\rm E}_8$ case.}. The sub-dominant ones are now of the form $\bar{g}_s^{2l/3}$. This will be important when we compute the functional forms of $\mathbb{F}_i(t)$ from {\bf Table \ref{firzacut3}} in section \ref{sec4.3}. Meanwhile, observing the fact that:
\bg\label{ennadaddy}
\ddot{\beta} = {\partial \beta \over \partial \bar{g}_s} \cdot{\partial^2 \bar{g}_s \over \partial t^2} + {\partial^2 \beta\over \partial \bar{g}_s^2} \left({\partial \bar{g}_s\over \partial t}\right)^2, \nd
we can compute the two functions $\vert\dot\beta\vert \vert {\rm log}~\bar{g}_s\vert$ and 
$\vert\ddot\beta\vert \vert {\rm log}~\bar{g}_s\vert$ in terms of powers of $\bar{g}_s$ and the non-perturbative term  ${\rm exp}\left(-{n_d(b, l)\over \bar{g}_s^{d/3}}\right)$. Plugging in the trans-series form of $\beta(t)$ from \eqref{ryanfan} in \eqref{ennadaddy}, we get: 

{\scriptsize
\bg\label{lovsik}
 \vert\dot\beta\vert \vert {\rm log}~\bar{g}_s\vert &= & \sum_{b, d} \sum_{l = 0}^\infty \left(\hat{c}^{(l)}_{b} \bar{g}_s^{b/3 + 2l/3 - 1} + 
\check{c}^{(l)}_{b} \bar{g}_s^{b/3 + 2l/3 - d/3 - 1}\right) \vert \bar{g}_s^\Delta - 1\vert {\rm exp}\left(-\sum_d {n_d(b, l)\over \bar{g}_s^{d/3}}\right) \cdot {\partial \bar{g}_s\over \partial t} \\
 \vert\ddot\beta\vert \vert {\rm log}~\bar{g}_s\vert &= & \sum_{b, d} \sum_{l = 0}^\infty \vert \bar{g}_s^\Delta - 1\vert {\rm exp}\left(-\sum_d {n_d(b, l)\over \bar{g}_s^{d/3}}\right)\Bigg\{
\left(\hat{c}^{(l)}_{b} \bar{g}_s^{b/3 + 2l/3 - 1} + 
\check{c}^{(l)}_{b} \bar{g}_s^{b/3 + 2l/3 - d/3 - 1}\right)  \cdot {\partial^2 \bar{g}_s\over \partial t^2}\nonumber\\
& +&  \left[\bar{g}_s^{b/3 + 2l/3 - 2}
\left(\hat{c}^{(l; 1)}_{b}  + 
\check{c}^{(l; 1)}_{b} \bar{g}_s^{- d/3}\right) 
+ \bar{g}_s^{b/3 + 2l/3 -d/3 - 2} \left(\hat{c}^{(l; 2)}_{b}  + 
\check{c}^{(l; 2)}_{b} \bar{g}_s^{- d/3}\right)\right]  \cdot \left({\partial \bar{g}_s\over \partial t}\right)^2\Bigg\},\nonumber
\nd}
where expectedly, both the expressions are small with the smallness being governed by $0 < g_s < 1$ as well as $\vert \bar{g}^\Delta_s - 1\vert$ (see {\bf figure \ref{logdelta}}). Moreover as predicted before, they are also expressed using polynomial powers of $\bar{g}_s$ in addition to the non-perturbative piece ${\rm exp}\left(-{n_d(b, l)\over \bar{g}_s^{d/3}}\right)$. The other parameters appearing in \eqref{lovsik} are defined in the following way:

{\scriptsize
\bg\label{serja}
&& \hat{c}^{(l)}_{b} = \left({b\over 3} + {2l\over 3}\right)c^{(l)}_{b}, ~~~~ \check{c}^{(l)}_{b} = {1\over 3} n_d(b, l) d c^{(l)}_{b}\\
&& \hat{c}^{(l;1)}_{b} = \left({b\over 3} + {2l\over 3} - 1\right)\hat{c}^{(l)}_{b}, ~~ \hat{c}^{(l;2)}_{b} = {1\over 3} n_d(b, l) d \hat{c}^{(l)}_{b}, ~~ \check{c}^{(l; 1)}_{b} = \left({b\over 3} + {2l\over 3} - {d\over 3} - 1\right)\check{c}^{(l)}_{b}, ~~ \check{c}^{(l; 2)}_{b} = {1\over 3} n_d(b, l) d \check{c}^{(l)}_{b}, \nonumber \nd}
which are by definition constant factors independent of $g_s$ with $c_b$ and $n_d$ defined earlier\footnote{Henceforth $n_d$ could either imply $n_d(b)$ or $n_d(b, l)$ depending on how generic we want.}. The signs of these coefficients do not matter at this stage and the two functions in \eqref{lovsik} are well-defined in the regime $0 \le g_s < 1$, and vanishing near $g_s \to 0$ (see the green curve in {\bf figure \ref{enmill}}).

\subsubsection{$g_s$ scalings of the curvature tensors in M-theory\label{sec4.2.4}}

However even with these aforementioned simplifications of choosing $\beta(t)$ etc., to determine the scaling of the curvature tensor, one needs to tread a bit more carefully. We will first compute the temporal derivative of ${\bf g}_{mn}$, where $(m, n) \in {\cal M}_4$, to get:
\bg\label{corsage1}
{\bf g}_{mn, 0}({\bf x}, y; g_s) \equiv -{\bf g}^{(1)}_{mn}({\bf x}, y) \left({g_s\over {\rm HH}_o}\right)^{\mathbb{A}(\alpha)} \pm 
{\bf g}^{(2)}_{mn}({\bf x}, y) \left({g_s\over {\rm HH}_o}\right)^{\mathbb{B}(\alpha)}, \nd
where the sign ambiguity is put in to take care of the sign ambiguity of the temporal derivative of $\alpha(t)$, and we have ignored the sub-leading corrections\footnote{This may be easily incorporated in $\mathbb{A}(\alpha)$ and $\mathbb{B}(\alpha)$ by changing $\alpha(t(g_s))$ to the following:
\bg\label{maharaja} \alpha(t(g_s)) ~ \rightarrow ~ \alpha(t(g_s)) + \vert {\rm F}_l^{(\alpha)}(t(\bar{g}_s))\vert, \nonumber \nd 
where ${\rm F}_l^{(\alpha)}(t(\bar{g}_s))$ takes the same form as in \eqref{ivbres} but with $f_l(t(\bar{g}_s) \to f^{(\alpha)}_l(t(\bar{g}_s)$
and $n_a(l) \to n_a^{(\alpha)}(l)$. A similar change could be incorporated to $\mathbb{A}(\beta)$ and $\mathbb{B}(\beta)$ by changing $\beta$ in the aforementioned way but with ${\rm F}_l^{(\beta)}(t(\bar{g}_s))$. We will address this in more details in section \ref{sec4.4}. \label{desjardintagre}} (including the non-perturbative ones alluded to above \eqref{ennadaddy}). This is also reflected in the definitions of $\mathbb{A}(\alpha)$ and $\mathbb{B}(\alpha)$, more appropriately in the latter, in the following way:
\bg\label{corsage2}
&& \mathbb{B}(\alpha) = -{2\over 3} + \alpha(t) - {\log\left(\mp \dot\alpha(t)\vert\log~\bar{g}_s\vert\right)\over \vert\log~\bar{g}_s\vert} \nonumber\\
&& \mathbb{A}(\alpha) = -{5\over 3} + \alpha(t) + \gamma_{1, 2}[\beta(t)] - 
{\log\left({2\over 3} -\alpha(t)\right)\over \vert\log~\bar{g}_s\vert} \ , \nd 
where $\gamma_{1, 2}[\beta(t)]$ $-$ which are only functions of $\beta(t)$ $-$ are defined in \eqref{ishena}; $\bar{g}_s \equiv {g_s\over {\rm H}(y){\rm H}_o({\bf x})} = \bar{g}_s[\beta(t)]$ from \eqref{covda}; and for the $SO(32)$ case, since $\alpha(t) = -\beta(t)$, $\dot\alpha(t) = -\dot\beta(t) = \vert\dot\beta(t)\vert$ because of an earlier constraint discussed after \eqref{sekdrem18}, the sign choice is negative in \eqref{corsage1} and positive in \eqref{corsage2}. See {\bf Table \ref{rindrag}} for explicit values. (The ${\rm E}_8 \times {\rm E}_8$ generalization of \eqref{corsage2} appears in \eqref{tifeni}.) In a similar vein:
\bg\label{corsage3}
{\bf g}_{\rho\sigma, 0}\equiv -{\bf g}^{(1)}_{\rho\sigma}({\bf x}, y) \left({g_s\over {\rm HH}_o}\right)^{\mathbb{A}(\beta)} \pm 
{\bf g}^{(2)}_{\rho\sigma}({\bf x}, y) \left({g_s\over {\rm HH}_o}\right)^{\mathbb{B}(\beta)}\nd
where $(\rho, \sigma) \in {\cal M}_2$. Now, since $0 < \beta(t) < {2\over 3}$, and $g_s << 1$, we can easily infer that $\vert\log~g_s\vert >> \log\left({2\over 3} \pm \beta(t)\right)$. Also since the derivatives are small, we expect $\vert\log~g_s\vert >> \log\big(\vert\dot\beta\vert\vert\log~g_s\vert\big)$. In this way, all the log corrections to \eqref{corsage2} appear to be small. However as shown in {\bf Table \ref{rindrag}} this doesn't imply that we can ignore them, at least not at this stage, so we should keep track of them. The temporal derivatives of the other components of the metric take the following form:
\bg\label{corsage4}
{\bf g}_{{\cal CD}, 0}({\bf x}, y; g_s) = \widetilde{\bf g}_{\cal CD}({\bf x}, y)\left({g_s\over {\rm HH}_o}\right)^{a_{\cal CD} - 1 + \gamma_{1, 2}(t)}, \nd
where $({\cal C, D}) \in {\bf R}^{2, 1} \times {\mathbb{T}^2\over {\cal G}}$, $\gamma_{1, 2}(t)$ as in \eqref{ishena}, and $a_{\cal CD}$ is the $g_s$ scaling of the respective metric components. As we see, there are no additional log corrections to these components at the dominant level compared to what we had in \eqref{corsage1} and \eqref{corsage3} (see also footnote \ref{desjardintagre}).

\begin{table}[tb]  
 \begin{center}
\renewcommand{\arraystretch}{1.5}
\begin{tabular}{|c||c||c|}\hline $\mathbb{F}_i(t)$  & $g_s$ scalings \\ \hline\hline
$\mathbb{F}_1(t)$ & $2\mathbb{A}(\alpha) + {8\over 3}, ~ 2\mathbb{B}(\alpha) + {8\over 3}, ~\mathbb{A}(\alpha) + \mathbb{B}(\alpha) + {8\over 3}$\\ \hline
$\mathbb{F}_2(t)$ & $\mathbb{A}(\alpha) + \mathbb{A}(\beta) + {8\over 3}, 
~\mathbb{A}(\alpha) + \mathbb{B}(\beta) + {8\over 3}, 
~ \mathbb{B}(\alpha) + \mathbb{A}(\beta) + {8\over 3}, 
~ \mathbb{B}(\alpha) + \mathbb{B}(\beta) + {8\over 3}$\\ \hline
$\mathbb{F}_3(t)$ & $\mathbb{A}(\beta), ~ \mathbb{B}(\beta)$ \\ \hline
$\mathbb{F}_4(t)$ & $\mathbb{A}(\alpha) + \beta(t) - \alpha(t), ~ \mathbb{B}(\alpha) + \beta(t) - \alpha(t)$\\ \hline
$\mathbb{F}_5(t)$ & $2\mathbb{A}(\beta) + {8\over 3}, ~ 2\mathbb{B}(\beta) + {8\over 3}, ~\mathbb{A}(\beta) + \mathbb{B}(\beta) + {8\over 3}$\\ \hline
$\mathbb{F}_6(t)$ & $\mathbb{A}(\alpha), ~ \mathbb{B}(\alpha)$ \\ \hline
$\mathbb{F}_7(t)$ & $\mathbb{A}(\beta) - \beta(t) + \alpha(t), ~ \mathbb{B}(\beta) - \beta(t) + \alpha(t)$\\
\hline
\end{tabular} 
\renewcommand{\arraystretch}{1}
\end{center}
 \caption[]{\Su The ${g_s\over {\rm H H}_o}$ scalings of the $\mathbb{F}_i(t)$ and the $\mathbb{C}_i[{\cal A}(\alpha)]$ factors appearing in {\bf Tables \ref{firzacut}} and {\bf \ref{firzacut2}} for the $SO(32)$ case. The definitions of $\bar{g}_s$ is in \eqref{covda}, $\gamma_{1, 2}$ are in \eqref{ishena}, $\gamma_{[3, .., 11]}$ are in \eqref{ishena2}, and $(\mathbb{A}(\sigma), \mathbb{B}(\sigma))$ are in \eqref{corsage2}. See {\bf Table \ref{rindrag}} for notations and {\bf Table \ref{privsocmey}} for the ${\rm E}_8 \times {\rm E}_8$ case.} 
  \label{firzacut3}
 \end{table}

The temporal derivatives of the metric components are important as they contribute to the curvature tensors. This is in general harder than what we had in \cite{desitter2, coherbeta, coherbeta2} because of the temporal derivatives of the exponents of $g_s$ themselves. Nevertheless we can work them out in enough details to be useful for the study of EOMs later in section \ref{sec6}. As an example, the  
curvature tensor $-$ and we will take one specific set of components for illustrative purpose $-$ scales in the following way:

{\footnotesize
\bg\label{moncrime}
{\bf R}_{mnpq}({\bf x}, y; g_s(t)) & = & {\bf R}^{(1)}_{mnpq}({\bf x}, y) \left({g_s\over {\rm HH}_o}\right)^{-{2\over 3} + 2\alpha(t) + 2\gamma_{1, 2}(t) -{2\log({2\over 3}-\alpha)\over \vert\log~\bar{g}_s\vert}}\\
& + & {\bf R}^{(2)}_{mnpq}({\bf x}, y)\left({g_s\over {\rm HH}_o}\right)^{-{2\over 3} +2\alpha(t) + \gamma_1(t) + \gamma_2(t) - {2\log({2\over 3} - \alpha(t))\over \vert\log~\bar{g}_s\vert}}\nonumber\\
& + & {\bf R}^{(3)}_{mnpq}({\bf x}, y) \left({g_s\over {\rm HH}_o}\right)^{{1\over 3}+ 2\alpha(t) + \gamma_{1, 2}(t) -{\log\left(\mp\dot\alpha({2\over 3}-\alpha)\vert\log~\bar{g}_s\vert\right)\over \vert\log~\bar{g}_s\vert}} \nonumber\\
& + & {\bf R}^{(4)}_{mnpq}({\bf x}, y) \left({g_s\over {\rm HH}_o}\right)^{-{2\over 3} + \alpha(t)} + 
{\bf R}^{(5)}_{mnpq}({\bf x}, y) \left({g_s\over {\rm HH}_o}\right)^{{4\over 3} + 2\alpha(t)}\nonumber\\
& + & {\bf R}^{(6)}_{mnpq}({\bf x}, y) \left({g_s\over {\rm HH}_o}\right)^{-{2\over 3} + 2\alpha(t) -\beta(t)} + {\bf R}^{(7)}_{mnpq}({\bf x}, y) \left({g_s\over {\rm HH}_o}\right)^{{4\over 3} + 2\alpha(t) - {2\log\left(\mp\dot\alpha\vert\log~\bar{g}_s\vert\right)\over \vert\log~\bar{g}_s\vert}}, \nonumber\nd}
where only the dominant scalings are shown, with the first entry for example is supposed to imply two different scalings, $-{2\over 3} + 2\alpha(t) + 2\gamma_1(t)-{2\log({2\over 3}-\alpha)\over \vert\log~\bar{g}_s\vert}$ and $-{2\over 3} + 2\alpha(t) + 2\gamma_2(t)-{2\log({2\over 3}-\alpha)\over \vert\log~\bar{g}_s\vert}$ as evident from \eqref{ishena} and \eqref{covda}. In {\bf Table \ref{firzacut}} the aforementioned scaling of the curvature component is expressed as:
\bg\label{vicrepe}
{\rm dom}\left(-{2\over 3} + \alpha(t), ~{4\over 3} + 2\alpha(t),~ -{2\over 3} + 2\alpha(t) - \beta(t), ~\mathbb{F}_1(t)\right)\nd
where $\mathbb{F}_1(t)$ may be read up from \eqref{moncrime} or from {\bf Table \ref{firzacut3}} (with a detailed derivation in section \ref{sec4.3}); and 
the notation for the {\it dominant} scaling being defined using  ${\rm dom}({\rm A}, {\rm B}, {\rm C}) = {\rm min}({\rm A}, {\rm B}, {\rm C}) = {\rm A}$ if 
${\rm A} < {\rm B}$ and ${\rm A} < {\rm C}$ (with no preferred ordering) within the temporal domain where $g_s < 1$. (For $g_s \ge 1$ ${\rm dom}({\rm A}, {\rm B}, {\rm C}) = {\rm max}({\rm A}, {\rm B}, {\rm C})$ within the appropriate temporal domain.) Using the aforementioned notations, the $g_s$ scalings of all the {\it on-shell} curvature 
tensors\footnote{The {\it on-shell} nature is a bit subtle here because we have used {\it off-shell} analysis, {\it i.e.} path integrals, to determine these so-called on-shell components. The remaining off-shell components are integrated away resulting in non-local counter-terms that was studied in great details in \cite{desitter2, joydeep}. We will discuss more on this aspect, following \cite{joydeep}, a bit later.} are given in {\bf Tables \ref{firzacut}}, {\bf \ref{firzacut5}} and 
{\bf \ref{firzacut2}}. Note that in computing these scalings, all permutations of the curvature indices are taken into account. For example ${\bf R}_{mn\rho\sigma}$ appearing in the second row of {\bf Table \ref{firzacut}} contains the set of six distinct components: ${\bf R}_{mn\rho\sigma}, {\bf R}_{m\rho n\sigma}, {\bf R}_{\rho\sigma mn}, {\bf R}_{\rho m \sigma n}, {\bf R}_{m\rho\sigma n}$, and ${\bf R}_{\rho mn \sigma}$ including other possibilities when $m \longleftrightarrow n, \rho \longleftrightarrow \sigma$.

\subsubsection{$g_s$ scalings of the G-flux components in M-theory \label{sec4.2.5}}

The G-flux components follow the definition advocated in \eqref{katusigel}, and here it might be more illuminating if we define them using the three-form fields. This will actually matter when we have a temporal derivative, for example as in ${\bf G}_{\rm 0ABD}$, and therefore it will be instructive to split-up the G-flux components in two sets: ${\bf G}_{\rm ABCD}({\bf x}, y; g_s)$ and ${\bf G}_{\rm 0ABD}({\bf x}, y; g_s)$ where $({\rm A, B}) \in {\bf R}^{2} \times {\cal M}_4 \times {\cal M}_2 \times {\mathbb{T}^2\over {\cal G}}$, ${\bf x} \in {\bf R}^2$ and $y \in {\cal M}_4 \times {\cal M}_2$. We will start by defining the three-form field components in the following way:
\bg\label{hannalat}
&& \langle {\bf C}_{\rm 0BD}\rangle_\sigma \equiv {\bf C}_{\rm 0BD}({\bf x}, y; g_s) = \sum_{k = 0}^\infty {\cal C}^{(k)}_{\rm 0BD}({\bf x}, y)\left({g_s\over {\rm H}(y){\rm H}_o({\bf x})}\right)^{l_{\rm 0B}^{\rm D}(t) + {2k\vert {\rm L}_{\rm 0B}^{\rm D}(k; t)\vert \over 3}}\nonumber\\
&& \langle{\bf C}_{\rm ABD}\rangle_\sigma \equiv {\bf C}_{\rm ABD}({\bf x}, y; g_s) = \sum_{k = 0}^\infty {\cal C}^{(k)}_{\rm ABD}({\bf x}, y)\left({g_s\over {\rm H}(y) {\rm H}_o({\bf x})}\right)^{l_{\rm AB}^{\rm D}(t) + {2k\vert {\rm L}_{\rm AB}^{\rm D}(k; t)\vert\over 3}},
\nd
where, compared to \cite{desitter2, coherbeta, coherbeta2}, we have made the scaling $l_{\rm AB}^{\rm D}(t)$ and $l_{\rm 0B}^{\rm D}(t)$ to be time-dependent, where $t \equiv t(g_s)$ as given in \eqref{covda}, and the orderings of ${\rm 0, A, B}$ and ${\rm D}$ in $l_{\rm AB}^{\rm D}$ or 
$l_{\rm 0B}^{\rm D}(t)$ are irrelevant\footnote{In other words, $l_{\rm AB}^{\rm D} = l_{\rm AD}^{\rm B} = l_{\rm BD}^{\rm A} = l_{\rm DA}^{\rm B} = ..$, {\it i.e.} the scaling remains the same for any permutation of the indices of a given three-form field.}. 
In a similar vein as $\beta(t)$ in \eqref{ryanfan}, we expect {\it dominant} non-perturbative corrections\footnote{By {\it dominant} we mean the corrections that appear directly for the $g_s$ scalings of the netric and the flux components that do not involve any ${\rm M}_p$ corrections. As such they will differ from the sub-dominant corrections to the metric and fluxes that will involve ${\rm M}_p$ corrections. \label{pogmomsi}} to appear in the definitions of $l_{\rm AB}^{\rm D}(t)$ and $l_{\rm 0B}^{\rm D}(t)$ in the following way:
\bg\label{ryanfan2}
&& l_{\rm 0B}^{\rm D}(t(g_s)) = \hat{\rm l}_{\rm 0B}^{\rm D} + 
\Delta \sum_{b, d} \sum_{l = 0}^\infty \left(l_{\rm 0B}^{\rm D}\right)^{l}_{b} \left( {g_s\over {\rm H}(y) {\rm H}({\bf x})}\right)^{{b\over 3} + {2l\over 3}}~{\rm exp}\left(-\sum_d {n_d(b, l)\over \bar{g}_s^{d/3}}\right)\\
&& l_{\rm AB}^{\rm D}(t(g_s)) = \hat{\rm l}_{\rm AB}^{\rm D} + 
\Delta \sum_{b, d} \sum_{l = 0}^\infty \left(l_{\rm AB}^{\rm D}\right)^{l}_{b} \left( {g_s\over {\rm H}(y) {\rm H}({\bf x})}\right)^{{b\over 3} + {2l\over 3}}~{\rm exp}\left(-\sum_d {m_d(b, l)\over \bar{g}_s^{d/3}}\right), \nonumber \nd
where $\hat{\rm l}_{\rm AB}^{\rm D}$ and $\hat{\rm l}_{\rm 0B}^{\rm D}$ are constants and so are the coefficients $\left(l_{\rm 0B}^{\rm D}\right)^{l}_{b}$ and $\left(l_{\rm AB}^{\rm D}\right)^{l}_{b}$. The coefficient $\Delta$ is the same one that appeared earlier in \eqref{ryanfan} with $0 < \Delta << 1$; and we expect $n_d$ to be generically different from $m_d$ $\forall d \in \mathbb{Z}$. The $l = 0$ form the dominant contributions and $l > 0$ are the sub-dominant ones.  

On the other hand the flux components will have {\it sub-dominant} non-perturbative corrections, just as in the metric components (see for example \eqref{tranpart}). These sub-dominant non-perturbative corrections may be absorbed in the definition of ${\rm L}_{\rm AB}^{\rm D}(k; t)$ and ${\rm L}_{\rm 0B}^{\rm D}(k; t)$, much like how we defined them for the metric components in \eqref{ivbres}\footnote{As cautioned below \eqref{tranpart}, even with an extension like in \eqref{ivbres} the generic scaling is much more non-trivial because it involves ${\rm M}_p$ corrections as described in footnote \ref{pogmomsi}. We will however not worry about them here and will address them later when we study the Schwinger-Dyson equations in section \ref{sec6}.}.
The modes in $k$ are defined using functions ${\rm L}_{\rm AB}^{\rm D}(k, t)$ and ${\rm L}_{\rm 0B}^{\rm D}(k, t)$ whose functional forms can be analyzed using the Schwinger-Dyson's equations \eqref{mariesole} as well as using flux quantizations, Bianchi identities and anomaly cancelations. This is intentional and resonates with the time-dependence of $\alpha_k(t)$ and $\beta_k(t)$ for the metric components of the $SO(32)$ case (and somewhat similar scalings for the ${\rm E}_8 \times {\rm E}_8$ case) in \eqref{gwenphone}. 
Note that the representations in \eqref{hannalat} and \eqref{ryanfan2} are valid for {\it both} $SO(32)$ and the ${\rm E}_8 \times {\rm E}_8$ theories. However the specifics of these two heterotic theories appear when we take derivatives, and especially the temporal derivatives, of the three-form fields. For example, using \eqref{hannalat} we can now define a part of the G-flux component, say $(d{\bf C})_{\rm 0ABD}$, for the $SO(32)$ theory in the following way:

{\scriptsize
\bg\label{secdrem18}
(d{\bf C})_{\rm 0ABD}(\zeta) & = & \sum_{k = 0}^\infty 
(d{\bf C})_{\rm 0ABD}^{(1, k)}({\bf x}, y)\left({g_s\over {\rm HH}_o}\right)^{l_{\rm AB}^{\rm D}(t) \mp
{{\rm log}\left(\big\vert\dot{l}_{\rm AB}^{\rm D}(t) 
+ {2k \vert \dot{\rm L}_{\rm AB}^{\rm D}(k; t)\vert\over 3}\big\vert \vert\log~\bar{g}_s\vert\right) \over \vert\log~\bar{g}_s\vert} + {2k\vert {\rm L}_{\rm AB}^{\rm D}(k; t)\vert\over 3}}\nonumber\\
&+ & 
\sum_{k = 0}^\infty (d{\bf C})_{\rm 0ABD}^{(2, k)}({\bf x}, y)\left({g_s\over {\rm HH}_o}\right)^{l_{\rm AB}^{\rm D}(t) - 1 + {\beta(t)\over 2} + 
{{\rm log}(2-\beta(t))\over \vert\log~\bar{g}_s\vert} \mp {\log\left(\big\vert l_{\rm AB}^{\rm D}(t) + {2k\vert {\rm L}_{\rm AB}^{\rm D}(k; t)\vert\over 3}\big\vert\right) \over\vert\log~\bar{g}_s\vert} + {2k\vert {\rm L}_{\rm AB}^{\rm D}(k; t)\vert\over 3}} \nonumber\\ 
&+ & 
\sum_{k = 0}^\infty (d{\bf C})_{\rm 0ABD}^{(3, k)}({\bf x}, y)\left({g_s\over {\rm HH}_o}\right)^{l_{\rm AB}^{\rm D}(t) + {1\over\vert\log~\bar{g}_s\vert}
{{\rm log}\left[{2-\beta(t)\over \vert\dot\beta(t)\vert\vert\log~\bar{g}_s\vert}\right]} 
\mp {\log\left(\big\vert l_{\rm AB}^{\rm D}(t) + {2k\vert {\rm L}_{\rm AB}^{\rm D}(k; t)\vert\over 3}\big\vert\right) \over\vert\log~\bar{g}_s\vert} + {2k\vert {\rm L}_{\rm AB}^{\rm D}(k; t)\vert\over 3}} \nonumber\\ 
& + & \sum_{k = 0}^\infty\left[(d{\bf C})_{\rm 0ABD}^{(4, k)}({\bf x}, y)\left({g_s\over {\rm HH}_o}\right)^{l_{\rm 0B}^{\rm D}(t) + {2k\vert {\rm L}_{\rm 0B}^{\rm D}(k; t)\vert\over 3}} +  (d{\bf C})_{\rm 0ABD}^{(5, k)}({\bf x}, y)\left({g_s\over {\rm HH}_o}\right)^{l_{\rm 0A}^{\rm D}(t) + {2k\vert {\rm L}_{\rm 0A}^{\rm D}(k; t)\vert\over 3}}\right] \nonumber\\
&+ &  \sum_{k = 0}^\infty(d{\bf C})_{\rm 0ABD}^{(6, k)}({\bf x}, y)\left({g_s\over {\rm HH}_o}\right)^{l_{\rm 0A}^{\rm B}(t) + {2k\vert {\rm L}_{\rm 0A}^{\rm B}(k; t)\vert\over 3}},
\nd}
where $\zeta \equiv ({\bf x}, y; g_s)$, $(d{\bf C})_{\rm 0ABD}^{(n, k)}({\bf x}, y)$ corresponds to appropriate derivative acting on the three-form, which in turn may be easily identified from the $g_s$ exponent\footnote{For example if the exponent is $l_{\rm AB}^{\rm D}$, then $(d{\bf C})_{\rm 0ABD}^{(n, k)}({\bf x}, y)$ would be proportional to $\partial_0 {\bf C}_{\rm ABD}({\bf x}, y)$ with $n = 1, 2$ and 3. A specific $n$ can then be assigned by looking at the log corrections to $l_{\rm AB}^{\rm D}$.}. The sign ambiguity in some of the terms above is to take care of the sign of $l_{\rm AB}^{\rm D}$, and $\dot{l}_{\rm AB}^{\rm D} \equiv \partial_0 l_{\rm AB}^{\rm D}$ is with respect to the conformal time. We can take $\dot{l}_{\rm AB}^{\rm D}$ to be monotonically decreasing function with time to be consistent with what we had for $\beta(t)$ before. On the other hand $(d{\bf C})_{\rm ABCD}$ takes the following form:

{\scriptsize
\bg\label{tatumey}
(d{\bf C})_{\rm ABCD}(\zeta) & = & \sum_{k = 0}^\infty (d{\bf C})_{\rm ABCD}^{(1, k)}({\bf x}, y)\left({g_s\over {\rm HH}_o}\right)^{l_{\rm BC}^{\rm D}(t)  + {2k \vert {\rm L}_{\rm BC}^{\rm D}(k; t)\vert\over 3}} +  \sum_{k = 0}^\infty (d{\bf C})_{\rm ABCD}^{(2, k)}({\bf x}, y)\left({g_s\over {\rm HH}_o}\right)^{l_{\rm AC}^{\rm D} + {2k\vert {\rm L}_{\rm AC}^{\rm D}(k; t)\vert \over 3}} \nonumber\\
& + & \sum_{k = 0}^\infty (d{\bf C})_{\rm ABCD}^{(3, k)}({\bf x}, y)\left({g_s\over {\rm HH}_o}\right)^{l_{\rm AB}^{\rm D}(t) + {2k\vert {\rm L}_{\rm AB}^{\rm D}(k; t)\vert\over 3}} +  \sum_{k = 0}^\infty (d{\bf C})_{\rm ABCD}^{(4, k)}({\bf x}, y)\left({g_s\over {\rm HH}_o}\right)^{l_{\rm AB}^{\rm C}(t) + {2k\vert {\rm L}_{\rm AB}^{\rm C}(k; t)\vert\over 3}}, \nonumber\\ \nd}
where $\zeta \equiv ({\bf x}, y; g_s)$.
This doesn't have the complications of \eqref{secdrem18} because of the absence of the temporal derivative. Note that all permutations and spatial/temporal derivatives are present inside the functional forms for $(d{\bf C})_{\rm ABCD}^{(n, k)}({\bf x}, y)$ and $(d{\bf C})^{(n, k)}_{\rm 0ABD}({\bf x}, y)$. In the presence of five-brane sources and quantum terms, the four-form G-flux ${\bf G}_4$ cannot simply be $d{\bf C}_3$: it should have contributions from all the sources and the quantum terms. This may be written as (see also \eqref{crawlquif2} and \eqref{malinechat}):
\bg\label{dietherapie}
{\bf G}_4 = d{\bf C}_3 -{c_2\over c_1} \hat{\mathbb{Y}}_4 - {c_3\over c_1} \ast_{11} \mathbb{Y}_7 + \hat{\rm N}_5 {\bf \Lambda}_4, \nd
where $c_i$ are ${\rm M}_p$ dependent constants, and $(\hat{\mathbb Y}_4, \mathbb{Y}_7)$ are the curvature four-form and quantum terms respectively as defined in eq. (4.219) in the second reference of \cite{coherbeta2}, ${\bf \Lambda}_4$ is a localized form (not globally defined) such that $d{\bf \Lambda}_4 \equiv {\bf \Lambda}_5$, and $\hat{\rm N}_5$ is the number of {\it dynamical} five-branes. These issues have been discussed in details in \cite{desitter2, coherbeta, coherbeta2}\footnote{See for example section 4.6 in the second reference of \cite{coherbeta2} where the quantum terms $\hat{\mathbb{Y}}_4$ and $\mathbb{Y}_7$ are carefully defined.} so we will be brief. However due to the presence of $\alpha(t)$ and $\beta(t)$ in the internal metric for the $SO(32)$ case, or $\hat\alpha(t)$ and $\hat\beta(t)$ for the ${\rm E}_8 \times {\rm E}_8$ case, the analysis is more involved than the one presented in \cite{desitter2, coherbeta, coherbeta2}. This means we can define the G-flux components using 
similar notation as in \cite{desitter2, coherbeta, coherbeta2}, but with a slight change to reflect the presence of log corrections at the lowest order, in the following way:
\bg\label{usher}
{\bf G}_{\cal ABCD}({\bf x}, y; g_s(t)) = \sum_{k = 0}^\infty {\cal G}^{(k)}_{\cal ABCD}({\bf x}, y) \left({g_s\over {\rm HH}_o}\right)^{l_{\cal AB}^{\cal CD}(t) + {2k\over 3}\vert {\rm L}_{\cal AB}^{\cal CD}(k; t)\vert}, \nd
where $({\cal A, B}) \in {\bf R}^{1, 2} \times {\cal M}_4 \times {\cal M}_2 \times {\mathbb{T}^2\over {\cal G}}$. We expect ${\cal G}^{( k)}_{\cal ABCD}({\bf x}, y)$ to be related to 
$(d{\bf C})_{\cal ABCD}^{(n, k)}({\bf x}, y)$ from \eqref{secdrem18} and \eqref{tatumey} for appropriate choice of $n$. To see how the $g_s$ scalings are matched we can compare the scalings of ${\bf G}_{\rm 0ABD}$ and $(d{\bf C})_{\rm 0ABD}$ in \eqref{dietherapie}. Choosing a specific ordering of
$l_{\rm 0A}^{\rm BD}(t)$ and $l_{\rm 0A}^{\rm BD}(k; t)$, we can relate them to $l_{\cal AB}^{\cal D}$ from \eqref{secdrem18} and \eqref{tatumey}. An example would be the following:

{\footnotesize
\bg\label{josymila}
&&l_{\rm 0A}^{\rm BD}(t) =  l_{\rm AB}^{\rm D}(t) \mp {{\rm log}(\vert \dot{l}_{\rm AB}^{{\rm D}}(t)\vert \vert\log~\bar{g}_s\vert) \over \vert\log~\bar{g}_s\vert}\\
&& \vert {\rm L}_{\rm 0A}^{\rm BD}(1; t)\vert = 
 {3\beta(t)\over 4} - {3\over 2} +  
{3{\rm log}(2-\beta(t))\over 2\vert\log~\bar{g}_s\vert} \pm 
 {3\over 2\vert\log~\bar{g}_s\vert}~\log\left[ {\vert \dot{l}_{\rm AB}^{{\rm D}}(t)\vert \vert\log~\bar{g}_s\vert \over  \vert l_{\rm AB}^{\rm D}(t)\vert}\right] \nonumber\\
&& \vert {\rm L}_{\rm 0A}^{\rm BD}(2; t)\vert = 
  {3\over 4\vert\log~\bar{g}_s\vert}~\log\left[{2-\beta(t)\over \vert\dot\beta(t)\vert\vert\log~\bar{g}_s\vert}\right] \pm 
  {3\over 4\vert\log~\bar{g}_s\vert}~\log\left[ {\vert \dot{l}_{\rm AB}^{{\rm D}}(t)\vert \vert\log~\bar{g}_s\vert \over  \vert l_{\rm AB}^{\rm D}(t)\vert}\right]  \nonumber\\
&& \vert {\rm L}_{\rm 0A}^{\rm BD}(3; t)\vert = {1\over 2}(l_{\rm 0B}^{\rm D}(t) - l_{\rm 0A}^{\rm BD}(t)), ~~
\vert {\rm L}_{\rm 0A}^{\rm BD}(4; t)\vert = {3\over 8}(l_{\rm 0A}^{\rm D}(t) - l_{\rm 0A}^{\rm BD}(t)), ~~\vert {\rm L}_{\rm 0A}^{\rm BD}(5; t)\vert = {3\over 10}(l_{\rm 0A}^{\rm B}(t) - l_{\rm 0A}^{\rm BD}(t)), \nonumber
\nd}
and so on; where $l_{\rm 0A}^{\rm BD}(t)$ appearing in the last three scalings can be read from the first line above. Our identifications in \eqref{josymila} is the simplest possible ones without worrying about possible overlapping ones\footnote{Looking at the functions forms for $d{\bf C}$ in \eqref{secdrem18} and \eqref{tatumey} one might consider that the following choice of the G-flux components:
\bg\label{usher2}
{\bf G}_{\cal ABCD}({\bf x}, y; g_s(t)) = \sum_{n = 1}^{\rm N}\sum_{k = 0}^\infty {\cal G}^{(n, k)}_{\cal ABCD}({\bf x}, y) \left({g_s\over {\rm HH}_o}\right)^{l_{\cal AB}^{\cal CD}(n; t) + {2k \over 3}\vert l_{\cal AB}^{\cal CD}(n, k; t)\vert}, \nd
would better suit to equate the coefficients in \eqref{dietherapie} 
with ${\rm N} \ge 4$ when all the legs of the fluxes are along the spatial direction as in \eqref{tatumey}, and ${\rm N} \ge 6$ when one of the leg is along the temporal direction as in \eqref{secdrem18}. A little thought will tell us that this complication is not necessary.
}. 
We also expect these scalings to be positive definite, but we do not impose any such constraints right now. In the limit of vanishing $c_2, c_3$ and $\hat{\rm N}_5$ in \eqref{dietherapie}, we expect the following identifications:
\bg\label{isahen}
&& {\cal G}_{\rm 0ABD}^{(l.{\rm mod}.7 + k)}({\bf x}, y) \equiv \left(d{\bf C}\right)^{(l.{\rm mod}.7 + 1; k)}_{\rm 0ABD}({\bf x}, y) \nonumber\\
&&  {\cal G}_{\rm ABCD}^{(l.{\rm mod}.5 + k)}({\bf x}, y) \equiv \left(d{\bf C}\right)^{(l.{\rm mod}.5 + 1; k)}_{\rm ABCD}({\bf x}, y)
\nd
for $l \ge 0$ and $k \ge 0$. Generically however the relation between 
${\cal G}^{(k)}_{\rm 0ABD}({\bf x}, y)$ and $\left(d{\bf C}\right)^{(n, k')}_{\rm 0ABD}$, with $k$ and $(n, k')$ related via \eqref{isahen}, is more involved then \eqref{isahen} because of \eqref{dietherapie}. We will soon incorporate the Bianchi identities to relate the G-flux components with the C-fields as well as the quantum series from \eqref{botsuga}. Note that for the simple minded analysis presented here, either \eqref{secdrem18} and \eqref{tatumey} or \eqref{usher} suffices, but a more elaborate analysis will involve additional ingredients (for example the sub-dominant non-perturbative corrections, dynamical five-branes et cetera). We will come back to this soon.

\subsection{Higher order corrections and scalings in $SO(32)$ and ${\rm E}_8 \times {\rm E}_8$ theories \label{sec4.4}}
The scalings that we discussed so far were mostly for the $SO(32)$ heterotic theory and in particular for the simpler case when we ignore the {\it sub-dominant} perturbative and the non-perturbative corrections (except for a brief digression in section \ref{sec4.2.2}). We did however take into account the {\it dominant} perturbative and non-perturbative corrections to the warp-factors $\alpha(t)$ and $\beta(t)$. Here we want to see what happens if we go beyond that for both the $SO(32)$ and the ${\rm E}_8 \times {\rm E}_8$ theories. 

Our starting point would be to consider the metric configurations in both $SO(32)$ and the ${\rm E}_8 \times {\rm E}_8$ cases. As we saw earlier, specifically  from {\bf Tables \ref{milleren1}} to {\bf \ref{milleren4}}, they may be derived from two different configurations in M-theory. For example in the $SO(32)$ case  
the M-theory metric discussed in section \ref{sec4.2.1} takes the following form:

{\footnotesize
\bg\label{sinistre}
ds^2 & = & \sum_{k = 0}^\infty h_k \left({g_s\over {\rm H}(y){\rm H}_o({\bf x})}\right)^{-{8\over 3} + {2k\vert f_k(t)\vert\over 3}} \sum_{\mu, \nu = 0}^2 {\bf g}_{\mu\nu}({\bf x}, y) dx^\mu dx^\nu\\
& + &\left({g_s\over {\rm H}(y){\rm H}_o({\bf x})}\right)^{4\over 3}\widetilde{\bf g}_{11, 11}({\bf x}, y) dx_{11}^2 + 
\sum_{k = 0}^\infty {\rm B}_k \left({g_s\over {\rm H}(y){\rm H}_o({\bf x})}\right)^{-{2\over 3} + \alpha(t) + {2k\vert\alpha_k(t)\vert\over 3}}
{\bf g}^{(k)}_{mn}({\bf x}, y) dy^m dy^n \nonumber\\
& + & \sum_{k = 0}^\infty \widetilde{h}_k\left({g_s\over {\rm H}(y){\rm H}_o({\bf x})}\right)^{{4\over 3} + {2k\vert \widetilde{f}_k(t)\vert\over 3}} 
{\bf g}_{33}({\bf x}, y) dx_3^2 +
\sum_{k = 0}^\infty {\rm A}_k \left({g_s\over {\rm H}(y){\rm H}_o({\bf x})}\right)^{-{2\over 3} + \beta(t) + {2k\vert\beta_k(t)\vert\over 3}}
{\bf g}^{(k)}_{\rho\sigma}({\bf x}, y) dy^\rho dy^\sigma, \nonumber \nd}
where $h_k$ and $\widetilde{h}_k$ would satisfy a constraint similar to the one shown in \eqref{englishgulab}, ${\bf x} \in {\bf R}^2$ and $y = (y^m, y^\alpha)$ with $y^m \in {\cal M}_4$ and $ y^\alpha \in {\cal M}_2$; and $({\rm A}_k, {\rm B}_k)$ are defined earlier in \eqref{sezoe70}. (We have kept the metric and flux components independent of $w^a \in {\mathbb{T}^2\over {\cal G}}$.) Note that we haven't added the non-perturbative corrections from section \ref{sec4.2.2} on the metric components (some parts of these corrections could be easily inserted by using \eqref{ivbres}, but a more generic picture will be presented later in section \ref{sec6}), because of our choice of the low energy M-theory metric. Additionally, we do not impose any of the simplifying features mentioned in section \ref{sec4.2.1}. This is intentional because we want to keep the system generic enough without violating the heterotic configurations resulting from the duality sequence
discussed in {\bf Tables \ref{milleren1}} to {\bf \ref{milleren4}}. 

The M-theory configuration for the ${\rm E}_8 \times {\rm E}_8$ case differs from \eqref{sinistre} not only in the choice of the internal space, but also in the choice of the fluxes. The latter will be dealt in section \ref{sec4.5}. The internal six-dimensional subspace is no longer a toroidal orbifold of the form ${\cal M}_4 \times {\mathbb{T}^2\over {\cal I}_2}$ but an orbifold of the form ${\cal M}_4 \times {\bf S}^1_{\theta_1} \times {{\bf S}_{\theta_2}^1\over {\cal I}_{\theta_2}}$. The metric takes the form:

{\scriptsize
\bg\label{sinistre00}
ds^2  &= & \sum_{k = 0}^\infty h_k \left({g_s\over {\rm H}(y){\rm H}_o({\bf x})}\right)^{-{8\over 3} + {2k\vert \hat{f}_k(t)\vert\over 3}} \sum_{\mu, \nu = 0}^2 {\bf g}_{\mu\nu}({\bf x}, y) dx^\mu dx^\nu
+ \left({g_s\over {\rm H}(y){\rm H}_o({\bf x})}\right)^{4\over 3}\widetilde{\bf g}_{11, 11}({\bf x}, y) dx_{11}^2 \\
&+ & 
\sum_{k = 0}^\infty {\rm C}_k \left({g_s\over {\rm H}(y){\rm H}_o({\bf x})}\right)^{-{2\over 3} + {2\hat\sigma(t)\over 3} + {2k\vert\hat\sigma_k(t)\vert\over 3}}
{\bf g}^{(k)}_{mn}({\bf x}, y) dy^m dy^n
+  \sum_{k = 0}^\infty \widetilde{h}_k\left({g_s\over {\rm H}(y){\rm H}_o({\bf x})}\right)^{{4\over 3}
+ {2k\vert \widetilde{\hat{f}}_k(t)\vert\over 3}} 
{\bf g}_{33}({\bf x}, y) dx_3^2  \nonumber\\
& + & 
\sum_{k = 0}^\infty\left[{\rm B}_k \left({g_s\over {\rm H}(y){\rm H}_o({\bf x})}\right)^{-{2\over 3} + {2\hat\beta(t)\over 3} + {2k\vert\hat\beta_k(t)\vert\over 3}}
{\bf g}^{(k)}_{\theta_2 \theta_2}({\bf x}, y) d\theta_2^2 + {\rm A}_k \left({g_s\over {\rm H}(y){\rm H}_o({\bf x})}\right)^{-{2\over 3} + {2\hat\alpha(t)\over 3} + {2k\vert\hat\alpha_k(t)\vert\over 3}}
{\bf g}^{(k)}_{\theta_1\theta_1}({\bf x}, y) d\theta_1^2\right], \nonumber \nd}
where $h_k$ and $\widetilde{h}_k$ would again satisfy a constraint similar to the one shown in \eqref{englishgulab}, ${\bf x} \in {\bf R}^2$ and $y = (y^m, \theta_1, \theta_2)$ with $y^m \in {\cal M}_4$ and $ (\theta_1, \theta_2) \in \Big({\bf S}^1_{\theta_1}, {{\bf S}_{\theta_2}^1\over {\cal I}_{\theta_2}}\Big)$. (As before, we have kept the metric and flux components independent of $w^a \in {\mathbb{T}^2\over {\cal G}}$ and avoided any simplifying procedures). In the last line of \eqref{sinistre00} the warp-factors are precisely the ${\rm F}_3(t)$ and ${\rm F}_1(t)$ discussed in {\bf Table \ref{milleren4}} with the condition that ${\rm F}_3(t) > {\rm F}_1(t)$ in the temporal domain $-{1\over \sqrt{\Lambda}} < t < -\epsilon_1$ and 
${\rm F}_1(t) = {\rm F}_3(t)$ in the domain $-\epsilon_1 \le t < 0$ in the string frame. There is also an intermediate blow-up phase where the M-theory metric takes the form \eqref{kingtut}. From there one could easily show that the heterotic dilaton, the distance between the two Horava-Witten walls and the axionic decay constant are all controlled by the warp-factors ${\rm F}_1(t)$ and ${\rm F}_3(t)$, as shown in the Einstein frame in \eqref{shamitagra}.

To proceed, we will make some simplifying choices for the internal metric configurations. For the $SO(32)$ and the ${\rm E}_8 \times {\rm E}_8$ cases we will assume ${\bf g}_{\rm MN}^{(k)}({\bf x}, y) = \mathbb{A}^{(k)}{\bf g}_{\rm MN}({\bf x}, y)$ where $({\rm M, N}) \in {\cal M}_4 \times {\mathbb{T}^2\over {\cal I}_2}$ for the $SO(32)$ theory; $({\rm M, N}) \in {\cal M}_4 \times {\bf S}^1_{\theta_1} \times {{\bf S}_{\theta_2}^1\over {\cal I}_{\theta_2}}$ for the ${\rm E}_8$ case; and $\mathbb{A}^{(k)}$ are constants that can be taken to be identity here. With this choice, 
the extension to include the sub-dominant perturbative corrections, and at least a subset of the sub-dominant non-perturbative ones, is surprisingly simple. For the $SO(32)$ case all we need to do is to re-express the two warp-factor ${\rm F}_1(t)$ and ${\rm F}_2(t)$, as well as the IIA coupling $g_s$, using $(\beta_e(t), \alpha_e(t))$ instead of $(\beta(t), \alpha(t))$ respectively in the following way:

{\footnotesize
\bg\label{khadsalam}
{\rm F}_1(t) = {\rm A}_0 \left({g_s \over {\rm H}(y) {\rm H}_o({\bf x})}\right)^{\beta_e(t)}, ~ {\rm F}_2(t) = {\rm B}_0 \left({g_s \over {\rm H}(y) {\rm H}_o({\bf x})}\right)^{\alpha_e(t)}, ~ {g_s \over {\rm H}(y) {\rm H}_o({\bf x})} = \left(\sqrt{\Lambda}\vert t \vert\right)^{2\over 2 - \beta_e(t)},\nd}
where $({\rm A}_0, {\rm B}_0)$ are positive constants defined in \eqref{sinistre}. The exponents $\alpha_e(t)$ and $\beta_e(t)$ contain all the sub-dominant perturbative corrections in the following way:

{\footnotesize
\bg\label{bdtran}
&& \beta_e(t) \equiv \sum_{k = 0}^\infty \beta_e(t, [k]) = \beta(t) - {1\over \vert\log~\bar{g}_s\vert}~\log\left[1 + \sum_{k = 1}^\infty {{\rm A}_k\over {\rm A}_0} \left({g_s \over {\rm H}(y) {\rm H}_o({\bf x})}\right)^{2k\vert\beta_k(t)\vert\over 3}\right] \nonumber\\
&& \alpha_e(t) \equiv \sum_{k = 0}^\infty \alpha_e(t, [k]) = \alpha(t) - {1\over \vert\log~\bar{g}_s\vert}~\log\left[1 + \sum_{k = 1}^\infty {{\rm B}_k\over {\rm B}_0} \left({g_s \over {\rm H}(y) {\rm H}_o({\bf x})}\right)^{2k\vert\alpha_k(t)\vert\over 3}\right], \nd}
where the order-by-order $k$-th term can be extracted by expanding the log functions on the RHS of \eqref{bdtran}. It is clear that $\beta_e(t, [0]) = \beta(t)$ and $\alpha_e(t, [0]) = \alpha(t)$. We can also insert a subset of the sub-dominant non-perturbative corrections using a reparametrization as in \eqref{ivbres}, namely:
\bg\label{ivbrestick}
\vert \sigma_k(t(\bar{g}_s))\vert ~ \to ~ \vert \sigma_k(t(\bar{g}_s))\vert + {3\over 2k}
\sum_{a \in \mathbb{Z}} {n_a(k) \over \bar{g}_s^{a/3} \vert {\rm log}~\bar{g}_s\vert}, \nd
where $\sigma_k(t) \equiv (\alpha_k(t), \beta_k(t))$ and $\bar{g}_s = 
{g_s\over {\rm H}(y) {\rm H}_o({\bf x})}$. Since $\alpha(t)$ and $\beta(t)$ already contain the dominant perturbative and non-perturbative corrections as seen from \eqref{ryanfan}, this provides a powerful way to  incorporate the additional $k$-dependent sub-dominant corrections. Additionally, the simpler forms for ${\rm F}_i(t)$ and $g_s$ in \eqref{khadsalam} suggest that we can simply replace:
\bg\label{bekensal}
\alpha(t) ~ \to ~ \alpha_e(t), ~~~~~ \beta(t) ~ \to ~ \beta_e(t), \nd 
in {\bf Tables \ref{rindrag}} to {\bf \ref{firzacut3}} to incorporate all the aforementioned corrections. (There would still be additional corrections coming from $\vert f_k(t)\vert$ factors in the metric \eqref{sinistre} that we will incorporate soon.) Moreover, expressing $\alpha_e(t)$ and $\beta_e(t)$ in terms of $\alpha_e(t, [k])$ and $\beta_e(t, [k])$ respectively provides the necessary $k$-dependent proliferation.

\begin{table}[tb]  
 \begin{center}
\resizebox{\columnwidth}{!}{%
 \renewcommand{\arraystretch}{5.0}
\begin{tabular}{|c||c||c||c||c|}\hline Theory &  $\langle{\bf g}_{00}\rangle_\sigma, \langle{\bf g}_{ij}\rangle_\sigma$ & $\langle{\bf g}_{33}\rangle_\sigma$ & Related to \\ \hline\hline
M-theory & $\left({g_s\over {\rm H}(y) {\rm H}({\bf x})}\right)^{-{8\over 3} + \zeta_e(t)}\eta_{ij}$ & $\left({g_s\over {\rm H}(y) {\rm H}({\bf x})}\right)^{{4\over 3} + \eta_e(t)}$ & $ \sum\limits_{\{{\bf s}\}}\left[{\mathbb{F}_{({\bf s}, \mu, \nu)}\over g_{({\bf s}, \mu, \nu)}^{1/l} {\rm M}_p^{2\kappa}}
\int_0^\infty d{\rm S} ~{e^{-{\rm S}/g_{({\bf s}, \mu, \nu)}^{1/l}} \over 1 - {\cal A}_{({\bf s}, \mu, \nu)}{\rm S}^l}\right]_{\Su {\rm P. V}}\int_{k_{\rm IR}}^\mu d^{11} k~ {\sigma_{\mu\nu}(k)\over a^2(k)}~\psi_k(x, y, w)$\\ \hline
Heterotic $SO(32)$ & \multicolumn{2}{l||}{$\sum\limits_{l = 0}^\infty\eta_{\mu\nu} b_l
\left({g_s\over {\rm H}(y){\rm H}({\bf x})_o}\right)^{-2 + \beta(t) +{2l\vert{\red\check\beta_l(t)}\vert\over 3}}\left[
\sum\limits_{k = 0}^\infty h_k 
\left({g_s\over {\rm H}(y){\rm H}({\bf x})_o}\right)^{2k\vert {\red\check{f}_k(t)}\vert \over 3}\right]^{1\over 2}$} & ${1\over \Lambda(t) \vert t \vert^2} = {\left(1 + {\check{\Lambda}(t)\over \Lambda}\right)^{-1}\over \Lambda \vert t \vert^2}$ \\ \hline
\end{tabular}}
\renewcommand{\arraystretch}{1}
\end{center}
 \caption[]{\Su Various components of the emergent space-time metric components and their connections to either the Borel-resummed values (in M-theory), or to the four-dimensional dynamical dark energy \eqref{marapaug} (in Heterotic $SO(32)$ theory). Here $(i, j) \in {\bf R}^2$, and $\sigma$ is associated to the corresponding Glauber-Sudarshan states in M-theory and Heterotic $SO(32)$ theory. The bare cosmological constant $\Lambda$ and the type IIA coupling $g_s$ are given by \eqref{montehan2} and \eqref{johnsonsteel} respectively. The explcit form for $\Lambda(t)$ from \eqref{marapaug} appears in \eqref{daisisaint2}. A similar table may be constructed for the ${\rm E}_8 \times {\rm E}_8$ case by following the duality chasings from {\bf Table \ref{milleren4}}, but we will not do so here.}
\label{scarstan25}
 \end{table}

Comparing \eqref{bdtran} with \eqref{paigest}, one might be tempted to 
relate $\vert\beta_k(t)\vert$ with $\vert{\red\check{\rm B}_k(t)}\vert$ directly instead of $\vert{\red{\rm B}_k(t)}\vert$ that we used in \eqref{paigest}, but the identification is bit more subtle. $\beta_k(t)$ and $\check{\rm B}_k(t)$ are defined in two different theories, M-theory and $SO(32)$ heterotic theory respectively, and they are connected by duality transformations (see {\bf Table \ref{scarstan25}}). This would imply $\check{\rm B}_k(t) = \beta_k(t) + {\cal O}(g_s, {\rm M}_p)$. We can express the  ${\cal O}(g_s, {\rm M}_p)$ corrections to $\check{\rm B}_k(t)$ in the following suggestive way:

{\footnotesize
\bg\label{stonnejess}
\check{\rm B}_k(t) = {\rm B}_k(t) + {\rm C}_k(t) = 
\beta_k(t) - {1\over \vert\log~\bar{g}_s\vert}~\log\left(\sum_{l = 0}^\infty \sum_{a, b}{d_{klab}\over \sqrt{2^l l!}}~\mathbb{H}_l\left[\bigg({g_s\over 
{\rm H}(y){\rm H}_o({\bf x})}\bigg)^a \cdot {1\over {\rm M}_p^b}\right]\right), \nd}
where $\mathbb{H}_l[z]$ is the $l$-th Hermite polynomial written in terms of the variable $z\equiv {\bar{g}^a_s\over {\rm M}_p^b}$ with the product defined over all possible choices of $(a, b) \in (\pm\mathbb{Z}, \pm\mathbb{Z})$; and $d_{klab}$ are constants. The Hermite functions can in principle reproduce any functions due to their completeness property, and in particular can reproduce function of the form $\left({g_s\over {\rm H}(y) {\rm H}({\bf x})}\right)^{{\rm C}_k(t(g_s))}$. Taking all of these carefully now, the dynamical dark energy from \eqref{marapaug} takes the following form:

{\footnotesize
\bg\label{daisisaint2}
\Lambda(t) = \Lambda\left(\sqrt{\Lambda}\vert t \vert\right)^{2\beta_e(t)-2\beta(t)\over 2-\beta_e(t)} \left[ 1 + \sum_{l = 1}^\infty b_l\left(\sqrt{\Lambda}\vert t \vert\right)^{4l\vert {\red\check{\rm B}_l(t)}\vert\over 3(2-\beta_e(t))}\right]^{-1}  \left[ 1 + \sum_{k = 1}^\infty h_k\left(\sqrt{\Lambda}\vert t \vert\right)^{4k\vert {\red\check{\rm F}_l(t)}\vert\over 3(2-\beta_e(t))}\right]^{-{1\over 2}}, \nd}
which is a closed form expression that generalizes our earlier result from \eqref{mmilan}, with $\Lambda$ given by \eqref{montehan2}. (The parameters in {\red red} have already been defined above and, in more details, earlier in section \ref{sec4.2.1}.) Note that, as it stands, \eqref{daisisaint2} doesn't have the pathologies associated with \eqref{daisisaint} because it reaches $\Lambda(t) = \Lambda$ as $t \to 0$. The slow decrease in $\Lambda(t)$ is clear from the above expression and therefore \eqref{daisisaint2} provides, probably for the first time, a description of the dynamical nature of the dark energy from a top-down theory.

One could also ask the question what if the cosmological constant is truly a {\it constant}, and the results of DESI-BAO \cite{desibao} suffers from some, yet unforeseen, statistical error. In such a case it will be worthwhile to compare the connection between $\bar{g}_s$ with 
the conformal time $t$ for the two cases, one with a dynamical dark energy and the other with a cosmological constant. The result may be presented as:

{\scriptsize
\bg\label{dollarlumemey}
{g_s\over {\rm H}(y) {\rm H}({\bf x})} = \begin{cases}~~ \left(\sqrt{\Lambda}\vert t\vert\right)^{2\over 2-\beta_e(t)}, ~~~~{\rm for}~\Lambda(t) = \Lambda + \check{\Lambda}(t) \\
~~~~~ \\
~~ \left(\sqrt{\Lambda}\vert t\vert\right)^{2\over 2-\beta(t) + {1\over 2\vert \log~\bar{g}_s\vert}~\log\Big[\Big(1+\sum\limits_{l = 1}^\infty b_l\bar{g}_s^{2l\vert{\red\check{\rm B}_l(t)}\vert/ 3}\Big)^2\Big(1+\sum\limits_{k = 1}^\infty h_k\bar{g}_s^{2k\vert{\red\check{\rm F}_k(t)}\vert/3}\Big)\Big]}, ~~~~{\rm for}~\Lambda(t) = \Lambda 
\end{cases}
\nd}
implying that a dynamical dark energy scenario in fact {\it simplifies} the computations! Furthermore, with a cosmological {\it constant}, the aforementioned identification between $\bar{g}_s$ in type IIA/M-theory and the conformal time $t$ depends heavily on the $(g_s, {\rm M}_p)$ corrections  accumulated during the duality chasing process. This is a bit of an unsatisfactory feature because why would a type IIA/M-theory result depend on what duality sequence we want to perform? It may nevertheless be possible that this is the right answer and the DESI-BAO results have (hitherto unforeseen) statistical errors. If this is the case, then our analysis is bound to become much more involved. In this paper, we will however not worry about the results coming from imposing a cosmological constant, and instead take the dynamical dark energy scenario seriously\footnote{Aesthetically, it is slightly disconcerting to assume that nature is controlled by a ``God-given" cosmological {\it constant} that remains unchanged during the whole period of the evolution of our universe. It is much more natural to assume a dynamical dark energy that is {\it decreasing} (but not increasing!) with time. Our analysis in \eqref{daisisaint2} does predict a decreasing $\Lambda(t)$ that eventually approaches a bare value of $\Lambda$ at $t = 0$. This bare value appears from Borel resumming instanton series from all the tunneling effects as shown in \eqref{montehan2}, and {\it not} from the zero-point energies (which are in fact canceled due to the underlying supersymmetric Minkowski background). See also \cite{dieterobert}  for some recent discussion on this.} via the first choice in \eqref{dollarlumemey}, {\it i.e.} the relation \eqref{johnsonsteel}. 

\begin{table}[tb]  
 \begin{center}
\renewcommand{\arraystretch}{1.5}
\begin{tabular}{|c||c||c|}\hline Metric components & Series controlled by & Temporal derivative\\ \hline\hline
${\bf g}_{\mu\nu}({\bf x}, y; g_s)$ & $\hat\zeta_e(t)$ & $\mathbb{C}_8(\hat\zeta_e), \mathbb{D}_e(\hat\zeta_e)$ \\ \hline
${\bf g}_{mn}({\bf x}, y; g_s)$ & $\hat\sigma_e(t)$ & $\mathbb{A}_8(\hat\sigma_e), \mathbb{B}_8(\hat\sigma_e)$ \\ \hline
${\bf g}_{\theta_1\theta_1}({\bf x}, y; g_s)$ & $\hat\alpha_e(t)$ & $\mathbb{A}_8(\hat\alpha_e), \mathbb{B}_8(\hat\alpha_e)$ \\ \hline
${\bf g}_{\theta_2\theta_2}({\bf x}, y; g_s)$ & $\hat\beta_e(t)$ & $\mathbb{A}_8(\hat\beta_e), \mathbb{B}_8(\hat\beta_e)$ \\ \hline
${\bf g}_{33}({\bf x}, y; g_s)$ & $\hat\eta_e(t)$ & $\mathbb{F}_8(\hat\eta_e), \mathbb{G}_e(\hat\eta_e)$ \\ \hline
${\bf g}_{11, 11}({\bf x}, y; g_s)$ & 0 & ${1\over 3} + \gamma_{1, 2}\big[{\hat\alpha_e(t) + \hat\beta_e(t)\over 2}\big]$ \\ \hline
\end{tabular}
\renewcommand{\arraystretch}{1}
\end{center}
 \caption[]{\Su M-theory metric components from \eqref{sinistre00} that lead to the ${\rm E}_8 \times {\rm E}_8$ theory in the presence of the higher-order sub-dominant corrections. The coordinates span: ${\bf x} \in {\bf R}^2$ and $y \in {\cal M}_4 \times {\bf S}^1_{\theta_1}\times {{\bf S}^1_{\theta_2}\over {\cal I}_{\theta_2}}$ as specified in \eqref{sinistre00} and in {\bf Table \ref{milleren4}}. The parameters in the second column are defined in  \eqref{bdtran2} and \eqref{masatwork}. The parameters in the third column are defined in \eqref{tifeni} and \eqref{vanandcar}.}
  \label{depravity}
 \end{table}

The story in the ${\rm E}_8 \times {\rm E}_8$ side is equally simple, modulo the $\vert\hat{f}_k(t)\vert$ factor in \eqref{sinistre00}. Due to the presence of three warp-factors ${\rm F}_i(t)$ with $i = 1, .., 3$, the ${\rm E}_8$ sides provides a more generic construction and one can extract the $SO(32)$ picture by making ${\rm F}_1(t) = {\rm F}_3(t)$. (There are of course some subtle differences in the choice of the metric and the internal spaces, but they don't change much of the story.) Following the same strategy as in \eqref{khadsalam}, we can express the warp-factors ${\rm F}_i(t)$ and the IIA coupling $g_s$ in the following way:
\bg\label{khadsalim}
&&{\rm F}_1(t) = {\rm A}_0 \left({g_s \over {\rm H}(y) {\rm H}_o({\bf x})}\right)^{\hat\alpha_e(t)}, ~~~~ {\rm F}_3(t) = {\rm B}_0 \left({g_s \over {\rm H}(y) {\rm H}_o({\bf x})}\right)^{\hat\beta_e(t)} \nonumber\\
&& {\rm F}_2(t) = {\rm C}_0 \left({g_s \over {\rm H}(y) {\rm H}_o({\bf x})}\right)^{\hat\sigma_e(t)},~~~~ {g_s \over {\rm H}(y) {\rm H}_o({\bf x})} = \left(\sqrt{\Lambda}\vert t \vert\right)^{4\over 4 - \hat\alpha_e(t) - \hat\beta_e(t)},\nd
where $({\rm A}_0, {\rm B}_0, {\rm C}_0)$ are the constant coefficients defined in \eqref{sezoe71} with $(\hat\alpha_e(t), \hat\beta_e(t), \hat\sigma_e(t))$ incorporate the sub-dominant corrections to $(\hat\alpha(t), \hat\beta(t), \hat\sigma(t))$ respectively in the following way which is similar to what we had in \eqref{bdtran}:

{\footnotesize
\bg\label{bdtran2}
&& \hat\beta_e(t) \equiv \sum_{k = 0}^\infty \hat\beta_e(t, [k]) = {2\hat\beta(t)\over 3} - {1\over \vert\log~\bar{g}_s\vert}~\log\left[1 + \sum_{k = 1}^\infty {{\rm B}_k\over {\rm B}_0} \left({g_s \over {\rm H}(y) {\rm H}_o({\bf x})}\right)^{2k\vert\hat\beta_k(t)\vert\over 3}\right] \nonumber\\
&& \hat\sigma_e(t) \equiv \sum_{k = 0}^\infty \hat\sigma_e(t, [k]) = {2\hat\sigma(t)\over 3} - {1\over \vert\log~\bar{g}_s\vert}~\log\left[1 + \sum_{k = 1}^\infty {{\rm C}_k\over {\rm C}_0} \left({g_s \over {\rm H}(y) {\rm H}_o({\bf x})}\right)^{2k\vert\hat\sigma_k(t)\vert\over 3}\right]\nonumber\\
&& \hat\alpha_e(t) \equiv \sum_{k = 0}^\infty \hat\alpha_e(t, [k]) = {2\hat\alpha(t)\over 3} - {1\over \vert\log~\bar{g}_s\vert}~\log\left[1 + \sum_{k = 1}^\infty {{\rm A}_k\over {\rm A}_0} \left({g_s \over {\rm H}(y) {\rm H}_o({\bf x})}\right)^{2k\vert\hat\alpha_k(t)\vert\over 3}\right],
\nd}
where again $\hat\alpha_e(t, [0]) = {2\hat\alpha(t)\over 3}, \hat\beta_e(t, [0]) = {2\hat\beta(t)\over 3}$ and $\hat\sigma_e(t, [0]) = {2\hat\sigma(t)\over 3}$. Interestingly, comparing the two $g_s$ values in \eqref{khadsalim} and \eqref{khadsalam}, it appears that if we make the following sequential changes to the results of the $SO(32)$ theory developed earlier (and shown in {\bf Tables \ref{rindrag}} to {\bf \ref{firzacut3}}):

{\scriptsize
\bg\label{tiffmont}
&& \beta(t) ~ \to ~ \beta_e(t) ~ \to ~ {\hat\alpha_e(t) + \hat\beta_e(t)\over 2} ~~ {\rm or}~~ \left({\hat\alpha_e(t)}, {\hat\beta_e(t)}\right), ~~~~~~~~ \alpha(t) \to \alpha_e(t) \to {\hat\sigma_e(t)}\\
&& \partial_0\left({g_s\over {\rm H}(y) {\rm H}_o({\bf x})}\right)
= \sum_{l = 1}^2 e_l \left({g_s\over {\rm H}(y) {\rm H}({\bf x})}\right)^{\gamma_l(\beta)} ~ \to ~ \sum_{l = 1}^2 e_l \left({g_s\over {\rm H}(y) {\rm H}({\bf x})}\right)^{\gamma_l(\beta_e)} ~ \to ~ 
\sum_{l = 1}^2 e_l \left({g_s\over {\rm H}(y) {\rm H}({\bf x})}\right)^{\gamma_l\left({\hat\alpha_e + \hat\beta_e\over 2}\right)}\nonumber\\
&& \partial^2_0\left({g_s\over {\rm H}(y) {\rm H}_o({\bf x})}\right)
= \sum_{l = 3}^{10} e_l \left({g_s\over {\rm H}(y) {\rm H}({\bf x})}\right)^{\gamma_l(\beta)} ~ \to ~ \sum_{l = 3}^{10} e_l \left({g_s\over {\rm H}(y) {\rm H}({\bf x})}\right)^{\gamma_l(\beta_e)} ~ \to ~ 
\sum_{l = 3}^{10} e_l \left({g_s\over {\rm H}(y) {\rm H}({\bf x})}\right)^{\gamma_l\left({\hat\alpha_e + \hat\beta_e\over 2}\right)}, \nonumber \nd}
where $e_l$'s are defined in \eqref{ishena} and \eqref{ishena2}, then we should be able to reproduce the results of the ${\rm E}_8 \times {\rm E}_8$ theory. The ambiguity in the first line above happens because $\hat\alpha_e(t)$ and $\hat\beta_e(t)$ enter the metric components and the string coupling $g_s$ in two different ways which, in turn, affects the scalings of the curvature components.  While this, and the identifications in \eqref{tiffmont}, will be generically true, there are still a few subtleties that we want to clarify in the following sections, sections \ref{sec4.3.1}, \ref{sec4.3.2} and \ref{curuva}, before we express the 
changes that need to be added to the {\bf Tables \ref{firzathai}} to {\bf \ref{firzacut103}} to get the scalings for the ${\rm E}_8$ case. 

For the case with G-fluxes, the story would be similar. Let us consider the G-flux ans\"atze given in \eqref{usher}. As for the case with the metric components, we could apply the simplifying choice ${\cal G}^{(k)}_{\cal ABCD}({\bf x}, y) = {\cal A}_{\scriptscriptstyle{\tiny{\cal (ABCD)}}}^{(k)} {\cal G}_{\cal ABCD}({\bf x}, y)$ where ${\cal A}_{\scriptscriptstyle{\tiny{\cal (ABCD)}}}^{(k)}$ are constants that depend on the choice of the flux components. (No summation or anti-symmetry is assumed for the constants 
${\cal A}_{\scriptscriptstyle{\tiny{\cal (ABCD)}}}^{(k)}$.) Plugging this in \eqref{usher}, we can rewrite this as:
\bg\label{andyrey}
{\bf G}_{\cal ABCD}({\bf x}, y; g_s) = {\cal G}_{\cal ABCD}({\bf x}, y) \left({g_s\over {\rm H}(y) {\rm H}_o({\bf x})}\right)^{\hat{l}_{e{\cal AB}}^{\cal CD}(t(g_s))}\nd
where $\hat{l}_{e{\cal AB}}^{\cal CD}(t(g_s))$ should contain all the perturbative and the non-perturbative $\bar{g}_s$ corrections, much like what we had for $(\alpha_e(t), \beta_e(t))$ for the $SO(32)$ and 
$(\hat\alpha_e(t), \hat\beta_e(t), \hat\sigma_e(t))$ for the ${\rm E}_8 \times {\rm E}_8$ cases. The scaling $l_{e{\cal AB}}^{\cal CD}(t(g_s))$ is now defined as:
\bg\label{malenbego}
\hat{l}_{e{\cal AB}}^{\cal CD}(t(g_s)) = l_{\cal AB}^{\cal CD}(t) 
- {1\over \vert\log~\bar{g}_s\vert}~\log\left[1 + \sum_{k = 1}^\infty {  
{\cal A}_{\scriptscriptstyle{\tiny{\cal (ABCD)}}}^{(k)}\over {\cal A}_{\scriptscriptstyle{\tiny{\cal (ABCD)}}}^{(0)}} \left({g_s \over {\rm H}(y) {\rm H}_o({\bf x})}\right)^{2k\vert {\rm L}_{\cal AB}^{\cal CD}(k; t)\vert\over 3}\right], \nd
where, as mentioned above, one may view ${\rm L}_{\cal AB}^{\cal CD}(k; t)$ to {\it contain} all the non-perturbative corrections in an additive way, much like what we had in \eqref{ivbrestick}. While the reformulation \eqref{andyrey} and \eqref{malenbego} works for both $SO(32)$ and the ${\rm E}_8 \times {\rm E}_8$ cases\footnote{Although, for illustrative purpose, we can use $\hat{l}_{e\cal AB}^{\cal CD}(t(g_s))$ for the ${\rm E}_8 \times {\rm E}_8$ case and $l_{e\cal AB}^{\cal CD}(t(g_s)$ for the $SO(32)$ case.} $-$ unlike the three-form fluxes in \eqref{secdrem18} and \eqref{tatumey} which are sensitive to which heterotic theory we consider, much like the metric components \eqref{bdtran} and \eqref{bdtran2} $-$ they are still {\it not} generic enough. There exist a more generic ans\"atze for both the flux and the metric components that we will discuss later when we study the Schwinger-Dyson equations. Meanwhile \eqref{malenbego} for G-fluxes and \eqref{bdtran} and \eqref{bdtran2} for the metric components for the two heterotic cases would suffice.

\subsubsection{One temporal-derivative acting on the metric components \label{sec4.3.1}}

One of the important subtleties is to quantify how the temporal derivatives act on the various metric components, starting with the one-derivative action here and, in the next sub-section, the two-derivatives' action. Clearly both these actions depend on the $\gamma_l$ factors, $l = 1, .., 10$, that we described in \eqref{tiffmont}. For the $SO(32)$ case, and in the absence of the sub-dominant corrections, the first-derivatives are  captured by $\mathbb{A}(\gamma)$ and $\mathbb{B}(\gamma)$ in \eqref{corsage2}, where $\gamma \equiv (\alpha, \beta)$. In the presence of the sub-dominant corrections, we will express the corresponding $SO(32)$ and ${\rm E}_8 \times {\rm E}_8$ as
$(\mathbb{A}(\gamma_e), \mathbb{B}(\gamma_e))$ and $(\mathbb{A}_8(\hat\gamma_e), \mathbb{B}_8(\hat\gamma_e))$ respectively (see also {\bf Table \ref{jessrog}}). They are defined as:
\bg\label{tifeni}
&& \mathbb{B}(\gamma_e) = -{2\over 3} + \gamma_e(t) - {\log\left(\mp \dot\gamma_e(t)\vert\log~\bar{g}_s\vert\right)\over \vert\log~\bar{g}_s\vert} \nonumber\\
&& \mathbb{B}_8(\hat\gamma_e) = -{2\over 3} + \hat\gamma_e(t) - {\log\left(\mp \dot{\hat\gamma}_e(t)\vert\log~\bar{g}_s\vert\right)\over \vert\log~\bar{g}_s\vert} \nonumber\\
&& \mathbb{A}(\gamma_e) = -{5\over 3} + \gamma_e(t) + \gamma_{1, 2}[\beta_e(t)] - 
{\log\left({2\over 3} -\gamma_e(t)\right)\over \vert\log~\bar{g}_s\vert} \nonumber\\
&&\mathbb{A}_8(\hat\gamma_e) = -{5\over 3} + \hat\gamma_e(t) + \gamma_{1, 2}\left[{\hat\alpha_e(t) + \hat\beta_e(t)\over 2}\right] - 
{\log\left({2\over 3} -\hat\gamma_e(t)\right)\over \vert\log~\bar{g}_s\vert}, \nd 
where $\hat\gamma_e \equiv ({\hat\sigma_e}, {\hat\alpha_e}, {\hat\beta_e})$. This follows exactly the replacement strategy on \eqref{corsage2} that we advocated in \eqref{tiffmont} namely, for the $SO(32)$ case the replacement is simple, whereas for the ${\rm E}_8 \times {\rm E}_8$ theory, the replacement is more non-trivial. Part of the non-triviality comes from the fact that a zero scaling in the $SO(32)$ side can either be represented as a zero scaling in the ${\rm E}_8 \times {\rm E}_8$ theory, or as:
\bg\label{unitymitford}
0\big\vert_{SO(32)} ~ \to ~ \pm \vert a \vert\left[{\hat\alpha_e(t) + \hat\beta_e(t) \over 2} - \left({\hat\alpha_e(t)}, {\hat\beta_e(t)}\right)\right]_{{\rm E}_8 \times {\rm E}_8}, \nd
with $a$ being any number. In section \ref{sec4.3} we will see numerous instances of this, for example in \eqref{chu8hath1}, \eqref{connorice} et cetera. This imples that the substitution method advocated in \eqref{tiffmont}, while correct, cannot always be used to predict the scalings in the ${\rm E}_8 \times {\rm E}_8$ side from the scalings in the $SO(32)$ case. (The other way around, {\it i.e.} predicting the scalings for $SO(32)$ case from the scalings in the ${\rm E}_8 \times {\rm E}_8$ case, is straightforward.)

The second set of subtleties come from the presence of the $(\vert f_k(t)\vert, \vert\widetilde{f}_k(t)\vert)$ and $(\vert \hat{f}_k(t)\vert, \vert \widetilde{\hat{f}}_k\vert)$ factors in \eqref{sinistre} and \eqref{sinistre00} respectively. They would contribute to the temporal derivatives of the metric components along $(\mu, \nu)$ and $(3, 3)$ directions. If we define two quantities:
\bg\label{masatwork}
&&\hat{\zeta}_e(t) = - {1 \over \vert\log~\bar{g}_s\vert}~\log\left[1 + \sum_{k = 1}^\infty {h_k\over h_0} \left({g_s \over {\rm H}(y) {\rm H}_o({\bf x})}\right)^{2k\vert\hat{f}_k(t)\vert\over 3}\right]\nonumber\\
&& \hat{\eta}_e(t) = - {1\over \vert\log~\bar{g}_s\vert}~\log\left[1 + \sum_{k = 1}^\infty {\widetilde{h}_k\over \widetilde{h}_0} \left({g_s \over {\rm H}(y) {\rm H}_o({\bf x})}\right)^{2k\vert\widetilde{\hat{f}}_k(t)\vert\over 3}\right], \nd
and a similar one $(\zeta_e(t), \eta_e(t))$ with $f_k(t)$ and $\widetilde{f}_k(t)$ for the $SO(32)$ case, then the temporal derivatives of the metric components along $(\mu, \nu) \in {\bf R}^{2, 1}$ and $w^3 \in w^a \in {\mathbb{T}^2\over {\cal G}}$ can be represented with scalings $(\mathbb{C}_8, \mathbb{D}_8)$ and $(\mathbb{F}_8, \mathbb{G}_8)$
respectively, where\footnote{In other words, ${\bf g}_{\mu\nu, 0}({\bf x}, y, g_s) = {\bf g}^{(1)}_{\mu\nu}({\bf x}, y) ~\bar{g}_s^{\mathbb{C}_8(\hat\zeta_e)} +{\bf g}^{(2)}_{\mu\nu}({\bf x}, y) ~\bar{g}_s^{\mathbb{D}_8(\hat\zeta_e)}$ and 
${\bf g}_{33, 0}({\bf x}, y, g_s) = {\bf g}^{(1)}_{33}({\bf x}, y) ~\bar{g}_s^{\mathbb{F}_8(\hat\eta_e)} +{\bf g}^{(2)}_{33}({\bf x}, y) ~\bar{g}_s^{\mathbb{G}_8(\hat\eta_e)}$ where $\bar{g}_s = {g_s\over {\rm H}(y){\rm H}({\bf x})}$ and $(\hat\zeta_e, \hat\eta_e)$ defined in \eqref{masatwork}. See also {\bf Table \ref{jessrog}} for more details.}:
\bg\label{vanandcar}
&& \mathbb{F}_8(\hat\eta_e) = +{4\over 3} + \hat\eta_e(t) - {\log\left(\mp \dot{\hat\eta}_e(t)\vert\log~\bar{g}_s\vert\right)\over \vert\log~\bar{g}_s\vert} \nonumber\\
&& \mathbb{C}_8(\hat\zeta_e) = -{8\over 3} + \hat\zeta_e(t) - {\log\left(\mp \dot{\hat\zeta}_e(t)\vert\log~\bar{g}_s\vert\right)\over \vert\log~\bar{g}_s\vert} \nonumber\\
&&\mathbb{G}_8(\hat\eta_e) = +{1\over 3} + \hat\eta_e(t) + \gamma_{1, 2}\left[{\hat\alpha_e(t) + \hat\beta_e(t)\over 2}\right] - 
{\log\left({4\over 3} +\hat\eta_e(t)\right)\over \vert\log~\bar{g}_s\vert}\nonumber\\
&&\mathbb{D}_8(\hat\zeta_e) = -{11\over 3} + \hat\zeta_e(t) + \gamma_{1, 2}\left[{\hat\alpha_e(t) + \hat\beta_e(t)\over 2}\right] - 
{\log\left({8\over 3} -\hat\zeta_e(t)\right)\over \vert\log~\bar{g}_s\vert}, \nd 
which is similar to what we have in \eqref{tifeni}. The difference being that the coefficients in \eqref{tifeni} are defined with respect to $\hat\gamma_e$ and $\gamma_e$ which take values in $(\hat\sigma_e, \hat\alpha_e, \hat\beta_e)$ and $(\alpha_e, \beta_e)$ respectively, while the coefficients in \eqref{vanandcar} are defined with respect to one set of choice from \eqref{masatwork}. For the $SO(32)$ case one can similarly define the coefficients as $(\mathbb{C}(\zeta_e), \mathbb{D}(\zeta_e))$ and $(\mathbb{F}(\eta_e), \mathbb{G}(\eta_e))$ with appropriately changing $\gamma_{1, 2}\Big[{\hat\alpha_e(t) + \hat\beta_e(t)\over 3}\Big]$ to $\gamma_{1, 2}[\beta_e(t)]$. See also {\bf Table \ref{depravity}} for a summary of the notations.

\begin{table}[tb]  
 \begin{center}
 \renewcommand{\arraystretch}{1.5}
\begin{tabular}{|c||c||c|}\hline Data  & $g_s$ representation  \\ \hline\hline
$\bar{g}_{s, 0}$ & $ e_1 ~\bar{g}_s^{\gamma_1\big[{\hat\alpha_e(t) + \hat\beta_e(t)\over 2}\big]} +e_2 ~\bar{g}_s^{\gamma_2\big[{\hat\alpha_e(t) + \hat\beta_e(t)\over 2}\big]}$\\ \hline
$\gamma_1$ & $1 - {\log\Big(\vert {\dot{\hat\alpha}_e(t) + \dot{\hat\beta}_e(t)\over 2}\vert \vert \log~\bar{g}_s\vert\Big)\over \vert\log~\bar{g}_s\vert} + {\log\Big(2 - {\hat\alpha_e(t) + \hat\beta_e(t)\over 2}\Big)\over \vert\log~\bar{g}_s\vert}$\\ \hline
$\gamma_2$ & ${\hat\alpha_e(t) + \hat\beta_e(t)\over 4} +{\log\Big(2 - {\hat\alpha_e(t) + \hat\beta_e(t)\over 2}\Big)\over \vert\log~\bar{g}_s\vert}$\\ \hline
${\bf g}_{mn, 0}({\bf x}, y; g_s)$ &  ${\bf g}^{(1)}_{mn}({\bf x}, y)~ \bar{g}_s^{\mathbb{A}_8(\hat\sigma_e)} +  
{\bf g}^{(2)}_{mn}({\bf x}, y)~\bar{g}_s^{\mathbb{B}_8(\hat\sigma_e)}$ \\ \hline 
${\bf g}_{\theta_1\theta_1, 0}({\bf x}, y; g_s)$ &  $  
{\bf g}^{(1)}_{\theta_1\theta_1}({\bf x}, y)~\bar{g}_s^{\mathbb{A}_8(\hat\alpha_e)} +  
{\bf g}^{(2)}_{\theta_1\theta_1}({\bf x}, y)~\bar{g}_s^{\mathbb{B}_8(\hat\alpha_e)}$ \\ \hline
${\bf g}_{\theta_2\theta_2, 0}({\bf x}, y; g_s)$ &  $  
{\bf g}^{(1)}_{\theta_2\theta_2}({\bf x}, y)~\bar{g}_s^{\mathbb{A}_8(\hat\beta_e)} +{\bf g}^{(2)}_{\theta_2\theta_2}({\bf x}, y)~\bar{g}_s^{\mathbb{B}_8(\hat\beta_e)}$ \\ \hline
${\bf g}_{\mu\nu, 0}({\bf x}, y; g_s)$ &  ${\bf g}^{(1)}_{\mu\nu}({\bf x}, y)~\bar{g}_s^{\mathbb{C}_8(\hat\zeta_e)} +  {\bf g}^{(2)}_{\mu\nu}({\bf x}, y)~\bar{g}_s^{\mathbb{D}_8(\hat\zeta_e)}$ \\
\hline
\end{tabular}
\renewcommand{\arraystretch}{1}
\end{center}
 \caption[]{\Su Data required to compute the first derivatives of the metric components with respect to the temporal coordinate for the ${\rm E}_8 \times {\rm E}_8$ case. As before ${\bf x} \in {\bf R}^2$, $y \in {\cal M}_4 \times {\bf S}^1_{\theta_1} \times {{\bf S}^1_{\theta_2}\over {\cal I}_{\theta_2}}$ and ${\bf g}_{\rm MN}^{(1, 2)}({\bf x}, y) \equiv c_{1,2} {\bf g}_{\rm MN}({\bf x}, y)$ with $c_{1, 2}$ independent of $({\bf x}, y)$ and $\bar{g}_s = {g_s\over {\rm H}(y) {\rm H}_o({\bf x})}$. The other parameters are defined in \eqref{bdtran2} and \eqref{masatwork}. See also {\bf Table \ref{depravity}}.} 
\label{jessrog}
 \end{table}

\begin{table}[tb]  
 \begin{center}
 \resizebox{\columnwidth}{!}{%
 \renewcommand{\arraystretch}{2.06}
\begin{tabular}{|c||c||c|}\hline Riemann tensors  & $g_s$ scalings  \\ \hline\hline
${\bf R}_{mnpq}$ &  ${\rm dom}\left(-{2\over 3} + \hat\sigma_e(t), ~\mathbb{F}_{1e}(t),~ {4\over 3} + 2\hat\sigma_e(t) - \hat\zeta_e(t),~ -{2\over 3} + 2\hat\sigma_e(t) - (\hat\alpha_e(t), \hat\beta_e(t))\right)$ \\ \hline 
${\bf R}_{mn\rho\sigma}$ &  ${\rm dom}\left(-{2\over 3} + \hat\sigma_e(t),~ -{2\over 3} + (\hat\alpha_e(t), \hat\beta_e(t)), ~\mathbb{F}_{2e}(t), ~{4\over 3} + \hat\sigma_e(t) + (\hat\alpha_e(t), \hat\beta_e(t)) - \hat\zeta_e(t), -{2\over 3} + \hat\sigma_e(t) \pm \hat\alpha_e(t) \mp \hat\beta_e(t)\right)$ \\ \hline 
${\bf R}_{\rho\sigma i0}$ &  ${\rm dom}\left(2 + (\hat\alpha_e(t), \hat\beta_e(t)) + \mathbb{C}_8(\hat\zeta_e) - \hat\zeta_e(t), ~2 + (\hat\alpha_e(t), \hat\beta_e(t)) + \mathbb{D}_8(\hat\zeta_e) - \hat\zeta_e(t), ~ \mathbb{F}_{3e}(t)\right)$ \\ \hline 
${\bf R}_{m\rho\sigma 0}$ &  ${\rm dom}\left(\mathbb{F}_{3e}(t), ~ \mathbb{F}_{4e}(t)\right)$ \\ \hline 
${\bf R}_{0\sigma\sigma\rho}$ &  ${\rm dom}~\mathbb{F}_{3e}(t)$ \\ \hline 
${\bf R}_{\rho\sigma\rho\sigma}$ &  ${\rm dom}\left(-{2\over 3} + (\hat\alpha_e(t), \hat\beta_e(t)), ~\mathbb{F}_{5e}(t), ~-{2\over 3} + 2\otimes (\hat\alpha_e(t), \hat\beta_e(t)) - \hat\sigma_e(t), ~{4\over 3} + 2\otimes(\hat\alpha_e(t), \hat\beta_e(t))-\hat\zeta_e(t)\right)$ \\ \hline
${\bf R}_{mn i0}$ &  ${\rm dom}\left(2 + \hat\sigma_e(t) + \mathbb{C}_8(\hat\zeta_e) - \hat\zeta_e(t),~2 + \hat\sigma_e(t) + \mathbb{D}_8(\hat\zeta_e) - \hat\zeta_e(t),~ \mathbb{F}_{6e}(t)\right)$ \\
\hline
\end{tabular}}
\renewcommand{\arraystretch}{1}
\end{center}
 \caption[]{\Su The ${g_s/{\rm H H}_o}$ scalings of the curvature tensors related to the ${\rm E}_8 \times {\rm E}_8$ theory assuming no dependence on ${\mathbb{T}^2/{\cal G}}$ directions, but including the sub-dominant corrections. Here $(m, n) \in {\cal M}_4, (\rho, \sigma) \in {\bf S}^1_{\theta_1} \times {{\bf S}^1_{\theta_2}/{\cal I}_{\theta_2}}, (i, j) \in {\bf R}^2$, $(a, b) \in {\mathbb{T}^2/{\cal G}}$, $\gamma_{1, 2}\big[{\hat\alpha_e(t) + \hat\beta_e(t)\over 3}\big]$ are defined from \eqref{ishena}, $(\hat\alpha_e(t), \hat\beta_e(t), \hat\sigma_e(t))$ are defined in \eqref{bdtran2}, $(\hat\zeta_e(t), \hat\eta_e(t))$ are defined in \eqref{masatwork}, $(\mathbb{F}_8(\hat\eta_e), \mathbb{G}_8(\hat\eta_e), \mathbb{C}_8(\hat\zeta_e), \mathbb{D}_8(\hat\zeta_e))$ are defined in \eqref{vanandcar}; $\mathbb{F}_{ie}(t)$ are 
 defined in {\bf Table \ref{privsocmey}}. In computing the scalings all permutations of the curvature indices are taken into account. One should compare the results here with the ones from {\bf Table \ref{firzacut}} which was done for the $SO(32)$ case without including the sub-dominant corrections.} 
\label{firzathai}
 \end{table}

\begin{table}[tb]  
 \begin{center}
\renewcommand{\arraystretch}{1.5}
\begin{tabular}{|c||c||c|}\hline Riemann tensors  & $g_s$ scalings  \\ \hline\hline
${\bf R}_{m\rho\sigma i}, {\bf R}_{i\sigma \sigma\rho}, {\bf R}_{m\sigma\sigma\rho}$ &  $-{2 \over 3} + (\hat\alpha_e(t), \hat\beta_e(t))$ \\ \hline 
${\bf R}_{mnpi}, {\bf R}_{mnp\sigma}, {\bf R}_{mn\sigma i}$ &  $-{2\over 3} + \hat\sigma_e(t)$ \\ \hline 
${\bf R}_{mabi}, {\bf R}_{\sigma ab i}, {\bf R}_{m\sigma a b}$ &  $~~{4\over 3} + (0, \hat\eta_e(t))$ \\ \hline 
${\bf R}_{m0i0}, {\bf R}_{mijk}, {\bf R}_{0\sigma i0}$ & $ -{8\over 3} + \hat\zeta_e(t)$ \\ \hline
${\bf R}_{\sigma ijk}, {\bf R}_{m\sigma ij}, {\bf R}_{0m 0\sigma}$ & $ -{8\over 3} + \hat\zeta_e(t)$ \\ \hline
\end{tabular}
\renewcommand{\arraystretch}{1}
\end{center}
 \caption[]{\Su The ${g_s\over {\rm H H}_o}$ scalings of the curvature tensors for the ${\rm E}_8 \times {\rm E}_8$ case that involve all the dominant and sub-dominant corrections. One may compare the scalings with the ones from {\bf Table \ref{firzacut5}} done for the simplified $SO(32)$ case.} 
\label{firzathai2}
 \end{table}

\begin{table}[tb]  
 \begin{center}
\resizebox{\columnwidth}{!}{%
\renewcommand{\arraystretch}{2.0}
\begin{tabular}{|c||c||c|}\hline Riemann tensors  & $g_s$ scalings \\ \hline\hline
${\bf R}_{0i0j}$ & ${\rm dom}\left(-{8\over 3} + \hat\zeta_e(t), -{14\over 3} + 2\hat\zeta_e(t) - \hat\sigma_e(t), -{14\over 3} + 2\hat\zeta_e(t) - (\hat\alpha_e(t), \hat\beta_e(t)), \mathbb{F}_{16e}(t)\right)$\\ \hline
${\bf R}_{0a0b}$ & ${\rm dom}\left({4\over 3} + (0, \hat\eta_e(t)), ~ -{2\over 3} + \hat\zeta_e(t) - \hat\sigma_e(t) + (0, \hat\eta_e(t)), ~ -{2\over 3} + \hat\zeta_e(t) - (\hat\alpha_e(t), \hat\beta_e(t)) + (0, \hat\eta_e(t)), ~
\mathbb{F}_{17e}(t)\right)$\\ \hline
${\bf R}_{0m0n}$ & ${\rm dom}\left(-{8\over 3} + \hat\zeta_e(t), ~-{2\over 3} + \hat\sigma_e(t),~\mathbb{F}_{14e}(t), ~ \mathbb{F}_{18e}(t),~ 
-{8\over 3} +\hat\zeta_e(t) + \hat\sigma_e(t) - (\hat\alpha_e(t), \hat\beta_e(t))\right)$ \\ \hline
${\bf R}_{0\rho 0\sigma}$ & ${\rm dom}\left(-{8\over 3} + \hat\zeta_e(t), ~-{2\over 3} + (\hat\alpha_e(t), \hat\beta_e(t)),~\mathbb{F}_{15e}(t), ~ \mathbb{F}_{20e}(t), ~-{8\over 3} + \hat\zeta_e(t) + (\hat\alpha_e(t), \hat\beta_e(t)) - \hat\sigma_e(t)\right)$ \\ \hline
\end{tabular}}
\renewcommand{\arraystretch}{1}
\end{center}
 \caption[]{\Su The ${g_s\over {\rm H H}_o}$ scalings of the curvature tensors that involve second derivative $\ddot{g}_s$ for the ${\rm E}_8 \times {\rm E}_8$ case. Again no dependence on the toroidal direction is assumed; and $\mathbb{F}_i(t)$ are defined in {\bf Table \ref{privsocmey}}. One should also compare the results to the ones from {\bf Table \ref{firzacut2}} which was done for the simplified $SO(32)$ case. Expectedly, the results for the generic ${\rm E}_8 \times {\rm E}_8$ case are more non-trivial.} 
  \label{firzathai3}
 \end{table}

\begin{table}[tb]  
 \begin{center}
 \resizebox{\columnwidth}{!}{%
 \renewcommand{\arraystretch}{2.3}
\begin{tabular}{|c||c||c|}\hline $\mathbb{F}_i(t)$  & $g_s$ scalings \\ \hline\hline
$\mathbb{F}_{1e}(t)$ & $2 \otimes \mathbb{A}_8(\hat\sigma_e) + {8\over 3} - \hat\zeta_e(t), ~ 2\otimes \mathbb{B}_8(\hat\sigma_e) + {8\over 3} - \hat\zeta_e(t), ~\mathbb{A}_8(\hat\sigma_e) + \mathbb{B}_8(\hat\sigma_e) + {8\over 3} - \hat\zeta_e(t)$\\ \hline
$\mathbb{F}_{2e}(t)$ & $\mathbb{A}_8(\hat\sigma_e) + \mathbb{A}_8(\hat\alpha_e, \hat\beta_e) + {8\over 3} - \hat\zeta_e(t), 
~\mathbb{A}_8(\hat\sigma_e) + \mathbb{B}_8(\hat\alpha_e, \hat\beta_e) + {8\over 3} - \hat\zeta_e(t), 
~ \mathbb{B}_8(\hat\sigma_e) + \mathbb{A}_8(\hat\alpha_e, \hat\beta_e) + {8\over 3} -\hat\zeta_e(t), 
~ \mathbb{B}_8(\hat\sigma_e) + \mathbb{B}_8(\hat\alpha_e, \hat\beta_e) + {8\over 3} - \hat\zeta_e(t)$\\ \hline
$\mathbb{F}_{3e}(t)$ & $\mathbb{A}_8(\hat\alpha_e, \hat\beta_e), ~ 
\mathbb{B}_8(\hat\alpha_e, \hat\beta_e)$ \\ \hline
$\mathbb{F}_{4e}(t)$ & $\mathbb{A}_8(\hat\sigma_e) + (\hat\alpha_e(t), \hat\beta_e(t)) - \hat\sigma_e(t), ~ \mathbb{B}_8(\hat\sigma_e) + (\hat\alpha_e(t), \hat\beta_e(t)) - \hat\sigma_e(t)$\\ \hline
$\mathbb{F}_{5e}(t)$ & $2\otimes \mathbb{A}_8(\hat\alpha_e, \hat\beta_e) + {8\over 3} - \hat\zeta_e(t), ~ 2\otimes \mathbb{B}_8(\hat\alpha_e, \hat\beta_e) + {8\over 3} - \hat\zeta_e(t), ~\mathbb{A}_8(\hat\alpha_e, \hat\beta_e) \oplus \mathbb{B}_8(\hat\alpha_e, \hat\beta_e) + {8\over 3} - \hat\zeta_e(t)$\\ \hline
$\mathbb{F}_{6e}(t)$ & $\mathbb{A}_8(\hat\sigma_e), ~ \mathbb{B}_8(\hat\sigma_e)$ \\ \hline
$\mathbb{F}_{7e}(t)$ & $\mathbb{A}_8(\hat\alpha_e, \hat\beta_e) - (\hat\alpha_e(t), \hat\beta_e(t)) + \hat\sigma_e(t), ~ \mathbb{B}_8(\hat\alpha_e, \hat\beta_e)- (\hat\alpha_e(t), \hat\beta_e(t)) + \hat\sigma_e(t)$\\
\hline
\end{tabular}}
\renewcommand{\arraystretch}{1}
\end{center}
 \caption[]{\Su The ${g_s\over {\rm H H}_o}$ scalings of the $\mathbb{F}_i(t)$  factors appearing in {\bf Tables \ref{firzathai}} and {\bf \ref{firzathai3}} for the ${\rm E}_8 \times {\rm E}_8$ case. 
 We have defined $\mathbb{N}_8(\Sigma_e) = (\mathbb{A}_8(\Sigma_e), \mathbb{B}_8(\Sigma_e)), \mathbb{M}_8(\hat\eta_e) = (\mathbb{F}_8(\hat\eta_e), \mathbb{G}_8(\hat\eta_e))$ and $\mathbb{P}_8(\hat\zeta_e) = (\mathbb{C}_8(\hat\zeta_e), \mathbb{D}_8(\hat\zeta_e))$ with $\Sigma_e(t) = \hat\sigma_e(t)$ or $\Sigma_e(t) = (\hat\alpha_e(t), \hat\beta_e(t))$. All other parameters are defined in section \ref{sec4.4} and in {\bf Table \ref{jessrog}}. Note the absence of $\mathbb{F}_{19e}(t)$ and $\mathbb{F}_{21e}(t)$, although their counterparts appear in the simplified $SO(32)$ case in {\bf Table \ref{firzacut3}}.} 
  \label{privsocmey}
 \end{table} 

With this we are ready to tackle the scalings of the curvature tensors that would now include all the higher order corrections and for a generic setting with three warp-factors. To get the results that we have in {\bf Tables \ref{firzacut}} to {\bf \ref{firzacut3}} we have to equate the two warp factors $\hat\alpha_e(t)$ and $\hat\beta_e(t)$ to $\beta_e(t)$ and make other changes mentioned earlier, including the switching off of the higher-order corrections. An example may be given in the following way:

{\scriptsize
\bg\label{nickid}
{\bf R}_{\rho\sigma i0}({\bf x}, y; g_s) & = & {\bf R}^{(1)}_{\rho\sigma i0}({\bf x}, y)\left({g_s\over {\rm H}(y) {\rm H}({\bf x})}\right)^{\mathbb{A}_8(\hat\alpha_e, \hat\beta_e)} + {\bf R}^{(2)}_{\rho\sigma i0}({\bf x}, y)\left({g_s\over {\rm H}(y) {\rm H}({\bf x})}\right)^{\mathbb{B}_8(\hat\alpha_e, \hat\beta_e)}\\
& + & {\bf R}^{(3)}_{\rho\sigma i0}({\bf x}, y)\left({g_s\over {\rm H}(y) {\rm H}({\bf x})}\right)^{2 + (\hat\alpha_e, \hat\beta_e) + \mathbb{C}_8(\hat\zeta_e) - \hat\zeta_e} + 
{\bf R}^{(4)}_{\rho\sigma i0}({\bf x}, y)\left({g_s\over {\rm H}(y) {\rm H}({\bf x})}\right)^{2 + (\hat\alpha_e, \hat\beta_e) + \mathbb{D}_8(\hat\zeta_e) - \hat\zeta_e}, \nonumber \nd}
where $(\hat\alpha_e, \hat\beta_e)$ suggests that we can choose between $\hat\alpha_e$ and $\hat\beta_e$ depending on $(\rho, \sigma)$ taking values $(\theta_1, \theta_1)$ or $(\theta_2, \theta_2)$ respectively. The other parameters are defined as follows: $(\mathbb{A}_8(\hat\alpha_e, \hat\beta_e), \mathbb{B}_8(\hat\alpha_e, \hat\beta_e))$ are given in \eqref{tifeni}, $\hat\zeta_e$ in \eqref{masatwork} and $(\mathbb{C}_8(\hat\zeta_e), \mathbb{D}_8(\hat\zeta_e))$ appear in \eqref{vanandcar}. The presence of $-\hat\zeta_e$ implies that we have to Taylor expand the series from \eqref{masatwork}. Note that, compared to 
the third row in {\bf Table \ref{firzacut}}, the scalings appear to be much more involved than $\mathbb{A}(\beta)$, $\mathbb{B}(\beta)$ and 
$-{5\over 3} + \beta(t) + \gamma_{1, 2}(\beta)$. This is true, but a simple substitution of the form discussed earlier would convert $\mathbb{A}_8(\hat\alpha_e, \hat\beta_e) \to \mathbb{A}(\beta)$ and  $\mathbb{B}_8(\hat\alpha_e, \hat\beta_e) \to \mathbb{B}(\beta)$. The third term appears from making $\hat\zeta_e = 1$ and converting ${\hat\alpha_e(t) + \hat\beta_e(t)\over 2} \to \beta_e(t) \to \beta(t)$. In {\bf Table \ref{firzathai}} we denote the scalings in \eqref{nickid} as:

{\footnotesize
\bg\label{aspbruke}
{\rm dom}\left(2 + (\hat\alpha_e(t), \hat\beta_e(t)) + \mathbb{C}_8(\hat\zeta_e) - \hat\zeta_e(t), ~2 + (\hat\alpha_e(t), \hat\beta_e(t)) + \mathbb{D}_8(\hat\zeta_e) - \hat\zeta_e(t), ~ \mathbb{F}_{3e}(t)\right), \nd}
where $\mathbb{F}_{3e}(t)$ is defined in {\bf Table \ref{privsocmey}}, and the other parameters are defined above. Note that the scalings presented in {\bf Tables \ref{firzathai}}, {\bf \ref{firzathai2}} and {\bf \ref{firzathai3}} are a generalizations of those presented in {\bf Tables \ref{firzacut}}, {\bf \ref{firzacut5}} and {\bf \ref{firzacut2}}, not only to include the ${\rm E}_8$ case with three warp-factors but also to include the sub-dominant corrections. Similarly the contents in {\bf Table \ref{privsocmey}} are a generalization of those in {\bf Table \ref{firzacut2}}. The reason for presenting the simpler cases first in 
{\bf Tables \ref{firzacut}}, {\bf \ref{firzacut5}} and {\bf \ref{firzacut2}} is two-fold: {\Su One}, the scalings and the physics are much more easier to interpret from the simpler tables than the generalized ones in {\bf Tables \ref{firzathai}}, {\bf \ref{firzathai2}} and {\bf \ref{firzathai3}}, and {\Su two}, we shall use the simpler tables to study the explicit EOMs in later sections. An example may be given directly from \eqref{aspbruke}. As it stands, it is slightly difficult to infer the explicit scalings of the curvature tensor ${\bf R}_{\rho\sigma i0}$ from \eqref{aspbruke}. To go to the simpler case, we first convert $(\hat\alpha_e(t), \hat\beta_e(t))$ to $\beta_e(t)$ which can then be converted to $\beta(t)$ by removing the sub-dominant corrections. The other parameters change in the following way:
\bg\label{ovricate}
&& \hat\zeta_e(t) \to 0, ~~~\mathbb{B}_8(\hat\alpha_e, \hat\beta_e) \to \mathbb{B}(\beta), ~~~\mathbb{A}_8(\hat\alpha_e, \hat\beta_e) \to \mathbb{A}(\beta)\nonumber\\
&& \mathbb{C}_8(\hat\zeta_e) \to -{2\over 3} + \beta(t) + \infty, ~~~
\mathbb{D}_8(\hat\zeta_e) \to -{11\over 3} + \gamma_{1, 2}(\beta), \nd
with the $\mathbb{C}_8(\hat\zeta_e)$ scaling dropping out and the remaining parameters conspiring to reproduce the scalings in the third rows of {\bf Tables \ref{firzacut}} and {\bf \ref{firzacut3}}. We will follow this strategy to compute the dominant behavior of $\mathbb{F}_{ie}(t)$ and $\mathbb{F}_i(t)$ factors in section \ref{sec4.3}. However before we do that, we need to figure out the generic scalings of the metric tensors once we allow second derivative with respect to the temporal coordinate for both the $SO(32)$ and the ${\rm E}_8 \times {\rm E}_8$ cases. In the following we study this.

\subsubsection{Two temporal-derivatives acting on the metric components \label{sec4.3.2}}

The behavior of the metric tensors for one derivatives of the temporal coordinate is summarized in {\bf Table \ref{jessrog}}. For two derivative, the story is slightly involved although a simpler study for the $SO(32)$ case, in the absence of the sub-dominant corrections, appears in \eqref{ishena2} using $\gamma_3[\beta], ..., \gamma_{10}[\beta]$ where the $\gamma_i[\beta]$ are defined in {\bf Table \ref{rindrag}}. For the generic ${\rm E}_8 \times {\rm E}_8$ case, we expect $\beta(t) \to \beta_e(t) \to {\hat\alpha_e(t) + \hat\beta_e(t)\over 2}$, and therefore:
\bg\label{alexpog}
\gamma_{[3,...,10]}[\beta(t)] ~ \to ~ \gamma_{[3,...,10]}[\beta_e(t)] ~ \to ~
\gamma_{[3,..., 10]}\Big[{\hat\alpha_e(t) + \hat\beta_e(t)\over 2}\Big], \nd
which we will use in the following to compute the changes to the scalings in {\bf Table \ref{firzacut3}} in the generic ${\rm E}_8 \times {\rm E}_8$ case. The first derivatives are defined using $\gamma_{1, 2}\Big[{\hat\alpha_e(t) + \hat\beta_e(t)\over 2}\Big]$, and if we take the metric components as ${\bf g}_{\rm MN}({\bf x}, y; g_s) = {\bf g}_{\rm MN}({\bf x}, y)~\bar{g}_s^{a_{({\rm MN})} + \Sigma_{e({\rm MN})}(t)}$, then:

{\footnotesize
\bg\label{libertine}
{\bf g}_{\rm MN, 0}({\bf x}, y; g_s) &= & {\bf g}^{(1)}_{\rm MN}({\bf x}, y) ~\left({g_s\over {\rm H}(y) {\rm H}_o({\bf x})}\right)^{a_{({\rm MN})} + \Sigma_{e({\rm MN})}(t) - {\log\left(\mp \dot\Sigma_{e({\rm MN})}(t) \vert\log~\bar{g}_s\vert \right)\over \vert\log~\bar{g}_s\vert}}\\
&+& {\bf g}^{(2)}_{\rm MN}({\bf x}, y) ~\left({g_s\over {\rm H}(y) {\rm H}_o({\bf x})}\right)^{a_{({\rm MN})} - 1 + \Sigma_{e({\rm MN})}(t) + \gamma_{1, 2}\Big[{\hat\alpha_e(t) + \hat\beta_e(t)\over 2}\Big] - {\log\vert a_{({\rm MN})} + \Sigma_{e({\rm MN})}(t)\vert \over \vert\log~\bar{g}_s\vert}}, \nonumber \nd}
where ${\bf g}^{(1, 2)}_{\rm MN}({\bf x}, y) = c_{1, 2}~{\bf g}_{\rm MN}({\bf x}, y)$ with a constant $c_{1, 2}$; $a_{({\rm MN})}$ are constants 
with $({\rm M, N}) \in {\bf R}^{1, 2} \times {\cal M}_4 \times {\bf S}^1_{\theta_1} \times {{\bf S}^1_{\theta_2}/{\cal I}_{\theta_2}} \times {\mathbb{T}^2\over {\cal G}}$ such that $a_{({\rm MN})} \equiv (a_{(\mu\nu)}, a_{(mn)}, a_{(\theta_i \theta_i)}, a_{(ab)})$ $= (-{8\over 3}, -{2\over 3},$ $ -{2\over 3}, +{4\over 3})$; and $\Sigma_{e({\rm MN})}(t) \equiv (\Sigma_{e(\mu\nu)}, \Sigma_{e(mn)}, \Sigma_{e(\theta_1 \theta_1)}, \Sigma_{e(\theta_2 \theta_2)}, \Sigma_{e(33)}, \Sigma_{e(11, 11)}) $  $= (\hat\zeta_e(t), \hat\sigma_e(t), $ $\hat\alpha_e(t), \hat\beta_e(t), \hat\eta_e(t), 0)$ in the same order as the coordinates being arranged. In the form \eqref{libertine}, one may easily map it to the various expansion parameters from {\bf Table \ref{jessrog}}. The second derivative now takes the following form:

{\scriptsize
\bg\label{howifjewls}
{\bf g}_{\rm MN, 00}({\bf x}, y; g_s) &=&  {\bf g}^{(1a)}_{\rm MN}({\bf x}, y) ~\bar{g}_s^{a_{({\rm MN})} + \Sigma_{e({\rm MN})}(t) - {2\log(\mp \dot\Sigma_{e({\rm MN})}(t)\vert\log~\bar{g}_s\vert)\over \vert\log~\bar{g}_s\vert}}\\
& + & {\bf g}^{(2a)}_{\rm MN}({\bf x}, y) ~\bar{g}_s^{a_{({\rm MN})} + \Sigma_{e({\rm MN})}(t) - {\log(\mp \ddot\Sigma_{e({\rm MN})}(t)\vert\log~\bar{g}_s\vert)\over \vert\log~\bar{g}_s\vert}} \nonumber\\
& + & {\bf g}^{(3a)}_{\rm MN}({\bf x}, y) ~\bar{g}_s^{a_{({\rm MN})} + \Sigma_{e({\rm MN})}(t) + \gamma_{1, 2} - 1- {\log(\mp \dot\Sigma_{e({\rm MN})}(t))\over \vert\log~\bar{g}_s\vert}}\nonumber\\
& + & {\bf g}^{(4a)}_{\rm MN}({\bf x}, y) ~\bar{g}_s^{a_{({\rm MN})} + \Sigma_{e({\rm MN})}(t) + \gamma_{1, 2} - 1- {\log(\mp \dot\Sigma_{e({\rm MN})}(t)\vert\log~\bar{g}_s\vert)\over \vert\log~\bar{g}_s\vert} - {\log(\vert a_{({\rm MN})} + \Sigma_{e({\rm MN})}(t)\vert)\over \vert\log~\bar{g}_s\vert}}\nonumber\\
& + & {\bf g}^{(5a)}_{\rm MN}({\bf x}, y) ~\bar{g}_s^{a_{({\rm MN})} + \Sigma_{e({\rm MN})}(t) + \gamma_{1, 2} - 1 - {\log(\vert a_{({\rm MN})} + \Sigma_{e({\rm MN})}(t)\vert) \over \vert\log~\bar{g}_s\vert} - {\log(\mp(\dot\Sigma_{e({\rm MN})}(t) + \dot\gamma_{1, 2})\vert\log~\bar{g}_s\vert)\over \vert\log~\bar{g}_s\vert}}\nonumber\\
& + & {\bf g}^{(6a)}_{\rm MN}({\bf x}, y) ~\bar{g}_s^{a_{({\rm MN})} + \Sigma_{e({\rm MN})}(t) + 2\gamma_{1, 2} - 2 - {\log(\vert a_{({\rm MN})} + \Sigma_{e({\rm MN})}(t)\vert) \over \vert\log~\bar{g}_s\vert} - {\log(\vert a_{({\rm MN})} + \Sigma_{e({\rm MN})}(t)  - 1\vert)\over \vert\log~\bar{g}_s\vert}}\nonumber
\nd}
where $\bar{g}_s = {g_s\over {\rm H}(y) {\rm H}_o({\bf x})}$, 
$\gamma_{1, 2} = \gamma_{1,2}\big[{\hat\alpha_e(t) + \hat\beta_e(t)\over 2}\big]$ and the superscript $a$ is used for later convenience. We can simplify a little bit the expressions in \eqref{howifjewls} by removing $\dot\gamma_{1,2}$ and replacing it by $\gamma_{[3,...,10]}$. This gives us the following expression:

{\footnotesize
\bg\label{selenzoe}
{\bf g}_{\rm MN, 00}({\bf x}, y; g_s) & = & \sum_{i = 1}^6 {\bf g}^{(i)}_{\rm MN}({\bf x}, y)~\left({g_s\over {\rm H}(y) {\rm H}_o({\bf x})}\right)^{\mathbb{A}_{i8}^{({\rm MN)}}(\Sigma_e(t))}\\
&=&  {\bf g}^{(1)}_{\rm MN}({\bf x}, y) ~\bar{g}_s^{a_{({\rm MN})} + \Sigma_{e({\rm MN})}(t) - {2\log(\mp \dot\Sigma_{e({\rm MN})}(t)\vert\log~\bar{g}_s\vert)\over \vert\log~\bar{g}_s\vert}}\nonumber\\
& + & {\bf g}^{(2)}_{\rm MN}({\bf x}, y) ~\bar{g}_s^{a_{({\rm MN})} + \Sigma_{e({\rm MN})}(t) - {\log(\mp \ddot\Sigma_{e({\rm MN})}(t)\vert\log~\bar{g}_s\vert)\over \vert\log~\bar{g}_s\vert}} \nonumber\\
& + & {\bf g}^{(3)}_{\rm MN}({\bf x}, y) ~\bar{g}_s^{a_{({\rm MN})} + \Sigma_{e({\rm MN})}(t) + \gamma_{1, 2} - 1- {\log(\mp \dot\Sigma_{e({\rm MN})}(t))\over \vert\log~\bar{g}_s\vert}}\nonumber\\
& + & {\bf g}^{(4)}_{\rm MN}({\bf x}, y) ~\bar{g}_s^{a_{({\rm MN})} + \Sigma_{e({\rm MN})}(t) + \gamma_{[3, ...., 10]} - 1 - {\log(\vert a_{({\rm MN})} + \Sigma_{e({\rm MN})}(t)\vert) \over \vert\log~\bar{g}_s\vert}} \nonumber\\
& + & {\bf g}^{(5)}_{\rm MN}({\bf x}, y) ~\bar{g}_s^{a_{({\rm MN})} + \Sigma_{e({\rm MN})}(t) + \gamma_{1, 2} - 1- {\log(\mp \dot\Sigma_{e({\rm MN})}(t)\vert\log~\bar{g}_s\vert)\over \vert\log~\bar{g}_s\vert} - {\log(\vert a_{({\rm MN})} + \Sigma_{e({\rm MN})}(t)\vert)\over \vert\log~\bar{g}_s\vert}}\nonumber\\
& + & {\bf g}^{(6)}_{\rm MN}({\bf x}, y) ~\bar{g}_s^{a_{({\rm MN})} + \Sigma_{e({\rm MN})}(t) + 2\gamma_{1, 2} - 2 - {\log(\vert a_{({\rm MN})} + \Sigma_{e({\rm MN})}(t)\vert) \over \vert\log~\bar{g}_s\vert} - {\log(\vert a_{({\rm MN})} + \Sigma_{e({\rm MN})}(t) - 1\vert)\over \vert\log~\bar{g}_s\vert}}\nonumber
\nd}
with the coefficients ${\bf g}^{(1, 2, 3)}_{\rm MN}({\bf x}, y) ={\bf g}^{(1a, 2a, 3a)}_{\rm MN}({\bf x}, y)$ and ${\bf g}_{\rm MN}^{(4)}({\bf x}, y), {\bf g}^{(5)}_{\rm MN}({\bf x}, y)$ and ${\bf g}_{\rm MN}^{(6)}({\bf x}, y)$ replacing ${\bf g}_{\rm MN}^{(5a)}({\bf x}, y), {\bf g}_{\rm MN}^{(4a)}({\bf x}, y)$ and ${\bf g}_{\rm MN}^{(6a)}({\bf x}, y)$ respectively. The other coefficient $\gamma_{[3, ...., 10]} = \gamma_{[3,..., 10]}\big[{\hat\alpha_e(t) + \hat\beta_e(t)\over 2}\big]$. It is interesting to note that, in the absence of the sub-dominant corrections, the scalings become very simple and are represented by:
\bg\label{gascon}
{\bf g}_{\rm MN, 00}({\bf x}, y; g_s) = {\bf g}^{(a_1)}_{\rm MN}({\bf x}, y) ~\bar{g}_s^{a_{({\rm MN)}} - 1 + \gamma_{[3, ..., 10]}} + 
{\bf g}^{(a_2)}_{\rm MN}({\bf x}, y) ~\bar{g}_s^{a_{({\rm MN)}} - 2 + 2\gamma_{1, 2}}, \nd
which is what appeared earlier and also in {\bf Table \ref{firzacut3}}. 
(The coefficients ${\bf g}_{\rm MN}^{(a_1)}({\bf x}, y)$ and ${\bf g}_{\rm MN}^{(a_2)}({\bf x}, y)$ are related to the coefficients ${\bf g}_{\rm MN}^{(4)}({\bf x}, y)$ and ${\bf g}_{\rm MN}^{(6)}({\bf x}, y)$ respectively.) Clearly now the story is more involved with lots of other terms that depend on the sub-dominant corrections, including their temporal derivatives. This is reflected, for example, in the expressions for the curvature tensor ${\bf R}_{0i0j}({\bf x}, y; g_s)$:

{\scriptsize
\bg\label{lusliber}
{\bf R}_{0i0j}({\bf x}, y; g_s) &=& \sum_{i = 1}^6{\bf R}^{(i)}_{0i0j}({\bf x}, y)~\left({g_s\over {\rm H}(y) {\rm H}_o({\bf x})}\right)^{\mathbb{A}_{i8}^{(ij)}(\hat\zeta_e)} + {\bf R}^{(7)}_{0i0j}({\bf x}, y)~\left({g_s\over {\rm H}(y) {\rm H}_o({\bf x})}\right)^{-{8\over 3} + \hat\zeta_e(t)} \nonumber\\
&+ & {\bf R}^{(8)}_{0i0j}({\bf x}, y)~\left({g_s\over {\rm H}(y) {\rm H}_o({\bf x})}\right)^{{8\over 3} -\hat\zeta_e(t) + 2\mathbb{D}_8(\hat\zeta_e)} +  {\bf R}^{(9)}_{0i0j}({\bf x}, y)~\left({g_s\over {\rm H}(y) {\rm H}_o({\bf x})}\right)^{{8\over 3} -\hat\zeta_e(t) + 2\mathbb{C}_8(\hat\zeta_e)}\nonumber\\
& + & {\bf R}^{(10)}_{0i0j}({\bf x}, y)~\left({g_s\over {\rm H}(y) {\rm H}_o({\bf x})}\right)^{{8\over 3} -\hat\zeta_e(t) + \mathbb{C}_8(\hat\zeta_e) + \mathbb{D}_8(\hat\zeta_e)}+  {\bf R}^{(11)}_{0i0j}({\bf x}, y)~\left({g_s\over {\rm H}(y) {\rm H}_o({\bf x})}\right)^{-{14\over 3} + 2\hat\zeta_e(t) - \hat\sigma_e(t)} \nonumber\\
& + & {\bf R}^{(12)}_{0i0j}({\bf x}, y)~\left({g_s\over {\rm H}(y) {\rm H}_o({\bf x})}\right)^{-{14\over 3} + 2\hat\zeta_e(t) - (\hat\alpha_e(t), \hat\beta_e(t))}
\nonumber
\nd}
where $\mathbb{A}_{i8}^{(ij)}(\hat\zeta_e)$ can be read from \eqref{selenzoe} and it has ${\bf \Su 16}$ distinct scalings, at least in the way we have represented the derivatives of $g_s$. $\mathbb{D}_8(\hat\zeta_e)$ on the other hand has ${\bf \Su 2}$ distinct scalings. Together therefore the curvature tensor ${\bf R}_{0i0j}({\bf x}, y; g_s)$ in \eqref{lusliber} may be expressed using at least ${\bf \Su 24}$ distinct scalings. In {\bf Table \ref{firzathai3}} we represent this as:
\bg\label{ryantell}
{\rm dom}\left(-{8\over 3} + \hat\zeta_e(t), -{14\over 3} + 2\hat\zeta_e(t) - \hat\sigma_e(t), -{14\over 3} + 2\hat\zeta_e(t) - (\hat\alpha_e(t), \hat\beta_e(t)), \mathbb{F}_{16e}(t)\right), \nd
where the expression for $\mathbb{F}_{16e}(t)$ appears in {\bf Table \ref{privsocmey}}. On the other hand the $g_s$ scalings of the curvature tensor ${\bf R}_{0a0b}({\bf x}, y; g_s)$ is bit more non-trivial. This takes the following form:

{\scriptsize
\bg\label{lusliber2}
{\bf R}_{0a0b}({\bf x}, y; g_s) &=& \sum_{i = 1}^6{\bf R}^{(i)}_{0a0b}({\bf x}, y)~\left({g_s\over {\rm H}(y) {\rm H}_o({\bf x})}\right)^{\mathbb{A}_{i8}^{(ab)}(\hat\eta_e)} + {\bf R}^{(7)}_{0a0b}({\bf x}, y)~\left({g_s\over {\rm H}(y) {\rm H}_o({\bf x})}\right)^{+{4\over 3} + (0, \hat\eta_e(t))} \nonumber\\
&+ & {\bf R}^{(8)}_{0a0b}({\bf x}, y)~\left({g_s\over {\rm H}(y) {\rm H}_o({\bf x})}\right)^{{8\over 3} -\hat\zeta_e(t) + \mathbb{C}_8(\hat\zeta_e) + \big(\mathbb{F}_8(\hat\eta_e), {1\over 3} + \gamma_{1, 2}\big[{\hat\alpha_e + \hat\beta_e\over 2}\big]\big)} \nonumber\\
&+ & {\bf R}^{(9)}_{0a0b}({\bf x}, y)~\left({g_s\over {\rm H}(y) {\rm H}_o({\bf x})}\right)^{{8\over 3} -\hat\zeta_e(t) + \mathbb{C}_8(\hat\zeta_e) + \big(\mathbb{G}_8(\hat\eta_e), {1\over 3} + \gamma_{1, 2}\big[{\hat\alpha_e + \hat\beta_e\over 2}\big]\big)} \nonumber\\
&+ & {\bf R}^{(10)}_{0a0b}({\bf x}, y)~\left({g_s\over {\rm H}(y) {\rm H}_o({\bf x})}\right)^{{8\over 3} -\hat\zeta_e(t) + \mathbb{D}_8(\hat\zeta_e) + \big(\mathbb{F}_8(\hat\eta_e), {1\over 3} + \gamma_{1, 2}\big[{\hat\alpha_e + \hat\beta_e\over 2}\big]\big)} \nonumber\\
&+ & {\bf R}^{(11)}_{0a0b}({\bf x}, y)~\left({g_s\over {\rm H}(y) {\rm H}_o({\bf x})}\right)^{{8\over 3} -\hat\zeta_e(t) + \mathbb{D}_8(\hat\zeta_e) + \big(\mathbb{G}_8(\hat\eta_e), {1\over 3} + \gamma_{1, 2}\big[{\hat\alpha_e + \hat\beta_e\over 2}\big]\big)}\nonumber\\
& + &  {\bf R}^{(12)}_{0a0b}({\bf x}, y)~\left({g_s\over {\rm H}(y) {\rm H}_o({\bf x})}\right)^{-{2\over 3} + \hat\zeta_e(t) - \hat\sigma_e(t) + (0, \hat\eta_e(t))}\nonumber\\
& + &  {\bf R}^{(13)}_{0a0b}({\bf x}, y)~\left({g_s\over {\rm H}(y) {\rm H}_o({\bf x})}\right)^{-{2\over 3} + \hat\zeta_e(t) - (\hat\alpha_e(t), \hat\beta_e(t)) + (0, \hat\eta_e(t))},
\nd}
where $(\mathbb{C}_8(\hat\zeta_e),\mathbb{D}_8(\hat\zeta_e),\mathbb{F}_8(\hat\eta_e),\mathbb{G}_8(\hat\eta_e))$ are defined in {\bf Table \ref{jessrog}}
and $\mathbb{A}_{i8}^{(ab)}(\hat\eta_e)$ can be read-off from \eqref{selenzoe}. The above scalings are represented in {\bf Table \ref{firzathai3}} in the following way:

{\scriptsize
\bg\label{pazthai}
{\rm dom}\left({4\over 3} + (0, \hat\eta_e(t)), ~ -{2\over 3} + \hat\zeta_e(t) - \hat\sigma_e(t) + (0, \hat\eta_e(t)), ~ -{2\over 3} + \hat\zeta_e(t) - (\hat\alpha_e(t), \hat\beta_e(t)) + (0, \hat\eta_e(t)), ~
\mathbb{F}_{17e}(t)\right), \nd}
where the form for $\mathbb{F}_{17e}(t)$ appears in {\bf Table \ref{privsocmey}}. In a similar vein one may work out the scalings of the curvature tensors ${\bf R}_{0m0n}({\bf x}, y; g_s)$ and ${\bf R}_{0\rho 0 \sigma}({\bf x}, y; g_s)$ where $(m, n) \in {\cal M}$ and $(\rho, \sigma) = \{(\theta_1, \theta_1), (\theta_2 \theta_2)\} \in {\bf S}^1_{\theta_1} \times {{\bf S}^1\over {\cal I}_{\theta_2}}$. All these results appear in {\bf Table \ref{firzathai3}} expressed using the $\mathbb{F}_{16e}(t), \mathbb{F}_{17e}(t), \mathbb{F}_{18e}(t)$ and $\mathbb{F}_{20e}(t)$ factors that are defined in {\bf Table \ref{privsocmey}}. Unfortunately these $\mathbb{F}_{ie}(t)$ factors, starting from $\mathbb{F}_{1e}(t)$ till $\mathbb{F}_{25e}(t)$ factors, are complicated functions involving logs and powers of the dominant and the sub-dominant corrections. Can they be simplified? In the following section we will discuss in detail under what conditions simplifications of the functional forms for the $\mathbb{F}_{ie}(t)$ factors may be attained. 

\subsubsection{One temporal derivative acting on the 3-form flux components \label{curuva}}

The G-flux components that are effected by one temporal derivative on the three-form flux components are typically of the form $(d{\bf C})_{\rm 0ABD}({\bf x}, y; g_s)$, which we dealt earlier in \eqref{secdrem18}. Since the three and the four-form fluxes are related by a more complicated relation involving quantum and five-brane terms $-$ as shown in \eqref{dietherapie} $-$ it will be easier to express the temporal derivative using only the three-form scalings from \eqref{hannalat}, {\it i.e.} $(l_{\rm 0B}^{\rm D}, {\rm L}_{\rm 0B}^{\rm D})$ and 
$(l_{\rm AB}^{\rm D}, {\rm L}_{\rm AB}^{\rm D})$. In the end, we can relate them to the G-flux scalings $l_{\cal AB}^{\cal CD}$ and ${\rm L}_{\cal AB}^{\cal CD}$ from \eqref{usher}, or even to $l_{e\cal AB}^{\cal CD}$ from \eqref{andyrey}, where 
$({\cal A, B}) \in {\bf R}^{2, 1} \times {\cal M}_6 \times {\mathbb{T}^2\over {\cal G}}$ with ${\cal M}_6$ taking different forms for the two heterotic cases. For example, if we take the ${\rm E}_8 \times {\rm E}_8$ case, ${\cal M}_6 = {\cal M}_4 \times {\bf S}^1_{\theta_1} \times {{\bf S}^1_{\theta_2}\over {\cal I}_{\theta_2}}$ and one temporal derivative on the three-form flux components leads to:

{\scriptsize
\bg\label{secdremgone}
(d{\bf C})_{\rm 0ABD}(\zeta) & = & \sum_{k = 0}^\infty 
(d{\bf C})_{\rm 0ABD}^{(1, k)}({\bf x}, y)~\bar{g}_s^{l_{\rm AB}^{\rm D}(t) \mp
{{\rm log}\left(\big\vert\dot{l}_{\rm AB}^{\rm D}(t) 
+ {2k \vert \dot{\rm L}_{\rm AB}^{\rm D}(k; t)\vert\over 3}\big\vert \vert\log~\bar{g}_s\vert\right) \over \vert\log~\bar{g}_s\vert} + {2k\vert {\rm L}_{\rm AB}^{\rm D}(k; t)\vert\over 3}}\nonumber\\
&+ & 
\sum_{k = 0}^\infty (d{\bf C})_{\rm 0ABD}^{(2, k)}({\bf x}, y)~\bar{g}_s^{l_{\rm AB}^{\rm D}(t) - 1 + {\hat\alpha_e(t) + \hat\beta_e(t)\over 4} + 
{{\rm log}(2-{\hat\alpha_e(t) + \hat\beta_e(t)\over 2})\over \vert\log~\bar{g}_s\vert} \mp {\log\left(\big\vert l_{\rm AB}^{\rm D}(t) + {2k\vert {\rm L}_{\rm AB}^{\rm D}(k; t)\vert\over 3}\big\vert\right) \over\vert\log~\bar{g}_s\vert} + {2k\vert {\rm L}_{\rm AB}^{\rm D}(k; t)\vert\over 3}} \nonumber\\ 
&+ & 
\sum_{k = 0}^\infty (d{\bf C})_{\rm 0ABD}^{(3, k)}({\bf x}, y)~\bar{g}_s^{l_{\rm AB}^{\rm D}(t) + {1\over\vert\log~\bar{g}_s\vert}
{{\rm log}\left[{4-\hat\alpha_e(t) - \hat\beta_e(t)\over \vert\dot{\hat\alpha}_e(t) + \dot{\hat\beta}_e(t)\vert\vert\log~\bar{g}_s\vert}\right]} 
\mp {\log\left(\big\vert l_{\rm AB}^{\rm D}(t) + {2k\vert {\rm L}_{\rm AB}^{\rm D}(k; t)\vert\over 3}\big\vert\right) \over\vert\log~\bar{g}_s\vert} + {2k\vert {\rm L}_{\rm AB}^{\rm D}(k; t)\vert\over 3}} \nonumber\\ 
& + & \sum_{k = 0}^\infty\left[(d{\bf C})_{\rm 0ABD}^{(4, k)}({\bf x}, y)~\bar{g}_s^{l_{\rm 0B}^{\rm D}(t) + {2k\vert {\rm L}_{\rm 0B}^{\rm D}(k; t)\vert\over 3}} +  (d{\bf C})_{\rm 0ABD}^{(5, k)}({\bf x}, y)\left({g_s\over {\rm HH}_o}\right)^{l_{\rm 0A}^{\rm D}(t) + {2k\vert {\rm L}_{\rm 0A}^{\rm D}(k; t)\vert\over 3}}\right] \nonumber\\
&+ &  \sum_{k = 0}^\infty(d{\bf C})_{\rm 0ABD}^{(6, k)}({\bf x}, y)~\bar{g}_s^{l_{\rm 0A}^{\rm B}(t) + {2k\vert {\rm L}_{\rm 0A}^{\rm B}(k; t)\vert\over 3}},
\nd}
where one may note that, since ${\partial \over \partial t} = \bar{g}_s^{\gamma_{1, 2}\big[{\hat\alpha_e(t) + \hat\beta_e(t)\over 2}\big]}~{\partial \over \partial \bar{g}_s}$, temporal derivatives like 
$\dot{l}_{\rm 0A}^{\rm B}$ et cetera in \eqref{secdremgone} will involve 
$(\hat\alpha_e(t), \hat\beta_e(t))$ because of \eqref{ryanfan2}. However there is a subtlety now. Since $\hat\alpha_e(t)$ and $\hat\beta_e(t)$ are themselves expressed as a series: $\hat\alpha_e(t) = \sum\limits_{l = 0}^\infty \hat\alpha_e(t; [l])$ and $\hat\beta_e(t) = \sum\limits_{l = 0}^\infty \hat\beta_e(t; [l])$, one may wonder if there is a need to relate the mode expansion in $l$ for $(\hat\alpha_e(t), \hat\beta_e(t))$ to the mode expansion in $k$ in \eqref{secdremgone}. The worrisome feature of such an identification is the ambiguity related to the mapping itself: what governs the mapping of the $l$-modings from $(\hat\alpha_e(t), \hat\beta_e(t))$ to the $k$-modings from the G-flux components? A little thought however will tell us that such identifications are unnecessary. One may simply use the whole of $\hat\alpha_e(t)$ and $\hat\beta_e(t)$ and identify the flux components by the procedure advocated in section \ref{sec4.2.5}. With this in mind, one may relate the $\bar{g}_s$ scalings of the three-form flux components to the ones from the four-form G-flux components in the following way:

{\footnotesize
\bg\label{josymila2}
&&l_{\rm 0A}^{\rm BD}(t) =  l_{\rm AB}^{\rm D}(t) \mp {{\rm log}(\vert \dot{l}_{\rm AB}^{{\rm D}}(t)\vert \vert\log~\bar{g}_s\vert) \over \vert\log~\bar{g}_s\vert}\\
&& \vert {\rm L}_{\rm 0A}^{\rm BD}(1; t)\vert = 
 {3(\hat\alpha_e(t) + \hat\beta_e(t))\over 8} - {3\over 2} +  
{3{\rm log}\big(2-{\hat\alpha_e(t) + \hat\beta_e(t)\over 2}\big)\over 2\vert\log~\bar{g}_s\vert} \pm 
 {3\over 2\vert\log~\bar{g}_s\vert}~\log\left[ {\vert \dot{l}_{\rm AB}^{{\rm D}}(t)\vert \vert\log~\bar{g}_s\vert \over  \vert l_{\rm AB}^{\rm D}(t)\vert}\right] \nonumber\\
&& \vert {\rm L}_{\rm 0A}^{\rm BD}(2; t)\vert = 
  {3\over 4\vert\log~\bar{g}_s\vert}~\log\left[{4-\hat\alpha_e(t) - \hat\beta_e(t)\over \vert\dot{\hat\alpha}_e(t) + \dot{\hat\beta}_e(t)\vert\vert\log~\bar{g}_s\vert}\right] \pm 
  {3\over 4\vert\log~\bar{g}_s\vert}~\log\left[ {\vert \dot{l}_{\rm AB}^{{\rm D}}(t)\vert \vert\log~\bar{g}_s\vert \over  \vert l_{\rm AB}^{\rm D}(t)\vert}\right]  \nonumber\\
&& \vert {\rm L}_{\rm 0A}^{\rm BD}(3; t)\vert = {1\over 2}(l_{\rm 0B}^{\rm D}(t) - l_{\rm 0A}^{\rm BD}(t)), ~~
\vert {\rm L}_{\rm 0A}^{\rm BD}(4; t)\vert = {3\over 8}(l_{\rm 0A}^{\rm D}(t) - l_{\rm 0A}^{\rm BD}(t)), ~~\vert {\rm L}_{\rm 0A}^{\rm BD}(5; t)\vert = {3\over 10}(l_{\rm 0A}^{\rm B}(t) - l_{\rm 0A}^{\rm BD}(t)), \nonumber
\nd}
which is precisely what we had earlier in \eqref{josymila}, with the exception that $\beta(t) \to {\hat\alpha_e(t) + \hat\beta_e(t)\over 2}$ and the derivatives defined accordingly. Clearly, the generalized $SO(32)$ case appears from \eqref{josymila2} by the standard replacement of $\hat\alpha_e(t) + \hat\beta_e(t)$ by $2\beta_e(t)$. These results are shown in {\bf Tables \ref{firzacut102}} and {\bf \ref{firzacut103}}. In section \ref{sec4.5} we will use all these results to quantify the Bianchi identities, flux quantizations and the anomaly cancellation for both the heterotic theories.

\begin{table}[tb]  
 \begin{center}
\renewcommand{\arraystretch}{2.4}
\resizebox{\columnwidth}{!}{%
\begin{tabular}{|c||c||c|}\hline ${\bf G}_{\rm 0ABD}$  & $(d{\bf C})_{\rm 0ABD}$ \\ \hline\hline
${\cal G}^{(0)}_{\rm 0ABD}({\bf x}, y)~\left({g_s\over {\rm HH}_o}\right)^{l_{\rm 0A}^{\rm BD}(t)}$ & $(d{\bf C})^{(1, 0)}_{\rm 0ABD}({\bf x}, y)~\left({g_s\over {\rm HH}_o}\right)^{l_{\rm AB}^{\rm D}(t)\mp
{{\rm log}\left(\vert\dot{l}_{\rm AB}^{\rm D}(t)\vert \vert\log~\bar{g}_s\vert\right) \over \vert\log~\bar{g}_s\vert}}$ \\ \hline
${\cal G}^{(1)}_{\rm 0ABD}({\bf x}, y)~\left({g_s\over {\rm HH}_o}\right)^{l_{\rm 0A}^{\rm BD}(t) + {2\over 3}\vert {\rm L}_{\rm 0A}^{\rm BD}(1; t)\vert}$ & $(d{\bf C})_{\rm 0ABD}^{(2, 0)}({\bf x}, y)\left({g_s\over {\rm HH}_o}\right)^{l_{\rm AB}^{\rm D}(t) - 1 + {\hat\alpha_e(t) + \hat\beta_e(t)\over 4} + 
{{\rm log}(4-{\hat\alpha_e(t) - \hat\beta_e(t)}) - \log~2\over \vert\log~\bar{g}_s\vert} \mp {\log\left(\vert l_{\rm AB}^{\rm D}(t)\vert\right) \over\vert\log~\bar{g}_s\vert}}$\\ \hline
${\cal G}^{(2)}_{\rm 0ABD}({\bf x}, y)~\left({g_s\over {\rm HH}_o}\right)^{l_{\rm 0A}^{\rm BD}(t) + {4\over 3}\vert {\rm L}_{\rm 0A}^{\rm BD}(2; t)\vert}$ & $(d{\bf C})_{\rm 0ABD}^{(3, 0)}({\bf x}, y)\left({g_s\over {\rm HH}_o}\right)^{l_{\rm AB}^{\rm D}(t) + {1\over\vert\log~\bar{g}_s\vert}
{{\rm log}\left[{4-{\hat\alpha_e(t) - \hat\beta_e(t)}\over \vert\dot{\hat\alpha}_e(t) + \dot{\hat\beta}_e(t)\vert\vert\log~\bar{g}_s\vert}\right]} 
\mp {\log\left(\vert l_{\rm AB}^{\rm D}(t)\vert\right) \over\vert\log~\bar{g}_s\vert}}$\\ \hline
${\cal G}^{(3)}_{\rm 0ABD}({\bf x}, y)~\left({g_s\over {\rm HH}_o}\right)^{l_{\rm 0A}^{\rm BD}(t) + 2\vert {\rm L}_{\rm 0A}^{\rm BD}(3; t)\vert}$ &
$(d{\bf C})_{\rm 0ABD}^{(4, 0)}({\bf x}, y)\left({g_s\over {\rm HH}_o}\right)^{l_{\rm 0B}^{\rm D}(t)}$ \\ \hline
${\cal G}^{(4)}_{\rm 0ABD}({\bf x}, y)~\left({g_s\over {\rm HH}_o}\right)^{l_{\rm 0A}^{\rm BD}(t) + {8\over 3}\vert {\rm L}_{\rm 0A}^{\rm BD}(4; t)\vert}$ &
$(d{\bf C})_{\rm 0ABD}^{(5, 0)}({\bf x}, y)\left({g_s\over {\rm HH}_o}\right)^{l_{\rm 0A}^{\rm D}(t)}$\\
\hline
\end{tabular}}
\renewcommand{\arraystretch}{1}
\end{center}
 \caption[]{\Su The mapping of the G-flux components ${\bf G}_{\rm 0ABD}$ from \eqref{usher} to $(d{\bf C})_{\rm 0ABD}$ from \eqref{secdrem18} with ${\rm H} {\rm H}_o \equiv {\rm H}(y) {\rm H}_o({\bf x})$ for the generalized ${\rm E}_8 \times {\rm E}_8$ case. To go to the generalized $SO(32)$ case and from there to the simplified $SO(32)$ case, one may simply replace ${\hat\alpha_e(t) + \hat\beta_e(t)}$ by $2\beta_e(t)$ and then by $2\beta(t)$ respectively. Note also that the temporal derivative, like $\dot{l}^{\rm C}_{\rm AB} \equiv {\partial{l}^{\rm C}_{\rm AB}\over \partial \bar{g}_s} \bar{g}_s^{\gamma_{1, 2}\big[{\hat\alpha_e(t) + \hat\beta_e(t)\over 2}\big]}$, so is sensitive to which theory we consider. Additionally, a sample of the mapping was shown earlier in \eqref{josymila} using a mod 7 identification from \eqref{isahen}. Here we do a more detailed mapping by including additional proliferation of components with respect to $k \in \mathbb{Z}$. One can easily see that there is a one-to-one correspondence and therefore a definition like \eqref{usher2} is not necessary.} 
  \label{firzacut102}
 \end{table}

\begin{table}[tb]  
 \begin{center}
\renewcommand{\arraystretch}{1.5}
\begin{tabular}{|c||c||c|}\hline ${\bf G}_{\rm ABCD}$  & $(d{\bf C})_{\rm ABCD}$ \\ \hline\hline
${\cal G}^{(0)}_{\rm ABCD}({\bf x}, y) ~\left({g_s\over {\rm HH}_o}\right)^{l_{\rm 0A}^{\rm BD}(t)}$ & $(d{\bf C})_{\rm ABCD}^{(1, 0)}({\bf x}, y)\left({g_s\over {\rm HH}_o}\right)^{l_{\rm BC}^{\rm D}(t)}$ \\ \hline
${\cal G}^{(1)}_{\rm ABCD}({\bf x}, y)~\left({g_s\over {\rm HH}_o}\right)^{l_{\rm AB}^{\rm CD}(t) + {2\over 3}\vert {\rm L}_{\rm AB}^{\rm CD}(1; t)\vert}$
& $(d{\bf C})_{\rm ABCD}^{(2, 0)}({\bf x}, y)\left({g_s\over {\rm HH}_o}\right)^{l_{\rm AC}^{\rm D}(t)}$ \\ \hline
${\cal G}^{(2)}_{\rm ABCD}({\bf x}, y)~\left({g_s\over {\rm HH}_o}\right)^{l_{\rm AB}^{\rm CD}(t) + {4\over 3}\vert {\rm L}_{\rm AB}^{\rm CD}(2; t)\vert}$
& $(d{\bf C})_{\rm ABCD}^{(3, 0)}({\bf x}, y)\left({g_s\over {\rm HH}_o}\right)^{l_{\rm AB}^{\rm D}(t)}$ \\ \hline
${\cal G}^{(3)}_{\rm ABCD}({\bf x}, y)~\left({g_s\over {\rm HH}_o}\right)^{l_{\rm AB}^{\rm CD}(t) + 2\vert {\rm L}_{\rm AB}^{\rm CD}(3; t)\vert}$
& $(d{\bf C})_{\rm ABCD}^{(4, 0)}({\bf x}, y)\left({g_s\over {\rm HH}_o}\right)^{l_{\rm AB}^{\rm C}(t)}$ \\ \hline
${\cal G}^{(4)}_{\rm ABCD}({\bf x}, y)~\left({g_s\over {\rm HH}_o}\right)^{l_{\rm AB}^{\rm CD}(t) + {8\over 3}\vert {\rm L}_{\rm AB}^{\rm CD}(4; t)\vert}$
& $(d{\bf C})_{\rm ABCD}^{(1, 1)}({\bf x}, y)\left({g_s\over {\rm HH}_o}\right)^{l_{\rm BC}^{\rm D}(t)  + {2\over 3}\vert {\rm L}_{\rm BC}^{\rm D}(1; t)\vert}$ \\ \hline
${\cal G}^{(5)}_{\rm ABCD}({\bf x}, y)~\left({g_s\over {\rm HH}_o}\right)^{l_{\rm AB}^{\rm CD}(t) + {10 \over 3}\vert {\rm L}_{\rm AB}^{\rm CD}(5; t)\vert}$
& $(d{\bf C})_{\rm ABCD}^{(2, 1)}({\bf x}, y)\left({g_s\over {\rm HH}_o}\right)^{l_{\rm AC}^{\rm D} + {2\over 3}\vert {\rm L}_{\rm AC}^{\rm D}(1; t)\vert}$ \\
\hline
\end{tabular}
\renewcommand{\arraystretch}{1}
\end{center}
 \caption[]{\Su The mapping of the G-flux components ${\bf G}_{\rm ABCD}$ from \eqref{usher} to $(d{\bf C})_{\rm ABCD}$ from \eqref{tatumey} for {\it all} the heterotic theories namely, the generalized ${\rm E}_8 \times {\rm E}_8$ and $SO(32)$ theories and the simplified $SO(32)$ theory. This mapping could also be implemented using a mod 5 identification from \eqref{isahen}. Here we do a more detailed mapping by including additional proliferation of components with respect to $k \in \mathbb{Z}$. Again, as in the {\bf Table \ref{firzacut102}}, one see that there is a one-to-one correspondence thus eliminating the necessity to implement a definition like \eqref{usher2}.} 
  \label{firzacut103}
 \end{table}

\begin{figure}[h]
\centering
\begin{tabular}{c}
\includegraphics[width=2in]{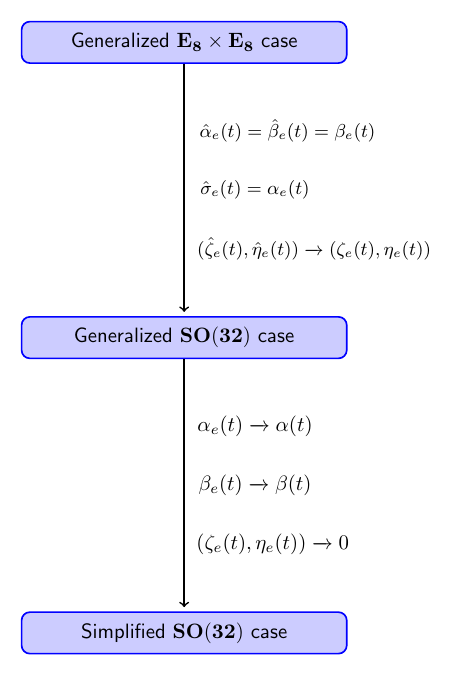}
\end{tabular}
\caption[]{Strategy used in section \ref{sec4.3} for computing the scalings of the $\mathbb{F}_{ie}(t)$ factors.}
\label{boxer}
\end{figure}

\subsection{$g_s$ scalings of and bounds on the $\mathbb{F}_{ie, i}(t)$  factors from {\bf Tables \ref{firzacut3}} and {\bf \ref{privsocmey}} \label{sec4.3}}

The log corrections to the $g_s$ scalings appearing in {\bf Tables \ref{firzathai}} and {\bf \ref{firzathai3}} depend on the $\mathbb{F}_{ie}(t)$ factors whose detailed forms are given in {\bf Table \ref{privsocmey}}. In the limit when we ignore the sub-dominant corrections, the curvature scalings can be simplified considerably and  they appear in {\bf Tables \ref{firzacut}} to {\bf \ref{firzacut3}}, expressed using the $\mathbb{F}_i(t)$ factors given in {\bf Table \ref{firzacut3}}. Our strategy in this section is to study the simplified scalings, {\it i.e.} ignoring the sub-dominant corrections in $\mathbb{F}_{ie}(t)$ but keeping the dominant perturbative and non-perturbative corrections. The reason for studying the simpler picture with $\mathbb{F}_i(t)$ is because most of the interesting and subtle manipulations may be directly seen from there,  and one may add in the sub-dominant corrections later. Once added, these $\mathbb{F}_{ie}(t)$ factors appear directly in the quantum scaling \eqref{botsuga}, and are typically constructed from the metric components which involve $\alpha_e(t)$ and $\beta_e(t)$ for the $SO(32)$ theory and $\hat\sigma_e(t), \hat\alpha_e(t)$ and $\hat\beta_e(t)$ for the ${\rm E}_8 \times {\rm E}_8$ theory. The appearance of the non-perturbative corrections is crucial: they are the dominant contributions in $\alpha_e(t)$ and $\beta_e(t)$ as shown in \eqref{ryanfan} (similarly in $\hat\sigma_e(t), \hat\alpha_e(t)$ and $\hat\beta_e(t)$), but are sub-dominant contributions in the metric components along ${\bf R}^{2, 1} \times {\mathbb{T}^2\over {\cal G}}$ in the sense of say \eqref{tranpart}. Alternatively, as emphasized above, the metric components along the internal six-dimensional base have {\it both} sub-dominant and dominant non-perturbative contributions. It is therefore important that we spell out the functional behavior of $\mathbb{F}_{ie}(t)$ as well as of $\mathbb{F}_i(t)$ in terms of $g_s$ carefully to check whether or not any relative minus signs appear in \eqref{botsuga}. Keeping this goal in mind, in this section, we will study the functional form and bounds on all  $\mathbb{F}_i(t)$ factors from the $\mathbb{F}_{ie}(t)$ factors.

\subsubsection{{Functional forms and bounds on} $\mathbb{F}_{1, 1e}(t)$  {and} $\mathbb{F}_{5, 5e}(t)$}

\vskip.1in

\noindent Our starting point will be $\mathbb{F}_{1e}(t)$ which appears in the first row of {\bf Table \ref{firzathai}} for the scaling of the curvature tensor ${\bf R}_{mnpq}$. Looking at {\bf Table \ref{privsocmey}}, it takes the following form:

{\footnotesize
\bg\label{lenasiri}
\mathbb{F}_{1e}(t) = \left(2 \otimes \mathbb{A}_8(\hat\sigma_e) + {8\over 3} - \hat\zeta_e(t), ~ 2\otimes \mathbb{B}_8(\hat\sigma_e) + {8\over 3} - \hat\zeta_e(t), ~\mathbb{A}_8(\hat\sigma_e) + \mathbb{B}_8(\hat\sigma_e) + {8\over 3} - \hat\zeta_e(t)\right), \nd}
where $({\bf A}_8(\hat\sigma_e), {\bf B}_8(\hat\sigma_e)$ are defined in \eqref{tifeni}, $\hat\sigma_e$ contains all the dominant and sub-dominant corrections and is defined in \eqref{bdtran2}, and $\hat\zeta_e$ contains the sub-dominant corrections to the $2+1$ dimensional spacetime metric and is defined in \eqref{masatwork}. The symbol $2\otimes \mathbb{A}_8(\hat\sigma_e)$ signifies two copies of $\mathbb{A}_8(\hat\sigma_e)$ when raised to powers of $\bar{g}_s$. The strategy that we shall follow here is shown in {\bf figure \ref{boxer}}: we first go to the generalized $SO(32)$, and from there go to the simplified $SO(32)$ case by following the steps shown in the figure. The simplified $SO(32)$ case provides a controlled laboratory to perform the requisite computations of the scales, and from there it will be easy to extract the exact scalings for the generalized ${\rm E}_8 \times {\rm E}_8$ and the $SO(32)$ cases.  

Let us see how this works explicitly. 
To go to the $SO(32)$ case, we first make $\hat\sigma_e(t) \to \alpha_e(t)$. This still contains the sub-dominant corrections not only for the warp-factor $\alpha(t)$, but also the sub-dominant corrections along the spacetime $2+1$ directions (denoted by $\hat\zeta_e(t) \to \zeta_e(t)$). If we switch off all the sub-dominant corrections, then $\alpha_e(t) \to \alpha(t)$ for the $SO(32)$ theory, and we can express $\mathbb{F}_1(t)$ in the following suggestive way:

{\footnotesize
\bg\label{pitpendu}
\mathbb{F}_1(t) & \equiv & \left(2\mathbb{A}(\alpha) + {8\over 3}, ~2\mathbb{B}(\alpha) + {8\over 3}, ~ \mathbb{A}(\alpha) + \mathbb{B}(\alpha) + {8\over 3}\right)\\
&=& \Bigg(-{{2\over 3} + 2\alpha(t) + 2\gamma_{1, 2}(t) -{2\log({2\over 3}-\alpha)\over \vert\log~\bar{g}_s\vert}},~~ {{4\over 3} + 2\alpha(t) - {2\log\left(\mp\dot\alpha\vert\log~\bar{g}_s\vert\right)\over \vert\log~\bar{g}_s\vert}}\nonumber\\ 
&& {{1\over 3}+ 2\alpha(t) + \gamma_{1, 2}(t) -{\log\left(\mp\dot\alpha({2\over 3}-\alpha)\vert\log~\bar{g}_s\vert\right)\over \vert\log~\bar{g}_s\vert}}, ~
-{2\over 3} + 2\alpha(t) + \gamma_1(t) + \gamma_2(t) - {2\log({2\over 3} -\alpha)\over \vert\log~\bar{g}_s\vert}\Bigg) 
\nonumber\\ 
&= & \Bigg({1\over 3} + 2\alpha(t) + {\beta(t)\over 2} + {1\over \vert\log~\bar{g}_s\vert}~\log\left[{2-\beta(t)\over \mp\dot{\alpha}(t)\vert\log~\bar{g}_s\vert({2\over 3} - \alpha(t))}\right], \nonumber\\ 
&& 
{1\over 3} + 2\alpha(t) + {\beta(t)\over 2} + {2\over \vert\log~\bar{g}_s\vert}~\log\left[{(2-\beta(t)) \over 
 ({2\over 3} - \alpha(t))\sqrt{\vert\dot\beta(t)\vert\vert\log~\bar{g}_s\vert}}\right],
\nonumber\\
&& {4\over 3} + 2\alpha(t) -{2\log(\mp\dot{\alpha}(t)\vert\log~\bar{g}_s\vert)\over \vert\log~\bar{g}_s\vert},~ 
{4\over 3} + 2\alpha(t) + {1\over \vert\log~\bar{g}_s\vert}~\log\left[{2-\beta(t)\over \mp\dot{\alpha}(t)\vert\dot{\beta}(t)\vert\vert\log~\bar{g}_s\vert^2({2\over 3} - \alpha(t))}\right], \nonumber\\
&& {4\over 3} + 2\alpha(t) + {2\over \vert\log~\bar{g}_s\vert}~\log\left[{2-\beta(t)\over \vert\dot{\beta}(t)\vert\vert\log~\bar{g}_s\vert({2\over 3} - \alpha(t))}\right],~
-{2\over 3} + 2\alpha(t) + \beta(t) + {2\over \vert\log~\bar{g}_s\vert}~\log\left[{2-\beta(t)\over {2\over 3} - \alpha(t)}\right]\Bigg), \nonumber \nd}
and since the log corrections are very small both here and in \eqref{ishena}, with ${1\over 3} \le \gamma_i \le 1$, the dominant contribution to $\mathbb{F}_1(t)$ appears solely from ${4\over 3} + 2\alpha(t), -{2\over 3} + 2\alpha(t) + \beta(t)$ and ${1\over 3} + 2\alpha(t) + {\beta(t)\over 2}$. This is what one would have expected from naive scaling arguments. However the log corrections, albeit small, are important as shown in {\bf Table \ref{rindrag}}, so we will carry them forward in our subsequent computations\footnote{This is an example of an additive small quantity that however cannot be ignored, although we can have some control on the dynamics by resorting to a definition like \eqref{tokio} for $\beta(t)$. See the discussions in section \ref{sec4.2.3}.}. 

Out of the three dominant scalings in $\mathbb{F}_1(t)$, the scaling coefficient $-{2\over 3} + 2\alpha(t) + \beta(t) = -{2\over 3} - \beta(t)$
becomes negative. (Recall that $\alpha(t) = -\beta(t)$.) This may be a bit more disconcerting than the other two terms. However this is {\it not} an issue because the term that actually contributes to the quantum scaling (as shown in {\bf Table \ref{firzacut6}}), takes the form:
\bg\label{kilie1}
{4\over 3} + \mathbb{F}_1(t) - 2\alpha(t) = \left({8\over 3}, ~{2\over 3} + \beta(t), ~{5\over 3} + {\beta(t)\over 2}\right), \nd
which are all positive definite as $0 < \beta(t) < {2\over 3}$. The log corrections then add deviations from the aforementioned dominant values. These may be quantified in the following way. Using \eqref{lovsik}, we note that:

{\scriptsize
\bg\label{ivbresbest}
&&\vert\dot\alpha(t)\vert\vert{\rm log}~\bar{g}_s\vert \to \sum_{b, d, d'} \left({g_s\over {\rm H}(y) {\rm H}_o({\bf x})}\right)^{{b\over 3} - 1 + {n_{d'}(b)\over \bar{g}_s^{d'/3}\vert{\rm log}~\bar{g}_s\vert} + \vert\gamma(t)\vert + \big(0, \Delta, \Delta - {d\over 3}, -{d\over 3}\big)}\\
&& \vert\dot\alpha(t)\vert^2\vert{\rm log}~\bar{g}_s\vert^2 \to
\sum_{b, b',..} \left({g_s\over {\rm H}(y) {\rm H}_o({\bf x})}\right)^{{b+b'\over 3} - 2 + {n_{d''}(b)\over \bar{g}_s^{d''/3}\vert{\rm log}~\bar{g}_s\vert} + {n_{d'''}(b)\over \bar{g}_s^{d'''/3}\vert{\rm log}~\bar{g}_s\vert}
+ 2\vert\gamma(t)\vert + \big(0, 2\Delta, \Delta, 2\Delta - {d + d'\over 3}, ..., \Delta - {d\over 3}\big)} \nonumber\\
&& ~~~~~~~~~~~~~~~~~~~~ < \sum_{b, d, d'} \left({g_s\over {\rm H}(y) {\rm H}_o({\bf x})}\right)^{{2b\over 3} - 2 + {2n_{d'}(b)\over \bar{g}_s^{d'/3}\vert{\rm log}~\bar{g}_s\vert} + 2\vert\gamma(t)\vert + \big(0, 2\Delta, \Delta, 2\Delta - {2d\over 3}, \Delta - {2d\over 3}, -{2d\over 3}, -{d\over 3}, 2\Delta - {d\over 3}, \Delta - {d\over 3}\big)}, \nonumber \nd}
where the summation over $(b, d, d', ..)$ should also include the coefficients 
$\hat{c}^{(0)}_{b}$ and $\check{c}^{(0)}_{b}$ from \eqref{lovsik} as we are taking the dominant $l = 0$ contributions. For the second case, we have shown the bound. {To avoid complicating the relations, we will not specify the coefficients and work with the bounded values wherever possible}. The terms in the brackets in \eqref{ivbresbest} should be assumed to contribute additively to the terms outside the brackets. For example, in the first case, there are four distinct $g_s$ scalings coming from $\big(0, \Delta, \Delta - {d\over 3}, -{d\over 3}\big)$ when added to the term ${b\over 3} - 1 + {n_d\over \bar{g}_s^{d/3}\vert{\rm log}~\bar{g}_s\vert} + \vert\gamma(t)\vert$ where $\gamma(t)$ is the dominant contribution that appears from the definition ${\partial \bar{g}_s\over \partial t} \propto \bar{g}_s^{\vert\gamma(t)\vert}$. This may be easily extracted from the dominant contribution in \eqref{stan22} and its clear that it takes a similar trans-series form as \eqref{ryanfan} (see \eqref{mcelhon})\footnote{We have expressed ${\partial \bar{g}_s\over \partial t}$ in terms of $\vert\gamma(t)\vert$ instead of $\vert\gamma_{1, 2}\vert$ from \eqref{ishena}. This is intentional because the expression in \eqref{ishena} (and also in \eqref{ishena2}) is a function of $\beta(t)$ and $\dot{\beta}(t)$, which are themselves functions of $\bar{g}_s$ as shown in \eqref{ryanfan}. To avoid this circular reasoning, we will express the derivative of $\bar{g}_s$ using $\vert\gamma(t)\vert$ which may be expressed solely in terms of $\bar{g}_s$ as in \eqref{mcelhon}. This way $\beta(t)$, $\gamma(t)$ {\it et cetera} will be in the same footings, and we can express $\gamma_{1, 2}(t)$ in terms of $\gamma(t)$ (as shown in {\bf Table \ref{munleena}}). \label{gemmaton}}. Additionally, since $b\ge 0$ and $0 \le d \le d_{\rm max}$ for $d_{\rm max} \in \mathbb{Z}$, we can define two  set  from \eqref{ivbresbest} in the following way:

{\scriptsize
\bg\label{standudh}
&& {\cal C}(\Delta; b, d) = \left({b\over 3}, \Delta + {b\over 3}, \Delta + {b-d\over 3}, {b-d\over 3}\right)\\
&& {\cal B}(\Delta; b, d) = \left( {2b\over 3} , 2\Delta + {2b\over 3}, \Delta +{2b\over 3}, 2\Delta + {2(b - d)\over 3}, \Delta + {2(b - d)\over 3}, {2(b - d)\over 3}, {2b - d\over 3}, 2\Delta + {2b - d\over 3}, \Delta + {2b - d\over 3}\right), \nonumber \nd}
so that the negative contributions would appear solely from the $-1$ and $-2$ as well as the $b - d$ exponents in \eqref{ivbresbest}. This might seem as a matter of concern because now the $g_s$ scalings of the curvature tensors would develop negative signs. The $-1$ and $-2$ factors, as will become clear below, pose no problem and so the main concern is on the $b -d$ factors. However as we discussed in sections \ref{sec4.2.3} and later in \ref{sec448}, and especially while dealing with \eqref{stan22}, the factor \eqref{hatpathanda} which here contributes to $\vert\gamma\vert$ now gives us:
\bg\label{clodkalua}
\vert\gamma\vert + {\cal C}(\Delta; b, d) \ge d_{\rm max}, ~~~
2\vert\gamma\vert + {\cal B}(\Delta; b, d) \ge 2d_{\rm max}, \nd
in the zero instanton sector, {\it i.e.} for $n_d = 0$,
implying that the aforementioned combinations remain positive definite, and therefore create no problems with the EFT. (For $n_d > 0$, the two combinations in \eqref{clodkalua} are always positive definite as we shall see from \eqref{mcelhon}.)
Now plugging \eqref{ivbresbest} and \eqref{standudh} in \eqref{pitpendu}, we see that the dominant contribution to $\mathbb{F}_1(t)$ changes to the following:

{\footnotesize
\bg\label{pitpendu0}
\mathbb{F}_1(t) \to \mathbb{F}_1(t; b, d, d') & \equiv & \bigg(-{2\over 3} + 2\alpha(t) + {\beta(t)\over 2} + {n_d\over \bar{g}_s^{d/3}\vert{\rm log}~\bar{g}_s\vert} + \vert\gamma(t)\vert + {\cal C}(\Delta; b, d'), \\
&- & {2\over 3} + 2\alpha(t) + {2n_d\over \bar{g}_s^{d/3}\vert{\rm log}~\bar{g}_s\vert} + 2\vert\gamma(t)\vert + {\cal B}(\Delta; b, d'), ~
-{2\over 3} + 2\alpha(t) + \beta(t)\bigg), \nonumber \nd}
where additional dominant non-perturbative terms would come from the trans-series form of $\beta(t)$ in \eqref{ryanfan}. The $(b, d, d')$ dependence of \eqref{pitpendu0} suggests that, once we exponentiate $\bar{g}_s$ by $\mathbb{F}_1(t; b, d, d')$ and insert the coefficients correctly, we have to sum over $b, d$ and $d'$. However even before we perform the summation, the dominant contribution \eqref{pitpendu0} is quite different from the naive expectation mentioned just below \eqref{pitpendu}. Consequently we expect \eqref{kilie1} to also change. This change is easy to quantify, and is given by:
\bg\label{kylieador}
{4\over 3} + \mathbb{F}_1(t; b, d, d') - 2\alpha(t) & = & 
\bigg({2\over 3} + {\beta(t)\over 2} + {n_d\over \bar{g}_s^{d/3}\vert{\rm log}~\bar{g}_s\vert} + \vert\gamma(t)\vert + {\cal C}(\Delta; b, d'), \\
& & {2\over 3} + {2n_d\over \bar{g}_s^{d/3}\vert{\rm log}~\bar{g}_s\vert} + 2\vert\gamma(t)\vert + {\cal B}(\Delta; b, d'), ~
{2\over 3} + \beta(t)\bigg), \nonumber \nd
which are all positive definite and in the limit $g_s \to 0$ the dominant contribution comes from ${2\over 3} + \beta(t) \to {2\over 3}$ because of our ans\"atze \eqref{ryanfan} for $\beta(t)$. For $n_d > 0$, it is clear that the two terms in \eqref{kylieador} are exponentially suppressed. For $n_d = 0$, and using \eqref{clodkalua}, it is easy to justify that ${2\over 3} + \beta(t)$ remains dominant as long as $d_{\rm max} > {\Delta\over 2}$. The latter is clearly possible because $d_{\rm max} \in \mathbb{Z}$.
Incidentally for vanishing $\beta(t)$, all the terms from \eqref{kilie1} would have been suppressed compared to the other terms in the first row of {\bf Table \ref{firzacut6}}. For the ${\rm E}_8 \times {\rm E}_8$ case, the analysis follows similar strategy as above with the dominant contribution coming from:

{\footnotesize
\bg\label{rusflo1}
{\rm dom}~\mathbb{F}_{1e}(t) = -{2\over 3} + 2\hat\sigma_e(t) - \hat\zeta_e(t) + {\hat\alpha_e(t) + \hat\beta_e(t)\over 2} + {2\over \vert\log~\bar{g}_s\vert}~\log\left({12 - 3\hat\alpha_e - 3\hat\beta_e\over 4 - 6\hat\sigma_e}\right), \nd}
and the contribution to the quantum series \eqref{botsuga} comes from ${4\over 3} + \mathbb{F}_{1e}(t) - 2\sigma_e(t)$. Ignoring the sub-dominant contributions $-$ as $(\hat\sigma_e(t), \hat\alpha_e(t), \hat\beta_e(t)) << 1$ and the log piece contributes a numerical factor of ${1\over 3}$ $-$ the scaling we now get is ${2\over 3} + {\hat\alpha(t) + \hat\beta(t)\over 2} - \hat\zeta_e(t)$. Plugging back the sub-dominant corrections, the scaling we expect respectively for the $SO(32)$ and the ${\rm E}_8 \times {\rm E}_8$ theory appears to be:
\bg\label{2cho1bora}
\boxed{{2\over 3} + \beta_e(t) - \zeta_e(t); ~~~~~ {2\over 3} + {\hat\alpha_e(t) + \hat\beta_e(t)\over 2} -\hat\zeta_e(t)} \nd
where $-\zeta_e(t)$ or $-\hat\zeta_e(t)$ would imply Taylor expanding the series in \eqref{masatwork} with an overall positive sign.
We can also ask for what choices of the parameters in \eqref{kylieador} would match-up with the scalings in \eqref{kilie1}. The answer is:
\bg\label{qingthic}
n_d = \gamma(t) = 0, ~~~~ b = d' + 3, \nd
which would pick up the dominant scaling from the two sets in \eqref{standudh}. One might worry that keeping vanishing $n_d$ would not give rise to the behavior we expect from \eqref{afleis} and from {\bf figure \ref{boudi}} using the functional form for $\beta(t)$ in \eqref{ryanfan}. This, as we saw earlier, is not the case, so with $n_d \ge 0$, the scalings in \eqref{kylieador} {\it almost} matches with the ones from \eqref{kilie1} except now that the first and the third scalings come with additional exponential suppressions from the non-perturbative terms. Unfortunately $\gamma = 0$ and $b = d' + 3$ cannot be imposed because of \eqref{hatpathanda} and the fact that the trans-series for $\beta(t)$ is defined for $b \ge 0$ and $0 \le d \le d_{\rm max}$. Thus \eqref{kylieador} remans the correct scalings compared to the simpler ones in \eqref{kilie1}.  
In a similar vein, $\mathbb{F}_{5e}(t)$ from {\bf Table \ref{privsocmey}} takes the following form:

{\footnotesize
\bg\label{ryanice}
\mathbb{F}_{5e}(t) = \left(2\otimes \mathbb{A}_8(\hat\alpha_e, \hat\beta_e) + {8\over 3} - \hat\zeta_e(t), ~ 2\otimes \mathbb{B}_8(\hat\alpha_e, \hat\beta_e) + {8\over 3} - \hat\zeta_e(t), ~\mathbb{A}_8(\hat\alpha_e, \hat\beta_e) \oplus \mathbb{B}_8(\hat\alpha_e, \hat\beta_e) + {8\over 3} - \hat\zeta_e(t)\right), \nonumber\\ \nd}
where all the parameters are defined in section \ref{sec4.4} with $(\hat\alpha_e(t), \hat\beta_e(t))$ related to whether we are taking $\theta_1$ or $\theta_2$ directions respectively. (Recall that the internal sub-manifold for the ${\rm E}_8$ case is ${\bf S}^1_{\theta_1} \times {{\bf S}^1_{\theta_2}\over {\cal I}_{\theta_2}}$.) Again we will take the simplified picture where $\hat\zeta_e(t) = 0$, and to go to the $SO(32)$ case, we shall replace $(\hat\alpha_e(t), \hat\beta_e(t)) \to (\hat\alpha(t), \hat\beta(t))$
by $\beta(t)$. This gives us:

{\footnotesize
\bg\label{pitpendu2}
\mathbb{F}_5(t) & \equiv & \left(2\mathbb{A}(\beta) + {8\over 3}, ~2\mathbb{B}(\beta) + {8\over 3}, ~ \mathbb{A}(\beta) + \mathbb{B}(\beta) + {8\over 3}\right) \\
&=& \Bigg(-{{2\over 3} + 2\beta(t) + 2\gamma_{1, 2}(t) -{2\log({2\over 3}-\beta)\over \vert\log~\bar{g}_s\vert}},~~ {{4\over 3} + 2\beta(t) - {2\log\left(\vert\dot\beta\vert\vert\log~\bar{g}_s\vert\right)\over \vert\log~\bar{g}_s\vert}}\nonumber\\ 
&& {{1\over 3}+ 2\beta(t) + \gamma_{1, 2}(t) -{\log\left(\vert\dot\beta\vert({2\over 3}-\beta)\vert\log~\bar{g}_s\vert\right)\over \vert\log~\bar{g}_s\vert}}, ~
-{2\over 3} + 2\beta(t) + \gamma_1(t) + \gamma_2(t) - {2\log({2\over 3} -\beta)\over \vert\log~\bar{g}_s\vert}\Bigg) 
\nonumber\\ 
&= & \Bigg({1\over 3} + {5\beta(t)\over 2} + {1\over \vert\log~\bar{g}_s\vert}~\log\left[{2-\beta(t)\over \vert\dot{\beta}(t)\vert\vert\log~\bar{g}_s\vert({2\over 3} - \beta(t))}\right], \nonumber\\ 
&& 
{1\over 3} + {5\beta(t)\over 2} + {2\over \vert\log~\bar{g}_s\vert}~\log\left[{(2-\beta(t)) \over 
 ({2\over 3} - \beta(t))\sqrt{\vert\dot\beta(t)\vert\vert\log~\bar{g}_s\vert}}\right],
\nonumber\\
&& {4\over 3} + 2\beta(t) -{2\log(\vert\dot{\beta}(t)\vert\vert\log~\bar{g}_s\vert)\over \vert\log~\bar{g}_s\vert},~ 
{4\over 3} + 2\beta(t) + {1\over \vert\log~\bar{g}_s\vert}~\log\left[{2-\beta(t)\over \vert\dot{\beta}(t)\vert^2\vert\log~\bar{g}_s\vert^2({2\over 3} - \beta(t))}\right], \nonumber\\
&& {4\over 3} + 2\beta(t) + {2\over \vert\log~\bar{g}_s\vert}~\log\left[{2-\beta(t)\over \vert\dot{\beta}(t)\vert\vert\log~\bar{g}_s\vert({2\over 3} - \beta(t))}\right],~
-{2\over 3} + 3\beta(t) + {2\over \vert\log~\bar{g}_s\vert}~\log\left[{2-\beta(t)\over {2\over 3} - \beta(t)}\right]\Bigg), \nonumber \nd}
which is the same as the replacement $\alpha(t)$ by $\beta(t)$, but keeping $\beta(t)$ intact, in \eqref{pitpendu}. Note that now, ignoring the log corrections, the dominant contribution appears to come from the three terms ${1\over 3} + {5\beta(t)\over 2}, 
{4\over 3} + 2\beta(t)$ and $-{2\over 3} + 3\beta(t)$. For $0 < \beta(t) < {2\over 3}$ there is a possibility that $-{2\over 3} + \beta(t)$ becomes negative. However what appears in the quantum scaling \eqref{botsuga} is the combination:
\bg\label{jhogren}
{4\over 3} + \mathbb{F}_5(t) - 2\beta(t) = \left({8\over 3}, ~ {2\over 3} + \beta(t), ~ {5\over 3} + {\beta(t)\over 2}\right), \nd
which is exactly the same as \eqref{kilie1}. The log terms will now add corrections to \eqref{jhogren} but not change the sign. To see this we will use \eqref{ivbresbest} and \eqref{standudh}, but replace $\beta(t)$ for $\alpha(t)$. This doesn't change anything because $\beta(t) = -\alpha(t)$, although the actual terms differ from what we had in \eqref{pitpendu0}. For example now we have:

{\footnotesize
\bg\label{pitpendu5}
\mathbb{F}_5(t) \to \mathbb{F}_5(t; b, d, d') & \equiv & \bigg(-{2\over 3} + {5\beta(t)\over 2} + {n_d\over \bar{g}_s^{d/3}\vert{\rm log}~\bar{g}_s\vert} + \vert\gamma(t)\vert + {\cal C}(\Delta; b, d'), \\
&- & {2\over 3} + 2\beta(t) + {2n_d\over \bar{g}_s^{d/3}\vert{\rm log}~\bar{g}_s\vert} + 2\vert\gamma(t)\vert + {\cal B}(\Delta; b, d'), ~
-{2\over 3} + 3\beta(t)\bigg), \nonumber \nd}
where ${\cal B}(\Delta; b, d')$ and ${\cal C}(\Delta; b, d')$ are defined in \eqref{standudh}. Interesting the replacement to go from \eqref{pitpendu0} to \eqref{pitpendu5} appears to be $\alpha(t) \to \beta(t)$ instead of $\alpha(t) \to -\beta(t)$. Plugging \eqref{pitpendu5} in \eqref{jhogren}, we get:
\bg\label{kylieador2}
{4\over 3} + \mathbb{F}_5(t; b, d, d') - 2\beta(t) & = & 
\bigg({2\over 3} + {\beta(t)\over 2} + {n_d\over \bar{g}_s^{d/3}\vert{\rm log}~\bar{g}_s\vert} + \vert\gamma(t)\vert + {\cal C}(\Delta; b, d'), \\
& & {2\over 3} + {2n_d\over \bar{g}_s^{d/3}\vert{\rm log}~\bar{g}_s\vert} + 2\vert\gamma(t)\vert + {\cal B}(\Delta; b, d'), ~
{2\over 3} + \beta(t)\bigg), \nonumber \nd
which matches exactly with \eqref{kylieador}. In fact imposing \eqref{qingthic}, we notice that \eqref{kylieador2} matches with \eqref{jhogren}. However as before, going with $n_d \ge 0$ resembles \eqref{jhogren} except that two of the terms are suppressed by the non-perturbative factors. Raising $\bar{g}_s$ by $\mathbb{F}_5(t; b, d, d')$ and then summing over $(b, d, d')$ appropriately gives us the final answer. The result is again positive definite and is dominated by ${2\over 3}$ at late time where $g_s \to 0$. For the ${\rm E}_8 \times {\rm E}_8$ case, the dominant contribution to $\mathbb{F}_{5e}(t)$ comes from:

{\scriptsize
\bg\label{metro3may}
{\rm dom}~\mathbb{F}_{5e}(t) = -{2\over 3} + 2\left(\hat\alpha_e(t), \hat\beta_e(t)\right) + {\hat\alpha_e(t) + \hat\beta_e(t)\over 2} - \hat\zeta_e(t) + {2\over \vert\log~\bar{g}_s\vert} ~{\log\big(4 -\hat\alpha_e - \hat\beta_e\big) - \log~2\over \log\big(2 - 3(\hat\alpha_e, \hat\beta_e)\big) - \log~3}, \nd}
where all the parameters are defined earlier in section \ref{sec4.4}. The contribution to \eqref{botsuga} now comes from ${4\over 3} +\mathbb{F}_{5e}(t) - 2(\hat\alpha_e(t), \hat\beta_e(t))$. Plugging \eqref{metro3may}, we again get \eqref{2cho1bora} as the dominant contributions from $\mathbb{F}_5(t)$ and $\mathbb{F}_{5e}(t)$ for the $SO(32)$ and the ${\rm E}_8 \times {\rm E}_8$ cases respectively.
Our analysis then shows that ${\bf R}_{\rho\sigma \rho\sigma}$ only contributes to the quantum series as positive powers of $g_s$ as evident from the sixth rows of {\bf Tables \ref{firzacut6}, \ref{lilalo1}} and {\bf \ref{lilalo2}}.

\subsubsection{{Functional form and bound on} $\mathbb{F}_{2, 2e}(t)$}

\vskip.1in

\noindent The next set is $\mathbb{F}_{2e}(t)$ and appears in the second row of {\bf Table \ref{firzathai}}. This is in general more non-trivial and involves various combinations of $\mathbb{A}_8(\hat\sigma_e), \mathbb{A}_8(\hat\alpha_e, \hat\beta_e)$ and the corresponding ones from $\mathbb{B}_8(\hat\sigma_e), \mathbb{B}_8(\hat\alpha_e, \hat\beta_e)$ given in \eqref{tifeni}. Looking at {\bf Table \ref{privsocmey}}, this takes the form:
\bg\label{russflow}
\mathbb{F}_{2e}(t) &= & \Big(\mathbb{A}_8(\hat\sigma_e) + \mathbb{A}_8(\hat\alpha_e, \hat\beta_e) + {8\over 3} - \hat\zeta_e(t),
~\mathbb{A}_8(\hat\sigma_e) + \mathbb{B}_8(\hat\alpha_e, \hat\beta_e) + {8\over 3} - \hat\zeta_e(t) \nonumber\\
&&  \mathbb{B}_8(\hat\sigma_e) + \mathbb{A}_8(\hat\alpha_e, \hat\beta_e) + {8\over 3} -\hat\zeta_e(t), 
~ \mathbb{B}_8(\hat\sigma_e) + \mathbb{B}(\hat\alpha_e, \hat\beta_e) + {8\over 3} - \hat\zeta_e(t)\Big), \nd
where all other parameters are defined in section \ref{sec4.4}. To go to the simpler case, we can remove all sub-dominant contributions, thus converting $\hat\sigma_e(t) \to \hat\sigma(t), \hat\alpha_e(t) \to \hat\alpha(t), \hat\beta_e(t) \to \hat\beta(t)$ and $\hat\zeta_e(t) \to 0$. From here it is easy to go to the $SO(32)$ case that appears in the second row of {\bf Table \ref{firzacut}}, with parameters defined in \eqref{rindrag}. (See {\bf figure \ref{boxer}}.) This takes the form:

{\footnotesize
\bg\label{bralesasl1}
\mathbb{F}_2(t) & = & \left(\mathbb{A}(\alpha) + \mathbb{A}(\beta) + {8\over 3},~ \mathbb{A}(\alpha) + \mathbb{B}(\beta) + {8\over 3},~ 
\mathbb{B}(\alpha) + \mathbb{A}(\beta) + {8\over 3}, ~
\mathbb{B}(\alpha) + \mathbb{B}(\beta) + {8\over 3}\right)\\
&=& \Bigg(-{2\over 3} + \alpha(t) + 2\beta(t) 
+ {1\over\vert\log~\bar{g}_s\vert}~\log\left[{(2-\beta(t))^2\over ({2\over 3} -\alpha(t))({2\over 3} - \beta(t))}\right], \nonumber\\
&& {1\over 3} + \alpha(t) + {3\beta(t)\over 2} + {1\over\vert\log~\bar{g}_s\vert}~\log\left[{(2-\beta(t))\over \vert\dot\beta(t)\vert\vert\log~\bar{g}_s\vert ({2\over 3} - \alpha(t))}\right], \nonumber\\
&& {1\over 3} + \alpha(t) + {3\beta(t)\over 2} + {1\over\vert\log~\bar{g}_s\vert}~\log\left[{(2-\beta(t))\over \mp\dot\alpha(t)\vert\log~\bar{g}_s\vert ({2\over 3} - \beta(t))}\right], \nonumber\\
&& {4\over 3} + \alpha(t) + \beta(t) + {1\over\vert\log~\bar{g}_s\vert} ~\log\left[{(2-\beta(t)) \over \mp\dot\alpha(t)\vert\dot\beta(t)\vert\vert\log~\bar{g}_s\vert^2({2\over 3} - \beta(t))}\right], \nonumber\\
&& {4\over 3} + \alpha(t) + \beta(t) + 
{2 \over \vert\log~\bar{g}_s\vert}~\log\left[{(2-\beta(t))\over 
\vert\dot\beta(t)\vert\vert\log~\bar{g}_s\vert\sqrt{({2\over 3}-\alpha(t))({2\over 3} -\beta(t))}}\right], \nonumber\\
&& {1\over 3} + \alpha(t) + {3\beta(t)\over 2} + 
{1\over \vert\log~\bar{g}_s\vert}~\log\left[{(2-\beta(t))^2\over 
\vert\dot\beta(t)\vert\vert\log~\bar{g}_s\vert({2\over 3}-\alpha(t))({2\over 3} -\beta(t))}\right], \nonumber\\
&& {4\over 3} + \alpha(t) + \beta(t) + {1\over \vert\log~\bar{g}_s\vert} ~\log\left[{2-\beta(t)\over f_o(t)({2\over 3} - \alpha(t))}\right],~
{4\over 3} + \alpha(t) + \beta(t) - {\log(\mp \dot\alpha(t) \vert\dot\beta(t)\vert\vert\log~\bar{g}_s\vert^2)\over\vert\log~\bar{g}_s\vert}\Bigg),  \nonumber
\nd}
leading to eight possible terms with various log dependences with $f_o(t)\equiv \vert\dot\beta(t)\vert^2 \vert\log~\bar{g}_g\vert^2$. Again, motivated from our earlier analysis, taking vanishing log corrections will give us enough hints as to how the final scalings would look like. Once we take that, the dominant contributions come from the three terms with scalings $-{2\over 3} + \alpha(t) + 2\beta(t), {1\over 3} + \alpha(t) + {3\beta(t)\over 2}$ and  ${4\over 3} + \alpha(t) + \beta(t))$. None of these terms are now worrisome when we take $\alpha(t) = -\beta(t)$ because of the positivity of $\beta(t)$ and because of $0 < \beta(t) < {2\over 3}$. However the contribution to \eqref{botsuga} (discussed in the following section) now becomes:
\bg\label{paltro}
{4\over 3} + \mathbb{F}_2(t) - \alpha(t) - \beta(t) = \left({8\over 3}, ~ 
{2\over 3} + \beta(t), ~ {5\over 3} + {\beta(t)\over 2}\right), \nd
which are all clearly positive definite. As before, we expect that the log terms now add 
small corrections to two of the aforementioned dominant scalings. This may be easily seen first from the fact that:

{\footnotesize
\bg\label{pitpendu6}
\mathbb{F}_2(t) \to \mathbb{F}_2(t; b, d, d') & \equiv & \bigg(-{2\over 3} + \alpha(t) + {3\beta(t)\over 2} + {n_d\over \bar{g}_s^{d/3}\vert{\rm log}~\bar{g}_s\vert} + \vert\gamma(t)\vert + {\cal C}(\Delta; b, d'), \\
&- & {2\over 3} + \alpha(t) + \beta(t) + {2n_d\over \bar{g}_s^{d/3}\vert{\rm log}~\bar{g}_s\vert} + 2\vert\gamma(t)\vert + {\cal B}(\Delta; b, d'), ~
-{2\over 3} + \alpha(t) + \beta(t)\bigg), \nonumber \nd}
where all the parameters appearing above are defined earlier. Since $\alpha(t)$ and $\beta(t)$ already have dominant non-perturbative corrections (see \eqref{ryanfan}), the first two terms in \eqref{pitpendu6} have additional non-perturbative corrections. Plugging \eqref{pitpendu6} in \eqref{bralesasl1}, we get:

\bg\label{kylieador3}
{4\over 3} + \mathbb{F}_2(t; b, d, d') - \alpha(t) - \beta(t) & = & 
\bigg({2\over 3} + {\beta(t)\over 2} + {n_d\over \bar{g}_s^{d/3}\vert{\rm log}~\bar{g}_s\vert} + \vert\gamma(t)\vert + {\cal C}(\Delta; b, d'), \\
& & {2\over 3} + {2n_d\over \bar{g}_s^{d/3}\vert{\rm log}~\bar{g}_s\vert} + 2\vert\gamma(t)\vert + {\cal B}(\Delta; b, d'), ~
{2\over 3} + \beta(t)\bigg), \nonumber \nd
which matches well with \eqref{kylieador2} and \eqref{kylieador}, and with $n_d \ge 0$ in \eqref{qingthic}, comes close to the scalings in \eqref{paltro} up to additional non-perturbative suppressions. For the ${\rm E}_8 \times {\rm E}_8$ case, the story is similar with the dominant contribution coming from:

{\scriptsize
\bg\label{russflow2}
{\rm dom}~\mathbb{F}_{2e}(t) = -{2\over 3} + \hat\sigma_e(t) -\hat\zeta_e(t)  + \left(\hat\alpha_e(t), \hat\beta_e(t)\right) + {\hat\alpha_e(t) + \hat\beta_e(t)\over 2} + {1\over \vert \log~\bar{g}_s\vert} ~\log\left[{\big(12 - 3\hat\alpha_e - 3\hat\beta_e\big)^2\over \big({4} - 6\hat\sigma_e\big)\big({4} - 6(\hat\alpha_e, \hat\beta_e)\big)}\right], \nonumber\\ \nd}
with the log piece sub-dominant because $(\hat\sigma_e(t), \hat\alpha_e, \hat\beta_e(t)) << 1$ and therefore only contributes to a numerical factor of ${2\over 3}$ to the overall $g_s$ scaling. The contribution to 
\eqref{botsuga} now comes from ${4\over 3} + \mathbb{F}_{2e}(t) - \hat\sigma_e(t) - (\hat\alpha_e(t), \hat\beta_e(t))$, which again gives us \eqref{2cho1bora}.
The other contributions to \eqref{botsuga} from the second rows of {\bf Table \ref{firzacut6}} and {\bf \ref{jennatara}} are clearly positive definite.

\begin{table}[tb]  
 \begin{center}
\renewcommand{\arraystretch}{1.5}
\begin{tabular}{|c||c||c|}\hline Riemann tensors  & Contributions to \eqref{botsuga}  \\ \hline\hline
${\bf R}_{mnpq}$ &  ${\rm dom}\left({2\over 3} - \alpha(t), ~{4\over 3} + \mathbb{F}_1(t) - 2\alpha(t),~ {8\over 3},~ {2\over 3}  - \beta(t)\right)$ \\ \hline 
${\bf R}_{mn\rho\sigma}$ &  ${\rm dom}\left({2\over 3} - \alpha(t),~ {2\over 3} - \beta(t), ~{4\over 3} + \mathbb{F}_2(t) -\alpha(t) - \beta(t), ~{8\over 3}\right)$ \\ \hline 
${\bf R}_{\rho\sigma i0}$ &  ${\rm dom}\left({5\over 3}  + \gamma_{1,2}(t), ~ {10\over 3} + \mathbb{F}_3(t) -\beta(t)\right)$ \\ \hline 
${\bf R}_{m\rho\sigma 0}$ &  ${\rm dom}\left({7\over 3} + \mathbb{F}_3(t) -\beta(t) - {\alpha(t)\over 2}, ~ {7\over 3} + \mathbb{F}_4(t) -\beta(t) -{\alpha(t)\over 2}\right)$ \\ \hline 
${\bf R}_{0\sigma\sigma\rho}$ &  ${\rm dom}\left({7\over 3} + \mathbb{F}_3(t) -{3\beta(t)\over 2}\right)$ \\ \hline 
${\bf R}_{\rho\sigma\rho\sigma}$ &  ${\rm dom}\left({2\over 3} - \beta(t), ~{4\over 3} + \mathbb{F}_5(t) -2\beta(t), ~{2\over 3}  - \alpha(t), ~{8\over 3}\right)$ \\ \hline
${\bf R}_{mn i0}$ &  ${\rm dom}\left({5\over 3} + \gamma_{1,2}(t), ~ {10\over 3} + \mathbb{F}_6(t) - \alpha(t)\right)$ \\
\hline
\end{tabular}
\renewcommand{\arraystretch}{1}
\end{center}
 \caption[]{\Su Contributions of the curvature tensors to \eqref{botsuga} for the simplified $SO(32)$ case from {\bf Table \ref{firzacut}}.} 
\label{firzacut6}
 \end{table}

\begin{table}[tb]  
 \begin{center}
\renewcommand{\arraystretch}{1.5}
\begin{tabular}{|c||c||c|}\hline Terms in $\mathbb{F}_i$  & Powers of $\bar{g}_s$  \\ \hline\hline
$\vert\ddot\beta\vert\vert\log~\bar{g}_s\vert$ &  $-2 + \Sigma(a_1, a_2, a_3, ...., a_{10})$ \\ \hline 
$\gamma_2 - 1$ &  $-1 + {n_d\over 2\bar{g}_s^{d/3} \vert\log~\bar{g}_s\vert} + {1\over 2}{\cal C}(\Delta; b, d)$ \\ \hline 
$\vert\dot\beta\vert\vert\log~\bar{g}_s\vert$ &  $-1 + {n_d\over \bar{g}_s^{d/3} \vert\log~\bar{g}_s\vert} + \vert\gamma\vert + {\cal C}(\Delta; b, d)$ \\ \hline 
$\gamma_1 - 1$ &  $-1 + {n_d\over \bar{g}_s^{d/3} \vert\log~\bar{g}_s\vert} + \vert\gamma\vert + {\cal C}(\Delta; b, d)$ \\ \hline 
$\vert\dot\beta\vert^2\vert\log~\bar{g}_s\vert^2$ &  $-2 + {2n_d\over \bar{g}_s^{d/3} \vert\log~\bar{g}_s\vert} + 2\vert\gamma\vert + {\cal B}(\Delta; b, d)$ \\ \hline 
$\vert\dot\beta\vert$ &  $-1 + {n_d\over \bar{g}_s^{d/3} \vert\log~\bar{g}_s\vert} + \vert\gamma\vert + {\vert\log~\Delta\vert\over \vert\log~\bar{g}_s\vert} + 
{\cal C}(0; b, d)$\\ \hline
$\vert\dot\beta\vert^2\vert\log~\bar{g}_s\vert$ &  $-2 + {n_d\over \bar{g}_s^{d/3} \vert\log~\bar{g}_s\vert} + 2\vert\gamma\vert + {\vert\log~\Delta\vert\over \vert\log~\bar{g}_s\vert} + 
{\cal F}(\Delta; b, d)$\\ \hline
\end{tabular}
\renewcommand{\arraystretch}{1}
\end{center}
 \caption[]{\Su $\gamma_{1, 2}$ and various combinations of the temporal derivatives of $\beta(t)$ for the simplified $SO(32)$ case, where $\beta(t)$ is given in \eqref{ryanfan}, expressed solely as powers of $\bar{g}_s$. The $\beta(t)$ dependent pieces take the form 
 $\vert\partial_t^n\beta\vert^p\vert\log~\bar{g}_s\vert^q$ with $(n, p, q) = (2,1,1), (1, 1, 1), (1, 2, 2), (1, 1, 0)$ and $(1, 2, 1)$. The parameters 
 $\gamma_{1, 2}, \gamma(t), ~\Sigma(a_1, a_2, a_3, ...., a_{10}), ~{\cal C}(\Delta; b, d), ~{\cal B}(\Delta; b, d)$ and ${\cal F}(\Delta; b, d)$ are defined in \eqref{ishena}, \eqref{mcelhon}, \eqref{nillchoke7}, \eqref{standudh} and \eqref{chinkpet} respectively. Similar series appear with $\alpha(t)$ because $\alpha(t) = -\beta(t)$.} 
\label{munleena}
 \end{table}

\begin{table}[tb]  
 \begin{center}
\renewcommand{\arraystretch}{1.5}
\begin{tabular}{|c||c||c|}\hline Riemann tensors  & Contributions to \eqref{botsuga} \\ \hline\hline
${\bf R}_{m\rho\sigma i}, {\bf R}_{mnpi}, {\bf R}_{mabi}, {\bf R}_{m0i0}, {\bf R}_{mijk}$ & ${5\over 3} - {\alpha(t)\over 2}$\\ \hline
${\bf R}_{i\sigma\sigma\rho}, {\bf R}_{mn\sigma i}, {\bf R}_{\sigma a b i}, {\bf R}_{0\sigma i 0}, {\bf R}_{\sigma ijk}$ & ${5\over 3} - {\beta(t)\over 2}$\\ \hline
$ {\bf R}_{m\sigma \sigma \rho}, {\bf R}_{mnp\sigma}, {\bf R}_{m\sigma ab}, {\bf R}_{m\sigma ij}, {\bf R}_{0m0\sigma}$ & ${2\over 3} - {\alpha(t)\over 2} - {\beta(t)\over 2}$\\ \hline
\end{tabular} 
\renewcommand{\arraystretch}{1}
\end{center}
 \caption[]{\Su Contributions of the curvature tensors to \eqref{botsuga} for the simplified $SO(32)$ case from {\bf Tables \ref{firzacut5}}.} 
  \label{firzacut8}
 \end{table} 

\begin{table}[tb]  
 \begin{center}
\renewcommand{\arraystretch}{1.5}
\begin{tabular}{|c||c||c|}\hline Riemann tensors  & Contributions to \eqref{botsuga} \\ \hline\hline
${\bf R}_{0i0j}$ & ${\rm dom}\left({8\over 3}, ~ {16\over 3} + \mathbb{F}_{16}(t), ~{2\over 3} - \alpha(t), 
~ {2\over 3} - \beta(t)\right)$ \\ \hline
${\bf R}_{0a0b}$ & ${\rm dom}\left({8\over 3}, ~{4\over 3} + \mathbb{F}_{17}(t), ~{2\over 3} - \alpha(t), 
~ {2\over 3} - \beta(t)\right)$ \\ \hline
${\bf R}_{0m0n}$ & ${\rm dom}\left({8\over 3}, ~{2\over 3} -\alpha(t),~\widetilde{\mathbb{F}}_{14}(t), ~ \widetilde{\mathbb{F}}_{18}(t),~\widetilde{\mathbb{F}}_{19}(t),~ 
{2\over 3} - \beta(t)\right)$ \\ \hline
${\bf R}_{0\rho 0\sigma}$ & ${\rm dom}\left({8\over 3}, ~{2\over 3} -\beta(t),~\widetilde{\mathbb{F}}_{15}(t), ~ \widetilde{\mathbb{F}}_{20}(t), ~\widetilde{\mathbb{F}}_{21}(t), ~{2\over 3} - \alpha(t)\right)$ \\ \hline
\end{tabular} 
\renewcommand{\arraystretch}{1}
\end{center}
 \caption[]{\Su Contributions of the curvature tensors for the simplified $SO(32)$ case to \eqref{botsuga} from {\bf Tables \ref{firzacut2}} with $\widetilde{\mathbb{F}}_{i}(t)$ defined for $i = 14, 18, 19$ as
 $\widetilde{\mathbb{F}}_{i}(t)= {10\over 3} + \mathbb{F}_i(t) - \alpha(t)$ and $\widetilde{\mathbb{F}}_{j}(t)$ defined for $j = 15, 20, 21$ as
 $\widetilde{\mathbb{F}}_{j}(t)= {10\over 3} + \mathbb{F}_j(t) - \beta(t)$.} 
  \label{firzacut9}
 \end{table}

\subsubsection{{Functional forms and bounds on} $\mathbb{F}_{3, 3e}(t)$  {and}  $\mathbb{F}_{6, 6e}(t)$}

\vskip.1in

\noindent Our next set is $\mathbb{F}_{3e}(t)$ and $\mathbb{F}_{6e}(t)$ whose functional forms appear respectively in the third and the sixth rows of {\bf Table \ref{privsocmey}}. $\mathbb{F}_{3e}(t)$ is responsible in providing the scalings of the curvature tensors ${\bf R}_{\rho\sigma i o}, {\bf R}_{m\rho \sigma 0}$ and ${\bf R}_{0\sigma \sigma \rho}$ (including all the respective permutations of the indices), whereas $\mathbb{F}_{6e}(t)$ is responsible for the scalings of ${\bf R}_{mni0}, {\bf R}_{mn\sigma 0}$ and 
${\bf R}_{mnp0}$ (again including all the respective permutations of the indices). $\mathbb{F}_{3e}(t)$ takes the following form:
\bg\label{sirvalentin}
\mathbb{F}_{3e}(t) = \left(\mathbb{A}_8(\hat\alpha_e(t), \hat\beta_e(t), 
\mathbb{B}_8(\hat\alpha_e(t), \hat\beta_e(t)\right), \nd
corresponding to the ${\rm E}_8 \times {\rm E}_8$ theory. One may easily read out the form of the above functions from \eqref{tifeni} and using our simplifying procedure, or referring to {\bf Table \ref{firzacut3}}, $\mathbb{F}_3(t)$ for the $SO(32)$ theory becomes: 
\bg\label{hkpilla}
\mathbb{F}_3(t) & = & \left(\mathbb{A}(\beta), ~ \mathbb{B}(\beta)\right)\nonumber\\
& = & \Bigg(-{2\over 3} + \beta(t) + {1\over \vert\log~\bar{g}_s\vert}~\log\left[{(2-\beta(t))\over \vert\dot\beta(t)\vert\vert\log~\bar{g}_s\vert({2\over 3} - \beta(t))}\right], \\
&-& {5\over 3} + {3\beta(t)\over 2} + {1\over\vert\log~\bar{g}_s\vert}~\log\left({2-\beta(t)\over {2\over 3} - \beta(t)}\right), ~ -{2\over 3} + \beta(t) - {\log(\vert\dot\beta(t)\vert \vert\log~\bar{g}_s\vert)\over \vert\log~\bar{g}_s\vert}\Bigg), \nonumber\nd
with the two dominant contributions given by $-{2\over 3} + \beta(t)$ and 
$-{5\over 3} + {3\beta(t)\over 2}$, once we ignore the log corrections. Since $0 < \beta(t) < {2\over 3}$, both these contributions are negative leading to strong curvature at late time. Looking at the actual contributions from the third, fourth and fifth rows in {\bf Table \ref{firzacut6}} to the quantum scaling \eqref{botsuga} we find that:

{\footnotesize
\bg\label{hkpilla2}
&& {10\over 3} + \mathbb{F}_3(t) - \beta(t) = \left({8\over 3}, ~{5\over 3} + {\beta(t)\over 2}\right) \\
&& {7\over 3} + \mathbb{F}_3(t) - \beta(t) -{\alpha(t)\over 2} = \left({5\over 3} - {\alpha(t)\over 2}, ~ {2\over 3} + {\beta(t)\over 2} -{\alpha(t)\over 2}\right), ~~~
{7\over 3} + \mathbb{F}_3(t) - {3\beta(t)\over 2} = \left({5\over 3} - {\beta(t)\over 2}, ~{2\over 3}\right), \nonumber \nd}
which are all positive definite. (The last term with a relative minus sign is positive because $0 < \beta(t) < {2\over 3}$.) This shows that the three set of curvature tensors, which would typically be very large, actually contribute as positive powers of $g_s$ to \eqref{botsuga}. The story in the presence of the log corrections may now be presented as:

{\footnotesize
\bg\label{pitpendu7}
&& \mathbb{F}_3(t) \to \mathbb{F}_3(t; b, d, d')  \equiv  \bigg(-{5\over 3} + \beta(t)  + {n_d\over \bar{g}_s^{d/3}\vert{\rm log}~\bar{g}_s\vert} + \vert\gamma(t)\vert + {\cal C}(\Delta; b, d'), 
~ -  {5\over 3} + {3\beta(t)\over 2}\bigg) \nonumber\\
&& {7\over 3} + \mathbb{F}_3(t; b, d, d') - {3\beta(t)\over 2}  =  
\bigg({2\over 3} - {\beta\over 2} + {n_d\over \bar{g}_s^{d/3}\vert{\rm log}~\bar{g}_s\vert} + \vert\gamma(t)\vert + {\cal C}(\Delta; b, d'), ~ {2\over 3}\bigg) \nonumber\\
&& {10\over 3} + \mathbb{F}_3(t; b, d, d') - \beta(t)  =  
\bigg({5\over 3} + {n_d\over \bar{g}_s^{d/3}\vert{\rm log}~\bar{g}_s\vert} + \vert\gamma(t)\vert + {\cal C}(\Delta; b, d'),
~ {5\over 3} + {\beta(t)\over 2}\bigg)\\
&& {7\over 3} + \mathbb{F}_3(t; b, d, d') - \beta(t) - {\alpha(t)\over 2} =  
\bigg({2\over 3} - {\alpha(t)\over 2} + {n_d\over \bar{g}_s^{d/3}\vert{\rm log}~\bar{g}_s\vert} + \vert\gamma(t)\vert + {\cal C}(\Delta; b, d'),~
 {2\over 3} + {\beta(t)\over 2} -{\alpha(t)\over 2}\bigg), \nonumber
\nd}
implying that the contributions to \eqref{botsuga} remains positive definite. Using \eqref{qingthic} we recover the scalings from \eqref{hkpilla2} although, as cautioned earlier, the correct scaling only comes from imposing $n_d \ge 0$ and keeping $b \ge 0$ and $0 \le d \le d_{\rm max}$. The condition 
$b = d' + 3$ is a very special one, and we are not required to impose because $b_{\rm min} = 0$ and all terms with $n_d > 0$ would automatically suppress the parts of the scalings in \eqref{pitpendu7} that accompany $n_d$. This way the dominant scaling still remains ${2\over 3}$. For the ${\rm E}_8 \times {\rm E}_8$ case, the dominant contribution comes from:

{\footnotesize
\bg\label{russflow3}
{\rm dom}~\mathbb{F}_{3e}(t) = -{5\over 3} + (\hat\alpha_e(t), \hat\beta_e(t)) + {\hat\alpha_e(t) + \hat\beta_e(t)\over 4} + {1\over \vert\log~\bar{g}_s\vert}~\log\left({12 - 3\hat\alpha_e - 3\hat\beta_e\over 4 - 6(\hat\alpha_e, \hat\beta_e)}\right), \nd}
where note the absence of the $\hat\zeta_e(t)$ factor compared to the other dominant contributions we saw so far. However the $\hat\zeta_e(t)$ factor does appear in the scaling of the quantum series \eqref{botsuga} that takes the form:
\bg\label{floanna}
&& {7\over 3} + \mathbb{F}_{3e}(t) - (\hat\alpha_e(t), \hat\beta_e(t)) - {\hat\sigma_e(t)\over 2} - {\hat\zeta_e(t)\over 2},\\
&& {10\over 3} + \mathbb{F}_{3e}(t) - (\hat\alpha_e(t), \hat\beta_e(t)) - \hat\zeta_e(t), ~~~~
{7\over 3} + \mathbb{F}_{3e}(t) - {3\over 2}(\hat\alpha_e(t), \hat\beta_e(t))  - {\hat\zeta_e(t)\over 2},\nonumber \nd
associated with the three curvature components ${\bf R}_{m\rho\sigma 0}, {\bf R}_{\rho\sigma i0}$ and ${\bf R}_{0\sigma\sigma \rho}$ respectively. Plugging the value of $\mathbb{F}_{3e}(t)$ from \eqref{russflow3} in 
\eqref{floanna} gives us the contributions to \eqref{botsuga} for the ${\rm E}_8 \times {\rm E}_8$ case. For the $SO(32)$ case, we can generalize the analysis from \eqref{pitpendu7} or alternatively, derive it from the ${\rm E}_8 \times {\rm E}_8$ case, to get:

{\scriptsize
\bg\label{chu8hath1}\boxed{
\begin{aligned}
& \left({5\over 3} - \zeta_e(t) + {\beta_e(t)\over 2}, ~~~ {2\over 3} - {\zeta_e(t)\over 2} + {\beta_e(t)\over 2} - {\alpha_e(t)\over 2}, ~~~ {2\over 3} - {\zeta_e(t)\over 2}\right); \\
& \left({5\over 3} - \hat\zeta_e(t) + {\frac{\hat\alpha_e(t) + \hat\beta_e(t)}{4}}, ~~~ {2\over 3} - {\hat\zeta_e(t)\over 2} + {\frac{\hat\alpha_e(t) + \hat\beta_e(t)}{4}} - {\hat\sigma_e\over 2}, ~~~ {2\over 3} - {\hat\zeta_e(t)\over 2} -{1\over 2}(\hat\alpha_e(t), \hat\beta_e(t)) + {\frac{\hat\alpha_e(t) + \hat\beta_e(t)}{4}}\right)
\end{aligned}}
\nonumber\\
\nd}
where the upper set is for the $SO(32)$ case and the lower set is for the ${\rm E}_8 \times {\rm E}_8$ case. These two scalings differ significantly from what we had in \eqref{2cho1bora}.
One expects similar story for $\mathbb{F}_{6e}(t)$ whose functional form takes the following values:
\bg\label{hkpgkmkkkbkb}
\mathbb{F}_{6e}(t) =  \left(\mathbb{A}_8(\hat\sigma_e), ~ \mathbb{B}_8(\hat\sigma_e)\right), \nd
as seen from {\bf Table \ref{privsocmey}}. The similarity to \eqref{sirvalentin} suggests that this case cannot be very different from what we had earlier. For the simplified scenario where we ignore the sub-dominant contributions, the $SO(32)$ case becomes:
\bg\label{hkpilla3}
\mathbb{F}_6(t) & = & \left(\mathbb{A}(\alpha), ~ \mathbb{B}(\alpha)\right)\nonumber\\
& = & \Bigg(-{2\over 3} + \alpha(t) + {1\over \vert\log~\bar{g}_s\vert}~\log\left[{(2-\beta(t))\over \vert\dot\beta(t)\vert\vert\log~\bar{g}_s\vert({2\over 3} - \alpha(t))}\right], \\
&-& {5\over 3} + \alpha(t) + {\beta(t)\over 2} + {1\over\vert\log~\bar{g}_s\vert}~\log\left({2-\beta(t)\over {2\over 3} - \alpha(t)}\right), ~ -{2\over 3} + \alpha(t) - {\log(\mp \dot\alpha(t) \vert\log~\bar{g}_s\vert)\over \vert\log~\bar{g}_s\vert}\Bigg), \nonumber\nd
whose dominant contributions come from $-{2\over 3} + \alpha(t)$ and 
$-{5\over 3} + \alpha(t) + {\beta(t)\over 2}$. Since $\alpha(t) = -\beta(t)$, all these become purely negative as $-{2\over 3} - \beta(t)$ and $-{5\over3} - {\beta(t)\over 2}$, signalling strong Riemann curvatures components. However looking at rows seven, eight and nine of {\bf Table \ref{firzacut6}}, and ignoring the log corrections, we find that the actual contributions to \eqref{botsuga} become:

{\footnotesize
\bg\label{hkpilla4}
&& {10\over 3} + \mathbb{F}_6(t) - \alpha(t) = \left({8\over 3}, ~{5\over 3} + {\beta(t)\over 2}\right) \\
&& {7\over 3} + \mathbb{F}_6(t) - {3\alpha(t)\over 2} = \left({5\over 3} -{\alpha(t)\over 2}, ~ {2\over 3} + {\beta(t)\over 2} -{\alpha(t)\over 2}\right), ~~~
{7\over 3} + \mathbb{F}_6(t) -\alpha(t) - {\beta(t)\over 2} = \left({5\over 3} - {\beta(t)\over 2}, ~{2\over 3}\right), \nonumber \nd}
which matches exactly with what we had in \eqref{hkpilla2}. In the presence of the log corrections, the scalings are very similar to \eqref{pitpendu7}, with some very minor difference. This may be quantified in the following way:

{\footnotesize
\bg\label{pitpendu8}
&& \mathbb{F}_6(t) \to \mathbb{F}_6(t; b, d, d')  \equiv  \bigg(-{5\over 3} + \alpha(t)  + {n_d\over \bar{g}_s^{d/3}\vert{\rm log}~\bar{g}_s\vert} + \vert\gamma(t)\vert + {\cal C}(\Delta; b, d'), 
~ -  {5\over 3} + \alpha(t) + {\beta(t)\over 2}\bigg) \nonumber\\
&& {10\over 3} + \mathbb{F}_6(t; b, d, d') - \alpha(t)  =  
\bigg({5\over 3} + {n_d\over \bar{g}_s^{d/3}\vert{\rm log}~\bar{g}_s\vert} + \vert\gamma(t)\vert + {\cal C}(\Delta; b, d'),
~ {5\over 3} + {\beta(t)\over 2}\bigg)\\
&& {7\over 3} + \mathbb{F}_6(t; b, d, d') - \alpha(t) - {\beta(t)\over 2}  =  
\bigg({2\over 3} - {\beta\over 2} + {n_d\over \bar{g}_s^{d/3}\vert{\rm log}~\bar{g}_s\vert} + \vert\gamma(t)\vert + {\cal C}(\Delta; b, d'), ~ {2\over 3}\bigg) \nonumber\\
&& {7\over 3} + \mathbb{F}_6(t; b, d, d') - {3\alpha(t)\over 2} =  
\bigg({2\over 3} - {\alpha(t)\over 2} + {n_d\over \bar{g}_s^{d/3}\vert{\rm log}~\bar{g}_s\vert} + \vert\gamma(t)\vert + {\cal C}(\Delta; b, d'),~
 {2\over 3} + {\beta(t)\over 2} -{\alpha(t)\over 2}\bigg), \nonumber
\nd}
which would match with \eqref{hkpilla4} only if we impose \eqref{qingthic}. With $n_d \ge 0$, {\it i.e.} including all terms with $n_d > 0$, the dominant scaling is controlled by 
${2\over 3}$ again, thus showing that the contributions of the curvature tensors to \eqref{botsuga} remain positive powers of $g_s$. For the ${\rm E}_8 \times {\rm E}_8$ case the dominant contribution comes from:
\bg\label{russflow3}
{\rm dom}~\mathbb{F}_{6e}(t) = -{5\over 3} + \hat\sigma_e(t) + {\hat\alpha_e(t) + \hat\beta_e(t)\over 4} + {1\over \vert\log~\bar{g}_s\vert}~\log\left({12 - 3\hat\alpha_e - 3\hat\beta_e\over 4 - 6\hat\sigma_e}\right), \nd
and the contributions to \eqref{botsuga} takes the form \eqref{floanna} with $\mathbb{F}_{6e}(t)$ replacing $\mathbb{F}_{3e}(t)$, and  $(\hat\alpha_e(t), \hat\beta_e(t)) \leftrightarrow \hat\sigma_e(t)$ but $\hat\zeta_e(t)$ unchanged. For the $SO(32)$ and the ${\rm E}_8 \times {\rm E}_8$ cases, they reproduce precisely \eqref{chu8hath1}.

\subsubsection{{Functional forms and bounds on} $\mathbb{F}_{4, 4e}(t)$ {and}  $\mathbb{F}_{7, 7e}(t)$}

\vskip.1in

\noindent Our next set is $\mathbb{F}_{4e}(t)$ and $\mathbb{F}_{7e}(t)$ which appear in the scaling exponents of ${\bf R}_{m\rho\sigma 0}$ and ${\bf R}_{mn\sigma 0}$ respectively. (In fact the other two exponents $\mathbb{F}_{3e}(t)$ and $\mathbb{F}_{6e}$ also appear respectively for these curvature tensors which we already discussed earlier.) These exponents appear in the fourth and the ninth rows in {\bf Table \ref{firzathai}}, and their function forms respectively in the fourth and the seventh rows of {\bf Table \ref{privsocmey}}. From there we see that $\mathbb{F}_{4e}(t)$ takes the following form:
\bg\label{kinmarsol}
\mathbb{F}_{4e}(t) &= & \left(\mathbb{A}_8(\hat\sigma_e) + (\hat\alpha_e(t), \hat\beta_e(t)) - \hat\sigma_e(t), ~ \mathbb{B}_8(\hat\sigma_e) + (\hat\alpha_e(t), \hat\beta_e(t)) - \hat\sigma_e(t)\right)\nonumber\\
& \xrightarrow{\text{${\bf D}_{16}$}} & \left(\mathbb{A}(\alpha) + \beta(t) - \alpha(t), ~ 
\mathbb{B}(\alpha) + \beta(t) - \alpha(t)\right) \\
& = & \Bigg(-{2\over 3} + \beta(t) + {1\over \vert\log~\bar{g}_s\vert}~\log\left[{2-\beta(t) \over \vert\dot\beta(t)\vert\vert\log~\bar{g}_s\vert({2\over 3} - \alpha(t))}\right], \nonumber\\
& & -{5\over 3} + {3\beta(t)\over 2} + {1\over\vert\log~\bar{g}_s\vert}~\log\left({2-\beta(t) \over {2\over 3} - \alpha(t)}\right), ~ -{2\over 3} + \beta(t) - {\log(\mp \dot\alpha(t)\vert\log~\bar{g}_s\vert) \over\vert\log~\bar{g}_s\vert}\Bigg),\nonumber
\nd
where the arrow denotes going from the ${\rm E}_8 \times {\rm E}_8$ case to the $SO(32)$ case in the simplified setting. For the $SO(32)$ side, 
we see that the dominant scalings in the absence of the log corrections are $-{2\over 3} + \beta(t)$ and $-{5\over 3} + {3\beta(t)\over 2}$, both of which are negative definite because $0 < \beta(t) < {2\over 3}$. This signal strong Riemann curvature tensor, but as before we should see how they contribute to the quantum scaling \eqref{botsuga} once log corrections are inserted in. Looking at the fourth row in {\bf Table \ref{firzacut6}} we see that the contributions are:

{\scriptsize
\bg\label{ablumm}
{7\over 3} + \mathbb{F}_4(t; b, d, d') - \beta(t) - {\alpha(t)\over 2} = \left({2\over 3} - {\alpha(t)\over 2} + {n_d\over \bar{g}_s^{d/3}\vert{\rm log}~\bar{g}_s\vert} + \vert\gamma(t)\vert + {\cal C}(\Delta; b, d'),
 ~ {2\over 3} + {\beta(t)\over 2} -{\alpha(t)\over 2}\right), \nd}
 and one may then sum over $(b, d, d')$ with appropriate coefficients once we raise them as powers of $\bar{g}_s$. Note that
both terms in \eqref{ablumm} are positive definite because $\alpha(t) = -\beta(t)$, implying that the specific curvature tensor (along with the permutation of the indices) contributes as positive powers of $g_s$ to \eqref{botsuga}. It is also easy to see that, for the ${\rm E}_8 \times {\rm E}_8$ case, the dominant contribution comes from:

{\footnotesize
\bg\label{marsolib}
{\rm dom}~\mathbb{F}_{4e}(t) = -{5\over 3} + {\hat\alpha_e(t) + \hat\beta_e(t)\over 4} + (\hat\alpha_e(t), \hat\beta_e(t)) + {1\over \vert\log~\bar{g}_s\vert}~\log\left({12 - 3\hat\alpha_e - 3\hat\beta_e\over 4 - 6\hat\sigma_e}\right), \nd}
which contributes to \eqref{botsuga} as ${7\over 3} + \mathbb{F}_{4e}(t) - (\hat\alpha_e(t), \hat\beta_e(t)) - {\hat\sigma_e(t)\over 2} - {\hat\zeta_e(t)\over 2}$. Note that the $\hat\zeta_e(t)$ factor enters through this in \eqref{botsuga}. The result we get however differs from \eqref{2cho1bora} in the placement of the warp-factors as:
\bg\label{duitrans}
\boxed{
{2\over 3} + {\beta_e(t)\over 2} - {\alpha_e(t)\over 2} - {\zeta_e(t)\over 2}; ~~~~ {2\over 3} + {\hat\alpha_e(t) + \hat\beta_e(t)\over 4} - {\hat\sigma_e(t)\over 2} - {\hat\zeta_e(t)\over 2}} \nd
for the $SO(32)$ and the ${\rm E}_8 \times {\rm E}_8$ cases respectively. Interestingly, for the $SO(32)$ case, if we make $\alpha_e(t) = - \beta_e(t)$ to keep the four-dimensional Newton's constant time-independent, we can recover the $SO(32)$ scaling from \eqref{2cho1bora}. For the ${\rm E}_8 \times {\rm E}_8$ case the relation is $\hat\sigma_e(t) = -{\hat\alpha_e(t) + 3\hat\beta_e(t)\over 4}$, which makes it different from the one in \eqref{2cho1bora}\footnote{Although for $\hat\alpha_e(t) = \hat\beta_e(t)$ the results would match with the $SO(32)$ case.}.
In a similar vein, the other exponent $\mathbb{F}_7(t)$ takes the following form:

{\footnotesize
\bg\label{kinmarsol2}
\mathbb{F}_{7e}(t) & = & \left(\mathbb{A}_8(\hat\alpha_e, \hat\beta_e) - (\hat\alpha_e(t), \hat\beta_e(t)) + \hat\sigma_e(t), ~ \mathbb{B}_8(\hat\alpha_e, \hat\beta_e)- (\hat\alpha_e(t), \hat\beta_e(t)) + \hat\sigma_e(t)\right)\nonumber\\
 & \xrightarrow{\text{${\bf D}_{16}$}} & \left(\mathbb{A}(\beta) - \beta(t) + \alpha(t), ~ 
\mathbb{B}(\beta) - \beta(t) + \alpha(t)\right) \\
& = & \Bigg(-{2\over 3} + \alpha(t) + {1\over \vert\log~\bar{g}_s\vert}~\log\left[{2-\beta(t) \over \vert\dot\beta(t)\vert\vert\log~\bar{g}_s\vert({2\over 3} - \beta(t))}\right], \nonumber\\
&& -{5\over 3} + \alpha(t) +  {\beta(t)\over 2} + {1\over\vert\log~\bar{g}_s\vert}~\log\left({2-\beta(t) \over {2\over 3} - \beta(t)}\right), ~ -{2\over 3} + \alpha(t) - {\log(\vert\dot\beta(t)\vert\vert\log~\bar{g}_s\vert) \over\vert\log~\bar{g}_s\vert}\Bigg),\nonumber
\nd}
with the dominant parts in the absence of log corrections being $-{2\over 3} + \alpha(t)$ and $-{5\over 3} + \alpha(t) + {\beta(t)\over 2}$, which are more negative now because $\alpha(t) = -\beta(t)$. However if we look at the contributions to the scaling \eqref{botsuga} from the ninth row in {\bf Table \ref{firzacut6}}, and insert in the log corrections, we find that:

{\footnotesize
\bg\label{clodgarcia}
{7\over 3} + \mathbb{F}_7(t; b, d, d') - \alpha(t) - {\beta(t)\over 2} = \left({2\over 3}, ~{2\over 3} - {\beta(t)\over 2} + {n_d\over \bar{g}_s^{d/3}\vert{\rm log}~\bar{g}_s\vert} + \vert\gamma(t)\vert + {\cal C}(\Delta; b, d')\right), \nd}
which are both positive definite because $0 < \beta(t) < {2\over 3}$, thus contributing as positive powers of $g_s$  to \eqref{botsuga} once we raise them as powers of $\bar{g}_s$ with appropriate coefficients. Again, the strong curvature tensor components, are relatively harmless in the quantum series \eqref{botsuga}.  Interestingly, \eqref{clodgarcia} becomes equal to \eqref{ablumm} only when $\alpha(t) = \beta(t)$. This equality would not keep four-dimensional Newton's constant time-independent therefore, while \eqref{clodgarcia} and \eqref{ablumm} would differ in their functional form, they would nevertheless be positive definite. We could also recover the dominant scalings in the absence of the log corrections by using \eqref{qingthic}, but this will clash with the expected behavior of $\beta(t)$. For the ${\rm E}_8 \times {\rm E}_8$ case, the dominant scaling becomes:
\bg\label{kgfgold}
{\rm dom}~\mathbb{F}_{7e}(t) = -{5\over 3} + {\hat\alpha_e(t) + \hat\beta_e(t)\over 4} + \hat\sigma_e(t) + {1\over \vert\log~\bar{g}_s\vert}~\log\left({12 - 3\hat\alpha_e - 3\hat\beta_e\over 4 - 6(\hat\alpha_e, \hat\beta_e)}\right), \nd
which enters the scaling in \eqref{botsuga} as ${7\over 3} + \mathbb{F}_{7e}(t) - \hat\sigma_e(t) - {1\over 2}(\hat\alpha_e, \hat\beta_e) - {\hat\zeta_e(t)\over 2}$. As before, the log piece simply contributes as a  $g_s$ independent numerical factor of ${1\over 3}$, and we can ignore it. The resulting scalings entering \eqref{botsuga} now becomes:
\bg\label{connorice}
\boxed{
{2\over 3} - {\zeta_e(t)\over 2}; ~~~~ {2\over 3} + {\hat\alpha_e(t) + \hat\beta_e(t)\over 4} - {1\over 2} (\hat\alpha_e(t), \hat\beta_e(t)) - {\hat\zeta_e(t)\over 2}} \nd
for the $SO(32)$ and the ${\rm E}_8 \times {\rm E}_8$ cases respectively.
Interesting, this case and the ones from \eqref{duitrans} are contained in the set of scalings from \eqref{chu8hath1}.

\subsubsection{{Functional forms and bounds on} $\mathbb{F}_{j,je,j+6,(j+6)e}(t)$}

\vskip.1in

\noindent Our next set is a set of four exponents, coming from the choice $j = 8$ and $j = 9$, will involve more complicated structures compared to what we encountered so far. Let us start with ${\rm F}_{8e}(t)$ that appears in the scaling of the curvature tensor ${\bf R}_{mnab}$. From {\bf Table \ref{privsocmey}}, this takes the following form:
\bg\label{lenasiribone}
\mathbb{F}_{8e}(t) & = & \Bigg({8\over 3} - \hat\zeta_e(t) + \mathbb{N}_8(\hat\sigma_e) + \left({1\over 3} + \gamma_{1, 2}\Big[{\hat\alpha_e(t) + \hat\beta_e(t)\over 2}\Big], \mathbb{F}_8(\hat\eta_e)\right), \nonumber\\
&&~{8\over 3} - \hat\zeta_e(t) + \mathbb{N}_8(\hat\sigma_e) + \left({1\over 3} + \gamma_{1, 2}\Big[{\hat\alpha_e(t) + \hat\beta_e(t)\over 2}\Big], \mathbb{G}_8(\hat\eta_e)\right)\Bigg), \nd
where $\mathbb{N}_8(\hat\sigma_e) = (\mathbb{A}_8(\hat\sigma_e), \mathbb{B}_8(\hat\sigma_e)$; and $(\mathbb{F}_{8}(\hat\eta_e), \mathbb{G}_8(\hat\eta_e))$ are defined in \eqref{vanandcar}. The latter two typically involve modes $\eta_e(t)$, defined in \eqref{masatwork}, and are related to the metric component along the $w^3$ direction (that forms the M-theory torus ${\mathbb{T}^2\over {\cal G}}$. The nested brackets in \eqref{lenasiribone} suggests that, for every choice of ${\rm N}_8(\hat\sigma_e)$ there are two possibilities from the two allowed choices of taking the temporal derivatives of ${\bf g}_{ab}({\bf x}, y; g_s)$ with $(w^a, w^b) \in {\mathbb{T}^2\over {\cal G}}$ in \eqref{sinistre00}. These may be explicitly written as:
\bg\label{lensiri1}
\mathbb{F}_{8e}(t) & = & \Bigg({8\over 3} - \hat\zeta_e(t) + \mathbb{A}_8(\hat\sigma_e) + \left({1\over 3} + \gamma_{1, 2}\Big[{\hat\alpha_e(t) + \hat\beta_e(t)\over 2}\Big], \mathbb{F}_8(\hat\eta_e)\right) \nonumber\\
&& ~~{8\over 3} - \hat\zeta_e(t) + \mathbb{A}_8(\hat\sigma_e) + \left({1\over 3} + \gamma_{1, 2}\Big[{\hat\alpha_e(t) + \hat\beta_e(t)\over 2}\Big], \mathbb{G}_8(\hat\eta_e)\right) \nonumber\\
&& ~~{8\over 3} - \hat\zeta_e(t) + \mathbb{B}_8(\hat\sigma_e) + \left({1\over 3} + \gamma_{1, 2}\Big[{\hat\alpha_e(t) + \hat\beta_e(t)\over 2}\Big], \mathbb{F}_8(\hat\eta_e)\right) \nonumber\\
&& ~~{8\over 3} - \hat\zeta_e(t) + \mathbb{B}_8(\hat\sigma_e) + \left({1\over 3} + \gamma_{1, 2}\Big[{\hat\alpha_e(t) + \hat\beta_e(t)\over 2}\Big], \mathbb{G}_8(\hat\eta_e)\right)\Bigg), \nd
thus leading to the eight possibilities. The story simplifies considerably once we ignore the sub-dominant corrections because $\mathbb{F}_8(\hat\eta_e) \to \infty$ as $\dot{\hat\eta}_e(t) \to 0$; and $\mathbb{G}_8(\hat\eta_e) = {1\over 3} + \hat\eta_e(t) + \gamma_{1, 2}\big[{\hat\alpha_e(t) + \hat\beta_e(t)\over 3}\big]$, because the log piece only provides a $g_s$ independent numerical factor of ${4\over 3}$.
Once we go to the simplified $SO(32)$ case from \eqref{lensiri1}, it is easy to infer that all these choices $-$ including the ones coming from $\mathbb{F}_{9e,14e,15e}(t)$ $-$ go as either 
$\{\mathbb{A}(\alpha), \mathbb{A}(\beta)\} + \gamma_{1, 2} + n$ or as 
$\{\mathbb{B}(\alpha), \mathbb{B}(\beta)\} + \gamma_{1, 2} + n$ with $n \in \pm\mathbb{Z}$ taking values $n = (-1, 3)$. This will be elaborated soon, but first let us start with $\mathbb{F}_8(t)$ whose functional form appears in the eighth row in 
{\bf Table \ref{firzacut6}} and contributes to the scaling of the curvature tensor ${\bf R}_{mnab}$ as shown in {\bf Table \ref{firzacut}}. This takes the following form:

{\footnotesize
\bg\label{garci1}
\mathbb{F}_8(t) & = & \left(3 + \mathbb{A}(\alpha) + \gamma_{1, 2}, ~ 3 + \mathbb{B}(\alpha) + \gamma_{1, 2}\right)\\
& = & \left(\mathbb{A}_1(\alpha) + \gamma_1 + 3, \mathbb{A}_1(\alpha) + \gamma_2 + 3, \mathbb{A}_2(\alpha) + \gamma_1 + 3, \mathbb{A}_2(\alpha) + \gamma_2 + 3, \mathbb{B}(\alpha) + \gamma_1 + 3, \mathbb{B}(\alpha) + \gamma_2 + 3\right), \nonumber\\
&=&\Bigg( {7\over 3} + \alpha(t) + {\beta(t)\over 2} + 
{1\over\vert\log~\bar{g}_s\vert}~\log\left[{(2-\beta(t))^2 \over \vert\dot\beta(t)\vert\vert\log~\bar{g}_s\vert({2\over 3} - \alpha(t))}\right], \nonumber\\
&& {10\over 3} + \alpha(t) + {1\over\vert\log~\bar{g}_s\vert}~\log\left[{2-\beta(t) \over \mp\dot\alpha(t)\vert\dot\beta(t)\vert\vert\log~\bar{g}_s\vert^2}\right], ~
{7\over 3} + \alpha(t) + {\beta(t)\over 2} + {1\over\vert\log~\bar{g}_s\vert} ~\log\left[{2-\beta(t)\over \mp \dot\alpha(t)\vert\log~\bar{g}_s\vert}\right], \nonumber\\
&& {10\over 3} + \alpha(t) + {2\over \vert\log~\bar{g}_s\vert}~\log\left[{2-\beta(t) \over \vert\dot\beta(t)\vert\vert\log~\bar{g}_s\vert\sqrt{{2\over 3} -\alpha(t)}}\right],~
{4\over 3} + \alpha(t) + \beta(t) + {2\over\vert\log~\bar{g}_s\vert}~\log\left[{2-\beta(t)\over \sqrt{{2\over 3} - \alpha(t)}}\right]\Bigg),\nonumber
\nd}
where $\mathbb{A}_1(\alpha)$ and $\mathbb{A}_2(\alpha)$ are the two values of $\mathbb{A}(\alpha)$ from \eqref{corsage2}, with $\gamma_1$ and $\gamma_2$ being the two values from \eqref{ishena}. The dominant contributions in the absence of the log corrections are now ${10\over 3} + \alpha(t), {7\over 3} + \alpha(t) + {\beta(t)\over 2}$ and ${4\over 3} + \alpha(t) + \beta(t)$. In fact they are all positive definite and the contributions to the quantum scaling \eqref{botsuga} once we insert in the log corrections become:
\bg\label{ishena3}
-{2\over 3} + \mathbb{F}_8(t; b, d, d') -\alpha(t) & = & \bigg({2\over 3} + {2n_d\over \bar{g}_s^{d/3}\vert{\rm log}~\bar{g}_s\vert} + 2\vert\gamma(t)\vert + {\cal B}(\Delta; b, d'), \\
&& {2\over 3} + {\beta(t)\over 2} + {n_d\over \bar{g}_s^{d/3}\vert{\rm log}~\bar{g}_s\vert} + \vert\gamma(t)\vert + {\cal C}(\Delta; b, d'), ~{2\over 3} + \beta(t)\bigg), \nonumber\nd
which, as mentioned above, are all positive definite because $0 < \beta(t) < {2\over 3}$; and ${\cal B}(\Delta; b, d')$ and ${\cal C}(\Delta; b, d')$ are defined in \eqref{standudh}. Thus both the curvature components as well as their contributions to \eqref{botsuga} are in positive powers of $g_s$. Extracting the results with the sub-dominant corrections for both $SO(32)$ and the ${\rm E}_8 \times {\rm E}_8$ cases is easy. The dominant contribution from \eqref{lensiri1} becomes:

{\footnotesize
\bg\label{lensiri2}
{\rm dom}~\mathbb{F}_{8e}(t) &=& \Bigg({4\over 3} -\hat\zeta_e(t) + \hat\sigma_e(t) + {\hat\alpha_e(t) + \hat\beta_e(t)\over 2} + {1\over \vert\log~\bar{g}_s\vert}\log\left[{(4\sqrt{3} -\sqrt{3}\hat\alpha_e - \sqrt{3}\hat\beta_e)^2\over 8 - 12\hat\sigma_e}\right], \\
&& ~~ {4\over 3} -\hat\zeta_e(t) + \hat\sigma_e(t) + \hat\eta_e(t) + {\hat\alpha_e(t) + \hat\beta_e(t)\over 2} +{1\over \vert\log~\bar{g}_s\vert}\log\left[{(12 -{3}\hat\alpha_e - {3}\hat\beta_e)^2\over (4-6\hat\sigma_e(t))(8 + 6\hat\eta_e)}\right]\Bigg), \nonumber
\nd}
which will contribute to \eqref{botsuga} as $-{2\over 3} + \mathbb{F}_{8e}(t) - \hat\sigma_e(t) - (0, \hat\eta_e(t))$, where the two choices come from the modes along the two directions of the M-theory torus ${\mathbb{T}^2\over {\cal G}}$. Putting everything together provides the same scaling as in \eqref{2cho1bora}
for the $SO(32)$ and the ${\rm E}_8 \times {\rm E}_8$ cases respectively.
In a similar vein, $\mathbb{F}_{14e}(t)$ takes the following form:
\bg\label{tariffcnn}
\mathbb{F}_{14e}(t) = \left({8\over 3} - \hat\zeta_e(t) + \mathbb{N}_8(\hat\sigma_e) + \mathbb{C}_8(\hat\zeta_e), ~{8\over 3} - \hat\zeta_e(t) + \mathbb{N}_8(\hat\sigma_e) + \mathbb{D}_8(\hat\zeta_e)\right), \nd
which differs from \eqref{lenasiribone} in the presence of $\mathbb{C}_8(\hat\zeta_e)$ and $\mathbb{D}_8(\hat\zeta_e)$ instead of $\mathbb{F}_8(\hat\eta_e)$ and $\mathbb{G}_8(\hat\eta_e)$. (See definitions in \eqref{vanandcar}.) Moreover the arguments of these function differ because now the temporal derivative act on the metric components along ${\bf R}^{2, 1}$ directions. From {\bf Table \ref{firzathai}} we see that $\mathbb{F}_{14e}(t)$ provides the scaling for the curvature tensor ${\bf R}_{mnij}$. We can also express \eqref{tariffcnn} in the following expanded  form:
\bg\label{tarifcnn2}
\mathbb{F}_{14e}(t) & = & \bigg({8\over 3} - \hat\zeta_e(t) + \mathbb{A}_8(\hat\sigma_e) + \mathbb{C}_8(\hat\zeta_e), 
~ ~ {8\over 3} - \hat\zeta_e(t) + \mathbb{A}_8(\hat\sigma_e) + \mathbb{D}_8(\hat\zeta_e) \nonumber\\
&& ~~ {8\over 3} - \hat\zeta_e(t) + \mathbb{B}_8(\hat\sigma_e) + \mathbb{C}_8(\hat\zeta_e),
~~ {8\over 3} - \hat\zeta_e(t) + \mathbb{B}_8(\hat\sigma_e) + \mathbb{D}_8(\hat\zeta_e)\bigg), \nd
where, using the definitions \eqref{vanandcar}, one can see that once we decouple the sub-leading corrections, {\it i.e.} make $\hat\zeta_e(t) \to 0$, $\mathbb{C}_8(\hat\zeta_e) \to 0$ and $\mathbb{D}_8(\hat\zeta_e) \to -{11\over 3} + \hat\zeta_e(t) + \gamma_{1,, 2}\big[{\hat\alpha_e + \hat\beta_e\over 3}\big]$ because the log piece only provides a $g_s$ independent scaling of ${8\over 3}$. Once we further transform to the simplified $SO(32)$ picture, this gives us
$\mathbb{F}_{14}(t)$ which differs from $\mathbb{F}_8(t)$ by an integer and takes the following form:

{\footnotesize
\bg\label{garci2}
\mathbb{F}_{14}(t) & = & \left(\mathbb{A}(\alpha) + \gamma_{1, 2} - 1, ~ \mathbb{B}(\alpha) + \gamma_{1, 2} - 1\right)\\
& = & \left(\mathbb{A}_1(\alpha) + \gamma_1 - 1, \mathbb{A}_1(\alpha) + \gamma_2 - 1, \mathbb{A}_2(\alpha) + \gamma_1 - 1, \mathbb{A}_2(\alpha) + \gamma_2 - 1, \mathbb{B}(\alpha) + \gamma_1 - 1 , \mathbb{B}(\alpha) + \gamma_2 - 1 \right), \nonumber\\
&=&\Bigg(-{5\over 3} + \alpha(t) + {\beta(t)\over 2} + 
{1\over\vert\log~\bar{g}_s\vert}~\log\left[{(2-\beta(t))^2 \over \vert\dot\beta(t)\vert\vert\log~\bar{g}_s\vert({2\over 3} - \alpha(t))}\right], \nonumber\\
&-& {2\over 3} + \alpha(t) + {1\over\vert\log~\bar{g}_s\vert}~\log\left[{2-\beta(t) \over \mp\dot\alpha(t)\vert\dot\beta(t)\vert\vert\log~\bar{g}_s\vert^2}\right], ~
-{5\over 3} + \alpha(t) + {\beta(t)\over 2} + {1\over\vert\log~\bar{g}_s\vert} ~\log\left[{2-\beta(t)\over \mp \dot\alpha(t)\vert\log~\bar{g}_s\vert}\right], \nonumber\\
&-& {2 \over 3} + \alpha(t) + {2\over \vert\log~\bar{g}_s\vert}~\log\left[{2-\beta(t) \over \vert\dot\beta(t)\vert\vert\log~\bar{g}_s\vert\sqrt{{2\over 3} -\alpha(t)}}\right],~
-{8\over 3} + \alpha(t) + \beta(t) + {2\over\vert\log~\bar{g}_s\vert}~\log\left[{2-\beta(t)\over \sqrt{{2\over 3} - \alpha(t)}}\right]\Bigg),\nonumber
\nd}
where now the dominant contributions in the absence of the log corrections are $-{5\over 3} + \alpha(t) + {\beta(t)\over 2}, -{2\over 3} + \alpha(t)$ and $ -{8\over 3} + \alpha(t) + \beta(t)$ which are all negative definite because $\alpha(t) = -\beta(t)$. These strong Riemann curvature components ${\bf R}_{mnij}$ and ${\bf R}_{0m0n}$ contribute as:
\bg\label{ishena4}
{10\over 3} + \mathbb{F}_{14}(t; b, d, d') - \alpha(t) & = & \bigg({2\over 3} + {2n_d\over \bar{g}_s^{d/3}\vert{\rm log}~\bar{g}_s\vert} + 2\vert\gamma(t)\vert + {\cal B}(\Delta; b, d'), \\
&& {2\over 3} + {\beta(t)\over 2} + {n_d\over \bar{g}_s^{d/3}\vert{\rm log}~\bar{g}_s\vert} + \vert\gamma(t)\vert + {\cal C}(\Delta; b, d'), ~{2\over 3} + \beta(t)\bigg), \nonumber\nd
in exactly the same way as \eqref{ishena3} to the quantum series  \eqref{botsuga} once we sum over $(b, d, d')$ when $\bar{g}_s$ is exponentiated by the terms in \eqref{ishena4}. The dominant contribution, at least in the simplfied context, is clearly ${2\over 3} + \beta(t)$, which is what we saw for the earlier cases too. For the generic ${\rm E}_8 \times {\rm E}_8$ case, the dominant contribution comes from:

{\footnotesize
\bg\label{sirilena3}
{\rm dom}~\mathbb{F}_{14e}(t) = -{8\over 3} + \hat\sigma_e(t) + {\hat\alpha_e(t) + \hat\beta_e(t)\over 2} + {1\over \vert\log~\bar{g}_s\vert}~\log\left[{(12 - 3\hat\alpha_e - 3\hat\beta_e)^2\over (4 - 6\hat\sigma_e(t))(16 - 6\hat\zeta_e)}\right], \nd}
from where the contribution to \eqref{botsuga} will come from ${10\over 3} + \mathbb{F}_{14e}(t) - \hat\sigma_e(t) - \hat\zeta_e(t)$. Plugging in the result from \eqref{sirilena3}, the generic scalings for the $SO(32)$ and the ${\rm E}_8 \times {\rm E}_8$ theory are precisely the ones from \eqref{2cho1bora}. 

The other two set are $\mathbb{F}_{9e}(t)$ and $\mathbb{F}_{15e}(t)$ whose functional forms appear respectively in the ninth and the fifteenth rows of {\bf Table \ref{firzathai}}. $\mathbb{F}_{9e}(t)$ which contributes to 
the $g_s$ scaling of the curvature tensor ${\bf R}_{\rho\sigma ab}$, takes the following form:
\bg\label{lenasiri4}
\mathbb{F}_{9e}(t) & = & \Bigg({8\over 3} - \hat\zeta_e(t) + \mathbb{N}_8(\hat\alpha_e, \hat\beta_e) + \bigg({1\over 3} + \gamma_{1, 2}\Big[{\hat\alpha_e(t) + \hat\beta_e(t)\over 2}\Big], ~\mathbb{F}_8(\hat\eta_e)\bigg), \nonumber\\
&& ~~{8\over 3} - \hat\zeta_e(t) + \mathbb{N}_8(\hat\alpha_e, \hat\beta_e) + \bigg({1\over 3} + \gamma_{1, 2}\Big[{\hat\alpha_e(t) + \hat\beta_e(t)\over 2}\Big], ~\mathbb{G}_8(\hat\eta_e)\bigg)\Bigg), \nd
which differs from $\mathbb{F}_{8e}(t)$ only in the choice of the argument for $\mathbb{N}_8$, where $\mathbb{N}_8(\hat\alpha_e, \hat\beta_e) = (\mathbb{A}_8(\hat\alpha_e, \hat\beta_e),\mathbb{B}_8(\hat\alpha_e, \hat\beta_e))$. In the limit we decouple the sub-dominant contribution as $\hat\eta_e(t) \to 0$, $\mathbb{F}_8(\hat\eta_e) \to 0$ and $\mathbb{G}_8(\hat\eta_e) \to {1\over 3} + \hat\eta_e(t) + \gamma_{1,2}\Big[{\hat\alpha_e + \hat\beta_e\over 2}\Big]$. Imposing further simplifications, and turning to the $SO(32)$ side, we get:

{\footnotesize
\bg\label{garci3}
\mathbb{F}_9(t) & = & \left(3 + \mathbb{A}(\beta) + \gamma_{1, 2}, ~ 3 + \mathbb{B}(\beta) + \gamma_{1, 2}\right)\\
& = & \left(\mathbb{A}_1(\beta) + \gamma_1 + 3, \mathbb{A}_1(\beta) + \gamma_2 + 3, \mathbb{A}_2(\beta) + \gamma_1 + 3, \mathbb{A}_2(\beta) + \gamma_2 + 3, \mathbb{B}(\beta) + \gamma_1 + 3, \mathbb{B}(\beta) + \gamma_2 + 3\right), \nonumber\\
&=&\Bigg( {7\over 3} + {3\beta(t)\over 2} + 
{1\over\vert\log~\bar{g}_s\vert}~\log\left[{(2-\beta(t))^2 \over \vert\dot\beta(t)\vert\vert\log~\bar{g}_s\vert({2\over 3} - \beta(t))}\right], \nonumber\\
&& {10\over 3} + \beta(t) + {2 \over\vert\log~\bar{g}_s\vert}~
\log\left[{\sqrt{2-\beta(t)} \over \vert\dot\beta(t)\vert\vert\log~\bar{g}_s\vert}\right], ~
{7\over 3} + {3\beta(t)\over 2} + {1\over\vert\log~\bar{g}_s\vert} ~\log\left[{2-\beta(t)\over \vert \dot\beta(t)\vert\vert\log~\bar{g}_s\vert}\right], \nonumber\\
&& {10\over 3} + \beta(t) + {2\over \vert\log~\bar{g}_s\vert}~\log\left[{2-\beta(t) \over \vert\dot\beta(t)\vert\vert\log~\bar{g}_s\vert\sqrt{{2\over 3} -\beta(t)}}\right],~
{4\over 3} + 2\beta(t) + {2\over\vert\log~\bar{g}_s\vert}~\log\left[{2-\beta(t)\over \sqrt{{2\over 3} - \beta(t)}}\right]\Bigg),\nonumber
\nd}
with dominant contributions in the absence of the log corrections are given by ${7\over 3} + {3\beta(t)\over 2}, {10\over 3} + \beta(t)$ and ${4\over 3} + 2\beta(t)$. These are all positive definite, and the contribution to \eqref{botsuga} appearing from the eleventh row of {\bf Table \ref{firzacut6}}, once we add back the log corrections, may be written in the following way:
\bg\label{sukkam}
-{2\over 3} + \mathbb{F}_9(t; b, d, d') - \beta(t) & = & \bigg({2\over 3} + {2n_d\over \bar{g}_s^{d/3}\vert{\rm log}~\bar{g}_s\vert} + 2\vert\gamma(t)\vert + {\cal B}(\Delta; b, d'), \\
&& {2\over 3} + {\beta(t)\over 2} + {n_d\over \bar{g}_s^{d/3}\vert{\rm log}~\bar{g}_s\vert} + \vert\gamma(t)\vert + {\cal C}(\Delta; b, d'), ~{2\over 3} + \beta(t)\bigg), \nonumber\nd
which are also positive definite and match exactly with \eqref{ishena3} and \eqref{ishena4}. The scalings in \eqref{sukkam} change slightly after we incorporate sub-dominant corrections from $\log(2-\beta(t))$ and 
$\log\big({2\over 3} - \beta(t)\big)$, which may be easily inferred from \eqref{garci3}. We will however leave them for the diligent reader and instead work out the dominant contribution for the ${\rm E}_8 \times {\rm E}_8$ case. This reproduces:

{\scriptsize
\bg\label{lenasiri5}
{\rm dom}~\mathbb{F}_{9e}(t) & = & \Bigg({4\over 3} - \hat\zeta_e(t) + (\hat\alpha_e(t), \hat\beta_e(t)) + {\hat\alpha_e(t) + \hat\beta_e(t)\over 2} + {1\over \vert\log~\bar{g}_s\vert}~\log\left[{(4\sqrt{3} -\sqrt{3}\hat\alpha_e - 
\sqrt{3}\hat\beta_e)^2\over 8 - 12\hat\sigma_e}\right], \\
&& ~~ {4\over 3} - \hat\zeta_e(t) + \hat\eta_e(t) + (\hat\alpha_e(t), \hat\beta_e(t)) + {\hat\alpha_e(t) + \hat\beta_e(t)\over 2} + {1\over \vert\log~\bar{g}_s\vert}~\log\left[{(12 -3\hat\alpha_e -3\hat\beta_e)^2\over (4 - 6(\hat\alpha_e, \hat\beta_e))(8 + 6\hat\eta_e)}\right]\Bigg), \nonumber
\nd}
from where the contribution to \eqref{botsuga} becomes $-{2\over 3} - (\hat\alpha_e(t), \hat\beta_e(t)) - (0, \hat\eta_e(t))$. Plugging everything in reproduces precisely \eqref{2cho1bora} for the $SO(32)$ and the ${\rm E}_8 \times {\rm E}_8$ cases up to the sub-dominant logarithmic corrections. 

The final set in our list is $\mathbb{F}_{15e}(t)$, and is an interesting one because it contributes to the $g_s$ scalings of both
${\bf R}_{\rho\sigma ij}$ and ${\bf R}_{0\rho 0 \sigma}$ as shown in {\bf Tables \ref{firzathai}} and {\bf \ref{firzathai3}}. From {\bf Table \ref{privsocmey}} we can easily read out its functional form as: 
\bg\label{sirilena6}
\mathbb{F}_{15e}(t) = \left({8\over 3} - \hat\zeta_e(t) + \mathbb{N}_8(\hat\alpha_e, \hat\beta_e) + \mathbb{C}_8(\hat\zeta_e), ~  
{8\over 3} - \hat\zeta_e(t) + \mathbb{N}_8(\hat\alpha_e, \hat\beta_e) + \mathbb{D}_8(\hat\zeta_e)\right), \nd
which is very similar to $\mathbb{F}_{14e}(t)$ except $\mathbb{N}_8$ appears with the argument $(\hat\alpha_e, \hat\beta_e)$. This would mean that most of the discussions we had earlier will go through unchanged, and the simplified $SO(32)$ case will take the form:

{\footnotesize
\bg\label{tososo}
\mathbb{F}_{15}(t) & = & \left(\mathbb{A}(\beta) + \gamma_{1, 2} - 1, ~ \mathbb{B}(\beta) + \gamma_{1, 2} - 1\right)\\
& = & \left(\mathbb{A}_1(\beta) + \gamma_1 - 1, \mathbb{A}_1(\beta) + \gamma_2 - 1, \mathbb{A}_2(\beta) + \gamma_1 - 1, \mathbb{A}_2(\beta) + \gamma_2 - 1, \mathbb{B}(\beta) + \gamma_1 - 1 , \mathbb{B}(\beta) + \gamma_2 - 1 \right), \nonumber\\
&=&\Bigg(-{5\over 3} + {3\beta(t)\over 2} + 
{1\over\vert\log~\bar{g}_s\vert}~\log\left[{(2-\beta(t))^2 \over \vert\dot\beta(t)\vert\vert\log~\bar{g}_s\vert({2\over 3} - \beta(t))}\right], \nonumber\\
&-& {2\over 3} + \beta(t) + {2\over\vert\log~\bar{g}_s\vert}~\log\left[{\sqrt{2-\beta(t)} \over \vert\dot\beta(t)\vert\vert\log~\bar{g}_s\vert}\right], ~
-{5\over 3} + {3\beta(t)\over 2} + {1\over\vert\log~\bar{g}_s\vert} ~\log\left[{2-\beta(t)\over \vert\dot\beta(t)\vert\vert\log~\bar{g}_s\vert}\right], \nonumber\\
&-& {2 \over 3} + \beta(t) + {2\over \vert\log~\bar{g}_s\vert}~\log\left[{2-\beta(t) \over \vert\dot\beta(t)\vert\vert\log~\bar{g}_s\vert\sqrt{{2\over 3} -\beta(t)}}\right],~
-{8\over 3} + 2\beta(t) + {2\over\vert\log~\bar{g}_s\vert}~\log\left[{2-\beta(t)\over \sqrt{{2\over 3} - \beta(t)}}\right]\Bigg),\nonumber
\nd}
with the dominant contribution coming from $-{5\over 3} + {3\beta(t)\over 2}, -{2\over 3} + \beta(t)$ and $-{8\over 3} + 2\beta(t)$, which are negative definite implying strong curvature components ${\bf R}_{\rho\sigma ij}$ and ${\bf R}_{0\rho 0 \sigma}$. These curvature components contribute as:
\bg\label{indiastoker}
{10\over 3} + \mathbb{F}_{15}(t; b, d, d') - \beta(t) & = & \bigg({2\over 3} + {2n_d\over \bar{g}_s^{d/3}\vert{\rm log}~\bar{g}_s\vert} + 2\vert\gamma(t)\vert + {\cal B}(\Delta; b, d'), \\
&& {2\over 3} + {\beta(t)\over 2} + {n_d\over \bar{g}_s^{d/3}\vert{\rm log}~\bar{g}_s\vert} + \vert\gamma(t)\vert + {\cal C}(\Delta; b, d'), ~{2\over 3} + \beta(t)\bigg), \nonumber\nd
matching exactly with \eqref{ishena3}, \eqref{ishena4} and \eqref{sukkam} and implying, as before, that these curvature coefficients only contribute as positive powers of $g_s$, despite being strong at weak $g_s$ coupling. Such a picture prevails in the most generic case with sub-dominant corrections too as may be easily seen starting with the ${\rm E}_8 \times {\rm E}_8$ case. The dominant contribution there becomes:

{\footnotesize
\bg\label{sirilena7}
{\rm dom}~\mathbb{F}_{15e}(t) = -{8\over 3} + (\hat\alpha_e(t), \hat\beta_e(t)) + {\hat\alpha_e(t) + \hat\beta_e(t)\over 2} + 
{1\over \vert\log~\bar{g}_s\vert}~\log\left[{(12 -{3}\hat\alpha_e -{3}\hat\beta_e)^2\over (4 - 6(\hat\alpha_e, \hat\beta_e))(16 - 6\hat\zeta_e)}\right]\nd}
which further contributes to \eqref{botsuga} as ${10\over 3} + \mathbb{F}_{15e}(t) - (\hat\alpha_e(t), \hat\beta_e(t)) - \hat\zeta_e(t)$. Putting everything together then gives us \eqref{2cho1bora} as the generic scalings for the $SO32)$ and the ${\rm E}_8 \times {\rm E}_8$ cases. Expectedly, all are positive definite.

\subsubsection{{Functional forms and bounds on} $\mathbb{F}_{l,le,(l+2),(l+2)e}(t)$}

\vskip.1in

\noindent The next set of terms with $l = 10, 11$ can be grouped together because of their similarities in their scalings. For example we can group $\mathbb{F}_{10e}(t)$ and $\mathbb{F}_{12e}(t)$ together, and similarly we can group $\mathbb{F}_{11e}(t)$ and $\mathbb{F}_{13e}(t)$ together. They are tied as:
\bg\label{spanmom}
\mathbb{F}_{(l +2)e}(t) = \mathbb{F}_{le}(t) - {8\over 3} - \left({4\over 3}, {4\over 3} + \hat\eta_e(t)\right) + \hat\zeta_e(t), \nd
for $l = 10, 11$. The two possibilities in the bracket denote the modes along the two toroidal directions of ${\mathbb{T}^2\over {\cal G}}$ in the metric \eqref{sinistre00}. The scaling $\mathbb{F}_{le}(t)$ can be expressed as:

{\footnotesize
\bg\label{sirilena8}
\mathbb{F}_{le}(t) &= & \Bigg({2\over 3} + \delta_{l, 10}\left(\mathbb{A}_8(\hat\sigma_e) - \hat\sigma_e(t)\right)+ \delta_{l, 11} \left(\mathbb{A}_8(\hat\alpha_e, \hat\beta_e) - (\hat\alpha_e(t), \hat\beta_e(t))\right) + \left({4\over 3}, {4\over 3} + \hat\eta_e(t)\right), \\
&& ~~ {2\over 3} + \delta_{l, 10}\left(\mathbb{B}_8(\hat\sigma_e) - \hat\sigma_e(t)\right)+ \delta_{l, 11} \left(\mathbb{B}_8(\hat\alpha_e, \hat\beta_e) - (\hat\alpha_e(t), \hat\beta_e(t))\right) + \left({4\over 3}, {4\over 3} + \hat\eta_e(t)\right)\Bigg), \nonumber\nd}
which allows us to choose between the various options. To go to the simplified picture, and in particular to the simplified version of $SO(32)$ theory, it is advisable to study the pairs of the scalings separately. In the $SO(32)$ case, the scalings are 
typically of the form $\mathbb{A}(\alpha) - \alpha(t) + n$ and $\mathbb{B}(\alpha) - \alpha(t) + n$ where $n = (2, -2)$. For the first case we take $\mathbb{F}_{10}(t)$ and $\mathbb{F}_{12}(t)$ which we express as:
\bg\label{leesob}
\mathbb{F}_{8 + 2j} & = & \left(\mathbb{A}(\alpha) - \alpha(t) + 2(3-2j), ~
\mathbb{B}(\alpha) - \alpha(t) + 2(3 - 2j)\right) \nonumber\\
&=& \Bigg({16\over 3} - 4j + {1\over \vert\log~\bar{g}_s\vert} ~\log\left[{2-\beta(t)\over \vert\dot\beta(t)\vert\vert\log~\bar{g}_s\vert({2\over 3} - \alpha(t))}\right], \\
&& {13\over 3} -4j + {\beta(t)\over 2} + {1\over\vert\log~\bar{g}_s\vert}~\log\left[{2-\beta(t)\over {2\over 3} - \alpha(t)}\right],~
{16\over 3} - 4j - {\log(\mp\dot\alpha(t) \vert\log~\bar{g}_s\vert)\over \vert\log~\bar{g}_s\vert}\Bigg), \nonumber
\nd
where we used $j = 1, 2$ to denote the scalings instead of $l$ and $l+2$. We notice that the dominant contributions in the absence of the log corrections are ${16\over 3} - 4j$ and ${13\over 3} - 4j + {\beta(t)\over 2}$. For $j = 1$, they are all positive definite, but for $j = 2$ they become negative implying strong curvature component ${\bf R}_{m0ij}$. Once we add in the dominant log corrections, the contributions to the quantum series \eqref{botsuga} from {\bf Table \ref{firzacut6}} become:

{\footnotesize
\bg\label{wickerman}
-{11\over 3} + 4j + \mathbb{F}_{8+2j}(t; b, d, d') - {\alpha(t)\over 2} = \left({2\over 3} + {\beta(t)\over 2} - {\alpha(t)\over 2}, ~ {2\over 3} - {\alpha(t)\over 2} + {n_d\over \bar{g}_s^{d/3}\vert{\rm log}~\bar{g}_s\vert} + \vert\gamma(t)\vert + {\cal C}(\Delta; b, d')\right), \nonumber\\ \nd}
which are positive definite and independent of $j$, implying finite contributions to \eqref{botsuga} from ${\bf R}_{0mab}$ as well as ${\bf R}_{m0ij}$. To go to the generic picture for both the $SO(32)$ and the ${\rm E}_8 \times {\rm E}_8$ cases, we can start by computing the dominant contribution for the latter. From \eqref{sirilena8} and \eqref{spanmom}, this becomes:

{\scriptsize
\bg\label{sirilena9}
{\rm dom}~\mathbb{F}_{(8+2j)e}(t) = {13\over 3} - 4j + {\hat\alpha_e(t) + \hat\beta_e(t)\over 4} + {1\over \vert\log~\bar{g}_s\vert}~\log\left({12 - 3\hat\alpha_e - 3\hat\beta_e\over 4 - 6\hat\sigma_e}\right) + \left(0, \hat\eta_e(t)\right)\delta_{j1} + \hat\zeta_e(t)\delta_{j2}, \nd}
where we can easily extract the two values for $j = 1$ and $j = 2$. To see how we can determine the contribution to \eqref{botsuga}, we can perform the following operation:
\bg\label{sirilena10}
-{11\over 3} + 4j + \mathbb{F}_{(8+2j)e}(t) - {\hat\zeta_e(t)\over 2} - {\hat\sigma_e(t)\over 2} - (0, \hat\eta_e(t)) \delta_{j1} - \hat\zeta_e(t) \delta_{j2}, \nd
with the two values of $j$ using \eqref{sirilena9}. This precisely gives us the same scalings as in \eqref{duitrans} for the $SO(32)$ and the ${\rm E}_8 \times {\rm E}_8$ cases irrespective of the values of $j$. 
In a similar vein the remaining simplified two components for the $SO(32)$ case are of the form 
$\mathbb{A}(\beta) - \beta(t) + n$ and $\mathbb{B}(\beta) - \beta(t) + n$ with $n = (2, -2)$. Their functional values are:
\bg\label{leeleesob}
\mathbb{F}_{9 + 2j} & = & \left(\mathbb{A}(\beta) - \beta(t) + 2(3-2j), ~
\mathbb{B}(\beta) - \beta(t) + 2(3 - 2j)\right) \nonumber\\
&=& \Bigg({16\over 3} - 4j + {1\over \vert\log~\bar{g}_s\vert} ~\log\left[{2-\beta(t)\over \vert\dot\beta(t)\vert\vert\log~\bar{g}_s\vert({2\over 3} - \beta(t))}\right], \\
&& {13\over 3} -4j + {\beta(t)\over 2} + {1\over\vert\log~\bar{g}_s\vert}~\log\left[{2-\beta(t)\over {2\over 3} - \beta(t)}\right],~
{16\over 3} - 4j - {\log(\vert\dot\beta(t)\vert \vert\log~\bar{g}_s\vert)\over \vert\log~\bar{g}_s\vert}\Bigg), \nonumber
\nd
with the dominant contributions in the absence of the log corrections being the same that we had earlier, namely ${16\over 3} - 4j$ and ${13\over 3} - 4j + {\beta(t)\over 2}$, implying now that the curvature component ${\bf R}_{\sigma 0 ij}$ becomes strong, but ${\bf R}_{0\sigma ab}$ remains weak. Once we add in the dominant log corrections, the contributions to \eqref{botsuga} from {\bf Table \ref{firzacut6}} become:

{\footnotesize
\bg\label{wickerman2}
-{11\over 3} + 4j + \mathbb{F}_{9+2j}(t; b, d, d') - {\beta(t)\over 2} = \left({2\over 3}, ~ {2\over 3} - {\beta(t)\over 2}+ {n_d\over \bar{g}_s^{d/3}\vert{\rm log}~\bar{g}_s\vert} + \vert\gamma(t)\vert + {\cal C}(\Delta; b, d')\right), \nd}
which are both positive definite and independent of $j$, but differ from 
\eqref{wickerman}. They become identical when $\alpha(t) = \beta(t)$, but such a choice will not keep the four-dimensional Newton's constant in the Heterotic side time-independent. Interestingly both the set of scalings \eqref{wickerman} and \eqref{wickerman2} match exactly with the ones from \eqref{ablumm} and \eqref{clodgarcia} respectively. The generic story is also similar to what we encountered earlier. The dominant contribution for the ${\rm E}_8 \times {\rm E}_8$ case becomes:

{\scriptsize
\bg\label{sirilena11}
{\rm dom}~\mathbb{F}_{(9+2j)e}(t) = {13\over 3} - 4j + {\hat\alpha_e(t) + \hat\beta_e(t)\over 4} + {1\over \vert\log~\bar{g}_s\vert}~\log\left({12 - 3\hat\alpha_e - 3\hat\beta_e\over 4 - 6(\hat\alpha_e, \hat\beta_e)}\right) + \left(0, \hat\eta_e(t)\right)\delta_{j1} + \hat\zeta_e(t)\delta_{j2}, \nd}
which differs from \eqref{sirilena9} only in the logarithmic factor. The contribution to \eqref{botsuga} is also very similar to \eqref{sirilena10} with only changes being $\mathbb{F}_{(8+2j)e}(t) \leftrightarrow \mathbb{F}_{(9+2j)e}(t)$ and  $\hat\sigma_e(t) \leftrightarrow (\hat\alpha_e(t), \hat\beta_e(t))$. This gives us precisely the scalings in \eqref{connorice} for the $SO(32)$ and the ${\rm E}_8 \times {\rm E}_8$ cases which are both positive definite.

\subsubsection{{Functional forms and bounds on} $\mathbb{F}^{(1)}_{16,16e}(t)$  {and} $\mathbb{F}^{(1)}_{17,17e}(t)$}

\vskip.1in

\noindent The next two set are slightly more involved compared to what we 
encountered so far because they involve second derivatives with respect to the temporal coordinate. They may typically be expressed in the following way:

{\scriptsize
\bg\label{khakee}
\mathbb{F}_{17e}(t) & = & \Bigg({8\over 3} - \hat\zeta_e(t) +\left({1\over 3} + \gamma_{1, 2}\big[{\hat\alpha_e(t) + \hat\beta_e(t)\over 2}\big], \mathbb{M}_8(\hat\eta_e)\right) + \mathbb{C}_8(\hat\zeta_e), \\
&& ~~{8\over 3} - \hat\zeta_e(t) +\left({1\over 3} + \gamma_{1, 2}\big[{\hat\alpha_e(t) + \hat\beta_e(t)\over 2}\big], \mathbb{M}_8(\hat\eta_e)\right) 
+ \mathbb{D}_8(\hat\zeta_e), \mathbb{A}_{i8}^{(ab)}(\hat\eta_e)\Bigg)\nonumber\\
\mathbb{F}_{16e}(t) &=& \Bigg({8\over 3} -\hat\zeta_e(t) + 2\otimes \mathbb{C}_8(\hat\zeta_e), ~{8\over 3} -\hat\zeta_e(t) + 2\otimes \mathbb{D}_8(\hat\zeta_e), ~{8\over 3} -\hat\zeta_e(t) +  \mathbb{C}_8(\hat\zeta_e) + \mathbb{D}_8(\hat\zeta_e), ~\mathbb{A}_{l8}^{(ij)}\Bigg), \nonumber \nd}
where $\mathbb{M}_8(\hat\eta_e) = (\mathbb{F}_8(\hat\eta_e), \mathbb{G}_8(\hat\eta_e))$ which, along with the other parameters appearing in \eqref{khakee}, are defined in {\bf Table \ref{jessrog}}.
They govern the $g_s$ scalings of 
${\bf R}_{0a0b}$ and ${\bf R}_{0i0j}$ respectively as shown in {\bf Table \ref{firzathai3}} and in \eqref{lusliber2} and \eqref{lusliber} respectively. Notice also the presence of $\mathbb{A}_{l8}^{(ab)}$ and $\mathbb{A}_{l8}^{(ij)}$ whose typical forms may be read-off from \eqref{selenzoe}. For convenience of discussion, we will split-up the set 
from \eqref{khakee} into two subsets: $\mathbb{F}^{(m)}_{(15 + l)e}(t)$ with $m = 1, l = (1, 2)$ and $m = 2, l = (1, 2)$ with the latter containing $(\mathbb{A}_{i8}^{(ij)}, \mathbb{A}_{i8}^{(ab)})$ and the former containing everything else from \eqref{khakee}.

To analyze the first subset from \eqref{khakee}, which we will do in this subsection, we will follow the same strategy as before, namely, go to the simplified $SO(32)$ case and study each and every terms. After which we will provide the answer for the generic case. In the simplified $SO(32)$ case, wherein we ignore the sub-dominant corrections, the first subset of terms from \eqref{khakee} take the form $2\gamma_{1, 2} + n$. The second subset, which we shall study in section \ref{scaruvill}, involve both $2\gamma_{1, 2} + n$ and $\gamma_{[3,...,10]} + m$ with $n = (-{14\over 3}, -{2\over 3})$ and 
$m = (-{11\over 3}, + {1\over 3})$ for all cases. 

Our starting set would be the ones from $\mathbb{F}_{16}(t)$ and $\mathbb{F}_{17}(t)$ that scales with respect to $\gamma_{1, 2}$. As before, we will express them using the parameter $j$ such that when $j = 1$ we get $\mathbb{F}_{16}(t)$ and when $j = 2$ we get $\mathbb{F}_{17}(t)$. With this in mind, the first subset of scalings become:
\bg\label{ellishky}
\mathbb{F}^{(1)}_{15 + j} &=& 2\gamma_{1, 2} + 4j - {26\over 3} \nonumber\\
& = & \left( 4j - {26\over 3} + 2\gamma_1, ~ 4j - {26\over 3} + 2\gamma_2, 
~4j - {26\over 3} + \gamma_1 + \gamma_2\right) \\
& = & \Bigg(4j - {20\over 3} + {2\over \vert\log~\bar{g}_s\vert}~\log\left[{2-\beta(t)\over \vert\dot\beta(t)\vert\vert\log~\bar{g}_s\vert}\right], ~ 4j -{26\over 3} +  \beta(t) + {2\log(2-\beta(t))\over\vert\log~\bar{g}_s\vert}, \nonumber\\
&& 4j - {23\over 3} + {\beta(t)\over 2} + {2\over\vert\log~\bar{g}_s\vert}~\log\left[{2-\beta(t)\over \sqrt{\vert\dot\beta(t)\vert\vert\log~\bar{g}_s\vert}}\right]\Bigg), \nonumber\nd
where the dominant contributions in the absence of the log corrections come from $4j - {20\over 3}, 4j - {26\over 3} + \beta(t)$ and $4j - {23\over 3} + {\beta(t)\over 2}$. For $j = 1$, all these are negative definite, whereas for $j = 2$, only one of the three is negative. Any negative entry means strong curvature components, and so both 
${\bf R}_{0i0j}$ and ${\bf R}_{0a0b}$ are strong and scale as negative powers of $g_s$. However the actual contributions to the quantum series  \eqref{botsuga} from {\bf Table \ref{firzacut9}}, once we add in the dominant log corrections, become:
\bg\label{sarkylie}
\mathbb{F}^{(1)}_{15 + j}(t; b, d, d') - 4j + {28\over 3} 
& = & \bigg({2\over 3} + {2n_d\over \bar{g}_s^{d/3}\vert{\rm log}~\bar{g}_s\vert} + 2\vert\gamma(t)\vert + {\cal B}(\Delta; b, d'), \\
&& {2\over 3} + {\beta(t)\over 2} + {n_d\over \bar{g}_s^{d/3}\vert{\rm log}~\bar{g}_s\vert} + \vert\gamma(t)\vert + {\cal C}(\Delta; b, d'), ~{2\over 3} + \beta(t)\bigg), \nonumber\nd
which match exactly with \eqref{indiastoker} and are all positive definite and independent of $j$, implying that the contributions are weak at late time as they all scale as positive powers of $g_s$. The sub-dominant log terms then add small corrections to \eqref{sarkylie}.

What happens in the generic case for the ${\rm E}_8 \times {\rm E}_8$ and $SO(32)$ sides? The story here is naturally involved, but we can use all the inputs from the simplified $SO(32)$ case discussed above to determine the dominant contributions from the first subset. For $\mathbb{F}^{(1)}_{17e}(t)$ the dominant contribution comes from:

{\footnotesize
\bg\label{manucox}
{\rm dom}~\mathbb{F}^{(1)}_{17e}(t) &= & 
-{2\over 3} + (0, \hat\eta_e(t)) + {\hat\alpha_e(t) + \hat\beta_e(t)\over 2} + 
{1\over \vert\log~\bar{g}_s\vert}~\log\bigg[{(12 -3\hat\alpha_e - 3\hat\beta_e)^2\over (8 + 6(0,\hat\eta_e))(16 - 6 \hat\zeta_e)}\bigg],\nd}
where the two choices come from $(0, \hat\eta_e(t))$. The contribution to \eqref{botsuga} will now appear from the operation ${4\over 3} + \mathbb{F}_{17e}^{(1)}(t) - \hat\zeta_e(t) - (0, \hat\eta_e(t))$. This gives us \eqref{2cho1bora} for the $SO(32)$ and the ${\rm E}_8 \times {\rm E}_8$ cases respectively. On the other hand, the dominant contribution from $\mathbb{F}_{16e}^{(1)}(t)$ becomes:

{\footnotesize
\bg\label{manucox2}
{\rm dom}~\mathbb{F}^{(1)}_{16e}(t) &= & 
-{14\over 3} + \hat\zeta_e(t) + {\hat\alpha_e(t) + \hat\beta_e(t)\over 2} + 
{2 \over \vert\log~\bar{g}_s\vert}~\log\bigg[{12 -{3}\hat\alpha_e - {3}\hat\beta_e\over 16 - 6\hat\zeta_e}\bigg], \nd}
which when added to ${16\over 3} - 2\hat\zeta_e(t)$ again reproduces \eqref{2cho1bora} as the contributions from the $SO(32)$ and the ${\rm E}_8 \times {\rm E}_8$ cases respectively.

\subsubsection{Functional forms and bounds on $\gamma$ and EFT criterion \label{sec448}}

Our next set would be the ones from $\mathbb{F}_{16}(t)$ and $\mathbb{F}_{17}(t)$ that scales with respect to $\gamma_{1, 2}$ with $\gamma_{[3,...,10]}$ defined in \eqref{ishena2}. These incorporate the action of the second derivative of $g_s$ and are crucial for the consistency of the Schwinger-Dyson's equations that we alluded to earlier and will discuss in more details later. 

As defined in \eqref{lovsik}, this would necessarily involve both first and the second derivative of $\bar{g}_s$ with respect to the conformal time, implying that the first derivative of $\gamma$ should enter the story. Recall that we have so-far defined ${\partial \bar{g}_s\over \partial t}$ in two different ways, namely as $\bar{g}_s^\gamma$ and as $\bar{g}_s^{\gamma_{1, 2}}$. We expect the former to be more complete representation of ${\partial \bar{g}_s\over \partial t}$ as it appears from say \eqref{stan22}, because the latter, as seen from say {\bf Table \ref{jessrog}}, is defined in terms of ${\partial \bar{g}_s\over \partial t}$. (In other words, $\gamma_{1, 2}$ should be expressible in terms of $\gamma$.) Additionally, again from \eqref{stan22}, we expect $\gamma$ to be at least bounded from below by $d_{\rm max}$. Can we represent $\gamma$ as a trans-series, {\it i.e.} as $\gamma(t(g_s))$, much like how we defined the parameters $\beta_e(t(g_s)), \alpha_e(t(g_s))$ for the $SO(32)$ case and $(\hat\sigma_e(t(g_s)), \hat\alpha_e(t(g_s)), \hat\beta_e(t(g_s))$ for the ${\rm E}_8 \times {\rm E}_8$ case? In this section, let us first discuss this, after which we will see its implication on $\mathbb{F}_{15+j}^{(2)}$ in section \ref{scaruvill}. Therefore to start, let us define:
\bg\label{mcelhon}
\vert\gamma(t(g_s))\vert = d_{\rm max} + \Delta \sum_{(b, d) \ge 0} \sum_{l = 0}^\infty h^{(l)}_{b} 
\left({g_s\over {\rm H}(y) {\rm H}({\bf x})}\right)^{{b\over 3} + {2l\over 3}} {\rm exp}\left(-\sum_d {n_d(b, l)\over \bar{g}_s^{d/3}}\right), \nd
which is the familiar trans-series form with $0 < \Delta << 1$ and constant coefficients $h^{(l)}_{b}$, except that we do not specify the bounds on $h^{(l)}_b$ right away. The coefficients $h^{(l)}_{b}$ may be derived by expanding the RHS of \eqref{stan22} as a Taylor series (for $l > 0$ we will have to insert \eqref{ryanfan} in \eqref{stan22} to equate the $h_{b}^{(l)}$ with $c_{b}^{(l)}$). Using \eqref{mcelhon}, the second derivative of $\bar{g}_s$ with respect to the conformal time takes the following form:

{\footnotesize
\bg\label{cataloo}
{\partial^2\bar{g}_s\over \partial t^2} & = & \partial_0\vert\gamma\vert~\log~\bar{g}_s~\bar{g}_s^{|\gamma|} + \vert\gamma\vert~\bar{g}_s^{2|\gamma|-1} \\
& = & \left[ 1 + 2\prod_{d'}\sum_{b'} \sum_{l' = 0}^\infty h^{(l')}_{b'} 
\bar{g}_s^{{b'\over 3} + {2l'\over 3} + {n_{d'}(b', l')\over \bar{g}_s^{d'/3} \vert\log~\bar{g}_s\vert}}\left(\bar{g}_s^\Delta - 1\right)\right]~\bar{g}_s^{2d_{\rm max}}\nonumber\\
&\times & \left[\prod_{d''}\sum_{b, d} \sum_{l = 0}^\infty h^{(l)}_{b} 
\bar{g}_s^{{b\over 3} + {2l\over 3} + {n_{d''}(b, l)\over \bar{g}_s^{d''/3} \vert\log~\bar{g}_s\vert} - 1}  
 \left(\Delta + {1\over 3}\bigg(b + 2l + {n_d(b, l) d\over \bar{g}_s^{d/3}}\bigg)\left(\bar{g}_s^\Delta - 1\right)\right) + 
 \bar{g}_s^{-1 - {\log~d_{\rm max}\over \vert\log~\bar{g}_s\vert}}\right],
\nonumber\nd}
where, if we take $b \ge 0$ and $0 \le (d,d', d'') \le (\tilde{d}_{\rm max}, \tilde{d}_{\rm max}, \tilde{d}_{\rm max})$, then the only negative exponent will come from the inverse $\bar{g}_s$ piece. (We do not relate $d_{\rm max}$ with $\tilde{d}_{\rm max}$ as yet.) This is much like what we encountered earlier, but now the story is a bit more involved because \eqref{cataloo} is a small piece in a much bigger structure in \eqref{lovsik}. Will this create an issue now?

To see this we need to express \eqref{cataloo} in terms of the dominant scalings. This would appear from the $l = 0$ part of the series and we will following the simplifying strategy adopted in writing \eqref{ivbresbest} to express the second derivative. What we now have is:
\bg\label{laelajohn}
{\partial^2\bar{g}_s\over \partial t^2} & \to & \left({g_s\over {\rm H}(y){\rm H}_o({\bf x})}\right)^{-1+ 2d_{\rm max}- {\log~d_{\rm max}\over \vert\log~\bar{g}_s\vert} }\\
&\oplus &\prod_{d'}\sum_{b,d} \left({g_s\over {\rm H}(y){\rm H}({\bf x})}\right)^{-1+ 2d_{\rm max}+{n_{d'}(b, 0)\over \bar{g}_s^{d'/3} \vert\log~\bar{g}_s\vert} + {\cal C}(\Delta; b, d)}\nonumber\\
&\oplus &\prod_{d', d''}\sum_{b, b', d}\left({g_s\over {\rm H}(y){\rm H}({\bf x})}\right)^{-1+ 2d_{\rm max}+{n_{d'}(b, 0)\over \bar{g}_s^{d'/3} \vert\log~\bar{g}_s\vert} +{n_{d''}(b', 0)\over \bar{g}_s^{d''/3} \vert\log~\bar{g}_s\vert}+ {\cal C}(\Delta;  b' + b, d)}\nonumber\\
&\oplus &\prod_{d', d''}\sum_{b, b', d}\left({g_s\over {\rm H}(y){\rm H}({\bf x})}\right)^{-1+ 2d_{\rm max}+{n_{d'}(b, 0)\over \bar{g}_s^{d'/3} \vert\log~\bar{g}_s\vert} +{n_{d''}(b', 0)\over \bar{g}_s^{d''/3} \vert\log~\bar{g}_s\vert}+ {\cal C}(\Delta; 3\Delta + b' + b, d)}, \nonumber \nd
where ${\cal C}(\Delta; b, d)$ is defined in \eqref{standudh}. In the second line we have $b \to b + b'$ and in the third line we have 
$b \to 3\Delta + b + b'$ inserted in the definition of ${\cal C}(\Delta; b, d)$ from \eqref{standudh}. The scaling representation in \eqref{laelajohn} doesn't show the coefficients, so before summing over $(b, b', d)$ we need to carefully insert them. Worrisome features like (a) the negativity of some terms in ${\cal C}(\Delta; b, d)$, and (b) the additional negative signs coming from the $-1$ exponents in \eqref{laelajohn}, are now alleviated. This justifies the inclusion of the constraint \eqref{clodkalua} in the zero instanton sector, {\it i.e.} for the $n_d = 0$ case.
 However, \eqref{laelajohn} is a small part of a bigger structure in \eqref{lovsik}, so we still need to work out the full scalings to see if there are any potential issues remaining.

The issues could appear when we try to evaluate the scalings of the various terms in the second relation in \eqref{lovsik}. We can express the scalings, without worrying too much about the coefficients, in the following way:
\bg\label{leylmilen}
\vert \ddot\beta\vert \vert \log~\bar{g}_s\vert & \to & \prod_{d'}\sum_{b, d}
\left({g_s\over {\rm H}(y){\rm H}({\bf x})}\right)^{-1 +{n_{d'}(b, 0)\over \bar{g}_s^{d'/3} \vert\log~\bar{g}_s\vert} + {\cal C}(\Delta; b, d)}\cdot {\partial^2\bar{g}_s\over \partial t^2} \nonumber\\
& \oplus & \prod_{d'}\sum_{b, d}
\left({g_s\over {\rm H}(y){\rm H}({\bf x})}\right)^{- 2 +{n_{d'}(b, 0)\over \bar{g}_s^{d'/3} \vert\log~\bar{g}_s\vert} + {\cal E}(\Delta; b, d)}\cdot \left({\partial\bar{g}_s\over \partial t}\right)^2, \nd
where one will have to introduce the scalings of ${\partial^2\bar{g}_s\over \partial t^2}$ from \eqref{laelajohn}, and 
$\left({\partial\bar{g}_s\over \partial t}\right)^2$ from $\bar{g}_s^{2\vert\gamma\vert}$. We have also defined ${\cal E}(\Delta; b, d)$, and ${\cal F}(\Delta; b, d)$ for later use, as:
\bg\label{chinkpet}
&&{\cal E}(\Delta; b, d) = \left({b\over 3}, \Delta + {b\over 3}, {b-d\over 3}, \Delta + {b - d\over 3}, {b-2d\over 3}, \Delta + {b-2d\over 3}\right)\\
&&{\cal F}(\Delta; b, d) = \left({2b\over 3}, \Delta + {2b\over 3}, {2(b-d)\over 3}, \Delta + {2(b - d)\over 3}, {2b-d\over 3}, \Delta + {2b-d\over 3}\right), \nonumber \nd
where we should view them as providing the upper bounds, much like how we defined the bound in \eqref{ivbresbest}. (Again,
we see from the definition of ${\cal E}(\Delta; b, d)$ that $b = d =0$ does create issues as anticipated earlier. However as we shall justify below, choices like $b > d$ or $b > 2d$ are not required to be imposed for the validity of the underlying EFT.) This being the case, the signs of the coefficients are not relevant and the scalings in \eqref{leylmilen} take the following form:

{\scriptsize
\bg\label{samthi}
\vert \ddot\beta\vert \vert \log~\bar{g}_s\vert 
& \to & \prod_{d'}\sum_{b, d}
\left({g_s\over {\rm H}(y){\rm H}({\bf x})}\right)^{-2 + 2d_{\rm max} +{n_{d'}\over \bar{g}_s^{d''/3} \vert\log~\bar{g}_s\vert} + {\cal E}(\Delta; b, d)}\\
& \oplus &\prod_{d'}\sum_{b, d}
\left({g_s\over {\rm H}(y){\rm H}({\bf x})}\right)^{-2+2d_{\rm max}+{n_{d'}(b, 0)\over \bar{g}_s^{d'/3} \vert\log~\bar{g}_s\vert}  - {\log~d_{\rm max}\over \vert\log~\bar{g}_s\vert}+ {\cal C}(\Delta; b, d)}\nonumber\\
& \oplus & \prod_{d', d''}\sum_{b, b', d}
\left({g_s\over {\rm H}(y){\rm H}({\bf x})}\right)^{-2+ 2d_{\rm max}+{b'\over 3} + {n_{d'}\over \bar{g}_s^{d'/3} \vert\log~\bar{g}_s\vert} +{n_{d''}\over \bar{g}_s^{d''/3} \vert\log~\bar{g}_s\vert}+ {\cal E}(\Delta; b, d)}\nonumber\\
& \oplus & \prod_{d', d''}\sum_{b, b', d}
\left({g_s\over {\rm H}(y){\rm H}({\bf x})}\right)^{-2+ 2d_{\rm max}+\Delta + {b'\over 3} + {n_{d'}\over \bar{g}_s^{d'/3} \vert\log~\bar{g}_s\vert} +{n_{d''}\over \bar{g}_s^{d''/3} \vert\log~\bar{g}_s\vert}+ {\cal E}(\Delta; b, d)}\nonumber\\
& \oplus & \prod_{d'', d'''}\sum_{b, b', d, d'}
\left({g_s\over {\rm H}(y){\rm H}({\bf x})}\right)^{-2+ 2d_{\rm max}+{n_{d''}\over \bar{g}_s^{d''/3} \vert\log~\bar{g}_s\vert} +{n_{d'''}\over \bar{g}_s^{d'''/3} \vert\log~\bar{g}_s\vert}+ {\cal C}(\Delta; b, d)+ {\cal C}(\Delta; b', d')}\nonumber\\
& \oplus & \prod_{d'', d''', d''''}\sum_{b, b', d, d'}
\left({g_s\over {\rm H}(y){\rm H}({\bf x})}\right)^{-2+ 2d_{\rm max}+{n_{d''}\over \bar{g}_s^{d''/3} \vert\log~\bar{g}_s\vert} +{n_{d'''}\over \bar{g}_s^{d'''/3} \vert\log~\bar{g}_s\vert}+{n_{d''''}\over \bar{g}_s^{d''''/3} \vert\log~\bar{g}_s\vert}+ {\cal C}_{12}(\Delta; b, b', d, d')}\nonumber\\
& \oplus & \prod_{d'', d''', d''''}\sum_{b, b', d, d'}
\left({g_s\over {\rm H}(y){\rm H}({\bf x})}\right)^{-2+ 2d_{\rm max}+{n_{d''}\over \bar{g}_s^{d''/3} \vert\log~\bar{g}_s\vert} +{n_{d'''}\over \bar{g}_s^{d'''/3} \vert\log~\bar{g}_s\vert}+{n_{d''''}\over \bar{g}_s^{d''''/3} \vert\log~\bar{g}_s\vert} + {\cal C}_{34}(\Delta; b, b', d, d')}\nonumber, \nd}
where the overall signs of ${\cal E}(\Delta; b, d)$, ${\cal C}_{12}(\Delta; b, b', d, d') $ and ${\cal C}_{34}(\Delta; b, b', d, d')$ 
in the exponents of \eqref{samthi}, as well as in \eqref{laelajohn} earlier, are now taken care of by 
the presence of $2d_{\rm max}$.
For convenience, we have also defined:
\bg\label{tapsipan}
&& {\cal C}_{12}(\Delta; b, b', d, d') \equiv {\cal C}(\Delta; b, d)+ {\cal C}(\Delta; b'+ b'', d') \nonumber\\
&& {\cal C}_{34}(\Delta; b, b', d, d') \equiv {\cal C}(\Delta; b, d)+ {\cal C}(\Delta; 3\Delta + b'+ b'', d'), \nd
where ${\cal C}(\Delta; b, d)$ is as in \eqref{standudh}.
As usual, the summing over $(b, d)$ and $(b'', d'')$ et cetera are done after we have inserted the appropriate coefficients. These are easily taken care of from what we discussed earlier so we will not elaborate further on this. Instead we define one more set of scalings in the following way:
\bg\label{delgad}
\vert \dot\beta\vert^2 \vert \log~\bar{g}_s\vert 
& \to & \prod_{d'}\sum_{b, d}
\left({g_s\over {\rm H}(y){\rm H}({\bf x})}\right)^{-2 +{n_{d'}\over \bar{g}_s^{d'/3} \vert\log~\bar{g}_s\vert} + 2\vert\gamma\vert + {\vert\log~\Delta\vert \over \vert\log~\bar{g}_s\vert} + {\cal F}(\Delta; b, d)}, \nd
where ${\cal F}(\Delta; b, d)$ is defined in \eqref{chinkpet}. The $-2$ factor is not a concern when it enters \eqref{botsuga}, and the negative signs in
${\cal F}(\Delta; b, d)$ are again taken care of $2\vert\gamma\vert$ factor. Note that the above is heavily suppressed by both the non-perturbative corrections, from $n_d> 0$ ({\it i.e.} when we are away from the zero instanton sector) and from $\gamma$ in \eqref{mcelhon} as well as by $\Delta$. 

Our analysis above, and from the explicit form for $\gamma$ in \eqref{mcelhon}, suggests that the EFT criterion governed by ${\partial \bar{g}_s\over \partial t} \propto \bar{g}_s^{\vert\gamma\vert}$ \cite{coherbeta2} is controlled by $d_{\rm max}$ {\it i.e.} the highest power of $\bar{g}_s$ appearing in the non-perturbative corrections to $\beta(t)$ 
in \eqref{ryanfan} for the $SO(32)$ case, or the highest power of $\bar{g}_s$ appearing in either $\hat\alpha(t)$ or $\hat\beta(t)$ in \eqref{horseman} for the ${\rm E}_8 \times {\rm E}_8$ case, {\it i.e.} 
$d_{\rm max} = {\rm dom}(d', d)$. On the other hand, if the highest powers are identical and are equal to $d_{\rm max}$, then it is this $d_{\rm max}$ that will govern the EFT criterion. Additionally, comparing \eqref{stan22} with \eqref{mcelhon}, we see that $\Delta \cdot h_b^{(l)}$ can be a finite integer and is not required to be arbitrarily small.

\subsubsection{{Functional forms and bounds on} $\mathbb{F}^{(2)}_{16,16e}(t)$  {and} $\mathbb{F}^{(2)}_{17,17e}(t)$ \label{scaruvill}}

With \eqref{samthi} and \eqref{delgad}, as well as \eqref{ivbresbest} earlier, we are almost ready to evaluate the remaining contributions from \eqref{khakee}
to the quantum series \eqref{botsuga}. One other definition that will considerably facilitate our analysis is the following:

{\scriptsize
\bg\label{nillchoke7}
\Sigma(a_1, a_2, ..., a_{10}) &\equiv & 2d_{\rm max} + {a_1b'\over 3} + 
{a_2n_{d}\over \bar{g}_s^{d/3} \vert\log~\bar{g}_s\vert} +{a_3n_{d'}\over \bar{g}_s^{d'/3} \vert\log~\bar{g}_s\vert}+{a_4 n_{d''}\over \bar{g}_s^{d''/3} \vert\log~\bar{g}_s\vert} -{a_5\log~d_{\rm max}\over \vert\log~\bar{g}_s\vert}\\ 
&+ & a_6 {\cal C}(\Delta; b, d)+ a_7 {\cal C}(\Delta; b', d') + a_8 {\cal C}(\Delta; b'+ b'', d')
+ a_9 {\cal C}(\Delta; 3\Delta + b'+ b'', d')+ a_{10}{\cal E}(\Delta; b, d), \nonumber \nd}
where $a_i$ can be 1 or 0. As an example, the first, second and the third exponents in \eqref{samthi} may be expressed as ${-2 + \Sigma(0, 1, 0, ..., 0, 1)}, -2 + \Sigma(0, 1, 0, 0, 1, 0, ..., 0)$ and $-2 + \Sigma(1, 1, 1, 0, ..., 0, 1)$ respectively. It is also clear that $\Sigma(a_1, ..., a_{10}) > 0$ because of the $2 d_{\rm max}$ factor. Question now is whether the remaining factors from \eqref{khakee}, namely $\mathbb{A}_{l8}^{(ab)}$ and $\mathbb{A}_{l8}^{(ij)}$, cause any concerns. To analyze this, 
we will start by $\mathbb{A}_{l8}^{(ab)}$ for the generic ${\rm E}_8 \times {\rm E}_8$ case which may be represented by the following set:

{\footnotesize
\bg\label{soplock1}
\mathbb{A}_{l8}^{(ab)} & = &\bigg\{
{4\over 3} + (0, \hat\eta_e(t)) - {2\log(\mp (0, \dot{\hat\eta}_{e}(t))\vert\log~\bar{g}_s\vert)\over \vert\log~\bar{g}_s\vert}; \\
&&~~{{4\over 3} + (0, \hat\eta_e(t))  - {\log(\mp (0, \ddot{\hat\eta}_{e}(t))\vert\log~\bar{g}_s\vert)\over \vert\log~\bar{g}_s\vert}}; \nonumber\\
&& ~~{1\over 3} + \gamma_{1, 2}\Big[{\hat\alpha_e(t) + \hat\beta_e(t)\over 2}\Big] + (0, \hat\eta_e(t))- {\log(\mp (0, \dot{\hat\eta}_{e}(t))\vert\log~\bar{g}_s\vert)\over\vert\log~\bar{g}_s\vert}; \nonumber\\
&& ~~{{1\over 3} + \gamma_{[3, ...., 10]}\Big[{\hat\alpha_e(t) + \hat\beta_e(t)\over 2}\Big] + (0, \hat\eta_e(t)) - {\log\left(\vert {4\over 3} + (0, \hat\eta_e(t))\vert\right) \over \vert\log~\bar{g}_s\vert}};  \nonumber\\
& & ~~{{1\over 3} + \gamma_{1, 2}\Big[{\hat\alpha_e(t) + \hat\beta_e(t)\over 2}\Big] +(0, \hat\eta_{e}(t)) - {\log\left(\mp (0, \dot{\hat\eta}_{e}(t))\vert\log~\bar{g}_s\vert\right)\over \vert\log~\bar{g}_s\vert} - {\log\left(\vert {4\over 3} + (0, \hat\eta_{e}(t))\vert\right)\over \vert\log~\bar{g}_s\vert}}; \nonumber\\
&& -{{2\over 3} + 2\gamma_{1, 2}\Big[{\hat\alpha_e(t) + \hat\beta_e(t)\over 2}\Big] +(0, \hat\eta_{e}(t)) - {\log\left(\vert {4\over 3} + (0, \hat\eta_{e}(t))\vert\right) \over \vert\log~\bar{g}_s\vert} - {\log\left(\vert {1\over 3} + (0, \hat\eta_{e}(t))\vert\right)\over \vert\log~\bar{g}_s\vert}}\bigg\}, \nonumber \nd}
with $l = 1, .., 6$ which one could identify from the RHS of the above expressions. The ordering of the terms follow the ordering originally presented in \eqref{selenzoe}. In a similar vein, the set of $\mathbb{A}_{l8}^{(ij)}$ may be represented by the following:
\bg\label{soplock2}
\mathbb{A}_{l8}^{(ij)} & = &\bigg\{
-{8\over 3} + \hat\zeta_e(t) - {2\log(\mp \dot{\hat\zeta}_{e}(t)\vert\log~\bar{g}_s\vert)\over \vert\log~\bar{g}_s\vert}; \\
&&~~~{-{8\over 3} + \hat\zeta_e(t)  - {\log(\mp \ddot{\hat\zeta}_{e}(t)\vert\log~\bar{g}_s\vert)\over \vert\log~\bar{g}_s\vert}}; \nonumber\\
&& ~~{-{11\over 3} + \gamma_{1, 2}\Big[{\hat\alpha_e(t) + \hat\beta_e(t)\over 2}\Big] + \hat\zeta_e(t)- {\log(\mp \dot{\hat\zeta}_{e}(t)\vert\log~\bar{g}_s\vert)\over\vert\log~\bar{g}_s\vert}}; \nonumber\\
&& ~~{-{11\over 3} + \gamma_{[3, ...., 10]}\Big[{\hat\alpha_e(t) + \hat\beta_e(t)\over 2}\Big] + \hat\zeta_e(t) - {\log\left(\vert {8\over 3} - \hat\zeta_e(t)\vert\right) \over \vert\log~\bar{g}_s\vert}};  \nonumber\\
&& ~~{-{11\over 3} + \gamma_{1, 2}\Big[{\hat\alpha_e(t) + \hat\beta_e(t)\over 2}\Big] + \hat\zeta_e(t) - {\log\left(\mp \dot{\hat\zeta}_{e}(t)\vert\log~\bar{g}_s\vert\right)\over \vert\log~\bar{g}_s\vert} - {\log\left(\vert {8\over 3}-\hat\zeta_{e}(t)\vert\right)\over \vert\log~\bar{g}_s\vert}}; \nonumber\\
&& ~-{{14\over 3} + 2\gamma_{1, 2}\Big[{\hat\alpha_e(t) + \hat\beta_e(t)\over 2}\Big]  + \hat\zeta_e(t) - {\log\left(\vert {8\over 3}- \hat\zeta_{e}(t)\vert\right) \over \vert\log~\bar{g}_s\vert} - {\log\left(\vert {11\over 3}- \hat\zeta_{e}(t)\vert\right)\over \vert\log~\bar{g}_s\vert}}\bigg\}, \nonumber \nd
where the complications come from $\gamma_{[3, ...., 10]}\Big[{\hat\alpha_e(t) + \hat\beta_e(t)\over 2}\Big]$ $-$ appearing in both \eqref{soplock1} and \eqref{soplock2} $-$ because of the presence of eight different terms. This may be read-off from \eqref{ishena2} by making the simple substitution of $\beta(t) \to {\hat\alpha_e(t) + \hat\beta_e(t)\over 2}$. In fact we can use the reverse transformation to go to the $SO(32)$ case, as we have done before. For the simplified $SO(32)$ case, the functional forms that we are looking for now are the following:
\bg\label{sakyliera}
\mathbb{F}^{(2)}_{15 + j} & = & \gamma_{[3,...,10]} + 4j - {23\over 3}\nonumber\\ 
&=& \Big(\gamma_3 + 4j - {23\over 3}, ~  \gamma_4 + 4j - {23\over 3}, ~ \gamma_5 + 4j - {23\over 3}, ~ 
\gamma_6 + 4j - {23\over 3},\nonumber\\
&& ~ \gamma_7 + 4j - {23\over 3}, ~ 
\gamma_8 + 4j - {23\over 3}, ~ \gamma_9 + 4j - {23\over 3}, ~ 
\gamma_{10} + 4j - {23\over 3}\Big)\\
&=& \Bigg({4j} - {23\over 3} + {\beta(t)\over 2} + {2\over \vert\log~\bar{g}_s\vert}~\log\left[{2-\beta(t)\over \sqrt{\vert\dot\beta(t)\vert\vert\log~\bar{g}_s\vert}}\right], \nonumber\\
&& 4j -{26\over 3} + \beta(t) + {2\log(2-\beta(t))\over\vert\log~\bar{g}_s\vert},~
4j -{26\over 3} + \beta(t) + {\log(2-\beta(t))\over\vert\log~\bar{g}_s\vert}, 
\nonumber\\
&& 4j - {20\over 3} + {1\over \vert\log~\bar{g}_s\vert}~\log\left[{2-\beta(t)\over \mp \ddot{\beta}(t)\vert\log~\bar{g}_s\vert}\right], ~
4j - {20\over 3} + {1\over \vert\log~\bar{g}_s\vert}~\log\left[{(2-\beta(t))^2\over\dot{\beta}^2(t)\vert\log~\bar{g}_s\vert}\right], 
\nonumber\\
&& 4j - {20\over 3} + {2\over \vert\log~\bar{g}_s\vert}~\log\left[{2-\beta(t)\over \vert\dot\beta(t)\vert\vert\log~\bar{g}_s\vert}\right], ~
4j - {23\over 3} + {\beta(t)\over 2} + \log\left[{(2-\beta(t))^2\over \vert\dot\beta(t)\vert\vert\log~\bar{g}_s\vert}\right], \nonumber\\
&&  4j -{23\over 3} + {\beta(t)\over 2} + {1\over\vert\log~\bar{g}_s\vert}~\log\left[{(2-\beta(t))^2\over \vert\dot\beta(t)\vert}\right]\Bigg), \nonumber
\nd
with the dominant contributions in the absence of the log corrections now coming from $4j - {23\over 3} + {\beta(t)\over 2}, 4j - {26\over 3}$ and $4j - {20\over 3}$, which are all negative definite for $j = 1$, and only one negative for $j = 2$. Nevertheless, these imply that the curvature tensors ${\bf R}_{0i0j}$ and ${\bf R}_{0a0b}$ are both strong at late time. In the presence of the log corrections the story is now more involved because of the presence of $\ddot\beta(t)$. Including them, 
$\mathbb{F}^{(2)}_{15 + j}$ from \eqref{sakyliera} takes the following form:

{\scriptsize
\bg\label{eyez69}
\mathbb{F}^{(2)}_{15 + j}(\xi) & = & \bigg(4j - {26\over 3} + {\beta(t)\over 2} +  {n_{d}\over \bar{g}_s^{d/3} \vert\log~\bar{g}_s\vert} + \vert\gamma\vert + {\cal C}(\Delta; b, d), ~
4j - {26\over 3} + \beta(t), \\
&& 4j - {26\over 3} + \Sigma(a_1, .., a_{10}),~  
4j - {26\over 3} +  {n_{d}\over \bar{g}_s^{d/3} \vert\log~\bar{g}_s\vert} + 2\vert\gamma\vert + {\vert\log~\Delta\vert\over \vert\log~\bar{g}_s\vert} + {\cal F}(\Delta; b, d), \nonumber\\ 
&& 4j - {26\over 3} +  {2n_{d}\over \bar{g}_s^{d/3} \vert\log~\bar{g}_s\vert} + 2\vert\gamma\vert  + {\cal B}(\Delta; b, d),~
4j - {26\over 3} + {\beta(t)\over 2} +  {n_{d}\over \bar{g}_s^{d/3} \vert\log~\bar{g}_s\vert} + \vert\gamma\vert + {\vert\log~\Delta\vert\over \vert\log~\bar{g}_s\vert} + {\cal C}(0; b, d)\bigg), \nonumber \nd}
where $\xi \equiv (t; b, b', b''; d, d', d'')$; ${\cal C}(\Delta; b, d)$ and ${\cal B}(\Delta; b, d)$ are defined in \eqref{standudh}; ${\cal F}(\Delta; b, d)$ in \eqref{chinkpet}; and 
$\Sigma(a_1, .., a_{10})$ in \eqref{nillchoke7} as shown in {\bf Table \ref{munleena}}. Note that the $a_i \in (1, 0)$ parameters in $\Sigma(a_1, .., a_{10})$ take all the values allowed from 
\eqref{samthi}. From {\bf Table \ref{firzacut9}}, and plugging in the scalings from \eqref{eyez69}, we see that the contribution takes the following form:

{\scriptsize
\bg\label{sarkylie2}
\mathbb{F}^{(2)}_{15 + j}(\xi) - 4j + {28\over 3} & = & \bigg({2\over 3} + {\beta(t)\over 2} +  {n_{d}\over \bar{g}_s^{d/3} \vert\log~\bar{g}_s\vert} + \vert\gamma\vert + {\cal C}(\Delta; b, d), ~
{2\over 3} + \beta(t), \\
&& {2\over 3} + \Sigma(a_1, .., a_{10}),~  
{2\over 3} +  {n_{d}\over \bar{g}_s^{d/3} \vert\log~\bar{g}_s\vert} + 2\vert\gamma\vert + {\vert\log~\Delta\vert\over \vert\log~\bar{g}_s\vert} + {\cal F}(\Delta; b, d), \nonumber\\ 
&& {2\over 3} +  {2n_{d}\over \bar{g}_s^{d/3} \vert\log~\bar{g}_s\vert} + 2\vert\gamma\vert  + {\cal B}(\Delta; b, d),~
{2\over 3} + {\beta(t)\over 2} +  {n_{d}\over \bar{g}_s^{d/3} \vert\log~\bar{g}_s\vert} + \vert\gamma\vert + {\vert\log~\Delta\vert\over \vert\log~\bar{g}_s\vert} + {\cal C}(0; b, d)\bigg), \nonumber \nd}
where three of the terms match 
exactly with the ones from \eqref{sarkylie}, however now there are many more terms. Interestingly all the terms except ${2\over 3} + \beta(t)$
are heavily suppressed by the non-perturbative contributions (including additional suppressions by $\Delta$ for some of the terms), once we raise them to powers of $\bar{g}_s$, implying again that the dominant contribution only comes from ${2\over 3} + \beta(t)$. 
The additional sub-dominant log terms add further small corrections to \eqref{sarkylie} and \eqref{sarkylie2}. Note that, despite certain similarities between \eqref{sarkylie} and \eqref{sarkylie2}, the sub-dominant log corrections to each set respectively are very different.

The story for the generic ${\rm E}_8 \times {\rm E}_8$ case may also be easily worked out from \eqref{soplock1} and \eqref{soplock2}, noting that $\mathbb{F}_{16e}^{(2)} = \{\mathbb{A}_{l8}^{(ij)}\}$ and $\mathbb{F}_{17e}^{(2)} = \{\mathbb{A}_{l8}^{(ab)}\}$ with $l = 1, .., 6$. The dominant contributions may be extracted from the following set of terms:
\bg\label{milensmit}
\mathbb{F}_{(15+j)e}^{(2)}(t) & = &\Bigg\{4j - {26\over 3} + 
2\gamma_{1, 2}\Big[{\hat\alpha_e(t) + \hat\beta_e(t)\over 2}\Big] +(2-j)\hat\zeta_e(t) + (j - 1)(0, \hat\eta_e(t)) \nonumber\\
&& -
{1\over \vert\log~\bar{g}_s\vert}~\log\left\vert {4\over 3} + (2-j)\Big({4\over 3} - \hat\zeta_e(t)\Big) + (j - 1)(0, \hat\eta_e(t))\right\vert \nonumber\\
&& - {1\over \vert\log~\bar{g}_s\vert}
\left\vert{1\over 3} + (2-j)\Big({10\over 3} - \hat\zeta_e(t)\Big) + (j - 1)(0, \hat\eta_e(t))\right\vert;  \nonumber\\
&& 4j - {23\over 3} + 
\gamma_{[3,..,10]}\Big[{\hat\alpha_e(t) + \hat\beta_e(t)\over 2}\Big] +(2-j)\hat\zeta_e(t)+ (j - 1)(0, \hat\eta_e(t))\nonumber\\
&& -
{1\over \vert\log~\bar{g}_s\vert}~\log\left\vert {4\over 3} + (2-j)\Big({4\over 3} - \hat\zeta_e(t)\Big) + (j - 1)(0, \hat\eta_e(t))\right\vert\Bigg\}, \nd
leading to two terms each for $j =1$ and $j = 2$. The contributions from the terms containing $\gamma_{1, 2}\Big[{\hat\alpha_e(t) + \hat\beta_e(t)\over 2}\Big]$ have already appeared in \eqref{manucox} and \eqref{manucox2}, up to the sub-dominant logarithmic corrections, so we will not discuss them here. We will instead concentrate only on the terms that contain $\gamma_{[3,..,10]}\Big[{\hat\alpha_e(t) + \hat\beta_e(t)\over 2}\Big]$. Looking at {\bf Table \ref{rindrag}}, we see that the only relevant terms are $\gamma_{4}\Big[{\hat\alpha_e(t) + \hat\beta_e(t)\over 2}\Big]$ and $\gamma_{5}\Big[{\hat\alpha_e(t) + \hat\beta_e(t)\over 2}\Big]$. Plugging this in \eqref{milensmit}, we get the following dominant contribution:

{\footnotesize
\bg\label{lilphil}
{\rm dom}~\mathbb{F}_{(15+j)e}^{(2)}(t) &= & 4j - {26\over 3} + 
{\hat\alpha_e(t) + \hat\beta_e(t)\over 2} +(2-j)\hat\zeta_e(t)+ (j - 1)(0, \hat\eta_e(t))\nonumber\\
& + & {1\over \vert\log~\bar{g}_s\vert}~\log\left[{ (4 - \hat\alpha_e(t) - \hat\beta_e(t))^l\over \big\vert {4.2^l\over 3} + 2^l(2-j)\big({4\over 3} - \hat\zeta_e(t)\big) + 2^l(j - 1)(0, \hat\eta_e(t))\big\vert}\right], \nd}
for $l= 1, 2$, which may be compared to the dominant contribution from \eqref{eyez69} for the simplified $SO(32)$ case. For the generic ${\rm E}_8 \times {\rm E}_8$ case, all we now need is to perform the operation:
\bg\label{mymylenjon}
{28\over 3} - 4j + \mathbb{F}_{(15+j)e}^{(2)}(t) + (j - 3) \hat\zeta_e(t) + (1-j)(0, \hat\eta_e(t)), \nd
with \eqref{lilphil} as the input. This gives us again \eqref{2cho1bora} as the contributions from this case to the quantum scaling \eqref{botsuga}, upto possible sub-dominant logarithmic corrections.

\vskip.1in

\subsubsection{{Functional forms and bounds on} $\mathbb{F}_{18, 20}(t)$  {and} $\mathbb{F}^{(1, 2)}_{18e, 20e}(t)$ \label{scaruvilla}}

\vskip.1in

\noindent Our next set of entries from {\bf Table \ref{privsocmey}} are
$\mathbb{F}_{18e}(t)$ and $\mathbb{F}_{20e}(t)$ that respectively contribute to the $g_s$ scalings of the curvature tensors ${\bf R}_{0m0n}$ and 
${\bf R}_{0\rho 0 \sigma}$ as shown in {\bf Table \ref{firzathai3}}. 
The generic form of these two scalings are:

{\footnotesize
\bg\label{koisherodinde}
\mathbb{F}_{18e}(t) &= & \left(2\otimes\mathbb{A}_8(\hat\sigma_e) -\hat\sigma_e(t) + {2\over 3}, ~ 
2\otimes\mathbb{B}_8(\hat\sigma_e) -\hat\sigma_e(t) + {2\over 3}, ~ \mathbb{A}_8(\hat\sigma_e) + \mathbb{B}_8(\hat\sigma_e) - \hat\sigma_e(t) + {2\over 3}, \mathbb{A}_{i8}^{(mn)}\right) \nonumber\\
\mathbb{F}_{20e}(t) &= & \Big(2\otimes\mathbb{A}_8(\hat\alpha_e, \hat\beta_e) -(\hat\alpha_e, \hat\beta_e(t)) + {2\over 3}, ~ 
2\otimes\mathbb{B}_8(\hat\alpha_e, \hat\beta_e) -(\hat\alpha_e, \hat\beta_e(t)) + {2\over 3}, ~ \mathbb{A}_8(\hat\alpha_e, \hat\beta_e) + \mathbb{B}_8(\hat\alpha_e, \hat\beta_e) \nonumber\\
&&~~ 
- (\hat\alpha_e, \hat\beta_e(t)) + {2\over 3}, \mathbb{A}_{i8}^{(\rho\sigma)}\Big), \nd}
where $\mathbb{A}_{i8}^{({\rm MN})}(t(g_s))$ with $i = 1, .., 6$ may be read from \eqref{selenzoe}. As in the previous section, we can split-up the set in \eqref{koisherodinde} as $\mathbb{F}_{18e}(t) = (\mathbb{F}_{18e}^{(1)}(t), \mathbb{F}_{18e}^{(2)}(t))$ and 
$\mathbb{F}_{20e}(t) = (\mathbb{F}_{20e}^{(1)}(t), \mathbb{F}_{20e}^{(2)}(t))$, with $\mathbb{F}_{18e}^{(2)}(t) = \{\mathbb{A}_{i8}^{(mn)}(t(g_s))\}$ and 
$\mathbb{F}_{20e}^{(2)}(t) = \{\mathbb{A}_{i8}^{(\rho\sigma)}(t(g_s))\}$, $i =1, .., 6$ and the remaining from \eqref{koisherodinde} in the other set $\mathbb{F}_{18e}^{(1)}(t)$ and $\mathbb{F}_{20e}^{(1)}(t)$ respectively. Once we go to the simplified $SO(32)$ case, $\mathbb{F}^{(1)}_{18e}(t)$ takes the following functional form:
\bg\label{sidalri}
\mathbb{F}_{18}(t) & = & \left(2\mathbb{A}(\alpha) - \alpha(t) + {2\over 3}, ~ 2\mathbb{B}(\alpha) - \alpha(t) + {2\over 3}, ~ \mathbb{A}(\alpha) + \mathbb{B}(\alpha) - \alpha + {2\over 3}\right)\\
&=& \Bigg(-{2\over 3} + \alpha(t) + {2\over \vert\log~\bar{g}_s\vert}~\log\left[{2-\beta(t) \over \vert\dot\beta(t)\vert\vert\log~\bar{g}_s\vert({2\over 3} - \alpha(t))}\right], \nonumber\\ 
&& -{5\over 3} + \alpha(t) + {\beta(t)\over 2} + {2\over\vert\log~\bar{g}_s\vert}~\log\left[{2-\beta(t) \over ({2\over 3} - \alpha(t))\sqrt{\vert\dot\beta(t)\vert\vert\log~\bar{g}_s\vert}}\right],
\nonumber\\
&& -{2\over 3} + \alpha(t) + {1\over\vert\log~\bar{g}_s\vert}~
\log\left[{2-\beta(t) \over \mp \dot\alpha(t) \vert \dot\beta(t)\vert\vert\log~\bar{g}_s\vert^2 ({2\over 3} - \alpha(t))}\right] \nonumber\\
&& -{5\over 3} + \alpha(t) + {\beta(t)\over 2} + {1\over\vert\log~\bar{g}_s\vert}~\log\left[{2-\beta(t)\over \mp \dot\alpha(t)\vert\log~\bar{g}_s\vert ({2\over 3} - \alpha(t))}\right], \nonumber\\
&& -{8\over 3} + \alpha(t) + \beta(t) + {2\over\vert\log~\bar{g}_s\vert}~\log\left[{2-\beta(t) \over {2\over 3} - \alpha(t)}\right], ~ -{2\over 3} + \alpha(t) - {2\log(\mp \dot\alpha(t)\vert\log~\bar{g}_s\vert) \over\vert\log~\bar{g}_s\vert}\Bigg), \nonumber
\nd
where note that we did not put a superscript of 1 because of our way of expressing the scalings for the simpler $SO(32)$ case as shown in row 18 of {\bf Table \ref{firzacut3}}. The dominant terms now being $-{2\over 3} + \alpha(t), -{5\over 3} + \alpha(t) + {\beta(t)\over 2}$ and $-{8\over 3} + \alpha(t) + \beta(t)$ in the absence of the log corrections. Using $\alpha(t) = -\beta(t)$, we see that all of them are negative definite implying that the curvature tensor ${\bf R}_{0m0n}$ becomes strong at late time. However the actual contributions to the quantum series \eqref{botsuga}, in the presence if the log corrections which are coming from {\bf Table \ref{firzacut9}} become:
\bg\label{selma}
{10\over 3} + \mathbb{F}_{18}(t) - \alpha(t) & = & \bigg({2\over 3} + {2n_d\over \bar{g}_s^{d/3}\vert{\rm log}~\bar{g}_s\vert} + 2\vert\gamma(t)\vert + {\cal B}(\Delta; b, d), \\
&& {2\over 3} + {\beta(t)\over 2} + {n_d\over \bar{g}_s^{d/3}\vert{\rm log}~\bar{g}_s\vert} + \vert\gamma(t)\vert + {\cal C}(\Delta; b, d), ~{2\over 3} + \beta(t)\bigg), \nonumber\nd
which are all positive definite, and in fact matches with what we had earlier in say \eqref{sarkylie2} and \eqref{indiastoker}. The dominant contribution is again ${2\over 3} + \beta(t)$ as the exponentially suppressed terms coming from $n_d > 0$ are expected to be very small when $g_s < 1$ and especially at late time. For the generalized ${\rm E}_8 \times {\rm E}_8$ case, the dominant contribution comes from:
\bg\label{sirilenakal}
{\rm dom}~\mathbb{F}^{(1)}_{18e}(t) = -{8\over 3} + \hat\sigma_e(t) + 
{\hat\alpha_e(t) + \hat\beta_e(t)\over 2} + {2\over \vert\log~\bar{g}_s\vert}~\log\bigg\vert{12 - 3\hat\alpha_e(t) - 3\hat\beta_e(t)\over 4 - 6\hat\sigma_e(t)}\bigg\vert, \nd
on which if we perform the operation ${10\over 3} + \mathbb{F}^{(1)}_{18e}(t) - \hat\zeta_e(t) - \hat\sigma_e(t)$ gives us the scalings from \eqref{2cho1bora} for the $SO(32)$ and the ${\rm E}_8 \times {\rm E}_8$ cases respectively that would enter \eqref{botsuga}.

Our next set from {\bf Table \ref{privsocmey}} is $\mathbb{F}^{(1)}_{20e}(t)$ which resembles $\mathbb{F}^{(1)}_{18e}(t)$ by exchanging $\hat\sigma_e(t)(t)$ with $(\hat\alpha_e(t), \hat\beta_e(t))$, keeping ${\hat\alpha_e(t) + \hat\beta_e(t)\over 2}$ in the definition of $\gamma_{1, 2}\big[{\hat\alpha_e(t) + \hat\beta_e(t)\over 2}\big]$ appeaing in $\mathbb{F}^{(1)}_{18e}(t)$ intact. Because of this, the functional forms for the set in $\mathbb{F}^{(1)}_{20e}(t)$ would be quite different. These in turn would govern the scalings of ${\bf R}_{0\rho 0\sigma}$. For the simplified $SO(32)$ case, the set is given by:
\bg\label{rycone1}
\mathbb{F}_{20}(t) & = & \left(2\mathbb{A}(\beta) - \beta(t) + {2\over 3}, ~ 2\mathbb{B}(\beta) - \beta(t) + {2\over 3}, ~ \mathbb{A}(\beta) + \mathbb{B}(\beta) - \beta + {2\over 3}\right)\\
&=& \Bigg(-{2\over 3} + \beta(t) + {2\over \vert\log~\bar{g}_s\vert}~\log\left[{2-\beta(t) \over \vert\dot\beta(t)\vert\vert\log~\bar{g}_s\vert({2\over 3} - \beta(t))}\right], \nonumber\\ 
&& -{5\over 3} + {3\beta(t)\over 2} + {2\over\vert\log~\bar{g}_s\vert}~\log\left[{2-\beta(t) \over ({2\over 3} - \beta(t))\sqrt{\vert\dot\beta(t)\vert\vert\log~\bar{g}_s\vert}}\right],
\nonumber\\
&& -{2\over 3} + \beta(t) + {1\over\vert\log~\bar{g}_s\vert}~
\log\left[{2-\beta(t) \over \vert \dot\beta(t)\vert^2\vert\log~\bar{g}_s\vert^2 ({2\over 3} - \beta(t))}\right] \nonumber\\
&& -{5\over 3} + {3\beta(t)\over 2} + {1\over\vert\log~\bar{g}_s\vert}~\log\left[{2-\beta(t)\over  \vert\dot\beta(t)\vert\vert\log~\bar{g}_s\vert ({2\over 3} - \beta(t))}\right], \nonumber\\
&& -{8\over 3} + 2\beta(t) + {2\over\vert\log~\bar{g}_s\vert}~\log\left[{2-\beta(t) \over {2\over 3} - \beta(t)}\right], ~ -{2\over 3} + \beta(t) - {2\log(\vert\dot\beta(t)\vert\vert\log~\bar{g}_s\vert) \over\vert\log~\bar{g}_s\vert}\Bigg), \nonumber
\nd
where again we did not have a superscript of 1 because of how it appears in row 20 of {\bf Table \ref{firzacut3}}. The dominant contributions now come from $-{2\over 3} + \beta(t), -{5\over 3} + {3\beta(t)\over 2}$ and $-{8\over 3} + 2\beta(t)$ in the absence of the log corrections, which are again all negative definite, implying strong curvature components ${\bf R}_{0\rho 0 \sigma}$ at late time. Once we carefully take the log corrections into account, the contributions to the quantum series \eqref{botsuga}, from {\bf Table \ref{firzacut9}}, are exactly similar to \eqref{selma}, {\it i.e.}:
\bg\label{selma2}
{10\over 3} + \mathbb{F}_{20}(t) - \beta(t) & = & \bigg({2\over 3} + {2n_d\over \bar{g}_s^{d/3}\vert{\rm log}~\bar{g}_s\vert} + 2\vert\gamma(t)\vert + {\cal B}(\Delta; b, d), \\
&& {2\over 3} + {\beta(t)\over 2} + {n_d\over \bar{g}_s^{d/3}\vert{\rm log}~\bar{g}_s\vert} + \vert\gamma(t)\vert + {\cal C}(\Delta; b, d), ~{2\over 3} + \beta(t)\bigg), \nonumber\nd
implying that the contributions become weaker as we go to late time. As before, despite the similarities between \eqref{selma} and \eqref{selma2}, the additional log corrections are in general different. (For example putting $\alpha(t) = -\beta(t)$ in \eqref{selma}, we can easily see that the log corrections are quite different from the log corrections in \eqref{selma2}. In other words \eqref{selma} has terms like $\log\left({6- 3\beta(t)\over 2 + 3\beta(t)}\right)$ whereas \eqref{selma2} has terms like $\log\left({6 - 3\beta(t)\over 2 - 3\beta(t)}\right)$.) For the generic ${\rm E}_8 \times {\rm E}_8$ case, wherein all the log corrections are inserted in, the story is equally simple. The dominant contribution comes from:

{\footnotesize
\bg\label{raffey}
{\rm dom}~\mathbb{F}^{(1)}_{20e}(t) = -{8\over 3} + (\hat\alpha_e(t), \hat\beta_e(t)) + {\hat\alpha_e(t) + \hat\beta_e(t)\over 2} + 
{2\over \vert\log~\bar{g}_s\vert}~\log\bigg\vert{12 - 3\hat\alpha_e(t) - 3\hat\beta_e(t)\over 4 - 6(\hat\alpha_e(t), \hat\beta_e(t))}\bigg\vert, \nd}
on which once we perform the operation ${10\over 3} + \mathbb{F}^{(1)}_{20e}(t) - \hat\zeta_e(t) - (\hat\alpha_e(t), \hat\beta_e(t))$, we reproduce the same scalings as in \eqref{2cho1bora} for the $SO(32)$ and the ${\rm E}_8 \times {\rm E}_8$ cases respectively.

The other two scalings 
$\mathbb{F}^{(2)}_{18e}(t)= \{\mathbb{A}_{i8}^{(mn)}(t(g_s))\}$ and 
$\mathbb{F}_{20e}^{(2)}(t) = \{\mathbb{A}_{i8}^{(\alpha\beta)}(t(g_s))\}$, $i =1, .., 6$, are necessary for the generic ${\rm E}_8 \times {\rm E}_8$ and $SO(32)$ cases as shown in {\bf Table \ref{privsocmey}} but do not appear in {\bf Table \ref{firzacut3}} for the simplified $SO(32)$ case. From \eqref{selenzoe}, we can express them as:

{\footnotesize
\bg\label{raffeyjones}
\mathbb{A}_{i8}^{({\rm MN})} & = & \Bigg\{
-{2\over 3} + \Sigma_{e({\rm MN})}(t) - {2\log(\mp \dot\Sigma_{e({\rm MN})}(t)\vert\log~\bar{g}_s\vert)\over \vert\log~\bar{g}_s\vert}; \\
&&~~-{2\over 3} + \Sigma_{e({\rm MN})}(t) - {\log(\mp \ddot\Sigma_{e({\rm MN})}(t)\vert\log~\bar{g}_s\vert)\over \vert\log~\bar{g}_s\vert}; \nonumber\\
&& ~~ -{5\over 3} + \Sigma_{e({\rm MN})}(t) + \gamma_{1, 2} - {\log(\mp \dot\Sigma_{e({\rm MN})}(t))\over \vert\log~\bar{g}_s\vert};\nonumber\\
&& ~~-{5\over 3} + \Sigma_{e({\rm MN})}(t) + \gamma_{[3, ...., 10]} - {\log\big\vert {2\over 3} - \Sigma_{e({\rm MN})}(t)\big\vert \over \vert\log~\bar{g}_s\vert}; \nonumber\\
&& ~~-{5\over 3} + \Sigma_{e({\rm MN})}(t) + \gamma_{1, 2} - {\log(\mp \dot\Sigma_{e({\rm MN})}(t)\vert\log~\bar{g}_s\vert)\over \vert\log~\bar{g}_s\vert} - {\log\big\vert {2\over 3} - \Sigma_{e({\rm MN})}(t)\big\vert\over \vert\log~\bar{g}_s\vert};\nonumber\\
&& ~~-{8\over 3}+ \Sigma_{e({\rm MN})}(t) + 2\gamma_{1, 2} - {\log\big\vert {2\over 3} - \Sigma_{e({\rm MN})}(t)\big\vert \over \vert\log~\bar{g}_s\vert} - {\log\big\vert {5\over 3} - \Sigma_{e({\rm MN})}(t)\big\vert\over \vert\log~\bar{g}_s\vert}\Bigg\}, \nonumber \nd}
where $\Sigma_{e(mn)}(t) = \hat\sigma_e(t)$ and $\Sigma_{e(\alpha\beta)} = (\hat\alpha_e(t), \hat\beta_e(t))$. The dominant contributions come from the fourth and the sixth terms in \eqref{raffeyjones}, the latter of which already appear in \eqref{sirilenakal} and \eqref{raffey} upto the sub-dominant logarithmic corrections, so we will ignore them. The remaining dominant contributions from the fourth term in \eqref{raffey} become:
\bg\label{tommraja}
{\rm dom}~\mathbb{F}_{(18+j)e}^{(2)}(t) &= & -{8\over 3} + 
{\hat\alpha_e(t) + \hat\beta_e(t)\over 2} + {1\over 2}\big(2-{j}\big)\hat\sigma_e(t)+ {j\over 2}(\hat\alpha_e(t), \hat\beta_e(t))\nonumber\\
& + & {1\over \vert\log~\bar{g}_s\vert}~\log\left[{ (4 - \hat\alpha_e(t) - \hat\beta_e(t))^l\over \big\vert {2.2^l\over 3} - 2^l(1-{j\over 2})\hat\sigma_e(t) - 2^{l-1} j(\hat\alpha_e, \hat\beta_e(t))\big\vert}\right], \nd
for $l = 1, 2$, and $j = 0, 2$. We can now perform the operation 
${10\over 3} +\mathbb{F}_{(18+j)e}^{(2)}(t) - \hat\zeta_e(t) - {1\over 2}(2-j) \hat\sigma_e(t) - {j\over 2} (\hat\alpha_e(t), \hat\beta_e(t))$, which will give us the same scalings as in \eqref{2cho1bora} for the $SO(32)$ and the ${\rm E}_8 \times {\rm E}_8$ cases respectively contributing to \eqref{botsuga}.

\subsubsection{Functional forms and bounds on $\mathbb{F}_{19}(t)$ and  $\mathbb{F}_{21}(t)$}

\vskip.1in

\noindent Our next set of scalings are a bit more non-trivial from what we encountered so far because they involve double temporal derivatives on the metric components ${\bf g}_{mn}$ and ${\bf g}_{\rho\sigma}$ respectively\footnote{It is easy to see from \eqref{corsage4} and \eqref{samthi} that the double temporal derivatives on the metric components ${\bf g}_{ab}, {\bf g}_{ij}$ and ${\bf g}_{00}$ do not lead to any complicated series.}. Moreover, the way we have constructed {\bf Table \ref{privsocmey}}, these scalings only appear for the simplified $SO(32)$ case. As evident from {\bf Table \ref{firzacut3}}, they would naturally involve derivatives of $\mathbb{A}(\alpha)$ and $\mathbb{A}(\beta)$ as well as of $\mathbb{B}(\alpha)$ and $\mathbb{B}(\beta)$. Such a consideration will lead to proliferation of terms whose condensed forms appear in {\bf Table \ref{firzacut3}}. To see this let us begin with $\mathbb{F}_{19}(t)$ whose explicit form becomes:

{\footnotesize
\bg\label{jayandasu}
\mathbb{F}_{19}(t) & = & \Bigg(\mathbb{B}(\alpha) - {\log(\mp\dot{\mathbb{B}}(\alpha)\vert\log~\bar{g}_s\vert)\over \vert\log~\bar{g}_s\vert}, ~ \mathbb{B}(\alpha) - 1 + \gamma_1(t) - {\log\vert\mathbb{B}(\alpha)\vert \over\vert\log~\bar{g}_s\vert}, ~ \mathbb{B}(\alpha) - 1 + \gamma_2(t) - {\log\vert\mathbb{B}(\alpha)\vert \over\vert\log~\bar{g}_s\vert}, \nonumber\\
&& \mathbb{A}_1(\alpha) - {\log(\mp\dot{\mathbb{A}}_1(\alpha)\vert\log~\bar{g}_s\vert)\over \vert\log~\bar{g}_s\vert}, ~ \mathbb{A}_1(\alpha) - 1 + \gamma_1(t) - {\log\vert\mathbb{A}_1(\alpha)\vert \over\vert\log~\bar{g}_s\vert}, ~ \mathbb{A}_1(\alpha) - 1 + \gamma_2(t) - {\log\vert\mathbb{A}_1(\alpha)\vert \over\vert\log~\bar{g}_s\vert}, \nonumber\\
&& \mathbb{A}_2(\alpha) - {\log(\mp\dot{\mathbb{A}}_2(\alpha)\vert\log~\bar{g}_s\vert)\over \vert\log~\bar{g}_s\vert}, ~ \mathbb{A}_2(\alpha) - 1 + \gamma_1(t) - {\log\vert\mathbb{A}_2(\alpha)\vert \over\vert\log~\bar{g}_s\vert}, ~ \mathbb{A}_2(\alpha) - 1 + \gamma_2(t) - {\log\vert\mathbb{A}_2(\alpha)\vert \over\vert\log~\bar{g}_s\vert}\Bigg), \nonumber\\
\nd}
where $\mathbb{A}_1(\alpha)$ and $\mathbb{A}_2(\alpha)$ are the two values of $\mathbb{A}(\alpha)$ from \eqref{corsage2} that are determined by the two values of $\gamma(t)$, {\it i.e.} $\gamma_1(t)$ and $\gamma_2(t)$, from \eqref{ishena} or \eqref{mcelhon}. The sign ambiguities inside the log terms in \eqref{jayandasu} are to take care of the signs of $\dot{\mathbb{A}}_i(\alpha)$ and $\dot{\mathbb{B}}(\alpha)$. Using the aforementioned definitions, $\dot{\mathbb{A}}_1(\alpha)$ takes the following form:

{\scriptsize
\bg\label{katmikpal}
\dot{\mathbb{A}}_1(\alpha) = \dot\alpha(t) + {1\over \vert\log~\bar{g}_s\vert}\left[{\vert\dot\beta(t)\vert \over 2-\beta(t)} + {\ddot\beta(t)\over \vert\dot\beta(t)\vert} + {c_1 \bar{g}_s^{\gamma_1 - 1} + c_2 \bar{g}_s^{\gamma_2 - 1}\over \vert\log~\bar{g}_s\vert}\left\{1 + 
\log\left({2-\beta(t)\over \vert\dot\beta(t)\vert\vert\log~\bar{g}_s\vert\vert({2\over 3} - \alpha(t))}\right)\right\} + {\ddot\alpha(t)\over {2\over 3} - \alpha(t)}\right], \nonumber\\ \nd}
from where it is easy to see that the first term contributes as 
$\log(\mp\dot\alpha(t) \vert\log~\bar{g}_s\vert)$ which appears to be smaller than 
$\vert\log~\bar{g}_s\vert)$ by assuming $\dot\alpha(t) = -\dot\beta(t)$ small. At the first sight, this is clearly consistent with our earlier assumption about the derivative of $\beta(t)$  $-$ recall that $\vert\dot\beta(t)\vert < {1\over \bar{g}_s}$ $-$ implying further that both $\log\left({\vert\dot\beta(t)\vert\over 2-\beta(t)}\right)$ and  $\log\left({\pm \dot\alpha(t)\over {2\over 3} - \alpha(t)}\right)$ are smaller than $\vert\log~\bar{g}_s\vert$. For the third term, we want 
$\log\left(\pm{\ddot\beta(t)\over \vert\dot\beta(t)\vert}\right) <\vert\log~\bar{g}_s\vert$. Since $\vert\dot\beta(t)\vert < {1\over \bar{g}_s}$, we want $\vert \ddot\beta(t)\vert$ to be of the same order of 
$\vert\dot\beta(t)\vert$ for consistency. Using the same arguments, the log terms inside the braces in \eqref{katmikpal} are clearly smaller than 
$\vert\log~\bar{g}_s\vert$ and therefore the size of the terms relies on $c_1 \bar{g}_s^{\gamma_1 -1} + c_2 \bar{g}_s^{\gamma_2 -1}$ with $c_1 = 1, c_2 = 2\sqrt{\Lambda}$. From \eqref{ishena} one can easily infer that 
$\gamma_1 - 1 = {1\over\vert\log~\bar{g}_s\vert}~\log\left({2-\beta(t)\over \vert\dot\beta(t)\vert\vert\log~\bar{g}_s\vert}\right)$ and 
$\gamma_2 - 1 = {\beta(t)\over 2} - 1 + {\log(2-\beta(t))\over\vert\log~\bar{g}_s\vert}$ with the possibility that $\gamma_1 > 0$ and $0 < \gamma_2 < 1$ (the latter is of course always true in our set-up) with $\gamma_1 > \gamma_2$. Such a consideration then suggests that a possible dominant contribution may appear from\footnote{We have assumed $\log\big(\sum\limits_{i = 1}^n a_i\big) = \log~a_1 + {\cal O}\left({1\over a_1}\right)$ if $a_1 > a_i~\forall i \ne 1$ with $a_1 \equiv{c_1 \bar{g}_s^{\gamma_1 - \gamma_2} + c_2\over \bar{g}_s^{\vert 1 - \gamma_2\vert} \vert\log~\bar{g}_s\vert} $, $n = 6$ in \eqref{katmikpal} and \eqref{paanvilla} and $n = 4$ in \eqref{metrottimey}. The bound discussed in \eqref{teresann} is therefore on ${\log~a_1\over \vert\log~\bar{g}_s\vert}$.}:
\bg\label{teresann}
{1\over \vert\log~\bar{g}_s\vert}~\log\left({c_1 \bar{g}_s^{\gamma_1 - \gamma_2} + c_2\over \bar{g}_s^{\vert 1 - \gamma_2\vert} \vert\log~\bar{g}_s\vert}\right) ~ < ~ \vert 1 - \gamma_2\vert, \nd
where we have assumed that $c_2 \equiv 2\sqrt{\Lambda} << 1$. Since the dominant term in $\gamma_2$ is ${\beta(t)\over 2}$, \eqref{teresann} is bounded from above by at least $1- {\beta(t)\over 2}$ (any additional sub-dominant log terms will add a small correction to it). The actual dominant contribution from \eqref{jayandasu} will be slightly different because of the addition of other terms\footnote{Here for example it is $\mathbb{A}_1(\alpha)$.}, and in particular from our earlier observations that the log terms {\it cannot} actually be ignored. In the absence of any log corrections the naive expectation would be a contribution $-{5\over 3} + \alpha(t) + {\beta(t)\over 2}$, much like what we had earlier. Unfortunately this is not quite right as there are other dominant contributions, even in the absence of log corrections, once we introduce the temporal derivatives of 
$\mathbb{A}_2(\alpha)$ and $\mathbb{B}(\alpha)$. The first one is given by:

{\footnotesize
\bg\label{paanvilla}
\dot{\mathbb{A}}_2(\alpha) = \dot\alpha(t) + {\dot\beta(t)\over 2} + {1\over \vert\log~\bar{g}_s\vert}\left[{c_1 \bar{g}_s^{\gamma_1 - 1} + c_2 \bar{g}_s^{\gamma_2 - 1} \over \vert\log~\bar{g}_s\vert}~\log\left({2-\beta(t)\over {2\over 3} - \alpha(t)}\right) + {\vert\dot\beta(t)\vert\over 2-\beta(t)} + {\dot\alpha(t)\over {2\over 3} -\alpha(t)}\right], \nd}
where we notice that the naive dominant contribution will come from an equivalent term like \eqref{teresann}, as all the other terms in \eqref{paanvilla} appear to be sub-dominant. This isn't quite right as we shall see soon. In a similar vein, $\dot{\mathbb{B}}(\alpha)$ may be expressed as:
\bg\label{metrottimey}
\dot{\mathbb{B}}(\alpha) = \dot\alpha(t) - {1\over \vert\log~\bar{g}_s\vert}\bigg[{\ddot\alpha(t)\over \dot\alpha(t)} +  
{c_1 \bar{g}_s^{\gamma_1 - 1} + c_2 \bar{g}_s^{\gamma_2 - 1} \over \vert\log~\bar{g}_s\vert}\big\{1 + \log\left(\mp\dot\alpha(t)\vert\log~\bar{g}_s\vert\right)\big\}\bigg], \nd
with again the naive dominant contribution appear to be provided by the bound in \eqref{teresann} as the others terms in \eqref{metrottimey} are sub-dominant. In the same vein it would appear that all the three terms: ${\log\vert\mathbb{A}_1(\alpha)\vert\over \vert\log~\bar{g}_s\vert},{\log\vert\mathbb{A}_2(\alpha)\vert\over \vert\log~\bar{g}_s\vert}$ and 
${\log\vert\mathbb{B}(\alpha)\vert\over \vert\log~\bar{g}_s\vert}$ in
\eqref{jayandasu} are sub-dominant. As cautioned earlier these naive expectations are not correct. To get the correct picture we will insert in all the log corrections carefully and express $\mathbb{F}_{19}(t)$ in the following way:

{\footnotesize
\bg\label{nicolodi}
\mathbb{F}_{19}(t) & = & -{1\over\vert\log~\bar{g}_s\vert}~\log\left[\mp\dot\alpha(t)\vert\log~\bar{g}_s\vert \pm {\ddot\alpha(t)\over \dot\alpha(t)} + 
{c_1 \bar{g}_s^{\gamma_1 - 1} + c_2 \bar{g}_s^{\gamma_2 - 1} \over \vert\log~\bar{g}_s\vert}\big\{1 + \log\left(\mp\dot\alpha(t)\vert\log~\bar{g}_s\vert\right)\big\}\right] 
-{2\over 3} + \alpha(t) \nonumber\\
&&- {\log(\mp\dot\alpha(t) \vert\log~\bar{g}_s\vert) \over \vert\log~\bar{g}_s\vert},
\nonumber\\
&&-{2\over 3} + \alpha(t) - {1\over\vert\log~\bar{g}_s\vert}~\log\left[{2\over 3} - \alpha(t) + {\log(\mp\dot\alpha(t) \vert\log~\bar{g}_s\vert) \over \vert\log~\bar{g}_s\vert}\right] + {1\over \vert\log~\bar{g}_s\vert}~\log\left[{2-\beta(t)\over \mp\dot\alpha(t)\vert\dot\beta(t)\vert\vert\log~\bar{g}_s\vert^2}\right], \nonumber\\
&&-{5\over 3} + \alpha(t) + {\beta(t)\over 2} - {1\over\vert\log~\bar{g}_s\vert}~\log\left[{2\over 3} - \alpha(t) + {\log(\mp\dot\alpha(t) \vert\log~\bar{g}_s\vert) \over \vert\log~\bar{g}_s\vert}\right] + {1\over \vert\log~\bar{g}_s\vert}~\log\left[{2-\beta(t)\over \mp\dot\alpha(t)\vert\log~\bar{g}_s\vert}\right], \nonumber\\
&& -{1\over \vert\log~\bar{g}_s\vert}~\log\Bigg[{\vert\dot\beta(t)\vert \over 2-\beta(t)} \mp {\ddot\beta(t)\over \vert\dot\beta(t)\vert} + {c_1\bar{g}_s^{\gamma_1 - 1} + c_2 \bar{g}_s^{\gamma_2 - 1}\over \vert\log~\bar{g}_s\vert}\left\{1 + 
\log\left({2-\beta(t)\over \vert\dot\beta(t)\vert\vert\log~\bar{g}_s\vert\vert({2\over 3} - \alpha(t))}\right)\right\}  \nonumber\\
&& \mp\dot\alpha(t) \vert\log~\bar{g}_s\vert \mp {\ddot\alpha(t)\over {2\over 3} - \alpha(t)}\Bigg] - {2\over 3} + \alpha(t) + 
{1\over \vert\log~\bar{g}_s\vert}~\log\left[{2-\beta(t)\over \vert\dot\beta(t)\vert\vert\log~\bar{g}_s\vert({2\over 3} -\alpha(t))}\right],  \nonumber\\
&& -{2\over 3} + \alpha(t)  -{1\over\vert\log~\bar{g}_s\vert}~\log\left[ {2\over 3} - \alpha(t) - 
{1\over \vert\log~\bar{g}_s\vert}~\log\left({2-\beta(t)\over \vert\dot\beta(t)\vert\vert\log~\bar{g}_s\vert({2\over 3} -\alpha(t))}\right)\right]
\nonumber\\
&& + {2\over\vert\log~\bar{g}_s\vert}~\log\left[ {2-\beta(t)\over \vert\dot\beta(t)\vert\vert\log~\bar{g}_s\vert\sqrt{{2\over 3} - \alpha(t)}}\right], \nonumber\\
&& -{5\over 3} + \alpha(t) + {\beta(t)\over 2} - {1\over\vert\log~\bar{g}_s\vert}~\log\left[{2\over 3} - \alpha(t) -{1 \over \vert\log~\bar{g}_s\vert}~\log\left({2-\beta(t)\over \vert\dot\beta(t)\vert \vert\log~\bar{g}_s\vert({2\over 3} - \alpha(t))}\right)\right] \nonumber\\
&& + {1\over \vert\log~\bar{g}_s\vert}~\log\left[{(2-\beta(t))^2\over \vert\dot\beta(t)\vert\vert\log~\bar{g}_s\vert({2\over 3}-\alpha(t))}\right], \nonumber\\
&& - \log\left[\left(\mp\dot\alpha(t) + {\vert\dot\beta(t)\vert\over 2}\right)\vert\log~\bar{g}_s\vert + {c_1 \bar{g}_s^{\gamma_1 - 1} + c_2 \bar{g}_s^{\gamma_2 - 1} \over \vert\log~\bar{g}_s\vert}~\log\left({2-\beta(t)\over {2\over 3} - \alpha(t)}\right) + {\vert\dot\beta(t)\vert\over 2-\beta(t)} \mp {\dot\alpha(t)\over {2\over 3} -\alpha(t)}\right] \nonumber\\
&& \times  {1\over \vert\log~\bar{g}_s\vert} -{5\over 3} + \alpha(t) + {\beta(t)\over 2} + {1\over \vert\log~\bar{g}_s\vert}~\log\left({2-\beta(t)\over {2\over 3} -\alpha(t)}\right), \nonumber\\
&& -{5\over 3} + \alpha(t) + {\beta(t)\over 2} - {1\over \vert\log~\bar{g}_s\vert}~\log\left[{5\over 3} - \alpha(t) - {\beta(t)\over 2} -{1 \over\vert\log~\bar{g}_s\vert}~\log\left({2-\beta(t)\over {2\over 3} - \alpha(t)}\right)\right] \nonumber\\
&& + {2\over\vert\log~\bar{g}_s\vert}~\log\left[{2-\beta(t)\over \sqrt{\vert\dot\beta(t)\vert\vert\log~\bar{g}_s\vert({2\over 3} - \alpha(t))}}\right], \nonumber\\
&& -{{\cal A}\over 3} - {1\over \vert\log~\bar{g}_s\vert}~\log\left[{5\over 3} - \alpha(t) - {\beta(t)\over 2} -{1 \over\vert\log~\bar{g}_s\vert}~\log\left({2-\beta(t)\over {2\over 3} - \alpha(t)}\right)\right]
+ {2\over\vert\log~\bar{g}_s\vert}~\log\left[{2-\beta(t)\over \sqrt{{2\over 3} - \alpha(t)}}\right], \nonumber\\
\nd}
where ${\cal A} \equiv 8-3(\alpha(t) + \beta(t))$. Ignoring the log corrections we see that the dominant contributions appear as $-{5\over 3} + \alpha(t) + {\beta(t)\over 2}, -{2\over 3} + \alpha(t)$ and $-{8\over 3} + \alpha(t) + \beta(t)$. They are all negative definite and therefore the curvature tensor ${\bf R}_{0m0n}$ would appear to become strong at late time. This is not an issue because the contribution of the curvature tensor to \eqref{botsuga}, from {\bf Table \ref{firzacut9}} and in the absence of the log corrections, actually becomes:
\bg\label{dariaN}
{10\over 3} + \mathbb{F}_{19}(t) - \alpha(t) = \left({8\over 3},~ {2\over 3} + \beta(t), ~{5\over 3} + {\beta(t)\over 2}\right), \nd
which are all positive definite. The actual story once we add in the log corrections is slightly different. To analyze this we will have to carefully work out all the terms in \eqref{nicolodi} and see how they scale. As a starter we note that:

{\scriptsize
\bg\label{dparker}
&& \vert\dot\alpha(t)\vert \left(c_1 \bar{g}_s^{\gamma_1 - 1} + c_2 \bar{g}_s^{\gamma_2 - 1}\right) = \vert\dot\alpha(t)\vert \left[c_1 \left({g_s\over {\rm H}(y){\rm H}_o({\bf x})}\right)^{-1 + {n_d\over \bar{g}_s^{d/3} \vert\log~\bar{g}_s\vert} + \vert\gamma(t)\vert + {\cal C}(\Delta; b, d)} + c_2 \left({g_s\over {\rm H}(y){\rm H}_o({\bf x})}\right)^{{\beta(t)\over 2} - 1}\right]\nonumber\\
&& \vert\dot\alpha(t)\vert \left(c_1 \bar{g}_s^{\gamma_1 - 1} + c_2 \bar{g}_s^{\gamma_2 - 1}\right) \log\left(\vert\dot\alpha(t)\vert\vert\log~\bar{g}_s\vert\right) = 
\vert\dot\alpha(t)\vert \left[\log\left({g_s\over {\rm H}(y){\rm H}_o({\bf x})}\right)^{-1 + \vert\gamma(t)\vert + {\cal C}(\Delta; b, d)} - {n_d\over \bar{g}_s^{d/3}}\right]\\
&& ~~~~~~~~~~~~~~~~~~~~~~~~~~~~~~~~~~~~~~~~~\times \left[c_1 \left({g_s\over {\rm H}(y){\rm H}_o({\bf x})}\right)^{-1 + {n_d\over \bar{g}_s^{d/3} \vert\log~\bar{g}_s\vert} + \vert\gamma(t)\vert + {\cal C}(\Delta; b, d)} + c_2 \left({g_s\over {\rm H}(y){\rm H}_o({\bf x})}\right)^{{\beta(t)\over 2} - 1}\right],\nonumber 
\nd}
which again confirms that $b \ge 2d$ for EFT to persist, but more importantly, both the contributions are highly subdominant because they are proportional to $\Delta << 1$. (Any worries that we could have 
terms like $\left({g_s\over {\rm H}(y){\rm H}_o({\bf x})}\right)^{{\beta(t)\over 2} - 1 - {d\over 3}}$ implying possible problems with EFT do not appear because of the following form for $\dot\alpha(t)$:
\bg\label{mirzap}
\vert\dot\alpha(t)\vert = \left({g_s\over {\rm H}(y){\rm H}_o({\bf x})}\right)^{-1 + {n_d\over \bar{g}_s^{d/3} \vert\log~\bar{g}_s\vert} + \vert\gamma(t)\vert + {\cal C}(0; b, d) - {\log~\Delta\over \vert\log~\bar{g}_s\vert}}, \nd
that absorbs all $\bar{g}_s^{-d/3}$ factors as $b \ge 2d$; where ${\cal C}(0; b, d)$ can be read-off from \eqref{standudh}.)
In fact all we require for $\Delta$ is to go to zero faster that $\bar{g}_s^{3}$ so as to ignore terms like \eqref{dparker} in \eqref{nicolodi}. We will however keep all terms so as to compare them carefully and see how they appear in the quantum scalings in \eqref{botsuga}. With this in mind, let us work out all the terms from \eqref{nicolodi}. The first one is:

{\footnotesize
\bg\label{sweetyme}
&& -{1\over\vert\log~\bar{g}_s\vert}~\log\left[\mp\dot\alpha(t)
\vert\log~\bar{g}_s\vert \pm {\ddot\alpha(t)\over \dot\alpha(t)} + 
{c_1 \bar{g}_s^{\gamma_1 - 1} + c_2 \bar{g}_s^{\gamma_2 - 1} \over \vert\log~\bar{g}_s\vert}\big\{1 + \log\left(\mp\dot\alpha(t)\vert\log~\bar{g}_s\vert\right)\big\}\right] 
-{2\over 3} + \alpha(t)\\
&&- {\log(\mp\dot\alpha(t) \vert\log~\bar{g}_s\vert) \over \vert\log~\bar{g}_s\vert} = \bigg(-{8\over 3} + \alpha(t) + {2n_d\over \bar{g}_s^{d/3} \vert\log~\bar{g}_s\vert} + 2\vert\gamma(t)\vert + {\cal B}(\Delta; b, d), -{8\over 3} + \alpha(t) + \Sigma(a_1, ..., a_9), \nonumber\\
&& -{8\over 3} + \alpha(t) + {2n_d\over \bar{g}_s^{d/3} \vert\log~\bar{g}_s\vert} + 2\vert\gamma(t)\vert 
+ {\cal C}(\Delta; b, d) + {\cal C}(0; b, d) - {\log~\Delta\over \vert\log~\bar{g}_s\vert}, \nonumber\\
&& -{8\over 3} + \alpha(t) + {2n_d\over \bar{g}_s^{d/3} \vert\log~\bar{g}_s\vert} + 2\vert\gamma(t)\vert 
+ {\cal C}(\Delta; b-d, d) + {\cal C}(0; b, d) - {\log~\Delta\over \vert\log~\bar{g}_s\vert}, \nonumber\\
&& -{8\over 3} + {n_d\over \bar{g}_s^{d/3} \vert\log~\bar{g}_s\vert} + \vert\gamma(t)\vert + {\cal C}(0; b-d, d) - {\log~\Delta\over \vert\log~\bar{g}_s\vert} - {\beta(t)\over 2}, 
-{8\over 3} + {n_d\over \bar{g}_s^{d/3} \vert\log~\bar{g}_s\vert} + \vert\gamma(t)\vert + {\cal C}(0; b, d)\nonumber\\ 
&& - {\log~\Delta\over \vert\log~\bar{g}_s\vert} - {\beta(t)\over 2}, 
-{8\over 3} + {n_d\over \bar{g}_s^{d/3} \vert\log~\bar{g}_s\vert} + \vert\gamma(t)\vert + {\cal C}(0; b, d) - {\log~\Delta\over \vert\log~\bar{g}_s\vert} - {\beta(t)\over 2}  - {\log\vert\log~\bar{g}_s\vert^{1-\vert\gamma(t)\vert - {\cal C}(\Delta; b, d)}\over \vert\log~\bar{g}_s\vert}\nonumber\\
&& -{8\over 3} + \alpha(t) + {2n_d\over \bar{g}_s^{d/3} \vert\log~\bar{g}_s\vert} + 2\vert\gamma(t)\vert 
+ {\cal C}(\Delta; b, d) + {\cal C}(0; b, d) - {\log~\Delta\over \vert\log~\bar{g}_s\vert} - {\log\vert\log~\bar{g}_s\vert^{1-\vert\gamma(t)\vert - {\cal C}(\Delta; b, d)}\over \vert\log~\bar{g}_s\vert}
\bigg), \nonumber \nd}
where again ${\cal C}(\Delta; b-d, d)$ etc can be read-off from \eqref{standudh}. The log.log pieces $-$ whose exponents are positive definite because both $\vert\gamma(t)\vert << 1$ and ${\cal C}(\Delta; b, d) << 1$ $-$ are harmless because they are eventually controlled by either of the following two limiting values:
\bg\label{obricate}
\lim_{\bar{g}_s \to 0} ~\bar{g}_s^{\vert{\rm A}\vert} ~\log~\bar{g}_s ~ \to ~ 0,~~~~~~\lim_{\bar{g}_s \to 0} ~{\log\vert \log~\bar{g}_s\vert \over \vert \log~\bar{g}_s\vert} ~ \to ~ 0, \nd
$\forall \vert{\rm A}\vert \in \mathbb{R}_+$ as may be easily seen from \eqref{dparker}. In fact, since all the terms in \eqref{sweetyme} either have $n_d$ or $\Delta$ or both, they are heavily suppressed much like what we saw earlier for other values of $\mathbb{F}_i$. In a similar vein the second term from \eqref{nicolodi} may be expressed in the following way:

{\footnotesize
\bg\label{chanells}
&&  -{2\over 3} + \alpha(t) - {1\over\vert\log~\bar{g}_s\vert}~\log\left[{2\over 3} - \alpha(t) + {\log(\mp\dot\alpha(t) \vert\log~\bar{g}_s\vert) \over \vert\log~\bar{g}_s\vert}\right] + {1\over \vert\log~\bar{g}_s\vert}~\log\left[{2-\beta(t)\over \mp\dot\alpha(t)\vert\dot\beta(t)\vert\vert\log~\bar{g}_s\vert^2}\right]
\nonumber\\
&& = \bigg(-{8\over 3} + \alpha(t) + {2n_d\over \bar{g}_s^{d/3} \vert\log~\bar{g}_s\vert} + 2\vert\gamma(t)\vert + {\cal B}(\Delta; b, d),\nonumber\\
&& -{8\over 3} + \alpha(t) + {2n_d\over \bar{g}_s^{d/3} \vert\log~\bar{g}_s\vert} + 2\vert\gamma(t)\vert 
+ {\cal C}(\Delta; b-d, d) + {\cal C}(0; b, d) - {\log~\Delta\over \vert\log~\bar{g}_s\vert},\\
&&  -{8\over 3} + \alpha(t) + {2n_d\over \bar{g}_s^{d/3} \vert\log~\bar{g}_s\vert} + 2\vert\gamma(t)\vert 
+ {\cal C}(\Delta; b, d) + {\cal C}(0; b, d) - {\log~\Delta\over \vert\log~\bar{g}_s\vert} - {\log\vert\log~\bar{g}_s\vert^{1-\vert\gamma(t)\vert - {\cal C}(\Delta; b, d)}\over \vert\log~\bar{g}_s\vert}\Bigg), \nonumber
\nd}
where the terms herein are a subset of the ones from \eqref{sweetyme}. The difference would come from terms like ${2 - 3\alpha(t)\over 6 - 3\beta(t)}$ and ${1\over 2 - \beta(t)}$ that we have ignored because of their relative suppression factors. Nevertheless, as in \eqref{sweetyme}, all the terms in \eqref{chanells} are heavily suppressed both by the $\Delta$  as well as by the non-perturbative factors. The remaining terms in \eqref{nicolodi} are all suppressed except for one term that we will discuss soon. The third, sixth, seventh and the eighth terms all go as $\bar{g}_s^{-{5\over 3} + \alpha(t) + {\beta(t)\over 2}}$, but with different suppression factors whose common factors appear to be $\vert\dot\beta(t)\vert\vert\log~\bar{g}_s\vert$ except for the seventh term. The suppression factor for the seventh term turns out to be:
\bg\label{jassher}
{3\over 2}{\vert\dot\beta(t)\vert}\vert\log~\bar{g}_s\vert + {c_1 \bar{g}_s^{\gamma_1 - 1} + c_2 \bar{g}_s^{\gamma_2 - 1} \over \vert\log~\bar{g}_s\vert}~\log\left({6-3\beta(t)\over {2} + 3\beta(t)}\right) + {8\vert\dot\beta(t)\vert\over (2-\beta(t))(2+3\beta(t))}, \nd
which when compared to the results from {\bf Table \ref{munleena}} show the non-perturbative suppressions for each of the terms in \eqref{jassher}. The fourth and the fifth terms that go as $\bar{g}_s^{-{2\over 3} + \alpha(t)}$ also allow a common suppression factor of 
$\vert\dot\beta(t)\vert\vert\log~\bar{g}_s\vert$. The only unsuppressed term is the ninth term that goes as:

{\scriptsize
\bg\label{jasmpenny}
\bar{g}_s^{-{8\over 3}}\left[{5\over 3} - \alpha(t) - {\beta(t)\over 2} -{1 \over\vert\log~\bar{g}_s\vert}~\log\left({2-\beta(t)\over {2\over 3} - \alpha(t)}\right)\right]
\left[{{{2\over 3} - \alpha(t)\over (2-\beta(t))^2}}\right] = 
{\bar{g}_s^{-{8\over 3}}\over 6}\left({5\over 3} - {3\beta(t)\over 2} - {\log~3\over \vert\log~\bar{g}_s}\right), \nd}
where other than the first term, all other terms in the bracket are suppressed. Looking at \eqref{dariaN}, we see that the dominant contribution is 
${2\over 3} + \beta(t)$. Incidentally, the dominant behavior matches with those from $\mathbb{F}_{14}(t)$ in \eqref{ishena4} and $\mathbb{F}_{18}(t)$ from \eqref{selma}, although the log corrections are different. This makes sense because all the three $\mathbb{F}_i(t)$, namely $\mathbb{F}_{14}(t), \mathbb{F}_{18}(t)$ and $\mathbb{F}_{19}(t)$ contribute to the $g_s$ scalings of ${\bf R}_{0m0n}$ as seen from {\bf Table \ref{firzacut9}}.

The last one in the set of $\mathbb{F}_i(t)$ is $\mathbb{F}_{21}(t)$, and it governs the $g_s$ scalings of the curvature tensor ${\bf R}_{0\rho 0 \sigma}$. The functional form for $\mathbb{F}_{21}(t)$ appears in {\bf Table \ref{firzacut3}}, and one would need explicit forms for $\dot{\mathbb{A}}_1(\beta),\dot{\mathbb{A}}_2(\beta)$ and $\dot{\mathbb{B}}(\beta)$. These may be easily worked out from \eqref{katmikpal}, \eqref{paanvilla} and \eqref{metrottimey} by replacing $\alpha(t)$ by $\beta(t)$ and keeping $\beta(t)$ therein unchanged. Plugging these in the functional form for $\mathbb{F}_{21}(t)$ in {\bf Table \ref{firzacut3}} gives us the following:

{\footnotesize
\bg\label{teneabre}
\mathbb{F}_{21}(t) & = &  -{1\over\vert\log~\bar{g}_s\vert}~\log\left[\vert\dot\beta(t)\vert\vert\log~\bar{g}_s\vert \mp {\ddot\beta(t)\over \vert\dot\beta(t)\vert} + 
{c_1 g_s^{\gamma_1 - 1} + c_2 g_s^{\gamma_2 - 1} \over \vert\log~\bar{g}_s\vert}\big\{1 + \log\left(\vert\dot\beta(t)\vert\vert\log~\bar{g}_s\vert\right)\big\}\right] 
-{2\over 3} + \beta(t) \nonumber\\
&&- {\log(\vert\dot\beta(t)\vert \vert\log~\bar{g}_s\vert) \over \vert\log~\bar{g}_s\vert},
\nonumber\\
&&-{2\over 3} + \beta(t) - {1\over\vert\log~\bar{g}_s\vert}~\log\left[{2\over 3} - \beta(t) + {\log(\vert\dot\beta(t)\vert \vert\log~\bar{g}_s\vert) \over \vert\log~\bar{g}_s\vert}\right] + {2 \over \vert\log~\bar{g}_s\vert}~\log\left[{\sqrt{2-\beta(t)}\over \vert\dot\beta(t)\vert\vert\log~\bar{g}_s\vert}\right], \nonumber\\
&&-{5\over 3} + {3\beta(t)\over 2} - {1\over\vert\log~\bar{g}_s\vert}~\log\left[{2\over 3} - \beta(t) + {\log(\vert\dot\beta(t)\vert \vert\log~\bar{g}_s\vert) \over \vert\log~\bar{g}_s\vert}\right] + {1\over \vert\log~\bar{g}_s\vert}~\log\left[{2-\beta(t)\over \vert\dot\beta(t)\vert\vert\log~\bar{g}_s\vert}\right], \nonumber\\
&& -{1\over \vert\log~\bar{g}_s\vert}~\log\Bigg[{\vert\dot\beta(t)\vert \over 2-\beta(t)} \mp {\ddot\beta(t)\over \vert\dot\beta(t)\vert} + {c_1g_s^{\gamma_1 - 1} + c_2 g_s^{\gamma_2 - 1}\over \vert\log~\bar{g}_s\vert}\left\{1 + 
\log\left({2-\beta(t)\over \vert\dot\beta(t)\vert\vert\log~\bar{g}_s\vert\vert({2\over 3} - \beta(t))}\right)\right\}  \nonumber\\
&& + \vert\dot\beta(t)\vert \vert\log~\bar{g}_s\vert \mp {\ddot\beta(t)\over {2\over 3} - \beta(t)}\Bigg] - {2\over 3} + \beta(t) + 
{1\over \vert\log~\bar{g}_s\vert}~\log\left[{2-\beta(t)\over \vert\dot\beta(t)\vert\vert\log~\bar{g}_s\vert({2\over 3} -\beta(t))}\right],  \nonumber\\
&& -{2\over 3} + \beta(t)  -{1\over\vert\log~\bar{g}_s\vert}~\log\left[ {2\over 3} - \beta(t) - 
{1\over \vert\log~\bar{g}_s\vert}~\log\left({2-\beta(t)\over \vert\dot\beta(t)\vert\vert\log~\bar{g}_s\vert({2\over 3} -\beta(t))}\right)\right]
\nonumber\\
&& + {2\over\vert\log~\bar{g}_s\vert}~\log\left[ {2-\beta(t)\over \vert\dot\beta(t)\vert\vert\log~\bar{g}_s\vert\sqrt{{2\over 3} - \beta(t)}}\right], \nonumber\\
&& -{5\over 3} + {3\beta(t)\over 2} - {1\over\vert\log~\bar{g}_s\vert}~\log\left[{2\over 3} - \beta(t) -{1 \over \vert\log~\bar{g}_s\vert}~\log\left({2-\beta(t)\over \vert\dot\beta(t)\vert \vert\log~\bar{g}_s\vert({2\over 3} - \beta(t))}\right)\right] \nonumber\\
&& + {1\over \vert\log~\bar{g}_s\vert}~\log\left[{(2-\beta(t))^2\over \vert\dot\beta(t)\vert\vert\log~\bar{g}_s\vert({2\over 3}-\beta(t))}\right], \nonumber\\
&& - \log\left[{3\over 2} \vert\dot\beta(t)\vert \vert\log~\bar{g}_s\vert + {c_1 g_s^{\gamma_1 - 1} + c_2 g_s^{\gamma_2 - 1} \over \vert\log~\bar{g}_s\vert}~\log\left({2-\beta(t)\over {2\over 3} - \beta(t)}\right) + {\vert\dot\beta(t)\vert\over 2-\beta(t)} +{\vert\dot\beta(t)\vert\over {2\over 3} -\beta(t)}\right] \nonumber\\
&& \times  {1\over \vert\log~\bar{g}_s\vert} -{5\over 3} +  {3\beta(t)\over 2} + {1\over \vert\log~\bar{g}_s\vert}~\log\left({2-\beta(t)\over {2\over 3} -\beta(t)}\right), \nonumber\\
&& -{5\over 3} + {3\beta(t)\over 2} - {1\over \vert\log~\bar{g}_s\vert}~\log\left[{5\over 3} - {3\beta(t)\over 2} -{1 \over\vert\log~\bar{g}_s\vert}~\log\left({2-\beta(t)\over {2\over 3} - \beta(t)}\right)\right] \nonumber\\
&& + {2\over\vert\log~\bar{g}_s\vert}~\log\left[{2-\beta(t)\over \sqrt{\vert\dot\beta(t)\vert\vert\log~\bar{g}_s\vert({2\over 3} - \beta(t))}}\right], \nonumber\\
&& -{8\over 3} + 2\beta(t) - {1\over \vert\log~\bar{g}_s\vert}~\log\left[{5\over 3} - {3\beta(t)\over 2} -{1 \over\vert\log~\bar{g}_s\vert}~\log\left({2-\beta(t)\over {2\over 3} - \beta(t)}\right)\right]
+ {2\over\vert\log~\bar{g}_s\vert}~\log\left[{2-\beta(t)\over \sqrt{{2\over 3} - \beta(t)}}\right], \nonumber\\
\nd}
where again the dominant contribution come from the last term of \eqref{teneabre} as $-{8\over 3} + 2\beta(t)$, which is negative definite because $0 < \beta(t) < {2\over 3}$. However the actual contributions to \eqref{botsuga}, may be seen from {\bf Table \ref{firzacut9}} as:
\bg\label{latilengt}
{10\over 3} + \mathbb{F}_{21}(t) - \beta(t) =  {2\over 3} + \beta(t), \nd
which is positive definite, implying that it becomes small at late time, and matching exactly with the ${2\over 3} + \beta(t)$ fact that we had earlier. Again, the log corrections to \eqref{latilengt} and \eqref{dariaN} would be very small but different.

\subsubsection{Functional forms and bounds on $\mathbb{F}_{22e}(t)$ till $\mathbb{F}_{25e}(t)$}

The remaining four scalings $\mathbb{F}_{22e}(t)$ till $\mathbb{F}_{25e}(t)$ appearing for the ${\rm E}_8 \times {\rm E}_8$ case do not appear explicitly for the simplified $SO(32)$ case. For the generic case, we will show that these scalings on the ${\rm E}_8$ side imply non-trivial changes in the $SO(32)$ side (but not for the simplified case, as mentioned above). Let us start with the scaling $\mathbb{F}_{22e}(t)$ which, from {\bf Table \ref{privsocmey}}, takes the following form:

{\scriptsize
\bg\label{sirilena13}
\mathbb{F}_{22e}(t) & = & \Bigg({8\over 3} - \hat\zeta_e(t) + 2\otimes \left({1\over 3} + \gamma_{1, 2}\big[{\hat\alpha_e(t) + \hat\beta_e(t)\over 2}\big], \mathbb{M}_8(\hat\eta_e)\right)\\
&& ~~{8\over 3} - \hat\zeta_e(t) + \left({1\over 3} + \gamma_{1, 2}\big[{\hat\alpha_e(t) + \hat\beta_e(t)\over 2}\big], \mathbb{F}_8(\hat\eta_e)\right) +\left({1\over 3} + \gamma_{1, 2}\big[{\hat\alpha_e(t) + \hat\beta_e(t)\over 2}\big], \mathbb{G}_8(\hat\eta_e)\right)\Bigg) \nonumber\\
& = & \Bigg({8\over 3} - \hat\zeta_e(t) + \left({1\over 3} + \gamma_{1, 2}\big[{\hat\alpha_e(t) + \hat\beta_e(t)\over 2}\big], \mathbb{F}_8(\hat\eta_e)\right) +\left({1\over 3} + \gamma_{1, 2}\big[{\hat\alpha_e(t) + \hat\beta_e(t)\over 2}\big], \mathbb{F}_8(\hat\eta_e)\right), \nonumber\\
&& ~~ {8\over 3} - \hat\zeta_e(t) + \left({1\over 3} + \gamma_{1, 2}\big[{\hat\alpha_e(t) + \hat\beta_e(t)\over 2}\big], \mathbb{F}_8(\hat\eta_e)\right) +\left({1\over 3} + \gamma_{1, 2}\big[{\hat\alpha_e(t) + \hat\beta_e(t)\over 2}\big], \mathbb{G}_8(\hat\eta_e)\right), \nonumber\\
&& ~~ {8\over 3} - \hat\zeta_e(t) + \left({1\over 3} + \gamma_{1, 2}\big[{\hat\alpha_e(t) + \hat\beta_e(t)\over 2}\big], \mathbb{G}_8(\hat\eta_e)\right) +\left({1\over 3} + \gamma_{1, 2}\big[{\hat\alpha_e(t) + \hat\beta_e(t)\over 2}\big], \mathbb{G}_8(\hat\eta_e)\right)\Bigg), \nonumber
\nd}
where $\mathbb{M}_8(\hat\eta_e) = (\mathbb{F}_8(\hat\eta_e),\mathbb{G}_8(\hat\eta_e))$. It is now easy to work out the contributions to the scaling by looking from third to the fifth line but, as motivated from our earlier analysis, it will suffice to look at the dominant contributions only. The dominant contributions to 
$\mathbb{F}_{22e}(t)$ come from the following three pieces:
\bg\label{sirilena14}
{\rm dom}~\mathbb{F}_{22e}(t) & = & \Bigg({10\over 3} - \hat\zeta_e(t) + 
{\hat\alpha_e(t) + \hat\beta_e(t)\over 2} + {2\log\left(2 - {\hat\alpha_e + \hat\beta_e\over 2}\right)\over \vert\log~\bar{g}_s\vert},\\
&& ~~{10\over 3} - \hat\zeta_e(t) + 
{\hat\alpha_e(t) + \hat\beta_e(t)\over 2} + {2\log\left(2 - {\hat\alpha_e + \hat\beta_e\over 2}\right)\over \vert\log~\bar{g}_s\vert} + \hat\eta_e(t) - {\log\left({4\over 3} + \hat\eta_e\right)\over \vert\log~\bar{g}_s\vert}, \nonumber\\
&& ~~{10\over 3} - \hat\zeta_e(t) + 
{\hat\alpha_e(t) + \hat\beta_e(t)\over 2} + {2\log\left(2 - {\hat\alpha_e + \hat\beta_e\over 2}\right)\over \vert\log~\bar{g}_s\vert} + 2\hat\eta_e(t) - {2\log\left({4\over 3} + \hat\eta_e\right)\over \vert\log~\bar{g}_s\vert}\Bigg), \nonumber \nd
that differ only by the sub-dominant terms governed by $\hat\eta_e(t)$. One may also easily see that, in the simplified case where we ignore the sub-dominant terms, the three scalings in \eqref{sirilena14} merge to: 
\bg\label{sirilena15}
{10\over 3} + 2\gamma_{1, 2}[\beta(t)] = {10\over 3} + \beta(t) + {2\log(2-\beta)\over\vert\log~\bar{g}_s\vert}, \nd
as it appears in {\bf Table \ref{firzacut}} for the curvature component ${\bf R}_{abab}$. On the other hand, the contribution to the quantum series \eqref{botsuga} comes from $-{8\over 3} + \mathbb{F}_{22e}(t) - 2(0, \hat\eta_e(t))$. Plugging in \eqref{sirilena14} gives us the same scalings as in \eqref{2cho1bora} for the $SO(32)$ and the ${\rm E}_8 \times {\rm E}_8$ cases.

The next item in our list is $\mathbb{F}_{23e}(t)$ which corresponds to the scalings of the curvature component ${\bf R}_{abi0}$. This may be represented in the following way:

{\footnotesize
\bg\label{sirilena16}
\mathbb{F}_{23e}(t) & = & \left({8\over 3} - \hat\zeta_e(t) + \left({4\over 3}, {4\over 3} + \hat\eta_e(t)\right) + \mathbb{C}_8(\hat\zeta_e), ~~{8\over 3} - \hat\zeta_e(t) + \left({4\over 3}, {4\over 3} + \hat\eta_e(t)\right) + \mathbb{D}_8(\hat\zeta_e)\right), \nd}
with the dominant contribution coming from the one containing $\mathbb{D}_8(\hat\zeta_e)$, with $\mathbb{D}_8(\hat\zeta_e)$ defined in \eqref{vanandcar}. This can actually be worked out carefully to give us:
\bg\label{sirilena17}
{\rm dom}~\mathbb{F}_{23e}(t) = {1\over 3} + (0, \hat\eta_e(t)) + {\hat\alpha_e(t) + \hat\beta_e(t)\over 4} + {1\over \vert\log~\bar{g}_s\vert}~\log\left({12-3\hat\alpha_e - 3\hat\beta_e\over 16 - 6\hat\zeta_e}\right), \nd
which should be compared to the elements in the fifteenth row of {\bf Table \ref{firzathai}} associated with the curvature tensor ${\bf R}_{abi0}$. Between the remaining ones therein, the dominant contribution would come from $\left({1\over 3} + \gamma_{1, 2}\Big[{\hat\alpha_e + \hat\beta_e\over 2}\Big], \mathbb{G}_8(\hat\eta_e)\right)$, resulting in a result very similar to \eqref{sirilena17} with the difference coming from the sub-dominant contributions (here governed by $\hat\eta_e(t)$ and 
$\hat\zeta_e(t)$). To get the contributions to \eqref{botsuga}, especially from \eqref{sirilena17}, we have to compute ${4\over 3} + \mathbb{F}_{23e}(t) - \hat\zeta_e(t) - (0, \hat\eta_e(t))$. This gives:
\bg\label{sirilena18}
\boxed{{5\over 3} - \zeta_e(t) + {\beta_e(t)\over 2}; ~~~~ {5\over 3} - \hat\zeta_e(t) + {\hat\alpha_e(t) + \hat\beta_e(t)\over 4}} \nd
which are the first terms in the two set of terms in \eqref{chu8hath1} associated with the $SO(32)$ and the ${\rm E}_8 \times {\rm E}_8$ cases respectively. Incidentally, the two remaining terms in the scalings of ${\bf R}_{abi0}$ also contribute as \eqref{sirilena18} upto sub-dominant terms. 

Another related term is $\mathbb{F}_{24e}(t)$ that provides the scaling for the curvature tensor ${\bf R}_{ijij}$. From {\bf Table \ref{privsocmey}} this takes the following form:
\bg\label{sirilena19}
\mathbb{F}_{24e}(t) = \left({8\over 3} - \hat\zeta_e(t) + \mathbb{P}_8(\hat\zeta_e) +\mathbb{P}_8(\hat\zeta_e), ~~ {8\over 3} - \hat\zeta_e(t) + \mathbb{C}_8(\hat\zeta_e) +\mathbb{D}_8(\hat\zeta_e)\right), \nd
where $\mathbb{P}_8(\hat\zeta_e) = \left(\mathbb{C}_8(\hat\zeta_e), \mathbb{D}_8(\hat\zeta_e)\right)$. From the form of $\mathbb{P}_8(\hat\zeta_e)$ in \eqref{vanandcar}, it is easy to see that only the term without $\mathbb{C}_8(\hat\zeta_e)$ would dominate the scaling. The dominant scaling becomes:
\bg\label{sirilena20}
{\rm dom}~\mathbb{F}_{24e}(t) = -{14\over 3} + \hat\zeta_e(t) + {\hat\alpha_e(t) + \hat\beta_e(t)\over 2} + {2\over \vert\log~\bar{g}_s\vert}~\log\left({12 - 3\hat\alpha_e - 3\hat\beta_e\over 16 - 6\hat\zeta_e}\right), \nd
which, once we remove the sub-dominant corrections and go to the simplified $SO(32)$ side, it becomes $-{14\over 3} + \beta(t) + {2\log(2 - \beta)\over\vert\log~\bar{g}_s\vert}$, exactly as in the entry $-{14\over 3} + 2\gamma_{1, 2}[\beta(t)]$ on the nineteenth row of {\bf Table \ref{firzacut}}. The contribution to \eqref{botsuga} now comes from 
${16\over 3} + \mathbb{F}_{24e}(t) - 2\hat\zeta_e(t)$, thus giving us the same scalings in \eqref{2cho1bora} for the $SO(32)$ and the ${\rm E}_8 \times {\rm E}_8$ cases respectively.

Our final scaling $\mathbb{F}_{25e}(t)$ for the curvature tensor ${\bf R}_{abij}$ is very similar to $\mathbb{F}_{8e}(t)$ from \eqref{lenasiri}, except $\mathbb{A}_8(\hat\sigma_e)$ and $\mathbb{B}_8(\hat\sigma_e)$ therein are replaced by $\mathbb{C}_8(\hat\zeta_e)$ and $\mathbb{D}_8(\hat\zeta_e)$ respectively. As such we expect their properties to be similar. Looking at the functional forms for $\mathbb{C}_8(\hat\zeta_e)$ and $\mathbb{F}_8(\hat\eta_e)$ in \eqref{vanandcar}, we see that the terms containing them would be non-perturbatively suppressed at late time. This implies that the dominant contribution becomes:
\bg\label{sirilena21}
{\rm dom}~\mathbb{F}_{25e}(t) &= & \bigg(3 - \hat\zeta_e(t) + \gamma_{1, 2}\Big[{\hat\alpha_e + \hat\beta_e\over 2}\Big] + \mathbb{D}_8(\hat\zeta_e), ~~{8\over 3} - \hat\zeta_e(t) + \mathbb{G}_8(\hat\eta_e) + \mathbb{D}(\hat\zeta_e)\bigg) \nonumber\\
& = & \Bigg(-{2\over 3} + {\hat\alpha_e(t) + \hat\beta_e(t)\over 2} + 
{1\over \vert\log~\bar{g}_s\vert}~\log\bigg[{(4\sqrt{3} -\sqrt{3}\hat\alpha_e - \sqrt{3}\hat\beta_e)^2\over 32 - 12\hat\zeta_e}\bigg], \\
&& ~~ -{2\over 3} + (0, \hat\eta_e(t)) + {\hat\alpha_e(t) + \hat\beta_e(t)\over 2} + 
{1\over \vert\log~\bar{g}_s\vert}~\log\bigg[{(12 -3\hat\alpha_e - 3\hat\beta_e)^2\over (8 + 6(0,\hat\eta_e))(16 - 6 \hat\zeta_e)}\bigg]\Bigg), \nonumber \nd
where one may see the similarity between the two terms up to sub-dominant corrections coming from $\hat\eta_e(t)$ and $\hat\zeta_e(t)$. They contribute to the scalings of the curvature tensor ${\bf R}_{abij}$, and from there one may extract the contribution to \eqref{botsuga} by performing the operation ${4\over 3} + \mathbb{F}_{25e}(t) - \hat\zeta_e(t) - (0, \hat\eta_e(t))$. This gives us the same scalings as in \eqref{2cho1bora} for the $SO(32)$ and the ${\rm E}_8 \times {\rm E}_8$ cases respectively.

This concludes our analysis of all the $\mathbb{F}_i(t)$ factors that are responsible in fixing parts of the $g_s$ scalings of the curvature tensors. (A summary appears in {\bf Table \ref{jennatara}}.) Our analysis reveals that, although many of the Riemann curvature tensors\footnote{Recall that they are computed as expectation values over the Glauber-Sudarshan states, as in \eqref{katusigel}, and are therefore in some sense {\it emergent} quantities.} become strong at late time, {\it i.e.} in the limit $\g_s \to 0$, their contributions to the quantum series \eqref{botsuga} $-$ which would naturally extend to the non-local and non-perturbative terms \cite{desitter2, coherbeta, coherbeta2} $-$ are generically small. In fact, inserting the log corrections, most of the $\mathbb{F}_i$ factors, if not all, provide dominant contributions of ${2\over 3} + {\hat\alpha_e(t) + \hat\beta_e(t)\over 2} - \hat\zeta_e(t)$ to \eqref{botsuga} for the ${\rm E}_8 \times {\rm E}_8$ case and $ {2\over 3} + \beta_e(t) - \zeta_e(t)$ for the $SO(32)$ case  with a few taking slightly different values as summarized in {\bf Table \ref{catdouble320}}. (The simplified $SO(32)$ case is summarized in {\bf Table \ref{cat320}}.) Despite this, the log corrections will turn out to be important in the subsequent analysis that we shall discuss below. This somewhat surprising result emphasizes once more that one has to carefully study the behavior of the Riemann tensors in the presence of the log corrections to ascertain their contributions to any quantum series, perturbative, non-perturbative or non-local ones. 

\begin{table}[tb]  
 \begin{center}
 \resizebox{\columnwidth}{!}{%
 \renewcommand{\arraystretch}{2.4}
\begin{tabular}{|c||c||c||c|}\hline $\mathbb{F}_{ie}(t)$  &  $SO(32)$ &
${\rm E}_8 \times {\rm E}_8$  \\ \hline\hline
$\mathbb{F}_{1e}(t)$ & $-{2\over 3} + 2\alpha_e(t) - \zeta_e(t) + {\beta_e(t)} + {2\over \vert\log~\bar{g}_s\vert}~\log\left({6 - 3\beta_e\over 2 - 3\alpha_e}\right)$ & $-{2\over 3} + 2\hat\sigma_e(t) - \hat\zeta_e(t) + {\hat\alpha_e(t) + \hat\beta_e(t)\over 2} + {2\over \vert\log~\bar{g}_s\vert}~\log\left({12 - 3\hat\alpha_e - 3\hat\beta_e\over 4 - 6\hat\sigma_e}\right)$\\ \hline
$\mathbb{F}_{2e}(t)$ & $-{2\over 3} + \alpha_e(t) -\zeta_e(t)  + 2\beta_e(t) + {1\over \vert \log~\bar{g}_s\vert} ~\log\left[{\big(6 -  3\beta_e\big)^2\over \big({2} - 3\alpha_e\big)\big({2} - 3\beta_e\big)}\right]
$ &  $ -{2\over 3} + \hat\sigma_e(t) -\hat\zeta_e(t)  + \left(\hat\alpha_e(t), \hat\beta_e(t)\right) + {\hat\alpha_e(t) + \hat\beta_e(t)\over 2} + {1\over \vert \log~\bar{g}_s\vert} ~\log\left[{\big(12 - 3\hat\alpha_e - 3\hat\beta_e\big)^2\over \big({4} - 6\hat\sigma_e\big)\big({4} - 6(\hat\alpha_e, \hat\beta_e)\big)}\right]$\\ \hline
$\mathbb{F}_{3e}(t) $ &  $-{5\over 3} + {3\beta_e(t)\over 2} + {1\over \vert\log~\bar{g}_s\vert}~\log\left({6 - 3\beta_e\over 2 - 3\beta_e}\right)$ &  $-{5\over 3} + (\hat\alpha_e(t), \hat\beta_e(t)) + {\hat\alpha_e(t) + \hat\beta_e(t)\over 4} + {1\over \vert\log~\bar{g}_s\vert}~\log\left({12 - 3\hat\alpha_e - 3\hat\beta_e\over 4 - 6(\hat\alpha_e, \hat\beta_e)}\right)$\\ \hline
$\mathbb{F}_{4e}(t) $ &  $-{5\over 3} + {3\beta_e(t)\over 2} +  {1\over \vert\log~\bar{g}_s\vert}~\log\left({6 - 3\beta_e\over 2 - 3\alpha_e}\right)$ &  $-{5\over 3} + {\hat\alpha_e(t) + \hat\beta_e(t)\over 4} + (\hat\alpha_e(t), \hat\beta_e(t)) + {1\over \vert\log~\bar{g}_s\vert}~\log\left({12 - 3\hat\alpha_e - 3\hat\beta_e\over 4 - 6\hat\sigma_e}\right)$ \\
\hline
\end{tabular}}
\renewcommand{\arraystretch}{1}
\end{center}
 \caption[]{\Su Summary of the dominant contributions of the $\mathbb{F}_{ie}(t)$ factors for the generalized $SO(32)$ and ${\rm E}_8 \times {\rm E}_8$ cases. From here one may easily extract the dominant contributions for the simplified $SO(32)$ case. Notice that the scalings are not the contributions to \eqref{botsuga} which, in turn are given in {\bf Tables \ref{catdouble320}} and {\bf \ref{cat320}}.} 
  \label{jennatara}
 \end{table}

\begin{table}[tb]  
 \begin{center}
 \resizebox{\columnwidth}{!}{%
 \renewcommand{\arraystretch}{2.5}
\begin{tabular}{|c||c||c||c||c|}\hline $\mathbb{F}_{ie}(t)$  &  $SO(32)$ &
${\rm E}_8 \times {\rm E}_8$ & Appears in \\ \hline\hline
$\mathbb{F}_{1e}(t), \mathbb{F}_{2e}(t), \mathbb{F}_{5e}(t), \mathbb{F}_{8e}(t), \mathbb{F}_{9e}(t)$ & ${2\over 3} + \beta_e(t) - \zeta_e(t)$ & ${2\over 3} + {\hat\alpha_e(t) + \hat\beta_e(t)\over 2} - \hat\zeta_e(t)$ & \eqref{2cho1bora} \\ \hline
$\mathbb{F}_{14e}(t), \mathbb{F}_{15e}(t), \mathbb{F}_{16e}(t), \mathbb{F}_{17e}(t)$ & ${2\over 3} + \beta_e(t) - \zeta_e(t)$ & ${2\over 3} + {\hat\alpha_e(t) + \hat\beta_e(t)\over 2} - \hat\zeta_e(t)$ & \eqref{2cho1bora} \\ \hline
$\mathbb{F}_{18e}(t), \mathbb{F}_{20e}(t), \mathbb{F}_{22e}(t), \mathbb{F}_{24e}(t), \mathbb{F}_{25e}(t)$ & ${2\over 3} + \beta_e(t) - \zeta_e(t)$ & ${2\over 3} + {\hat\alpha_e(t) + \hat\beta_e(t)\over 2} - \hat\zeta_e(t)$ & \eqref{2cho1bora} \\ \hline
$\mathbb{F}^{(1)}_{3e}(t), \mathbb{F}^{(1)}_{6e}(t)$ & ${5\over 3} + {\beta_e\over 2} - \zeta_e(t)$ & ${5\over 3} + {\hat\alpha_e(t) + \hat\beta_e(t)\over 4} - \hat\zeta_e(t)$ & \eqref{chu8hath1}\\ \hline
$\mathbb{F}^{(2)}_{3e}(t), \mathbb{F}^{(2)}_{6e}(t)$ & ${2\over 3} + {\beta_e\over 2} - {\alpha_e(t)\over 2} - {\zeta_e(t)\over 2}$ & ${2\over 3} + {\hat\alpha_e(t) + \hat\beta_e(t)\over 4} -{\hat\sigma_e(t)\over 2} - {\hat\zeta_e(t)\over 2}$ & \eqref{chu8hath1}\\ \hline
$\mathbb{F}^{(3)}_{3e}(t), \mathbb{F}^{(3)}_{6e}(t)$ & ${2\over 3}  - {\zeta_e(t)\over 2}$ & ${2\over 3} + {\hat\alpha_e(t) + \hat\beta_e(t)\over 4} -{1\over 2} (\hat\alpha_e(t), \hat\beta_e(t)) - {\hat\zeta_e(t)\over 2}$ & \eqref{chu8hath1}\\ \hline
$\mathbb{F}_{4e}(t), \mathbb{F}_{10e}(t), \mathbb{F}_{12e}(t)$ & ${2\over 3} +{\beta_e(t)\over 2} - {\alpha_e(t)\over 2} - {\zeta_e(t)\over 2}$ & ${2\over 3} + {\hat\alpha_e(t) + \hat\beta_e(t)\over 2}  - {\hat\sigma_e(t)\over 2} - {\hat\zeta_e(t)\over 2}$ & \eqref{duitrans}\\
\hline
\end{tabular}}
\renewcommand{\arraystretch}{1}
\end{center}
 \caption[]{\Su Dominant contributions of the $\mathbb{F}_{ie}(t)$ factors for the generalized $SO(32)$ and ${\rm E}_8 \times {\rm E}_8$ cases to the quantum scalings in \eqref{botsuga}. All parameters appearing here are linked to the equations that describe them. The functional forms for the $\mathbb{F}_{ie}(t)$ factors are provided in 
 {\bf Table \ref{privsocmey}}. One may compare these scalings with the simplified ones for the $SO(32)$ case appearing in {\bf Table \ref{cat320}}.} 
  \label{catdouble320}
 \end{table} 


\begin{table}[tb]  
 \begin{center}
\renewcommand{\arraystretch}{1.75}
\begin{tabular}{|c||c||c|}\hline $\mathbb{F}_i(t)$  & Dominant Contributions to \eqref{botsuga} \\ \hline\hline
$\mathbb{F}_{7}(t), \mathbb{F}_{11}(t), \mathbb{F}_{13}(t)$ & ${2\over 3}$ \\ \hline
$\mathbb{F}_1(t), \mathbb{F}_2(t), \mathbb{F}_4(t), \mathbb{F}_5(t)$ & ${2\over 3} + \beta(t)$ \\ \hline
$\mathbb{F}_8(t), \mathbb{F}_9(t), \mathbb{F}_{10}(t), \mathbb{F}_{12}(t)$ & ${2\over 3} + \beta(t)$ \\ \hline
$\mathbb{F}_{14}(t), \mathbb{F}_{15}(t), \mathbb{F}_{16}(t), \mathbb{F}_{17}(t)$ & ${2\over 3} + \beta(t)$ \\ \hline
$\mathbb{F}_{18}(t), \mathbb{F}_{19}(t), \mathbb{F}_{20}(t), \mathbb{F}_{21}(t)$ & ${2\over 3} + \beta(t)$ \\ \hline
$\mathbb{F}_{3}(t), \mathbb{F}_{6}(t)$ & ${2\over 3}, {2\over 3} + \beta(t), {5\over 3} + {\beta(t)\over 2}$ \\ \hline
\end{tabular} 
\renewcommand{\arraystretch}{1}
\end{center}
 \caption[]{\Su Dominant contributions of the $\mathbb{F}_i(t)$ factors for the $SO(32)$ case once the log corrections are inserted in but ignoring the sub-dominant contributions. Detailed derivations of these factors have appeared in section \ref{sec4.3} which the readers may look-up for more details.} 
  \label{cat320}
 \end{table} 


\begin{table}[tb]  
 \begin{center}
 \resizebox{\columnwidth}{!}{%
 \renewcommand{\arraystretch}{2.7}
\begin{tabular}{|c||c||c||c|}\hline Riemann tensors & Additive factors & Contributions to \eqref{botsuga}  \\ \hline\hline
${\bf R}_{mnpq}$ & ${4\over 3} - 2\hat\sigma_e(t)$ & ${\rm dom}\left({2\over 3} - \hat\sigma_e(t), ~{2\over 3} + {\hat\alpha_e(t) + \hat\beta_e(t)\over 2} - \hat\zeta_s(t),~ {8\over 3} - \hat\zeta_e(t),~ {2\over 3} - (\hat\alpha_e(t), \hat\beta_e(t))\right)$ \\ \hline 
${\bf R}_{mn\rho\sigma}$ & ${4\over 3} - \hat\sigma_e(t) - (\hat\alpha_e(t), \hat\beta_e(t))$ & ${\rm dom}\left({2\over 3} - \hat\sigma_e(t),~ {2\over 3}- (\hat\alpha_e(t), \hat\beta_e(t)), ~{2\over 3} + {\hat\alpha_e(t) + \hat\beta_e(t)\over 2} -\hat\zeta_e(t), ~ {8\over 3} - \hat\zeta_e(t),~ {2\over 3} \pm \hat\alpha_e(t) \mp \hat\beta_e(t) - (\hat\alpha_e(t), \hat\beta_e(t))\right)$ \\ \hline 
${\bf R}_{\rho\sigma i0}$ & $ {10\over 3} - (\hat\alpha_e(t), \hat\beta_e(t)) - \hat\zeta_e(t) $ &  ${5\over 3} - \hat\zeta_e(t) + {\hat\alpha_e(t) + \hat\beta_e(t)\over 4}$ \\ \hline 
${\bf R}_{m\rho\sigma 0}$ & $ {7\over 3} - (\hat\alpha_e(t), \hat\beta_e(t)) - {\hat\sigma_e(t)\over 2} -{\hat\zeta_e(t)\over 2}$ &  $ {2\over 3} + {\hat\alpha_e(t) + \hat\beta_e(t) \over 4} - {\hat\sigma_e(t) \over 2} - {\hat\zeta_e(t) \over 2} $ \\ \hline 
${\bf R}_{0\sigma\sigma\rho}$ & $ {7\over 3} - {\hat\zeta_e(t)\over 2} -{3\over 2}(\hat\alpha_e(t), \hat\beta_e(t)) $&  ${2\over 3} - {\hat\zeta_e(t)\over 2} - {1\over 2}(\hat\alpha_e(t), \hat\beta_e(t)) + {\hat\alpha_e(t) + \hat\beta_e(t) \over 4}$ \\ \hline 
${\bf R}_{\rho\sigma\rho\sigma}$ & ${4\over 3} - 2(\hat\alpha_e(t), \hat\beta_e(t))$  &  ${\rm dom}\left({2\over 3} - (\hat\alpha_e(t), \hat\beta_e(t)), ~{2\over 3} + {\hat\alpha_e(t) + \hat\beta_e(t)\over 2} - \hat\zeta_e(t), ~{2\over 3} - \hat\sigma_e(t), ~ {8\over 3} - \hat\zeta_e(t)\right)$ \\ \hline
${\bf R}_{mn i0}$ & $ {10\over 3} - \hat\sigma_e(t) - \hat\zeta_e(t)$  &  ${5\over 3} - \hat\zeta_e(t) + {\hat\alpha_e(t) + \hat\beta_e(t)\over 4}$ \\
\hline
\end{tabular}}
\renewcommand{\arraystretch}{1}
\end{center}
 \caption[]{\Su The contributions from the curvature tensors related to the generalized ${\rm E}_8 \times {\rm E}_8$ theory to the quantum scalings in \eqref{botsuga}, assuming no dependence on ${\mathbb{T}^2/{\cal G}}$ directions, and excluding all the logarithmic corrections. Here as in {\bf Table \ref{firzathai}}, $(m, n) \in {\cal M}_4, (\rho, \sigma) \in {\bf S}^1_{\theta_1} \times {{\bf S}^1_{\theta_2}/{\cal I}_{\theta_2}}, (i, j) \in {\bf R}^2$, $(a, b) \in {\mathbb{T}^2/{\cal G}}$, $\gamma_{1, 2}\big[{\hat\alpha_e(t) + \hat\beta_e(t)\over 3}\big]$ are defined from \eqref{ishena}, $(\hat\alpha_e(t), \hat\beta_e(t), \hat\sigma_e(t))$ are defined in \eqref{bdtran2}, and $(\hat\zeta_e(t), \hat\eta_e(t))$ are defined in \eqref{masatwork}. In computing the scalings all permutations of the curvature indices are taken into account. Note the positivity of all the elements in the third column.} 
\label{lilalo1}
 \end{table}

\begin{table}[tb]  
 \begin{center}
 \resizebox{\columnwidth}{!}{%
 \renewcommand{\arraystretch}{3.0}
\begin{tabular}{|c||c||c||c|}\hline Riemann tensors & Additive factors & Contributions to \eqref{botsuga}  \\ \hline\hline
${\bf R}_{m\rho\sigma i}$ & ${7\over 3} - {\hat\sigma_e(t)\over 2} - (\hat\alpha_e(t), \hat\beta_e(t)) - {\hat\zeta_e(t)\over 2}$ & ${5\over 3} - {\hat\sigma_e(t)\over 2} - {\hat\zeta_e(t)\over 2}$ \\ \hline 
${\bf R}_{i\sigma\sigma\rho}$ & ${7\over 3} - {\hat\zeta_e(t)\over 2} - {3\over 2}(\hat\alpha_e(t), \hat\beta_e(t))$ & ${5\over 3} - {\hat\zeta_e(t)\over 2} - {1\over 2}(\hat\alpha_e(t), \hat\beta_e(t))$ \\ \hline
${\bf R}_{m\sigma\sigma\rho}$ & ${4\over 3} - {\hat\sigma_e(t)\over 2} - {3\over 2}(\hat\alpha_e(t), \hat\beta_e(t))$ & ${2\over 3} - {\hat\sigma_e(t)\over 2} - {1\over 2}(\hat\alpha_e(t), \hat\beta_e(t))$ \\ \hline
${\bf R}_{mnpi}$ & ${7\over 3} - {3\hat\sigma_e(t)\over 2} - {\hat\zeta_e(t)\over 2}$ & ${5\over 3} - {\hat\sigma_e(t)\over 2} - {\hat\zeta_e(t)\over 2}$ \\ \hline
${\bf R}_{mnp\sigma}$ & ${4\over 3} - {3\hat\sigma_e(t)\over 2} - {1\over 2}(\hat\alpha_e(t), \hat\beta_e(t))$ & ${2\over 3} - {\hat\sigma_e(t)\over 2} - {1\over 2}(\hat\alpha_e(t), \hat\beta_e(t))$ \\ \hline
${\bf R}_{mn\sigma i}$ & ${7\over 3} - {\hat\sigma_e(t)} - {1\over 2}(\hat\alpha_e(t), \hat\beta_e(t)) - {\hat\zeta_e(t)\over 2}$ & ${5\over 3} - {\hat\zeta_e(t)\over 2} - {1\over 2}(\hat\alpha_e(t), \hat\beta_e(t))$ \\ \hline
${\bf R}_{mabi}$ & ${1\over 3} - {\hat\sigma_e(t)\over 2} - {\hat\zeta_e(t)\over 2} - (0, \hat\eta_e(t))$ & ${5\over 3} - {\hat\sigma_e(t)\over 2} - {\hat\zeta_e(t)\over 2}$ \\
\hline
\end{tabular}}
\renewcommand{\arraystretch}{1}
\end{center}
 \caption[]{\Su Contributions of the remaining Riemann tensors from {\bf Tables \ref{firzathai2}} and {\bf \ref{firzathai3}} to the quantum scalings in \eqref{botsuga} for the ${\rm E}_8 \times {\rm E}_8$ case. All other parameters are as the ones described in {\bf Table \ref{lilalo1}}. Note that, as in the previous table, all the curvature tensors contribute as {\it positive}  powers of $\bar{g}_s$, implying that they do not violate the underlying EFT. Interestingly, many of the curvature tensors (at least till row 12), provide unique contributions to the quantum scalings in \eqref{botsuga}.}
\label{lilalo3}
 \end{table}

\begin{table}[tb]  
 \begin{center}
 \resizebox{\columnwidth}{!}{%
 \renewcommand{\arraystretch}{1.7}
\begin{tabular}{|c||c||c||c|}\hline Riemann tensors & Additive factors & Contributions to \eqref{botsuga}  \\ \hline\hline
${\bf R}_{mnpq}$ & ${4\over 3} - 2\alpha_e(t)$ & ${\rm dom}\left({2\over 3} - \alpha_e(t), ~{2\over 3} + \beta_e(t) - \zeta_e(t),~ {8\over 3} - \zeta_e(t),~ {2\over 3} - \beta_e(t)\right)$ \\ \hline 
${\bf R}_{mn\rho\sigma}$ & ${4\over 3} - \alpha_e(t) - \beta_e(t)$ & ${\rm dom}\left({2\over 3} - \alpha_e(t),~ {2\over 3}- \beta_e(t), ~{2\over 3} + {\beta_e(t)} - \zeta_e(t), ~ {8\over 3} -\zeta_e(t)\right)$ \\ \hline 
${\bf R}_{\rho\sigma i0}$ & $ {10\over 3} - \beta_e(t) - \zeta_e(t) $ &  ${5\over 3} - \zeta_e(t) + {\beta_e(t)\over 2}$ \\ \hline 
${\bf R}_{m\rho\sigma 0}$ & $ {7\over 3} - \beta_e(t) - {\alpha_e(t)\over 2} - {\zeta_e(t)\over 2} $ &  ${2\over 3} + {\beta_e(t) \over 2} - {\alpha_e(t) \over 2} - {\zeta_e(t) \over 2}$ \\ \hline 
${\bf R}_{0\sigma\sigma\rho}$ & $ {7\over 3} - {\zeta_e(t)\over 2} - {3\beta_e(t)\over 2} $&  ${2\over 3} - {\zeta_e(t)\over 2}$ \\ \hline 
${\bf R}_{\rho\sigma\rho\sigma}$ & ${4\over 3} - 2\beta_e(t)$  &  ${\rm dom}\left({2\over 3} - \beta_e(t), ~{2\over 3} + \beta_e(t) - \zeta_e(t), ~{2\over 3} - \alpha_e(t), ~ {8\over 3} - \zeta_e(t)\right)$ \\ \hline
${\bf R}_{mn i0}$ & $ {10\over 3} - \alpha_e(t) - \zeta_e(t)$  &  ${5\over 3} - \zeta_e(t) + {\beta_e(t)\over 2}$ \\
\hline
\end{tabular}}
\renewcommand{\arraystretch}{1}
\end{center}
 \caption[]{\Su The contributions from the curvature tensors related to the generalized $SO(32)$ theory to the quantum scalings in \eqref{botsuga}, assuming no dependence on ${\mathbb{T}^2/{\cal G}}$ directions, and excluding all the logarithmic corrections. Here as in {\bf Table \ref{firzathai}}, $(m, n) \in {\cal M}_4, (\rho, \sigma) \in \mathbb{T}^2/{\cal I}_{\rho\sigma}, (i, j) \in {\bf R}^2$, $(a, b) \in {\mathbb{T}^2/{\cal G}}$, $\gamma_{1, 2}\big[{\hat\alpha_e(t) + \hat\beta_e(t)\over 3}\big]$ are defined from \eqref{ishena}, $(\hat\alpha_e(t), \hat\beta_e(t), \hat\sigma_e(t))$ are defined in \eqref{bdtran2}, and $(\hat\zeta_e(t), \hat\eta_e(t))$ are defined in \eqref{masatwork}. In computing the scalings, all permutations of the curvature indices are taken into account. Comparing the scalings to the ones in {\bf Table \ref{lilalo1}}, we see certain simplifications coming from \eqref{tiffmont}. Note again the positivity of all the elements in the third column.} 
\label{lilalo2}
 \end{table}

\begin{table}[tb]  
 \begin{center}
 \resizebox{\columnwidth}{!}{%
 \renewcommand{\arraystretch}{2.44}
\begin{tabular}{|c||c||c||c|}\hline Riemann tensors & Additive factors & Contributions to \eqref{botsuga}  \\ \hline\hline
${\bf R}_{m\rho\sigma i}$ & ${7\over 3} - {\alpha_e(t)\over 2} - \beta_e(t) - {\zeta_e(t)\over 2}$ & ${5\over 3} - {\alpha_e(t)\over 2} - {\zeta_e(t)\over 2}$ \\ \hline 
${\bf R}_{i\sigma\sigma\rho}$ & ${7\over 3} - {\zeta_e(t)\over 2} - {3\beta_e(t)\over 2}$ & ${5\over 3} - {\zeta_e(t)\over 2} - {\beta_e(t)\over 2}$ \\ \hline
${\bf R}_{m\sigma\sigma\rho}$ & ${4\over 3} - {\alpha_e(t)\over 2} - {3\beta_e(t)\over 2}$ & ${2\over 3} - {\alpha_e(t)\over 2} - {\beta_e(t)\over 2}$ \\ \hline
${\bf R}_{mnpi}$ & ${7\over 3} - {3\alpha_e(t)\over 2} - {\zeta_e(t)\over 2}$ & ${5\over 3} - {\alpha_e(t)\over 2} - {\zeta_e(t)\over 2}$ \\ \hline
${\bf R}_{mnp\sigma}$ & ${4\over 3} - {3\alpha_e(t)\over 2} - {\beta_e(t)\over 2}$ & ${2\over 3} - {\alpha_e(t)\over 2} - {\beta_e(t)\over 2}$ \\ \hline
${\bf R}_{mn\sigma i}$ & ${7\over 3} - {\alpha_e(t)} - {\beta_e(t)\over 2} - {\zeta_e(t)\over 2}$ & ${5\over 3} - {\zeta_e(t)\over 2} - {\beta_e(t)\over 2}$ \\ \hline
${\bf R}_{mabi}$ & ${1\over 3} - {\alpha_e(t)\over 2} - {\zeta_e(t)\over 2} - (0, \eta_e(t))$ & ${5\over 3} - {\alpha_e(t)\over 2} - {\zeta_e(t)\over 2}$ \\
\hline
\end{tabular}}
\renewcommand{\arraystretch}{1}
\end{center}
 \caption[]{\Su Contributions of the remaining Riemann tensors from {\bf Tables \ref{firzathai2}} and {\bf \ref{firzathai3}} to the quantum scalings in \eqref{botsuga} for the generalized $SO(32)$ case. All other parameters are as the ones described in {\bf Table \ref{lilalo2}}. Note that, as in the previous table and for the ${\rm E}_8 \times {\rm E}_8$ case, all the curvature tensors contribute as {\it positive}  powers of $\bar{g}_s$, implying that they do not violate the underlying EFT. In a similar vein as in {\bf Table \ref{lilalo3}}, many of the curvature tensors (at least till row 12), provide unique contributions to the quantum scalings in \eqref{botsuga}.}
\label{lilalo4}
 \end{table}

\begin{table}[tb]  
 \begin{center}
 \resizebox{\columnwidth}{!}{%
 \renewcommand{\arraystretch}{2.2}
\begin{tabular}{|c||c||c||c|}\hline G-fluxes & Contributions from ${\rm E}_8 \times {\rm E}_8$ & Contributions from $SO(32)$ \\ \hline\hline
${\bf G}_{mnpq}$ & $\hat{l}_{emn}^{pq}(t) + {4\over 3} - 2\hat\sigma_e(t)$ & $l_{emn}^{pq}(t) + {4\over 3} - 2\alpha_e(t)$ \\ \hline 
${\bf G}_{mnp\alpha}$ & $\hat{l}_{emn}^{p\alpha}(t) + {4\over 3} - {3\hat\sigma_e(t)\over 2} - {1\over 2}(\hat\alpha_e(t), \hat\beta_e(t))$ & $l_{emn}^{p\alpha}(t) + {4\over 3} - {3\alpha_e(t)\over 2} - {\beta_e(t)\over 2}$ \\ \hline 
${\bf G}_{mnpa}$ & $\hat{l}_{emn}^{pa} + {1\over 3} -{3\hat\sigma_e(t)\over 2} - {1\over 2} (0, \hat\eta_e(t))$
& ${l}_{emn}^{pa} + {1\over 3} -{3\alpha_e(t)\over 2} - {1\over 2} (0, \eta_e(t))$\\ \hline
${\bf G}_{mn\alpha\beta}$ & $\hat{l}_{emn}^{\alpha\beta}(t) + {4\over 3} - {\hat\sigma_e(t)} - {\hat\alpha_e(t) +\hat\beta_e(t)\over 2}$ & $l_{emn}^{\alpha\beta}(t) + {4\over 3} - {\alpha_e(t)} - {\beta_e(t)}$ \\ \hline 
${\bf G}_{mn\alpha a}$ & $\hat{l}_{emn}^{\alpha a}(t) + {1\over 3} - {\hat\sigma_e(t)} - {1\over 2}(\hat\alpha_e(t), \hat\beta_e(t)) - {1\over 2}(0, \hat\eta_e(t))$ & $l_{emn}^{\alpha a}(t) + {1\over 3} - {\alpha_e(t)} - {\beta_e(t)\over 2} - {1\over 2} (0, \eta_e(t))$ \\ \hline 
${\bf G}_{m\alpha\beta a}$ & $\hat{l}_{e\alpha\beta}^{ma}(t) + {1\over 3} - {\hat\sigma_e(t)\over 2} - {\hat\alpha_e(t) + \hat\beta_e(t)\over 2} - {1\over 2}(0, \hat\eta_e(t))$ & $l_{e\alpha\beta}^{ma}(t) + {1\over 3} - {\alpha_e(t)\over 2} - {\beta_e(t)} - {1\over 2} (0, \eta_e(t))$ \\ \hline 
${\bf G}_{0ijm}$ & $\hat{l}_{eij}^{0m}(t) + {13\over 3} - {\hat\sigma_e(t)\over 2} - {3\hat\zeta_e(t)\over 2}$ & $l_{eij}^{0m}(t) + {13\over 3} - {\alpha_e(t)\over 2} - {3\zeta_e(t)\over 2}$ \\
\hline 
\end{tabular}}
\renewcommand{\arraystretch}{1}
\end{center}
 \caption[]{\Su Contributions from the G-flux components to the quantum scalings in \eqref{botsuga}. Note the appearance of ${\red -{2\over 3}}$ factors for G-flux components ${\bf G}_{{\rm MN}ab}$ where ${\rm M, N} \in {\cal M}_4 \times {\cal M}_2$. Because of the negative sign, for $\hat{l}_{e{\rm MN}}^{ab} = l_{e{\rm MN}}^{ab} = 0$, there would be no EFT description possible as described in \cite{desitter2, coherbeta, coherbeta2, joydeep}.}
\label{lilaloo}
 \end{table}

\begin{table}[tb]  
 \begin{center}
 \resizebox{\columnwidth}{!}{%
 \renewcommand{\arraystretch}{2.2}
\begin{tabular}{|c||c||c||c|}\hline G-fluxes & Contributions from ${\rm E}_8 \times {\rm E}_8$ & Contributions from $SO(32)$ \\ \hline\hline
${\bf G}_{ma\alpha i}$ & $\hat{l}_{ema}^{\alpha i}(t) + {4\over 3} -{\hat\sigma_e(t)\over 2} - {1\over 2}(\hat\alpha_e(t), \hat\beta_e(t)) - {1\over 2}(0, \hat\eta_e(t))  - {\hat\zeta_e(t)\over 2}$ & $l_{ema}^{\alpha i}(t) + {4\over 3} -{\alpha_e(t)\over 2} - {\beta_e(t)\over 2} - {1\over 2}(0, \eta_e(t)) - {\zeta_e(t)\over 2}$ \\ \hline 
${\bf G}_{mnij}$ & $\hat{l}_{emn}^{ij}(t) + {10\over 3} -{\hat\sigma_e(t)}  - {\hat\zeta_e(t)}$ & $l_{emn}^{ij}(t) + {10\over 3} -{\alpha_e(t)} -  {\zeta_e(t)}$ \\ \hline 
${\bf G}_{m\alpha ij}$ & $\hat{l}_{em\alpha}^{ij}(t) + {10\over 3} -{\hat\sigma_e(t)\over 2} -{1\over 2}(\hat\alpha_e(t), \hat\beta_e(t))  - {\hat\zeta_e(t)}$ & $l_{em\alpha}^{ij}(t) + {10\over 3} -{\alpha_e(t)\over 2} - {\beta_e(t)\over 2} -  {\zeta_e(t)}$ \\ \hline 
${\bf G}_{\alpha\beta ij}$ & $\hat{l}_{e\alpha\beta}^{ij}(t) + {10\over 3} -{\hat\alpha_e(t) +\hat\beta_e(t)\over 2}  - {\hat\zeta_e(t)}$ & $l_{e\alpha\beta}^{ij}(t) + {10\over 3}  - {\beta_e(t)} -  {\zeta_e(t)}$ \\ \hline 
${\bf G}_{maij}$ & $\hat{l}_{ema}^{ij}(t) + {7\over 3} -{\hat\sigma_e(t)\over 2} -{1\over 2}(0, \hat\eta_e(t))  - {\hat\zeta_e(t)}$ & $l_{ema}^{ij}(t) + {7\over 3} -{\alpha_e(t)\over 2} - {1\over 2} (0, \eta_e(t)) -  {\zeta_e(t)}$ \\ \hline 
${\bf G}_{\alpha aij}$ & $\hat{l}_{e\alpha a}^{ij}(t) + {7\over 3} -{1\over 2}(\hat\alpha_e(t), \hat\beta_e(t)) -{1\over 2}(0, \hat\eta_e(t))  - {\hat\zeta_e(t)}$ & $l_{e\alpha a}^{ij}(t) + {7\over 3} -{\beta_e(t)\over 2} - {1\over 2} (0, \eta_e(t)) -  {\zeta_e(t)}$ \\ \hline
${\bf G}_{abij}$ & $\hat{l}_{eab}^{ij}(t) + {4\over 3} - {\hat\eta_e(t)\over 2}  - {\hat\zeta_e(t)}$ & $l_{eab}^{ij}(t) + {4\over 3} - {\eta_e(t)\over 2} -  {\zeta_e(t)}$ \\
\hline
\end{tabular}}
\renewcommand{\arraystretch}{1}
\end{center}
 \caption[]{\Su Remaining contributions from the G-flux components to the quantum scalings in \eqref{botsuga}.}
\label{morganL1}
 \end{table}


\begin{table}[tb]  
 \begin{center}
 \resizebox{\columnwidth}{!}{%
 \renewcommand{\arraystretch}{4.1}
\begin{tabular}{|c||c||c||c|}\hline Riemann Tensors  &  ${\rm E}_8 \times {\rm E}_8$ &
$SO(32)$ \\ \hline\hline
${\bf R}_{ijij}, {\bf R}_{abij}, {\bf R}_{mnij}, {\bf R}_{\rho\sigma ij}, {\bf R}_{mnab}$ & {} & {} \\
${\bf R}_{\rho\sigma ab}, {\bf R}_{abab}, {\bf R}_{\rho\sigma\rho\sigma}$ & ${\rm dom}\left({8\over 3} - \hat\zeta_e(t), {2\over 3} - \hat\sigma_e(t), {2\over 3} - (\hat\alpha_e(t), \hat\beta_e(t)), {2\over 3} + {\hat\alpha_e(t) + \hat\beta_e(t)\over 2} - \hat\zeta_e(t)\right)$ & ${\rm dom}\left({8\over 3} - \zeta_e(t), {2\over 3} - \alpha_e(t), {2\over 3} - \beta_e(t), {2\over 3} + {\beta_e(t)} - \hat\zeta_e(t)\right)$\\
${\bf R}_{mnpq}, {\bf R}_{0i0j}, {\bf R}_{0a0b}, {\bf R}_{0m0n}, {\bf R}_{0\rho 0\sigma}$ & {} & {} \\ \cline{1-1}\cline{2-2}
${\bf R}_{mn\rho\sigma}$ & ${\rm dom}\left({2\over 3} - \hat\sigma_e(t),~ {2\over 3}- (\hat\alpha_e(t), \hat\beta_e(t)), ~{2\over 3} + {\hat\alpha_e(t) + \hat\beta_e(t)\over 2} -\hat\zeta_e(t), ~ {8\over 3} - \hat\zeta_e(t),~ {2\over 3} \pm \hat\alpha_e(t) \mp \hat\beta_e(t) - (\hat\alpha_e(t), \hat\beta_e(t))\right)$ & {} \\ \hline
${\bf R}_{\rho\sigma i0}, {\bf R}_{mni0}, {\bf R}_{abi0}, {\bf R}_{ijk0}$ & ${5\over 3} + {\hat\alpha_e(t) + \hat\beta_e(t)\over 4} - \hat\zeta_e(t)$ & ${5\over 3} + {\beta_e(t)\over 2} - \zeta_e(t)$ \\ \hline
${\bf R}_{m\rho\sigma i}, {\bf R}_{mnpi}, {\bf R}_{mabi}, {\bf R}_{mijk}, {\bf R}_{m0i0}$ & ${5\over 3} + {\hat\sigma_e(t)\over 2} - {\hat\zeta_e(t)\over 2}$ & ${5\over 3} + {\alpha_e(t)\over 2} - {\zeta_e(t)\over 2}$ \\ \hline
${\bf R}_{i\sigma\sigma \rho}, {\bf R}_{mn\sigma i}, {\bf R}_{\sigma ab i}, {\bf R}_{0\sigma i0}, {\bf R}_{\sigma ijk}$ & ${5\over 3} - {1\over 2}(\hat\alpha_e(t), \hat\beta_e(t)) - {\hat\zeta_e(t)\over 2}$  & ${5\over 3} - {\beta_e(t)\over 2} - {\hat\zeta_e(t)\over 2}$\\
\hline
\end{tabular}}
\renewcommand{\arraystretch}{1}
\end{center}
 \caption[]{\Su Collecting the contributions from {\bf Tables \ref{lilalo1}}, {\bf \ref{lilalo3}}, {\bf \ref{lilalo2}} and {\bf \ref{lilalo4}} to \eqref{botsuga}. Note that all the scalings are positive definite so there is no violation of EFT in both the Heterotic theories.}
\label{scarstan23}
 \end{table}

\begin{table}[tb]  
 \begin{center}
\renewcommand{\arraystretch}{1.5}
\begin{tabular}{|c||c||c|}\hline Riemann tensors  & Contributions to \eqref{botsuga} \\ \hline\hline
$ {\bf R}_{ijk0},  {\bf R}_{abi0}$ & ${5\over 3} + \gamma_{1,2}(t)$\\ \hline
${\bf R}_{\alpha\beta i0}, {\bf R}_{mni0}, {\bf R}_{m\alpha\beta i}, {\bf R}_{mnpi}, {\bf R}_{m0i0}, {\bf R}_{mijk}, {\bf R}_{mabi}$ & ${5\over 3} + 
{\beta(t)\over 2}$\\ \hline
${\bf R}_{i\alpha\alpha\beta}, {\bf R}_{mn\alpha i}, {\bf R}_{\alpha 0i0}, {\bf R}_{\alpha ijk}, 
{\bf R}_{\alpha ab i}$ & ${5\over 3} - {\beta(t) \over 2}$\\ \hline
${\bf R}_{mnp0}, {\bf R}_{m\alpha\beta 0}, {\bf R}_{m0ij}, {\bf R}_{0mab}$ & ${2\over 3} + {\beta(t)}$\\ \hline
${\bf R}_{mn\alpha 0}, {\bf R}_{0\alpha\alpha\beta}, {\bf R}_{\alpha 0ij}, {\bf R}_{0\alpha ab}$ & ${2\over 3}$\\ \hline
${\bf R}_{mnp\alpha}, {\bf R}_{m\alpha\alpha\beta}, {\bf R}_{m\alpha ab}, {\bf R}_{m\alpha ij}, 
{\bf R}_{0m0\alpha}$ & ${2\over 3}$\\ \hline
${\bf R}_{0\alpha0\beta}, {\bf R}_{ijij}, {\bf R}_{0i0j},  {\bf R}_{abij}, {\bf R}_{0a0b}, {\bf R}_{\alpha\beta \alpha\beta}, {\bf R}_{0m0n}$ & ${\rm dom}\left({2\over 3} -\alpha(t), {2\over 3} - \beta(t),  {8\over 3}\right)$\\ \hline
${\bf R}_{mnpq}, {\bf R}_{mn\alpha\beta}, {\bf R}_{\alpha\beta ab},  {\bf R}_{abab}, {\bf R}_{mnij}, {\bf R}_{\alpha\beta ij}, {\bf R}_{mnab}$ & ${\rm dom}\left({2\over 3} -\alpha(t), {2\over 3} - \beta(t), {8\over 3}\right)$\\ \hline
 \end{tabular} 
\renewcommand{\arraystretch}{1}
\end{center}
 \caption[]{\Su The ${g_s\over {\rm H H}_o}$ scalings of the curvature tensors as they appear in \eqref{botsuga} for the simplified $SO(32)$ heterotic theory. We have used $\alpha(t) = - \beta(t)$ to further simplify the analysis.} 
  \label{firzacut4}
 \end{table} 

\begin{table}[tb]  
 \begin{center}
\resizebox{\columnwidth}{!}{%
\renewcommand{\arraystretch}{2.3}
\begin{tabular}{|c||c||c||c||c|}\hline ${\bf G}_{\rm ABCD}$ tensors & Class & {\rm coeff}$\left({\rm G}^{\rm ABCD}\right)$ & Contributions to \eqref{botsuga} \\ \hline\hline
${\bf G}_{mnpq}$ & $\mathbb{G}_9$ & $a^{-8/3} {\rm H}^{-16/3} {\rm F}_2^{-4}$ & $l_{mn}^{pq} + {4\over 3} - 2\alpha(t)$ \\ \hline 
${\bf G}_{mnp\alpha}$ & $\mathbb{G}_9$ &$a^{-8/3} {\rm H}^{-16/3} {\rm F}_1^{-1} {\rm F}_2^{-3}$ & $l_{mn}^{p\alpha} + {4\over 3} - {3\alpha(t)\over 2} - {\beta(t)\over 2}$ \\ \hline 
${\bf G}_{mnpa}$ & $\mathbb{G}_{10}$ & $a^{-2/3} {\rm H}^{-16/3} {\rm F}_2^{-3}$ & $l_{mn}^{pa} + {1\over 3} - {3\alpha_o\over 2}$ \\ \hline 
${\bf G}_{mn\alpha\beta}$ & $\mathbb{G}_9$ & $a^{-8/3} {\rm H}^{-16/3} {\rm F}_1^{-2} {\rm F}_2^{-2}$ & $l_{mn}^{\alpha\beta} + {1\over 3} - {\alpha(t)\over 2} - {\beta(t)}$ \\ \hline 
${\bf G}_{0ijm}$ & $\mathbb{G}_{6}$ &$a^{-26/3} {\rm H}^{20/3} {\rm F}_2^{-1}$ & $l_{0i}^{jm} + {13\over 3} - {\alpha_o\over 2}$ 
\\ \hline 
${\bf G}_{0ij\alpha}$ & $\mathbb{G}_{6}$ &$a^{-26/3} {\rm H}^{20/3} {\rm F}_1^{-1}$ & $l_{0i}^{j\alpha} + {13\over 3}  - 
{\beta(t)\over 2}$ \\ \hline 
${\bf G}_{0ija}$ & $\mathbb{G}_{7}$ &$a^{-20/3}$ & $l_{0i}^{ja} + {10\over 3}$ \\
\hline 
\end{tabular}}
\renewcommand{\arraystretch}{1}
\end{center}
 \caption[]{\Su G-Fluxes and their ${g_s\over {\rm HH_o}}$ scalings shown on the fourth column contributing to \eqref{botsuga} for the simplified $SO(32)$ case. The second column depicts the classes shown in {\bf Table \ref{lokeys}}, and the third column depicts the precise coefficients of the G-fluxes entering \eqref{botsuga} in terms of the warp-factors $({\rm H}(y), {\rm F}_i(t))$ and the type IIB FRLW metric parameter $a(t)$ from \eqref{viomyer1}.}
  \label{fchumbon5}
 \end{table}

\begin{table}[tb]  
 \begin{center}
\resizebox{\columnwidth}{!}{%
\renewcommand{\arraystretch}{2.1}
\begin{tabular}{|c||c||c||c||c|}\hline ${\bf G}_{\rm ABCD}$ tensors & Class & {\rm coeff}$\left({\bf G}^{\rm ABCD}\right)$ & Contributions to \eqref{botsuga} \\ \hline\hline
${\bf G}_{mnp\mu}$ & $\mathbb{G}_1, \mathbb{G}_{12}$ & $a^{-14/3} {\rm H}^{-4/3}  {\rm F}_2^{-3}$ & $l_{mn}^{p\mu} + {7\over 3} - {3\alpha(t)\over 2}$ \\ \hline 
${\bf G}_{m\alpha\beta\mu}$ & $\mathbb{G}_1, \mathbb{G}_{12}$ & $a^{-14/3} {\rm H}^{-4/3} {\rm F}_1^{-2} {\rm F}_2^{-1}$ & $l_{m\alpha}^{\beta\mu} + {7\over 3} - {\alpha(t)\over 2} - {\beta(t)}$ \\ \hline 
${\bf G}_{mn\alpha\mu}$ & $\mathbb{G}_1, \mathbb{G}_{12}$ & $a^{-14/3} {\rm H}^{-4/3} {\rm F}_1^{-1} {\rm F}_2^{-2}$ & $l_{mn}^{\alpha\mu} + {7\over 3} - {\alpha(t)} - {\beta(t)\over 2}$ \\ \hline 
${\bf G}_{mna\mu}$ & $\mathbb{G}_2, \mathbb{G}_{13}$ & $a^{-8/3} {\rm H}^{-4/3} {\rm F}_2^{-2}$ & $l_{mn}^{a\mu} + {4\over 3} - {\alpha(t)}$ \\ \hline 
${\bf G}_{abm\mu}$ & $\mathbb{G}_3, \mathbb{G}_{14}$ &$a^{-2/3} {\rm H}^{-4/3} {\rm F}_2^{-1}$ & $l_{ab}^{m\mu} + {1\over 3} - {\alpha(t)\over 2}$ \\ \hline 
${\bf G}_{\alpha\beta a\mu}$ & $\mathbb{G}_2, \mathbb{G}_{13}$ & $a^{-8/3} {\rm H}^{-4/3} {\rm F}_1^{-2}$ & $l_{\alpha\beta}^{a\mu} + {4\over 3}  - {\beta(t)\over 2}$ \\ \hline 
${\bf G}_{ab\alpha\mu}$ & $\mathbb{G}_3, \mathbb{G}_{14}$ &$a^{-2/3} {\rm H}^{-4/3} {\rm F}_1^{-1}$ & $l_{ab}^{\alpha\mu} + {1\over 3}  - {\beta(t)\over 2}$ \\
\hline 
\end{tabular}}
\renewcommand{\arraystretch}{1}
\end{center}
 \caption[]{\Su G-Fluxes and their ${g_s\over {\rm HH_o}}$ scalings shown in the fourth column, similar to what we had in {\bf Table \ref{fchumbon5}}, contributing to \eqref{botsuga} for the simplified $SO(32)$ case. The difference is in the second column which depicts the components of fluxes each of them are now being represented by two different classes from {\bf Table \ref{lokeys}}.}
  \label{fchumbon6}
 \end{table}


\subsection{$g_s$ scalings of \eqref{botsuga}, constraints and revisiting the EFT criteria \label{sec4.55}}

With all the results collected so far, we are now ready for the main set of computations in the paper. These computations involve (a) studying the scalings of the quantum series \eqref{botsuga} for all the heterotic theories, (b) analyzing the Bianchi identities, flux quantizations and anomaly cancellations and (c) studying the Schwinger-Dyson equations for all the on and off-shell metric components. In this section we will study the scalings of the perturbative quantum series \eqref{botsuga} with special emphasis on the logarithmic corrections that we discussed in much details in section \ref{sec4.3}. In the following, let us start from this.

\subsubsection{The uncanceled log corrections to the curvature tensors \label{sec4.5.1}}

The logarithmic corrections studied in section \ref{sec4.3} may be divided into two sets: one that involve $(\alpha_e(t), \beta_e(t))$ and 
$(\hat\alpha_e(t), \hat\beta_e(t), \hat\sigma_e(t))$ for the generalized $SO(32)$ and the ${\rm E}_8 \times {\rm E}_8$ theories respectively, and the other that involve their corresponding temporal derivatives. Both of these corrections to the $\mathbb{F}_{ie}(t)$ factors, discussed in section \ref{sec4.3}, are sub-dominant\footnote{Meaning that the temporally invariant parts largely dominate over the temporally dependent 
warp-factors.}, but that doesn't mean that they would also contribute sub-dominantly to the quantum series \eqref{botsuga}. The surprising feature of our analysis is that, such logarithmic terms do contribute non-trivially and excluding them would not give us the full answer when we study the corresponding Schwinger-Dyson equations. To see this, we will have to go back to the analysis in section \ref{sec4.3} and look at the contributions from each of the $\mathbb{F}_{ie}(t)$ terms to the curvature tensors more carefully. 

The logarithmic corrections to the $\mathbb{F}_{ie}(t)$ factors, at least the ones involving $(\alpha_e(t), \beta_e(t))$ and 
$(\hat\alpha_e(t), \hat\beta_e(t), \hat\sigma_e(t))$ for the generalized $SO(32)$ and the ${\rm E}_8 \times {\rm E}_8$ theories respectively, are carefully tabulated in {\bf Table \ref{jennatara}}. While they appear as sub-dominant contributions in $\mathbb{F}_{ie}(t)$, they can in fact contribute non-trivially to the Riemann tensors, and therefore to the quantum scaling of \eqref{botsuga}. On the other hand, the logarithmic corrections involving derivatives of these factors only contribute sub-dominantly to the Riemann tensors and therefore also sub-dominantly to \eqref{botsuga}. 

There are {\it four} distinct terms that contribute to the logarithmic corrections for the $SO(32)$ and the ${\rm E}_8 \times {\rm E}_8$ cases respectively. These may be easily extracted from {\bf Table \ref{jennatara}}, and we collect them here for the two cases as sets 
${\cal S}_{32}$ and ${\cal S}_8$ respectively whose elements are:

{\footnotesize
\bg\label{emilxgulab}
&& {\cal S}_{32} = \left(1, ~ {4-6\alpha_e(t)\over 6 - 3\beta_e(t)}, ~ {4 - 6\beta_e(t)\over 6 - 3\beta_e(t)}, ~ {16 - 6\zeta_e(t)\over 6 - 3\beta_e(t)}, ~ {8 + 6(0, \eta_e(t))\over 6 - 3\beta_e(t)}\right)\\
&& {\cal S}_8 = \left(1, ~ {4-6\hat\sigma_e(t) \over 12 - 3\hat\alpha_e(t) - 3\hat\beta_e(t)}, ~ {4 - 6(\hat\alpha_e(t), \hat\beta_e(t))\over 12 - 3\hat\alpha_e(t) - 3\hat\beta_e(t)}, ~ {16 - 6\hat\zeta_e(t)\over 12 - 3\hat\alpha_e(t) - 3\hat\beta_e(t)}, ~ {8 + 6(0, \hat\eta_e(t))\over 12 - 3\hat\alpha_e(t) - 3\hat\beta_e(t)}\right),\nonumber \nd}
where all the parameters appearing in \eqref{emilxgulab} have been defined in section \ref{sec4.4}. By definition, these parameters accommodate most, if not all, of the higher order perturbative and non-perturbative corrections. We can now extract the logarithmic corrections from the two sets above by following some simple mathematical operations. Define a set ${\cal S}$ with commuting elements and the associated operations on it as:

{\footnotesize
\bg\label{dollarhaa}
{\cal S} = (a, b, c, d), ~~~~ {\cal S} \ast {\cal S} \equiv (a^2, b^2, c^2, d^2), ~~~~ {\cal S} \otimes {\cal S} \equiv (a^2, b^2, c^2, d^2, ab, ac, ad, bc, bd, cd), \nd}
with $ab = ba$ are taken to be one element. Using \eqref{dollarhaa}, the terms contributing to the logarithmic corrections for the $SO(32)$ and the ${\rm E}_8 \times {\rm E}_8$ cases will respectively come from the following two sets:
\bg\label{creeprani}
{\cal S}_{32} \otimes {\cal S}_{32} - {\cal S}_{32} \ast {\cal S}_{32}, ~~~~ {\rm and}~~~~ {\cal S}_{8} \otimes {\cal S}_{8} - {\cal S}_{8} \ast {\cal S}_{8}, \nd
thus keeping only the cross-terms. The logarithmic corrections may now be quantified by using the following strategy. First, define two column matrices, $\mathbb{M}_{32}$ and $\mathbb{M}_8$, by taking the elements from the two sets in \eqref{creeprani} for the $SO(32)$ and the ${\rm E}_8 \times {\rm E}_8$ cases respectively. The ten elements from each of the two sets in \eqref{creeprani} would form the ten elements of the column matrices $\mathbb{M}_{32}$ and $\mathbb{M}_8$. If we define another column matrix $\mathbb{N}_n$ whose elements are taken from the set ${\cal S}_n \equiv (a_n, b_n, c_n, d_n, ..., j_n)$, with $(a_n, b_n, ..., j_n) \in (\mathbb{Z}, \mathbb{Z}, ..., \mathbb{Z})$, then the logarithmic corrections may be expressed as:

{\scriptsize
\bg\label{crawlranipa}
-{1\over \vert\log~\bar{g}_s\vert}~\log\left(\mathbb{N}_n^\top \mathbb{M}_{32}\right) &= & -{1\over \vert\log~\bar{g}_s\vert}~\log\Bigg[a_n
\left( {4-6\alpha_e(t)\over 6 - 3\beta_e(t)}\right) + b_n\left({4 - 6\beta_e(t)\over 6 - 3\beta_e(t)}\right)+ ... \\
&+& e_n\left({(4-6\alpha_e(t))(4 - 6\beta_e(t))\over (6 - 3\beta_e(t))^2}\right) + ...+   j_n\left({(16 - 6\zeta_e(t)) (8 + 6(0, \eta_e(t)))\over (6 - 3\beta_e(t))^2}\right) \Bigg], \nonumber\\
-{1\over \vert\log~\bar{g}_s\vert}~\log\left(\mathbb{N}_n^\top \mathbb{M}_{8}\right) &= & -{1\over \vert\log~\bar{g}_s\vert}~\log\Bigg[a_n\left({4-6\hat\sigma_e(t) \over 12 - 3\hat\alpha_e(t) - 3\hat\beta_e(t)}\right) + b_n\left({4 - 6(\hat\alpha_e(t), \hat\beta_e(t))\over 12 - 3\hat\alpha_e(t) - 3\hat\beta_e(t)}\right) + ... \nonumber\\
& + & e_n\left({(4-6\hat\sigma_e(t)) (4 - 6(\hat\alpha_e(t), \hat\beta_e(t)))\over (12 - 3\hat\alpha_e(t) - 3\hat\beta_e(t))^2}\right) + ... + j_n \left({(16 - 6\hat\zeta_e(t))(8 + 6(0, \hat\eta_e(t)))\over (12 - 3\hat\alpha_e(t) - 3\hat\beta_e(t))^2}\right)\Bigg]\nonumber
\nd}
where from \eqref{botsuga} we see that $n \in \mathbb{Z}$ and $1 \le n \le 41$. Looking at the form of \eqref{crawlranipa} and recalling the fact from section \ref{sec4.4} that $(\alpha_e(t), \beta_e(t)) << 1$ and $(\hat\alpha_e(t), \hat\beta_e(t), \hat\sigma_e(t)) << 1$, the temporal dependent terms in the log factors should be highly sub-dominant. Moreover the elements of the column matrix $\mathbb{N}_n$ is typically proportional to $\delta_{1m(n)}$ where $m(n)$ is the $m$-th element in the $n$-th column matrix $\mathbb{N}_n$ with $1 \le m(n) \le 10$ and $1 \le n \le 41$. Thus it would appear that the individual logarithmic terms can only contribute sub-dominant $g_s$ dependent terms, and therefore should be ignored. Unfortunately following such an argument would lead to serious errors! We will see an example of this soon, but before that let us extend the logarithmic corrections further by incorporating temporal derivatives of the warp-factors. These may be quantified by again defining two sets, much like the ones in \eqref{emilxgulab}, in the following way:

{\footnotesize
\bg\label{emilxgulab2}
&& {\cal T}_{32} = \left(1, ~ {\dot{\alpha}_e(t)\vert\log~\bar{g}_s\vert\over 2 - \dot{\beta}_e(t)}, ~ {\dot\beta_e(t)\vert\log~\bar{g}_s\vert\over 2 - \dot\beta_e(t)}, ~ {\dot\zeta_e(t)\vert\log~\bar{g}_s\vert\over 2 - \dot\beta_e(t)}, ~ {\dot\eta_e(t)\vert\log~\bar{g}_s\vert\over 2 - \dot\beta_e(t)}\right)\\
&& {\cal T}_8 = \left(1, ~ {\dot{\hat\sigma}_e(t)\vert\log~\bar{g}_s\vert\over 4 - \dot{\hat\alpha}_e(t) - \dot{\hat\beta}_e(t)}, ~ {(\dot{\hat\alpha}_e(t), \dot{\hat\beta}_e(t))\vert\log~\bar{g}_s\vert\over 4 - \dot{\hat\alpha}_e(t) - \dot{\hat\beta}_e(t)}, ~ {\dot{\hat\zeta}_e(t) \vert\log~\bar{g}_s\vert 
\over 4 - \dot{\hat\alpha}_e(t) - \dot{\hat\beta}_e(t)}, ~ { \dot{\hat\eta}_e(t)\vert\log~\bar{g}_s\vert\over 4 - \dot{\hat\alpha}_e(t) - \dot{\hat\beta}_e(t)}\right),\nonumber \nd}
where, as it stands, they are all sub-dominant contributions. However we will continue to keep track of them in the same way as we did earlier. In fact, following the same strategy as in \eqref{creeprani}, we can define two column matrices $\mathbb{Q}_{32}$ and $\mathbb{Q}_8$ from \eqref{emilxgulab2}. To complete the story, 
we need another column matrix $\mathbb{P}_n$, much like $\mathbb{N}_n$ before, whose elements are taken from the set $\widetilde{\cal S}_n \equiv (\tilde{a}_n, \tilde{b}_n, \tilde{c}_n, \tilde{d}_n, ..., \tilde{j}_n)$, with $(\tilde{a}_n, \tilde{b}_n, \tilde{c}_n, \tilde{d}_n, ..., \tilde{j}_n)\in (\mathbb{Z}, \mathbb{Z}, ..., \mathbb{Z})$ and $1 \le n \le 41$. Putting everything together now leads to the following generic logarithmic corrections:

{\scriptsize
\bg\label{lotusada}
&& {\rm dom}\left[-{1\over \vert\log~\bar{g}_s\vert}~\log\left(\mathbb{N}_n^\top \mathbb{M}_{8}\right),~-{1\over \vert\log~\bar{g}_s\vert}~\log\left(\mathbb{P}_n^\top \mathbb{Q}_{8}\right),~-{1\over \vert\log~\bar{g}_s\vert}~\log\left(
\mathbb{N}_n^{'\top} \mathbb{M}_{8}\cdot\mathbb{P}_n^{'\top} \mathbb{Q}_{8}\right)\right],\nonumber\\
&& {\rm dom}\left[-{1\over \vert\log~\bar{g}_s\vert}~\log\left(\mathbb{N}_n^\top \mathbb{M}_{32}\right), ~-{1\over \vert\log~\bar{g}_s\vert}~\log\left(\mathbb{P}_n^\top \mathbb{Q}_{32}\right), ~-{1\over \vert\log~\bar{g}_s\vert}~\log\left(
\mathbb{N}_n^{'\top} \mathbb{M}_{32}\cdot\mathbb{P}_n^{'\top} \mathbb{Q}_{32}
\right)\right], \nonumber\\ \nd}
to the ${\rm E}_8 \times {\rm E}_8$ and $SO(32)$ cases respectively. Note that $(\mathbb{N}'_n, \mathbb{P}'_n)$ are column matrices with constant (but not identical!) elements as in $(\mathbb{N}_n, \mathbb{P}_n)$. As expected, the dominant contributions would typically come from $\mathbb{N}_n$, while the sub-dominant ones would come from $\mathbb{P}_n$ and from the combination of $\mathbb{N}'_n$ and $\mathbb{P}'_n$. In the following, let us construct an example to see how this may work out in practice. 

The example that we have in mind is the computation of the Ricci tensor ${\bf R}_{0n}$, where $y^n \in {\cal M}_4$. Looking at {\bf Tables \ref{niksmit11}}, {\bf \ref{niksmit12}} and {\bf \ref{niksmit120}} we see that it scales as $\left({g_s\over {\rm H}(y){\rm H}({\bf x})}\right)^{-1 + {\hat\alpha_e(t) + \hat\beta_e(t)\over 4}}$ for the generalized ${\rm E}_8 \times {\rm E}_8$ case, $\left({g_s\over {\rm H}(y){\rm H}({\bf x})}\right)^{-1 + {\beta_e(t)\over 2}}$ for the generalized $SO(32)$ case, and as $\left({g_s\over {\rm H}(y){\rm H}({\bf x})}\right)^{-1 + {\beta(t)\over 2}}$ for the simplified $SO(32)$ case. This is the correct answer only if we ignore the logarithmic corrections. In the presence of the logarithmic corrections, the scaling picks up extra factors in the following way:

{\scriptsize
\bg\label{lulup}
{\bf R}_{0n}({\bf x}, y; g_s) &=&\sum_{i = 1}^4 {\bf R}_{0n}^{(n_i)}({\bf x}, y)\left[\bar{a}_{i} + \mathbb{N}_{n_i}^\top \mathbb{M}_8(\bar{g}_s)\otimes\left(1 + \bar{b}_{i}(4 - \hat\alpha_e(t) - \hat\beta_e(t))\right)\right]\left({g_s\over {\rm H}(y){\rm H}({\bf x})}\right)^{-1 + {\hat\alpha_e(t) + \hat\beta_e(t)\over 4}}\nonumber\\
&+&\sum_{i = 1}^4{\bf R}_{0n}^{(m_i)}({\bf x}, y)\left[\bar{c}_i +  \mathbb{P}_{m_i}^\top \mathbb{Q}_8(\bar{g}_s)\otimes\left(1 + \bar{d}_i(4 - \dot{\hat\alpha}_e(t) - \dot{\hat\beta}_e(t))\right)\right]\left({g_s\over {\rm H}(y){\rm H}({\bf x})}\right)^{{\cal A}_{m_i}(t)} \nonumber\\ 
&+& \sum_{i = 1}^4 {\bf R}_{0n}^{(l_i)}({\bf x}, y)\Big[\bar{e}_i + \mathbb{N}_{l_i}^{'\top} \mathbb{M}_8(\bar{g}_s)\cdot
\mathbb{P}_{l_i}^{'\top} \mathbb{Q}_8(\bar{g}_s)\otimes\left(1 + \bar{f}_{i}(4 - \hat\alpha_e(t) - \hat\beta_e(t))\right)\left(1 + \bar{g}_i(4 - \dot{\hat\alpha}_e(t) - \dot{\hat\beta}_e(t))\right)  \nonumber\\
&\times &\left({g_s\over {\rm H}(y){\rm H}({\bf x})}\right)^{{\cal B}_{l_i}(t)}, \nd}
for the generalized ${\rm E}_8 \times {\rm E}_8$ case and $(\bar{a}_i, \bar{b}_i, ..., \bar{g}_i)$ are constants. The $\otimes$ product signifies the following: the constant column matrices multiplying the individual terms are now allowed to have different elements. For example, in the first line the column matrix $\mathbb{N}_{n_i}$ can have different set of ten elements while accompanying $\bar{b}_{i}(4 - \hat\alpha_e(t) - \hat\beta_e(t))$. Thus:

{\scriptsize
\bg\label{farsidedeal}
\mathbb{N}_{n_i}^\top \mathbb{M}_8(\bar{g}_s)\otimes\left(1 + \bar{b}_{i}(4 - \hat\alpha_e(t) - \hat\beta_e(t))\right) \equiv \mathbb{N}_{n_i}(\{\zeta\})^\top \mathbb{M}_8(\bar{g}_s) + \bar{b}_i \mathbb{N}_{n_i}(\{\zeta'\})^\top \mathbb{M}_8(\bar{g}_s)(4 - \hat\alpha_e(t) - \hat\beta_e(t)), \nd}
where $\{\zeta\}$ and $\{\zeta'\}$ are the two set of ten constant integers. In a similar way, in the third line, all four terms will be accompanied by 
$\mathbb{N}_{l_i}(\{\zeta_k\})^{'\top}\mathbb{M}_8(\bar{g}_s)\cdot\mathbb{P}_{m_i}(\{\zeta_k\})^{'\top}\mathbb{Q}_8(\bar{g}_s)$ with $k = 1, ..., 4$ respectively. The $g_s$ scalings in the second and the fourth lines, namely ${\cal A}_{m_i}(t(g_s))$ and ${\cal B}_{l_i}(t(g_s))$, may be easily worked out following the scalings of the Riemann tensors tabulated earlier. Furthermore, due to the subdominant nature of the derivative terms, we expect $\bar{c}_i = \bar{e}_i = 0$. For the generalized and the simplified $SO(32)$ cases, one may easily work out the scalings of ${\bf R}_{0n}({\bf x}, y; g_s)$ by following \eqref{lotusada} and {\bf Tables \ref{niksmit12}} and {\bf \ref{niksmit120}}. In fact for the former, {\it i.e.} the generalized $SO(32)$ case, we do have an explicit functional form for ${\bf R}_{0n}({\bf x}, y; g_s)$ given by\footnote{See eq. (4.99) in the first reference of \cite{desitter2}.}:
\bg\label{palmroyale}
{\bf R}_{0n}({\bf x}, y; g_s) &=& -2\left({\dot{\rm F}_1(t)\over {\rm F}_1(t)} +{\dot{\rm F}_2(t)\over {\rm F}_2(t)}\right) {\partial_n {\rm H}(y)\over {\rm H}(y)} \\
& = & 2\bigg[(\dot\alpha_e(t) + \dot\beta_e(t))\vert\log~\bar{g}_s\vert - {\alpha_e(t) + \beta_e(t)\over 2 - \beta_e(t)}~\left({g_s\over {\rm H}(y){\rm H}({\bf x})}\right)^{-1 + {\beta_e(t)\over 2}}\bigg] {\partial_n {\rm H}(y)\over {\rm H}(y)}, \nonumber \nd
which, as one may easily see, matches precisely with \eqref{lulup} with 
${\bf R}_{0n}^{(n_i)}({\bf x}, y) = {\partial_n {\rm H}(y)\over {\rm H}(y)}$ and ${\rm H}_o({\bf x}) = 1$. The other parameters from \eqref{lulup} are identified in the following way:
\bg\label{julmore}
&&\mathbb{N}'_{l_i} = \mathbb{P}'_{l_i} = \{0\}, ~~ \bar{b}_i = \bar{c}_i = \bar{e}_i = \bar{f}_i = \bar{g}_i = 0 \nonumber\\
&& \bar{a}_1 = -{8\over 3}, ~~\bar{a}_j = 0, ~~\mathbb{N}_{n_p} = \{0\}, ~~ \forall ~~(j, p) \ge (2, 3) \nonumber\\
&& \mathbb{N}_{n_1} = \big(\begin{matrix} 2 & 0 & 0 &... & 0\end{matrix}\big), ~~~\mathbb{N}_{n_2} = \big(\begin{matrix} 0 & 2 & 0 &... & 0\end{matrix}\big), ~~~ {\cal A}_{m_i} = {\cal B}_{l_i} = 0 \nonumber\\
&&  \mathbb{P}_{m_1} = \big(\begin{matrix} 2 & 0 & 0 &... & 0\end{matrix}\big), ~~~\mathbb{P}_{m_2} = \big(\begin{matrix} 0 & 2 & 0 &... & 0\end{matrix}\big), ~~~ \mathbb{P}_{m_l} = \{0\}, ~~\forall ~~ l \ge 3, \nd
expectedly confirming the fact that most of the parameters in \eqref{lulup} vanish\footnote{One may also verify from the first reference and eq. (4.99) of \cite{desitter2} that ${\bf R}_{n\rho}({\bf x}, y; g_s) = -{8\partial_n{\rm H}(y) \partial_\rho{\rm H}(y)\over {\rm H}^2(y)}$. This matches with the scalings in {\bf Tables \ref{niksmit9}}, {\bf \ref{niksmit10}} and {\bf \ref{niksmit100}}, implying further that all the parameters in \eqref{lulup} vanish to provide zero $g_s$ scaling. This remains true for all the five participating curvature tensors.}. Comparing with \eqref{claudhostess}, \eqref{hatpathanda} and {\bf figure \ref{logdelta}}, the first term (involving temporal derivatives of the warp-factors) in \eqref{palmroyale} is clearly sub-dominant compared to the second term and therefore may be ignored. Such a choice will also help us to deal with Ostrogradsky's instabilities as we shall soon see. Taking all the aforementioned arguments into account implies that the logarithmic corrections to the scalings of the curvature tensors may be succinctly captured by inserting the following factors in the $g_s$ scalings from {\bf Tables \ref{lilalo1}}, {\bf \ref{lilalo3}}, {\bf \ref{lilalo2}} and 
{\bf \ref{lilalo4}}:

{\footnotesize
\bg\label{marianbrick}
{\bf L_{N}}[\bar{a}_i, \bar{b}_i, \mathbb{N}_i; \mathbb{M}_{8(32)}] \equiv
\bar{a}_{i} + \mathbb{N}_{i}^\top \mathbb{M}_{8(32)}(\bar{g}_s)\otimes
~\begin{cases} 1 + \bar{b}_i(2 - \beta_e(t)) ~~~~~~~~~~~~~ {\rm for}~SO(32)\\
~~~~\\
1 + \bar{b}_{i}(4 - \hat\alpha_e(t) - \hat\beta_e(t))~~~ {\rm for}~~ {\rm E}_8 \times {\rm E}_8 \end{cases}, \nd}
where $1 \le i \le 41$, $\mathbb{M}_{8(32)} = \mathbb{M}_8$ or $\mathbb{M}_{32}$ for ${\rm E}_8 \times {\rm E}_8$ or $SO(32)$ respectively, and $(\bar{a}_i, \bar{b}_i) \in (\mathbb{R}, \mathbb{R})$ (although for most cases we expect $\bar{b}_i = 0$). Such a modification will then directly influence the Riemann tensors and the quantum series in \eqref{botsuga}, and subsequently therefore to the Schwinger-Dyson equations.

\subsubsection{Further corrections to the metric and the flux components \label{sec4.5.2}}

There would also be additional corrections to the metric and the flux components over and above the corrections that we have incorporated so far. However before analyzing this let us summarize the story that has been developed till now. Starting with section \ref{sec4.2.1}, we have discussed the perturbative corrections to the metric components, followed by the non-perturbative extensions in section \ref{sec4.2.2} (see for example \eqref{tranpart} and \eqref{ivbres}). Such corrections also enter the definitions of the warp-factors for both $SO(32)$ and ${\rm E}_8 \times {\rm E}_8$ theories as shown in \eqref{ryanfan} and \eqref{horseman} respectively. Their consistency with the underlying axionic cosmology is discussed in section \ref{sec4.2.3}. Similar story is developed for the G-flux components in section \ref{sec4.2.5}. All the aforementioned corrections may be neatly packaged as $(\alpha_e(t), \beta_e(t))$ for the $SO(32)$ theory, as shown in \eqref{bdtran} and \eqref{ivbrestick}; and as $(\hat\alpha_e(t), \hat\beta_e(t), \hat\sigma_e(t))$ for the ${\rm E}_8 \times {\rm E}_8$ theory, as shown in \eqref{bdtran2}. The corresponding G-flux components also get neatly packaged in a way shown in \eqref{andyrey}.

Despite all the aforementioned developments, the story is not complete: there would still be additional corrections to the metric and the flux components as alluded to above. These additional corrections would come from the non-perturbative effects related to the ${\rm M}_p$ scalings, and from the mixed non-perturbative effects from both the $g_s$ and the ${\rm M}_p$ scalings. As with the logarithmic corrections discussed in the previous section, including these additional non-perturbative effects  is bound to make the subsequent analysis rather challenging. Nevertheless we will show that there is a way to analyze the system and, from there, extract useful outcomes. 

We expect both the metric and the flux components to absorb the additional corrections. Some aspect of these corrections have been described earlier in \cite{coherbeta2} (see the second reference therein). Here we want to do a more detailed study regarding the $g_s$ scalings et cetera. The metric components, under the inclusion of all the aforementioned corrections, now take the following form:

{\scriptsize
\bg\label{katepathorabar}
\langle {\bf g}_{\rm AB}\rangle_\sigma \equiv {\bf g}_{\rm AB}({\bf x}, y; g_s) = \sum_{k = 0}^\infty {\bf g}^{(k)}_{\rm AB}({\bf x}, y)\left({g_s\over {\rm H}(y){\rm H}({\bf x})}\right)^{a^{({\rm AB})}(t) + {2k\vert f^{({\rm AB})}_k(t)\vert\over 3}} ~{\rm exp}\left[-\sum_{n, m}c_{nm}^{({\rm AB})}(k) {\left({\rm M}^2_p {\rm X}^2\right)^n\over \bar{g}_s^{m/3}}\right], \nd}
where ${\rm X}^2 \equiv \eta_{\rm AB} {\rm X}^{\rm A} {\rm X}^{\rm B}$ is the {unwarped} distance such that ${\rm X} \in {\bf R}^2 \times {\cal M}_4 \times {\cal M}_2$ with ${\cal M}_2$ taking two different orbifold limits for the two heterotic cases, {\it i.e.} ${\cal M}_2 = {\mathbb{T}^2\over {\cal I}_2}$ for the $SO(32)$ case and ${\cal M}_2 = {\bf S}^1_{\theta_1} \times {{\bf S}^1_{\theta_2}\over {\cal I}_{\theta_2}}$ for the ${\rm E}_8 \times {\rm E}_8$ case. The {\it unwarped} distance is necessary here because a warped distance would have included the information of the metric itself on the RHS of \eqref{katepathorabar} which would, in turn, have made a representation like \eqref{katepathorabar} rather hard to deal with. The remaining parameters are $a^{({\rm AB})}(t)$ and $f^{({\rm AB})}_k(t)$ and the constant $c^{({\rm AB})}_{nm}(k)$. The parameter $a^{({\rm AB})}(t)$ is defined as:

{\footnotesize
\bg\label{karenmach}
a^{({\rm AB})}(t) = \begin{cases} \left(a^{({\mu\nu})}(t), a^{(mn)}(t), a^{(\rho\sigma)}(t), a^{(ab)}(t)\right) = \left(-{8\over 3}, -{2\over 3} + \alpha(t), -{2\over 3} + \beta(t), {4\over 3}\right)\\
~~~~~\\
\left(a^{(\mu\nu)}(t), a^{(mn)}(t), a^{(\theta_1\theta_1)}(t), a^{(\theta_2\theta_2)}(t), a^{(ab)}(t)\right) = \big(-{8\over 3}, -{2\over 3} + \hat\sigma(t), -{2\over 3} + \hat\alpha(t), -{2\over 3} + \hat\beta(t), {4\over 3}\big) \end{cases} \nd}
for the $SO(32)$ and the ${\rm E}_8 \times {\rm E}_8$ cases respectively.
The other parameter $f^{({\rm AB})}_k(t)$ form the additional perturbative corrections that we have discussed earlier (see sections \ref{sec4.2.1} and \ref{sec4.2.2}). Finally the constant parameter $c^{({\rm AB})}_{nm}(k)$ that enters the exponential factor in \eqref{katepathorabar} provides the following distribution of the non-perturbative effects:

{\scriptsize
\bg\label{marmatrix}
{\rm exp}\left[-\sum_{n, m}c_{nm}^{({\rm AB})}(k) {\left({\rm M}^2_p {\rm X}^2\right)^n\over \bar{g}_s^{m/3}}\right] = {\rm exp}\left[-\sum_{ m = 0}^\infty {c_{0m}^{({\rm AB})}(k)\over \bar{g}_s^{m/3}} -\sum_{n = 0}^\infty c_{n0}^{({\rm AB})}(k) \left({\rm M}^2_p {\rm X}^2\right)^n  
-\sum_{(n, m)\ge 1}^\infty c_{nm}^{({\rm AB})}(k) {\left({\rm M}^2_p {\rm X}^2\right)^n\over \bar{g}_s^{m/3}}\right], \nonumber\\ \nd}
where the first term is clearly related to $n_a(l)$ appearing in \eqref{tranpart}, but the other two terms are new. In fact the third term is a mixture of both the first and the second terms. Unfortunately such a term is the main source of problem for us because we cannot use a manipulation like \eqref{ivbres} to absorb it in the definition of $(\zeta_e(t), \alpha_e(t), \beta_e(t), \eta_e(t))$ as in \eqref{bdtran} or in the definition of
$(\hat\zeta_e(t), \hat\sigma_e(t), \hat\alpha_e(t), \hat\beta_e(t), \hat\eta_e(t))$ as in \eqref{bdtran2} for the $SO(32)$ and the ${\rm E}_8 \times {\rm E}_8$ cases respectively. How should we then proceed?

\begin{figure}[h]
\centering
\begin{tabular}{c}
\includegraphics[width=4in]{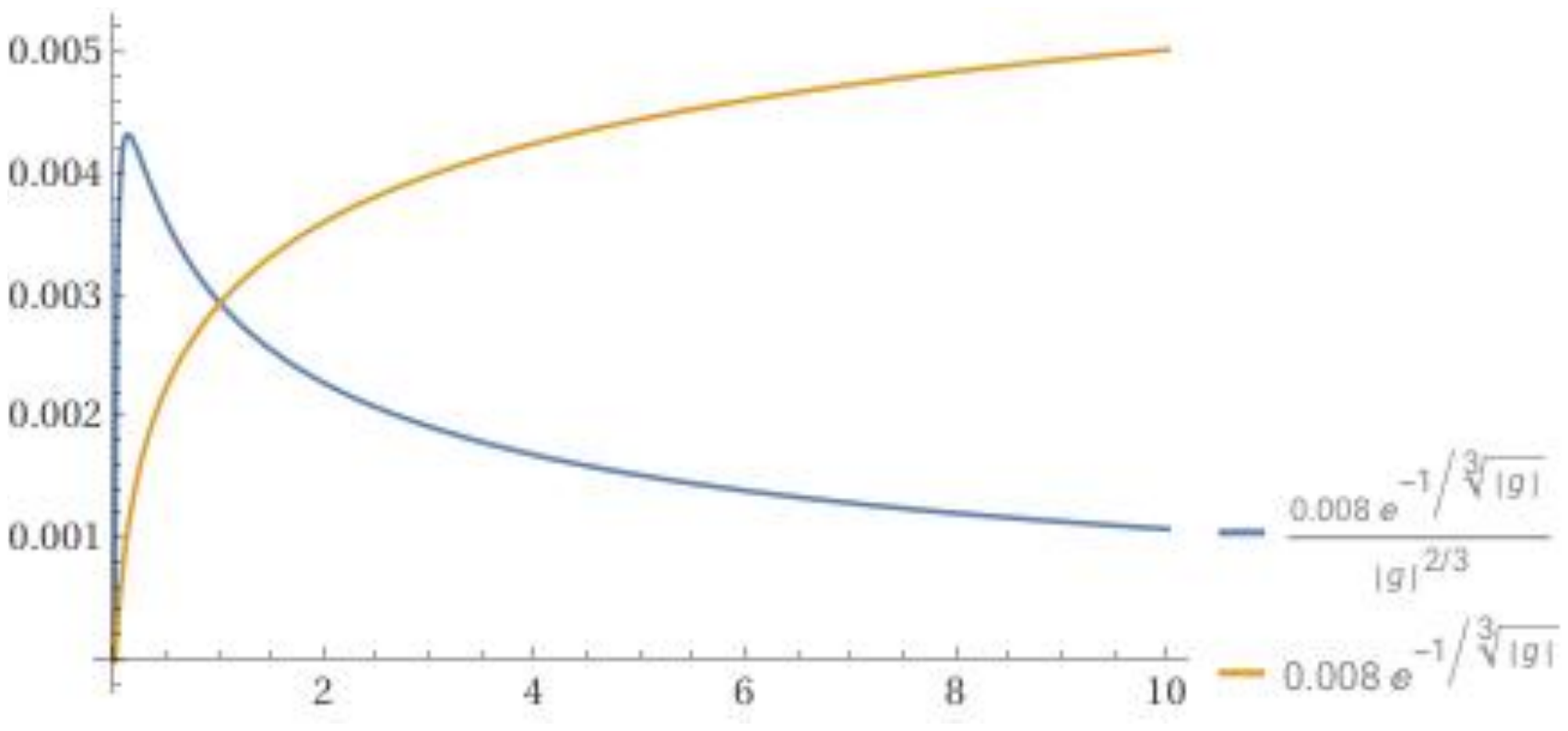}
\end{tabular}
\vskip-.7in
\caption[]{Behavior of the non-perturbative terms ${\rm exp}\left(-{1\over g^{1/3}}\right)$ (in orange) and ${1\over g^{2/3}}{\rm exp}\left(-{1\over g^{1/3}}\right)$ (in blue) as $g \equiv \bar{g}_s \to 0$. Expectedly, the exponential suppression factor always dominates near the origin irrespective of any accompanying factors.}
\label{npbehavior}
\end{figure}

One possibility would be to express the mixed term, {\it i.e.} the third term in \eqref{marmatrix} as a {\it product} of two terms: one, as a $g_s$ dependent term and the other, as a ${\rm M}_p{\rm X}$ dependent term. Such a realization appears from the so-called Mehler formula\footnote{The Mehler formula appears from the generalization of the standard representation of the Hermite polynomials, namely: $\sum\limits_{n\ge 0} \mathbb{H}_n({\rm X}) {u^n\over n!} = {\rm exp}\left(2{\rm X} u - u^2\right)$ in the following way:
$$\sum\limits_{n\ge 0} \mathbb{H}_n({\rm X})\mathbb{H}_n({\rm Y})~{u^n\over n!} = (1 - 4u^2)~{\rm exp}\left[{4{\rm XY}u - 4({\rm X}^2 + {\rm Y}^2)u^2\over 1 - 4u^2}\right]$$
\noindent where $u = {\rho\over 2}$ and $\vert\rho\vert < 1$. Details about the proof of the Mehler formula appears in \cite{mehlerproof}, and one may even take $\rho = e^{-i\alpha}$ to generalize it further (see the last reference in \cite{mehlerproof}).} 
\cite{mehler}:

{\scriptsize
\bg\label{misskalchul}
{\rm exp}\left[-{2\rho f_1(g_s) f_2({\rm M}_p{\rm X})\over 1 -\rho^2}\right] = 
\sqrt{1-\rho^2} \sum_{n = 0}^\infty {\left(-\rho/ 2\right)^n\over n!}~\mathbb{H}_n(f_1(g_s)) ~{\rm exp}\left[{\rho^2 f_1^2(g_s)\over {1-\rho^2}}\right] ~\mathbb{H}_n(f_2({\rm M}_p{\rm X})~
{\rm exp}\left[{\rho^2 f_2^2({\rm M}_p{\rm X})\over {1-\rho^2}}\right],
\nonumber\\ \nd}
with $\vert\rho\vert < 1$ and $\mathbb{H}_n(f)$ are the Hermite polynomials expressed using $f$ functions. We can choose $\rho = \sqrt{2} - 1$, which would keep ${2\rho \over 1 - \rho^2} = 1$ and therefore simplify \eqref{misskalchul}. At the first glance the Mehler formula does exactly what we wanted: it splits the exponential of a mixed function into a sum over product of two functions. A careful look however shows that there is a subtle problem in implementing \eqref{misskalchul} for the present case. To see this, let us use the Mehler formula to express the third term in \eqref{marmatrix} using the Hermite polynomials as:
\bg\label{crawlrani}
{\rm exp}\left[-\sum_{(n, m)\ge 1}^\infty c_{nm}^{({\rm AB})}(k) {\left({\rm M}^2_p {\rm X}^2\right)^n\over \bar{g}_s^{m/3}}\right]
& = & \prod_{(n, m)\ge 1} \sqrt{1-\rho^2(n, m)} \sum_{l = 0}^\infty {(-\rho(n, m)/2)^l\over l!}\nonumber \\
&\times& 
~\mathbb{H}_l({\rm M}_p^{2n}{\rm X}^{2n})~{\rm exp}\left[{\rho^2(n, m) ({\rm M}_p^{2} {\rm X}^{2})^{2n}\over 1 - \rho^2(n, m)}\right] \\
&\times& ~\mathbb{H}_l\left({1\over \bar{g}_s^{m/3}}\right)~{\rm exp}\left[{\rho^2(n, m)\over (1-\rho^2(n, m)) \bar{g}_s^{2m/3}}\right], \nonumber
\nd
where $\rho(n, m) = {\sqrt{1 + (c_{nm}^{({\rm AB})})^2} - 1\over c_{nm}^{({\rm AB})}}$ and $c_{nm}^{({\rm AB})}(k) \ge 0$ otherwise the non-perturbative terms in \eqref{katepathorabar} and \eqref{marmatrix} cannot be properly defined (note that $(n, m)$ are not Lorentz indices). The LHS of the equality in \eqref{crawlrani} is a well-defined function for any values of $(\bar{g}_s, {\rm M}_p, {\rm X})$ and, in particular, in the limit of $\bar{g}_s \to 0$ and ${\rm M}_p{\rm X} \to \infty$. Unfortunately in this limit the RHS is not well-defined: both the exponential pieces as well as the Hermite polynomials blow-up when we take $\bar{g}_s \to 0$ and ${\rm M}_p{\rm X} \to \infty$. We can alternatively define two quantities:
\bg\label{candalice}
{\rm Z}_m \equiv {\rm exp}\left(-{1\over \bar{g}_s^{m/3}}\right), ~~~~~~~
{\rm W}_n \equiv {\rm exp}\left[-\left({\rm M}_p^2{\rm X}^2\right)^n  \right], \nd
which are both well defined functions and lie within $0 < ({\rm Z}_m, {\rm W}_n) < 1$ for all values of $(\bar{g}_s, {\rm M}_p, {\rm X})$. (See also {\bf figure \ref{npbehavior}}.)
Using \eqref{candalice}, we can express the LHS of \eqref{crawlrani} in the following suggestive way that illustrates the underlying problem:

{\footnotesize
\bg\label{amarcandy}
\prod_{(n, m) \ge 1} \left({\rm Z}_m\right)^{c_{nm}^{({\rm AB})}(k)\vert\log~{\rm W}_n\vert} = \prod_{(n, m)\ge 1} {\rm exp}\left[- c_{nm}^{({\rm AB})}(k)\vert\log~{\rm Z}_m\vert\vert\log~{\rm W}_n\vert\right], \nd}
where the logarithmic terms blow-up when $({\rm Z}_m, {\rm W}_n) \to (0, 0)$, keeping \eqref{amarcandy} well-defined. In fact there is no perturbative expansion of the exponential piece in \eqref{amarcandy} because if $\bar{g}_s \to \epsilon^a$ and ${\rm M}_p^2 {\rm X}^2 \to \epsilon^{-b}$, then we require $c_{nm}^{({\rm AB})}(k) \to \epsilon^\gamma$ where $\gamma > bn + {am\over 3}$ and $\epsilon \to 0$. For arbitrary $(n, m)$ this amounts to $\gamma \to \infty$ and therefore $c_{nm}^{({\rm AB})}(k) = 0$. Thus, expectedly, there is no perturbative limit possible and it appears that the Mehler decomposition works in the reverse limit where $\bar{g}_s >> 1$ and ${\rm M}_p^2 {\rm X}^2 << 1$.
Nevertheless the Mehler decomposition, including the choice of the good coordinates \eqref{candalice}, provide enough hints to decompose \eqref{amarcandy}, or equivalently the LHS of \eqref{crawlrani}, in the following way:

{\scriptsize
\bg\label{mistmean}
\prod_{(n, m)\ge 1} {\rm exp}\left[ -c_{nm}^{({\rm AB})}(k)\vert\log~{\rm Z}_m\vert\vert\log~{\rm W}_n\vert\right] = \prod_{(n, m)\ge 1} \sum_{l = 0}^\infty d^{({\rm AB})}_l\left(k;{1\over \vert\log~{\rm Z}_m\vert}\right) f^{({\rm AB})}_l\left(k;{1\over \vert\log~{\rm W}_n\vert}\right) ~\mathbb{H}_l({\rm Z}_m) \mathbb{H}_l({\rm W}_n), \nonumber\\ \nd}
where both sides of \eqref{mistmean} are well defined because of our choice of the coordinates from \eqref{candalice} and the fact that 
${1\over \vert\log~{\rm Z}_m\vert}$ and ${1\over \vert\log~{\rm W}_n\vert}$ are perturbative in $\bar{g}_s$ and ${1\over {\rm M}_p^2 {\rm X}^2}$ respectively. The RHS of \eqref{mistmean} has some similarities with the Mehler form \eqref{crawlrani} as both contains the Hermite polynomials accompanied by two functions, however now we do have a well-defined expansion that does not blow-up. Moreover, looking at the explicit form for the Hermite polynomials:
\bg\label{steelmisty}
\mathbb{H}_l({\rm Z}_m) = l! \sum_{p = 0}^{\lfloor {l\over 2}\rfloor} {(-1)^p\over k!(l - 2p)!}~\left(2{\rm Z}_m\right)^{l-2p} =l! \sum_{p = 0}^{\lfloor {l\over 2}\rfloor} {(-1)^p 2^{l-2p}\over p!(l - 2p)!}~{\rm exp}\left(-{\vert l-2p\vert\over \bar{g}_s^{m/3}}\right), \nd
we see that this is precisely an instanton series in $\bar{g}^{m/3}_s$! Similar instanton series appear for the Hermite polynomials $\mathbb{H}_l({\rm W}_n)$ in ${\rm M}_p^2 {\rm X}^2$. The perturbative functions $d^{({\rm AB})}_l\left(k;{1\over \vert\log~{\rm Z}_m\vert}\right)$ and $f^{({\rm AB})}_l\left(k;{1\over \vert\log~{\rm W}_n\vert}\right)$
can be viewed as functions contributing to the fluctuation determinants around the two instanton series respectively. Thus the mixed term in the non-perturbative contributions to the metric from \eqref{katepathorabar} and \eqref{marmatrix} may also be interpreted as coming from instanton series with the fluctuation determinants inserted in. This means that we can rewrite the $g_s$ scalings of the metric components in \eqref{katepathorabar} as $\left({g_s\over {\rm H}(y){\rm H}_o({\bf x})}\right)^{a^{({\rm AB})}(t) - \sigma^{({\rm AB})}(t) + \Sigma_e^{({\rm AB})}(t)}$, where $\Sigma_e^{({\rm AB})}(t)$ is given by:

{\scriptsize
\bg\label{meanermis}
&&\Sigma_e^{({\rm AB})}(t) = \sigma^{({\rm AB})}(t) -{1\over \vert\log~\bar{g}_s\vert}\left[1 + \sum_{k = 1}^\infty {{\rm A}_k^{({\rm AB})}\over {\rm A}_0^{({\rm AB})}}\left({g_s\over {\rm H}(y){\rm H}_o({\bf x})}\right)^{2k \vert {\cal B}_k^{({\rm AB})}(t)\vert\over 3}\right]\\
&& \vert{\cal B}_k^{({\rm AB})}(t)\vert = \vert f_k^{({\rm AB})}(t)\vert + {3\over 2k}\sum_{n = 0}^\infty {\vert c_{0n}^{({\rm AB})}(k)\over \bar{g}_s^{n/3}\vert\log~\bar{g}_s\vert} - {3\over 2k\vert\log~\bar{g}_s\vert}\sum_{m\ge 1}\log \left[{\rm dom}\left( \left\{d^{({\rm AB})}_l\left(k;{1\over \vert\log~{\rm Z}_m\vert}\right) ~\mathbb{H}_l({\rm Z}_m)\right\}_{\forall l\ge 0} \right)\right],\nonumber
\nd}
where $\sigma^{({\rm AB})}(t) = (\sigma^{(\mu\nu)}, \sigma^{(mn)}, \sigma^{(\theta_1\theta_1)}, \sigma^{(\theta_2\theta_2)}, \sigma^{(ab)}) = (0, \hat\sigma(t), \hat\alpha(t), \hat\beta(t), 0)$ for the ${\rm E}_8 \times {\rm E}_8$ case and $(\sigma^{(\mu\nu)}, \sigma^{(mn)}, \sigma^{(\rho\sigma)}, \sigma^{(ab)}) = (0, \alpha(t), \beta(t), 0)$ for the $SO(32)$ case. The decomposition \eqref{meanermis} is facilitated by allowing a slightly simplifying ans\"atze 
for the spatial part of the metric components:

{\scriptsize
\bg\label{ghoritagra}
&& {\bf g}_{\rm AB}^{(k)}({\bf x}, y) ~{\rm exp}\left[-\sum_{n=0}^\infty c_{n0}^{({\rm AB})}(k) ({\rm M}_p^2 {\rm X}^2)^n\right]\prod_{n\ge 1} {\rm dom}\left( \left\{f^{({\rm AB})}_l\left(k;{1\over \vert\log~{\rm W}_n\vert}\right) ~\mathbb{H}_l({\rm W}_n)\right\}_{\forall l\ge 0} \right) \\
& \equiv & {\rm A}^{({\rm AB})}_k ~{\bf g}_{\rm AB}({\bf x}, y) ~{\rm exp}\left[-\sum_{n=0}^\infty c_{n0}^{({\rm AB})} ({\rm M}_p^2 {\rm X}^2)^n\right] \prod_{n\ge 1}{\rm dom}\left( \left\{f^{({\rm AB})}_l\left({1\over \vert\log~{\rm W}_n\vert}\right) ~\mathbb{H}_l({\rm W}_n)\right\}_{\forall l\ge 0} \right) \equiv {\rm A}^{({\rm AB})}_k ~\check{\bf g}_{\rm AB}({\bf x}, y),\nonumber
\nd}
which is possible with the appropriate choice of the flux components and ${\rm X} \in ({\bf x}, y)$. The ans\"atze \eqref{ghoritagra} has the obvious advantage of splitting the spatial and the temporal parts of the metric in the way specified in \eqref{meanermis}. Note that, despite the relative minus sign, the RHS of the second relation in \eqref{meanermis} remains positive definite. Thus 
$\Sigma_e^{({\rm AB})}(t) = (\hat\zeta_e(t), \hat\sigma_e(t), \hat\alpha_e(t), \hat\beta_e(t), \hat\eta_e(t))$ for the ${\rm E}_8 \times {\rm E}_8$ case, and $\Sigma_e^{({\rm AB})}(t) = (\zeta_e(t), \alpha_e(t), \beta_e(t), \eta_e(t))$ for the $SO(32)$ case, with the input from \eqref{meanermis}, now contain the most generic possible cases in our set-up. In a similar vein, one can develop the scalings for the G-flux components and show the corresponding genericity for the $\hat{l}_{e{\rm AB}}^{\rm CD}$ and ${l}_{e{\rm AB}}^{\rm CD}$ scalings for the ${\rm E}_8 \times {\rm E}_8$ and the $SO(32)$ cases respectively.

\subsubsection{Ostrogradsky instabilities and higher order interactions \label{ostro}}

There is yet another subtlety that needs to be addressed before we compute the $g_s$ scaling from \eqref{botsuga}, and has to do with Ostrogradsky's instabilities. 
To see this, we should first note that $n_0 > 0$ in \eqref{botsuga} is not the only place where multiple temporal derivatives appear. They also appear inside the curvature tensors ${\bf R}_{\rm ABCD}({\bf x}, y; g_s)$ as well as inside the flux tensors ${\bf G}_{0{\rm MNP}}({\bf x}, y; g_s)$, including their powers, where $({\rm A, B}) \in {\bf R}^{2,1} \times {\cal M}_4 \times {\cal M}_2 \times {\mathbb{T}^2\over {\cal G}}$ and $({\rm M, N}) \in {\bf R}^{2} \times {\cal M}_4 \times {\cal M}_2 \times {\mathbb{T}^2\over {\cal G}}$ with ${\cal M}_2$ as usual taking two different forms to contain the $SO(32)$ and the ${\rm E}_8 \times {\rm E}_8$ cases. Because of these multiple temporal derivatives, Ostrogradsky's instabilities (or ghosts) could in principle appear. However 
there are a few points that make the story different from the usual system with multiple temporal derivatives.

\vskip.1in
\noindent $\bullet$ Existence of the underlying supersymmety at the vacuum level, {\it i.e.} at the level of supersymmetric Minkowski minima, guarantees the absence of any kinds of instabilities. Moreover the two-dimensional conformal field theory origin of string theory and the absence of any minima other than the aforementioned Minkowski ones remove the possibilities of ghost modes. 

\vskip.1in
\noindent $\bullet$ The moduli stabilization at the vacuum Minkowski level, and the dynamical duality sequences presented in {\bf Tables \ref{milleren1}, \ref{milleren2}, \ref{milleren3}, \ref{millerenoo}} and {\bf \ref{milleren4}} using seed M-theory emergent metric configurations (from Glauber-Sudarshan states) justify the complete absence of any instabilities.

\vskip.1in 
\noindent Aside from the above generic discussions, which clearly tells us that any constructions that appear from string theory will generically have no Ostrogradsky instabilities, there are more specific reasons pertaining to the our Glauber-Sudarshan states that also suggest the absence of such instabilities. These reasoning may be expressed in the following way.

\vskip.1in
\noindent $\bullet$ Computations using the Glauber-Sudarshan states essentially require us to impose IR cut-offs because of the underlying UV/IR mixing \cite{borel2}. The energy scale is $k_{\rm IR} < k < \mu$ as shown in \eqref{greeneve}, and therefore the theory is bounded from below as well as from the top. Moreover the dynamical moduli stabilization discussed in \cite{coherbeta, coherbeta2} concurs with the moduli stabilization at the vacuum level, thus removing any instabilities.

\vskip.1in
\noindent $\bullet$ The total on-shell Hamiltonian, including all possible ghost contributions, always {\it vanish} in both the Minkowski minimum as well as in the emergent de Sitter or quasi de Sitter states \cite{wdwpaper}. The off-shell Hamiltonian does not have to vanish, but the Hamiltonian as an operator again always annihilates any quantum state. Thus generation and propagation of instabilities cannot occur in the usual ways seen in the field theories.

\vskip.1in
\noindent $\bullet$ The total action is expressed in a trans-series form (see \eqref{kimkarol}), and therefore expectation values computed using this always {\it converge}. This is unlike usual perturbative actions that show factorial growths \cite{joydeep, wdwpaper}.

\vskip.1in
\noindent $\bullet$ The interaction terms, whose perturbative form is captured in \eqref{botsuga}, \eqref{botsuga2.0} or \eqref{fahingsha10}, and the full trans-series form appearing in \eqref{kimkarol}, have sub-dominant contributions to the total action as they are suppressed both perturbatively and non-perturbatively by powers of ${\rm M}_p$. The emergent action, and specifically the interaction parts, have additional suppressions from the neighboring Glauber-Sudarshan states of the on-shell ones \cite{wdwpaper}. These suppressions suggest that instabilities cannot be generated at least in the usual ways from the higher derivative terms. 

\vskip.1in

\noindent $\bullet$ All the on-shell metric and the flux components may be expressed as functions over a given Cauchy slice (or a given spatial slice) with overall $g_s$ scalings that control their temporal behaviors.

\vskip.1in

\noindent The aforementioned seven points spell out the key differences between what we have now from string-theory and M-theory, and the standard higher derivative theories that do not originate from string or M theories. In particular, 
the first and the second points that we made above are important. The higher order terms in \eqref{botsuga}, \eqref{botsuga2.0} or \eqref{fahingsha10}, and in the full trans-series form \eqref{kimkarol}, appear with coefficients that are carefully controlled right from the supersymmetric Minkowski level. Additionally, because of the lower bounds in the momenta, we are essentially cutting off all momenta smaller than $k_{\rm IR}$ in theory and thus preventing the theory to develop any ghost or phantom states\footnote{Of course such a procedure does not remove other ghosts like Faddeev-Popov or Batalin-Vilkovisky which are, in fact, essential in their own way for the consistency of the system \cite{wdwpaper}. Additional ghosts like the Boulware-Deser ghosts do not exist in the system because in the energy scale $k_{\rm IR} < k < \mu$ there are no massive graviton states.}. Note that, as explained in detail in \cite{borel2}, the IR cut-off is {\it essential} not only for the nodal diagrams to make sense, but also to comply with the underlying UV/IR mixing. 

While the above argument provides enough reason to justify the absence of any Ostrogradsky instabilities in the system, one would still like to see if there are other possible arguments. One other argument is related to the short-distance behavior of the theory. It is known that M-theory is well-defined at arbitrary short distances, despite the fact that our knowledge of the short-distance degrees of freedom is completely lacking. Under RG flow we do not expect the low energy theory to develop pathologies like Ostrogradsky instabilities. However we are also performing Borel resummations at every energy scale and, as discussed after \eqref{ckcigaret}, it is not clear whether Borel resummation is compatible with the RG flow. This means any instabilities should be verified at every energy scale. However again, because of the stringy origin and because of the IR cut-off, we expect instability like the Ostrogradsky instabilities should not appear. 

Yet another argument comes from looking at the Wheeler-De Witt (WdW) equation for the emergent states. The solution of such an equation gives us the envelope wave-functional over all the on and off-shell Glauber-Sudarshan states (see \cite{wdwpaper} for more details on this). We expect the wave-functional to peak near the on-shell Glauber-Sudarshan states. This suggest that the metric configurations in the close neighborhood to take the following dominant behavior\footnote{The product structure in \eqref{sarakalu} is easy to explain when it takes a polynomial form. In the exponential form, this is guaranteed using the Mehler decomposition as shown in \eqref{misskalchul}.}:
\bg\label{sarakalu}
{\bf g}_{\rm AB}({\bf x}, y; g_s) \equiv {\bf g}_{\rm AB}^{(0)}({\bf x}, y; g_s) = \check{\bf g}_{\rm AB}({\bf x}, y)\left({g_s\over {\rm H}(y){\rm H}({\bf x})}\right)^{\Sigma^{(0)}_{e{\rm AB}}(t(g_s))}, \nd
where $\check{\bf g}_{\rm AB}({\bf x}, y)$ is a function defined over a given spatial slice (parametrized by the coordinates $({\bf x}, y)$ with ${\bf x} \in {\bf R}^2$ and $y\in  {\cal M}_4 \times {\cal M}_2 \times {\mathbb{T}^2\over {\cal G}}$; and $\Sigma^{(0)}_{e{\rm AB}}(t(g_s))$ captures the $g_s$ scaling. In the off-shell case, they are completely arbitrary, but in the on-shell case they are defined from \eqref{meanermis} and \eqref{ghoritagra} as:

{\scriptsize
\bg\label{sarahchor}
&& \Sigma^{(0)}_{e{\rm AB}}(t(g_s)) = a^{({\rm AB})}(t(g_s)) - \sigma^{({\rm AB})}(t(g_s)) + \Sigma_e^{({\rm AB})}(t(g_s))\\
&&\check{\bf g}_{\rm AB}({\bf x}, y)=  {\bf g}_{\rm AB}({\bf x}, y) ~{\rm exp}\left[-\sum_{n=0}^\infty c_{n0}^{({\rm AB})} ({\rm M}_p^2 {\rm X}^2)^n\right] \prod_{n\ge 1}{\rm dom}\left( \left\{f^{({\rm AB})}_l\left({1\over \vert\log~{\rm W}_n\vert}\right) ~\mathbb{H}_l({\rm W}_n)\right\}_{\forall l\ge 0} \right)\nonumber \nd}
where all the parameters appearing in \eqref{sarahchor} have been defined earlier in section \ref{sec4.5.2}. One may equivalently express the off-shell four-form G-flux components in a similar way, namely as four-forms defined on a spatial slice with overall $g_s$ scalings. It is now easy to see that the $n$-th order temporal derivatives act on the metric components as:
\bg\label{ennakhela}
\partial_0^n {\bf g}_{\rm AB}({\bf x}, y; g_s) \equiv {\bf g}_{\rm AB}^{(n)}({\bf x}, y; g_s) = \check{\bf g}_{\rm AB}({\bf x}, y)\left({g_s\over {\rm H}(y){\rm H}({\bf x})}\right)^{\Sigma^{(n)}_{e{\rm AB}}(t(g_s))}, \nd
which has exactly the same form as \eqref{sarakalu} up to a different $g_s$ scaling. Thus on a given spatial slice all the temporal derivatives of the metric components are simply proportional to the original metric components. The proportionality factors are controlled by ${\Sigma^{(n)}_{e{\rm AB}}(t(g_s))}$ that have the following recurrence relation:

{\scriptsize
\bg\label{daniniash}
{\Sigma^{(n)}_{e{\rm AB}}(t)} = {\rm dom}\left( {\Sigma^{(n-1)}_{e{\rm AB}}(t)} - {\log\left(\pm\dot\Sigma^{(n-1)}_{e{\rm AB}}(t)\vert\log~\bar{g}_s\vert\right) \over\vert\log~\bar{g}_s\vert}, ~\gamma_{1, 2} - 1 + \Sigma^{(n-1)}_{e{\rm AB}}(t) - {\log\left(\Sigma^{(n-1)}_{e{\rm AB}}(t)\right)\over \vert\log~\bar{g}_s\vert}  \right), \nd}
where $\gamma_{1, 2}$ equals $\gamma_{1, 2}[\beta_e(t)]$ for the $SO(32)$ case or equals $\gamma_{1, 2}\left[{\hat\alpha_e(t) + \hat\beta_e(t)\over 2}\right]$ for the ${\rm E}_8 \times {\rm E}_8$ case. One may choose 
$\Sigma^{(n-1)}_{e{\rm AB}}(t)$ appropriately for the two heterotic cases. It is also easily verify that, in the on-shell case, for $n = 1$ we get the results from {\bf Table \ref{jessrog}}.

Our little exercise above suggests that the higher order temporal derivatives of the metric (and also flux) components are not independent but are related to the zeroth order component \eqref{sarakalu}. In other words, for the case with the metric components, we expect:
\bg\label{penpax}
{\bf g}^{(n)}_{\rm AB}({\bf x}, y; g_s) = {\bf g}^{(0)}_{\rm AB}({\bf x}, y; g_s) \left({g_s\over {\rm H}(y){\rm H}({\bf x})}\right)^{\Sigma^{(n)}_{e{\rm AB}}(t(g_s)) -\Sigma^{(0)}_{e{\rm AB}}(t(g_s))}, \nd
and are therefore not independent degrees of freedom, so should be inserted in any definition of the Hamiltonian using Lagrange multipliers. Such a procedure does generically remove any Ostrogradsky instabilities \cite{woodard}, but since we are imposing IR cut-offs in the nodal diagrams \cite{borel2}, and because of the underlying stringy origin of our model, the issue is not pertinent for the present case.

\subsubsection{The quantum scalings in the generalized ${\rm E}_8 \times {\rm E}_8$ case \label{sec4.5.3}}

With all the results at hand now, we are ready to work out the scaling of the perturbative quantum series in \eqref{botsuga} for the ${\rm E}_8 \times {\rm E}_8$ case. However before moving ahead, we should remind the readers of an upcoming complication resulting from the existence of higher order temporal derivatives denoted by $\{\partial_0^{n_0}\}$ in \eqref{botsuga}. Since the $g_s$ scalings of the curvature and fluxes themselves have temporal dependence, higher order temporal derivatives would significantly change the final scaling of the quantum series in \eqref{botsuga}. Because of this we will have to tread very carefully here. As a starter therefore we will keep $n_0 = 0$ and then study the $n_0 > 0$ case separately. Note that none of the $n_i$, that control the spatial and temporal derivatives, are taken to be {\it negative} here. Negative $n_i$, or equivalently inverse derivatives, would signal non-localities which we shall deal separately later (see also \cite{desitter2, coherbeta} for some more details on this). Looking at {\bf Tables \ref{lilaloo}}, {\bf \ref{morganL1}} and {\bf \ref{scarstan23}}, the quantum scaling of \eqref{botsuga} with $n_0 = 0$ $-$ which we denote as $\theta_{nl}^{(0)}$ $-$ becomes:

{\scriptsize
\bg\label{brittbaba}
\theta^{(0)}_{nl}  &= &  {\rm dom}\left({8\over 3} - \hat\zeta_e(t), {2\over 3} - \hat\sigma_e(t), {2\over 3} - (\hat\alpha_e(t), \hat\beta_e(t)), {2\over 3} + {\hat\alpha_e(t) + \hat\beta_e(t)\over 2} - \hat\zeta_e(t)\right) \sum_{i =1}^{13}l_i + \left({4\over 3} - {\hat\zeta_e(t)\over 2}\right)n_1 \nonumber\\
&+&  {\rm dom}\left({2\over 3} - \hat\sigma_e(t),~ {2\over 3}- (\hat\alpha_e(t), \hat\beta_e(t)), ~{2\over 3} + {\hat\alpha_e(t) + \hat\beta_e(t)\over 2} -\hat\zeta_e(t), ~ {8\over 3} - \hat\zeta_e(t),~ {2\over 3} \pm \hat\alpha_e(t) \mp \hat\beta_e(t) - (\hat\alpha_e(t), \hat\beta_e(t))\right) l_{14}\nonumber\\
&+& \left( {5\over 3} + {\hat\alpha_e(t) + \hat\beta_e(t)\over 4} - \hat\zeta_e(t)\right)\sum_{j = 15}^{18} l_j  + \left( {5\over 3} - {\hat\sigma_e(t)\over 2} - {\hat\zeta_e(t)\over 2}\right) \sum_{k = 19}^{23} l_k + \left({5\over 3} - {1\over 2}(\hat\alpha_e(t), \hat\beta_e(t)) - {\hat\zeta_e(t)\over 2}\right) \sum_{p = 24}^{28} l_p \nonumber\\
&+& \left( {2\over 3}  - {\hat\sigma_e(t)\over 2} - {1\over 2}(\hat\alpha_e(t), \hat\beta_e(t))\right) \sum_{q=29}^{33} l_q +
\left( {2\over 3} - {1\over 2}(\hat\alpha_e(t), \hat\beta_e(t)) + {\hat\alpha_e(t) + \hat\beta_e(t)\over 4} - {\hat\zeta_e(t)\over 2}\right) \sum_{r=34}^{37} l_r \nonumber\\
&+ & \left( {2\over 3} + {\hat\alpha_e(t) + \hat\beta_e(t)\over 4} - {\hat\sigma_e(t)\over 2} - {\hat\zeta_e(t)\over 2}\right) \sum_{s = 38}^{41} l_s {\red -} {1\over \vert\log~\bar{g}_s\vert} \sum_{i = 1}^{41} {\rm log}~ {\bf \red L_{N}}[\bar{a}_i, \bar{b}_i, \mathbb{N}_i; \mathbb{M}_{8}]
+ \left(\hat{l}_{emn}^{pq}(t) + {4\over 3} - 2\hat\sigma_e(t)\right)l_{42}\nonumber\\
&+& \left(\hat{l}_{emn}^{p\alpha}(t) + {4\over 3} - {3\hat\sigma_e(t)\over 2} - {1\over 2}(\hat\alpha_e(t), \hat\beta_e(t))\right)l_{43}  
+ \left(\hat{l}_{emn}^{pa} + {1\over 3} -{3\hat\sigma_e(t)\over 2} - {1\over 2} (0, \hat\eta_e(t))\right)l_{44} \nonumber\\ 
&+ & \left( \hat{l}_{emn}^{\alpha\beta}(t) + {4\over 3} - {\hat\sigma_e(t)} - (\hat\alpha_e(t), \hat\beta_e(t))\right)l_{45} 
+ \left(\hat{l}_{emn}^{\alpha a}(t) + {1\over 3} - {\hat\sigma_e(t)} - {1\over 2}(\hat\alpha_e(t), \hat\beta_e(t)) - {1\over 2}(0, \hat\eta_e(t))\right)l_{46} \nonumber\\
&+ & \left(\hat{l}_{e\alpha\beta}^{ma}(t) + {1\over 3} - {\hat\sigma_e(t)\over 2} - (\hat\alpha_e(t), \hat\beta_e(t)) - {1\over 2}(0, \hat\eta_e(t))\right)l_{47}
+ \left(\hat{l}_{eij}^{0m}(t) + {13\over 3} - {\hat\sigma_e(t)\over 2} - {3\hat\zeta_e(t)\over 2}\right)l_{48} \nonumber\\
&+ & \left(\hat{l}_{eij}^{0\alpha}(t) + {13\over 3} - {1\over 2}(\hat\alpha_e(t), \hat\beta_e(t)) - {3\hat\zeta_e(t)\over 2}\right)l_{49}
+ \left( \hat{l}_{emn}^{ab}(t) {\red - {2\over 3}} - {\hat\sigma_e(t)}  - {\hat\eta_e(t)\over 2}\right)l_{50}\nonumber\\
&+ &  \left(\hat{l}_{e\alpha\beta}^{ab}(t) {\red - {2\over 3}} - (\hat\alpha_e(t), \hat\beta_e(t))  - {\hat\eta_e(t)\over 2}\right)l_{51} 
+ \left(\hat{l}_{em\alpha}^{ab}(t) {\red - {2\over 3}} - {\hat\sigma_e(t)\over 2} - {1\over 2}(\hat\alpha_e(t), \hat\beta_e(t))  - {\hat\eta_e(t)\over 2}\right)l_{52}\nonumber\\
&+ &\left( \hat{l}_{emn}^{pi}(t) + {7\over 3} - {3\hat\sigma_e(t)\over 2} - {\hat\zeta_e(t)\over 2}\right)l_{53} 
+ \left(\hat{l}_{e\alpha\beta}^{mi}(t) + {7\over 3} - {\hat\sigma_e(t)\over 2} - (\hat\alpha_e(t), \hat\beta_e(t))  - {\hat\zeta_e(t)\over 2}\right)l_{54} \nonumber\\
&+ &\left( \hat{l}_{emn}^{\alpha i}(t) + {7\over 3} - {\hat\sigma_e(t)} - {1\over 2}(\hat\alpha_e(t), \hat\beta_e(t))  - {\hat\zeta_e(t)\over 2}\right)l_{55}
+ \left(\hat{l}_{emn}^{ai}(t) + {4\over 3} - {\hat\sigma_e(t)} - {1\over 2}(0, \hat\eta_e(t))  - {\hat\zeta_e(t)\over 2}\right)l_{56} \nonumber\\
&+ &\left( \hat{l}_{eab}^{mi}(t) + {1\over 3} - {\hat\sigma_e(t)\over 2} - {\hat\eta_e(t)\over 2}  - {\hat\zeta_e(t)\over 2}\right)l_{57}
+ \left(\hat{l}_{e\alpha\beta}^{ai}(t) + {4\over 3} - (\hat\alpha_e(t), \hat\beta_e(t)) - {1\over 2}(0, \hat\eta_e(t))  - {\hat\zeta_e(t)\over 2}\right)l_{58}\nonumber\\
&+ & \left(\hat{l}_{eab}^{\alpha i}(t) + {1\over 3} - {1\over 2}(\hat\alpha_e(t), \hat\beta_e(t)) - {\hat\eta_e(t)\over 2}  - {\hat\zeta_e(t)\over 2}\right)l_{59} + \left(\hat{l}_{emn}^{ij}(t) + {10\over 3} -{\hat\sigma_e(t)}  - {\hat\zeta_e(t)}\right)l_{61} \nonumber\\
& + & \left(\hat{l}_{ema}^{\alpha i}(t) + {4\over 3} -{\hat\sigma_e(t)\over 2} - {1\over 2}(\hat\alpha_e(t), \hat\beta_e(t)) - {1\over 2}(0, \hat\eta_e(t))  - {\hat\zeta_e(t)\over 2}\right)l_{60}
+ \left(\hat{l}_{emn}^{0p} + {7\over 3} -{3\hat\sigma_e(t)\over 2} - {\hat\zeta_e(t)\over 2}\right)l_{68}  \nonumber\\
& + & \left(\hat{l}_{e\alpha\beta}^{ij}(t) + {10\over 3} -(\hat\alpha_e(t), \hat\beta_e(t))  - {\hat\zeta_e(t)}\right)l_{63}
+ \left(\hat{l}_{ema}^{ij}(t) + {7\over 3} -{\hat\sigma_e(t)\over 2} -{1\over 2}(0, \hat\eta_e(t))  - {\hat\zeta_e(t)}\right)l_{64}\nonumber\\
&+&\left( \hat{l}_{e\alpha a}^{ij}(t) + {7\over 3} -{1\over 2}(\hat\alpha_e(t), \hat\beta_e(t)) -{1\over 2}(0, \hat\eta_e(t))  - {\hat\zeta_e(t)}\right)l_{65} 
+\left( \hat{l}_{eab}^{ij}(t) + {4\over 3} -{\hat\eta_e(t)\over 2}  - {\hat\zeta_e(t)}\right)l_{66} \nonumber\\
&+& \left( \hat{l}_{eij}^{0a}(t) + {10\over 3} -{1\over 2}(0, \hat\eta_e(t))  - {3\hat\zeta_e(t)\over 2}\right)l_{67}
+ \left(\hat{l}_{em\alpha}^{ij}(t) + {10\over 3} -{\hat\sigma_e(t)\over 2} -{1\over 2}(\hat\alpha_e(t), \hat\beta_e(t))  - {\hat\zeta_e(t)}\right)l_{62}
\nonumber\\
&+& \left( \hat{l}_{emn}^{0\alpha} + {7\over 3} -{\hat\sigma_e(t)} -{1\over 2} (\hat\alpha_e(t), \hat\beta_e(t)) - {\hat\zeta_e(t)\over 2}\right)l_{69}
+ \left(\hat{l}_{e\alpha\beta}^{0m} + {7\over 3} -{\hat\sigma_e(t)\over 2} - (\hat\alpha_e(t), \hat\beta_e(t)) - {\hat\zeta_e(t)\over 2}\right)l_{70} \nonumber\\
&+ & \left( \hat{l}_{eab}^{0m} + {1\over 3} -{\hat\sigma_e(t)\over 2} -{\hat\eta_e(t)\over 2} - {\hat\zeta_e(t)\over 2}\right)l_{71} 
+ \left(\hat{l}_{eab}^{0\alpha} + {1\over 3} -{1\over 2}(\hat\alpha_e(t), \hat\beta_e(t)) -{\hat\eta_e(t)\over 2} - {\hat\zeta_e(t)\over 2}\right)l_{72}
\nonumber\\
& + & \left( \hat{l}_{emn}^{0a} + {4\over 3} -{\hat\sigma_e(t)} -{1\over 2}(0, \hat\eta_e(t)) - {\hat\zeta_e(t)\over 2}\right)l_{73}
+\left( \hat{l}_{e0m}^{ia} + {7\over 3} - {\hat\sigma_e(t)\over 2} -{1\over 2}(0, \hat\eta_e(t)) - {\hat\zeta_e(t)}\right)l_{80}
\nonumber\\
&+ & \left( \hat{l}_{e\alpha\beta}^{0a} + {4\over 3} -(\hat\alpha_e(t), \hat\beta_e(t)) -{1\over 2}(0, \hat\eta_e(t)) - {\hat\zeta_e(t)\over 2}\right)l_{75}
+ \left(\hat{l}_{e0\alpha}^{ia} + {7\over 3} - {1\over 2}(\hat\alpha_e(t), \hat\beta_e(t)) -{1\over 2}(0, \hat\eta_e(t)) - {\hat\zeta_e(t)}\right)l_{81}
 \nonumber\\
&+& \left( \hat{l}_{e0m}^{\alpha i} + {10\over 3} -{\hat\sigma_e(t)\over 2} - {1\over 2} (\hat\alpha_e(t), \hat\beta_e(t)) - {\hat\zeta_e(t)}\right)l_{77}
+ \left(\hat{l}_{e\alpha\beta}^{0i} + {10\over 3} -(\hat\alpha_e(t), \hat\beta_e(t)) - {\hat\zeta_e(t)}\right)l_{78} \nonumber\\
& + & \left( \hat{l}_{eab}^{0i} + {4\over 3} -{\hat\eta_e(t)\over 2} - {\hat\zeta_e(t)}\right)l_{79} + \left(\hat{l}_{e0m}^{\alpha a} + {4\over 3} -{\hat\sigma_e(t)\over 2} - {1\over 2} (\hat\alpha_e(t), \hat\beta_e(t)) -{1\over 2}(0, \hat\eta_e(t)) - {\hat\zeta_e(t)\over 2}\right)l_{74}  \nonumber\\
& + & \left(\hat{l}_{emn}^{0i} + {10\over 3} -{\hat\sigma_e(t)} - {\hat\zeta_e(t)}\right)l_{76} + \left({1\over 3} - {\hat\sigma_e(t)\over 2}\right)n_2  + \left({1\over 3} - {\hat\alpha_e(t)\over 2}\right)n_{\theta_1}+ \left({1\over 3} - {\hat\beta_e(t)\over 2}\right)n_{\theta_2},
\nd}
where $n_1$ is the number of derivatives along ${\bf R}^2$, $n_2$ is the number of derivatives along ${\cal M}_4$, $n_{\theta_1}$ is the number of derivatives along ${\bf S}^1_{\theta_1}$ and $n_{\theta_2}$ is the number of derivatives along ${{\bf S}^1_{\theta_2} \over {\cal I}_{\theta_2}}$. The relative {\red minus} sign appearing on the fifth line is completely harmless because the ${\bf L_N}$ factor from \eqref{marianbrick} when raised by $\bar{g}_s$ always produces positive $g_s$ scalings. Similarly, the negative signs appearing on the ninth and the tenth lines are also harmless as long as $\hat{l}^{ab}_{e{\rm MN}}$, where $({\rm M, N}) \in {\cal M}_4 \times{\bf S}^1_{\theta_1} \times{{\bf S}^1_{\theta_2} \over {\cal I}_{\theta_2}}$, remains non-zero and positive. Thus the $g_s$ scaling \eqref{britt} shows why an EFT description can be given for the ${\rm E}_8 \times {\rm E}_8$ case. Question is what happens when we allow non-zero temporal derivatives in \eqref{botsuga}, {\it i.e.} when we allow $n_0 > 0$? 

The $n_0 > 0$ case is bit more subtle because in \eqref{botsuga} we  expect $({\bf g}^{-1})^{n+m} \partial_0^{2m} \ne ({\bf g}^{-1})^m \partial_0^{2m} ({\bf g}^{-1})^{n}$. Devising a covariant derivative ${\cal D}_0$ such that ${\cal D}_0 {\bf g}_{\rm AB} = 0$ unfortunately doesn't work because it is very hard to construct covariantly constant metric components in a temporally varying background such as ours. For this case we expect:

{\footnotesize
\bg\label{cortadams}
\left({\bf g}^{-1}\right)^{n_0/2} {\partial}_0^{n_0} \left[\left({\bf g}^{-1}\right)^{(n, l)}{\cal Q}_{nl}^{(0)}\right]  = 
\left({\bf g}^{-1}\right)^{(n, l) + {n_0/2}} {\partial}_0^{n_0} {\cal Q}^{(0)}_{nl} + \left({\bf g}^{-1}\right)^{n_0/2} {\cal Q}_{nl}^{(0)}
{\partial}_0^{n_0}\left[\left({\bf g}^{-1}\right)^{(n, l)}\right] +.., \nd}
where ${\cal Q}_{nl}^{(0)}$ is the un-contracted quantum series and equals to $\mathbb{Q}^{(\{l_i\},n_i)}_{\rm T}$ in \eqref{botsuga} with $n_0 = 0$ when contracted by the inverse metric components\footnote{In other words $\left({\bf g}^{-1}\right)^{(n, l)}{\cal Q}_{nl}^{(0)} =\mathbb{Q}^{(\{l_i\},n_i)}_{\rm T}\big\vert_{n_0 = 0}$ in \eqref{botsuga}. See also \eqref{botsuga2.0} and \eqref{fahingsha10} later.}. The dotted terms, as discussed in the paragraph below \eqref{botsuga}, are the additional pieces that can in principle be generated once we take the curvature related corrections from \eqref{botsuga}. In other words, the dotted terms in \eqref{cortadams} are {\it contained} within \eqref{botsuga}, implying that:
\bg\label{julconfchatoor}
\left({\bf g}^{-1}\right)^{(n, l) + {n_0/2}} {\partial}_0^{n_0} {\cal Q}^{(0)}_{nl} \subset{
\left({\bf g}^{-1}\right)^{n_0/2} {\partial}_0^{n_0} \left[\left({\bf g}^{-1}\right)^{(n, l)}{\cal Q}_{nl}^{(0)}\right]}. \nd
This means using the LHS of \eqref{cortadams} guarantees the encompassment of many more terms that would have otherwise been absent if we had used the LHS of \eqref{julconfchatoor}.
 For such a scenario, the case with $n_0 > 0$, the LHS of \eqref{cortadams} 
may be presented as a recurrence relation of the form:
\bg\label{daniniash2}
\theta_{nl} \equiv \theta^{(n_0)}_{nl}  & = & {\rm dom}\Bigg[\theta^{(n_0-1)}_{nl}+{4\over 3} - {\hat\zeta_e(t)\over 2} - {\log\Big(\pm\dot\theta^{(n_0-1)}_{nl}\vert\log~\bar{g}_s\vert\Big) \over\vert\log~\bar{g}_s\vert}, \\
&&~~~~~~~~ 
~\gamma_{1, 2} + {1\over 3} - {\hat\zeta_e(t)\over 2} + \theta^{(n_0-1)}_{nl} - {\log\Big(\theta^{(n_0-1)}_{nl}\Big)\over \vert\log~\bar{g}_s\vert}  \Bigg], \nonumber\nd
which is reminiscent of \eqref{daniniash} discussed earlier. The above relation, while not in contradiction with EFT expectedly doesn't  match with the LHS of \eqref{julconfchatoor}. This is because of at least two complications, one involving $\log\left(\theta^{(n_0-1)}_{nl}\right)$ and the other involving $\log\left(\pm\dot\theta^{(n_0-1)}_{nl}\right)$. While both can be ignored because of their sub-dominant natures, the LHS of \eqref{cortadams} still continues to capture the $\bar{g}_s$ scalings with multiple temporal derivatives. The dominant contribution from the first term is only a positive numerical factor (with sub-dominant $g_s$ corrections) and the second term involves derivatives of the sub-dominant corrections $\hat\zeta_e(t), \hat\sigma_e(t)$ et cetera. This means only the second line, excluding the log piece, in \eqref{daniniash2} matters thus implying a better analytic control of the LHS of \eqref{cortadams}. Putting in these conditions provide the following remarkably simple solution for $\theta_{nl}$ from the recurrence relation \eqref{daniniash2}:
\bg\label{cortanaeve}
\theta_{nl} \equiv \theta_{nl}^{(n_0)} = \theta_{nl}^{(0)} + n_0 ~{\rm dom}\left(1, {\hat\alpha_e(t) + \hat\beta_e(t)\over 4}\right) - {n_0 \hat\zeta_e(t)\over 2} + {n_0 \over 3}, \nd
which is expectedly always positive definite\footnote{Recall that $\hat\zeta_e(t) << 1$ for $-{1\over \sqrt{\Lambda}} < t < 0$, where $\Lambda$ is the bare cosmological constant from \eqref{marapaug}.} and therefore shows no contradiction with the underlying EFT. Note that, as long as we ignore the log corrections, the quantum scaling remains simple. However one could, with some efforts, incorporate the log corrections from \eqref{cortadams} and \eqref{daniniash2}, but we shall not do it here and leave it as an exercise for our diligent readers. 

\subsubsection{The quantum scalings in the generalized $SO(32)$ case \label{sec4.5.4}}

The story for the generalized $SO(32)$ case is similar to what we had in \eqref{brittbaba} and \eqref{cortanaeve}, but with some key differences. 
Looking at {\bf Tables \ref{lilaloo}}, {\bf \ref{morganL1}} and {\bf \ref{scarstan23}}, the quantum scaling of \eqref{botsuga} for this case  becomes:

{\scriptsize
\bg\label{brittbaba2}
\theta_{nl}  &= &  {\rm dom}\left({8\over 3} - \zeta_e(t), {2\over 3} - \alpha_e(t), {2\over 3} - \beta_e(t), {2\over 3} + {\beta_e(t)} - \zeta_e(t)\right) \sum_{i =1}^{14}l_i + \left({4\over 3} - {\zeta_e(t)\over 2}\right)n_1 \nonumber\\
&+& \left( {5\over 3} + {\beta_e(t)\over 2} - \zeta_e(t)\right)\sum_{j = 15}^{18} l_j  + \left( {5\over 3} - {\alpha_e(t)\over 2} - {\zeta_e(t)\over 2}\right) \sum_{k = 19}^{23} l_k + \left({5\over 3} - {\beta_e(t)\over 2} - {\zeta_e(t)\over 2}\right) \sum_{p = 24}^{28} l_p \nonumber\\
&+& \left( {2\over 3}  - {\alpha_e(t)\over 2} - {\beta_e(t)\over 2}\right) \sum_{q=29}^{33} l_q +
\left( {2\over 3} - {\zeta_e(t)\over 2}\right) \sum_{r=34}^{37} l_r 
+ n_0 ~{\rm dom}\left(1, {\beta_e(t)\over 2}\right) + {n_0 \zeta_e(t)\over 2} + {n_0 \over 3}
\nonumber\\
&+ & \left( {2\over 3} + {\beta_e(t)\over 2} - {\alpha_e(t)\over 2} - {\zeta_e(t)\over 2}\right) \sum_{s = 38}^{41} l_s {\red -} {1\over \vert\log~\bar{g}_s\vert} \sum_{i = 1}^{41} {\rm log}~ {\bf \red L_{N}}[\bar{a}_i, \bar{b}_i, \mathbb{N}_i; \mathbb{M}_{32}]
+ \left({l}_{emn}^{pq}(t) + {4\over 3} - 2\alpha_e(t)\right)l_{42}\nonumber\\
&+& \left({l}_{emn}^{p\alpha}(t) + {4\over 3} - {3\alpha_e(t)\over 2} - {\beta_e(t)\over 2}\right)l_{43}  
+ \left({l}_{emn}^{pa} + {1\over 3} -{3\alpha_e(t)\over 2} - {1\over 2} (0, \eta_e(t))\right)l_{44} \nonumber\\ 
&+ & \left({l}_{emn}^{\alpha\beta}(t) + {4\over 3} - {\alpha_e(t)} - \beta_e(t)\right)l_{45} 
+ \left({l}_{emn}^{\alpha a}(t) + {1\over 3} - {\alpha_e(t)} - {\beta_e(t)\over 2} - {1\over 2}(0, \eta_e(t))\right)l_{46} \nonumber\\
&+ & \left({l}_{e\alpha\beta}^{ma}(t) + {1\over 3} - {\alpha_e(t)\over 2} - \beta_e(t) - {1\over 2}(0, \eta_e(t))\right)l_{47}
+ \left({l}_{eij}^{0m}(t) + {13\over 3} - {\alpha_e(t)\over 2} - {3\zeta_e(t)\over 2}\right)l_{48} \nonumber\\
&+ & \left({l}_{eij}^{0\alpha}(t) + {13\over 3} - {\beta_e(t)\over 2} - {3\zeta_e(t)\over 2}\right)l_{49}
+ \left({l}_{emn}^{ab}(t) {\red - {2\over 3}} - {\alpha_e(t)}  - {\eta_e(t)\over 2}\right)l_{50}\nonumber\\
&+ &  \left({l}_{e\alpha\beta}^{ab}(t) {\red - {2\over 3}} - \beta_e(t)  - {\eta_e(t)\over 2}\right)l_{51} 
+ \left({l}_{em\alpha}^{ab}(t) {\red - {2\over 3}} - {\alpha_e(t)\over 2} - {\beta_e(t)\over 2}  - {\eta_e(t)\over 2}\right)l_{52}\nonumber\\
&+ &\left({l}_{emn}^{pi}(t) + {7\over 3} - {3\alpha_e(t)\over 2} - {\zeta_e(t)\over 2}\right)l_{53} 
+ \left({l}_{e\alpha\beta}^{mi}(t) + {7\over 3} - {\alpha_e(t)\over 2} - \beta_e(t)  - {\zeta_e(t)\over 2}\right)l_{54} \nonumber\\
&+ &\left({l}_{emn}^{\alpha i}(t) + {7\over 3} - {\alpha_e(t)} - {\beta_e(t)\over 2}  - {\zeta_e(t)\over 2}\right)l_{55}
+ \left({l}_{emn}^{ai}(t) + {4\over 3} - {\alpha_e(t)} - {1\over 2}(0, \eta_e(t))  - {\zeta_e(t)\over 2}\right)l_{56} \nonumber\\
&+ &\left({l}_{eab}^{mi}(t) + {1\over 3} - {\alpha_e(t)\over 2} - {\eta_e(t)\over 2}  - {\zeta_e(t)\over 2}\right)l_{57}
+ \left({l}_{e\alpha\beta}^{ai}(t) + {4\over 3} - \beta_e(t) - {1\over 2}(0, \eta_e(t))  - {\zeta_e(t)\over 2}\right)l_{58}\nonumber\\
&+ & \left({l}_{eab}^{\alpha i}(t) + {1\over 3} - {\beta_e(t)\over 2} - {\eta_e(t)\over 2}  - {\zeta_e(t)\over 2}\right)l_{59} + \left({l}_{emn}^{ij}(t) + {10\over 3} -{\alpha_e(t)}  - {\zeta_e(t)}\right)l_{61} \nonumber\\
& + & \left({l}_{ema}^{\alpha i}(t) + {4\over 3} -{\alpha_e(t)\over 2} - {\beta_e(t)\over 2} - {1\over 2}(0, \eta_e(t))  - {\zeta_e(t)\over 2}\right)l_{60}
+ \left({l}_{emn}^{0p} + {7\over 3} -{3\alpha_e(t)\over 2} - {\zeta_e(t)\over 2}\right)l_{68}  \nonumber\\
& + & \left({l}_{e\alpha\beta}^{ij}(t) + {10\over 3} -\beta_e(t)  - {\zeta_e(t)}\right)l_{63}
+ \left({l}_{ema}^{ij}(t) + {7\over 3} -{\alpha_e(t)\over 2} -{1\over 2}(0, \eta_e(t))  - {\zeta_e(t)}\right)l_{64}\nonumber\\
&+&\left({l}_{e\alpha a}^{ij}(t) + {7\over 3} -{\beta_e(t)\over 2} -{1\over 2}(0, \eta_e(t))  - {\zeta_e(t)}\right)l_{65} 
+\left({l}_{eab}^{ij}(t) + {4\over 3} -{\eta_e(t)\over 2}  - {\zeta_e(t)}\right)l_{66} \nonumber\\
&+& \left({l}_{eij}^{0a}(t) + {10\over 3} -{1\over 2}(0, \eta_e(t))  - {3\zeta_e(t)\over 2}\right)l_{67}
+ \left({l}_{em\alpha}^{ij}(t) + {10\over 3} -{\alpha_e(t)\over 2} -{\beta_e(t)\over 2}  - {\zeta_e(t)}\right)l_{62}
\nonumber\\
&+& \left({l}_{emn}^{0\alpha} + {7\over 3} -{\alpha_e(t)} -{\beta_e(t)\over 2} - {\zeta_e(t)\over 2}\right)l_{69}
+ \left({l}_{e\alpha\beta}^{0m} + {7\over 3} -{\alpha_e(t)\over 2} - \beta_e(t) - {\zeta_e(t)\over 2}\right)l_{70} \nonumber\\
&+ & \left({l}_{eab}^{0m} + {1\over 3} -{\alpha_e(t)\over 2} -{\eta_e(t)\over 2} - {\zeta_e(t)\over 2}\right)l_{71} 
+ \left({l}_{eab}^{0\alpha} + {1\over 3} -{\beta_e(t)\over 2} -{ \eta_e(t)\over 2} - {\zeta_e(t)\over 2}\right)l_{72}
\nonumber\\
& + & \left({l}_{emn}^{0a} + {4\over 3} -{\alpha_e(t)} -{1\over 2}(0, \eta_e(t)) - {\zeta_e(t)\over 2}\right)l_{73}
+\left({l}_{e0m}^{ia} + {7\over 3} - {\alpha_e(t)\over 2} -{1\over 2}(0, \eta_e(t)) - {\zeta_e(t)}\right)l_{80}
\nonumber\\
&+ & \left({l}_{e\alpha\beta}^{0a} + {4\over 3} -\beta_e(t) -{1\over 2}(0, \eta_e(t)) - {\zeta_e(t)\over 2}\right)l_{75}
+ \left({l}_{e0\alpha}^{ia} + {7\over 3} - {\beta_e(t)\over 2} -{1\over 2}(0, \eta_e(t)) - {\zeta_e(t)}\right)l_{81}
 \nonumber\\
&+& \left({l}_{e0m}^{\alpha i} + {10\over 3} -{\alpha_e(t)\over 2} - {\beta_e(t)\over 2} - {\zeta_e(t)}\right)l_{77}
+ \left({l}_{e\alpha\beta}^{0i} + {10\over 3} -\beta_e(t) - {\zeta_e(t)}\right)l_{78} \nonumber\\
& + & \left({l}_{eab}^{0i} + {4\over 3} -{\eta_e(t)\over 2} - {\zeta_e(t)}\right)l_{79} + \left({l}_{e0m}^{\alpha a} + {4\over 3} -{\alpha_e(t)\over 2} - {\beta_e(t)\over 2} -{1\over 2}(0, \eta_e(t)) - {\zeta_e(t)\over 2}\right)l_{74}  \nonumber\\
& + & \left({l}_{emn}^{0i} + {10\over 3} -{\alpha_e(t)} - {\zeta_e(t)}\right)l_{76} + \left({1\over 3} - {\alpha_e(t)\over 2}\right)n_2  + \left({1\over 3} - {\beta_e(t)\over 2}\right)n_3,
\nd}
where $n_1$ relates derivatives along ${\bf R}^2$, $n_2$ relates derivatives along ${\cal M}_4$ and $n_3$ relates derivatives along ${\cal M}_2 = {\mathbb{T}^2\over {\cal I}_2}$. Note that we have incorporated the consequence from the temporal derivatives from \eqref{cortanaeve} in  the third line. Expectedly \eqref{brittbaba2} is simpler than \eqref{brittbaba}, but does contain all the nuances from all the higher order perturbative and non-perturbative corrections to the metric and the flux components. Moreover, $\theta_{nl}$ is strictly a positive definite quantity despite the presence of a relative minus sign on the fourth line.
This appears from \eqref{marianbrick} but, because of its logarithmic form, is harmless and thus only contribute positive scalings. The other relative minus signs on the eighth and the ninth lines are again harmless as they are controlled by $l_{e{\rm MN}}^{ab} > {2\over 3}$ with $({\rm M, N}) \in {\cal M}_4 \times {\mathbb{T}^2\over {\cal I}_2}$.

\subsubsection{The quantum scalings in the simplified $SO(32)$ case \label{sec4.5.5}}

Our next venture is to compute the scaling for the simplified $SO(32)$ case. This may be easily derived from \eqref{brittbaba2} by making $\zeta_e(t) = \eta_e(t) = 0$ and $(\alpha_e(t), \beta_e(t)) = (\alpha(t), \beta(t))$, but it is instructive to explicitly show it here. One of the use of the simplified story is it's application to the more precise form of the Schwinger-Dyson equation as shown in later sections. With these inputs, and using the results from {\bf Tables \ref{firzacut6}}, {\bf \ref{firzacut8}}, {\bf \ref{firzacut9}}, {\bf \ref{firzacut4}}, {\bf \ref{fchumbon5}} and {\bf \ref{fchumbon6}}, the ${g_s\over {\rm HH}_o}$ scaling of the quantum series \eqref{botsuga} takes the following form:

{\footnotesize
\bg\label{fahingsha8}
\theta_{nl} &= & {\rm dom}\left({2\over 3} + \beta(t), {2\over 3} - \beta(t), {8\over 3}\right) \sum_{i = 1}^{14} l_i + \left({5\over 3} + \gamma_{1, 2}(t)\right) \sum_{j = 15}^{16} l_j 
+ \left({1\over 3} + {\beta(t)\over 2}\right)n_1  \nonumber\\
& + & \left({5\over 3} + {\beta(t)\over 2}\right) \sum_{k = 17}^{23} l_k  + 
\left({5\over 3} - {\beta(t)\over 2}\right) \sum_{m = 24}^{28} l_m
+ {2\over 3} \sum_{n = 29}^{37} l_n  
+ \left({2\over 3} + {\beta(t)}\right) \sum_{p = 38}^{41} l_p \nonumber\\ 
&+ & \left({1\over 3} + \gamma_{1, 2}(t)\right)n_0 + {4n_3\over 3} + \left(l_{\alpha\beta}^{ai} + {4\over 3} - \beta(t)\right)l_{58} + \left({1\over 3} - {\beta(t)\over 2}\right)n_2 \nonumber\\
&+& \left(l_{mn}^{pq} + {4\over 3} - 2\alpha(t)\right)l_{42} + \left(l_{mn}^{p\alpha} + {4\over 3} 
-{3\alpha(t)\over 2}  - {\beta(t) \over 2}\right)l_{43} +
\left(l_{mn}^{pa} + {1\over 3} - {3\alpha(t)\over 2} \right)l_{44}\nonumber\\ 
&+& \left(l_{mn}^{\alpha\beta} + {4\over 3} - \alpha(t) - \beta(t)\right)l_{45} 
+ \left(l_{mn}^{\alpha a} + {1\over 3} - \alpha(t) - {\beta(t)\over 2}\right)l_{46}
+  \left(l_{ma}^{\alpha\beta} + {1\over 3} - {\alpha(t)\over 2} - \beta(t)\right)l_{47} \nonumber\\
&+ &  \left(l_{ij}^{0m} + {13\over 3} -{\alpha(t)\over 2}\right)l_{48} 
+ \left(l_{ij}^{0\alpha} + {13\over 3} - {\beta(t)\over 2}\right)l_{49} + \left(l_{mn}^{ab} {- {2\over 3}} - \alpha(t)\right){l_{50}}
+   \left(l_{\alpha\beta}^{ab} {- {2\over 3}} -\beta(t)\right){l_{51}}\nonumber\\
&+&  \left(l_{m\alpha}^{ab} {- {2\over 3}} - {\alpha(t)\over 2} - {\beta(t)\over 2}\right){l_{52}} + \left(l_{mn}^{pi} + {7\over 3} - {3\alpha(t)\over 2}\right)l_{53} 
+   \left(l_{m\alpha}^{\beta i} + {7\over 3} - {\alpha(t)\over 2} - \beta(t)\right)l_{54}\nonumber\\
& + & \left(l_{mn}^{\alpha i} + {7\over 3} - \alpha(t) - {\beta(t)\over 2}\right)l_{55} 
+ \left(l_{mn}^{ai} + {4\over 3} - \alpha(t)\right)l_{56}
+  \left(l_{ab}^{mi} + {1\over 3} - {\alpha(t)\over 2} \right)l_{57} \nonumber\\ 
& + &  \left(l_{ab}^{\alpha i} + {1\over 3} - {\beta(t)\over 2} \right)l_{59} + \left(l_{m\alpha}^{ai} + {4\over 3} 
-{\alpha(t)\over 2} - {\beta(t)\over 2}\right) l_{60} 
+ \left(l_{mn}^{ij} + {10\over 3} -\alpha(t)\right)l_{61}\nonumber\\
&+& \left(l_{m\alpha}^{ij} + {10\over 3} - {\alpha(t)\over 2} - {\beta(t)\over 2} \right)l_{62} + \left(l_{\alpha\beta}^{ij} + {10\over 3} -\beta(t)\right)l_{63} 
+ \left(l_{ma}^{ij} + {7\over 3} -{\alpha(t)\over 2}\right)l_{64}\nonumber\\
& + &  \left(l_{\alpha a}^{ij} + {7\over 3} -{\beta(t)\over 2}\right)l_{65} 
+ \left(l_{ab}^{ij} + {4\over 3}\right)l_{66} + \left(l_{ij}^{0a} + {10\over 3}\right)l_{67} 
+  \left(l_{0m}^{np} + {7\over 3} - {3\alpha(t)\over 2} \right)l_{68}\nonumber\\ 
& + &  \left(l_{0m}^{n \alpha} + {7\over 3} - \alpha(t) - {\beta(t)\over 2}\right)l_{69} + \left(l_{0m}^{\alpha\beta} + {7\over 3} - {\alpha(t)\over 2} - \beta(t)\right)l_{70} 
+ \left(l_{0m}^{ab} + {1\over 3} - {\alpha(t)\over 2}\right)l_{71} \nonumber\\
&+& \left(l_{0\alpha}^{ab} + {1\over 3} -  {\beta(t)\over 2}\right)l_{72} + \left(l_{0m}^{na} + {4\over 3} - \alpha(t)\right)l_{73} 
+ \left(l_{0m}^{\alpha a} + {4\over 3} - {\alpha(t)\over 2} - {\beta(t)\over 2}\right)l_{74}\nonumber\\
 &+ & \left(l_{0\alpha}^{\beta a} + {4\over 3} - \beta(t)\right)l_{75} 
+ \left(l_{0m}^{ni} + {10\over 3} -\alpha(t)\right)l_{76}
+ \left(l_{0m}^{\alpha i} + {10\over 3} -{\alpha(t)\over 2} - {\beta(t)\over 2}\right)l_{77} \nonumber\\
&+ & \left(l_{0a}^{bi} + {4\over 3}\right)l_{78}
+  \left(l_{0\alpha}^{\beta i} + {10\over 3} - \beta(t)\right)l_{79} 
+ \left(l_{0m}^{ia} + {7\over 3} - {\alpha(t)\over 2}\right)l_{80} + \left(l_{0\alpha}^{ia} + {7\over 3} - {\beta(t)\over 2}\right)l_{81},  
\nd}
where $l_i$, with $i = 1, ..., 41$ are the powers of the curvature tensors in \eqref{fahingsha8} and $(\gamma_1, \gamma_2)$ are the two values of the $g_s$ exponents in \eqref{ishena}. As before, this is clearly positive definite and therefore maintains the EFT status.

\begin{table}[tb]  
 \begin{center}
 \resizebox{\columnwidth}{!}{%
 \renewcommand{\arraystretch}{1.9}
\begin{tabular}{|c|c|c||c|c|c|c|c|}\hline G-fluxes & ${g_s\over {\rm H}(y){\rm H}_o({\bf x})}$ scalings & Lower bounds & G-fluxes & ${g_s\over {\rm H}(y){\rm H}_o({\bf x})}$ scalings & Lower bounds\\ \hline\hline
${\bf G}_{mnpq}$ & $\hat{l}_{emn}^{pq}(t)$ & $-1^\pm$ &  ${\bf G}_{\alpha\beta ij}$ & $\hat{l}_{e\alpha\beta}^{ij}(t) $ & $-3^\pm$ \\ \hline 
${\bf G}_{mnp\alpha}$ & $\hat{l}_{emn}^{p\alpha}(t)$ & $-1^\pm$ & ${\bf G}_{maij}$ & $\hat{l}_{ema}^{ij}(t) $ & $-2^\pm$ \\ \hline
${\bf G}_{mnpa}$ & $\hat{l}_{emn}^{pa}$
& $~~0^\pm$ & ${\bf G}_{\alpha aij}$ & $\hat{l}_{e\alpha a}^{ij}(t) $ & $-2^\pm$ \\ \hline
${\bf G}_{mn\alpha\beta}$ & $\hat{l}_{emn}^{\alpha\beta}$ & $-1^\pm$ &
${\bf G}_{abij}$ & $\hat{l}_{eab}^{ij}(t)$ & $-1^\pm$ \\ \hline
${\bf G}_{mn\alpha a}$ & $\hat{l}_{emn}^{\alpha a}(t)$ & $~~0^\pm$ &
${\bf G}_{0ija}$ & $\hat{l}_{eij}^{0a}(t)$ & $-3^\pm$ \\ \hline
${\bf G}_{m\alpha\beta a}$ & $\hat{l}_{e\alpha\beta}^{ma}(t)$ & $~~0^\pm$ &
${\bf G}_{0mnp}$ & $\hat{l}_{emn}^{0p} $
& $-2^\pm$\\ \hline
${\bf G}_{0ijm}$ & $\hat{l}_{eij}^{0m}(t) $ & $-4^\pm$ &  ${\bf G}_{0mn\alpha}$ & $\hat{l}_{emn}^{0\alpha} $
& $-2^\pm$\\
\hline 
\end{tabular}}
\renewcommand{\arraystretch}{1}
\end{center}
 \caption[]{\Su Lower bounds on the various ${g_s\over {\rm H}(y){\rm H}_o({\bf x})}$ scalings of the G-flux components.}
\label{micandi0_0}
 \end{table}



\begin{table}[tb]  
 \begin{center}
 \resizebox{\columnwidth}{!}{%
 \renewcommand{\arraystretch}{3.0}
\begin{tabular}{|c||c||c||c|}\hline fluxes & dual fluxes & Scalings for ${\rm E}_8 \times {\rm E}_8$ & Scalings for $SO(32)$ \\ \hline\hline
${\bf G}_{mnpq}$ & ${\bf G}_{0ij\alpha\beta ab}$   &  $\hat{l}_{emn}^{pq}(t) -2 + {3\hat\zeta_e(t)\over 2} - 2\hat\sigma_e(t) + {\hat\alpha_e(t) + \hat\beta_e(t)\over 2} + {\hat\eta_e(t)\over 2}$ & $l_{emn}^{pq}(t)-2 + {3\zeta_e(t)\over 2} - 2\alpha_e(t) + {\beta_e(t)} + {\eta_e(t)\over 2}$ \\ \hline 
${\bf G}_{mnp\alpha}$ & ${\bf G}_{0ijq\beta ab}$  & $\hat{l}_{emn}^{p\alpha}(t) -2 + {3\hat\zeta_e(t)\over 2} -\hat\sigma_e(t) + {\hat\alpha_e(t) + \hat\beta_e(t)\over 2} - (\hat\alpha_e(t), \hat\beta_e(t)) + {\hat\eta_e(t)\over 2}$ & $l_{emn}^{p\alpha}(t)-2 + {3\zeta_e(t)\over 2} -\alpha_e(t) +  {\eta_e(t)\over 2}$ \\ \hline 
${\bf G}_{mnpa}$ & ${\bf G}_{0ijq\alpha\beta b}$  & $\hat{l}_{emn}^{pa} 
-4 + {3\hat\zeta_e(t)\over 2} -\hat\sigma_e(t) + {\hat\alpha_e(t) + \hat\beta_e(t)\over 2} + \big({\hat\eta_e(t)\over 2}, -{\hat\eta_e(t)\over 2}\big)$
& ${l}_{emn}^{pa} -4 + {3\zeta_e(t)\over 2} -\alpha_e(t) + {\beta_e(t)} + \big({\eta_e(t)\over 2}, -{eta_e(t)\over 2}\big)$\\ \hline
${\bf G}_{mn\alpha\beta}$ & ${\bf G}_{0ijpqab}$  & $\hat{l}_{emn}^{\alpha\beta}(t)- 2 + {3\hat\zeta_e(t)\over 2} + {\hat\alpha_e(t) + \hat\beta_e(t)\over 2} + {\hat\eta_e(t)\over 2}$ & $l_{emn}^{\alpha\beta}(t)- 2 + {3\zeta_e(t)\over 2} + {\beta_e(t)} + {\eta_e(t)\over 2}$ \\ \hline 
${\bf G}_{mn\alpha a}$ & ${\bf G}_{0ijpq\beta b}$ & $\hat{l}_{emn}^{\alpha a}(t)-4 + {3\hat\zeta_e(t)\over 2} + {\hat\alpha_e(t) + \hat\beta_e(t)\over 2} - (\hat\alpha_e(t), \hat\beta_e(t)) + \big({\hat\eta_e(t)\over 2}, -{\hat\eta_e(t)\over 2}\big)$& $l_{emn}^{\alpha a}(t)-4 + {3\zeta_e(t)\over 2} +  \big({\eta_e(t)\over 2}, -{\eta_e(t)\over 2}\big)$ \\
\hline 
\end{tabular}}
\renewcommand{\arraystretch}{1}
\end{center}
 \caption[]{\Su The ${g_s\over {\rm H}(y){\rm H}_o({\bf x})}$ scalings of the dual fluxes for both the ${\rm E}_8 \times {\rm E}_8$ and $SO(32)$ cases. Note that the 
 contributions from these dual G-flux components to the quantum scalings in \eqref{botsuga2.0} remain exactly the same as the respective contributions to the quantum scalings in \eqref{botsuga}. This somewhat surprising feature is explained in section \ref{duldul}.}
\label{fridaylin1}
 \end{table}

\begin{table}[tb]  
 \begin{center}
 \resizebox{\columnwidth}{!}{%
 \renewcommand{\arraystretch}{2.8}
\begin{tabular}{|c||c||c||c|}\hline fluxes & dual fluxes & Scalings for ${\rm E}_8 \times {\rm E}_8$ & Scalings for $SO(32)$ \\ \hline\hline
${\bf G}_{ma\alpha i}$  & ${\bf G}_{0jnpqb\beta}$ & $\hat{l}_{ema}^{\alpha i}(t)-2 + {\hat\zeta_e(t)\over 2} + \hat\sigma_e(t) + {\hat\alpha_e(t) + \hat\beta_e(t)\over 2} -(\hat\alpha_e(t), \hat\beta_e(t)) +\big({\hat\eta_e(t)\over2},  -{\hat\eta_e(t)\over 2}\big)$ & $l_{ema}^{\alpha i}(t)-2 + {\zeta_e(t)\over 2} + \alpha_e(t)  +\big({\eta_e(t)\over2},  -{\eta_e(t)\over 2}\big)$ \\ \hline 
${\bf G}_{mnij}$ & ${\bf G}_{0pq\alpha\beta ab}$ & $\hat{l}_{emn}^{ij}(t) 
+ 2 -{\hat\zeta_e(t)\over 2}  + {\hat\alpha_e(t) + \hat\beta_e(t)\over 2}  +{\hat\eta_e(t)\over 2}$ & $l_{emn}^{ij}(t)+ 2 -{\zeta_e(t)\over 2}  + {\beta_e(t)}  +{\eta_e(t)\over 2}$ \\ \hline 
${\bf G}_{m\alpha ij}$ & ${\bf G}_{0npq\beta ab}$ & $\hat{l}_{em\alpha}^{ij}(t)+2 - {\hat\zeta_e(t)\over 2} + \hat\sigma_e(t) + {\hat\alpha_e(t) + \hat\beta_e(t)\over 2} -(\hat\alpha_e(t), \hat\beta_e(t)) +{\hat\eta_e(t)\over 2}$ & $l_{em\alpha}^{ij}(t)+2 - {\zeta_e(t)\over 2} + \alpha_e(t)  +{\eta_e(t)\over 2}$ \\ \hline 
${\bf G}_{\alpha\beta ij}$ & ${\bf G}_{0mnpqab}$  & $\hat{l}_{e\alpha\beta}^{ij}(t)+2 - {\hat\zeta_e(t)\over 2} + 2\hat\sigma_e(t) - {\hat\alpha_e(t) + \hat\beta_e(t)\over 2}  +{\hat\eta_e(t)\over 2}$ & $l_{e\alpha\beta}^{ij}(t)+2 - {\zeta_e(t)\over 2} + 2\alpha_e(t) - {\beta_e(t)}  +{\eta_e(t)\over 2}$ \\ \hline 
${\bf G}_{maij}$ & ${\bf G}_{0npqb\alpha\beta}$ & $\hat{l}_{ema}^{ij}(t) - {\hat\zeta_e(t)\over 2} + \hat\sigma_e(t) + {\hat\alpha_e(t) + \hat\beta_e(t)\over 2}  +\big({\hat\eta_e(t)\over 2}, -{\hat\eta_e(t)\over 2}\big)$ & $l_{ema}^{ij}(t)- {\zeta_e(t)\over 2} + \alpha_e(t) + {\beta_e(t)}  +\big({\eta_e(t)\over 2}, -{\eta_e(t)\over 2}\big)$ \\
\hline
\end{tabular}}
\renewcommand{\arraystretch}{1}
\end{center}
 \caption[]{\Su Scalings of the remaining components of the dual fluxes.}
\label{fridaylin2}
 \end{table}

\subsubsection{The quantum scalings in the dualized version of \eqref{botsuga} \label{duldul}}

One may also write the quantum series in terms of the {\it dual} of the four-forms, {\it i.e.} seven-forms. Such a construction becomes necessary to study Bianchi identities, flux quantizations and anomaly cancellations which we shall discuss soon. Putting in the seven-forms modify \eqref{botsuga} in the following way:

{\scriptsize
\bg\label{botsuga2.0}
\widetilde{\mathbb{Q}}_{\rm T}^{(\{l_i\}, n_i)} &= & \left[{\bf g}^{-1}\right] \prod_{i = 0}^3 \left[\partial\right]^{n_i} 
\prod_{{\rm k} = 1}^{41} \left({\bf R}_{\rm A_k B_k C_k D_k}\right)^{l_{\rm k}} \prod_{{\rm r} = 42}^{81} 
\left({\bf G}_{\rm A_r B_r C_r D_r E_r F_r J_r}\right)^{l_{\rm r}}\\
& = & {\bf g}^{m_i m'_i}.... {\bf g}^{j_k j'_k} 
\{\partial_m^{n_1}\} \{\partial_\alpha^{n_2}\} \{\partial_i^{n_3}\}\{\partial_0^{n_0}\}
\left({\bf R}_{mnpq}\right)^{l_1} \left({\bf R}_{mnab}\right)^{l_2}\left({\bf R}_{\alpha\beta ab}\right)^{l_3}\left({\bf R}_{abab}\right)^{l_4} \nonumber\\
&\times& \left({\bf R}_{mnij}\right)^{l_5}\left({\bf R}_{\alpha\beta ij}\right)^{l_6}
\left({\bf R}_{0m0n}\right)^{l_7}\left({\bf R}_{0\alpha 0\beta}\right)^{l_8}\left({\bf R}_{ijij}\right)^{l_9}
\left({\bf R}_{0i0j}\right)^{l_{10}}\left({\bf R}_{abij}\right)^{l_{11}}
\nonumber\\
& \times & \left({\bf R}_{0a0b}\right)^{l_{12}}\left({\bf R}_{\alpha\beta\alpha\beta}\right)^{l_{13}}\left({\bf R}_{mn\alpha\beta}\right)^{l_{14}}\left({\bf R}_{ijk0}\right)^{l_{15}}\left({\bf R}_{abi0}\right)^{l_{16}}\left({\bf R}_{\alpha\beta i0}\right)^{l_{17}}
\left({\bf R}_{mni0}\right)^{l_{18}}
\nonumber\\
& \times & \left({\bf R}_{m\alpha\beta i}\right)^{l_{19}}\left({\bf R}_{mnpi}\right)^{l_{20}}
\left({\bf R}_{m0i0}\right)^{l_{21}}\left({\bf R}_{mijk}\right)^{l_{22}}
\left({\bf R}_{mabi}\right)^{l_{23}}\left({\bf R}_{i\alpha\alpha\beta}\right)^{l_{24}}
\left({\bf R}_{mn\alpha i}\right)^{l_{25}}
\nonumber\\
&\times& \left({\bf R}_{\alpha 0i0}\right)^{l_{26}}\left({\bf R}_{\alpha ijk}\right)^{l_{27}}
\left({\bf R}_{\alpha abi}\right)^{l_{28}}\left({\bf R}_{mnp\alpha}\right)^{l_{29}}
\left({\bf R}_{m\alpha\alpha\beta}\right)^{l_{30}}\left({\bf R}_{m\alpha ab}\right)^{l_{31}}
\left({\bf R}_{m\alpha ij}\right)^{l_{32}}\nonumber\\
&\times& \left({\bf R}_{0m0\alpha}\right)^{l_{33}} 
\left({\bf R}_{mn\alpha 0}\right)^{l_{34}}
\left({\bf R}_{0\alpha\alpha\beta}\right)^{l_{35}}
\left({\bf R}_{\alpha 0ij}\right)^{l_{36}}\left({\bf R}_{0\alpha ab}\right)^{l_{37}}\left({\bf R}_{mnp0}\right)^{l_{38}}
\left({\bf R}_{m\alpha\beta 0}\right)^{l_{39}}\nonumber\\
&\times&\left({\bf R}_{m0ij}\right)^{l_{40}}
\left({\bf R}_{0mab}\right)^{l_{41}}\left({\bf G}_{0ij\alpha\beta ab}\right)^{l_{42}}\left({\bf G}_{0ijq\beta ab}\right)^{l_{43}}
\left({\bf G}_{0ijq\alpha\beta b}\right)^{l_{44}}\left({\bf G}_{0ijpqab}\right)^{l_{45}}
\nonumber\\
&\times&\left({\bf G}_{0ijpq\beta b}\right)^{l_{46}}\left({\bf G}_{0ijnpqb}\right)^{l_{47}}\left({\bf G}_{npq\alpha\beta ab}\right)^{l_{48}} 
\left({\bf G}_{mnpq\beta ab}\right)^{l_{49}}
\left({\bf G}_{0ijpq\alpha\beta}\right)^{l_{50}}\left({\bf G}_{0ijmnpq}\right)^{l_{51}}
 \nonumber\\
&\times&\left({\bf G}_{0ijnpq\beta}\right)^{l_{52}} \left({\bf G}_{0jq\alpha\beta ab}\right)^{l_{53}}\left({\bf G}_{0jnpq ab}\right)^{l_{54}}\left({\bf G}_{0jpq\beta ab}\right)^{l_{55}} 
\left({\bf G}_{0jpq\alpha\beta b}\right)^{l_{56}}\left({\bf G}_{0jnpq\alpha\beta}\right)^{l_{57}} \nonumber\\
&\times&\left({\bf G}_{0jbmnpq}\right)^{l_{58}}
\left({\bf G}_{0j\beta mnpq}\right)^{l_{59}} \left({\bf G}_{0jnpqb\beta}\right)^{60}\left({\bf G}_{0pq\alpha\beta ab}\right)^{l_{61}}\left({\bf G}_{0npq\beta ab}\right)^{l_{62}} 
\left({\bf G}_{0mnpqab}\right)^{l_{63}}
 \nonumber\\
&\times&\left({\bf G}_{0npqb\alpha\beta}\right)^{l_{64}}\left({\bf G}_{0\beta bmnpq}\right)^{l_{65}}
\left({\bf G}_{0\alpha\beta mnpq}\right)^{l_{66}} \left({\bf G}_{\alpha\beta mnpqb}\right)^{l_{67}}\left({\bf G}_{ijq\alpha\beta ab}\right)^{l_{68}}
\left({\bf G}_{ijpq\beta ab}\right)^{l_{69}} 
\nonumber\\
&\times&\left({\bf G}_{ijnpqab}\right)^{l_{70}}\left({\bf G}_{ijnpq\alpha\beta}\right)^{l_{71}}
\left({\bf G}_{ij\beta mnpq}\right)^{l_{72}}\left({\bf G}_{ij\alpha\beta pqb}\right)^{l_{73}}
\left({\bf G}_{ijnpq\beta b}\right)^{l_{74}}\left({\bf G}_{ijmnpqb}\right)^{l_{75}}\nonumber\\
&\times & \left({\bf G}_{jpq\alpha\beta ab}\right)^{l_{76}} \left({\bf G}_{jnpq\beta ab}\right)^{l_{77}} 
\left({\bf G}_{jmnpqab}\right)^{l_{78}} \left({\bf G}_{jmnpq\alpha\beta}\right)^{l_{79}}\left({\bf G}_{jnpq\alpha\beta b}\right)^{l_{80}}
\left({\bf G}_{jmnpq\beta b}\right)^{l_{81}},\nonumber \nd}
where as before $(m, n) \in {\cal M}_4, (i, j) \in {\bf R}^2$ and $ (a, b) \in {\mathbb{T}^2\over {\cal G}}$.  The remaining mapping is $(\alpha, \beta) \in {\mathbb{T}^2\over {\cal I}_2}$ or ${\bf S}^1_{\theta_1} \times {{\bf S}^1_{\theta_2}\over {\cal I}_{\theta_2}}$ depending on whether we consider $SO(32)$ case or the ${\rm E}_8 \times {\rm E}_8$ case respectively. The dual seven-form is defined in the usual way:

{\scriptsize
\bg\label{amqueen}
{\bf G}_7 &=& {\bf G}_{\rm M_1 M_2 M_3.... M_7}~ dx^{{\rm M}_1} \wedge dx^{{\rm M}_2} \wedge... \wedge dx^{{\rm M}_7}\\
&=& {1\over 7!} {\bf G}_{\rm A'B'C'D'} \sqrt{-{\bf g}_{11}} ~{\bf g}^{\rm A'A} {\bf g}^{\rm B'B} {\bf g}^{\rm C'C}  {\bf g}^{\rm D'D} ~\epsilon_{\rm ABCDM_1M_2M_3M_4M_5M_6M_7} ~dx^{{\rm M}_1} \wedge dx^{{\rm M}_2} \wedge... \wedge dx^{{\rm M}_7}, \nonumber\nd}
where without loss of generalities we used similar notations for the form and its dual. Note that the duality has to be computed with respect to the full non-perturbative completions of the metric and the flux components as described in section \ref{sec4.5.2}. It is easy to see that:
\bg\label{lindavendri}
\sqrt{-{\bf g}_{11}} = \begin{cases} -{14\over 3} + {3\zeta_e(t)\over 2}  + 2\alpha_e(t) + \beta_e(t) + {\eta_e(t)\over 2}  ~~~~~~~~~~~~~ {\rm for}~SO(32)\\
~~~~\\
-{14\over 3} + {3\hat\zeta_e(t)\over 2}  + 2\hat\sigma_e(t) + {\hat\alpha_e(t) + \hat\beta_e(t)\over 2} + {\hat\eta_e(t)\over 2}~~~~~~~ {\rm for}~~ {\rm E}_8 \times {\rm E}_8 \end{cases}, \nd
which when inserted in \eqref{amqueen} provides the scalings of the dual forms. The results are provided in {\bf Tables \ref{fridaylin1}} and {\bf \ref{fridaylin2}}. Looking at the entries in the aforementioned tables and comparing it with \eqref{amqueen}, it is easy to confirm that, due to the following identity between a four-form ${\bf G}_{\rm ABCD}$ and its dual ${\bf G}_{\rm M_1M_2...M_7}$:
\bg\label{altexkalu}
&& {\bf G}_{a_1 \ldots a_7}
= \frac{1}{4!}\,
\epsilon_{a_1 \ldots a_7 b_1 b_2 b_3 b_4}\,
{\bf G}^{b_1 b_2 b_3 b_4} \nonumber\\
&& |{\bf G}_7|^2
\;=\;
\frac{1}{7!}\,{\bf G}_{\rm M_1 \ldots M_7}{\bf G}^{\rm M_1 \ldots M_7}
\;=\;
\frac{1}{4!}\,{\bf G}_{\rm A_1 \ldots A_4}{\bf G}^{\rm A_1 \ldots A_4}
\;=\;
|{\bf G}_4|^2, \nd
where ${\bf F}_{ab} \equiv {\bf F}_{\rm MN} ~e^{\rm M}_a e^{\rm N}_b$ with $e^{\rm M}_a$ being the vielbein,
 the scaling of the quantum series in \eqref{botsuga2.0} {\it will exactly be the same as} the scaling of the original quantum series in \eqref{botsuga}. This is consistent with the type IIB case studied in \cite{desitter2}.

Note that in writing \eqref{fahingsha8} we have not mentioned the fermionic contributions. Fermionic terms may be incorporated by generalizing the definitions of the metric and the flux terms themselves. We will discuss more on this soon, but it will turn out that all the 
fermionic contributions, from both the curvature as well as the G-flux components, are sub-dominant. Despite this, and as mentioned above, the ${g_s\over {\rm HH}_o}$ scaling of \eqref{fahingsha8} as in \eqref{botsuga} is not simple, in particular we see that now there are too many relative {\it minus} signs. This is clearly not good, as uncontrolled relative signs would signify a breakdown of EFT description (see \cite{coherbeta2} for details on this). The scalings of the curvature and the derivative terms, from the first three lines of \eqref{botsuga} immediately puts the following bounds on the values of $\alpha(t), \beta(t)$ and $\gamma_{1, 2}(t)$:
\bg\label{macharr}
\alpha(t) < +{2\over3}, ~~~~ 0 < \beta(t) < + {2\over 3}, ~~~~ \gamma_{1, 2} > -{1\over 3} \nd
implying that the dominant scalings of ${\rm F}_1$ and ${\rm F}_2$ for the $SO(32)$ theory cannot exceed the aforementioned bounds, and $\gamma_{1, 2}(t) \ge 0$ because the scalings jump as $\pm {\mathbb{Z}\over 3}$. This is of course consistent with what we have been taking so far, but now the domain of validity of $(\alpha(t), \beta(t))$ has a stronger footing from the EFT criterion. For the 
${\rm E}_8 \times {\rm E}_8$ theory the story remains similar. All we need is to replace $\alpha(t) \to \hat\sigma(t), \gamma_{1, 2}(t) \to \hat\gamma_{1, 2}(t)$ and $\beta(t) \to (\hat\alpha(t), \hat\beta(t))$ in \eqref{botsuga} and also in all the subsequent tables. The replacement of 
$\beta(t) \to (\hat\alpha(t), \hat\beta(t))$ in \eqref{botsuga} means that we are looking at dominant of the two quantities $\left({\mathbb{Z}\over 3} - \hat\alpha(t)\right)$ and $\left({\mathbb{Z}\over 3} - \hat\beta(t)\right)$. For $\hat\alpha(t) > \hat\beta(t)$, it is the former that dominates, although at late time $\hat\alpha(t) = \hat\beta(t)$ and the system reduces to the two warp-factor case of \eqref{botsuga}. Similarly one may see from the flux contributions, the replacement becomes:

{\footnotesize
\bg\label{mcascanning}
l_{\rm AB}^{\rm CD} + {\mathbb{Z}\over 3} - a_1 \alpha(t) - a_2\beta(t) ~ \to ~ {\rm dom}\left(l_{\rm AB}^{\rm CD} + {\mathbb{Z}\over 3} - a_1 \hat\sigma(t) - a_2\hat\alpha(t), ~l_{\rm AB}^{\rm CD} + {\mathbb{Z}\over 3} - a_1 \hat\sigma(t) - a_2\hat\beta(t)\right), \nd}
where $a_i \in \mathbb{Z}$. Again an equality between $\hat\alpha(t)$ and $\hat\beta(t)$ will bring us back to \eqref{botsuga}. Similar replacements extend to all the tables contributing to \eqref{botsuga}.

Interestingly, by demanding the metric components to be {\it independent} of the coordinates of ${\cal M}_2$, {\it i.e.} independent of the $(\alpha,\beta)$ directions, it removes all the scalings with $-\beta(t)$ in the curvature and the derivative terms. The $-\beta(t)$ terms now only appear in the G-flux scalings. Again, similar story will be repeated for the ${\rm E}_8 \times {\rm E}_8$ case, where $\hat\alpha(t)$ and $\hat\beta(t)$ will show up in the G-flux scalings. For the $SO(32)$ case, 
realizing the internal six-manifold as a toroidal fibration over a four-dimensional base, much like what we had in  \cite{DRS} except now that the base and the toroidal fibration have temporal dependences, would not lead to any late-time singularities as we noticed before. Similar set-up albeit with a different topology would appear in the ${\rm E}_8 \times {\rm E}_8$ case. All these consistent compactifications would lead to a new non-K\"ahler six-manifolds in the cosmological setting. We will elaborate more on this issue in the following section.

\newpage

\begin{table}[tb]  
 \begin{center}
 \resizebox{\columnwidth}{!}{%
 \renewcommand{\arraystretch}{3.4}
}
\renewcommand{\arraystretch}{1}
\end{center}
 \caption[]{\Su Comparing the $g_s$ scalings of the two-form ${\bf R}^{a_0b_0}_{[mn]} dy^{m} \wedge dy^{n}$ as well as the Ricci tensor ${\bf R}_{mn}$ for $(m, n) \in {\cal M}_4$ for the generalized ${\rm E}_8 \times {\rm E}_8$ case.}
\label{niksmit1}
 \end{table}

\begin{table}[tb]  
 \begin{center}
 \resizebox{\columnwidth}{!}{%
 \renewcommand{\arraystretch}{3.3}
\begin{tabular}{|c||c||c||c|}\hline Riemann tensors & Contributions to ${\bf R}^{a_0b_0}_{[mn]}$ & Contributions to ${\bf R}_{mn}$\\ \hline\hline
${\bf R}_{mni0}$ & $1 + {\alpha}_e(t) - \zeta_e(t) + {\beta_e(t)\over 2}$ & $......$ \\ \hline 
${\bf R}_{mnp0}$ & ${{\alpha}_e(t)\over 2} - {\zeta_e(t)\over 2} + {\beta_e(t)\over 2}$ & $......$ \\ \hline 
${\bf R}_{mn\sigma 0}$ & ${{\alpha}_e(t)} - {\zeta_e(t)\over 2}$ & $......$ \\ \hline 
${\bf R}_{mnpi}$ & $1 + {{\alpha}_e(t)\over 2} - {\zeta_e(t)\over 2}$ & $......$ \\ \hline 
${\bf R}_{mnp\sigma}$ & ${{\alpha}_e(t)\over 2} - {\beta_e(t)\over 2}$ & $......$ \\ \hline 
${\bf R}_{mn\sigma i}$ & $1 + {{\alpha}_e(t)\over 2} - {\beta_e(t)\over 2} - {\zeta_e(t)\over 2}$ & $......$ \\ \hline 
${\bf R}_{mnpq}, {\bf R}_{mnab}$ & ${\rm dom}\left(0, \alpha_e(t)-\beta_e(t), 2 + \alpha_e(t) - \zeta_e(t), \alpha_e(t)+ {\beta_e(t)} - \zeta_e(t)\right)$ &${\rm dom}\left(0, \alpha_e(t)-\beta_e(t), 2 + \alpha_e(t) - \zeta_e(t), \alpha_e(t)+ {\beta_e(t)} - \zeta_e(t)\right)$\\ \hline 
${\bf R}_{mnij}, {\bf R}_{m0n0}, {\bf R}_{mn\rho\sigma}$ & ${\rm dom}\left(0, \alpha_e(t)-\beta_e(t), 2 + \alpha_e(t) - \zeta_e(t), \alpha_e(t)+ {\beta_e(t)} - \zeta_e(t)\right)$ &${\rm dom}\left(0, \alpha_e(t)-\beta_e(t), 2 + \alpha_e(t) - \zeta_e(t), \alpha_e(t)+ {\beta_e(t)} - \zeta_e(t)\right)$\\ \hline
\end{tabular}}
\renewcommand{\arraystretch}{1}
\end{center}
 \caption[]{\Su Comparing the $g_s$ scalings of the two-form ${\bf R}^{a_0b_0}_{[mn]} dy^{m} \wedge dy^{n}$ as well as the Ricci tensor ${\bf R}_{mn}$ for $(m, n) \in {\cal M}_4$ for the generalized $SO(32)$ case.}
\label{niksmit2}
 \end{table} 

\begin{table}[tb]  
 \begin{center}
 \resizebox{\columnwidth}{!}{%
 \renewcommand{\arraystretch}{4.5}
%
}
\renewcommand{\arraystretch}{1}
\end{center}
 \caption[]{\Su Comparing the $g_s$ scalings of the two-form ${\bf R}^{a_0b_0}_{[\alpha\beta]} dy^{\alpha} \wedge dy^{\beta}$ as well as the Ricci tensor ${\bf R}_{\alpha\beta}$ for $(\alpha, \beta) \in {\cal M}_2$ for the generalized ${\rm E}_8 \times {\rm E}_8$ case.}
\label{niksmit3}
 \end{table}

\begin{table}[tb]  
 \begin{center}
 \resizebox{\columnwidth}{!}{%
 \renewcommand{\arraystretch}{3}
%
}
\renewcommand{\arraystretch}{1}
\end{center}
 \caption[]{\Su Comparing the $g_s$ scalings of the two-form ${\bf R}^{a_0b_0}_{[\alpha\beta]} dy^{\alpha} \wedge dy^{\beta}$ as well as the Ricci tensor ${\bf R}_{\alpha\beta}$ for $(\alpha, \beta) \in {\cal M}_2$ for the generalized 
 for the generalized $SO(32)$ case.}
\label{niksmit4}
 \end{table}

\begin{table}[tb]  
 \begin{center}
\renewcommand{\arraystretch}{1.5}
%
\renewcommand{\arraystretch}{1}
\end{center}
 \caption[]{\Su Comparing the $g_s$ scalings of the two-form ${\bf R}^{a_0b_0}_{[mn]} dy^{m} \wedge dy^{n}$ as well as the Ricci tensor ${\bf R}_{mn}$ for $(m, n) \in {\cal M}_4$ for the simplified $SO(32)$ case. For vanishing $\beta(t)$, both the two-form and the Ricci tensor have dominant scalings of 0 confirming the $g_s$ scalings in \cite{desitter2}. For non-vanishing $\beta(t)$, the dominant contributions for both the form and the tensor come from $\left({g_s\over {\rm HH}_o}\right)^{-2\beta(t)}$ with $g_s << 1$.}
  \label{fchumbon1}
 \end{table}

\begin{table}[tb]  
 \begin{center}
\renewcommand{\arraystretch}{1.5}
\begin{tabular}{|c||c||c|}\hline Riemann curvatures & Contributions to ${\bf R}^{a_0b_0}_{[\alpha\beta]}$ & Contributions to ${\bf R}_{\alpha\beta}$
 \\ \hline\hline
${\bf R}_{\alpha\beta i0}$ & $1 + {3\beta(t)\over 2}$ & $......$\\ \hline
${\bf R}_{\alpha\beta \alpha m}$ & ${\beta(t)}$ & $......$  \\ \hline
${\bf R}_{\alpha\beta m0}$ & $2\beta(t)$ & $......$ \\ \hline
${\bf R}_{\alpha\beta\alpha 0}$ & ${\beta(t)}$ & $......$ \\ \hline
${\bf R}_{\alpha\beta m i}$ & $1 + {3\beta(t)\over 2}$ & $......$\\ \hline
${\bf R}_{\alpha\beta\alpha i}$ & $1 +  {\beta(t)\over 2}$ & $......$ \\ \hline
${\bf R}_{\alpha\beta mn}, {\bf R}_{\alpha\beta ab}$  
& ${\rm dom}\left(0, 2 + \beta(t), 2\beta(t)\right)$& ${\rm dom}\left(0, 2 + \beta(t), 2\beta(t)\right)$\\ \hline
${\bf R}_{\alpha\beta\alpha\beta}, {\bf R}_{\alpha\beta ij}, {\bf R}_{\alpha 0\beta 0}$ 
& ${\rm dom}\left(0, 2 + \beta(t), 2\beta(t)\right)$ & ${\rm dom}\left(0, 2 + \beta(t), 2\beta(t)\right)$ \\ \hline
\end{tabular}
\renewcommand{\arraystretch}{1}
\end{center}
 \caption[]{\Su Comparing the $g_s$ scalings of the two-form ${\bf R}^{a_0b_0}_{[\alpha\beta]} dy^{\alpha} \wedge dy^{\beta}$ as well as the Ricci tensor ${\bf R}_{\alpha\beta}$ for $(\alpha, \beta) \in {\cal M}_2$ for the simplified $SO(32)$ case. For vanishing  $\beta(t)$, both the two-form and the Ricci tensor have dominant scalings of 0 confirming the $g_s$ scalings in \cite{desitter2}. For non-vanishing $\beta(t)$, the dominant contributions for both the form and the tensor again come from $\left({g_s\over {\rm HH}_o}\right)^{0}$ with $g_s << 1$.}
  \label{fchumbon2}
 \end{table}

\begin{table}[tb]  
 \begin{center}
 \resizebox{\columnwidth}{!}{%
 \renewcommand{\arraystretch}{5.0}
}
\renewcommand{\arraystretch}{1}
\end{center}
 \caption[]{\Su Comparing the $g_s$ scalings of the two-form ${\bf R}^{a_0b_0}_{[ab]} dw^{a} \wedge dw^{b}$ as well as the Ricci tensor ${\bf R}_{ab}$ for $(a, b) \in {\mathbb{T}^2\over {\cal G}}$ for the generalized ${\rm E}_8 \times {\rm E}_8$ case.}
\label{niksmit5}
 \end{table}

\begin{table}[tb]  
 \begin{center}
 \resizebox{\columnwidth}{!}{%
 \renewcommand{\arraystretch}{4.0}
%
}
\renewcommand{\arraystretch}{1}
\end{center}
 \caption[]{\Su Comparing the $g_s$ scalings of the two-form ${\bf R}^{a_0b_0}_{[ab]} dw^{a} \wedge dw^{b}$ as well as the Ricci tensor ${\bf R}_{ab}$ for $(a, b) \in {\mathbb{T}^2\over {\cal G}}$ for the generalized $SO(32)$ case.}
\label{niksmit6}
 \end{table}

\begin{table}[tb]  
 \begin{center}
 \resizebox{\columnwidth}{!}{%
 \renewcommand{\arraystretch}{3.2}
%
}
\renewcommand{\arraystretch}{1}
\end{center}
 \caption[]{\Su Comparing the $g_s$ scalings of the two-form ${\bf R}^{a_0b_0}_{[ij]} dx^{i} \wedge dx^{j}$ as well as the Ricci tensor ${\bf R}_{ij}$ for $(i, j) \in {\bf R}^2$ for the generalized ${\rm E}_8 \times {\rm E}_8$ case.}
\label{niksmit7}
 \end{table}

\begin{table}[tb]  
 \begin{center}
 \resizebox{\columnwidth}{!}{%
 \renewcommand{\arraystretch}{3}
%
}
\renewcommand{\arraystretch}{1}
\end{center}
 \caption[]{\Su Comparing the $g_s$ scalings of the two-form ${\bf R}^{a_0b_0}_{[ij]} dx^{i} \wedge dx^{j}$ as well as the Ricci tensor ${\bf R}_{ij}$ for $(i, j) \in {\bf R}^2$ for the generalized $SO(32)$ case.}
\label{niksmit8}
 \end{table}

\begin{table}[tb]  
 \begin{center}
\renewcommand{\arraystretch}{1.7}
%
\renewcommand{\arraystretch}{1}
\end{center}
 \caption[]{\Su Comparing the $g_s$ scalings of the two-form ${\bf R}^{a_0b_0}_{[ab]} dw^a \wedge dw^b$ as well as the Ricci tensor ${\bf R}_{ab}$ for the simplified $SO(32)$ case and for $(a, b) \in {\mathbb{T}^2\over {\cal G}}$. For vanishing $\beta(t)$, the dominant scaling is $\bar{g}_s^2$ \cite{desitter2}, otherwise it is $\bar{g}_s^{2-\beta(t)}$.}
  \label{fchumbon3}
 \end{table}

\begin{table}[tb]  
 \begin{center}
  \renewcommand{\arraystretch}{1.7}
%
\renewcommand{\arraystretch}{1}
\end{center}
 \caption[]{\Su Comparing the $g_s$ scalings of the two-form ${\bf R}^{a_0b_0}_{[ij]} dx^{i} \wedge dx^{j}$ as well as the Ricci tensor ${\bf R}_{ij}$ for $(i, j) \in {\bf R}^2$ for the simplified $SO(32)$ case. For vanishing $\beta(t)$, the dominant scaling is $\bar{g}_s^{-2}$, otherwise it remains to be $\bar{g}_s^{-2-\beta(t)}$.}
\label{niksmit88}
 \end{table}

\begin{table}[tb]  
 \begin{center}
 \resizebox{\columnwidth}{!}{%
 \renewcommand{\arraystretch}{6}
%
}
\renewcommand{\arraystretch}{1}
\end{center}
 \caption[]{\Su Comparing the $g_s$ scalings of the two-form ${\bf R}^{a_0b_0}_{[m\rho]} dy^{m} \wedge dy^{\rho}$ as well as the Ricci tensor ${\bf R}_{m\rho}$ for $m \in {\cal M}_4$, $\rho \in {\cal M}_2$ for the generalized ${\rm E}_8 \times {\rm E}_8$ case.}
\label{niksmit9}
 \end{table}

\begin{table}[tb]  
 \begin{center}
 \resizebox{\columnwidth}{!}{%
 \renewcommand{\arraystretch}{3.1}
%
}
\renewcommand{\arraystretch}{1}
\end{center}
 \caption[]{\Su Comparing the $g_s$ scalings of the two-form ${\bf R}^{a_0b_0}_{[m\rho]} dy^{m} \wedge dy^{\rho}$ as well as the Ricci tensor ${\bf R}_{m\rho}$ for $m \in {\cal M}_4$, $\rho \in {\cal M}_2$ for the generalized $SO(32)$ case.}
\label{niksmit10}
 \end{table}
 
\begin{table}[tb]  
 \begin{center}
 \resizebox{\columnwidth}{!}{%
 \renewcommand{\arraystretch}{3.5}
%
}
\renewcommand{\arraystretch}{1}
\end{center}
 \caption[]{\Su Comparing the $g_s$ scalings of the two-form ${\bf R}^{a_0b_0}_{[0n]} dx^{0} \wedge dy^{n}$ as well as the Ricci tensor ${\bf R}_{0m}$ for $n \in {\cal M}_4$ for the generalized ${\rm E}_8 \times {\rm E}_8$ case.}
\label{niksmit11}
 \end{table}

\begin{table}[tb]  
 \begin{center}
 \resizebox{\columnwidth}{!}{%
 \renewcommand{\arraystretch}{3.3}
%
}
\renewcommand{\arraystretch}{1}
\end{center}
 \caption[]{\Su Comparing the $g_s$ scalings of the two-form ${\bf R}^{a_0b_0}_{[0n]} dx^{0} \wedge dy^{n}$ as well as the Ricci tensor ${\bf R}_{0m}$ for $n \in {\cal M}_4$ for the generalized $SO(32)$ case.}
\label{niksmit12}
 \end{table}

\begin{table}[tb]  
 \begin{center}
 \renewcommand{\arraystretch}{2.0}
%
\renewcommand{\arraystretch}{1}
\end{center}
 \caption[]{\Su Comparing the $g_s$ scalings of the two-form ${\bf R}^{a_0b_0}_{[m\rho]} dy^{m} \wedge dy^{\rho}$ as well as the Ricci tensor ${\bf R}_{m\rho}$ for $m \in {\cal M}_4$, $\rho \in {\cal M}_2$ for the simplified $SO(32)$ case. For vanishing $\beta(t)$, the dominant scaling for the 2-form is $\bar{g}_s^0$, whereas for non-vanishing $\beta(t)$, it becomes $\bar{g}_s^{-\beta(t)}$.}
\label{niksmit100}
 \end{table}

\begin{table}[tb]  
 \begin{center}
 \renewcommand{\arraystretch}{2}

\renewcommand{\arraystretch}{1}
\end{center}
 \caption[]{\Su Comparing the $g_s$ scalings of the two-form ${\bf R}^{a_0b_0}_{[0n]} dx^{0} \wedge dy^{n}$ as well as the Ricci tensor ${\bf R}_{0m}$ for $n \in {\cal M}_4$ for the simplified $SO(32)$ case. For non-vanishing $\beta(t)$, the dominant scaling for the 2-form is 
 $\bar{g}_s^{-1 - {2\beta(t)\over 2}}$, whereas for vanishing $\beta(t)$ it is $\bar{g}_s^{-1}$.}
\label{niksmit120}
 \end{table}

\begin{table}[tb]  
 \begin{center}
 \resizebox{\columnwidth}{!}{%
 \renewcommand{\arraystretch}{4.1}
%
}
\renewcommand{\arraystretch}{1}
\end{center}
 \caption[]{\Su The $\bar{g}_s \equiv {g_s\over {\rm H}(y) {\rm H}_o({\bf x})}$ scalings for the Ricci tensor ${\bf R}_{00}$ for all the heterotic theories, including the simplified $SO(32)$ theory. For the latter, the dominant scaling is $\bar{g}_s^{-2-\beta(t)}$.}
\label{niksmit13}
 \end{table}

\begin{table}[tb]  
 \begin{center}
 \resizebox{\columnwidth}{!}{%
 \renewcommand{\arraystretch}{2.7}
%
}
\renewcommand{\arraystretch}{1}
\end{center}
 \caption[]{\Su The $\bar{g}_s \equiv {g_s\over {\rm H}(y) {\rm H}_o({\bf x})}$ scalings for the two form ${\bf R}_{[ia]}^{a_ob_o} d{\bf x}^i \wedge dw^a$ for $i \in {\bf R}^2$ and $a \in {\mathbb{T}^2\over {\cal G}}$ for all the heterotic theories. For the simplified $SO(32)$ case, the dominant scaling becomes $\bar{g}_s^{-\beta(t)}$. Note that there are no contributions to the Ricci tensor ${\bf R}_{ia}$.}
\label{niksmit14}
 \end{table}

\begin{table}[tb]  
 \begin{center}
 \resizebox{\columnwidth}{!}{%
 \renewcommand{\arraystretch}{2.3}
}
\renewcommand{\arraystretch}{1}
\end{center}
 \caption[]{\Su The $\bar{g}_s \equiv {g_s\over {\rm H}(y) {\rm H}_o({\bf x})}$ scalings for the two form ${\bf R}_{[a0]}^{a_ob_o} d{\bf x}^0 \wedge dw^a$ for $a \in {\mathbb{T}^2\over {\cal G}}$ for all the heterotic theories. For the simplified $SO(32)$ case, the dominant scaling becomes $\bar{g}_s^{-\beta(t)}$. Note that there are no contributions to the Ricci tensor ${\bf R}_{a0}$.}
\label{niksmit15}
 \end{table}

\begin{table}[tb]  
 \begin{center}
 \resizebox{\columnwidth}{!}{%
 \renewcommand{\arraystretch}{2.3}
%
}
\renewcommand{\arraystretch}{1}
\end{center}
 \caption[]{\Su The $\bar{g}_s \equiv {g_s\over {\rm H}(y) {\rm H}_o({\bf x})}$ scalings for the two form ${\bf R}_{[ma]}^{a_ob_o} dy^m \wedge dw^a$ for $m \in {\cal M}_4$ and $a \in {\mathbb{T}^2\over {\cal G}}$ for all the heterotic theories. For the simplified $SO(32)$ case, the dominant scaling becomes $\bar{g}_s^{-\beta(t)}$. Note that there are no contributions to the Ricci tensor ${\bf R}_{ma}$.}
\label{niksmit16}
 \end{table}

\begin{table}[tb]  
 \begin{center}
 \resizebox{\columnwidth}{!}{%
 \renewcommand{\arraystretch}{2.3}
}
\renewcommand{\arraystretch}{1}
\end{center}
 \caption[]{\Su The $\bar{g}_s \equiv {g_s\over {\rm H}(y) {\rm H}_o({\bf x})}$ scalings for the two form ${\bf R}_{[\alpha a]}^{a_ob_o} dy^\alpha \wedge dw^a$ for $\alpha \in {\cal M}_2$ and $a \in {\mathbb{T}^2\over {\cal G}}$ for all the heterotic theories. For the simplified $SO(32)$ case, the dominant scaling becomes $\bar{g}_s^{1-{\beta(t)\over 2}}$. Note that there are no contributions to the Ricci tensor ${\bf R}_{\alpha a}$.}
\label{niksmit17}
 \end{table}

\begin{table}[tb]  
 \begin{center}
 \resizebox{\columnwidth}{!}{%
 \renewcommand{\arraystretch}{3.2}
}
\renewcommand{\arraystretch}{1}
\end{center}
 \caption[]{\Su Comparing the $g_s$ scalings of the two-form ${\bf R}^{a_0b_0}_{[0i]} dx^{0} \wedge dx^{i}$ as well as the Ricci tensor ${\bf R}_{0i}$ for $i \in {\bf R}^2$ for the generalized ${\rm E}_8 \times {\rm E}_8$ case.}
\label{niksmit18}
 \end{table}

\begin{table}[tb]  
 \begin{center}
 \resizebox{\columnwidth}{!}{%
 \renewcommand{\arraystretch}{2.6}
%
}
\renewcommand{\arraystretch}{1}
\end{center}
 \caption[]{\Su Comparing the $g_s$ scalings of the two-form ${\bf R}^{a_0b_0}_{[0i]} dx^{0} \wedge dx^{i}$ as well as the Ricci tensor ${\bf R}_{0i}$ for $i \in {\bf R}^2$ for the generalized $SO(32)$ case.}
\label{niksmit19}
 \end{table}

\begin{table}[tb]  
 \begin{center}
 \resizebox{\columnwidth}{!}{%
 \renewcommand{\arraystretch}{1.5}
%
}
\renewcommand{\arraystretch}{1}
\end{center}
 \caption[]{\Su Comparing the $g_s$ scalings of the two-form ${\bf R}^{a_0b_0}_{[0i]} dx^{0} \wedge dx^{i}$ as well as the Ricci tensor ${\bf R}_{0i}$ for $i \in {\bf R}^2$ for the simplified $SO(32)$ case. The dominant scaling for the 2-form is $\bar{g}_s^{-2-\beta(t)}$.}
\label{niksmit20}
 \end{table} 

\begin{table}[tb]  
 \begin{center}
 \resizebox{\columnwidth}{!}{%
 \renewcommand{\arraystretch}{3.5}
%
}
\renewcommand{\arraystretch}{1}
\end{center}
 \caption[]{\Su Comparing the $g_s$ scalings of the two-form ${\bf R}^{a_0b_0}_{[im]} dx^{i} \wedge dy^{m}$ as well as the Ricci tensor ${\bf R}_{im}$ for $i \in {\bf R}^2$  and $m \in {\cal M}_4$ for the generalized ${\rm E}_8 \times {\rm E}_8$ case.}
\label{niksmit21}
 \end{table} 

 \begin{table}[tb]  
 \begin{center}
 \resizebox{\columnwidth}{!}{%
 \renewcommand{\arraystretch}{3.5}
%
}
\renewcommand{\arraystretch}{1}
\end{center}
 \caption[]{\Su Comparing the $g_s$ scalings of the two-form ${\bf R}^{a_0b_0}_{[im]} dx^{i} \wedge dy^{m}$ as well as the Ricci tensor ${\bf R}_{im}$ for $i \in {\bf R}^2$  and $m \in {\cal M}_4$ for the generalized $SO(32)$ case.}
\label{niksmit22}
 \end{table} 

\begin{table}[tb]  
 \begin{center}
 \resizebox{\columnwidth}{!}{%
 \renewcommand{\arraystretch}{3.5}
%
}
\renewcommand{\arraystretch}{1}
\end{center}
 \caption[]{\Su Comparing the $g_s$ scalings of the two-form ${\bf R}^{a_0b_0}_{[i\rho]} dx^{i} \wedge dy^{\rho}$ as well as the Ricci tensor ${\bf R}_{im}$ for $i \in {\bf R}^2$  and $\rho \in {\cal M}_2$ for the generalized ${\rm E}_8 \times {\rm E}_8$ case.}
\label{niksmit23}
 \end{table} 

\begin{table}[tb]  
 \begin{center}
 \resizebox{\columnwidth}{!}{%
 \renewcommand{\arraystretch}{3.2}
%
}
\renewcommand{\arraystretch}{1}
\end{center}
 \caption[]{\Su Comparing the $g_s$ scalings of the two-form ${\bf R}^{a_0b_0}_{[i\rho]} dx^{i} \wedge dy^{\rho}$ as well as the Ricci tensor ${\bf R}_{im}$ for $i \in {\bf R}^2$  and $\rho \in {\cal M}_2$ for the generalized $SO(32)$ case.}
\label{niksmit232}
 \end{table} 

\begin{table}[tb]  
 \begin{center}
 \resizebox{\columnwidth}{!}{%
 \renewcommand{\arraystretch}{3.6}
%
}
\renewcommand{\arraystretch}{1}
\end{center}
 \caption[]{\Su Comparing the $g_s$ scalings of the two-form ${\bf R}^{a_0b_0}_{[0\alpha]} dx^{0} \wedge dy^{\alpha}$ as well as the Ricci tensor ${\bf R}_{0\alpha}$ for $\alpha \in {\cal M}_2$ for the generalized ${\rm E}_8 \times {\rm E}_8$ case.}
\label{niksmit24}
 \end{table}

\begin{table}[tb]  
 \begin{center}
 \resizebox{\columnwidth}{!}{%
 \renewcommand{\arraystretch}{3.4}
%
}
\renewcommand{\arraystretch}{1}
\end{center}
 \caption[]{\Su Comparing the $g_s$ scalings of the two-form ${\bf R}^{a_0b_0}_{[0\alpha]} dx^{0} \wedge dy^{\alpha}$ as well as the Ricci tensor ${\bf R}_{0\alpha}$ for $\alpha \in {\cal M}_2$ for the generalized $SO(32)$ case.}
\label{niksmit25}
 \end{table}

\begin{table}[tb]  
 \begin{center}
 \resizebox{\columnwidth}{!}{%
 \renewcommand{\arraystretch}{1.5}
%
}
\renewcommand{\arraystretch}{1}
\end{center}
 \caption[]{\Su Comparing the $g_s$ scalings of the two-form ${\bf R}^{a_0b_0}_{[im]} dx^{i} \wedge dy^{m}$ as well as the Ricci tensor ${\bf R}_{im}$ for $i \in {\bf R}^2$  and $m \in {\cal M}_4$ for the simplified $SO(32)$ case. The dominant scaling for the 2-form is $\bar{g}_s^{-1-{3\beta(t)\over 2}}.$}
\label{niksmit220}
 \end{table} 

\begin{table}[tb]  
 \begin{center}
 \resizebox{\columnwidth}{!}{%
 \renewcommand{\arraystretch}{1.5}
%
}
\renewcommand{\arraystretch}{1}
\end{center}
 \caption[]{\Su Comparing the $g_s$ scalings of the two-form ${\bf R}^{a_0b_0}_{[i\rho]} dx^{i} \wedge dy^{\rho}$ as well as the Ricci tensor ${\bf R}_{im}$ for $i \in {\bf R}^2$  and $\rho \in {\cal M}_2$ for the simplified $SO(32)$ case. The dominant scaling for the 2-form is
 $\bar{g}_s^{-1 +{\beta(t)\over 2}}.$}
\label{niksmit231}
 \end{table} 

\begin{table}[tb]  
 \begin{center}
 \resizebox{\columnwidth}{!}{%
 \renewcommand{\arraystretch}{1.5}
%
}
\renewcommand{\arraystretch}{1}
\end{center}
 \caption[]{\Su Comparing the $g_s$ scalings of the two-form ${\bf R}^{a_0b_0}_{[0\alpha]} dx^{0} \wedge dy^{\alpha}$ as well as the Ricci tensor ${\bf R}_{0\alpha}$ for $\alpha \in {\cal M}_2$ for the simplified $SO(32)$ case. The dominant scaling for the 2-form is $\bar{g}_s^{-1 - {\beta(t)\over 2}}$.}
\label{niksmit250}
 \end{table}




\section{Fermionic extension of \eqref{botsuga} and non-perturbative effects \label{servant}}

To study the Rarita-Schwinger fermions we need to define certain aspects of the story first, namely, what do we mean by the fermions, vielbeins and the Gamma matrices in the excited state, compared to their well-known presence in the Minkowski (solitonic) vacuum sector. Some of this was discussed in section \ref{creepquif} earlier, but here we want to also comment on the specific representations of the Gamma matrices that we shall use for our computations. First $-$ exactly in the way we expressed the on-shell bosonic degrees of freedom as expectation values in \eqref{katusigel} $-$ the Rarita-Schwinger fermions should also be expressed in the same light, {\it i.e.} as $\langle\Psi_{\rm A}\rangle_\sigma \equiv\Psi_{\rm A} =\Psi_{\rm A}({\bf x}, y; g_s(t))$ thus forming part of the {\it emergent} degrees of freedom. There is however a subtlety now when we try to decompose the metric components in terms of the vielbeins:

{\footnotesize
\bg\label{maecomics}
\langle{\bf g}_{\rm AB}\rangle_\sigma = {\bf g}_{\rm AB} = \langle {\bf e}^a_{\rm A} {\bf e}^b_{\rm B}\rangle_\sigma \eta_{ab} = 
\langle {\bf e}^a_{\rm A}\rangle_\sigma \langle {\bf e}^b_{\rm B}\rangle_\sigma \eta_{ab} + \sum\limits_{\sigma'}  \langle {\bf e}^a_{\rm A} {\bf e}^b_{\rm B}\rangle_{(\sigma'|\sigma)} \eta_{ab} + ... \equiv \hat{\bf e}^a_{\rm A} \hat{\bf e}^b_{\rm B} \eta_{ab} \nd}
where $({\rm A, B}) \in {\bf R}^{2, 1} \times {\cal M}_4 \times {\cal M}_2 \times \xoxo$. The first equality in \eqref{maecomics} is from \eqref{katusigel}; the second equality is from the standard decomposition of the metric components in terms of the vielbeins at the solitonic, {\it i.e.} Minkowski minimum, level; the third equality, with the dotted terms, is from the intermediate Glauber-Sudarshan states when we allow for the resolution of the identity as in \cite{joydeep}.  
The subtlety with the above decomposition in \eqref{maecomics} should be clear: the emergent metric is not completely captured by ${\bf e}^a_{\rm A} e^b_{\rm B} \eta_{ab}$, where ${\bf e}^a_{\rm A} \equiv \langle {\bf e}^a_{\rm A}\rangle_\sigma$, because of the intermediate Glauber-Sudarshan states. We can still write the emergent metric in terms of some {\it effective} vielbeins $\hat{\bf e}^a_{\rm A}$ that is constructed in a way shown above. Using the above formalism, one could also try to define the emergent Gamma matrices in the following way: 
\bg\label{maeargo}
\langle\Gamma_{\rm A}\rangle_\sigma = \Gamma_{\rm A} = \langle {\bf e}^a_{\rm A}\Gamma_a\rangle_\sigma =\langle {\bf e}^a_{\rm A}\rangle_\sigma \Gamma_a + \sum_{\sigma'}\langle {\bf e}^a_{\rm A}\Gamma_a\rangle_{(\sigma'|\sigma)} + ..., \nd
with the dotted terms as usual coming from the resolution of identity \cite{joydeep}. The above is {\it almost}  $\hat{\bf e}^a_{\rm A}\Gamma_a$ but not quite as the identification is not exactly compatible with \eqref{maecomics}, unless the Wheeler-de Witt wavefunction is sharply localized along the on-shell Glauber-Sudarshan state $\vert\sigma\rangle$ (see \cite{wdwpaper}). Therefore instead of \eqref{maeargo}, we will simply define the emergent Gamma matrices to be $\Gamma_{\rm A} = \hat{\bf e}^a_{\rm A}\Gamma_a$. This way the Gamma matrices will satisfy:
\bg\label{meghann}
[\Gamma_{\rm A}, \Gamma_{\rm B}]_+ = 2 {\bf g}_{\rm AB} \mathbb{I}, ~~ \Gamma_0^\dagger = + \Gamma_0, \Gamma^\dagger_{\rm A'} = - \Gamma_{\rm A'}~~~ \implies ~~(\Gamma^0)^2 \propto \xxy^{{8\over 3} - \hat\zeta_e(t)}\nd
with  ${\rm A'} \in {\bf R}^{2} \times {\cal M}_4 \times {\cal M}_2 \times \xoxo$, thus fitting perfectly with the Clifford algebra and allowing inherent temporal (and spatial) dependence for the ${\rm E}_8 \times {\rm E}_8$ case. (The $SO(32)$ case is by replacing $\hat\zeta_e(t)$ by $\zeta_e(t)$.) Note two things: {\Su one}, we have used the {\it mostly  minus} signature to allow for \eqref{meghann} just for this section. We will resort back to the mostly plus signature from next section as we have been following before. And {\Su two}, using the fact that the resolution of identity in \eqref{maecomics} should not change the $\bar{g}_s$ scalings of the metric components, we can fix the scalings of $\hat{\bf e}^a_{\rm A}$. Our aim now would be express the fermionic degrees of freedom by extending the definitions of the metric and flux degrees of freedom, much like \eqref{charlot} that extends the metric to incorporate the Gamma matrices.

A possible extension of the G-flux components, by incorporating eleven-dimensional gravitino, was studied in the second reference of \cite{coherbeta}. Here we want to generalize the picture further by first introducing a
matrix valued operator of the form:
\bg\label{darlene}
{\cal D}_{\rm M} \equiv \big[1 - (1-i)\delta_{\rm M0}\big]\Gamma_{\rm M} +  i\mathbb{I}~{\rm M}_p^{-1}{\partial}_{\rm M}, \nd
with the imaginary part scaling as the vielbein\footnote{By this we will always mean the effective vielbein unless mentioned otherwise.} (and thus would have the corresponding ${g_s\over {\rm HH}_o}$ scaling), whereas the real part would simply be the derivative. (We expect the covariant derivative ${\rm D}_{\rm M}$ to be automatically generated from the curvature and the flux terms in the series 
\eqref{botsuga}.) This distinction between the real and the imaginary parts comes from the purely imaginary representations of the eleven-dimensional Gamma matrices for the spatial directions. Such a representation is possible at least with the mostly minus signature and may be easily worked out with the matrix products of the Pauli matrices. This is a Majorana representation\footnote{The Hamiltonian density for the Majorana gravitino takes the form 
${\cal H} \equiv i \Psi_{\rm M}^\dagger \Gamma^0 \Gamma^{\rm MNP} {\rm D}_{\rm N} \Psi_{\rm P}$, 
where ${\rm D}_{\rm N}$ is the covariant derivative. There is no mass term and the hermiticity condition 
leads to $$\Gamma_0 \Gamma_{\rm MNP} + \Gamma_{\rm MNP}^\dagger\Gamma_0^\dagger = 0,$$
\noindent where ${\rm sign}(\Gamma^2_0) = +1$ because of our choice of the mostly minus signature. If there was a mass term for the gravitino then the reality of the Hamiltonian would imply 
$\Gamma_0^\dagger = +\Gamma_0$. We will however continue with this choice. This would imply 
$\Gamma^\dagger_{\rm M} = -\Gamma_{\rm M}, \Gamma_0^\dagger = + \Gamma_0$ and ${\partial}^\dagger_{\rm M, 0} = - {\partial}_{\rm M, 0}$. In this sense $\Gamma_{\rm M}, i\Gamma_0$ and ${\partial}_{\rm M, 0}$ are purely {\it imaginary}. Interestingly, one could ask if we can allow 
$\Gamma_{\rm M, 0}^\dagger = - \Gamma_{\rm M, 0}$, {\it i.e.} allow all Gamma matrices to be anti-Hermitian. This turns out to be {\it possible}, but only for Hermitian space-time.}.
The reason for choosing the specific form \eqref{darlene} is to allow for the four possibilities of the form\footnote{With $\Gamma_0$ replaced by $i\Gamma_0$ in this section.}:

{\footnotesize
\bg\label{maciel} 
&&\left({\rm Re}~{\cal D}{\cal D}^\dagger\right)_{\rm MA} = \left({\rm Re}~{\cal D}^\dagger{\cal D}\right)_{\rm MA} = -\Gamma_{\rm M} \Gamma_{\rm A} - {\rm M}_p^{-2}{\partial}_{\rm M} {\partial}_{\rm A}\\
&& \left({\rm Im}~{\cal D}{\cal D}^\dagger\right)_{\rm MA} = - \left({\rm Im}~{\cal D}^\dagger{\cal D}\right)_{\rm MA} = 
{\rm M}_p^{-1}\left(\Gamma_{\rm M} {\partial}_{\rm A} -  \Gamma_{\rm A} {\partial}_{\rm M}\right) -{\rm M}_p^{-1}[{\partial}_{\rm M}, \Gamma_{\rm A}]  \nonumber\\
&&\left({\rm Re}~{\cal D}^2\right)_{\rm MA} = \Gamma_{\rm M} \Gamma_{\rm A} - {\rm M}_p^{-2}{\partial}_{\rm M} {\partial}_{\rm A}, ~~~~ \left({\rm Im}~{\cal D}^2\right)_{\rm MA} = {\rm M}_p^{-1}\left(\Gamma_{\rm M} {\partial}_{\rm A} +   \Gamma_{\rm A}{\partial}_{\rm M}\right) + {\rm M}_p^{-1}[{\partial}_{\rm M}, \Gamma_{\rm A}], \nonumber \nd}
where we cannot generically use the fact that ${\rm D}_{\rm N}\Gamma_{\rm M} = 0$ to simplify further because the vielbeins are {\it not necessarily} covariantly constants with ${\rm D}_{\rm N}$ being the covariant derivative (with respect to metrics and the fluxes)\footnote{Recall that, in a time-dependent background, we do not necessarily expect the vielbeins to be covariantly constants. This may be easily checked by taking the usual definition of the spin-connection $-$ which is typically derived by invoking covariant constancy of the vielbeins in a time-{\it independent} background $-$ but now using the vielbeins $\hat{\bf e}^a_{\rm M}$ and study the $\bar{g}_s$ scalings of the various terms in the equation ${\rm D}_{\rm M} \hat{\bf e}^a_{\rm N} = 0$. Such a analysis will lead to a highly constrained system with no non-trivial solution. Moreover, in a time-dependent background using ${\rm D}_{\rm N}$ in \eqref{darlene} will complicate the definition by introducing a spin-connection resulting in non-trivial $\bar{g}_s$ scalings of the various terms in \eqref{darlene}.}. Note that the ${\rm Re}$ and ${\rm Im}$ connotations are used for the elements in the matrix representation of ${\cal D}_{\rm M}$ and not to its eigenvalues (in other words we are not imposing hermiticity. Thus if we make the real parts of \eqref{maciel} symmetric, then they would also be hermitian). Using \eqref{maciel}, let us start by defining the following Rarita-Schwinger field $\lambda_{\rm M}$ as\footnote{Other possible definition exists, namely $\lambda_{\rm M} =  {{\cal O}^{(1){\rm A}}_{\rm A}} \Psi_{\rm M} = [{\rm tr}~{\cal O}^{(1)}] \Psi_{\rm M}$, but we won't use it here.}:

{\footnotesize
\bg\label{pmlaren}
\lambda_{\rm M} =  {\cal O}^{(1)}_{\rm MA} \Psi^{\rm A} 
\equiv  \sum_{\{k_i\}}\Big(a_{k_1} {\rm Re}~{\cal D}^{2k_1}  + 
b_{k_2} {\rm Im}~{\cal D}^{2k_2} + d_{k_3} {\rm Re}~\vert{\cal D}\vert^{2k_3}  
+ e_{k_4} {\rm Im}~\vert{\cal D}\vert^{2k_4}\Big)_{\rm MA} \Psi^{\rm A}, \nd}
with the Gamma-matrices in the purely imaginary representation and $\Psi_{\rm M}$ being the eleven-dimensional gravitino. Since the ordering of ${\rm Im}~\vert{\cal D}\vert^2$ is important, we will simply choose the convention $\vert{\cal D}\vert^2 = {\cal D} {\cal D}^\dagger$. It is easy to see that, for ${k_i} = 1$ with $(a_1, b_1, d_1, e_1)$ kept arbitrary, \eqref{pmlaren} 
gives us:
\bg\label{lalaren}
\lambda_{\rm M} &=&  2(a_1 - d_1) \left(\Psi_{\rm M} + \Gamma_{\rm MA} \Psi^{\rm A}\right) - (a_1 + d_1) 
{\rm M}_p^{-2}
{\partial}_{\rm M} {\partial}_{\rm A} \Psi^{\rm A}\nonumber\\
 &+ & (b_1 + e_1){\rm M}_p^{-1} \Gamma_{\rm M} {\partial}_{\rm A} \Psi^{\rm A}
+ (b_1 - e_1){\rm M}_p^{-1} \Big(\Gamma_{\rm A} {\partial}_{\rm M} + [\partial_{\rm M}, \Gamma_{\rm A}]\Big) \Psi^{\rm A}, \nd  
where we see the usefulness of choosing the complex operator in \eqref{darlene}: it allows us to isolate the actions of the various real operators on the gravitino. At the lowest order, since we do not expect second-order derivatives on the gravitino, we can impose $a_1 = - d_1 = {1\over 4}$, giving us the following simplified expression:

{\footnotesize
\bg\label{pamnu}
\lambda_{\rm M} = \Psi_{\rm M} + \Gamma_{\rm MA} \Psi^{\rm A} 
+ (b_1 + e_1) {\rm M}_p^{-1} \Gamma_{\rm M} {\partial}_{\rm A} \Psi^{\rm A}
+ (b_1 - e_1) {\rm M}_p^{-1} \Big(\Gamma_{\rm A} {\partial}_{\rm M} + [\partial_{\rm M}, \Gamma_{\rm A}]\Big)\Psi^{\rm A}. \nd}
The definition \eqref{pmlaren}, while useful, relies on specific choice of $(a_1, d_1)$ to get rid of the second-derivative terms. There is also an ambiguity in the ordering of the indices ${\rm M}$ and ${\rm A}$ 
in \eqref{pmlaren}. Symmetrizing over these indices would still keep the second derivative terms at the lowest order, while anti-symmetrization gives:

{\footnotesize
\bg\label{pmlaren2}
&& \left({\rm Re}~{\cal D}^2\right)_{[{\rm MA}]} =  - \left({\rm Re}~\vert{\cal D}\vert^2\right)_{[{\rm MA}]} = 
\Gamma_{\rm MA}\\
&& \left({\rm Im}~{\cal D}^2\right)_{[{\rm MA}]} =[\partial_{\rm M}, \Gamma_{\rm A}] -[\partial_{\rm A}, \Gamma_{\rm M}], ~~~~~ 
\left({\rm Im}~\vert{\cal D}\vert^2\right)_{[{\rm MA}]} = [\partial_{\rm A}, \Gamma_{\rm M}]_+ - [\partial_{\rm M}, \Gamma_{\rm A}]_+, \nonumber\nd}
thus simplifying slightly the real parts of the expression \eqref{pmlaren}, where $[\cdot,\cdot]_+$ is the anti-commutator. Now defining $\lambda_{\rm M} =  {\cal O}^{(1)}_{\rm [MA]} \Psi^{\rm A}$,
at lowest order when $k_i = 1$, the contributions from each of the terms in \eqref{pmlaren} is now quite different, and the Rarita-Schwinger fermion takes the form:

{\scriptsize
\bg\label{pmlaren3}
\lambda_{\rm M} = \left(a_1 - d_1\right) \Gamma_{\rm MA} \Psi^{\rm A} + 
{\rm M}_p^{-1}\Big(b_1 [\partial_{\rm M}, \Gamma_{\rm A}] - e_1 
[\partial_{\rm M}, \Gamma_{\rm A}]_+\Big)\Psi^{\rm A} 
- {\rm M}_p^{-1}\Big(b_1 [\partial_{\rm A}, \Gamma_{\rm M}] + e_1 
[\partial_{\rm A}, \Gamma_{\rm M}]_+\Big)\Psi^{\rm A}, \nd}
for any values of $(a_1, b_1, d_1, e_1)$. Compared to \eqref{pamnu}, slight simplification has occurred from the absence of the second-order derivatives on the gravitino. Further simplifications to \eqref{pamnu} (and also \eqref{pmlaren3}) can happen if the vielbeins become covariantly constant. Unfortunately, as mentioned earlier, this is not guaranteed for temporally varying background so we will avoid using it here. The aim of the above exercise then is to see whether we can further augment the metric definition in \eqref{charlot} by fermionic terms constructed out of $\lambda_{\rm N}$. In other words, can we extend $\hat{\bf g}_{\rm CD}$ from \eqref{charlot} to the following:
\bg\label{wendy}
\hat{\bf g}_{\rm CD} = {\bf g}_{({\rm CD})}\mathbb{I} + c_2\Gamma_{\rm CD} + c_3 \bar{\lambda}_{\rm C} 
\lambda_{\rm D}, \nd
where $(c_2, c_3)$ are dimensionless constants? Such a fermionic extension would be possible if we could generate the kinetic term of the gravitino by plugging in \eqref{wendy} to \eqref{botsuga}. One possibility 
would be to look for a term of the form:

{\scriptsize
\bg\label{ruthL}
\hat{\bf g}^{\rm CD} {\bf R}_{\rm 0C0D} = {\bf g}^{\rm CD} \hat{\bf g}_{\rm CD, 0} {\bf g}_{00, 0} {\bf g}^{00} + .... & = &  \xxy^{{8\over 3} - \hat\zeta_e(t) + {\rm dom}\left({\cal A}_{e{\rm CD}}(t), {\cal B}_{e{\rm CD}}(t)\right) + {\rm dom}\left(\mathbb{C}_8(\hat\zeta_e), \mathbb{D}_8(\hat\zeta_e)\right)}\nonumber\\
& \times & \bar{\lambda}_{\rm C}({\bf x}, y, w)~ \lambda_{\rm D}({\bf x}, y, w)~ {\bf g}^{\rm CD}({\bf x}, y, w) + ..., \nd}
for the ${\rm E}_8 \times {\rm E}_8$ case,  
where we have taken $t_{\rm C}$ is the dominant scaling\footnote{In other words, 
$\lambda_{\rm C}({\bf x}, y, w; g_s) = \left({g_s\over {\rm HH}_o}\right)^{t_{\rm C} + \Lambda_{e{\rm C}}(t)} \lambda_{\rm C}({\bf x}, y, w)$. The constraint on $t_{\rm C}$ is that the dominant scaling of $\hat{\bf g}_{\rm CD}$ cannot exceed the dominant scaling of the metric factor ${\bf g}_{({\rm CD})} \equiv {\bf g}_{\rm CD}$ in 
\eqref{wendy} otherwise we will have problems with the EFT. \label{versailles}} of $\lambda_{\rm C}$ from \eqref{pmlaren}. The other parameters appearing in \eqref{ruthL} are defined as follows. The $\left(\mathbb{C}_8(\hat\zeta_e), \mathbb{D}_8(\hat\zeta_e)\right)$ have been defined in {\bf Table \ref{jessrog}}, and $\left({\cal A}_{e{\rm CD}}(t), {\cal B}_{e{\rm CD}}(t)\right)$ take the following form:

{\footnotesize
\bg\label{korbenijohnson}
&& {\cal A}_{e{\rm CD}} = t_{\rm CD} - a_{\rm CD} + \Lambda_{e{\rm CD}}(t) - \hat\Sigma_{e{\rm CD}}(t) - {\log\vert \dot{\Lambda}_{e{\rm CD}}(t)~\log~\bar{g}_s\vert \over \vert\log~\bar{g}_s\vert} \\
&&{\cal B}_{e{\rm CD}} = t_{\rm CD} - a_{\rm CD} + \Lambda_{e{\rm CD}}(t) - \hat\Sigma_{e{\rm CD}}(t) + \gamma_{1, 2}\left[{\hat\alpha_e(t) + \hat\beta_e(t)\over 2}\right] - 1 - {\log\vert t_{\rm CD} + \Lambda_{e{\rm CD}}(t)\vert \over \vert\log~\bar{g}_s\vert} \nonumber\\
&& t_{\rm CD} = t_{\rm C} + t_{\rm D}, ~~ \Lambda_{e{\rm CD}}(t) = 
\Lambda_{e{\rm C}}(t) + \Lambda_{e{\rm D}}(t), ~~~ 
{\bf g}_{\rm CD}({\bf x}, y, w; g_s) = \tilde{\bf g}_{\rm CD}({\bf x}, y, w) 
\xxy^{a_{\rm CD} + \hat\Sigma_{e{\rm CD}}(t)}, \nonumber \nd}
with $a_{\rm CD} \equiv a^{({\rm CD})} - \sigma^{({\rm CD})}$ from \eqref{meanermis} and \eqref{karenmach} being the dominant scalings; $({\rm C, D}) \in {\bf R}^{2, 1} \times {\cal M}_4 \times {\cal M}_2 \times \xoxo$, being the dominant scaling and $\hat\Sigma_{e{\rm CD}}(t) = (\hat\zeta_e(t), \hat\alpha_e(t),\hat\beta_e(t), \hat\sigma_e(t),$ $\hat\eta_e(t)) = \Sigma^{({\rm CD})}_e(t)$ from \eqref{meanermis} depending on which components of the metric we take in \eqref{ruthL}. Note that, the ${g_s\over {\rm HH}_o}$ scaling shown above does not include the hidden scalings that would come from the Gamma functions (for example 
there would be an additional scaling of $\left({g_s\over {\rm HH}_o}\right)^{{4\over 3} - {\hat\zeta_e(t)\over 2}}$ from $\Gamma^0$ appearing in \eqref{ruthL}). There are also other terms coming from the definition \eqref{wendy} that we denoted above by the dotted terms. Ignoring these subtleties,  and using the definition \eqref{pmlaren}, we find:
\bg\label{langmore}
{\bf g}^{\rm CD} \bar{\lambda}_{\rm C} \lambda_{\rm D} + ... &= & 
f_1 \bar{\Psi}_{[{\rm R}}\Gamma_{\rm A} \Gamma_{\rm C} \Gamma_{\rm B} \Gamma_{\rm D} \Psi_{{\rm S}]} {\bf g}^{\rm AR} {\bf g}^{\rm BS} {\bf g}^{\rm CD}\\
&+ &  f_2 {\rm M}_p^{-1} \bar{\Psi}_{\rm M} \Gamma^{\rm MNP} {\rm D}_{\rm N} \Psi_{\rm P} + 
f_3 {\rm M}_p^{-2} \bar{\Psi}_{[{\rm A}}\Gamma_{\rm BC} {\rm D}_{\rm H} {\rm D}_{\rm E} \Psi_{{\rm F}]} 
{\bf g}^{\rm AB}{\bf g}^{\rm CH}{\bf g}^{\rm EF} + ....., \nonumber \nd
where $f_i$ are dimensionless constants and the dotted terms\footnote{Some of the dotted terms, appearing from \eqref{botsuga}, are used to convert the normal derivatives to covariant derivatives.} are additional derivative terms of ${\cal O}\left({\rm M}_p^{-3}\right)$ including the commutator and the anti-commutator terms. At the low energy supergravity level we only expect the second term so the existence of higher derivative terms at the two-gravitino level would lead to numerous issues ranging from the consistency with supersymmetry to the presence of ghosts. Fortunately the first and the third terms vanish due to anti-symmetry, but the system is more constrained and is not clear if we can reproduce 
all the fermionic interactions at the supergravity level (even torsional correction to the spin-connection doesn't seem to appear naturally here).
Using either \eqref{pamnu} or \eqref{pmlaren3} simplifies some of the dotted terms in \eqref{langmore} but does not substantially change the outcome. 

\subsection{Construction of the fermionic bilinear terms \label{wistoon}}

Our little exercise above tells us that the modification of the metric as in \eqref{wendy} cannot be that simple. This is because the fermionic extension of the metric leads to a term of the following form:
\bg\label{savwatson}
\left(\overline\lambda \lambda\right)_{\rm MN} = \left(\Psi^\dagger \big[{\cal O}^{(1)}\big]^\dagger \Gamma^0 \big[{\cal O}^{(1)}\big]\Psi\right)_{\rm MN}, \nd
where ${\cal O}^{(1)}$ is as defined in \eqref{pmlaren}, is not generic enough to reproduce all the interactions correctly. 
However what we did above isn't without merit: both the operator \eqref{darlene} and the Rarita-Schwinger fermion $\lambda_{\rm N}$ in \eqref{pmlaren} will be useful in the following way. Choosing the coefficients
in \eqref{pmlaren} appropriately, we can modify the metric \eqref{charlot} by adding fermionic contributions
as:
\bg\label{akash}
\hat{\bf g}_{\rm MN} \equiv \hat{\bf e}_{\rm M}^a \hat{\bf e}_{\rm N}^b\Big[\eta_{ab} \mathbb{I}  + c_2 \Gamma_{[ab]}\Big] + c_3 \left(\bar\lambda {\cal O} \lambda\right)_{\rm MN}, \nd
where the Lorentz indices over the Rarita-Schwinger fermion as well as the operator ${\cal O}$ will have to be inserted in and $c_3$ is a dimensionless parameter\footnote{We expect $c_3$ to be a real number. Any factors of $i = \sqrt{-1}$ to keep the fermionic terms real will be automatically inserted in as in  
\eqref{langmore} above.}. For the present case we  can quantify the fermionic contribution in the following way\footnote{Henceforth $\Gamma_{\rm A}$ defined in the usual way so that ${\cal D}_{\rm A} + {\cal D}^\dagger_{\rm A} = 2i\partial_{\rm A}$ and ${\cal D}_{\rm A} - {\cal D}^\dagger_{\rm A} = 2(-i)^{\delta_{\rm A0}}\Gamma_{\rm A}$, unless mentioned otherwise.}:

{\footnotesize
\bg\label{cascova}
&&\Big(\bar\lambda {\cal O} \lambda\Big)_{\rm MN} \equiv \mathbb{X}_{\rm MN} = \sum_{k_i} {c_{k_1k_2}} \left[\bar\Psi \cdot \left({\cal D} - {\cal D}^\dagger\right)^{k_1} \cdot 
\left({\cal D} + {\cal D}^\dagger\right)^{k_2}\cdot \Psi\right]_{\rm MN}\\
&&~ + \sum_{k_i} {\tilde{c}_{k_3k_4k_5k_6}} \left[\bar\Psi \cdot \left({\cal D} - {\cal D}^\dagger\right)^{k_3} \cdot 
\left({\cal D} + {\cal D}^\dagger\right)^{k_4}\cdot \Psi~
\bar\Psi \cdot \left({\cal D} - {\cal D}^\dagger\right)^{k_5} \cdot 
\left({\cal D} + {\cal D}^\dagger\right)^{k_6}\cdot \Psi \right]_{\rm MN} + ...., \nonumber \nd}
where the dotted terms are higher order fermionic interactions. The above series\footnote{An alternative viewpoint for the series \eqref{cascova} will be to define $\lambda_{\rm N} = \Psi_{\rm N}$ but now make the operator ${\cal O}$ more complicated. This way we could avoid non-trivial Fierz orderings arising from the choice \eqref{pmlaren}. Moreover, it should be understood that in \eqref{cascova} we can allow all permutations of the operator $\left({\cal D} - {\cal D}^\dagger\right)^{k_i} \cdot 
\left({\cal D} + {\cal D}^\dagger\right)^{k_j}$. For example for $k_i = k_j = 2$, we should include intermediate operators like $ c_{22}\left({\cal D} - {\cal D}^\dagger\right)^{2} \cdot 
\left({\cal D} + {\cal D}^\dagger\right)^{2},~~c_{1111}\left({\cal D} - {\cal D}^\dagger\right)\cdot  
\left({\cal D} + {\cal D}^\dagger\right) \cdot\left({\cal D} - {\cal D}^\dagger\right) \cdot 
\left({\cal D} + {\cal D}^\dagger\right),~~c_{121}\left({\cal D} - {\cal D}^\dagger\right) \cdot 
\left({\cal D} + {\cal D}^\dagger\right)^{2}\left({\cal D} - {\cal D}^\dagger\right)$ and all other permutations with different coefficients. The coefficients $c_{k_ik_j} \equiv c_{\{k_ik_j\}}$ can then be fixed by imposing eleven-dimensional supersymmetry. As discussed later, such complications may be necessary to get the generic interactions of the Rarita-Schwinger fermions from \eqref{botsuga}. \label{leanne}}, which is clearly much more general than \eqref{savwatson}, would make sense 
if $\sum\limits_i k_i \in 2\mathbb{Z}$. Also to avoid generating mass terms for the gravitino we will keep
$k_2 > 0$. 
On the other hand, low energy supersymmetry will make $c_{0n} = \tilde{c}_{0n0m} = ... =  0$, so to avoid this we can demand $k_1 > 0$ too. (This will however not imply generically $k_{2n+1} > 0$ for $n \ge 1$.)
The {\it dots} between the fermions and the operators signify the placement of the Lorentz indices. The coefficients $\hat{c}_{k_1....k_{2n}}$ are dimensionless numbers but are not necessarily real (because the Gamma matrices are in the purely imaginary representations and because of our choice \eqref{darlene}). The gravitinos are taken to be dimensionless, and therefore the dimensions are solely carried by the derivatives\footnote{There is minor subtlety here that needs some clarification. We can keep all the bosonic and the fermionic fields dimensionless and provide the ${\rm M}_p$ scalings of the terms in the action via the derivative counting. While this works well for the bosonic fields ({\it i.e.} the metric and the three-form fields), the gravitino kinetic term is suppressed by ${\rm M}_p$ instead of ${\rm M}_p^2$. As such at large wave-lengths it would appear to be more dominant than the bosonic term. (Even in \eqref{cascova} if we keep $k_2 > 0$, but $k_4 = k _6 = 0$, the second term will appear to be more dominant than the first term at large wave-lengths.) This is actually {\it not} a problem because we can always redefine the gravitino as 
$\Psi^{\rm here}_{\rm N} = {\rm M}_p^{-1/2}\Psi^{\rm HW}_{\rm N}$ where HW refers to \cite{horava}. Using dimensionless fields, the supersymmetry transformations at the solitonic Minkowski take the following form:
\bg\label{sust}
&&\delta \hat{e}^a_{\rm N} = {1\over 2} \bar\eta \Gamma^a \Psi_{\rm N} + ..., ~~~
\delta {\bf C}_{\rm MNP} = -{\sqrt{2}\over 8} \bar\eta \Gamma_{[{\rm MN}}\Psi_{{\rm P}]} + ...\nonumber\\
&& \delta \Psi_{\rm N} = {\rm M}_p^{-1} {\rm D}_{\rm N} \eta + {\sqrt{2} {\rm M}_p^{-1}\over 288}
\left(\Gamma_{\rm N}^{~~{\rm PQRS}} - 8 \delta^{\rm P}_{\rm N} \Gamma^{\rm QRS}\right) \eta {\bf G}_{\rm PQRS} + .... \nonumber \nd
where the dotted terms involve higher order terms necessary to keep the action with the quantum corrections \eqref{botsuga} supersymmetric, including terms that {\it supercovariantize}  the G-flux and the spin-connection. The Grassmann Majorana parameter $\eta$ is also kept dimensionless here, and therefore explicit ${\rm M}_p$ dependences appear in the supersymmetry transformations. If one uses the dimensionful gravitino, then $\eta$ will have a dimension of $-{1\over 2}$ as may be easily seen from the above transformations. \label{susie}}. Hence the higher order terms are suppressed by powers of ${\rm M}_p$ as one would have expected. Taking 
$c_{k_1k_2} = c_{31} = i$ and contracting 
\eqref{cascova} by ${\bf g}^{\rm MN}$
reproduces 
exactly the kinetic term for the gravitino with the full covariant derivative structure appearing from the curvature and flux corrections in \eqref{botsuga}. Constraints from supersymmetry may now be easily imposed by choosing appropriate values for the coefficients $c_{k_1...k_{2n}}$. For example, consistency with low energy supersymmetry implies that all $c_{k_1k_2} = 0$ except $c_{31}$. Interestingly, the series 
\eqref{cascova} matches exactly with the all order fermionic series predicted in \cite{evanfermion} once we dimensionally reduce the theory to IIB. In fact, since the seven and the three-branes' action come from the 
M-theory action near the singularities of the eight-manifold, the series \eqref{cascova} would directly predict the all-order fermionic terms on the D3 and D7 world volumes, thus confirming the results of 
\cite{evanfermion}.  

\begin{table}[tb]  
 \begin{center}
\resizebox{\columnwidth}{!}{%
\renewcommand{\arraystretch}{4.5}
}
\renewcommand{\arraystretch}{1}
\end{center}
 \caption[]{\Su Some lowest order contributions to $\hat{\bf g}_{\rm MN}$ where $({\rm M, N}) \in {\cal M}_4 \times {\cal M}_2$, with their corresponding $\bar{g}_s$ scalings following the notations in \eqref{tikiming}. Note that the scaling of $\bar{\Psi}_{\rm M} \Psi_{\rm A} \bar{\Psi}^{\rm A} \Psi_{\rm N}$ is the same as the one shown in the third row.} 
  \label{lusliberti}
 \end{table}

Let us try to make the aforementioned statements a bit more quantitative. With the fermionic modification of the metric of the form \eqref{akash} and the fermionic term taking the series structure \eqref{cascova}, we can easily work out the possible interactions involving gravitino from the curvature corrections in 
\eqref{botsuga}. For this we will have to specify some $\bar{g}_s$ scaling of the gravitinos. Taking the cue from footnote \ref{versailles}, we can express the gravitino scaling as:
\bg\label{tikiming}
\Psi_{\rm C}({\bf x}, y, w; g_s) & = & \sum_{k \in \mathbb{Z}} \Psi^{(k)}_{\rm C}({\bf x}, y, w) \xxy^{l_{\rm C} + \Lambda_{\rm C}(t) + {2k\over 3}|\widetilde\Lambda_{\rm C}(k; t)| + ...}\nonumber\\
&= & \widetilde{\Psi}_{\rm C}({\bf x}, y, w) \xxy^{l_{\rm C} + \hat\Lambda_{e{\rm C}}(t)}, \nd
where $l_{\rm C}$ is the dominant scaling and, as a first take, we expect $l_{\rm C} + l_{\rm D} + {5\over 3}$ to be of the same order as $a_{\rm CD}$, where $a_{\rm CD}$ is the dominant scaling of the on-shell metric components\footnote{A more precise determination for both on and off-shell cases will be discussed a bit later.}   ${\bf g}_{\rm CD}$.  
The dotted terms are all the possible non-perturbative and mixed corrections, which are summed over appropriately in $\hat\Lambda_{e{\rm C}}(t)$ much like how we collected all the non-perturbative corrections of the metric components in\footnote{We shall sometime use $\hat\Sigma_{e{\rm CD}}(t)$ to describe the same set of components when it becomes necessary to specify the orientations of the metric components.} $\hat\Sigma_e(t) \equiv (\hat\zeta_e(t), \hat\alpha_e(t), \hat\beta_e(t), \hat\sigma_e(t), \hat\eta_e(t))$. We have also defined, as in previously for the metric and the flux components, $\Psi_{\rm C}^{(k)}({\bf x}, y, w) = {\cal A}_\psi^{(k)}\widetilde\Psi_{\rm C}({\bf x}, y, w)$ with ${\bf x} \in {\bf R}^2, y \in {\cal M}_4 \times {\cal M}_2$ and $w \in \xoxo$.

There is however one subtlety that appears when we look at the scalings in {\bf Table \ref{lusliberti}}: not only does the number of positive scalings increase, but also does the number of $l_{\rm M}$'s, ${\rm M} \in {\cal M}_4 \times {\cal M}_2$, in equal proportions. The first row in {\bf Table \ref{lusliberti}} then provides the following constraints on $l_{m}$ and 
$l_\alpha$:
\bg\label{liberti}
l_m \ge -{7\over 6}^\pm, ~~~~ l_\alpha \ge -{7\over 6}^\pm, \nd
where the $\pm$ superscripts may be easily worked out from the contents in {\bf Table \ref{lusliberti}}. Taking $l_{\rm M} = -{7\over 6}^\pm$ now leads to the following issue. Consider a generic contribution to $\hat{\bf g}_{\rm MN}$ of the form $\bar{\Psi}_{\rm M}\Psi_{\rm N}(\bar\Psi_{\rm A}\Psi^{\rm A})^{{\cal N} -1}$ with ${\cal N} \ge 2$. It has no ${\rm M}_p$ suppressions, because of our choice of the dimensionless fermions, and appears from \eqref{cascova} by taking $k_i = 0$. This scales as:
\bg\label{merritt}
2{\cal N}\left(-{7\over 6}\right)^\pm + {4\over 3}^\pm + 2^\pm({\cal N} -1) = -{1\over 3}(2+{\cal N})^\pm, \nd
where the first term comes from the $\bar{g}_s$ scalings of the $2{\cal N}$ Rarita-Schwinger fermions, the second term comes from $\Gamma^0$ and the third term comes from the ${\cal N} -1$ fermionic condensates. We see that, as ${\cal N}$ increases, the $\bar{g}_s$ scaling of the condensate becomes more {\it negative} and therefore upstages the $\bar{g}_s$ scalings of the metric components ${\bf g}_{\rm MN}$ for $({\rm M, N}) \in {\cal M}_4 \times {\cal M}_2$. This means the choice \eqref{liberti} would make the contributions of fermionic condensates to $\hat{\bf g}_{\rm MN}$ dominate, unless\footnote{If $l_{\rm M}$ denotes the scaling of the fermions appearing in $\bar{\Psi}_{\rm M}\Psi_{\rm N}(\bar\Psi_{\rm A}\Psi^{\rm A})^{{\cal N} -1}$, where ${\cal N} \ge 2$, then $l_{\rm M}$ satisfies the following condition:
\bg 2l_{\rm M} + {4\over 3} + (2{\cal N} -2)l_{\rm M} + 2({\cal N} - 1) \ge -{2\over 3}, \nonumber \nd
where the first and the third terms are from the fermion products, the second terms is from $\Gamma^0$, the fourth term is from all the remaining Gamma matrices, and we have ignored the $\pm$ superscripts. The above inequality leads to $l_{\rm M} \ge -1$.} $l_{\rm M} > -1$, implying:
\begin{empheq}[box={\mybluebox[5pt]}]{equation}
\Psi_{\rm M}({\bf x}, y; g_s(t)) = \sum_{l \ge 1}\sum_{k_l = 0}^\infty \Psi^{(k, l)}_{{\rm M}}({\bf x}, y) \left({g_s\over {\rm HH}_o}\right)^{-1^\pm + {l\over 3} + {2k_l\over 3}\vert \widetilde{\hat\Lambda}_{\rm M}(k_l; t)\vert} 
\label{roseQ}
\end{empheq}
where $\widetilde{\hat\Lambda}_{\rm M}(k_l; t)$ contains all the perturbative and the non-perturbative corrections that would eventually be represented by $\hat\Lambda_{e{\rm M}}(t)$ in \eqref{tikiming}.
(For ${\rm C} \in {\mathbb{T}^2\over {\cal G}}$, and to avoid clashing with the EFT, $l_a \ge {2\over 3}$.) The form \eqref{roseQ} doesn't exactly fix the $\bar{g}_s$ scalings of the fermionic components because we expect a specific choice of $l$ in the series above to show up. We will tighten the lower bound further when we discuss the fermionic extensions of the G-flux components. Interestingly however, due to the definition of the generalized metric \eqref{akash}, we now expect on-shell cross-term components of the form  $\hat{\bf g}_{{\rm M}a}, \hat{\bf g}_{\mu{\rm M}}$ and $\hat{\bf g}_{\mu a}$ with ${\rm M} \in {\cal M}_4 \times {\cal M}_2, a \in \xoxo$ and $\mu \in {\bf R}^{2, 1}$.
Of course, on one hand, these cross-terms are only in the fermionic sector, scaling as appropriate 
powers of ${g_s\over {\rm HH}_o}$. On the other hand, they would imply the possibility of the existence of cross-term energy momentum tensors of the form ${\bf T}_{{\rm M}a}, {\bf T}_{\mu{\rm M}}$ and ${\bf T}_{\mu a}$ which may be constructed from \eqref{botsuga} {\it without} involving cross-term on-shell metric components ${\bf g}_{{\rm M}a}, {\bf g}_{\mu{\rm M}}$ and ${\bf g}_{\mu a}$. The existence of these energy momentum tensors will become immensely useful when we study the Schwinger-Dyson equations for the cross-term metric components later. We should also remember that, existence of the generalized metric components from the fermionic sector, would imply adding new curvature polynomials to an already complicated perturbative series \eqref{botsuga}. 
To avoid such un-necessary complications at this stage, and to simplify the cross-term structure, we will define the gravitinos only over the internal six-manifold 
${\cal M}_4 \times {\cal M}_2$ in this section. This means in \eqref{tikiming}, $\Psi_{\rm M} \equiv \Psi_{\rm M}\left(y, {g_s\over {\rm HH}_o}\right)$ with a dominant scaling $l_{\rm M}$. In the series \eqref{cascova} it would imply the absence of the indices $(w^a, x^i)$ in the expansion for $\left({\cal D} + {\cal D}^\dagger\right)^{k_{2n}}$
with $n \in \mathbb{Z}$. The only possibility is the appearance of the indices $(w^a, x^i)$ in the expansion 
for $\left({\cal D} - {\cal D}^\dagger\right)^{k_{2n + 1}}$. At the lowest order of fermions in the series \eqref{cascova}, it {will} allow generations of cross-terms $\hat{\bf g}_{{\rm M}a}, 
\hat{\bf g}_{{\rm M}\mu}$ and $\hat{\bf g}_{a\mu}$ for the ${\rm E}_8 \times {\rm E}_8$ case of the form\footnote{With $c_2 = 0$ in \eqref{akash}, and since we restricted the indices of $\Psi_{\rm M}$ to lie within ${\cal M}_4 \times {\cal M}_2$, terms like $c_{11}\overline\Psi_{\rm P}\Gamma^{\rm P} \partial_{\rm C} \Psi_{\rm D}$ do not appear, although we can allow $c_{31} \overline\Psi_{[{\rm C}} \Gamma^{\rm CD} \Gamma_{|{\rm E}|} \partial_{{\rm D}]} \Psi_{\rm M}$ with ${\rm E} \in {\bf R}^{2, 1} \times \xoxo$ for the metric components $\hat{\bf g}_{{\rm M}a}$ and $\hat{\bf g}_{{\rm M}\mu}$. However such combinations will not give us metric components $\hat{\bf g}_{a\mu}$. Thus generically at lowest order the fermionic combination that can produce all the three set of cross-term metric components is \eqref{lillee}.}:
\begin{table}[tb]  
 \begin{center}
\resizebox{\columnwidth}{!}{%
\renewcommand{\arraystretch}{3.5}
\begin{tabular}{|c||c||c|}\hline Cross-terms & Forms & ${g_s\over {\rm HH}_o}$ scalings \\ \hline\hline
 $\left(\bar\lambda{\cal O}\lambda\right)^{(1)}_{{\rm M}a}$ & ${c_{51}\over 64{\rm M}_p} ~\bar\Psi_{[\rm P} \Gamma^{\rm PQR}\Gamma_{\vert{\rm M}a\vert} \partial_{\rm Q}\Psi_{\rm R]}$ & ${8\over 3} - {\hat\zeta_e(t)\over 2} - {1\over 2}(\delta_{{\rm Q}n} + \delta_{{\rm R}p}){\hat\sigma_e(t)} - {1\over 2}(\hat\alpha_e(t), \hat\beta_e(t))(\delta_{{\rm Q}\alpha} + \delta_{{\rm R}\beta}) + {1\over 2}(0, \hat\eta_e(t)) + {\rm dom}\left({\cal A}_l, {\cal B}_l, {\cal C}_l\right)$  \\ \hline 
 $\left(\bar\lambda{\cal O}\lambda\right)^{(1)}_{{\rm M}\mu}$ & ${c_{51}\over 64{\rm M}_p} ~\bar\Psi_{[\rm P} \Gamma^{\rm PQR}\Gamma_{\vert{\rm M}\mu\vert} \partial_{\rm Q}\Psi_{\rm R]}$ & ${2\over 3} - {1\over 2}(\delta_{{\rm Q}n} + \delta_{{\rm R}p}){\hat\sigma_e(t)} - {1\over 2}(\hat\alpha_e(t), \hat\beta_e(t))(\delta_{{\rm Q}\alpha} + \delta_{{\rm R}\beta}) + {\rm dom}\left({\cal A}_l, {\cal B}_l, {\cal C}_l\right)$  \\ \hline 
$\left(\bar\lambda{\cal O}\lambda\right)^{(1)}_{a\mu}$ & ${c_{51}\over 64{\rm M}_p} ~\bar\Psi_{[\rm P} \Gamma^{\rm PQR}\Gamma_{\vert a\mu\vert} \partial_{\rm Q}\Psi_{\rm R]}$ & ${5\over 3}  - {1\over 2}(\delta_{{\rm Q}n} + \delta_{{\rm R}p}){\hat\sigma_e(t)} - {1\over 2}(\hat\alpha_e(t), \hat\beta_e(t))(\delta_{{\rm Q}\alpha} + \delta_{{\rm R}\beta})+ {1\over 2}(0, \hat\eta_e(t)) + {\rm dom}\left({\cal A}_l, {\cal B}_l, {\cal C}_l\right)$  \\ \hline 
 \end{tabular}}
\renewcommand{\arraystretch}{1}
\end{center}
 \caption[]{ \Su The precise ${g_s\over {\rm H}(y){\rm H}_o({\bf x})}$ scalings of the various terms of \eqref{lillee} contributing to the cross-term metric components in \eqref{akash}. We have also defined 
 ${\cal A}_l = l_m + l_n + \hat\Lambda_{em}(t)+ \hat\Lambda_{en}(t),  
 {\cal B}_l = l_m + l_{\alpha, \beta} + \hat\Lambda_{em}(t)+ \hat\Lambda_{e(\alpha, \beta)}(t)$ and ${\cal C}_l = l_{\alpha, \beta} + l_{\beta, \alpha} + \hat\Lambda_{e(\alpha, \beta)}(t)+ \hat\Lambda_{e(\beta, \alpha)}(t)$ using the parameters from \eqref{tikiming}.} 
  \label{rozechat}
 \end{table}

 {\footnotesize
\bg\label{lillee}
\left(\bar\lambda{\cal O}\lambda\right)^{(1)}_{\rm CD} = {c_{51}\over 64{\rm M}_p} ~\bar\Psi_{[\rm P} \Gamma^{\rm PQR}\Gamma_{\vert{\rm CD}\vert} \partial_{\rm Q}\Psi_{\rm R]} = \begin{cases}
~{1\over {\rm M}_p}\sum\limits_{k} \hat{\bf g}_{[{\rm M}a]}^{(k)}(y)\left({g_s\over {\rm HH}_o}\right)^{l_{\rm P} + l_{\rm R} + {2\over 3}(k|\widetilde\Lambda_{\rm P}(k;t)|+.. + 4)+ ...}\\
~{1\over {\rm M}_p}\sum\limits_{k} \hat{\bf g}_{[{\rm M}\mu]}^{(k)}(y)\left({g_s\over {\rm HH}_o}\right)^{l_{\rm P} + l_{\rm R} + {2\over 3}(k|\widetilde\Lambda_{\rm P}(k;t)|+..+ 1)+...}\\
~{1\over {\rm M}_p}\sum\limits_{k} \hat{\bf g}_{[a\mu]}^{(k)}(y)\left({g_s\over {\rm HH}_o}\right)^{l_{\rm P} + l_{\rm R} + {1\over 3} (2k|\widetilde\Lambda_{\rm P}(k;t)|+..+ 5)+...}
\end{cases} ~
\nd}
where $c_{51}$ is a non-zero constant, $\left({\rm P, Q, R}\right) \in {\cal M}_4 \times {\cal M}_2$ and the dominant scaling $l_{\rm P} + l_{\rm Q}$
could variously be $l_m + l_n, l_m + l_{\alpha, \beta}$ or $l_{\alpha, \beta} + l_{\beta, \alpha}$ depending on which submanifold we define the gravitinos. The dotted terms are the additional perturbative and the non-perturbative corrections, whose precise forms appear in {\bf Table \ref{rozechat}}. Note that the symmetric part of \eqref{lillee} vanishes, so it is only the anti-symmetric part which remains non-zero. The raising and lowering of the indices in \eqref{lillee} are done by the symmetric part of \eqref{akash} and therefore many of the curvature contributions using \eqref{lillee} could vanish. On the other hand, at higher orders in the fermionic expansion of \eqref{cascova} we could in principle get symmetric pieces by putting the free indices across the fermion bilinears. An example would be:

{\footnotesize
\bg\label{lillee1}
\left(\bar\lambda{\cal O}\lambda\right)^{(2)}_{\rm CD} = {\tilde{c}_{2020}\over 64} ~\bar\Psi_{[{\rm P}} \Gamma^{\rm PQ}\Gamma_{\vert{\rm C}\vert}\Psi_{{\rm Q}]} \bar\Psi_{[{\rm R}} \Gamma^{\rm RS}\Gamma_{\vert{\rm D}\vert}\Psi_{{\rm S}]}= \begin{cases}
~\sum\limits_{k} \hat{\bf g}_{a\mu}^{(k)}\left({g_s\over {\rm HH}_o}\right)^{l_{\rm P} + l_{\rm Q} + l_{\rm R} + l_{\rm S} + {2\over 3} (k|\widetilde\Lambda_{\rm P}(k;t)|+..+ 5)}\\
~\sum\limits_{k} \hat{\bf g}_{{\rm M}\mu}^{(k)}\left({g_s\over {\rm HH}_o}\right)^{l_{\rm P} + l_{\rm Q} + l_{\rm R} + l_{\rm S}+ {1\over 3}(2k|\widetilde\Lambda_{\rm P}(k;t)|+..+ 7)}\\
~\sum\limits_{k} \hat{\bf g}_{{\rm M}a}^{(k)}\left({g_s\over {\rm HH}_o}\right)^{l_{\rm P} + l_{\rm Q} + l_{\rm R} + l_{\rm S} + {1\over 3}(2k|\widetilde\Lambda_{\rm P}(k;t)|+..+ 13)}
\end{cases} ~
\nd}
where $c_{2020}$ is another constant, and the dominant scalings are distributed over the six-manifold ${\cal M}_4 \times {\cal M}_2$. The dotted terms are additional corrections whose precise forms are given in in {\bf Table \ref{rozechat2}}. In the absence of any derivatives, there appears no ${\rm M}_p$ suppressions. This isn't a problem as discussed in footnote \ref{susie}, and with every additional derivative the suppression factor becomes ${1\over {\rm M}_p} \left({g_s\over {\rm HH}_o}\right)^{{1\over 3} - {\hat\sigma_e(t)\over 2}}$. Thus all higher order  fermionic contributions to the metric in \eqref{akash} from \eqref{cascova} are highly subdominant in the limit $g_s < 1$. 

\begin{table}[tb]  
 \begin{center}
\resizebox{\columnwidth}{!}{%
\renewcommand{\arraystretch}{5.5}
}
\renewcommand{\arraystretch}{1}
\end{center}
 \caption[]{ \Su The precise ${g_s\over {\rm H}(y){\rm H}_o({\bf x})}$ scalings of the various terms of \eqref{lillee1} contributing to the cross-term metric components in \eqref{akash}. We have also defined 
 ${\cal A}_l = l_m + l_n + \hat\Lambda_{em}(t)+ \hat\Lambda_{en}(t),  
 {\cal B}_l = l_m + l_{\alpha, \beta} + \hat\Lambda_{em}(t)+ \hat\Lambda_{e(\alpha, \beta)}(t), {\cal C}_l = l_{\alpha, \beta} + l_{\beta, \alpha} + \hat\Lambda_{e(\alpha, \beta)}(t)+ \hat\Lambda_{e(\beta, \alpha)}(t)$  and ${\cal A}'_l = l_{m'} + l_{n'} + \hat\Lambda_{em'}(t)+ \hat\Lambda_{en'}(t),  
 {\cal B}'_l = l_{m'} + l_{\alpha', \beta'} + \hat\Lambda_{em'}(t)+ \hat\Lambda_{e(\alpha', \beta')}(t), {\cal C}'_l = l_{\alpha', \beta'} + l_{\beta', \alpha'} + \hat\Lambda_{e(\alpha', \beta')}(t)+ \hat\Lambda_{e(\beta', \alpha')}(t)$ using the parameters from \eqref{tikiming}.} 
  \label{rozechat2}
 \end{table}

\subsection{First look at the fermionic extension of metric components in \eqref{botsuga} \label{olivewatt}}

The two examples in \eqref{lillee} and \eqref{lillee1}, despite their relative suppressions, suggest that the quantum series in \eqref{botsuga} may be further augmented by powers of the fermionic bilinears. There are of course multiple possible ways of expressing such a power series of the bilinear terms, and most would be phenomenological,  but if we put constraints from low energy supersymmetry then at least we might provide some structure to such  a series\footnote{There is of course no supersymmetry at the level of the excited states, but since the emergent degrees of freedom appear from the expectation values of the actual bosonic and fermionic degrees of freedom, the supersymmetry should be manifest at the vacuum Minkowski (or the solitonic) level. This is the supersymmetry that we are alluding to. More details on how the {\it supergravity} action controls the dynamics of the emergent degrees of freedom via the Schwinger-Dyson equations will become clearer in section \ref{sec6.1}. See also \cite{joydeep}.}. One of the simplest possibility would be to express the fermionic contributions as:

{\footnotesize
\bg\label{dragun1}
\mathbb{Q}_{\rm met} \equiv 
\sum_{n} c_{(n)} ~{\rm tr} \left(\bar\lambda {\cal O} \lambda\right)^n
= c_{(1)}
\left(\bar\lambda {\cal O} \lambda\right)^{\rm C}_{\rm C} + c_{(2)}
\left(\bar\lambda {\cal O} \lambda\right)^{\rm C}_{\rm D} \left(\bar\lambda {\cal O} \lambda\right)^{\rm D}_{\rm C}
+ c_{(3)} \left(\bar\lambda {\cal O} \lambda\right)^{\rm C}_{\rm D} 
\left(\bar\lambda {\cal O} \lambda\right)^{\rm D}_{\rm E} \left(\bar\lambda {\cal O} \lambda\right)^{\rm E}_{\rm C}
+ ..., \nonumber\\ \nd}
where $c_{(n)}$ are constants independent of $g_s$ or ${\rm M}_p$. These coefficients could in principle be fixed by supersymmetry (in the time independent case), although such higher powers of fermionic terms have not been studied in the literature so far. At the lowest order in the fermionic bilinears, if we define 
$\bar\lambda {\cal O}\lambda \equiv \mathbb{X}$ as in \eqref{cascova}, then one may easily see:

{\footnotesize
\bg\label{lili}
{\rm tr}~\mathbb{X} = {\bf g}^{\rm CD}\left(\bar\lambda {\cal O} \lambda\right)_{\rm CD} &=& 
{c_{31}\over 16{\rm M}_p} \bar{\Psi}_{\rm M} \Gamma^{\rm MNP} \partial_{\rm N} \Psi_{\rm P} + 
{1\over 64 {\rm M}_p}\bar{\Psi}_{\rm M}\Gamma^{\rm MNP} \left(c_{411}\Gamma^{\rm Q} \partial_{[{\rm N}} \Gamma_{{\rm Q}]}
- c_{312}\partial_{[{\rm N}} \Gamma_{{\rm Q}]}\Gamma^{\rm Q}\right)\Psi_{\rm P}\nonumber\\
& + & {1\over 256 {\rm M}_p} \bar{\Psi}_{\rm M}\Gamma^{\rm MNPRS}\left(c_{512}\partial_{[{\rm R}} \Gamma_{{\rm S}]} \Gamma_{\rm N} + c_{611}\Gamma_{\rm N} \partial_{[{\rm R}} \Gamma_{{\rm S}]}\right)\Psi_{\rm P} + ... + {\cal O}\left(\Psi_{\rm M}^4\right), \nd} 
where we have used \eqref{cascova} and the coefficients $c_{nm}$ are defined there\footnote{There is an interesting subtlety that we should point out here. From \eqref{cascova} there appears possibilities of the form ${c_{31}\over 16{\rm M}_p} \bar{\Psi}_{\rm M} \Gamma^{\rm MNP} \partial_{\rm N} \Psi_{\rm P}$ and ${c_{51}\over 64{\rm M}_p} \bar{\Psi}_{\rm M} \Gamma^{\rm MNP}\Gamma^{\rm Q}\Gamma_{\rm Q} \partial_{\rm N} \Psi_{\rm P}$. Using the fact that $\Gamma^{\rm Q} \Gamma_{\rm Q} = 11$, it would appear that we can absorb the second interaction in the first. Alternatively, we could split $c_{31} = (a_1 + a_2) c_{31}$, with $a_1 + a_2 = 1$, such that a subset of the terms in \eqref{lili} takes the form:
\bg\label{oma2tag}
{a_1 c_{31}\over 16{\rm M}_p} \bar{\Psi}_{\rm M} \Gamma^{\rm MNP} \partial_{\rm N} \Psi_{\rm P} + \left({a_2c_{31}\over 16{\rm M}_p} + {c_{51}\over 64{\rm M}_p}\right)\bar{\Psi}_{\rm M} \Gamma^{\rm MNP} \partial_{\rm N}\Gamma^{\rm Q}\Gamma_{\rm Q} \Psi_{\rm P}, \nonumber \nd
\noindent showing that the first term is now a specific fraction of $c_{31}$, whereas the second terms is a combination of two terms. Looking at \eqref{susycond} we can then impose the following conditions on the coefficients:
\bg\label{oma2tag2}
c_{31} a_1 = -8i, ~~~ 4c_{31} a_2 + c_{51} = -32i, \nonumber \nd
with the requirement that $a_1 + a_2 = 1$ as mentioned earlier. Such a way of distributing the coefficients does not fix them uniquely, but shows that the inherent ambiguity of inserting powers of $\Gamma^{\rm Q}\Gamma_{\rm Q}$ does not alter any of the physical results. As such we can ignore these nuances and simply distribute the Gamma matrices from ${\cal D} - {\cal D}^\dagger$ in a way they appear in \eqref{cascova}.
Similar relations may be invoked for the other terms in the series \eqref{lili} without worrying about the repercussions from powers of $\Gamma^{\rm Q}\Gamma_{\rm Q}$ etc. \label{leanne2}}. (The dotted terms are additional permutations with appropriate coefficients as discussed in footnote \ref{leanne}.)
All the three set of terms are to the same order in ${\rm M}_p$, which is good, and if we ignore the 
${\cal O}\left(\Psi_{\rm M}^4\right)$ terms for the time being, we see that the second and the third set of fermionic bilinear terms combine with the first term to provide the spin-connection 
$\omega_{\rm N}^{ab}\Gamma_{ab}$ necessary to get the covariant derivative structure. This happens for (see also footnote \ref{leanne2}):
\bg\label{susycond}
c_{31} = -8 i, ~~~~~ c_{411} = c_{312} = 32 i, ~~~~~ c_{512} = c_{611} = -128 i, \nd
which are also the conditions that come from imposing low energy supersymmetry. Interestingly, the 
${\cal O}\left(\Psi_{\rm M}^4\right)$ terms that we ignored above would contribute a 
${\cal O}\left(\Psi_{\rm M}^2\right)$ {\it correction} to the spin-connection itself. One such possibility would be an interaction of the form:
\bg\label{dragun2}
{\tilde{c}_{3030}\over 64} ~\bar{\Psi}_{\rm M} \Gamma^{\rm MNP} \Psi_{\rm P} \cdot \bar{\Psi}^{\rm Q} 
\Gamma_{\rm QNR} \Psi^{\rm R}, \nd
which, for $\tilde{c}_{3030} = - 4i$, precisely provides the necessary contribution to {\it supercovariant} connection\footnote{A supercovariant connection, that typically appears at low energies in the eleven-dimensional supergravity action, is an extension of the usual spin-connection $\omega^{ab}_{\rm M}$ to:
\bg\label{skiler}
\hat{\omega}_{\rm M}^{ab}\Gamma_{ab} \equiv \omega^{ab}_{\rm M}\Gamma_{ab} + 
{1\over 8} \bar\Psi^{\rm Q}\Gamma_{\rm QMR} \Psi^{\rm R}, \nonumber \nd
whose supersymmetry variation does not involve derivatives of the infinitesimal Grassmann parameter. As such it doesn't involve any extra derivative factor. \label{supspin}}. Note however the absence of a ${\rm M}_p$ suppression which is because of our choice of dimensionless fermions. This could be easily compensated by giving appropriate dimensions to the fermions so that the ${\rm M}_p$ scalings match over 
\eqref{lili} and \eqref{dragun2}.

The fermionic bilinear series \eqref{dragun1} can also be represented using the generalized metric 
$\hat{\bf g}_{\rm CD}$ in \eqref{akash} using metric products.
For example, in the absence of the fermionic terms, metric product like ${\bf g}^{\rm CD} {\bf g}_{\rm CD}$ contributes nothing to \eqref{botsuga}. However now we could use \eqref{akash} to express \eqref{dragun1} in the following suggestive way:

{\footnotesize
\bg\label{dragun3}
\mathbb{Q}_{\rm met} &\equiv & \sum_{n = 1}^\infty c_{(n)} {\rm tr}~\mathbb{X}^n =  c_{(1)} {\rm tr}~\mathbb{X} + c_{(2)} {\rm tr}~\mathbb{X}^2 + 
c_{(3)} {\rm tr}~\mathbb{X}^3 + c_{(4)} {\rm tr}~\mathbb{X}^4 + c_{(5)} {\rm tr}~\mathbb{X}^5 + .... \nonumber\\
& = & c_{(1)} {\bf g}^{\rm CD} \left(\hat{\bf g}_{\rm CD} - {\bf g}_{\rm CD}\right) + 
c_{(2)}{\bf g}^{\rm CD}{\bf g}^{\rm EF}\left(\hat{\bf g}_{\rm CE} - {\bf g}_{\rm CE}\right)\left(\hat{\bf g}_{\rm DF}- {\bf g}_{\rm DF}\right)\\
&+& 
c_{(3)} {\bf g}^{\rm CD}{\bf g}^{\rm EF} {\bf g}^{\rm GH}\left(\hat{\bf g}_{\rm DE} - {\bf g}_{\rm DE}\right)
\left(\hat{\bf g}_{\rm FG}- {\bf g}_{\rm FG}\right)\left(\hat{\bf g}_{\rm CH}- {\bf g}_{\rm CH}\right) + 
{\cal O}\left(\left(\hat{\bf g}_{\rm CD} - {\bf g}_{\rm CD}\right)^4\right)
\nonumber \nd}
with the matrix $\mathbb{X}$ defined as $\mathbb{X} = \bar\lambda {\cal O}\lambda$ and $c_2 = 0$ in \eqref{akash}. 
By construction, while the metric products keep only the fermionic factors in the series, every term of the series is an infinite series by itself because of the definition of the matrix $\mathbb{X}$ in \eqref{cascova}. Thus the system has enough coefficients so that supersymmetric constraints do not make this a over-constrained system.

Our aforementioned construction is encouraging but as it stands, this still cannot be the complete story. For example, since ${\rm tr}~\mathbb{X}^p \ne \left({\rm tr}~\mathbb{X}\right)^p$, quantum terms like 
$\left({\rm tr}~\mathbb{X}\right)^p$ or $\left({\rm tr}~\mathbb{X}^p\right)^q \left({\rm tr}~\mathbb{X}^m\right)^n$ with $(m, n, p, q) \in \mathbb{Z}_+$ are absent in \eqref{dragun3}. Thus instead of \eqref{dragun3}, or 
equivalently \eqref{dragun1}, we can propose the following quantum series:
\bg\label{elmoss}
\mathbb{Q}^{(2)}_{\rm met} \equiv \sum_{\{p_i\}} c_{p_1p_2.....p_\infty} \prod_{i = 1}^\infty {\rm tr}~\mathbb{X}^{p_i} = 
\sum_{m, n, .., s}  c_{mnpq...s} {\rm tr}~\mathbb{X}^m \cdot {\rm tr}~\mathbb{X}^n \cdot {\rm tr}~\mathbb{X}^p \cdot {\rm tr}~\mathbb{X}^q \cdot \cdot \cdot {\rm tr}~\mathbb{X}^s, \nonumber\\ \nd
for which \eqref{dragun1} and \eqref{dragun3} are subsets with the coefficient $c_{(n)}$ identified to $c_{n00...0}$ in \eqref{elmoss}.  However, while on one hand \eqref{elmoss} provides a more complete picture of the fermionic interactions, on the other hand, \eqref{elmoss} appears to be harder to deal with at first sight. Is there a way to express it using the generalized metric \eqref{akash}? 

One possibility would be to rewrite \eqref{elmoss} as a series in powers of ${\rm tr}~\mathbb{X}^p$ with 
$p \in \mathbb{Z}$. Keeping in mind that such a series should reproduce \eqref{dragun3} at least at the lowest order, let us propose the following way to re-express \eqref{elmoss}:

{\footnotesize
\bg\label{elmoss2} 
\mathbb{Q}^{(2)}_{\rm met} &\equiv & \sum_{k_p}\left( \sum_{p} c^{(p)}_{k_p} \left[{\rm tr}\left(1 + \mathbb{X}\right)^p\right]^{k_p}\right)\nonumber\\
& = &  \sum_{k_1} c^{(1)}_{k_1} \left({\rm D} + {\rm tr}~\mathbb{X}\right)^{k_1} + 
\sum_{k_2} c^{(2)}_{k_2} \left({\rm D} + 2 {\rm tr}~\mathbb{X} + {\rm tr}~\mathbb{X}^2\right)^{k_2} + 
\sum_{k_3} c^{(3)}_{k_3} \left({\rm D} + 3 {\rm tr}~\mathbb{X} + 3{\rm tr}~\mathbb{X}^2 + {\rm tr}~\mathbb{X}^3\right)^{k_3} + ... , \nonumber\\ \nd}
where ${\rm D}$ is the space-time dimensions, and one may easily see that all powers of 
${\rm tr}~\mathbb{X}^p$ can in-principle appear in \eqref{elmoss2}. In fact, with a little effort, we can relate the coefficients $c_{mnp...s}$ from \eqref{elmoss} with $c^{(p)}_{k_p}$ from \eqref{elmoss2}. This is now purely a problem of combinatorics, and is easy to see that to generate any term from \eqref{elmoss} we
need at least the series with a coefficient $c^{(p_o)}_{k_{p_o}}$ for $p_o \ge {\rm dom}(m, n, p, q, ...s)$ in \eqref{elmoss2}\footnote{For small values of $(m, n, p, .., s)$ it is not too hard to work out the combinatoric factors that relate the coefficients of \eqref{elmoss} with the ones from \eqref{elmoss2}. As an example, for $m = n = p = ... = s = 0$ and $m = 1, n = p = ...= s = 0$ we see that:
\bg\label{sweetpea}
c_{000...0} =\sum_{\{k_p\}}\left(\sum_p c^{(p)}_{k_p} {\rm D}^{k_p}\right), ~~~~
c_{100...0} = \sum_{\{k_p\}}\left(\sum_p p k_p c^{(p)}_{k_p} {\rm D}^{k_p - 1}\right), \nonumber\nd
where the summations are ordered as above. Since \eqref{lili} with \eqref{susycond} already capture the 
supersymmetric kinetic term for the fermions, with no additional constant factors, we can impose
$c_{000...0} = 0$ and $c_{100...0} = 1$ with the former removing any needs for adding a cosmological constant term. This is of course what we should expect at the low energy supergravity level.}. If we absorb the factor of ${\rm D}$ in the definition of $c^{(p_o)}_{k_{p_o}}$, then the combinatoric elements may be arranged as powers of ${\rm D}^{-1}$ in addition to the ${\rm M}_p$ suppressions of the fermionic bilinears. All of these informations are clearly in the definition of the generalized metric \eqref{akash}, and therefore, after the dust settles, \eqref{elmoss2} may succinctly be expressed as powers of the generalized metric in the following way:

{\footnotesize
\bg\label{dragun4}
\mathbb{Q}^{(2)}_{\rm met} &\equiv & \sum_{k_p} \left(\sum_{p} c^{(p)}_{k_p} \left[{\rm tr}\left(1 + \mathbb{X}\right)^p\right]^{k_p}\right)
 =  \sum_{k_1} c^{(1)}_{k_1} \left(\hat{\bf g}^{\rm C}_{\rm C}\right)^{k_1} + \sum_{k_2} c^{(2)}_{k_2} 
\left(\hat{\bf g}^{\rm C}_{\rm D} \hat{\bf g}^{\rm D}_{\rm C}\right)^{k_2} + 
\sum_{k_3} c^{(3)}_{k_3} 
\left(\hat{\bf g}^{\rm C}_{\rm D} \hat{\bf g}^{\rm D}_{\rm E} \hat{\bf g}^{\rm E}_{\rm C}\right)^{k_3} + ....
\nonumber\\
& = &  \sum_{k_1} c^{(1)}_{k_1}\left({\bf g}^{\rm CD} \hat{\bf g}_{\rm CD}\right)^{k_1} + 
\sum_{k_2} c^{(2)}_{k_2}\left({\bf g}^{\rm CD}{\bf g}^{\rm EF}\hat{\bf g}_{\rm CE} \hat{\bf g}_{\rm DF}\right)^{k_2} + 
\sum_{k_3} c^{(3)}_{k_3} \left({\bf g}^{\rm CD}{\bf g}^{\rm EF}{\bf g}^{\rm GH}\hat{\bf g}_{\rm DE} \hat{\bf g}_{\rm FG} \hat{\bf g}_{\rm CH}\right)^{k_3} +..., \nonumber\\ \nd}
which would provide non-trivial contributions from \eqref{akash} because of \eqref{cascova}. 
The second line of \eqref{dragun4} is not just the fact that  
the indices are raised or lowered by the symmetric bosonic part of the metric in \eqref{akash}, but a more direct demonstration that a generic fermionic interaction of the form \eqref{elmoss}, may be expressed completely in terms of the generalized metric \eqref{akash} as in \eqref{dragun4}. 

\subsection{A trans-series extension of the fermionic quantum series \label{rajatomy}}

The aforementioned rewriting of the quantum series \eqref{elmoss} in terms of the generalized metric may not be the only utility of the series \eqref{elmoss2}. Due to the summation structure of \eqref{elmoss2}, or its equivalent form \eqref{dragun4}, there could arise the possibility of {\it summing} the series and express it in a more non-perturbative (or in a trans-series) form. The latter however requires a redefinition of ${\rm tr}~\mathbb{X}$ in a more suggestive format so as to fit in with possible non-perturbative and non-local forms.  Let us then define ${\rm tr}~\mathbb{X}$ in the following two ways (which we will call ${\rm tr}~\mathbb{X}_i$ to keep them distinct from the earlier {\it perturbative} definition):
\bg\label{boncek}
&& {\rm tr}~\mathbb{X}_1(x, y) \equiv {\rm M}_p^{6}\int^y d^6z \sqrt{{\bf g}_6(x, z)} {\bf g}^{\rm MN}(x, z) \Big(\bar\lambda(x, z) {\cal O}(x, z)\lambda(x, z)\Big)_{\rm MN}\\
&& {\rm tr}~\mathbb{X}_2(x, y) \equiv {\rm M}_p^{6}\int d^6z \sqrt{{\bf g}_6(x, z)} f(y - z){\bf g}^{\rm MN}(x, z) \Big(\bar\lambda(x, z) {\cal O}(x, z)\lambda(x, z)\Big)_{\rm MN}, \nonumber \nd  
where $f(y - z)$ is the non-locality function defined in \cite{desitter2, coherbeta} with $z^{\rm M} \in {\cal M}_4 \times {\cal M}_2$ and $x \in {\bf R}^{1, 2}$; and we have assumed the fermionic components 
$\Psi_{\rm M}(z)$ to be functions of only the internal six-manifold ${\cal M}_4 \times {\cal M}_2$. Note that the operator ${\cal O} = {\cal O}(x, z)$, despite being constructed out of fermionic bi-linears, could in-principle be a function of all the coordinates (except the toroidal ones). In general however the integrals in \eqref{boncek} can extend over the whole eight-manifold but since we kept everything to be independent of the toroidal directions, we can get 
{\it delocalized} instanton effects from the toroidal integrals (see \cite{coherbeta} for more details).

The rest of the procedure is somewhat similar to what we had in \cite{joydeep}. If we do not involve the integrals, then \eqref{elmoss2} is basically what we can get with every term therein being suppressed by positive powers of $\bar{g}_s$, {\it i.e.}
$\xxy^{+{\rm ive}}$. To bring \eqref{elmoss2} in a trans-series (or convergent) form, we take the definition 
\eqref{boncek} and rewrite the coefficients $c_{k_p}^{(p)}$ as:
\bg\label{hassholy}
c_{k_p}^{(p)} & \equiv & \sum_{l\ge 1} {d^{(p)}_l (-l)^{k_p} \over k_p!} = \sum_{l \ge 1} \mathbb{M}_{k_p l} d^{(p)}_l \nonumber\\
& = & {d_1^{(p)}(-1)^{k_p}\over k_p!} + {d_2^{(p)}(-2)^{k_p}\over k_p!} +{d_3^{(p)}(-3)^{k_p}\over k_p!} + ...+ {d_n^{(p)}(-n)^{k_p}\over k_p!} + .. \nd 
which should be understood as though we are replacing all the 
$c_{k_p}^{(p)}$ coefficients in \eqref{elmoss2} by the series from \eqref{hassholy}. One might worry that this will make \eqref{hassholy} much more difficult to deal with, but as we shall show, the reverse will be true. The matrix $\mathbb{M}$, with components $\mathbb{M}_{k_pl}$, can be represented as:
\bg\label{omapant}
\mathbb{M} \equiv \mathbb{M}_{k_pl} = \left(\begin{matrix} -1 & -2 & -3 & -4& -5& -6 & -7 & -8 & ....\\
~ & ~ & ~ & ~ & ~ & ~ & ~ & ~ & ~ \\
~{1\over 2} & ~2 & ~{9\over 2} & ~8 & ~{25\over 2} & ~18 & ~{49\over 2} & ~32 &.... \\
~ & ~ & ~ & ~ & ~ & ~ & ~ & ~ & ~\\
-{1\over 6} & -{4\over 3} & -{9\over 2} & -{32\over 3} & -{125\over 6} & -36 & -{343\over 6} & -{256\over 3} &....\\
~ & ~ & ~ & ~ & ~ & ~ & ~ & ~ &~\\
~{1\over 24} & ~{2\over 3} & ~{27\over 8} & ~{32\over 3} & ~{625\over 24}  &~ 54 & ~{2401\over 24} & ~{512\over 3} &.... \\
~ & ~ & ~ & ~ & ~ & ~ & ~ & ~ &~\\
-{1\over 120} & -{4\over 15} & -{81\over 40} & -{128\over 15} & -{625\over 24} & -{324\over 5} & 
-{16807\over 120} & -{4096\over 15} &....\\
~ & ~ & ~ & ~ & ~ & ~ & ~ & ~ &~\\
~{1\over 720} & ~{4\over 45} & ~{81\over 80} & ~{256\over 45} & ~{3125\over 144} & ~{324\over 5} & ~{117649\over 720} & ~{16384\over 45} &....\\
~ & ~ & ~ & ~ & ~ & ~ & ~ & ~ &~  \\
-{1\over 5040} & -{8\over 315} & -{243\over 560} & -{1024\over 315} & -{15625\over 1008} & 
-{1944\over 35} & -{117649\over 720} & -{131072\over 315} & ...\\
~ & ~ & ~ & ~ & ~ & ~ & ~ & ~ &~ \\
~{1\over 40320} & ~{2\over 315} & ~ {729\over 4480} & ~{512\over 315}  & ~{78125\over 8064} & ~{1458\over 35} & ~{823543\over 5760} & ~{131072\over 315} & ...\\
~ & ~ & ~ & ~ & ~ & ~ & ~ & ~ &~ \\.... & .... & .... & .... & .... & .... & .... & .... & ....  \end{matrix}\right), \nd
which is basically a ${\rm N} \times {\rm N}$ matrix with ${\rm N} \to \infty$. Solution would exist if and only if the matrix ${\rm M}$ has an inverse which, in turn, would imply that the determinant of ${\rm M}$ should be {\it non-zero}. Since the matrix ${\rm M}$ from \eqref{omapant} is an $\infty \times \infty$ dimensional matrix, it would appear that the determinant (and consequently the inverse) will be impossible to compute. Herein however we encounter a pleasant surprise: the determinant of the matrix \eqref{omapant} is always $\pm 1$, with the sign depending on the ranks as given in 
 {\bf Table \ref{lilluth2}}. This implies:
\bg\label{benedetta1}
{\rm det}~\mathbb{M}_{{\rm N} \times {\rm N}} = \begin{cases} 
~{\rm cos}~{\rm N}\pi, ~~~~~~~~~~~~~ {\rm N} \in 2\mathbb{Z}\\
~{\rm sin}~\left({\rm N} + {1\over 2}\right)\pi, ~~~~ {\rm N} \in 2\mathbb{Z} + 1
\end{cases}
\nd 
which would at least provide a reason for the inverse of matrix $\mathbb{M}$ to exist. However \eqref{benedetta1} is not enough because we have to actually {\it compute} the inverse of $\mathbb{M}$ to determine $d_l^{(p)}$ in terms of $c_{k_p}^{(p)}$. This unfortunately is a challenging exercise because the rank ${\rm N}$ of $\mathbb{M}$ could be arbitrarily large (and even infinite). How then should we proceed? One possibility would be to start with {\it finite} ${\rm N}$ and determine the inverse of the matrix \eqref{hassholy}. Such a procedure could in-principle suggest us some patterns which we could use to express the inverse when ${\rm N} \to \infty$. A good starting point would be to take small even and odd ranks, for example ${\rm N} = 5, 6, 7, 8$ and determine the inverse of \eqref{hassholy}. This gives:

{\footnotesize
\bg\label{reshish}
\mathbb{M}^{-1}_{7 \times 7} = \left(
\right)\nonumber\\ \nd}
for the four cases. Expectedly the inverses exist, which is good, but we do notice some repeating patterns: they are shown in {\red red} in the four examples of \eqref{hkpgkmkkkbkb}. In fact one may easily check that the aforementioned reddened patterns repeat for any ${\rm N}$, no matter how large the rank is. This means $d_l^{(p)} = \sum\limits_{k_p} (\mathbb{M}^{-1})_{lk_p} c_{k_p}^{(p)}$, which when written in terms of the components take the following form:
\bg\label{kettleman2}
\left(%
\right), \nd
thus providing a precise way to map the set of coefficients $c_{k_p}^{(p)}$ to the set of coefficients $d_l^{(p)}$. Unfortunately looking at the inverse matrix in \eqref{kettleman2}, we immediately see a problem: in the limit ${\rm N} \to \infty$, many of the terms in the matrix blow up! This means in the mapping:
\bg\label{blsjun18}
d_l^{(p)} = \mathbb{M}^{-1}_{lk_p} c^{(p)}_{k_p}, \nd
$d_l^{(p)}$ cannot be finite numbers if $c_{k_p}^{(p)}$ appearing in \eqref{elmoss2} are {\it finite} numbers. Thus only if the set $c_{k_p}^{(p)}$ becomes {\it arbitrarily small}, then there is a chance that the set of coefficients $d_l^{(p)}$ becomes finite. This observation may in fact suggest a way out of the conundrum. First, however let us ask why do we, in the first place, get large values for $d_l^{(p)}$ with finite set of values of $c_{k_p}^{(p)}$? The answer is simple but instructive and it relies on the fact that we are trying to express a {\it perturbative} series in terms of an exponential one. Such a construction naturally leads to issues because of the following reasons.

Consider the polynomial series in \eqref{elmoss2}. Since every $c_{k_p}^{(p)}$ in \eqref{elmoss2} is expressed in terms of a linear combinations of $d_l^{(p)}$, the proposal \eqref{hassholy} basically converts every term of \eqref{elmoss2} to an exponential series (more detail on this will be given soon). For example, taking only the first term in \eqref{elmoss2} and using \eqref{hassholy} then gives us\footnote{For a proof, see eq. (3.58) to eq. (3.60) in the first reference of \cite{coherbeta}.}:
\bg\label{brunlope}
\sum_{k_1}c^{(1)}_{k_1} \left({\rm D} + {\rm tr}~\mathbb{X}_i\right)^{k_1} =  \sum_{l \ge 1}
d_l^{(1)} \Big[{\rm exp}\left(-l{\rm D} - l{\rm tr}~\mathbb{X}_i\right) - 1\Big], \nd
with ${\rm D}$ being the space-time dimension and $\mathbb{X}_i$ for $i= 1, 2$ are given by \eqref{boncek}. In the limit where 
${\rm D} + {\rm tr}~\mathbb{X}_i << 1$, the exponential term on the RHS of \eqref{brunlope} goes to zero faster than any polynomial series and therefore for finite $c_{k_p}^{(p)}$, the coefficients $d_l^{(p)}$ in the exponential series has to be arbitrarily large to produce the same finite result as the one from the polynomial series. On the other hand, in the limit ${\rm D} + {\rm tr}~\mathbb{X}_i >>  1$, the polynomial series on the LHS of \eqref{brunlope} would only make sense for $c_{k_p}^{(p)} \to 0$. Such a choice produces finite values for $d_l^{(p)}$ from \eqref{kettleman2}. For the present case, since ${\rm D} = 6, 8$ and ${\rm tr}~\mathbb{X}_i$ for $i = 1, 2$ from \eqref{boncek} are not necessarily restricted to take small values, the LHS of \eqref{brunlope} would make sense for $c_{k_p}^{(p)} \to 0$, {\it i.e.} with almost zero radius of convergence. Plugging in such a choice in \eqref{kettleman2} gives us finite values for $d_l^{(p)}$. As we shall also see later, it is the trans-series form for the interactions that makes sense here, and therefore it will only be the exponential series and not the polynomial one from \eqref{elmoss2} that we should seriously consider to capture the real dynamics of the system. The procedure \eqref{hassholy} and the matrix transformation \eqref{kettleman2} should then be viewed as a {\it trick} that converts polynomial series to  exponential ones. 

There are still a few subtleties regarding the summation process that we should point out here. First one should be clear: our summation procedure is {\it not} a Borel resummation. There are no obvious factorial (or Gevrey) growths accompanying the polynomial series, so the usual Borel-\'Ecalle procedure cannot be applied here. To appreciate this let us look at one of the term in the polynomial series on the LHS of \eqref{brunlope}, say $({\rm D} + {\rm tr}~\mathbb{X}_i)^q$, with 
$q \in \{k_1\}$. Plugging this in the path-integral will produce a perturbative series of the form:
\bg\label{melismey}
\sum_{\rm N} {g^{\rm N}\over {\rm N}!} \left(1 + {1\over {\rm D}}{\rm tr}~\mathbb{X}_i\right)^{q{\rm N}}, \nd
where $g = c_q^{(1)} {\rm D}^{q}$ is the coupling constant. From the binomial expansion we can ascertain the growths at every order in ${\rm N}$ which, in general, could be of Gevrey type \cite{gevrey, borelborel, borel2}. Measuring any expectation values using the aforementioned path-integral will now involve the Borel-\'Ecalle resummation \cite{borel2}\footnote{Note the key difference. The Borel-\'Ecalle resummation appears once we take every piece of the polynomial interactions, appearing on the LHS of \eqref{brunlope}, and use (each of) them in the path-integral to compute the corrections to certain expectation values. The resummation procedure then provides us the Borel-\'Ecalle resummed corrections to the expectation values and {\it not} to the action directly.}. However such resummations do not directly tell us how the action should behave. The trans-series form of the action, that appears as $\hat{\bf S}_{\rm tot}({\bf \Xi})$ in \eqref{kimkarol}, needs to be determined from various other consistency checks emanating specifically from demanding correct instanton behaviors, convergence etc.

Secondly, and as mentioned earlier, the coefficients $c_{k_p}^{(p)}$ need to have zero radii of convergences for the series on the LHS of \eqref{brunlope} (with ${\rm tr}~\mathbb{X}_i$ defined in \eqref{boncek}) to make sense. To see the consistency of the above statement, we will have to study the $\bar{g}_s$ scalings of the two functions in \eqref{boncek}. The scalings are easy to compute and they typically take the following form:
\bg\label{brlopes}
\left({g_s\over {{\rm H}(y){\rm H}_o({\bf x})}}\right)^{\theta(t)-2 +2\hat\sigma_e(t) +{\hat\alpha_e(t) + \hat\beta_e(t)\over 2}} =\left({g_s\over {{\rm H}(y){\rm H}_o({\bf x})}}\right)^{-(2^\pm - \theta)}, \nd
where $\theta$ will eventually be related to $\theta_{nl}$, which is the quantum scaling given by \eqref{cortanaeve} and \eqref{brittbaba} once we express \eqref{botsuga} using the generalized metric and fluxes. The $\pm$ superscript is there to indicate the small difference from $2$ from the presence of $(\hat\sigma_e(t), \hat\alpha_e(t), \hat\beta_e(t))$. (These details are not important now, and will become clearer as we move along.) For the present case we can take $\theta(t)$ to be the scaling of the fermionic condensate. For $\theta(t) < 2^\pm$, the series is non-perturbative in $\bar{g}_s$, whereas for $\theta(t) > 2^\pm$ the series become perturbative in $\bar{g}_s$. The rest of the details are similar to what appeared in eq. (6.28) and eq. (6.29) in \cite{joydeep}. After the dust settles, 
plugging \eqref{hassholy} in \eqref{elmoss2} with the definition \eqref{boncek}, we get:
\begin{table}[tb]  
 \begin{center}
\renewcommand{\arraystretch}{1.5}
\begin{tabular}{|c||c||c||c||c||c||c||c|| c||c|}\hline ${\rm N}$  & 2 & 3 & 4 & 5 & 6 & 7 & 8 & ... \\ \hline\hline
${\rm det}$ & $-1$ & $+1$ & $+1$ & $-1$ & $-1$ & $+1$ & $+1$ &... \\ \hline 
 \end{tabular}
\renewcommand{\arraystretch}{1}
\end{center}
 \caption[]{\Su The determinants of $\mathbb{M}_{{\rm N} \times {\rm N}}$ which only take values $\pm 1$ depending on what rank ${\rm N}$ we take. The generic result then is \eqref{benedetta1} $\forall {\rm N} \in \mathbb{Z}$.} 
  \label{lilluth2}
 \end{table}

{\footnotesize
\bg\label{cheril}
&& \sum_{k_1}c^{(1)}_{k_1} \left({\rm D} + {\rm tr}~\mathbb{X}_i\right)^{k_1} =  \sum_{l \ge 1}
d_l^{(1)} \Big[{\rm exp}\left(-l{\rm D} - l{\rm tr}~\mathbb{X}_i\right) - 1\Big] \\
&& \sum_{k_2}c^{(2)}_{k_2} \left({\rm D} + 2 {\rm tr}~\mathbb{X}_i + {\rm tr}~\mathbb{X}_i^2\right)^{k_2} =  
\sum_{l \ge 1} 
d_l^{(2)} \Big[{\rm exp}\left(-l{\rm D} - 2l{\rm tr}~\mathbb{X}_i - l{\rm tr}~\mathbb{X}_i^2\right) - 1\Big] \nonumber\\
&& \sum_{k_3}c^{(3)}_{k_3} \left({\rm D} + 3 {\rm tr}~\mathbb{X}_i + 3{\rm tr}~\mathbb{X}_i^2 + {\rm tr}~\mathbb{X}_i^3\right)^{k_3} =  
\sum_{l \ge 1} 
d_l^{(3)} \Big[{\rm exp}\left(-l{\rm D} - 3l {\rm tr}~\mathbb{X}_i - 3l{\rm tr}~\mathbb{X}_i^2 
-l {\rm tr}~\mathbb{X}_i^3\right) - 1\Big], \nonumber \nd}
with {\it finite} $d_l^{(p)}$;
where $i = 1, 2$, and ${\rm D}$ is the internal dimension that we can take it to be $8$ or $6$ depending on whether we include the toroidal directions or not. All the series above appear convergent due to the ${\rm exp}(-l{\rm D})$ and ${\rm exp}\left(-l {\rm tr}~\mathbb{X}^{2q}\right)$ suppressions for $(l, q) \in \mathbb{Z}_+$, no matter what sign ${\rm tr}~\mathbb{X}^{2q-1}$ carries. The reason is the following. For $\mathbb{X}_i = -\vert\mathbb{X}_i\vert$, the trace of the odd-powers of $\mathbb{X}_i$ become negative definite\footnote{Recall from \eqref{boncek}, that the traces are computed by contracting the fermion condensates with the positive-definite inverse metric components of the internal space ${\cal M}_4 \times {\cal M}_2$. As such the signs are determined by the signs of the condensates, {\it i.e.} $\mathbb{X}_i$ only.}, {\it i.e.} ${\rm tr}~\mathbb{X}^{2q-1} = -\vert{\rm tr}~\mathbb{X}^{2q-1}\vert$. For $\vert{\rm tr}~\mathbb{X}_i\vert > {\rm D}$, and $\vert{\rm tr}~\mathbb{X}^{2q+1}\vert >\vert{\rm tr}~\mathbb{X}^{2q}\vert$, the polynomial series will go as $-\left(\vert{\rm tr}~\mathbb{X}_i\vert - {\rm D}\right)$, {\it et cetera}. For such a case we can easily see that:
\bg\label{lastanniv}
c_{k_p}^{(p)} & \equiv & \sum_{l\ge 1} {d^{(p)}_l (+l)^{k_p} \over k_p!}  ~~\implies ~~
\left(\begin{matrix} d_1^{(p)}\\ ~\\ d_2^{(p)}\\ ~ \\d_3^{(p)}\\ ~\\...\\ ~\\ d_{\rm N}^{(p)} 
\end{matrix}\right) = \left(\begin{matrix}
{{\rm N}} & ~ ... & ~ ... & ~ ... & ~ ... &~ \mp{\rm N}\\
~ & ~ & ~ & ~ & ~ &  ~  \\
~... & ~...& ~... & ~...& ~...& ~ ... \\
~ & ~ & ~ & ~ & ~ & ~ &   ~\\
~... & ~...& ~... & ~...&  ~...&~ ... \\
 ~ & ~ & ~ & ~ & ~ &  ~ \\
\pm {{\rm N}\over {\rm N - 1}} & ~ ... &  ~... & ~... & ~ ... & ~ -{\rm N} \\
  ~ & ~ & ~ & ~ & ~ & ~ & ~ \\
\mp{1\over {\rm N}} &  ~... &  ~ ... & ~ ... & ~- {{\rm N} - 1\over 2} & ~ \pm 1\\
 \end{matrix}\right)
\left(\begin{matrix}
c_1^{(p)}\\ ~\\ c_2^{(p)}\\~\\c_3^{(p)}\\ ~\\... \\ ~\\c_{\rm N}^{(p)} 
\end{matrix}\right), \nonumber \\ \nd
from where we can infer that the inverse matrix basically takes the same form as the one we had in \eqref{kettleman2}, with some small alterations in signs. (The large ${\rm N}$ issue is still there as one may see from \eqref{lastanniv}.) The above analysis then reveals that 
we can again express \eqref{elmoss2} in terms of a convergent series, irrespective of the sign of ${\rm tr}~\mathbb{X}_i$. This suggests replacing \eqref{cheril} by the following set of convergent series:

{\footnotesize
\bg\label{cheril2}
&& \sum_{k_1}c^{(1)}_{k_1} \left({\rm D} + {\rm tr}~\mathbb{X}_i\right)^{k_1} =  \sum_{l \ge 1}
d_l^{(1)} \Big[{\rm exp}\left(\pm l{\rm D} -l\vert{\rm tr}~\mathbb{X}_i\vert\right) - 1\Big] \\
&& \sum_{k_2}c^{(2)}_{k_2} \left({\rm D} + 2 {\rm tr}~\mathbb{X}_i + {\rm tr}~\mathbb{X}_i^2\right)^{k_2} =  
\sum_{l \ge 1} 
d_l^{(2)} \Big[{\rm exp}\left(\pm l{\rm D} -l\vert 2{\rm tr}~\mathbb{X}_i + {\rm tr}~\mathbb{X}_i^2\vert\right) - 1\Big] \nonumber\\
&& \sum_{k_3}c^{(3)}_{k_3} \left({\rm D} + 3 {\rm tr}~\mathbb{X}_i + 3{\rm tr}~\mathbb{X}_i^2 + {\rm tr}~\mathbb{X}_i^3\right)^{k_3} =  
\sum_{l \ge 1} 
d_l^{(3)} \Big[{\rm exp}\left(\pm l{\rm D}-l\vert 3 {\rm tr}~\mathbb{X}_i + 3{\rm tr}~\mathbb{X}_i^2 
+ {\rm tr}~\mathbb{X}_i^3\vert\right) - 1\Big], \nonumber \nd}
where the $\pm$ ambiguity comes from the $\pm l$ choices in \eqref{lastanniv} and \eqref{kettleman2} for $\mathbb{X}_i < 0$ and $\mathbb{X}_i > 0$ respectively. This ambiguity is harmless because 
${\rm exp}(\pm l{\rm D})$ can always be absorbed in the definition of $d_l^{(p)}$ without altering its sign. Moreover, $\mathbb{X}_i^p$ for $p \in \mathbb{Z}_+$ are further suppressed by ${\rm M}_p$ so they do not contribute beyond certain order. This means, for most computations, the $l = 1$ term from the first line of \eqref{cheril} would suffice giving us:

{\scriptsize
\bg\label{paige}
\mathbb{Q}_{\rm nloc} = \beta \int d^{11} x\sqrt{-{\bf g}_{11}(x, y)}
~{\rm exp}\left[-{\rm M}_p^6\int d^6z \sqrt{{\bf g}_6(x, z)} \left\vert f(y - z) 
{\bf g}^{\rm MN}(x, z) \left(\bar{\lambda}(x, z){\cal O}(x, z)\lambda(x, z)\right)_{\rm MN}\right\vert\right],  \nd} 
as the non-perturbative and non-local contribution from the ${\rm 5}$-brane instantons wrapping the internal six-manifold ${\cal M}_4 \times {\cal M}_2$. (The traditional non-perturbative M5-brane instanton contribution can be easily got from \eqref{cheril} by using ${\rm tr}~\mathbb{X}_1$ instead of ${\rm tr}~\mathbb{X}_2$.) Note that 
$\beta = d_1^{(1)} e^{\mp{\rm D}} {\rm M}_p^{11}$, and we have not shown the cosmological constant piece 
$-d_1^{(1)} {\rm M}_p^{11}  \int d^{11} x \sqrt{-{\bf g}_{11}(x, y, w)}$ in \eqref{paige}. The reason is that, and as also shown in \cite{coherbeta}, such a term does not contribute to the Schwinger-Dyson's equations\footnote{Another reason is that, if we take the fluctuation determinants accompanying each of the instanton saddles, the $(-1)$ pieces appearing in \eqref{cheril2} can be absorbed in contributions from the zero instanton sector \cite{joydeep}. We will have more to say on this a little later.\label{hayleyhuh}}. Thus the above contribution is in addition to any perturbative fermionic contributions that we can have from \eqref{dragun4}. Note also that the exponential piece has an explicit dependence on $x \in {\bf R}^{2, 1}$ but the integral is over $z \in {\cal M}_4 \times {\cal M}_2$, which would mean that the ${g_s\over {\rm HH}_o}$ dependence can be carried over to the bulk eleven-dimensional integral. This is easily reflected in the computation of the energy-momentum tensor, which takes the following explicit form:

{\footnotesize
\bg\label{mca300}
\mathbb{T}_{\rm CE}(z_1, z_2) & = & -{2\over \sqrt{-{\bf g}_{11}(z_1, z_2)}}~{\delta \mathbb{Q}_{\rm nloc} \over 
\delta {\bf g}^{\rm CE}(z_1, z_2)}\\
& = & \beta \mathbb{V}_2 {\rm M}_p^{-11} {\bf g}_{\rm CE}(z_1, z_2)\left(e^{-{\rm tr}~\mathbb{X}_2(z_1, z_2)} - e^{\pm{\rm D}}\right) 
+ 2\beta {\rm M}_p^{-5} \mathbb{V}_2 \int d^6 y~f(y - z_2) e^{-{\rm tr}~\mathbb{X}_2(z_1, y)}\nonumber\\
&\times& {\delta\over \delta {\bf g}^{\rm CE}(z_1, z_2)} 
\Big(\sqrt{{\bf g}_6(z_1, z_2)} 
{\bf g}^{\rm MN}(z_1, z_2) \left\vert\left(\bar{\lambda}(z_1, z_2){\cal O}(z_1, z_2)\lambda(z_1, z_1)\right)_{\rm MN}\right\vert
\Big) \sqrt{{\bf g}_{11}(z_1, y)\over {\bf g}_{11}(z_1, z_2)}, \nonumber \nd}
where we have taken $f(y-z_2) > 0$ in the domian of the integral, {\it i.e.} over the six-manifold ${\cal M}_4 \times {\cal M}_2$; $\mathbb{V}_2$ is the un-warped dimensionless volume of ${\mathbb{T}^2\over {\cal G}}$ and 
$(z_1, z_2) \in ({\bf R}^{2, 1}, {\cal M}_4 \times {\cal M}_2)$. Note two things: one, we have inserted back the cosmological constant piece in the second line of \eqref{mca300}, and two, the ${\rm M}_p$ scaling of the first term cancels out but the ${\rm M}_p$ scaling of the second term is more non-trivial, being
suppressed by ${\rm M}_p^{\sigma_{2\mathbb{X}} - 6}$, where $\sigma_{2\mathbb{X}}$ is the scaling of ${\rm tr}~\mathbb{X}_2$. Similarly, the ${g_s\over {\rm HH}_o}$ scaling of the first term is much simpler compared to the corresponding scaling of the second term. Putting everything together, the energy-momentum tensor from the non-local fermionic counter-terms scale as:
\bg\label{britt}
\mathbb{T}_{\rm CE}(z_1, z_2)  &= &\left({g_s\over {\rm HH}_o}\right)^{a_{\rm CE} + \hat\Sigma_{e{\rm CD}}}
\left[{\rm exp}\left(-{\rm M}_p^{6-\sigma_{2\mathbb{X}}}\left({g_s\over {\rm HH}_o}\right)^{\theta_{2\mathbb{X}} - 2^\pm}\right) - e^{\rm D}\right]\\
& + & {\rm M}_p^{6-\sigma_{2\mathbb{X}}}\left({g_s\over {\rm HH}_o}\right)^{\theta_{2\mathbb{X}} - 2^\pm + a_{\rm CE} + \hat\Sigma_{e{\rm CD}}} {\rm exp}\left(-{\rm M}_p^{6-\sigma_{2\mathbb{X}}}\left({g_s\over {\rm HH}_o}\right)^{\theta_{2\mathbb{X}} - 2^\pm}\right), \nonumber \nd
where $\left({g_s\over {\rm HH}_o}\right)^{a_{\rm CE}}$, with $a_{\rm CE} = a^{({\rm CE})} - \sigma^{({\rm CE})}$ from \eqref{meanermis} and \eqref{karenmach}, is the dominant scaling of the metric component 
${\bf g}_{\rm CE}$ (and by definition this is also the dominant scaling of the generalized metric component
$\hat{\bf g}_{\rm CE}$); $\hat\Sigma_{e{\rm CE}} = \Sigma_e^{({\rm CE})}$ from \eqref{meanermis}; and the superscript $2^\pm$ has already been explained after \eqref{brlopes}. The scalings in \eqref{britt} are computed assuming the non-locality function 
$f(y - z_2)$ in \eqref{mca300} has no explicit $g_s$ or ${\rm M}_p$ dependences. With this assumption, the 
result in \eqref{britt} is unsurprisingly very similar to the ones we got in \cite{coherbeta} and 
\cite{coherbeta2} from the bosonic counter-parts. The key differences are in the explicit factors appearing for the respective scalings.
Interestingly, and as also observed in \cite{coherbeta, coherbeta2}, the non-locality function integrates out as:
\bg\label{tagchin}
\int d^6 y \sqrt{-{\bf g}_{11}(z_1, y)}~f(y - z_2)~e^{-{\rm tr}~\mathbb{X}_2(z_1, y)} = 
\left({g_s\over {\rm HH}_o}\right)^{-{14\over 3}^\pm} \check{\mathbb{F}}_2(z_1, z_2; g_s, {\rm M}_p), \nd
giving rise to a perfectly {\it local} correction to the energy-momentum tensor. The $\pm$ superscript appears from \eqref{lindavendri}, and the extra $g_s$ scaling that we see in \eqref{tagchin} is actually harmless: it is cancelled by the metric determinant  in \eqref{mca300}.  In a similar vein, the
${\rm tr}~\mathbb{X}_1$ terms in \eqref{boncek} will contribute additional corrections to the energy-momentum tensor. 

There is however an annoying factor of metric in the expression for the energy-momentum tensor from the first terms in \eqref{mca300} and \eqref{britt}. Of course, as mentioned earlier, such a term doesn't contribute to the Schwinger-Dyson's equations, so we can in principle ignore it for the present computations. Note that, in the limit ${g_s\over {\rm HH}_o} \to 0$ we can do slightly better, if we combine the results from the fermionic and the bosonic contributions. The bosonic contributions appear from localized and de-localized 
brane instantons (both from the standard as well as the non-local instantons). Imposing supersymmetry will 
kill off the cosmological constant pieces altogether, giving us:  
\bg\label{gooplab}
\sum_k c_k + \sum_{p, l\ge 1} d_l^{(p)} + .... = 0, \nd
where the first term is from the bosonic contributions appearing in eq (3.43) of \cite{coherbeta3}, and the dotted terms are additional non-perturbative (and non-local) contributions. The sum over $p \in \mathbb{Z}$ appears because we can add all the contributions in \eqref{cheril2}. Observe that the toroidal volume factor 
$\mathbb{V}_2$ doesn't appear in \eqref{gooplab} because we have assumed the bosonic contributions to also be independent of the toroidal directions. The additional term proportional to the metric has a coefficient:
\bg\label{foxdavax}
\sum_k c_k e^{-k\vert \mathbb{G}^{(1)}(z_1, z_2)\vert} + \sum_{l \ge 1} d_l^{(1)} e^{-l\vert {\rm D} + {\rm tr}~\mathbb{X}_2(z_1, z_2)\vert} + ....., \nd
where the first term with $\mathbb{G}^{(1)}(z_1, z_2)$ is again from the bosonic contributions in eq. (3.43) of \cite{coherbeta3}, but now the dotted terms will involve contributions from higher $p$ in \eqref{cheril} as well as from other non-perturbative and non-local contributions. This clearly doesn't vanish. Interestingly, when ${g_s\over {\rm HH}_o} \to 0$, if $\mathbb{G}^{(q)}(z_1, z_2)$ scales as positive power of $g_s$, it vanishes and therefore the term proportional to $c_k$ in \eqref{foxdavax} cancels out from the $c_k$ piece in \eqref{gooplab}, but this doesn't appear to be the case with the fermionic piece. Therefore, while \eqref{gooplab} appears to cancel the extra cosmological constant pieces, this cannot be the full story. In the following we will show that the real reason for the consistency of the exponential series in \eqref{cheril2} lies in the fluctuation determinants around the instanton series as alluded to earlier in footnote \ref{hayleyhuh}.

\subsection{More on the fermionic extension of the metric components in \eqref{botsuga} \label{tomysellpiz}}

The series in \eqref{paige}, or the energy-momentum tensor derived from it in \eqref{mca300}, is clearly non-perturbative in ${\rm M}_p$, and contrary to naive expectations, can remain non-perturbative to arbitrary orders in the fermionic expansion. The reason is not too hard to see. Any ${\rm M}_p$ suppressions should appear from the derivative countings, and if we take {\it localized} fermions (much like the localized G-flux components ${\bf G}_{{\rm MN}ab}$ that we took in \cite{desitter2, coherbeta, coherbeta2}), then every derivative will bring down two powers of ${\rm M}_p$ to counteract one power of 
${\rm M}_p$ suppression from each derivative. Net result is a gain of one positive power of ${\rm M}_p$ from each derivative, thus extending the non-perturbative behavior to arbitrary orders. In the language of 
\eqref{britt}, this implies that $\sigma_{2\mathbb{X}}$ can become {\it negative}. 
The series in \eqref{mca300} is also non-perturbative in 
${g_s\over {\rm HH}_o}$, coming from the determinant factor, and will continue to contribute non-perturbatively at every order in ${g_s\over {\rm HH}_o}$ to the energy-momentum tensor much like what we had in \cite{desitter2, coherbeta, coherbeta2}. As an example, since $a_{\mu\nu} = -{8\over 3}$ from \eqref{meanermis} for 
$(\mu, \nu) \in {\bf R}^{2, 1}$, the quantum scaling of the fermionic term should be $\theta_{2\mathbb{X}} = {8\over 3}^\pm$ to contribute as $\left({g_s\over {\rm HH}_o}\right)^{-2^\pm}$ to the space-time Einstein's equation, where the $\pm$ superscripts correspond to small corrections discussed earlier. In fact $\theta_{2\mathbb{X}} = {8\over 3}^\pm$ contributes non-perturbatively to {\it all} the Einstein's equations. In \cite{desitter2} we saw that such a scaling for \eqref{botsuga} necessitates the introduction of quartic order curvature corrections. We now see that the fermionic corrections to \eqref{botsuga} doesn't change the conclusion. This will be elaborated more when we encounter the Schwinger-Dyson's equations later. 

Some more subtlety ensues. Replacing the metric by a generalized metric means that the curvature components should also get fermionic contributions. We encountered such issues earlier in 
\eqref{ruthL}, but our aim therein was to reproduce terms of the low energy supergravity. Since this is no longer an issue, the curvature terms can reproduce additional interactions of the fermions with the metric degrees of freedom. (For simplicity we will also take localized fermions, so that the traces do not involve integrals over ${\cal M}_4 \times {\cal M}_2$. This way ${\rm tr}~\mathbb{X}_i$ from \eqref{boncek} will simply be ${\rm tr}~\mathbb{X}$ from \eqref{lili}.) To see the consistency of the series \eqref{akash} and \eqref{cascova}, let us consider a few curvature terms from 
the quantum series \eqref{botsuga} by replacing the metric therein by the generalized metric \eqref{akash}:
\bg\label{daril} 
\mathbb{Q}_{\rm curv} \equiv q_1\hat{\bf g}^{\rm CD} \hat{\bf g}^{00} \hat{\bf R}_{\rm 0C0D}  + 
q_2\hat{\bf g}^{\rm CD} \hat{\bf g}^{\rm EF} \hat{\bf R}_{\rm CEDF} + 
q_3\hat{\bf g}^{\rm CD} \hat{\bf g}^{\rm EF}\left(\hat{\bf g}^{00}\right)^2 \hat{\bf R}_{\rm 0C0D}
\hat{\bf R}_{\rm 0E0F}  + ..  \nd  
where $q_i$ are dimensionless constants and  the indices are raised or lowered using the metric ${\bf g}_{\rm CD}$. The dotted terms are the higher order curvature contributions from \eqref{botsuga}. The generalized Riemann curvature $\hat{\bf R}_{\rm 0C0D}$, or more generically $\hat{\bf R}_{\rm ABCD}$, is defined in the same way as the usual Riemann curvature except now we are allowed to use the generalized metric $\hat{\bf g}_{\rm AB}$ from \eqref{akash}. Using this, one may 
easily see that $\mathbb{Q}_{\rm curv}$ from \eqref{daril} decomposes in the following way:

{\footnotesize
\bg\label{darilebu}
\mathbb{Q}^{(1)}_{\rm curv} &=& q_1 \hat{\bf g}^{\rm CD}\hat{\bf g}^{00} \hat{\bf g}_{{\rm CD},0} \hat{\bf g}_{00, 0} 
\hat{\bf g}^{00} + q_2\hat{\bf g}^{\rm CD} \hat{\bf g}^{\rm EF} \hat{\bf g}_{{\rm CD}, 0} \hat{\bf g}_{{\rm EF}, 0} \hat{\bf g}^{00} + q_3\hat{\bf g}^{\rm CD} \hat{\bf g}^{\rm EF}\left(\hat{\bf g}^{00}\right)^4 \left(\hat{\bf g}_{00, 0}\right)^2 
 \hat{\bf g}_{{\rm CD}, 0}  \hat{\bf g}_{{\rm EF}, 0} + 
.... \nonumber\\
& = & q_1c_3 {\bf g}^{00}\left({\bf g}^{\rm CD} + c_3 \left(\bar\lambda \theta\lambda\right)^{\rm CD}\right)
\left(\bar\lambda\theta \lambda\right)_{{\rm CD}, 0} + 
{\bf g}^{00}\left(q_2 + q_3 \left({\bf g}^{00}\right)^3 \left({\bf g}_{00, 0}\right)^2\right)\left({\bf g}^{\rm CD} + c_3 \left(\bar\lambda \theta\lambda\right)^{\rm CD}\right) \nonumber\\
&\times & \left({\bf g}^{\rm EF} + c_3 \left(\bar\lambda \theta\lambda\right)^{\rm EF}\right)
\left({\bf g}_{\rm CD, 0} + c_3 \left(\bar\lambda \theta\lambda\right)_{\rm CD, 0}\right)
\left({\bf g}_{\rm EF, 0} + c_3 \left(\bar\lambda \theta\lambda\right)_{\rm EF, 0}\right) + ..., \nd}
where we have restricted ourselves to the temporal derivatives for simplicity (denoted by 
$\mathbb{Q}_{\rm curv}^{(1)}$), and ignored the $c_2$ factor in \eqref{akash}. The dotted terms will contain other temporal derivatives, and also other possible curvature terms, that could be constructed along the same lines. 
Note that we have taken ${\bf g}_{00}$ instead of $\hat{\bf g}_{00}$ in the second line of \eqref{darilebu}. 
This is because to quadratic orders in fermions $\Psi_{\rm M}$ there are no contributions and the contributions only begin from the 
quartic order as we saw in \eqref{lillee} and \eqref{lillee1}.  The temporal and the spatial derivatives are now slightly more non-trivial. This is because of the following identities:
\bg\label{belluchi}
&& {\bf g}^{\rm CD} \left(\bar\lambda {\cal O} \lambda\right)_{\rm CD, M} = 
\left({\rm tr}~\mathbb{X}\right)_{,\rm M} - {\bf g}^{\rm CD}_{,\rm M} \mathbb{X}_{\rm CD}\\
&&{\bf g}^{\rm CD}\left(\bar\lambda {\cal O} \lambda\right)_{\rm CD, 0} \equiv {\bf g}^{\rm CD}\mathbb{X}_{\rm CD, 0} = {\rm tr}~{\mathbb{X}}\cdot\xxy^{{\rm dom}(\mathbb{A}_{2\mathbb{X}},~\mathbb{B}_{2\mathbb{X}},~  \widetilde{\mathbb{A}}_8(\hat\Sigma_e), ~ \widetilde{\mathbb{B}}_8(\hat\Sigma_{e}))}, \nonumber \nd
where  $\mathbb{X}_{\rm CD}(x, y) \equiv \tilde{\mathbb{X}}_{\rm CD}({\bf x}, y) \xxy^{\theta_{2\mathbb{X}}}$ which scales as $\theta_{2\mathbb{X}}$; and ${\rm tr}~\mathbb{X} \equiv {\bf g}^{\rm CD} \mathbb{X}_{\rm CD}$. The scalings in the second line of \eqref{belluchi}, that involve $\mathbb{A}_{2\mathbb{X}}, \mathbb{B}_{2\mathbb{X}}, \widetilde{\mathbb{A}}_8(\hat\Sigma_e)$ and $\widetilde{\mathbb{B}}_8(\hat\Sigma_e)$, are all separated by positive signs relative to each other.  
These parameters appearing in \eqref{belluchi} are defined as follows:
\bg\label{kellymac}
&& \widetilde{\mathbb{B}}_8(\hat\Sigma_e) =  - 1 + \gamma_{1,2}\bigg[{\hat\alpha_e(t) + \hat\beta_e(t)\over 2}\bigg] - {\log \vert a_{\rm CD} + {\hat\Sigma}_{e{\rm CD}}(t)\vert \over \vert\log~\bar{g}_s\vert} \\
&&\mathbb{B}_{2\mathbb{X}}= - 1 + \gamma_{1,2}\bigg[{\hat\alpha_e(t) + \hat\beta_e(t)\over 2}\bigg] - {\log \vert\theta_{2\mathbb{X}}(t) - a_{\rm CD} - {\hat\Sigma}_{e{\rm CD}}(t)\vert \over \vert\log~\bar{g}_s\vert} \nonumber\\
&& \widetilde{\mathbb{A}}_8(\hat\Sigma_e) =  - {\log \vert \dot{\hat\Sigma}_{e{\rm CD}}(t) \log~\bar{g}_s\vert \over \vert\log~\bar{g}_s\vert}, ~~ \mathbb{A}_{2\mathbb{X}} =  - {\log\vert (\dot{\theta}_{2\mathbb{X}}(t) - \dot{\hat\Sigma}_{e{\rm CD}}(t)) \log~\bar{g}_s\vert \over \vert\log~\bar{g}_s\vert}, \nonumber
\nd
which may be compared to the scalings appearing in {\bf Table \ref{jessrog}}. Plugging \eqref{kellymac} in \eqref{belluchi}, we note that, compared to the spatial derivatives, the temporal derivatives and especially derivatives like $({\rm tr}~\mathbb{X}), 0$ are typically related to ${\rm tr}~{\mathbb{X}}$ accompanied by non-trivial $\bar{g}_s$ scalings. 
 This means that a temporal derivative will always act as scalings by factors of  $\xxy^{{\rm dom}(\mathbb{A}_{2\mathbb{X}},~\mathbb{B}_{2\mathbb{X}},~  \widetilde{\mathbb{A}}_8(\hat\Sigma_e), ~ \widetilde{\mathbb{B}}_8(\hat\Sigma_{e}))}$, but this will not be the case with spatial derivatives. For our purpose, we can simply the coefficients \eqref{kellymac} slightly by observing that $a_{\rm CD}$ are constants and $\hat\Sigma_{e{\rm CD}}(t) << 1$. For such a case, the logarithmic pieces appearing for $\widetilde{\mathbb{B}}_8(\hat\Sigma_e)$ and $\mathbb{B}_{2\mathbb{X}}$ are dominated by $a_{\rm CD}$, so are basically constant and do not contribute non-trivially to the scalings. Similar for the other two cases in \eqref{kellymac}, both $\dot{\theta}_{2\mathbb{X}}(t)$ and $ \dot{\hat\Sigma}_{e{\rm CD}}(t)$ are very small so that we can approximate \eqref{kellymac} to the following:
\bg\label{kellymac2}
\left(\widetilde{\mathbb{A}}_8(\hat\Sigma_e), \mathbb{A}_{2\mathbb{X}}\right) ~ \to ~ \left(\infty, \infty\right), ~~~
\left(\widetilde{\mathbb{B}}_8(\hat\Sigma_e), \mathbb{B}_{2\mathbb{X}}\right) ~\to ~  \gamma_{1,2}\bigg[{\hat\alpha_e(t) + \hat\beta_e(t)\over 2}\bigg] -1, \nd
implying that $\xxy^{\widetilde{\mathbb{A}}_8(\hat\Sigma_e), ~\mathbb{A}_{2\mathbb{X}}} \to 0$ as $\bar{g}_s << 1$. With the aforementioned simplification, one 
may easily see that the temporal derivative pieces in \eqref{darilebu} take the following suggestive form:
\bg\label{dixlin}
\mathbb{Q}^{(1)}_{\rm curv} &= & q_1{\bf g}^{00}\left({\rm tr}~\mathbb{X} +  {\rm tr}~\mathbb{X}^2\right) 
\xxy^{\gamma_o - 1} \\
& + & {\bf g}^{00}\left[q_2 + q_3 {\bf g}^{00}\xxy^{2\gamma_o - 2}\right] 
\xxy^{\gamma_o - 1} \left({\rm D} + 2{\rm tr}~\mathbb{X} + 
{\rm tr}~\mathbb{X}^2\right)^2 + .., \nonumber \nd
where we kept $c_3 = 1$ and defined $\gamma_o$ as $\gamma_o \equiv \gamma_{1,2}\Big[{\hat\alpha_e(t) + \hat\beta_e(t)\over 2}\Big]$ to avoid clutter. The dotted terms are higher curvature contributions that involve temporal derivatives of the metric from the curvature polynomials. Combining the quantum series from 
\eqref{elmoss2} and the ones from \eqref{dixlin} above, we see that the coefficients change in the following way:
\bg\label{runy}
&& c_1^{(\pm)} \to c_1^{(\pm)} \mp q_1 {\bf g}^{00} \xxy^{\gamma_o - 1} \pm ...
\nonumber\\
&&c_2^{(2)} \to c_2^{(2)} + {\bf g}^{00}\left(q_2 + q_3 {\bf g}^{00}\xxy^{2\gamma_o - 2}\right)\xxy^{\gamma_o - 1} + ..., \nd
where $c_1^{(+)} \equiv c_1^{(1)}$ and $c_1^{(-)} \equiv c_1^{(2)}$. The dotted terms are the additional
corrections to the same order in ${\rm tr}~\mathbb{X}^p$. Note two things: One, the changes to the coefficients in 
\eqref{elmoss2} are not simply renormalization of the coefficients because they involve ${\bf g}_{00}(x, y)$
fields, so they typically involve interactions. And two, since ${\bf g}^{00}$ scales as $\xxy^{{8\over 3} - \hat\zeta_e(t)}$, the shifts in the coefficients $c_1^{(\pm)}$ and $c_2^{(2)}$ are small and therefore they do not interfere with the mapping \eqref{blsjun18}. On the other hand the spatial (and subsequently the internal) derivative parts of \eqref{daril} will be more non-trivial. We can even combine the spatial, internal and the temporal derivatives together to express the quantum series in a more condensed format. Looking at \eqref{daril} one may easily see that we can express it in the following alternative form:
\bg\label{taylewod}
\mathbb{Q}_{\rm curv} &= & \sum C_{\{l_i\}\{p_j\}\{n_k\}} \prod_i \left(\hat{\bf g}^{\rm C_i D_i}\right)^{l_i} 
\prod_j \left(\hat{\bf g}_{{}_{\rm E_jF_j, M_j}}\right)^{p_j} \prod_k \left(\hat{\bf g}_{{}_{\rm G_k H_k, N_kP_k}}\right)^{n_k}\\
& = &  \sum C_{\{l_i\}\{p_j\}\{n_k\}} \prod_i \left({\bf g}^{\rm C_i D_i} + \left(\bar\lambda {\cal O} \lambda\right)^{\rm C_i D_i}\right)^{l_i} 
\prod_j \left({\bf g}_{{}_{\rm E_jF_j, M_j}} + \left(\bar\lambda {\cal O} \lambda\right)_{\rm E_j F_J, M_j}\right)^{p_j} 
\nonumber\\
&& ~~~~ \times \prod_k \left({\bf g}_{{}_{\rm G_k H_k, N_kP_k}} + \left(\bar\lambda {\cal O} \lambda\right)_{\rm G_k H_k, N_kP_k} \right)^{n_k}\nonumber \nd
where now $\left({\rm C_i, D_i, E_j, F_j, G_k, H_k}\right) \in {\bf R}^{2, 1} \times {\cal M}_4 \times {\cal M}_2 \times {\mathbb{T}^2\over {\cal G}}$ and $\left({\rm M_j, N_k, P_k}\right) \in {\bf R}^{2, 1} \times {\cal M}_4 \times {\cal M}_2$.  It should also be clear that although we expect $\sum l_i > \sum p_j + \sum n_k$, Lorentz invariance of \eqref{taylewod} would relate all the $l_i$ to some combinations of all $p_j$ and $n_k$  for this to make sense. (Since the curvature tensor involve two derivatives, we expect the middle term in $p_j$ to reflect that.) Additionally, the series in \eqref{dixlin}, and consequently \eqref{runy}, are a subset of 
\eqref{taylewod}. This means the $g_s$ scaling of \eqref{taylewod} is dominated by the $g_s$ scalings of the powers of the Riemann curvature terms because a further rewriting of \eqref{taylewod} as:

{\footnotesize
\bg\label{hostel}
\mathbb{Q}_{\rm curv} & \equiv & \prod_j \hat{\bf g}^{\rm E_j F_j} \prod_i \left(\hat{\bf R}_{\rm A_i B_i C_i D_i}\right)^{l_i} =  
\prod_j {\bf g}^{\rm E_j F_j} \prod_i \left({\bf R}_{\rm A_i B_i C_i D_i}\right)^{l_i} \\ 
&+ & 
\prod_j {\bf g}^{\rm K_j L_j} \sum {\cal C}_{\{r_p\}...\{q_l\}} \prod_{p, t, ..., l} 
\left({\bf g}_{{}_{\rm A_pB_p, Q_p}}\right)^{r_p} \left({\bf g}_{{}_{\rm I_t J_t, R_t S_t}}\right)^{s_t} 
\left(\mathbb{X}_{\rm C_i D_i}\right)^{m_i} \left(\mathbb{X}_{\rm E_j F_j, M_j}\right)^{n_j}
\left(\mathbb{X}_{\rm G_l H_l, N_l P_l}\right)^{q_l}, \nonumber \nd}
where ${\cal C}_{\{r_p\}...\{q_l\}}$ are the coefficients that may be derived from $C_{\{l_i\}\{p_j\}\{n_k\}}$
in \eqref{taylewod}, would imply that the series in powers and derivatives of $\mathbb{X}_{\rm C_kD_k}$ are sub-dominant as they all scale as subdominant powers of ${g_s\over {\rm HH}_o}$ $-$ compared to the corresponding powers and derivatives of the ${\bf g}_{\rm C_kD_k}$ metric components $-$ from \eqref{tikiming}
and footnote \ref{versailles}. To see how this works precisely, let us look at the scaling of the curvature component $\hat{\bf R}_{mnpq}$ and compare this with the corresponding scaling from {\bf Tables \ref{firzathai}} and {\bf \ref{lilalo1}}. First, using the gravitino scaling from \eqref{tikiming} let us consider the series from 
\eqref{cascova} by keeping $k_{2n} = 0$ with $n > 1$. This way, with dimensionless fermions, there will be no ${\rm M}_p$ suppressions (or alternatively, ${\rm M}_p$ suppressed higher order terms with dimensionful fermions) except for $k_2 > 0$. 
For the $g_s$ scaling of $\mathbb{X}_{mn}$ it is clear that one of the condition for not going beyond the first term of 
\eqref{cascova} is that the fully contracted fermionc bi-linear term, $\bar\Psi_{\rm M}{\slashed{\partial}}\Psi^{\rm M}$, scales as:

{\scriptsize
\bg\label{kelimacdo}
{\rm scale}\left(\bar\Psi_{\rm M}{\slashed{\partial}}\Psi^{\rm M}\right) = \begin{cases}
~l_m + l_n + {7\over 3} + \hat\Lambda_{em}(t) + \hat\Lambda_{en}(t) - {\hat\zeta_e(t)\over 2} - {\rm dom}\left({3\hat\sigma_e(t)\over 2}, \hat\sigma_e(t) + {1\over 2}(\hat\alpha_e(t), \hat\beta_e(t))\right)\\
~~~~~\\
~ l_\alpha + l_\beta + {7\over 3} + \hat\Lambda_{e\alpha}(t) + \hat\Lambda_{e\beta}(t) - {\hat\zeta_e(t)\over 2} - {\rm dom}\left({3\over 2}(\hat\alpha_e(t), \hat\beta_e(t)), {\hat\sigma_e(t)\over 2} + (\hat\alpha_e(t), \hat\beta_e(t))\right)    
\end{cases}
\nd}
which are both positive definite because of the fermionic scaling as in \eqref{roseQ}.  The constraints on $(\hat\alpha(t), \hat\beta(t), \hat\sigma(t))$ from {\bf Table \ref{milleren4}} do not change any of the results. However if we would have chosen the scalings from \eqref{liberti}, then \eqref{kelimacdo} would have lead to the possibility of relative minus signs {\it of the same order} resulting in issues with EFT. This is at least avoided by the choice \eqref{roseQ}.

It is also important to comment that, the way we have tried to represent the fermions is through metric and $-$ as we show in section \ref{annypyar} $-$ through the G-flux components. As such the typical fermionic contribution should be sub-dominant as:
\bg\label{hamraaz2}
\mathbb{X}_{mn}(x, y) = \widetilde{\mathbb{X}}_{mn}({\bf x}, y) 
\left({g_s\over {\rm HH}_o}\right)^{{\rm dom}\left(l_m + l_n + {5\over 3}^\pm, ~l_m + l_n + l_{\rm M} + l_{\rm N} + {10\over 3}^\pm, ...\right)}, \nd
where $({\rm M, N}) \in {\cal M}_4 \times {\cal M}_2$, and the dotted terms are the higher fermionic condensates' contributions from \eqref{cascova}. It is clear from \eqref{roseQ} that the dominant contribution from the scaling comes from the first term in \eqref{hamraaz2}. The result \eqref{hamraaz2} can be further generalized to the following transformation to $\hat\sigma_e(t), \hat\alpha_e(t), \hat\beta_e(t)$:
\bg\label{vegjun29}
&& \hat\sigma_e(t) ~ \to ~ \hat\sigma_e(t) + {\rm dom}\left(0, {1\over 3}^\pm + {2l\over 3}, 0^\pm + {4l\over 3}, ...\right) \nonumber\\
&& (\hat\alpha_e(t), \hat\beta_e(t)) ~ \to ~(\hat\alpha_e(t), \hat\beta_e(t)) + {\rm dom}\left(0, {1\over 3}^\pm + {2l\over 3}, 0^\pm + {4l\over 3}, ...\right), \nd
where $l > 0$ and we have used \eqref{roseQ}. The other parameters, namely $(\hat\zeta_e(t), \hat\eta_e(t))$ remain unchanged because of our choice of the Rarita-Schwinger fermions to lie on the internal six-manifold ${\cal M}_4 \times {\cal M}_2$. It is also clear from \eqref{vegjun29} that the fermionic extension to the metric in \eqref{akash} and \eqref{cascova} do not upstage the $\bar{g}_s$ scalings of the metric components. This is good because it would mean the curvature scalings in {\bf Tables \ref{firzathai}}, {\bf \ref{firzathai2}}, {\bf \ref{firzathai3}} and {\bf \ref{privsocmey}} associated with ${\bf R}_{\rm ABCD}$ could still be regarded as the dominant contributions. To evaluate the sub-dominant contributions coming from the generalized metric one may easily replace $(\hat\sigma_e(t), \hat\alpha_e(t), \hat\beta_e(t))$ by \eqref{vegjun29} to produce the $\bar{g}_s$ scalings of the generalized curvature components $\hat{\bf R}_{\rm ABCD}$.

Let us make a few more observations on the series \eqref{hostel}, and specifically on the second line of 
\eqref{hostel}.  First, of course, Lorentz invariance is maintained for each term in the above series. And again, as before, the terms in powers of $n_j$ should reflect the fact that a curvature tensor has two derivatives. However, while the temporal derivatives could change the coefficients of \eqref{elmoss2} as in \eqref{runy}, this doesn't appear to be the case with \eqref{hostel} even if we use the first identity in 
\eqref{belluchi}. Thus if we restrict 
$({\rm M_j, N_l, P_l}) \in {\bf R}^2 \times {\cal M}_4 \times {\cal M}_2$, then they would give rise to additional derivative interactions of the fermions. The coefficients 
${\cal C}_{\{r_p\}...\{q_l\}}$ (or equivalently $C_{\{l_i\}\{p_j\}\{n_k\}}$ from \eqref{taylewod}) could be fixed 
using supersymmetry constraints, much like how we did previously.

To conclude: the fermionic corrections cannot change the dominant scaling of the quantum series in \eqref{botsuga} as long as the fermionic scalings are given by \eqref{roseQ} with $l > 0$. This however doesn't completely fix the scalings because of our attempt to analyze the fermionic interaction {\it only} through the fermionic extensions of the bosonic degrees of freedom. Our attempt here does provide a consistent picture but doesn't exactly give us a way to get the $\bar{g}_s$ scalings perturbatively. On the other hand non-perturbatively, {\it i.e.} for example using \eqref{britt} in the EOMs, would give us  $\theta_{2\mathbb{X}} = {8\over 3}$ as one might expect but that is only the result for the condensate \eqref{cascova}. To get at least the lower bounds on the scalings one will have to study the fermionic extension of the G-flux components. However before we dwell on this, there is still one more subtlety remaining with our study of the fermionic degrees of freedom, namely, the presence of {\it localized} fermions and their effects on the condensate. This is what we turn to next.

\subsection{Further subtleties with localized fermions and fermionic bilinears \label{tomiraaj}}

All our aforementioned conclusions were based on the assumption that $k_{2n} = 0$ for $n > 1$ in \eqref{cascova}. Switching on the higher order derivatives would not change the conclusions because of the ${\rm M}_p$ suppressions of the derivative terms. However subtleties arise when we allow {\it localized} fermions because the derivative actions would bring down extra powers of ${\rm M}_p$. In fact, as observed earlier, each derivative brings down at least two powers of ${\rm M}_p$, which would cancel against one power of ${\rm M}_p$ from the derivative scaling, thus allowing a net gain of one {\it positive} power of ${\rm M}_p$ from the derivative action. This fortunately is not a problem as long as we use the fermionic series as in \eqref{boncek}: the non-perturbative exponential suppressions of the positive powers of ${\rm M}_p$ as in \eqref{cheril} and 
\eqref{paige} allow well-defined behavior at late time and also when ${\rm M}_p \to \infty$. Perturbatively, however the issue still persists: localized fermions in \eqref{cascova} would create problems when 
${\rm M}_p \to \infty$. Is there a way out of this?

The issue appears when the localization is along a sub-manifold of the six-dimensional base ${\cal M}_4 \times {\cal M}_2$ as we have restricted the derivative operator ${\cal D}_{\rm M}$ from \eqref{darlene} to be along the six-manifold base. The problem at hand is now more technical compared to what we had earlier. To proceed, let us first make the following re-definition of the fermions $\Psi_{\rm M}$ and the coefficients $c_{k_1k_2}$ from \eqref{cascova}:
\bg\label{fakhost}
\Psi_{\rm M}(y_m, y_\alpha; g_s) = \Psi_{\rm M}(y_m, g_s) ~e^{-{\rm M}_p^2 y_\alpha^2}, ~~~~~~
c_{k_1 k_2} \equiv c_{|k_1|k_2}, \nd
where $|k_1|$ implies that we have kept $k_1$ fixed but will vary $k_2$ in \eqref{cascova}; and 
$(y_m, y_\alpha) \in ({\cal M}_4, {\cal M}_2)$ such that $y_\alpha^2 \equiv {\bf g}_{\alpha\beta} y^\alpha 
y^\beta$. We can also make similar re-definition of the coefficient $c_{|k_3|k_4|k_5|k_6}$ but will not do so here. We will also take the simpler case where the derivatives appearing in \eqref{cascova} are fully contracted, meaning that:
\bg\label{censor}
\left({\cal D} + {\cal D}^\dagger\right)^{2k_2} \equiv {\rm M}_p^{-2k_2}\sum_b f_{bk_2} \left(\partial^m\partial_m\right)^{k_2 - b} 
\left(\partial^\alpha \partial_\alpha\right)^b, \nd
with $0 \le b \le k_2$, $(y^m, y^\alpha) \in ({\cal M}_4 \times {\cal M}_2)$, $f_{bk_2}$ are the combinatorial coefficients;  and the indices are raised or lowered using the warped metric ${\bf g}_{\rm MN}$. As mentioned above we will also concentrate on the first term in \eqref{cascova}. 
Using \eqref{fakhost} and \eqref{censor} in \eqref{cascova}, the first term takes the following form:
\bg\label{gangubai}
\mathbb{X}^{(1, k_1)}_{\rm MN} &\equiv &\sum_{k_2 = 2}^{2\mathbb{Z}} \sum_{b = 0}^{k_2} {\rm M}_p^{-2k_2} f_{bk_2} c_{|k_1|k_2}{\rm exp}\left(-{\rm M}_p^2 {\bf g}_{\alpha\beta} y^\alpha y^\beta\right)\left(\partial^\alpha \partial_\alpha\right)^{b}{\rm exp}\left(-{\rm M}_p^2 {\bf g}_{\alpha\beta} y^\alpha y^\beta\right)\nonumber\\
&&~~~~~~\times \left[\bar\Psi(y^m, g_s) \cdot \left({\cal D} - {\cal D}^\dagger\right)^{|k_1|} \left(\partial^m \partial_m\right)^{k_2 - b} \cdot \Psi(y^m, g_s)\right]_{\rm MN}, \nd
where we have replaced $k_2$ in \eqref{cascova} by $2k_2$ to comply with the fully contracted case 
\eqref{censor}.   We will also take $k_2 \in 2\mathbb{Z}$ to simplify the subsequent analysis. Such a constraint is not necessary, except to avoid clutter so we will stick with that. It is now easy to see that, for all $b \le {k_2\over 2}$, the series in \eqref{gangubai} remains 
perturbative in ${\rm M}_p$ for ${\rm M}_p \to \infty$. The interesting case is when ${k_2\over 2} < b \le k_2$. The derivatives can act in multiple ways, giving rise to various powers of ${\rm M}_p$. For example, 
\eqref{gangubai} takes the form:
\bg\label{gangubai2}
\mathbb{X}^{(1, k_1)}_{\rm MN} &\equiv &\sum_{k_2 = 2}^{2\mathbb{Z}} c_{|k_1|k_2}\sum_{b = k_2/2}^{k_2} f_{bk_2}
 \left[\sum_{h = 0}^{2b-1} {\rm M}_p^{-2k_2 + 4b - 2h} 
f_h^{(b, k_2)}(y^\alpha, g_s) + {\cal O}\left({1\over {\rm M}_p^2}\right)\right] e^{-2{\rm M}_p^2 y_\alpha^2}\nonumber\\
&&~~~~~~\times \left[\bar\Psi(y^m, g_s) \cdot \left({\cal D} - {\cal D}^\dagger\right)^{|k_1|} \left(\partial^m \partial_m\right)^{k_2 - b} \cdot \Psi(y^m, g_s)\right]_{\rm MN}, \nd
where $f_h^{(b, k_2)}(y^\alpha, g_s)$ with ${k_2\over 2} < b \le k_2, 0 \le h \le 2b - 1$ are set of functions that may be easily derived from \eqref{gangubai} by the derivative actions on the exponential, $y^\alpha$ as well as on the warped-metric ${\bf g}_{\alpha\beta}$. Note the dependence of $f_h^{(b, k_2)}(y^\alpha, g_s)$ on ${g_s\over {\rm HH}_o}$, and also the overall dependence on {\it positive} powers of ${\rm M}_p$. They both appear from the derivative actions, as expected. For $k_2 = 2$ there are only two values of $b$ that are relevant for us and they are $b = 1, 2$. This gives us:

{\footnotesize
\bg\label{thehunt}
\mathbb{X}_{\rm MN}^{(1, k_1, 2)} &=& c_{|k_1|2} f_{12}\left[f_0^{(1, 2)}(y^\alpha, g_s) + {\bar{f}_1^{(1, 2)}(y^\alpha, g_s)\over {\rm M}_p^2} + {\cal O}\left({1\over {\rm M}_p^4}\right)\right]\\
&\times&  e^{-2{\rm M}_p^2 y_\alpha^2} 
\left[\bar\Psi(y^m, g_s) \cdot \left({\cal D} - {\cal D}^\dagger\right)^{|k_1|} \left(\partial^m \partial_m\right) \cdot \Psi(y^m, g_s)\right]_{\rm MN} \nonumber\\
&+ & c_{|k_1|2} f_{22}\left[{\rm M}_p^4 f_0^{(2, 2)}(y^\alpha, g_s) + {\rm M}_p^2 f_1^{(2, 2)}(y^\alpha, g_s) + 
f_2^{(2, 2)}(y^\alpha, g_s) + {\bar{f}_3^{(2, 2)}(y^\alpha, g_s) \over {\rm M}_p^2} + 
{\cal O}\left({1\over {\rm M}_p^4}\right)\right] \nonumber\\
& \times & e^{-2{\rm M}_p^2 y_\alpha^2} 
\left[\bar\Psi(y^m, g_s) \cdot \left({\cal D} - {\cal D}^\dagger\right)^{|k_1|} \cdot \Psi(y^m, g_s)\right]_{\rm MN}
\nonumber \nd}
where $\mathbb{X}^{(1, k_1, k_2)}$ is the further sub-division of \eqref{gangubai2} over $k_2 \in 2\mathbb{Z}$. The barred dimensionful functions $\bar{f}_1^{(1, 2)}(y^\alpha, g_s)$ and $\bar{f}_3^{(2, 2)}(y^\alpha, g_s)$ denote corrections to ${f}_1^{(1, 2)}(y^\alpha, g_s)$ and ${f}_3^{(2, 2)}(y^\alpha, g_s)$ respectively from the
corresponding perturbative series. (Recall that the perturbative series appear from $b < {k_2\over 2}$.) In a similar vein, for $k_2 = 4$, $b$ takes the values $b = 2, 3, 4$, and consequently the series \eqref{gangubai2} takes the following form:

{\footnotesize
\bg\label{thehunt2}
\mathbb{X}_{\rm MN}^{(1, k_1, 4)} &=& c_{|k_1|4} f_{24}\left[f_0^{(2, 4)}(y^\alpha, g_s) + {\bar{f}_1^{(2, 4)}(y^\alpha, g_s)\over {\rm M}_p^2}  + {\bar{f}_2^{(2, 4)}(y^\alpha, g_s)\over {\rm M}_p^4}  
+ {\bar{f}_3^{(2, 4)}(y^\alpha, g_s)\over {\rm M}_p^6}  + {\cal O}\left({1\over {\rm M}_p^8}\right)\right]\nonumber\\
&\times&  e^{-2{\rm M}_p^2 y_\alpha^2} 
\left[\bar\Psi(y^m, g_s) \cdot \left({\cal D} - {\cal D}^\dagger\right)^{|k_1|} \left(\partial^m \partial_m\right)^2 \cdot \Psi(y^m, g_s)\right]_{\rm MN} \nonumber\\
&+ & c_{|k_1|4} f_{34}\Bigg[{\rm M}_p^4 f_0^{(3, 4)}(y^\alpha, g_s) + {\rm M}_p^2 f_1^{(3, 4)}(y^\alpha, g_s) + 
f_2^{(3, 4)}(y^\alpha, g_s) + {\bar{f}_3^{(3, 4)}(y^\alpha, g_s) \over {\rm M}_p^2} 
+ {\bar{f}_4^{(3, 4)}(y^\alpha, g_s) \over {\rm M}_p^4} \nonumber\\
&+ & {\bar{f}_5^{(3, 4)}(y^\alpha, g_s) \over {\rm M}_p^6} + 
{\cal O}\left({1\over {\rm M}_p^8}\right)\Bigg]
 e^{-2{\rm M}_p^2 y_\alpha^2} 
\left[\bar\Psi(y^m, g_s) \cdot \left({\cal D} - {\cal D}^\dagger\right)^{|k_1|}  \left(\partial^m \partial_m\right)\cdot \Psi(y^m, g_s)\right]_{\rm MN} \nonumber\\
&+ & c_{|k_1|4} f_{44}\Bigg[{\rm M}_p^8 f_0^{(4, 4)}(y^\alpha, g_s) + {\rm M}_p^6 f_1^{(4, 4)}(y^\alpha, g_s) + 
{\rm M}_p^4 f_2^{(4, 4)}(y^\alpha, g_s) + {\rm M}_p^2 {f}_3^{(4, 4)}(y^\alpha, g_s)  \nonumber\\
& + &  
{f}_4^{(4, 4)}(y^\alpha, g_s) + {\bar{f}_5^{(4, 4)}(y^\alpha, g_s) \over {\rm M}_p^2} + {\bar{f}_6^{(4, 4)}(y^\alpha, g_s) \over {\rm M}_p^4} +
  {\bar{f}_7^{(4, 4)}(y^\alpha, g_s) \over {\rm M}_p^6} + 
{\cal O}\left({1\over {\rm M}_p^8}\right)\Bigg] \\
& \times &  e^{-2{\rm M}_p^2 y_\alpha^2} 
\left[\bar\Psi(y^m, g_s) \cdot \left({\cal D} - {\cal D}^\dagger\right)^{|k_1|}\cdot \Psi(y^m, g_s)\right]_{\rm MN},
\nonumber \nd}
where again all terms are suppressed by $e^{-2{\rm M}_p^2 y_\alpha^2}$, but the coefficients of the positive powers of ${\rm M}_p$ are quite non-trivial: they involve the dimensionful functions $f_h^{(b, k_2)}$ as well as well as products of the fermion bi-linears. Summing such a series would be a difficult 
exercise\footnote{One might be tempted to use the redefinition of the coefficients as in \eqref{hassholy}. Unfortunately, a simple redefinition like: 
\bg\label{friday}
c_{|k_1| k_2} \equiv \sum_{l \ge 1} 2^{|k_1|} d_l^{(|k_1|)}{\left(-2l{\rm M}_p\right)^{2k_2}\over (2k_2)!}, \nonumber\nd 
is not effective as before because of the presence of functions like $f_h^{(b, k_2)}(y^\alpha, g_s)$. However the series could still be summed, albeit {\it not} as a Borel resummation, and one may show that it takes the trans-series form \eqref{gilpinbeti}. The situation here is simplified by the presence of $e^{-2{\rm M}^2_py^2_\alpha}$ accompanying every term with positive powers of ${\rm M}_p$.}, but one can at least expect the following trans-series form for $\mathbb{X}^{(1)}_{\rm MN}$:

{\footnotesize
\bg\label{gilpinbeti}
\mathbb{X}_{\rm MN}^{(1)} &= & \sum_{k_1, k_2} \mathbb{X}_{\rm MN}^{(1, k_1, k_2)} \\
&= & e^{-2{\rm M}_p^2 y_\alpha^2} \sum_{k_1} 
\left[\bar\Psi(y^m, g_s) \cdot \left({\cal D} - {\cal D}^\dagger\right)^{|k_1|}
\left(\sum_{k = 1}^\infty \bar{c}_{kk_1}e^{-k {\rm M}_p^2 {\cal O}^{(k)}(y^\alpha, g_s, \partial_m)}
+ {\cal O}\left({1\over {\rm M}_p^2}\right)\right)\cdot \Psi(y^m, g_s)\right]_{\rm MN}, \nonumber \nd}
where ${\cal O}^{(k)}(y^\alpha, g_s, \partial_m)$ are set of operators that may be extracted from \eqref{thehunt} and \eqref{thehunt2} and $\bar{c}_{kk_1}$ are dimensionless constants. The above form of the trans-series, as one would have expected, is well-defined in the presence of localized fermions when 
${\rm M}_p \to \infty$. One should also keep track of the ${g_s\over {\rm HH}_o}$ scalings that would appear from the internal metric components but they would not change the form of the trans-series.

\subsection{Fermionic extension of the G-flux components, EOMs and \eqref{botsuga} \label{annyypyar}}

Having constructed the non-perturbative extension \eqref{gilpinbeti}, it is now time to see how G-flux components may have fermionic extensions. 
Our analysis so far has given us not only a way to integrate fermions inside the metric using the series \eqref{cascova}, but also possible trans-series extension once non-perturbative (and non-local) effects are taken into account. There are four set of series that appear from the aforementioned considerations.

\vskip.1in

\noindent $\bullet$ Pertubative in ${g_s\over {\rm HH}_o}$ and perturbative in ${\rm M}_p$. This is basically the fermionc extension \eqref{cascova} in \eqref{akash}. 

\vskip.1in

\noindent $\bullet$ Perturbative in ${g_s\over {\rm HH}_o}$, but non-perturbative in ${\rm M}_p$. This would be similar to \eqref{gilpinbeti} when we keep the $g_s$ expansion perturbative.

\vskip.1in

\noindent $\bullet$ Non-perturbative in ${g_s\over {\rm HH}_o}$ but perturbative in ${\rm M}_p$. This could also include non-local terms like \eqref{boncek}. Keeping ${\rm M}_p$ perturbative, this would lead to either \eqref{cheril} or \eqref{paige}. 

\vskip.1in

\noindent $\bullet$ Non-perturbative in both ${g_s\over {\rm HH}_o}$ and ${\rm M}_p$. This would be similar to \eqref{cheril} or \eqref{paige} when we take localized fermions like \eqref{fakhost} into account in \eqref{boncek}, thus naturally extending this to the non-local cases. 

\vskip.1in

\noindent In a similar vein we can therefore extend the G-flux components to allow for the Rarita-Schwinger fermions. Using the series in \eqref{cascova}, we can make the following proposal:
\bg\label{vendlinda}
\hat{\bf G}_{\rm ABCD} = {\bf G}_{\rm ABCD} + c_4 {\rm M}_p^{-1} \left(\bar\lambda {\cal O}\lambda\right)_{[{\rm ABCD}]}, \nd
where $c_4$ is a dimensionless constant, and we have used anti-symmetry with the fermionic indices so as to comply with the total anti-symmetry of the G-flux components. The latter is not necessary and we can resort to a generic rank four-tensor $\hat{\bf G}_{\rm ABCD}$, as we have done for the rank two metric case in \eqref{akash}, but we will avoid these complications here. Note two things: {\Su one}, is that generically $({\rm A, B,..}) \in 
{\bf R}^{2, 1} \times {\cal M}_4 \times {\cal M}_2 \times {\mathbb{T}^2\over {\cal G}}$, but since the fermions are defined only over the internal six-manifold, the indices on the fermonic part of \eqref{vendlinda}
need to be arranged appropriately. And {\Su two}, the fermionic extension is made directly to four-form ${\bf G}_4$ flux components ${\bf G}_{\rm ABCD}$ although we could also do it for the three-form ${\bf C}_3$ flux components ${\bf C}_{\rm ABC}$. But since ${\bf G}_4 \ne d{\bf C}_3$ as shown in \eqref{dietherapie}, this is not a contradiction. (Additionally, we rarely need to use ${\bf C}_3$ explicitly in our analysis except in very few cases that we shall discuss when we study the flux EOMs.) It is therefore easy to see that:
\bg\label{vendlinda2}
\hat{\bf G}_{\rm ABCD} = {\bf G}_{\rm ABCD} + {1\over 64} c_4 {\rm M}_p^{-1} \Big({c_{60}} \bar\Psi^{\rm M} 
\Gamma_{\rm ABMNCD} \Psi^{\rm N} + 16 {c_{20}} \bar\Psi_{[{\rm A}} \Gamma_{\rm BC} \Psi_{{\rm D}]} + .....\Big),\nd
where the dotted terms are additional fermionic interactions from \eqref{cascova}, and the coefficients 
$c_{k_10}$ are the coefficients $c_{k_1 k_2}$ in \eqref{cascova} for $k_2 = 0$. Thus the terms in 
\eqref{vendlinda} are the ones with no additional ${\rm M}_p$ suppressions, and are exactly the ones that have appeared earlier in \cite{bergshoeff} and in \cite{evanfermion}. Not surprisingly, we now see that  there can be higher order fermionic contributions to the G-flux components. If we take dimensionful fermions, all higher order fermionic terms will be perturbatively suppressed by ${\rm M}_p$. 

The fermionic contributions to the G-flux components should also be sub-dominant when we compare their $g_s$ scalings. Let us take components ${\bf G}_{\rm MNPQ}$, whose $\bar{g}_s$ scalings appear in \eqref{andyrey}. In section \ref{sec8.2} we will compute the dominant contribution to the scalings, and the answer therein appears in \eqref{salgill3}: the dominant scalings for all the components of ${\bf G}_{\rm MNPQ}$ are $0^\pm$. Now looking at the dominant scalings of the fermionic components appearing in \eqref{roseQ}, it is easy to infer that:
\bg\label{taglaundri}
6\left(-{1\over 3} + {\hat\sigma_e(t)\over 2}\right) + 2\left(-1^\pm + {l\over 3}\right) + 2\left({2\over 3} - (\hat\alpha_e(t), \hat\beta_e(t))\right) \ge 0, \nd
for the G-flux with components $\hat{\bf G}_{mnpq}$  from the fermionic contribution $\bar\Psi^r\Gamma_{mnpqrs}\Psi^s$ in \eqref{vendlinda2} 
with $(m, n, p, q) \in {\cal M}_4$ and $(r, s) \in {\cal M}_2$. The first term in \eqref{taglaundri} comes from all the six Gamma matrices; the second term comes from the two Rarita-Schwinger fermions; and the last term comes from the inverse metric components. Solving \eqref{taglaundri}
suggests that the lower bound for $l$ in \eqref{roseQ} should be $l \ge 4$, implying that we can rewrite \eqref{roseQ} as:
\begin{empheq}[box={\mybluebox[5pt]}]{equation}
\Psi_{\rm M}({\bf x}, y; g_s(t)) = \sum_{l \ge 0}\sum_{k_l = 0}^\infty \Psi^{(k, l)}_{{\rm M}}({\bf x}, y) \left({g_s\over {\rm HH}_o}\right)^{{1\over 3}^\pm + {l\over 3} + {2k_l\over 3}\vert \widetilde{\hat\Lambda}_{\rm M}(k_l; t)\vert} 
\label{roseQ2}
\end{empheq}
with $l$ now taking the lower bound of $l \ge 0$. The dotted terms in \eqref{vendlinda2} that involve powers of derivatives $\partial_{\rm Q}\partial^{\rm Q}$, or powers of ${\rm tr}~\mathbb{X}$ cannot change the conclusion that $l_{\rm C} > 0$ in \eqref{tikiming}.

Our result in \eqref{roseQ2} is based on the fact that the fermionic scalings in the generalized metric and flux components, \eqref{akash} and \eqref{vendlinda} respectively, should not upstage the $\bar{g}_s$ scalings of the bosonic degrees of freedom. This is a reasonable assumption so that in \eqref{botsuga} the system continues to allow for the same EFT description that we had earlier. Of course the dominance of the fermionic terms are allowed within reason, namely that they shouldn't create relative minus signs\footnote{The issue is deeper. If $l_{\rm C} < 0$, then powers of fermionic bilinears can produce higher powers of inverse $\bar{g}_s$. They should be properly summed up in exponential series otherwise we will have conflicts with both EFT and the Schwinger-Dyson equations. But this is not the main reason. Since de Sitter or quasi de Sitter spacetime is controlled by the bosonic degrees of freedom, fermionic dominance will imply that such backgrounds are {\it fermionic condensates}. While this opens up a fascinating new possibility, such a viewpoint is unfortunately plagued with innumerable phenomenological issues. As such we will continue with \eqref{roseQ2}. \label{cascovbhalo}} in the $g_s$ scalings of \eqref{botsuga}. The latter, if allowed, would imply new curvature terms, and therefore new $g_s$ scalings (that we have consistently ignored) making the analysis more difficult. We will therefore continue to keep the fermionic contributions sub-dominant. 

In view of the above comments it is therefore instructive to see how the fermionic contributions influence the quantum series in \eqref{botsuga}. We will only look at the perturbative series as one can get the non-perturbative (including the non-local) series by appropriately bringing it in the trans-series form. The quantum series then becomes:

{\footnotesize
\bg\label{fahingsha10}
\hat{\mathbb{Q}}_{\rm T}^{(\{l_i\}, n_i)} =  \left[\hat{\bf g}^{-1}\right] \prod_{i = 0}^3 \left[\partial\right]^{n_i} 
\left(\hat{\bf g}^{\rm C_1}_{\rm C_2} \hat{\bf g}^{\rm C_2}_{\rm C_3} .....\hat{\bf g}^{{\rm C}_{n_4}}_{\rm C_1}
\right)^{n_5}
\prod_{{\rm k} = 1}^{41} \left(\hat{\bf R}_{\rm A_k B_k C_k D_k}\right)^{l_{\rm k}} \prod_{{\rm r} = 42}^{81} 
\left(\hat{\bf G}_{\rm A_r B_r C_r D_r}\right)^{l_{\rm r}}, \nd}
which differs from \eqref{botsuga} by the presence of generalized metric $\hat{\bf g}_{\rm CD}$ as well as generalized curvature $\hat{\bf R}_{\rm ABCD}$ and generalized G-flux components 
$\hat{\bf G}_{\rm ABCD}$ whose forms have been discussed earlier. Note that we have used the generalized metric components for inverse metric also in \eqref{fahingsha10}. This is necessary to get terms like \eqref{ozark} and \eqref{rooth} to appear consistently. The fermionic EOMs can now be derived from 
\eqref{fahingsha10} via:
\bg\label{tagmukh}
{\delta\over \delta \bar{\Psi}_{\rm M}(z_1, z_2)} \int d^{11}x \sqrt{-{\bf g}_{11}(x, y, w^a)}\left(
\hat{\mathbb{Q}}_{\rm T}^{(\{l_i\}, n_i)}(x, y, w^a) + .....\right) = 0, \nd
where $(z_1, z_2) \in {\bf R}^{2,1}, {\cal M}_4 \times {\cal M}_2$, and the dotted terms are the non-perturbative, non-local and the topological contributions. We will the discuss the latter contributions in the following section wherein the first term with $\hat{\mathbb{Q}}_{\rm T}^{(\{l_i\}, n_i)}(x, y, w^a)$ will be the zero instanton sector and the dotted terms will form the rest of a trans-series. Taking the Hermitian conjugate of \eqref{tagmukh} will give us the other set of Rarita-Schwinger fermionic equations.


\begin{table}[tb] 
 \begin{center}
\renewcommand{\arraystretch}{1.5}
\begin{tabular}{|c||c||c||c||c||c||c||c|}\hline Class & ${\Su{\bf G}_{\rm ABCD}}$ & Class & ${\Su{\bf G}_{\rm ABCD}}$ & Class & ${\Su{\bf C}_{\rm ABC}}$ & Class & ${\Su{\bf C}_{\rm ABC}}$ \\ \hline\hline
$\mathbb{G}_1$ & ${\bf G}_{\rm 0MNP}$ & $\mathbb{G}_{10}$ & ${\bf G}_{{\rm MNP}a}$ & $\mathbb{X}_1$ & 
${\bf C}_{\rm MNP}$ & $\mathbb{X}_{10}$ & ${\bf C}_{0ia}$ \\ \hline
$\mathbb{G}_2$ & ${\bf G}_{{\rm 0NP}a}$ &$\mathbb{G}_{11}$ & ${\bf G}_{{\rm MN}ab}$ 
& $\mathbb{X}_2$ & ${\bf C}_{\rm 0MN}$ & $\mathbb{X}_{11}$ & ${\bf C}_{ij{\rm M}}$ \\ \hline
$\mathbb{G}_3$ & ${\bf G}_{{\rm 0N}ab}$ & $\mathbb{G}_{12}$ & ${\bf G}_{{\rm MNP}i}$ & $\mathbb{X}_3$ & ${\bf C}_{{\rm NP}a}$ & $\mathbb{X}_{12}$ & ${\bf C}_{0ij}$ \\ \hline
$\mathbb{G}_4$ & ${\bf G}_{{\rm 0MN}i}$ & $\mathbb{G}_{13}$ & ${\bf G}_{{\rm MN}ai}$ & $\mathbb{X}_4$ & ${\bf C}_{{\rm 0P}a}$ & $\mathbb{X}_{13}$ & ${\bf C}_{ija}$ \\ \hline
$\mathbb{G}_5$ & ${\bf G}_{{\rm 0N}ia}$ & $\mathbb{G}_{14}$ & ${\bf G}_{{\rm M}abi}$ & $\mathbb{X}_5$ & ${\bf C}_{{\rm N}ab}$ & $\mathbb{X}_{14}$ & ${\bf C}_{abi}$ \\ \hline
$\mathbb{G}_6$ & ${\bf G}_{0ij{\rm M}}$ & $\mathbb{G}_{15}$ & ${\bf G}_{{\rm MN}ij}$ & $\mathbb{X}_6$ & ${\bf C}_{0ab}$ & $....$ & $.....$ \\ \hline
$\mathbb{G}_7$ & ${\bf G}_{0ija}$ & $\mathbb{G}_{16}$ & ${\bf G}_{{\rm M}aij}$ & $\mathbb{X}_7$ & ${\bf C}_{{\rm MN}i}$ & $....$ & $......$ \\ \hline 
$\mathbb{G}_8$ & ${\bf G}_{0abi}$ & $\mathbb{G}_{17}$ & ${\bf G}_{abij}$  & $\mathbb{X}_8$ & ${\bf C}_{{\rm 0N}i}$ 
& $....$ & $.....$ \\ \hline
$\mathbb{G}_9$ & ${\bf G}_{\rm MNPQ}$ & $....$ & $.....$ & $\mathbb{X}_9$ & ${\bf C}_{{\rm N}ia}$ & $....$ & $.....$\\ \hline 
\end{tabular}
\renewcommand{\arraystretch}{1}
\end{center}
 \caption[]{\Su Identifications of the ${\rm G}_i$ and the $\mathbb{X}_i$ tensors ($\mathbb{X}_1$ and $\mathbb{X}_2$ not to be confused with \eqref{boncek}) to the four and three-form flux components  ${\bf G}_{\rm ABCD}$ and ${\bf C}_{\rm ABC}$ respectively. These flux components appear in the EOMs and the Bianchi identities 
discussed in section \ref{sec7s} and \ref{sec4.5} respectively. Note that $({\rm A, B}) \in {\bf R}^{2, 1} \times {\cal M}_4 \times {\cal M}_2 \times {\mathbb{T}^2\over {\cal G}}$.}
  \label{lokeys}
 \end{table}


\section{Schwinger-Dyson's equations and existence of a de Sitter state \label{sec6}}

Our analysis with the fermionic terms gave us a way to extend the quantum series \eqref{botsuga} in a reasonable way. We also saw how to incorporate non-perturbative and non-local effects by summing over certain series appropriately. Such a procedure matches well with \cite{desitter2, coherbeta, coherbeta2}, but the analysis presented for extracting the non-perturbative effects is not exactly a Borel resummation as also mentioned therein. In this section we would like to derive the corresponding Schwinger-Dyson's equations for the metric and fluxes by incorporating these non-perturbative and non-local effects. However before delving into this, let us clarify how the equations look like for the 
truly off-shell and the off-shell-behaving-like-the-on-shell states for our case.  

\subsection{Schwinger-Dyson's equations for the off-shell and the on-shell states \label{sec6.1}}

To study the dynamics of the ``off-shell" and the ``on-shell" states, with the latter also appearing from an off-shell analysis, {\it i.e.} from a path-integral, we need to go back to the Schwinger-Dyson's equations that we discussed in \eqref{mariesole}. The typical Schwinger-Dyson's equations for the on-shell components of the metric, the fluxes and the fermions take the following form:

{\footnotesize
\bg\label{chelmey}
\left\langle {\delta\left(\check{\bf S}_{\rm tot}({\bf \Xi, \Upsilon}_g) - \check{\bf S}_{\rm ghost}({\bf \Xi, \Upsilon}_g)\right)\over \delta {\bf \Xi}}\right\rangle_\sigma &= &
 {\delta\check{\bf S}_{\rm tot}(\langle{\bf \Xi}\rangle_\sigma)\over \delta \left\langle{\bf \Xi}\right\rangle_\sigma} +\sum_{\sigma'}  \left\langle {\delta\left(\check{\bf S}_{\rm tot}({\bf \Xi, \Upsilon}_g) - \check{\bf S}_{\rm ghost}({\bf \Xi, \Upsilon}_g)\right)\over \delta {\bf \Xi}}\right\rangle_{(\sigma\vert\sigma')} \nonumber\\ 
 &=& 
-\left\langle {\delta\check{\bf S}_{\rm ghost}({\bf \Xi, \Upsilon}_g)\over \delta {\bf \Xi}}\right\rangle_\sigma - \left\langle {\delta\over \delta {\bf \Xi}}~{\rm log}\left(\mathbb{D}^\dagger(\sigma)\mathbb{D}(\sigma)\right)\right\rangle_\sigma \nd}
where the difference from \eqref{mariesole} is the usage of $\check{\bf S}_{\rm tot}$ instead of $\hat{\bf S}_{\rm tot}$, although we will continue to use $\check{\bf S}_{\rm tot} \approx \hat{\bf S}_{\rm tot}$ (see \eqref{monfluent}). We have also used the result from footnote \ref{desclaud}. However unlike \eqref{monfluent} and footnote \ref{loicry}, we can now define $\hat{\bf S}_{\rm tot}({\bf \Xi, \Upsilon}_g)$ in the following way:
\bg\label{tomiraja}
\hat{\bf S}_{\rm tot}({\bf \Xi, \Upsilon}_g)={\bf S}_{\rm kin}({\bf \Xi})+ {\bf S}_{\rm NP}({\bf \Xi}) + {\bf S}_{\rm nloc}({\bf \Xi}) + \hat{\bf S}_{\rm ghost}({\bf \Xi, \Upsilon}_g), \nd
thus including a convergent series of non-local interactions. Note also the presence of the functional derivative of $\mathbb{D}^\dagger(\sigma) \mathbb{D}(\sigma)$ with respect to ${\bf \Xi}$. This is because of the dependence of $\mathbb{D}(\sigma)$ on the on-shell degrees of freedom. The metric components, whose EOMs appear above, can be expressed as in \eqref{ebrowning} which we quote here again: 

{\footnotesize
\bg\label{ebrowning0}
\langle {\bf g}_{\rm AB} \rangle_{\sigma} &\equiv & {\int [{\cal D} {\bf \Xi}] [{\cal D}
{\bf \Upsilon}_{g}]~e^{-\hat{\bf S}_{\rm tot}({\bf \Xi, \Upsilon}_g)}~ \mathbb{D}^\dagger({\alpha}, {\beta}, {\gamma}; {\bf \Xi})~ {\bf g}_{\rm AB}(x, y, z)~
\mathbb{D}({\alpha}, {\beta}, {\gamma}; {\bf \Xi}) \over 
\int [{\cal D} {\bf \Xi}] [{\cal D}
{\bf \Upsilon}_{g}]~e^{-\hat{\bf S}_{\rm tot}( {\bf \Xi, \Upsilon}_g)} ~\mathbb{D}^\dagger({\alpha}, {\beta}, {\gamma}; {\bf \Xi}) 
\mathbb{D}({\alpha}, {\beta}, {\gamma}; {\bf \Xi})}, \nd}
where $\mathbb{D}(\sigma) \equiv \mathbb{D}(\alpha, \beta, \gamma; {\bf \Xi})$ is only the linear part, with ${\bf \Xi} = ({\bf g}_{\rm AB}, {\bf C}_{\rm ABD}, {\bf \Psi}_{\rm A}, \overline{\bf \Psi}_{\rm A})$; and we used $\check{\bf S}_{\rm tot} \approx \hat{\bf S}_{\rm tot}$. The {\it on-shell} states are now defined from the above off-shell computation when the components of the metric, fluxes and the Rarita-Schwinger fermions lie in the set ${\bf \Xi}$. All other components that do not lie in the set ${\bf \Xi}$ are the truly off-shell states\footnote{Note the {\it absence} of the off-shell degrees of freedom from the measure and from the action in \eqref{ebrowning0}. As discussed in \cite{joydeep}, their effects are absorbed in the definition of 
$\hat{\bf S}_{\rm tot}({\bf \Xi, \Upsilon}_g)$ from \eqref{tomiraja}. We will elaborate on this a bit more soon.}. The off-shell and the on-shell states are then defined from \eqref{ebrowning0} in the following way:
\bg\label{angwhiteT}
&& \langle {\bf g}_{\rm 0A'}\rangle_\sigma ~ = ~ \langle {\bf g}_{\rm M'a}\rangle_\sigma ~ = ~ \langle {\bf g}_{m\rho} \rangle_\sigma ~ = ~ \langle {\bf g}_{i{\rm M}}\rangle_\sigma ~ = ~... ~ = ~ 0\nonumber\\
&&\langle {\bf g}_{\rm AB}\rangle_\sigma = \big({\bf g}_{00}(y, g_s),~ {\bf g}_{ij}(y, g_s), ~{\bf g}_{mn}(y, g_s), ~{\bf g}_{\rho\sigma}(y, g_s), ~{\bf g}_{ab}(y, g_s)\big), \nd
where ${\rm A'} \in {\bf R}^2 \times {\cal M}_4 \times {\cal M}_2 \times {\mathbb{T}^2\over {\cal G}}, ~ {\rm M'} \in {\bf R}^2 \times {\cal M}_4 \times {\cal M}_2, ~ {\rm M} \in {\cal M}_4 \times {\cal M}_2$. In the path-integral computation of the metric components, as in \eqref{ebrowning} and \eqref{ebrowning0}, the truly off-shell components are also the ones where the corresponding vacua remain the un-shifted interacting vacua $\vert\Omega\rangle$. This make them automatically vanish as one-point functions in any interacting theory vanish due to conservation of momenta\footnote{There might be some confusion in the nomenclature here. Absence of the emergent off-shell degrees of freedom does not mean the absence of off-shell states for the emergent on-shell degrees of freedom. The latter continues to arise in the path-integrals in the usual ways thus leading to the derivation of the underlying Wheeler-de Witt equation as shown in \cite{wdwpaper}.}. Question then is how to determine the corresponding Schwinger-Dyson's equations for the truly off-shell states?

To determine the Schwinger-Dyson's equations we should remember that while the expectation values of the off-shell states vanish according to \eqref{angwhiteT}, the off-shell states would themselves propagate in loops in the Minkowski background. In the path-integral formalism, we can integrate them out giving rise to non-local interactions. In the process however we can determine possible Schwinger-Dyson's equations of the form:

{\footnotesize
\bg\label{lilshank}
\Big\langle \lambda_3\big({\bf R}_{\rm A''B''} - {\bf T}^{\rm loc}_{\rm A''B''} - {\bf T}^{\rm nloc}_{\rm A''B''} +  .. \big)\Big\rangle_\sigma = 
{\bf R}_{\rm A''B''}(\langle{\bf \Xi}\rangle_\sigma) - {\bf T}^{\rm loc}_{\rm A''B''}(\langle{{\bf \Xi}}\rangle_\sigma)- {\bf T}^{\rm nloc}_{\rm A''B''}(\langle{{\bf \Xi}}\rangle_\sigma)= 0, \nd}
where $\langle {\bf g}_{\rm A''B''}\rangle_\sigma = 0$ with $({\rm A'', B''})$ spanning the components in the first line of \eqref{angwhiteT}, and 
${\bf T}^{\rm loc}_{\rm A''B''}$ is the localized contribution that include the perturbative and the non-perturbative contributions\footnote{Depending how we arrange the resurgence trans-series.}, and $\lambda_3$ is the remnant of the determinant of the eleven-dimensional metric. The dotted terms involve other contributions including Faddeev-Popov and Batalin-Vilkovisky ghosts (denoted by the set ${\bf \Upsilon}_g$). Note that 
neither ${\bf R}_{\rm A''B''}(\langle{\bf \Xi}\rangle_\sigma)$ nor the local and the non-local energy momentum tensors are functions of ${\bf g}_{\rm A''B''}$. The expectation values themselves are now taken over an action that is expressed via the local and the non-local interactions, including ghosts, constructed solely in terms of the on-shell degrees of freedom. To see how this works, and in particular the consistency of \eqref{lilshank}, let us consider one such off-shell component ${\bf g}_{\rm 0A'}$ with ${\rm A'} \in {\bf R}^2 \times {\cal M}_4 \times {\cal M}_2 \times {\mathbb{T}^2\over {\cal G}}$. Taking the expectation value $\langle \sigma \vert \sigma \rangle$ and shifting the component ${\bf g}_{\rm 0A'}$ by a small amount leads to the expected Schwinger-Dyson's equation of the form:

{\scriptsize
\bg\label{tgrlbahlu}
\int \left[{\cal D}\hat{\bf \Xi}\right]\int \left[{\cal D}{\bf g}_{\rm 0A'}\right] ~e^{-{\bf S}^t_{\rm tot}(\hat{\bf \Xi}, {\bf g}_{\rm 0A'})} 
\left({\bf R}_{\rm 0A'}({\bf \Xi}, {\bf g}_{\rm 0A'}) - {1\over 2} {\bf g}_{\rm 0A'} {\bf R}({\bf \Xi}, {\bf g}_{\rm 0A'}) - {\bf T}_{\rm 0A'}({\bf \Xi}, {\bf g}_{\rm 0A'}) + ...\right) \mathbb{D}^\dagger(\sigma) \mathbb{D}(\sigma) = 0, \nonumber\\
\nd}
where  $\hat{\bf \Xi} = ({\bf \Xi, \Upsilon}_g)$; ${\bf S}^t_{\rm tot}(\hat{\bf \Xi}, {\bf g}_{\rm 0A'}) \equiv {\bf S}_{\rm tot}(\hat{\bf \Xi}, {\bf g}_{\rm 0A'}) - {1\over 2}\log[-{\rm det}~{\bf g}_{11}({\bf \Xi, g}_{\rm 0A'})] $, ${\bf S}_{\rm tot}(\hat{\bf \Xi}, {\bf g}_{\rm 0A'}) = {\bf S}_{\rm kin}({\bf \Xi}, {\bf g}_{\rm 0A'}) + {\bf S}_{\rm pert}({\bf \Xi}, {\bf g}_{\rm 0A'}) + {\bf S}_{\rm ghost}(\hat{\bf \Xi}, {\bf g}_{\rm 0A'})$. (While ${\bf S}_{\rm tot}(\hat{\bf \Xi}, {\bf g}_{\rm 0A'})$ does not depend on the eleven-dimensional coordinates, the log term, namely $\log[-{\rm det}~{\bf g}_{11}({\bf \Xi, g}_{\rm 0A'})] \equiv \log[-{\rm det}~{\bf g}_{11}({\bf \Xi, g}_{\rm 0A'}; x, y, w)]$  does depend on them with $x \in {\bf R}^{2, 1}, y \in {\cal M}_4 \times {\cal M}_2, w \in \xoxo$. This means 
${\bf S}^t_{\rm tot}(\hat{\bf \Xi}, {\bf g}_{\rm 0A'}) \equiv {\bf S}^t_{\rm tot}(\hat{\bf \Xi}, {\bf g}_{\rm 0A'}; x, y, w)$.) We have ignored other off-shell components for simplicity and 
${\bf \Xi} = ({\bf g}_{\rm AB}, {\bf C}_{\rm ABD}, {\bf \Psi}_{\rm A}, \overline{\bf \Psi}_{\rm A})$ is the set of all  allowed on-shell degrees of freedom as they appear, for example, in the second line of \eqref{angwhiteT} for the metric components. The dotted terms are related to the Faddeev-Popov and other ghosts with ${\bf T}_{\rm 0A'}({\bf \Xi}, {\bf g}_{\rm 0A'}) \equiv {\bf T}^{\rm pert}_{\rm 0A'}({\bf \Xi}, {\bf g}_{\rm 0A'})$. Note that $\mathbb{D}(\sigma) \equiv \mathbb{D}(\sigma, {\bf \Xi})$ does not depend on the off-shell components as emphasized earlier, but $-{\bf g}_{11}({\bf \Xi, g}_{\rm 0A'})$ does, where ${\bf \Xi}$ for the determinant case should be understood as all the on-shell metric components. We can use a congruence transformation to express the determinant as:
\bg\label{lilykhel}
{\bf g}_{11}({\bf \Xi, g}_{\rm 0A'}) = \mathbb{V}^\top({\bf \Xi, g}_{\rm 0A'}){\bf g}_{11,d}({\bf \Xi, g}_{\rm 0A'})\mathbb{V}({\bf \Xi, g}_{\rm 0A'}), \nd
where $\mathbb{V}({\bf \Xi, g}_{\rm 0A'})$ is an invertible matrix and 
${\bf g}_{11,d}({\bf \Xi, g}_{\rm 0A'})$ is a diagonal matrix\footnote{Previously we had used $-{\bf g}_{11}$ to denote the determinant of the metric with only the on-shell components. Now that we allow off-shell components also, we denote the matrix using ${\bf g}_{11}({\bf \Xi, g}_{\rm 0A'})$ and its determinant using ${\rm det}~{\bf g}_{11}({\bf \Xi, g}_{\rm 0A'})$. Similar notations are used for the diagonal one.}. One of the advantage of expressing the determinant of the metric using the diagonal one is that, on every term of the diagonal matrix, the on and off-shell metric components mix non-trivially. This will be useful momentarily when
we perform the integral over ${\bf g}_{\rm 0A'}$. To do this, let us divide the total action and the tensors in the following suggestive way:

{\footnotesize
\bg\label{stanfordtag}
&&{\bf R}_{\rm 0A'}({\bf \Xi}, {\bf g}_{\rm 0A'}) = \hat{\bf R}_{\rm 0A'}({\bf \Xi}) + 
\widetilde{\bf R}_{\rm 0A'}({\bf \Xi}, {\bf g}_{\rm 0A'}), ~~ 
{\bf R}({\bf \Xi}, {\bf g}_{\rm 0A'}) = \hat{\bf R}({\bf \Xi}) + 
\widetilde{\bf R}({\bf \Xi}, {\bf g}_{\rm 0A'})\\
&& {\bf T}_{\rm 0A'}({\bf \Xi}, {\bf g}_{\rm 0A'}) = \hat{\bf T}_{\rm 0A'}({\bf \Xi}) + 
\widetilde{\bf T}_{\rm 0A'}({\bf \Xi}, {\bf g}_{\rm 0A'}), ~~ 
{\bf S}_{\rm tot}({\bf \Xi}, {\bf g}_{\rm 0A'}, {\bf \Upsilon}_g) = \overline{\bf S}_{\rm tot}(\hat{\bf \Xi}) + 
\widetilde{\bf S}_{\rm tot}(\hat{\bf \Xi}, {\bf g}_{\rm 0A'}), \nonumber \nd}
which basically split them up into pieces that depend only on $({\bf \Xi, \Upsilon}_g) \equiv \hat{\bf \Xi}$ and pieces that depend on both $({\bf \Xi, \Upsilon}_g)$ and ${\bf g}_{\rm 0A'}$. (The distribution of 
${\bf R}({\bf \Xi}, {\bf g}_{\rm 0A'})$ in terms of $\hat{\bf R}({\bf \Xi})$  and $\widetilde{\bf R}({\bf \Xi}, {\bf g}_{\rm 0A'})$ should be understood only as a functional distribution so that we are not obligated to provide meanings to the two parts.) 
Such a splitting in \eqref{stanfordtag} would split the Schwinger-Dyson's equation in \eqref{tgrlbahlu} in the following way:

{\footnotesize
\bg\label{violetgrm}
0 &= & \int \left[{\cal D} \hat{\bf \Xi}\right] ~e^{-\overline{\bf S}_{\rm tot}({\bf \Xi, \Upsilon}_g)} \left(\hat{\bf R}_{\rm 0A'}({\bf \Xi}) - \hat{\bf T}_{\rm 0A'}({\bf \Xi})\right) 
\mathbb{D}^\dagger(\sigma) \mathbb{D}(\sigma)
\int \left[{\cal D} {\bf g}_{\rm 0A'}\right] ~e^{-\widetilde{\bf S}^t_{\rm tot}({\bf \Xi, g}_{\rm 0A'}, {\bf \Upsilon}_g)} \\
&+& \int \left[{\cal D}\hat{\bf \Xi}\right]  ~e^{-\overline{\bf S}_{\rm tot}(\hat{\bf \Xi})}
\mathbb{D}^\dagger(\sigma) \mathbb{D}(\sigma)
\int \left[{\cal D}{\bf g}_{\rm 0A'}\right] ~e^{-\widetilde{\bf S}^t_{\rm tot}(\hat{\bf \Xi}, {\bf g}_{\rm 0A'})} 
\left(\widetilde{\bf G}_{\rm 0A'}({\bf \Xi, g}_{\rm 0A'}) - \widetilde{\bf T}_{\rm 0A'}({\bf \Xi, g}_{\rm 0A'}) + ...\right) \nonumber
\nd}
where $\widetilde{\bf G}_{\rm 0A'}({\bf \Xi, g}_{\rm 0A'}) =\widetilde{\bf R}_{\rm 0A'}({\bf \Xi, g}_{\rm 0A'}) - {1\over 2} {\bf g}_{\rm 0A'} {\bf R}({\bf \Xi, g}_{\rm 0A'})$; and the off-shell action is defined as $\widetilde{\bf S}^t_{\rm tot}({\bf \Xi, g}_{\rm 0A'}, {\bf \Upsilon}_g) $ $\equiv$
$\widetilde {\bf S}_{\rm tot}({\bf \Xi, g}_{\rm 0A'}, {\bf \Upsilon}_g)$ $-{1\over 2}\log[-{\rm det}~{\bf g}_{11, d}({\bf \Xi, g}_{\rm 0A'})]$. Observe that we did not use the $\hat{\bf R}({\bf \Xi})$ and $\widetilde{\bf R}({\bf \Xi, g}_{\rm 0A'})$ decomposition in \eqref{violetgrm}. This is because the remaining integral of the form:
\bg\label{smitvanand}
\int \left[{\cal D} \hat{\bf \Xi}\right] ~e^{-\overline{\bf S}_{\rm tot}({\bf \Xi, \Upsilon}_g)} \hat{\bf R}({\bf \Xi})  
\mathbb{D}^\dagger(\sigma) \mathbb{D}(\sigma)
\int \left[{\cal D} {\bf g}_{\rm 0A'}\right] ~e^{-\widetilde{\bf S}^t_{\rm tot}({\bf \Xi, g}_{\rm 0A'}, {\bf \Upsilon}_g)} {\bf g}_{\rm 0A'} \stackrel{?}{=} 0, \nd
only vanishes in special cases, say for example if we have one cross-term component, and generically the integral in \eqref{smitvanand} is non-zero. The dotted terms  in \eqref{violetgrm} are now the ghosts {\it and} the non-perturbative contributions. Note two things: {\Su one}, the advantage of using the diagonal matrix is to put all the complications to the off-shell side so that we can integrate them out\footnote{The determinant however remains unchanged.}; and {\Su two}, the other term in the second line of \eqref{chelmey}, {\it i.e.} the term related to the functional derivative of $\mathbb{D}^\dagger(\sigma) \mathbb{D}(\sigma)$, does not contribute to the Schwinger-Dyson's equation. Finally, the two integrals that contribute to \eqref{violetgrm} may be extracted from the following series of integrals:

{\footnotesize
\bg\label{bhmeymelismey}
&&  \int \left[{\cal D} {\bf g}_{\rm 0A'}\right]~e^{-\widetilde{\bf S}_{\rm tot}(\hat{\bf \Xi}, {\bf g}_{\rm 0A'})} =  e^{-{\bf S}_{\rm tot}^{\rm nloc}(\hat{\bf \Xi})}\\
&&  \int \left[{\cal D} {\bf g}_{\rm 0A'}\right]~e^{-\widetilde{\bf S}^t_{\rm tot}(\hat{\bf \Xi}, {\bf g}_{\rm 0A'})} = \lambda_1({\bf \Xi}) \lambda_2(\hat{\bf \Xi}; x, y, w) e^{-{\bf S}_{\rm tot}^{\rm nloc}(\hat{\bf \Xi})}\nonumber\\
&& \int \left[{\cal D}{\bf g}_{\rm 0A'}\right] ~e^{-\widetilde{\bf S}^t_{\rm tot}(\hat{\bf \Xi}, {\bf g}_{\rm 0A'})} 
\left(\widetilde{\bf G}_{\rm 0A'}({\bf \Xi, g}_{\rm 0A'}) - \widetilde{\bf T}_{\rm 0A'}({\bf \Xi, g}_{\rm 0A'})\right) =\lambda_1({\bf \Xi}) \lambda_2(\hat{\bf \Xi}; x, y, w)e^{-{\bf S}_{\rm tot}^{\rm nloc}(\hat{\bf \Xi})}~{\bf T}^{\rm nloc}_{\rm 0A'}({\bf \Xi}),\nonumber \nd}
where $\lambda_1({\bf \Xi})$ and $\lambda_2(\hat{\bf \Xi}; x, y, w)$
are the remnants of the determinant after we have integrated out the off-shell field ${\bf g}_{\rm 0A'}$, and we expect $\lambda_1({\bf \Xi})$ to be local. Both $\lambda_2(\hat{\bf \Xi}; x, y, w)$ and ${\bf S}_{\rm tot}^{\rm nloc}(\hat{\bf \Xi})$ are non-local. The former is also a function of the coordinates coming from integrating out composite operators. This may be ascertained from the following explicit forms of the two functions $\lambda_1({\bf \Xi})$ and $\lambda_2(\hat{\bf \Xi}; x, y, w)$:

{\footnotesize
\bg\label{rusiflow}
\lambda_1({\bf \Xi}) = \sqrt{-{\bf g}_{11}({\bf \Xi})}, ~~~~ 
\lambda_2(\hat{\bf \Xi}; x, y, w) = {\int \left[{\cal D} {\bf g}_{\rm 0A'}\right]~e^{-\widetilde{\bf S}_{\rm tot}(\hat{\bf \Xi}, {\bf g}_{\rm 0A'})}~{\bf \Psi}({\bf \Xi}, {\bf g}_{\rm 0A'}; x, y, w) \over\int \left[{\cal D} {\bf g}_{\rm 0A'}\right]~e^{-\widetilde{\bf S}_{\rm tot}(\hat{\bf \Xi}, {\bf g}_{\rm 0A'})}}, \nd}
where ${\bf \Psi}({\bf \Xi}, {\bf g}_{\rm 0A'}; x, y, w)$ is the composite operator\footnote{This may be easily derived from \eqref{lilykhel} after removing the piece $\lambda_1({\bf \Xi})$, but as we shall soon see, the explicit form for ${\bf \Psi}({\bf \Xi}, {\bf g}_{\rm 0A'}; x, y, w)$ will not be required at least to study the ``on-shell" EOMs.} that may be expressed as a series for more efficient computation. From \eqref{rusiflow} it should be clear why $\lambda_2(\hat{\bf \Xi}; x, y, w)$ is non-local. The non-local action ${\bf S}_{\rm tot}^{\rm nloc}(\hat{\bf \Xi})$ and the non-local energy-momentum tensor term ${\bf T}^{\rm nloc}_{\rm 0A'}({\bf \Xi})$  are now functions of $\hat{\bf \Xi}$ and ${\bf \Xi}$ respectively (the second relation in \eqref{bhmeymelismey} could be taken as the definition of the non-local energy-momentum tensor). The non-locality\footnote{It is easy to see how by integrating out the massless off-shell states we get a non-local action. For concreteness we shall integrate out a massless scalar field $\varphi(x)$ in $3+1$ dimensions in the following path-integral structure:
\bg\label{beatmik}
&&\int \left[{\cal D}\hat{\bf \Xi}\right]~{\rm exp}\left[{-{\bf S}(\hat{\bf \Xi})}\right] \int \left[{\cal D}\varphi\right] ~{\rm exp}\left[{-\int d^4 x ~\varphi(x) \square \varphi(x)-\int d^4 x ~{\rm C}(\hat{\bf \Xi}(x))\varphi^2(x)}\right] \nonumber\\
&& = \int \left[{\cal D}\hat{\bf \Xi}\right]~{\rm exp}\left[{-{\bf S}(\hat{\bf \Xi}) + \log\left(1 + \sum\limits_{p = 1}^\infty \prod\limits_{i = 1}^p \int d^4 x_i ~{1 \over \square_i}\cdot{\rm C}_i(\hat{\bf \Xi}(x_i))\right)}\right] \nonumber \nd
which clearly leads to a non-local action expressed completely in terms of
the degrees of freedom $\hat{\bf \Xi}$ (for concreteness, we can view the set of $\hat{\bf \Xi}$ to be the same that we had used above). Note that ${\rm C}_i(\hat{\bf \Xi}(x_i))$ are related to ${\rm C}(\hat{\bf \Xi}(x_i))$ upto multiplicative factors. The series in log may now be expanded and resummed as an exponential series in a way we discussed earlier. Since this is just a toy model, we will not spend more time elaborating on this here.}
of course comes from integrating out the massless off-shell state ${\bf g}_{\rm 0A'}$, and the above derivation provides a concrete proof of the non-local counter-terms that we introduced in \cite{desitter2, coherbeta, coherbeta2}. Combining \eqref{bhmeymelismey} with \eqref{violetgrm}, we get the following Schwinger-Dyson's equation:
\bg\label{butterflyeffect}
&& \int \big[{\cal D} \hat{\bf \Xi}\big] ~e^{-\overline{\bf S}_{\rm tot}(\hat{\bf \Xi})-{\bf S}_{\rm tot}^{\rm nloc}(\hat{\bf \Xi})} \lambda_3(\hat{\bf \Xi})\left(\hat{\bf R}_{\rm 0A'}({\bf \Xi}) - \hat{\bf T}_{\rm 0A'}({\bf \Xi}) -{\bf T}^{\rm nloc}_{\rm 0A'}({\bf \Xi}) + ... \right) 
\mathbb{D}^\dagger(\sigma) \mathbb{D}(\sigma) \nonumber\\
&& =  \Big\langle \lambda_3(\hat{\bf \Xi})\big(\hat{\bf R}_{\rm 0A'}({\bf \Xi}) - \hat{\bf T}_{\rm 0A'}({\bf \Xi}) -{\bf T}^{\rm nloc}_{\rm 0A'}({\bf \Xi}) + ...\big) \Big\rangle_\sigma = 0, \nd
where $\lambda_3(\hat{\bf \Xi}) \equiv \lambda_3(\hat{\bf \Xi}; x, y, z) =\lambda_1({\bf \Xi}) \lambda_2(\hat{\bf \Xi}; x, y, w)$. \eqref{butterflyeffect}
reproduces the LHS of \eqref{lilshank} with the $\lambda_3({\bf \Xi})$ factor, but it doesn't tell us yet how to get the RHS of \eqref{lilshank}. To get the RHS of \eqref{lilshank} from \eqref{butterflyeffect} we have to look into the dotted terms $-$ that involve the ghosts and the non-perturbative contributions $-$ more carefully. 
We shall however ignore the non-perturbative contributions for the time-being and concentrate on the ghosts only. The ghost action appears in ${\bf S}_{\rm tot}(\hat{\bf \Xi}, {\bf g}_{\rm 0A'})$, so we can make a finer splitting in \eqref{stanfordtag} as:
\bg\label{kiegrey}
\overline{\bf S}_{\rm tot}(\hat{\bf \Xi}) = \overline{\bf S}^{(1)}_{\rm tot}({\bf \Xi}) 
+ \overline{\bf S}_{\rm ghost}(\hat{\bf \Xi}), ~~\widetilde{\bf S}_{\rm tot}(\hat{\bf \Xi}, {\bf g}_{\rm 0A'}) = \widetilde{\bf S}^{(1)}_{\rm tot}({\bf \Xi, g}_{\rm 0A'}) 
+ \widetilde{\bf S}_{\rm ghost}(\hat{\bf \Xi}, {\bf g}_{\rm 0A'}), \nd
such that the non-local action ${\bf S}^{\rm nloc}_{\rm tot}(\hat{\bf \Xi})$ will have the appropriate contributions from the ghost sector. In fact some parts of the dotted terms in the second line of \eqref{violetgrm} will then be expressed only in terms of ${\delta\widetilde{\bf S}_{\rm ghost}(\hat{\bf \Xi}, {\bf g}_{\rm 0A'})\over \delta {\bf g}^{\rm 0A'}}$ such that we can define the following effective action from the ghost sector:

{\footnotesize
\bg\label{kiekal}
\int \left[{\cal D}{\bf g}_{\rm 0A'}\right] ~e^{-\widetilde{\bf S}^{(1)}_{\rm tot}({\bf \Xi, g}_{\rm 0A'}) - \widetilde{\bf S}_{\rm ghost}(\hat{\bf \Xi}, {\bf g}_{\rm 0A'})}
~  {\delta\widetilde{\bf S}_{\rm ghost}(\hat{\bf \Xi}, {\bf g}_{\rm 0A'})\over \delta {\bf g}^{\rm 0A'}} 
\equiv ~\lambda_3(\hat{\bf \Xi}) e^{-{\bf S}_{\rm tot}^{\rm nloc}(\hat{\bf \Xi})}~[{\bf T}^{\rm nloc}_{\rm ghost}]_{\rm 0A'}(\hat{\bf \Xi}), \nd}
which confirms the aforementioned terms containing the appropriate ghost contributions. (Note the absence of ${1\over 2}\log[-{\rm det}~{\bf g}_{11, d}({\bf \Xi, g}_{\rm 0A'})]$ in the first line of \eqref{kiekal}.) Some of the dotted terms in \eqref{butterflyeffect} now come from the non-local energy-momentum tensor $[{\bf T}^{\rm nloc}_{\rm ghost}]_{\rm 0A'}(\hat{\bf \Xi})$, implying that the local contributions from $\overline{\bf S}_{\rm ghost}(\hat{\bf \Xi})$ to \eqref{butterflyeffect} are present only through the total action 
$\overline{\bf S}_{\rm tot}(\hat{\bf \Xi}) + {\bf S}_{\rm tot}^{\rm nloc}(\hat{\bf \Xi})$. Plugging the contributions from \eqref{kiekal} into \eqref{butterflyeffect} gives us the following cleaner form from the Schwinger-Dyson equation \eqref{butterflyeffect}:

{\footnotesize
\bg\label{hardi}
&& \Big\langle \lambda_3(\hat{\bf \Xi})\big(\hat{\bf R}_{\rm 0A'}({\bf \Xi}) - \hat{\bf T}_{\rm 0A'}({\bf \Xi}) -{\rm T}^{\rm nloc}_{\rm 0A'}({\bf \Xi}) + [{\bf T}^{\rm nloc}_{\rm ghost}]_{\rm 0A'}(\hat{\bf \Xi}) + ..\big)\Big\rangle_\sigma\nonumber\\
&= & \lambda_3(\langle{\bf \Xi}\rangle_\sigma)\Big(  {\bf R}_{\rm 0A'}(\langle{\bf \Xi}\rangle_\sigma) - {\bf T}_{\rm 0A'} (\langle{\bf \Xi}\rangle_\sigma) - {\bf T}^{\rm nloc}_{\rm 0A'} (\langle{\bf \Xi}\rangle_\sigma) + ..\Big)+ \langle\lambda_3(\hat{\bf \Xi})[{\bf T}^{\rm nloc}_{\rm ghost}]_{\rm 0A'}(\hat{\bf \Xi})\rangle_\sigma\nonumber\\
&+& \sum_{\sigma' \ne \sigma} \left[\langle\lambda_3(\hat{\bf \Xi})\hat{\bf R}_{\rm 0A'}({\bf \Xi})\rangle_{(\sigma |\sigma')}  - \langle\lambda_3({\bf \Xi})\hat{\bf T}_{\rm 0A'}({\bf \Xi})\rangle_{(\sigma |\sigma')} - \langle\lambda_3({\bf \Xi})\hat{\bf T}^{\rm nloc}_{\rm 0A'}({\bf \Xi})\rangle_{(\sigma |\sigma')}+..\right] 
= 0, \nd}
where to motivate $\lambda_3(\langle{\bf \Xi}\rangle_\sigma)$ we first write this in the split form as: $\lambda_3(\hat{\bf \Xi}) \equiv \overline\lambda_3({\bf \Xi}) +\widetilde\lambda_3(\hat{\bf \Xi})$, and since $\langle{\bf \Upsilon}_g\rangle_\sigma = 0$, $\widetilde\lambda_3(\langle\hat{\bf \Xi}\rangle_\sigma) =\widetilde\lambda_3(\langle{\bf \Xi}\rangle_\sigma, \langle{\bf \Upsilon}_g\rangle_\sigma)=\widetilde\lambda_3(\langle{\bf \Xi}\rangle_\sigma)$. Together they reproduce $\lambda_3(\langle{\bf \Xi}\rangle_\sigma)$ in the second line of \eqref{hardi}, and the sum on the third line is over all the other off-shell Glauber-Sudarshan states to be defined momentarily (note $\lambda_3(\hat{\bf \Xi})$ and not just $\widetilde\lambda_3(\hat{\bf \Xi})$ therein) . Additionally,
any ghost dependence of ${\rm T}^{\rm nloc}_{\rm 0A'}(\hat{\bf \Xi})$ from the second relation in \eqref{bhmeymelismey}, {\it i.e.} 
${\rm T}^{\rm nloc}_{\rm 0A'}(\hat{\bf \Xi}) \to {\rm T}^{\rm nloc}_{\rm 0A'}({\bf \Xi}) + \widetilde{\rm T}^{\rm nloc}_{\rm 0A'}(\hat{\bf \Xi})$, the second term $\widetilde{\rm T}^{\rm nloc}_{\rm 0A'}(\hat{\bf \Xi})$ 
can be easily absorbed in the definition of $[{\bf T}^{\rm nloc}_{\rm ghost}]_{\rm 0A'}(\hat{\bf \Xi})$ keeping \eqref{hardi} unchanged. The other parameters are defined as follows:  $\sigma'$ denotes a set of the intermediate states 
$\{\vert\sigma'\rangle\} \ne \vert\sigma\rangle$ that would appear in the path-integral computation as shown in \cite{coherbeta2}, thus motivating the splitting in the equation \eqref{hardi}; and the dotted terms are the non-perturbative contributions (that we haven't determined yet). The expectation value of the set ${\bf \Xi}$, {\it i.e.} $\langle {\bf \Xi}\rangle_\sigma$, follows the definition laid out in \eqref{vkrieps} and therefore can be replaced by their warped-values. A natural split of \eqref{hardi} could easily be argued to be the following, since $\lambda_3(\langle{\bf \Xi}\rangle_\sigma) \ne 0$:

{\scriptsize
\bg\label{nikclau}
&& {\Su  {\bf R}_{\rm 0A'}(\langle{\bf \Xi}\rangle_\sigma) - {\bf T}_{\rm 0A'} (\langle{\bf \Xi}\rangle_\sigma) - {\bf T}^{\rm nloc}_{\rm 0A'} (\langle{\bf \Xi}\rangle_\sigma) + .. = 0}\\
&& {\Su \sum_{\sigma' \ne \sigma} \Big[\langle\lambda_3(\hat{\bf \Xi})\hat{\bf R}_{\rm 0A'}({\bf \Xi})\rangle_{(\sigma |\sigma')}  - \langle\lambda_3(\hat{\bf \Xi})\hat{\bf T}_{\rm 0A'}({\bf \Xi})\rangle_{(\sigma |\sigma')} - \langle\lambda_3(\hat{\bf \Xi})\hat{\bf T}^{\rm nloc}_{\rm 0A'}({\bf \Xi})\rangle_{(\sigma |\sigma')}\Big] + \langle\lambda_3(\hat{\bf \Xi})[{\bf T}^{\rm nloc}_{\rm ghost}]_{\rm 0A'}(\hat{\bf \Xi})\rangle_\sigma + ..
= 0}, \nonumber \nd}
which is {\it almost} exactly\footnote{We still need to derive the non-perturbative contributions, which are denoted by the dotted terms in both the equations of \eqref{nikclau}. They will be studied in section \ref{grace}.} as predicted in \eqref{lilshank}, and may be compared to an equivalent result derived in section 4.1.6 in the first reference of \cite{desitter2}, using a slightly different method. The first equation in \eqref{nikclau} is expressed completely in terms of the on-shell degrees of freedom and is therefore an ``on-shell" EOM {\it independent of the ghosts}\footnote{This is crucial and helps us to make sense of the {\it split} in \eqref{nikclau}: an EOM that determines the dynamics of the on-shell degrees of freedom cannot have  ghost contributions appearing explicitly through the energy-momentum tensor. Thus the ghost contributions should appear via the second equation in \eqref{nikclau} which involve intermediate Glauber-Sudarshan states $\vert\sigma'\rangle$.}, although a trace of the ghost contribution still remains in the computation of $\langle{\bf \Xi}\rangle_\sigma$ as a path-integral (for example as in \eqref{ebrowning0}). The ghosts appear more explicitly in the second equation in \eqref{nikclau}, and should be viewed as the defining equation for the non-local energy-momentum tensor from the ghosts. In fact, since we expect the ghosts to show up in the loops, or here, in the intermediate states $\{\vert\sigma'\rangle\}$, we can make the following splitting:

{\footnotesize
\bg\label{neela}
\langle\lambda_3(\hat{\bf \Xi})[{\bf T}^{\rm nloc}_{\rm ghost}]_{\rm 0A'}(\hat{\bf \Xi})\rangle_\sigma 
=\lambda_3(\langle{\bf \Xi}\rangle_\sigma)[{\bf T}^{\rm nloc}_{\rm ghost}]_{\rm 0A'}(\langle{\bf \Xi}\rangle_\sigma)
+  \sum_{\sigma' \ne \sigma} \langle\lambda_3(\hat{\bf \Xi})[{\bf T}^{\rm nloc}_{\rm ghost}]_{\rm 0A'}(\hat{\bf \Xi})\rangle_{(\sigma|\sigma')}, \nd}
and keep $[{\bf T}^{\rm nloc}_{\rm ghost}]_{\rm 0A'}(\langle{\bf \Xi}\rangle_\sigma) = 0$, so that the on-shell EOM does not get effected. This way the second equation in \eqref{nikclau} will be determined completely in terms of the intermediate states 
$\{\vert\sigma'\rangle\}$, and the first equation in \eqref{nikclau} continues to provide the ``on-shell" behavior.

On the other hand, for the on-shell degrees of freedom like $\langle{\bf g}_{\rm AB}\rangle_\sigma$ from \eqref{angwhiteT} which are the ones from the set ${\bf \Xi}$, the story is almost similar but with some minor differences. Making a similar split between barred and tilde variables, the Schwinger-Dyson's equation now gives us the following:

{\footnotesize
\bg\label{rposner}
0 &= & \int \left[{\cal D} \hat{\bf \Xi}\right] ~e^{-\overline{\bf S}_{\rm tot}(\hat{\bf \Xi})} \left(\hat{\bf R}_{\rm AB}({\bf \Xi}) - {1\over 2} {\bf g}_{\rm AB}\hat{\bf R}({\bf \Xi})- \hat{\bf T}_{\rm AB}({\bf \Xi})\right) 
\mathbb{D}^\dagger(\sigma) \mathbb{D}(\sigma)
\int \left[{\cal D} {\bf g}_{\rm 0A'}\right] ~e^{-\widetilde{\bf S}^t_{\rm tot}(\hat{\bf \Xi}, {\bf g}_{\rm 0A'})} \nonumber\\
&+& \int \left[{\cal D}\hat{\bf \Xi}\right]  ~e^{-\overline{\bf S}_{\rm tot}(\hat{\bf \Xi})}
\mathbb{D}^\dagger(\sigma) \mathbb{D}(\sigma)
\int \left[{\cal D}{\bf g}_{\rm 0A'}\right] ~e^{-\widetilde{\bf S}^t_{\rm tot}(\hat{\bf \Xi}, {\bf g}_{\rm 0A'})} 
\left(\widetilde{\bf G}_{\rm AB}({\bf \Xi, g}_{\rm 0A'}) - \widetilde{\bf T}_{\rm AB}({\bf \Xi, g}_{\rm 0A'}) + ...\right) \nonumber\\
\nd}
where $\widetilde{\bf G}_{\rm AB}({\bf \Xi, g}_{\rm 0A'})$ is the Einstein tensor; the action entering the integral over the off-shell state is $\widetilde{\bf S}^t_{\rm tot}({\bf \Xi, g}_{\rm 0A'}, {\bf \Upsilon}_g) $ $\equiv$
$\widetilde {\bf S}_{\rm tot}({\bf \Xi, g}_{\rm 0A'}, {\bf \Upsilon}_g)$ $-{1\over 2}\log[-{\rm det}~{\bf g}_{11, d}({\bf \Xi, g}_{\rm 0A'})]$; and the dotted terms are again the ghosts and the non-perturbative contributions. There is however a difference now: a part of the path-integral in the second line may be expressed as a total derivative in the following way:

{\footnotesize
\bg\label{milenthom}
\int \left[{\cal D}{\bf g}_{\rm 0A'}\right] ~e^{-\widetilde{\bf S}^t_{\rm tot}(\hat{\bf \Xi}, {\bf g}_{\rm 0A'})} 
\left(\widetilde{\bf G}_{\rm AB}(\hat{\bf \Xi}, {\bf g}_{\rm 0A'}) - 
\widetilde{\bf T}_{\rm AB}(\hat{\bf \Xi}, {\bf g}_{\rm 0A'})\right) = -{\partial\over \partial {\bf g}_{\rm AB}}\int 
\left[{\cal D}{\bf g}_{\rm 0A'}\right] ~e^{-\widetilde{\bf S}_{\rm tot}(\hat{\bf \Xi}, {\bf g}_{\rm 0A'})}, \nd}
where note the differences from the second line in \eqref{rposner}: we have expressed the Einstein plus the energy-momentum tensors; and the action (with the off-shell state) using respectively $\hat{\bf \Xi} = ({\bf \Xi, \Upsilon}_g)$ (instead of ${\bf \Xi}$) on LHS and 
$\widetilde{\bf S}_{\rm tot}(\hat{\bf \Xi}, {\bf g}_{\rm 0A'})$ (instead of $\widetilde{\bf S}^t_{\rm tot}(\hat{\bf \Xi}, {\bf g}_{\rm 0A'})$) on RHS by taking contributions from the dotted terms involving ghosts. Such a relation, like the one in \eqref{milenthom}, does not work for the second line in \eqref{violetgrm}, because the Einstein tensors (as well as the corresponding energy-momentum tensors) in the first and the second lines of \eqref{violetgrm} both appeared\footnote{Note however the caveat in \eqref{smitvanand}.} from $\widetilde{\bf S}_{\rm tot}(\hat{\bf \Xi}, {\bf g}_{\rm 0A'})$, whereas the corresponding tensors in \eqref{rposner} appear from $\overline{\bf S}_{\rm tot}(\hat{\bf \Xi})$ and $\widetilde{\bf S}_{\rm tot}(\hat{\bf \Xi}, {\bf g}_{\rm 0A'})$ respectively. Now using the result from the first equation in \eqref{bhmeymelismey}, it is easy to infer from the fact that $\lambda_3(\langle{\bf \Xi}\rangle_\sigma) \ne 0$:
\bg\label{coffiemarieso}
&& {\Su {\bf R}_{\rm AB}(\langle{\bf \Xi}\rangle_\sigma)-{1\over 2}{\bf g}_{\rm AB}{\bf R}(\langle{\bf \Xi}\rangle_\sigma)- {\bf T}_{\rm AB} (\langle{\bf \Xi}\rangle_\sigma) - {\bf T}^{\rm nloc}_{\rm AB} (\langle{\bf \Xi}\rangle_\sigma) + .. = 0} \\
&& {\Su \sum_{\sigma' \ne \sigma} \Big[\langle\lambda_3(\hat{\bf \Xi})\hat{\bf R}_{\rm AB}({\bf \Xi})\rangle_{(\sigma |\sigma')} -{1\over 2} \langle  \lambda_3(\hat{\bf \Xi}){\bf g}_{\rm AB}\hat{\bf R}({\bf \Xi})\rangle_{(\sigma |\sigma')}}\nonumber\\
&& {\Su - \langle\lambda_3(\hat{\bf \Xi})\hat{\bf T}_{\rm AB}({\bf \Xi})\rangle_{(\sigma |\sigma')} 
- \langle\lambda_3(\hat{\bf \Xi})\hat{\bf T}^{\rm nloc}_{\rm AB}({\bf \Xi})\rangle_{(\sigma |\sigma')} - \langle\lambda_3(\hat{\bf \Xi})[{\bf T}^{\rm nloc}_{\rm ghost}]_{\rm AB}(\hat{\bf \Xi})\rangle\Big] = 0}, \nonumber \nd
where the set $\{\vert\sigma'\rangle\}$ are the intermediate Glauber-Sudarshan states, and the dotted terms are the non-perturbative contributions. Looking at the first set of equations in \eqref{nikclau} and \eqref{coffiemarieso} for the off-shell and the on-shell degrees of freedom respectively, we find that they are {\it exactly} the EOMs advocated in \cite{desitter2, coherbeta, coherbeta2} but now for the {\it emergent degrees of freedom}! Similar story will unfold for the flux and the fermionic degrees of freedom. For the fermions, this was discussed briefly in section \ref{servant}, and the flux case will be studied in section \ref{sec7s}. The action governing the expectation values is not $\hat{\bf S}_{\rm tot}(\hat{\bf \Xi})$ from \eqref{tomiraja} yet, rather it is now:
\bg\label{tomipizzal}
\overline{\bf S}_{\rm tot}(\hat{\bf \Xi}) + {\bf S}^{\rm nloc}_{\rm tot}(\hat{\bf \Xi}) = {\bf S}_{\rm kin}({\bf \Xi}) + {\bf S}_{\rm pert}({\bf \Xi}) + {\bf S}^{\rm nloc}_{\rm pert}({\bf \Xi}) + \hat{\bf S}_{\rm ghost}(\hat{\bf \Xi}), \nd
with only {\it perturbative} interactions. Using the concepts of Borel-\'Ecalle resummations and the summation techniques from section \ref{servant}, we can replace ${\bf S}_{\rm pert}({\bf \Xi})$ and $ {\bf S}^{\rm nloc}_{\rm pert}({\bf \Xi})$ by their non-perturbative and the resummed values respectively. This will convert the action to $\hat{\bf S}_{\rm tot}(\hat{\bf \Xi})$ from \eqref{tomiraja}, which can now be used to compute the expectation values. In fact such a procedure will remove the dotted terms in \eqref{nikclau} and \eqref{coffiemarieso}, and replace the energy-momentum tensors by their full non-perturbative values. Applying further the renormalization techniques from \cite{wdwpaper} will convert $\hat{\bf S}_{\rm tot}(\hat{\bf \Xi})$ to $\check{\bf S}_{\rm tot}(\hat{\bf \Xi})$, although we will never use the latter here. 

\subsection{Comparing gauging with the emergent metric conditions \eqref{angwhiteT} \label{lenamyfrench}}

\begin{table}[h]  
 \begin{center} 
\resizebox{\columnwidth}{!}{%
 \renewcommand{\arraystretch}{2.5}
\begin{tabular}{|c||c||c||c||c||c|}\hline 
\textbf{Block} & \textbf{Components} & \textbf{Count} & \textbf{Point} & \textbf{Patch (generic)} & \textbf{Global (conditions)}\\ \hline
Time--time                 & ${\bf g}_{00}$                 & $1$  & --- & --- & --- \\ \hline
Spatial ${\bf R}^2$            & ${\bf g}_{ij}$ (sym.)          & $3$  & --- & --- & --- \\ \hline
$\mathcal{M}_4$            & ${\bf g}_{mn}$ (sym.)          & $10$ & --- & --- & --- \\ \hline
$\mathcal{M}_2$            & ${\bf g}_{\alpha\beta}$ (sym.) & $3$  & --- & --- & --- \\ \hline
Fiber ${\mathbb{T}^2\over {\cal G}}$                     & ${\bf g}_{ab}$ (sym.)          & $3$  & --- & --- & \parbox[t]{6.8cm}{Reduced (not zero) by ${\rm SL}(2,\mathbb{Z})$ and torus diffeos to area \& complex-structure moduli} \\ \hline
\textbf{Time--space }     & ${\bf g}_{0i},\, {\bf g}_{0m},\, {\bf g}_{0\alpha}$ & $\mathbf{8}$ &
$\checkmark$ & $\checkmark$ \;\parbox[t]{5.5cm}{(synchronous / Gaussian normal gauge near a regular time slice)} &
\parbox[t]{6.4cm}{$\checkmark$ if a \emph{global} synchronous foliation exists (often obstructed by caustics/horizons)} \\ \hline
\textbf{Spatial cross }   & ${\bf g}_{im},\, {\bf g}_{i\alpha},\, {\bf g}_{m\alpha}$ & $\mathbf{20}$ &
$\checkmark$ &
$\times$ \;\parbox[t]{5.5cm}{(in general cannot all be removed)} &
\parbox[t]{6.4cm}{$\checkmark$ iff the $9$D base is an {orthogonal (possibly multiply warped) product} aligned with ${\bf R}^2,\mathcal{M}_4,\mathcal{M}_2$ (distributions orthogonal \& integrable)} \\ \hline
\textbf{Base--fiber}       & ${\bf g}_{0a},\, {\bf g}_{ia},\, {\bf g}_{ma},\, {\bf g}_{\alpha a}$ & $\mathbf{18}$ &
$\checkmark$ &
\parbox[t]{5.5cm}{Only $2$ indep.\ combinations (use $y^a\!\to y^a+\xi^a(x)$); e.g.\ set ${\bf g}_{0a}=0$} &
\parbox[t]{6.4cm}{$\checkmark$ all $18$ iff the $\mathbb{T}^2$ bundle is trivial \emph{and} KK field strengths/holonomies vanish: ${\rm F}^a{}_{\rm MN}=0$ and Wilson lines trivial} \\ \hline
\end{tabular}}
\renewcommand{\arraystretch}{1}
\end{center}
 \caption[]{\Su Comparing \eqref{angwhiteT} with gauging of the metric \eqref{virgbhalotag}. ``Point'' means at a single spacetime point (local inertial frame); ``Patch'' means on a generic open neighborhood (no special geometric/topological assumptions); ``Global'' lists the condition under which the zeroing can be done everywhere. At a {\it point} (Riemann normal coordinates in full $11$D) all off-diagonals vanish and the diagonal blocks take Minkowski/Euclidean values; this normalizes the metric at that point only, not on a region. On a {\it generic patch}, we can always kill the $8$ base time--space mixings via synchronous gauge and use the two fiber diffeomorphisms to set, say, $g_{0a}=0$; that is $8+2=10$ components guaranteed. More requires geometric conditions (orthogonal product for spatial cross blocks; trivial KK data for base--fiber).
{\it Globally}, the best-case scenario (block-diagonal in the torus) needs a trivial $\mathbb{T}^2$ fibration with ${\rm F}^a=d{\rm A}^a=0$ and trivial Wilson lines. A global synchronous slicing is an additional, independent requirement to set all ${\bf g}_{0*}$ to zero everywhere.}
\label{indigtagoutside}
 \end{table} 

Let us take a short detour to compare the result developed in section \ref{sec6.1} $-$ namely with the emergent metric components \eqref{angwhiteT} where the off-diagonal components naturally vanish because of our choice of $\mathbb{D}(\sigma, {\bf \Xi})$ $-$ with a possible eleven-dimensional metric configurations with {\it all} components non-zero and ask how many of the metric components can be {\it gauged} to zero for the latter configuration. We can generate the latter configuration by choosing another displacement operator $\mathbb{D}'(\sigma', {\bf \Xi}')$ which span all the allowed degrees of freedom (we denote them by ${\bf \Xi}'$ and the corresponding Glauber-Sudarshan states by $\vert\sigma'\rangle$).
In other words we are looking at a metric configuration of the form:
\bg\label{virgbhalotag}
ds^2 = \tilde{\bf g}_{\rm MN}(y) dy^{\rm M} dy^{\rm N} + \tilde{\bf g}_{ab}(y) \Big(dw^a + {\rm A}^a_{\rm M}(y) dy^{\rm M}\Big) \Big(dw^b + {\rm A}^b_{\rm N}(y) dy^{\rm N}\Big), \nd
where $({\rm M, N}) \in (0, i, m, \alpha)$ label the 9 ``base'' coordinates and $a, b = 1, 2$ the eleven-dimensional torus fiber. The difference now is that the compact internal manifolds $\mathcal M_4, \mathcal M_2$ and ${\mathbb{T}^2\over \mathcal G}$ are non-trivially fibered over $2+1$ dimensional non-compact spacetime ${\bf R}^{2, 1}$.
The mixed components  are precisely the two KK gauge fields ${\rm A}^a_{\rm M}(y)$ for the ${\rm U(1) \times U(1)}$ isometries of the fiber torus. It should also be clear from our notation that: 
\bg\label{emininvirg}
&& \langle {\bf g}_{\rm MN}(y)\rangle_{\sigma'} \equiv {\bf g}_{\rm MN}(y) = \tilde{\bf g}_{\rm MN}(y) + \tilde{\bf g}_{ab}(y){\rm A}^a_{\rm M} {\rm A}^b_{\rm N}(y) \nonumber\\
&& \langle{\bf g}_{ab}(y)\rangle_{\sigma'} \equiv {\bf g}_{ab}(y) = \tilde{\bf g}_{ab}(y), ~~\langle{\bf g}_{a{\rm M}}(y)\rangle_{\sigma'} \equiv {\bf g}_{a{\rm M}} = \tilde{\bf g}_{ab}(y) {\rm A}^b_{\rm M}(y), \nd
from where it is easy to see that there are 
$1+3+10+3+3= {\bf 20}$ diagonal block components that are not gaugeable to zero except at a single point where they can be trivialized. If we don't include the fiber, there are in total ${\bf 8}$ time-space components of the form ${\bf g}_{0{\rm M'}}(y)$ and ${\bf 20}$ spatial-cross components of the form ${\bf g}_{i{\rm M}''}(y)$ where ${\rm M}' = (i, m, \alpha)$ and ${\rm M}'' = (m, \alpha)$. Including fiber, there are ${\bf 18}$ fiber-base components of the form ${\bf g}_{a{\rm M}}(y)$. Together there are $8 + 20 + 18 = {\bf 46}$ off-diagonal components from a total of ${\bf 66}$ metric components.

It is easy to see that at a single point (Riemann normal coordinates in full eleven-dimension) we can make all ${\bf 46}$ off–diagonal components vanish. The  
diagonal ones become Minkowski/Euclidean values at the point (but are not zero). However on a generic open patch (with no special geometric/topological assumptions), the time–space mixing components on the nine-dimensional base assuming synchronous/Gaussian normal gauge around a smooth time slice, we can make the set of ${\bf 8}$ components to vanish, {\it i.e.}:
\bg\label{peribindu1}
{\bf g}_{0i}(y)= {\bf g}_{0m}(y)={\bf g}_{0\alpha}(y)=0,
\nd
however this breaks at caustics or at horizons. For the base-fiber mixing components ${\bf g}_{a{\rm M}}(y)$ there are only two fiber reparametrization gauges $\xi^a(x)$ (one per ${\rm U(1)}$ of the fiber torus), so at most two independent combinations of the ${\bf 18}$ components can be set to zero everywhere on the patch. A common choice (temporal KK gauge) is:
\bg\label{peribindu2}
{\bf g}_{0a}(y) =0 \quad (a=1,2),
\nd
thus implying ${\bf 2}$ components that may be set to zero as mentioned above. The remaining spatial-cross components ${\bf g}_{im}(y), {\bf g}_{i\alpha}(y)$ and ${\bf g}_{m\alpha}(y)$ cannot all be eliminated on a patch in general. They vanish on a patch iff the nine-dimensional base metric admits an orthogonal (possibly warped) product aligned with ${\bf R}^2$, $\mathcal{M}_4$, and $\mathcal{M}_2$ implying integrable or mutually orthogonal distributions. Therefore on a general patch we can kill at least $8 + 2 = {\bf 10}$ of the off-diagonal components and maybe more if the base geometry is an orthogonal product.

Globally the story is slightly more non-trivial. If the ${\mathbb{T}^2 \over \mathcal G}$ fibration over the base is topologically trivial and the KK field strengths/holonomies vanish, {\it i.e.}:
\bg\label{peribindu3}
{\rm F}^a_{\rm MN}=0, \nd
and all Wilson lines are trivial,
then we can set all ${\bf 18}$ base–fiber off–diagonal components to zero globally, {\it i.e.}
${\bf g}_{a{\rm M}}=0$. In general with a non-trivial ${\rm A}^a_{\rm M}(y)$ the condition \eqref{peribindu3} is hard to satisfy, so generically it is difficult to kill the base-fiber components. On the other hand, 
time–space mixing components ${\bf g}_{0i}, {\bf g}_{0m}$ and  ${\bf g}_{0\alpha}$ can be set to zero globally only if the spacetime admits a global Gaussian normal foliation (global time function with non-intersecting normal geodesics). For four-dimensional de Sitter this is a bit more subtle. On one hand, the congruence of co-moving world-lines (constant angular coordinates on ${\bf S}^3$ from \eqref{peribindu00}) is
hypersurface-orthogonal (vorticity $\omega=0$) and shear-free ($\sigma=0$). Consequently
Raychaudhuri’s equation reduces to:
\bg\label{peribindu6}
\frac{d\theta}{d\tau}
= -\frac{1}{3}\,\theta^{2} - \sigma^{2} + \omega^{2} - {\bf R}_{\mu\nu}\,u^{\mu}u^{\nu} = -{1\over 3} \theta^2 -{\bf R}_{\mu\nu}\,u^{\mu}u^{\nu},
\nd
where we expect this evolution to never drive $\theta$ to $-\infty$ in finite proper time, so the orthogonal geodesics do not focus to form caustics. Geometrically the $\partial_t$ flow lines never cross; each co-moving observer sits at fixed $(\chi,\theta,\phi)$ on the ${\bf S}^3$
for the global de Sitter \eqref{peribindu00}. On the other hand, the de Sitter space that we study here are {\it transient} with a finite temporal domain $-{1\over \sqrt\Lambda} < t < 0$ in flat-slicing, so the
evolution of $\theta$ by Raychaudhuri's equation \eqref{peribindu6} may become non-trivial\footnote{For example for a four-dimensional transient spacetime decaying by Coleman--De Luccia bubbles (first-order 
transition), 
a single, global synchronous foliation for the entire spacetime typically fails because of the following reasons.
Normal geodesics shot orthogonally from a putative ``initial'' slice can intersect at bubble walls or within the interior (caustics), breaking Gaussian normal coordinates. Or, different bubbles nucleated at different times lead to multiple intersecting congruences implying no single time function $t$ with lapse and shift:
$({\rm N}=1, {\rm N}^{i}=0)$ that covers everything smoothly. In practice, one could use \emph{patchwise} synchronous charts (separating the exterior region, bubble-interior open slices, etc.) and match across the wall but that may not always guarantee global foliations.}. Thus the transient de Sitter spacetime that we want to study using \eqref{virgbhalotag} {\it may not} allow the aforementioned 
globally, so this is not guaranteed. 

Finally, the spatial cross–blocks components ${\bf g}_{im}, {\bf g}_{i\alpha}$, and ${\bf g}_{m\alpha}$ vanish globally iff the nine-dimensional base metric is a (possibly multiply warped) orthogonal product aligned with the chosen factorization. Otherwise, they cannot all be gauged away. Therefore globally, we can kill at most ${\bf 2}$ components ${\bf g}_{0a}=0$ from the two fiber gauges, and 
potentially up to ${\bf 18}$ components ${\bf g}_{a{\rm M}}$ if KK data is trivial (which is of course unlikely in our case).
Plus potentially ${\bf 8}$ components of the form ${\bf g}_{0\ast}$ if a global synchronous slicing exists (which is again unlikely for the transient background that we want to study). The remaining
${\bf 20}$ spatial cross terms, namely ${\bf g}_{im}, {\bf g}_{i\alpha}$ and ${\bf g}_{m\alpha}$ are zero globally only if the base is an orthogonal product.  All these results are summarized in {\bf Table \ref{indigtagoutside}}.

Our above discussion suggests that taking an alternative configuration like \eqref{virgbhalotag} and then trying to resort to the requisite block-diagonal form by gauging doesn't help. In fact the transient nature of our background forbids any kind of simplifications that could result from gauging various metric components. The best way to get rid of the off-diagonal metric components for the emergent background is to choose the displacement operator in the way described earlier. The off-diagonal components appearing in the loops are then integrated away as in section \ref{sec6.1} leading us to the non-local interactions.

\subsection{Further analysis of the energy-momentum tensors and the EOMs \label{sammipaaador}}

The analysis of the EOMs for the on-shell and the off-shell metric components have shown the importance of the proposal that the de-Sitter or the quasi de-Sitter states can only be consistently studied using the Glauber-Sudarshan states. In fact one may easily see that the cross-term EOMs in \eqref{nikclau} can {\it only} be derived using the path-integral approach over the excited state $\vert\sigma\rangle$ discussed above and clearly {\it not from any vacuum configuration}. For the on-shell degrees of freedom, the story may be summarized by the following sequence:
\bg\label{duilotine}
{\bf S}_{\rm tot}(\hat{\bf \Xi}, {\bf g}_{\rm 0A'}, ..)~ \xrightarrow{\rm integ}~ \overline{\bf S}_{\rm tot}(\hat{\bf \Xi}) + {\bf S}^{\rm nloc}_{\rm tot}(\hat{\bf \Xi})~ \xrightarrow{\rm resum} ~\hat{\bf S}_{\rm tot}(\hat{\bf \Xi})~ \xrightarrow{\rm renorm} ~\check{\bf S}_{\rm tot}(\hat{\bf \Xi}), \nd
where ${\bf g}_{\rm 0A'}$ and the dotted terms form the set of off-shell metric components (see \eqref{tgrlbahlu}); and $\hat{\bf \Xi} = ({\bf \Xi, \Upsilon}_g)$. Eventually $-$ after we isolate the ghost EOMs as in second equations in \eqref{nikclau} and \eqref{coffiemarieso} and assume $\check{\bf S}_{\rm tot}(\hat{\bf \Xi}) \approx \hat{\bf S}_{\rm tot}(\hat{\bf \Xi})$ $-$ what decides the emergent EOMs for the on-shell degrees of freedom ${\bf \Xi}$ is the equation ${\delta\hat{\bf S}_{\rm tot}(\langle{\bf \Xi}\rangle_\sigma) \over \delta \langle{\bf \Xi}\rangle_\sigma} = 0$ leading us to the first equation in \eqref{coffiemarieso} but now with full non-perturbative completion.

However two questions still remain. {\Su One}, what is the precise form of $\hat{\bf S}_{\rm tot}(\langle{\bf \Xi}\rangle_\sigma)$? And {\Su two}, how do we determine the $g_s$ scalings of the energy-momentum tensors ${\bf T}_{\rm 0A'}(\langle{\bf \Xi}\rangle_\sigma)$ and ${\bf T}^{\rm nloc}_{\rm 0A'}(\langle{\bf \Xi}\rangle_\sigma)$?  The latter is necessary to quantify the quantum contributions, both for the perturbative and the non-perturbative sectors, in the cross-term EOMs \eqref{nikclau}. Additionally, we have cheated at two places: \textcolor{blue}{one}, going from \eqref{hardi} to \eqref{nikclau}\footnote{And also in the first equality of \eqref{chelmey}.} we have taken the following resolution of the identity:
\bg\label{capquebec}
{\bf 1} = \sum_{\sigma, \sigma^\ast} {\vert\sigma\rangle\langle\sigma\vert\over \langle\sigma\vert\sigma\rangle} ~ \to ~ \int d^2\sigma ~{\vert\sigma\rangle\langle\sigma\vert\over \langle\sigma\vert\sigma\rangle}\nd
which is not exactly true \cite{joydeep}; and \textcolor{blue}{two}, we have ignored the contribution ${\delta\over \delta {\bf g}^{\rm AB}}~{\rm log}\left(\mathbb{D}^\dagger(\sigma)\mathbb{D}(\sigma)\right)$ to \eqref{rposner}. While such contributions are absent in \eqref{violetgrm} because of the dependence of $\mathbb{D}(\sigma)$ only on the on-shell variables ${\bf \Xi}$, they are present for the on-shell EOMs. Inserting them back, how do they change the outcome? 

Let us start by answering the second question, namely the $g_s$ scalings of the energy-momentum tensors. In fact this has already been discussed in the first reference of \cite{desitter2}, but here we will re-derive this in a slightly different way. We will discuss the answer to the first question in the next section.

At the level of the excited state there are no off-shell degrees of freedom because of \eqref{angwhiteT}, so clearly an interaction of the form ${\bf g}_{\rm 0A'} {\bf T}^{\rm 0A'}$ vanish despite the possibility of a non-zero ${\bf T}_{\rm 0A'}(\langle{\bf \Xi}\rangle_\sigma)$. How can we then determine the $\bar{g}_s$ scaling of the cross-term energy-momentum tensor? The answer comes rather miraculously from the analysis that we presented in section \ref{servant}, namely that we can go for more generalized form of the metric as given in \eqref{akash}. For the off-shell degrees of freedom we are in fact looking at an interaction of the form:
\bg\label{omaolivia}
\hat{\bf g}_{\rm A'B'} {\bf T}^{\rm A'B'}(\langle{\bf \Xi}\rangle_\sigma)
= c_2\Gamma_{\rm A'B'}{\bf T}^{\rm A'B'}(\langle{\bf \Xi}\rangle_\sigma)+ c_3(\bar\lambda {\cal O}\lambda)_{\rm A'B'} {\bf T}^{\rm A'B'}(\langle{\bf \Xi}\rangle_\sigma), \nd
where $({\rm A', B'})$ span the coordinate range specified in the first line of \eqref{angwhiteT}. The interaction \eqref{omaolivia} clearly appears from \eqref{fahingsha10}, and since from \eqref{roseQ2} the $\bar{g}_s$ scalings of the Rarita-Schwinger fermions are sub-dominant, the $\bar{g}_s$ scaling of ${\bf T}^{\rm A'B'}(\langle{\bf \Xi}\rangle_\sigma)$ can be easily inferred from the first term in \eqref{omaolivia}. A list of the $\bar{g}_s$ scalings of the cross-term energy-momentum tensors appear in {\bf Table \ref{omabiratag}}. The answer presented here matches well with the ones given in the first reference of \cite{desitter2} despite the fact that we are now using completely different techniques. Moreover, {\bf Table \ref{omabiratag}} captures an exhaustive list of the cross-term energy-momentum tensors.

\begin{table}[tb]  
 \begin{center}
\resizebox{\columnwidth}{!}{%
\renewcommand{\arraystretch}{3.5}
\begin{tabular}{|c|c||c|c|}\hline Energy-Momentum Tensors & ${g_s\over {\rm HH}_o}$ Scalings & Energy-Momentum Tensors & ${g_s\over {\rm HH}_o}$ scalings \\ \hline\hline
${\bf T}_{0i}$ & $\theta_{nl}(t) - {8\over 3} + \hat\zeta_e(t)$ & ${\bf T}_{i\alpha}$ & $\theta_{nl}(t) - {5\over 3} + {\hat\zeta_e(t)\over 2} + {1\over 2}(\hat\alpha_e(t), \hat\beta_e(t))$ \\ \hline
${\bf T}_{0m}$ & $\theta_{nl}(t) - {5\over 3} + {\hat\zeta_e(t)\over 2} + {\hat\sigma_e(t)\over 2}$ & ${\bf T}_{ia}$ & $\theta_{nl}(t) - {2\over 3} + {\hat\zeta_e(t)\over 2} + {1\over 2}(0, \hat\eta_e(t))$ \\ \hline
${\bf T}_{0\alpha}$ & $\theta_{nl}(t) - {5\over 3} + {\hat\zeta_e(t)\over 2} + {1\over 2}(\hat\alpha_e(t), \hat\beta_e(t))$ & ${\bf T}_{m\alpha}$ & $\theta_{nl}(t) - {2\over 3} + {\hat\sigma_e(t)\over 2} + {1\over 2}(\hat\alpha_e(t), \hat\beta_e(t))$ \\ \hline
${\bf T}_{0a}$ & $\theta_{nl}(t) - {2\over 3} + {\hat\zeta_e(t)\over 2} + {1\over 2}(0, \hat\eta_e(t))$ & ${\bf T}_{ma}$ & $\theta_{nl}(t) + {1\over 3} + {\hat\sigma_e(t)\over 2} +{1\over 2}(0, \hat\eta_e(t))$ \\ \hline
${\bf T}_{im}$ & $\theta_{nl}(t) - {5\over 3} + {\hat\zeta_e(t)\over 2} + {\hat\sigma_e(t)\over 2}$ & ${\bf T}_{\alpha a}$ & $\theta_{nl}(t) + {1\over 3} +{1\over 2}(\hat\alpha_e(t), \hat\beta_e(t))+{1\over 2}(0, \hat\eta_e(t))$ \\ \hline
\end{tabular}}
\renewcommand{\arraystretch}{1}
\end{center}
 \caption[]{\Su ${g_s\over {\rm H}(y){\rm H}_o({\bf x})}$ scalings of the cross-term energy-momentum tensors with $\theta_{nl}(t)$ being the scalings of the quantum series \eqref{fahingsha10}. The vielbeins used in the computation of the scalings are the effective vielbeins appearing in \eqref{maecomics}.} 
  \label{omabiratag}
 \end{table}

 The other two questions, namely the modification of \eqref{capquebec} and the insertion of ${\delta\over \delta {\bf g}^{\rm AB}}~{\rm log}\left(\mathbb{D}^\dagger(\sigma)\mathbb{D}(\sigma)\right)$, are easy to answer. The first one has already been discussed in section 6.1 of \cite{joydeep} with the modification to the resolution of identity in the presence of the intermediate Glauber-Sudarshan states is presented in eq. (6.11) of \cite{joydeep}. Interested readers may want to look up the derivation presented therein. 

 For the second one, we may easily see that insertion of ${\delta\over \delta {\bf g}^{\rm AB}}~{\rm log}\left(\mathbb{D}^\dagger(\sigma)\mathbb{D}(\sigma)\right)$ will only effect the ghost EOMs, namely the second set of equations in \eqref{nikclau} and \eqref{coffiemarieso} respectively. Some aspect of the derivation has appeared in \cite{coherbeta} and \cite{coherbeta2}, so we will not dwell on this any further here\footnote{Plus the effects are only on the ghost EOMs which we won't study here.}. Interested readers may again look up the aforementioned references. 
 
\subsection{The trans-series form of the Schwinger-Dyson equations \label{trans1}}

With all the discussion of the EOMs and the energy momentum tensors, we are now ready to answer the first question raised in the previous section, namely what is the precise form for the effective action 
$\hat{\bf S}_{\rm tot}(\hat{\bf \Xi})$? This, in some sense, has been discussed in \cite{joydeep}: the action may be expressed as a sum over the instanton saddles along with their fluctuation determinants thus forming a resurgent transseries. The new thing here would be to express it in the form where all the instanton corrections are clearly laid out. Recall that there are at least two kinds of instantons at play here: the BBS \cite{bbs} and the KKLT \cite{kklt} instantons, in addition to other possibilities. The effective action \eqref{tomiraja} can be expressed as sum over all the instantons saddles and their fluctuation determinants in the following way\footnote{Note that the action is defined over a given solitonic vacuum which is the Minkowski minimum for us. Generically for other solitons or solitonic vacua the story goes in the following way. Pick the topological sector that contains the soliton, treat the soliton configuration as the {\it vacuum} of that sector, and build a trans-series whose 0-instanton piece is the perturbative expansion about the soliton, while nonzero exponential sectors encode instanton-type processes in that solitonic background (kink–antikink tunneling, soliton decay, etc.). Thus around any M-theory branes $-$ which are solitonic solutions $-$ we can construct the action as a trans-series there. The story is clearly very involved, so to avoid such complications, we will not worry too much about the branes for the time-being.} :
\bg\label{kimkarol}
&& \hat{\bf S}_{\rm tot}({\bf \Xi}) \equiv \hat{\bf S}_{\rm tot}(\hat{\bf \Xi})\big\vert_{{\bf \Upsilon}_g= 0} = {\bf S}_{\rm kin}({\bf \Xi}) + {\bf S}_{\rm NP}({\bf \Xi}) + {\bf S}_{\rm nloc}({\bf \Xi})\\ 
&& = {\rm M}_p^{9}
\int d^{11}{\rm X} \sqrt{-{\bf g}_{11}({\rm X})} 
~{\bf R}_{11}({\rm X}) +{\rm M}_p^9\int\Big({\bf G}_4({\rm X}) \wedge \ast_{11}{\bf G}_4({\rm X})+   
{\bf C}_3({\rm X}) \wedge {\bf G}_4({\rm X}) \wedge {\bf G}_4({\rm X)}\Big)\nonumber\\
&&  + ~{\rm M}_p^3\int {\bf C}_3({\rm X}) \wedge \mathbb{X}_8({\rm X})
+ {\rm M}_p^{11}\int d^{11} {\rm X} \sqrt{-{\bf g}_{11}({\rm X})}~
\sum_d\sum_{{\rm S}_d = 0}^\infty h_{{\rm S}_d}~{\bf Q}_{\rm pert}(\bar{c}({\rm S}_d); {\bf \Xi}({\rm X}))\nonumber\\
&& \times ~{\rm exp}\left(-{\rm S}_d {\rm M}_p^d \int_0^{\rm Y} d^d{\rm Y'} \sqrt{{\bf g}_d({\rm Y}', x)} \big\vert {\bf Q}_{\rm pert}(\hat{c}({\rm S}_d); {\bf \Xi}({\rm Y}', x))\big\vert \right)\nonumber\\
&&+ ~{\bf S}_{\rm fermions+antifermions} + {\bf S}_{\rm mixed}+ 
{\rm M}_p^{11}\int d^{11}{\rm X} \sqrt{-{\bf g}_{11}({\rm X})} \nonumber\\
&& \times 
\sum_d\sum_{{\rm P}_d = 0}^\infty b_{{\rm P}_d}~ {\rm exp}\left({-{\rm P}_d{\rm M}_p^d\int_{{\cal M}_d} d^d{\rm Y'} 
\sqrt{{\bf g}_d({\rm Y}', x)}\big\vert \mathbb{F}({\rm Y} -{\rm Y'}; t) {\bf Q}_{\rm pert}(\tilde{c}({\rm P}_d); {\bf \Xi}({\rm Y}', x))\big\vert}\right)\nonumber\\
&&\times \left[{\bf Q}_{\rm pert}(\check{c}({\rm P}_d); {\bf \Xi}({\rm X})) + 
{\rm M}_p^d\int_{{\cal M}_d} d^d{\rm Y''} \sqrt{{\bf g}_d({\rm Y}'', x)} ~{\bf Q}_{\rm pert}(\grave{c}({\rm P}_d); {\bf \Xi}({\rm Y}'', x)) \mathbb{F}({\rm Y}-{\rm Y}''; t)\right]  \nonumber, 
\nd
where ${\rm X} \equiv (x, {\rm Y})$ with $x \in {\bf R}^{2, 1}$ and ${\rm Y} \subset {\cal M}_4 \times {\cal M}_2 \times \xoxo$; and the summations over $d$ in the third and the sixth lines are defined for $1\le d \le 10$ with $d \in \mathbb{Z}$. The perturbative series accompanying each of the non-perturbative terms is arranged to capture the fluctuation determinant around the ${\rm S}$-instantons. Thus:
\bg\label{braz2tag}
{\bf Q}_{\rm pert}(c({\rm S}); {\bf \Xi}({\rm X})) = \sum_{\{l_i\}, n_j}\mathbb{Q}_{\rm T}^{(\{l_i\}, n_j)}(c({\rm S}; \{l_i\}, n_j); {\bf \Xi}({\rm X})), \nd
where $\mathbb{Q}_{\rm T}^{(\{l_i\}, n_j)}(c({\rm S}; \{l_i\}, n_j); {\bf \Xi}({\rm X}))$ is given by \eqref{botsuga} with $c({\rm S}; \{l_i\}, n_j)$ denoting the coefficients of the various terms in the series.
Note that $d = 6$ in \eqref{kimkarol} corresponds to both the BBS \cite{bbs} and the KKLT \cite{kklt} instantons: the BBS instantons are M5-brane instantons wrapped on a six-cycle of the internal manifold, whereas the KKLT instantons are M5-brane instantons wrapped on $\Sigma_4 \times \xoxo$ with $\Sigma_4$ being a four-cycle in ${\cal M}_4 \times {\cal M}_2$. For $d = 3$ we will have the M2-brane instantons wrapping three cycle $\Sigma_3$ in the internal eight-manifold ${\cal M}_4 \times {\cal M}_2 \times \xoxo$. Some of these configuration can break the four-dimensional isometries (see details in section \ref{kalulight}). Similarly with $8 < d \le 10$ we can have instanton configurations that could also break the four-dimensional de Sitter isometries in the dual heterotic side. For quasi de Sitter states, such configurations could in principle be allowed, although we will not study them here. For us, we will restrict $d$ to lie in the range $1\le d \le 8$ within the internal eight-manifold. The fermionic and the mixed interactions appear from the analysis that we presented in section \ref{servant}. The procedure would be to replace the metric and the flux components by generalized metric and flux components, {\it i.e.} the ones with fermionic extensions as in \eqref{akash} and \eqref{vendlinda} with ${\bf Q}_{\rm pert}(c, {\bf \Xi})$ replaced by \eqref{fahingsha10} and $c \equiv (\bar{c}({\rm S}_d), \hat{c}({\rm S}_d), \tilde{c}({\rm P}_d), \check{c}({\rm P}_d), \grave{c}({\rm P}_d))$. Some of these coefficients could in principle be fixed by (a) comparing the action with the known low-energy effective action at the supergravity level, {\it i.e.} at the zero instanton sector and up to quartic orders in the curvatures, and (b) by demanding the absence of Ostrogradsky's (and other potential) instabilities. The generic picture, as presented here, is not known but as we shall see that in our analysis somewhat surprisingly the details about the coefficients are {\it not} necessary at least to study the $\bar{g}_s$ scalings of the various terms from \eqref{botsuga} or \eqref{fahingsha10}. 

The M2-brane and the M5-brane {\it instantons} appearing in the effective action \eqref{kimkarol} are easy to interpret in the excited state when we use 
$\hat{\bf S}_{\rm tot}(\langle{\bf \Xi}\rangle_\sigma)$. So are the fermionic degrees of freedom: they being the emergent degrees of freedom $\langle\Psi_{\rm A}\rangle_\sigma$ as discussed in section \ref{servant}. Interestingly, the M2-branes and the M5-branes also become Glauber-Sudarshan states in ways described in \cite{dileep}. This makes sense in the general framework where the degrees of freedom that are responsible to create an accelerating four-dimensional universe are {\it emergent} degrees of freedom. The action governing the dynamics of 
$\langle{\bf \Xi}\rangle_\sigma$ is simply $\hat{\bf S}_{\rm tot}(\langle{\bf \Xi}\rangle_\sigma)$ (which actually appears from $\check{\bf S}_{\rm tot}(\langle{\bf \Xi}\rangle_\sigma)$ when the Wheeler-de Witt wave-function becomes localized along the ``classical'' trajectory \cite{wdwpaper}) from where the EOMs takes the form:
\bg\label{trottagra}
{\delta\check{\bf S}_{\rm tot}(\langle{\bf \Xi}\rangle_\sigma) \over \delta \langle{\bf \Xi}\rangle_\sigma} = 0 ~~~~ \implies ~~~~ {\delta\hat{\bf S}_{\rm tot}(\langle{\bf \Xi}\rangle_\sigma) \over \delta \langle{\bf \Xi}\rangle_\sigma} = 0, \nd
the later of which we will study in the rest of the paper. Recall that 
$\langle{\bf \Xi}\rangle_\sigma = ({\bf g}_{\rm AB}, {\bf C}_{\rm ABD},$ $\Psi_{\rm A}, \bar\Psi_{\rm A})$, and therefore there are multiple set of equations which need to be solved to demonstrate the consistency of the four-dimensional de Sitter or the quasi de Sitter backgrounds. This will be a rather challenging exercise because of the interconnectedness of the equations so we will have to tread very carefully. The strategy would then be the following: in the remainder of this and the next sub-sections we will lay out the basic structure of the on-shell metric EOMs. In sections \ref{sec7s} and \ref{sec4.5} we will provide a detailed study of the G-flux EOMs, Bianchi identities, anomaly cancellations and flux quantizations. In section \ref{secmetric}, we will return back to the on-shell metric EOMs, including the cross-terms EOMs whose energy-momentum tensors were laid out in {\bf Table \ref{omabiratag}}, and demonstrate consistency in the presence of the full trans-series form of the effective action.  

The on-shell EOMs should be sourced by both perturbative and non-perturbative effects, including the corresponding non-local ones. They may be derived from \eqref{trottagra} in the following way.
Choosing $\hat{\bf S}_{\rm tot}(\langle{\bf\Xi}\rangle_\sigma)$ from \eqref{kimkarol} and $\langle{\bf\Xi}\rangle_\sigma = \langle {\bf g}_{\rm AB}\rangle_\sigma \equiv {\bf g}_{\rm AB}$, the EOM that we get by plugging them in \eqref{trottagra} may be expressed as:

{\footnotesize
\bg\label{mootcha}
{\bf R}_{\rm AB}(\langle {\bf \Xi}({\rm X})\rangle_\sigma) - 
{1\over 2} \langle {\bf g}_{\rm AB}({\rm X})\rangle_\sigma {\bf R}(\langle {\bf \Xi}({\rm X})\rangle_\sigma) &= & {\bf T}_{\rm AB}^{(f)}(\langle {\bf \Xi}({\rm X})\rangle_\sigma) + \sum_{i = 1}^4
{\bf T}_{\rm AB}^{({\rm pert};i)}(\langle {\bf \Xi}({\rm X})\rangle_\sigma)\nonumber\\ 
&+ & \sum_{j = 1}^2{\bf T}_{\rm AB}^{({\rm BBS}; j)}(\langle {\bf \Xi}({\rm X})\rangle_\sigma)+ \sum_{k = 1}^2
{\bf T}_{\rm AB}^{({\rm KKLT}; k)}(\langle {\bf \Xi}({\rm X})\rangle_\sigma)\nonumber\\
&+ & \sum_{l = 1}^2 {\bf T}_{\rm AB}^{({\rm NP}; l)}(\langle {\bf \Xi}({\rm X})\rangle_\sigma), \nd}
where the first term is the G-flux contribution, the second set of terms is all from the perturbative effects, the third set is from the BBS instantons, the fourth set is from the KKLT type instantons, and the final set is from other set of instantons or more generic non-perturbative effects. One may further quantify them in the following way. The first ``classical" contribution is from the G-fluxes. They are given by:

{\scriptsize
\bg\label{malishell1}
{\bf T}_{\rm AB}^{{\red(f)}}(\langle {\bf \Xi}({\rm X})\rangle_\sigma) = 
{2 \over \sqrt{{\bf g}_{11}(\langle{\bf \Xi}({\rm X})\rangle_\sigma)}} {\delta\over \delta \langle{\bf g}^{\rm AB}({\rm X})\rangle_\sigma} \Big({\bf G}_4( \langle{\bf \Xi}({\rm X})\rangle_\sigma) \wedge \ast_{11}{\bf G}_4( \langle{\bf \Xi}({\rm X})\rangle_\sigma)\Big), \nd}
whose $\bar{g}_s$ scaling behavior is completely controlled by the $\bar{g}_s$ scalings of the G-flux components and the metric components that enter\footnote{With $\sqrt{-{\bf g}_{11}(\langle{\bf \Xi}({\rm X})\rangle_\sigma)}$ hiding inside $\ast_{11}$.} the Hodge star $\ast_{11}$. Note that we are not taking the fermionic extension of the G-fluxes to simplify the ensuing analysis (implying that we will be ignoring the fermions for the time being). The perturbative contributions, on the other hand, come from the following set of terms:

{\scriptsize
\bg\label{malishell2}
{\bf T}_{\rm AB}^{{\red({\rm pert};1)}}(\langle {\bf \Xi}({\rm X})\rangle_\sigma) &= &  - ~{2{\rm M}_p^2\over \sqrt{{\bf g}_{11}(\langle{\bf \Xi}({\rm X})\rangle_\sigma)}}\sum_d
\sum_{{\rm S}_d= 0}^\infty h_{{\rm S}_d} {\delta\over \delta \langle{\bf g}^{\rm AB}({\rm X})\rangle_\sigma}
\left[\sqrt{{\bf g}_{11}(\langle{\bf \Xi}({\rm X})\rangle_\sigma)}{\bf Q}_{\rm pert}(\bar{c}_d({\rm S}_d); \langle{\bf \Xi}({\rm X})\rangle_\sigma)\right] \nonumber\\
& & \times  ~{\rm exp}\left(-{\rm S}_d {\rm M}_p^d \int_0^{\rm Y} d^d{\rm Y}'\sqrt{{\bf g}_d(\langle {\bf \Xi}({\rm Y}', x)\rangle_\sigma)} \big\vert {\bf Q}_{\rm pert}(\hat{c}_d({\rm S}_d); \langle{\bf \Xi}({\rm Y}', x)\rangle_\sigma)\big\vert \right)\\
{\bf T}_{\rm AB}^{{\red({\rm pert};2)}}(\langle {\bf \Xi}({\rm X})\rangle_\sigma) &=& -{{\rm M}_p^2\over \sqrt{{\bf g}_{11}(\langle {\bf \Xi}({\rm X})\rangle_\sigma)}} \sum_d\sum_{{\rm P}_d = 0}^\infty b_{{\rm P}_d}{\delta\over \delta \langle{\bf g}^{\rm AB}({\rm X})\rangle_\sigma}
\left(\sqrt{{\bf g}_{11}(\langle{\bf \Xi}({\rm X})\rangle_\sigma)}{\bf Q}_{\rm pert}(\check{c}_d({\rm P}_d); \langle{\bf \Xi}({\rm X})\rangle_\sigma)\right)\nonumber\\
&& \times ~{\rm exp}\left({-{\rm P}_d{\rm M}_p^d\int_{{\cal M}_d} d^d{\rm Y}' 
\sqrt{{\bf g}_d(\langle{\bf \Xi}({\rm Y}', x)\rangle_\sigma)}\big\vert \mathbb{F}({\rm Y} -{\rm Y}'; t) {\bf Q}_{\rm pert}(\tilde{c}_d({\rm P}_d); \langle{\bf \Xi}({\rm Y}', x)\rangle_\sigma)\big\vert}\right)\nonumber\\
{\bf T}_{\rm AB}^{{\red({\rm pert};3)}}(\langle {\bf \Xi}({\rm X})\rangle_\sigma)& = & ~ - \sum_d {\rm M}_p^{d+2} \langle {\bf g}_{\rm AB}({\rm X})\rangle_\sigma \nonumber\\
&& \times ~ \sum_{{\rm P}_d = 0}^\infty\int_{{\cal M}_d} d^d{\rm Y}' \sqrt{{\bf g}_d(\langle{\bf \Xi}({\rm Y}', x)\rangle_\sigma)} ~{\bf Q}_{\rm pert}(\grave{c}_d({\rm P}_d); \langle{\bf \Xi}({\rm Y}', x)\rangle_\sigma)~\times \mathbb{F}({\rm Y}-{\rm Y}'; t)\nonumber\\
&& ~\times ~{\rm exp}\left({-{\rm P}_d{\rm M}_p^d\int_{{\cal M}_d} d^d{\rm Y}'' 
\sqrt{{\bf g}_d(\langle{\bf \Xi}({\rm Y}'', x)\rangle_\sigma)}\big\vert \mathbb{F}({\rm Y} -{\rm Y}''; t) {\bf Q}_{\rm pert}(\tilde{c}_d({\rm P}_d); \langle{\bf \Xi}({\rm Y}'', x)\rangle_\sigma)\big\vert}\right)\nonumber\\
{\bf T}_{\rm AB}^{{\red({\rm pert};4)}}(\langle {\bf \Xi}({\rm X})\rangle_\sigma) &= & -{\rm M}_p^{10} \int d^8{\rm Y}' \sqrt{\bf g_{11}(\langle {\bf \Xi}({\rm Y}', x)\rangle_\sigma) \over {\bf g}_{11}(\langle {\bf \Xi}({\rm X})\rangle_\sigma)} \nonumber\\
&&\times ~ \sum_d\sum_{{\rm P}_d = 0}^\infty {\delta \over \delta \langle {\bf g}_{\rm AB}({\rm X})\rangle_\sigma} \left(\sqrt{{\bf g}_d(\langle{\bf \Xi}({\rm X})\rangle_\sigma)} ~{\bf Q}_{\rm pert}(\grave{c}_d({\rm P}_d); \langle{\bf \Xi}({\rm X})\rangle_\sigma)\right)
~\times\mathbb{F}({\rm Y}'-{\rm Y}; t)\nonumber\\
&& \times ~{\rm exp}\left({-{\rm P}_d{\rm M}_p^d\int_{{\cal M}_d} d^d{\rm Y}'' 
\sqrt{{\bf g}_d(\langle{\bf \Xi}({\rm Y}'', x)\rangle_\sigma)}\big\vert \mathbb{F}({\rm Y}' - {\rm Y}''; t) {\bf Q}_{\rm pert}(\tilde{c}_d({\rm P}_d); \langle{\bf \Xi}({\rm Y}'', x)\rangle_\sigma)\big\vert}\right)\nonumber
\nd}
where ${\rm X} = (x, {\rm Y})$ with $x \in {\bf R}^{2, 1}$ and ${\rm Y} \subset{{\cal M}_4 \times {\cal M}_2 \times \xoxo}$. The first term in \eqref{malishell1}, {\it i.e.} $i = 1$, is from local interactions and the remaining ones, {\it i.e.} $i = 2, 3, 4$, are from non-local interactions controlled by the non-locality function $\mathbb{F}({\rm Y} - {\rm Y}'; t)$. Note that, despite the presence of $\mathbb{F}({\rm Y} - {\rm Y}'; t)$, the contributions are completely local\footnote{We are avoiding non-localities with respect to time as they would lead to much more complicated dynamics. See \cite{desitter2} for details on this.} because of the integral over ${\rm Y}'$. The sum over $d$, as mentioned earlier, could typically involve all possibilities $1 \le d \le 10$, although we will concentrate mostly on $ 1 \le d \le 8$ which follows the natural distribution of the compact and the non-compact spaces in our M-theory set-up. The exponential suppressions for $({\rm S}_d, {\rm P}_d) \ge (1, 1)$ denote the {\it perturbative} contributions from the non-zero instanton sectors coming exclusively from the corresponding fluctuation determinants. These are distinct from the {\it non-perturbative} contributions which are of three kinds: BBS \cite{bbs}, KKLT \cite{kklt} and generic NP types. The contributions from the BBS instantons are:

{\scriptsize
\bg\label{teelmolish1}
{\bf T}_{\rm AB}^{{\red({\rm BBS}; 1)}}(\langle {\bf \Xi}({\rm X})\rangle_\sigma) & = &  ~{\rm M}_p^{10} \int d^8{\rm Y}' \sqrt{ {\bf g}_{11}(\langle{\bf \Xi}({\rm Y}', x)\rangle_\sigma) \over{\bf g}_{11}(\langle{\bf \Xi}({\rm X})\rangle_\sigma)}~\sum_{{s} = 1}^\infty ~{s} h_{{s}} ~
{\bf Q}_{\rm pert}(\bar{c}_6({s}); \langle{\bf \Xi}({\rm Y'}, x)\rangle_\sigma)\\
&& \times ~ {\delta\over \delta \langle{\bf g}^{\rm AB}({\rm X})\rangle_\sigma}
\left[\sqrt{{\bf g}_{6}(\langle{\bf \Xi}({\rm X})\rangle_\sigma)}{\bf Q}_{\rm pert}(\hat{c}_6({s}); \langle{\bf \Xi}({\rm X})\rangle_\sigma)\right]
~\Theta({\rm Y}'-{\rm Y})\nonumber\\
&& \times ~{\rm exp}\left(-{s} {\rm M}_p^6 \int_0^{{\rm Y}'} d^6{\rm Y}'' \sqrt{{\bf g}_6(\langle {\bf \Xi}({\rm Y}'', x)\rangle_\sigma)} \big\vert {\bf Q}_{\rm pert}(\hat{c}_6({s}); \langle{\bf \Xi}({\rm Y}'', x)\rangle_\sigma)\big\vert \right) \nonumber\\
{\bf T}_{\rm AB}^{{\red({\rm BBS}; 2)}}(\langle {\bf \Xi}({\rm X})\rangle_\sigma)&=& {\rm M}_p^{10} \int d^8{\rm Y}' \sqrt{ {\bf g}_{11}(\langle{\bf \Xi}({\rm Y}', x)\rangle_\sigma) \over{\bf g}_{11}(\langle{\bf \Xi}({\rm X})\rangle_\sigma)}~\mathbb{F}({\rm Y'-Y}; t) \nonumber\\
&&\times ~\sum_{{p} = 1}^\infty {p}b_{p}~
{\delta\over \delta \langle{\bf g}^{\rm AB}({\rm X})\rangle_\sigma}
\left[\sqrt{{\bf g}_{6}(\langle{\bf \Xi}({\rm X})\rangle_\sigma)}{\bf Q}_{\rm pert}(\tilde{c}_6(p); \langle{\bf \Xi}({\rm X})\rangle_\sigma)\right] 
~\times \Big[{\bf Q}_{\rm pert}(\check{c}_6({p}); \langle{\bf \Xi}({\rm Y}', x)\rangle_\sigma) \nonumber\\
&& + ~
{\rm M}_p^6\int_{{\cal M}_6} d^6{\rm Y}''\sqrt{{\bf g}_6(\langle {\bf \Xi}({\rm Y}'', x)\rangle_\sigma)} ~{\bf Q}_{\rm pert}(\grave{c}_6({ p}); \langle{\bf \Xi}({\rm Y}'', x)\rangle_\sigma) \mathbb{F}({\rm Y}'-{\rm Y}''; t)\Big]\nonumber\\
&& \times ~ {\rm exp}\left({-{p}{\rm M}_p^6\int_{{\cal M}_6} d^6y''' 
\sqrt{{\bf g}_6(\langle{\bf \Xi}({\rm Y}''', x)\rangle_\sigma)}\big\vert \mathbb{F}({\rm Y}' -{\rm Y}'''; t) {\bf Q}_{\rm pert}(\tilde{c}_6({p}); \langle{\bf \Xi}({\rm Y}''', x)\rangle_\sigma)\big\vert}\right),\nonumber 
\nd}
where the first one is the sum over contributions from the five-brane instantons wrapped on ${\cal M}_4 \times {\cal M}_2$ along with their fluctuation determinants. The second one is similar contributions but from the non-local sector indicated by the presence of $\mathbb{F}({\rm Y}' -{\rm Y}''')$. Again, the non-locality factors are all appropriately integrated away so that we always get local results. In a similar vein, the KKLT type instantons contribute as:

{\scriptsize
\bg\label{teelmolish2}
{\bf T}_{\rm AB}^{{\red({\rm KKLT}; 1)}}(\langle {\bf \Xi}({\rm X})\rangle_\sigma) & = & ~{\rm M}_p^{10} \int d^8{\rm Y}' \sqrt{ {\bf g}_{11}(\langle{\bf \Xi}({\rm Y}', x)\rangle_\sigma) \over{\bf g}_{11}(\langle{\bf \Xi}({\rm X})\rangle_\sigma)}~\sum_{{s} = 1}^\infty ~{s} h_{s} ~
{\bf Q}_{\rm pert}(\bar{c}_3(s); \langle{\bf \Xi}({\rm Y'}, x)\rangle_\sigma)\\
&& \times ~ {\delta\over \delta \langle{\bf g}^{\rm AB}({\rm X})\rangle_\sigma}
\left[\sqrt{{\bf g}_{6}(\langle{\bf \Xi}({\rm X})\rangle_\sigma)}{\bf Q}_{\rm pert}(\hat{c}_3({s}); \langle{\bf \Xi}({\rm X})\rangle_\sigma)\right]
~\Theta({\rm Y}'-{\rm Y})\nonumber\\
&& \times ~{\rm exp}\left(-{s} {\rm M}_p^6 \int_0^{{\rm Y}'} d^4{\rm U}'' d^2w'' \sqrt{{\bf g}_6(\langle {\bf \Xi}({\rm U}'', w'', x)\rangle_\sigma)} \big\vert {\bf Q}_{\rm pert}(\hat{c}_3({s}); \langle{\bf \Xi}({\rm U}'', w'', x)\rangle_\sigma)\big\vert \right)\nonumber\\
{\bf T}_{\rm AB}^{{\red({\rm KKLT}; 2)}}(\langle {\bf \Xi}({\rm X})\rangle_\sigma) & = & {\rm M}_p^{10} \int d^8{\rm Y}' \sqrt{ {\bf g}_{11}(\langle{\bf \Xi}({\rm Y}', x)\rangle_\sigma) \over{\bf g}_{11}(\langle{\bf \Xi}({\rm X})\rangle_\sigma)}~\mathbb{F}({\rm Y'-Y}; t)\nonumber\\
&& \times ~\sum_{{p} = 1}^\infty { p}b_{{p}}~
{\delta\over \delta \langle{\bf g}^{\rm AB}({\rm X})\rangle_\sigma}
\left[\sqrt{{\bf g}_{6}(\langle{\bf \Xi}({\rm X})\rangle_\sigma)}{\bf Q}_{\rm pert}(\tilde{c}_3({p}); \langle{\bf \Xi}({\rm X})\rangle_\sigma)\right]
\times \Big[{\bf Q}_{\rm pert}(\check{c}_3({p}); \langle{\bf \Xi}({\rm Y}', x)\rangle_\sigma) \nonumber\\
&&+ ~ 
{\rm M}_p^6\int_{\Sigma_4 \times \xoxo} d^6{\rm Y}''\sqrt{{\bf g}_6(\langle {\bf \Xi}({\rm Y}'', x)\rangle_\sigma)} ~{\bf Q}_{\rm pert}(\grave{c}_3({p}); \langle{\bf \Xi}({\rm Y}'', x)\rangle_\sigma) \mathbb{F}({\rm Y}'-{\rm Y}''; t)\Big]\nonumber\\
&& ~ \times ~ {\rm exp}\left({-{p}{\rm M}_p^6\int_{\Sigma_4 \times \xoxo} d^6{\rm Y}''' 
\sqrt{{\bf g}_6(\langle{\bf \Xi}({\rm Y}''', x)\rangle_\sigma)}\big\vert \mathbb{F}({\rm Y}' -{\rm Y}'''; t) {\bf Q}_{\rm pert}(\tilde{c}_3({p}); \langle{\bf \Xi}({\rm Y}''', x)\rangle_\sigma)\big\vert}\right),\nonumber \nd}
where the first one is the sum over contributions from the M5-brane instantons wrapped on a six-cycle $\Sigma_4 \times \xoxo$ where $\Sigma_4$ is a four-cycle inside the six-manifold {\it i.e.} $\Sigma_4 \subset{{\cal M}_4 \times {\cal M}_2}$, along with their fluctuation determinants. The second one is similar contributions but from the non-local sector indicated by the presence of $\mathbb{F}({\rm Y}' -{\rm Y}''')$. Again, as mentioned earlier, the non-locality factors are all appropriately integrated away so that we always get local results. It should also be clear that, if ${\cal M}_4$ and ${\cal M}_2$ do not allow odd cycles, then certain types of KKLT type instantons would not exist although the BBS type instantons from \eqref{teelmolish1} would still continue to exist and hence, contribute. The remaining contributions come from generic instantons or non-perturbative terms of the following form:

{\scriptsize
\bg\label{teelmolish3}
{\bf T}_{\rm AB}^{{\red({\rm NP}; 1)}}(\langle {\bf \Xi}({\rm X})\rangle_\sigma) & = &  {\rm M}_p^{10}  \int d^8{\rm Y}' \sqrt{ {\bf g}_{11}(\langle{\bf \Xi}({\rm Y}', x)\rangle_\sigma) \over{\bf g}_{11}(\langle{\bf \Xi}({\rm X})\rangle_\sigma)}~\sum_{d \ne 6}\sum_{{\rm S}_d = 0}^\infty ~{\rm S}_d h_{{\rm S}_d} ~
{\bf Q}_{\rm pert}(\bar{c}_d({\rm S}_d); \langle{\bf \Xi}({\rm Y'}, x)\rangle_\sigma)\nonumber\\
&& \times ~ {\delta\over \delta \langle{\bf g}^{\rm AB}({\rm X})\rangle_\sigma}
\left[\sqrt{{\bf g}_{d}(\langle{\bf \Xi}({\rm X})\rangle_\sigma)}{\bf Q}_{\rm pert}(\hat{c}_d({\rm S}_d); \langle{\bf \Xi}({\rm X})\rangle_\sigma)\right]
~\Theta({\rm Y}'-{\rm Y})\\
&& \times ~{\rm exp}\left(-{\rm S}_d {\rm M}_p^d \int_0^{{\rm Y}'} d^d{\rm Y}'' \sqrt{{\bf g}_d(\langle {\bf \Xi}({\rm Y}'', x)\rangle_\sigma)} \big\vert {\bf Q}_{\rm pert}(\hat{c}_d({\rm S}_d); \langle{\bf \Xi}({\rm Y}'', x)\rangle_\sigma)\big\vert \right)\nonumber\\
{\bf T}_{\rm AB}^{{\red({\rm NP}; 2)}}(\langle {\bf \Xi}({\rm X})\rangle_\sigma) & = & 
{\rm M}_p^{10} \int d^8{\rm Y}' \sqrt{ {\bf g}_{11}(\langle{\bf \Xi}({\rm Y}', x)\rangle_\sigma) \over{\bf g}_{11}(\langle{\bf \Xi}({\rm X})\rangle_\sigma)}~\mathbb{F}({\rm Y'-Y}; t)\nonumber\\
&& \times ~\sum_{d \ne 6}\sum_{{\rm P}_d = 1}^\infty {\rm P}_db_{{\rm P}_d}~
{\delta\over \delta \langle{\bf g}^{\rm AB}({\rm X})\rangle_\sigma}
\left[\sqrt{{\bf g}_{d}(\langle{\bf \Xi}({\rm X})\rangle_\sigma)}{\bf Q}_{\rm pert}(\tilde{c}_d({\rm P}_d); \langle{\bf \Xi}({\rm X})\rangle_\sigma)\right] \nonumber\\
&& \times ~\Big[{\bf Q}_{\rm pert}(\check{c}_d({\rm P}_d); \langle{\bf \Xi}({\rm Y}', x)\rangle_\sigma) \nonumber\\
&& + ~ 
{\rm M}_p^d\int_{{\cal M}_d} d^d{\rm Y}''\sqrt{{\bf g}_d(\langle {\bf \Xi}({\rm Y}'', x)\rangle_\sigma)} ~{\bf Q}_{\rm pert}(\grave{c}_d({\rm P}_d); \langle{\bf \Xi}({\rm Y}'', x)\rangle_\sigma) \mathbb{F}({\rm Y}'-{\rm Y}''; t)\Big]\nonumber\\
&&\times ~ {\rm exp}\left({-{\rm P}_d{\rm M}_p^d\int_{{\cal M}_d} d^dy''' 
\sqrt{{\bf g}_d(\langle{\bf \Xi}({\rm Y}''', x)\rangle_\sigma)}\big\vert \mathbb{F}({\rm Y}' -{\rm Y}'''; t) {\bf Q}_{\rm pert}(\tilde{c}_d({\rm P}_d); \langle{\bf \Xi}({\rm Y}''', x)\rangle_\sigma)\big\vert}\right), \nonumber
\nd}
where ${\cal M}_d \subset{{\cal M}_4 \times {\cal M}_2 \times \xoxo}$, implying that we are looking at configurations with $1 \le d \le 8$. We have also assumed that $\xoxo$ does not allow a one-cycle. In fact, in the absence of any globally defined odd-cycles, we expect $d = 2, 4$ and $d = 8$ cycles to contribute here, in addition to the ones from \eqref{teelmolish1} and \eqref{teelmolish2}. Plugging \eqref{malishell1}, \eqref{malishell2}, \eqref{teelmolish1}, \eqref{teelmolish2} and \eqref{teelmolish3} in \eqref{mootcha} provides the following EOM from the action \eqref{kimkarol}:

{\scriptsize
\bg\label{tranishonal3}
&& {\bf R}_{\rm AB}(\langle {\bf \Xi}({\rm X})\rangle_\sigma) - 
{1\over 2} \langle {\bf g}_{\rm AB}({\rm X})\rangle_\sigma {\bf R}(\langle {\bf \Xi}({\rm X})\rangle_\sigma)\\ 
&& = ~
{2 \over \sqrt{{\bf g}_{11}(\langle{\bf \Xi}({\rm X})\rangle_\sigma)}} {\delta\over \delta \langle{\bf g}^{\rm AB}({\rm X})\rangle_\sigma} \left(\sqrt{{\bf g}_{11}(\langle{\bf \Xi}({\rm X})\rangle_\sigma)} {\bf G}_4( \langle{\bf \Xi}({\rm X})\rangle_\sigma) \wedge \ast_{11}{\bf G}_4( \langle{\bf \Xi}({\rm X})\rangle_\sigma)\right)\nonumber\\
&& - ~{2{\rm M}_p^2\over \sqrt{{\bf g}_{11}(\langle{\bf \Xi}({\rm X})\rangle_\sigma)}}\sum_d
\sum_{{\rm S}_d= 0}^\infty h_{{\rm S}_d} {\delta\over \delta \langle{\bf g}^{\rm AB}({\rm X})\rangle_\sigma}
\left[\sqrt{{\bf g}_{11}(\langle{\bf \Xi}({\rm X})\rangle_\sigma)}{\bf Q}_{\rm pert}(\bar{c}_d({\rm S}_d); \langle{\bf \Xi}({\rm X})\rangle_\sigma)\right] \nonumber\\
&& \times ~{\rm exp}\left(-{\rm S}_d {\rm M}_p^d \int_0^{\rm Y} d^d{\rm Y}'\sqrt{{\bf g}_d(\langle {\bf \Xi}({\rm Y}', x)\rangle_\sigma)} \big\vert {\bf Q}_{\rm pert}(\hat{c}_d({\rm S}_d); \langle{\bf \Xi}({\rm Y}', x)\rangle_\sigma)\big\vert \right)\nonumber\\
&& + ~{\rm M}_p^{10} \int d^8{\rm Y}' \sqrt{ {\bf g}_{11}(\langle{\bf \Xi}({\rm Y}', x)\rangle_\sigma) \over{\bf g}_{11}(\langle{\bf \Xi}({\rm X})\rangle_\sigma)}~\sum_{s = 1}^\infty ~{s} h_{{s}} ~
{\bf Q}_{\rm pert}(\bar{c}_6({s}); \langle{\bf \Xi}({\rm Y'}, x)\rangle_\sigma)\nonumber\\
&& \times ~ {\delta\over \delta \langle{\bf g}^{\rm AB}({\rm X})\rangle_\sigma}
\left[\sqrt{{\bf g}_{6}(\langle{\bf \Xi}({\rm X})\rangle_\sigma)}{\bf Q}_{\rm pert}(\hat{c}_6({s}); \langle{\bf \Xi}({\rm X})\rangle_\sigma)\right]
~\Theta({\rm Y}'-{\rm Y})\nonumber\\
&& \times ~{\rm exp}\left(-{s} {\rm M}_p^6 \int_0^{{\rm Y}'} d^6{\rm Y}'' \sqrt{{\bf g}_6(\langle {\bf \Xi}({\rm Y}'', x)\rangle_\sigma)} \big\vert {\bf Q}_{\rm pert}(\hat{c}_6({s}); \langle{\bf \Xi}({\rm Y}'', x)\rangle_\sigma)\big\vert \right)\nonumber\\
&& + ~{\rm M}_p^{10} \int d^8{\rm Y}' \sqrt{ {\bf g}_{11}(\langle{\bf \Xi}({\rm Y}', x)\rangle_\sigma) \over{\bf g}_{11}(\langle{\bf \Xi}({\rm X})\rangle_\sigma)}~\sum_{{s} = 1}^\infty ~{s} h_{s} ~
{\bf Q}_{\rm pert}(\bar{c}_3(s); \langle{\bf \Xi}({\rm Y'}, x)\rangle_\sigma)\nonumber\\
&& \times ~ {\delta\over \delta \langle{\bf g}^{\rm AB}({\rm X})\rangle_\sigma}
\left[\sqrt{{\bf g}_{6}(\langle{\bf \Xi}({\rm X})\rangle_\sigma)}{\bf Q}_{\rm pert}(\hat{c}_3({s}); \langle{\bf \Xi}({\rm X})\rangle_\sigma)\right]
~\Theta({\rm Y}'-{\rm Y})\nonumber\\
&& \times ~{\rm exp}\left(-{s} {\rm M}_p^6 \int_0^{{\rm Y}'} d^4{\rm U}'' d^2w'' \sqrt{{\bf g}_6(\langle {\bf \Xi}({\rm U}'', w'', x)\rangle_\sigma)} \big\vert {\bf Q}_{\rm pert}(\hat{c}_3({s}); \langle{\bf \Xi}({\rm U}'', w'', x)\rangle_\sigma)\big\vert \right)\nonumber\\
&& +{\rm M}_p^{10}  \int d^8{\rm Y}' \sqrt{ {\bf g}_{11}(\langle{\bf \Xi}({\rm Y}', x)\rangle_\sigma) \over{\bf g}_{11}(\langle{\bf \Xi}({\rm X})\rangle_\sigma)}~\sum_{d \ne 6}\sum_{{\rm S}_d = 0}^\infty ~{\rm S}_d h_{{\rm S}_d} ~
{\bf Q}_{\rm pert}(\bar{c}_d({\rm S}_d); \langle{\bf \Xi}({\rm Y'}, x)\rangle_\sigma)\nonumber\\
&& \times ~ {\delta\over \delta \langle{\bf g}^{\rm AB}({\rm X})\rangle_\sigma}
\left[\sqrt{{\bf g}_{d}(\langle{\bf \Xi}({\rm X})\rangle_\sigma)}{\bf Q}_{\rm pert}(\hat{c}_d({\rm S}_d); \langle{\bf \Xi}({\rm X})\rangle_\sigma)\right]
~\Theta({\rm Y}'-{\rm Y})\nonumber\\
&& \times ~{\rm exp}\left(-{\rm S}_d {\rm M}_p^d \int_0^{{\rm Y}'} d^d{\rm Y}'' \sqrt{{\bf g}_d(\langle {\bf \Xi}({\rm Y}'', x)\rangle_\sigma)} \big\vert {\bf Q}_{\rm pert}(\hat{c}_d({\rm S}_d); \langle{\bf \Xi}({\rm Y}'', x)\rangle_\sigma)\big\vert \right)\nonumber\\
&& -{{\rm M}_p^2\over \sqrt{{\bf g}_{11}(\langle {\bf \Xi}({\rm X})\rangle_\sigma)}} \sum_d\sum_{{\rm P}_d = 0}^\infty b_{{\rm P}_d}{\delta\over \delta \langle{\bf g}^{\rm AB}({\rm X})\rangle_\sigma}
\left(\sqrt{{\bf g}_{11}(\langle{\bf \Xi}({\rm X})\rangle_\sigma)}{\bf Q}_{\rm pert}(\check{c}_d({\rm P}_d); \langle{\bf \Xi}({\rm X})\rangle_\sigma)\right)\nonumber\\
&& \times ~{\rm exp}\left({-{\rm P}_d{\rm M}_p^d\int_{{\cal M}_d} d^d{\rm Y}' 
\sqrt{{\bf g}_d(\langle{\bf \Xi}({\rm Y}', x)\rangle_\sigma)}\big\vert \mathbb{F}({\rm Y} -{\rm Y}') {\bf Q}_{\rm pert}(\tilde{c}_d({\rm P}_d); \langle{\bf \Xi}({\rm Y}', x)\rangle_\sigma)\big\vert}\right)\nonumber\\
&& ~ - \sum_d {\rm M}_p^{d+2}\sum_{{\rm P}_d = 0}^\infty \langle {\bf g}_{\rm AB}({\rm X})\rangle_\sigma 
\int_{{\cal M}_d} d^d{\rm Y}' \sqrt{{\bf g}_d(\langle{\bf \Xi}({\rm Y}', x)\rangle_\sigma)} ~{\bf Q}_{\rm pert}(\grave{c}_d({\rm P}_d); \langle{\bf \Xi}({\rm Y}', x)\rangle_\sigma) \mathbb{F}({\rm Y}-{\rm Y}'; t)\nonumber\\
&& ~\times ~{\rm exp}\left({-{\rm P}_d{\rm M}_p^d\int_{{\cal M}_d} d^d{\rm Y}'' 
\sqrt{{\bf g}_d(\langle{\bf \Xi}({\rm Y}'', x)\rangle_\sigma)}\big\vert \mathbb{F}({\rm Y} -{\rm Y}''; t) {\bf Q}_{\rm pert}(\tilde{c}_d({\rm P}_d); \langle{\bf \Xi}({\rm Y}'', x)\rangle_\sigma)\big\vert}\right)\nonumber\\
&& -{\rm M}_p^{10} \int d^8{\rm Y}' \sqrt{\bf g_{11}(\langle {\bf \Xi}({\rm Y}', x)\rangle_\sigma \over {\bf g}_{11}(\langle {\bf \Xi}({\rm X})\rangle_\sigma}\sum_d\sum_{{\rm P}_d = 0}^\infty {\delta \over \delta \langle {\bf g}_{\rm AB}({\rm X})\rangle_\sigma} \left(\sqrt{{\bf g}_d(\langle{\bf \Xi}({\rm X})\rangle_\sigma)} ~{\bf Q}_{\rm pert}(\grave{c}_d({\rm P}_d); \langle{\bf \Xi}({\rm X})\rangle_\sigma)\right) \mathbb{F}({\rm Y}'-{\rm Y}; t)\nonumber\\
&& ~\times ~{\rm exp}\left({-{\rm P}_d{\rm M}_p^d\int_{{\cal M}_d} d^d{\rm Y}'' 
\sqrt{{\bf g}_d(\langle{\bf \Xi}({\rm Y}'', x)\rangle_\sigma)}\big\vert \mathbb{F}({\rm Y}' - {\rm Y}''; t) {\bf Q}_{\rm pert}(\tilde{c}_d({\rm P}_d); \langle{\bf \Xi}({\rm Y}'', x)\rangle_\sigma)\big\vert}\right)\nonumber\\
&& ~ + ~ {\rm M}_p^{10} \int d^8{\rm Y}' \sqrt{ {\bf g}_{11}(\langle{\bf \Xi}({\rm Y}', x)\rangle_\sigma) \over{\bf g}_{11}(\langle{\bf \Xi}({\rm X})\rangle_\sigma)}~\mathbb{F}({\rm Y'-Y}; t)\sum_{{p} = 1}^\infty {p}b_{p}~
{\delta\over \delta \langle{\bf g}^{\rm AB}({\rm X})\rangle_\sigma}
\left[\sqrt{{\bf g}_{6}(\langle{\bf \Xi}({\rm X})\rangle_\sigma)}{\bf Q}_{\rm pert}(\tilde{c}_6(p); \langle{\bf \Xi}({\rm X})\rangle_\sigma)\right] \nonumber\\
&& ~\times \left[{\bf Q}_{\rm pert}(\check{c}_6({p}); \langle{\bf \Xi}({\rm Y}', x)\rangle_\sigma) + 
{\rm M}_p^6\int_{{\cal M}_6} d^6{\rm Y}''\sqrt{{\bf g}_6(\langle {\bf \Xi}({\rm Y}'', x)\rangle_\sigma)} ~{\bf Q}_{\rm pert}(\grave{c}_6({ p}); \langle{\bf \Xi}({\rm Y}'', x)\rangle_\sigma) \mathbb{F}({\rm Y}'-{\rm Y}''; t)\right]\nonumber\\
&& ~ \times ~ {\rm exp}\left({-{p}{\rm M}_p^6\int_{{\cal M}_6} d^6y''' 
\sqrt{{\bf g}_6(\langle{\bf \Xi}({\rm Y}''', x)\rangle_\sigma)}\big\vert \mathbb{F}({\rm Y}' -{\rm Y}'''; t) {\bf Q}_{\rm pert}(\tilde{c}_6({p}); \langle{\bf \Xi}({\rm Y}''', x)\rangle_\sigma)\big\vert}\right)\nonumber\\
&& ~ {\rm M}_p^{10} \int d^8{\rm Y}' \sqrt{ {\bf g}_{11}(\langle{\bf \Xi}({\rm Y}', x)\rangle_\sigma) \over{\bf g}_{11}(\langle{\bf \Xi}({\rm X})\rangle_\sigma)}~\mathbb{F}({\rm Y'-Y}; t)
~\sum_{{p} = 1}^\infty { p}b_{{p}}~
{\delta\over \delta \langle{\bf g}^{\rm AB}({\rm X})\rangle_\sigma}
\left[\sqrt{{\bf g}_{6}(\langle{\bf \Xi}({\rm X})\rangle_\sigma)}{\bf Q}_{\rm pert}(\tilde{c}_3({p}); \langle{\bf \Xi}({\rm X})\rangle_\sigma)\right] \nonumber\\
&& \times \Big[{\bf Q}_{\rm pert}(\check{c}_3({p}); \langle{\bf \Xi}({\rm Y}', x)\rangle_\sigma) +  
{\rm M}_p^6\int_{\Sigma_4 \times \xoxo} d^6{\rm Y}''\sqrt{{\bf g}_6(\langle {\bf \Xi}({\rm Y}'', x)\rangle_\sigma)} ~{\bf Q}_{\rm pert}(\grave{c}_3({p}); \langle{\bf \Xi}({\rm Y}'', x)\rangle_\sigma) \mathbb{F}({\rm Y}'-{\rm Y}''; t)\Big]\nonumber\\
&& ~ \times ~ {\rm exp}\left({-{p}{\rm M}_p^6\int_{\Sigma_4 \times \xoxo} d^6{\rm Y}''' 
\sqrt{{\bf g}_6(\langle{\bf \Xi}({\rm Y}''', x)\rangle_\sigma)}\big\vert \mathbb{F}({\rm Y}' -{\rm Y}'''; t) {\bf Q}_{\rm pert}(\tilde{c}_3({p}); \langle{\bf \Xi}({\rm Y}''', x)\rangle_\sigma)\big\vert}\right)
+ ~ {\rm M}_p^{10} \int d^8{\rm Y}' \nonumber\\
&& \times ~\sqrt{ {\bf g}_{11}(\langle{\bf \Xi}({\rm Y}', x)\rangle_\sigma) \over{\bf g}_{11}(\langle{\bf \Xi}({\rm X})\rangle_\sigma)}\mathbb{F}({\rm Y'-Y}; t)\sum_{d \ne 6}\sum_{{\rm P}_d = 1}^\infty {\rm P}_db_{{\rm P}_d}~
{\delta\over \delta \langle{\bf g}^{\rm AB}({\rm X})\rangle_\sigma}
\left[\sqrt{{\bf g}_{d}(\langle{\bf \Xi}({\rm X})\rangle_\sigma)}{\bf Q}_{\rm pert}(\tilde{c}_d({\rm P}_d); \langle{\bf \Xi}({\rm X})\rangle_\sigma)\right] \nonumber\\
&& ~\times \left[{\bf Q}_{\rm pert}(\check{c}_d({\rm P}_d); \langle{\bf \Xi}({\rm Y}', x)\rangle_\sigma) + 
{\rm M}_p^d\int_{{\cal M}_d} d^d{\rm Y}''\sqrt{{\bf g}_d(\langle {\bf \Xi}({\rm Y}'', x)\rangle_\sigma)} ~{\bf Q}_{\rm pert}(\grave{c}_d({\rm P}_d); \langle{\bf \Xi}({\rm Y}'', x)\rangle_\sigma) \mathbb{F}({\rm Y}'-{\rm Y}''; t)\right]\nonumber\\
&& ~ \times ~ {\rm exp}\left({-{\rm P}_d{\rm M}_p^d\int_{{\cal M}_d} d^dy''' 
\sqrt{{\bf g}_d(\langle{\bf \Xi}({\rm Y}''', x)\rangle_\sigma)}\big\vert \mathbb{F}({\rm Y}' -{\rm Y}'''; t) {\bf Q}_{\rm pert}(\tilde{c}_d({\rm P}_d); \langle{\bf \Xi}({\rm Y}''', x)\rangle_\sigma)\big\vert}\right),\nonumber
\nd}
which controls the dynamics of the metric components pertaining to the de-Sitter and the quasi de-Sitter configurations. In fact, we can now easily incorporate the fermionic contributions by replacing ${\bf g}_{\rm AB}$ by the generalized metric \eqref{akash} everywhere {\it except} in ${\delta\over \delta \langle{\bf g}^{\rm AB}({\rm X})\rangle_\sigma}$, G-flux components by \eqref{vendlinda} and ${\bf Q}_{\rm pert}(c, \langle {\bf \Xi}\rangle_\sigma)$ from \eqref{botsuga} by \eqref{fahingsha10}.

\subsection{A note on the five-brane instantons and the trans-series action \label{kalulight}}

From the supersymmetric Minkowski vacuum point of view, 
the BBS instantons \cite{bbs}, as mentioned above, are again configurations of M5-branes instantons wrapping the internal six-cycle of the ambient eight-manifold. The typical action of such an instanton at the supersymmetric vacuum Minkowski level is:
\bg\label{sadapa}
{\bf S}_{\rm M5} = 2\pi {\rm M}_p^6\int_{\Sigma_6} d^6y\sqrt{{\bf g}_6} + i\int_{\Sigma_6} {\bf C}_6 + ..., \nd
where ${\bf g}$ is the pullback metric and the dotted terms are the ${\cal O}({\bf R}^4, {\bf G}^2 {\bf R}^4, {\bf R}^6, ..)$ corrections.
The six-form term with integral of ${\bf C}_6$ over a six-manifold is not gauge-invariant, but here we will interpret as the ``charge'' term.
In addition to that there are interactions involving three-form $\widetilde{\bf H}_3 \equiv d\widetilde{\bf B}_2 - {\bf C}_3$, where $\widetilde{\bf B}_2$ is the two-form of the world-volume tensor multiplet and ${\bf C}_3$ is the usual M-theory three-form. 
Interestingly, the full multiplet structure of the five-brane instanton (or even for the usual M5-brane) can be derived from the {\it bulk} fields. This is easy to see for the standard M5-brane in the following way (the instanton discussion is similar). By dimensionally reducing it to IIA and then T-dualizing orthogonal to the NS5-brane converts it to a Taub-NUT space in the type IIB side. The two-form $\widetilde{\bf B}_2$ can be extracted from the IIB four-form ${\bf C}_4$ as ${\bf C}_4 = \widetilde{\bf B}_2 \wedge \Omega_2$, where $\Omega_2$ is the normalizable harmonic two-form on a Taub-NUT space \cite{imamura}. In fact a careful study shows that the complete M5-brane action may be derived from the bulk action in IIB \cite{imamura}, thus confirming our claim that the information about the world-volume multiplet is hiding in the bulk action itself. Now using the fact that ${\rm M}_p^6 = (g_s^{(0)})^{-2} {\rm M}^6_s$, the contribution of the five-brane instanton in the path-integral is:
\bg\label{sadapa2}
{\rm exp}\left(-{\bf S}_{\rm M5}\right) \sim {\rm exp}\left[-{{\rm const}\cdot {\rm Vol}(\Sigma_6)\over (g_s^{(0)})^2} - \delta {\bf  S}_{\rm M5}\left({\bf R}^4, {\bf G}^2 {\bf R}^4, {\bf R}^6, ..\right)\right], \nd
where $g_s^{(0)}$ is the IIA string coupling at the supersymmetric Minkowski vacuum. Unfortunately the action expressed in the form \eqref{sadapa}, and it's path-integral representation from \eqref{sadapa2}, is not the full story. Due to the asymptotic nature of the higher order interaction terms, one needs to express the effective action as a trans-series form. In other words:
\bg\label{kalameBM}
{\bf S}_{\rm eff} = {\bf S}_{\rm pert} + \sum_{n = 1}^\infty \sum_{k = 0}^\infty c_{n, k} \big(g_s^{(0)}\big)^k ~{\rm exp}\left(-{\bf S}_{\rm M5}\right), \nd
where $c_{n, k}$ captures the perturbative fluctuations over the 
$n$-instantons sector. Exponentiating \eqref{kalameBM} in the path-integral would now involve {\it double exponentials} which would guarantee convergence. One could also incorporate some level of non-locality to express the trans-series action as (see also \cite{maximE}):

{\footnotesize
\bg\label{kalameBM2}
{\bf S}_{\rm eff} = {\rm M}_p^{11}\int d^{11}y \sqrt{-{\bf g}_{11}(y)}~{\cal O}(y) \sum_{n = 0}^\infty {\rm exp}\left(-n{\rm M}_p^6\int d^6 y'\sqrt{{\bf g}_6(y')} \mathbb{F}(y - y') {\cal O}'(y')\right), \nd}
with ${\cal O}(y)$ and ${\cal O}'(y)$ are generic operators with all order perturbative interactions at the far IR as in \eqref{botsuga} or \eqref{fahingsha10}, but now attuned to the supersymmetric Minkowski vacuum. At the level of the excited non-supersymmetric Glauber-Sudarshan state, the action for the five-brane instantons is as given in \eqref{kimkarol} via saddle points in a trans-series form with the corresponding fluctuation determinants. The full action with the higher derivative corrections are captured $-$ for both the local and the non-local cases $-$ by ${\bf Q}_{\rm pert}(c; {\bf \Xi}(x, y))$ from \eqref{botsuga}, or more appropriately, from \eqref{fahingsha10}. In fact the latter, {\it i.e.} \eqref{fahingsha10}, can be used to even construct the topological interactions on the world-volume using the Gamma-matrices from \eqref{indig2thi}. 

The KKLT instantons at the vacuum Minkowski level are M5-brane instantons wrapped on $\Sigma_4 \times \xoxo$ with $\Sigma_4$ being a four-cycle in ${\cal M}_4 \times {\cal M}_2$. The KKLT instantons correspond to the IIB {\it dual} of the D3-brane instantons wrapped on $\Sigma_4$. The BBS instantons dualize to Taub-NUT instantons in the IIB side oriented along $3+1$ dimensions and continue as gravitational instantons in the $SO(32)$ side. The KKLT instantons, on the other hand, continue as heterotic five-brane instantons in the $SO(32)$ heterotic side. The ${\rm E}_8\times {\rm E}_8$ story is slightly more complicated because of the intermediate blow-up that we elaborated in  {\bf Table \ref{milleren4}} and in \eqref{dija}, but the story should be similar at least for the case with the BBS instantons. Other configurations are not possible because we have assumed that $\xoxo$ allows no globally defined one-cycle. At the level of the excited non-supersymmetric Glauber-Sudarshan state we expect similar configurations because the duality sequence simulates the duality sequence of the vacuum solitonic states albeit with non-trivial ${\rm M}_p$ corrections as shown in section \ref{sec2.3}.

For $d = 3$ we will have the M2-brane instantons wrapping three cycle $\Sigma_3$ in the internal eight-manifold ${\cal M}_4 \times {\cal M}_2 \times \xoxo$. For $\Sigma_3 \subset{{\cal M}_4 \times {\cal M}_2}$, they become D3-brane instantons in the IIB side wrapping $\Sigma_3$ and oriented along ${\bf R}_3$ direction along $3+1$ dimensional space-time. As such they break the four-dimensional isometries. In the heterotic $SO(32)$ case, depending on the orientation of $\Sigma_3$, we can get five-brane or string instantons, again breaking the four-dimensional isometries. For M2-brane instantons oriented along 
${\bf S}^1 \times \xoxo$, where ${\bf S}^1$ is a one-cycle in ${\cal M}_4 \times {\cal M}_2$, they lead to interesting configuration of {\it perturbative} KK fluctuations in the heterotic $SO(32)$ side.

The instanton actions presented in \eqref{kimkarol} do contain the required higher order curvature and flux corrections, including parts of the Born-Infeld like terms. The metric and the flux components appearing in the action for the instantons, {\it i.e.} the exponential pieces in \eqref{kimkarol}, should actually be viewed as pull-back components but since the instantons are {\it wrapped} around spaces that share the dynamical motions of the ambient spaces, the world-volume coordinates coincide, at least, for the brane instantons. Other instantons that do not have the usual brane representations, typically have higher order interactions as in \eqref{kimkarol}.

The topological terms also work out correctly if we take the ${1\over \sqrt{|{\bf g}_d|}}$ from \eqref{indig2thi}. The Gamma-matrices from say the generalized metric \eqref{akash} when inserted in \eqref{fahingsha10}
can give rise to the required Levi-Civita symbol in $d$-dimensions to allow for the following conversion:
\bg\label{kitange}
\sqrt{|{\bf g}_d|} ~\hat{\bf g}^{\rm M_1M_2}.....\hat{\bf g}^{{\rm M}_{d-1}{\rm M}_d} ~{\bf G}_{{\rm M_1 M_2....M}_d} ~ \to ~ \epsilon^{{\rm M_1M_2....M}_{d-1}{\rm M}_d} ~{\bf G}_{{\rm M_1 M_2 ....M}_{d-1}{\rm M}_d}, \nd
including other non-topological terms. The charges of the brane instantons can be extracted from such terms by integrating over a manifold with a boundary. In general in the presence of gauge invariant field strengths and curvature tensors, we expect:
\bg\label{kangelss}
&& \int_{{\cal M}_d} {\bf G}_d = \int_{\partial{\cal M}_d} {\bf C}_{d-1}\nonumber\\
&& \int_{{\cal M}_{12}} {\bf G}_4 \wedge \mathbb{X}_8 = \int_{\partial{\cal M}_{12}} {\bf C}_3 \wedge \mathbb{X}_8 \nonumber\\
&& \int_{{\cal M}_{12}} {\bf G}_4 \wedge {\bf G}_4 \wedge {\bf G}_4 = \int_{\partial{\cal M}_{12}} {\bf C}_3 \wedge {\bf G}_4 \wedge {\bf G}_4, \nd
where the last two suggest a F-theory origin of the M-theory interactions with $\partial{\cal M}_{12} \equiv {\cal M}_{11}$ (see \cite{FMS} for more details). The first one however appears slightly ambiguous because it suggests $-$ say for the BBS instantons $-$ a {\it seven-manifold} whose boundary is the six-manifold ${\cal M}_4 \times {\cal M}_2$. There could in principle be multiple choices for the seven-manifold, but if we insist on the dominant contribution, it can single out a unique seven-manifold. The results for all the brane-instantons discussed here are succinctly summarized by {\bf Table \ref{gilsalaam}}. These will be very useful when we study the Schwinger-Dyson equations for the metric components in section \ref{secmetric}. 

\begin{table}[tb]  
 \begin{center}
\resizebox{\columnwidth}{!}{%
\renewcommand{\arraystretch}{1.4}
}
\renewcommand{\arraystretch}{1}
\end{center}
 \caption[]{\Su The ${g_s\over {\rm H}(y){\rm H}_o({\bf x})}$ scalings of the charges of the various brane-instantons appearing in our construction. Other instantons that do not have brane representations are not shown here. In computing the $\bar{g}_s$ scalings we have assumed that both ${\cal M}_2$ and $\xoxo$ do not allow globally defined one-cycles, although ${\cal M}_4$ does. The $\bar{g}_s$ scalings for the dual fluxes are derived from {\bf Tables \ref{fridaylin1}} and {\bf \ref{fridaylin2}}, whereas the results in the fifth column are derived from the EOMs and the Bianchi identities of the fluxes in section \ref{sec7s} and \ref{sec4.5}. The $\pm$ superscripts can be easily derived from the analysis presented here, but we will not do so. The charges for the various instantons may be inferred from the dominant $\bar{g}_s$ scalings appearing in the fifth column. Somewhat interestingly, and as we shall discuss in section \ref{secmetric}, the dominant $\bar{g}_s$ scalings of the instanton charges (using \eqref{kangelss}) appear to scale in the {\it same} way as their respective Born-Infeld parts (see however \eqref{mandalamey}).}
 \label{gilsalaam}
 \end{table}


\subsection{A short detour on the wormhole-dressed far IR M-theory action \label{trans2}}

In section \ref{trans00}, we discussed using simple examples how creation of baby universes could influence the trans-series action in our universe. Our conclusion therein was that the action gets {\it dressed} appropriately by wormhole effects which are additional non-instanton like saddles in the path-integral. In this section we will see how this might effect the action \eqref{kimkarol}. Since \eqref{kimkarol} forms the basis of our construction of the de Sitter excited states including the emergent fluxes {\it et cetera}, it is important to know how much of \eqref{kimkarol} would change once the wormhole dressing is inserted in. We will start by first giving a non-technical summary, along the lines of section \ref{trans00} but attuned to \eqref{kimkarol}, and then provide a more quantitative analysis.

\subsubsection{Non-technical summary of the wormhole-dressed action \label{trans01}}

Let us summarize, in intuitive terms, what changes occur in the effective description of M-theory when we include the contributions of Euclidean wormholes in the path integral. The original eleven-dimensional action already includes classical gravitational dynamics, gauge field interactions, topological terms, and non-perturbative effects from Euclidean branes. Wormholes do not simply add one more interaction on top of this; they fundamentally alter how the theory organizes its couplings and how distant points in spacetime, or even different instanton sectors, communicate with one another. To see this, first recall from \eqref{kimkarol} that the original M-theory action contains four main ingredients that may be summarized in the following way.

\vskip.1in

\noindent $\bullet$ \textbf{Core dynamics:} Einstein gravity and the kinetic term for the four-form field strength ${\bf G}_4$, including the kinetic term for the Rarita-Schwinger fermions.

\vskip.1in
\noindent $\bullet$ \textbf{Topological couplings:} Terms like ${\bf C}_3 \wedge {\bf G}_4 \wedge {\bf G}_4$ and ${\bf C}_3 \wedge \mathbb{X}_8$ encode important global and curvature effects, including charge quantization and tadpole constraints.

\vskip.1in

\noindent $\bullet$ \textbf{Non-perturbative (NP) contributions:} These arise from Euclidean M2 and M5 branes (instantons) wrapping compact cycles, leading to exponential terms weighted by brane-instanton actions; as well as other possible NP effects.

\vskip.1in

\noindent $\bullet$ \textbf{Pre-existing nonlocal interactions:} Even without wormholes, fields can interact at a distance through long-range propagation, described by kernels that connect separated spacetime points. They appear from integrating out the off-shell metric components.

\vskip.1in

\noindent Once we introduce the wormholes, question is: what do the wormholes change? It turns out that the wormholes produce three qualitative effects on this structure:

\paragraph{$\bullet$ Fluctuating couplings (“$\alpha$-parameters”):}
Wormholes induce tiny fluctuations in what we normally think of as fixed coupling constants. Gravitational couplings, topological coefficients, instanton weights, and even nonlocal kernels can vary slightly from one “baby-universe sector” to another. In a given sector they are fixed but shifted; if we average over sectors, we integrate over these fluctuations. This is mathematically described by introducing Gaussian random variables $\alpha$ that shift each coupling as we saw in section \ref{trans00}.

\paragraph{$\bullet$ Nonlocal connections (“bilocals”):}
After integrating out the wormhole degrees of freedom, the effective action acquires new terms that directly link field operators at two separate points in spacetime. These are known as \emph{bilocal terms}, and they represent weak, long-distance correlations induced by wormholes. Such terms are absent in a strictly local quantum field theory but are natural in a quantum gravity path integral.

\paragraph{$\bullet$ Instanton cross-couplings:}
Wormholes can also connect distinct Euclidean branes or even different instanton events. This results in current-current type terms on their worldvolumes, allowing one instanton to influence another. Physically, this means that the usual exponential suppression factors for instantons are no longer independent—they become coupled through wormhole effects.

\vskip.1in

\noindent The above three effects basically summarizes what one would expect from introducing wormholes connecting baby universes. However there are two equivalent viewpoints which form the two complementary ways to describe these wormhole effects:

\vskip.1in

\noindent $\bullet$ \textbf{$\alpha$-ensemble picture:} Treat every coupling and NP weight as shifted by a small random parameter $\alpha$. A fixed $\alpha$ corresponds to a specific “baby-universe state,” while integrating over $\alpha$ averages over all possible wormhole contributions.

\vskip.1in

\noindent $\bullet$ \textbf{Bilocal picture:} Instead of shifting couplings, one can include explicit bilocal terms in the action. These terms connect operators or brane currents at distinct points. The two pictures are mathematically equivalent: integrating over $\alpha$ reproduces the bilocals.

\vskip.1in

\noindent This brings us to the crucial question of what 
effects do these wormholes play on the non-perturbative sector of our model. For example the 
NP (both local and non-local instanton) sectors of \eqref{kimkarol} is particularly sensitive to wormholes. These may be summarized as:
\vskip.1in

\noindent $\bullet$ The coefficients (weights) of instanton contributions shift and can mix across different instanton types.

\vskip.1in

 \noindent $\bullet$ The exponents in the instanton actions gain additional quadratic terms that represent cross-couplings with other instantons.

\vskip.1in
 
 \noindent $\bullet$ The pre-existing nonlocal kernel becomes a dressed kernel, $\mathbb{F}_{\rm eff}({\rm Y-Y'})$, which now includes wormhole contributions.
\vskip.1in
\noindent As a result, multi-instanton interactions that were previously absent now naturally arise, and the structure of the trans-series (the infinite series over instanton sectors) becomes richer and more entangled. Expectedly the full partition function of the theory is again best organized as a \emph{trans-series}, a systematic sum over all instanton numbers and wormhole sectors. In this language:
\vskip.1in
\noindent $\bullet$  Fixing an $\alpha$-state gives a conventional theory with shifted couplings and no explicit bilocals.
\vskip.1in
\noindent  $\bullet$ Averaging over $\alpha$ gives a theory with explicit non-localities and operator correlations induced by wormholes.

\vskip.1in

\noindent Note however that there are a few constraints and selection rules. For example, not every coupling or correlation is permitted. Gauge invariance, flux quantization, tadpole cancellation conditions, and supersymmetry all restrict which operator pairs or instanton sectors can be connected by wormholes. These selection rules ensure the consistency of the theory even in the presence of nonlocal terms. Putting everything together, 
the final “wormhole-dressed” action is just the original M-theory action, but with three new kinds of ingredients added:
\vskip.1in
\noindent $\bullet$ {Bulk bilocal terms} linking fields at distant spacetime points.
\vskip.1in
\noindent $\bullet$  {Worldvolume bilocal terms} linking currents on brane instantons.
\vskip.1in
\noindent $\bullet$ {Dressed coefficients and kernels} encoding fluctuations and mixing induced by wormholes.
\vskip.1in

\noindent Conceptually, this action tells us that wormholes do not merely add new interactions. Instead, they reshape the structure of the theory itself — couplings become state-dependent, sectors that were previously independent become entangled, and long-distance correlations emerge naturally. The resulting formulation gives us a unified way to interpolate between “fixed-coupling, factorizing” quantum gravity and “ensemble-averaged, weakly nonlocal” descriptions of M-theory.

\subsubsection{A more quantitative analysis of the wormhole effects on \eqref{kimkarol} \label{trans1b}}

To start, let us quickly recall the field content appearing in \eqref{kimkarol} which we represent
 collectively by ${\bf \Xi}$: metric ${\bf g}_{11}$, three-form ${\bf C}_3$ with field strength ${\bf G}_4=d{\bf C}_3$, and the Rarita-Schwinger fermions ${\bf \Psi}_{\rm M}, \bar{\bf\Psi}_{\rm M}$ (with all off-shell fields components integrated out). World-volume coordinates on Euclidean M-branes/instantons are ${\rm Y}$, with induced metric ${\bf g}_d$.  Nonlocal bulk–brane couplings are represented by kernels like $\mathbb{F}({\rm Y}-{\rm Y}';t)$. Additionally, although not directly appearing in \eqref{kimkarol}, composite bulk operators built from ${\bf \Xi}$ will be collected in ${\cal O}_A$; and the world-volume currents/densities will be collected into a vector $\vec{\cal J}^{(d)}$. We use ${\rm M}_p$ for the eleven-dimensional Planck scale.

The question that we want to investigate here is what the wormholes would do quantitatively to the action \eqref{kimkarol}.
Euclidean wormholes contribute amplitudes connecting separate asymptotic regions, effectively introducing new “moduli” (the baby-universe data) that randomize couplings. Concretely, one may treat \emph{each} coupling (local, nonlocal, and NP weights) as shifted by a random variable $\alpha$. Fixing $\alpha$ selects a “baby-universe state”; integrating over $\alpha$ averages over wormhole sectors.

Question is how should we quantify $\alpha$, or more appropriately, $\alpha_{\rm A}$ where the subscript ${\rm A}$ is related to an operator $\mathcal O_{\rm A}$ constructed from the on-shell degrees of freedom ${\bf \Xi}$? In the process, it should also become clear as to why 
 $\alpha_{\rm A}$ shifts every coupling in the theory. To answer these questions, let us express $\hat{\bf S}_{\rm tot}({\bf \Xi})$ from \eqref{kimkarol} in the following alternative way:
\begin{equation}\label{ryandaisi}
\hat{\bf S}_{\rm tot}({\bf \Xi}) \;=\; \sum_{\rm A} g_{\rm A} \int d^{11}{\rm X} \, \mathcal O_{\rm A}({\rm X}) \;+\; \cdots \, ,
\end{equation}
where each $g_{\rm A}$ is the coupling multiplying a local operator $\mathcal O_{\rm A}({\rm X})$ from \eqref{kimkarol} and the dots denote contributions from the branes, fermions et cetera. The interesting thing about the Euclidean wormhole is that it generates an additional bilocal term in the path integral of the form:
\begin{equation}\label{bringback}
{\bf S}_{\rm wh}({\bf \Xi}) \;=\; \frac12 \sum_{\rm A,B} \int d^{11}{\rm X}\, d^{11}{\rm Y} \; \mathcal O_{\rm A}({\rm X})\, \Delta^{(\mathrm{wh})}_{\rm AB}({\rm X,Y})\, \mathcal O_{\rm B}({\rm Y}) \, ,
\end{equation}
where $\Delta^{(\mathrm{wh})}_{\rm AB}({\rm X,Y})$ is the wormhole kernel encoding the baby-universe exchange. This term is nonlocal and quadratic in the operators $\mathcal O_{\rm A}$, with the non-locality coming from the kernel factor. At this stage one can make the follow useful Hubbard-Stratonovich \cite{hubbard} transformation:
\begin{equation}
e^{-{\bf S}_{\rm wh}({\bf \Xi})} 
\;=\; \int \mathcal D\alpha \; 
\exp\!\Big[ - \bar{\bf S}_{\rm wh}[{\bf \Xi}, \alpha] \Big] \, ,
\end{equation}
which not only rewrites ${\bf S}_{\rm wh}({\bf \Xi})$ in terms of another function $\bar{\bf S}_{\rm wh}[{\bf \Xi}, \alpha]$ but also helps to {\it linearize} the quadratic function  by introducing the auxiliary fields $\alpha_{\rm A}({\rm X})$. This is clear from the following form for $\bar{\bf S}_{\rm wh}[{\bf \Xi}, \alpha]$:

{\footnotesize
\bg\label{waywardmey}
\bar{\bf S}_{\rm wh}[{\bf \Xi},\alpha] \;=\; \frac12 \sum_{\rm A,B}\int d^{11}{\rm X}\, d^{11}{\rm Y}\; 
\alpha_{\rm A}({\rm X})\, \Big(\Delta^{(\mathrm{wh})}\Big)^{-1}_{\rm AB}({\rm X,Y})\, \alpha_{\rm B}({\rm Y})
\;-\; \sum_{\rm A} \int d^{11}{\rm X}\; \alpha_{\rm A}({\rm X})\, \mathcal O_{\rm A}({\rm X}) \, ,
\nd}
where the Hubbard-Stratonovich transformation is done most efficiently by rotating the auxiliary contour (a Lefschetz-thimble choice) to absorb an unnecessary factor of $\sqrt{-1}$. Plugging this in the path-integral we get:

{\scriptsize
\bg\label{ninarnun}
&& \int \mathcal D{\bf \Xi} ~{\rm exp}\left[-\hat{\bf S}_{\rm tot}({\bf \Xi}) - {\bf S}_{\rm wh}({\bf \Xi})\right]\\
= && \int\mathcal D{\bf \Xi} ~\mathcal D\alpha ~{\rm exp}\left[-\frac12 \sum_{\rm A,B}\int d^{11}{\rm X}\, d^{11}{\rm Y}\; 
\alpha_{\rm A}({\rm X})\, \Big(\Delta^{(\mathrm{wh})}\Big)^{-1}_{\rm AB}({\rm X,Y})\, \alpha_{\rm B}({\rm Y})\right] ~{\rm exp}\left[-\sum_{\rm A}(g_{\rm A} - \alpha_{\rm A})\int d^{11}{\rm X}~\mathcal O_{\rm A}({\rm X})\right],\nonumber \nd}
where the second term here is crucial and it holds the key to what we have been emphasizing so far, namely, that the presence of the wormholes does two things simultaneously: one, it {\it shifts} all the couplings present in the action $\hat{\bf S}_{\rm tot}({\bf \Xi})$ in \eqref{kimkarol}, and two, it {\it integrates} over all the coupling shifts.  In other words:
\begin{equation}
g_{\rm A} \;\longrightarrow\; g_{\rm A} - \alpha_{\rm A}({\rm X}) \, ,
\end{equation}
followed by the integration procedure. As emphasized above, 
this holds for every operator $\mathcal O_{\rm A}$ present in the theory, and is the direct origin of the universal $\alpha$-shift of couplings. If one conditions on a fixed baby-universe state (i.e. does not integrate over $\alpha$), then each universe in the multiverse simply has effective couplings:
\begin{equation}\label{vacage71}
g_{\rm A}^{\rm eff} \;=\; g_{\rm A} - \alpha_{\rm A} \, .
\end{equation}
If instead one integrates over $\alpha$ with the Gaussian weight $P[\alpha] \sim \exp\left[-\tfrac12 \alpha (\Delta^{(\mathrm{wh})})^{-1} \alpha\right]$, then the theory becomes an ensemble average over couplings:
\begin{equation}
\langle \mathcal O \rangle = \int \mathcal D\alpha \; P[\alpha] \; \langle \mathcal O \rangle_{g - \alpha} \, ,
\end{equation}
where $P[\alpha]$ can be termed as the induced probability distribution of $\alpha$ and $\langle ..\rangle_{g-\alpha}$ signifies the path integral with an action that allows couplings shifted by $\alpha_{\rm A}$. In the special case where the wormhole kernel factorizes as $\Delta^{(\mathrm{wh})}_{\rm AB}({\rm X,Y}) \simeq c_{\rm AB}$, the $\alpha_{\rm A}({\rm X})$ reduce to spacetime constants $\alpha_{\rm A}$, reproducing Coleman's original picture in which the couplings of each universe are shifted by random constants determined by the baby-universe sector. In all cases, the essential reason the $\alpha_{\rm A}({\rm X})$ shift \emph{every} coupling is that the linear term they generate through the Hubbard--Stratonovich transformation has exactly the same structure as the original coupling terms in the action.

Our aforementioned analysis still leaves a few questions unanswered. For example, we saw that all the coupling constants $g_{\rm A}$ change by the wormhole factors $\alpha_{\rm A}$ respectively, but we did not analyze the situation when the coupling constants $g_{\rm A} = g_{\rm A}(\mu)$ are functions of the energy scale $\mu$. To see what happens for this case,  
fix a renormalization scale $\mu$ and write the (Euclidean) EFT as:
\begin{equation}
\hat{\bf S}_{\rm tot}(\mu)\;=\;\sum_{\rm A} g_{\rm A}(\mu)\,\int d^{11}{\rm X}\,\mathcal O_{\rm A}({\rm X})\,.
\end{equation}
Including wormhole effects and performing the Hubbard--Stratonovich (HS) linearization with an
imaginary contour for the auxiliary fields (so no explicit $i$ appears in the linear term),
conditioning on a fixed HS configuration $\alpha\equiv\{\alpha_A\}$ is equivalent to a
\emph{universal shift of couplings} at the same scale, {\it i.e.}:

{\footnotesize
\begin{equation}\label{vacage70}
{\;
g_{\rm A}^{\rm eff}(\mu;\alpha)\;=\;g_{\rm A}(\mu)\;-\;\alpha_{\rm A}
\;}\qquad\Rightarrow\qquad
\hat{\bf S}_{\rm tot}^{(\alpha)}(\mu)\;=\;\sum_{\rm A} \Big(g_{\rm A}(\mu)-\alpha_{\rm A}\Big)\!\int d^{11}{\rm X}\,\mathcal O_{\rm A}({\rm X}).
\end{equation}}
(With the opposite sign convention for the bilocal kernel one would obtain $g_A^{\rm eff}=g_A+\alpha_A$; the physics is unchanged provided the convention is used consistently.)
Once $\alpha$ is fixed, {\it i.e.} once we condition on a baby–universe sector, the EFT runs in the usual way with the effective initial condition:
\begin{equation}
\mu\,\frac{d}{d\mu}g_{\rm A}^{\rm eff}(\mu;\alpha)\;=\;\beta_{\rm A}\!\Big(g^{\rm eff}_{\rm A_1}(\mu;\alpha),g^{\rm eff}_{\rm A_2}(\mu;\alpha), ..g^{\rm eff}_{\rm A_n}(\mu;\alpha)\Big)\, ,
\end{equation}
with $n \in \mathbb{Z}_+$ denoting the number of couplings at that scale.
In the zero–mode (Coleman) approximation the $\alpha_{\rm }$ are scale–independent constants; with the full kernel, one may keep $\alpha_{\rm A}({\rm X})$ slow–varying in $X$, in which case the ``couplings'' become background fields at the coarse–graining scale. On the other hand, 
if instead of conditioning we \emph{average} over $\alpha$ with the normalized Gaussian prior $P[\alpha]\;\propto\;\exp\!\Big[-\tfrac12\,\alpha\,(\Delta^{(\mathrm{wh})})^{-1}\alpha\Big]$
then observables are ensemble averages over shifted theories:
\begin{equation}
\langle\mathcal X\rangle\;=\;\int\! \mathcal D\alpha\;P[\alpha]\;\langle\mathcal X\rangle_{\{g_A(\mu)-\alpha_A\}}\,.
\end{equation}
In the zero–mode limit $\,\Delta^{(\mathrm{wh})}_{\rm AB}({\rm X,Y})\!\to\!c_{\rm AB}$, the $\alpha_{\rm A}$ become spacetime constants and therefore reduces to a shift by constant parameters at the chosen scale $\mu$.

Yet another issue is the presence of the {$(\det\Delta)^{1/2}$} normalization factor when we use the Hubbard--Stratonovich (HS) decomposition on the wormhole--induced bilocal \eqref{bringback} in Euclidean signature
with a positive (or suitably regulated) kernel $\Delta_{\rm AB}$. The HS identity reads:
\begin{equation}
\label{eq:HS-identity}
\int \mathcal D\alpha\;
\exp\!\left[-\frac{1}{2}\,\alpha\,\Delta^{-1}\alpha+\alpha\,\mathcal O\right]
=(\det 2\pi\,\Delta)^{1/2}\;
\exp\!\left[+\frac{1}{2}\,\mathcal O\,\Delta\,\mathcal O\right],
\end{equation}
where contractions are shorthand for
$\alpha\,\Delta^{-1}\alpha=\sum\limits_{\rm A,B}\!\int\! d^{11}{\rm X}\,d^{11}{\rm Y}\; \alpha_{\rm A}({\rm X})\,(\Delta^{-1})_{\rm AB}({\rm X,Y})\,\alpha_{\rm B}({\rm Y})$
and
$\alpha\,\mathcal O=\sum\limits_{\rm A}\!\int\! d^{11}{\rm X}\;\alpha_{\rm A}({\rm X})\,\mathcal O_{\rm }({\rm X})$. The apparent ``extra factor'' $(\det 2\pi\,\Delta)^{1/2}$ is simply the Gaussian normalization. There are a few standard ways to handle it, depending on whether $\Delta$ depends on dynamical fields. For example if the kernel 
$\Delta_{\rm AB}$ is field-independent, then we can absorb it in the definition of a normalized measure. In other words, we can define the measure as:
\begin{equation}\label{2tagmahal}
d\mu_\Delta(\alpha)\;\equiv\; \frac{\mathcal D\alpha}{(\det 2\pi\,\Delta)^{1/2}}\;
\exp\!\left[-\frac{1}{2}\,\alpha\,\Delta^{-1}\alpha\right],
\end{equation}
where the contraction of the $\alpha_{\rm A}$ and $\Delta_{\rm AB}$ has been defined earlier. Using \eqref{2tagmahal} in 
\eqref{eq:HS-identity} will imply the following path-integral representation:
\begin{equation}
\exp\!\left[+\frac{1}{2}\,\mathcal O\,\Delta\,\mathcal O\right]
=\int d\mu_\Delta(\alpha)\; e^{\alpha\,\mathcal O},
\end{equation}
implying that if $\Delta$ is a $c$--number kernel (independent of the dynamical fields ${\bf \Xi}$), then $(\det 2\pi\,\Delta)^{1/2}$ is an overall constant and cancels in normalized observables. Thus one simply works with the normalized measure $d\mu_\Delta$.

On the other hand, 
if $\Delta=\Delta[{\bf \Xi}]$ depends on the fields (e.g.\ through the background metric, fluxes, or moduli), the normalization is not a harmless constant. One writes:
\begin{equation}\label{tagalore}
(\det 2\pi\,\Delta[{\bf \Xi}])^{1/2}
=\exp\!\left[\frac{1}{2}\,{\rm Tr}~\ln~ \Delta[{\bf \Xi}] + \frac{1}{2}\,{\rm Tr}~\ln(2\pi)\right],
\end{equation}
with the assumption that it 
is the result of a convergent Gaussian over real variables, which requires the kernel $\Delta({\bf \Xi})$ (or the quadratic form it comes from) to be positive definite\footnote{If $\Delta[{\bf \Xi}]$ is self-adjoint and negative definite, the naive real Gaussian underlying $(\det 2\pi\,\Delta[{\bf \Xi}])^{1/2}=\exp\!\{\tfrac{1}{2}\,{\rm Tr}\ln \Delta[{\bf \Xi}]+\tfrac{1}{2}\,{\rm Tr}\ln(2\pi)\}$ is divergent; the correct definition uses a contour rotation (steepest–descent/Lefschetz) so that the fluctuation integral runs along $i\mathbb R$. Let $\{\lambda_i\}$ be the eigenvalues of $\Delta({\bf \Xi})$ in finite dimension. If all $\lambda_i<0$, then choosing the logarithm across the negative real axis yields:
$$
\frac12\,{\rm Tr}~\ln ~\Delta({\bf \Xi}) \;=\; \frac12\sum_i \ln ~\lambda_i({\bf \Xi})
\;=\; \frac12\sum_i\!\big(\ln~|\lambda_i({\bf \Xi})|+i\pi\big)
\;=\; \frac12\,{\rm Tr}~\ln~|\Delta({\bf \Xi})| \;+\; i\,\frac{\pi}{2}\,N,
$$
and hence
$
(\det 2\pi\,\Delta({\bf \Xi}))^{1/2}\;=\;e^{\,i\frac{\pi}{2}N}\,(\det 2\pi\,|\Delta|({\bf \Xi}))^{1/2}.
$
More generally, if $\Delta({\bf \Xi})$ has mixed sign with $n_-(\Delta)$ negative eigenvalues (the Morse index), then:
$$
\frac12\,{\rm Tr}~\ln ~\Delta({\bf \Xi}) \;=\; \frac12\,{\rm Tr}~\ln~|\Delta({\bf \Xi})| \;+\; i\,\frac{\pi}{2}\,n_-(\Delta),
\qquad
(\det 2\pi\,\Delta({\bf \Xi})^{1/2}\;=\; e^{\,i\frac{\pi}{2}n_-(\Delta)}\,(\det 2\pi\,|\Delta({\bf \Xi})|)^{1/2}.
$$
In the functional (infinite-dimensional) case one regularizes the magnitude with a zeta/heat-kernel prescription and encodes the phase by the spectral asymmetry (the $\eta$-invariant): 
$$
{\rm Im}\,{\rm Tr}~\ln ~\Delta({\bf \Xi}) \;=\; \frac{\pi}{2}\,\eta_\Delta(0),
\qquad 
\eta_\Delta(s)\;=\;\sum_i {\rm sign}(\lambda_i)\,|\lambda_i|^{-s}\ \xrightarrow{\ \text{a.c.}\ }\ s=0,
$$
where the operation $a.c$ stands for analytically continue $\eta_\Delta(s)$ from its domain of convergence to a neighborhood of $s=0$ and then evaluate at $s=0$. This way the final form becomes:
$$
(\det 2\pi\,\Delta({\bf \Xi}))^{1/2}
\;=\;
\exp\!\left[\frac12\,{\rm Tr}~\ln~|\Delta({\bf \Xi})|+\frac12\,{\rm Tr}\ln(2\pi)\right]\;
\exp\!\Big(\,i\,\frac{\pi}{2}\,\eta_\Delta(0)\Big).
$$
Physically, a negative (or mixed-sign) spectrum signals an unstable/tunneling direction of the saddle; the Gaussian is then defined on the steepest-descent contour, giving the magnitude from $|\Delta({\bf \Xi})|$ and a phase $e^{i\pi/2}$ per negative mode (or, in the continuum, $e^{i\pi\,\eta_\Delta(0)/2}$). Zero modes, when present, are removed from $\det$ (denoted by a prime) and treated via collective coordinates, and in many applications ratios of determinants are used so that overall constants like $\tfrac12\,{\rm Tr}~\ln(2\pi)$ and any background-independent phase factors cancel.}. \eqref{tagalore} then 
contributes:
\begin{equation}
+\;\frac{1}{2}\,{\rm Tr}\ln \Delta[{\bf \Xi}]
\end{equation}
to the Euclidean effective action. In practice one evaluates ${\rm Tr}~\ln~ \Delta[{\bf \Xi}]$ with heat--kernel/Seeley--DeWitt methods: the UV--divergent local pieces renormalize $\int\!\sqrt {\bf g}$, $\int\!\sqrt {\bf g}\,{\bf R}$, $\int\!\sqrt {\bf g}\,{\bf R}^2,\dots$, while any finite nonlocal remainder is a genuine wormhole--induced correction. If instead $\Delta$ depends only on fixed wormhole moduli (not on ${\bf \Xi}$), the determinant can be absorbed into the wormhole measure or into sector coefficients (e.g.\ $h_{{\rm S}_d},b_{{\rm P}_d}$ in \eqref{kimkarol}).

Finally, if $\Delta({\bf \Xi})$ has zero modes, one may replace $\det$ by the reduced determinant $\det'$ on the subspace orthogonal to zero modes, and treat the latter by introducing collective coordinates with the appropriate Jacobian (Faddeev--Popov or moduli--space measure). The HS identity then carries $(\det' 2\pi\,\Delta({\bf \Xi}))^{1/2}$, and the zero--mode volume factor is accounted for separately.

(The completely analogous statements hold for current--sector HS linearization: for a current bilocal $\tfrac{1}{2}\,J\,\mathbb K\,J$ one introduces auxiliary $\lambda$ with prior $\exp[-\tfrac12\,\lambda\,\mathbb K^{-1}\lambda]$; a field--independent $\mathbb K$ gives a removable $(\det 2\pi\,\mathbb K)^{1/2}$, while a field--dependent $\mathbb K[{\bf \Xi}]$ contributes $+\tfrac12\,{\rm Tr}~\ln ~\mathbb K[{\bf \Xi}]$ to the effective action (again with reduced determinants and collective coordinates if zero modes are present). We will not discuss this further here and a more detailed account will be presented elsewhere.)

This brings us to the final set of questions: what happens if the wormhole contributions are more generic than the bilocal term \eqref{bringback}? The analysis turns out to be surprisingly tractable with the aid of some manipulations whose validity will be discussed soon. To start, we will assume that the bilocal term in \eqref{bringback} may be replaced by a more generic function which convert the partition function to the following:
\bg\label{vacage}
\mathcal Z=\int \mathcal D{\bf \Xi}\;e^{-\hat{\bf S}_{\rm tot}[{\bf \Xi}]}\,~\mathcal F[\mathcal O],
\nd
which takes a product form in a similar vein to \eqref{ninarnun}. In fact to bring \eqref{vacage} to a form similar to \eqref{ninarnun}, we can define:
\bg\label{abianchi1}
\hat{\bf S}_{\rm tot}[\Xi]=\sum_{\rm A} g_{\rm A}(\mu)\int d^{11}{\rm X}\,\mathcal O_{\rm A}({\rm X}),\qquad
\mathcal F[\mathcal O]\equiv e^{-\nu[\mathcal O]},
\nd
at a given scale $\mu$. (This is the far IR scale and we shall keep it fixed as we have been doing so far.) Question now is how should we proceed with $\mathcal F[\mathcal O]$? The answer lies in the following interesting manipulation of $e^{-\nu[\mathcal O]}$, namely:

{\footnotesize
\bg\label{vacage2}
e^{-\nu[\mathcal O]}
=\int \mathcal D\alpha\;P[\alpha]\;
\exp\!\Big(\int d^{11}X\,\alpha_{\rm A}({\rm X})\,\mathcal O_{\rm A}({\rm X})\Big)
\equiv \big\langle e^{\int \alpha\cdot\mathcal O}\big\rangle_{P[\alpha]},
~~~ \int \mathcal D\alpha\,P[\alpha]=1,
\nd}
with $\langle F[\alpha]\rangle_{P[\alpha]}\equiv \int \mathcal D\alpha\,P[\alpha]\,F[\alpha]$. \eqref{vacage2}, as one may easily identify, is a generalized Laplace transform over auxiliary fields $\alpha_{\rm A}({\rm X})$ with a normalized weight $P[\alpha]$. Plugging \eqref{vacage2} in the partition function \eqref{vacage}, immediately produces:
\bg\label{vacage3}
\mathcal Z &=& \int\mathcal D{\bf \Xi}~\mathcal D\alpha ~P[\alpha] ~{\rm exp}\left[-\hat{\bf S}_{\rm tot}[{\bf \Xi}] + \int d^{11}{\rm X} ~\alpha_{\rm A}({\rm X}) \mathcal O_{\rm A}({\rm X})\right] \nonumber\\
&=&\Big\langle \int \mathcal D{\bf \Xi}\;
\exp\!\Big[-\sum_A \big(g_A(\mu)-\alpha_A\big)\!\int d^{11}X\,\mathcal O_A(X)\Big]
\Big\rangle_{P[\alpha]},
\nd
where in the second line we have 
interchanged the integrals to express it in a desired form. Hence, for a fixed auxiliary configuration $\alpha$, we see that the effective coupling becomes:
\bg\label{vacage4}
g_{\rm A}^{\rm eff}(\mu;\alpha)=g_{\rm A}(\mu)-\alpha_{\rm A},
\nd
which is almost similar to what we had in \eqref{vacage71} or, more appropriately, in \eqref{vacage70}. However now the difference is for the choice of $P[\alpha]$: this is no longer a simple Gaussian function as evident from the Laplace transformation \eqref{vacage2}. Because of this, the wormholes start introducing more complicated interactions than the bilocal ones. A way to see this is the following. Taking \eqref{vacage2}, 
expand the exponential and take the average term-by-term:

{\scriptsize
\bg\label{vacage5}
e^{-\nu[\mathcal O]}
&= &\sum_{n\ge 0}\frac{1}{n!}~\Bigg\langle\left(\!\int \alpha\cdot\mathcal O\right)^{\!n}\Bigg\rangle_{P[\alpha]}\\
&= & 1+\int d^{11}{\rm X}\,\mu_{\rm A}\,\mathcal O_{\rm A}
+\frac12\int d^{11}{\rm X}\,d^{11}{\rm Y}\,\langle \alpha_{\rm A}\alpha_{\rm B}\rangle\,\mathcal O_{\rm A}\mathcal O_{\rm B} +\frac{1}{3!}\int d^{11}{\rm X}\,d^{11}{\rm Y}\,d^{11}{\rm Z}\,
\langle \alpha_{\rm A}\alpha_{\rm B}\alpha_{\rm C}\rangle\,
\mathcal O_{\rm A}\mathcal O_{\rm B}\mathcal O_{\rm C}+\cdots, \nonumber
\nd}
with $\mu_{\rm A}\equiv\langle\alpha_{\rm A}\rangle$ and $\langle\cdots\rangle$ denoting moments w.r.t.\ $P[\alpha]$. It is now easy to see that for $P[\alpha]$ a Gaussian function, $\mu_{\rm A} = 0$ and the first non-zero value comes from $\langle \alpha_{\rm A} \alpha_{\rm B}\rangle_{P[\alpha]} \equiv \Delta_{\rm AB}$. In fact all the {\it even} moments would be non-zero, thus reproducing \eqref{bringback} once we identify $\Delta_{\rm AB}$ with $\Delta^{\rm wh}_{\rm AB}$. However for generic choice of $P[\alpha]$ this is in general not true as many be easily inferred from \eqref{vacage5}. In fact we can make this even more precise by 
defining $W[{\rm J}]$ and ${\rm K}^{(n)}_{{\rm A_1\cdots A}_n}({\rm X_1,\ldots,X}_n)$ as:
\bg\label{vacage6}
&& W[{\rm J}]\equiv \ln~\int \mathcal D\alpha ~P[\alpha]~ {\rm exp}\left(\int d^{11}{\rm X}~{\rm J}_{\rm A}({\rm X})\alpha_{\rm A}({\rm X})\right)\\
&& {\rm K}^{(n)}_{{\rm A_1\cdots A}_n}({\rm X_1,\ldots,X}_n)
=\left.\frac{\delta^n W[{\rm J}]}{\delta {\rm J}_{\rm A_1}({\rm X_1})\cdots \delta {\rm J}_{{\rm A}_n}({\rm X}_n)}\right|_{\rm J=0}
=\langle \alpha_{\rm A_1}({\rm X_1})\cdots\alpha_{{\rm A}_n}({\rm X}_n)\rangle_c,\nonumber
\nd
where ${\rm J}_{\rm A}$ is the standard {\it source} term that we introduce to facilitate path integral computations (and put ${\rm J}_{\rm A} = 0$ at the end). Together they are useful to quantify the individual terms in \eqref{vacage5} in the following way:
\bg\label{vacage7}
e^{-\nu[\mathcal O]}
=\exp\!\left[
\sum_{n\ge 1}\frac{1}{n!}
\int d^{11}{\rm X_1} \cdots d^{11}{{\rm X}_n}~
{\rm K}^{(n)}_{{\rm A_1\cdots A}_n}({\rm X_1,\ldots,X}_n)\,
\prod_{i=1}^n \mathcal O_{{\rm A}_i}({\rm X}_i)
\right],
\nd
which is our generalization of the bilocal term from \eqref{bringback}. Using \eqref{vacage7}, it is easy to justify what we said earlier, namely, 
if $P[\alpha]$ is Gaussian, all ${\rm K}^{(n>2)}=0$ and the only nontrivial term is the bilocal from ${\rm K}^{(2)}$. Denoting ${\rm K}^{(2)}$ by $\Delta^{(\mathrm{wh})}$ reproduces the special case from \eqref{bringback}. For non-Gaussian $P[\alpha]$, ${\rm K}^{(n\ge3)}\neq 0$ and the full multi-local tower is present. However despite the complicated nature of \eqref{vacage7}, our earlier conclusion that the effective coupling is the shifted one from \eqref{vacage4} continues to remain valid. 

\subsubsection{Wormhole kernel, integration contours and Lefschetz thimbles \label{thimbles}}

Our aforementioned conclusion relies heavily on the fact that $\mathcal F[\mathcal O]$ from \eqref{abianchi1} can be represented by the Laplace transformed form \eqref{vacage2}. Can this be always true? The answer is {\it not always} and this is where more subtleties enter the story.

The identity \eqref{vacage2} is valid only when $\exp(-\nu[\mathcal O])$ is the (generalized) Laplace transform (moment–generating functional) of some probability measure $P[\alpha]$ over auxiliary fields~$\alpha$. If we define ${\rm F}[{\rm J}]\equiv \exp(-\nu[{\rm J}])$ with 
   ${\rm F}[0]=1$, and ${\rm F[J]}$  finite and continuous for sources ${\rm J}$ near $0$ and impose the conditions that 
the characteristic functional $\Phi[\varphi]\equiv {\rm F}[i\,\varphi]=\exp(-\nu[i\,\varphi])$ is positive–definite, {\it i.e.}\ for any test function $\varphi_k$ and complex numbers $c_k$,
  $
  \sum\limits_{k,\ell} c_k^{\ast}c_\ell\,\Phi[\varphi_k-\varphi_\ell]\ge 0,
  $
  and $\Phi$ continuous with $\Phi[0]=1$, then there exists a probability measure $P[\alpha]$ with
  $
  \Phi[\varphi]=\langle \exp(i\!\int \varphi\!\cdot\!\alpha)\rangle.
  $
  Additionally, if ${\rm F[J]}$ is also finite near ${\rm J}=0$, the Laplace (real–source) representation exists.
(For a full Euclidean field theory of $\alpha$ one may also ask for reflection positivity, etc., but that is not required just for the integral identity.)

\begin{figure}[h]
\centering
\begin{tabular}{c}
\includegraphics[width=4in]{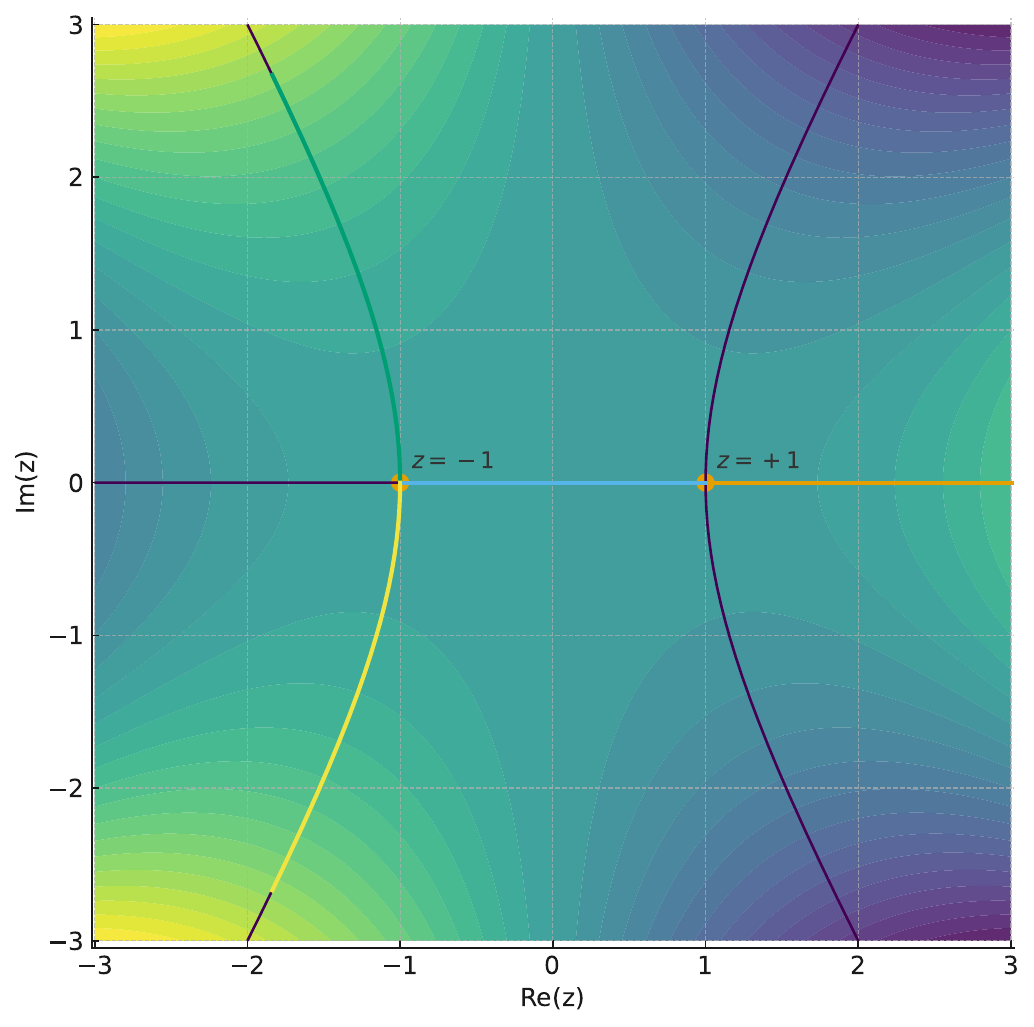}
\end{tabular}
\caption[]{Thimbles for the simple toy model in \eqref{stepsilver2}. The saddles at $z = \pm 1$ are shown. Also the Lefschetz thimbles obtained by integrating the (reverse) gradient-flow ${dz\over d\tau} = -\overline{{\rm S}'(z)}$ from each saddle along its steepest-descent directions.}
\label{thimbleL}
\end{figure}

\begin{figure}[h]
\centering
\begin{tabular}{c}
\includegraphics[width=3in]{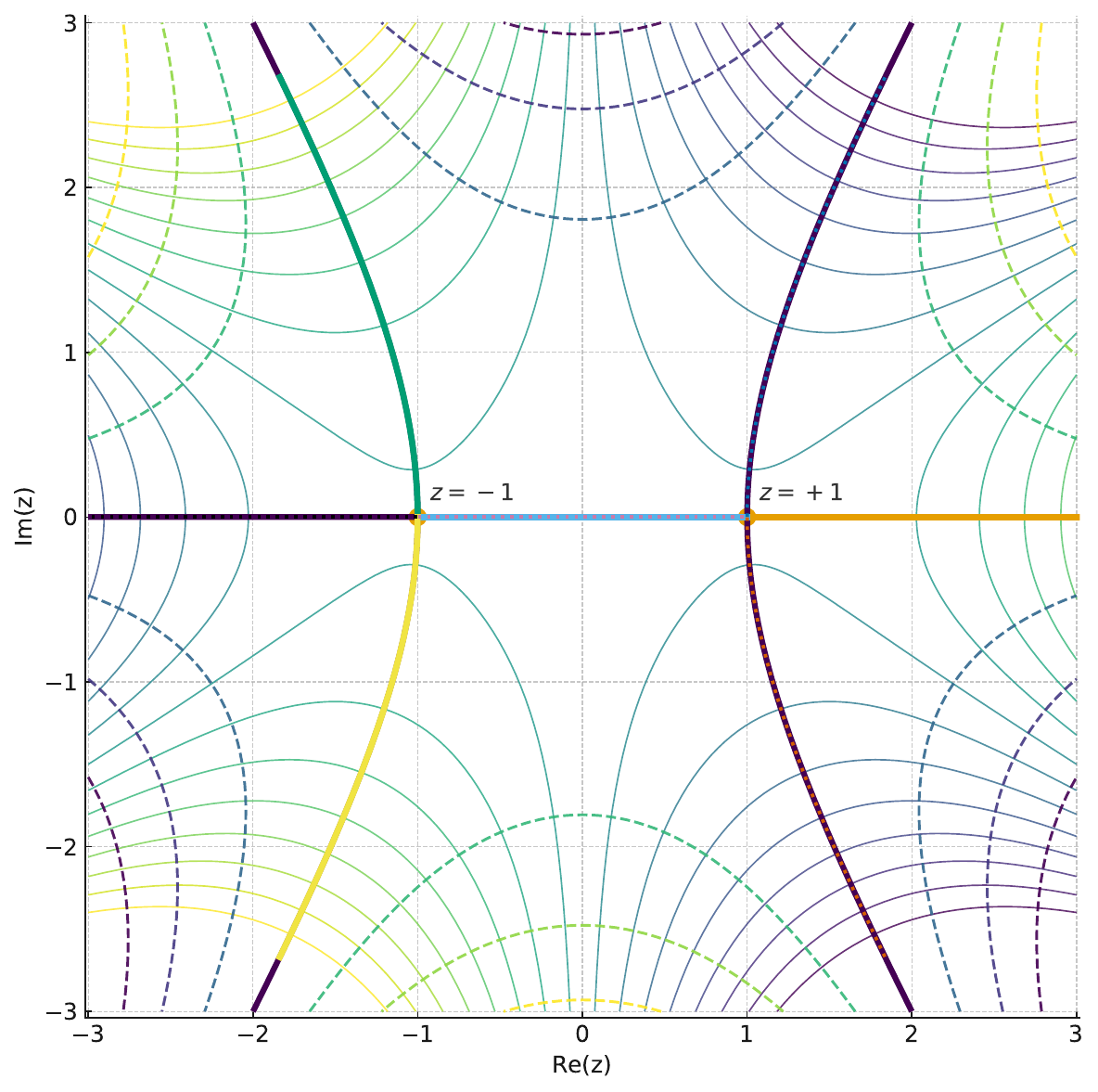}
\end{tabular}
\caption[]{A clearer contour projection of the previous {\bf figure \ref{thimbleL}}. The ${\rm Re}~{\rm S}(z)$ contours are represented by thin solid lines; the ${\rm Im}~{\rm S}(z)$ constant-phase contours are represented by  dashed lines; the ${\rm Im}~{\rm S}(z) = 0$ is shown by the extra thick curve; the thimbles are represented by very thick solid curves leaving the saddles; the anti-thimbles are represented by dotted curves; and the saddles at $z = \pm 1$ are marked with dots and labels.}
\label{greencontour}
\end{figure}

If the aforementioned conditions fail then there are at least two ways to deal with the situation. The {\Su first} way is to
 allow for a complex (signed) weight and/or integrate along a deformed contour (e.g.\ a Lefschetz thimble, see {\bf figure \ref{thimbleL}}). In other words, define:
\bg\label{isapet1}
  \exp(-\nu[\mathcal O])=\int_{\mathcal C}\mathcal D\alpha\, P[\alpha]\;\exp\!\Big(\int \alpha\!\cdot\!\mathcal O\Big), \nd
  where $\mathcal C$ is the contour (as example will be given below using {\bf figure \ref{greencontour}}). \eqref{isapet1} is 
  exact, but positivity is lost (because of possible sign problem). The 
  {\Su second} way is to 
   define ${\rm W[J]}\equiv \ln ~{\rm F[J]}=-\nu[{\rm J}]$. Then the  connected $n$-point kernels are ${\rm K}^{(n)}=\big.\delta^n {\rm W}/\delta {\rm J}^n\big|_{\rm J=0}$ as before, and therefore \eqref{vacage7} provides a fully general construct that avoids constructing $P[\alpha]$.

\begin{figure}[h]
\centering
\begin{tabular}{c}
\includegraphics[width=4in]{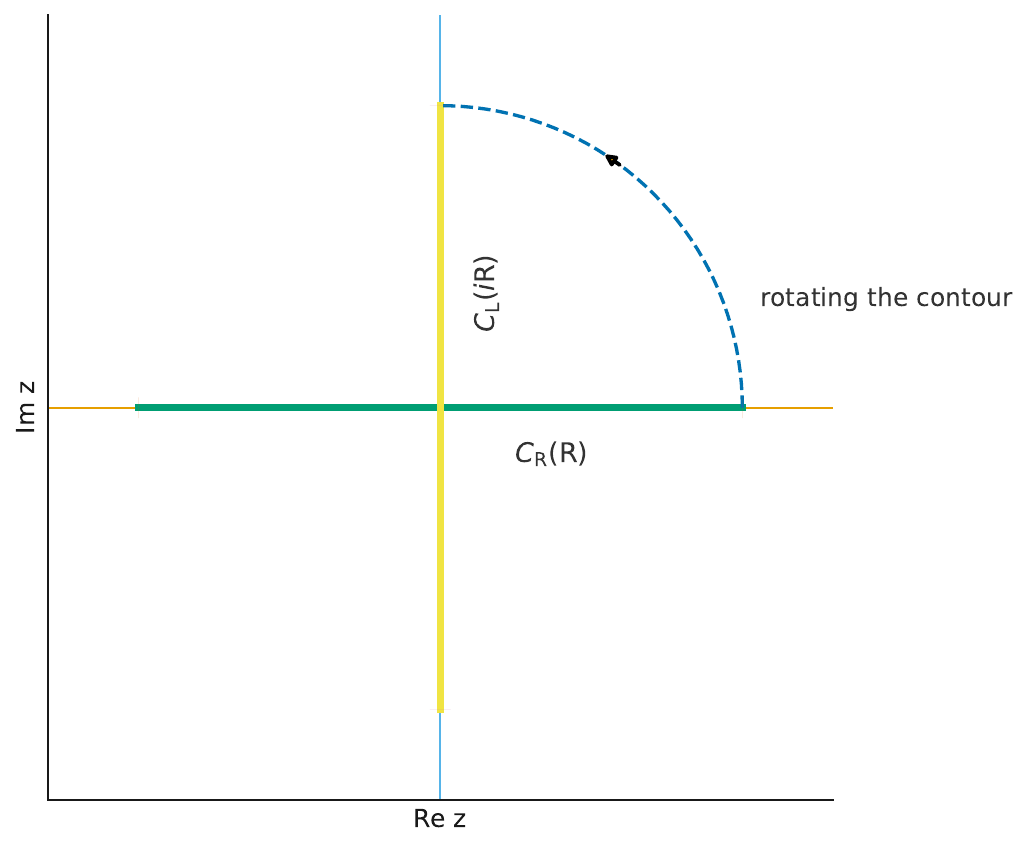}
\end{tabular}
\caption[]{Rotation of the contour to the imaginary axis for the case when any eigenvalue of the kernel in ${\rm S}_\alpha[\alpha]$ from \eqref{isapet2} is negative definite.}
\label{contourrotation}
\end{figure}

Therefore the bottom line is the following.
If $\exp(-\nu[\mathcal O])$ is a bona fide Laplace transform of a positive measure, we can represent wormhole effects as an average over sectors with shifted couplings: at fixed $\alpha$, we can replace $g_A$ by $g_A-\alpha_A$, compute observables, and then average over $\alpha$ with $P[\alpha]$ (Gaussian or not). If not, either use a complex/contour measure \eqref{isapet1} or work directly with the connected multi–local kernels ${\rm K}^{(n)}$ from \eqref{vacage7}, which is always valid.

In our discussion above we briefly mentioned the appearance of Lefschetz thimbles (as shown in {\bf figure \ref{thimbleL}}), so the question is 
why do they appear here? Let us clarify this in some details now. We will
start from the HS/Laplace representation but with a generic choice of $P[\alpha]$ as:
\bg\label{isapet2}
P[\alpha]\propto e^{-{\rm S}_\alpha[\alpha]}\, \qquad
e^{-\nu[\mathcal O]}
=\int \mathcal D\alpha\; 
\exp\!\Big(-{\rm S}_\alpha[\alpha] + \int d^{11}{\rm X} \alpha_{\rm A}({\rm X})\,\mathcal O_{\rm A}({\rm X})\Big), \nd
where ${\rm S}_\alpha[\alpha]$ is a generic function of $\alpha_{\rm A}$. 
When $P[\alpha]$ is Gaussian, ${\rm S}_\alpha[\alpha]$ is quadratic. After diagonalizing the kernel, the functional integral factorizes into independent one–dimensional Gaussians over the mode amplitudes. If an eigenvalue is negative, the real contour diverges. In that case one rotates that mode’s contour to $i\mathbb R$ (steepest descent). This is shown in {\bf figure \ref{contourrotation}} and this procedure essentially takes care of the Gaussian case. 

For non–Gaussian $P[\alpha]$, the auxiliary action ${\rm S}_\alpha[\alpha]$ contains nonlinear terms. After a mode decomposition (or even without fully diagonalizing), each integration variable reduces to a complex integral of the form:
\bg\label{isapet3}
{\rm I(J)}\;=\;\int_{C} dz\;\exp\!\big[-{\rm S}(z;{\rm J})\big],
\qquad
{\rm S}(z;{\rm J})\;=\;\tfrac12\,\kappa\,z^2 - {\rm J}\,z \;+\; \text{nonlinear terms},
\nd
where $z$ represents a (possibly complexified) mode of $\alpha$, and the source ${\rm J}$ is built from $\mathcal O_A$ in the following standard way:
\bg\label{isapet4}
{\rm J} \;=\; \sum_{\rm A} \int d^{11}{\rm X}~ u_{\rm A}({\rm X})\,\mathcal O_{\rm A}({\rm X}), 
\nd
where $u_{\rm A}$, or more appropriately $u_{\rm A}^{(n)}$, are the eigenmodes for the $\alpha_{\rm A}$ field. To see how \eqref{isapet4} appears,
choose any (discrete) mode basis $u_{\rm A}^{(n)}({\rm X})$ for the $\alpha_{\rm A}$-field, and expand
$\alpha_{\rm A}({\rm X})\;=\;\sum\limits_{n} a_n\,u_{\rm A}^{(n)}({\rm X})$.
Inserting it in 
$\int d^{11}{\rm X}\;\alpha_{\rm A}({\rm X})\,\mathcal O_{\rm A}({\rm X})$ expresses it as $\sum\limits_n a_n\,{\rm J}_n$, 
so that the integral over the $n$-th amplitude $a_n$ appears as a one-variable integral with linear source ${\rm J}_n$ as shown by:
\bg\label{isapet5}
{\rm J}_n \;=\; \sum_{\rm A} \int d^{11}{\rm X}\; u_{\rm A}^{(n)}({\rm X})\,\mathcal O_{\rm A}({\rm X}),
\nd
this reproducing \eqref{isapet4} with the assumption that ${\rm J} \to {\rm J}_n$ and $z \to a_n$. Note that 
for a continuous spectrum of modes, we can replace $\sum\limits_n\to\int dn$ and $\delta^{mn}\to\delta(m-n)$ in the standard way.


The key point now is the following. 
The original contour $C$ (often the real line and shown in {\bf figure \ref{thimbleL}}) need not be convergent. Picard–Lefschetz theory states that one can deform $C$ into a finite sum of Lefschetz thimbles $\mathcal J_\sigma$ (steepest–descent manifolds) passing through the saddles $z_\sigma$ of ${\rm S}(z; {\rm J})$ (i.e.\ solutions of $\partial {\rm S}(z; {\rm J})/\partial z=0$). In other words:
\bg\label{stepsilver}
\int_{C} dz\,e^{-{\rm S}(z; {\rm J})}
\;=\;
\sum_{\sigma} n_\sigma(C)\;
\int_{\mathcal J_\sigma} dz\,e^{-{\rm S}(z; {\rm J})}\,,
\nd
where the integers $n_\sigma(C)$ are intersection numbers encoding which saddles contribute for the chosen $C$. Along each $\mathcal J_\sigma$, $\mathrm{Im}\,{\rm S}(z; {\rm J})$ is constant and $\mathrm{Re}\,{\rm S}(z; {\rm J})\to +\infty$ at the ends, so the thimble integral converges.
(In the language of $\alpha$ integral, 
for non–Gaussian $P[\alpha]$, the $\alpha$–integral is defined by deforming the original contour into steepest–descent thimbles through the complex saddles of the auxiliary action ${\rm S}_\alpha[\alpha]$ with the linear source coming from $\int \alpha\!\cdot\!\mathcal O$.)

\begin{figure}[h]
\centering
\begin{tabular}{c}
\includegraphics[width=5in]{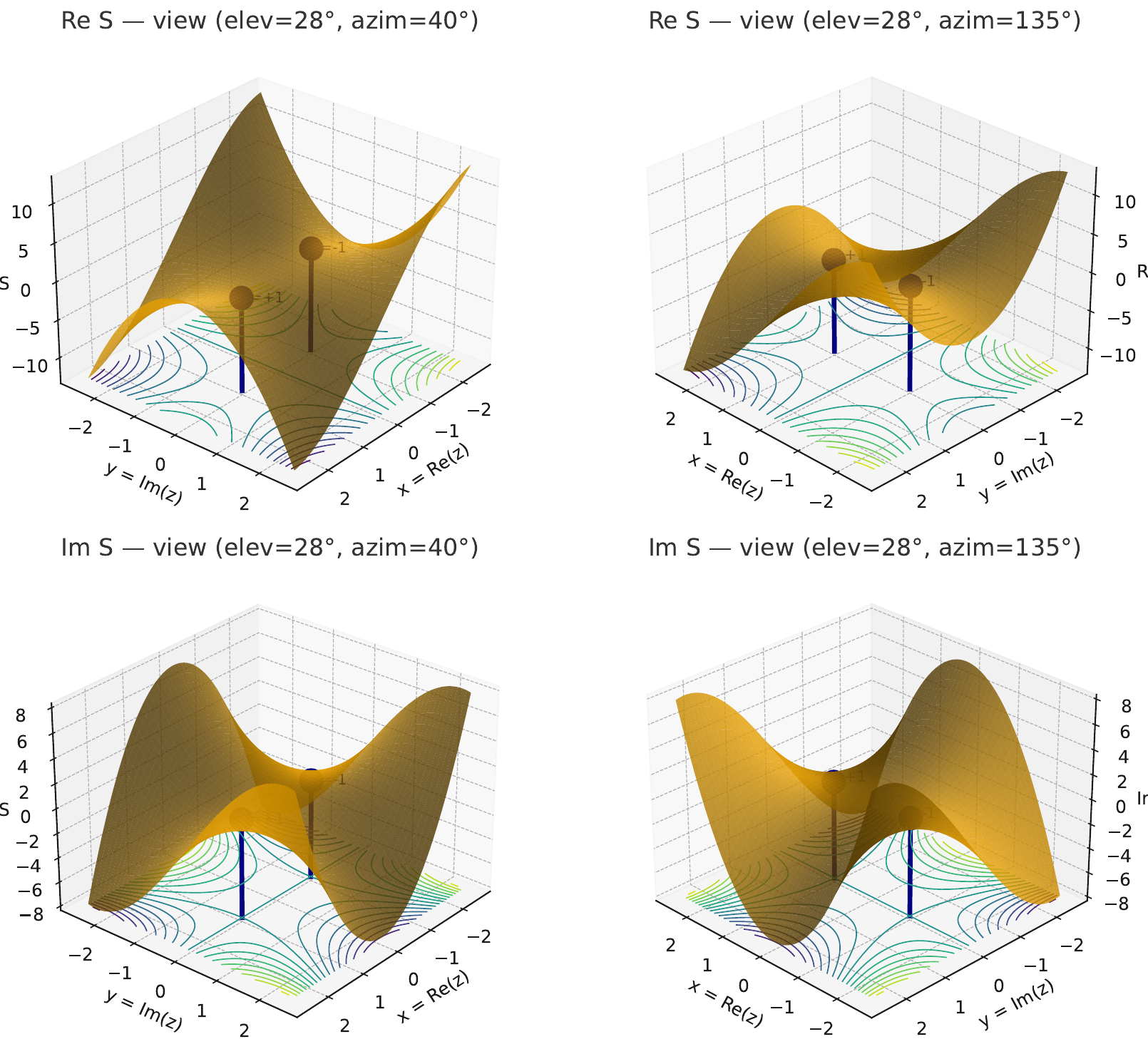}
\end{tabular}
\caption[]{The $3d$ plots for ${\rm Re}~{\rm S}(z)$ and ${\rm Im}~{\rm S}(z)$ from \eqref{stepsilver2} with the saddles and the contours shown clearly. The dark blue pins in each of the plots show the two saddle points.}
\label{z33figure}
\end{figure}

Let us now see how this works out with a simple toy model. We will define ${\rm S}(z; {\rm J})$ with no quadratic piece but with a cubic and a linear piece. We will start by taking ${\rm J} = 1$ for simplicity. The action that we have in mind is the following:
\bg\label{stepsilver2}
{\rm S}(z; 1) \equiv {\rm S}(z)=\frac{z^3}{3}-z,\nd
which is the standard Airy–type cubic capturing the local geometry near a cubic turning point. This is the model, shown in {\bf figure \ref{z33figure}}, whose integration contour is described in {\bf figures \ref{thimbleL}} and {\bf \ref{greencontour}}. The axes
“$\mathrm{Re}\,z$ / $\mathrm{Im}\,z$” denote the complex plane for a single mode $z$ of the auxiliary field $\alpha$. In the full functional integral we have a product over modes; the {\bf figure \ref{thimbleL}} depicts one factor.
The saddles $z_\star=\pm 1$ solve:
\bg\label{stepsilver3}
{\rm S}'(z)=z^2-1=0, 
\nd
implying that each saddle has its thimble $\mathcal J(\pm1)$, which are the steepest descent manifolds. They are depicted by the colored 
curves\footnote{It is easy to see where the curves in {\bf figures \ref{thimbleL}} and {\bf \ref{greencontour}} come from. Using the steepest-descent equation and the notations from \eqref{mumbmeyaxion} and \eqref{mumbmeykitag}, we obtain
    $\frac{dz}{d\tau}
     = -(\bar z^{2}-1)
     = -(x^{2}-y^{2}-1) + 2 i x y$. This yields, for the real and the imaginary parts, the set of equations:
$ \frac{dx}{d\tau} = -x^{2} + y^{2} + 1$ and 
$    \frac{dy}{d\tau} = 2 x y$ which are two first-order equations that generate \emph{all} the flow lines in the complex
plane, including those that do not touch the real axis. In fact
the real axis $y=0$ is an invariant sub-manifold of the flow because
    $\left.\frac{dy}{d\tau}\right|_{y=0}
= 2 x y = 0$ implying that, if a trajectory begins on $y=0$, it remains on $y=0$ for all $\tau$.  
Conversely, if $y(\tau_{0})\neq 0$ at some initial time $\tau_{0}$, the
trajectory can never cross the real axis.  This explains why many flow lines
remain entirely in the upper or lower half-plane: they are simply other
integral curves of the same gradient--flow equations. Interestingly, 
eliminating $\tau$ gives the differential equation 
    $\frac{dy}{dx}
    = \frac{2 x y}{-x^{2} + y^{2} + 1}$ for the shape of each flow line.    Solutions of this first-order equation, with initial condition $(x_{0},y_{0})$,
generate all gradient--flow curves, including those that do not lie on the real
axis.  The special flow lines that begin at the critical points $z=\pm 1$
are the Lefschetz thimbles.} in {\bf figures \ref{thimbleL}} and {\bf \ref{greencontour}}. 

Now defining the Morse function $h(z) \equiv {\rm Re}~{\rm S}(z)$, 
a Lefschetz thimble $\mathcal J_\sigma$ attached to a saddle $z_\sigma$
(where ${\rm S}'(z) = 0$) is the set of points reached by the downward gradient flow of $h(z) \equiv {\rm Re}~{\rm S}(z)$
that emanates from that saddle satisfying ${dz\over d\tau} = -\overline{{\rm S}'(z)}$, where $\tau$ is just an auxiliary ``flow time” used to define the steepest-descent (Picard–Lefschetz) gradient flow in the complexified variable(s) (thus should not be confused with the physical time in our problem). Along this flow:
\bg\label{stepsilver4}
{d\over d\tau}~{\rm Re}~{\rm S}(z) = - \vert {\rm S}'(z)\vert^2 \le 0, ~~~~ {d\over d\tau}~{\rm Im}~{\rm S}(z) = 0, \nd
thus a thimble $\mathcal J_\sigma$ attached to a critical point $z_\sigma$ (where $\partial {\rm S} = 0$)
is the set of points that flow into $z_\sigma$
 as $\tau \to +\infty$
under this equation (stable manifold). Practically, to trace the curve outward from the saddle, one may integrate the reverse (``upward") flow ${\partial z\over \partial \tau} = + \overline{{\rm S}'(z)}$
starting arbitrarily close to $z_\sigma$.

\begin{figure}[h]
\centering
\begin{tabular}{c}
\includegraphics[width=3in]{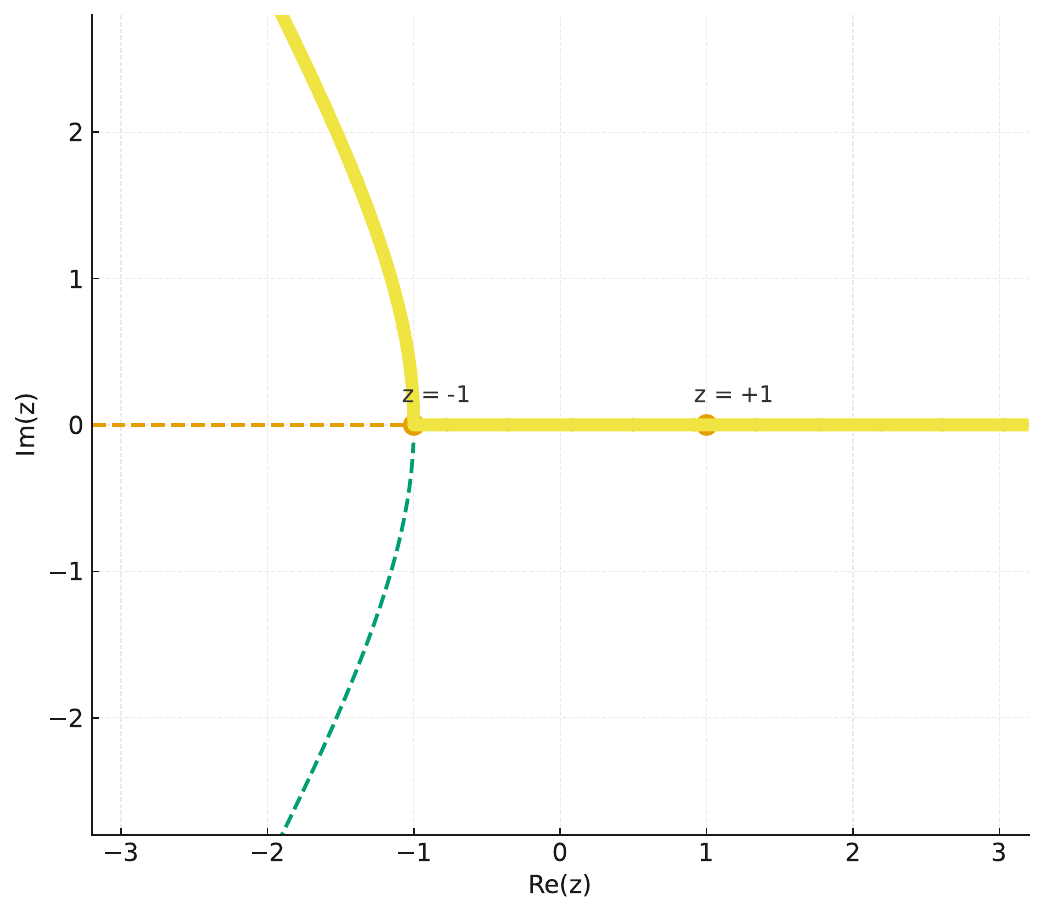}
\end{tabular}
\caption[]{The contour of integration using only the upper branch at $z=-1$: it comes in from $-\infty$ along the upper hyperbola $y=+\sqrt{3(x^2-1)}$, reaches $z=-1$, and then continues along the real axis through $z=+1$ to $+\infty$. The dashed curves in the background show the full ${\rm Im}~{\rm S}(z) = 0$
set (real axis + both hyperbola branches) for context.}
\label{contourinteg}
\end{figure}

Our above discussion implies that each thimble is a downward gradient flow of $\mathrm{Re}\,{\rm S}(z)$ emanating from its saddle; along it $\mathrm{Im}\,{\rm S}(z)$ is constant and the ends lie in Stokes sectors where $e^{-{\rm S}(z)}$ decays. For the cubic, the asymptotic directions are separated by $120^\circ$ because at large $|z|$, where $z \equiv {\rm R}~e^{i\theta}$, we have:
\bg\label{stepisabel}
\mathrm{Re}\,{\rm S}(z)\sim \frac{{\rm R}^3}{3}\cos(3\theta), ~~~~~ {\rm Im}~{\rm S}(z) \sim {{\rm R}^3\over 3}~\sin(3\theta) = {\rm const} \equiv c, \nd
implying that $\theta$ adjusts as ${\rm R}$ changes keeping ${\rm exp}\left(-i{\rm Im}~{\rm S}(z)\right) = {\rm exp}(-ic)$ a constant phase factor along the thimble. (Note that if we kept $\theta$ fixed, the phase would not be a constant -- that's exactly why thimble is a {\it curved} path and not a straight ray.) We can take $c = 0$ which would provide two curves: the real axis $y = 0$ and two hyperbola branches $y^2 = 3(x^2-1)$. But a thimble is not every ${\rm Im}~{\rm S}(z)= 0$ curve through the saddle: it is the subset of curves on which ${\rm Re}~{\rm S}(z)$ increases. Schematically, the contour deforms into a sum of thimbles:
\bg\label{chuckmey}
C \;\simeq\; \mathcal J(+1)\ \pm\ \mathcal J(-1),
\nd
with the relative sign determined by the intersection numbers between $C$ and the upward cycles (unstable manifolds). These signs may jump when a parameter crosses a Stokes line. For us we will only consider the relative positive sign. The contour of integration is shown in {\bf figure \ref{contourinteg}}.

Once we include a linear source,
${\rm S}(z;{\rm J})=\frac{z^3}{3}-{\rm J}\,z$, 
the saddles satisfy ${\rm S}'(z;{\rm J})=z^2-{\rm J}=0$ giving us
$
z_\star=\pm\sqrt{\rm J},
$.
Therefore the thimbles rotate with $\arg {\rm J}$ and the set of contributing saddles can change (also known as the Stokes phenomenon). In the HS application, ${\rm J}$ is the projection of $\mathcal O$ onto the mode
${\rm J}=\int d^{11}{\rm X}~ u({\rm X})\,\mathcal O({\rm X})$, much like what we had earlier.

\begin{figure}[h]
\centering
\begin{tabular}{c}
\includegraphics[width=4in]{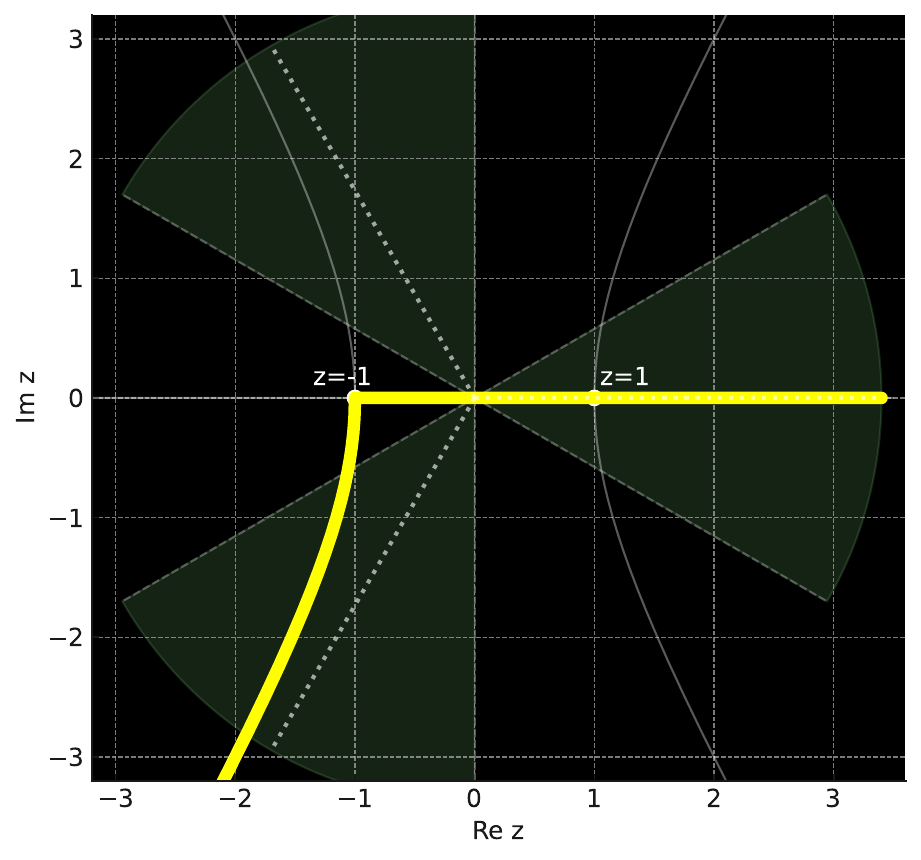}
\end{tabular}
\caption[]{Another allowed contour of integration using only the lower branch at $z=-1$: it comes in from $-\infty$ along the lower hyperbola $y=-\sqrt{3(x^2-1)}$, reaches $z=-1$, and then continues along the real axis through $z=+1$ to $+\infty$. The decay wedges from {\bf figure \ref{decaywedge}} is plotted in the background for convenience.}
\label{airycontour}
\end{figure}

The rest of the argument to compute the integral in \eqref{isapet3} is rather standard. 
Considering the holomorphic action ${\rm S}(z; 1) \equiv {\rm S}(z)$ with a nondegenerate saddle $z_\sigma$ so that ${\rm S}'(z_\sigma)=0$ and ${\rm S}''(z_\sigma)\neq 0$, we can expand about $z_\sigma$ with $z=z_\sigma+w$ and express \eqref{isapet3} as:
\bg\label{isapet51}
{\rm S}(z_\sigma+w)
= {\rm S}_\sigma + \frac{1}{2}\,\lambda\,w^2 + \frac{a}{3!}\,w^3 + \frac{b}{4!}\,w^4 + \cdots, \nd
with 
$\lambda \equiv S''(z_\sigma), a\equiv S^{(3)}(z_\sigma)$ and $b\equiv S^{(4)}(z_\sigma)$.
On the Lefschetz thimble $\mathcal J_\sigma$ through $z_\sigma$, $\mathrm{Im}\,{\rm S}(z)$ is constant and $\mathrm{Re}\,{\rm S}(z)$ increases away from the saddle (see {\bf figures \ref{contourinteg}} and {\bf \ref{airycontour}}), so the local Gaussian $\exp(-\tfrac12\lambda w^2)$ decays. Choosing the branch so that the thimble is mapped to $t\in\mathbb R$ by $w=t/\sqrt{\lambda}$, one finds:
\bg\label{isapet55}
\int_{\mathcal J_\sigma}\! dz\,e^{-{\rm S}(z)}
= e^{-{\rm S}_\sigma}\,\frac{1}{\sqrt{\lambda}}
\int_{\mathbb R}\! dt\,
\exp\!\left[-\frac{t^2}{2}\;-\;\frac{a}{3!\,\lambda^{3/2}}\,t^3\;-\;\frac{b}{4!\,\lambda^{2}}\,t^4\;+\;\cdots\right],
\nd
where $\lambda \equiv S''(z_\sigma)$ as defined earlier.
Using Gaussian moments $\int_{\mathbb R} dt\,e^{-t^2/2}=\sqrt{2\pi}$, $\langle t^{2n}\rangle=(2n{-}1)!!$, $\langle t^{2n+1}\rangle=0$, the first terms of the steepest–descent expansion are:
\bg\label{isapet56}
\int_{\mathcal J_\sigma}\! dz\,e^{-{\rm S}(z)}
= e^{-{\rm S}_\sigma}\,\sqrt{\frac{2\pi}{\lambda}}\,
\left[\,1\;+\;\left(\frac{5a^2}{24\,\lambda^3}-\frac{b}{8\,\lambda^2}\right)\;+\;\cdots\right],
\nd
where the dotted terms are higher orders in inverses of $\lambda$. We can re-express \eqref{isapet56} using the definitions of $\lambda, a$ and $b$ proposed above. Putting everything together provides the following expression for the integral over the thimble $\mathcal J_\sigma$:
\bg\label{isapet57}
\int_{\mathcal J_\sigma}\! dz\,e^{-{\rm S}(z)}
= e^{-{\rm S}(z_\sigma)}\,\sqrt{\frac{2\pi}{{\rm S}''(z_\sigma)}}\,
\left[\,1\;+\;\frac{5\,[{\rm S}^{(3)}(z_\sigma)]^2 - 3\,{\rm S}''(z_\sigma)\,{\rm S}^{(4)}(z_\sigma)}{24\,[{\rm S}''(z_\sigma)]^3}\;+\;\cdots\right],
\nd
where the term in the square bracket may be expressed as 1 + loop corrections. \eqref{isapet57} then provides us with
the essential ingredients to extract the answer from \eqref{stepsilver} once we take $n_\pm = 1$ and view  
the square root factor in \eqref{isapet57} to carry the thimble’s phase/orientation:
\begin{equation}
\sqrt{\frac{1}{{\rm S}''(z_\sigma)}} \;=\; \frac{e^{-\,\tfrac{i}{2}\arg {\rm S}''(z_\sigma)}}{\sqrt{|{\rm S}''(z_\sigma)|}}\times(\text{orientation sign}).
\end{equation}
Therefore summing the contributing thimbles yields the exact result for that mode. The full $\alpha$ functional integral is the product (or, in many variables, a multidimensional thimble integral) over all modes.

To conclude, 
$P[\alpha]$ defines the complex action ${\rm S}_\alpha[\alpha]$ entering the weight $e^{-{\rm S}_\alpha[\alpha]}$ in \eqref{isapet2}. The operator
$\mathcal O_{\rm A}$ appears as the linear source $\int d^{11}{\rm X}~ \alpha_{\rm A}({\rm X})\,\mathcal O_{\rm A}({\rm X})$ that shifts the saddles (and thus determines which thimbles contribute).
The $\alpha_{\rm A}({\rm X})$ are the integration variables; after mode decomposition $\alpha_{\rm A}({\rm X})=\sum\limits_n a_n\,u_{\rm A}^{(n)}({\rm X})$, each amplitude $a_n$ is integrated along its thimble(s).
{\bf Figures \ref{contourinteg}} and {\bf \ref{airycontour}} are  faithful one–mode representations of the non–Gaussian case in \eqref{isapet3}: the original “real” $\alpha$–contour is replaced by the oriented sum of steepest–descent thimbles through the complex saddles of ${\rm S}_\alpha[\alpha] - \int \alpha\!\cdot\!\mathcal O$, ensuring convergence and yielding a well–defined (generally complex) weight; the intersection numbers encode which saddles actually contribute for the chosen original contour.



\section{Schwinger-Dyson equations for the global and localized fluxes \label{sec7s}}

Having got all the metric equations of motion, it is now time to study the Schwinger-Dyson equations corresponding to all the on-shell flux components appearing in \eqref{botsuga}. This will eventually help us to study the Bianchi identities, flux quantizations and anomaly cancellations, which we shall elaborate in section \ref{sec4.5}. In the process, the importance of the dual formalism, as in \eqref{botsuga2.0},  will become useful. The flux ans\"atze that we took in \eqref{usher}, can now be split into global and local fluxes in the following way:
\bg\label{viorosse}
{\cal G}^{(k)}_{\cal ABCD}({\bf x}, y) = {\cal G}^{(k; {\rm global})}_{\cal ABCD}({\bf x}, y) + {\cal F}^{(k)}_{\cal AB}({\bf x}, y) \Omega^{(k)}_{\cal CD}(y), \nd
where $({\cal A, B, C, D}) \in {\bf R}^{1, 2} \times {\cal M}_4 \times {\cal M}_2 \times {\mathbb{T}^2\over {\cal G}}$ although we expect $\Omega^{(k)}_{\rm CD}(y)$ to be localized forms in the internal space
${\cal M}_4 \times {\cal M}_2 \times {\mathbb{T}^2\over {\cal G}}$. For this section and the next, we will however not distinguish between the global and the localized fluxes, but the distinction will become important when we study the flux quantizations and anomaly cancellations in section \ref{secanomaly}.

To proceed we will follow closely the analysis presented for the type IIB case in the second reference of \cite{coherbeta2} starting from section 4.3 onwards. The EOMs have already been laid out in eq. (4.18), (4.19), (4.26) and (4.27) in the second reference of \cite{coherbeta2} (which we shall refer to separately as \cite{coherbeta3}). Recalling the fact that all the flux and the metric components are emergent quantities extracted from the expectation values as in \eqref{katusigel} (see details in section \ref{creepquif}), the corresponding EOMs can be rewritten for convenience from \cite{coherbeta3} in the following way:
\bg\label{crawlquif}
\partial_{{\rm N}_k}\big(\mathbb{T}_7^{(f)}\big)_{\rm N_1.....N_7} &= &  b_1~
\partial_{{\rm N}_k}\big(\mathbb{T}_7^{(q)}\big)_{\rm N_1.....N_7} + 
b_2 \big(\mathbb{T}^{(k)}_8\big)_{{\rm N_1.....N_7}{\rm N}_k} \\
&+& b_3 \big(\mathbb{X}^{(k)}_8\big)_{{\rm N_1.....N_7}{\rm N}_k} + 
b_4 \big({\bf \Lambda}_8\big)_{{\rm N_1.....N_7}{\rm N}_k} + 
b_5~ {\overbracket[1pt][7pt]{\mathbb{T}_0 \partial_{{\rm N}_k}\big(\mathbb{T}}}{}_7^{(np)}\big)_{\rm N_1.....N_7},\nonumber\nd
where $b_i$ are constants, and the superscripts are defined as follows: $f$ is for fluxes, $q$ is for quantum ({\it i.e.} the zero instanton sector of \eqref{kimkarol}), $np$ is for non-perturbative ({\it i.e.} the non-zero instanton sector of \eqref{kimkarol}) and $k$ is for the direction of the derivative. The seven and the eight-form tensors are defined in the following way \cite{coherbeta3}:
\bg\label{olivecostamey}
&&\big(\mathbb{T}_7^{(q)}\big)_{\rm N_1 ...... N_7} = \sqrt{-{\bf g}_{11}} \left(\mathbb{Y}_4\right)^{\rm ABCD}~
\epsilon_{{\rm ABCDN_1 N_2.... N_7}}\\
&&\big(\mathbb{T}_7^{(f)}\big)_{\rm N_1 ...... N_7} = \sqrt{-{\bf g}_{11}} ~{\bf G}^{\rm ABCD}~
\epsilon_{{\rm ABCDN_1 N_2.... N_7}} \nonumber\\ 
&&\big(\mathbb{T}^{(k)}_8\big)_{{\rm N_1 ...... N_7}{\rm N}_k} = {\bf G}_{[{\rm N_1 N_2 N_3 N_4}} 
{\bf G}_{{\rm N_5 N_6 N_7}{\rm N}_k]} \nonumber\\
&&\big(\mathbb{X}^{(k)}_8\big)_{{\rm N_1 ...... N_7}{\rm N}_k} = 
{1\over 3\cdot 2^9 \cdot \pi^4} \left({\rm tr} ~\mathbb{R}_{\rm tot}^4 
- {1\over 4} \left({\rm tr}~\mathbb{R}_{\rm tot}^2\right)^2\right)_{{\rm N_1 N_2.....N_7}{\rm N}_k} \nonumber\\
&& {\overbracket[1pt][7pt]{\mathbb{T}_0 \partial_{{\rm N}_k}\big(\mathbb{T}}}{}_7^{(np)}\big)_{\rm N_1.....N_7} =
\int d^8z_2 \sqrt{{\bf g}_8(z_2)}~
\mathbb{T}_0(x, z_2) \partial_{[{\rm N}_k} \big(\mathbb{T}_7^{np}\big)_{{\rm N_1.....N_7}]}(x, z, z_2), \nonumber \nd
where all the tensors appearing in \eqref{olivecostamey} are defined carefully in section 4.3 of \cite{coherbeta3}. The only difference is that the forms for $\mathbb{T}_0$ and $\mathbb{T}^{np}_7$ are more complicated than the ones that have appeared in eq. (4.28) of \cite{coherbeta3}. This is because the non-zero instanton sector of the trans-series action in \eqref{kimkarol} is much more involved than the non-perturbative action that we took in \cite{coherbeta3}. For the brevity of the analysis, we will not dwell on this sector in detail here and leave a more elaborate analysis for future work. We will concentrate mostly on the zero instanton sector, {\it i.e.} the sector involving \eqref{botsuga}, \eqref{botsuga2.0} and \eqref{fahingsha10}. Moreover, when taking the derivatives in \eqref{crawlquif} we will for example not take the mixed pieces with coefficients $c_{nm}$ in \eqref{marmatrix} for the both the flux and the metric components, although we will continue to take $c_{0m}$ and $c_{n0}$ pieces. 

\subsection{EOMs for the flux components ${\bf G}_{0ij{\rm M}}$ and ${\bf G}_{0ija}$ \label{blackbond1}}

Let us start by using \eqref{crawlquif} to study the EOMs, {\it i.e.} the Schwinger-Dyson equations for the flux components ${\bf G}_{0ij{\rm M}}$, where $y^{\rm M} \in {\cal M}_4 \times {\cal M}_2$ and ${\bf G}_{0ija}$ where $w^a\in {\mathbb{T}^2\over {\cal G}}$. This means we will be looking at the flux components ${\bf G}_{0ijm}, {\bf G}_{0ij\theta_1}, {\bf G}_{0ij\theta_2}$ and ${\bf G}_{0ija}$ for the ${\rm E}_8 \times {\rm E}_8$ case which implies ${\cal M}_2 = {\bf S}^1_{\theta_1} \times {{\bf S}^1_{\theta_2}\over {\cal I}_{\theta_2}}$. Since the generalized and the simplified $SO(32)$ cases can be derived easily from the ${\rm E}_8 \times {\rm E}_8$ case, we will not discuss them in detail here. The curvature forms, that enter in the definition for the eight-form $\mathbb{X}^{(m, 0, i)}_8$ can be easily extracted specifically from 
{\bf Tables \ref{niksmit1}, \ref{niksmit3}, \ref{niksmit4}, \ref{niksmit5}, \ref{niksmit9}, \ref{niksmit11},  \ref{niksmit15}, \ref{niksmit16}, \ref{niksmit17}, \ref{niksmit21}, \ref{niksmit23}}, {\bf \ref{niksmit24}} till {\bf \ref{niksmit250}} and, using them, we can express the relevant $\mathbb{R}_{\rm tot}$ in the following way (see also \cite{coherbeta3}):
\bg\label{devonkalu}
\mathbb{R}_{\rm tot} &= & {\bf R}_{[0i]} dx^0 \wedge dx^i + {\bf R}_{[i\alpha]} dx^i \wedge  dy^\alpha + {\bf R}_{[im]} dx^i \wedge dy^m\nonumber\\
& + & {\bf R}_{[0m]} dx^0 \wedge dy^m + {\bf R}_{[0\alpha]} dx^0 \wedge  dy^\alpha + {\bf R}_{[0 a]} dx^0 \wedge dw^a\nonumber\\
&+& {\bf R}_{[mn]} dy^m \wedge dy^n + {\bf R}_{[\alpha\beta]} dy^\alpha\wedge dy^\beta  + {\bf R}_{[ab]} dw^a \wedge dw^b \nonumber\\
& + & {\bf R}_{[m\alpha]} dy^m \wedge dy^\alpha+ {\bf R}_{[ma]} dy^m \wedge dw^a + {\bf R}_{[\alpha a]} dy^\alpha \wedge dw^a,
\nd
where we have suppressed the internal indices. The $g_s$ scalings of each of the curvature forms may be easily extracted from the aforementioned tables. As we can see there are a wide range of scalings for each of the curvature forms, and therefore it would make sense to find the {\it dominant} contribution to the eight-form $\mathbb{X}^{(m, 0, i)}_8$. This gives us:
\bg\label{devlee}
&& {\rm dom}~\Big[\mathbb{X}^{(i)}_8({\bf x}, y, w; g_s)\Big]_{npq\alpha\beta abi}= \left({g_s\over {\rm H}(y){\rm H}_o({\bf x})}\right)^{1 + \hat\chi^{(3)}_e(t)} \left[\widetilde{\mathbb{X}}^{(i)}_8({\bf x}, y, w)\right]_{npq\alpha\beta abi}\nonumber\\
&& {\rm dom}~\Big[\mathbb{X}^{(0)}_8({\bf x}, y, w; g_s)\Big]_{0npq\alpha\beta ab}= \left({g_s\over {\rm H}(y){\rm H}_o({\bf x})}\right)^{1 + \hat\chi^{(2)}_e(t)} \left[\widetilde{\mathbb{X}}^{(0)}_8({\bf x}, y, w)\right]_{0npq\alpha\beta ab}\\
&& {\rm dom}~\Big[\mathbb{X}^{(m)}_8({\bf x}, y, w; g_s)\Big]_{mnpq\alpha\beta ab}= \left({g_s\over {\rm H}(y){\rm H}_o({\bf x})}\right)^{2 + \hat\chi^{(1)}_e(t)} \left[\widetilde{\mathbb{X}}^{(m)}_8({\bf x}, y, w)\right]_{mnpq\alpha\beta ab}, \nonumber \nd
where $\hat\chi^{(n)}_e(t) \equiv \hat\chi^{(n)}_e(\hat\alpha_e(t), \hat\beta_e(t), \hat\sigma_e(t), \hat\zeta_e(t), \hat\eta_e(t))$ whose explicit value may be easily computed from (a) the $g_s$ scalings appearing in {\bf Tables \ref{niksmit1}, \ref{niksmit3}, \ref{niksmit4}, \ref{niksmit5}, \ref{niksmit9}, \ref{niksmit11},  \ref{niksmit15}, \ref{niksmit16}, \ref{niksmit17}, \ref{niksmit21}, \ref{niksmit23}}, {\bf \ref{niksmit24}}, till \eqref{niksmit250} and from (b) the uncanceled logarithmic corrections discussed in section \ref{sec4.5.1}. Finally, we can express the eight-form ${\bf \Lambda}_8$ associated with the {\it dynamical} M2 branes\footnote{Note that the M2 branes, and including other branes, are also taken as Glauber-Sudarshan states in the way discussed in \cite{dileep}. Thus they all have some quantum widths quantified by ${\bf \Lambda}_8^{(k)}(y, w)$.} in the following suggestive way:
\bg\label{samthigap}
{\bf \Lambda}_8({\bf x}, y, w; g_s) = \sum_{k = 0}^\infty {\bf \Lambda}_8^{(k)}(y, w)\left({g_s\over {\rm H}(y){\rm H}_o({\bf x})}\right)^{{2k|j_k|\over 3}} \to \left({g_s\over {\rm H}(y){\rm H}_o({\bf x})}\right)^{\hat{j}_e(t)} \widetilde{\bf \Lambda}_8(y, w), \nd
where $\hat{j}_e(t)$ contains all the perturbative and non-perturbative corrections to the bare scaling (see section \ref{sec4.5.2}). We have also taken $\hat{j}_e(t) > 0$. This will become clearer soon. Putting everything together, 
we can express the scalings of the tensors appearing in \eqref{crawlquif} as in {\bf Table \ref{crecraqwiff1}}.

\begin{table}[H]  
 \begin{center}
\resizebox{\columnwidth}{!}{%
\renewcommand{\arraystretch}{2.3}
}
\renewcommand{\arraystretch}{1}
\end{center}
 \caption[]{\Su ${g_s\over {\rm H}(y){\rm H}_o({\bf x})}$ scalings of the various terms of \eqref{crawlquif} contributing to the EOM of ${\bf G}_{0ijm}$. The quantum scaling $\theta_{nl}$ is given by \eqref{brittbaba}. The lower bounds, consistent with the underlying EFT, are shown in the last column. In the second row, we also denote the bound on $\theta_{nl}$.} 
  \label{crecraqwiff1}
 \end{table}
From \eqref{crawlquif}, one would expect that all the elements in the third column of {\bf Table \ref{crecraqwiff1}} to be equal to each other. However at the lowest order, as discussed in \cite{coherbeta3}, this does not happen. At the lowest order, where:

{\footnotesize
\bg\label{shwarmamey}
(\hat\alpha_e(t), \hat\beta_e(t), \hat\sigma_e(t), \hat{l}_{e{\rm AB}}^{\rm CD}(t), \hat\chi_e^{(1)}(t), \hat{j}_e(t), \hat\zeta_e(t), \hat\eta_e(t)) = \left({2\hat\alpha(t)\over 3}, {2\hat\beta(t)\over 3}, {2\hat\sigma(t)\over 3}, \hat{l}_{\rm AB}^{\rm CD}, \hat\chi^{(1)}(t), \hat{j}(t), 0, 0\right), \nd}
we expect an identification of the form given by eq. (4.36) in \cite{coherbeta3}, wherein we used over-brackets and under-brackets to allow for the necessary sub-identifications. In {\bf Table \ref{crecraqwiff1}} we can make the following identifications:
\bg\label{cata00}
{\rm Row}~ 1 = {\rm Row}~ 2 = {\rm Row}~ 9, ~~~~~ {\rm Row}~ 3 = {\rm Row}~ 6 = {\rm Row}~9, \nd
which are implemented at the lowest orders. At higher orders we expect all the rows in \eqref{cata00} to be identified to each other. Moreover from {\bf Table \ref{micandi0_0}} the scalings of the flux components are bounded from below as:
\bg\label{tundegolab}
\hat{l}_{e[mn}^{[pq} \oplus \hat{l}_{e\alpha\beta]}^{ab]} \ge 0, \nd
which can be easily verified from all the scalings participating in \eqref{tundegolab}.
Imposing \eqref{cata00} then provides the following values for the scalings of the flux components and the quantum terms for the derivatives acting along $y^m$ directions:

{\footnotesize
\bg\label{brettbess}
\hat{l}_{0i}^{jm}(t) = -4 + {\rm dom}~\hat{j}(t) + {\hat\beta(t) - \hat\alpha(t)\over 6}, ~~~\theta_{nl}(t) = {2\over 3} + 2~{\rm dom}~\hat{j}(t) + {5\hat\beta(t) - \hat\alpha(t)\over 6}, \nd}
with $\theta_{nl}(t)$ as given by \eqref{brittbaba}\footnote{Interestingly, comparing the elements of the first two rows in the third column of {\bf Table \ref{crecraqwiff1}} we find that $\theta_{nl} = 2\left(\hat{l}_{e0i}^{jm} + {13\over 3} - {\hat\sigma_e(t)\over 2} - {3\hat\zeta_e(t)\over 2}\right)$. This is precisely how the contribution of ${\bf G}_{0ijm}$ appears in \eqref{brittbaba} with $l_{48} = 2$. In fact this observation is generic: such identifications hold for all components of the G-fluxes. This was also observed for the IIB case in \cite{coherbeta3}. \label{fasermaxe}}. Since all the temporally dependent parameters are very small, we see that the scaling of the flux component ${\bf G}_{0ijm}$ differs slightly from the scaling for the same flux components in the type IIB case from \cite{coherbeta3}.
At early time, within the temporal bound $-{1\over \sqrt{\Lambda}} < t < 0$ ($\Lambda$ being the bare cosmological constant), we know that $\hat\alpha(t) > \hat\beta(t) > {\hat\alpha(t)\over 9}$ and so in the domain $\hat\beta(t) < \hat\alpha(t) < 9\hat\beta(t)$ one might try to identify ${\rm dom}~\hat{j}(t)$ with ${\hat\alpha(t) - \hat\beta(t)\over 6}$. From our ans\"atze \eqref{samthigap}
such a behavior is not supported at late time, where we expect $\hat\alpha(t) = \hat\beta(t)$, because it makes ${\rm dom}~\hat{j}(t) = 0$. Furthermore, an integral over the compact eight-manifold would suggest that the integral of the temporally invariant part of ${\bf \Lambda}_8$ vanishes, {\it i.e.} $\int_{{\cal M}_8} {\bf \Lambda}_8^{(0)} = 0$ with ${\cal M}_8 = {\cal M}_4 \times {\cal M}_2 \times {\mathbb{T}^2\over {\cal G}}$. On the other hand, the under-bracket identification in eq. (4.26) of \cite{coherbeta3} provides the following scalings:

{\footnotesize
\bg\label{brettbess2}
\hat{l}_{[mn}^{[pq}(t) \oplus \hat{l}_{\alpha\beta]}^{ab]}(t) = 2 + \hat\chi^{(1)}(t), ~~~~ \theta_{nl}(t) = {8\over 3} + \hat\chi^{(1)}(t) + {7\hat\beta(t) - 3\hat\alpha(t)\over 12},~~~~\hat{j}(t) = 2 + \hat\chi^{(1)}(t), \nd}
which again differ from the values $l_{mn}^{pq} = l_{\alpha\beta}^{ab} = 1$ (and the permutation of the indices) from \cite{coherbeta3} by small quantities. Note that we have extended the under-bracket identification from eq. (4.26) in \cite{coherbeta3} to include the dynamical M2 branes also, as reflected in \eqref{cata00}. The relation between $\hat{j}(t)$ and $\hat{\chi}^{(1)}(t)$ confirms that $\hat{j}(t) > 0$. An integral over the eight-manifold ${\cal M}_8$ now provides precisely the anomaly cancellation condition of 
\cite{becker2}.  Additionally, the quantum scaling $\theta_{nl}$ when compared to \eqref{brittbaba} provides the necessary contributing factors from \eqref{botsuga}. 

An alternative way to address this problem is to look at the bounds on the flux components imposed by the EFT constraints as given in {\bf Table \ref{micandi0_0}}. The identification then takes the following form:
\bg\label{zoeitfront}
{\rm Row}~1 = {\rm Row}~2 = {\rm Row}~3 = {\rm Row}~9, \nd
which would scale as $\left({g_s\over {\rm H}(y){\rm H}_o({\bf x})}\right)^{0^\pm}$ with the fluxes scaling as in \eqref{tundegolab} and \eqref{brettbess}. This gives us $\theta_{nl} \ge {2\over 3}^\pm$. At the order $\left({g_s\over {\rm H}(y){\rm H}_o({\bf x})}\right)^{2^\pm}$ we expect the identification \eqref{zoeitfront} to extend to include Row 6 also, giving us $\theta_{nl} \ge {8\over 3}^\pm$. Beyond this order, the non-perturbative corrections enter the story.

When the derivatives act along the temporal direction in \eqref{crawlquif}, the story gets a bit more involved because of the inherent temporal dependence of the scalings appearing in the first two rows of {\bf Table \ref{crecraqwiff1}}. If we define:
\bg\label{vespodana}
&&{\rm L}_{0i}^{jm} \equiv \hat{l}_{e0i}^{jm} + 4 - {3\hat\zeta_e(t)\over 2} + \hat\sigma_e(t) + {\hat\alpha_e(t) + \hat\beta_e(t)\over 2} + {\hat\eta_e(t)\over 2} \nonumber\\
&&\Theta_{0i}^{jm} \equiv \theta_{nl} - \hat{l}_{e0i}^{jm} - {14\over 3} + {3\hat\zeta_e(t)\over 2}  + 2\hat\sigma_e(t) + {\hat\alpha_e(t) + \hat\beta_e(t)\over 2} + {\hat\eta_e(t)\over 2}, \nd
then the temporal derivatives on them may be worked out using the strategy laid out earlier. This is shown in {\bf Table \ref{ccqwiff0}} using the $\gamma_{1, 2}\big[{\hat\alpha_e(t) + \hat\beta_e(t)\over 2}\big]$ function from {\bf Table \ref{jessrog}}.

\begin{table}[H]  
 \begin{center}
\resizebox{\columnwidth}{!}{%
\renewcommand{\arraystretch}{1.5}
\begin{tabular}{|c||c||c|}\hline Rows & ${\bf G}_{0ijm}$ tensors &  ${g_s\over {\rm HH}_o}$ scaling \\ \hline\hline
1 & $\partial_0\left(\sqrt{-{\bf g}_{11}} {\bf G}_{0i'j'm'}{\bf g}^{00} {\bf g}^{i'i}{\bf g}^{j'j} {\bf g}^{m'm}
\epsilon_{0ijmnpq\alpha\beta ab}\right)$ & ${\rm dom}\left({\rm L}_{0i}^{jm} -{\log\left(\pm \dot{\rm L}_{0i}^{jm}\vert\log~\bar{g}_s\vert\right)\over \vert\log~\bar{g}_s\vert}, {\rm L}_{0i}^{jm} - 1 +\gamma_{1, 2}\big[{\hat\alpha_e(t) + \hat\beta_e(t)\over 2}\big] - {\log\vert{\rm L}_{0i}^{jm}\vert\over \vert\log~\bar{g}_s\vert}\right)$ \\ \hline 
2 & $\partial_0\left(\sqrt{-{\bf g}_{11}} \mathbb{Y}_4^{0ijm} 
\epsilon_{0ijmnpq\alpha\beta ab}\right)$ & ${\rm dom}\left({\Theta}_{0i}^{jm} -{\log\left(\pm \dot{\Theta}_{0i}^{jm}\vert\log~\bar{g}_s\vert\right)\over \vert\log~\bar{g}_s\vert}, {\Theta}_{0i}^{jm} - 1 +\gamma_{1, 2}\big[{\hat\alpha_e(t) + \hat\beta_e(t)\over 2}\big] - {\log\vert{\Theta}_{0i}^{jm}\vert\over \vert\log~\bar{g}_s\vert}\right)$ \\ \hline 
\end{tabular}}
\renewcommand{\arraystretch}{1}
\end{center}
 \caption[]{\Su ${g_s\over {\rm H}(y){\rm H}_o({\bf x})}$ scalings of the temporal derivatives of the first two terms in \eqref{crawlquif} using the definitions from \eqref{vespodana}.} 
  \label{ccqwiff0}
 \end{table}

\noindent Since $\gamma_1 = 1 + {\cal O}(\log)$ and $\gamma_2 = {\hat\alpha_e(t) + \hat\beta_e(t)\over 4} + {\cal O}(\log)$, the dominant scalings in the two rows of {\bf Table \ref{ccqwiff0}} are respectively ${\rm L}_{0i}^{jm} - 1$ and $\Theta_{0i}^{jm} - 1$. Interestingly there are no M2 branes contributing to the EOM from \eqref{crawlquif} because they would involve M2 branes wrapping one-cycles on ${\cal M}_4$, and additionally breaking the four-dimensional isometry in the Heterotic side.
The bound on the scalings of the flux components now becomes:
\bg\label{tundegolab2}
\hat{l}_{e[0n}^{[pq} \oplus \hat{l}_{e\alpha\beta]}^{ab]} \ge -1, \nd
which may be read up from {\bf Table \ref{micandi0_0}}. At order $\left({g_s\over {\rm H}(y){\rm H}_o({\bf x})}\right)^{-1^\pm}$, the two rows in {\bf Table \ref{ccqwiff0}} is identified with the scaling in Row 4 of {\bf Table \ref{crecraqwiff1}}. Finally at order  $\left({g_s\over {\rm H}(y){\rm H}_o({\bf x})}\right)^{+1^\pm}$, the scaling identifications become:
\bg\label{arabicsar}
\left({\rm Row}~1 = {\rm Row}~2\right)_{\bf Table ~\ref{ccqwiff0}} = \left({\rm Row}~4 = {\rm Row}~6\right)_{\bf Table ~\ref{crecraqwiff1}}, \nd
consistent with the lower bounds on the scalings of the flux components from {\bf Table \ref{micandi0_0}}. For both cases, $\theta_{nl}$ continues to have the lower bounds of ${2\over 3}^\pm$ and ${8\over 3}^\pm$ seen earlier. Beyond these orders, we expect the non-perturbative corrections to add in. 

When the derivatives act along the spatial ${\bf R}^2$ directions, Rows 1 and 2 from {\bf Table \ref{crecraqwiff1}} do not contribute because of the underlying duality sequence as given in {\bf Table \ref{milleren4}}. Since the flux products from Row 5 in {\bf Table \ref{crecraqwiff1}} are bounded by $-1^\pm$, they cannot contribute and the spatial parts should vanish. At order $\left({g_s\over {\rm H}(y){\rm H}_o({\bf x})}\right)^{+1^\pm}$ the equality between rows 5 and 8. (In fact once we allow for spatial dependence, as in \cite{coherbeta3}, at this order the equality will extend to rows 1 and 2.)

In a similar vein the tensors appearing in the EOM associated with ${\bf G}_{0ij\alpha}$ take the form given by {\bf Table \ref{crecraqwiff2}}.
\begin{table}[tb]  
 \begin{center}
\resizebox{\columnwidth}{!}{%
\renewcommand{\arraystretch}{2.3}
}
\renewcommand{\arraystretch}{1}
\end{center}
 \caption[]{\Su ${g_s\over {\rm H}(y){\rm H}_o({\bf x})}$ scalings of the various terms of \eqref{crawlquif} contributing to the EOM of ${\bf G}_{0ij\alpha}$. The quantum scaling $\theta_{nl}$ is given by \eqref{brittbaba}.} 
  \label{crecraqwiff2}
 \end{table}
We see that there are minor changes in some of the rows compared to the ones in {\bf Table \ref{crecraqwiff1}}. Naively, this would mean that the results from \eqref{brettbess} and \eqref{brettbess2} should not change much. However from the duality sequence in {\bf Table \ref{milleren4}}, we can only allow the metric and the flux components to be functions of the coordinates of ${\cal M}_4$ and $g_s$, implying that the $(\alpha, i)$ derivative terms ({\it i.e.} the first two terms and the last term) in \eqref{crawlquif} should vanish. For the $\alpha$ derivative pieces, looking at {\bf Table \ref{crecraqwiff2}}, suggests that the flux product terms in Row 3 should vanish at order $\left({g_s\over {\rm H}(y){\rm H}_o({\bf x})}\right)^{0^\pm}$. At order $\left({g_s\over {\rm H}(y){\rm H}_o({\bf x})}\right)^{+2^\pm}$, we can equate rows 3 and 6 with 9. This leads to \eqref{brettbess2} without the $\theta_{nl}$ term. Not surprisingly, this is also consistent with the anomaly cancellation condition of \cite{becker2}. On the other hand, with the spatial derivative along ${\bf R}^2$, the first and the second rows of {\bf Table \ref{crecraqwiff2}} do not contribute and to order $\left({g_s\over {\rm H}(y){\rm H}_o({\bf x})}\right)^{-1^\pm}$, the flux products in Row 5 should vanish. At order $\left({g_s\over {\rm H}(y){\rm H}_o({\bf x})}\right)^{+1^\pm}$, we expect equality between rows 5 and 8.

More important here is the temporal derivative, and as we saw in {\bf Table \ref{ccqwiff0}}, the scalings in the first two rows of {\bf Table \ref{crecraqwiff2}} shifts by $-1$. This implies that the lower bounds are now $-1^\pm$ for the first two rows, and therefore to order 
$\left({g_s\over {\rm H}(y){\rm H}_o({\bf x})}\right)^{-1^\pm}$, we can equate them to the fourth row in {\bf Table \ref{crecraqwiff2}} with $\theta_{nl} \ge {2\over 3}^\pm$ in \eqref{brittbaba}. Finally to order
$\left({g_s\over {\rm H}(y){\rm H}_o({\bf x})}\right)^{+1^\pm}$, we can allow the following equality from the fourth column of {\bf Table \ref{crecraqwiff2}}:
\bg\label{arabsera}
-1^\pm + {\rm Row}~ 1 = -1^\pm + {\rm Row}~2 = {\rm Row} ~4 = {\rm Row}~7, \nd
implying that the quantum series \eqref{botsuga} now contributes for 
$\theta_{nl} \ge {8\over 3}^\pm$. Beyond these orders, the non-perturbative terms should also start participating.

Our final set of flux components are the ones with legs along the toroidal directions. They are of the form ${\bf G}_{0ija}$. We can again collect all the tensors contributing to the EOMs of ${\bf G}_{0ija}$ from 
\eqref{crawlquif} as in {\bf Table \ref{ccquiff3}}.
\begin{table}[tb]  
 \begin{center}
\resizebox{\columnwidth}{!}{%
\renewcommand{\arraystretch}{2.3}
}
\renewcommand{\arraystretch}{1}
\end{center}
 \caption[]{ \Su ${g_s\over {\rm H}(y){\rm H}_o({\bf x})}$ scalings of the various terms of \eqref{crawlquif} contributing to the EOM of ${\bf G}_{0ija}$.} 
  \label{ccquiff3}
 \end{table}
Since all fields and flux components are independent of the $w^a \in {\mathbb{T}^2\over {\cal G}}$ directions, the first two rows in {\bf Table \ref{ccquiff3}} do not contribute to \eqref{crawlquif} when we take the derivatives along the toroidal directions. At order 
$\left({g_s\over {\rm H}(y){\rm H}_o({\bf x})}\right)^{0^\pm}$, we can equate rows 3 and 9 with ${\rm dom}~j = 0$. Once we go to order 
$\left({g_s\over {\rm H}(y){\rm H}_o({\bf x})}\right)^{+2^\pm}$, we can allow the following identifications:
\bg\label{maimandi}
{\rm Row}~ 3 = {\rm Row}~6 = {\rm Row}~9, \nd
giving us back the full anomaly cancellation condition from \cite{becker2}. Note that there are no non-perturbative corrections to the anomaly cancellation condition because of the form of the EOM in \eqref{crawlquif}. The analysis remains somewhat similar when we take the derivatives along the spatial ${\bf R}^2$ directions. The difference being the absence of the M2 brane contributions. At order $\left({g_s\over {\rm H}(y){\rm H}_o({\bf x})}\right)^{0^\pm}$ we expect an equality between rows 5 and 8. Below this order, the product of the fluxes vanish. 

The story changes quite a bit once we take the temporal derivatives. The scalings in the first two rows in {\bf Table \ref{ccquiff3}} are now bounded from below by $-2^\pm$. Again the M2 branes do not participate, and so to order $\left({g_s\over {\rm H}(y){\rm H}_o({\bf x})}\right)^{-2^\pm}$, we can equate the first two rows with the fourth row in {\bf Table \ref{ccquiff3}} with $\theta_{nl} \ge {2\over 3}^\pm$ in \eqref{brittbaba}. When we go to order $\left({g_s\over {\rm H}(y){\rm H}_o({\bf x})}\right)^{0^\pm}$, we expect:
\bg\label{arabsera2}
-1^\pm + {\rm Row}~ 1 = -1^\pm + {\rm Row}~2 = {\rm Row} ~4 = {\rm Row}~7, \nd
in {\bf Table \ref{ccquiff3}} with $\theta_{nl} \ge {8\over 3}^\pm$. The non-perturbative corrections should start contributing from this order. We should also compare this and the earlier results for the scalings to the ones from \cite{coherbeta3}. Here we have only investigated the consequences from the bounds imposed on the flux scalings from {\bf Table \ref{micandi0_0}}, but did not study the exact scalings of the flux components. Once we do that, the results would be closer to what we had in \cite{coherbeta3}. Unfortunately due to the sheer number of flux components appearing here, this is not an easy exercise, and therefore we will address the precise flux scalings in a separate work. Here we can discuss the range of possible values. For example, using the analysis presented earlier, let us express the actual scalings as:
\bg\label{argotrani}
\hat{l}_{e0i}^{jm} = -4^\pm + \vert x_1\vert, ~~~~ \hat{l}_{e0i}^{j\alpha} = -4^\pm + \vert x_2\vert, ~~~~ \hat{l}_{e0i}^{ja} = -3^\pm + \vert x_3\vert,\nd
where the lower bounds are taken from {\bf Table \ref{micandi0_0}}. Now comparing all the results from {\bf Tables \ref{crecraqwiff1}, \ref{ccqwiff0}, \ref{crecraqwiff2}} and {\bf \ref{ccquiff3}}, one can easily see that:
\bg\label{argothai}
0^\pm \le (x_1, x_2, x_3) \le 2^\pm, \nd
providing us a range for the scalings that are not only consistent with the underlying EFT but also with all the EOMs. In \cite{coherbeta3} we took $x_1 = x_2 = 0$ and $x_3 = 1$, but the analysis is more technically challenging now because of the presence of small parameters like 
$\hat\alpha_e(t), \hat\beta_e(t), \hat\sigma_e(t), \hat\zeta_e(t)$ and $\hat\eta_e(t)$. Nevertheless we can make the following ans\"atze for the G-flux components:
\begin{empheq}[box={\mybluebox[5pt]}]{align}
& {\bf G}_{0ija}({\bf x}, y; g_s(t)) = \sum_{l = 0}^6\sum_{k_l = 0}^\infty {\cal G}^{(k, l)}_{0ija}({\bf x}, y) \left({g_s\over {\rm HH}_o}\right)^{-3^\pm + {l\over 3} + {2k_l\over 3}\vert {\rm L}_{0i}^{ja}(k_l; t)\vert}\nonumber\\
&{\bf G}_{0ij{\rm M}}({\bf x}, y; g_s(t)) = \sum_{l = 0}^6\sum_{k_l = 0}^\infty {\cal G}^{(k, l)}_{0ij{\rm M}}({\bf x}, y) \left({g_s\over {\rm HH}_o}\right)^{-4^\pm + {l\over 3} + {2k_l\over 3}\vert {\rm L}_{\rm 0i}^{j{\rm M}}(k_l; t)\vert}
\label{usher1}
\end{empheq}
where ${\rm L}_{\rm 0i}^{j{\rm M}}(k_l; t)$ and ${\rm L}_{\rm 0i}^{ja}(k_l; t)$ contain all the perturbative and the non-perturbative corrections as given by \eqref{andyrey} and \eqref{malenbego}, including the ones discussed in section \ref{sec4.5.2}.

Note that the ans\"atze presented in \eqref{usher1} is not only consistent with the underlying EFT, but would also solve the EOMs with, at least, one of the choice of $l$. All the small corrections, coming from the inclusion of $(\hat\alpha_e(t), \hat\beta_e(t), \hat\sigma_e(t), \hat\eta_e(t), \hat\zeta_e(t))$ and the log terms, are included in the $\pm$ superscripts. They may be easily quantified from {\bf Table \ref{micandi0_0}}, {\bf Tables \ref{niksmit1}} to {\bf \ref{niksmit250}}, and the analysis of section \ref{sec4.5.1}. This is the closest that we can get to quantifying precisely the flux components in the heterotic ${\rm E}_8 \times {\rm E}_8$. (The $SO(32)$ case can be easily worked out from here.)
Therefore, in the following, our strategy would be to map out the range for the scalings of the flux components and from there express the behavior of the flux components as a series. Herein awaits a surprise as we shall soon see.

\subsection{EOMs for the flux components ${\bf G}_{\rm MNPQ}$ and ${\bf G}_{\rm 0MNP}$ \label{blackbond2}}

The flux components ${\bf G}_{\rm MNPQ}$ and ${\bf G}_{\rm 0MNP}$ for $({\rm M, N}) \in {\cal M}_4 \times {\cal M}_2$ would now be ${\bf G}_{mnpq}$, ${\bf G}_{mnp\alpha}$, $ 
{\bf G}_{mn\alpha\beta}$, $ {\bf G}_{0mnp}$, ${\bf G}_{0mn\alpha}$ and ${\bf G}_{0m\alpha\beta}$ . The tensors participating in the EOM for all the flux components from \eqref{crawlquif}, may be expressed as in {\bf Tables \ref{ccrani1}, \ref{ccrani2}} and {\bf \ref{ccrani3}}. 
\begin{table}[tb]  
 \begin{center}
\resizebox{\columnwidth}{!}{%
\renewcommand{\arraystretch}{2.5}
\begin{tabular}{|c||c||c||c||c|}\hline Rows & ${\bf G}_{mnpq}$ tensors & Forms & ${g_s\over {\rm HH}_o}$ scalings & Lower bounds \\ \hline\hline
1 & $\big(\mathbb{T}_7^{(f)}\big)_{0ij\alpha\beta ab}$ & $\sqrt{-{\bf g}_{11}} {\bf G}_{m'n'p'q'}{\bf g}^{m'm} {\bf g}^{n'n}{\bf g}^{p'p} {\bf g}^{q'q}
\epsilon_{mnpq0ij\alpha\beta ab}$ & $\hat{l}_{emn}^{pq} - 2 + {3\hat\zeta_e(t)\over 2} - 2\hat\sigma_e(t) + {\hat\alpha_e(t) + \hat\beta_e(t)\over 2}  + {\hat\eta_e(t)\over 2}$ & $-3^\pm$ \\ \hline 
2 & $\big(\mathbb{T}_7^{(q)}\big)_{0ij\alpha\beta ab}$ & $\sqrt{-{\bf g}_{11}} \mathbb{Y}_4^{mnpq} 
\epsilon_{mnpq0ij\alpha\beta ab}$ & $\theta_{nl} - \hat{l}_{emn}^{pq} - {14\over 3} + {3\hat\zeta_e(t)\over 2}  + 2\hat\sigma_e(t) + {\hat\alpha_e(t) + \hat\beta_e(t)\over 2} + {\hat\eta_e(t)\over 2}$ & $-3^\pm\big({2\over 3}^\pm\big)$ \\ \hline 
3 & $\big(\mathbb{T}^{(m)}_8\big)_{m0ij\alpha\beta ab}$ & ${\bf G}_{[m0ij} {\bf G}_{\alpha\beta ab]}$ & $ \hat{l}_{e[m0}^{[ij} \oplus \hat{l}_{e\alpha\beta]}^{ab]}$ & $-3^\pm$
\\ \hline 
4 & $\big(\mathbb{X}^{(m)}_8\big)_{m0ij\alpha\beta ab}$ & ${\rm dom} \left({\rm tr} ~\mathbb{R}_{\rm tot}^4 
- {1\over 4} \left({\rm tr}~\mathbb{R}_{\rm tot}^2\right)^2\right)_{m0ij\alpha\beta ab}$ & $-1 + \hat\chi^{(8)}_e(t)$ & $-1^\pm$ \\ \hline
\end{tabular}}
\renewcommand{\arraystretch}{1}
\end{center}
 \caption[]{ \Su ${g_s\over {\rm H}(y){\rm H}_o({\bf x})}$ scalings of the various terms of \eqref{crawlquif} contributing to the EOM of ${\bf G}_{mnpq}$.} 
  \label{ccrani1}
 \end{table}
 The $\mathbb{X}_8$ polynomials are now computed using the curvature two-forms from {\bf Table \ref{niksmit1}} to {\bf Table \ref{niksmit250}}. They take the form:

 {\footnotesize
\bg\label{devlee2}
&& {\rm dom}~\Big[\mathbb{X}^{(m)}_8({\bf x}, y, w; g_s)\Big]_{m0ij\alpha\beta ab}= \left({g_s\over {\rm H}(y){\rm H}_o({\bf x})}\right)^{-1 + \hat\chi^{(8)}_e(t)} \left[\widetilde{\mathbb{X}}^{(m)}_8({\bf x}, y, w)\right]_{m0ij\alpha\beta ab}\nonumber\\
&& {\rm dom}~\Big[\mathbb{X}^{(m)}_8({\bf x}, y, w; g_s)\Big]_{0ijmq\beta ab}= \left({g_s\over {\rm H}(y){\rm H}_o({\bf x})}\right)^{-1 + \hat\chi^{(9)}_e(t)} \left[\widetilde{\mathbb{X}}^{(m)}_8({\bf x}, y, w)\right]_{0ijmq\beta ab}\nonumber\\
&& {\rm dom}~\Big[\mathbb{X}^{(\alpha)}_8({\bf x}, y, w; g_s)\Big]_{0ijq\alpha\beta ab}= \left({g_s\over {\rm H}(y){\rm H}_o({\bf x})}\right)^{-1 + \hat\chi^{(10)}_e(t)} \left[\widetilde{\mathbb{X}}^{(m)}_8({\bf x}, y, w)\right]_{0ijq\alpha\beta ab}\nonumber\\
&& {\rm dom}~\Big[\mathbb{X}^{(m)}_8({\bf x}, y, w; g_s)\Big]_{0ijmpq ab}= \left({g_s\over {\rm H}(y){\rm H}_o({\bf x})}\right)^{-1 + \hat\chi^{(11)}_e(t)} \left[\widetilde{\mathbb{X}}^{(m)}_8({\bf x}, y, w)\right]_{0ijmpq ab}\nonumber\\
&& {\rm dom}~\Big[\mathbb{X}^{(\alpha)}_8({\bf x}, y, w; g_s)\Big]_{0ijpq\alpha ab}= \left({g_s\over {\rm H}(y){\rm H}_o({\bf x})}\right)^{-1 + \hat\chi^{(12)}_e(t)} \left[\widetilde{\mathbb{X}}^{(m)}_8({\bf x}, y, w)\right]_{0ijpq\alpha ab},
\nd}
which may be compared to \eqref{devlee} earlier. For the ${\bf G}_{mnpq}$ flux components, it is easy to see from {\bf Table \ref{ccrani1}} that to order $\left({g_s\over {\rm H}(y) {\rm H}_o({\bf x})}\right)^{-3^\pm}$, the first three rows may be equated to each other for $\theta_{nl} \ge {2\over 3}^\pm$ in \eqref{brittbaba}. Once we go to order $\left({g_s\over {\rm H}(y) {\rm H}_o({\bf x})}\right)^{-1^\pm}$, all four rows get equated with $\theta_{nl} \ge {8\over 3}^\pm$. Note that in eq. (4.46) of \cite{coherbeta3}, the identifications are done differently. This is because therein we had taken $l_{0i}^{jm} = -4, l_{mn}^{ab} = 1$ and $l_{mn}^{pq} = 1$. Here, as mentioned earlier, we are only looking at the lower bounds of the scalings of the G-flux components. Comparing the first and the third rows in {\bf Table \ref{ccrani1}} produces the lower bound of $\hat{l}_{emn}^{pq} \ge -1^\pm$ confirming the result from {\bf Table \ref{micandi0_0}}. This will continue to be the case for the other two set of components.

When we take the flux components ${\bf G}_{mnp\alpha}$, which would be ${\bf G}_{mnp\theta_1}$ and ${\bf G}_{mnp\theta_2}$, the tensors contributing to \eqref{crawlquif} is now given by {\bf Table \ref{ccrani2}}. 
\begin{table}[tb]  
 \begin{center}
\resizebox{\columnwidth}{!}{%
\renewcommand{\arraystretch}{2.5}
\begin{tabular}{|c||c||c||c||c|}\hline Rows & ${\bf G}_{mnp\alpha}$ tensors & Forms & ${g_s\over {\rm HH}_o}$ scalings & Lower bounds \\ \hline\hline
1 & $\big(\mathbb{T}_7^{(f)}\big)_{0ijq\beta ab}$ & $\sqrt{-{\bf g}_{11}} {\bf G}_{m'n'p'\alpha'}{\bf g}^{m'm} {\bf g}^{n'n}{\bf g}^{p'p} {\bf g}^{\alpha'\alpha}
\epsilon_{mnpq0ij\alpha\beta ab}$ & $\hat{l}_{emn}^{p\alpha} - 2 + {3\hat\zeta_e(t)\over 2} + 2\hat\sigma_e(t) + {\hat\alpha_e(t) + \hat\beta_e(t)\over 2} - (\hat\alpha_e(t), \hat\beta_e(t))  + {\hat\eta_e(t)\over 2}$ & $-3^\pm$ \\ \hline 
2 & $\big(\mathbb{T}_7^{(q)}\big)_{0ijq\beta ab}$ & $\sqrt{-{\bf g}_{11}} \mathbb{Y}_4^{mnp\alpha} 
\epsilon_{mnpq0ij\alpha\beta ab}$ & $\theta_{nl} - \hat{l}_{emn}^{p\alpha} - {14\over 3} + {3\hat\zeta_e(t)\over 2}  + 2\hat\sigma_e(t) + {\hat\alpha_e(t) + \hat\beta_e(t)\over 2} + {\hat\eta_e(t)\over 2}$ & $-3^\pm\big({2\over 3}^\pm\big)$ \\ \hline 
3 & $\big(\mathbb{T}^{(m)}_8\big)_{m0ijq\beta ab}$ & ${\bf G}_{[m0ij} {\bf G}_{q\beta ab]}$ & $ \hat{l}_{e[m0}^{[ij} \oplus \hat{l}_{eq\beta]}^{ab]}$ & $-3^\pm$
\\ \hline 
4 & $\big(\mathbb{T}^{(\alpha)}_8\big)_{0ijq\alpha\beta ab}$ & ${\bf G}_{[0ijq} {\bf G}_{\alpha\beta ab]}$ & $ \hat{l}_{e[0i}^{[jq} \oplus \hat{l}_{e\alpha\beta]}^{ab]}$ & $-3^\pm$
\\ \hline 
5 & $\big(\mathbb{X}^{(m)}_8\big)_{m0ijq\beta ab}$ & ${\rm dom} \left({\rm tr} ~\mathbb{R}_{\rm tot}^4 
- {1\over 4} \left({\rm tr}~\mathbb{R}_{\rm tot}^2\right)^2\right)_{m0ijq\beta ab}$ & $-1 + \hat\chi^{(9)}_e(t)$ & $-1^\pm$ \\ \hline
6 & $\big(\mathbb{X}^{(\alpha)}_8\big)_{0ijq\alpha\beta ab}$ & ${\rm dom} \left({\rm tr} ~\mathbb{R}_{\rm tot}^4 
- {1\over 4} \left({\rm tr}~\mathbb{R}_{\rm tot}^2\right)^2\right)_{0ijq\alpha\beta ab}$ & $-1 + \hat\chi^{(10)}_e(t)$ & $-1^\pm$ \\ \hline
\end{tabular}}
\renewcommand{\arraystretch}{1}
\end{center}
 \caption[]{ \Su ${g_s\over {\rm H}(y){\rm H}_o({\bf x})}$ scalings of the various terms of \eqref{crawlquif} contributing to the EOM of ${\bf G}_{mnp\alpha}$.} 
  \label{ccrani2}
 \end{table}
Despite the difference in the scalings for the various rows in {\bf Table \ref{ccrani2}}, the story is very similar to the one from {\bf Table \ref{ccrani1}}. The identifications for the derivatives along ${\cal M}_4$ directions remain exactly the same as before. For derivatives along ${\cal M}_2$ directions, the first two rows in {\bf Table \ref{ccrani2}} do not contribute. At order $\left({g_s\over {\rm H}(y){\rm H}_o({\bf x})}\right)^{-3^\pm}$, the flux products from the fourth row vanish, and only at order $\left({g_s\over {\rm H}(y){\rm H}_o({\bf x})}\right)^{-1^\pm}$, the scalings from fourth and the sixth rows in {\bf Table \ref{ccrani2}} get identified.

For the ${\bf G}_{mn\alpha\beta}$ flux components, the matching of the scales from {\bf Table \ref{ccrani3}} remains similar to what we saw for 
the case with ${\bf G}_{mnp\alpha}$ in {\bf Table \ref{ccrani2}}. For the derivatives along ${\cal M}_4$ directions, the rows 1, 2, 3 and 5 may be identified with $\theta_{nl} \ge {8\over 3}^\pm$. For derivatives along ${\cal M}_2$, there is a sub-identification as before. It is also easy to confirm, by equating the first two rows in {\bf  Tables \ref{ccrani1}, \ref{ccrani2}} and {\bf \ref{ccrani3}}, that:
\bg\label{sarakallu}
\theta_{nl} = 2\hat{l}_{e{\rm MN}}^{\rm PQ} + {8\over 3} -(2 + \delta_{{\rm P}p} + \delta_{{\rm Q}q})\hat\sigma_e(t) - (\hat\alpha_e(t), \hat\beta_e(t))( \delta_{{\rm P}\alpha} + \delta_{{\rm Q}\beta}), \nd
which matches exactly with \eqref{brittbaba} for $l_{42} = l_{44} = l_{45} = 2$. This is an important consistency check that one may extract from \eqref{crawlquif}. In a similar vein,
\begin{table}[tb]  
 \begin{center}
\resizebox{\columnwidth}{!}{%
\renewcommand{\arraystretch}{2.5}
\begin{tabular}{|c||c||c||c||c|}\hline Rows & ${\bf G}_{mn\alpha\beta}$ tensors & Forms & ${g_s\over {\rm HH}_o}$ scalings & Lower bounds \\ \hline\hline
1 & $\big(\mathbb{T}_7^{(f)}\big)_{0ijpqab}$ & $\sqrt{-{\bf g}_{11}} {\bf G}_{m'n'\alpha'\beta'}{\bf g}^{m'm} {\bf g}^{n'n}{\bf g}^{\alpha'\alpha}{\bf g}^{\beta'\beta}
\epsilon_{mnpq0ij\alpha\beta ab}$ & $\hat{l}_{emn}^{\alpha\beta} - 2 + {3\hat\zeta_e(t)\over 2} + {\hat\alpha_e(t) + \hat\beta_e(t)\over 2} - (\hat\alpha_e(t), \hat\beta_e(t))  + {\hat\eta_e(t)\over 2}$ & $-3^\pm$ \\ \hline 
2 & $\big(\mathbb{T}_7^{(q)}\big)_{0ijpqab}$ & $\sqrt{-{\bf g}_{11}} \mathbb{Y}_4^{mn\alpha\beta} 
\epsilon_{mnpq0ij\alpha\beta ab}$ & $\theta_{nl} - \hat{l}_{emn}^{\alpha\beta} - {14\over 3} + {3\hat\zeta_e(t)\over 2}  + 2\hat\sigma_e(t) + {\hat\alpha_e(t) + \hat\beta_e(t)\over 2} + {\hat\eta_e(t)\over 2}$ & $-3^\pm\big({2\over 3}^\pm\big)$ \\ \hline 
3 & $\big(\mathbb{T}^{(m)}_8\big)_{m0ijpqab}$ & ${\bf G}_{[m0ij} {\bf G}_{pqab]}$ & $ \hat{l}_{e[m0}^{[ij} \oplus \hat{l}_{epq]}^{ab]}$ & $-3^\pm$
\\ \hline 
4 & $\big(\mathbb{T}^{(\alpha)}_8\big)_{0ijpq\alpha ab}$ & ${\bf G}_{[0ijp} {\bf G}_{q\alpha ab]}$ & $ \hat{l}_{e[0i}^{[jp} \oplus \hat{l}_{eq\alpha]}^{ab]}$ & $-3^\pm$
\\ \hline 
5 & $\big(\mathbb{X}^{(m)}_8\big)_{m0ijpqab}$ & ${\rm dom} \left({\rm tr} ~\mathbb{R}_{\rm tot}^4 
- {1\over 4} \left({\rm tr}~\mathbb{R}_{\rm tot}^2\right)^2\right)_{m0ijpq ab}$ & $-1 + \hat\chi^{(11)}_e(t)$ & $-1^\pm$ \\ \hline
6 & $\big(\mathbb{X}^{(\alpha)}_8\big)_{0ijpq\alpha ab}$ & ${\rm dom} \left({\rm tr} ~\mathbb{R}_{\rm tot}^4 
- {1\over 4} \left({\rm tr}~\mathbb{R}_{\rm tot}^2\right)^2\right)_{0ijpq\alpha ab}$ & $-1 + \hat\chi^{(12)}_e(t)$ & $-1^\pm$ \\ \hline
\end{tabular}}
\renewcommand{\arraystretch}{1}
\end{center}
 \caption[]{ \Su ${g_s\over {\rm H}(y){\rm H}_o({\bf x})}$ scalings of the various terms of \eqref{crawlquif} contributing to the EOM of ${\bf G}_{mn\alpha\beta}$.} 
  \label{ccrani3}
 \end{table}
to determine the
exact scalings we can follow the procedure as laid out earlier, namely, define the three scalings in the following way:
\bg\label{selfhath}
\hat{l}_{emn}^{pq} = -1^\pm + \vert x_4\vert, ~~~ \hat{l}_{emn}^{p\alpha} = -1^\pm + \vert x_5\vert, ~~~ \hat{l}_{emn}^{\alpha\beta} = -1^\pm + \vert x_6\vert, \nd
where $x_i = x_i(t(g_s))$ because of their dependence on the small parameters like 
$\hat\alpha_e(t)$, $ \hat\beta_e(t)$, $ \hat\sigma_e(t)$, $ \hat\zeta_e(t)$ and $\hat\eta_e(t)$. Now using the results from {\bf Tables \ref{ccrani1}, \ref{ccrani2}} and {\bf \ref{ccrani3}}, it is not too hard to infer that:
\bg\label{argoador}
0^\pm \le (x_4, x_5, x_6) \le 2^\pm, \nd
which is similar to what we had in \eqref{argothai}. However due to the $\pm$ superscripts, they are not necessarily required to be identical. In \cite{coherbeta3} we had taken $x_4 = x_5 = x_6 = 2$, but here it is much harder to determine the precise values. Nevertheless, as in \eqref{usher1}, we can still express the ans\"atze for the ${\bf G}_{\rm MNPQ}$ flux components as:
\begin{empheq}[box={\mybluebox[5pt]}]{equation}
{\bf G}_{{\rm MNPQ}}({\bf x}, y; g_s(t)) = \sum_{l = 0}^6\sum_{k_l = 0}^\infty {\cal G}^{(k, l)}_{{\rm MNPQ}}({\bf x}, y) \left({g_s\over {\rm HH}_o}\right)^{-1^\pm + {l\over 3} + {2k_l\over 3}\vert {\rm L}_{\rm MN}^{\rm PQ}(k_l; t)\vert} 
\label{usher0}
\end{empheq}
where ${\rm L}_{\rm MN}^{\rm PQ}(k_l; t)$ contains all the perturbative and the non-perturbative corrections discussed in \eqref{andyrey}, \eqref{malenbego} and section \ref{sec4.5.1}.

The temporal derivatives would appear once we take the flux components to be ${\bf G}_{\rm 0MNP}$ with $({\rm M, N}) \in {\cal M}_4 \times {\cal M}_2$. This means we are looking at flux components of the form ${\bf G}_{0mnp}, {\bf G}_{0mn\alpha}$ and ${\bf G}_{0m\alpha\beta}$. For ${\bf G}_{0mnp}$, the tensors contributing to \eqref{crawlquif} may be collected in {\bf Table \ref{ccrani4}}. 

\begin{table}[tb]  
 \begin{center}
\resizebox{\columnwidth}{!}{%
\renewcommand{\arraystretch}{2.5}
\begin{tabular}{|c||c||c||c||c|}\hline Rows & ${\bf G}_{0mnp}$ tensors & Forms & ${g_s\over {\rm HH}_o}$ scalings & Lower bounds \\ \hline\hline
1 & $\big(\mathbb{T}_7^{(f)}\big)_{ijq\alpha\beta ab}$ & $\sqrt{-{\bf g}_{11}} {\bf G}_{0m'n'p'}{\bf g}^{00} {\bf g}^{m'm}{\bf g}^{n'n}{\bf g}^{p'p}
\epsilon_{mnpq0ij\alpha\beta ab}$ & $\hat{l}_{e0m}^{np} + {\hat\zeta_e(t)\over 2} -\hat\sigma_e(t) + {\hat\alpha_e(t) + \hat\beta_e(t)\over 2}  + {\hat\eta_e(t)\over 2}$ & $-2^\pm$ \\ \hline 
2 & $\big(\mathbb{T}_7^{(q)}\big)_{ijq\alpha\beta ab}$ & $\sqrt{-{\bf g}_{11}} \mathbb{Y}_4^{0mnp} 
\epsilon_{mnpq0ij\alpha\beta ab}$ & $\theta_{nl} - \hat{l}_{e0m}^{np} - {14\over 3} + {3\hat\zeta_e(t)\over 2}  + 2\hat\sigma_e(t) + {\hat\alpha_e(t) + \hat\beta_e(t)\over 2} + {\hat\eta_e(t)\over 2}$ & $-2^\pm\big({2\over 3}^\pm\big)$ \\ \hline 
3 & $\big(\mathbb{T}^{(m)}_8\big)_{mijq\alpha\beta ab}$ & ${\bf G}_{[mqij} {\bf G}_{\alpha\beta ab]}$ & $ l_{e[mq}^{[ij} \oplus l_{e\alpha\beta]}^{ab]}$ & $-2^\pm$
\\ \hline 
4 & $\big(\mathbb{T}^{(0)}_8\big)_{0ijq\alpha\beta ab}$ & ${\bf G}_{[0ijq} {\bf G}_{\alpha\beta ab]}$ & $ l_{e[0i}^{[jq} \oplus l_{e\alpha\beta]}^{ab]}$ & $-3^\pm$
\\ \hline 
5 & $\big(\mathbb{X}^{(m)}_8\big)_{mijq\alpha\beta ab}$ & ${\rm dom} \left({\rm tr} ~\mathbb{R}_{\rm tot}^4 
- {1\over 4} \left({\rm tr}~\mathbb{R}_{\rm tot}^2\right)^2\right)_{mijq \alpha\beta ab}$ & $0 + \hat\chi^{(13)}_e(t)$ & $0^\pm$ \\ \hline
6 & $\big(\mathbb{X}^{(0)}_8\big)_{0ijq\alpha\beta ab}$ & ${\rm dom} \left({\rm tr} ~\mathbb{R}_{\rm tot}^4 
- {1\over 4} \left({\rm tr}~\mathbb{R}_{\rm tot}^2\right)^2\right)_{0ijq\alpha\beta ab}$ & $-1 + \hat\chi^{(14)}_e(t)$ & $-1^\pm$ \\ \hline
\end{tabular}}
\renewcommand{\arraystretch}{1}
\end{center}
 \caption[]{ \Su ${g_s\over {\rm H}(y){\rm H}_o({\bf x})}$ scalings of the various terms of \eqref{crawlquif} contributing to the EOM of ${\bf G}_{0mnp}$.} 
  \label{ccrani4}
 \end{table}

When the derivatives act along the ${\cal M}_4$ directions, The first three rows of {\bf Table \ref{ccrani4}} may be identified to order $\left({g_s\over {\rm H}(y){\rm H}_o({\bf x})}\right)^{-2^\pm}$ with $\theta_{nl} \ge {2\over 3}^\pm$. Once we go to order $\left({g_s\over {\rm H}(y){\rm H}_o({\bf x})}\right)^{0^\pm}$ we can identify the first three rows with the fifth row of {\bf Table \ref{ccrani4}}. Once we look at the temporal derivatives, rows 1, 2, 4 and 6 get identified at 
order $\left({g_s\over {\rm H}(y){\rm H}_o({\bf x})}\right)^{-1^\pm}$ with $\theta_{nl} \ge {8\over 3}^\pm$. In a similar vein, we can work out the details for the flux components ${\bf G}_{0mn\alpha}$ and ${\bf G}_{0m\alpha\beta}$ whose scalings of the tensors contributing to \eqref{crawlquif} are shown in {\bf Tables \ref{ccrani5}} and {\bf \ref{ccrani52}} respectively. For both the flux components ${\bf G}_{0mn\alpha}$ and ${\bf G}_{0m\alpha\beta}$ the scalings are expectedly similar to what we have for ${\bf G}_{0mnp}$, however now we also have derivatives along the ${\cal M}_2$ directions as shown in {\bf Tables \ref{ccrani5}} and {\bf \ref{ccrani52}}. Moreover, comparing the first two rows in {\bf Tables \ref{ccrani4}, \ref{ccrani5}} and {\bf \ref{ccrani52}}, we find that:
\bg\label{quinnbaba}
\theta_{nl} = 2\hat{l}_{e{\rm 0M}}^{\rm NP} + {14\over 3} - \hat\zeta_e(t) - \hat\sigma_e(t)\left(1 + \delta_{{\rm N}n} + \delta_{{\rm P}p}\right) - (\hat\alpha_e(t), \hat\beta_e(t))\left( \delta_{{\rm N}\alpha} + \delta_{{\rm P}\beta}\right), \nd
matching exactly with \eqref{brittbaba} for $l_{68} = l_{69} = l_{70} = 2$. This matching suggests that the two ways of getting the various terms in \eqref{brittbaba}, one being direct contraction using the vielbeins and the other being the flux EOM approach, are deeply interconnected. Another consistency check appears when we compare the first and the third rows in {\bf Tables \ref{ccrani4}, \ref{ccrani5}} and {\bf \ref{ccrani52}}: the bounds become $\hat{l}_{e{\rm 0M}}^{\rm NP} \ge -2^\pm$ consistent with {\bf Table \ref{micandi0_0}}. (Interestingly, if we compare the first rows with the fifth rows in the aforementioned tables, we recover the same lower bounds provided we replace $\hat{l}_{e{\rm 0M}}^{\rm NP} \to\hat{l}_{e{\rm 0M}}^{\rm NP} - 1$ because of the temporal derivatives.) What we now need are the precise behaviors of the scalings $\hat{l}_{e{\rm 0M}}^{\rm NP}$. For this,
if we now denote the actual scalings as:
\bg\label{selfhath2}
\hat{l}_{e0m}^{np} = -2^\pm + \vert x_7\vert, ~~~ \hat{l}_{e0m}^{n\alpha} = -2^\pm + \vert x_8\vert, ~~~ \hat{l}_{e0m}^{\alpha\beta} = -2^\pm + \vert x_9\vert, \nd
where again the superscripts denote small corrections from the presence of $\hat\alpha_e(t)$, $ \hat\beta_e(t)$, $ \hat\sigma_e(t)$, $ \hat\zeta_e(t)$ and $\hat\eta_e(t)$, then {\bf Tables \ref{ccrani4}, \ref{ccrani5}} and {\bf \ref{ccrani52}} gives us:
\bg\label{argoador2}
0^\pm \le (x_7, x_8, x_9) \le 2^\pm, \nd
which is exactly what we had in \eqref{argoador} and \eqref{argothai}. In \cite{coherbeta3}, it was observed that $x_7 = x_8 = x_9 = 2$ appeared to have a better fit with the available data, although $x_7 = x_8 = x_9 = 0$ was also a viable possibility. Here however we will make a more generic ans\"atze for the G-flux components ${\bf G}_{0{\rm MNP}}$ with $({\rm M, N}) \in {\cal M}_4 \times {\cal M}_2$:
\begin{empheq}[box={\mybluebox[5pt]}]{equation}
{\bf G}_{0{\rm MNP}}({\bf x}, y; g_s(t)) = \sum_{l = 0}^6\sum_{k_l = 0}^\infty {\cal G}^{(k, l)}_{0{\rm MNP}}({\bf x}, y) \left({g_s\over {\rm HH}_o}\right)^{-2^\pm + {l\over 3} + {2k_l\over 3}\vert {\rm L}_{\rm 0M}^{\rm NP}(k_l; t)\vert} 
\label{usher2}
\end{empheq}
where, from \eqref{andyrey} and \eqref{malenbego}, ${\rm L}_{\rm 0M AB}^{\rm NP}(k_l; t)$ contains all the perturbative and non-perturbative corrections including the ones discussed in section \ref{sec4.5.2}.

\begin{table}[tb]   
 \begin{center}
\resizebox{\columnwidth}{!}{%
\renewcommand{\arraystretch}{2.5}
}
\renewcommand{\arraystretch}{1}
\end{center}
 \caption[]{ \Su ${g_s\over {\rm H}(y){\rm H}_o({\bf x})}$ scalings of the various terms of \eqref{crawlquif} contributing to the EOM of ${\bf G}_{0mn\alpha}$.} 
  \label{ccrani5}
 \end{table}

\begin{table}[tb]  
 \begin{center}
\resizebox{\columnwidth}{!}{%
\renewcommand{\arraystretch}{2.5}
%
}
\renewcommand{\arraystretch}{1}
\end{center}
 \caption[]{ \Su ${g_s\over {\rm H}(y){\rm H}_o({\bf x})}$ scalings of the various terms of \eqref{crawlquif} contributing to the EOM of ${\bf G}_{0m\alpha\beta}$.} 
  \label{ccrani52}
 \end{table}

\subsection{EOMs for the flux components ${\bf G}_{{\rm MNP}a}$ and ${\bf G}_{{\rm 0NP}a}$ \label{blackbond3}} 

Our next set of G-fluxes are of the form ${\bf G}_{{\rm MNP}a}$, and they cover flux components ${\bf G}_{mnpa}, {\bf G}_{mn\alpha a}$ and ${\bf G}_{m\alpha\beta a}$. The tensors contributing to their EOMs in \eqref{crawlquif} are collected in {\bf Tables \ref{ccrani6}} and {\bf \ref{ccrani7}}, the latter being more generic. Comparing the first two rows in {\bf Table \ref{ccrani7}}, we find that:
\bg\label{diormea}
\theta_{nl} = 2\hat{l}_{e{\rm MN}}^{{\rm P}a} + {2\over 3} - \hat\sigma_e(t)\left(1 + \delta_{{\rm N}n} + \delta_{{\rm P}p}\right) - 
(\hat\alpha_e(t), \hat\beta_e(t))\left( \delta_{{\rm N}\alpha} + \delta_{{\rm P}\beta}\right), \nd
which again matches with \eqref{brittbaba} for $l_{44} = l_{46}  = l_{47} = 2$. Another consistency check appears from comparing the first and the third rows in {\bf Tables \ref{ccrani6}} and {\bf \ref{ccrani7}}: they give us $\hat{l}_{e{\rm MN}}^{{\rm P}a} \ge 0^\pm$ which matches with the one from {\bf Table \ref{micandi0_0}}. Comparing now the EOMs from \eqref{crawlquif}, it is clear that 
the relevant derivative actions are from the derivatives along directions ${\cal M}_4$, and we expect that to order $\left({g_s\over {\rm H}(y){\rm H}_o({\bf x})}\right)^{-2^\pm}$ rows 1, 2, 3 and 5 in {\bf Table \ref{ccrani7}} should match-up. At order $\left({g_s\over {\rm H}(y){\rm H}_o({\bf x})}\right)^{-4^\pm}$
the spatial parts of the flux product should vanish. The derivatives acting along the toroidal ${\mathbb{T}^2\over {\cal G}}$ directions imply that we can equate rows 4 and 6 at order $\left({g_s\over {\rm H}(y){\rm H}_o({\bf x})}\right)^{-1^\pm}$, below which again we expect vanishing of the flux products in fourth row of {\bf Table \ref{ccrani7}}. If we now denote the precise scalings of the flux components in the following way:
\bg\label{selfhath3}
\hat{l}_{emn}^{pa} = 0^\pm + \vert x_{10}\vert, ~~~ \hat{l}_{emn}^{\alpha a} = 0^\pm + \vert x_{11}\vert, ~~~ \hat{l}_{em\alpha}^{\beta a} = 0^\pm + \vert x_{12}\vert, \nd
where again the superscripts denote small corrections from the presence of $\hat\alpha_e(t)$, $ \hat\beta_e(t)$, $ \hat\sigma_e(t)$, $ \hat\zeta_e(t)$ and $\hat\eta_e(t)$. Looking at {\bf Tables \ref{micandi0_0}, \ref{ccrani6}} and {\bf \ref{ccrani7}} gives us:
\bg\label{argoador3}
0^\pm \le (x_{10}, x_{11}, x_{12}) \le 2^\pm, \nd
precisely matching with \eqref{argothai}, \eqref{argoador} and \eqref{argoador2}. In \cite{coherbeta3} we took $x_{10} = x_{11} = x_{12} 
 = 1$, but now we want to keep it more generic so as to include the small parameters discussed earlier. This gives us the following ans\"atze for the flux components:
\begin{empheq}[box={\mybluebox[5pt]}]{equation}
{\bf G}_{{\rm MNP}a}({\bf x}, y; g_s(t)) = \sum_{l = 0}^6\sum_{k_l = 0}^\infty {\cal G}^{(k, l)}_{{\rm MNP}a}({\bf x}, y) \left({g_s\over {\rm HH}_o}\right)^{0^\pm + {l\over 3} + {2k_l\over 3}\vert {\rm L}_{\rm MN}^{{\rm P}a}(k_l; t)\vert} 
\label{usher3}
\end{empheq} 
which suggests that, even for $l = 0$ in \eqref{usher3}, the dominant parts of the flux components {\it cannot} be time-independent because of the presence of the small parameters $\hat\alpha_e(t)$, $ \hat\beta_e(t)$, $ \hat\sigma_e(t)$, $ \hat\zeta_e(t)$ and $\hat\eta_e(t)$
\begin{table}[tb]  
 \begin{center}
\resizebox{\columnwidth}{!}{%
\renewcommand{\arraystretch}{2.5}
}
\renewcommand{\arraystretch}{1}
\end{center}
 \caption[]{ \Su ${g_s\over {\rm H}(y){\rm H}_o({\bf x})}$ scalings of the various terms of \eqref{crawlquif} contributing to the EOM of ${\bf G}_{mnpa}$.} 
  \label{ccrani6}
 \end{table}

 \begin{table}[tb]  
 \begin{center}
\resizebox{\columnwidth}{!}{%
\renewcommand{\arraystretch}{3.5}
%
}
\renewcommand{\arraystretch}{1}
\end{center}
 \caption[]{ \Su ${g_s\over {\rm H}(y){\rm H}_o({\bf x})}$ scalings of the various terms of \eqref{crawlquif} contributing to the EOM of ${\bf G}_{{\rm MNP}a}$. This includes ${\bf G}_{mnpa}, {\bf G}_{mn\alpha a}$ and ${\bf G}_{m\alpha\beta a}$ flux components.} 
  \label{ccrani7}
 \end{table}

\begin{table}[tb]  
 \begin{center}
\resizebox{\columnwidth}{!}{%
\renewcommand{\arraystretch}{3.5}
%
}
\renewcommand{\arraystretch}{1}
\end{center}
 \caption[]{ \Su ${g_s\over {\rm H}(y){\rm H}_o({\bf x})}$ scalings of the various terms of \eqref{crawlquif} contributing to the EOM of ${\bf G}_{0{\rm NP}a}$. This includes ${\bf G}_{0npa}, {\bf G}_{0n\alpha a}$ and ${\bf G}_{0\alpha\beta a}$ flux components.} 
  \label{ccrani8}
 \end{table}

The story gets slightly more involved for the next set of G-fluxes ${\bf G}_{0{\rm NP}a}$ which cover ${\bf G}_{0npa}, {\bf G}_{0n\alpha a}$ and ${\bf G}_{0\alpha\beta a}$ flux components. The relevant derivative actions are now along the internal ${\cal M}_4$ and the temporal directions. The $\mathbb{X}_8$ polynomials are constructed from the two forms given in {\bf Tables \ref{niksmit1}} to {\bf \ref{niksmit250}}, as shown in {\bf Table \ref{ccrani8}}. From {\bf Table \ref{micandi0_0}}, we see that $\hat{l}_{e0{\rm N}}^{{\rm P}a}$ is bounded from below by $-1^\pm$. Such lower bounds also appear from comparing the first and the third rows in {\bf Table \ref{ccrani8}}, or comparing the first and the fifth rows, keeping track of the fact that the scaling in the first row has to reduce by 1 to comply with the temporal-derivative action. This means we can express the precise scalings of the relevant flux components as:
\bg\label{selfhath4}
\hat{l}_{e0n}^{pa} = -1^\pm + \vert x_{13}\vert, ~~~ \hat{l}_{e0n}^{\alpha a} = -1^\pm + \vert x_{14}\vert, ~~~ \hat{l}_{e0\alpha}^{\beta a} = -1^\pm + \vert x_{15}\vert, \nd
where again the superscripts denote small corrections from the presence of $\hat\alpha_e(t)$, $ \hat\beta_e(t)$, $ \hat\sigma_e(t)$, $ \hat\zeta_e(t)$ and $\hat\eta_e(t)$. Looking at {\bf Tables \ref{micandi0_0}}, and {\bf \ref{ccrani8}} tells us that the action of the temporal derivatives reduce the scalings in Row 1 by 1, thus matching with the scalings in Row 6. This then gives us:
\bg\label{argoador3}
0^\pm \le (x_{13}, x_{14}, x_{15}) \le 2^\pm, \nd
exactly as before. In \cite{coherbeta3} we discussed the cases where 
$x_{13} = x_{14} = x_{15} = 0$ and $x_{13} = x_{14} = x_{15} = 2$. Here, as before, we want to be generic, and propose the following ans\"atze for the flux components ${\bf G}_{0{\rm NP}a}$:
\begin{empheq}[box={\mybluebox[5pt]}]{equation}
{\bf G}_{0{\rm NP}a}({\bf x}, y; g_s(t)) = \sum_{l = 0}^6\sum_{k_l = 0}^\infty {\cal G}^{(k, l)}_{0{\rm NP}a}({\bf x}, y) \left({g_s\over {\rm HH}_o}\right)^{-1^\pm + {l\over 3} + {2k_l\over 3}\vert {\rm L}_{\rm 0N}^{{\rm P}a}(k_l; t)\vert} 
\label{usher4}
\end{empheq} 
where even for $l = 3$ we do not have a time-independent part of the flux components because of the presence of the small parameters $\hat\alpha_e(t)$, $ \hat\beta_e(t)$, $ \hat\sigma_e(t)$, $ \hat\zeta_e(t)$ and $\hat\eta_e(t)$.

\subsection{EOMs for the flux components ${\bf G}_{{\rm MN}ab}$ and ${\bf G}_{{\rm 0N}ab}$ \label{blackbond4}}

For the G-fluxes of the form ${\bf G}_{{\rm MN}ab}$, we are dealing with the flux components ${\bf G}_{mnab}, {\bf G}_{m\alpha ab}$ and ${\bf G}_{\alpha\beta ab}$. As before, comparing the first two rows in {\bf Table \ref{ccrani9}} we find that:

{\footnotesize
\bg\label{oniston}
\theta_{nl} = 2\hat{l}_{e{\rm MN}}^{ab} - {4\over 3} - \left(\delta_{{\rm M}m} + \delta_{{\rm N}n}\right)\hat\sigma_e(t) -\left(\delta_{{\rm M}\alpha} + \delta_{{\rm N}\beta}\right)(\hat\alpha_e(t), \hat\beta_e(t))
- {\hat\eta_e(t)}, \nd}
which matches with \eqref{brittbaba} for the three choices of the flux components with $l_{50} = l_{51} = l_{52} = 2$. Clearly at 
$\left({g_s\over {\rm H}(y){\rm H}_o({\bf x})}\right)^{-5^\pm}$ the scalings in the first three rows of {\bf Table \ref{ccrani9}} match up when we take the derivatives along the ${\cal M}_4 \times {\cal M}_2$ directions with the bounds presented in {\bf Table \ref{micandi0_0}}. This lower bound can also be extracted from {\bf Table \ref{ccrani9}} by equating the first and the third rows, which gives us $\hat{l}_{e{\rm MN}}^{ab} \ge +1^\pm$. The matching with the fifth row of {\bf Table \ref{ccrani9}} happens when we go to the order 
$\left({g_s\over {\rm H}(y){\rm H}_o({\bf x})}\right)^{-3^\pm}$ where the scaling of the $\mathbb{X}_8$ polynomials are determined from the curvature forms computed in {\bf Tables \ref{niksmit1}} to {\bf \ref{niksmit250}}. For the derivatives along the toroidal directions, the flux products should vanish at order $\left({g_s\over {\rm H}(y){\rm H}_o({\bf x})}\right)^{-4^\pm}$. However at order $\left({g_s\over {\rm H}(y){\rm H}_o({\bf x})}\right)^{-2^\pm}$, rows 1, 2, 4 and 6 should match up if we consider the scalings given in {\bf Table \ref{micandi0_0}}. 
The exact scalings may now be represented as:
\bg\label{selfhath5}
\hat{l}_{emn}^{ab} = +1^\pm + \vert x_{16}\vert, ~~~ \hat{l}_{em\alpha}^{ab} = +1^\pm + \vert x_{17}\vert, ~~~ \hat{l}_{e\alpha\beta}^{ab} = +1^\pm + \vert x_{18}\vert, \nd
where again the superscripts denote small corrections from the presence of $\hat\alpha_e(t)$, $ \hat\beta_e(t)$, $ \hat\sigma_e(t)$, $ \hat\zeta_e(t)$ and $\hat\eta_e(t)$. Looking at {\bf Tables \ref{micandi0_0}}, and {\bf \ref{ccrani9}} then gives us:
\bg\label{argoador4}
0^\pm \le (x_{16}, x_{17}, x_{18}) \le 2^\pm, \nd
exactly as before. In \cite{coherbeta3} we took $x_{16} = x_{17} = x_{18} = 0$ so that the scalings for all the three flux components were $+1$. Here we want to keep it more generic and therefore we take the following ans\"atze:
\begin{empheq}[box={\mybluebox[5pt]}]{equation}
{\bf G}_{{\rm MN}ab}({\bf x}, y; g_s(t)) = \sum_{l = 0}^6\sum_{k_l = 0}^\infty {\cal G}^{(k, l)}_{{\rm MN}ab}({\bf x}, y) \left({g_s\over {\rm HH}_o}\right)^{+1^\pm + {l\over 3} + {2k_l\over 3}\vert {\rm L}_{\rm MN}^{ab}(k_l; t)\vert} 
\label{usher5}
\end{empheq} 
which is interesting because this is the first set of flux components whose lower bounds are non-zero and positive definite integers. Note that the positive scalings of $+1^\pm$ is important otherwise there would be a serious conflict with the underlying EFT as seen from \eqref{brittbaba}.

\begin{table}[tb]  
 \begin{center}
\resizebox{\columnwidth}{!}{%
\renewcommand{\arraystretch}{3.5}
\begin{tabular}{|c||c||c||c||c|}\hline Rows & ${\bf G}_{{\rm MN}ab}$ tensors & Forms & ${g_s\over {\rm HH}_o}$ scalings & Lower bounds \\ \hline\hline
1 & $\big(\mathbb{T}_7^{(f)}\big)_{0ij{\rm PQRS}}$ & $\sqrt{-{\bf g}_{11}} {\bf G}_{{\rm M'}a'b'}{\bf g}^{\rm M'M} {\bf g}^{\rm N'N}{\bf g}^{\rm a'a}{\bf g}^{b'b}
\epsilon_{0ij{\rm MNPQRS}ab}$ & $\hat{l}_{e{\rm MN}}^{ab} - 6 + {3\hat\zeta_e(t)\over 2} + (2- \delta_{{\rm M}m} - \delta_{{\rm N}n}){\hat\sigma_e(t)} + {\hat\alpha_e(t) + \hat\beta_e(t)\over 2} - (\hat\alpha_e(t), \hat\beta_e(t))(\delta_{{\rm N}\alpha} + \delta_{{\rm M}\beta})  -{\hat\eta_e(t)\over 2}$ & $-5^\pm$ \\ \hline 
2 & $\big(\mathbb{T}_7^{(q)}\big)_{0ij{\rm PQRS} }$ & $\sqrt{-{\bf g}_{11}} \mathbb{Y}_4^{{\rm MN}ab} 
\epsilon_{0ij{\rm MNPQRS}ab}$ & $\theta_{nl} - \hat{l}_{e{\rm MN}}^{ab} - {14\over 3} + {3\hat\zeta_e(t)\over 2}  + 2\hat\sigma_e(t) + {\hat\alpha_e(t) + \hat\beta_e(t)\over 2} + {\hat\eta_e(t)\over 2}$ & $-5^\pm\big({2\over 3}^\pm\big)$ \\ \hline 
3 & $\big(\mathbb{T}^{({\rm M})}_8\big)_{0ij{\rm MPQRS}}$ & ${\bf G}_{[0ij{\rm M}} {\bf G}_{{\rm PQRS}]}$ & $ l_{e[0i}^{[j{\rm M}} \oplus l_{e{\rm PQ}]}^{{\rm RS}]}$ & $-5^\pm$
\\ \hline 
4 & $\big(\mathbb{T}^{(a)}_8\big)_{0ij{\rm PQRS} a}$ & ${\bf G}_{[0ij{\rm P}} {\bf G}_{{\rm QRS} b]}$ & $ l_{e[0i}^{[j{\rm Q}} \oplus l_{e{\rm RS}]}^{ab]}$ & $-4^\pm$
\\ \hline 
5 & $\big(\mathbb{X}^{({\rm M})}_8\big)_{0ij{\rm MPQRS} }$ & ${\rm dom} \left({\rm tr} ~\mathbb{R}_{\rm tot}^4 
- {1\over 4} \left({\rm tr}~\mathbb{R}_{\rm tot}^2\right)^2\right)_{0ij{\rm MPQRS}}$ & $-3 + \hat\chi^{(36)}_e(t), ~~-3 + \hat\chi^{(38)}_e(t),~~ -3 + \hat\chi^{(40)}_e(t)$ & $-3^\pm$ \\ \hline
6 & $\big(\mathbb{X}^{(a)}_8\big)_{0ij{\rm PQRS} a}$ & ${\rm dom} \left({\rm tr} ~\mathbb{R}_{\rm tot}^4 
- {1\over 4} \left({\rm tr}~\mathbb{R}_{\rm tot}^2\right)^2\right)_{0ij{\rm PQRS} a}$ & $-2 + \hat\chi^{(37)}_e(t), ~~
 -2 + \hat\chi^{(39)}_e(t), ~~-2 + \hat\chi^{(41)}_e(t)$ & $-2^\pm$ \\ \hline
\end{tabular}}
\renewcommand{\arraystretch}{1}
\end{center}
 \caption[]{ \Su ${g_s\over {\rm H}(y){\rm H}_o({\bf x})}$ scalings of the various terms of \eqref{crawlquif} contributing to the EOM of ${\bf G}_{{\rm MN}ab}$. This includes ${\bf G}_{mnab}, {\bf G}_{m\alpha ab}$ and ${\bf G}_{\alpha\beta ab}$ flux components.} 
  \label{ccrani9}
 \end{table}

Our next set in this list are the G-fluxes ${\bf G}_{{\rm 0N}ab}$ which involve the flux components ${\bf G}_{0nab}$ and ${\bf G}_{0\alpha ab}$.
We can start by making one check: comparing rows 1 and 2 of {\bf Table \ref{ccrani10}} gives us the following form for $\theta_{nl}$:
\bg\label{viorosse1}
\theta_{nl} = 2\hat{l}_{e{\rm 0N}}^{ab} + {2\over 3} - \hat\zeta_e(t) - 
\delta_{{\rm N}n} \hat\sigma_e(t) - (\hat\alpha_e(t), \hat\beta_e(t)) - \hat\eta_e(t), \nd
which matches exactly with \eqref{brittbaba} for $l_{71} = l_{72} = 2$. 
One may also make the usual matching of the rows in {\bf Table \ref{ccrani10}}, keeping in mind that now temporal derivatives are involved. The $\mathbb{X}_8$ polynomials are computed from the curvature forms given in {\bf Tables \ref{niksmit1}} to {\bf \ref{niksmit250}}. The lower bounds from {\bf Table \ref{micandi0_0}} appear by equating rows 1 and 3 or rows 1 and 5, with a careful consideration for the latter because of the temporal derivative. Comparing the derivatives along the internal ${\cal M}_4 \times {\cal M}_2$ directions we see that, for $\hat{l}_{e{\rm oN}}^{ab} \ge 0$ and to order $\left({g_s\over {\rm H}(y){\rm H}_o({\bf x})}\right)^{-4^\pm}$, the first three rows in {\bf Table \ref{ccrani10}} match up. Going beyond that, and to order $\left({g_s\over {\rm H}(y){\rm H}_o({\bf x})}\right)^{-2^\pm}$, we can match it up with the sixth row. Taking the temporal derivatives into account, the scalings in the first row reduces by 1 which fits very well with the scalings in the eighth row that shows a $-1$ reduction in the $g_s$ scalings. Thus to order 
$\left({g_s\over {\rm H}(y){\rm H}_o({\bf x})}\right)^{-3^\pm}$ rows 1, 2, 5 and 8 can be matched up. When the derivatives act along the toroidal directions, the matching happens at order $\left({g_s\over {\rm H}(y){\rm H}_o({\bf x})}\right)^{-1^\pm}$ between rows 4 and 7, below which the flux product vanishes. If we now denote the precise scaling as:
\bg\label{selfhath6}
\hat{l}_{e0n}^{ab} = 0^\pm + \vert x_{19}\vert, ~~~~~\hat{l}_{e0\alpha}^{ab} = 0^\pm + \vert x_{20}\vert, \nd
where again the superscripts denote small corrections from the presence of $\hat\alpha_e(t)$, $ \hat\beta_e(t)$, $ \hat\sigma_e(t)$, $ \hat\zeta_e(t)$ and $\hat\eta_e(t)$. Looking at {\bf Tables \ref{micandi0_0}}, and {\bf \ref{ccrani10}} then gives us:
\bg\label{argoador5}
0^\pm \le (x_{19}, x_{20}) \le 2^\pm, \nd
expectedly matching with earlier results. In \cite{coherbeta3} similar bounds were observed (without the $\pm$ superscripts because of the analysis being in type IIB), but now we can provide a more precise determination of the G-flux components. Our ans\"atze becomes:
\begin{empheq}[box={\mybluebox[5pt]}]{equation}
{\bf G}_{{\rm 0N}ab}({\bf x}, y; g_s(t)) = \sum_{l = 0}^6\sum_{k_l = 0}^\infty {\cal G}^{(k, l)}_{{\rm 0N}ab}({\bf x}, y) \left({g_s\over {\rm HH}_o}\right)^{0^\pm + {l\over 3} + {2k_l\over 3}\vert {\rm L}_{\rm 0N}^{ab}(k_l; t)\vert} 
\label{usher6}
\end{empheq} 
which tells us that, even for $l = k = 0$, the flux components do not have  time-independent pieces. This is similar to what we had observed for the ${\bf G}_{{\rm MNP}a}$ flux components in \eqref{usher3}.

\begin{table}[tb]  
 \begin{center}
\resizebox{\columnwidth}{!}{%
\renewcommand{\arraystretch}{3.5}
}
\renewcommand{\arraystretch}{1}
\end{center}
 \caption[]{ \Su ${g_s\over {\rm H}(y){\rm H}_o({\bf x})}$ scalings of the various terms of \eqref{crawlquif} contributing to the EOM of ${\bf G}_{0{\rm N}ab}$. This includes ${\bf G}_{0nab}$, and $ {\bf G}_{0\alpha ab}$  flux components.} 
  \label{ccrani10}
 \end{table}

\subsection{EOMs for the flux components ${\bf G}_{{\rm MN}ai}, {\bf G}_{{\rm 0N}ai}$ and ${\bf G}_{{\rm M}abi}$ \label{blackbond5}}

Our next set of flux components will involve at least one spatial direction {\it i.e.} we now allow flux components with at least one leg along the spatial ${\bf R}^2$ directions. This is of course not the first time we have involved the spatial components. In \eqref{usher1} we did encounter ${\bf G}_{0ij{\rm M}}$ and ${\bf G}_{0ija}$ flux components. Here we will do a systematic study of the other possible choices of such flux components.

\begin{table}[tb]  
 \begin{center}
\resizebox{\columnwidth}{!}{%
\renewcommand{\arraystretch}{3.5}
}
\renewcommand{\arraystretch}{1}
\end{center}
 \caption[]{ \Su ${g_s\over {\rm H}(y){\rm H}_o({\bf x})}$ scalings of the various terms of \eqref{crawlquif} contributing to the EOM of ${\bf G}_{{\rm MN}ai}$. This includes ${\bf G}_{mnai}, {\bf G}_{m\alpha a i}$, and $ {\bf G}_{\alpha\beta ai}$  flux components.} 
  \label{ccrani11}
 \end{table}

Our first set of G-fluxes are of the form ${\bf G}_{{\rm MN}ai}$ with flux components ${\bf G}_{mnai}, {\bf G}_{m\alpha ai}$ and ${\bf G}_{\alpha\beta ai}$. The tensors contributing to the ${\bf G}_{{\rm MN}ai}$ EOMs are given in {\bf Table \ref{ccrani11}}. Comparing the first two rows of the table, provides the following form for $\theta_{nl}$:

{\footnotesize
\bg\label{ajcandare}
\theta_{nl} = 2\hat{l}_{e{\rm MN}}^{ai} + {8\over 3} - \hat\zeta_e(t) - 
\left(\delta_{{\rm M}m} + \delta_{{\rm N}n}\right)\hat\sigma_e(t) - (\hat\alpha_e(t), \hat\beta_e(t)) \left(\delta_{{\rm M}\alpha} + \delta_{{\rm N}\beta}\right) - (0, \hat\eta_e(t)), \nd}
which matches with \eqref{brittbaba} for $l_{56} = l_{58} = l_{60} = 2$. 
The derivative action that is crucial now is the one along the ${\cal M}_4$ directions. To order $\left({g_s\over {\rm H}(y){\rm H}_o({\bf x})}\right)^{-3^\pm}$, with $\hat{l}_{e{\rm MN}}^{ai} \ge -1^\pm$, we can see that the first three rows of {\bf Table \ref{ccrani11}} match up. The lower bounds appear as before by comparing the scalings in rows 1 and 3 of {\bf Table \ref{ccrani11}}, thus matching with the ones from {\bf Table \ref{micandi0_0}}. With the same lower bound, and to order $\left({g_s\over {\rm H}(y){\rm H}_o({\bf x})}\right)^{-1^\pm}$, we can further match up the scalings with the sixth row. On the other hand, if we take the derivatives along the toroidal ${\mathbb{T}^2\over {\cal G}}$ or the spatial ${\bf R}^2$ directions, the matching becomes Row 4 = Row 7 at order $\left({g_s\over {\rm H}(y){\rm H}_o({\bf x})}\right)^{0^\pm}$, and Row 5 = Row 8 at order $\left({g_s\over {\rm H}(y){\rm H}_o({\bf x})}\right)^{-2^\pm}$ respectively. Below these orders, we expect the flux products to respectively vanish. If we now denote the exact scalings as:
\bg\label{selfhath7}
\hat{l}_{emn}^{ai} = -1^\pm + \vert x_{21}\vert, ~~~ \hat{l}_{em\alpha}^{ai} = -1^\pm + \vert x_{22}\vert, ~~~ \hat{l}_{e\alpha\beta}^{ai} = -1^\pm + \vert x_{23}\vert, \nd
where again the superscripts denote small corrections from the presence of $\hat\alpha_e(t)$, $ \hat\beta_e(t)$, $ \hat\sigma_e(t)$, $ \hat\zeta_e(t)$ and $\hat\eta_e(t)$, then looking at {\bf Tables \ref{micandi0_0}}, and {\bf \ref{ccrani11}} gives us:
\bg\label{argoador6}
0^\pm \le (x_{21}, x_{22}, x_{23}) \le 2^\pm, \nd
exactly as before. As in \cite{coherbeta3} we notice that the bounds on the scalings in {\bf Table \ref{ccrani11}} matches well with the bounds on the scalings in {\bf Table \ref{ccrani8}} for the G-fluxes ${\bf G}_{{\rm 0NP}a}$. This is somewhat expected because the flux components 
in ${\bf G}_{{\rm 0NP}a}$ and ${\bf G}_{{\rm MN}ai}$ appear to differ by $x^i \leftrightarrow x^0$ in their indices. Despite this, the matching in {\bf Table \ref{ccrani8}} is done differently for derivatives along the temporal direction compared to the matching in {\bf Table \ref{ccrani11}} for derivatives along the spatial ${\bf R}^2$ directions.  
Nevertheless the flux ans\"atze takes the following form:
\begin{empheq}[box={\mybluebox[5pt]}]{equation}
{\bf G}_{{\rm MN}ai}({\bf x}, y; g_s(t)) = \sum_{l = 0}^6\sum_{k_l = 0}^\infty {\cal G}^{(k, l)}_{{\rm MN}ai}({\bf x}, y) \left({g_s\over {\rm HH}_o}\right)^{-1^\pm + {l\over 3} + {2k_l\over 3}\vert {\rm L}_{\rm MN}^{ai}(k_l; t)\vert} 
\label{usher7}
\end{empheq} 
where the scaling seems to match with \eqref{usher4}. This however doesn't mean that the two kinds of fluxes would behave in similar ways. The choice of $l$ in \eqref{usher7} and \eqref{usher4} could be different and thus be responsible in creating the necessary difference.

A slightly more involved case is with our next set of G-fluxes of the form  ${\bf G}_{{\rm 0N}ai}$ whose flux components are ${\bf G}_{0nai}$ and ${\bf G}_{0\alpha ai}$. The tensors contributing to the EOMs from \eqref{crawlquif} are collected in {\bf Table \ref{ccrani12}}. Interestingly, now we have four different derivatives' action: along spatial ${\bf R}^2$, temporal, internal ${\cal M}_4$ and toroidal ${\mathbb{T}^2\over {\cal G}}$. Comparing the first two rows in {\bf Table \ref{ccrani12}} provides our first consistency check:
\bg\label{aapplediesl}
\theta_{nl} = 2\hat{l}_{\rm 0N}^{ai} +{14\over 3} -2\hat\zeta_e(t) - \hat\sigma_e(t) \delta_{{\rm N}n} - (\hat\alpha_e(t), \hat\beta_e(t))\delta_{{\rm N}\alpha} - (0, \hat\eta_e(t)), \nd 
which matches with \eqref{brittbaba} for $l_{80} = l_{81} =2$. The second consistency check comes from comparing the first row with the third row in {\bf Table \ref{ccrani12}}, or comparing the first row with the sixth
row $-$ with the due consideration that the scaling in the first row has to reduce by 1 to comply with the temporal derivative $-$ thus giving us $\hat{l}_{e{\rm 0N}}^{ai} \ge -2^\pm$.

\begin{table}[tb]  
 \begin{center}
\resizebox{\columnwidth}{!}{%
\renewcommand{\arraystretch}{3.5}
}
\renewcommand{\arraystretch}{1}
\end{center}
 \caption[]{ \Su ${g_s\over {\rm H}(y){\rm H}_o({\bf x})}$ scalings of the various terms of \eqref{crawlquif} contributing to the EOM of ${\bf G}_{{\rm 0N}ai}$. This includes ${\bf G}_{0nai}$, and $ {\bf G}_{0\alpha ai}$  flux components.} 
  \label{ccrani12}
 \end{table}

 Other comparisons can also be made from looking at various orders in 
 ${g_s\over {\rm H}(y){\rm H}_o({\bf x})}$ between the rows that follow the procedure laid out earlier. We will not do this here, instead we will work out the upper bounds on the scalings $\hat{l}_{e{\rm 0N}}^{ai}$. This is where the importance of the $\mathbb{X}_8$ polynomials enter the story, which are computed from the curvature forms given in {\bf Tables \ref{niksmit1}} till {\bf \ref{niksmit250}}. Comparing the first row with the seventh row in {\bf Table \ref{ccrani12}} gives us $\hat{l}_{e{\rm 0N}}^{ai} \le 0^\pm$. Exactly the same lower bounds appear from comparing the first row with tenth row, once we keep track of the temporal-derivative action. If we now denote the exact scalings as:
\bg\label{selfhath8}
\hat{l}_{e0n}^{ai} = -2^\pm + \vert x_{24}\vert, ~~~ \hat{l}_{e0\alpha}^{ai} = -2^\pm + \vert x_{25}\vert,\nd
where again the superscripts denote small corrections from the presence of $\hat\alpha_e(t)$, $ \hat\beta_e(t)$, $ \hat\sigma_e(t)$, $ \hat\zeta_e(t)$ and $\hat\eta_e(t)$, then looking at {\bf Tables \ref{micandi0_0}}, and {\bf \ref{ccrani12}} gives us:
\bg\label{argoador7}
0^\pm \le (x_{24}, x_{25}) \le 2^\pm, \nd 
which is the same interval as before. In \cite{coherbeta3} we encountered similar result except the result was precisely between 0 and 2. Now due to the $\pm$ superscripts, the exact analysis is a bit hard to perform. Nevertheless we can provide the following ans\"atze for the flux components:
\begin{empheq}[box={\mybluebox[5pt]}]{equation}
{\bf G}_{{\rm 0N}ai}({\bf x}, y; g_s(t)) = \sum_{l = 0}^6\sum_{k_l = 0}^\infty {\cal G}^{(k, l)}_{{\rm 0N}ai}({\bf x}, y) \left({g_s\over {\rm HH}_o}\right)^{-2^\pm + {l\over 3} + {2k_l\over 3}\vert {\rm L}_{\rm 0N}^{ai}(k_l; t)\vert} 
\label{usher8}
\end{empheq} 
where we notice that even for $l = 6$, the flux components can never have temporally independent pieces. 

\begin{table}[tb]  
 \begin{center}
\resizebox{\columnwidth}{!}{%
\renewcommand{\arraystretch}{3.5}
}
\renewcommand{\arraystretch}{1}
\end{center}
 \caption[]{ \Su ${g_s\over {\rm H}(y){\rm H}_o({\bf x})}$ scalings of the various terms of \eqref{crawlquif} contributing to the EOM of ${\bf G}_{{\rm M}abi}$. This includes ${\bf G}_{mabi}$, and $ {\bf G}_{\alpha abi}$  flux components.} 
  \label{ccrani13}
 \end{table}

Our final set of G-fluxes are of the form ${\bf G}_{{\rm M}abi}$ that allow flux components of the form ${\bf G}_{mabi}$ and ${\bf G}_{\alpha abi}$. All the usual consistency checks can be made. For example comparing row 1 with 3 and 6 in {\bf Table \ref{ccrani13}} provides the lower and the upper bounds of $0^\pm \le \hat{l}_{e{\rm M}a}^{bi} \le 2^\pm$. Comparing the first two rows in {\bf Table \ref{ccrani13}} gives us:
\bg\label{whitbrit}
\theta_{nl} = 2\hat{l}_{e{\rm M}a}^{bi} + {2\over 3} - \hat\zeta_e(t) - \hat\sigma_e(t)\delta_{{\rm M}m} - (\hat\alpha_e(t), \hat\beta_e(t)) \delta_{{\rm M}\alpha} - \hat\eta_e(t), \nd
which expectedly matches with \eqref{brittbaba} for $l_{57} = l_{59} = 2$.
 If we now denote the exact scalings as:
\bg\label{selfhath9}
\hat{l}_{ema}^{bi} = 0^\pm + \vert x_{26}\vert, ~~~ \hat{l}_{e\alpha a}^{bi} = 0^\pm + \vert x_{27}\vert,\nd
where again the superscripts denote small corrections from the presence of $\hat\alpha_e(t)$, $ \hat\beta_e(t)$, $ \hat\sigma_e(t)$, $ \hat\zeta_e(t)$ and $\hat\eta_e(t)$, then looking at {\bf Tables \ref{micandi0_0}}, and {\bf \ref{ccrani12}} gives us:
\bg\label{argoador4}
0^\pm \le (x_{26}, x_{27}) \le 2^\pm, \nd 
exactly as before. The ans\"atze for the G-flux components can now be given by the following expression:
\begin{empheq}[box={\mybluebox[5pt]}]{equation}
{\bf G}_{{\rm M}abi}({\bf x}, y; g_s(t)) = \sum_{l = 0}^6\sum_{k_l = 0}^\infty {\cal G}^{(k, l)}_{{\rm M}abi}({\bf x}, y) \left({g_s\over {\rm HH}_o}\right)^{0^\pm + {l\over 3} + {2k_l\over 3}\vert {\rm L}_{{\rm M}a}^{bi}(k_l; t)\vert} 
\label{usher9}
\end{empheq} 
which shows that even for $l = 0$, we do not expect any time-independent pieces from the G-flux components.

\subsection{EOMs for the flux components ${\bf G}_{{\rm MNP}i}, {\bf G}_{{\rm 0MN}i}$ and ${\bf G}_{0abi}$ \label{blackbond6}}

\begin{table}[tb]  
 \begin{center}
\resizebox{\columnwidth}{!}{%
\renewcommand{\arraystretch}{3.5}
\begin{tabular}{|c||c||c||c||c|}\hline Rows & ${\bf G}_{{\rm MNP}i}$ tensors & Forms & ${g_s\over {\rm HH}_o}$ scalings & Lower bounds \\ \hline\hline
1 & $\big(\mathbb{T}_7^{(f)}\big)_{0j{\rm QRS}ab}$ & $\sqrt{-{\bf g}_{11}} {\bf G}_{{\rm M'N'P'}i'}{\bf g}^{\rm M'M}{\bf g}^{\rm N'N} {\bf g}^{\rm P'P}{\bf g}^{i'i}
\epsilon_{0ij{\rm MNPQRS}ab}$ & $\hat{l}_{e{\rm MN}}^{{\rm P}i} + {\hat\zeta_e(t)\over 2} + (1  - \delta_{{\rm N}n}- \delta_{{\rm P}p}){\hat\sigma_e(t)} + {\hat\alpha_e(t) + \hat\beta_e(t)\over 2} - (\hat\alpha_e(t), \hat\beta_e(t))\left(\delta_{{\rm N}\alpha}- \delta_{{\rm P}\beta}\right) + {\hat\eta_e(t)\over 2}$ & $-2^\pm$ \\ \hline 
2 & $\big(\mathbb{T}_7^{(q)}\big)_{0j{\rm QRS}ab}$ & $\sqrt{-{\bf g}_{11}} \mathbb{Y}_4^{{\rm MNP}i} 
\epsilon_{0ij{\rm MNPQRS}ab}$ & $\theta_{nl} - \hat{l}_{e{\rm MN}}^{{\rm P}i} - {14\over 3} + {3\hat\zeta_e(t)\over 2}  + 2\hat\sigma_e(t) + {\hat\alpha_e(t) + \hat\beta_e(t)\over 2} + {\hat\eta_e(t)\over 2}$ & $-2^\pm\big({2\over 3}^\pm\big)$ \\ \hline 
3 & $\big(\mathbb{T}^{({\rm M})}_8\big)_{0j{\rm MQRS}ab}$ & ${\bf G}_{[0j{\rm MQ}} {\bf G}_{{\rm RS}ab]}$ & $ l_{e[0j}^{[{\rm MQ}} \oplus l_{e{\rm RS}]}^{ab]}$ & $-2^\pm$
\\ \hline 
4 & $\big(\mathbb{T}^{(i)}_8\big)_{0ij{\rm QRS}ab}$ & ${\bf G}_{[0ij{\rm Q}} {\bf G}_{{\rm RS}ab]}$ & $ l_{e[0i}^{[j{\rm Q}} \oplus l_{e{\rm RS}]}^{ab]}$ & $-3^\pm$
\\ \hline 
5 & $\big(\mathbb{X}^{({\rm M})}_8\big)_{0j{\rm MQRS}ab}$ & ${\rm dom} \left({\rm tr} ~\mathbb{R}_{\rm tot}^4 
- {1\over 4} \left({\rm tr}~\mathbb{R}_{\rm tot}^2\right)^2\right)_{0j{\rm MQRS}ab}$ & $0 + \hat\chi^{(71)}_e(t),~~0 + \hat\chi^{(73)}_e(t),~~0 + \hat\chi^{(75)}_e(t)$ & $0^\pm$ \\ \hline
6 & $\big(\mathbb{X}^{(i)}_8\big)_{0ij{\rm QRS}ab}$ & ${\rm dom} \left({\rm tr} ~\mathbb{R}_{\rm tot}^4 
- {1\over 4} \left({\rm tr}~\mathbb{R}_{\rm tot}^2\right)^2\right)_{0ij{\rm QRS}ab}$ & $-1 + \hat\chi^{(72)}_e(t), ~~
 -1 + \hat\chi^{(74)}_e(t),~~-1 + \hat\chi^{(76)}_e(t)$ & $-1^\pm$ \\ \hline
 \end{tabular}}
\renewcommand{\arraystretch}{1}
\end{center}
 \caption[]{ \Su ${g_s\over {\rm H}(y){\rm H}_o({\bf x})}$ scalings of the various terms of \eqref{crawlquif} contributing to the EOM of ${\bf G}_{{\rm MNP}i}$. This includes ${\bf G}_{mnpi}, {\bf G}_{mn\alpha i}$, and $ {\bf G}_{m\alpha\beta i}$  flux components.} 
  \label{ccrani14}
 \end{table}

Our next set of G-fluxes are of the form ${\bf G}_{{\rm MNP}i}$ whose flux components are ${\bf G}_{mnpi}, {\bf G}_{mn\alpha i}$ and ${\bf G}_{m\alpha\beta i}$. The tensors contributing to the EOMs are collected in {\bf Table \ref{ccrani14}}. All the usual consistency checks may be performed in the way discussed earlier. For example, comparing the first two rows of {\bf Table \ref{ccrani14}} provides the following value for $\theta_{nl}$:
\bg\label{secwekeep}
\theta_{nl} = 2\hat{l}_{e{\rm MN}}^{{\rm P}i} + {14\over 3} - \hat\zeta_e(t) - \left(1 + \delta_{{\rm N}n} + \delta_{{\rm P}p}\right)\hat\sigma_e(t) - (\hat\alpha_e(t), \hat\beta_e(t))\left( \delta_{{\rm N}\alpha} + \delta_{{\rm P}\beta}\right), \nd 
which matches precisely with \eqref{brittbaba} for $l_{53} = l_{54} = l_{55}= 2$. Similarly, comparing the first row with the third row provides the lower bounds $\hat{l}_{e{\rm MN}}^{{\rm P}i} \ge -2^\pm$, and comparing the first row with the fifth row provides the upper bounds $\hat{l}_{e{\rm MN}}^{{\rm P}i} \le 0^\pm$. If we now denote the exact scalings as:
\bg\label{selfhath10}
\hat{l}_{emn}^{pi} = -2^\pm + \vert x_{28}\vert, ~~~ \hat{l}_{em\alpha }^{pi} = -2^\pm + \vert x_{29}\vert,~~~ \hat{l}_{e\alpha\beta }^{pi} = -2^\pm + \vert x_{30}\vert\nd
where again the superscripts denote small corrections from the presence of $\hat\alpha_e(t)$, $ \hat\beta_e(t)$, $ \hat\sigma_e(t)$, $ \hat\zeta_e(t)$ and $\hat\eta_e(t)$, then looking at {\bf Tables \ref{micandi0_0}}, and {\bf \ref{ccrani14}} gives us:
\bg\label{argoador9}
0^\pm \le (x_{28}, x_{29}, x_{30}) \le 2^\pm, \nd 
exactly spanning the same range as before. In \cite{coherbeta3} we saw similar story but without the $\pm$ superscripts. The ans\"atze for the G-flux components can now be given by the following expression:
\begin{empheq}[box={\mybluebox[5pt]}]{equation}
{\bf G}_{{\rm MNP}i}({\bf x}, y; g_s(t)) = \sum_{l = 0}^6\sum_{k_l = 0}^\infty {\cal G}^{(k, l)}_{{\rm MNP}i}({\bf x}, y) \left({g_s\over {\rm HH}_o}\right)^{-2^\pm + {l\over 3} + {2k_l\over 3}\vert {\rm L}_{{\rm MN}}^{{\rm P}i}(k_l; t)\vert} 
\label{usher10}
\end{empheq} 
where even for $l = 6$ in the series \eqref{usher10} we do not expect  time-independent pieces.

\begin{table}[tb]  
 \begin{center}
\resizebox{\columnwidth}{!}{%
\renewcommand{\arraystretch}{3.5}
}
\renewcommand{\arraystretch}{1}
\end{center}
 \caption[]{ \Su ${g_s\over {\rm H}(y){\rm H}_o({\bf x})}$ scalings of the various terms of \eqref{crawlquif} contributing to the EOM of ${\bf G}_{{\rm 0MN}i}$. This includes ${\bf G}_{0mni}, {\bf G}_{0m\alpha i}$, and $ {\bf G}_{0\alpha\beta i}$  flux components.} 
  \label{ccrani15}
 \end{table}

We now go to the set of G-fluxes are of the form ${\bf G}_{{\rm 0MN}i}$ whose flux components are ${\bf G}_{0mni}, {\bf G}_{0m\alpha i}$ and ${\bf G}_{0\alpha\beta i}$. The tensors contributing to the EOMs are collected in {\bf Table \ref{ccrani15}}. Again, all the usual consistency checks may be performed in the way discussed earlier. For example, comparing the first two rows of {\bf Table \ref{ccrani15}} provides the value for $\theta_{nl}$ as:
\bg\label{secwekeep2}
\theta_{nl} = 2\hat{l}_{e{\rm MN}}^{ij} + {20\over 3} - 2\hat\zeta_e(t) - \hat\sigma_e(t)\left(\delta_{{\rm M}m} + \delta_{{\rm N}n}\right) - (\hat\alpha_e(t), \hat\beta_e(t))\left( \delta_{{\rm M}\alpha} + \delta_{{\rm N}\beta}\right), \nd 
which matches precisely with \eqref{brittbaba} for $l_{76} = l_{77} = l_{78}= 2$. Similarly, comparing the first row with the third row provides the lower bounds $\hat{l}_{e{\rm 0M}}^{{\rm N}i} \ge -3^\pm$, and comparing the first row with the sixth row provides the upper bounds $\hat{l}_{e{\rm 0M}}^{{\rm N}i} \le -1^\pm$. If we now denote the exact scalings as:
\bg\label{selfhath11}
\hat{l}_{e0m}^{ni} = -3^\pm + \vert x_{31}\vert, ~~~ \hat{l}_{e0m}^{\alpha i} = -3^\pm + \vert x_{32}\vert,~~~ \hat{l}_{e0\alpha}^{\beta i} = -3^\pm + \vert x_{33}\vert\nd
where again the superscripts denote small corrections from the presence of $\hat\alpha_e(t)$, $ \hat\beta_e(t)$, $ \hat\sigma_e(t)$, $ \hat\zeta_e(t)$ and $\hat\eta_e(t)$, then looking at {\bf Tables \ref{micandi0_0}}, and {\bf \ref{ccrani15}} gives us:
\bg\label{argoador10}
0^\pm \le (x_{31}, x_{32}, x_{33}) \le 2^\pm, \nd 
exactly spanning the same range as before. In \cite{coherbeta3} we saw similar story but without the $\pm$ superscripts. The ans\"atze for the G-flux components can now be given by the following expression:
\begin{empheq}[box={\mybluebox[5pt]}]{equation}
{\bf G}_{{\rm 0MN}i}({\bf x}, y; g_s(t)) = \sum_{l = 0}^6\sum_{k_l = 0}^\infty {\cal G}^{(k, l)}_{{\rm 0MN}i}({\bf x}, y) \left({g_s\over {\rm HH}_o}\right)^{-3^\pm + {l\over 3} + {2k_l\over 3}\vert {\rm L}_{{\rm 0M}}^{{\rm N}i}(k_l; t)\vert} 
\label{usher11}
\end{empheq} 
where we notice that for the whole range $0 \le l \le 6$ in the series \eqref{usher11} we do not expect  time-independent pieces.

\begin{table}[tb]  
 \begin{center}
\resizebox{\columnwidth}{!}{%
\renewcommand{\arraystretch}{3.5}
}
\renewcommand{\arraystretch}{1}
\end{center}
 \caption[]{ \Su ${g_s\over {\rm H}(y){\rm H}_o({\bf x})}$ scalings of the various terms of \eqref{crawlquif} contributing to the EOM of ${\bf G}_{0abi}$.} 
  \label{ccrani16}
 \end{table}

Our final set of G-flux is of the form ${\bf G}_{0abi}$. The tensors contributing to the EOMs are collected in {\bf Table \ref{ccrani16}}. We can now make all the usual consistency checks in the way discussed earlier. For example, comparing the first two rows of {\bf Table \ref{ccrani16}} provides the following value for $\theta_{nl}$:
\bg\label{secwekeep3}
\theta_{nl} = 2\hat{l}_{e{\rm 0}a}^{bi} + {8\over 3} - 2\hat\zeta_e(t) - \hat\eta_e(t), \nd 
which matches precisely with \eqref{brittbaba} for $l_{79} = 2$. Similarly, comparing the first row with the fifth row provides $-$ keeping due consideration to the fact that the scalings in the first row should reduce by 1 $-$ provides the lower bound $\hat{l}_{e{\rm 0}a}^{bi} \ge -1^\pm$, and comparing the first row with the eighth row  $-$ again keeping due consideration to the fact that the scaling in the first rwo reduces by 1 $-$ provides the upper bound $\hat{l}_{e{\rm 0}a}^{bi} \le +1^\pm$. If we now denote the exact scalings as:
\bg\label{selfhath12}
\hat{l}_{e0a}^{bi} = -1^\pm + \vert x_{34}\vert, \nd
where again the superscripts denote small corrections from the presence of $\hat\alpha_e(t)$, $ \hat\beta_e(t)$, $ \hat\sigma_e(t)$, $ \hat\zeta_e(t)$ and $\hat\eta_e(t)$, then looking at {\bf Tables \ref{micandi0_0}}, and {\bf \ref{ccrani16}} gives us:
\bg\label{argoador11}
0^\pm \le x_{34} \le 2^\pm, \nd 
exactly spanning the same range as before. In \cite{coherbeta3} we saw similar story but without the $\pm$ superscripts. The ans\"atze for the G-flux component can now be given by the following expression:
\begin{empheq}[box={\mybluebox[5pt]}]{equation}
{\bf G}_{{\rm 0}abi}({\bf x}, y; g_s(t)) = \sum_{l = 0}^6\sum_{k_l = 0}^\infty {\cal G}^{(k, l)}_{{\rm 0}abi}({\bf x}, y) \left({g_s\over {\rm HH}_o}\right)^{-1^\pm + {l\over 3} + {2k_l\over 3}\vert {\rm L}_{{\rm 0}a}^{bi}(k_l; t)\vert} 
\label{usher12}
\end{empheq} 
where we notice that even for $l = 1$ in the series \eqref{usher11} we do not expect  time-independent pieces.

\subsection{EOMs for the flux components ${\bf G}_{{\rm MN}ij}, {\bf G}_{{\rm M}aij}$ and ${\bf G}_{abij}$ \label{blackbond7}}

\begin{table}[tb]  
 \begin{center}
\resizebox{\columnwidth}{!}{%
\renewcommand{\arraystretch}{3.5}
\begin{tabular}{|c||c||c||c||c|}\hline Rows & ${\bf G}_{{\rm MN}ij}$ tensors & Forms & ${g_s\over {\rm HH}_o}$ scalings & Lower bounds \\ \hline\hline 
1 & $\big(\mathbb{T}_7^{(f)}\big)_{0{\rm PQRS}ab}$ & $\sqrt{-{\bf g}_{11}} {\bf G}_{{\rm M'N'}i'j'}{\bf g}^{\rm M'M}{\bf g}^{\rm N'N} {\bf g}^{\rm i'i}{\bf g}^{j'j}
\epsilon_{0ij{\rm MNPQRS}ab}$ & $\hat{l}_{e{\rm MN}}^{ij} + 2 -{\hat\zeta_e(t)\over 2} + (2  - \delta_{{\rm M}m}- \delta_{{\rm N}n}){\hat\sigma_e(t)} + {\hat\alpha_e(t) + \hat\beta_e(t)\over 2} - (\hat\alpha_e(t), \hat\beta_e(t))\left(\delta_{{\rm M}\alpha}- \delta_{{\rm N}\beta}\right) + {\hat\eta_e(t)\over 2}$ & $-1^\pm$ \\ \hline 
2 & $\big(\mathbb{T}_7^{(q)}\big)_{0{\rm PQRS}ab}$ & $\sqrt{-{\bf g}_{11}} \mathbb{Y}_4^{{\rm MN}ij} 
\epsilon_{0ij{\rm MNPQRS}ab}$ & $\theta_{nl} - \hat{l}_{e{\rm MN}}^{ij} - {14\over 3} + {3\hat\zeta_e(t)\over 2}  + 2\hat\sigma_e(t) + {\hat\alpha_e(t) + \hat\beta_e(t)\over 2} + {\hat\eta_e(t)\over 2}$ & $-1^\pm\big({2\over 3}^\pm\big)$ \\ \hline 
3 & $\big(\mathbb{T}^{({\rm M})}_8\big)_{0{\rm MPQRS}ab}$ & ${\bf G}_{[0{\rm MPQ}} {\bf G}_{{\rm RS}ab]}$ & $ l_{e[0{\rm M}}^{[{\rm PQ}} \oplus l_{e{\rm RS}]}^{ab]}$ & $-1^\pm$
\\ \hline 
4 & $\big(\mathbb{T}^{(i)}_8\big)_{0i{\rm PQRS}ab}$ & ${\bf G}_{[0i{\rm PQ}} {\bf G}_{{\rm RS}ab]}$ & $ l_{e[0i}^{[{\rm PQ}} \oplus l_{e{\rm RS}]}^{ab]}$ & $-2^\pm$
\\ \hline 
5 & $\big(\mathbb{X}^{({\rm M})}_8\big)_{0{\rm MPQRS}ab}$ & ${\rm dom} \left({\rm tr} ~\mathbb{R}_{\rm tot}^4 
- {1\over 4} \left({\rm tr}~\mathbb{R}_{\rm tot}^2\right)^2\right)_{0{\rm MPQRS}ab}$ & $1 + \hat\chi^{(89)}_e(t),~~1 + \hat\chi^{(91)}_e(t),~~1 + \hat\chi^{(93)}_e(t)$ & $+1^\pm$ \\ \hline
6 & $\big(\mathbb{X}^{(i)}_8\big)_{0i{\rm PQRS}ab}$ & ${\rm dom} \left({\rm tr} ~\mathbb{R}_{\rm tot}^4 
- {1\over 4} \left({\rm tr}~\mathbb{R}_{\rm tot}^2\right)^2\right)_{0i{\rm PQRS}ab}$ & $0 + \hat\chi^{(90)}_e(t), ~~
 0 + \hat\chi^{(92)}_e(t),~~0 + \hat\chi^{(94)}_e(t)$ & $0^\pm$ \\ \hline
 \end{tabular}}
\renewcommand{\arraystretch}{1}
\end{center}
 \caption[]{ \Su ${g_s\over {\rm H}(y){\rm H}_o({\bf x})}$ scalings of the various terms of \eqref{crawlquif} contributing to the EOM of ${\bf G}_{{\rm MN}ij}$. This includes ${\bf G}_{mnij}, {\bf G}_{m\alpha ij}$, and $ {\bf G}_{\alpha\beta ij}$  flux components.} 
  \label{ccrani17}
 \end{table}

In this last section we will deal with the G-fluxes of the form ${\bf G}_{{\rm MN}ij}, {\bf G}_{{\rm M}aij}$ and ${\bf G}_{abij}$ that have two legs along the spatial ${\bf R}^2$ directions. Our first set is
${\bf G}_{{\rm MN}ij}$ that allows flux components of the form
${\bf G}_{mnij}, {\bf G}_{m\alpha ij}$ and ${\bf G}_{\alpha\beta ij}$.
The tensors contributing to the EOMs are now collected in {\bf Table \ref{ccrani17}}. Again, all the usual consistency checks may be performed in the way discussed earlier. For example, comparing the first two rows of {\bf Table \ref{ccrani17}} provides the following value for $\theta_{nl}$:
\bg\label{secwekeep4}
\theta_{nl} = 2\hat{l}_{e{\rm MN}}^{ij} + {20\over 3} - 2\hat\zeta_e(t) - \left(\delta_{{\rm M}m} + \delta_{{\rm N}n}\right)\hat\sigma_e(t) - (\hat\alpha_e(t), \hat\beta_e(t))\left( \delta_{{\rm M}\alpha} + \delta_{{\rm N}\beta}\right), \nd 
which matches precisely with \eqref{brittbaba} for $l_{61} = l_{62} = l_{63}= 2$. Note the similarity with \eqref{secwekeep2}. Similarly, comparing the first row with the third row provides the lower bounds $\hat{l}_{e{\rm MN}}^{ij} \ge -3^\pm$, and comparing the first row with the fifth row provides the upper bounds $\hat{l}_{e{\rm MN}}^{ij} \le -1^\pm$. If we now denote the exact scalings as:
\bg\label{selfhath13}
\hat{l}_{emn}^{ij} = -3^\pm + \vert x_{35}\vert, ~~~ \hat{l}_{em\alpha }^{ij} = -3^\pm + \vert x_{36}\vert,~~~ \hat{l}_{e\alpha\beta }^{ij} = -3^\pm + \vert x_{37}\vert\nd
where again the superscripts denote small corrections from the presence of $\hat\alpha_e(t)$, $ \hat\beta_e(t)$, $ \hat\sigma_e(t)$, $ \hat\zeta_e(t)$ and $\hat\eta_e(t)$, then looking at {\bf Tables \ref{micandi0_0}}, and {\bf \ref{ccrani17}} gives us:
\bg\label{argoador12}
0^\pm \le (x_{35}, x_{36}, x_{37}) \le 2^\pm, \nd 
exactly spanning the same range as before. In \cite{coherbeta3} we saw similar story but without the $\pm$ superscripts. The ans\"atze for the G-flux components can now be given by the following expression:
\begin{empheq}[box={\mybluebox[5pt]}]{equation}
{\bf G}_{{\rm MN}ij}({\bf x}, y; g_s(t)) = \sum_{l = 0}^6\sum_{k_l = 0}^\infty {\cal G}^{(k, l)}_{{\rm MN}ij}({\bf x}, y) \left({g_s\over {\rm HH}_o}\right)^{-3^\pm + {l\over 3} + {2k_l\over 3}\vert {\rm L}_{{\rm MN}}^{ij}(k_l; t)\vert} 
\label{usher13}
\end{empheq} 
where for any values of $l$ in the range $0 \le l \le 6$ in the series \eqref{usher11} we do not expect  time-independent pieces. Note the similarity of \eqref{usher13} with \eqref{usher11} which is expected because for both cases we are dealing with ${\bf R}^{2, 1}$ directions.

\begin{table}[tb]  
 \begin{center}
\resizebox{\columnwidth}{!}{%
\renewcommand{\arraystretch}{3.5}
}
\renewcommand{\arraystretch}{1}
\end{center}
 \caption[]{ \Su ${g_s\over {\rm H}(y){\rm H}_o({\bf x})}$ scalings of the various terms of \eqref{crawlquif} contributing to the EOM of ${\bf G}_{{\rm M}aij}$. This includes ${\bf G}_{maij}$ and $ {\bf G}_{{\rm M}\alpha ij}$  flux components.} 
  \label{ccrani18}
 \end{table}

Our next set if ${\bf G}_{{\rm M}aij}$ which allows components of the form
${\bf G}_{maij}$ and ${\bf G}_{\alpha aij}$. The tensors contributing to the EOMs from \eqref{crawlquif} are collected in {\bf Table \ref{ccrani18}}. Comparing the first row with the third and sixth rows of {\bf Table \ref{ccrani18}} easily gives us the bounds $-2^\pm \le \hat{l}_{e{\rm M}a}^{ij} \le 0$. Similarly, comparing the first and the second rows gives us:
\bg\label{secwekeep5}
\theta_{nl} = 2\hat{l}_{e{\rm M}a}^{ij} + {14\over 3} - 2\hat\zeta_e(t) - \hat\sigma_e(t)\delta_{{\rm M}m} - (\hat\alpha_e(t), \hat\beta_e(t))\delta_{{\rm M}\alpha} -(0, \hat\eta_e(t)), \nd
which matches with \eqref{brittbaba} for $l_{64} = l_{65} = 2$. Rest of the analysis follows the pattern laid out earlier. If we denote the exact scalings of the flux components as:
\bg\label{hoasholly}
\hat{l}_{ema}^{ij} = -2^\pm + \vert x_{38}\vert, ~~~~ \hat{l}_{e\alpha a}^{ij} = -2^\pm + \vert x_{39}\vert, ~~~~ {\rm with}~~~ 0^\pm \le (x_{38}, x_{39}) \le 2^\pm, 
\nd
then from \eqref{hoasholly} one can provide the precise ans\"atze for the G-flux components ${\bf G}_{{\rm M}aij}$ in the following way:
\begin{empheq}[box={\mybluebox[5pt]}]{equation}
{\bf G}_{{\rm M}aij}({\bf x}, y; g_s(t)) = \sum_{l = 0}^6\sum_{k_l = 0}^\infty {\cal G}^{(k, l)}_{{\rm M}aij}({\bf x}, y) \left({g_s\over {\rm HH}_o}\right)^{-2^\pm + {l\over 3} + {2k_l\over 3}\vert {\rm L}_{{\rm M}a}^{ij}(k_l; t)\vert} 
\label{usher14}
\end{empheq}

\begin{table}[tb]  
 \begin{center}
\resizebox{\columnwidth}{!}{%
\renewcommand{\arraystretch}{3.5}
\begin{tabular}{|c||c||c||c||c|}\hline Rows & ${\bf G}_{abij}$ tensors & Forms & ${g_s\over {\rm HH}_o}$ scalings & Lower bounds \\ \hline\hline
1 & $\big(\mathbb{T}_7^{(f)}\big)_{0{\rm MNPQRS}}$ & $\sqrt{-{\bf g}_{11}} {\bf G}_{a'b'i'j'}{\bf g}^{a'a}{\bf g}^{b'b} {\bf g}^{i'i}{\bf g}^{j'j}
\epsilon_{0ij{\rm MNPQRS}ab}$ & $\hat{l}_{eab}^{ij} - 2 - {\hat\zeta_e(t)\over 2} + 2 {\hat\sigma_e(t)} + {\hat\alpha_e(t) + \hat\beta_e(t)\over 2} - {\hat\eta_e(t)\over 2}$ & $-3^\pm$ \\ \hline 
2 & $\big(\mathbb{T}_7^{(q)}\big)_{0{\rm MNPQRS}}$ & $\sqrt{-{\bf g}_{11}} \mathbb{Y}_4^{abij} 
\epsilon_{0ij{\rm MNPQRS}ab}$ & $\theta_{nl} - \hat{l}_{eab}^{ij} - {14\over 3} + {3\hat\zeta_e(t)\over 2}  + 2\hat\sigma_e(t) + {\hat\alpha_e(t) + \hat\beta_e(t)\over 2} + {\hat\eta_e(t)\over 2}$ & $-3^\pm\big({2\over 3}^\pm\big)$ \\ \hline 
3 & $\big(\mathbb{T}^{(a)}_8\big)_{0{\rm MNPQRS}a}$ & ${\bf G}_{[0{\rm MNP}} {\bf G}_{{\rm QRS}a]}$ & $ l_{e[0{\rm M}}^{[{\rm NP}} \oplus l_{e{\rm QR}]}^{{\rm S}a]}$ & $-2^\pm$
\\ \hline 
4 & $\big(\mathbb{T}^{(i)}_8\big)_{0i{\rm MNPQRS}}$ & ${\bf G}_{[0i{\rm MN}}{\bf G}_{{\rm PQRS}]}$ & $ l_{e[0i}^{[{\rm MN}} \oplus l_{e{\rm PQ}]}^{{\rm RS}]}$ & $-4^\pm$
\\ \hline 
5 & $\big(\mathbb{X}^{(a)}_8\big)_{0{\rm MNPQRS}a}$ & ${\rm dom} \left({\rm tr} ~\mathbb{R}_{\rm tot}^4 
- {1\over 4} \left({\rm tr}~\mathbb{R}_{\rm tot}^2\right)^2\right)_{0{\rm MNPQRS}a}$ & $0 + \hat\chi^{(101)}_e(t)$ & $0^\pm$ \\ \hline
6 & $\big(\mathbb{X}^{(i)}_8\big)_{0i{\rm MNPQRS}}$ & ${\rm dom} \left({\rm tr} ~\mathbb{R}_{\rm tot}^4 
- {1\over 4} \left({\rm tr}~\mathbb{R}_{\rm tot}^2\right)^2\right)_{0i{\rm MNPQRS}}$ & $-2 + \hat\chi^{(102)}_e(t)$ & $-2^\pm$ \\ \hline
 \end{tabular}}
\renewcommand{\arraystretch}{1}
\end{center}
 \caption[]{ \Su ${g_s\over {\rm H}(y){\rm H}_o({\bf x})}$ scalings of the various terms of \eqref{crawlquif} contributing to the EOM of ${\bf G}_{abij}$.} 
  \label{ccrani19}
 \end{table}

Our last G-flux component is ${\bf G}_{abij}$ with $(a, b) \in {\mathbb{T}^2\over {\cal G}}$. The tensors contributing to the EOMs from \eqref{crawlquif} are collected in {\bf Table \ref{ccrani19}}. However we now face a few issues that prohibit us to use the scheme laid out earlier. For example, since we expect the fields to be independent of the toroidal ${\mathbb{T}^2\over {\cal G}}$  and the spatial ${\bf R}^2$ directions, we cannot determine the lower and the upper bounds on $\hat{l}_{eab}^{ij}$ by comparing the rows in {\bf Table \ref{ccrani19}}. However using alternative techniques, as given in {\bf Table \ref{micandi0_0}}, provide us with $\hat{l}_{eab}^{ij} \ge -1^\pm$. We could also compare the first two rows in {\bf Table \ref{ccrani19}} to get:
\bg\label{hoasmeye}
\theta_{nl} = 2\hat{l}_{eab}^{ij} + {8\over 3} - 2\hat\zeta_e(t) - \hat\eta_e(t), \nd
which precisely matches with \eqref{brittbaba} for $l_{66} = 2$. Now, knowing the lower bound from {\bf Table \ref{micandi0_0}} in fact suggests that the exact scaling may be expressed as:
\bg\label{kaluhoas}
\hat{l}_{eab}^{ij} = -1^\pm + \vert x_{40}\vert, \nd
but now, since the upper bound is unknown, we cannot determine the range for $x_{40}$. This however does not prohibit us to propose the following ans\"atze for the exact form for the flux component:
\begin{empheq}[box={\mybluebox[5pt]}]{equation}
{\bf G}_{abij}({\bf x}, y; g_s(t)) = \sum_{l = 0}^6\sum_{k_l = 0}^\infty {\cal G}^{(k, l)}_{abij}({\bf x}, y) \left({g_s\over {\rm HH}_o}\right)^{-1^\pm + {l\over 3} + {2k_l\over 3}\vert {\rm L}_{ab}^{ij}(k_l; t)\vert} 
\label{usher15}
\end{empheq}

\begin{figure}[h]
\centering
\begin{tabular}{c}
\includegraphics[width=6.2in]{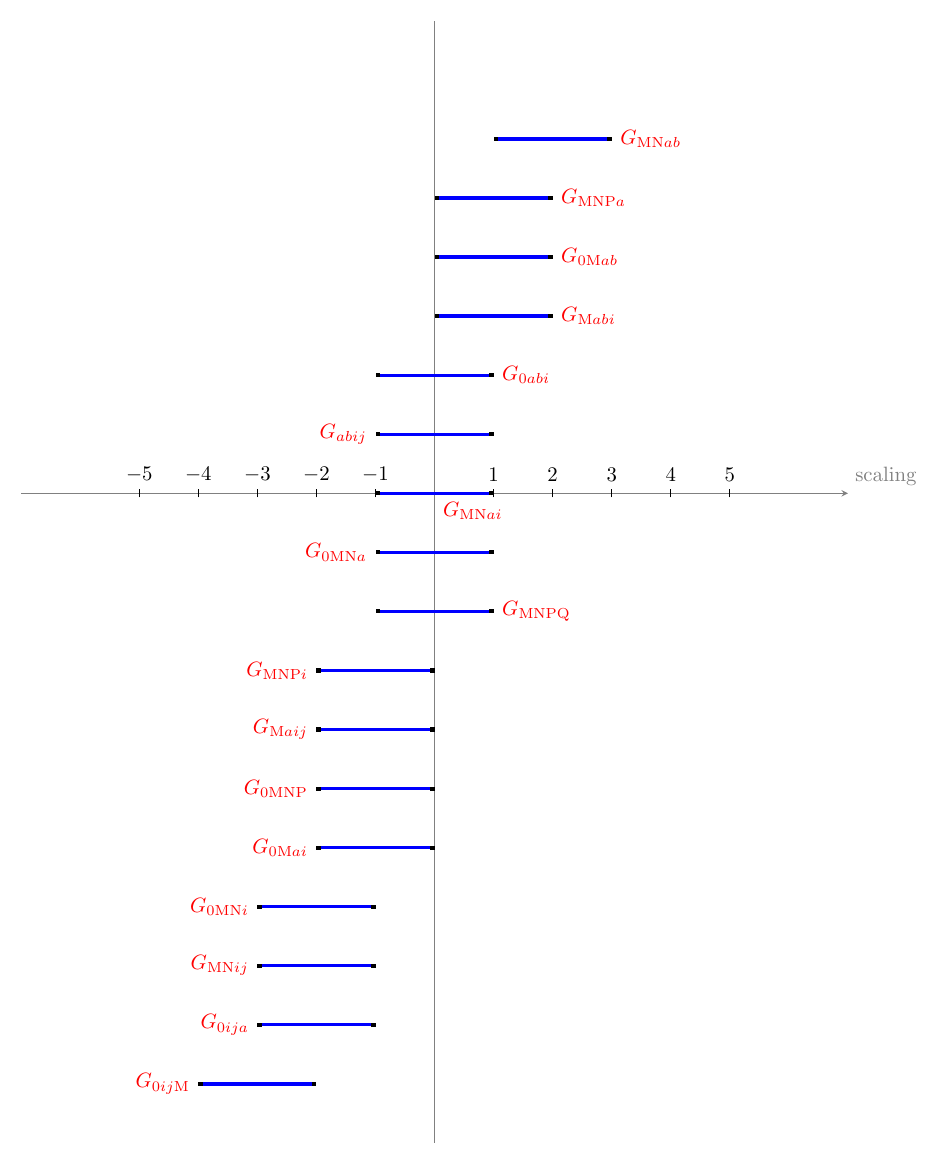}
\end{tabular}
\vskip-.1in
\caption[]{The flux EOMs from \eqref{crawlquif} do not exactly fix the scalings precisely, rather they provide ranges of validities. For all the on-shell flux components entering \eqref{crawlquif}, the ranges are shown here. The y-axis do not have any significance, however as we go up along the y-axis, we see the gradual shifts of the flux components towards positive scalings.}
\label{fluxbehavior1}
\end{figure}


\section{Dual forms and the Bianchi identities for the G-flux components\label{sec4.5}}

In the previous section we managed to determine the  form of the on-shell G-fluxes for all the 40 components that appear in \eqref{botsuga}. However the results were presented in terms of a parameter $l \in \mathbb{Z}$ that takes integer values within the range $0 \le l \le 6$. The question is whether we can quantify the precise value of $l$ for all the flux components. The answer turns out to be in the affirmative if we take the Bianchi identities for all the flux components carefully. In the following we shall elaborate the story in detail. 

To proceed we should note that Bianchi identities are derived from the EOMs of the {\it dual} fluxes, {\it i.e.} the EOMs for the dual seven-form fluxes. The scalings of the various components of the dual fluxes are collected in {\bf Tables \ref{fridaylin1}} and {\bf \ref{fridaylin2}}.
The precise EOMs for the dual fluxes have already been derived in section 4.6 of \cite{coherbeta3}, so we can simply quote the answer here. The EOMs for the dual forms may now be expressed as:

{\footnotesize
\bg\label{crawlquif2}
\partial_{{\rm N}_k}\big(\hat{\mathbb{T}}_4^{(f)}\big)_{\rm N_1.....N_4} =   c_1
\partial_{{\rm N}_k}\big(\hat{\mathbb{T}}_4^{(q)}\big)_{\rm N_1.....N_4} + 
c_2\partial_{{\rm N}_k}\big(\hat{\mathbb{Y}}_4\big)_{{\rm N_1.....N_4}} 
 + 
c_3 \hat{\mathbb{N}}_5\big({\bf \Lambda}_5\big)_{{\rm N_1.....N_4}{\rm N}_k}, \nd}
which is similar to \eqref{crawlquif}, but now expressed using four-forms. The $c_i$, similar to $b_i$ in \eqref{crawlquif}, are $g_s$ independent coefficients (see \cite{coherbeta3}). Note that we did not explicitly write the non-perturbative terms. These are absorbed in the definition of $\hat{\mathbb{T}}_4^{(q)}$. The second term on the RHS of the equality in \eqref{crawlquif2} may be argued from the ${\bf C}_3 \wedge \mathbb{Z}_8$ coupling, where $\mathbb{Z}_8 = b_2 {\bf G}_4 \wedge {\bf G}_4 + b_3 \mathbb{X}_8$ with $(b_2, b_3)$ as in \eqref{crawlquif}, by the following chain of dualities\footnote{Recall that for a $p$-form $\omega_p$ and a $q$-form $\eta_q$ in $d$ dimensions $\ast\omega_p \wedge \ast\eta_q = (-1)^{p(d-q)} \omega_p \wedge \eta_q$, with $p+q = d$.}:
\bg\label{malinechat}
{\bf C}_3 \wedge \mathbb{Z}_8 \to {\bf G}_4 \wedge \mathbb{Z}_7 = 
\ast {\bf G}_7 \wedge \ast \hat{\mathbb{Y}}_4 = {\bf G}_7 \wedge \hat{\mathbb{Y}}_4, \nd
\noindent ignoring signs and boundary terms, and we have taken $\mathbb{Z}_8 = d\mathbb{Z}_7$ locally. Defining ${\bf G}_7 = d{\bf C}_6 + ..$, we can reproduce the required term in \eqref{crawlquif2}. However there is a small subtlety here. On one hand $\mathbb{Z}_7$ is not a invariant topological form because we have ignored the boundary terms, so $\ast \hat{\mathbb{Y}}_4$ should reflect this. On the other hand ${\rm tr}~\mathbb{R}_{\rm tot} \wedge \mathbb{R}_{\rm tot}$ is a well-defined topologically invariant four-form that could contribute to $\hat{\mathbb{Y}}_4$ in \eqref{malinechat}, the other being the gauge invariant ${\bf G}_4$. The latter is already been accounted for in \eqref{crawlquif2}, so we will express $\hat{\mathbb{Y}}_4$ as in \eqref{dietherapie} using ${\rm tr}~\mathbb{R}_{\rm tot} \wedge \mathbb{R}_{\rm tot}$. (Such a choice will lead to some issue later on that we shall discuss in section \ref{sec8.8}.)  Meanwhile, one might get a little worried by the fact that, since ${\rm tr}~\mathbb{R}_{\rm tot} \wedge \mathbb{R}_{\rm tot}$ is a {\it locally} exact form, {\it i.e.} ${\rm tr}~\mathbb{R}_{\rm tot} \wedge \mathbb{R}_{\rm tot} \equiv d \Omega_3$,
$d({\rm tr}~\mathbb{R}_{\rm tot} \wedge \mathbb{R}_{\rm tot}) = 0$ and the RHS of \eqref{crawlquif2} is simply sourced by five-branes in the absence of the quantum corrections. This is not correct because of a small subtlety, and in fact lies at the heart of the problem. The precise form of ${\rm tr}~\mathbb{R}_{\rm tot} \wedge \mathbb{R}_{\rm tot}$ and the action of the exterior derivative is as follows:
\bg\label{argomeythi}
&& {\rm tr}~\mathbb{R}_{\rm tot} \wedge \mathbb{R}_{\rm tot}({\bf x}, y, w^a; g_s) = 
{\rm tr}~{\bf R}_{\rm tot} \wedge {\bf R}_{\rm tot}({\bf x}, y, x_3; g_s)\Theta(x_{11} - w_o) \\
&& d\left({\rm tr}~\mathbb{R}_{\rm tot} \wedge \mathbb{R}_{\rm tot}\right) = d\left( {\rm tr}~{\bf R}_{\rm tot} \wedge {\bf R}_{\rm tot}\Theta(x_{11}-w_o)\right) = {\rm tr}~{\bf R}_{\rm tot} \wedge {\bf R}_{\rm tot} \wedge \delta(x_{11}-w_o) dx_{11}, \nonumber\nd
which is also precisely the way the curvature terms manifested from the O8-planes in the type IIA dual picture of \cite{jatkar1} (see for example eqn. (2.13) - (2.16) there). \eqref{argomeythi} also implies that on a five-manifold ${\cal M}_5$ with a boundary, we expect:
\bg\label{capitainemey}
\int_{{\cal M}_5} d({\rm tr}~\mathbb{R}_{\rm tot} \wedge \mathbb{R}_{\rm tot}) = \int_{\partial{\cal M}_5} {\rm tr}~\mathbb{R}_{\rm tot} \wedge \mathbb{R}_{\rm tot}, \nd
which is the manifestation of the fact that ${\rm tr}~\mathbb{R}_{\rm tot} \wedge \mathbb{R}_{\rm tot}$ is not globally exact. Plus the fact that the exterior derivative action produces localized four-form as in \eqref{argomeythi} provides additional justification. Therefore
with these at hand, the tensors appearing in \eqref{crawlquif2} are now defined in the following way:
\bg\label{valentinpen}
&& \big(\hat{\mathbb{T}}_4^{(f)}\big)_{\rm N_1N_2 N_3N_4} = {\bf G}_{\rm N_1 N_2 N_3 N_4} \nonumber\\
&& \big(\hat{\mathbb{Y}}_4\big)_{{\rm N_1N_2 N_3N_4}} = \left({\rm tr}~\mathbb{R}_{\rm tot} \wedge \mathbb{R}_{\rm tot}\right)_{\rm N_1 N_2 N_3 N_4} \nonumber\\
&& (\hat{\mathbb{T}}_4^{(q)}\big)_{\rm N_1N_2N_3N_4} = \left(\ast \mathbb{Y}_7\right)_{\rm N_1N_2 N_3N_4} = \sqrt{-{\bf g}_{11}} \mathbb{Y}_7^{\rm ABCDEFG} ~\epsilon_{\rm ABCDEFG N_1N_2N_3N_4}, \nd
where the scalings for $\mathbb{Y}_7$ $-$ which also include the non-perturbative corrections via trans-series form \eqref{kimkarol} of the action $-$ will be inferred from \eqref{botsuga2.0} and from {\bf Tables \ref{fridaylin1}} and {\bf \ref{fridaylin2}}. The number of five-branes $\hat{\mathbb{N}}_5 \equiv \hat{\mathbb{N}}_5(\bar{g}_s)$, is a {\it dynamical} quantity that would scale with respect to $\bar{g}_s$ in the following way:
\bg\label{carolpenl}
\hat{\mathbb{N}}_5(\bar{g}_s) = \hat{\mathbb{N}}_5^{(0)}\left({g_s\over {\rm H}(y) {\rm H}_o({\bf x})}\right)^{\hat{h}_e(t)}, \nd
where $\hat{h}_e(t)$ contains all the perturbative and non-perturbative corrections to the bare scaling\footnote{There is a small subtlety here: the choice of $\hat{h}_e(t)$ would depend on the {\it orientations} of the five-branes as shown in {\bf Tables \ref{ccrani201}, \ref{ccrani203}} and {\bf \ref{ccrani205}}. This will become clearer as we move along.}. This is similar to $\hat{j}_e(t)$ for the dynamical M2-branes from \eqref{samthigap}. The difference from \eqref{samthigap} is that we keep ${\bf \Lambda}_5$ to be $g_s$ independent and put all the $g_s$ dependence on $\hat{\mathbb{N}}_5$. A democratic formulation, where we put all the temporal dependence for the dynamical M2-branes on $\hat{\mathbb{N}}_2$ and keep ${\bf \Lambda}_8$ to be $g_s$ independent, exists but is not required here.

\subsection{Bianchi identities for the flux components ${\bf G}_{0ij{\rm M}}$ and ${\bf G}_{0ija}$ \label{cchase1}}

\begin{table}[tb]  
 \begin{center}
\resizebox{\columnwidth}{!}{%
\renewcommand{\arraystretch}{2.5}
\begin{tabular}{|c||c||c||c|}\hline Rows & ${\bf G}_{0ij{\rm M}}$ tensors & Forms & ${g_s\over {\rm HH}_o}$ scalings \\ \hline\hline
1 & $\big(\hat{\mathbb{T}}_4^{(f)}\big)_{0ij{\rm M}}$ & ${\bf G}_{0ij{\rm M}}$ & $-4^\pm + {l\over 3}$  \\ \hline 
2 & $\big(\hat{\mathbb{T}}_4^{(q)}\big)_{0ij{\rm M}}$ & $\sqrt{-{\bf g}_{11}} \mathbb{Y}_7^{{\rm NPQRS}ab} 
\epsilon_{0ij{\rm MNPQRS}ab}$ & $\theta_{nl}  - {14\over 3} - {l\over 3} + {3\hat\zeta_e(t)} + \delta_{{\rm M}m} \hat\sigma_e(t) + (\hat\alpha_e(t), \hat\beta_e(t))\delta_{{\rm M}\alpha} $  \\ \hline 
3 & $\big(\hat{\mathbb{Y}}_4\big)_{0ij{\rm M}}$ & ${\rm dom} \left({\rm tr} ~\mathbb{R}_{\rm tot} \wedge \mathbb{R}_{\rm tot}\right)_{0ij{\rm M}}$ & $-3^\pm$  \\ \hline
 \end{tabular}}
\renewcommand{\arraystretch}{1}
\end{center}
 \caption[]{ \Su ${g_s\over {\rm H}(y){\rm H}_o({\bf x})}$ scalings of the various terms of \eqref{crawlquif2} contributing to the Bianchi identity of ${\bf G}_{0ij{\rm M}}$. To compute the scaling in the second row, we use the dual flux scalings from rows 7 and 8 in {\bf Table \ref{fridaylin1}}. Note that the dynamical M5-branes do not contribute.} 
  \label{ccrani20}
 \end{table}

\begin{figure}[h]
\centering
\begin{tabular}{c}
\includegraphics[width=4.5in]{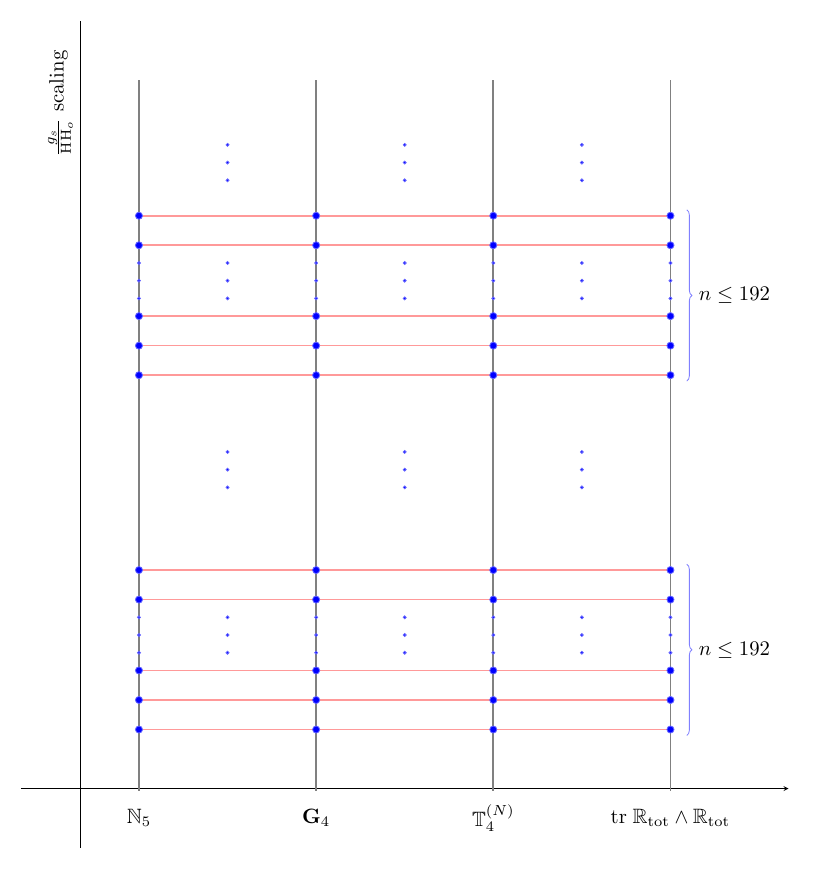}
\end{tabular}
\vskip-.1in
\caption[]{Typical ways the Bianchi identities \eqref{crawlquif2} are satisfied to all orders in $\bar{g}_s \equiv {g_s\over {\rm H}(y) {\rm H}_o({\bf x})}$. The quantum terms, that in principle could include both perturbative and non-perturbative contributions via the trans-series \eqref{kimkarol}, are jointly denoted by $\mathbb{T}_4^{(N)} \equiv (\hat{\mathbb{T}}_4^{(q)})_{\rm N_1..N_4} dy^{\rm N_1}\wedge..\wedge dy^{\rm N_4}$ as given in \eqref{valentinpen}.
Unlike the case with the EOMs from \eqref{crawlquif} $-$ which may also be inferred from {\bf figure \ref{eombehavior}} $-$ all terms in \eqref{crawlquif2} can be arranged to scale in exactly the same way right from the lowest order, including the matching of the $\pm$ superscripts. In fact this matching suggests that the terms appearing in \eqref{crawlquif2} may be grouped in copies of $n$, where $n \le 192$. Note that the {\it spacings} between the various terms in the individual groups may not be identical. These and other details will be further elaborated in section \ref{sec8.8}.}
\label{bianchibehavior}
\end{figure}

\begin{figure}[h]
\centering
\begin{tabular}{c}
\includegraphics[width=6in]{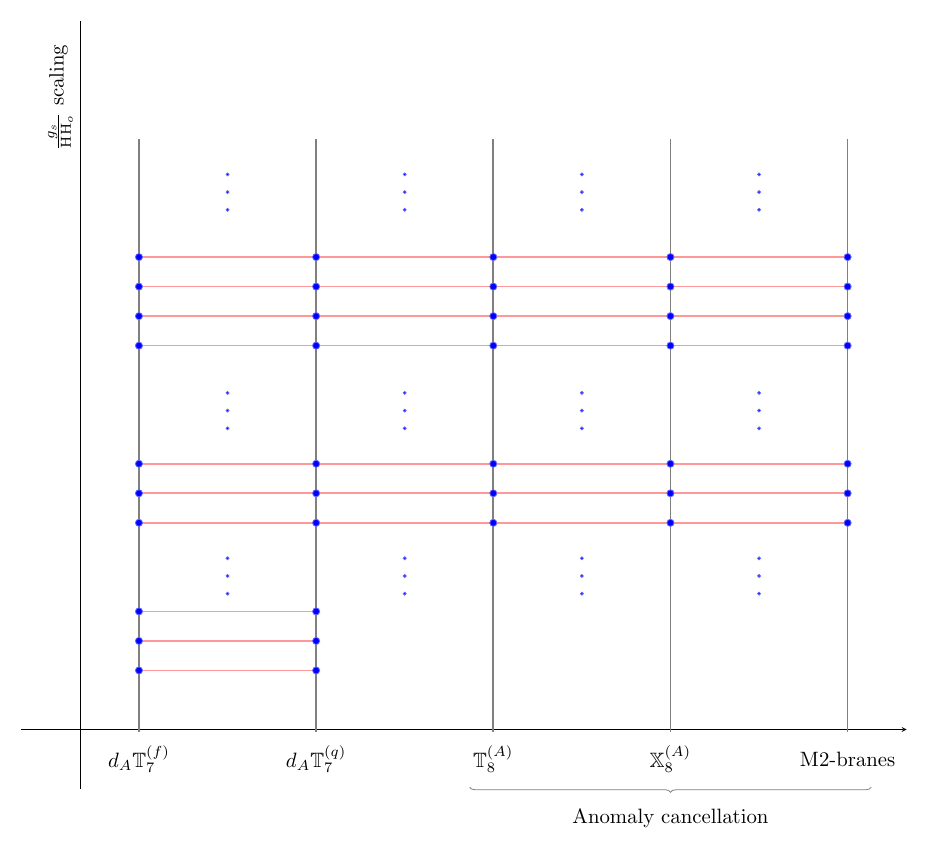}
\end{tabular}
\vskip-.1in
\caption[]{Typical ways the flux EOMs \eqref{crawlquif} are satisfied to all orders in $\bar{g}_s \equiv {g_s\over {\rm H}(y) {\rm H}_o({\bf x})}$. At the lowest order, the $\bar{g}_s$ scalings of the flux and the quantum terms, namely $\mathbb{T}_7^{(f)}$ and $\mathbb{T}_7^{(q)}$ respectively from \eqref{olivecostamey}, are matched and as we go to higher orders in $\bar{g}_s$, the matching extends to include all the other terms in \eqref{crawlquif}. The lowest order matching between  $\mathbb{T}_7^{(f)}$ and $\mathbb{T}_7^{(q)}$ fits consistently with \eqref{brittbaba} for the ${\rm E}_8 \times {\rm E}_8$ case as discussed in footnote \ref{fasermaxe}. Interestingly, a subset of the terms, that include the flux products $\mathbb{T}_8^{({\rm A})}$, the $\mathbb{X}_8^{{\rm A})}$ polynomial, and the number of dynamical M2-branes, are responsible for the anomaly cancellation as discussed here and will also be elaborated in section \ref{secanomaly}. The subscript and the superscript ${\rm A}$ are related to the directions along which the corresponding flux components are oriented. The quantum term $\mathbb{T}_7^{(q)}$ can contain both the perturbative and the non-perturbative corrections as a trans-series given in \eqref{kimkarol}.}
\label{eombehavior}
\end{figure}

We will start by computing the Bianchi identities of the G-fluxes of the form ${\bf G}_{0ij{\rm M}}$ whose components are ${\bf G}_{0ijm}$ and ${\bf G}_{0ij\alpha}$. The tensor contributing to \eqref{crawlquif2} are collected in {\bf Table \ref{ccrani20}}. Comparing the $g_s$ scalings in the three rows immediately fixes both $l$ and $\theta_{nl}$ to be of the following form:
\bg\label{salome1}
l = 3, ~~~~~ \theta_{nl} = {8\over 3}^\pm, \nd
implying that $x_1 = x_2 = 1^\pm$ in \eqref{argotrani}. In fact as we shall see for all the 40 on-shell fluxes, $x_i = 1^\pm$, and therefore the functional form for ${\bf G}_{0ij{\rm M}}$ as given before in \eqref{usher1} becomes:
\begin{empheq}[box={\mybluebox[5pt]}]{align}
{\bf G}_{0ij{\rm M}}({\bf x}, y; g_s(t)) = \sum_{k = 0}^\infty {\cal G}^{(k)}_{0ij{\rm M}}({\bf x}, y) \left({g_s\over {\rm HH}_o}\right)^{-3^\pm  + {2k\over 3}\vert {\rm L}_{\rm 0i}^{j{\rm M}}(k; t)\vert}
\label{salgill1}
\end{empheq}
where ${\rm L}_{\rm 0i}^{j{\rm M}}(k; t)$ contains all the sub-dominant perturbative and the non-perturbative corrections (see section \ref{sec8.8} for more details on this). The functional form \eqref{salgill1} now solves both the EOMs \eqref{crawlquif} as well as the Bianchi identities \eqref{crawlquif2}. It is interesting that the Bianchi identities help us fix the precise scalings to $-3^\pm$ which lie half-way in the range $-4^\pm \le \hat{l}_{e0i}^{j{\rm M}} \le -2^\pm$. This is shown in {\bf figure \ref{fluxbehavior2}}. The contents of {\bf Table \ref{crecraqwiff1}} now change to the one shown in {\bf Table \ref{crecraqwiff101}}. At order $\left({g_s\over {\rm H}(y) {\rm H}_o({\bf x})}\right)^{1^\pm}$, the first row of {\bf Table \ref{crecraqwiff101}} can be equated to the second row. At order $\left({g_s\over {\rm H}(y) {\rm H}_o({\bf x})}\right)^{2^\pm}$, and when we take derivatives along the ${\cal M}_4$ directions, rows 3, 6 and 9 also start participating and they match up giving us $\hat{j}_e = 2^\pm$ and the anomaly cancellation condition with the dynamical M2-branes.  Beyond this, or to the same order, the non-perturbative effects enter the picture and we realize the full EOM from \eqref{crawlquif}. Similarly the temporal and spatial ${\bf R}^2$ derivatives' action start at 
order $\left({g_s\over {\rm H}(y) {\rm H}_o({\bf x})}\right)^{1^\pm}$, and continues upward to accommodate other terms of \eqref{crawlquif} except the M2-brane term. Finally, comparing the first two rows of {\bf Table \ref{crecraqwiff101}} now gives us $\theta_{nl} = {8\over 3}^\pm$ in \eqref{crawlquif} which matches precisely with $\theta_{nl} = {8\over 3}^\pm$ that we got from the Bianchi identity \eqref{crawlquif2} in \eqref{salome1}. In {\bf figures \ref{bianchibehavior}} and {\bf \ref{eombehavior}}, typical ways how the Bianchi identities and the EOMs are satisfied in our case are shown. More elaborations will be given in section \ref{sec8.8}.

\begin{table}[tb]  
 \begin{center}
\resizebox{\columnwidth}{!}{%
\renewcommand{\arraystretch}{2.3}
}
\renewcommand{\arraystretch}{1}
\end{center}
 \caption[]{\Su ${g_s\over {\rm H}(y){\rm H}_o({\bf x})}$ scalings of the various terms from {\bf Table \ref{crecraqwiff1}} once we use \eqref{salgill1} and $x_i = 1^\pm$ in \eqref{argothai} and other related equations.} 
  \label{crecraqwiff101}
 \end{table}

\begin{table}[tb]  
 \begin{center}
\resizebox{\columnwidth}{!}{%
\renewcommand{\arraystretch}{2.5}
%
}
\renewcommand{\arraystretch}{1}
\end{center}
 \caption[]{ \Su ${g_s\over {\rm H}(y){\rm H}_o({\bf x})}$ scalings of the various terms of \eqref{crawlquif2} contributing to the Bianchi identity of ${\bf G}_{0ija}$. To compute the scaling in the second row, we use the dual flux scalings from row 8 in {\bf Table \ref{fridaylin2}}. Note that the dynamical M5-branes again do not contribute.} 
  \label{ccrani200}
 \end{table}

For the ${\bf G}_{0ija}$ flux components, the tensors contributing to the Bianch identity \eqref{crawlquif2} are collected in {\bf Table \ref{ccrani200}}. Comparing the three rows we immediate get the same results as in \eqref{salome1}. This tells us that the G-flux components now take the form as:
\begin{empheq}[box={\mybluebox[5pt]}]{align}
{\bf G}_{0ija}({\bf x}, y; g_s(t)) = \sum_{k = 0}^\infty {\cal G}^{(k)}_{0ija}({\bf x}, y) \left({g_s\over {\rm HH}_o}\right)^{-2^\pm  + {2k\over 3}\vert {\rm L}_{0i}^{ja}(k; t)\vert}
\label{salgill2}
\end{empheq}
from what we had in \eqref{usher1}. Again we see that the Bianchi identity helps us to choose the precise scaling which, in turn, lies half-way between the range $-3^\pm \le \hat{l}_{e0i}^{ja} \le -1^\pm$ as shown in {\bf figure \ref{fluxbehavior2}}. The contents of {\bf Table \ref{ccquiff3}} now change to {\bf Table \ref{ccquiff30}}.

\begin{table}[tb]  
 \begin{center}
\resizebox{\columnwidth}{!}{%
\renewcommand{\arraystretch}{2.3}
\begin{tabular}{|c||c||c||c|}\hline Rows & ${\bf G}_{0ija}$ tensors & Forms & ${g_s\over {\rm HH}_o}$ scalings  \\ \hline\hline
1 & $\big(\mathbb{T}_7^{(f)}\big)_{{\rm MNPQRS} b}$ & $\sqrt{-{\bf g}_{11}} {\bf G}_{0i'j'a'}{\bf g}^{00} {\bf g}^{i'i}{\bf g}^{j'j} {\bf g}^{a'a}
\epsilon_{0ij{\rm MNPQRS} ab}$  & $0^\pm$ \\ \hline 
2 & $\big(\mathbb{T}_7^{(q)}\big)_{{\rm MNPQRS} b}$ & $\sqrt{-{\bf g}_{11}} \mathbb{Y}_4^{0ija} 
\epsilon_{0ij{\rm MNPQRS}ab}$ & $\theta_{nl} - {8\over 3}^\pm$ \\ \hline 
3 & $\big(\mathbb{T}^{(a)}_8\big)_{{\rm MNPQRS} ab}$ & ${\bf G}_{[{\rm MNPQ}} {\bf G}_{{\rm RS} ab]}$ & $ 2^\pm$
\\ \hline 
4 & $\big(\mathbb{T}^{(0)}_8\big)_{0{\rm MNPQRS} b}$ & ${\bf G}_{[0{\rm MNP}} {\bf G}_{{\rm Q}\alpha\beta b]}$ & $ 0^\pm$
\\ \hline 
5 & $\big(\mathbb{T}^{(i)}_8\big)_{{\rm MNPQRS} ib}$ & ${\bf G}_{[{\rm MNPQ}} {\bf G}_{{\rm RS} ib]}$ & $ 0^\pm$
\\ \hline 
6 & $\big(\mathbb{X}^{(a)}_8\big)_{{\rm MNPQRS} ab}$ & ${\rm dom} \left({\rm tr} ~\mathbb{R}_{\rm tot}^4 
- {1\over 4} \left({\rm tr}~\mathbb{R}_{\rm tot}^2\right)^2\right)_{{\rm MNPQRS} ab}$  & $2^\pm$ \\ \hline
7 & $\big(\mathbb{X}^{(0)}_8\big)_{{\rm MNPQRS} 0b}$ & ${\rm dom} \left({\rm tr} ~\mathbb{R}_{\rm tot}^4 
- {1\over 4} \left({\rm tr}~\mathbb{R}_{\rm tot}^2\right)^2\right)_{{\rm MNPQRS} 0b}$  & $0^\pm$ \\ \hline
8 & $\big(\mathbb{X}^{(i)}_8\big)_{{\rm MNPQRS} ib}$ & ${\rm dom} \left({\rm tr} ~\mathbb{R}_{\rm tot}^4 
- {1\over 4} \left({\rm tr}~\mathbb{R}_{\rm tot}^2\right)^2\right)_{{\rm MNPQRS} ib}$  & $0^\pm$ \\ \hline
9 & $\big({\bf \Lambda}_8\big)_{{\rm MNPQRS} ab}$ & $\left({\bf \Lambda}_8({\bf x}, y, w; g_s)\right)_{{\rm MNPQRS} ab}$
& $\hat{j}_e(t)$ \\ \hline
\end{tabular}}
\renewcommand{\arraystretch}{1}
\end{center}
 \caption[]{ \Su ${g_s\over {\rm H}(y){\rm H}_o({\bf x})}$ scalings of the various terms from {\bf Table \ref{ccquiff3}} now using the exact scaling for ${\bf G}_{0ija}$ and other flux components.} 
  \label{ccquiff30}
 \end{table}

At order $\left({g_s\over {\rm H}(y){\rm H}_o({\bf x})}\right)^{-1^\pm}$, once we take the temporal derivative in \eqref{crawlquif}, we can balance the first two rows of {\bf Table \ref{ccquiff30}}. This gives us $\theta_{nl} = {8\over 3}^\pm$, thus matching with what we have from \eqref{salome1}. At order $\left({g_s\over {\rm H}(y){\rm H}_o({\bf x})}\right)^{2^\pm}$, we can match rows 3, 6 and 9, giving us $\hat{j}_e(t) = 2^\pm$ for the dynamical M2 branes. This matches precisely with the what we had from {\bf Table \ref{crecraqwiff101}} as well as the anomaly cancellation condition. At order $\left({g_s\over {\rm H}(y){\rm H}_o({\bf x})}\right)^{0^\pm}$, with the temporal derivative, the first two rows at next to leading orders can now match with rows 4 and 7 at the lowest orders. Similar story can be constructed for the derivatives along the spatial ${\bf R}^2$ directions.

\subsection{Bianchi identities for the flux components ${\bf G}_{\rm MNPQ}$ and ${\bf G}_{\rm 0MNP}$ \label{sec8.2}} 

\begin{table}[tb]  
 \begin{center}
\resizebox{\columnwidth}{!}{%
\renewcommand{\arraystretch}{2.5}
\begin{tabular}{|c||c||c||c|}\hline Rows & ${\bf G}_{\rm MNPQ}$ tensors & Forms & ${g_s\over {\rm HH}_o}$ scalings \\ \hline\hline
1 & $\big(\hat{\mathbb{T}}_4^{(f)}\big)_{\rm MNPQ}$ & ${\bf G}_{\rm MNPQ}$ & $-1^\pm + {l\over 3}$  \\ \hline 
2 & $\big(\hat{\mathbb{T}}_4^{(q)}\big)_{\rm MNPQ}$ & $\sqrt{-{\bf g}_{11}} \mathbb{Y}_7^{0ij{\rm RS}ab} 
\epsilon_{0ij{\rm MNPQRS}ab}$ & $\theta_{nl}  - {5\over 3} - {l\over 3} + (2 + \delta_{{\rm P}p} + \delta_{{\rm Q}q})\hat\sigma_e(t) + (\hat\alpha_e(t), \hat\beta_e(t))(\delta_{{\rm P}\alpha} + \delta_{{\rm Q}\beta})$  \\ \hline 
3 & $\big(\hat{\mathbb{Y}}_4\big)_{\rm MNPQ}$ & ${\rm dom} \left({\rm tr} ~\mathbb{R}_{\rm tot} \wedge \mathbb{R}_{\rm tot}\right)_{{\rm MNPQ}}$ & $0^\pm$  \\ \hline
4 & $\big(\widetilde{\bf \Lambda}_5\big)_{\rm MNPQS}$ & $\hat{\mathbb{N}}_5\left({\bf \Lambda}_5\right)_{\rm MNPQS}$ & $\hat{h}_e(t)$\\ \hline
 \end{tabular}}
\renewcommand{\arraystretch}{1}
\end{center}
 \caption[]{ \Su ${g_s\over {\rm H}(y){\rm H}_o({\bf x})}$ scalings of the various terms of \eqref{crawlquif2} contributing to the Bianchi identity of ${\bf G}_{\rm MNPQ}$. To compute the scaling in the second row, we use the dual flux scalings from rows 1, 2 and 4 in {\bf Table \ref{fridaylin1}}. Note that the dynamical M5-branes do participate with $g_s$ scalings $\hat{h}_e(t)$.} 
  \label{ccrani201}
 \end{table}

For the ${\bf G}_{\rm MNPQ}$ flux components, the Bianchi identities \eqref{crawlquif2} tell us that the five-branes should now contribute. The tensors contributing to the Bianchi identities are collected in {\bf Table \ref{ccrani201}}. From here we note that for the ${\bf G}_{mnpq}$ flux components we require M5-branes to be oriented along the spatial $x^{0, i, j}$ directions and wrapping three-cycles ${\bf S}^1_\alpha \times {\mathbb{T}^2\over {\cal G}}$, where ${\bf S}^1_\alpha \in {\cal M}_2$. For ${\bf G}_{mn\alpha\beta}$, the five-branes wrap one-cycles in ${\cal M}_4$. For ${\bf G}_{mnp\alpha}$, they can either wrap one-cycles in ${\cal M}_2$ or in ${\cal M}_4$.
These five-branes may or may not be stable depending on whether ${\cal M}_2$ and ${\cal M}_4$ allow topological one-cycles or not\footnote{Recall that the internal manifold ${\cal M}_4 \times {\cal M}_2$ is generically non-K\"ahler and non-complex.}, so we need to tread a little carefully here. Comparing the first three rows of {\bf Table \ref{ccrani201}} again gives us \eqref{salome1}. Comparing with the fourth row tells us that the aforementioned five-branes are dynamical only because of the $\pm$ superscript in $\hat{h}_e(t) = 0^\pm$. All in all, this fixes the functional form of the G-flux components to be:
\begin{empheq}[box={\mybluebox[5pt]}]{align}
{\bf G}_{\rm MNPQ}({\bf x}, y; g_s(t)) = \sum_{k = 0}^\infty {\cal G}^{(k)}_{\rm MNPQ}({\bf x}, y) \left({g_s\over {\rm HH}_o}\right)^{0^\pm  + {2k\over 3}\vert {\rm L}_{\rm MN}^{\rm PQ}(k; t)\vert}
\label{salgill3}
\end{empheq}
again demonstrating that the Bianchi identities fix the scalings right at the half-way in the range $-1\le \hat{l}_{e{\rm MN}}^{\rm PQ} \le 1$. This is shown in {\bf figure \ref{fluxbehavior2}}. Note that if ${\rm S} = \alpha \in {\cal M}_2$ in the fourth row of {\bf Table \ref{ccrani201}}, ${\bf \Lambda}_5 = 0$ from \eqref{crawlquif2}, so these five-branes do not contribute. On the other hand, if ${\rm S} = m \in {\cal M}_4$, then the contributions are as described above. 

\begin{table}[tb]  
 \begin{center}
\resizebox{\columnwidth}{!}{%
\renewcommand{\arraystretch}{2.5}
\begin{tabular}{|c||c||c||c|}\hline Rows & ${\bf G}_{\rm MNPQ}$ tensors & Forms & ${g_s\over {\rm HH}_o}$ scalings  \\ \hline\hline
1 & $\big(\mathbb{T}_7^{(f)}\big)_{0ij{\rm RS} ab}$ & $\sqrt{-{\bf g}_{11}} {\bf G}_{\rm M'N'P'Q'}{\bf g}^{\rm M'M} {\bf g}^{\rm N'N}{\bf g}^{\rm P'P} {\bf g}^{\rm Q'Q}
\epsilon_{0ij{\rm MNPQRS}ab}$ & $-2^\pm$ \\ \hline 
2 & $\big(\mathbb{T}_7^{(q)}\big)_{0ij{\rm RS} ab}$ & $\sqrt{-{\bf g}_{11}} \mathbb{Y}_4^{\rm MNPQ} 
\epsilon_{0ij{\rm MNPQRS}ab}$ & $\theta_{nl}  - {14\over 3}^\pm$ \\ \hline 
3 & $\big(\mathbb{T}^{({\rm M})}_8\big)_{0ij{\rm MRS} ab}$ & ${\bf G}_{[0ij{\rm M}} {\bf G}_{{\rm RS}ab]}$ & $ -1^\pm$
\\ \hline 
4 & $\big(\mathbb{X}^{({\rm M})}_8\big)_{0ij{\rm MRS} ab}$ & ${\rm dom} \left({\rm tr} ~\mathbb{R}_{\rm tot}^4 
- {1\over 4} \left({\rm tr}~\mathbb{R}_{\rm tot}^2\right)^2\right)_{0ij{\rm MRS} ab}$ & $-1^\pm$ \\ \hline
\end{tabular}}
\renewcommand{\arraystretch}{1}
\end{center}
 \caption[]{ \Su ${g_s\over {\rm H}(y){\rm H}_o({\bf x})}$ scalings of the various terms from {\bf Table \ref{ccrani1}, \ref{ccrani2}} and {\bf \ref{ccrani3}} now using the exact scalings of ${\bf G}_{\rm MNPQ}$ from \eqref{salgill3} and other flux components.} 
  \label{ccrani101}
 \end{table}

Now using the exact scalings of the G-fluxes, we can easily see that the contents of {\bf Tables \ref{ccrani1}, \ref{ccrani2}} and {\bf \ref{ccrani3}} change to the ones given in {\bf Table \ref{ccrani101}}. At order $\left({g_s\over {\rm H}(y){\rm H}_o({\bf x})}\right)^{-2^\pm}$, we can equate the first two rows of 
{\bf Table \ref{ccrani101}}, giving us $\theta_{nl} = {8\over 3}^\pm$, exactly as we got from the Bianchi identities. At order $\left({g_s\over {\rm H}(y){\rm H}_o({\bf x})}\right)^{-1^\pm}$, we expect all four rows to match-up with next-to-leading order contributions from the first two rows and leading order contributions from the other two rows.

\begin{table}[tb]  
 \begin{center}
\resizebox{\columnwidth}{!}{%
\renewcommand{\arraystretch}{3}
\begin{tabular}{|c||c||c||c|}\hline Rows & ${\bf G}_{{\rm 0MNP}}$ tensors & Forms & ${g_s\over {\rm HH}_o}$ scalings \\ \hline\hline
1 & $\big(\hat{\mathbb{T}}_4^{(f)}\big)_{{\rm 0MNP}}$ & ${\bf G}_{{\rm 0MNP}}$ & $-2^\pm + {l\over 3}$  \\ \hline 
2 & $\big(\hat{\mathbb{T}}_4^{(q)}\big)_{{\rm 0MNP}}$ & $\sqrt{-{\bf g}_{11}} \mathbb{Y}_7^{ij{\rm QRS}ab} 
\epsilon_{0ij{\rm MNPQRS}ab}$ & $\theta_{nl}  - {8\over 3} - {l\over 3}  +\hat\zeta_e(t) + (1 +\delta_{{\rm N}n} + \delta_{{\rm P}p})\hat\sigma_e(t) + (\hat\alpha_e(t), \hat\beta_e(t))(\delta_{{\rm N}\alpha} + \delta_{{\rm P}\beta})$  \\ \hline 
3 & $\big(\hat{\mathbb{Y}}_4\big)_{{\rm 0MNP}}$ & ${\rm dom} \left({\rm tr} ~\mathbb{R}_{\rm tot} \wedge \mathbb{R}_{\rm tot}\right)_{{\rm 0MNP}}$ & $-1^\pm$  \\ \hline
 \end{tabular}}
\renewcommand{\arraystretch}{1}
\end{center}
 \caption[]{ \Su ${g_s\over {\rm H}(y){\rm H}_o({\bf x})}$ scalings of the various terms of \eqref{crawlquif2} contributing to the Bianchi identity of ${\bf G}_{{\rm 0MNP}}$. To compute the scaling in the second row, we use the dual flux scalings from rows 9, 10 and 11 in {\bf Table \ref{fridaylin2}}. Note that now the dynamical M5-branes do not participate.} 
  \label{ccrani202}
 \end{table}

The story for the flux components ${\bf G}_{\rm 0MNP}$ is similar since the derivatives all act in the same way in \eqref{crawlquif2}. Comparing the three rows in {\bf Table \ref{ccrani202}} we immediately recover \eqref{salome1}, giving us the follow precise scaling of the flux components:
\begin{empheq}[box={\mybluebox[5pt]}]{align}
{\bf G}_{\rm 0MNP}({\bf x}, y; g_s(t)) = \sum_{k = 0}^\infty {\cal G}^{(k)}_{\rm 0MNP}({\bf x}, y) \left({g_s\over {\rm HH}_o}\right)^{-1^\pm  + {2k\over 3}\vert {\rm L}_{\rm 0M}^{\rm NP}(k; t)\vert}
\label{salgill4}
\end{empheq}
which takes the value half-way between the range $-2^\pm \le \hat{l}_{e{\rm 0M}}^{\rm NP} \le 0^\pm$. This is shown in {\bf figure \ref{fluxbehavior2}}. The contents of {\bf Tables \ref{ccrani4}, \ref{ccrani5}} and {\bf \ref{ccrani52}} change to the one given in {\bf Table \ref{ccrani521}}. 
\begin{table}[tb]   
 \begin{center}
\resizebox{\columnwidth}{!}{%
\renewcommand{\arraystretch}{2.5}
\begin{tabular}{|c||c||c||c|}\hline Rows & ${\bf G}_{\rm 0MNP}$ tensors & Forms & ${g_s\over {\rm HH}_o}$ scalings  \\ \hline\hline
1 & $\big(\mathbb{T}_7^{(f)}\big)_{ij{\rm QRS} ab}$ & $\sqrt{-{\bf g}_{11}} {\bf G}_{\rm 0M'N'P'}{\bf g}^{00} {\bf g}^{\rm M'M}{\bf g}^{\rm N'N}{\bf g}^{\rm P'P}
\epsilon_{0ij{\rm MNPQRS}ab}$ & $-1^\pm$ \\ \hline 
2 & $\big(\mathbb{T}_7^{(q)}\big)_{ij{\rm QRS} ab}$ & $\sqrt{-{\bf g}_{11}} \mathbb{Y}_4^{{\rm 0MNP}} 
\epsilon_{0ij{\rm MNPQRS} ab}$ & $\theta_{nl} - {11\over 3}^\pm$ \\ \hline 
3 & $\big(\mathbb{T}^{({\rm M})}_8\big)_{ij{\rm MQRS} ab}$ & ${\bf G}_{[ij{\rm MQ}} {\bf G}_{{\rm RS}ab]}$ & $ ~~0^\pm$
\\ \hline 
4 & $\big(\mathbb{T}^{(0)}_8\big)_{0ij{\rm QRS}ab}$ & ${\bf G}_{[0ij{\rm Q}} {\bf G}_{{\rm RS} ab]}$ & $ -1^\pm$
\\ \hline 
5 & $\big(\mathbb{X}^{({\rm M})}_8\big)_{ij{\rm MQRS} ab}$ & ${\rm dom} \left({\rm tr} ~\mathbb{R}_{\rm tot}^4 
- {1\over 4} \left({\rm tr}~\mathbb{R}_{\rm tot}^2\right)^2\right)_{ijq{\rm MQRS} ab}$  & $~~0^\pm$ \\ \hline
6 & $\big(\mathbb{X}^{(0)}_8\big)_{0ij{\rm QRS} ab}$ & ${\rm dom} \left({\rm tr} ~\mathbb{R}_{\rm tot}^4 
- {1\over 4} \left({\rm tr}~\mathbb{R}_{\rm tot}^2\right)^2\right)_{0ij{\rm QRS} ab}$  & $-1^\pm$ \\ \hline
\end{tabular}}
\renewcommand{\arraystretch}{1}
\end{center}
 \caption[]{ \Su ${g_s\over {\rm H}(y){\rm H}_o({\bf x})}$ scalings of the various terms from {\bf Tables \ref{ccrani4}, \ref{ccrani5}} and {\bf \ref{ccrani52}} now using the exact scalings of ${\bf G}_{\rm 0MNP}$ from \eqref{salgill4} and other flux components.} 
  \label{ccrani521}
 \end{table}
To order $\xxy^{-1^\pm}$ the first two rows in {\bf Table \ref{ccrani521}} match up to give $\theta_{nl} = {8\over 3}^\pm$, thus matching with what we expect from the Bianchi identities. At order 
$\xxy^{0^\pm}$, we can further match the NLO scalings with rows 3 and 5 at the leading order for derivatives along ${\cal M}_4$. For the temporal derivative, the NLO matching happens with rows 4 and 6 at order $\xxy^{-1^\pm}$.

\subsection{Bianchi identities for the flux components ${\bf G}_{{\rm MNP}a}$ and ${\bf G}_{{\rm 0NP}a}$ \label{cchase3}}

\begin{table}[tb]  
 \begin{center}
\resizebox{\columnwidth}{!}{%
\renewcommand{\arraystretch}{3}
\begin{tabular}{|c||c||c||c|}\hline Rows & ${\bf G}_{{\rm MNP}a}$ tensors & Forms & ${g_s\over {\rm HH}_o}$ scalings \\ \hline\hline
1 & $\big(\hat{\mathbb{T}}_4^{(f)}\big)_{{\rm MNP}a}$ & ${\bf G}_{{\rm MNP}a}$ & $0^\pm + {l\over 3}$  \\ \hline 
2 & $\big(\hat{\mathbb{T}}_4^{(q)}\big)_{{\rm MNP}a}$ & $\sqrt{-{\bf g}_{11}} \mathbb{Y}_7^{0ij{\rm QRS}b} 
\epsilon_{0ij{\rm MNPQRS}ab}$ & $\theta_{nl}  - {2\over 3} - {l\over 3}  + (1 + \delta_{{\rm N}n} + \delta_{{\rm P}p})\hat\sigma_e(t) + (\hat\alpha_e(t), \hat\beta_e(t))(\delta_{{\rm N}\alpha} + \delta_{{\rm P}\beta}) + (0, \hat\eta_e(t))$  \\ \hline 
3 & $\big(\hat{\mathbb{Y}}_4\big)_{{\rm MNP}a}$ & ${\rm dom} \left({\rm tr} ~\mathbb{R}_{\rm tot} \wedge \mathbb{R}_{\rm tot}\right)_{{\rm MNP}a}$ & $1^\pm$  \\ \hline
4 & $\big(\widetilde{\bf \Lambda}_5\big)_{{\rm MNPS}a}$ & $\hat{\mathbb{N}}_5\left({\bf \Lambda}_5\right)_{{\rm MNPS}a}$ & $\hat{l}_e(t)$\\ \hline
 \end{tabular}}
\renewcommand{\arraystretch}{1}
\end{center}
 \caption[]{ \Su ${g_s\over {\rm H}(y){\rm H}_o({\bf x})}$ scalings of the various terms of \eqref{crawlquif2} contributing to the Bianchi identity of ${\bf G}_{{\rm MNP}a}$. To compute the scaling in the second row, we use the dual flux scalings from rows 3, 5 and 6 in {\bf Table \ref{fridaylin1}}. Note that now the dynamical M5-branes do  participate with $g_s$ scalings $\hat{l}_e(t)$.} 
  \label{ccrani203}
 \end{table}

For the ${\bf G}_{{\rm MNP}a}$ flux components, the tensors contributing  to the Bianchi identities are shown in {\bf Table \ref{ccrani203}}. We see that the M5-branes now contribute. These fivebranes have topologies of the form ${\bf R}^{2, 1} \times {\bf S}^2_{{\cal M}_4}\times {\bf S}^1_{\xoxo}, {\bf R}^{2, 1} \times {\cal M}_2 \times{\bf S}^1_{\xoxo}$ and 
${\bf R}^{2, 1} \times {\bf S}^1_{{\cal M}_4} \times {\bf S}^1_{{\cal M}_2} \times{\bf S}^1_{\xoxo}$ where ${\bf S}^n_{{\cal M}_d}$ denotes a $n$-cycle in the manifold ${\cal M}_d$ of dimension $d$. It is easy to see that, if we compare all the four rows in {\bf Table \ref{ccrani203}}, we recover \eqref{salome1}, with $\hat{l}_e(t) = 1^\pm$ implying the dynamical nature of the aforementioned five-branes and the flux components taking the following precise form:
\begin{empheq}[box={\mybluebox[5pt]}]{align}
{\bf G}_{{\rm MNP}a}({\bf x}, y; g_s(t)) = \sum_{k = 0}^\infty {\cal G}^{(k)}_{{\rm MNP}a}({\bf x}, y) \left({g_s\over {\rm HH}_o}\right)^{1^\pm  + {2k\over 3}\vert {\rm L}_{\rm MN}^{{\rm P}a}(k; t)\vert}
\label{salgill5}
\end{empheq}
as shown in {\bf figure \ref{fluxbehavior2}}. The contents of {\bf Table \ref{ccrani7}} now change to the ones given in {\bf Table \ref{ccrani77}}.
\begin{table}[tb]  
 \begin{center}
\resizebox{\columnwidth}{!}{%
\renewcommand{\arraystretch}{3.5}
\begin{tabular}{|c||c||c||c||c|}\hline Rows & ${\bf G}_{{\rm MNP}a}$ tensors & Forms & ${g_s\over {\rm HH}_o}$ scalings  \\ \hline\hline
1 & $\big(\mathbb{T}_7^{(f)}\big)_{0ij{\rm QRS}b}$ & $\sqrt{-{\bf g}_{11}} {\bf G}_{{\rm M'N'P'}a'}{\bf g}^{\rm M'M} {\bf g}^{\rm N'N}{\bf g}^{\rm P'P}{\bf g}^{a'a}
\epsilon_{0ij{\rm MNPQRS}ab}$ & $-3^\pm$ \\ \hline 
2 & $\big(\mathbb{T}_7^{(q)}\big)_{0ij{\rm QRS} b}$ & $\sqrt{-{\bf g}_{11}} \mathbb{Y}_4^{{\rm MNP}a} 
\epsilon_{0ij{\rm MNPQRS}ab}$ & $\theta_{nl}  - {17\over 3}^\pm$ \\ \hline 
3 & $\big(\mathbb{T}^{({\rm M})}_8\big)_{0ij{\rm MQRS} b}$ & ${\bf G}_{[0ij{\rm M}} {\bf G}_{{\rm QRS} b]}$  & $-2^\pm$
\\ \hline 
4 & $\big(\mathbb{T}^{(a)}_8\big)_{0ij{\rm QRS} ab}$ & ${\bf G}_{[0ij{\rm Q}} {\bf G}_{{\rm RS} ab]}$  & $-1^\pm$
\\ \hline 
5 & $\big(\mathbb{X}^{({\rm M})}_8\big)_{0ij{\rm MQRS} b}$ & ${\rm dom} \left({\rm tr} ~\mathbb{R}_{\rm tot}^4 
- {1\over 4} \left({\rm tr}~\mathbb{R}_{\rm tot}^2\right)^2\right)_{0ij{\rm MQRS} b}$ & $-2^\pm, ~~-2^\pm,~~ -2^\pm$ \\ \hline
6 & $\big(\mathbb{X}^{(a)}_8\big)_{0ij{\rm QRS} ab}$ & ${\rm dom} \left({\rm tr} ~\mathbb{R}_{\rm tot}^4 
- {1\over 4} \left({\rm tr}~\mathbb{R}_{\rm tot}^2\right)^2\right)_{0ij{\rm QRS} ab}$ & $-1\pm, ~~-1^\pm, ~~-1^\pm$ \\ \hline
\end{tabular}}
\renewcommand{\arraystretch}{1}
\end{center}
 \caption[]{ \Su ${g_s\over {\rm H}(y){\rm H}_o({\bf x})}$ scalings of the various terms from {\bf Table \ref{ccrani7}} once we take the precise scalings of the ${\bf G}_{{\rm MNP}a}$ from \eqref{salgill5} and other flux components.} 
  \label{ccrani77}
 \end{table}
To order $\xxy^{-3^\pm}$ the first two rows of {\bf Table \ref{ccrani77}} match up giving us $\theta_{nl} = {8\over 3}^\pm$, which is same as what we had from the Bianchi identities. To order $\xxy^{-2^\pm}$ we can further match the NLO scalings from the first two rows with rows 3 and 5 once we take derivatives along ${\cal M}_4$ directions. For derivatives along ${\mathbb{T}^2\over {\cal G}}$ directions, the match is only between rows 4 and 6 at order $\xxy^{-1^\pm}$ because the first two rows do not contribute.

\begin{table}[tb]  
 \begin{center}
\resizebox{\columnwidth}{!}{%
\renewcommand{\arraystretch}{3}
\begin{tabular}{|c||c||c||c|}\hline Rows & ${\bf G}_{{\rm 0NP}a}$ tensors & Forms & ${g_s\over {\rm HH}_o}$ scalings \\ \hline\hline
1 & $\big(\hat{\mathbb{T}}_4^{(f)}\big)_{{\rm 0NP}a}$ & ${\bf G}_{{\rm 0NP}a}$ & $-1^\pm + {l\over 3}$  \\ \hline 
2 & $\big(\hat{\mathbb{T}}_4^{(q)}\big)_{{\rm 0NP}a}$ & $\sqrt{-{\bf g}_{11}} \mathbb{Y}_7^{ij{\rm MQRS}b} 
\epsilon_{0ij{\rm MNPQRS}ab}$ & $\theta_{nl}  - {5\over 3} - {l\over 3} + +\hat\zeta_e(t) + (\delta_{{\rm N}n} + \delta_{{\rm P}p})\hat\sigma_e(t) + (\hat\alpha_e(t), \hat\beta_e(t))(\delta_{{\rm N}\alpha} + \delta_{{\rm P}\beta}) + (0, \hat\eta_e(t))$  \\ \hline 
3 & $\big(\hat{\mathbb{Y}}_4\big)_{{\rm 0NP}a}$ & ${\rm dom} \left({\rm tr} ~\mathbb{R}_{\rm tot} \wedge \mathbb{R}_{\rm tot}\right)_{{\rm 0NP}a}$ & $0^\pm$  \\ \hline
 \end{tabular}}
\renewcommand{\arraystretch}{1}
\end{center}
 \caption[]{ \Su ${g_s\over {\rm H}(y){\rm H}_o({\bf x})}$ scalings of the various terms of \eqref{crawlquif2} contributing to the Bianchi identity of ${\bf G}_{{\rm 0NP}a}$. To compute the scaling in the second row, we use the dual flux scalings from rows 14, 15 and 16 in {\bf Table \ref{fridaylin2}}. Note that now the dynamical M5-branes do not participate.} 
  \label{ccrani204}
 \end{table}

For the ${\bf G}_{{\rm 0NP}a}$ flux components, the tensors contributing to the Bianchi identities \eqref{crawlquif2} are collected in {\bf Table \ref{ccrani204}}. Again comparing the three rows, we recover \eqref{salome1}, and therefore the flux components take the following precise form:
\begin{empheq}[box={\mybluebox[5pt]}]{align}
{\bf G}_{{\rm 0NP}a}({\bf x}, y; g_s(t)) = \sum_{k = 0}^\infty {\cal G}^{(k)}_{{\rm 0NP}a}({\bf x}, y) \left({g_s\over {\rm HH}_o}\right)^{0^\pm  + {2k\over 3}\vert {\rm L}_{\rm 0N}^{{\rm P}a}(k; t)\vert}
\label{salgill6}
\end{empheq}
as shown in {\bf figure \ref{fluxbehavior2}}. Note that, as discussed earlier, the $0^\pm$ scalings at the leading order do not imply time-independence precisely because of the $\pm$ superscripts. The contents of {\bf Table \ref{ccrani8}} change to the ones given in {\bf Table \ref{ccrani88}}.
\begin{table}[tb]  
 \begin{center}
\resizebox{\columnwidth}{!}{%
\renewcommand{\arraystretch}{3.0}
\begin{tabular}{|c||c||c||c||c|}\hline Rows & ${\bf G}_{0{\rm NP}a}$ tensors & Forms & ${g_s\over {\rm HH}_o}$ scalings  \\ \hline\hline
1 & $\big(\mathbb{T}_7^{(f)}\big)_{ij{\rm MQRS}b}$ & $\sqrt{-{\bf g}_{11}} {\bf G}_{{\rm 0N'P'}a'}{\bf g}^{\rm 00} {\bf g}^{\rm N'N}{\bf g}^{\rm P'P}{\bf g}^{a'a}
\epsilon_{0ij{\rm MNPQRS}ab}$  & $-2^\pm$ \\ \hline 
2 & $\big(\mathbb{T}_7^{(q)}\big)_{ij{\rm MQRS} b}$ & $\sqrt{-{\bf g}_{11}} \mathbb{Y}_4^{0{\rm NP}a} 
\epsilon_{0ij{\rm MNPQRS}ab}$ & $\theta_{nl}  - {14\over 3}^\pm$ \\ \hline 
3 & $\big(\mathbb{T}^{({\rm M})}_8\big)_{ij{\rm MNQRS} b}$ & ${\bf G}_{[ij{\rm MN}} {\bf G}_{{\rm QRS} b]}$ & $-1^\pm$
\\ \hline 
4 & $\big(\mathbb{T}^{(a)}_8\big)_{ij{\rm MQRS} ab}$ & ${\bf G}_{[ij{\rm MQ}} {\bf G}_{{\rm RS} ab]}$  & $~~0^\pm$
\\ \hline 
5 & $\big(\mathbb{T}^{(0)}_8\big)_{0ij{\rm MQRS} b}$ & ${\bf G}_{[0ij{\rm M}} {\bf G}_{{\rm QRS} b]}$  & $-2^\pm$
\\ \hline 
6 & $\big(\mathbb{X}^{({\rm M})}_8\big)_{ij{\rm MNQRS} b}$ & ${\rm dom} \left({\rm tr} ~\mathbb{R}_{\rm tot}^4 
- {1\over 4} \left({\rm tr}~\mathbb{R}_{\rm tot}^2\right)^2\right)_{ij{\rm MNQRS} b}$ & $-1^\pm, ~~-1^\pm,~~ -1^\pm$ \\ \hline
7 & $\big(\mathbb{X}^{(a)}_8\big)_{ij{\rm MQRS} ab}$ & ${\rm dom} \left({\rm tr} ~\mathbb{R}_{\rm tot}^4 
- {1\over 4} \left({\rm tr}~\mathbb{R}_{\rm tot}^2\right)^2\right)_{ij{\rm MQRS} ab}$ & $0^\pm, ~~0^\pm, ~~0^\pm$ \\ \hline
8 & $\big(\mathbb{X}^{(0)}_8\big)_{0ij{\rm MQRS} b}$ & ${\rm dom} \left({\rm tr} ~\mathbb{R}_{\rm tot}^4 
- {1\over 4} \left({\rm tr}~\mathbb{R}_{\rm tot}^2\right)^2\right)_{0ij{\rm MQRS} b}$ & $-2^\pm, ~~-2^\pm, ~~-2^\pm$  \\ \hline
 \end{tabular}}
\renewcommand{\arraystretch}{1}
\end{center}
 \caption[]{ \Su ${g_s\over {\rm H}(y){\rm H}_o({\bf x})}$ scalings of the various terms from {\bf Table \ref{ccrani8}} one the contribution from the exact scalings of the ${\bf G}_{0{\rm NP}a}$ from \eqref{salgill6} and other flux components are added in.} 
  \label{ccrani88}
 \end{table}
 Comparing the first two rows in {\bf Table \ref{ccrani88}}, we immediate get $\theta_{nl} = {8\over 3}^\pm$, consistent with what we have from the Bianchi identities. With temporal derivatives, at order $\xxy^{-2^\pm}$, rows 1 and 2 at NLO match up with rows 5 and 8 at leading order. With derivatives along ${\cal M}_4$, the match is with rows 3 and 6 at leading order. With derivatives along ${\mathbb{T}^2\over {\cal G}}$, the match is only between rows 4 and 6, because the first two rows do not contribute.

\subsection{Bianchi identities for the flux components ${\bf G}_{{\rm MN}ab}$ and ${\bf G}_{{\rm 0N}ab}$ \label{cchase4}} 

\begin{table}[tb]  
 \begin{center}
\resizebox{\columnwidth}{!}{%
\renewcommand{\arraystretch}{3}
\begin{tabular}{|c||c||c||c|}\hline Rows & ${\bf G}_{{\rm MN}ab}$ tensors & Forms & ${g_s\over {\rm HH}_o}$ scalings \\ \hline\hline
1 & $\big(\hat{\mathbb{T}}_4^{(f)}\big)_{{\rm MN}ab}$ & ${\bf G}_{{\rm MN}ab}$ & $1^\pm + {l\over 3}$  \\ \hline 
2 & $\big(\hat{\mathbb{T}}_4^{(q)}\big)_{{\rm MN}ab}$ & $\sqrt{-{\bf g}_{11}} \mathbb{Y}_7^{0ij{\rm PQRS}} 
\epsilon_{0ij{\rm MNPQRS}ab}$ & $\theta_{nl}  + {1\over 3} - {l\over 3}  + (\delta_{{\rm M}m} + \delta_{{\rm N}n})\hat\sigma_e(t) + (\hat\alpha_e(t), \hat\beta_e(t))(\delta_{{\rm M}\alpha} + \delta_{{\rm N}\beta}) + \hat\eta_e(t)$  \\ \hline 
3 & $\big(\hat{\mathbb{Y}}_4\big)_{{\rm MN}ab}$ & ${\rm dom} \left({\rm tr} ~\mathbb{R}_{\rm tot} \wedge \mathbb{R}_{\rm tot}\right)_{{\rm MN}ab}$ & $2^\pm$  \\ \hline
4 & $\big(\widetilde{\bf \Lambda}_5\big)_{{\rm MNS}ab}$ & $\hat{\mathbb{N}}_5\left({\bf \Lambda}_5\right)_{{\rm MNS}ab}$ & $\hat{p}_e(t)$\\ \hline
 \end{tabular}}
\renewcommand{\arraystretch}{1}
\end{center}
 \caption[]{ \Su ${g_s\over {\rm H}(y){\rm H}_o({\bf x})}$ scalings of the various terms of \eqref{crawlquif2} contributing to the Bianchi identity of ${\bf G}_{{\rm MN}ab}$. To compute the scaling in the second row, we use the dual flux scalings from rows 9, 10 and 11 in {\bf Table \ref{fridaylin1}}. Note that now the dynamical M5-branes do  participate with $g_s$ scaling $\hat{p}_e(t)$.} 
  \label{ccrani205}
 \end{table}

The tensors contributing to the Bianchi identities for the G-flux components of the form ${\bf G}_{{\rm MN}ab}$ are collected in {\bf Table \ref{ccrani205}}. Due to the orientation of the flux components, the five-branes do contribute. They have the topologies of the form ${\bf R}^{2, 1} \times {\bf S}^3_{{\cal M}_4}, {\bf R}^{2, 1} \times {\bf S}^1_{{\cal M}_4} \times {\cal M}_2$ and ${\bf R}^{2, 1} \times {\bf S}^2_{{\cal M}_4} \times {\bf S}^1_{{\cal M}_4}$, different from the orientations of the five-branes for the case with flux components ${\bf G}_{{\rm MNP}a}$. Comparing the all the four rows in {\bf Table \ref{ccrani205}}, we see that $\hat{p}_e(t) = 2^\pm$. Consequently the G-flux components now have the following precise scalings:
\begin{empheq}[box={\mybluebox[5pt]}]{align}
{\bf G}_{{\rm MN}ab}({\bf x}, y; g_s(t)) = \sum_{k = 0}^\infty {\cal G}^{(k)}_{{\rm MN}ab}({\bf x}, y) \left({g_s\over {\rm HH}_o}\right)^{2^\pm  + {2k\over 3}\vert {\rm L}_{\rm MN}^{ab}(k; t)\vert}
\label{salgill7}
\end{empheq}
fixing them right at the half-way point in the range $1^\pm \le \hat{l}_{e{\rm MN}}^{ab} \le 3^\pm$ as shown in {\bf Figure \ref{fluxbehavior2}}. Since the $\pm$ superscripts only add very small corrections to the existing scalings, the positive exponents suggest that these flux components decouple at late time where $\bar{g}_s \to 0$. The contents of {\bf Table \ref{ccrani9}} now change to the ones shown in 
{\bf Table \ref{ccrani99}}.
\begin{table}[tb]  
 \begin{center}
\resizebox{\columnwidth}{!}{%
\renewcommand{\arraystretch}{3.0}
\begin{tabular}{|c||c||c||c|}\hline Rows & ${\bf G}_{{\rm MN}ab}$ tensors & Forms & ${g_s\over {\rm HH}_o}$ scalings  \\ \hline\hline
1 & $\big(\mathbb{T}_7^{(f)}\big)_{0ij{\rm PQRS}}$ & $\sqrt{-{\bf g}_{11}} {\bf G}_{{\rm M'}a'b'}{\bf g}^{\rm M'M} {\bf g}^{\rm N'N}{\bf g}^{\rm a'a}{\bf g}^{b'b}
\epsilon_{0ij{\rm MNPQRS}ab}$ & $-4^\pm$ \\ \hline 
2 & $\big(\mathbb{T}_7^{(q)}\big)_{0ij{\rm PQRS} }$ & $\sqrt{-{\bf g}_{11}} \mathbb{Y}_4^{{\rm MN}ab} 
\epsilon_{0ij{\rm MNPQRS}ab}$ & $\theta_{nl} - {20\over 3}^\pm $ \\ \hline 
3 & $\big(\mathbb{T}^{({\rm M})}_8\big)_{0ij{\rm MPQRS}}$ & ${\bf G}_{[0ij{\rm M}} {\bf G}_{{\rm PQRS}]}$  & $-3^\pm$
\\ \hline 
4 & $\big(\mathbb{T}^{(a)}_8\big)_{0ij{\rm PQRS} a}$ & ${\bf G}_{[0ij{\rm P}} {\bf G}_{{\rm QRS} b]}$ & $-2^\pm$
\\ \hline 
5 & $\big(\mathbb{X}^{({\rm M})}_8\big)_{0ij{\rm MPQRS} }$ & ${\rm dom} \left({\rm tr} ~\mathbb{R}_{\rm tot}^4 
- {1\over 4} \left({\rm tr}~\mathbb{R}_{\rm tot}^2\right)^2\right)_{0ij{\rm MPQRS}}$ & $-3^\pm, ~~-3^\pm, ~~-3^\pm$  \\ \hline
6 & $\big(\mathbb{X}^{(a)}_8\big)_{0ij{\rm PQRS} a}$ & ${\rm dom} \left({\rm tr} ~\mathbb{R}_{\rm tot}^4 
- {1\over 4} \left({\rm tr}~\mathbb{R}_{\rm tot}^2\right)^2\right)_{0ij{\rm PQRS} a}$ & $-2^\pm, ~~-2^\pm, ~~-2^\pm$ \\ \hline
\end{tabular}}
\renewcommand{\arraystretch}{1}
\end{center}
 \caption[]{ \Su ${g_s\over {\rm H}(y){\rm H}_o({\bf x})}$ scalings of the various terms from {\bf Table \ref{ccrani9}} one we insert the exact scalings of ${\bf G}_{{\rm MN}ab}$ from \eqref{salgill7} and other flux components.} 
  \label{ccrani99}
 \end{table}
 Comparing the first two rows of {\bf Table \ref{ccrani99}} at order 
 $\xxy^{-4^\pm}$ again gives us $\theta_{nl} = {8\over 3}^\pm$, consistent with what we got from the Bianchi identities. Looking at derivatives along the ${\cal M}_4$ directions, rows 1 and 2 at NLO match up with rows 3 and 5 at leading order of $\xxy^{-3^\pm}$. Looking at derivatives along $\xoxo$ directions, the match is only between rows 4 and 6 at order $\xxy^{-2^\pm}$, as the first two rows do not contribute.

\begin{table}[tb]  
 \begin{center}
\resizebox{\columnwidth}{!}{%
\renewcommand{\arraystretch}{3}
\begin{tabular}{|c||c||c||c|}\hline Rows & ${\bf G}_{{\rm 0N}ab}$ tensors & Forms & ${g_s\over {\rm HH}_o}$ scalings \\ \hline\hline
1 & $\big(\hat{\mathbb{T}}_4^{(f)}\big)_{{\rm 0N}ab}$ & ${\bf G}_{{\rm 0N}ab}$ & $0^\pm + {l\over 3}$  \\ \hline 
2 & $\big(\hat{\mathbb{T}}_4^{(q)}\big)_{{\rm 0N}ab}$ & $\sqrt{-{\bf g}_{11}} \mathbb{Y}_7^{ij{\rm MPQRS}} 
\epsilon_{0ij{\rm MNPQRS}ab}$ & $\theta_{nl}  - {2\over 3} - {l\over 3}  + \hat\zeta_e(t) + \delta_{{\rm N}n}\hat\sigma_e(t) + (\hat\alpha_e(t), \hat\beta_e(t))\delta_{{\rm N}\alpha} + \hat\eta_e(t)$  \\ \hline 
3 & $\big(\hat{\mathbb{Y}}_4\big)_{{\rm 0N}ab}$ & ${\rm dom} \left({\rm tr} ~\mathbb{R}_{\rm tot} \wedge \mathbb{R}_{\rm tot}\right)_{{\rm 0N}ab}$ & $1^\pm$  \\ \hline
 \end{tabular}}
\renewcommand{\arraystretch}{1}
\end{center}
 \caption[]{ \Su ${g_s\over {\rm H}(y){\rm H}_o({\bf x})}$ scalings of the various terms of \eqref{crawlquif2} contributing to the Bianchi identity of ${\bf G}_{{\rm 0N}ab}$. To compute the scaling in the second row, we use the dual flux scalings from rows 12 and 13 in {\bf Table \ref{fridaylin2}}. Note that now the dynamical M5-branes do  not participate.} 
  \label{ccrani206}
 \end{table}

The story for the ${\bf G}_{{\rm 0N}ab}$ flux components is slightly different because the five-branes do not contribute to the Bianchi identities. The tensors contributing to \eqref{crawlquif2} are now collected in {\bf Table \ref{ccrani206}}. All the usual checks can be made and we recover \eqref{salome1} as before. The precise scalings of the flux components now take the following form:
\begin{empheq}[box={\mybluebox[5pt]}]{align}
{\bf G}_{{\rm 0N}ab}({\bf x}, y; g_s(t)) = \sum_{k = 0}^\infty {\cal G}^{(k)}_{{\rm 0N}ab}({\bf x}, y) \left({g_s\over {\rm HH}_o}\right)^{1^\pm  + {2k\over 3}\vert {\rm L}_{\rm 0N}^{ab}(k; t)\vert}
\label{salgill8}
\end{empheq}
fixing them right at the half-way point in the range $0^\pm \le \hat{l}_{e{\rm 0N}}^{ab} \le 2^\pm$ as shown in {\bf figure \ref{fluxbehavior2}}. Because of the positive exponents, these flux components decouple at late time when $\bar{g}_s \to 0$. The contents of {\bf Table \ref{ccrani10}} now change to the ones shown in {\bf Table \ref{ccrani1010}}.
\begin{table}[tb]  
 \begin{center}
\resizebox{\columnwidth}{!}{%
\renewcommand{\arraystretch}{3.0}
\begin{tabular}{|c||c||c||c||c|}\hline Rows & ${\bf G}_{0{\rm N}ab}$ tensors & Forms & ${g_s\over {\rm HH}_o}$ scalings  \\ \hline\hline
1 & $\big(\mathbb{T}_7^{(f)}\big)_{ij{\rm MPQRS}}$ & $\sqrt{-{\bf g}_{11}} {\bf G}_{{\rm 0N'}a'b'}{\bf g}^{\rm 00} {\bf g}^{\rm N'N}{\bf g}^{\rm a'a}{\bf g}^{b'b}
\epsilon_{0ij{\rm MNPQRS}ab}$  & $-3^\pm$ \\ \hline 
2 & $\big(\mathbb{T}_7^{(q)}\big)_{ij{\rm MQRS} b}$ & $\sqrt{-{\bf g}_{11}} \mathbb{Y}_4^{0{\rm N}ab} 
\epsilon_{0ij{\rm MNPQRS}ab}$ & $\theta_{nl}  - {17\over 3}^\pm $ \\ \hline 
3 & $\big(\mathbb{T}^{({\rm N})}_8\big)_{ij{\rm MNPQRS}}$ & ${\bf G}_{[ij{\rm MN}} {\bf G}_{{\rm PQRS}]}$ & $-2^\pm$
\\ \hline 
4 & $\big(\mathbb{T}^{(a)}_8\big)_{ij{\rm MPQRS} a}$ & ${\bf G}_{[ij{\rm MP}} {\bf G}_{{\rm QRS} a]}$  & $-1^\pm$
\\ \hline 
5 & $\big(\mathbb{T}^{(0)}_8\big)_{0ij{\rm MPQRS}}$ & ${\bf G}_{[0ij{\rm M}} {\bf G}_{{\rm PQRS}]}$  & $-3^\pm$
\\ \hline 
6 & $\big(\mathbb{X}^{({\rm N})}_8\big)_{ij{\rm MNPQRS}}$ & ${\rm dom} \left({\rm tr} ~\mathbb{R}_{\rm tot}^4 
- {1\over 4} \left({\rm tr}~\mathbb{R}_{\rm tot}^2\right)^2\right)_{ij{\rm MNPQRS} }$ & $-2^\pm , ~~-2^\pm$ \\ \hline
7 & $\big(\mathbb{X}^{(a)}_8\big)_{ij{\rm MPQRS} a}$ & ${\rm dom} \left({\rm tr} ~\mathbb{R}_{\rm tot}^4 
- {1\over 4} \left({\rm tr}~\mathbb{R}_{\rm tot}^2\right)^2\right)_{ij{\rm MPQRS} a}$ & $-1^\pm, ~~
 -1^\pm$ \\ \hline
8 & $\big(\mathbb{X}^{(0)}_8\big)_{0ij{\rm MPQRS}}$ & ${\rm dom} \left({\rm tr} ~\mathbb{R}_{\rm tot}^4 
- {1\over 4} \left({\rm tr}~\mathbb{R}_{\rm tot}^2\right)^2\right)_{0ij{\rm MPQRS}}$ & $-3^\pm , ~~
 -3^\pm$ \\ \hline
 \end{tabular}}
\renewcommand{\arraystretch}{1}
\end{center}
 \caption[]{ \Su ${g_s\over {\rm H}(y){\rm H}_o({\bf x})}$ scalings of the various terms from {\bf Table \ref{ccrani10}} once the contributions of ${\bf G}_{0{\rm N}ab}$ from \eqref{salgill8} and other flux components are inserted in.} 
  \label{ccrani1010}
 \end{table}
Looking at temporal derivative in \eqref{crawlquif}, the first two rows in {\bf Table \eqref{ccrani1010}} at NLO will only match up with rows 5 and 8 at leading order $\xxy^{-3^\pm}$, whereas at order $\xxy^{-4^\pm}$ only the first two rows match with temporal derivative. With derivatives along ${\cal M}_4$ directions, the match is with rows 3 and 6 at leading order $\xxy^{-2^\pm}$. With derivatives along the toroidal $\xoxo$ directions, the match is only between rows 4 and 7 at order $\xxy^{-1^\pm}$.

\subsection{Bianchi identities for the flux components ${\bf G}_{{\rm MN}ai}, {\bf G}_{{\rm 0N}ai}$ and ${\bf G}_{{\rm M}abi}$ \label{cchase5}} 

\begin{table}[tb]  
 \begin{center}
\resizebox{\columnwidth}{!}{%
\renewcommand{\arraystretch}{3}
\begin{tabular}{|c||c||c||c|}\hline Rows & ${\bf G}_{{\rm MN}ai}$ tensors & Forms & ${g_s\over {\rm HH}_o}$ scalings \\ \hline\hline
1 & $\big(\hat{\mathbb{T}}_4^{(f)}\big)_{{\rm MN}ai}$ & ${\bf G}_{{\rm MN}ai}$ & $-1^\pm + {l\over 3}$  \\ \hline 
2 & $\big(\hat{\mathbb{T}}_4^{(q)}\big)_{{\rm MN}ai}$ & $\sqrt{-{\bf g}_{11}} \mathbb{Y}_7^{0j{\rm PQRS}b} 
\epsilon_{0ij{\rm MNPQRS}ab}$ & $\theta_{nl}  - {5\over 3} - {l\over 3}  + \hat\zeta_e(t) + (\delta_{{\rm M}m} + \delta_{{\rm N}n})\hat\sigma_e(t) + (\hat\alpha_e(t), \hat\beta_e(t))(\delta_{{\rm M}\alpha}+\delta_{{\rm N}\beta}) + (0, \hat\eta_e(t))$  \\ \hline 
3 & $\big(\hat{\mathbb{Y}}_4\big)_{{\rm MN}ai}$ & ${\rm dom} \left({\rm tr} ~\mathbb{R}_{\rm tot} \wedge \mathbb{R}_{\rm tot}\right)_{{\rm MN}ai}$ & $0^\pm$  \\ \hline
 \end{tabular}}
\renewcommand{\arraystretch}{1}
\end{center}
 \caption[]{ \Su ${g_s\over {\rm H}(y){\rm H}_o({\bf x})}$ scalings of the various terms of \eqref{crawlquif2} contributing to the Bianchi identity of ${\bf G}_{{\rm MN}ai}$. To compute the scaling in the second row, we use the dual flux scalings from rows 15 and 17 in {\bf Table \ref{fridaylin1}} and row 1 in {\bf Table \ref{fridaylin2}}. Note that the dynamical M5-branes do  not participate.} 
  \label{ccrani207}
 \end{table}

The tensors contributing to the Bianchi identities for flux components ${\bf G}_{{\rm MNP}i}$ are collected in {\bf Table \ref{ccrani207}}. All the usual checks can be performed and equating the three rows in {\bf Table \ref{ccrani207}} we get back \eqref{salome1}. The precise flux scalings then take the following form:
\begin{empheq}[box={\mybluebox[5pt]}]{align}
{\bf G}_{{\rm MN}ai}({\bf x}, y; g_s(t)) = \sum_{k = 0}^\infty {\cal G}^{(k)}_{{\rm MN}ai}({\bf x}, y) \left({g_s\over {\rm HH}_o}\right)^{0^\pm  + {2k\over 3}\vert {\rm L}_{\rm MN}^{ai}(k; t)\vert}
\label{salgill9}
\end{empheq}
thus putting us right at the half-way point in the range $-1^\pm \le \hat{l}_{e{\rm MN}}^{ai} \le +1^\pm$. The leading order $0^\pm$ exponent suggests that the flux components are never time-independent. The contents of {\bf Table \ref{ccrani11}} change to the ones given in {\bf Table \ref{ccrani1111}}.
\begin{table}[tb]  
 \begin{center}
\resizebox{\columnwidth}{!}{%
\renewcommand{\arraystretch}{3.0}
\begin{tabular}{|c||c||c||c||c|}\hline Rows & ${\bf G}_{{\rm MN}ai}$ tensors & Forms & ${g_s\over {\rm HH}_o}$ scalings  \\ \hline\hline
1 & $\big(\mathbb{T}_7^{(f)}\big)_{0j{\rm PQRS}b}$ & $\sqrt{-{\bf g}_{11}} {\bf G}_{{\rm M'N'}a'i'}{\bf g}^{\rm M'M} {\bf g}^{\rm N'N}{\bf g}^{\rm a'a}{\bf g}^{i'i}
\epsilon_{0ij{\rm MNPQRS}ab}$  & $-2^\pm$ \\ \hline 
2 & $\big(\mathbb{T}_7^{(q)}\big)_{0j{\rm PQRS} b}$ & $\sqrt{-{\bf g}_{11}} \mathbb{Y}_4^{{\rm MN}ai} 
\epsilon_{0ij{\rm MNPQRS}ab}$ & $\theta_{nl}  - {14\over 3}^\pm$ \\ \hline 
3 & $\big(\mathbb{T}^{({\rm M})}_8\big)_{0j{\rm MPQRS}b}$ & ${\bf G}_{[0j{\rm MP}} {\bf G}_{{\rm QRS}b]}$  & $-1^\pm$
\\ \hline 
4 & $\big(\mathbb{T}^{(a)}_8\big)_{0j{\rm PQRS} ab}$ & ${\bf G}_{[0j{\rm PQ}} {\bf G}_{{\rm RS} ab]}$  & $~~0^\pm$
\\ \hline 
5 & $\big(\mathbb{T}^{(i)}_8\big)_{0ij{\rm PQRS}b}$ & ${\bf G}_{[0ij{\rm P}} {\bf G}_{{\rm QRS}b]}$  & $-2^\pm$
\\ \hline 
6 & $\big(\mathbb{X}^{({\rm M})}_8\big)_{0j{\rm MPQRS}b}$ & ${\rm dom} \left({\rm tr} ~\mathbb{R}_{\rm tot}^4 
- {1\over 4} \left({\rm tr}~\mathbb{R}_{\rm tot}^2\right)^2\right)_{0j{\rm MPQRS}b}$ & $-1^\pm ,~~-1^\pm , ~~-1^\pm$ \\ \hline
7 & $\big(\mathbb{X}^{(a)}_8\big)_{0j{\rm PQRS} ab}$ & ${\rm dom} \left({\rm tr} ~\mathbb{R}_{\rm tot}^4 
- {1\over 4} \left({\rm tr}~\mathbb{R}_{\rm tot}^2\right)^2\right)_{0j{\rm PQRS}ab}$ & $0^\pm , ~~0^\pm ,~~0^\pm$ \\ \hline
8 & $\big(\mathbb{X}^{(i)}_8\big)_{0ij{\rm PQRS}b}$ & ${\rm dom} \left({\rm tr} ~\mathbb{R}_{\rm tot}^4 
- {1\over 4} \left({\rm tr}~\mathbb{R}_{\rm tot}^2\right)^2\right)_{0ij{\rm PQRS}b}$ & $-2^\pm, ~~
 -2^\pm ,~~
 -2^\pm$ \\ \hline
 \end{tabular}}
\renewcommand{\arraystretch}{1}
\end{center}
 \caption[]{ \Su ${g_s\over {\rm H}(y){\rm H}_o({\bf x})}$ scalings of the various terms from {\bf Table \ref{ccrani11}} when we take the contribution from \eqref{salgill9} and other flux components.} 
  \label{ccrani1111}
 \end{table}
Comparing the first two rows in {\bf Table \ref{ccrani1111}} give us, as before, $\theta_{nl} = {8\over 3}^\pm$. At NLO they match up with rows 3 and 6 at leading order $\xxy^{-1^\pm}$ when we take derivatives along ${\cal M}_4$ directions. For derivatives along $\xoxo$ and ${\bf R}^2$ directions the matching happens between rows (4, 7) at leading order 
$\xxy^{0^\pm}$ and rows (5, 8) at leading order $\xxy^{-2^\pm}$ respectively without the involvement of the first two rows.

\begin{table}[tb]  
 \begin{center}
\resizebox{\columnwidth}{!}{%
\renewcommand{\arraystretch}{3}
}
\renewcommand{\arraystretch}{1}
\end{center}
 \caption[]{ \Su ${g_s\over {\rm H}(y){\rm H}_o({\bf x})}$ scalings of the various terms of \eqref{crawlquif2} contributing to the Bianchi identity of ${\bf G}_{{\rm 0N}ai}$. To compute the scaling in the second row, we use the dual flux scalings from rows 21 and 22 in {\bf Table \ref{fridaylin2}}. Note that the dynamical M5-branes do  not participate.} 
  \label{ccrani208}
 \end{table}

\begin{table}[tb]  
 \begin{center}
\resizebox{\columnwidth}{!}{%
\renewcommand{\arraystretch}{3}
%
}
\renewcommand{\arraystretch}{1}
\end{center}
 \caption[]{ \Su ${g_s\over {\rm H}(y){\rm H}_o({\bf x})}$ scalings of the various terms of \eqref{crawlquif2} contributing to the Bianchi identity of ${\bf G}_{{\rm M}abi}$. To compute the scaling in the second row, we use the dual flux scalings from rows 16 and 18 in {\bf Table \ref{fridaylin1}}. The dynamical M5-branes do  not participate.} 
  \label{ccrani209}
 \end{table}

For the remaining two flux components ${\bf G}_{{\rm 0N}ai}$ and ${\bf G}_{{\rm M}abi}$, the tensors contributing to the Bianchi identities \eqref{crawlquif2} are collected in {\bf Tables \ref{ccrani208}} and {\bf \ref{ccrani209}}. When we compare and equate all the rows in each of these tables, we not only recover \eqref{salome1}, but also get the following exact scalings for the two set of flux components:
\begin{empheq}[box={\mybluebox[5pt]}]{align}
&{\bf G}_{{\rm 0N}ai}({\bf x}, y; g_s(t)) = \sum_{k = 0}^\infty {\cal G}^{(k)}_{{\rm 0N}ai}({\bf x}, y) \left({g_s\over {\rm HH}_o}\right)^{-1^\pm  + {2k\over 3}\vert {\rm L}_{\rm 0N}^{ai}(k; t)\vert}
\nonumber\\
&{\bf G}_{{\rm M}abi}({\bf x}, y; g_s(t)) = \sum_{k = 0}^\infty {\cal G}^{(k)}_{{\rm M}abi}({\bf x}, y) \left({g_s\over {\rm HH}_o}\right)^{1^\pm  + {2k\over 3}\vert {\rm L}_{{\rm M}a}^{bi}(k; t)\vert}
\label{salgill10}
\end{empheq}
which are unsurprisingly the half-way points in the two ranges $-2^\pm \le \hat{l}_{e{\rm 0N}}^{ai} \le 0^\pm$ and $0^\pm \le \hat{l}_{e{\rm M}a}^{bi} \le 2^\pm$ respectively. Changes to the contents of the {\bf Tables \ref{ccrani12}} and {\bf \ref{ccrani13}} are shown in {\bf Tables \ref{ccrani1212}} and {\bf \ref{ccrani1313}}. As before, one may easily verify the relations between various rows in each of the two tables, including the fact that the scalings of the quantum terms remain at $\theta_{nl} = {8\over 3}^\pm$.

\begin{table}[tb]  
 \begin{center}
\resizebox{\columnwidth}{!}{%
\renewcommand{\arraystretch}{2.5}
}
\renewcommand{\arraystretch}{1}
\end{center}
 \caption[]{ \Su ${g_s\over {\rm H}(y){\rm H}_o({\bf x})}$ scalings of the various terms from {\bf Table \ref{ccrani12}} when we take \eqref{salgill10} and other flux components into account.} 
  \label{ccrani1212}
 \end{table}

\begin{table}[tb]  
 \begin{center}
\resizebox{\columnwidth}{!}{%
\renewcommand{\arraystretch}{3.0}
%
}
\renewcommand{\arraystretch}{1}
\end{center}
 \caption[]{ \Su ${g_s\over {\rm H}(y){\rm H}_o({\bf x})}$ scalings of the various terms from {\bf Table \ref{ccrani13}} once we take  contributions from \eqref{salgill10} and other flux components.} 
  \label{ccrani1313}
 \end{table}

\subsection{Bianchi identities for the flux components ${\bf G}_{{\rm 0}abi}, {\bf G}_{{\rm 0MN}i}$ and ${\bf G}_{{\rm MNP}i}$ \label{cchase6}} 

\begin{table}[tb]  
 \begin{center}
\resizebox{\columnwidth}{!}{%
\renewcommand{\arraystretch}{2.4}
%
}
\renewcommand{\arraystretch}{1}
\end{center}
 \caption[]{ \Su ${g_s\over {\rm H}(y){\rm H}_o({\bf x})}$ scalings of the various terms of \eqref{crawlquif2} contributing to the Bianchi identity of ${\bf G}_{{\rm 0}abi}$. To compute the scaling in the second row, we use the dual flux scalings from row 20 in {\bf Table \ref{fridaylin2}}. The dynamical M5-branes do  not participate.} 
  \label{ccrani212}
 \end{table}

\begin{table}[tb]  
 \begin{center}
\resizebox{\columnwidth}{!}{%
\renewcommand{\arraystretch}{3}
%
}
\renewcommand{\arraystretch}{1}
\end{center}
 \caption[]{ \Su ${g_s\over {\rm H}(y){\rm H}_o({\bf x})}$ scalings of the various terms of \eqref{crawlquif2} contributing to the Bianchi identity of ${\bf G}_{{\rm 0MN}i}$. To compute the scaling in the second row, we use the dual flux scalings from rows 17, 18 and 19 in {\bf Table \ref{fridaylin2}}. The dynamical M5-branes do  not participate.} 
  \label{ccrani211}
 \end{table}

\begin{table}[tb]  
 \begin{center}
\resizebox{\columnwidth}{!}{%
\renewcommand{\arraystretch}{3}
%
}
\renewcommand{\arraystretch}{1}
\end{center}
 \caption[]{ \Su ${g_s\over {\rm H}(y){\rm H}_o({\bf x})}$ scalings of the various terms of \eqref{crawlquif2} contributing to the Bianchi identity of ${\bf G}_{{\rm MNP}i}$. To compute the scaling in the second row, we use the dual flux scalings from rows 12, 13 and 14 in {\bf Table \ref{fridaylin1}}. The dynamical M5-branes do  not participate.} 
  \label{ccrani210}
 \end{table}

For the three flux components, ${\bf G}_{{\rm 0}abi}, {\bf G}_{{\rm 0MN}i}$ and ${\bf G}_{{\rm MNP}i}$, the tensors contributing to the Bianchi identities are collected in {\bf Tables \ref{ccrani212}, \ref{ccrani211}} and {\bf \ref{ccrani210}} respectively. All checks may be easily performed as before, and using them we can express the precise scalings of these flux components in the following way:
\begin{empheq}[box={\mybluebox[5pt]}]{align}
&{\bf G}_{{\rm 0}abi}({\bf x}, y; g_s(t)) = \sum_{k = 0}^\infty {\cal G}^{(k)}_{{\rm 0}abi}({\bf x}, y) \left({g_s\over {\rm HH}_o}\right)^{0^\pm  + {2k\over 3}\vert {\rm L}_{0a}^{bi}(k; t)\vert}\nonumber\\
&{\bf G}_{{\rm 0MN}i}({\bf x}, y; g_s(t)) = \sum_{k = 0}^\infty {\cal G}^{(k)}_{{\rm 0MN}i}({\bf x}, y) \left({g_s\over {\rm HH}_o}\right)^{-2^\pm  + {2k\over 3}\vert {\rm L}_{{\rm 0M}}^{{\rm N}i}(k; t)\vert}\nonumber\\
&{\bf G}_{{\rm MNP}i}({\bf x}, y; g_s(t)) = \sum_{k = 0}^\infty {\cal G}^{(k)}_{{\rm MNP}i}({\bf x}, y) \left({g_s\over {\rm HH}_o}\right)^{-1^\pm  + {2k\over 3}\vert {\rm L}_{\rm MN}^{{\rm P}i}(k; t)\vert}
\label{salgill11}
\end{empheq}
which again confirms the fact that the Bianchi identities fix the scalings of the flux components right at the half-way points in the ranges $-2^\pm \le \hat{l}_{e{\rm MN}}^{{\rm P}i} \le 0^\pm, -3^\pm \le \hat{l}_{e{\rm 0M}}^{{\rm N}i} \le -1^\pm$ and $-1^\pm \le \hat{l}_{e{\rm 0}a}^{bi} \le +1^\pm$ respectively. Including all the exact scalings from \eqref{salgill11} and other flux components, the contents of {\bf Tables \ref{ccrani16}, \ref{ccrani15}} and {\bf \ref{ccrani14}} changes to the ones given in {\bf Tables \ref{ccrani1616}, \ref{ccrani1515}} and {\bf \ref{ccrani1414}} respectively. It is now easy to match the various scalings to \eqref{crawlquif} for derivatives acting along either ${\cal M}_4, \xoxo$, temporal or ${\bf R}^2$ directions. Needless to say, the quantum scalings again give us $\theta_{nl} = {8\over 3}^\pm$ thus matching with the ones we got from \eqref{salome1}.

\begin{table}[tb]  
 \begin{center}
\resizebox{\columnwidth}{!}{%
\renewcommand{\arraystretch}{3.0}
}
\renewcommand{\arraystretch}{1}
\end{center}
 \caption[]{ \Su ${g_s\over {\rm H}(y){\rm H}_o({\bf x})}$ scalings of the various terms from {\bf Table \ref{ccrani16}} once we take the precise contributions from \eqref{salgill11} and other flux terms.} 
  \label{ccrani1616}
 \end{table}

\begin{table}[tb]  
 \begin{center}
\resizebox{\columnwidth}{!}{%
\renewcommand{\arraystretch}{3.0}
%
}
\renewcommand{\arraystretch}{1}
\end{center}
 \caption[]{ \Su ${g_s\over {\rm H}(y){\rm H}_o({\bf x})}$ scalings of the various terms from {\bf Table \ref{ccrani15}} once we take the exact contributions from \eqref{salgill11} and other flux components.} 
  \label{ccrani1515}
 \end{table}

\begin{table}[tb]  
 \begin{center}
\resizebox{\columnwidth}{!}{%
\renewcommand{\arraystretch}{3.0}
%
}
\renewcommand{\arraystretch}{1}
\end{center}
 \caption[]{ \Su ${g_s\over {\rm H}(y){\rm H}_o({\bf x})}$ scalings of the various terms from {\bf Table \ref{ccrani14}} once we take the precise scalings from \eqref{salgill11} and other flux components.} 
  \label{ccrani1414}
 \end{table}

\subsection{Bianchi identities for the flux components ${\bf G}_{abij}, {\bf G}_{{\rm M}aij}$ and ${\bf G}_{{\rm MN}ij}$ \label{cchase7}}

\begin{table}[tb]  
 \begin{center}
\resizebox{\columnwidth}{!}{%
\renewcommand{\arraystretch}{2.4}
%
}
\renewcommand{\arraystretch}{1}
\end{center}
 \caption[]{ \Su ${g_s\over {\rm H}(y){\rm H}_o({\bf x})}$ scalings of the various terms of \eqref{crawlquif2} contributing to the Bianchi identity of ${\bf G}_{abij}$. To compute the scaling in the second row, we use the dual flux scalings from row 7 in {\bf Table \ref{fridaylin2}}. The dynamical M5-branes do  not participate.} 
  \label{ccrani215}
 \end{table}

\begin{table}[tb]  
 \begin{center}
\resizebox{\columnwidth}{!}{%
\renewcommand{\arraystretch}{3}
%
}
\renewcommand{\arraystretch}{1}
\end{center}
 \caption[]{ \Su ${g_s\over {\rm H}(y){\rm H}_o({\bf x})}$ scalings of the various terms of \eqref{crawlquif2} contributing to the Bianchi identity of ${\bf G}_{{\rm M}aij}$. To compute the scaling in the second row, we use the dual flux scalings from rows 5 and 6 in {\bf Table \ref{fridaylin2}}. The dynamical M5-branes do  not participate.} 
  \label{ccrani214}
 \end{table}

\begin{table}[tb]  
 \begin{center}
\resizebox{\columnwidth}{!}{%
\renewcommand{\arraystretch}{3}
%
}
\renewcommand{\arraystretch}{1}
\end{center}
 \caption[]{ \Su ${g_s\over {\rm H}(y){\rm H}_o({\bf x})}$ scalings of the various terms of \eqref{crawlquif2} contributing to the Bianchi identity of ${\bf G}_{{\rm MN}ij}$. To compute the scaling in the second row, we use the dual flux scalings from rows 2, 3 and 4 in {\bf Table \ref{fridaylin2}}. The dynamical M5-branes do  not participate.} 
  \label{ccrani213}
 \end{table}

\begin{figure}[h]
\centering
\begin{tabular}{c}
\includegraphics[width=6.2in]{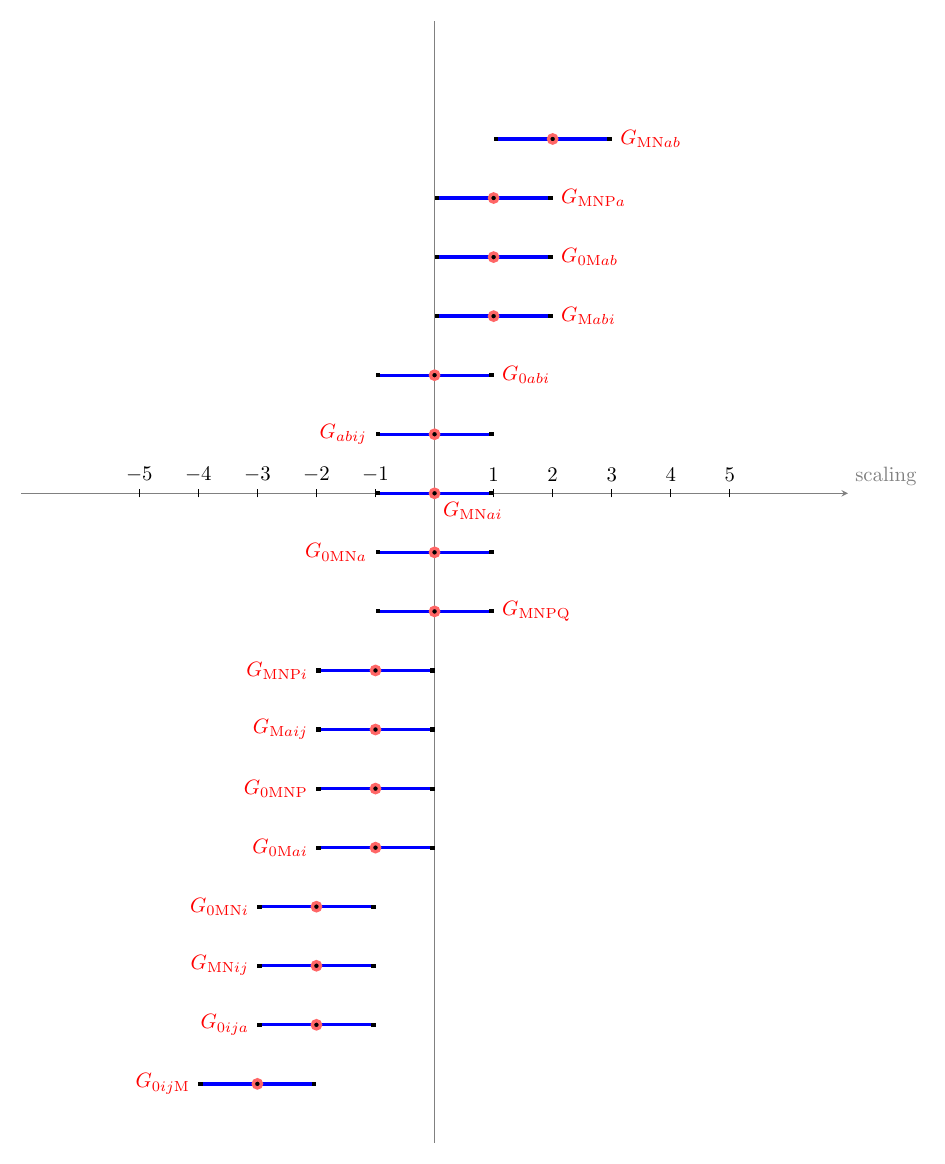}
\end{tabular}
\vskip-.1in
\caption[]{Modification to {\bf figure \ref{fluxbehavior1}} once we carefully take into account all the Bianchi identities. They fix the precise scalings of all the flux components always at the half-way points $-$ shown in blue/red dots in the figure. Note that, because of the $\pm$ corrections to the scalings, the flux components with leading order $0^\pm$ scalings are not time-independent.}
\label{fluxbehavior2}
\end{figure}

For our last three flux components, namely ${\bf G}_{abij}, {\bf G}_{{\rm M}aij}$ and ${\bf G}_{{\rm M}aij}$, the tensors contributing to the Bianchi identities \eqref{crawlquif2} are collected in {\bf Tables \ref{ccrani215}, \ref{ccrani214}} and {\bf \ref{ccrani213}}. Comparing the first two rows in each of the three tables easily give us \eqref{salome1}. The exact scalings of the three flux components now become:
\begin{empheq}[box={\mybluebox[5pt]}]{align}
&{\bf G}_{abij}({\bf x}, y; g_s(t)) = \sum_{k = 0}^\infty {\cal G}^{(k)}_{abij}({\bf x}, y) \left({g_s\over {\rm HH}_o}\right)^{0^\pm  + {2k\over 3}\vert {\rm L}_{ab}^{ij}(k; t)\vert}\nonumber\\
&{\bf G}_{{\rm M}aij}({\bf x}, y; g_s(t)) = \sum_{k = 0}^\infty {\cal G}^{(k)}_{{\rm M}aij}({\bf x}, y) \left({g_s\over {\rm HH}_o}\right)^{-1^\pm  + {2k\over 3}\vert {\rm L}_{{\rm M}a}^{ij}(k; t)\vert}\nonumber\\
&{\bf G}_{{\rm MN}ij}({\bf x}, y; g_s(t)) = \sum_{k = 0}^\infty {\cal G}^{(k)}_{{\rm MN}ij}({\bf x}, y) \left({g_s\over {\rm HH}_o}\right)^{-2^\pm  + {2k\over 3}\vert {\rm L}_{\rm MN}^{ij}(k; t)\vert}
\label{salgill12}
\end{empheq}
which again confirms the fact that the Bianchi identities fix the scalings of the flux components right at the half-way points (see {\bf figure \ref{fluxbehavior2}}) in the ranges $-1^\pm \le \hat{l}_{eab}^{ij} \le 0^\pm, -2^\pm \le \hat{l}_{e{\rm M}a}^{ij} \le 0^\pm$ and $-3^\pm \le \hat{l}_{e{\rm MN}}^{ij} \le -1^\pm$ respectively. Including all the exact scalings from \eqref{salgill12} and other flux components, the contents of {\bf Tables \ref{ccrani19}, \ref{ccrani18}} and {\bf \ref{ccrani17}} changes to the ones given in {\bf Tables \ref{ccrani1919}, \ref{ccrani1818}} and {\bf \ref{ccrani1717}} respectively. It is now easy to match the various scalings with \eqref{crawlquif} for derivatives acting along either ${\cal M}_4, \xoxo$, temporal or ${\bf R}^2$ directions at different NLOs and LOs. Needless to say, the quantum scalings again give us $\theta_{nl} = {8\over 3}^\pm$ thus matching with the ones we got from \eqref{salome1}.

\begin{table}[tb]  
 \begin{center}
\resizebox{\columnwidth}{!}{%
\renewcommand{\arraystretch}{3.0}
}
\renewcommand{\arraystretch}{1}
\end{center}
 \caption[]{ \Su ${g_s\over {\rm H}(y){\rm H}_o({\bf x})}$ scalings of the various terms from {\bf Table \ref{ccrani19}} once we take the exact scalings from \eqref{salgill12} and other flux components.} 
  \label{ccrani1919}
 \end{table}

\begin{table}[tb]  
 \begin{center}
\resizebox{\columnwidth}{!}{%
\renewcommand{\arraystretch}{3.0}
%
}
\renewcommand{\arraystretch}{1}
\end{center}
 \caption[]{ \Su ${g_s\over {\rm H}(y){\rm H}_o({\bf x})}$ scalings of the various terms from {\bf Table \ref{ccrani18}} once we consider the exact scalings from \eqref{salgill12} and other flux components.} 
  \label{ccrani1818}
 \end{table}

\begin{table}[tb]  
 \begin{center}
\resizebox{\columnwidth}{!}{%
\renewcommand{\arraystretch}{3.0}
%
}
\renewcommand{\arraystretch}{1}
\end{center}
 \caption[]{ \Su ${g_s\over {\rm H}(y){\rm H}_o({\bf x})}$ scalings of the various terms from {\bf Table \ref{ccrani17}} once we consider the exact scalings from \eqref{salgill12} and other flux components.} 
  \label{ccrani1717}
 \end{table}


\subsection{Conclusion and summary of the flux EOMs and the Bianchi identities \label{sec8.8}}

In section \ref{sec7s} and here we studied in detail the Schwinger-Dyson equations for the fluxes and the Bianchi identities respectively. The results are summarized in {\bf figures \ref{fluxbehavior1}} and {\bf \ref{fluxbehavior2}} whose explicit forms may be inferred from \eqref{salgill1}, \eqref{salgill2}, \eqref{salgill3}, \eqref{salgill4}, \eqref{salgill5}, \eqref{salgill6}, \eqref{salgill7}, \eqref{salgill8}, \eqref{salgill9}, \eqref{salgill10}, \eqref{salgill11} and \eqref{salgill12}. The $\pm$ superscripts on each of the leading order scalings can be explicitly computed, but the analysis is technically challenging because it involves the corrections to the $\mathbb{X}_8$ polynomials coming from the small parameters $(\hat\zeta_e(t), \hat\alpha_e(t), \hat\beta_e(t), \hat\sigma_e(t), \hat\eta_e(t))$ via the curvature forms. The small corrections to the curvature forms have been meticulously collected in {\bf Tables \ref{niksmit1}} to {\bf \ref{niksmit250}} so, with a little hard work, they can be computed.
We will however not work these details out here but leave them for the diligent readers to complete the story. Despite this, many precise results have emerged from our analysis. Assuming $0^\pm > 0$, we see from {\bf figure \ref{fluxbehavior2}} that slightly more than half of the flux components have positive scalings at leading orders. This means they would decouple at late time when $\bar{g}_s \to 0$. Moreover, both the dynamical M2 and M5-branes have positive scalings implying that at late time they too should decouple from the flux dynamics (meaning that their effects on the flux EOMs should be minimal or vanishing). This makes sense in light of the fact that at late time, in the ${\rm E}_8 \times {\rm E}_8$ heterotic case, the dynamics take the form as depicted in {\bf figures \ref{f11f33}} and {\bf \ref{axdecay}}. The $SO(32)$ case can be easily analyzed from the ${\rm E}_8 \times {\rm E}_8$ case using the simplifying strategy given in {\bf figure \ref{boxer}}.

Before ending this section let us clarify one more thing related to the $\pm$ superscripts on the $\bar{g}_s$ scalings of the various terms. The crucial question is when we match the $\bar{g}_s$ scalings of the various terms in (say) the EOMs, the dominant scalings match up, but how do we know that the $\pm$ pieces also match up? As an example, consider two terms in the flux EOM that go as $a_1^\pm$ and $a_2^\pm$. If $a_1 = a_2$, the dominant terms match up, but since the $\pm$ superscripts are controlled by small numbers, how can we justify the {\it exact} matching?  

This is where the Bianchi identities that we studied here becomes immensely useful. However looking at the various tables for the Bianchi identities, namely {\bf Tables \ref{ccrani20}, \ref{ccrani200}, \ref{ccrani201}, \ref{ccrani202}, \ref{ccrani203}, \ref{ccrani204}, \ref{ccrani205}, \ref{ccrani206}, \ref{ccrani207}, \ref{ccrani208}, \ref{ccrani209}, \ref{ccrani212}, \ref{ccrani211}, \ref{ccrani210}, \ref{ccrani215}, \ref{ccrani214}} and {\bf \ref{ccrani213}},  one might be concerned by the presence of the $\pm$ superscripts there. This is actually {\it not} a problem. To see this let us consider one such table, namely {\bf Table \ref{ccrani203}}. The scaling of the third row is fixed from our choice of the scalings of the on-shell metric components. This means the $\pm$ superscript of $1^\pm$ is known precisely in terms of the small parameters $(\hat\zeta_e(t), \hat\sigma_e(t), \hat\alpha_e(t), \hat\beta_e(t), \hat\eta_e(t))$. Equating Row 1 to Row 3 fixes $l$ precisely. To quantify this let us express the equality between the $\bar{g}_s$ scalings between Row 1 and Row 3 as:
\bg\label{comiccontag}
{\bf G}_{[{\rm ABCD}]} \propto {\bf R}_{[{\rm AB}]} {\bf R}_{[{\rm CD}]} + e_1
{\bf R}_{[{\rm AC}]} {\bf R}_{[{\rm BD}]} + e_2 {\bf R}_{[{\rm AD}]} {\bf R}_{[{\rm BC}]}, \nd
where $e_i = \pm$ to fix the relative signs and we have suppressed the vielbein indices. Looking at the $\bar{g}_s$ scalings of the curvature two-forms from {\bf Tables \ref{niksmit1}} till {\bf \ref{niksmit250}}, we see that each of the curvature two forms has at most {\it eight} possible choices of the $\bar{g}_s$ scalings once we express them using the small parameters $(\hat\zeta_e(t), \hat\sigma_e(t), \hat\alpha_e(t), \hat\beta_e(t), \hat\eta_e(t))$. However this condensed way of expressing the $\bar{g}_s$ scalings of the metric components uses the approximation ${\bf g}_{\rm AB}^{(k)}({\bf x}, y) = \mathbb{A}^{(k)} {\bf g}_{\rm AB}({\bf x}, y)$, so that we can express the full perturbative and the non-perturbative series as in \eqref{bdtran2} shown in section \ref{sec4.4}. (Similar story is with the G-flux components expressed as \eqref{andyrey}.) If we do not use this approximation then the analysis becomes very hard to track, as every individual terms in the series need to be dealt separately. However in such a scenario, we can make the following qualitative observations. The ${\rm tr}~{\bf R} \wedge {\bf R}$ piece of the Bianchi identity \eqref{crawlquif2}, whose explicit form is shown on the RHS of \eqref{comiccontag}, basically controls the scalings of fluxes, branes and quantum effects entering \eqref{crawlquif2}. From the $\bar{g}_s$ scalings of the curvature forms, we can easily infer that the $\bar{g}_s$ scalings of ${\rm tr}~{\bf R} \wedge {\bf R}$ can be grouped as copies of $n$ scalings where $n \le 192$. In other words, each of these 192 $\bar{g}_s$ scalings would form the {\it lowest} order scalings of a given sequence whose higher order terms are controlled by $(\hat\alpha_k(t), \hat\beta_k(t), \hat\sigma_k(t), \hat\zeta_k(t), \hat\eta_k(t))$ $-$ and their non-perturbative extensions as in \eqref{ivbrestick} $-$ for all $k \in \mathbb{Z}$ assuming no overlaps. For the G-flux components, the $\bar{g}_s$ scalings are controlled by lowest order bare scalings $l_{\rm AB}^{\rm CD}(t)$ and the higher order ${\rm L}_{\rm AB}^{\rm CD}(p; t)$ for all $p \in \mathbb{Z}$ as shown in \eqref{usher}. (Under the aforementioned approximation this may be summed up as in \eqref{ivbrestick}, but here we will refrain from taking any approximation.) Now the $\bar{g}_s$ scalings of the fluxes in Row 1 can be easily matched with the $\bar{g}_s$ scalings of  ${\rm tr}~{\bf R} \wedge {\bf R}$  from Row 3 if we take:
\bg\label{huhhuh}
k \equiv p ~~ {\rm mod}~n,~~~~~ n \le 192, \nd
which is easily achieved because of the non-trivial functions ${\rm L}_{\rm AB}^{\rm CD}(p; t)$ appearing in \eqref{usher}. Now making Row 2 = Row 3, will fix $\theta_{nl}$ completely. Finally equating everything to Row 4 will determine the sequence of scalings for the dynamical M5-branes whose lowest order, and hence the dominant one, appears in say {\bf Table \ref{ccrani203}}. Once we extend the aforementioned criteria to all the tables, any ambiguities related to the $\pm$ superscripts dissolve away and we get the Bianchi identities precisely. The non-perturbative corrections, coming from the trans-series form of the action \eqref{kimkarol}, would now be quantified by {\it different} values for the quantum scalings $\theta_{nl}$ associated with the different instanton numbers. In {\bf figure \ref{bianchibehavior}}, we denote how the Bianchi identities are satisfied to all orders in 
${g_s\over {\rm H}(y){\rm H}_o({\bf x})}$, both at the perturbative and at the non-perturbative levels. 

We can now extend the arguments to the EOMs whose details appear in 
{\bf Tables \ref{crecraqwiff101}, \ref{ccquiff30}, \ref{ccrani101}, \ref{ccrani521}, \ref{ccrani77}, \ref{ccrani88}, \ref{ccrani99}, \ref{ccrani1010}, \ref{ccrani1111}, \ref{ccrani1212}, \ref{ccrani1313}, \ref{ccrani1616}, \ref{ccrani1515}, \ref{ccrani1414}, \ref{ccrani1919}, \ref{ccrani1818}} and {\bf \ref{ccrani1717}}. Taking the Hodge dual of \eqref{dietherapie} and then the exterior derivative, gives us the following equation:
\bg\label{dietherapie2}
d\ast_{11}{\bf G}_4 = d\ast_{11} d{\bf C}_3 -{c_2\over c_1} d\ast_{11}  \left({\rm tr}~\mathbb{R}_{\rm tot} \wedge \mathbb{R}_{\rm tot}\right) - {c_3\over c_1} d\mathbb{Y}_7 + d\ast_{11}\big(\hat{\rm N}_5 {\bf \Lambda}_4\big), \nd
which doesn't exactly match with \eqref{crawlquif}. This was of course already anticipated in \eqref{malinechat} because of ignoring the boundary terms. However once we take the Hodge dual we cannot ignore the boundary contributions. To see this let us consider the boundary contribution at the first chain of duality in \eqref{malinechat} be $\mathbb{T}_b = d({\bf C}_3 \wedge \mathbb{Z}_7) \equiv d t_b$. Since $\mathbb{T}_b$ is an exact form, $d\mathbb{T}_b = 0$, so we can ignore it in \eqref{crawlquif2}. However once we take the Hodge dual, and then subsequently the derivative, we are dealing with $d\ast_{11} \mathbb{T}_b$ which cannot be ignored. Thus we see that:
\bg\label{omaoliviet}
b_2 {\bf G}_4 \wedge {\bf G}_4 + b_3 \mathbb{X}_8 \subset{{c_2\over c_1}d\ast_{11}  \left({\rm tr}~\mathbb{R}_{\rm tot} \wedge \mathbb{R}_{\rm tot}\right)}, \nd
up to overall signs and $(b_2, b_3)$ are as in \eqref{crawlquif}. Plugging \eqref{omaoliviet} in \eqref{dietherapie2} would reproduce the EOMs from the Bianchi identities, implying that solving the Bianchi identities is equivalent to solving the EOMs. Moreover, both the terms on the LHS of \eqref{omaoliviet} scale in the same way with respect to $\bar{g}_s$ which may be inferred from all the aforementioned Bianchi identity tables. The RHS of \eqref{omaoliviet}, on the other hand, scales in the same way as four-form flux components as evident from {\bf Table \ref{starR}}.

\begin{table}[tb]  
 \begin{center}
\resizebox{\columnwidth}{!}{%
\renewcommand{\arraystretch}{3.5}
}
\renewcommand{\arraystretch}{1}
\end{center}
 \caption[]{ \Su The $\bar{g}_s$ scalings of $\ast {\rm tr}\big(\mathbb{R}_{\rm tot} \wedge \mathbb{R}_{\rm tot}\big)_{\rm ABCD}$ for all the cases considered here. Note the precise matching with the seven-form flux components $\mathbb{T}_7^{(f)}$ from \eqref{olivecostamey}.} 
  \label{starR}
 \end{table}

Our discussion above, and from the scalings of the quantum series in \eqref{brittbaba} for the ${\rm E}_8 \times {\rm E}_8$ case, the first two rows of {\bf Tables \ref{crecraqwiff101}, \ref{ccquiff30}, \ref{ccrani101}} till {\bf \ref{ccrani1717}} always match-up at the lowest orders in $\bar{g}_s$ (see {\bf figure \ref{eombehavior}} and footnote \ref{fasermaxe}). As we go to the higher orders in $\bar{g}_s$ we expect the remaining terms from \eqref{crawlquif} to also match-up. However the question is how we know that the matching is precise in the sense that the $\pm$ superscripts also line up precisely. This can be motivated from the following two observations.

\vskip.1in

\noindent $\bullet$ From our earlier discussions we saw that the matching of the $\bar{g}_s$ scalings for the Bianchi identities are precise even with the $\pm$ superscripts as shown in {\bf figure \ref{bianchibehavior}}. 

\vskip.1in

\begin{figure}[h]
\centering
\begin{tabular}{c}
\includegraphics[width=6in]{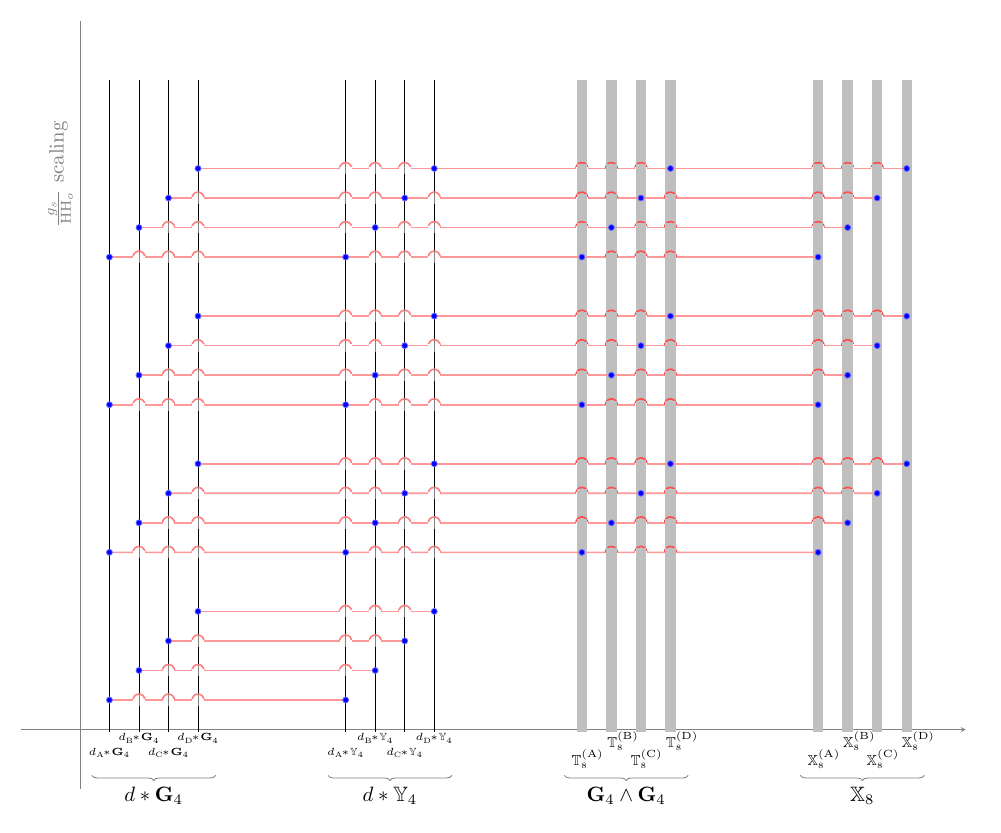}
\end{tabular}
\vskip-.1in
\caption[]{Matching the $\bar{g}_s$ scalings to all orders for the EOMs given in \eqref{crawlquif}. This differs from {\bf figure \ref{eombehavior}} by the appearance of all components of the EOMs in one set-up. All the parameters appearing here are defined in \eqref{olivecostamey}. For convenience we have absorbed the non-perturbative contributions, {\it i.e.} the contributions from the non-zero instanton sectors, in the definition of $\ast\mathbb{Y}_4$ instead of writing them separately as in the fifth line of \eqref{olivecostamey}.}
\label{eomBB}
\end{figure}

\noindent $\bullet$ Matching of $\bar{g}_s$ scalings in the Bianchi identities strongly suggests that at the order the $\bar{g}_s$ scalings of the LHS of \eqref{omaoliviet} match-up, the first two rows in each of the {\bf Tables \ref{crecraqwiff101}, \ref{ccquiff30}, \ref{ccrani101}} till {\bf \ref{ccrani1717}} would also respectively line-up precisely in a way shown in {\bf figure \ref{eombehavior}} and, in a more elaborate way, in {\bf figure \ref{eomBB}}.

\vskip.1in

\noindent Putting everything together, our analysis in sections \ref{sec7s} and \ref{sec4.5} suggests that we can solve both the EOMs and the Bianchi identities precisely to all orders in the $\bar{g}_s$ both perturbatively and non-perturbatively at the far IR.

\section{Flux quantizations and anomaly cancellations for the global fluxes \label{secanomaly}}

Our aim in this section is to address two related concepts, flux quantizations and anomaly cancellation, using the curvature scalings and Bianchi identities that we developed in the earlier sections. Our approach here will be simple, in the sense that we will not consider the full trans-series form of the action from \eqref{kimkarol}. Instead we will concentrate on the zero instanton sector of the action, {\it i.e.} the $s = b_p = 0$ sector of the action \eqref{kimkarol}. In the following, our aim to address the flux quantizations first. But before we go into this, let us clarify s few subtleties related to the temporal dependence of $\mathbb{X}_8$ polynomial that we studied earlier.

\subsection{Revisiting time-dependent $\mathbb{X}_8$ and the Chern-Weil transgression \label{jesskibhaluu}}

The time dependence of $\mathbb{X}_8$ polynomial is crucial to analyze the EOMs as well as the Bianchi-identities (although for the latter it is the first Pontryagin density that appears, they are related by the chain of dualities in \eqref{malinechat}). The question that we want to ask here is value of the temporal derivative of $\mathbb{X}_8$ integrated over the eight-manifold $\mathcal M_8 = \mathcal M_4 \times \mathcal M_2 \times {\mathbb{T}^2\over {\cal G}}$, {\it i.e.}:
\bg\label{jayfollows}
{\partial \over \partial t}\int_{\mathcal M_8} \mathbb{X}_8({\bf x}, y; \bar{g}_s) \equiv \bar{g}_{s, 0} {\partial \over \partial \bar{g}_s}\int_{\mathcal M_8} \mathbb{X}_8({\bf x}, y; \bar{g}_s), \nd
where $\bar{g}_{s, 0}$ is defined in {\bf Table \ref{jessrog}}. The story becomes more subtle because, while 
our earlier $\bar{g}_s$–scalings describe how the \emph{pointwise} $8$–form density changes, the Chern–Weil
transgression guarantees that these $\bar{g}_s$, or more appropriately the $t$–variations, are exact:
\bg\label{maymey}
\partial_t \mathbb{X}_8(\mathbb{R}(t)) \;=\; d\,\Theta_7(\omega(t),\partial_t\omega(t)),
\nd
where $\mathbb{R}(t)$ is the curvature two-form and $\Theta_7(\omega(t),\partial_t\omega(t))$ is the Chern-Weil seven-form (defined up to $d$ of a $6$-form). \eqref{maymey} would imply 
that they integrate to zero on a closed $\mathcal M_8$. Thus $\int_{\mathcal M_8}\mathbb{X}_8$ is $t$–independent even though
$\mathbb{X}_8({\bf x}, y;t)$ itself is $t$–dependent.

To see how \eqref{maymey} may appear, let $\omega(t)$ be a smooth one–parameter family of spin connections and
$\mathbb{R}(t) \;\equiv\; d\omega(t) + \omega(t)\wedge\omega(t)$
its curvature 2–form. 
The $t$–variation of the curvature is\footnote{We will henceforth represent, in this section, all the time-dependent n-forms $\mathbb{F}_n(x, y; t)$ as $\mathbb{F}_n(t)$ to avoid clutter.}:
\begin{equation}
\partial_t \mathbb{R}(t) \;=\; {\rm D}_t \dot\omega(t) \;\equiv\; d\dot\omega(t) + [\omega(t),\dot\omega(t)],
\label{eq:dRt}
\end{equation}
with ${\rm D}_t$ the covariant exterior derivative. We will take the simple Bianchi identity (without worrying about the fluxes etc) to be ${\rm D}_t \mathbb{R}(t)=0$. Using the
invariance of the trace, $\operatorname{tr}([X,Y]\cdots)=0$, one has
$\operatorname{tr}\big(D_t (\cdots)\big) \;=\; d\,\operatorname{tr}(\cdots)$ and we can easily show that for 
the monomial $\operatorname{tr} \mathbb{R}^{n}(t)$:
\begin{equation}
\partial_t \operatorname{tr} \mathbb{R}^{n}(t)
\;=\; n\,\operatorname{tr}\!\big[(\partial_t \mathbb{R}(t))\,R_t^{\,n-1}\big]
\;=\; n\,\operatorname{tr}\!\big[D_t \dot\omega_t \, \mathbb{R}^{n-1}(t)\big]
\;=\; n\, d\,\operatorname{tr}\!\big[\dot\omega_t\, \mathbb{R}^{n-1}(t)\big],
\label{eq:var-trRn}
\end{equation}
where the last step uses the trace identity above and ${\rm D}_t \mathbb{R}^{n-1}(t)=0$ (by Bianchi identity). Notice how the temporal derivative converts to a (locally) exact form. In particular, \eqref{eq:var-trRn} implies the following identities:
\bg\label{may3mey}
&& \partial_t \operatorname{tr} \mathbb{R}^{4}(t) \;=\; 4\, d\,\operatorname{tr}\!\big[\dot\omega_t\, \mathbb{R}^{3}(t)\big],
~~
\partial_t \operatorname{tr} \mathbb{R}^{2}(t) \;=\; 2\, d\,\operatorname{tr}\!\big[\dot\omega_t\, \mathbb{R}(t)\big]\nonumber\\
&& \partial_t \big[\operatorname{tr} \mathbb{R}^{2}(t)\big]^{\!2}
\;= \; 4\, d\big[ \operatorname{tr} \mathbb{R}^{2}(t)\cdot\operatorname{tr}\!\big(\dot\omega(t)\, \mathbb{R}(t)\big)\big],
\nd
where we again used the fact that $\operatorname{tr} \mathbb{R}^{2}(t)$ is closed {\it i.e.} $d\,\operatorname{tr} \mathbb{R}^{2}(t)=0$ by
the Bianchi identity. \eqref{may3mey} has all the necessary ingredients to compute the temporal variation of $\mathbb{X}_8$ polynomial. Putting everything together, we get:
\bg\label{lilybring}
\partial_t \mathbb{X}_8(\mathbb{R}(t)) &=&  \partial_t \operatorname{tr} \mathbb{R}^{4}(t)
   \;-\; \frac{1}{4}\,\partial_t \big[\operatorname{tr} \mathbb{R}^{2}(t)\big]^{2} \nonumber\\
&=& d\,\Big[
  4\,\operatorname{tr}\!\big(\dot\omega(t)\, \mathbb{R}^{3}(t)\big)
  \;-\; \big(\operatorname{tr} \mathbb{R}^{2}(t)\big)\,\operatorname{tr}\!\big(\dot\omega(t)\, \mathbb{R}(t)\big)
\Big],
\nd
showing us that the temporal derivative of the $\mathbb{X}_8(t)$ polynomial is an exact form given in terms of a transgression seven-form 
as in \eqref{maymey}. The explicit structure of the seven-form may be easily seen from \eqref{lilybring} as:
\bg\label{keiraship}
\Theta_7(\omega_t,\dot\omega_t)
\;=\;
4\,\operatorname{tr}\!\big[\dot\omega_t\, \mathbb{R}^{3}(t)\big]
\;-\;
\big[\operatorname{tr} \mathbb{R}^{2}(t)\big]\,\operatorname{tr}\big[\dot\omega_t\, \mathbb{R}(t)\big]\,,
\nd
which is defined up to addition of an exact $d\Lambda_6$, implying that any such shift leaves
$d\Theta_7$ (and hence $\partial_t \mathbb{X}_8(t)$) unchanged.

Let us make a few remarks related to our above discussion. {\Su One}, 
the derivation is a special case of the general Chern–Weil transgression formula for any
invariant polynomial $\mathbb{P}(\mathbb{R})$ whose temporal derivative yields:
\begin{equation}
\partial_t \mathbb{P}[\mathbb{R}(t)] \;=\; d\,\mathbb{T}_{\mathbb{P}}[\omega(t),\dot\omega(t)],
\qquad
\mathbb{T}_{\operatorname{tr} \mathbb{R}^n} = n\,\operatorname{tr}[\dot\omega(t)\,\mathbb{R}^{n-1}(t)].
\end{equation}
{\Su Two}, different overall normalizations of $\mathbb{X}_8$ (e.g.\ $\mathbb{X}_8 = \tfrac1{192}(p_1^2-4p_2)$) simply
rescale $\Theta_7$ by the same constant; the exactness statement is unchanged; and {\Su three}, 
replacing the spin connection by a general gauge connection $\mathcal A(t)$ with curvature
$\mathbb{F}(t)=d\mathcal A(t)+ \mathcal A^2(t)$ yields the identical proof with $\omega(t) \to \mathcal A(t)$, $\mathbb{R}(t) \to \mathbb{F}(t)$.

\eqref{maymey} with \eqref{keiraship} suggests that while $\mathbb{X}_8(t)$ could easily be time-dependent, the integral of $\mathbb{X}_8(t)$ over a {\it smooth} eight-manifold $\mathcal M_8$ can only be time-independent. In other words, the Chern-Weil transgression formula suggests that:
\bg\label{harkapil}
\mathbb{X}_8(\Delta t) = \mathbb{X}_8(0) + d\big[\Theta_7(t) \Delta t\big], \nd
where $\Theta_7(t)$ is a {\it globally defined} seven-form if the manifold $\mathcal M_8$ does not have any boundaries, or orbifold singularities. For the two heterotic cases that we study in this paper, namely ${\rm E}_8 \times {\rm E}_8$ and $SO(32)$, $\mathbb{X}_8(0) = 0$ and the eight-manifold $\mathcal M_8$ takes the form $\mathcal M_4 \times \mathcal M_2 \times {\mathbb{T}^2\over \mathcal G}$ with $\mathcal M_2$ given locally by the following orbifolded structure:
\bg\label{garkamel}
\mathcal M_2 = {\bf S}^1 \times {{\bf S}^1\over \mathbb{Z}_2}, ~~~~ 
\mathcal M_2 = {\mathbb{T}^2\over \mathbb{Z}_2}, \nd
respectively. (One may see this from the first rows in {\bf Tables \ref{milleren4}} and {\bf \ref{milleren2}}.) In the presence of the orbifold singularities, the Chern-Weil transgression formula now gives us:
\bg\label{brazjessrog}
\frac{\partial}{\partial t}\int_{\mathcal M_8} \mathbb{X}_8(t)
\;=\; \int_{\partial \mathcal M_8}\Theta_7\!\big[\omega(t),\dot\omega(t)\big]
\;=\; \sum_{i=1}^{\rm N} \int_{\mathcal W_7^{(i)}} \Theta_7(\omega(t),\dot\omega(t)), \nd
where ${\rm N}$ is the number of orbifold seven-planes (ignoring the $2+1$ dimensional non-compact directions) with geometry given by $\mathcal W_7^{(i)}$. For the ${\rm E}_8 \times {\rm E}_8$ theory, this suggests localized forms on the Horava-Witten walls $\mathcal W_7^{(i)}\big\vert_{i = 1, 2} \times {\bf R}^{2, 1}$; and for the $SO(32)$ case, the localized forms are on the four O7-planes in the intermediate type IIB set-up (see {\bf Table \ref{milleren2}}). In a  similar vein,  ${\rm tr}~\mathbb{R}_{\rm tot} \wedge \mathbb{R}_{\rm tot}$ in \eqref{capitainemey} is defined over a four-manifold which is a boundary of a given five-manifold. 

Before ending this section let us clear up one more subtlety related to the presence of dynamical M2-branes. (Similar story will hold for dynamical M5-branes so won't analyze this separately.) The subtlety is the following. 
For an M2 source the transverse (internal) metric is singular: the warp factor
${\rm H}(r)$ diverges at $r=0$. This means the Levi–Civita connection is not smooth at the brane
position(s), so one cannot apply the global Chern–Weil argument \emph{directly} on $\mathcal M_8$. Does this now create new surfaces with localized seven-form $\Theta_7$? To answer this we need to carefully look at the local geometries of the M2-branes near the singularities.

Let $\Sigma(t)=\{p_i(t)\}$ be the (time–dependent) set of M2-branes' locations in $\mathcal M_8$. (For the present analysis we will ignore the fact that $\mathcal M_8$ has orbifold singularities from the start.)
We can excise small geodesic eight-balls $\mathcal B_{8i}(\varepsilon)$ around each $p_i(t)$ and work on the punctured
manifold:
\begin{equation}
\mathcal M_8^\ast(t)\;=\;\mathcal M_8\setminus\bigcup_i \mathcal B_{8i}(\varepsilon),\qquad
\partial \mathcal M_8^\ast(t)=\bigsqcup_i \mathcal S^7_i(\varepsilon),
\end{equation}
where $\mathcal S^7_i$ is the $i$-th boundary seven-sphere. 
On $\mathcal M_8^\ast(t)$ the connection is smooth, so by Chern–Weil transgression
we expect:
\begin{equation}
\partial_t \mathbb{X}_8[\mathbb{R}(t)] \;=\; d\,\Theta_7[\omega(t),\dot\omega(t)],
\end{equation}
on $\mathcal M^\ast_8(t)$. (The proof that we gave earlier essentially goes through in the same way for the punctured manifold $\mathcal M^\ast_8(t)$.)
Taking the time derivative of the \emph{excision integral} and using Stokes’ theorem now gives:
\begin{equation}
\frac{\partial}{\partial t}\int_{\mathcal M_8^\ast(t)}\!\mathbb{X}_8(t)
\;=\;\sum_i\int_{\mathcal S^7_i(\varepsilon)} \Theta_7[\omega(t),\dot\omega(t)]
\;+\; \underbrace{\int_{\partial \mathcal M_8^\ast(t)} \iota_{v}\,\mathbb{X}_8(t)}_{\text{motion of the small spheres}},
\label{eq:excision-dt}
\end{equation}
where in the last term, $v$ is the boundary velocity field induced by the motion of $p_i(t)$ and possible change of
the excision surfaces. (We are ignoring the fact that the number of the M2-branes could also change with time.)
Near an isolated M2-brane, the transverse metric is (up to smooth diffeomorphism) \emph{conformally flat}.
Two key facts then hold. {\Su One}, on $\mathcal S^7(\varepsilon)$ the induced metric is (after a trivial
rescaling) the standard round metric. The Levi–Civita connection on a round odd sphere has
\emph{vanishing gravitational Chern–Simons invariant}, implying that:
\begin{equation}
\int_{\mathcal S^7(\varepsilon)}\!\Theta_7\big[\omega_{\mathcal S^7}(t),\dot\omega_{\mathcal S^7}(t)\big] \;=\; 0, 
\end{equation}
which may also be equivalently argued in the following way:
because $\mathcal B_8$ is contractible, the Pontryagin classes vanish on it; and therefore by transgression,
the Chern-Simons integral on its boundary must vanish. And {\Su two}, the second term in \eqref{eq:excision-dt}
depends on how the small spheres are dragged as the brane moves. Locally $\mathbb{X}_8=d\Theta_7$ on
$\mathcal B_8\setminus\{0\}$; using Cartan’s homotopy formula:
\bg\label{jcmeymalis}
\mathcal{L}_v \mathbb{X}_8(t) =d\big[\iota_v \mathbb{X}_8(t)\big]+\iota_v\big[d\mathbb{X}_8(t)\big], \nd
where $\mathcal{L}_v$ is the Lie-derivative,
and the closure of $\mathbb{X}_8(t)$ {\it i.e.} $d\mathbb{X}_8(t)=0$, one can trade the motion term for another Chern-Simons boundary integral on $\mathcal S^7(\varepsilon)$. This 
vanishes by argument number one. Thus the time derivative of the excised integral is zero. Therefore, taking $\varepsilon\to0$ (the excised volume tends to zero and does not contribute), we obtain the same result that the integral of $\mathbb{X}_8(t)$ cannot be time-dependent
for ordinary, isolated M2-brane singularities. Combining everything together tells us that the temporal dependence only comes from the presence of orbifold singularities as in \eqref{brazjessrog}.

\subsection{Flux quantizations for the global fluxes over internal manifolds \label{kolamey}}

Flux quantizations in a system where both the metric and the flux components have temporal dependence are non-trivial to define. Therefore it would be useful for us to define exactly what is meant by quantizations here. In the definition of the G-flux components in terms of global and localized pieces, it is the time-independent parts of the global components that would typically be quantized. In other words, the quantization procedure should be independent of $\bar{g}_s$ and, in particular, should reproduce the Witten's quantization condition \cite{wittenquantize}. To see how this may work out in our case, let us look at all the internal flux components carefully.

\subsubsection{Flux quantizations for ${\bf G}_{\rm MNPQ}$ components \label{sutton1}}

Our first internal flux components are the ${\bf G}_{\rm MNPQ}$ components where $({\rm M, N}) \in {\cal M}_4 \times {\cal M}_2$. The Bianchi identities governing their dynamics are given in \eqref{crawlquif2} and in {\bf Table \ref{ccrani201}}. The terms contrbuting to the EOMs \eqref{crawlquif} are collected in {\bf Table \ref{ccrani101}}. The explicit form for the G-flux components is given by \eqref{salgill3}, where we see that the leading order $\bar{g}_s$ scalings are $0^\pm$. In fact the five-branes contributing to the dynamics also scale as $0^\pm$ as evident from {\bf Table \ref{ccrani201}}. The five-branes have a topology of the form:
\bg\label{sirenmey}
{\bf R}^{2, 1} \times {\bf S}^1_{{\cal M}_d} \times \xoxo, \nd
where $d = 4, 2$, and the 1-cycles are allowed because of the non-K\"ahlerity of the internal manifold whose local topology is 
${\cal M}_4 \times {\cal M}_2$.
Moreover, since we have taken $0^\pm > 0$, both the flux components and the five-branes are now dynamical. The puzzling aspect however is the presence of the quantum series $(\mathbb{T}^{(q)})_{\rm MNPQ}$ given in terms of $\ast\mathbb{Y}_7$ in \eqref{valentinpen}. This would appear to change the flux quantization condition of Witten \cite{wittenquantize}. Is there a way out of the conundrum?

The answer was already known in some sense from \cite{coherbeta3}. The {\it localized} fluxes from \eqref{viorosse} can in fact absorb the quantum corrections so that the global fluxes would continue to satisfy the required quantization conditions. To see that, let us consider a five-manifold ${\bf \Sigma}_5$ whose boundary is a four-dimensional manifold ${\bf \Sigma}_4$. (This way we avoid the issues from section \ref{jesskibhaluu}.) In other words, $\partial{\bf \Sigma}_5 = {\bf \Sigma}_4$. Integrating \eqref{crawlquif2} over the five-manifold gives us:

{\footnotesize
\bg\label{hwlexador}
\int_{{\bf \Sigma}_4} {\bf G}_{\rm MNPQ}~dy^{\rm M} \wedge dy^{\rm N} \wedge dy^{\rm P} \wedge dy^{\rm Q}  & = & c_1  \int_{{\bf \Sigma}_4} \big(\mathbb{T}^{(q)}\big)_{\rm MNPQ}~dy^{\rm M} \wedge dy^{\rm N} \wedge dy^{\rm P} \wedge dy^{\rm Q}\\
&+& c_2\int_{{\bf \Sigma}_4} \big({\rm tr}~\mathbb{R}_{\rm tot} \wedge \mathbb{R}_{\rm tot}\big)_{\rm MNPQ}~dy^{\rm M} \wedge dy^{\rm N} \wedge dy^{\rm P} \wedge dy^{\rm Q} + c_3 \hat{\mathbb{N}}^{(1)}_5, \nonumber
\nd}
where the $\bar{g}_s$ dependence of each terms cancel out as evident from {\bf Table \ref{ccrani201}}! (From {\bf Table \ref{ccrani201}} we can easily see that all terms in \eqref{hwlexador} scale as $\xxy^{0^\pm}$ at leading order.) $\hat{\mathbb{N}}^{(1)}_5$ denotes the number of dynamical five-branes with topologies \eqref{sirenmey}. However now there is an additional piece from quantum corrections with a coefficient $c_1$, where $c_i$ are $\bar{g}_s$ independent coefficients. This can be balanced by the localized fluxes from \eqref{viorosse} in the following way:

{\scriptsize
\bg\label{perumei}
\int_{{\bf \Sigma}_4} {\cal F}^{(k)}_{\rm MN} \Omega^{(k)}_{\rm PQ} 
\xxy^{0^\pm + {2k\over 3}\vert{\rm L}_{\rm MN}^{\rm PQ}(k; t)\vert} 
dy^{[{\rm MNPQ}]} = 
\int_{{\bf \Sigma}_4} \big(\hat{\mathbb{T}}^{(q;k)}\big)_{\rm MNPQ} 
\xxy^{\theta_{nl}(k) - {8\over 3}^\pm} dy^{[{\rm MNPQ}]}, \nonumber\\ \nd}
which gives us $\theta_{nl} = {8\over 3}^\pm +{2k\over 3}\vert{\rm L}_{\rm MN}^{\rm PQ}(k; t)\vert$, consistent with both EOMs and the Bianchi identities. (We have defined $dy^{[{\rm MNPQ}]} \equiv dy^{\rm M} \wedge dy^{\rm N} \wedge dy^{\rm P} \wedge dy^{\rm Q}$.)   Note that \eqref{perumei} is an integrated condition, so locally the EOMs and the Bianchi identities will continue to have both the localized and the global fluxes. Now combining \eqref{viorosse}, \eqref{hwlexador} and \eqref{perumei}, we get the following flux quantization condition:
\bg\label{kilievio}
\int_{{\bf \Sigma}_4} {\cal G}^{(k; {\rm global})}_{\rm MNPQ}~ dy^{[{\rm MNPQ}]} = c_2\int_{{\bf \Sigma}_4} \big({\rm tr}~\mathbb{R}^{(k)}_{\rm tot} \wedge \mathbb{R}^{(k)}_{\rm tot}\big)_{\rm MNPQ}~
dy^{[{\rm MNPQ}]} + c_3 \hat{\mathbb{N}}^{(k; 1)}_5,\nd
which is consistent with Witten's flux quantization condition \cite{wittenquantize}, but now we see that it is valid at every order in $k \in \mathbb{Z}$. Thus even though the system has inherent temporal dependence, there appears to be a well-defined flux quantization procedure with dynamical five-branes.

\subsubsection{Flux quantizations for ${\bf G}_{{\rm MNP}a}$  
components \label{sutton2}}

The story for the other set of flux components, namely  ${\bf G}_{{\rm MNP}a}$ components, is somewhat similar to what we developed above. The Bianchi identities will produce an integral condition like \eqref{hwlexador}, with the dynamical five-branes for the ${\bf G}_{{\rm MNP}a}$ case are typically oriented in the following way:
\bg\label{essexshap}
{\bf R}^{2, 1} \times {\bf S}^2_{{\cal M}_4}\times {\bf S}^1_{\xoxo},~~ {\bf R}^{2, 1} \times {\cal M}_2 \times{\bf S}^1_{\xoxo},~~{\bf R}^{2, 1} \times {\bf S}^1_{{\cal M}_4} \times {\bf S}^1_{{\cal M}_2} \times{\bf S}^1_{\xoxo}\nd
as discussed earlier. The worrisome feature, unlike the previous case, is the five-branes wrapping a 1-cycle of the toroidal internal space 
$\xoxo$. For ${\cal G} = \mathbb{I}$, there is no issue but if ${\cal G}$ is a modulus operation without fixed points, then one might worry about the absence of topological 1-cycles. Fortunately this is not an issue for us because, while locally the eight-manifold has a topology of the form ${\cal M}_4 \times {\cal M}_2 \times \xoxo$, globally however it is a non-K\"ahler eight manifold with $\xoxo$ being a local fibration over a six-dimensional base (with local topology) ${\cal M}_4 \times {\cal M}_2$. These five-branes have a dominant scalings of $\hat{l}_e = 1^\pm$ as evident from comparing all the rows in {\bf Table \ref{ccrani203}}. Looking at the precise scalings of the flux components in \eqref{salgill5}, we see that the gauge fluxes now balance the quantum series in the following way:

{\scriptsize
\bg\label{perumei2}
\int_{{\bf \Sigma}_4} {\cal F}^{(k)}_{[{\rm MN}} \Omega^{(k)}_{{\rm P}a]} 
\xxy^{1^\pm + {2k\over 3}\vert{\rm L}_{\rm MN}^{{\rm P}a}(k; t)\vert} 
dy^{[{\rm MNP}a]} = 
\int_{{\bf \Sigma}_4} \big(\hat{\mathbb{T}}^{(q;k)}\big)_{{\rm MNP}a} 
\xxy^{\theta_{nl}(k) - {5\over 3}^\pm} dy^{[{\rm MNP}a]}, \nonumber\\ \nd}
which should be compared with \eqref{perumei}: the distributions of the $\bar{g}_s$ scalings are quite different. Nevertheless we again get  
$\theta_{nl} = {8\over 3}^\pm +{2k\over 3}\vert{\rm L}_{\rm MN}^{{\rm P}a}(k; t)\vert$, consistent with both EOMs and the Bianchi identities. Note that now we can have localized forms $\Omega_{\rm MN}$ and $\Omega_{{\rm M}a}$, the latter being allowed because of the non-K\"ahlerity of the underlying eight-manifold. Combining everything together the flux quantization condition takes the form:
\bg\label{kilievio2}
\int_{{\bf \Sigma}'_4} {\cal G}^{(k; {\rm global})}_{{\rm MNP}a}~ dy^{[{\rm MNP}a]} = c_2\int_{{\bf \Sigma}'_4} \big({\rm tr}~\mathbb{R}^{(k)}_{\rm tot} \wedge \mathbb{R}^{(k)}_{\rm tot}\big)_{{\rm MNP}a}~
dy^{[{\rm MNP}a]} + c_3 \hat{\mathbb{N}}^{(k; 2)}_5,\nd
where $\hat{\mathbb{N}}^{(k; 2)}_5$ is the number of five-branes with the orientations \eqref{essexshap}. The result \eqref{kilievio2} is not only independent of time but also consistent with Witten flux-quantization rule \cite{wittenquantize}.

\subsubsection{Flux quantizations for ${\bf G}_{{\rm MN}ab}$ 
components \label{sutton3}}

The final set of G-fluxes whose quantization scheme needs to be worked out is the ${\bf G}_{{\rm MN}ab}$ flux components. These flux components are interesting because they have to be necessarily time-dependent otherwise an EFT would not exist. This is most evident from the quantum scalings \eqref{brittbaba} and \eqref{brittbaba2} for the ${\rm E}_8 \times {\rm E}_8$ and the $SO(32)$ cases respectively. Happily, looking at the precise scalings for the flux components in \eqref{salgill7}, we see that the leading order scalings are $2^\pm$, and are therefore consistent with the EFT requirements. Moreover, from {\bf Table \ref{ccrani205}}, the dynamical five-branes also scale as $\hat{p}_e(t) = 2^\pm$, and they have the following topologies:
\bg\label{MMartarmey}
{\bf R}^{2, 1} \times {\bf S}^3_{{\cal M}_4}, ~~ {\bf R}^{2, 1} \times {\bf S}^1_{{\cal M}_4} \times {\cal M}_2, ~~ {\bf R}^{2, 1} \times {\bf S}^2_{{\cal M}_4} \times {\bf S}^1_{{\cal M}_4}, \nd
which show that they wrap 3-cycles in the internal manifold ${\cal M}_4 \times {\cal M}_2$. As discussed earlier, this is allowed because of the non-K\"ahlerity of the internal space. 

The quantization conditions may now be derived as before by integrating \eqref{crawlquif2} over a five manifold ${\bf \Sigma}''_5$ with a boundary ${\bf \Sigma}''_4$. The gauge fluxes now balance the quantum series in the following way:

{\scriptsize
\bg\label{perumei3}
\int_{{\bf \Sigma}''_4} {\cal F}^{(k)}_{[{\rm MN}} \Omega^{(k)}_{ab]} 
\xxy^{2^\pm + {2k\over 3}\vert{\rm L}_{\rm MN}^{ab}(k; t)\vert} 
dy^{[{\rm MN}ab]} = 
\int_{{\bf \Sigma}''_4} \big(\hat{\mathbb{T}}^{(q;k)}\big)_{{\rm MN}ab} 
\xxy^{\theta_{nl}(k) - {2\over 3}^\pm} dy^{[{\rm MN}ab]}, \nonumber\\ \nd}
which differs from \eqref{perumei} and \eqref{perumei2} in the distributions of the $\bar{g}_s$ factors. Nevertheless, the quantum terms contributing to \eqref{perumei3} scales as $\theta_{nl} ={8\over 3}^\pm +{2k\over 3}\vert{\rm L}_{\rm MN}^{ab}(k; t)\vert$, similar to the leading order terms that we had before and differing only to the sub-leading orders. Note that we are now looking at localized forms $\Omega_{ab}, \Omega_{{\rm M}a}$ and $\Omega_{\rm MN}$, and consequently \eqref{perumei3} is only an integrated condition. The EOMs \eqref{crawlquif} and the Bianchi identities \eqref{crawlquif2} continue to have both the localized and the global fluxes. Combining everything together the flux quantization condition takes the form:
\bg\label{kilievio3}
\int_{{\bf \Sigma}''_4} {\cal G}^{(k; {\rm global})}_{{\rm MN}ab}~ dy^{[{\rm MN}ab]} = c_2\int_{{\bf \Sigma}''_4} \big({\rm tr}~\mathbb{R}^{(k)}_{\rm tot} \wedge \mathbb{R}^{(k)}_{\rm tot}\big)_{{\rm MN}ab}~
dy^{[{\rm MN}ab]} + c_3 \hat{\mathbb{N}}^{(k; 3)}_5,\nd
where $\hat{\mathbb{N}}^{(k; 3)}_5$ is the number of five-branes with the orientations \eqref{MMartarmey}. This is expectedly consistent with Witten's flux quantization rule \cite{wittenquantize}. 

The flux quantization conditions in \eqref{kilievio}, \eqref{kilievio2} and \eqref{kilievio3}, while succinctly provides a consistent picture for the global fluxes, however raises questions for the localized fluxes. Shouldn't the localized fluxes also be quantized over some well-defined 2-cycles? The answer is in general {\it no}, because the localized fluxes generically dualize to abelian fluxes in the heterotic side that are combinations of the gauge fluxes and NS two-forms. There is no reason why such combinations should be quantized, so typically, 
\eqref{perumei}, \eqref{perumei2} and \eqref{perumei3} are all there is that can be said about the two-form fluxes ${\cal F}_{\rm MN}, {\cal F}_{ab}$ and ${\cal F}_{{\rm M}a}$.

On the other hand, existence of localized forms $\Omega_{\rm MN}, \Omega_{{\rm M}a}$ and $\Omega_{ab}$ suggests that one could allow 2-forms ${\cal F}_{\mu{\rm M}}, {\cal F}_{\mu\nu}$ and ${\cal F}_{\mu a}$ where $(\mu, \nu) \in {\bf R}^{2, 1}$. However there is no integrated condition on these fluxes because we do not expect flux quantizations over non-compact spaces. 

\subsection{Anomaly cancellations with global fluxes, branes and curvatures \label{suttonjen}}

Our final set of analysis related to the fluxes would be anomaly cancellations with two- and five-branes. Over compact non-K\"ahler internal eight manifold, the aforementioned anomaly cancellations become {\it necessary} for the consistency of the construction. However because of the underlying temporal dependence of the fluxes, metric and the branes, such an analysis could become non-trivial and therefore needs to be approached carefully.

\subsubsection{Anomaly cancellations with dynamical M2-branes and 
fluxes \label{suttonjen1}}

Our first case is the anomaly cancellations with dynamical M2-branes and fluxes. These cancellations are absolutely essential to avoid the tadpoles in the theory. A brief discussion on the cancellation procedure is given earlier in {\bf Tables \ref{crecraqwiff101}} and {\bf \ref{ccquiff30}}, and here we want to elaborate the story in detail.  

\begin{table}[tb]  
 \begin{center}
\resizebox{\columnwidth}{!}{%
\renewcommand{\arraystretch}{2.3}
}
\renewcommand{\arraystretch}{1}
\end{center}
 \caption[]{\Su Anomaly cancellations with M2-branes and fluxes when we consider the EOMs associated with ${\bf G}_{0ij{\rm M}}$.} 
  \label{anomaly1}
 \end{table}

\begin{table}[tb]  
 \begin{center}
\resizebox{\columnwidth}{!}{%
\renewcommand{\arraystretch}{2.3}
%
}
\renewcommand{\arraystretch}{1}
\end{center}
 \caption[]{ \Su Anomaly cancellations with M2-branes and fluxes when we consider the EOMs associated with ${\bf G}_{0ija}$.} 
  \label{anomaly2}
 \end{table}

\begin{table}[tb]  
 \begin{center}
 \resizebox{\columnwidth}{!}{%
 \renewcommand{\arraystretch}{5.0}
%
}
\renewcommand{\arraystretch}{1}
\end{center}
 \caption[]{\Su $\bar{g}_s$ scalings of the ``on-shell" curvature tensors contributing to the Einstein tensors for the generalized ${\rm E}_8 \times {\rm E}_8$ case, collected from {\bf Tables \ref{niksmit1}, \ref{niksmit3}, \ref{niksmit5}, \ref{niksmit7}} and {\bf \ref{niksmit13}}. Note that the scalings do not include the log corrections that we derived in section \ref{sec4.5.1}. These corrections have to be added in.}
\label{niksmit007}
 \end{table} 

\begin{table}[tb]  
 \begin{center}
 \renewcommand{\arraystretch}{1.0}
\begin{tabular}{|c||c|}\hline  Contributions to Einstein Tensors & 
${g_s\over {\rm H}(y) {\rm H}_o({\bf x})}$ Scalings \\ \hline
${\bf R}_{m\rho}, {\bf R}_{im}, {\bf R}_{i\rho}$ & $0$ \\ \hline
${\bf R}_{0n}, {\bf R}_{0\alpha}, {\bf R}_{0i}$ & $ - 1 + {\hat\alpha_e(t) + \hat\beta_e(t)\over 4}$ \\ \hline
${\bf R}_{ia}, {\bf R}_{0a}, {\bf R}_{ma}, {\bf R}_{\alpha a}$ & $\infty$\\ \hline
\end{tabular}
\renewcommand{\arraystretch}{1}
\end{center}
 \caption[]{\Su $\bar{g}_s$ scalings of the cross-term curvature tensors contributing to the Einstein tensors for the generalized ${\rm E}_8 \times {\rm E}_8$ case, collected from {\bf Tables \ref{niksmit9}, \ref{niksmit11}, \ref{niksmit18}, \ref{niksmit21}, \ref{niksmit23}} and {\bf \ref{niksmit24}}. Note that the scalings do not include the log corrections that we derived in section \ref{sec4.5.1}. These corrections have to be added in.}
\label{niksmit008}
 \end{table}

In {\bf Tables \ref{anomaly1}} and {\bf \ref{anomaly2}} we have tabulated the relevant parts that contribute to the anomaly cancellations. Both these set appear when we study EOMs \eqref{crawlquif} associated with 
G-flux components ${\bf G}_{0ij{\rm M}}$ and ${\bf G}_{0ija}$. Integrating \eqref{crawlquif} over the eight-manifold ${\cal M}_8$ whose local topology is ${\cal M}_4 \times {\cal M}_2 \times \xoxo$, we see that:
\bg\label{indig2tagre}
\int_{{\cal M}_8} \mathbb{T}_8^{(m, a)} + {b_4\over b_2}\int_{{\cal M}_8} {\bf \Lambda}_8 = -{b_3\over b_2} \int_{{\cal M}_8} \mathbb{X}_8^{(m, a)}, \nd
where $b_i$ are the $\bar{g}_s$ independent coefficients from \eqref{crawlquif}. The $\mathbb{X}_8^{(m, a)}$ remains time-dependent because of the orbifold singularities in $\mathcal M_2$ (see discussion in section \ref{jesskibhaluu}). From {\bf Tables \ref{anomaly1}} and {\bf \ref{anomaly2}}, both sides of the equation \eqref{indig2tagre} scale as $\xxy^{2^\pm}$ and therefore the $\bar{g}_s$ dependence scale out resulting in a time-independent equation reminiscent of the anomaly cancellation formula in \cite{becker2, DRS}. Note two things: {\Su One}, because of the underlying non-K\"ahlerity of the eight-manifold, the integral of $\mathbb{X}^{(m, a)}_8$ over the eight-manifold does not give the Euler-characteristics of the eight-manifold. And {\Su two}, the non-perturbative and non-local effects have not entered \eqref{indig2tagre}. This is because the contributions from the non-perturbative effects to \eqref{crawlquif} appear as total derivatives.

\subsubsection{Anomaly cancellations with dynamical M5-branes and curvatures \label{suttonjen2}}

Our next issue is the anomaly cancellation with the dynamical M5-branes. These five-branes are clearly different from the five-brane instantons that we took in section \ref{kalulight} and in {\bf Table \ref{gilsalaam}}. The five-branes appear more prominently in the EOMs for the dual fluxes, or alternatively in the Bianchi identities \eqref{crawlquif2}.

The story for the anomaly cancellation with dynamical M5-branes now follows a rather straightforwardly from the Bianchi identity \eqref{crawlquif2} with some interesting subtleties that we shall point out as we go along. The first subtlety comes from dealing with ${\rm tr}~\mathbb{R}_{\rm tot} \wedge \mathbb{R}_{\rm tot}$. This has been explained carefully in \eqref{argomeythi}, and using it we can express the Bianchi identity \eqref{crawlquif2} in the following way:
\bg\label{tiptipBP}
d{\bf G}_4 = c_2{\rm tr}~{\bf R}_{\rm tot} \wedge {\bf R}_{\rm tot} \wedge 
\sum_{i = 1}^2 \delta(x_{11} - w_i) dx_{11} + c_1 d\ast\mathbb{Y}_7 + c_3 \hat{\mathbb{N}}_5, \nd
where the derivation of the curvature term follows \eqref{malinechat}, $\mathbb{Y}_7$ is the full set of perturbative and the non-perturbative quantum corrections discussed in details earlier, $c_i$ are numerical constants, and $\hat{\mathbb{N}}_5 = \hat{\mathbb{N}}_5(\bar{g}_s)$ is the number of {\it dynamical} five-branes whose behavior for the various orientations of the five-branes were derived in section \ref{sec4.5}. The sum is over the two Horava-Witten walls that appears right at the onset of the duality chain in {\bf Table \ref{milleren4}}. 

One of the useful outcome of our study of the Bianchi identities in section \ref{sec4.5} is that the $\bar{g}_s$ dependence of all the terms appearing in \eqref{tiptipBP} {\it cancels} out. This means, once \eqref{tiptipBP} is stripped off the $\bar{g}_s$ dependence, the resulting equation becomes completely time {\it independent}. This is good, but raises the following puzzle. If we claim that \eqref{tiptipBP} can lead to the heterotic anomaly cancellation condition then $-$ because of the quantum term $\mathbb{Y}_7$ $-$ the anomaly cancellation condition should get corrected by loop and non-perturbative effects. This conclusion seems counterintuitive to the belief that anomalies are 1-loop effects that {\it do not} receive additional quantum corrections.

A way out of the conundrum is to view the four-form flux ${\bf G}_4$ to be made of a global and a localized pieces in a way described in \eqref{viorosse}. We can then express \eqref{tiptipBP} as:

{\footnotesize
\bg\label{tiptip2}
d{\bf G}_4^{({\rm global})} + d{\bf G}_4^{({\rm local})} = c_2 {\rm tr}~{\bf R}_{\rm tot} \wedge {\bf R}_{\rm tot} \wedge 
\sum_{i = 1}^2 \delta(x_{11} - w_i) dx_{11} + c_1 d\ast\mathbb{Y}_7 + c_3 \hat{\mathbb{N}}_5, \nd}
with the local term now balancing the quantum piece $\mathbb{Y}_7$. We can also turn the argument around to define the localized flux using the quantum terms. In fact these localized fluxes become the abelian gauge fluxes and their enhancement to the full non-perturbative group (${\rm E}_8 \times {\rm E}_8$ or $SO(32)$) happen via wrapped two-branes on the internal cycle. The story is elaborate and has been described earlier in \cite{DRS} so we will not dwell on it here. Instead we will express \eqref{tiptip2} as:
\bg\label{tiptip3}
&&  d{\bf G}_4^{({\rm local})} = c_1 d\ast\mathbb{Y}_7 \nonumber\\
&& d{\bf G}_4^{({\rm global})}  = c_2 {\rm tr}~{\bf R}_{\rm tot} \wedge {\bf R}_{\rm tot} \wedge 
\sum_{i = 1}^2 \delta(x_{11} - w_i) dx_{11}  + c_3 \hat{\mathbb{N}}_5, \nd
where the second relation almost seems like the expected anomaly condition except that it is missing the gauge fluxes, provided we assume that ${\bf G}_4^{({\rm global})}$ reduces to the heterotic three-form. Unfortunately the story is not that straightforward. The ${\bf G}_4^{({\rm global})}$ actually reduces to the NS-NS ${\bf H}_3$ and RR ${\bf F}_3$ three-forms in the type IIB side. Under a duality transformation $-$ see \cite{DRS} for the $SO(32)$ case $-$ we expect the NS-NS three form ${\bf H}_3$ to dissolve in the geometry and consequently convert the internal six-manifold to a non-K\"ahler one \cite{DRS}. The RR three form ${\bf F}_3$ eventually becomes the heterotic three-form ${\cal H}$. The curvature two-form in the heterotic side is now measured with respect to the torsional connection:
\bg\label{tommiraja}
(\omega_+)_{{\rm M}}^{ab} = \omega_{{\rm M}}^{ab} + {1\over 2} {\cal H}_{{\rm M}}^{ab}, \nd
and {\it not} the standard Einstein connection $\omega_{{\rm M}}^{ab}$. The one-form ${\cal H}_{\rm M}^{ab}$ is constructed out of the heterotic three-form ${\cal H}_{\rm MNP}$ as ${\cal H}_{\rm M}^{ab} = {\cal H}_{\rm MNP} e^{{\rm N}a} e^{{\rm P}b}$,
with all the parameters defined in the heterotic side (and not in the M-theory side)\footnote{This in particular means that the vielbein $e_{\rm N}^a$ is the standard one extracted from the heterotic metric and not the generalized vielbein $\hat{\bf e}_{\rm N}^a$ that we defined in \eqref{maecomics} for M-theory.}. Thus the duality transformation to the heterotic side converts the LHS of the second equation in \eqref{tiptip3} to $d{\cal H}$, and the curvature four-form on the RHS to the curvature four-form now defined with respect to the torsional connection \eqref{tommiraja}. Finally, the dynamical five-branes become heterotic five-branes which are {\it small instantons} for both ${\rm E}_8 \times {\rm E}_8$ and $SO(32)$ theories \cite{wittensmall}. Putting everything together converts the second equation in \eqref{tiptip3} to:
\bg\label{tiptip4}
d{\cal H} = {\alpha'\over 4}\left({\rm tr}~{\bf R}_+ \wedge {\bf R}_+ - {1\over 30} {\rm tr}~{\bf F} \wedge {\bf F}\right), \nd
where  ${\rm tr}~{\bf F} \wedge {\bf F}$ captures the instanton number; and the two-form ${\bf R}_+ \equiv {\bf R}_+(\omega_+)$, which would mean that ${\cal H}$ appears on {\it both} sides of the equation. To determine ${\cal H}$ we will have to solve the anomaly equation iteratively, order by order in $\alpha'$. Some of these details have been discussed in \cite{DRS}, which the readers may want to look up. A more precise derivation of how exactly the second equation in \eqref{tiptip3} converts to \eqref{tiptip4} will be discussed elsewhere.
 
\section{Schwinger-Dyson equations for the on/off-shell metric components \label{secmetric}}

Having discussed all the flux EOMs, it is now time to study the EOMs corresponding to the on and the off-shell metric components. Before starting let us remind the readers that by ``off-shell" metric components, we specifically mean the cross-term components appearing in the first line of \eqref{angwhiteT}. Consequently, the ``on-shell" metric components would be the ones appearing in the second line of \eqref{angwhiteT}. Both these set of metric components are {\it emergent} in the sense discussed earlier. In fact, as emphasized many times in the text, all the degrees of freedom $-$ including metric, flux and fermionic components $-$ are emergent degrees of freedom here. Since all these components are computed using path-integrals they are all truly off-shell, and by ``on-shell" and ``off-shell" we will mean those off-shell degrees of freedom that satisfy the corresponding Schwinger-Dyson equations, namely:

\vskip.1in

\noindent $\bullet$ The first set of equations in \eqref{coffiemarieso} and \eqref{nikclau} respectively for the metric components.

\vskip.1in

\noindent $\bullet$ The set of equations from \eqref{crawlquif} for the flux components.

\vskip.1in

\noindent $\bullet$ The set of equations from \eqref{tagmukh} for the Rarita-Schwinger fermionic components.

\vskip.1in

\noindent The above discussion makes it clear that the emergent off-shell degrees of freedom are mostly for the metric components $-$ satisfying the first set of equations in \eqref{nikclau} $-$ because for the flux and the fermionic components we take all the allowed possibilities thus making them all emergent on-shell degrees of freedom. Our aim for this section is to solve the first set of equations from \eqref{coffiemarieso} and \eqref{nikclau} respectively. For brevity, we will only study the generalized ${\rm E}_8 \times {\rm E}_8$ case, as the other two cases, namely the generalized $SO(32)$ and the simplified $SO(32)$ cases, may be easily extracted from the generalized ${\rm E}_8 \times {\rm E}_8$ case following {\bf figure \ref{boxer}}.

To proceed, we will need some minor modification of \eqref{brittbaba} by inserting the dominant scalings of the G-flux components. Recall that, for the on-shell metric components ${\bf g}_{\rm AB}$, we typically expressed the $\bar{g}_s$ scalings as $\Sigma^{(0)}_{e{\rm AB}}(t)$ in \eqref{sarakalu}, with   $\Sigma^{(0)}_{e{\rm AB}}(t)$ defined in \eqref{sarahchor} (see also \eqref{katepathorabar}). We can express the $\bar{g}_s$ scalings of the G-flux components, appearing as $\hat{l}_{e{\cal AB}}^{\cal CD}$ in \eqref{andyrey} and \eqref{malenbego}, as $\hat{l}_{e{\cal AB}}^{\cal CD} = a_{\cal AB}^{\cal CD} +\bar{l}_{e{\cal AB}}^{\cal CD}$, where $a_{\cal AB}^{\cal CD}$ are the dominant scalings of the flux components and $\bar{l}_{e{\cal AB}}^{\cal CD}$ are the small corrections, including the $\pm$ superscripts, derived in section \ref{sec4.5}. Inserting them in \eqref{brittbaba} changes it to the following:

{\scriptsize
\bg\label{brittbaba007}
\theta_{nl}  &= &  {\rm dom}\left({8\over 3} - \hat\zeta_e(t), {2\over 3} - \hat\sigma_e(t), {2\over 3} - (\hat\alpha_e(t), \hat\beta_e(t)), {2\over 3} + {\hat\alpha_e(t) + \hat\beta_e(t)\over 2} - \hat\zeta_e(t)\right) \sum_{i =1}^{13}l_i + \left({4\over 3} - {\hat\zeta_e(t)\over 2}\right)n_1 \nonumber\\
&+&  {\rm dom}\left({2\over 3} - \hat\sigma_e(t),~ {2\over 3}- (\hat\alpha_e(t), \hat\beta_e(t)), ~{2\over 3} + {\hat\alpha_e(t) + \hat\beta_e(t)\over 2} -\hat\zeta_e(t), ~ {8\over 3} - \hat\zeta_e(t),~ {2\over 3} \pm \hat\alpha_e(t) \mp \hat\beta_e(t) - (\hat\alpha_e(t), \hat\beta_e(t))\right) l_{14}\nonumber\\
&+& \left( {5\over 3} + {\hat\alpha_e(t) + \hat\beta_e(t)\over 4} - \hat\zeta_e(t)\right)\sum_{j = 15}^{18} l_j  + \left( {5\over 3} - {\hat\sigma_e(t)\over 2} - {\hat\zeta_e(t)\over 2}\right) \sum_{k = 19}^{23} l_k + \left({5\over 3} - {1\over 2}(\hat\alpha_e(t), \hat\beta_e(t)) - {\hat\zeta_e(t)\over 2}\right) \sum_{p = 24}^{28} l_p \nonumber\\
&+& \left( {2\over 3}  - {\hat\sigma_e(t)\over 2} - {1\over 2}(\hat\alpha_e(t), \hat\beta_e(t))\right) \sum_{q=29}^{33} l_q +
\left( {2\over 3} - {1\over 2}(\hat\alpha_e(t), \hat\beta_e(t)) + {\hat\alpha_e(t) + \hat\beta_e(t)\over 4} - {\hat\zeta_e(t)\over 2}\right) \sum_{r=34}^{37} l_r \nonumber\\
&+ & \left( {2\over 3} + {\hat\alpha_e(t) + \hat\beta_e(t)\over 4} - {\hat\sigma_e(t)\over 2} - {\hat\zeta_e(t)\over 2}\right) \sum_{s = 38}^{41} l_s {\red -} {1\over \vert\log~\bar{g}_s\vert} \sum_{i = 1}^{41} {\rm log}~ {\bf \red L_{N}}[\bar{a}_i, \bar{b}_i, \mathbb{N}_i; \mathbb{M}_{8}]
+ \left({4\over 3} +\bar{l}_{emn}^{pq}(t)- 2\hat\sigma_e(t)\right)l_{42}\nonumber\\
&+& \left({4\over 3} +\bar{l}_{emn}^{p\alpha}(t)- {3\hat\sigma_e(t)\over 2} - {1\over 2}(\hat\alpha_e(t), \hat\beta_e(t))\right)l_{43}  
+ \left({4\over 3} +\bar{l}_{emn}^{pa} -{3\hat\sigma_e(t)\over 2} - {1\over 2} (0, \hat\eta_e(t))\right)l_{44} \nonumber\\ 
&+ & \left({4\over 3} +\bar{l}_{emn}^{\alpha\beta}(t) - {\hat\sigma_e(t)} - (\hat\alpha_e(t), \hat\beta_e(t))\right)l_{45} 
+ \left( {4\over 3} +  \bar{l}_{emn}^{\alpha a}(t) - {\hat\sigma_e(t)} - {1\over 2}(\hat\alpha_e(t), \hat\beta_e(t)) - {1\over 2}(0, \hat\eta_e(t))\right)l_{46} \nonumber\\
&+ & \left({4\over 3} + \bar{l}_{e\alpha\beta}^{ma}(t) - {\hat\sigma_e(t)\over 2} - (\hat\alpha_e(t), \hat\beta_e(t)) - {1\over 2}(0, \hat\eta_e(t))\right)l_{47}
+ \left({4\over 3} +  \bar{l}_{eij}^{0m}(t) - {\hat\sigma_e(t)\over 2} - {3\hat\zeta_e(t)\over 2}\right)l_{48} \nonumber\\
&+ & \left({4\over 3} + \bar{l}_{eij}^{0\alpha}(t) - {1\over 2}(\hat\alpha_e(t), \hat\beta_e(t)) - {3\hat\zeta_e(t)\over 2}\right)l_{49}
+ \left({4\over 3} +  \bar{l}_{emn}^{ab}(t) - {\hat\sigma_e(t)}  - {\hat\eta_e(t)\over 2}\right)l_{50} +n_0 ~{\rm dom}\left(1, {\hat\alpha_e(t) + \hat\beta_e(t)\over 4}\right)\nonumber\\
&+ &  \left({4\over 3} + \bar{l}_{e\alpha\beta}^{ab}(t) - (\hat\alpha_e(t), \hat\beta_e(t))  - {\hat\eta_e(t)\over 2}\right)l_{51} 
+ \left({4\over 3} + \bar{l}_{em\alpha}^{ab}(t) - {\hat\sigma_e(t)\over 2} - {1\over 2}(\hat\alpha_e(t), \hat\beta_e(t))  - {\hat\eta_e(t)\over 2}\right)l_{52}\nonumber\\
&+ &\left( {4\over 3} + \bar{l}_{emn}^{pi}(t) - {3\hat\sigma_e(t)\over 2} - {\hat\zeta_e(t)\over 2}\right)l_{53} 
+ \left({4\over 3} + \bar{l}_{e\alpha\beta}^{mi}(t) - {\hat\sigma_e(t)\over 2} - (\hat\alpha_e(t), \hat\beta_e(t))  - {\hat\zeta_e(t)\over 2}\right)l_{54} \nonumber\\
&+ &\left( {4\over 3} + \bar{l}_{emn}^{\alpha i}(t) - {\hat\sigma_e(t)} - {1\over 2}(\hat\alpha_e(t), \hat\beta_e(t))  - {\hat\zeta_e(t)\over 2}\right)l_{55}
+ \left({4\over 3} + \bar{l}_{emn}^{ai}(t) - {\hat\sigma_e(t)} - {1\over 2}(0, \hat\eta_e(t))  - {\hat\zeta_e(t)\over 2}\right)l_{56} \nonumber\\
&+ &\left( {4\over 3} + \bar{l}_{eab}^{mi}(t) - {\hat\sigma_e(t)\over 2} - {\hat\eta_e(t)\over 2}  - {\hat\zeta_e(t)\over 2}\right)l_{57}
+ \left({4\over 3} + \bar{l}_{e\alpha\beta}^{ai}(t)  - (\hat\alpha_e(t), \hat\beta_e(t)) - {1\over 2}(0, \hat\eta_e(t))  - {\hat\zeta_e(t)\over 2}\right)l_{58}\nonumber\\
&+ & \left({4\over 3} + \bar{l}_{eab}^{\alpha i}(t) - {1\over 2}(\hat\alpha_e(t), \hat\beta_e(t)) - {\hat\eta_e(t)\over 2}  - {\hat\zeta_e(t)\over 2}\right)l_{59} + \left({4\over 3} + \bar{l}_{emn}^{ij}(t)  -{\hat\sigma_e(t)}  - {\hat\zeta_e(t)}\right)l_{61} \nonumber\\
& + & \left({4\over 3} + \bar{l}_{ema}^{\alpha i}(t) -{\hat\sigma_e(t)\over 2} - {1\over 2}(\hat\alpha_e(t), \hat\beta_e(t)) - {1\over 2}(0, \hat\eta_e(t))  - {\hat\zeta_e(t)\over 2}\right)l_{60}
+ \left({4\over 3} + \bar{l}_{emn}^{0p}  -{3\hat\sigma_e(t)\over 2} - {\hat\zeta_e(t)\over 2}\right)l_{68}  \nonumber\\
& + & \left({4\over 3} + \bar{l}_{e\alpha\beta}^{ij}(t) -(\hat\alpha_e(t), \hat\beta_e(t))  - {\hat\zeta_e(t)}\right)l_{63}
+ \left({4\over 3} + \bar{l}_{ema}^{ij}(t) -{\hat\sigma_e(t)\over 2} -{1\over 2}(0, \hat\eta_e(t))  - {\hat\zeta_e(t)}\right)l_{64}\nonumber\\
&+&\left( {4\over 3} + \bar{l}_{e\alpha a}^{ij}(t) -{1\over 2}(\hat\alpha_e(t), \hat\beta_e(t)) -{1\over 2}(0, \hat\eta_e(t))  - {\hat\zeta_e(t)}\right)l_{65} 
+\left({4\over 3} + \bar{l}_{eab}^{ij}(t) -{\hat\eta_e(t)\over 2}  - {\hat\zeta_e(t)}\right)l_{66} \nonumber\\
&+& \left({4\over 3} +  \bar{l}_{eij}^{0a}(t) -{1\over 2}(0, \hat\eta_e(t))  - {3\hat\zeta_e(t)\over 2}\right)l_{67}
+ \left({4\over 3} + \bar{l}_{em\alpha}^{ij}(t) -{\hat\sigma_e(t)\over 2} -{1\over 2}(\hat\alpha_e(t), \hat\beta_e(t))  - {\hat\zeta_e(t)}\right)l_{62}
\nonumber\\
&+& \left({4\over 3} + \bar{l}_{emn}^{0\alpha} -{\hat\sigma_e(t)} -{1\over 2} (\hat\alpha_e(t), \hat\beta_e(t)) - {\hat\zeta_e(t)\over 2}\right)l_{69}
+ \left({4\over 3} + \bar{l}_{e\alpha\beta}^{0m} -{\hat\sigma_e(t)\over 2} - (\hat\alpha_e(t), \hat\beta_e(t)) - {\hat\zeta_e(t)\over 2}\right)l_{70} \nonumber\\
&+ & \left( {4\over 3} + \bar{l}_{eab}^{0m}  -{\hat\sigma_e(t)\over 2} -{\hat\eta_e(t)\over 2} - {\hat\zeta_e(t)\over 2}\right)l_{71} 
+ \left({4\over 3} + \bar{l}_{eab}^{0\alpha} -{1\over 2}(\hat\alpha_e(t), \hat\beta_e(t)) -{\hat\eta_e(t)\over 2} - {\hat\zeta_e(t)\over 2}\right)l_{72}  + {n_0 \over 3}\nonumber\\
& + & \left( {4\over 3} + \bar{l}_{emn}^{0a} -{\hat\sigma_e(t)} -{1\over 2}(0, \hat\eta_e(t)) - {\hat\zeta_e(t)\over 2}\right)l_{73}
+\left( {4\over 3} + \bar{l}_{e0m}^{ia} - {\hat\sigma_e(t)\over 2} -{1\over 2}(0, \hat\eta_e(t)) - {\hat\zeta_e(t)}\right)l_{80} - {n_0 \hat\zeta_e(t)\over 2}\nonumber\\
&+ & \left( {4\over 3} + \bar{l}_{e\alpha\beta}^{0a}  -(\hat\alpha_e(t), \hat\beta_e(t)) -{1\over 2}(0, \hat\eta_e(t)) - {\hat\zeta_e(t)\over 2}\right)l_{75}
+ \left({4\over 3} + \bar{l}_{e0\alpha}^{ia} - {1\over 2}(\hat\alpha_e(t), \hat\beta_e(t)) -{1\over 2}(0, \hat\eta_e(t)) - {\hat\zeta_e(t)}\right)l_{81}
 \nonumber\\
&+& \left({4\over 3} +  \bar{l}_{e0m}^{\alpha i}  -{\hat\sigma_e(t)\over 2} - {1\over 2} (\hat\alpha_e(t), \hat\beta_e(t)) - {\hat\zeta_e(t)}\right)l_{77}
+ \left({4\over 3} + \bar{l}_{e\alpha\beta}^{0i}  -(\hat\alpha_e(t), \hat\beta_e(t)) - {\hat\zeta_e(t)}\right)l_{78} \nonumber\\
& + & \left( {4\over 3} + \bar{l}_{eab}^{0i} -{\hat\eta_e(t)\over 2} - {\hat\zeta_e(t)}\right)l_{79} + \left({4\over 3} + \bar{l}_{e0m}^{\alpha a} -{\hat\sigma_e(t)\over 2} - {1\over 2} (\hat\alpha_e(t), \hat\beta_e(t)) -{1\over 2}(0, \hat\eta_e(t)) - {\hat\zeta_e(t)\over 2}\right)l_{74}  \nonumber\\
& + & \left({4\over 3} + \bar{l}_{emn}^{0i} -{\hat\sigma_e(t)} - {\hat\zeta_e(t)}\right)l_{76} + \left({1\over 3} - {\hat\sigma_e(t)\over 2}\right)n_2  + \left({1\over 3} - {\hat\alpha_e(t)\over 2}\right)n_{\theta_1}+ \left({1\over 3} - {\hat\beta_e(t)\over 2}\right)n_{\theta_2}, \nd}
where we notice something rather surprising: all the G-flux components have the {\it same} dominant scalings of ${4\over 3}$. Moreover, compared to \eqref{brittbaba}, ${\bf G}_{{\rm MN}ab}$ flux components no longer have any relative {\it minus} signs. In fact the signs of all the dominant scalings are positive definite, thus justifying the existence of the EFT at the far IR. With this we are ready to tackle the EOMs of all the on and the off-shell metric components.

\subsection{EOMs for the emergent on-shell metric components ${\bf g}_{mn}$ \label{eomgmn}}

The dynamics of the on-shell emergent metric components $\langle {\bf g}_{mn}\rangle_\sigma \equiv {\bf g}_{mn}$ is controlled by the EOM given in \eqref{mootcha} with the energy-momentum tensors coming from \eqref{malishell1}, \eqref{malishell2}, \eqref{teelmolish1}, \eqref{teelmolish2} and \eqref{teelmolish3}. The $\bar{g}_s$ scalings of the set of twelve energy-momentum tensors are given in {\bf Table \ref{niksmit0071}}. Looking at Row 1 we see that the dominant scaling of ${\bf G}_{mn}$ is $0^\pm$. A more precise answer can be extracted from {\bf Table \ref{niksmit007}} and is given by:
\bg\label{sushiador}
{\rm scale}({\bf G}_{mn}) & = & {\rm dom}\Big(0, \hat\sigma_e(t) - (\hat\alpha_e(t), \hat\beta_e(t)), 2 + \hat\sigma_e(t) - \hat\zeta_e(t), \\
&& ~~ \hat\sigma_e(t) + {\hat\alpha_e(t) + \hat\beta_e(t)\over 2} - \hat\zeta_e(t), \hat\sigma_e(t) - (\hat\alpha_e(t), \hat\beta_e(t)) \pm \hat\alpha_e(t) \mp \hat\beta_e(t)\Big), \nonumber \nd
showing the range from $0^\pm$ to $2^\pm$ with the $\pm$ superscripts related to the small parameters appearing in \eqref{sushiador}. Comparing this to Row 2 of {\bf Table \ref{niksmit0071}}, we see that the fluxes do not scale in the right way to contribute to ${\bf G}_{mn}$, {\it i.e.} to the LHS of \eqref{mootcha}. They do however appear to contribute at higher orders. This is a consequence of the no-go theorems studied in \cite{GMN}, which state that the fluxes, branes and O-planes cannot provide positive energy to the system and therefore cannot be used to generate a de Sitter solution. For the present case, we see that the no-go theorems extend to the emergent quantities also: the emergent fluxes, branes et cetera cannot contribute to \eqref{mootcha} to the lowest orders.

\begin{table}[h]  
 \begin{center} 
\resizebox{\columnwidth}{!}{%
 \renewcommand{\arraystretch}{4.9}
}
\renewcommand{\arraystretch}{1}
\end{center}
 \caption[]{\Su The $\bar{g}_s$ scalings of the various terms in \eqref{mootcha} contributing to the EOM for the metric components ${\bf g}_{mn}$. We have also defined $\mathbb{U}_d({\rm Y};t) \equiv \sqrt{{\bf g}_d({\rm Y};t)} {\bf Q}_{\rm pert}({\rm Y};t)$ and $\mathbb{Q}_d({\rm Y}; t) \equiv {\bf Q}_{\rm pert}({\rm Y};t) + \int d^d{\rm Y'} \mathbb{U}_d({\rm Y'};t) \mathbb{F}({\rm Y - Y'}; t)$. To avoid clutter, we have defined all the quantities at a specific point in ${\bf R}^2$, so that ${\bf x}$ does not appear explicitly here. $\theta_d$ can be derived from the volume of the $d$ dimensional wrapped manifold, and $a_{\rm F}$ is the $\bar{g}_s$ scaling of the non-locality function $\mathbb{F}({\rm Y-Y'}; t(\bar{g}_s))$.}
\label{niksmit0071}
 \end{table} 

Let us now come to Row 3 which includes contributions from perturbative quantum corrections. The fourth column provides the $\bar{g}_s$ scaling of the perturbative contributions from the first term in \eqref{malishell2}. If we ignore the exponential piece for the time being, we see that they typically scale as $\xxy^{\theta_{nl} - {2\over 3} + \hat\sigma_e(t)}$ with $\theta_{nl}$ given by \eqref{brittbaba007}. Thus comparing rows 1 and 3 in {\bf Table \ref{niksmit0071}}, we have:
\bg\label{cyclistmey}
\theta_{nl} - {2\over 3} + \hat\sigma_e(t) = \begin{cases}~2 + \hat\sigma_e(t) - \hat\zeta_e(t)\\ ~~~\\
~\hat\sigma_e(t) - (\hat\alpha_e(t), \hat\beta_e(t)) \pm \hat\alpha_e(t) \mp \hat\beta_e(t)\\ ~~~\\
~\hat\sigma_e(t) + {\hat\alpha_e(t)+ \hat\beta_e(t)\over 2} - \hat\zeta_e(t)\\ ~~~\\
~\hat\sigma_e(t) - (\hat\alpha_e(t), \hat\beta_e(t))\\~~~\\
~0  \end{cases} \nd
which the RHS is arranged so that the dominant scaling shows up on the bottom, and the highest sub-dominant one shows up at the top. The next three terms in between are arranged in one specific way because we haven't imposed any specific relation between $\hat\alpha_e(t), \hat\beta_e(t)$ and $\hat\zeta_e(t)$. Solving for $\theta_{nl}$ in \eqref{cyclistmey} gives us:
\bg\label{cyclebaaz}
\theta_{nl}  = \begin{cases} ~{8\over 3} - \hat\zeta_e(t)\\ ~~~\\
~{2\over 3} - (\hat\alpha_e(t), \hat\beta_e(t)) \pm \hat\alpha_e(t) \mp \hat\beta_e(t)\\ ~~~\\
~{2\over 3} + {\hat\alpha_e(t)+ \hat\beta_e(t)\over 2} - \hat\zeta_e(t)\\ 
~~~\\
~ {2\over 3} - (\hat\alpha_e(t), \hat\beta_e(t))\\~~~\\
~{2\over 3} - \hat\sigma_e(t)  \end{cases} \nd
which may now be compared to the value of $\theta_{nl}$ from \eqref{brittbaba007}. This precisely gives us $l_{14} = 1$, implying that there are {\it no nontrivial perturbative terms from \eqref{botsuga} that would contribute to ${\bf T}_{mn}^{{\red({\rm pert; 1})}}$}. Fluxes, branes, O-planes and perturbative quantum effects are all red herrings in the problem. This is again expected from the no-go conditions of \cite{GMN}. 

The above conclusion ignores the exponential piece that goes as 
${\rm exp}\left(-{\bar{g}_s^{\theta_{nl}}\over \bar{g}_s^{\theta_d}}\right)$, where $\theta_{nl}$ is given by \eqref{cyclebaaz} and $\theta_d$ can be derived from the volume of the sub-manifold ${\cal M}_d$. For ${2\over 3}^\pm < \theta_d < {8\over 3}^\pm$ and for $\theta_d > {8\over 3}^\pm$, further suppressions can come from the exponential piece. For $\theta_d = {8\over 3}^\pm = {8\over 3} - \hat\zeta_e$, we could still have suppressions coming from the spatial contributions to the exponential piece, implying once again that the perturbative terms are truly inconsequential. We will discuss more about the exponential factor a little later. Meanwhile the question is: what about the perturbative contributions coming from the non-local counterterms as given in rows 3, 4 and 5 of {\bf Table \ref{niksmit0071}}?  To study them, let us first define the non-locality function $\mathbb{F}({\rm Y - Y'}; t)$ as:
\bg\label{omameys}
\mathbb{F}({\rm Y - Y'}; t(\bar{g}_s)) = \xxy^{a_{\rm F}(t)} {\bf F}({\rm Y-Y'}), \nd
where $a_{\rm F}(t)$ could in principle have either sign, although a negative sign would imply that the non-localities {\it grow} at late time. However despite that, ${\bf T}_{mn}^{{\red({\rm pert; 2})}}$ suffers the same fate as
${\bf T}_{mn}^{{\red({\rm pert; 1})}}$. For the remaining two cases, namely for ${\bf T}_{mn}^{{\red({\rm pert; 3})}}$ and ${\bf T}_{mn}^{{\red({\rm pert; 4})}}$, the quantum scaling $\theta_{nl}$ takes the following form:
\bg\label{cyclebaaz2}
\theta_{nl}  = \begin{cases} ~{8\over 3} - \hat\zeta_e(t) + \theta_d(t) - a_{\rm F}(t)\\ ~~~\\
~{2\over 3} - (\hat\alpha_e(t), \hat\beta_e(t)) \pm \hat\alpha_e(t) \mp \hat\beta_e(t)+ \theta_d(t) - a_{\rm F}(t)\\ ~~~\\
~{2\over 3} + {\hat\alpha_e(t)+ \hat\beta_e(t)\over 2} 
- \hat\zeta_e(t)+ \theta_d(t) - a_{\rm F}(t)\\ ~~~\\
~{2\over 3} - (\hat\alpha_e(t), \hat\beta_e(t))+ \theta_d(t) - a_{\rm F}(t)\\~~~\\
~{2\over 3} - \hat\sigma_e(t)+ \theta_d(t) - a_{\rm F}(t)\end{cases} \nd
which seem to give a non-trivial contribution. However a careful look suggest that this may not be the case. To see this, first note that 
since $1 \le d \le 8$ over the internal space, ${\cal M}_d \subset{{\cal M}_4 \times {\cal M}_2 \times \xoxo}$ and consequently $\theta_d(t) \le 2^\pm$. This immediate puts a constraint on $a_{\rm F}(t)$ to be $a_{\rm F}(t) \le 2^\pm$, where the superscripts as usual deal with the small corrections discussed earlier. The aforementioned two energy-momentum tensors, as clear from their $\bar{g}_s$ scalings, are not exactly perturbative because of the ${1\over \bar{g}_s^{\theta_d}}$ factors. There are three cases to consider.
\vskip.1in
\noindent $\bullet$ For $a_{\rm F}(t) = 0$ and $\theta_{nl}(t) < {2\over 3}^\pm + \theta_d(t)$ we are no longer in the perturbative regime unless 
${\cal M}_d \subset{\xoxo \times \Sigma_4}$ where $\Sigma_4$ is a four-cycle in ${\cal M}_4 \times {\cal M}_2$. In the perturbative regime, there are no solutions as clear from \eqref{cyclebaaz2}.

\vskip.1in
\noindent $\bullet$ For $a_{\rm F}(t) = -\vert a_{\rm F}(t)\vert$ and $\theta_{nl}(t) < {2\over 3}^\pm + \vert a_{\rm F}(t)\vert + \theta_d(t)$, we are again away from the perturbative regime. There is also more stronger constraint on ${\cal M}_d$. Moreover, with negative $a_{\rm F}(t)$, the non-locality increases with time. In the non-perturbative regime solutions could exist, but perturbatively we find no solutions.
\vskip.1in
\noindent $\bullet$ For positive $a_{\rm F}(t)$, perturbative regime comes from ${\cal M}_d \subset{\xoxo \times \Sigma_4}$ with $\Sigma_4$ being a four-cycle in ${\cal M}_4 \times {\cal M}_2$. Again, comparing it to \eqref{cyclebaaz2}, we see that there are no solutions as expected. Non-perturbatively with ${\cal M}_d \subset{{\cal M}_4 \times {\cal M}_2}$, solutions could exist.

\vskip.1in
\noindent All in all, the above discussions suggest that perturbative energy-momentum tensors on the RHS of \eqref{mootcha}  do not lead to any solutions, thus confirming the no-go theorems of \cite{GMN}.

Let us now come to our first set of genuine non-perturbative contributions coming from the BBS \cite{bbs} instantons which are M5-brane instantons wrapped on the six-manifold ${\cal M}_4 \times {\cal M}_2$. There are two kinds of contributions here: ${\bf T}_{mn}^{{\red({\rm BBS; 1})}}$ from local M5-brane instantons and 
${\bf T}_{mn}^{{\red({\rm BBS; 2})}}$ from non-local M5-brane instantons (although the non-locality factor is always integrated away). Therefore using \eqref{kimkarol} and \eqref{teelmolish1}, the $\bar{g}_s$ scalings are captured in rows 7 and 8 of {\bf Table \ref{niksmit0071}} respectively. For the first case, comparing rows 1 and 7 gives us:
\bg\label{cyclebaaz3}
\theta_{nl}  = \begin{cases} ~{8\over 3} - \hat\zeta_e(t) + 2 - 2\hat\sigma_e(t) - {\hat\alpha_e(t)+ \hat\beta_e(t)\over 2} \\ ~~~\\
~{2\over 3} - (\hat\alpha_e(t), \hat\beta_e(t)) \pm \hat\alpha_e(t) \mp \hat\beta_e(t)+ 2 - 2\hat\sigma_e(t) - {\hat\alpha_e(t) + \hat\beta_e(t)\over 2}\\ ~~~\\
~{2\over 3} + {\hat\alpha_e(t)+ \hat\beta_e(t)\over 2} 
- \hat\zeta_e(t)+ 2 - 2\hat\sigma_e(t) - {\hat\alpha_e(t)+ \hat\beta_e(t)\over 2}\\ ~~~\\
~{2\over 3} - (\hat\alpha_e(t), \hat\beta_e(t))+ 2 - 2\hat\sigma_e(t) - {\hat\alpha_e(t)+ \hat\beta_e(t)\over 2}\\~~~\\
~{2\over 3} - \hat\sigma_e(t)+ 2 - 2\hat\sigma_e(t) - {\hat\alpha_e(t)+ \hat\beta_e(t)\over 2}\end{cases} \nd
which implies that a factor of $2 - 2\hat\sigma_e(t) - {\hat\alpha_e(t)+ \hat\beta_e(t)\over 2}$ is added to \eqref{cyclebaaz}. This may now be compared to the value of $\theta_{nl}$ from \eqref{brittbaba007}. The result we now have is:
\bg\label{isaferr}
l_{14} = 1, ~~~~\sum_{q= 29}^{33} l_q = 2, ~~~~ \sum_{i = 1}^{13} l_i = 1, \nd
where $l_q = 2$ should be assumed to come from summing over the two choices of scalings in \eqref{brittbaba007}: ${2\over 3} - {\hat\sigma_e(t)\over 2} - {\hat\alpha_e(t)\over 2}$ and ${2\over 3} - {\hat\sigma_e(t)\over 2} - {\hat\beta_e(t)\over 2}$ picked up from ${1\over 2}(\hat\alpha_e(t), \hat\beta_e(t))$; and for $l_i = 1$ we are choosing ${2\over 3} - \hat\sigma_e(t)$ from the first line in \eqref{brittbaba007}.  Considering this, \eqref{isaferr} implies that we have {\it quartic} order in the curvature tensors, meaning that the BBS instantons contribute quartic order curvature terms to drive the acceleration of the universe. These quartic terms could in-principle come from \eqref{rooth} with $b_1 = b_2'=b_3'= 0$, $(b_2, b_3) \ne 0$ and $k = 1$, thus confirming a prediction of \cite{sothi}. Note that, for $1 \le l_i \le 13$, we are taking the contribution ${2\over 3} - \hat\sigma_e(t)$, with the assumption that the other terms in the collection would contribute at a different order. Two other set of possibilities from \eqref{brittbaba007} are of the following form:

{\footnotesize
\bg\label{isaferr2}
\Big(l_{14} = 1, ~~\sum_{q= 29}^{33} l_q = 2, ~~ n_2 = 2\Big),~~~~
\Big(l_{14} = 1, ~~ \sum_{i = 1}^{13} l_i = 1, ~~ n_2 = 2, ~~n_{\theta_1} = n_{\theta_2} = 1\Big)\nd}
with $n_2$ being the number of derivatives along ${\cal M}_4$, $n_{\theta_i}$ being the number of derivatives along ${\cal M}_2$, and $l_i = 1$ chooses the factor ${2\over 3} - \hat\sigma_e(t)$ in \eqref{botsuga}, are also allowed\footnote{While a term like $\square {\bf R}^3$ is allowed at the supersymmetric vacuum level, and contributes to the supersymmetric completion of ${\bf R}^4$ type of terms \cite{gatpurali}, a term like $\square_m\square_\alpha {\bf R}^2 \subset{\square^4 {\bf R}^2}$ most likely breaks supersymmetry. As such it will appear here with zero coefficient.} and so are other possibilities from the flux sector once we lay down the relation between $\bar{l}_{e{\rm AB}}^{\rm CD}$ and $(\hat\zeta_e(t), \hat\sigma_e(t), \hat\alpha_e(t), \hat\beta_e(t), \hat\eta_e(t))$ by working out the $\pm$ superscripts in sections \ref{sec7s} and \ref{sec4.5}. In fact an exhaustive list of quantum corrections coming from the BBS instantons can be worked out once we ascertain the aforementioned connections. Clearly since $(l_i, n_i) \in (+\mathbb{Z}, +\mathbb{Z})$, the list of quantum corrections are {\it finite}, which is a further confirmation of the fact that  an EFT with a quasi de Sitter excited state exists in the landscape.

The second type of BBS instantons has to do with the non-local effects, whose contributions to the energy-momentum tensors are given in row 8 of 
{\bf Table \ref{niksmit0071}}. There are in fact two kinds of contributions from the quantum series \eqref{botsuga}: the first one allows a scaling similar to \eqref{cyclebaaz3} except that the RHS will also have a factor of $-a_{\rm F}$ coming from \eqref{omameys}. This is similar to \eqref{cyclebaaz2} with $\theta_d \equiv 2 - 2\hat\sigma_e(t) - {\hat\alpha_e(t)+ \hat\beta_e(t)\over 2}$. For vanishing $a_{\rm F}$, solutions are similar to \eqref{isaferr} and \eqref{isaferr2}, whereas for $0 < a_{\rm F} < 2^\pm$, solutions would not exist if $a_{\rm F}$ is closer to $2^\pm$, and for $a_{\rm F} > 2^\pm$ no solutions exist. For $a_{\rm F} = -|a_{\rm F}|$, solutions would exist if and only if the values for $a_{\rm F}$ lie in the subset of the terms in \eqref{botsuga} with $l_i \in \mathbb{Z}^+$. On the other hand, for the second type of contributions, the quantum scaling becomes:
\bg\label{cyclebaaz4}
\theta_{nl}  = \begin{cases} ~{8\over 3} - \hat\zeta_e(t) + 4 - 4\hat\sigma_e(t) - \hat\alpha_e(t)- \hat\beta_e(t)-2a_{\rm F}(t)\\ ~~~\\
~{2\over 3} - (\hat\alpha_e(t), \hat\beta_e(t)) \pm \hat\alpha_e(t) \mp \hat\beta_e(t)+ 4 - 4\hat\sigma_e(t) - \hat\alpha_e(t)- \hat\beta_e(t)-2a_{\rm F}(t)\\ ~~~\\
~{2\over 3} + {\hat\alpha_e(t)+ \hat\beta_e(t)\over 2} 
- \hat\zeta_e(t)+ 4 - 4\hat\sigma_e(t) - \hat\alpha_e(t)- \hat\beta_e(t)-2a_{\rm F}(t)\\ ~~~\\
~{2\over 3} - (\hat\alpha_e(t), \hat\beta_e(t))+ 4 - 4\hat\sigma_e(t) - \hat\alpha_e(t)- \hat\beta_e(t)-2a_{\rm F}(t)\\~~~\\
~{2\over 3} - \hat\sigma_e(t)+ 4 - 4\hat\sigma_e(t) - \hat\alpha_e(t)- \hat\beta_e(t)-2a_{\rm F}(t)\end{cases} \nd
which is clearly more non-trivial than \eqref{cyclebaaz3}. The analysis for various choices of $a_{\rm F}$ can be discussed as before, but here we will study the $a_{\rm F} = 0$ only. For such a case, at least the following two sets of solutions are possible:
\bg\label{isaferr3}
\Big(l_{14} = 1, ~~\sum_{q= 29}^{33} l_q = 4, ~~ \sum_{i = 1}^{13} l_i = 2\Big), ~~~~ \Big( l_{14} = 1, ~~\sum_{q= 29}^{33} l_q = 4, ~~ n_2 = 4\Big), \nd
from \eqref{brittbaba007}, in addition to a {\it finite} set of other possibilities. For the choice $l_q = 4$, we can choose two copies each of ${2\over 3} - {\hat\sigma_e(t)\over 2} - {\hat\alpha_e(t)\over 2}$  
and ${2\over 3} - {\hat\sigma_e(t)\over 2} - {\hat\beta_e(t)\over 2}$ respectively
(An exhaustive list can again be constructed, provided the 
relation between $\bar{l}_{e{\rm AB}}^{\rm CD}$ and $(\hat\zeta_e(t), \hat\sigma_e(t), \hat\alpha_e(t), \hat\beta_e(t), \hat\eta_e(t))$ are spelled out from sections \ref{sec7s} and \ref{sec4.5}, but we will not do so here.) Our analysis suggests the quantum terms contributing from the non-local BBS type instantons are fifth and seventh order in curvatures; or with mixed flux, curvature and derivative terms\footnote{Note that terms like ${\bf R}^7$ and $\square^2{\bf R}^5$, including other mixed terms with fluxes that allow fourteen derivatives, have not yet been analyzed in the literature related to their supersymmetric completion. However from the excited state point of view we see that they appear naturally to allow for accelerated backgrounds. Thus our study here should be viewed as a prediction for the existence of such terms, despite the fact that they are suppressed by ${\rm M}_p^{-14}$.}. If we also include the fermions, the story can become more interesting, but we will leave this for future works. Note two things: {\Su one}, all the quantum contributions are countably finite in number confirming the fact that a well-defined EFT is in play here; and {\Su two}, all these terms are heavily suppressed by ${\rm M}_p$ so the dominant contributions would come from the quartic contributions studied before. 

There is yet another point related to the BBS instantons that needs mentioning at this stage before we go to the study of the KKLT instantons. Comparing the results from {\bf Table \ref{gilsalaam}}, and from {\bf Tables \ref{fridaylin1}} and {\bf \ref{fridaylin2}}, we can easily infer that the dominant charges ({\it i.e.} the Chern-Simons term from \eqref{sadapa}) of the BBS instantons scale with respect to $\bar{g}_s$ as $\bar{g}_s^{\theta_{\rm BBS}}$, where $\theta_{\rm BBS}$ is given by:

{\footnotesize
\bg\label{mandalamey}
\theta_{\rm BBS} = \underbrace{-2 + 2\hat\sigma_e(t) + {\hat\alpha_e(t)+ \hat\beta_e(t)\over 2}}_{\text{\rm Born-Infeld part}} - \underbrace{ {\hat\zeta_e(t)\over 2} - {\hat\eta_e(t)\over 2} + ...}_{\text{\rm Small correction}}, \nd}
and the dotted term comes from $\bar{l}_{e{\rm AB}}^{\rm CD}$ factors from \eqref{brittbaba007}. \eqref{mandalamey}
differs from the Born-Infeld part\footnote{See \eqref{sadapa} for the supersymmetric Minkowski case. The present case, which is for the non-supersymmetric Glauber-Sudarshan state, the form of the BBS instanton action would remain similar to \eqref{sadapa} except that the terms therein are being replaced by the expectation values.} by a very small, but non-zero, amount. The smallness can be inferred from the fact that $(\hat\zeta_e(t), \hat\eta_e(t), \hat\alpha_e(t), \hat\beta_e(t))$ are all very small. Interestingly, and according to \eqref{englishgulab}, 
$\hat\zeta_e(t) + \hat\eta_e(t) \approx 0$ but because of the $\bar{g}_s$ and ${\rm M}_p$ corrections one cannot make $\hat\zeta_e(t) = -\hat\eta_e(t)$.  Thus for the  generalized ${\rm E}_8 \times {\rm E}_8$ case, the correction appearing in \eqref{mandalamey} is genuinely non-zero and small. Similar story extends to the generalized $SO(32)$ case.

What does this mean? One conclusion would be that the BBS instanton charges\footnote{A word of caution. By ``charges" we mean the value of the Chern-Simons term from \eqref{sadapa} and not the actual charges of the instantons as the latter are not globally conserved ones. This is evident from the Bianchi identity that we studied earlier.} get {\it renormalized} by the small correction from \eqref{mandalamey}. These corrections are clearly $\bar{g}_s$ dependent, but they do not contribute to the energy-momentum tensors because of their topological nature\footnote{In fact since the Chern-Simons term is purely topological $-$ whether or not we express it in a gauge invariant way \eqref{kangelss} $-$ the form \eqref{mandalamey} suggests that the some imprint of the Born-Infeld (BI) part is still present. The deviation away from it is consistent with how every term in the trans-series deviates from the scaling of the BI part by quantum effects.}. They do however contribute to the Bianchi identities as given in \eqref{crawlquif2} via the $\ast\mathbb{Y}_7$ factor in \eqref{valentinpen}. Since, and as we saw in sections \ref{sec4.5} and \ref{secanomaly}, the Bianchi identities and flux quantizations are satisfied to all orders, both perturbatively and non-perturbatively, the charges of five-branes and the five-brane instantons are not violated. In fact the localized fluxes from \eqref{viorosse} can {\it absorb} the instanton charges in a way studied in section \ref{kolamey}.

Coming back to the KKLT instanton case \cite{kklt}, which deals with five-brane instantons wrapped on $\Sigma_4\times \xoxo$, where $\Sigma_4$ is a four-cycle in ${\cal M}_4 \times {\cal M}_2$, {\it i.e.} 
$\Sigma_4 \subset{{\cal M}_4 \times {\cal M}_2}$, the problem appears early on when we compute the volume of the wrapped manifold. This scales as $\bar{g}_s^{-\theta_6}$, where $\theta_6 = -2\hat\sigma_e(t) - {\hat\eta_e(t)\over 2}$ if $\Sigma_4 = {\cal M}_4$ or $\theta_6 =-\hat\sigma_e(t) - {\hat\alpha_e(t) + \hat\beta_e(t)\over 2} - {\hat\eta_e(t)\over 2}$, if $\Sigma_4 = \Sigma_2 \times {\cal M}_2$, where $\Sigma_2 \in {\cal M}_4$ is a two-cycle. Comparing rows 1 and 9 in {\bf Table \ref{niksmit0071}}, we get the following scalings for the quantum term \eqref{botsuga}:
\bg\label{cyclebaaz6}
\theta_{nl}  = \begin{cases} ~{8\over 3} - \hat\zeta_e(t)- \hat\sigma_e(t) - {\hat\eta_e(t)\over 2} - \begin{cases} ~\hat\sigma_e(t)\\~~~\\
~{\hat\alpha_e(t) + \hat\beta_e(t)\over 2}
\end{cases}\\
~{2\over 3} - (\hat\alpha_e(t), \hat\beta_e(t)) \pm \hat\alpha_e(t) \mp \hat\beta_e(t)- \hat\sigma_e(t) - {\hat\eta_e(t)\over 2} - \begin{cases} ~\hat\sigma_e(t)\\~~~\\
~{\hat\alpha_e(t) + \hat\beta_e(t)\over 2}
\end{cases}\\ 
~{2\over 3} + {\hat\alpha_e(t)+ \hat\beta_e(t)\over 2} - \hat\zeta_e(t)  
- \hat\sigma_e(t) - {\hat\eta_e(t)\over 2} - \begin{cases} ~\hat\sigma_e(t)\\~~~\\
~{\hat\alpha_e(t) + \hat\beta_e(t)\over 2}
\end{cases}\\ 
~ {2\over 3} - (\hat\alpha_e(t), \hat\beta_e(t))- \hat\sigma_e(t) - {\hat\eta_e(t)\over 2} - \begin{cases} ~\hat\sigma_e(t)\\~~~\\
~{\hat\alpha_e(t) + \hat\beta_e(t)\over 2}
\end{cases}\\
~{2\over 3} - 2\hat\sigma_e(t) - {\hat\eta_e(t)\over 2} - \begin{cases} ~\hat\sigma_e(t)\\ ~~~\\
~{\hat\alpha_e(t) + \hat\beta_e(t)\over 2}
\end{cases}
\end{cases} \nd
where the ambiguity lies on which five-branes we consider and what wrapping manifolds. As before we should now compare \eqref{cyclebaaz6} to the terms appearing in \eqref{brittbaba007}. Here is where we face our first problem: some of the terms in \eqref{cyclebaaz6} appear from $l_{14} =1$ in \eqref{brittbaba007}, but the remaining terms do not appear to match with any set of terms from \eqref{brittbaba007}. The problematic terms are the ones that arise from $\theta_6$ above, as there appears no counterparts of them in \eqref{brittbaba007}. In fact the presence of ${\hat\eta_e(t)\over 2}$ is one of the main problem because this factor never shows up in the scalings of the curvature components in \eqref{brittbaba007} (see the first five lines therein). Even the derivative terms do not have this factor as we have taken the fields to be independent of the toroidal direction. The only way ${\hat\eta_e(t)\over 2}$ would appear is from the flux sector. Unfortunately the dominant scalings of all the flux components are ${4\over 3}$, and therefore they cannot accommodate the scalings in \eqref{cyclebaaz6}. We could consider the fermionic sector, as studied in section \ref{servant}, and use \eqref{botsuga2.0} instead. However since the fermionic scalings are sub-dominant compared to their metric and flux counterparts $-$ as seen from \eqref{akash}, \eqref{vendlinda} and \eqref{roseQ2} $-$ it is not clear whether they could accommodate the scalings required from \eqref{cyclebaaz6} simply because the dominant factors (which are always positive) in \eqref{brittbaba007} can only {\it increase} so that the fermions continue to contribute sub-dominantly\footnote{An interesting possibility is to allow for fermionic dominance as discussed briefly in footnote \ref{cascovbhalo}, implying that all bosonic degrees of freedom are secretly fermionic condensates. Such a scenario would allow the KKLT instantons to contribute from the flux sector to accommodate the ${\hat\eta_e(t)\over 2}$ scaling, but the issues pointed out in footnote \ref{cascovbhalo} remains. It will be interesting to analyze this scenario carefully.} to \eqref{brittbaba007}.
On the other hand if we take the non-local effects into account, as given in row 10 of {\bf Table \ref{niksmit0071}}, and compare it to row 1, we get:
\bg\label{cyclebaaz7}
\theta_{nl}  = \begin{cases} ~{8\over 3} - \hat\zeta_e(t)- n\hat\sigma_e(t) - {n\hat\eta_e(t)\over 2}- na_{\rm F} - \begin{cases} ~n\hat\sigma_e(t)\\~~~\\
~{n\hat\alpha_e(t) + n\hat\beta_e(t)\over 2}
\end{cases}\\
~{2\over 3} - (\hat\alpha_e(t), \hat\beta_e(t)) \pm \hat\alpha_e(t) \mp \hat\beta_e(t)- n\hat\sigma_e(t) - {n\hat\eta_e(t)\over 2}- na_{\rm F}- \begin{cases} ~n\hat\sigma_e(t)\\~~~\\
~{n\hat\alpha_e(t) + n\hat\beta_e(t)\over 2}
\end{cases}\\ 
~{2\over 3} + {\hat\alpha_e(t)+ \hat\beta_e(t)\over 2} - \hat\zeta_e(t)  
- n\hat\sigma_e(t) - {n\hat\eta_e(t)\over 2}- na_{\rm F}- \begin{cases} ~n\hat\sigma_e(t)\\~~~\\
~{n\hat\alpha_e(t) + n\hat\beta_e(t)\over 2}
\end{cases}\\ 
~ {2\over 3} - (\hat\alpha_e(t), \hat\beta_e(t))- n\hat\sigma_e(t) - {n\hat\eta_e(t)\over 2}- na_{\rm F}- \begin{cases} ~n\hat\sigma_e(t)\\~~~\\
~{n\hat\alpha_e(t) + n\hat\beta_e(t)\over 2}
\end{cases}\\
~{2\over 3} - (n+1)\hat\sigma_e(t) - {n\hat\eta_e(t)\over 2} - na_{\rm F} -\begin{cases} ~n\hat\sigma_e(t)\\ ~~~\\
~{n\hat\alpha_e(t) + n\hat\beta_e(t)\over 2}
\end{cases}
\end{cases} \nd
where $n = 1, 2$ depending on which fluctuation determinants we choose around the non-local KKLT instantons. For $a_{\rm F} > 0$ there appears to be no solutions whereas for $a_{\rm F} \le -{2\over 3}$ solutions could in principle exist for either values of $n$. Again, the presence of ${n\hat\eta_e(t)\over 2}$ implies that no curvature terms from \eqref{brittbaba007} or \eqref{botsuga} can contribute. Solutions might exist only from the flux sector depending on the appropriate value of $a_{\rm F}$. Despite this, there is something unnatural about the aforementioned cases: solutions seem to exist for cases where the non-localities increase with time and become much more prominent at late time. Additionally, one may easily check that the dominant charges of the KKLT instantons scale with respect to $\bar{g}_s$ as $\bar{g}_s^{\theta_{\rm KKLT}}$, where $\theta_{\rm KKLT}$ takes the following form:
\bg\label{wiilsonT}
\theta_{\rm KKLT} = \begin{cases}  \underbrace{2\hat\sigma_e(t) + {\hat\eta_e(t)\over 2}}_{\text{\rm Born-Infeld part}} - \underbrace{ {\hat\zeta_e(t)\over 2} - {\hat\alpha_e(t) + \hat\beta_e(t)\over 2} + ...}_{\text{\rm Small correction}}\\~~~~\\
\underbrace{\hat\sigma_e(t) + {\hat\eta_e(t)\over 2} + {\hat\alpha_e(t) + \hat\beta_e(t)\over 2}}_{\text{\rm Born-Infeld part}} - \underbrace{ {\hat\zeta_e(t)\over 2} - {\hat\sigma_e(t)} + ...}_{\text{\rm Small correction}}\end{cases}
\nd
where the two choices are from the two possible orientations of the five-brane instantons, and the dotted terms come from the $\bar{l}_{e{\rm AB}}^{\rm CD}$ factors in \eqref{brittbaba007}. Of course, as mentioned earlier, one should remember that the instantons carry no global charges, and therefore these charges are {\it absorbed} by the Bianchi identities that we studied in section \ref{sec4.5}. Note that, unlike the case with the BBS instantons in \eqref{mandalamey}, the small corrections cannot be argued to be arbitrarily small because there exists no duality frame where the aforementioned combinations, with appropriate $\bar{g}_s$ and ${\rm M}_p$ corrections (see section \ref{sec4.2.1}), vanish.

Before moving ahead with the remaining non-perturbative effects from 
{\bf Table \ref{niksmit0071}}, let us clarify a few subtleties with the five-brane instantons' contributions (especially with the BBS instantons' contributions). {\Su First}, we note from rows 7 and 8 of {\bf Table \ref{niksmit0071}}, that the quantum scalings are captured by $\theta_{nl}$ from the perturbative fluctuations and $(\theta'_{nl}, \theta''_{nl})$ from the non-perturbative, hence the genuine five-branes instantons, terms. If we restrict ourselves to only the sector with the perturbative fluctuations, then these quantum contributions are heavly suppressed by the instanton terms. On the other hand, if we restrict 
ourselves to the genuine non-perturbative sectors, then these quantum corrections would indeed contribute non-trivially on the RHS of the Schwinger-Dyson equations \eqref{mootcha}. Thus \eqref{cyclebaaz3} and \eqref{cyclebaaz4} (including \eqref{cyclebaaz6} and \eqref{cyclebaaz7}) should be thought of as coming from the non-perturbative sectors only.

{\Su Secondly}, we have been ignoring the exponential factors accompanying the quantum scalings from rows 7, 8, 9 and 10 in {\bf Table \ref{niksmit0071}}. It is now time to take this into considerations and see whether the inclusion would have any influence on our earlier results. The typical relation that we have in mind is:
\bg\label{daniniHA}
{\bar{g}_s^{\theta_{nl}(t) - {2\over 3} + \hat\sigma_e(t)}\over \bar{g}_s^{\theta_d(t)}}~{\rm exp}\left[-|c_1|{\bar{g_s}^{\theta_{nl}(t)}\over \bar{g}_s^{\theta_d}(t)}\right] = c_2\bar{g}_s^{a(t)}, \nd
where $c_i$ are constants, and $\theta_{nl}(t)$ is the scaling coming purely from the non-perturbative sector. (One may easily check that the exponential suppression will otherwise kill that term completely, confirming the point we made earlier.) To solve \eqref{daniniHA}, let us  make the following manipulations:
\bg\label{daninawilson}
\theta_{nl}(t) = \theta_d(t) + z(t) ~~ \implies ~~ \bar{g}_s^{z(t)} ~{\rm exp}\left[-\vert c_1\vert\bar{g}_s^{z(t)}\right] = c_2\bar{g}_s^{b(t)},
\nd where $z(t)$ is another parameter, not to be confused with any coordinates used earlier; and $b(t) = a(t) + {2\over 3} - \hat\sigma_e(t)$. For simplicity we will take $c_1 = 1, c_2 = c$, and using this, the solution to \eqref{daninawilson} may be determined by first writing $z(t) = b(t) + \delta(t)$ to convert \eqref{daninawilson} to the following equation:
\bg\label{oxmon}
\log~u(t) = \log~c + \bar{g}_s^b(t) u(t), \nd 
where $u(t) \equiv \bar{g}_s^{\delta(t)}$. Since $0 < \bar{g}_s < 1$ and 
$0 < b(t) < 1$, we can solve \eqref{oxmon} iteratively to express $u(t)$ as $u(t) = c + \bar{g}_s^{b(t)} c + ...$, which would imply the following solution:
\bg\label{oxmonjol}
z(t) = b(t) + {\log~c\over \log~\bar{g}_s} + {c\bar{g}_s^{b(t)}\over \log~\bar{g}_s} + ..., \nd
where we can easily see that the solution simplifies if we make $c = 1$. One may also solve \eqref{daninawilson} exactly by using the Lambert $\mathbb{W}(-x)$ function. The exact solution becomes:

{\scriptsize
\bg\label{shonaliag}
z(t) &= & b(t) + {\mathbb{W}(-\bar{g}_s^b(t))\over \vert\log~\bar{g}_s\vert}\\
&= & \begin{cases}~b(t) + {1\over \vert\log~\bar{g}_s\vert}\sum\limits_{k = 1}^\infty {(-\bar{g}_s^{b(t)})^k(-k)^{k - 1}\over k!} ~~~~~~~ {\rm for}~ e|\bar{g}_s^{b(t)}| <1, ~|-1+\bar{g}^{1, b(t)}_s| < 1\\~~~~\\
~{\log(-1+\bar{g}_s^{b(t)}) + {\sum\limits_{k = 1}^\infty (-1)^{k+1} k^{-1}(-1+ \bar{g}_s^{b(t)})^{-k} + \sum\limits_{p = 1}^\infty (-\bar{g}_s^{b(t)})^p(-p)^p(pp!)^{-1} } \over \log(-1+\bar{g}_s) - {\sum\limits_{l = 1}^\infty (-1)^l l^{-1}(-1+\bar{g}_s)^{-l}}} ~~~{\rm for}~e|\bar{g}_s^{b(t)}| <1,~ |-1+\bar{g}^{1, b(t)}_s|>1
\end{cases}\nonumber
\nd}
which consistently reproduces \eqref{oxmonjol} for $c = 1$ from the first choice whose validity regime is $\vert\bar{g}_s^{b(t)}\vert < {1\over e}$ and $\vert-1+\bar{g}^{1, b(t)}_s\vert < 1$. The second choice in \eqref{shonaliag} is not realized here, and we have shown the plot of $z(t)$ in {\bf figure \ref{llotus}} for the range of allowed values for $\bar{g}_s$ and $b(t)$. One may now combine \eqref{daninawilson} and \eqref{shonaliag} to express the quantum scaling as:

\begin{figure}%
        \centering
        {\includegraphics[scale=0.40]{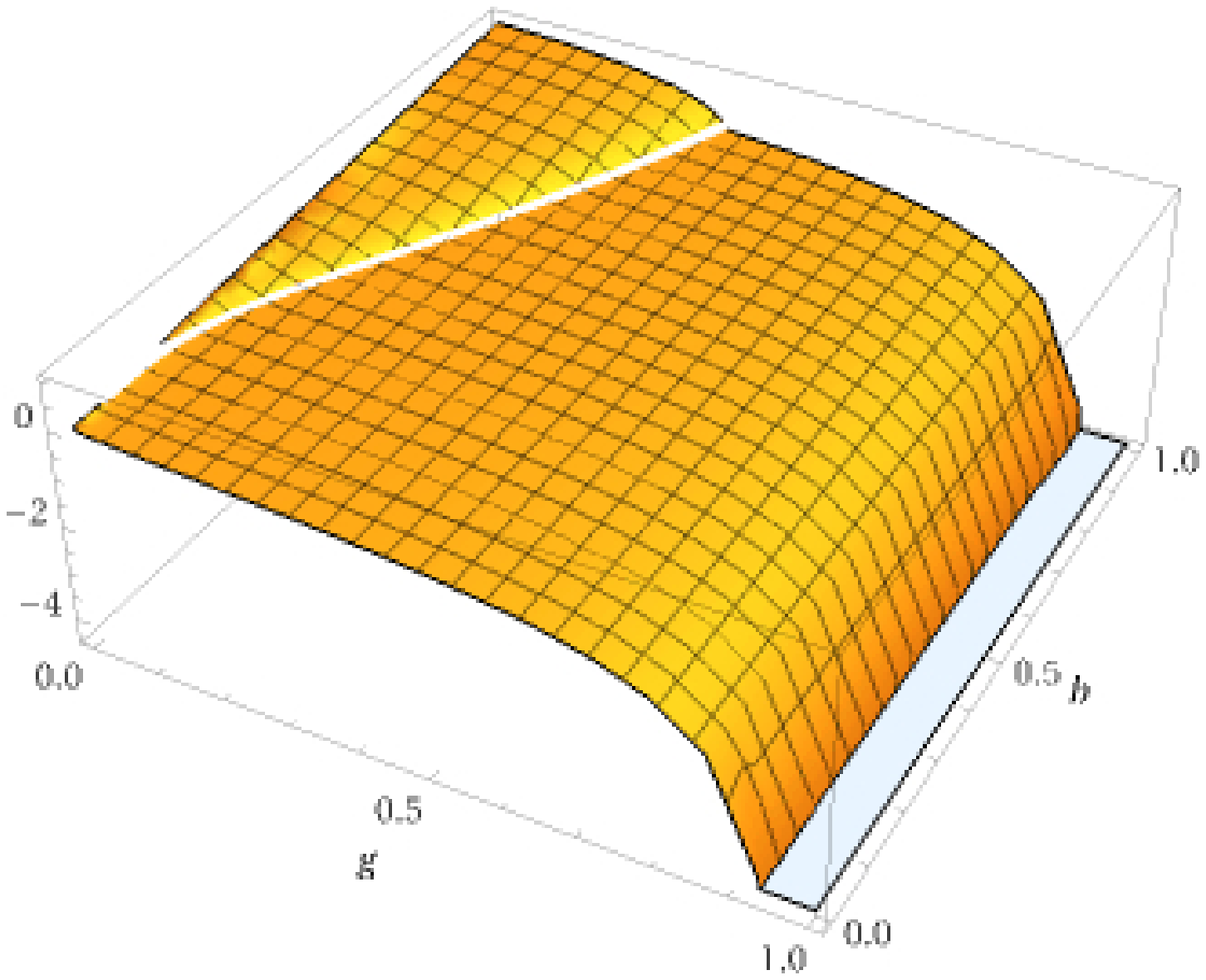}}%
        \qquad \qquad
        {\includegraphics[scale=0.40]{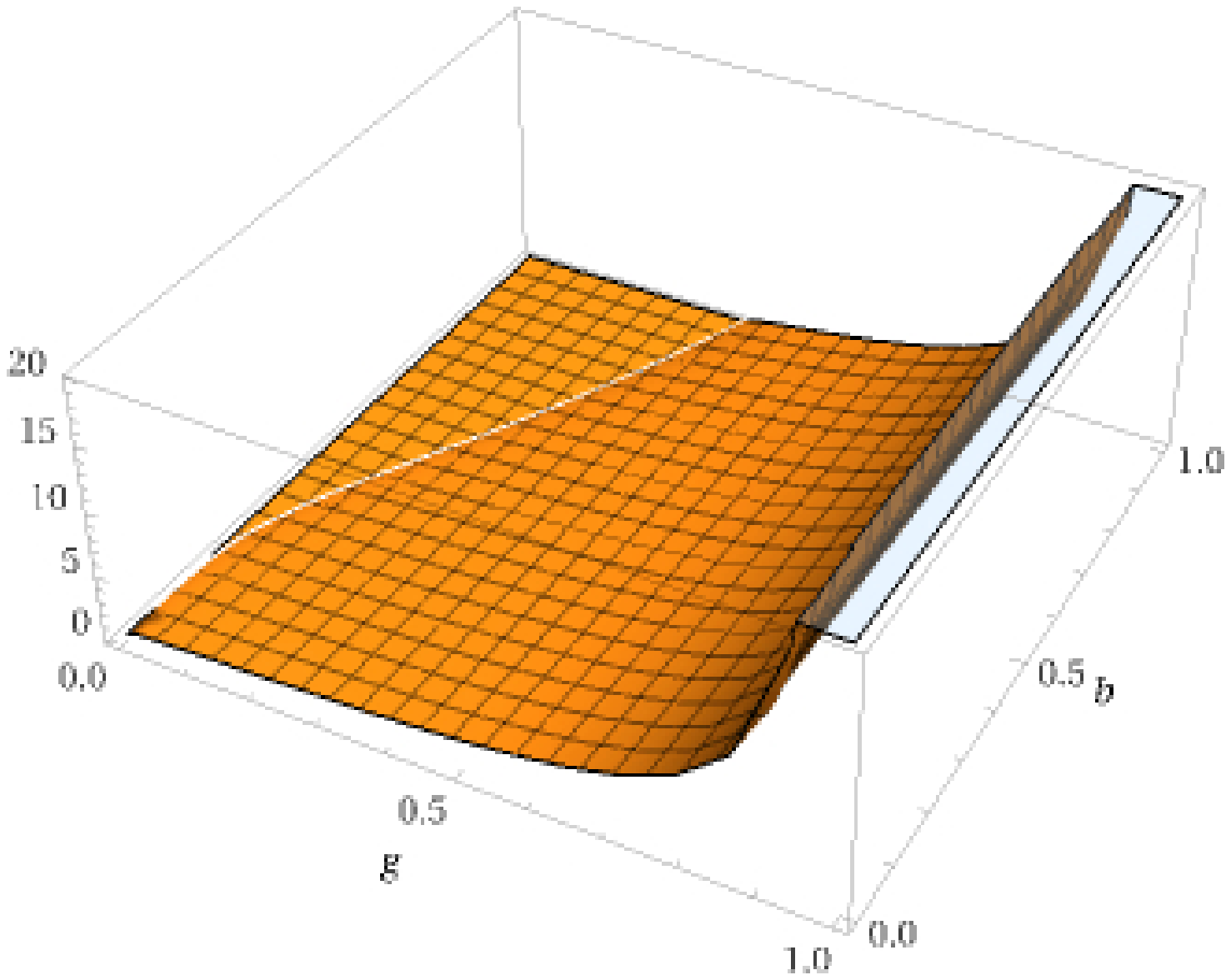}}%
        \caption{Plot of the real (top) and the imaginary (bottom) parts of the function $z(t)$ in \eqref{shonaliag} for $g_s \equiv g$ and $b(t) \equiv b$ ranging between $0 < g < 1$ and $0 < b < 1$. Expectedly, the real part, shown in the top, changes very little for a given choice of $b(t)$ within $0 < g < 1$. The imaginary part, shown in the bottom, is not used here but depicted for completeness.}%
        \label{llotus}%
    \end{figure}

{\footnotesize
\bg\label{brettbhalo}
\bar{g}_s^{\theta_{nl}(t)} =  \bar{g}_s^{\theta_d(t) + b(t)}~{\rm exp}\left[\sum\limits_{k = 1}^\infty {(\bar{g}_s^{b(t)})^k k^{k - 1}\over k!}\right]
 =  \bar{g}_s^{\theta_d(t) + b(t)} \sum_{{\rm N} = 0}^\infty {1\over {\rm N}!} \left[\sum\limits_{k = 1}^\infty {(\bar{g}_s^{b(t)})^k k^{k - 1}\over k!}\right]^{\rm N}, \nd}
which is now expanded {\it perturbatively} because $0 < \bar{g}_s < 1$ and $0 < b(t) < 1$. The series in \eqref{brettbhalo} suggests that the dominant contribution to $\theta_{nl}(t)$ remains $\theta_d(t) + b(t) = \theta_d + a(t) + {2\over 3} - \hat\sigma_e(t)$ which is exactly what appeared in \eqref{cyclebaaz3} for the BBS instantons and \eqref{cyclebaaz6} for the KKLT instantons. For the former, {\it i.e.} for the BBS instantons, the series in \eqref{brettbhalo} suggests higher order curvature contributions which are highly suppressed by the inverse powers of ${\rm M}_p$. This reinforces what we said earlier: {\Su the BBS instantons dominantly contribute quartic order curvature terms to drive the acceleration of the universe}. 

Let us now clarify the remaining contributions from other possible instantons appearing in rows 11 and 12 of {\bf Table \ref{niksmit0071}}. These instantons could be the M2-brane instantons, or the bound states of two and five-brane instantons, or even more exotic varieties that appear from the {\it Stokes lines}\footnote{For more details on the {\it Stokes lines} and how they contribute to the non-perturbative effects, the readers may refer to \cite{ecalle, unsal, dorigoni}. Some of these exotic non-perturbative contributions could come from non-BPS brane instantons, while others may not have such simple interpretations. This is already an interesting direction of research but will unfortunately deviate us away from our present analysis. We will elaborate on this in future works.} in the trans-series representation of the action \eqref{kimkarol}. The two-brane instantons, wrapping three-cycle $\Sigma_3 \in {\cal M}_4 \times {\cal M}_2$, will typically break the four-dimensional isometries (assuming they survive in the  ${\rm E}_8 \times {\rm E}_8$ heterotic side) and therefore we will not consider them here. Other types of two-brane instantons, wrapping three-cycle 
$\xoxo \times {\bf S}^1$ where ${\bf S}^1 \in {\cal M}_4$, possibly become string instantons in the type IIB side but they again appear to violate the four-dimensional isometries. Thus we will not consider them here too, although one could verify the relations between scalings of the Born-Infeld parts and the two-brane charges for the two set of two-brane instantons. The Born-Infeld parts scale as $-1^\pm$ and $1^\pm$ respectively\footnote{Or more precisely as $-1 + {3\hat\sigma_e(t)\over 2}$ and $1 + {\hat\sigma_e(t)\over 2} + {\hat\eta_e(t)\over 2}$ if we only consider the sub-manifolds within ${\cal M}_4$.}, and the charges $-$ whose dominant scalings come from the flux components ${\bf G}_{{\rm MNP}i}$ and ${\bf G}_{{\rm M}abi}$ respectively $-$ scale in exactly the same way up to small corrections similar to what we saw in \eqref{mandalamey} and \eqref{wiilsonT} earlier. Finally, we could also have space-time filling two-branes instantons (in Euclidean space) wrapping ${\bf R}^{2, 1}$. For such a configuration we expect the quantum scaling to be:
\bg\label{cyclebaaz8}
\theta_{nl}  = \begin{cases} ~{8\over 3} - \hat\zeta_e(t) +{4} - {3\hat\zeta_e(t)\over 2} \\ ~~~\\
~{2\over 3} - (\hat\alpha_e(t), \hat\beta_e(t)) \pm \hat\alpha_e(t) \mp \hat\beta_e(t)+{4} - {3\hat\zeta_e(t)\over 2}\\ ~~~\\
~{2\over 3} + {\hat\alpha_e(t)+ \hat\beta_e(t)\over 2} 
- \hat\zeta_e(t)+{4} - {3\hat\zeta_e(t)\over 2}  \\ ~~~\\
~{2\over 3} - (\hat\alpha_e(t), \hat\beta_e(t))+{4} - {3\hat\zeta_e(t)\over 2}\\~~~\\
~{2\over 3} - \hat\sigma_e(t)+ {4} - {3\hat\zeta_e(t)\over 2}\end{cases} \nd
which may now be compared to \eqref{brittbaba007}. However before we do that let us interpret such a configuration. In M-theory such a configuration should lead to tunneling events, and the usual jump of the four-form fluxes. However now comparing the charge term, {\it i.e.} the Chern-Simons piece, we see that it differs from the Born-Infeld term by substantial amount compared to what we saw in \eqref{mandalamey} and \eqref{wiilsonT} where the differences were very small. The Born-Infeld term scales as $-4 + {3\hat\zeta_e(t)\over 2}$, whereas the charge term scales as $-3^\pm$. The difference is now $-1^\pm$ and not a small correction. Moreover, comparing $\theta_{nl}$ to \eqref{brittbaba007} we find that:
\bg\label{palkimey}
\Big(l_{14} = 1, ~~~\sum\limits_{i = 1}^{13} l_i = 1, ~~~ n_1//n_0= 1\Big), ~~~\Big(l_{14} = 1, ~~~n_1//n_0 = 3\Big), \nd
where we have chosen ${8\over 3} - \hat\zeta_e(t)$ from the first line in \eqref{brittbaba007}; and $n_1//n_0 = l$ implies $n_1 = l$ or $n_0 = l$ for $l \in +\mathbb{Z}$. Unfortunately neither of the two choices in \eqref{palkimey} can lead to Lorentz invariant interactions on the two-branes, thus signaling that contributions to the acceleration of the universe at least cannot come from the choices in \eqref{palkimey}.

What about more exotic varieties which do not have simple brane-instanton representations? One possibility would be an instanton configuration covering the full eight manifold ${\cal M}_4 \times {\cal M}_2 \times \xoxo$. For such a case $\theta_{nl}$ takes the form:
\bg\label{cyclebaaz80}
\theta_{nl}  = \begin{cases} ~{8\over 3} - \hat\zeta_e(t) +{2\over 3} - 2\hat\sigma_e(t) - {\hat\alpha_e(t) + \hat\beta_e(t)\over 2} - {\hat\eta_e(t)\over 2}\\ ~~~\\
~{2\over 3} - (\hat\alpha_e(t), \hat\beta_e(t)) \pm \hat\alpha_e(t) \mp \hat\beta_e(t)+{2\over 3} - 2\hat\sigma_e(t) - {\hat\alpha_e(t) + \hat\beta_e(t)\over 2} - {\hat\eta_e(t)\over 2}\\ ~~~\\
~{2\over 3} + {\hat\alpha_e(t)+ \hat\beta_e(t)\over 2} 
- \hat\zeta_e(t)+{2\over 3} - 2\hat\sigma_e(t) - {\hat\alpha_e(t) + \hat\beta_e(t)\over 2} - {\hat\eta_e(t)\over 2}\\ ~~~\\
~{2\over 3} - (\hat\alpha_e(t), \hat\beta_e(t))+{2\over 3} - 2\hat\sigma_e(t) - {\hat\alpha_e(t) + \hat\beta_e(t)\over 2} - {\hat\eta_e(t)\over 2}\\~~~\\
~{2\over 3} - \hat\sigma_e(t)+ {2\over 3} - 2\hat\sigma_e(t) - {\hat\alpha_e(t) + \hat\beta_e(t)\over 2} - {\hat\eta_e(t)\over 2}\end{cases} \nd
which when compared to \eqref{brittbaba007} we see the following issues. The presence of $\hat\eta_e(t)$ means that it cannot come from any of the curvature terms in \eqref{botsuga}. Even if we approximate $\hat\eta_e(t) \approx -\hat\zeta_e(t)$, this doesn't help. If we choose $l_{14} =1$, there appears no simple solution to the system. If we look only into the flux sector in \eqref{brittbaba007}, and because of the dominant scalings of ${4\over 3}^\pm$ for all the terms therein, it appears that there are no simple solutions there too. 

The above object could probably be related to a non-BPS seven-brane instanton from the supersymmetric Minkowski vacuum point of view, or to a more exotic variety. Unfortunately, as we saw above, such configurations do not seem to drive the acceleration of the universe. What about a space-filling eight-brane instanton configuration\footnote{From supersymmetric Minkowski vacuum point of view, this is a non-BPS eight-brane instanton configuration that breaks supersymmetry spontaneously much like how an anti-brane does \cite{evanfermion}. From the Glauber-Sudarshan state point of view, this is an emergent configuration much like any other configurations that we studied here. Since the supersymmetry is already broken spontaneously by the Glauber-Sudarshan state, the eight-brane instanton configuration probably decays to a more stable configuration within the allowed temporal domain of $-{1\over \sqrt{\Lambda}} < t < 0$ ($\Lambda$ being the bare cosmological constant from \eqref{marapaug}) if the decay happens faster than the aforementioned interval. The story is more involved now because the decay, which is typically studied using tachyon condensation on such objects \cite{sentachyon}, is more complicated because we have no knowledge of how the open string tachyons behave in an accelerated background. Even the knowledge of the quantization of a string on such a background is a challenging problem. This is an interesting topic to study in itself, and our brief analysis has raised many questions whose answers demand more careful analysis. We will therefore leave it for a future investigation.}
wrapping 
${\rm Euc}({\bf R}^3) \times {\cal M}_4 \times {\cal M}_2$?  Introducing such a configuration at the level of the action \eqref{kimkarol} as a trans-series form is straightforward: we simply need to extend the internal integral from eight dimensions to nine-dimensions (see also \cite{maximE}). For such a configuration, $\theta_{nl}$ takes the following form:
\bg\label{cyclebaaz88}
\theta_{nl}  = \begin{cases} ~{8\over 3} - \hat\zeta_e(t) +{6} - 2\hat\sigma_e(t) - {\hat\alpha_e(t) + \hat\beta_e(t)\over 2}- {3\hat\zeta_e(t)\over 2}\\ ~~~\\
~{2\over 3} - (\hat\alpha_e(t), \hat\beta_e(t)) \pm \hat\alpha_e(t) \mp \hat\beta_e(t)+{6} - 2\hat\sigma_e(t) - {\hat\alpha_e(t) + \hat\beta_e(t)\over 2}- {3\hat\zeta_e(t)\over 2}\\ ~~~\\
~{2\over 3} + {\hat\alpha_e(t)+ \hat\beta_e(t)\over 2} 
- \hat\zeta_e(t)+{6} - 2\hat\sigma_e(t) - {\hat\alpha_e(t) + \hat\beta_e(t)\over 2}- {3\hat\zeta_e(t)\over 2}\\ ~~~\\
~{2\over 3} - (\hat\alpha_e(t), \hat\beta_e(t))+{6} - 2\hat\sigma_e(t) - {\hat\alpha_e(t) + \hat\beta_e(t)\over 2}- {3\hat\zeta_e(t)\over 2}\\~~~\\
~{2\over 3} - \hat\sigma_e(t)+ {6} - 2\hat\sigma_e(t) - {\hat\alpha_e(t) + \hat\beta_e(t)\over 2} - {3\hat\zeta_e(t)\over 2} \end{cases} \nd
which may now be compared to \eqref{brittbaba007}. It is clear that $l_{14} = 1$ continues to provide part of the scalings in \eqref{cyclebaaz88}, so the question is whether the remaining pieces can be mapped to some set of non-negative integer values of $l_r$. In the sector of the curvature and derivatives, we can find the following two sets of solutions:
\bg\label{cascovaaram}
\Big(n_1=1, ~\sum_{p = 24}^{28}l_p = 2, ~\sum_{i = 1}^{13}l_i = 3\Big), ~~~
\Big(n_1 = 3, ~\sum_{q=29}^{33} l_q=2, ~\sum_{i=1}^{13}l_i = 2\Big), \nd
where we choose ${2\over 3} - \hat\sigma_e(t)$ from coefficient of $l_i$ in the first line of \eqref{brittbaba007}, and the two allowed choices from both $l_p \in \mathbb{Z}^+$ and $l_q \in \mathbb{Z}^+$ for $24\le p\le 28$ and $29\le q\le 33$ in respectively the third and the fourth lines of \eqref{brittbaba007}. Looking at \eqref{cascovaaram}
we see that $n_1$, which is the number of derivatives acting along ${\bf R}^2$, is odd. This is unfortunate because it implies that we cannot have  Lorentz invariant interaction terms from the two choices in \eqref{cascovaaram}. It is possible that solutions exist if we include the non-locality factor $a_{\rm F}$; or include the flux and the fermionic sectors. However since the scalings in \eqref{cyclebaaz88} starts from ${20\over 3}^\pm$, the contributions from the world-volume of these instantons can only be sub-dominant compared to the contributions from the BBS instantons discussed earlier. Thus the BBS instantons continue to provide the dominant contributions to drive the acceleration of our universe.

\subsection{EOMs for the emergent on-shell metric components ${\bf g}_{\alpha\beta}, {\bf g}_{ab}$ and ${\bf g}_{\mu\nu}$ \label{eomgbeta}}

The EOMs for the emergent metric configuration ${\bf g}_{\alpha\beta}$ now follow similar story as for the emergent metric configuration ${\bf g}_{mn}$ discussed in section \ref{eomgmn} but with minor changes as detailed in {\bf Table \ref{niksmit0072}}. 

\begin{table}[h]  
 \begin{center} 
\resizebox{\columnwidth}{!}{%
 \renewcommand{\arraystretch}{5.3}
}
\renewcommand{\arraystretch}{1}
\end{center}
 \caption[]{\Su The $\bar{g}_s$ scalings of the various terms in \eqref{mootcha} contributing to the EOM for the metric components ${\bf g}_{\alpha\beta}$. The other functions and parameters like  $\mathbb{U}_d({\rm Y};t), \mathbb{Q}_d({\rm Y}; t), \theta_d$ and $a_{\rm F}$ have been defined earlier in {\bf Table \ref{niksmit0071}}. As before, to avoid clutter, we have defined all the quantities at a specific point in ${\bf R}^2$, so that ${\bf x}\in {\bf R}^2$ does not appear explicitly here.}
\label{niksmit0072}
 \end{table} 

Looking at the first row in {\bf Table \ref{niksmit0072}} we see that the Einstein tensor ${\bf G}_{\alpha\beta}$ has a dominant scaling of $0^\pm$. A more elaborate scalings of the various ingredients in ${\bf G}_{\alpha\beta}$ may be extracted from {\bf Table \ref{niksmit007}} and is given by:
\bg\label{lotustommy}
{\rm scale}({\bf G}_{\alpha\beta}) &=& {\rm dom}\Big(0, (\hat\alpha_e(t), \hat\beta_e(t)) - \hat\sigma_e(t), 2 + (\hat\alpha_e(t), \hat\beta_e(t)) -\hat\zeta_e(t) \nonumber\\
&& ~~ (\hat\alpha_e(t), \hat\beta_e(t)) + {\hat\alpha_e(t) +\hat\beta_e(t)\over 2} - \hat\zeta_e(t), \pm\hat\alpha_e(t) \mp \hat\beta_e(t)\Big), \nd
which may be compared to \eqref{sushiador}. From Row 2 in {\bf Table \ref{niksmit0072}} we see that the energy momentum tensors from the flux components show a dominant scaling of $\bar{g}_s^{2^\pm}$ and therefore they cannot balance the Einstein tensor ${\bf G}_{\alpha\beta}$. This is of course consistent with the no-go conditions from \cite{GMN}.

Let us now come to Row 3 which includes contributions from perturbative quantum corrections. The fourth column provides the $\bar{g}_s$ scaling of the perturbative contributions from the first term in \eqref{malishell2}. If we ignore the exponential piece for the time being, we see that they typically scale as $\xxy^{\theta_{nl} - {2\over 3} + (\hat\alpha_e(t), \hat\beta_e(t))}$ with $\theta_{nl}$ given by \eqref{brittbaba007}. Thus comparing rows 1 and 3 in {\bf Table \ref{niksmit0072}}, we have:

{\scriptsize
\bg\label{bicycle1}
\theta_{nl} - {2\over 3} + (\hat\alpha_e(t), \hat\beta_e(t)) = \begin{cases}~2 +(\hat\alpha_e(t), \hat\beta_e(t)) - \hat\zeta_e(t)\\ ~~~\\
\pm \hat\alpha_e(t) \mp \hat\beta_e(t)\\ ~~~\\
~(\hat\alpha_e(t), \hat\beta_e(t))+ {\hat\alpha_e(t)+ \hat\beta_e(t)\over 2} - \hat\zeta_e(t)\\ ~~~\\
~(\hat\alpha_e(t), \hat\beta_e(t)) - \hat\sigma_e(t)\\~~~\\
~0  \end{cases} \to 
\theta_{nl}  = \begin{cases} ~{8\over 3} - \hat\zeta_e(t)\\ ~~~\\
~{2\over 3} - (\hat\alpha_e(t), \hat\beta_e(t)) \pm \hat\alpha_e(t) \mp \hat\beta_e(t)\\ ~~~\\
~{2\over 3} + {\hat\alpha_e(t)+ \hat\beta_e(t)\over 2} - \hat\zeta_e(t)\\ 
~~~\\
~ {2\over 3} - (\hat\alpha_e(t), \hat\beta_e(t))\\~~~\\
~{2\over 3} - \hat\sigma_e(t)  \end{cases} 
\nd}
which is precisely what we got for $\theta_{nl}$ in \eqref{cyclebaaz}. Thus even though the scalings of the Einstein tensor ${\bf G}_{\alpha\beta}$ are different from the ones for ${\bf G}_{mn}$, the quantum scalings of the perturbative corrections remain exactly identical: we have $l_{14} =1$ in \eqref{brittbaba007}. This implies that the perturbative correction from Row 3 cannot solve the EOM.

In fact once $\theta_{nl}$ takes the form as in \eqref{bicycle1}, which is exactly similar to \eqref{cyclebaaz}, rest of the conclusions remain similar to the case with ${\bf g}_{mn}$ metric configuration. In particular this means that all the perturbative terms, given in Rows 3, 4, 5 and 6 of {\bf Table \ref{niksmit0072}}, are red herrings in the problem and they cannot be used to drive the acceleration of our universe. For the BBS instanton case from row 7, we have:

{\footnotesize
\bg\label{bicycle2}
\theta_{nl} - {2\over 3} + (\hat\alpha_e(t), \hat\beta_e(t)) - 2 + 2\hat\sigma_e(t) + {\hat\alpha_e(t) + \hat\beta_e(t)\over 2} = \begin{cases}~2 +(\hat\alpha_e(t), \hat\beta_e(t)) - \hat\zeta_e(t)\\ ~~~\\
\pm \hat\alpha_e(t) \mp \hat\beta_e(t)\\ ~~~\\
~(\hat\alpha_e(t), \hat\beta_e(t))+ {\hat\alpha_e(t)+ \hat\beta_e(t)\over 2} - \hat\zeta_e(t)\\ ~~~\\
~(\hat\alpha_e(t), \hat\beta_e(t)) - \hat\sigma_e(t)\\~~~\\
~0  \end{cases} \nd}
leading to \eqref{cyclebaaz3}, implying that the contributions come as before from \eqref{isaferr} and \eqref{isaferr2}, {\it i.e.} from the quartic curvature corrections on the world-volume of the BBS instantons. As we saw therein, this conclusion survives even if we incorporate the exponential corrections. The rest of the analysis, that includes the effects of other instantons and non-perturbative objects, is identical to what we had with the metric configuration ${\bf g}_{mn}$.

\begin{table}[h]  
 \begin{center} 
\resizebox{\columnwidth}{!}{%
 \renewcommand{\arraystretch}{5.3}
}
\renewcommand{\arraystretch}{1}
\end{center}
 \caption[]{\Su The $\bar{g}_s$ scalings of the various terms in \eqref{mootcha} contributing to the EOM for the metric components ${\bf g}_{ab}$. As in {\bf Table \ref{niksmit0072}}, the other functions and parameters like  $\mathbb{U}_d({\rm Y};t), \mathbb{Q}_d({\rm Y}; t), \theta_d$ and $a_{\rm F}$ have been defined earlier in {\bf Table \ref{niksmit0071}}. Similarly, to avoid clutter, we have defined all the quantities at a specific point in ${\bf R}^2$, so that ${\bf x}\in {\bf R}^2$ does not appear explicitly here.}
\label{niksmit0073}
 \end{table} 

For the emergent metric configuration ${\bf g}_{ab}$, the scalings of the various terms contributing to the EOM is given in {\bf Table \ref{niksmit0073}}. From Row 1 we see that the dominant scaling of the emergent Einstein tensor ${\bf G}_{ab}$ scales as $\bar{g}_s^{2^\pm}$. A more accurate analysis from {\bf Table \ref{niksmit007}} gives us the following values:

{\footnotesize
\bg\label{pakimey}
{\rm scale}({\bf G}_{ab}) &= & {\rm dom}\Big(2 + (0, \hat\eta_e(t)) - (\hat\alpha_e(t), \hat\beta_e(t)), 2 + (0, \hat\eta_e(t)) - \hat\sigma_e(t), 4 + (0, \hat\eta_e(t)) - \hat\zeta_e(t)\\ 
&& 2 + (0, \hat\eta_e(t)) + {\hat\alpha_e(t) + \hat\beta_e(t)\over 2} -\hat\zeta_e(t), 2 + (0, \hat\eta_e(t)) - (\hat\alpha_e(t), \hat\beta_e(t)) \pm\hat\alpha_e(t) \mp \hat\beta_e(t)\Big), \nonumber \nd}
which ranges from $2^\pm$ to $4^\pm$. Comparing this to Row 2 we see that the energy-momentum tensor from the fluxes, which has a dominant scaling of $4^\pm$ cannot balance the EOM \eqref{mootcha}. This is of course another confirmation of the no-go theorems of \cite{GMN}. The perturbative contribution from Rows 3 to 6 are captured by the quantum scaling $\theta_{nl}$ from \eqref{brittbaba007}, which takes the form:

{\scriptsize
\bg\label{bicyclelll}
\theta_{nl} + {4\over 3} + (0, \hat\eta_e(t)) = \begin{cases}~4 + (0, \hat\eta_e(t)) - \hat\zeta_e(t)\\ ~~~\\
 ~2 + (0, \hat\eta_e(t)) - (\hat\alpha_e(t), \hat\beta_e(t)) \pm\hat\alpha_e(t) \mp \hat\beta_e(t)\\ ~~~\\
~2 + (0, \hat\eta_e(t)) + {\hat\alpha_e(t) + \hat\beta_e(t)\over 2} -\hat\zeta_e(t)\\ ~~~\\
~2 + (0, \hat\eta_e(t)) - (\hat\alpha_e(t), \hat\beta_e(t))\\~~~\\
~2 + (0, \hat\eta_e(t)) - \hat\sigma_e(t)\end{cases} \to 
\theta_{nl}  = \begin{cases} ~{8\over 3} - \hat\zeta_e(t)\\ ~~~\\
~{2\over 3} - (\hat\alpha_e(t), \hat\beta_e(t)) \pm \hat\alpha_e(t) \mp \hat\beta_e(t)\\ ~~~\\
~{2\over 3} + {\hat\alpha_e(t)+ \hat\beta_e(t)\over 2} - \hat\zeta_e(t)\\ 
~~~\\
~ {2\over 3} - (\hat\alpha_e(t), \hat\beta_e(t))\\~~~\\
~{2\over 3} - \hat\sigma_e(t)  \end{cases} 
\nd}
which matches exactly with what we had in \eqref{cyclebaaz} and \eqref{bicycle1}, and is therefore related to $l_{14} = 1$ in \eqref{brittbaba007}. This implies that the perturbative terms are again red herrings in the problem. The contribution from the BBS instantons now gives us the following scaling equation:

{\footnotesize
\bg\label{bicycleMMM}
\theta_{nl} + {4\over 3} + (0, \hat\eta_e(t)) -2 + 2\hat\sigma_e(t) + {\hat\alpha_e(t) + \hat\beta_e(t)\over 2} = \begin{cases}~4 + (0, \hat\eta_e(t)) - \hat\zeta_e(t)\\ ~~~\\
 ~2 + (0, \hat\eta_e(t)) - (\hat\alpha_e(t), \hat\beta_e(t)) \pm\hat\alpha_e(t) \mp \hat\beta_e(t)\\ ~~~\\
~2 + (0, \hat\eta_e(t)) + {\hat\alpha_e(t) + \hat\beta_e(t)\over 2} -\hat\zeta_e(t)\\ ~~~\\
~2 + (0, \hat\eta_e(t)) - (\hat\alpha_e(t), \hat\beta_e(t))\\~~~\\
~2 + (0, \hat\eta_e(t)) - \hat\sigma_e(t)\end{cases} \nd}
leading us again to the same values for $\theta_{nl}$ that we had in \eqref{cyclebaaz3}, consequently allowing us to the choices \eqref{isaferr} and \eqref{isaferr2}. This is good because it suggests that it is the dominant quartic curvature terms on the world-volume of the BBS instantons are again responsible for driving the acceleration of the universe. From here, and for the remaining non-perturbative contributions, the story follows the path laid out in section \ref{eomgmn}.

\begin{table}[h]  
 \begin{center} 
\resizebox{\columnwidth}{!}{%
 \renewcommand{\arraystretch}{5.1}
}
\renewcommand{\arraystretch}{1}
\end{center}
 \caption[]{\Su The $\bar{g}_s$ scalings of the various terms in \eqref{mootcha} contributing to the EOM for the metric components ${\bf g}_{\mu\nu}$. As in {\bf Table \ref{niksmit0072}}, the other functions and parameters like  $\mathbb{U}_d({\rm Y};t), \mathbb{Q}_d({\rm Y}; t), \theta_d$ and $a_{\rm F}$ have been defined earlier in {\bf Table \ref{niksmit0071}}. Similarly, to avoid clutter, we have defined all the quantities at a specific point in ${\bf R}^2$, so that ${\bf x}\in {\bf R}^2$ does not appear explicitly here.}
\label{niksmit0100}
 \end{table} 

The story for the emergent metric configuration ${\bf g}_{\mu\nu} \equiv \langle{\bf g}_{\mu\nu}\rangle_\sigma$ is very similar to what we studied for the other components with some minor but crucial changes. These have been shown in {\bf Table \ref{niksmit0100}}, and in the following we will elaborate on some of the details. 

Looking at the first row in {\bf Table \ref{niksmit0100}}, we see that the emergent Einstein tensor scales as $\xxy^{-2^\pm}$. This is the dominant scaling, with $\pm$ denoting the small corrections. We can write the exact form of the Einstein tensor using {\bf Table \ref{niksmit007}}. The result is:

{\footnotesize
\bg\label{chibonita}
{\rm scale}({\bf G}_{\mu\nu}) &=& {\rm dom}\Big(-2-\hat\sigma_e(t) + \hat\zeta_e(t), -2 - (\hat\alpha_e(t), \hat\beta_e(t)) + \hat\zeta_e(t),\nonumber\\ 
 && -2 + {\hat\alpha_e(t) + \hat\beta_e(t)\over 2},
-2 - (\hat\alpha_e(t), \hat\beta_e(t)) \pm \hat\alpha_e(t) \mp \hat\beta_e(t) + \hat\zeta_e(t), 0\Big), \nd}
which confirms our earlier conclusion that the dominant scale is $-2^\pm$. Looking at the second row in {\bf Table \ref{niksmit0100}}, we see that the fluxes scale as $\xxy^{0^\pm}$ so clearly cannot balance the Einstein tensor. This again fits well with the no-go theorems \cite{GMN}. The perturbative contribution from Rows 3 to 6 are captured by the quantum scaling $\theta_{nl}$ from \eqref{brittbaba007}, which takes the form:

{\scriptsize
\bg\label{bicilite88}
\theta_{nl} - {8\over 3} + \hat\zeta_e(t) = \begin{cases}~~~0\\ ~~~\\
 ~-2 + \hat\zeta_e(t) - (\hat\alpha_e(t), \hat\beta_e(t)) \pm\hat\alpha_e(t) \mp \hat\beta_e(t)\\ ~~~\\
~-2 + {\hat\alpha_e(t) + \hat\beta_e(t)\over 2}\\ ~~~\\
~-2 + \hat\zeta_e(t) - (\hat\alpha_e(t), \hat\beta_e(t))\\~~~\\
~-2 + \hat\zeta_e(t) - \hat\sigma_e(t)\end{cases} \to 
\theta_{nl}  = \begin{cases} ~{8\over 3} - \hat\zeta_e(t)\\ ~~~\\
~{2\over 3} - (\hat\alpha_e(t), \hat\beta_e(t)) \pm \hat\alpha_e(t) \mp \hat\beta_e(t)\\ ~~~\\
~{2\over 3} + {\hat\alpha_e(t)+ \hat\beta_e(t)\over 2} - \hat\zeta_e(t)\\ 
~~~\\
~ {2\over 3} - (\hat\alpha_e(t), \hat\beta_e(t))\\~~~\\
~{2\over 3} - \hat\sigma_e(t)  \end{cases} 
\nd}
leading us unsurprisingly to the {\it same} $\theta_{nl}$ that we got in \eqref{cyclebaaz}, \eqref{bicycle1} and \eqref{bicyclelll}. Therefore, in a similar vein, comparing \eqref{bicilite88} with \eqref{brittbaba007} gives us $l_{14} = 1$ which is a term linear in the curvature. As such this is identified with the kinetic term, and therefore the perturbative series do not provide non-trivial terms to balance the Einstein tensor. This is yet again a confirmation of the no-go theorems \cite{GMN} that predicted that both the fluxes and the perturbative terms cannot drive the acceleration of our universe. The first non-trivial contribution comes from the BBS instantons, and for them we have the following equation:

{\footnotesize
\bg\label{holudlili}
\theta_{nl} - {8\over 3} + \hat\zeta_e(t)-2 + 2\hat\sigma_e(t) + {\hat\alpha_e(t) + \hat\beta_e(t)\over 2} = \begin{cases}~~~0\\ ~~~\\
 ~-2 + \hat\zeta_e(t) - (\hat\alpha_e(t), \hat\beta_e(t)) \pm\hat\alpha_e(t) \mp \hat\beta_e(t)\\ ~~~\\
~-2 + {\hat\alpha_e(t) + \hat\beta_e(t)\over 2}\\ ~~~\\
~-2 + \hat\zeta_e(t) - (\hat\alpha_e(t), \hat\beta_e(t))\\~~~\\
~-2 + \hat\zeta_e(t) - \hat\sigma_e(t)\end{cases}\nd }
leading us back to the same values for $\theta_{nl}$ that we had in \eqref{cyclebaaz3} and elsewhere for the other metric components, consequently allowing us to the choices \eqref{isaferr} and \eqref{isaferr2}. This is again good because it suggests that it is the dominant quartic curvature terms on the world-volume of the BBS instantons are again responsible for driving the acceleration of the universe. From here, and for the remaining non-perturbative contributions $-$ that include the KKLT instantons and other exotic instantons including their non-local extensions $-$ the story follows the path laid out in section \ref{eomgmn}.

The analysis presented in {\bf Tables \ref{niksmit0071}, \ref{niksmit0072}, \ref{niksmit0073}} and {\bf \ref{niksmit0100}} suggests that the $\bar{g}_s$ scalings of the quantum terms related to the perturbative and the non-perturbative effects for all the {\it on-shell} metric components are essentially similar. This means once we study the 
$\bar{g}_s$ scaling behavior of any given emergent {\it on-shell} component $-$ say ${\bf g}_{mn}$ $-$ we are essentially predicting the behavior of all other emergent on-shell metric components. This is of course expected from the underlying consistency of the model, but such similarities do not mean that the spatial components of the on-shell metric components would behave in identical ways.

\subsection{Counting the number of solutions with instanton embeddings \label{counting}}

In the study of the number of contributing terms, by equating the scalings of the Einstein tensor with the scalings $\theta_{nl}$ from the quantum series in \eqref{brittbaba007}, we always took $l_{14} = 1$. While this suggests a clear one-to-one correspondence of the subset of $\theta_{nl}$ (from the aforementioned mapping) to the set of terms in \eqref{brittbaba007}, it doesn't prohibit the possibility of the existence of other contributions. In this section we will try to see what other allowed solutions are possible within the curvature and the derivative sectors of \eqref{botsuga}. The equation that we have in mind is the following:

{\scriptsize
\bg\label{melisrain}
&&  x_1 \left({4\over 3} - {\hat\zeta_e(t)\over 2}\right) +
x_2 \left( {5\over 3} + {\hat\alpha_e(t) + \hat\beta_e(t)\over 4} - \hat\zeta_e(t)\right) + x_3 \left({5\over 3} - {\hat\alpha_e(t)\over 2} - {\hat\zeta_e(t)\over 2}\right) 
+ x_4 \left({5\over 3} - {\hat\beta_e(t)\over 2} - {\hat\zeta_e(t)\over 2}\right) \nonumber\\
&+& x_5\left({2\over 3} - {\hat\sigma_e(t)\over 2} - {\hat\alpha_e(t)\over 2}\right) + x_6\left({2\over 3} - {\hat\sigma_e(t)\over 2} - {\hat\beta_e(t)\over 2}\right)
+ x_7
\left( {2\over 3} - {\hat\alpha_e(t)\over 2} + {\hat\alpha_e(t) + \hat\beta_e(t)\over 4} - {\hat\zeta_e(t)\over 2}\right) \nonumber\\
&+ & x_8
\left( {2\over 3} - {\hat\beta_e(t)\over 2} + {\hat\alpha_e(t) + \hat\beta_e(t)\over 4} - {\hat\zeta_e(t)\over 2}\right)
+ x_9\left( {2\over 3} + {\hat\alpha_e(t) + \hat\beta_e(t)\over 4} - {\hat\sigma_e(t)\over 2} - {\hat\zeta_e(t)\over 2}\right)
+ x_{10}\left({8\over 3} - \hat\zeta_e(t)\right)\nonumber\\ 
&+& x_{11}\left({2\over 3} - \hat\sigma_e(t)\right) + x_{12}\left({2\over 3} - \hat\alpha_e(t)\right) + x_{13}\left({2\over 3} - \hat\beta_e(t)\right) + x_{14}\left({2\over 3} + {\hat\alpha_e(t) + \hat\beta_e(t)\over 2} - \hat\zeta_e(t)\right)\nonumber\\
& + & x_{15}\left({2\over 3} - 2\hat\alpha_e(t) + \hat\beta_e(t)\right)
+
x_{16}\left({2\over 3} - 2\hat\beta_e(t) + \hat\alpha_e(t)\right)
+ x_{17}\left( {5\over 3} - {\hat\sigma_e(t)\over 2} - {\hat\zeta_e(t)\over 2}\right) \nonumber\\
& = &~~ \begin{cases} ~{8\over 3} - \hat\zeta_e(t) + m_1\theta_d(t) - m_2a_{\rm F}(t)\\ ~~~\\
~{2\over 3} - 2\hat\alpha_e(t)+ \hat\beta_e(t)) + m_1\theta_d(t) - m_2a_{\rm F}(t)\\ ~~~\\
~{2\over 3} - 2\hat\beta_e(t) + \hat\alpha_e(t)) + m_1\theta_d(t) - m_2a_{\rm F}(t)\\ ~~~\\
~{2\over 3} + {\hat\alpha_e(t)+ \hat\beta_e(t)\over 2} 
- \hat\zeta_e(t)+ m_1\theta_d(t) - m_2a_{\rm F}(t)\\ ~~~\\
~{2\over 3} - \hat\beta_e(t)+ m_1\theta_d(t) - m_2a_{\rm F}(t)\\~~~\\
~{2\over 3} - \hat\alpha_e(t) + m_1\theta_d(t) - m_2a_{\rm F}(t)\\~~~\\
~{2\over 3} - \hat\sigma_e(t)+ m_1\theta_d(t) - m_2a_{\rm F}(t)\end{cases} 
\nd}
where $1 \le m_1 \le 2$ and $0\le m_2 \le 2$ with $m_i \in \mathbb{Z}^+$ and $\theta_d$ is related to the scaling of the inverse volume of the sub-manifold on which we have our $d$-dimensional instanton configuration (be it a brane-instanton or a more exotic variety). The $x_i$ defined in \eqref{melisrain} are non-negative integers, {\it i.e.} 
$x_i \in \mathbb{Z}^+$, and are defined in the following way:

{\footnotesize
\bg\label{nahirmey}
&& x_1 = n_1;~~~ x_2 \in \sum_{j = 15}^{18} l_j; ~~~ (x_3, x_4) \in \sum_{p = 24}^{28} l_p; ~~~ (x_5, x_6) \in\sum_{q=29}^{33} l_q \\ 
&& (x_7, x_8) \in \sum_{r=34}^{37} l_r;~~ 
x_9 \in\sum_{s = 38}^{41} l_s;~~~ (x_{15}, x_{16})\in l_{14}; ~~~ x_{17} \in \sum_{k = 19}^{23} l_k; ~~~ (x_{10},..., x_{14}) \in (l_{14},~\sum_{i =1}^{13}l_i), \nonumber \nd}
thus relating them to the factors appearing in \eqref{botsuga} and \eqref{brittbaba007}. Our aim is to find the value of the 17-tuple that solves \eqref{melisrain} for every row on the RHS. This is a non-trivial exercise, so the strategy would be to choose a particular row on the RHS of \eqref{melisrain} and try to find the set of 17-tuples. Once we collect the full set of 17-tuples associated to all the 7 rows on the RHS, we can select the subset of the solution set for the system. Note two things: {\Su one}, $(x_{10}, .., x_{13}) \in \mathbb{Z}^+$ could signify either curvatures or derivatives, {\Su two}, we have not considered the fluxes in \eqref{melisrain} simply because we have not worked out the $\hat{l}_{e{\rm AB}}^{\rm CD}$ factors (which may easily be worked out from the curvature two-forms, but we don't do it here). However to check whether we can have a {\it finite} number of solutions to an equation like \eqref{melisrain} (with the fluxes included in), all we need are the dominant scalings of the flux components entering \eqref{brittbaba007}. We will denote the flux components by $l_p$ with $42\le p\le 81$ and $p \in \mathbb{Z}^+$. For the BBS instantons,  where $m_1 = 1, m_2 = 0$ and $\theta_d = 2 -2\hat\sigma_e(t) - {\hat\alpha_e(t) + \hat\beta_e(t)\over 2}$, this gives us the following two constraints on $x_i$ and $l_p$:
\bg\label{falooda}
4x_1 + 5\bigg(\sum_{i = 2}^4 x_i + x_{17}\bigg) + 2\bigg(\sum_{j = 5}^9 x_j + \sum_{l= 11}^{16} x_l\bigg) + 8x_{10} + \sum_{p = 42}^{81}l_p =
\begin{cases}~8\\~~~\\~14 \end{cases} \nd
where the RHS denote the only two dominant choices from \eqref{melisrain}. Interestingly the equation \eqref{falooda} clearly has a finite number of solutions, because of the positivity constraints from both sides of the equation, but it would be instructive to find {\it all} the possible solutions. For this, it will be advisable to group some of the $x_i$ and $l_p$ into manageable sets, for example by defining them as $\sum\limits_{i = 2}^4 x_i + x_{17} \equiv {\rm B}$, 
$\sum\limits_{j = 5}^9 x_j + \sum\limits_{l= 11}^{16} x_l \equiv {\rm C}$ and 
$\sum\limits_{p = 42}^{81}l_p \equiv {\rm L}$, to express \eqref{falooda} as:
\bg\label{falooda2}
4x_1 + 5{\rm B} + 2{\rm C} + 8x_{10} + {\rm L} \equiv {\rm N} =  \begin{cases}~8\\~~~\\~14 \end{cases} \nd
and ask the question as to how many non-negative integer solutions are possible for the 5-tuple $(x_1, {\rm B}, {\rm C}, x_{10}, {\rm L})$. The answer is ${\bf 12}$ for ${\rm N} = 8$ and ${\bf 40}$ for ${\rm N} = 14$. They may be expressed as the following exhaustive list of 5-tuples:
\bg\label{sapnamistimey}
{\rm N} = {\bf 8}: && (0,0,0,0,8),
(0,0,0,1,0),
(0,0,1,0,6),
(0,0,2,0,4),
(0,0,3,0,2),
(0,0,4,0,0)\nonumber\\
&& (0,1,0,0,3),
(0,1,1,0,1),
(1,0,0,0,4),
(1,0,1,0,2),
(1,0,2,0,0),
(2,0,0,0,0)\nonumber\\
{\rm N} = {\bf 14}: && (0,0,0,0,14),
(0,0,0,1,6),
(0,0,1,0,12),
(0,0,1,1,4),
(0,0,2,0,10),
(0,0,2,1,2)\nonumber\\
&& (0,0,3,0,8),
(0,0,4,0,6),
(0,0,5,0,4),
(0,0,6,0,2),
(0,0,7,0,0),
(0,1,0,0,9) \nonumber\\
&& (0,1,1,0,7),
(0,1,2,0,5),
(0,2,0,0,4),
(0,2,1,0,2),
(1,0,0,0,10),
(1,0,0,1,2) \nonumber\\
&& (1,0,1,0,8),
(1,0,1,1,0),
(1,0,2,0,6),
(1,0,3,0,4),
(1,0,4,0,2),
(1,0,5,0,0) \nonumber\\
&& (1,1,0,0,5),
(1,1,1,0,3),
(1,1,2,0,1),
(1,2,0,0,0),
(2,0,0,0,6),
(2,0,1,0,4)\nonumber\\
&& (2,0,2,0,2),
(2,0,3,0,0),
(2,1,0,0,1),
(3,0,0,0,2),
(3,0,1,0,0), (0,0,3,1,0) \nonumber\\
&& (0,1,0,1,1), (0,1,3,0,3), (0,1,4,0,1), (0,2,2,0,0),
\nd
which provides the necessary {\it bound} on the number of possible choices. In fact if $\hat\sigma_e(t) = \hat\alpha_e(t) = \hat\beta_e(t) = \hat\zeta_e(t) = \hat\eta_e(t) = 0$, \eqref{sapnamistimey} would have been the allowed choices. However now that we have non-zero values of these parameters (which as we saw earlier are all very small), the actual number of choices gets further constrained. To see this we will concentrate on the lowest row on the RHS of \eqref{melisrain}, {\it i.e.}
take ${8\over 3} - 3\hat\sigma_e(t) - {\hat\alpha_e(t) + \hat\beta_e(t)\over 2}$ for the BBS instanton case and ask what are number of possible choices for the 17-tuple (where we concentrate only on the curvature and the derivative sectors of \eqref{botsuga} and \eqref{brittbaba007}). The answer turns out to be a unique one and is given by:
\bg\label{sapnamisti1}
x_5 = x_6 = 1, ~~~ x_{11} = 2, ~~~ x_i = 0~~~{\rm for} ~i \ne (5, 6, 11),\nd
which, because of the ambiguity in the definition of $(x_{10}, .., x_{13}) \in \mathbb{Z}^+$, is related to the three choices in \eqref{isaferr} and \eqref{isaferr2}. In a similar vein we can ask how many solutions are possible for the 17-tuple if the RHS of \eqref{melisrain} takes the value 
${8\over 3} -2\hat\sigma_e(t) - {3\hat\alpha_e(t)\over 2} - {\hat\beta_e(t)\over 2}$. Now there are {\it two} sets of solutions:
\bg\label{sapnamitha}
\Big(x_5 = 3, x_6 = 1, x_i = 0\Big), ~~~~~ \Big(x_5 = x_6 = x_{11} = x_{12}= 1, x_i = 0\Big), \nd
where the first solution does not consistently fit with \eqref{sapnamisti1}, and therefore it is only the second solution in \eqref{sapnamitha} that makes sense here. Unsurprisingly, this also matches with the three solutions from \eqref{isaferr} and \eqref{isaferr2}. Similarly, in  the third row from the bottom on the RHS of \eqref{melisrain} where we have ${8\over 3} - 2\hat\sigma_e(t) - {\hat\alpha_e(t)\over 2} - {3\hat\beta_e(t)\over 2}$, we can ask how many solutions for the 17-tuple are allowed. The answer is again {\it two} sets:
\bg\label{sapmit}
\Big(x_5 = 1, x_6 = 3, x_i = 0\Big), ~~~~~ \Big(x_5 = x_6 = x_{11} = x_{13}= 1, x_i = 0\Big), \nd
out of which the second set appears to be consistent with our choice \eqref{sapnamisti1} and consequently also with \eqref{isaferr} and \eqref{isaferr2}. One may go further and query about the next row on the RHS of \eqref{melisrain}. This goes as ${8\over 3} - 2\hat\sigma_e(t) - \hat\zeta_e(t)$, and we can ask how many 17-tuples are possible with this choice. The answer turns out to be {\it five} sets:

{\footnotesize
\bg\label{projapotinil}
&& \Big(x_5 = x_6 = x_{11} = x_{14} = 1, x_i = 0\Big), ~~~ 
\Big(x_6 = x_7 = x_9 = x_{11} = 1, x_i = 0\Big)\\
&& \Big(x_5 = x_8 = x_9 = x_{11} =1, x_i = 0\Big), ~~ \Big(x_7 = x_8 = 1, x_{11} = 2, x_i = 0\Big), ~~ \Big(x_5 = x_6 = 1, x_9 = 2, x_i = 0\Big), \nonumber \nd}
out of which it's only the first one that has $x_5 = x_6 = 1$ and fits well with \eqref{sapnamisti1}. The last one, although allows $x_5 = x_6 = 1$, unfortunately doesn't fit with the structure in \eqref{brittbaba007} and is consequently discarded. Needless to say, the first choice reproduces \eqref{isaferr} and \eqref{isaferr2}. Our analysis also suggests that individually there are possibilities of multiple choices for the 17-tuple, but we should only look for solution that fits appropriately with \eqref{brittbaba007} and \eqref{sapnamisti1}. In a similar vein, our next row goes as ${8\over 3} - {5\hat\beta_e(t)\over 2} + {\hat\alpha_e(t)\over 2} - 2\hat\sigma_e(t)$, and we can ask how many solutions are possible for the 17-tuple. The answer turns out to be a {\it unique} one:
\bg\label{amirani}
x_5 = x_6 = x_{11} = x_{16} = 1, ~~~ x_i = 0~~~~{\rm for}~~ i \ne 5, 6, 11, 16, \nd
which naturally fits well with \eqref{sapnamisti1} and \eqref{brittbaba007}, and consequently reproduces \eqref{isaferr} and \eqref{isaferr2}. If we flip $\hat\alpha_e(t) \leftrightarrow \hat\beta_e(t)$, we expect the uniqueness of the 17-tuple to survive. This is indeed the case for the next row with element ${8\over 3} - {5\hat\alpha_e(t)\over 2} + {\hat\beta_e(t)\over 2} - 2\hat\sigma_e(t)$, whose unique solution for the 17-tuple turns out to be:
\bg\label{amirani2}
x_5 = x_6 = x_{11} = x_{15} = 1, ~~~ x_i = 0~~~~{\rm for}~~ i \ne 5, 6, 11, 15, \nd
fitting well with both \eqref{brittbaba007} and \eqref{sapnamisti1}. Our final row, which is the top row on the RHS of \eqref{melisrain}, has the element ${14\over 3} - 2\hat\sigma_e(t) - \hat\zeta_e(t) - {\hat\alpha_e(t) + \hat\beta_e(t)\over 2}$ and we can ask how many solutions for the 17-tuple are allowed with the aforementioned choice. There are now {\it three} possibilities:
\bg\label{jacobs}
&&\Big(x_5=x_6=x_{10}=x_{11}=1, x_i=0\Big)\nonumber\\ &&\Big(x_5=x_6=x_{11}=1,x_1=2,x_i=0\Big), ~~\Big(x_3=x_4=1,x_{11}=2, x_i =0\Big),\nd
where the first two sets fit with \eqref{melisrain} and \eqref{brittbaba007} and in fact, due to the aforementioned ambiguity between the scalings of some curvature terms and the derivatives, they lead to identical scalings. The third set from \eqref{jacobs} do not match with \eqref{melisrain} and is therefore discarded.
\begin{table}[h]  
 \begin{center} 
\resizebox{\columnwidth}{!}{%
 \renewcommand{\arraystretch}{1.3}
\begin{tabular}{|c||c||c|}\hline $(x_1,{\rm B,C}, x_{10})$ & \text{check} & \text{ realizations}\\ \hline
(0,0,7,0)& $2\cdot7=14$ & $\binom{17}{10}=19448$\\ \hline 
(0,0,3,1)& $2\cdot3+8=14$ & $\binom{13}{10}=286$\\ \hline 
(0,2,2,0)& $5\cdot2+2\cdot2=14$ & $\binom{5}{3}\binom{12}{10}=10\cdot66=660$\\ \hline
(1,0,5,0)& $4+2\cdot5=14$ & $\binom{15}{10}=3003$\\ \hline
(1,0,1,1)& $4+2+8=14$ & $\binom{11}{10}=11$\\ \hline
(1,2,0,0)& $4+10=14$ & $\binom{5}{3}=10$\\ \hline
(2,0,3,0)& $8+6=14$ & $\binom{13}{10}=286$\\ \hline
(3,0,1,0)& $12+2=14$ & $\binom{11}{10}=11$ \\ \hline
\end{tabular}}
\renewcommand{\arraystretch}{1}
\end{center}
 \caption[]{\Su Finding the total number of nonnegative integer solutions for $4x_1 + 5{\rm B} + 2{\rm C} + 8x_{10} = 14$ where 
 ${\rm B} \equiv \sum\limits_{i = 2}^4 x_i + x_{17}$ and  
${\rm C} \equiv \sum\limits_{j = 5}^9 x_j + \sum\limits_{l= 11}^{16} x_l$. The number of nonnegative integer solution is 23715.}
\label{canposttag}
 \end{table} 

It is interesting to note that we are almost getting a unique set for the 17-tuple once we take into account all the rows on the RHS of \eqref{melisrain} for the BBS instantons. In fact the reduction of the possible choices comes from the existence of the small parameters 
$(\hat\zeta_e(t), \hat\sigma_e(t), \hat\alpha_e(t), \hat\beta_e(t), \hat\eta_e(t))$ that basically converts the number of choices with 5-tuples from \eqref{sapnamistimey} to substantially small sets of choices. The 5-tuple $(x_1, {\rm B, C}, x_{10}, {\rm L})$ contains the flux contributions from \eqref{botsuga} and \eqref{brittbaba007}, but since we are looking at 4-tuple\footnote{Recall that we are only analyzing the problem in the curvature and the derivative sectors of \eqref{botsuga} and \eqref{brittbaba007}.}, or more appropriately 17-tuple in the notations of $x_i$, it would be interesting to compute how many such 17-tuples are possible if we did not have the small parameters $(\hat\zeta_e(t), \hat\sigma_e(t), \hat\alpha_e(t), \hat\beta_e(t), \hat\eta_e(t))$. Using the definitions of ${\rm B}$ and ${\rm C}$ as
${\rm B} \equiv \sum\limits_{i = 2}^4 x_i + x_{17}$ and  
${\rm C} \equiv \sum\limits_{j = 5}^9 x_j + \sum\limits_{l= 11}^{16} x_l$, we are basically asking how many nonnegative integer solutions are possible for given values of ${\rm B}$ and ${\rm C}$. For this subset, the number is:
\bg\label{timhorttag}
\binom{{\rm B}+3}{3}\cdot\binom{{\rm C}+10}{10}, \nd
which may then be summed over ${\rm B}$ and ${\rm C}$. Following \eqref{sapnamistimey}, our question reduces to finding the number of nonnegative integer solutions for the two sets of equations: $4x_1 + 5{\rm B} + 2{\rm C} + 8x_{10} = 8$ and $4x_1 + 5{\rm B} + 2{\rm C} + 8x_{10} = 14$ and then use \eqref{timhorttag} to find the number of solutions for the 17-tuple. The detailed analysis for the second case is worked out in {\bf Table \ref{canposttag}}. The answer is:
\bg\label{canpostshoi}
{\rm N} = 8:~~{\bf 1069}, ~~~~~{\rm N} = 14:~~{\bf 23715}, \nd
and therefore it is remarkable that the inclusion of $(\hat\zeta_e(t), \hat\sigma_e(t), \hat\alpha_e(t), \hat\beta_e(t), \hat\eta_e(t))$ simply reduces this to an almost unique set of 17-tuple reflected by \eqref{isaferr} and \eqref{isaferr2} (the apparent ambiguity leading to the three sets therein is explained earlier).

Let us now consider the BBS instanton case with non-localities. In \eqref{melisrain} we will study the scenario with $m_1 = m_2 = 2$ and $a_{\rm F} = -\vert a_{\rm F}\vert \equiv -{x_{18}\over 3}$. For simplicity we will start with the topmost row on the RHS of \eqref{melisrain}, and ask how many non-negative integer solutions are possible. The equation that we want to solve is:
\bg\label{timcanpost}
{20\over 3} -\hat\zeta_e(t) - 4\hat\sigma_e(t) - \hat\alpha_e(t) - \hat\beta_e(t) = \sum_{i = 1}^{17} x_i c_i - {2\over 3} x_{18}, \nd
where $c_i \equiv c_i(\hat\zeta_e(t),\hat\alpha_e(t),\hat\beta_e(t), \hat\sigma_e(t))$ are the coefficients appearing on the LHS of \eqref{melisrain}. For the special case where $\hat\zeta_e(t)=\hat\alpha_e(t)=\hat\beta_e(t)=\hat\sigma_e(t) = 0$, one may easily check that because of the relative {\it minus} sign, there are {\it infinite} number of non-negative integer solutions for the 18-tuple $(x_1, x_2, ..., x_{18})$. Once we switch on the set of small parameters $(\hat\zeta_e(t),\hat\alpha_e(t),\hat\beta_e(t), \hat\sigma_e(t))$, the number of possible solutions for the 18-tuple drastically reduces to sets of 24 18-tuples. They are listed as follows:

{\scriptsize
\bg\label{desimetro}
(x_1, x_2, ...., x_{17}, x_{18}): && 
(0, 0, 0, 0, 2, 2, 0, 0, 0, 0, 1, 0, 0, 0, 0, 0, 2, 0),~~(0, 0, 0, 0, 0, 2, 0, 0, 0, 0, 2, 1, 0, 0, 0, 0, 2, 0)\nonumber\\
&&(0, 0, 1, 0, 1, 2, 0, 0, 0, 0, 2, 0, 0, 0, 0, 0, 1, 0),~~
(0, 0, 0, 0, 2, 0, 0, 0, 0, 0, 2, 0, 1, 0, 0, 0, 2, 0)\nonumber\\
&&(0, 0, 0, 1, 2, 1, 0, 0, 0, 0, 2, 0, 0, 0, 0, 0, 1, 0),~~
(2, 0, 0, 0, 2, 2, 0, 0, 0, 0, 2, 0, 0, 0, 0, 0, 0, 0)\nonumber\\
&&(0, 0, 0, 0, 0, 0, 0, 0, 0, 0, 3, 1, 1, 0, 0, 0, 2, 0), ~~
(0, 0, 0, 0, 0, 0, 0, 0, 0, 0, 3, 0, 0, 0, 1, 1, 2, 0)\nonumber\\
&&(0, 0, 0, 1, 0, 1, 0, 0, 0, 0, 3, 1, 0, 0, 0, 0, 1, 0), ~~
(0, 0, 2, 0, 0, 2, 0, 0, 0, 0, 3, 0, 0, 0, 0, 0, 0, 0)\nonumber\\
&&(2, 0, 0, 0, 0, 2, 0, 0, 0, 0, 3, 1, 0, 0, 0, 0, 0, 0), ~~
(0, 0, 1, 0, 1, 0, 0, 0, 0, 0, 3, 0, 1, 0, 0, 0, 1, 0)\nonumber\\
&&(0, 0, 1, 1, 1, 1, 0, 0, 0, 0, 3, 0, 0, 0, 0, 0, 0, 0), ~~
(0, 0, 0, 2, 2, 0, 0, 0, 0, 0, 3, 0, 0, 0, 0, 0, 0, 0)\nonumber\\
&&(2, 0, 0, 0, 2, 0, 0, 0, 0, 0, 3, 0, 1, 0, 0, 0, 0, 0), ~~
(0, 0, 0, 2, 0, 0, 0, 0, 0, 0, 4, 1, 0, 0, 0, 0, 0, 0)\nonumber\\
&&(0, 0, 2, 0, 0, 0, 0, 0, 0, 0, 4, 0, 1, 0, 0, 0, 0, 0), ~~
(2, 0, 0, 0, 0, 0, 0, 0, 0, 0, 4, 1, 1, 0, 0, 0, 0, 0)\nonumber\\
&&(2, 0, 0, 0, 0, 0, 0, 0, 0, 0, 4, 0, 0, 0, 1, 1, 0, 0), ~~
{\red(0, 0, 0, 0, 2, 2, 0, 0, 0, 1, 2, 0, 0, 0, 0, 0, 0, 0)}\nonumber\\
&&(0, 0, 0, 0, 0, 2, 0, 0, 0, 1, 3, 1, 0, 0, 0, 0, 0, 0), ~~
(0, 0, 0, 0, 2, 0, 0, 0, 0, 1, 3, 0, 1, 0, 0, 0, 0, 0)\nonumber\\
&&(0, 0, 0, 0, 0, 0, 0, 0, 0, 1, 4, 1, 1, 0, 0, 0, 0, 0), ~~
(0, 0, 0, 0, 0, 0, 0, 0, 0, 1, 4, 0, 0, 0, 1, 1, 0, 0)\nonumber\\ \nd}
where we see that $x_{18} = 3\vert a_{\rm F}\vert \in \{0, 1, 2\}$, implying that the temporal dependence of the non-locality function cannot exceed $\xxy^{-{2\over 3}}$; and we will discuss the importance of the solution given in {\red red} soon when we collect all the solutions from \eqref{melisrain}. Question however is, what made the infinite sets of solutions to reduce down to only 24 sets? To answer this, we need to  
look at the various equations that emanate from \eqref{timcanpost}. The first one is when we compare with the constant term on the LHS of \eqref{timcanpost}:
\bg\label{futsi1}
4x_1 + 5{\rm B} + 2{\rm C} + 8x_{10} - 2x_{18} = 20, \nd
where $({\rm B, C})$ are the same ones that appeared in \eqref{falooda2}. Due to the relative minus sign, this is exactly the equation that allows for an infinite number of non-negative integer solutions for the 18-tuple. If we flip the sign of the $x_{18}$ term, then all terms in \eqref{futsi1} are positive definite. It is easy to infer that all the non-negative integers appearing in \eqref{futsi1} are bounded from above by: $x_1 \le 5, {\rm B} \le 4, {\rm C} \le 10, x_{10} \le 2$ and $x_{18} \le 10$. Together they give rise to {\bf 71} 5-tuples.
We can then use \eqref{timhorttag} to find the number of solutions for the 18-tuple. 

Our present aim however is to continue with the relative sign in \eqref{timcanpost}, and ask how does the number of solutions get reduced to the 24 18-tuples in \eqref{desimetro}. One hint comes from comparing the $\hat\zeta_e(t)$ terms on both sides of the equation \eqref{timcanpost}. This gives us:
\bg\label{corytara}
x_1 + 2(x_2 + x_{10} + x_{14}) + \sum_{i = 3}^9 x_i-x_5-x_6 + x_{17} = 2, \nd
which immediately tells us that the 10-tuple should have finite number of solutions and are bounded from above as $(x_1, x_3, x_4, x_7, x_8, x_9, x_{17}) \le 2$ and $(x_2, x_{10}, x_{14}) \le 1$. Now comparing the $\hat\sigma_e(t)$ on both sides of \eqref{timcanpost} gives us:
\bg\label{corybritt}
\sum_{i = 5}^6 x_i + x_9 + 2x_{11} - x_{17} = 8, \nd
which should be alarming because of the relative minus sign, but is actually harmless because $x_{17}$ is bounded from above as $x_{17} \le 2$ because of \eqref{corytara}. Combining \eqref{corytara} and \eqref{corybritt} tells us that $x_{11} \le 4$ and that the only 5-tuple that needs to be fixed is $(x_{12}, x_{13}, x_{15}, x_{16}, x_{18})$. Now comparing $\hat\alpha_e(t)$ and $\hat\beta_e(t)$ on both sides of \eqref{timcanpost} give us respectively the following set of equations:
\bg\label{britcotivane}
&&{\rm A}:~x_2 - 2x_3 - 2x_5 - x_7 + x_8 + x_9 - 4 x_{12} + 2x_{14} - 8x_{15} + 4 x_{16} = -4 \nonumber\\
&&{\rm B}:~x_2 - 2x_4 - 2x_6 + x_7 - x_8 + x_9 - 4 x_{13} + 2x_{14} + 4x_{15} - 8 x_{16} = -4, \nd
from where, combining ${\rm A}$ and ${\rm B}$ as ${\rm A+B}$ and ${\rm A-B}$, immediately tells us that the 4-tuple are bounded from above as $(x_{12}, x_{13}, x_{15}, x_{16}) \le 2$. Finally, plugging in the bounds of $x_i, 1\le i \le 17$ in \eqref{futsi1} tells us that $x_{18}$ is also bounded from above. 

Our above analysis suggests something interesting. In certain cases even if we have relative minus signs (here for example it is with $x_{18})$), the very presence of the small parameters $(\hat\zeta_e(t), \hat\sigma_e(t), \hat\alpha_e(t), \hat\beta_e(t), \hat\eta_e(t))$ can render the number of allowed solutions to be finite, thus restoring EFT. We can see whether this trend continues if we go for the other rows on the RHS of \eqref{melisrain}. The equation that we have in mind is:
\bg\label{timcanpost2}
{14\over 3} - 5\hat\sigma_e(t) - \hat\alpha_e(t) - \hat\beta_e(t) = \sum_{i = 1}^{17} x_i c_i - {2\over 3} x_{18}, \nd
where $c_i$ can be extracted from the LHS of \eqref{melisrain}. Now since there is no $\hat\zeta_e(t)$ on the LHS of \eqref{timcanpost2}, the RHS of \eqref{corytara} becomes zero. This implies that all the $x_i$ appearing in \eqref{corytara} {\it vanish}. Comparing $\hat\alpha_e(t)$ and $\hat\beta_e(t)$ on both sides of \eqref{timcanpost2}, we recover \eqref{britcotivane} except that now all the $x_i$ appearing in \eqref{corytara} are put to zero in here. Similarly, comparing $\hat\sigma_e(t)$, we recover \eqref{corybritt} except with 10 instead of 8 on the RHS. Performing similar manipulations as above, we immediately get $x_{18} = 0$, implying that $a_{\rm F} = 0$, and the 18-tuple takes the following {\it five} values:

{\scriptsize
\bg\label{ctdekhai}
(x_1, x_2,..., x_{17}, x_{18}):&& {\red(0,0,0,0,2,2,0,0,0,0,3,0,0,0,0,0,0,0)},~~
(0,0,0,0,0,2,0,0,0,0,4,1,0,0,0,0,0,0)\nonumber\\
&& (0,0,0,0,2,0,0,0,0,0,4,0,1,0,0,0,0,0), ~~
(0,0,0,0,0,0,0,0,0,0,5,1,1,0,0,0,0,0)\nonumber\\
&&(0,0,0,0,0,0,0,0,0,0,5,0,0,0,1,1,0,0), \nd}
suggesting that the number of solutions has reduced quite a bit, but more importantly, the non-locality factor vanishes instead of $-{2\over 3}$ before. The importance of the solution in {\red red} here and elsewhere below will be elaborated soon. In a similar vein, our next equation from the second row bottom-up in \eqref{melisrain} is:
\bg\label{timcanpost3}
{14\over 3} - 4\hat\sigma_e(t) - 2\hat\alpha_e(t) - \hat\beta_e(t) = \sum_{i = 1}^{17} x_i c_i - {2\over 3} x_{18}, \nd
where, as in \eqref{timcanpost2} because of the absence of $\hat\zeta_e(t)$ on the LHS of \eqref{timcanpost3}, all the $x_i$ appearing on the LHS of \eqref{corytara} vanishes reducing to a small system comprised of the 8-tuple $(x_5, x_6, x_{11}, x_{12}, x_{13},$
$x_{15}, x_{16}, x_{18})$. The other constraints coming from matching $(\hat\sigma_e(t), \hat\alpha_e(t), \hat\beta_e(t))$ from both sides of 
\eqref{timcanpost3} immediately makes $x_{18} \equiv 3|a_{\rm F}| = 0$ and provides the following {\it eleven} solutions for the 18-tuple:

{\scriptsize
\bg\label{lotusmagic}
(x_1, x_2, ..., x_{17}, x_{18}): && (0,0,0,0,0,0,0,0,0,0,4,2,1,0,0,0,0,0),~~
(0,0,0,0,0,0,0,0,0,0,4,0,2,0,1,0,0,0)\nonumber\\
&& (0,0,0,0,0,0,0,0,0,0,4,1,0,0,1,1,0,0), ~~
(0,0,0,0,0,2,0,0,0,0,3,2,0,0,0,0,0,0)\nonumber\\
&& (0,0,0,0,0,2,0,0,0,0,3,0,1,0,1,0,0,0), ~~
(0,0,0,0,0,4,0,0,0,0,2,0,0,0,1,0,0,0)\nonumber\\
&& (0,0,0,0,2,0,0,0,0,0,3,1,1,0,0,0,0,0), ~~
(0,0,0,0,2,0,0,0,0,0,3,0,0,0,1,1,0,0)\nonumber\\
&& {\red(0,0,0,0,2,2,0,0,0,0,2,1,0,0,0,0,0,0)}, ~~
(0,0,0,0,4,0,0,0,0,0,2,0,1,0,0,0,0,0)\nonumber\\
&& (0,0,0,0,4,2,0,0,0,0,1,0,0,0,0,0,0,0), \nd}
where the number of solutions for the 18-tuple has increased a bit from what we had in \eqref{ctdekhai}, but the pattern is similar. The scaling of the non-locality function also consistently vanishes. Similarly, if we now flip $\hat\alpha_e(t)$ with $\hat\beta_e(t)$ and vice versa on the LHS of \eqref{timcanpost3}, we still get {\it eleven} solutions for the 18-tuple. They are given by:

{\scriptsize
\bg\label{lotusmagic2}
(x_1, x_2, ..., x_{17}, x_{18}): && (0,0,0,0,0,0,0,0,0,0,4,0,1,0,1,1,0,0),~~
(0,0,0,0,0,0,0,0,0,0,4,1,2,0,0,0,0,0)\nonumber\\
&& (0,0,0,0,0,0,0,0,0,0,4,2,0,0,0,1,0,0), ~~
(0,0,0,0,0,2,0,0,0,0,3,0,0,0,1,1,0,0)\nonumber\\
&& (0,0,0,0,0,2,0,0,0,0,3,1,1,0,0,0,0,0), ~~
(0,0,0,0,0,4,0,0,0,0,2,1,0,0,0,0,0,0)\nonumber\\
&& (0,0,0,0,2,0,0,0,0,0,3,0,2,0,0,0,0,0), ~~
(0,0,0,0,2,0,0,0,0,0,3,1,0,0,0,1,0,0)\nonumber\\
&& {\red(0,0,0,0,2,2,0,0,0,0,2,0,1,0,0,0,0,0)}, ~~
(0,0,0,0,2,4,0,0,0,0,1,0,0,0,0,0,0,0)\nonumber\\
&& (0,0,0,0,4,0,0,0,0,0,2,0,0,0,0,1,0,0), \nd}
which differs slightly by the placement of the integers from \eqref{lotusmagic}. The scaling of the non-locality function vanishes, {\it i.e.}
$x_{18} = 3|a_{\rm F}| = 0$, expectedly as before because the constraints have not changed too much. We can ask whether this continues to be the case if we take other rows in \eqref{melisrain}. It turns out that if we consider the middle row on the RHS of \eqref{melisrain} $x_{18} = 3|a_{\rm F}| \ne 0$. Specifically, we are now dealing with the following equation:
\bg\label{timcanpost5}
{14\over 3} - 4\hat\sigma_e(t) -\hat\zeta_e(t) - {\hat\alpha_e(t)\over 2} - {\hat\beta_e(t)\over 2} = \sum_{i = 1}^{17} x_i c_i - {2\over 3} x_{18}, \nd
where we note that $\hat\zeta_e(t)$ has appeared on the LHS. Inclusion of this would significantly change the outcome, and in particular we now no longer expect $x_{18} = 3|a_{\rm F}| = 0$ as mentioned above. Comparing $\hat\zeta_e(t), \hat\sigma_e(t), \hat\alpha_e(t), \hat\beta_e(t)$ and the constant term on both sides of \eqref{timcanpost5} now lead to {\it twenty-five}
possible solutions for the 18-tuple:

{\scriptsize
\bg\label{tundegulab}
(x_1, x_2, ..., x_{17}, x_{18}): &&(0, 0, 0, 0, 2, 2, 0, 0, 2, 0, 1, 0, 0, 0, 0, 0, 0, 0), ~~
(0, 0, 0, 0, 0, 2, 0, 0, 2, 0, 2, 1, 0, 0, 0, 0, 0, 0)\nonumber\\
&&(0, 0, 0, 0, 1, 1, 0, 0, 0, 0, 2, 0, 0, 0, 0, 0, 2, 2), ~~
(0, 0, 0, 0, 1, 2, 1, 0, 1, 0, 2, 0, 0, 0, 0, 0, 0, 0)\nonumber\\
&&(0, 0, 0, 0, 2, 0, 0, 0, 2, 0, 2, 0, 1, 0, 0, 0, 0, 0),~~
(0, 0, 0, 0, 2, 1, 0, 1, 1, 0, 2, 0, 0, 0, 0, 0, 0, 0)\nonumber\\
&&{\red(0, 0, 0, 0, 2, 2, 0, 0, 0, 0, 2, 0, 0, 1, 0, 0, 0, 0)}, ~~
(0, 0, 0, 0, 0, 0, 0, 0, 2, 0, 3, 1, 1, 0, 0, 0, 0, 0)\nonumber\\
&&(0, 0, 0, 0, 0, 0, 0, 0, 2, 0, 3, 0, 0, 0, 1, 1, 0, 0),~~
(0, 0, 0, 0, 0, 1, 0, 1, 1, 0, 3, 1, 0, 0, 0, 0, 0, 0)\nonumber\\
&&(0, 0, 1, 0, 0, 1, 0, 0, 0, 0, 3, 0, 0, 0, 0, 0, 1, 2),~~
(0, 0, 0, 0, 0, 2, 0, 0, 0, 0, 3, 1, 0, 1, 0, 0, 0, 0)\nonumber\\
&&(0, 0, 0, 0, 0, 2, 2, 0, 0, 0, 3, 0, 0, 0, 0, 0, 0, 0),~~
(0, 0, 0, 0, 1, 0, 1, 0, 1, 0, 3, 0, 1, 0, 0, 0, 0, 0)\nonumber\\
&&(0, 0, 0, 1, 1, 0, 0, 0, 0, 0, 3, 0, 0, 0, 0, 0, 1, 2),~~
(0, 0, 0, 0, 1, 1, 1, 1, 0, 0, 3, 0, 0, 0, 0, 0, 0, 0)\nonumber\\
&&(2, 0, 0, 0, 1, 1, 0, 0, 0, 0, 3, 0, 0, 0, 0, 0, 0, 2),~~
(0, 0, 0, 0, 2, 0, 0, 0, 0, 0, 3, 0, 1, 1, 0, 0, 0, 0)\nonumber\\
&&(0, 0, 0, 0, 2, 0, 0, 2, 0, 0, 3, 0, 0, 0, 0, 0, 0, 0),~~
(0, 0, 0, 0, 0, 0, 0, 0, 0, 0, 4, 1, 1, 1, 0, 0, 0, 0)\nonumber\\
&&(0, 0, 0, 0, 0, 0, 0, 2, 0, 0, 4, 1, 0, 0, 0, 0, 0, 0),~~
(0, 0, 0, 0, 0, 0, 2, 0, 0, 0, 4, 0, 1, 0, 0, 0, 0, 0)\nonumber\\
&&(0, 0, 0, 0, 0, 0, 0, 0, 0, 0, 4, 0, 0, 1, 1, 1, 0, 0),~~
(0, 0, 1, 1, 0, 0, 0, 0, 0, 0, 4, 0, 0, 0, 0, 0, 0, 2)\nonumber\\
&&(0, 0, 0, 0, 1, 1, 0, 0, 0, 1, 3, 0, 0, 0, 0, 0, 0, 2) \nd}
where we see that $x_8 = 3|a_{\rm F}| = \{0, 2, 3, 4\}$ implying that dominant scaling of the non-local terms if $\xxy^{-{4\over 3}}$
which is bigger than what we observed for \eqref{desimetro}. Interestingly \eqref{desimetro} is also the first one where $\hat\zeta_e(t)$ appeared on both sides of the equation \eqref{timcanpost}. This it seems that whenever we have $\hat\zeta_e(t)$ dependence on both sides of the equations from \eqref{melisrain}, the scaling of the non-locality factor $x_{18} = 3|a_{\rm F}|$ becomes non-zero. We can see whether this remains the case for the following equation:
\bg\label{timcanpost6}
{14\over 3} - 4\hat\sigma_e(t) - 3\hat\beta_e(t) = \sum_{i = 1}^{17} x_i c_i - {2\over 3} x_{18}, \nd
where not only $\hat\zeta_e(t)$ is absence, but also is $\hat\alpha_e(t)$. The constraints are now a bit stronger compared to the earlier ones, and one may easily convince oneself that $x_{18} = 3|a_{\rm F}| = 0$, and so no temporal dependence of the non-locality function \eqref{omameys}. There are now {\it nine} solutions to the 18-tuple, which may be listed as:

{\scriptsize
\bg\label{omameykotana}
(x_1, x_2, ..., x_{17}, x_{18}): && (0, 0, 0, 0, 0, 6, 0, 0, 0, 0, 1, 0, 0, 0, 0, 0, 0, 0), ~~
(0, 0, 0, 0, 0, 4, 0, 0, 0, 0, 2, 0, 1, 0, 0, 0, 0, 0)\nonumber\\
&&{\red(0, 0, 0, 0, 2, 2, 0, 0, 0, 0, 2, 0, 0, 0, 0, 1, 0, 0)}, ~~
(0, 0, 0, 0, 0, 2, 0, 0, 0, 0, 3, 0, 2, 0, 0, 0, 0, 0)\nonumber\\
&&(0, 0, 0, 0, 0, 2, 0, 0, 0, 0, 3, 1, 0, 0, 0, 1, 0, 0), ~~
(0, 0, 0, 0, 2, 0, 0, 0, 0, 0, 3, 0, 1, 0, 0, 1, 0, 0)\nonumber\\
&&(0, 0, 0, 0, 0, 0, 0, 0, 0, 0, 4, 0, 3, 0, 0, 0, 0, 0), ~~
(0, 0, 0, 0, 0, 0, 0, 0, 0, 0, 4, 1, 1, 0, 0, 1, 0, 0)\nonumber\\
&&(0, 0, 0, 0, 0, 0, 0, 0, 0, 0, 4, 0, 0, 0, 1, 2, 0, 0), \nd}
which fortunately has the expected pattern that we saw earlier. Question now is what happens if we replace $\hat\beta_e(t)$ by $\hat\alpha_e(t)$ on the LHS of \eqref{timcanpost6}. As before, the $\hat\zeta_e(t)$ will force $(x_1,.., x_4, x_7,..,x_{10}, x_{14}, x_{17})$ to vanish. The remaining equations provide the following {\it nine} solutions for the 18-tuple:

{\scriptsize
\bg\label{paanyue}
(x_1, x_2,...,x_{17}, x_{18}): && (0,0,0,0,0,0,0,0,0,0,4,0,0,0,2,1,0,0),~~
(0,0,0,0,0,0,0,0,0,0,4,1,1,0,1,0,0,0)\nonumber\\
&&
(0,0,0,0,0,0,0,0,0,0,4,3,0,0,0,0,0,0), ~~
(0,0,0,0,0,2,0,0,0,0,3,1,0,0,1,0,0,0)\nonumber\\
&& 
(0,0,0,0,2,0,0,0,0,0,3,0,1,0,1,0,0,0),~~
(0,0,0,0,2,0,0,0,0,0,3,2,0,0,0,0,0,0)\nonumber\\
&&
{\red(0,0,0,0,2,2,0,0,0,0,2,0,0,0,1,0,0,0)},~~
(0,0,0,0,4,0,0,0,0,0,2,1,0,0,0,0,0,0)\nonumber\\
&&
(0,0,0,0,6,0,0,0,0,0,1,0,0,0,0,0,0,0), \nd}
which clearly suggests a vanishing scaling of the non-locality function \eqref{omameys}, {\it i.e.}  $x_{18} = 2|a_{\rm F}| = 0$. The other solutions follow similar pattern as in \eqref{omameykotana}; and with \eqref{paanyue} this completes the collection of all the solution sets of 18-tuples that arise from the seven rows on the RHS of \eqref{melisrain}. In terms of increasing order of rows on the RHS of \eqref{melisrain}, the solution sets are: \eqref{ctdekhai}, \eqref{lotusmagic}, \eqref{lotusmagic2}, \eqref{tundegulab}, \eqref{omameykotana}, \eqref{paanyue} and \eqref{desimetro}. Comparing all the sets of solutions, we notice that the only solutions from each set
that solves all the EOMs in \eqref{melisrain} are the ones shown in {\red red} in \eqref{ctdekhai}, \eqref{lotusmagic}, \eqref{lotusmagic2}, \eqref{tundegulab}, \eqref{omameykotana}, \eqref{paanyue} and \eqref{desimetro}. These may be listed in the following way:

{\footnotesize
\bg\label{delrey}
 (0,0,0,0,2,2,0,0,0,0,3,0,0,0,0,0,0,0) &= & {14\over 3} - 5\hat\sigma_e(t) - \hat\alpha_e(t) - \hat\beta_e(t)
 \nonumber\\
 (0,0,0,0,2,2,0,0,0,0,2,1,0,0,0,0,0,0) & = & {14\over 3} - 4\hat\sigma_e(t) - 2\hat\alpha_e(t) - \hat\beta_e(t) 
\nonumber\\
(0,0,0,0,2,2,0,0,0,0,2,0,1,0,0,0,0,0)& = & {14\over 3} - 4\hat\sigma_e(t) - 2\hat\beta_e(t) - \hat\alpha_e(t)
\nonumber\\
(0,0,0,0,2,2,0,0,0,0,2,0,0,1,0,0,0,0) & = &
{14\over 3} - 4\hat\sigma_e(t) -\hat\zeta_e(t) - {\hat\alpha_e(t)\over 2} - {\hat\beta_e(t)\over 2} \nonumber\\
(0, 0, 0, 0, 2, 2, 0, 0, 0, 0, 2, 0, 0, 0, 1, 0, 0, 0) & = & 
{14\over 3} - 4\hat\sigma_e(t) - 3\hat\alpha_e(t)  \nonumber\\
(0, 0, 0, 0, 2, 2, 0, 0, 0, 0, 2, 0, 0, 0, 0, 1, 0, 0) & = &
{14\over 3} - 4\hat\sigma_e(t) - 3\hat\beta_e(t)  \nonumber\\
(0,0,0,0,2,2,0,0,0,1,2,0,0,0,0,0,0,0) & = &
{20\over 3} -\hat\zeta_e(t) - 4\hat\sigma_e(t) - \hat\alpha_e(t) - \hat\beta_e(t), \nd}
from where it is easy to see that all these solutions satisfy $x_{18} = 3|a_{\rm F}| = 0$, but more importantly, $x_5 = x_6 = 2$ for all cases and $x_{11} = 2$ for all cases except one. The remaining ones each have 1 distributed respectively as $x_{10} =1, x_{11} = 1, x_{12} = 1, x_{13} = 1, x_{14} = 1, x_{15} = 1$ and $x_{16} =1$. The distribution of 1's suggests that $l_{14} =1$ in \eqref{brittbaba007}. The distribution of 2's from $x_5= x_6 = x_{11} = 2$ suggests that $l_q=4$ and we choose $2\left({2\over 3} - \hat\sigma_e(t)\right)$ from either $l_i = 2$ 
or $n_2 = 4$ in \eqref{brittbaba007}. Thus the $\bar{g}_s$ scalings of  quantum terms from the non-local BBS instantons contributing to the acceleration of the universe come from:
\bg\label{almadel}
a_{\rm F} = 0, ~\Big(l_{14} = 1, ~~ \sum_{q = 29}^{33} l_q = 4, ~~\sum_{1 = 1}^{13} l_i = 2\Big), ~ \Big(l_{14} = 1, ~~ \sum_{q = 29}^{33} l_q = 4, ~~n_2 = 4\Big), \nd
which confirms what we had earlier in \eqref{isaferr3} using simple counting. Our result in \eqref{almadel} is a consequence of a  more exhaustive search that not only confirms our earlier analysis, but also tells us that the non-locality factor should always be $\bar{g}_s$ independent in \eqref{omameys}, and therefore cannot be made arbitrarily complicated. Question is: what happens if we flip the sign of the $x_{18}$ term? Could this allow non-zero values for $x_{18}$ for all the seven cases from \eqref{melisrain}? To see this let us revisit \eqref{timcanpost5}, but now with a relative {\it plus} sign, {\it i.e.} express the equation as:
\bg\label{timcanpost52}
{14\over 3} - 4\hat\sigma_e(t) -\hat\zeta_e(t) - {\hat\alpha_e(t)\over 2} - {\hat\beta_e(t)\over 2} = \sum_{i = 1}^{17} x_i c_i + {2\over 3} x_{18}, \nd
and ask how many solutions are possible for the 18-tuple, and whether there are some with non-zero values of $x_{18}$. The answer turns out to be {\it nineteen} non-negative integer solutions:

{\scriptsize
\bg\label{chototite}
(x_1, x_2, ..., x_{17}, x_{18}): && (0,0,0,0,0,0,0,0,0,0,4,0,0,1,1,1,0,0),~~
(0,0,0,0,0,0,0,0,0,0,4,1,1,1,0,0,0,0)\nonumber\\
&&
(0,0,0,0,0,0,0,0,2,0,3,0,0,0,1,1,0,0),~~
(0,0,0,0,0,0,0,0,2,0,3,1,1,0,0,0,0,0)\nonumber\\
&&
(0,0,0,0,0,0,0,2,0,0,4,1,0,0,0,0,0,0), ~~
(0,0,0,0,0,0,2,0,0,0,4,0,1,0,0,0,0,0)\nonumber\\
&&
(0,0,0,0,0,1,0,1,1,0,3,1,0,0,0,0,0,0),~~
(0,0,0,0,0,2,0,0,0,0,3,1,0,1,0,0,0,0)\nonumber\\
&&
(0,0,0,0,0,2,0,0,2,0,2,1,0,0,0,0,0,0), ~~
(0,0,0,0,0,2,2,0,0,0,3,0,0,0,0,0,0,0)\nonumber\\
&&
(0,0,0,0,1,0,1,0,1,0,3,0,1,0,0,0,0,0), ~~
(0,0,0,0,1,1,1,1,0,0,3,0,0,0,0,0,0,0)\nonumber\\
&&
(0,0,0,0,1,2,1,0,1,0,2,0,0,0,0,0,0,0), ~~
(0,0,0,0,2,0,0,0,0,0,3,0,1,1,0,0,0,0)\nonumber\\
&&
(0,0,0,0,2,0,0,0,2,0,2,0,1,0,0,0,0,0), ~~
(0,0,0,0,2,0,0,2,0,0,3,0,0,0,0,0,0,0)\nonumber\\
&&
(0,0,0,0,2,1,0,1,1,0,2,0,0,0,0,0,0,0), ~~
(0,0,0,0,2,2,0,0,0,0,2,0,0,1,0,0,0,0)\nonumber\\
&&
(0,0,0,0,2,2,0,0,2,0,1,0,0,0,0,0,0,0),\nd}
but all with vanishing $x_{18}$. Comparing with \eqref{tundegulab} we see that they overlap with the nineteen vanishing $x_{18}$ solutions therein. We can now ask the same questions for all the remaining six equations on the RHS of \eqref{melisrain}, but the answers are always the same: solutions exist only for $x_{18} = 3|a_{\rm F}|=0$ and expectedly overlap with the subsets of the ones with vanishing $x_{18}$ earlier. The contributions are from the {\it septic} order curvature terms on the world-volume of the non-local BBS instantons.

What about the KKLT five-brane instantons, namely the five-brane instantons wrapping the toroidal direction $\xoxo$ and four-cycles on the base ${\cal M}_4 \times {\cal M}_2$? Comparing the LHS of \eqref{melisrain} and the RHS of \eqref{cyclebaaz7}, it is easy to rule out any solutions because of the $\hat\eta_e(t)$ parameter appearing only on one side. This is clear from the following equation:

{\scriptsize
\bg\label{melisrain3}
&&  x_1 \left({4\over 3} - {\hat\zeta_e(t)\over 2}\right) +
x_2 \left( {5\over 3} + {\hat\alpha_e(t) + \hat\beta_e(t)\over 4} - \hat\zeta_e(t)\right) + x_3 \left({5\over 3} - {\hat\alpha_e(t)\over 2} - {\hat\zeta_e(t)\over 2}\right) 
+ x_4 \left({5\over 3} - {\hat\beta_e(t)\over 2} - {\hat\zeta_e(t)\over 2}\right) \nonumber\\
&+& x_5\left({2\over 3} - {\hat\sigma_e(t)\over 2} - {\hat\alpha_e(t)\over 2}\right) + x_6\left({2\over 3} - {\hat\sigma_e(t)\over 2} - {\hat\beta_e(t)\over 2}\right)
+ x_7
\left( {2\over 3} - {\hat\alpha_e(t)\over 2} + {\hat\alpha_e(t) + \hat\beta_e(t)\over 4} - {\hat\zeta_e(t)\over 2}\right) \nonumber\\
&+ & x_8
\left( {2\over 3} - {\hat\beta_e(t)\over 2} + {\hat\alpha_e(t) + \hat\beta_e(t)\over 4} - {\hat\zeta_e(t)\over 2}\right)
+ x_9\left( {2\over 3} + {\hat\alpha_e(t) + \hat\beta_e(t)\over 4} - {\hat\sigma_e(t)\over 2} - {\hat\zeta_e(t)\over 2}\right)
+ x_{10}\left({8\over 3} - \hat\zeta_e(t)\right)\nonumber\\ 
&+& x_{11}\left({2\over 3} - \hat\sigma_e(t)\right) + x_{12}\left({2\over 3} - \hat\alpha_e(t)\right) + x_{13}\left({2\over 3} - \hat\beta_e(t)\right) + x_{14}\left({2\over 3} + {\hat\alpha_e(t) + \hat\beta_e(t)\over 2} - \hat\zeta_e(t)\right)\nonumber\\
& + & x_{15}\left({2\over 3} - 2\hat\alpha_e(t) + \hat\beta_e(t)\right)
+
x_{16}\left({2\over 3} - 2\hat\beta_e(t) + \hat\alpha_e(t)\right)
+ x_{17}\left( {5\over 3} + {\hat\sigma_e(t)\over 2} - {\hat\zeta_e(t)\over 2}\right) \nonumber\\
& = & \begin{cases} ~{8\over 3} - \hat\zeta_e(t)- n\hat\sigma_e(t) - {n\hat\eta_e(t)\over 2}- na_{\rm F} - \begin{cases} ~n\hat\sigma_e(t)\\~~~\\
~{n\hat\alpha_e(t) + n\hat\beta_e(t)\over 2}
\end{cases}\\
~{2\over 3} - (\hat\alpha_e(t), \hat\beta_e(t)) \pm \hat\alpha_e(t) \mp \hat\beta_e(t)- n\hat\sigma_e(t) - {n\hat\eta_e(t)\over 2}- na_{\rm F}- \begin{cases} ~n\hat\sigma_e(t)\\~~~\\
~{n\hat\alpha_e(t) + n\hat\beta_e(t)\over 2}
\end{cases}\\ 
~{2\over 3} + {\hat\alpha_e(t)+ \hat\beta_e(t)\over 2} - \hat\zeta_e(t)  
- n\hat\sigma_e(t) - {n\hat\eta_e(t)\over 2}- na_{\rm F}- \begin{cases} ~n\hat\sigma_e(t)\\~~~\\
~{n\hat\alpha_e(t) + n\hat\beta_e(t)\over 2}
\end{cases}\\ 
~ {2\over 3} - (\hat\alpha_e(t), \hat\beta_e(t))- n\hat\sigma_e(t) - {n\hat\eta_e(t)\over 2}- na_{\rm F}- \begin{cases} ~n\hat\sigma_e(t)\\~~~\\
~{n\hat\alpha_e(t) + n\hat\beta_e(t)\over 2}
\end{cases}\\
~{2\over 3} - (n+1)\hat\sigma_e(t) - {n\hat\eta_e(t)\over 2} - na_{\rm F} -\begin{cases} ~n\hat\sigma_e(t)\\ ~~~\\
~{n\hat\alpha_e(t) + n\hat\beta_e(t)\over 2}
\end{cases}
\end{cases} \nd}
where it is easy to see that for $(\hat\eta_e(t), a_{\rm F}) > 0$ no solutions exist. In fact even if we make $\hat\eta_e(t) = 0$ but keep $a_{\rm F} > 0$, there are no solutions at least in the curvature and the derivative sectors of \eqref{botsuga} and \eqref{brittbaba007}. The situation does not improve if we keep $\hat\eta_e(t) = a_{\rm F} = 0$.  Solutions fail to arise there too. So it seems solutions could only exist if $\hat\eta_e(t) = 0$ and $a_{\rm F} = -|a_{\rm F}| = -{x_{18}\over 3}$. To verify this we will choose $n = 2$ on the RHS of \eqref{melisrain3}, and track the rows from the bottom to the top. It is easy to see that:
\bg\label{oliveoma1}
{2\over 3} - 3\hat\sigma_e(t) - \hat\alpha_e(t) - \hat\beta_e(t) = 
\begin{cases} ~(0,0,0,0,0,0,0,0,0,0,3,0,0,0,1,1,0,4)\\
~(0,0,0,0,0,0,0,0,0,0,3,1,1,0,0,0,0,4)\\
~(0,0,0,0,0,2,0,0,0,0,2,1,0,0,0,0,0,4)\\
~{\red(0,0,0,0,2,2,0,0,0,0,1,0,0,0,0,0,0,4)}\\
~(0,0,0,0,2,0,0,0,0,0,2,0,1,0,0,0,0,4)
\end{cases}\nd
suggesting that the scaling of the non-locality function \eqref{omameys} should be $x_{18} = 3|a_{\rm F}| = 4$. The fact that it comes with an overall minus sign helps us to get solutions for the 18-tuple (at least for this row in \eqref{melisrain3}). The importance of the solution in {\red red} here, and in the following, will be elaborated soon. For the same row, but with a different wrapping of the five-brane instantons on ${\cal M}_4 \times {\cal M}_2$, we get:
\bg\label{oliveoma2}
{2\over 3} - 5\hat\sigma_e(t) =  {\red(0,0,0,0,0,0,0,0,0,0,5,0,0,0,0,0,0,4)}, \nd
which again suggests $x_{18} = 3|a_{\rm F}| = 4$. The result is now $x_{11} = 5$, and this appears to be the only allowed solution. For the next row there are two possibilities for each of the two possible embeddings of the instantons. For the first embedding, the first possibility gives us the following choices:
\bg\label{oliveoma3}
{2\over 3} - 2\hat\sigma_e(t) - 2\hat\alpha_e(t) - \hat\beta_e(t) = \begin{cases} ~(0,0,0,0,0,0,0,0,0,0,2,0,2,0,1,0,0,4)\\
~(0,0,0,0,0,0,0,0,0,0,2,1,0,0,1,1,0,4)\\
~(0,0,0,0,0,0,0,0,0,0,2,2,1,0,0,0,0,4)\\
~(0,0,0,0,0,2,0,0,0,0,1,0,1,0,1,0,0,4)\\
~(0,0,0,0,0,2,0,0,0,0,1,2,0,0,0,0,0,4)\\
~(0,0,0,0,0,4,0,0,0,0,0,0,0,0,1,0,0,4)\\
~(0,0,0,0,2,0,0,0,0,0,1,0,0,0,1,1,0,4)\\
~(0,0,0,0,2,0,0,0,0,0,1,1,1,0,0,0,0,4)\\
~{\red(0,0,0,0,2,2,0,0,0,0,0,1,0,0,0,0,0,4)}\\
~(0,0,0,0,4,0,0,0,0,0,0,0,1,0,0,0,0,4) \end{cases} \nd
which are {\it ten} possible solutions all with the non-locality factor 
$x_{18} = 3|a_{\rm F}| = 4$. The second possibility with the first embedding now gives us the following:
\bg\label{oliveoma4}
{2\over 3} - 2\hat\sigma_e(t) - 2\hat\beta_e(t) - \hat\alpha_e(t) = \begin{cases} ~ (0,0,0,0,0,0,0,0,0,0,2,0,1,0,1,1,0,4)\\
~(0,0,0,0,0,0,0,0,0,0,2,1,2,0,0,0,0,4)\\
~(0,0,0,0,0,0,0,0,0,0,2,2,0,0,0,1,0,4)\\
~(0,0,0,0,0,2,0,0,0,0,1,0,0,0,1,1,0,4)\\
~(0,0,0,0,0,2,0,0,0,0,1,1,1,0,0,0,0,4)\\
~(0,0,0,0,0,4,0,0,0,0,0,1,0,0,0,0,0,4)\\
~(0,0,0,0,2,0,0,0,0,0,1,0,2,0,0,0,0,4)\\
~(0,0,0,0,2,0,0,0,0,0,1,1,0,0,0,1,0,4)\\
~{\red(0,0,0,0,2,2,0,0,0,0,0,0,1,0,0,0,0,4)}\\
~(0,0,0,0,4,0,0,0,0,0,0,0,0,0,0,1,0,4)\end{cases} \nd
with the {\it ten} solutions differing slightly from what we had in \eqref{oliveoma3}. The non-locality factor remains the same and one may easily verify these solutions by plugging them on the LHS of \eqref{melisrain3}. For the second kind of embedding, the two possibilities give us:
\bg\label{oliveoma5}
\theta_{nl} = \begin{cases}~{2\over 3} - 4\hat\sigma_e(t) - \hat\alpha_e(t) = \begin{cases}~
{\red(0,0,0,0,0,0,0,0,0,0,4,1,0,0,0,0,0,4)}\\
~(0,0,0,0,2,0,0,0,0,0,3,0,0,0,0,0,0,4)\end{cases}\\~~\\
~{2\over 3} - 4\hat\sigma_e(t) - \hat\beta_e(t) = \begin{cases}~ {\red(0,0,0,0,0,0,0,0,0,0,4,0,1,0,0,0,0,4)}\\
~(0,0,0,0,0,2,0,0,0,0,3,0,0,0,0,0,0,4) \end{cases}
\end{cases} \nd
both with the same non-locality factors of $x_{18} = 3|a_{\rm F}| = 4$. We see that the two sets of solutions differ slightly in the placement of the factors over the 18-tuples. But these differences will be crucial and we will elaborate on them soon. Meanwhile, our next item on the list is the one that has $\hat\zeta_e(t)$ on the RHS of \eqref{melisrain3}. For the first kind of embedding, this gives us the following set:

{\footnotesize
\bg\label{olivelaal}
{2\over 3} - 2\hat\sigma_e(t) - {\hat\alpha_e(t)\over 2} - {\hat\beta_e(t)\over 2} - \hat\zeta_e(t) = \begin{cases}~ 
(0, 0, 0, 0, 0, 2, 0, 0, 2, 0, 0, 1, 0, 0, 0, 0, 0, 4)\\
~(0, 0, 0, 0, 1, 1, 0, 0, 0, 0, 0, 0, 0, 0, 0, 0, 2, 6)\\
~(0, 0, 0, 0, 1, 2, 1, 0, 1, 0, 0, 0, 0, 0, 0, 0, 0, 4)\\
~(0, 0, 0, 0, 2, 0, 0, 0, 2, 0, 0, 0, 1, 0, 0, 0, 0, 4)\\
~(0, 0, 0, 0, 2, 1, 0, 1, 1, 0, 0, 0, 0, 0, 0, 0, 0, 4)\\
~{\red(0, 0, 0, 0, 2, 2, 0, 0, 0, 0, 0, 0, 0, 1, 0, 0, 0, 4)}\\
~(0, 0, 0, 0, 0, 0, 0, 0, 2, 0, 1, 1, 1, 0, 0, 0, 0, 4)\\
~(0, 0, 0, 0, 0, 0, 0, 0, 2, 0, 1, 0, 0, 0, 1, 1, 0, 4)\\
~(0, 0, 0, 0, 0, 1, 0, 1, 1, 0, 1, 1, 0, 0, 0, 0, 0, 4)\\
~(0, 0, 1, 0, 0, 1, 0, 0, 0, 0, 1, 0, 0, 0, 0, 0, 1, 6)\\
~(0, 0, 0, 0, 0, 2, 0, 0, 0, 0, 1, 1, 0, 1, 0, 0, 0, 4)\\
~(0, 0, 0, 0, 0, 2, 2, 0, 0, 0, 1, 0, 0, 0, 0, 0, 0, 4)\\
~(0, 0, 0, 0, 1, 0, 1, 0, 1, 0, 1, 0, 1, 0, 0, 0, 0, 4)\\
~(0, 0, 0, 1, 1, 0, 0, 0, 0, 0, 1, 0, 0, 0, 0, 0, 1, 6)\\
~(0, 0, 0, 0, 1, 1, 1, 1, 0, 0, 1, 0, 0, 0, 0, 0, 0, 4)\\
~(2, 0, 0, 0, 1, 1, 0, 0, 0, 0, 1, 0, 0, 0, 0, 0, 0, 6)\\
~(0, 0, 0, 0, 2, 0, 0, 0, 0, 0, 1, 0, 1, 1, 0, 0, 0, 4)\\
~(0, 0, 0, 0, 2, 0, 0, 2, 0, 0, 1, 0, 0, 0, 0, 0, 0, 4)\\
~(0, 0, 0, 0, 0, 0, 0, 0, 0, 0, 2, 1, 1, 1, 0, 0, 0, 4)\\
~(0, 0, 0, 0, 0, 0, 0, 2, 0, 0, 2, 1, 0, 0, 0, 0, 0, 4)\\
~(0, 0, 0, 0, 0, 0, 2, 0, 0, 0, 2, 0, 1, 0, 0, 0, 0, 4)\\
~(0, 0, 0, 0, 0, 0, 0, 0, 0, 0, 2, 0, 0, 1, 1, 1, 0, 4)\\
~(0, 0, 1, 1, 0, 0, 0, 0, 0, 0, 2, 0, 0, 0, 0, 0, 0, 6)\\
(~0, 0, 0, 0, 1, 1, 0, 0, 0, 1, 1, 0, 0, 0, 0, 0, 0, 6)\end{cases} \nd}
which are in total {\it twenty one} solutions for the 18-tuple, most with the non-locality factor of $x_{18} = 3|a_{\rm F}| = 4$, but two with non-locality factor of $x_{18} = 3|a_{\rm F}| = 6$. For the second kind of embedding, we get {\it two} solutions for the 18-tuple:

{\footnotesize
\bg\label{olivered}
{2\over 3} - 4\hat\sigma_e(t) + {\hat\alpha_e(t) + \hat\beta_e(t)\over 2} - \hat\zeta_e(t) = \begin{cases}~ {\red(0,0,0,0,0,0,0,0,0,0,4,0,0,1,0,0,0,4)}\\
~(0,0,0,0,0,0,0,0,2,0,3,0,0,0,0,0,0,4)\end{cases} \nd}
with the non-locality factor $x_{18} = 3|a_{\rm F}| = 4$ as before. For the second row from the top on the RHS of \eqref{melisrain}, and for the first kind of embedding of the five-brane instantons, we find that there are four possible choices of the $\theta_{nl}$ scalings from \eqref{brittbaba007} out of which two of them have already been dealt in \eqref{oliveoma3} and \eqref{oliveoma4}. This leaves only two distinct scalings, where the first one gives the following {\it eight} solutions:
\bg\label{olivered2}
{2\over 3} - 2\hat\sigma_e(t) - 3\hat\alpha_e(t) = \begin{cases}~
(0,0,0,0,0,0,0,0,0,0,2,0,0,0,2,1,0,4)\\
~(0,0,0,0,0,0,0,0,0,0,2,1,1,0,1,0,0,4)\\
~(0,0,0,0,0,0,0,0,0,0,2,3,0,0,0,0,0,4)\\
~(0,0,0,0,0,2,0,0,0,0,1,1,0,0,1,0,0,4)\\
~(0,0,0,0,2,0,0,0,0,0,1,0,1,0,1,0,0,4)\\
~(0,0,0,0,2,0,0,0,0,0,1,2,0,0,0,0,0,4)\\
~{\red(0,0,0,0,2,2,0,0,0,0,0,0,0,0,1,0,0,4)}\\
~(0,0,0,0,4,0,0,0,0,0,0,1,0,0,0,0,0,4)\end{cases} \nd
complete with non-locality factor $x_{18} = 3|a_{\rm F}| = 4$. The second one appears by replacing $\hat\alpha_e(t)$ by $\hat\beta_e(t)$ on the LHS of \eqref{olivered2}. Making this change leads to the following {\it eight} solutions for the 18-tuple:
\bg\label{olivered3}
{2\over 3} - 2\hat\sigma_e(t) - 3\hat\beta_e(t) = \begin{cases}~
(0,0,0,0,0,0,0,0,0,0,2,0,0,0,1,2,0,4)\\
~(0,0,0,0,0,0,0,0,0,0,2,0,3,0,0,0,0,4)\\
~(0,0,0,0,0,0,0,0,0,0,2,1,1,0,0,1,0,4)\\
~(0,0,0,0,0,2,0,0,0,0,1,0,2,0,0,0,0,4)\\
~(0,0,0,0,0,2,0,0,0,0,1,1,0,0,0,1,0,4)\\
~(0,0,0,0,0,4,0,0,0,0,0,0,1,0,0,0,0,4)\\
~{\red(0,0,0,0,2,2,0,0,0,0,0,0,0,0,0,1,0,4)}\\
~(0,0,0,0,2,0,0,0,0,0,1,0,1,0,0,1,0,4)
\end{cases} \nd
again complete with the non-locality factor $x_{18} = 3|a_{\rm F}| = 4$. For the second kind of embedding, we will again have four scalings for $\theta_{nl}$ from \eqref{brittbaba}. Out of which two of them have already been dealt with in \eqref{oliveoma5}. Thus there are only two distinct scalings. Interestingly, each of them only has {\it one} solution for the 18-tuple:

{\footnotesize
\bg\label{olivered5}
\theta_{nl} = \begin{cases} ~{2\over 3} - 4\hat\sigma_e(t) - 2\hat\alpha_e(t) + \hat\beta_e(t) = {\red(0,0,0,0,0,0,0,0,0,0,4,0,0,0,1,0,0,4)}\\~~\\
~{2\over 3} - 4\hat\sigma_e(t) - 2\hat\beta_e(t) + \hat\alpha_e(t) = 
{\red(0,0,0,0,0,0,0,0,0,0,4,0,0,0,0,1,0,4)}\end{cases}\nd}
with the same  non-locality factor $x_{18} = 3|a_{\rm F}| = 4$. The importance of the solutions in {\red red} will be elaborated soon. Our final set appears in the topmost row on the RHS of \eqref{melisrain3}. In fact this is the second scaling that has $\hat\zeta_e(t)$ explicitly on the RHS of \eqref{melisrain3}. Solving this leads to {\it forty six} solutions for the 18-tuple:

{\footnotesize
\bg\label{olivered10}
{8\over 3} - 2\hat\sigma_e(t) - \hat\alpha_e(t) - \hat\beta_e(t) - \hat\zeta_e(t) = \begin{cases}~ (0, 0, 0, 0, 0, 2, 0, 0, 0, 0, 0, 1, 0, 0, 0, 0, 2, 4)\\
~(0, 0, 0, 0, 0, 3, 1, 0, 1, 0, 0, 1, 0, 0, 0, 0, 0, 2)\\
~(0, 0, 0, 0, 1, 1, 0, 0, 2, 0, 0, 1, 1, 0, 0, 0, 0, 2)\\
~(0, 0, 0, 0, 1, 1, 0, 0, 2, 0, 0, 0, 0, 0, 1, 1, 0, 2)\\
~(0, 0, 0, 0, 1, 2, 0, 1, 1, 0, 0, 1, 0, 0, 0, 0, 0, 2)\\
~(0, 0, 1, 0, 1, 2, 0, 0, 0, 0, 0, 0, 0, 0, 0, 0, 1, 4)\\
~(0, 0, 0, 0, 1, 3, 0, 0, 0, 0, 0, 1, 0, 1, 0, 0, 0, 2)\\
~(0, 0, 0, 0, 1, 3, 2, 0, 0, 0, 0, 0, 0, 0, 0, 0, 0, 2)\\
~(0, 0, 0, 0, 2, 0, 0, 0, 0, 0, 0, 0, 1, 0, 0, 0, 2, 4)\\
~(0, 0, 0, 0, 2, 1, 1, 0, 1, 0, 0, 0, 1, 0, 0, 0, 0, 2)\\
~(0, 0, 0, 1, 2, 1, 0, 0, 0, 0, 0, 0, 0, 0, 0, 0, 1, 4)\\
~(0, 0, 0, 0, 2, 2, 1, 1, 0, 0, 0, 0, 0, 0, 0, 0, 0, 2)\\
~(2, 0, 0, 0, 2, 2, 0, 0, 0, 0, 0, 0, 0, 0, 0, 0, 0, 4)\\
~(0, 0, 0, 0, 3, 0, 0, 1, 1, 0, 0, 0, 1, 0, 0, 0, 0, 2)\\
~(0, 0, 0, 0, 3, 1, 0, 0, 0, 0, 0, 0, 1, 1, 0, 0, 0, 2)\\
~(0, 0, 0, 0, 3, 1, 0, 2, 0, 0, 0, 0, 0, 0, 0, 0, 0, 2)\\
~(0, 0, 0, 0, 0, 0, 0, 0, 0, 0, 1, 1, 1, 0, 0, 0, 2, 4)\\
~(0, 0, 0, 0, 0, 0, 0, 0, 0, 0, 1, 0, 0, 0, 1, 1, 2, 4)\\
~(0, 0, 0, 0, 0, 1, 1, 0, 1, 0, 1, 1, 1, 0, 0, 0, 0, 2)\\
~(0, 0, 0, 0, 0, 1, 1, 0, 1, 0, 1, 0, 0, 0, 1, 1, 0, 2)\\
~(0, 0, 0, 1, 0, 1, 0, 0, 0, 0, 1, 1, 0, 0, 0, 0, 1, 4)\\
~(0, 0, 0, 0, 0, 2, 1, 1, 0, 0, 1, 1, 0, 0, 0, 0, 0, 2)\\
~(0, 0, 2, 0, 0, 2, 0, 0, 0, 0, 1, 0, 0, 0, 0, 0, 0, 4)\\
~(2, 0, 0, 0, 0, 2, 0, 0, 0, 0, 1, 1, 0, 0, 0, 0, 0, 4)\\
~(0, 0, 0, 0, 1, 0, 0, 1, 1, 0, 1, 1, 1, 0, 0, 0, 0, 2)\\
~(0, 0, 0, 0, 1, 0, 0, 1, 1, 0, 1, 0, 0, 0, 1, 1, 0, 2)\\
~{\red(0, 0, 0, 0, 2, 2, 0, 0, 0, 1, 0, 0, 0, 0, 0, 0, 0, 4)}\\
~(0, 0, 1, 0, 1, 0, 0, 0, 0, 0, 1, 0, 1, 0, 0, 0, 1, 4)\\
~(0, 0, 0, 0, 1, 1, 0, 0, 0, 0, 1, 1, 1, 1, 0, 0, 0, 2)\\
~(0, 0, 0, 0, 1, 1, 0, 2, 0, 0, 1, 1, 0, 0, 0, 0, 0, 2)\\
~(0, 0, 0, 0, 1, 1, 2, 0, 0, 0, 1, 0, 1, 0, 0, 0, 0, 2)\\
~(0, 0, 0, 0, 1, 1, 0, 0, 0, 0, 1, 0, 0, 1, 1, 1, 0, 2)\\
~(0, 0, 1, 1, 1, 1, 0, 0, 0, 0, 1, 0, 0, 0, 0, 0, 0, 4)\\
~(0, 0, 0, 0, 2, 0, 1, 1, 0, 0, 1, 0, 1, 0, 0, 0, 0, 2)\\
~(0, 0, 0, 2, 2, 0, 0, 0, 0, 0, 1, 0, 0, 0, 0, 0, 0, 4)\\
~(2, 0, 0, 0, 2, 0, 0, 0, 0, 0, 1, 0, 1, 0, 0, 0, 0, 4)\\
~(0, 0, 0, 0, 0, 0, 1, 1, 0, 0, 2, 1, 1, 0, 0, 0, 0, 2)\\
~(0, 0, 0, 0, 0, 0, 0, 0, 0, 1, 2, 0, 0, 0, 1, 1, 0, 4)\\
~(0, 0, 0, 0, 0, 0, 1, 1, 0, 0, 2, 0, 0, 0, 1, 1, 0, 2)\\
~(0, 0, 0, 2, 0, 0, 0, 0, 0, 0, 2, 1, 0, 0, 0, 0, 0, 4)\\
~(0, 0, 2, 0, 0, 0, 0, 0, 0, 0, 2, 0, 1, 0, 0, 0, 0, 4)\\
~(2, 0, 0, 0, 0, 0, 0, 0, 0, 0, 2, 1, 1, 0, 0, 0, 0, 4)\\
~(2, 0, 0, 0, 0, 0, 0, 0, 0, 0, 2, 0, 0, 0, 1, 1, 0, 4)\\
~(0, 0, 0, 0, 0, 2, 0, 0, 0, 1, 1, 1, 0, 0, 0, 0, 0, 4)\\
~(0, 0, 0, 0, 2, 0, 0, 0, 0, 1, 1, 0, 1, 0, 0, 0, 0, 4)\\
~(0, 0, 0, 0, 0, 0, 0, 0, 0, 1, 2, 1, 1, 0, 0, 0, 0, 4)
\end{cases} \nd}
where we now see that the non-locality factor can be $x_{18} = 3|a_{\rm F}| = 4$ and $x_{18} = 3|a_{\rm F}| = 2$. There is one special case marked in {\red red} that we will elaborate soon. Meanwhile, for the other possible embedding of the KKLT instantons, we find that:
\bg\label{olivered22}
{8\over 3} - 4\hat\sigma_e(t) - \hat\zeta_e(t) = \begin{cases}~ (0, 0, 0, 0, 1, 1, 0, 0, 2, 0, 2, 0, 0, 0, 0, 0, 0, 2)\\
~(0, 0, 0, 0, 0, 0, 0, 0, 0, 0, 3, 0, 0, 0, 0, 0, 2, 4)\\
~(0, 0, 0, 0, 0, 1, 1, 0, 1, 0, 3, 0, 0, 0, 0, 0, 0, 2)\\
~(0, 0, 0, 0, 1, 0, 0, 1, 1, 0, 3, 0, 0, 0, 0, 0, 0, 2)\\
~{\red(0, 0, 0, 0, 0, 0, 0, 0, 0, 1, 4, 0, 0, 0, 0, 0, 0, 4)}\\
~(0, 0, 0, 0, 1, 1, 0, 0, 0, 0, 3, 0, 0, 1, 0, 0, 0, 2)\\
~(0, 0, 0, 0, 0, 0, 1, 1, 0, 0, 4, 0, 0, 0, 0, 0, 0, 2)\\
~(2, 0, 0, 0, 0, 0, 0, 0, 0, 0, 4, 0, 0, 0, 0, 0, 0, 4)
\end{cases} \nd
totaling to {\it eight} non-negative integer solutions for the 18-tuple. Note that only two of them have the non-locality factor of $x_{18} = 3|a_{\rm F}| = 4$. The remaining are either with $x_{18} = 3|a_{\rm F}| = 2$ or with $x_{18} = 3|a_{\rm F}| = 6$.

Let us now collect all the sets of solutions marked in {\red red}. We will first go with the KKLT instantons that wrap the toroidal direction $\xoxo$ and the four-cycle ${\cal M}_4$ on the base. The solutions for this embedding are given in \eqref{oliveoma2}, \eqref{oliveoma5}, \eqref{olivered}, \eqref{olivered5} and \eqref{olivered22}. They are represented here as:

{\footnotesize
\bg\label{oliveomachu}
&& (0,0,0,0,0,0,0,0,0,{\red 1},4,0,0,0,0,0,0,4) = {8\over 3} - 4\hat\sigma_e(t) -\hat\zeta_e(t)\nonumber\\
&& (0,0,0,0,0,0,0,0,0,0,5,0,0,0,0,0,0,4) = {2\over 3} - 5\hat\sigma_e(t)\nonumber\\
&& (0,0,0,0,0,0,0,0,0,0,4,{\red 1},0,0,0,0,0,4) = {2\over 3} - 4\hat\sigma_e(t) - \hat\alpha_e(t)\nonumber\\
&& (0,0,0,0,0,0,0,0,0,0,4,0,{\red 1},0,0,0,0,4) = {2\over 3} - 4\hat\sigma_e(t) - \hat\beta_e(t)\nonumber\\
&& (0,0,0,0,0,0,0,0,0,0,4,0,0,{\red 1},0,0,0,4) = {2\over 3} - 4\hat\sigma_e(t) + {\hat\alpha_e(t) - \hat\beta_e(t)\over 2} - \hat\zeta_e(t)\nonumber\\
&& (0,0,0,0,0,0,0,0,0,0,4,0,0,0,{\red 1},0,0,4) = {2\over 3} - 4\hat\sigma_e(t) -2\hat\alpha_e(t) + \hat\beta_e(t)\nonumber\\
&& (0,0,0,0,0,0,0,0,0,0,4,0,0,0,0,{\red 1},0,4) = {2\over 3} - 4\hat\sigma_e(t) -2\hat\beta_e(t) + \hat\alpha_e(t), \nd}
where if we write $5 = 4 + 1$, the distribution of ${\red 1}$ diagonally across \eqref{oliveomachu} suggests that there is a pattern. Looking at \eqref{brittbaba007} it is easy to see that the solutions in \eqref{oliveomachu} can be captured by the following choices of the exponents in \eqref{botsuga}:
\bg\label{evabsamtag}
a_{\rm F} = -{4\over 3}, ~~~~\Big(l_{14} = 1, ~~\sum_{i = 1}^{13} l_i = 4\Big), ~~~~\Big(l_{14} = 1, ~~n_2 = 8\Big), \nd
where we choose ${2\over 3} - \hat\sigma_e(t)$ from the coefficients of
$l_i\in\mathbb{Z}^+$ with $1 \le i \le 13$ in \eqref{brittbaba007}. \eqref{evabsamtag}  implies that the quantum terms from the non-local KKLT five-brane instantons $-$ with world-volume along $\xoxo\times {\cal M}_4$ $-$ contribute ${\bf R}^5$ and 
$\square^4 {\bf R}$ terms to drive the acceleration of our universe provided $a_{\rm F} = -{4\over 3}$ in \eqref{omameys}. Similarly for the instantons wrapping $\xoxo \times {\cal M}_2 \times {\bf \Sigma}_2$, where ${\bf \Sigma}_2$ is a two-cycle inside ${\cal M}_4$ {\it i.e.}  ${\bf \Sigma}_2 \in {\cal M}_4$, we get:

{\footnotesize
\bg\label{olivechu1}
&& (0,0,0,0,2,2,0,0,0,{\red 1},0,0,0,0,0,0,0,4)={8\over 3} - 2\hat\sigma_e(t) - \hat\alpha_e(t) - \hat\beta_e(t) - \hat\zeta_e(t)\nonumber\\
&& (0,0,0,0,2,2,0,0,0,0,{\red 1},0,0,0,0,0,0,4)={2\over 3} - 3\hat\sigma_e(t) - \hat\alpha_e(t) - \hat\beta_e(t)\nonumber\\
&& (0,0,0,0,2,2,0,0,0,0,0,{\red 1},0,0,0,0,0,4)={2\over 3} - 2\hat\sigma_e(t) - 2\hat\alpha_e(t) - \hat\beta_e(t)\nonumber\\
&& (0,0,0,0,2,2,0,0,0,0,0,0,{\red 1},0,0,0,0,4)={2\over 3} - 2\hat\sigma_e(t) - 2\hat\beta_e(t) - \hat\alpha_e(t)\nonumber\\
&& (0,0,0,0,2,2,0,0,0,0,0,0,0,{\red 1},0,0,0,4)={2\over 3} - 2\hat\sigma_e(t) - {\hat\alpha_e(t)\over 2} - {\hat\beta_e(t)\over 2} - \hat\zeta_e(t)\nonumber\\
&& (0,0,0,0,2,2,0,0,0,0,0,0,0,0,{\red 1},0,0,4)={2\over 3} - 2\hat\sigma_e(t) - 3\hat\alpha_e(t)\nonumber\\
&& (0,0,0,0,2,2,0,0,0,0,0,0,0,0,0,{\red 1},0,4)={2\over 3} - 2\hat\sigma_e(t) - 3\hat\beta_e(t), \nd}
where the distribution of ${\red 1}$ suggests similar pattern as in \eqref{evabsamtag}, whereas the distributions of 2's as $x_5 = x_6 = 2$ suggest a different choice of the $l_j$ from \eqref{brittbaba007}. It is not too hard to see that the solutions in \eqref{olivechu1} may be captured by:
\bg\label{olivechu2}
a_{\rm F} = -{4\over 3}, ~~~~ l_{14} = 1, ~~~~ \sum_{q = 29}^{33} l_q = 4, \nd
which again accounts for ${\bf R}^5$ contributions from the non-local KKLT type instantons wrapping $\xoxo \times {\cal M}_2 \times {\bf \Sigma}_2$. As we saw earlier, solutions with $a_{\rm F} \ge 0$ are not possible for both the allowed embeddings of the non-local KKLT type instantons if we are in the curvatures and the derivatives sector. Including fluxes or fermions could possibly change the outcomes. We will not investigate those cases here and leave them for our diligent readers to complete. 

There are however a few unsatisfactory features of the KKLT instanton configurations that we study here. First is the presence of negative $a_{\rm F}$ which implies, from \eqref{omameys}, that the non-localities would increase with respect to the conformal time. For the BBS instantons this was not the case: the non-locality factor $a_{\rm F}$ was zero, so there was no temporal dependence in \eqref{omameys}. The negative $a_{\rm F}$ for the KKLT case is necessary to compensate for the higher dominant scalings in \eqref{brittbaba007} because of the absence of any relative minus signs with the dominant pieces. 

Secondly, to get solutions in the sector with curvatures and derivative, we had to make $\hat\eta_e(t) = 0$. This is problematic from the viewpoint of the duality chasings that we studied in section \ref{sec4.2.1}. The duality chasing requires us to impose a constraint of the form \eqref{englishgulab} implying that $\hat\zeta_e(t) + \hat\eta_e(t) \approx 0$ where the approximation is due to $g_s$ and ${\rm M}_p$ corrections. If we take this at face-value and impose $\hat\eta_e(t) = -\hat\zeta_e(t)$ on the RHS of \eqref{melisrain3}, we can easily show that there are no solutions possible. To see this let us pick one such term, namely ${2\over 3} - 5\hat\sigma_e(t) - \hat\eta_e(t) = {2\over 3} - 5\hat\sigma_e(t) + \hat\zeta_e(t)$ from the lowest row on the RHS of \eqref{melisrain3}. Comparing this to the LHS of \eqref{melisrain3} gives us the following relation between $x_i$:
\bg\label{blluebook2}
-x_1 -2x_2 - x_3 -x_4- x_7 - x_8 - x_9 -2x_{10} -2x_{14} - x_{17} = +1, \nd
which appears by comparing the $\hat\zeta_e(t)$ coefficients on both sides of \eqref{melisrain3}. It is clear from \eqref{blluebook2} that there are no non-negative integer solutions for $x_i$ possible. Thus the solution set is empty. In fact this line of computation goes through for most of the cases on the RHS of \eqref{melisrain3}. While this argument may serve as a nail in the coffin, it is by no means a conclusive one.
It is possible that in the intermediate IIB duality frame:
\bg\label{bluebook}
\hat\eta'_e(t) + \hat\zeta'_e(t) = 0, ~~~~\hat\eta'_e(t) = \hat\eta_e(t) + {\cal O}(\bar{g}_s, {\rm M}_p), ~~ \hat\zeta'_e(t) = \hat\zeta_e(t) + {\cal O}(\bar{g}_s, {\rm M}_p), \nd
so that we could continue assuming $\hat\eta_e(t) \ne \hat\zeta_e(t)$ in the M-theory frame. Thus 
even if we have $\hat\eta_e(t) = 0$, it does not immediately lead to some isometry breaking in the dual frames, and in particular in the heterotic dual frames. However keeping $\hat\eta_e(t) = 0$ will impose much stronger constraints on $\hat\zeta_e(t)$ already at the level of the M-theory seed configuration. More importantly, there are no solutions possible with KKLT instantons {\it without} involving the non-locality term \eqref{omameys} at least in the curvature and the derivative sectors of \eqref{botsuga} and \eqref{brittbaba007}. Solutions could still exist in the fluxes and the fermionic sectors from \eqref{botsuga2.0}, but we won't study them here.

We can also ask what happens with the space-time filling two-brane instantons wrapping the Euclidean three-dimensions in M-theory. The scalings of the quantum terms are given in \eqref{cyclebaaz8} and, if we add in some non-locality by hand, we can ask how many solutions are possible in the curvatures and the derivatives sectors of \eqref{botsuga}. Taking the lowest row on the RHS of \eqref{cyclebaaz8}, the answer turns out to be {\rm three} solutions for the 18-tuple:

{\footnotesize
\bg\label{polkamey}
{14\over 3} - \hat\sigma_e(t) - {3\hat\zeta_e(t)\over 2} = 
\sum_{i = 1}^{17} x_ic_i - {2x_{18}\over 3} = \begin{cases} ~ (1, 0, 0, 0, 0, 0, 0, 0, 0, 0, 0, 0, 0, 0, 0, 0, 2, 0)\\
~(3, 0, 0, 0, 0, 0, 0, 0, 0, 0, 1, 0, 0, 0, 0, 0, 0, 0)\\
~(1, 0, 0, 0, 0, 0, 0, 0, 0, 1, 1, 0, 0, 0, 0, 0, 0, 0)\end{cases} \nd}
where all the three solutions are problematic because $x_1 = 1, 3$, and therefore lead to Lorentz violating interactions as mentioned earlier in \eqref{palkimey}. This may be seen from the fact that the coefficient $c_1$ of $x_1$ is $c_1 = {4\over 3} - {\hat\zeta_e(t)\over 2}$ and therefore cannot be replaced by $c_{10} = 2c_1$ which is related to curvature\footnote{Recall that the powers of the curvatures appearing in \eqref{botsuga} have to be non-negative integers.}. Flipping the sign of the $x_{18}$ term produces more solutions, but the problem remains:

{\footnotesize
\bg\label{polkamey2}
{14\over 3} - \hat\sigma_e(t) - {3\hat\zeta_e(t)\over 2} = 
\sum_{i = 1}^{17} x_ic_i + {2x_{18}\over 3} = \begin{cases} ~
(1,0,0,0,0,0,0,0,0,1,1,0,0,0,0,0,0,0)\\
~(1,0,0,0,0,0,1,1,0,0,1,0,0,0,0,0,0,2)\\
~(1,0,0,0,0,1,1,0,1,0,0,0,0,0,0,0,0,2)\\
~(1,0,0,0,1,0,0,1,1,0,0,0,0,0,0,0,0,2)\\
~(1,0,0,0,1,1,0,0,0,0,0,0,0,1,0,0,0,2)\\
~(3,0,0,0,0,0,0,0,0,0,1,0,0,0,0,0,0,0)\end{cases}\nd}
where we now have {\it six} solutions, but $x_1$ continues to have odd values of 1 and 3. We see similar problem with the space-time filling eight-brane instanton configurations wrapping ${\rm Euc}({\bf R}^3) \times {\cal M}_4 \times {\cal M}_2$, whose quantum scalings are given by \eqref{cyclebaaz88}. The exhaustive number of solutions for the 18-tuple is now:

{\scriptsize
\bg\label{cascokhel}
{20\over 3} - 3\hat\sigma_e(t) - {\hat\alpha_e(t) + \hat\beta_e(t)\over 2} - {3\hat\zeta_e(t)\over 2} = \sum_{i = 1}^{17}c_i x_i + {n x_{18}\over 3} = \begin{cases} ~ (1,0,1,1,0,0,0,0,0,0,3,0,0,0,0,0,0,0)\\
~(3,0,0,0,1,1,0,0,0,0,2,0,0,0,0,0,0,0)\end{cases} \nd}
as predicted in \eqref{cascovaaram}, where $n \in \mathbb{Z}$. Thus no matter whether or not we include the non-localities (with either sign), \eqref{cascokhel} implies that $x_1= 1, 3$, so no Lorentz invariant operators are possible. What about seven-brane instanton configurations filling the internal eight-manifold ${\cal M}_4 \times {\cal M}_2 \times \xoxo$? Using \eqref{cyclebaaz80} we did not find any solutions. Extending to include non-localities, we find:

{\scriptsize
\bg\label{cascamukh}
{4\over 3} - 3\hat\sigma_e(t) - {\hat\alpha_e(t) + \hat\beta_e(t)\over 2} = \sum_{i = 1}^{17} c_i x_i + {nx_{18}\over 3} = \begin{cases}~ (0,0,0,0,1,1,0,0,0,0,2,0,0,0,0,0,0,4),~n = -1\\
~{\red(0,0,0,0,1,1,0,0,0,0,2,0,0,0,0,0,0,2)},~n = -2\\
~(0,0,0,0,1,1,0,0,0,0,2,0,0,0,0,0,0,1), ~n = -4\end{cases}\nd}
which confirms what we had earlier, namely solutions do not exist if the non-locality factor $x_{18} = 3|a_{\rm F}|$ vanishes and/or if $\hat\eta_e(t)$ is non-zero. Switching on $x_{18}$ and keeping $\hat\eta_e(t) = 0$, we have one solution for each of the three allowed choices for $n \in \mathbb{Z}$. No other possibilities are allowed. Similar story happens for the next row on the RHS of \eqref{cyclebaaz80}. The solutions are now given by the following sets:

{\footnotesize
\bg\label{cascamukh2}
{4\over 3} - {3\hat\alpha_e(t)\over 2} - 2\hat\sigma_e(t) - {\hat\beta_e(t)\over 2} = \begin{cases} ~n=-4 \begin{cases}~ (0,0,0,0,1,1,0,0,0,0,1,1,0,0,0,0,0,1)\\
~(0,0,0,0,3,1,0,0,0,0,0,0,0,0,0,0,0,1) \end{cases}\\
~n = -2 \begin{cases}~ {\red(0,0,0,0,1,1,0,0,0,0,1,1,0,0,0,0,0,2)}\\
~(0,0,0,0,3,1,0,0,0,0,0,0,0,0,0,0,0,2)\end{cases}\\
~n = -1 \begin{cases}~ (0,0,0,0,1,1,0,0,0,0,1,1,0,0,0,0,0,4)\\
~(0,0,0,0,3,1,0,0,0,0,0,0,0,0,0,0,0,4)\end{cases} \end{cases} \nd}
where the proliferation with respect to the non-locality factors do not effect the values of the 17-tuple. Once we replace $\hat\alpha_e(t)$ by $\hat\beta_e(t)$ and vive-versa, the solutions we get are:

{\footnotesize
\bg\label{cascamukh3}
{4\over 3} - {3\hat\beta_e(t)\over 2} - 2\hat\sigma_e(t) - {\hat\alpha_e(t)\over 2} = \begin{cases} ~n=-4 \begin{cases}~ (0,0,0,0,1,1,0,0,0,0,1,0,1,0,0,0,0,1)\\
~(0,0,0,0,1,3,0,0,0,0,0,0,0,0,0,0,0,1) \end{cases}\\
~n = -2 \begin{cases}~ {\red(0,0,0,0,1,1,0,0,0,0,1,0,1,0,0,0,0,2)}\\
~(0,0,0,0,1,3,0,0,0,0,0,0,0,0,0,0,0,2)\end{cases}\\
~n = -1 \begin{cases}~ (0,0,0,0,1,1,0,0,0,0,1,0,1,0,0,0,0,4)\\
~(0,0,0,0,1,3,0,0,0,0,0,0,0,0,0,0,0,4)\end{cases} \end{cases} \nd}
which show similar proliferation as \eqref{cascamukh2}. The only difference is in the distribution of 1's and 3's. The next case is interesting because $\hat\alpha_e(t)$ and $\hat\beta_e(t)$ are eliminated. The resulting equation and it's solutions are:
\bg\label{cascamukh5}
{4\over 3} - 2\hat\sigma_e(t) - \hat\zeta_e(t) = \begin{cases}
~nx_{18} = -12~~(0,0,0,0,0,0,0,0,0,0,3,0,0,0,0,0,2)\\
~nx_{18} = -8 \begin{cases} ~(0,0,0,0,0,0,0,0,0,1,2,0,0,0,0,0,0)\\
~ (2, 0, 0, 0, 0, 0, 0, 0, 0, 0, 2, 0, 0, 0, 0, 0, 0)\end{cases}\\
~nx_{18} = -4 \begin{cases}~(0,0,0,0,0,0,1,1,0,0,2,0,0,0,0,0,0)\\
~(0, 0, 0, 0, 0, 1, 1, 0, 1, 0, 1, 0, 0, 0, 0, 0, 0)\\
~(0, 0, 0, 0, 1, 0, 0, 1, 1, 0, 1, 0, 0, 0, 0, 0, 0)\\
~{\red(0, 0, 0, 0, 1, 1, 0, 0, 0, 0, 1, 0, 0, 1, 0, 0, 0)}\\
~ (0, 0, 0, 0, 1, 1, 0, 0, 2, 0, 0, 0, 0, 0, 0, 0, 0)\end{cases} \end{cases} \nd
where the distribution of the 17-tuples are shown on the right for given values of $n x_{18}$. Since $n \in \mathbb{Z}^-$ and $x_{18} \in \mathbb{Z}^+$, we can easily figure out the values of the two-tuple $(n, x_{18})$ for any given value of the 17-tuple\footnote{For example with $nx_{18} = -12$, we have $(n, x_{18}) = (-1, 12), (-2, 6), (-3, 4), (-4, 3), (-6, 2), (-12, 1)$. Associated with each there would be the corresponding 17-tuple shown in \eqref{cascamukh5}.}. For the next row on the RHS of \eqref{cyclebaaz80} there are four choices out of which two have been discussed in \eqref{cascamukh2} and \eqref{cascamukh3}. The remaining two, and their solutions, are represented by:

{\scriptsize
\bg\label{cascamukh45} 
\theta_{nl} = \begin{cases}~{4\over 3}-2\hat\sigma_e(t) - {5\hat\alpha_e(t)\over 2} + {\hat\beta_e(t)\over 2} = {\red(0, 0, 0, 0, 1, 1, 0, 0, 0, 0, 1, 0, 0, 0, 1, 0, 0)}, ~nx_{18} = -4\\
~{4\over 3}-2\hat\sigma_e(t) - {5\hat\beta_e(t)\over 2} + {\hat\alpha_e(t)\over 2} = {\red(0, 0, 0, 0, 1, 1, 0, 0, 0, 0, 1, 0, 0, 0, 0, 1, 0)}, ~nx_{18} = -4\end{cases} \nd}
with the assumption that $n\in \mathbb{Z}^-$ and $x_{18} \in \mathbb{Z}^+$ as before. Finally for the last case, which is the topmost row on the RHS of \eqref{cyclebaaz80}, for the brevity of the number of solutions we will only consider $n = -2$. There are now {\it twenty-four} non-negative integer solutions keeping $\hat\eta_e(t) = 0$. They may be listed as:

{\footnotesize
\bg\label{meloniroll}
{10\over 3} - 2\hat\sigma_e(t) -\hat\zeta_e(t) - {\hat\alpha_e(t) + \hat\beta_e(t)\over 2} = \begin{cases}
~(0, 0, 0, 0, 0, 2, 0, 0, 2, 0, 0, 1, 0, 0, 0, 0, 0, 0)\\
~(0, 0, 0, 0, 1, 1, 0, 0, 0, 0, 0, 0, 0, 0, 0, 0, 2, 2)\\
~(0, 0, 0, 0, 1, 2, 1, 0, 1, 0, 0, 0, 0, 0, 0, 0, 0, 0)\\
~(0, 0, 0, 0, 2, 0, 0, 0, 2, 0, 0, 0, 1, 0, 0, 0, 0, 0)\\
~(0, 0, 0, 0, 2, 1, 0, 1, 1, 0, 0, 0, 0, 0, 0, 0, 0, 0)\\
~(0, 0, 0, 0, 2, 2, 0, 0, 0, 0, 0, 0, 0, 1, 0, 0, 0, 0)\\
~(0, 0, 0, 0, 0, 0, 0, 0, 2, 0, 1, 1, 1, 0, 0, 0, 0, 0)\\
~(0, 0, 0, 0, 0, 0, 0, 0, 2, 0, 1, 0, 0, 0, 1, 1, 0, 0)\\
~(0, 0, 0, 0, 0, 1, 0, 1, 1, 0, 1, 1, 0, 0, 0, 0, 0, 0)\\
~(0, 0, 1, 0, 0, 1, 0, 0, 0, 0, 1, 0, 0, 0, 0, 0, 1, 2)\\
~(0, 0, 0, 0, 0, 2, 0, 0, 0, 0, 1, 1, 0, 1, 0, 0, 0, 0)\\
~{\red(0, 0, 0, 0, 1, 1, 0, 0, 0, 1, 1, 0, 0, 0, 0, 0, 0, 2)}\\
~(0, 0, 0, 0, 0, 2, 2, 0, 0, 0, 1, 0, 0, 0, 0, 0, 0, 0)\\
~(0, 0, 0, 0, 1, 0, 1, 0, 1, 0, 1, 0, 1, 0, 0, 0, 0, 0)\\
~(0, 0, 0, 1, 1, 0, 0, 0, 0, 0, 1, 0, 0, 0, 0, 0, 1, 2)\\
~(0, 0, 0, 0, 1, 1, 1, 1, 0, 0, 1, 0, 0, 0, 0, 0, 0, 0)\\
~(2, 0, 0, 0, 1, 1, 0, 0, 0, 0, 1, 0, 0, 0, 0, 0, 0, 2)\\
~(0, 0, 0, 0, 2, 0, 0, 0, 0, 0, 1, 0, 1, 1, 0, 0, 0, 0)\\
~(0, 0, 0, 0, 2, 0, 0, 2, 0, 0, 1, 0, 0, 0, 0, 0, 0, 0)\\
~(0, 0, 0, 0, 0, 0, 0, 0, 0, 0, 2, 1, 1, 1, 0, 0, 0, 0)\\
~(0, 0, 0, 0, 0, 0, 0, 2, 0, 0, 2, 1, 0, 0, 0, 0, 0, 0)\\
~(0, 0, 0, 0, 0, 0, 2, 0, 0, 0, 2, 0, 1, 0, 0, 0, 0, 0)\\
~(0, 0, 0, 0, 0, 0, 0, 0, 0, 0, 2, 0, 0, 1, 1, 1, 0, 0)\\
~(0, 0, 1, 1, 0, 0, 0, 0, 0, 0, 2, 0, 0, 0, 0, 0, 0, 2)
\end{cases} \nd}
showing that now there are many cases with $x_{18} = 3|a_{\rm F}| = 0$. However these cases do not match with the ones from the other scalings and therefore should be discarded. Collecting all the ones in {\red red}
from \eqref{cascamukh} to \eqref{meloniroll}, we get the following pattern of distribution of the 18-tuple:

{\footnotesize
\bg\label{meloni2}
&& (0,0,0,0,1,1,0,0,0,{\red 1},1,0,0,0,0,0,0,2) = {10\over 3} - 2\hat\sigma_e(t) -\hat\zeta_e(t) - {\hat\alpha_e(t) + \hat\beta_e(t)\over 2} \nonumber\\
&& (0,0,0,0,1,1,0,0,0,0,2,0,0,0,0,0,0,2) = {4\over 3} - 3\hat\sigma_e(t) - {\hat\alpha_e(t) + \hat\beta_e(t)\over 2}\nonumber\\
&& (0,0,0,0,1,1,0,0,0,0,1,{\red 1},0,0,0,0,0,2) = {4\over 3} - {3\hat\alpha_e(t)\over 2} - 2\hat\sigma_e(t) - {\hat\beta_e(t)\over 2} \nonumber\\
&& (0,0,0,0,1,1,0,0,0,0,1,0,{\red 1},0,0,0,0,2) = {4\over 3} - {3\hat\beta_e(t)\over 2} - 2\hat\sigma_e(t) - {\hat\alpha_e(t)\over 2}\nonumber\\
&& (0, 0, 0, 0, 1, 1, 0, 0, 0, 0, 1, 0, 0, {\red 1}, 0, 0, 0, 2) = {4\over 3} - 2\hat\sigma_e(t) - \hat\zeta_e(t)\nonumber\\
&& (0, 0, 0, 0, 1, 1, 0, 0, 0, 0, 1, 0, 0, 0, {\red 1}, 0, 0,2) = {4\over 3}-2\hat\sigma_e(t) - {5\hat\alpha_e(t)\over 2} + {\hat\beta_e(t)\over 2} \nonumber\\
&& (0, 0, 0, 0, 1, 1, 0, 0, 0, 0, 1, 0, 0, 0, 0, {\red 1}, 0,2) = {4\over 3}-2\hat\sigma_e(t) - {5\hat\beta_e(t)\over 2} + {\hat\alpha_e(t)\over 2}\nd}
where the distribution of ${\red 1}$ shows a pattern, familiar from our earlier examples. It should also be clear that we are dealing with {\it quartic} order in curvatures because the quantum terms contributing here come from:

{\footnotesize
\bg\label{nihartle}
a_{\rm F} = -{2\over 3}, ~~\Big(\sum_{q=29}^{33} l_q = 2, ~~l_{14} =1, ~~ \sum_{i = 1}^{13} l_i = 1\Big), ~~ \Big(\sum_{q=29}^{33} l_q = 2, ~~l_{14} =1, ~~ n_2 = 2\Big), \nd}
where it is understood that we are choosing ${2\over 3} - \hat\sigma_e(t)$ from the first line in \eqref{brittbaba007}, corresponding to the choice $x_{11} = 1$. The quantum terms are typically of the form ${\bf R}^4$ and $\square {\bf R}^3$, similar to what we had for the BBS instanton case. The difference is that, now we have $a_{\rm F} = -{2\over 3}$ and we have kept $\hat\eta_e(t) = 0$. For the BBS case none of the aforementioned extra features were necessary. The solution given in \eqref{nihartle} suffers from the same problems that we faced for the KKLT case, at least in the curvatures and the derivatives sector of \eqref{botsuga} $-$ see the discussion after \eqref{olivechu2} $-$ and therefore comparing all the solutions from the other instantons and non-perturbative objects we conclude that it is the BBS instantons that provide the {\it dominant} quantum terms to drive the acceleration of our universe. 

\subsection{EOMs for the cross-term metric components and non-localities \label{grace}}

After the detailed study of the on-shell metric degrees of freedom, it is time now to discuss the cross-term metric components. These are off-shell degrees of freedom (in the usual sense) and are therefore integrated away as studied in detail in section \ref{sec6.1}. However this integration procedure does not imply that EOMs related to these degrees of freedom won't exist. Again, in the same section \ref{sec6.1} we managed to derive these EOMs and the results are presented in \eqref{nikclau} as two sets of equations that are given completely in terms of the emergent Ricci tensor ${\bf R}_{\rm 0A'}(\langle{\bf\Xi}\rangle_\sigma)$, the emergent energy-momentum tensors (both local and non-local ones), and the remnant of the determinant factor from say \eqref{lilykhel}. However the energy-momentum tensor 
discussed in \eqref{violetgrm}, or the final ones after integration in \eqref{nikclau}, only deal with the perturbative effects so it is time now to elaborate on the non-perturbative contributions. Our aim here is to first revisit the equation:
\bg\label{violetgrm3}
&& \int \left[{\cal D} \hat{\bf \Xi}\right] \int \left[{\cal D} {\bf g}_{\rm A'B'}\right] ~e^{-{\bf S}^t_{\rm tot}({\bf \Xi, g}_{\rm A'B'}, {\bf \Upsilon}_g)}~{\bf T}_{\rm A'B'}({\bf \Xi}, {\bf g}_{\rm A'B'}) = \\
&& \int \left[{\cal D} \hat{\bf \Xi}\right] ~e^{-\overline{\bf S}_{\rm tot}({\bf \Xi, \Upsilon}_g)} \left(- \hat{\bf T}_{\rm A'B'}({\bf \Xi})\right) 
\mathbb{D}^\dagger(\sigma) \mathbb{D}(\sigma)
\int \left[{\cal D} {\bf g}_{\rm A'B'}\right] ~e^{-\widetilde{\bf S}^t_{\rm tot}({\bf \Xi, g}_{\rm A'B'}, {\bf \Upsilon}_g)} \nonumber\\
&+& \int \left[{\cal D}\hat{\bf \Xi}\right]  ~e^{-\overline{\bf S}_{\rm tot}(\hat{\bf \Xi})}
\mathbb{D}^\dagger(\sigma) \mathbb{D}(\sigma)
\int \left[{\cal D}{\bf g}_{\rm A'B'}\right] ~e^{-\widetilde{\bf S}^t_{\rm tot}(\hat{\bf \Xi}, {\bf g}_{\rm A'B'})} 
\left( - \widetilde{\bf T}_{\rm A'B'}({\bf \Xi, g}_{\rm A'B'}) + ...\right) \nonumber
\nd
where we are using the same notations as in section \ref{sec6.1}, namely 
$\hat{\bf \Xi} = ({\bf \Xi}, {\bf \Upsilon}_g)$, with the latter being the ghosts; and the bar and the tilde decomposition of the action is defined in \eqref{stanfordtag}; and figure out the precise way we could quantify the emergent energy-momentum tensors. (The small difference now is that we take ${\bf g}_{\rm A'B'}$ as our off-shell degrees of freedom instead of just ${\bf g}_{\rm 0A'}$ taken earlier.) The superscript $t$ denotes the addition of the space-time dependent determinant factor $-{1\over 2}\log[-{\rm det}~{\bf g}_{11, d}({\bf \Xi, g}_{\rm A'B'}; x, y, w)]$ as described below \eqref{violetgrm}. The minus signs on the RHS of \eqref{violetgrm3} are simply for convenience.

To study the non-perturbative contributions to the emergent energy-momentum tensors, we will have to construct an action similar to the trans-series form of the action advocated in \eqref{kimkarol}. Recall that \eqref{kimkarol} only deals with the emergent ``on-shell" degrees of freedom, and so if we want to study ${\bf T}_{\rm A'B'}({\bf \Xi}, {\bf g}_{\rm A'B'})$ we will have to start anew and construct the corresponding action that involves the off-shell states. After inserting this in the path-integral, we can then integrate out the off-shell states. In other words, what are looking for is an equivalent of \eqref{tgrlbahlu}, namely:

{\scriptsize
\bg\label{tgrlbahlu3}
\int \left[{\cal D}\hat{\bf \Xi}\right]\int \left[{\cal D}{\bf g}_{\rm A'B'}\right] ~e^{-{\bf S}^t_{\rm tot}(\hat{\bf \Xi}, {\bf g}_{\rm A'B'})} 
\left({\bf R}_{\rm A'B'}({\bf \Xi}, {\bf g}_{\rm A'B'}) - {1\over 2} {\bf g}_{\rm A'B'} {\bf R}({\bf \Xi}, {\bf g}_{\rm A'B'}) + {\bf T}_{\rm A'B'}({\bf \Xi}, {\bf g}_{\rm A'B'})\right) \mathbb{D}^\dagger(\sigma) \mathbb{D}(\sigma) = 0, \nonumber\\
\nd}
with the assumption that the energy-momentum tensor decomposes as in \eqref{violetgrm3}. The appearance of ${\bf S}_{\rm tot}(\hat{\bf \Xi}, {\bf g}_{\rm A'B'})$ holds the key, because both the sets of equations in \eqref{nikclau} and \eqref{coffiemarieso} are derived using this. This is clear from the following decomposition:

{\scriptsize
\bg\label{palkidi}
{\bf S}_{\rm tot}(\hat{\bf \Xi}, {\bf g}_{\rm A'B'}) \to \begin{cases}
\Bigg\langle {\delta{\bf S}_{\rm tot}(\hat{\bf \Xi}, {\bf g}_{\rm A'B'})\over \delta {\bf g}^{\rm AB}}\Bigg\rangle_\sigma = \Big\langle 
\sqrt{-{\rm det}~{\bf g}_{11, d}({\bf\Xi, g}_{\rm A'B'})}\Big({\bf G}_{\rm AB}({\bf\Xi, g}_{\rm A'B'}) - {\bf T}_{\rm AB}({\bf\Xi, g}_{\rm A'B'})\Big)\Big\rangle_\sigma\\~~~\\
\Bigg\langle {\delta{\bf S}_{\rm tot}(\hat{\bf \Xi}, {\bf g}_{\rm A'B'})\over \delta {\bf g}^{\rm A'B'}}\Bigg\rangle_\sigma = \Big\langle 
\sqrt{-{\rm det}~{\bf g}_{11, d}({\bf\Xi, g}_{\rm A'B'})}\Big({\bf G}_{\rm A'B'}({\bf\Xi, g}_{\rm A'B'}) - {\bf T}_{\rm A'B'}({\bf\Xi, g}_{\rm A'B'})\Big)\Big\rangle_\sigma
\end{cases} \nd}
from where in fact when we equate the two rows to zero, the two sets of equations \eqref{coffiemarieso} and \eqref{nikclau} respectively emerge naturally. Our aim here is to quantify the energy-momentum tensor ${\bf T}_{\rm A'B'}({\bf\Xi, g}_{\rm A'B'})$ appearing in the second row of \eqref{palkidi}. For this let us decompose the action in the following way:

{\footnotesize
\bg\label{erinburn}
{\bf S}_{\rm tot}(\hat{\bf \Xi}, {\bf g}_{\rm A'B'}) \to 
{\bf S}'_{\rm tot}(\hat{\bf \Xi}, {\bf g}_{\rm A'B'})= 
{\bf S}^{\rm kin}_{\rm tot}(\hat{\bf \Xi}, {\bf g}_{\rm A'B'}) + 
{\bf S}^{\rm sors}_{\rm tot}({\bf \Xi}, {\bf g}_{\rm A'B'}) +
{\bf S}^{\rm ghosts}_{\rm tot}(\hat{\bf \Xi}, {\bf g}_{\rm A'B'}), \nd}
which is a slightly different distribution than what we took in section \ref{sec6.1} earlier, but can be easily mapped to each other. (``sors" denote the non-perturbative sources.) Note however that the action written in terms of the ``on-shell" components and ghosts, as in \eqref{tomiraja}, is only {\it after} we have integrated away the ``off-shell" components and then (a) Borel resummed the resulting Gevrey series to generate the non-perturbative parts and (b) resummed according to the technique elaborated in section \ref{rajatomy} to generate the non-local terms. On the other hand, in the action \ref{erinburn}, the first term on the RHS deals with the kinetic terms for all the on and off-shell degrees of freedom including the ghosts, the second term deals with the sources and the third term deals with the interactions between the degrees of freedom, sources and the ghosts {\it all with their respective non-perturbative completions}. Thus, instead of the sequence \eqref{duilotine}, we are now looking at a slightly different sequence of the form:
\bg\label{duilotine}
{\bf S}_{\rm tot}(\hat{\bf \Xi}, {\bf g}_{\rm A'B'}, ..)~ \xrightarrow{\rm NP~comp}{\bf S}'_{\rm tot}(\hat{\bf \Xi}, {\bf g}_{\rm A'B'})\xrightarrow{\rm integ} ~\hat{\bf S}_{\rm tot}(\hat{\bf \Xi})~ \xrightarrow{\rm renorm} ~\check{\bf S}_{\rm tot}(\hat{\bf \Xi}), \nd
where in the intermediate step we have allowed the full non-perturbative completion of the action both in terms of the on- and the off-shell degrees of freedom. (In fact ${\bf S}'_{\rm tot}(\hat{\bf \Xi}, {\bf g}_{\rm A'B'})$ is precisely the effective action that we will see at the supersymmetric warped Minkowski level when we impose no constraints on the degrees of freedom.) Thus once we integrate out the off-shell degrees of freedom, we recover the same action $\hat{\bf S}_{\rm tot}(\hat{\bf \Xi})$ that we had in \eqref{tomiraja} and \eqref{kimkarol}. One advantage of using ${\bf S}'_{\rm tot}(\hat{\bf \Xi}, {\bf g}_{\rm A'B'})$, instead of $\overline{\bf S}_{\rm tot}(\hat{\bf \Xi})$ and 
$\widetilde{\bf S}_{\rm tot}(\hat{\bf \Xi}, {\bf g}_{\rm 0A'})$ from \eqref{stanfordtag}, is that we don't have to go though the intermediate 
resummation process and instead land directly to $\hat{\bf S}_{\rm tot}(\hat{\bf \Xi})$ after integration. Both this procedure and the previous one advocated in section \ref{sec6.1} are {\it identical} in the sense that both lead to the same effective action, but in the second case we are dealing with the non-perturbative form from an early stage. Therefore, using ${\bf S}'_{\rm tot}(\hat{\bf \Xi}, {\bf g}_{\rm A'B'})$, 
our aim here is to study the second term {\it i.e.} the action corresponding to the non-perturbative sources ${\bf S}^{\rm sors}_{\rm tot}({\bf \Xi}, {\bf g}_{\rm A'B'})$ . This takes the following form:

{\scriptsize
\bg\label{kidcascaman}
&&{\bf S}^{\rm sors}_{\rm tot}({\bf \Xi}, {\bf g}_{\rm A'B'}) =
{\rm M}_p^{11}\int d^{11} {\rm X} \sqrt{-{\rm det}~{\bf g}_{11,d}({\bf \Xi}({\rm X}), {\bf g}_{\rm A'B'}({\rm X})}~
\sum_d\sum_{\{{\rm S}^{(i)}_d\} = 0}^\infty h_{\{{\rm S}^{(i)}_d\}}\Big[~{\bf Q}^{(1)}_{\rm pert}({c}_1(\{{\rm S}^{(i)}_d\}); {\bf \Xi}({\rm X})) \nonumber\\
&&~~~~~+~{\bf g}^{\rm A'B'}({\rm X}) {\bf Q}^{(2)}_{\rm A'B'}({c}_2(\{{\rm S}^{(i)}_d\});{\bf \Xi}({\rm X}) + ~{\bf Q}^{(3)}_{\rm pert}({c}_3(\{{\rm S}^{(i)}_d\}); {\bf \Xi}({\rm X}), \{{\bf g}_{\rm A'B'}({\rm X})\})\Big]\nonumber\\
&&~~~~~ \otimes \Bigg[~{\rm exp}\left(-{\rm S}^{(1)}_d {\rm M}_p^d \int_0^{\rm Y} d^d{\rm Y'} \sqrt{|\det~{\bf g}_d({\bf\Xi}({\rm Y}', x), {\bf g}_{\rm A'B'}({\rm Y}', x))|} ~~\big\vert {\bf Q}^{(1)}_{\rm pert}(\hat{c}_1({\rm S}^{(1)}_d); {\bf \Xi}({\rm Y}', x))\big\vert \right)\\
&&~~~~~\times  ~{\rm exp}\left(-{\rm S}^{(2)}_d {\rm M}_p^d \int_0^{\rm Y} d^d{\rm Y'} \sqrt{|\det~{\bf g}_d({\bf\Xi}({\rm Y}', x), {\bf g}_{\rm A'B'}({\rm Y}', x))|}~~\big\vert
{\bf g}^{\rm A'B'}({\rm Y'}, x) {\bf Q}^{(2)}_{\rm A'B'}(\hat{c}_2({\rm S}^{(2)}_d);{\bf \Xi}({\rm Y'}, x))\big\vert\right) \nonumber\\
&&~~~~~\times  ~{\rm exp}\left(-{\rm S}^{(3)}_d {\rm M}_p^d \int_0^{\rm Y} d^d{\rm Y'} \sqrt{|\det~{\bf g}_d({\bf\Xi}({\rm Y}', x), {\bf g}_{\rm A'B'}({\rm Y}', x))|} ~~\big\vert
 {\bf Q}^{(3)}_{\rm pert}(\hat{c}_3({\rm S}^{(3)}_d); {\bf \Xi}({\rm Y'}, x), \{{\bf g}_{\rm A'B'}({\rm Y'}, x)\})\big\vert\right)\Bigg], \nonumber\nd}
 where the symbol $\otimes$ fixes the coefficients of the quantum series of the perturbative fluctuations around any given instanton saddle. As an example, consider one of the perturbative quantum series 
 ${\bf Q}^{(1)}_{\rm pert}({c}_1(\{{\rm S}^{(i)}_d\}); {\bf \Xi}({\rm X})) ={\bf Q}^{(1)}_{\rm pert}({c}_1({\rm S}^{(1)}_d, {\rm S}^{(2)}_d, 
{\rm S}^{(3)}_d); {\bf \Xi}({\rm X}))$. Depending on which combination of the non-perturbative instanton sector it attaches to, the coefficients ${c}_1({\rm S}^{(1)}_d, {\rm S}^{(2)}_d, 
{\rm S}^{(3)}_d)$ will be determined. We can denote this symbolically in the following suggestive way:

{\scriptsize
\bg\label{pauorwg}
{\bf S}_{\rm tot}^{\rm sors} & = & \left({\bf Q}^{(1)} + {\bf g}{\bf Q}^{(2)} + {\bf Q}^{(3)}\right) \otimes \left(1 + e^{{\bf Q}^{(1)}}+ e^{2{\bf Q}^{(1)}} +...\right) 
\left(1 + e^{{\bf gQ}^{(2)}}+ e^{2{\bf gQ}^{(2)}} +...\right) \left(1 + e^{{\bf Q}^{(3)}}+ e^{2{\bf Q}^{(3)}} +...\right), \nonumber\\
& = & \left({\bf Q}^{(1)} + {\bf g}{\bf Q}^{(2)} + {\bf Q}^{(3)}\right) \otimes \Big(1 + {\cal S}_1 + {\cal S}_2 + {\cal S}_3 + {\Su{\cal S}_1 {\cal S}_2} + {\cal S}_2 {\cal S}_3+ {\cal S}_3{\cal S}_1 + {\Su{\cal S}_1  {\cal S}_2 {\cal S}_3}\Big), \nd}
with the $\otimes$ fixing the fluctuation determinants around any combinations of the instanton saddles; and ${\cal S}_i$ is related to the instanton series associated with ${\bf Q}^{(i)}$. As an example, for the terms highlighted in {\Su blue}, the coefficients for the fluctuation determinants will be $c_i({\rm S}_d^{(1)}, {\rm S}_d^{(2)}, 0)$ and $c_i({\rm S}_d^{(1)}, {\rm S}_d^{(2)}, {\rm S}_d^{(3)})$ respectively. The other parameters appearing in \eqref{kidcascaman} are defined in the following way. The determinant of the metric of the sub-manifold ${\cal M}_d$ on which we have the wrapped instantons, may or may not depend on the cross-term metric components. This needs to be verified in the case-by-case basis. The perturbative fluctuations are classified into three sets of quantum series: The first and the second sets, namely:

{\footnotesize
\bg\label{daistailfula}
({\bf Q}^{(1)}_{\rm pert}({c}_1(\{{\rm S}^{(i)}_d\}); {\bf \Xi}({\rm X})),{\bf Q}^{(1)}_{\rm pert}(\hat{c}_1({\rm S}^{(1)}_d)); {\bf \Xi}({\rm X}))) ~~ {\rm and} ~~({\bf Q}^{(2)}_{\rm A'B'}({c}_2(\{{\rm S}^{(i)}_d\});{\bf \Xi}({\rm X})),{\bf Q}^{(2)}_{\rm A'B'}(\hat{c}_2({\rm S}^{(2)}_d);{\bf \Xi}({\rm X}))), \nonumber\\ \nd}
are only functions of the ``on-shell" degrees of freedom ${\bf \Xi}$. (Note that the first terms in each of the two sets are expressed using  
all the three ${\rm S}_d^{(i)}$ with $i = 1, 2, 3$. This is important for the fluctuations to be properly quantified here.) Whereas the third set, namely:

{\footnotesize
\bg\label{tailordes2}
({\bf Q}^{(3)}_{\rm pert}({c}_3(\{{\rm S}^{(i)}_d\}); {\bf \Xi}({\rm X}), \{{\bf g}_{\rm A'B'}({\rm X})\}),{\bf Q}^{(3)}_{\rm pert}(\hat{c}_3({\rm S}^{(i)}_d); {\bf \Xi}({\rm X}), \{{\bf g}_{\rm A'B'}({\rm X})\})), \nd}
is a function of at least two different ``off-shell" components. Together they quantify the full quantum interactions over a given Minkowski minimum (or the corresponding solitonic solution). Note also that \eqref{kidcascaman} is the most generic way we can represent the quantum corrections in the model over a given minimum or a soliton.

Our aim now is to compute the emergent energy-momentum tensor by using the action \eqref{kidcascaman} in the path integral and integrate out the off-shell degrees of freedom. Since this is going to be an elaborate exercise, let us rewrite the action \eqref{kidcascaman} in a more manageable format without losing any information. To this effect, let 
 ${\rm X}=({\rm Y},x)$ and ${\rm Z}=({\rm Y_Z},x_{\rm Z})$ and we define the following quantities:

{\scriptsize 
\bg\label{Tking}
&& \mathcal E_i^{(d,\{{\rm S}_d\})}({\rm X})
=\exp\!\Big[-\,{\rm S}^{(i)}_d\,{\rm M}_p^d\,{\rm I}^{(i)}_d({\rm X})\Big], ~~\det~{\bf g}_d({\rm Y'},x) =\det~{\bf g}_d({\bf \Xi}({\rm Y'},x), {\bf g}_{\rm A'B'}({\rm Y'}, x))\\
&& \mathcal F({\rm X})
= {\bf Q}^{(1)}_{\rm pert}\!\big(c_1(\{{\rm S}_d\};{\bf \Xi}({\rm X})\big)
 + {\bf g}^{\rm A'B'}({\rm X})\,{\bf Q}^{(2)}_{\rm A'B'}\!\big(c_2(\{{\rm S}_d\}; {\bf\Xi}({\rm X})\big)
 + {\bf Q}^{(3)}_{\rm pert}\!\big(c_3(\{{\rm S}_d\}) ;{\bf \Xi}({\rm X}),\{{\bf g}_{\rm A'B'}({\rm X})\}\big)\nonumber\\ 
&&{\rm I}^{(i)}_d({\rm X})
=\int_{0}^{\rm Y} d^d{\rm Y'}\;\sqrt{|\det~{\bf g}_d({\rm Y'},x)|}\;\big|\hat{\mathcal Q}_i({\rm Y'},x)\big|, ~~~~~ (\hat{\mathcal Q}_1({\rm X}), \hat{\mathcal Q}_2({\rm X}, \hat{\mathcal Q}_3({\rm X})) = \nonumber\\
&&
\Big({\bf Q}^{(1)}_{\rm pert}\!\big(\hat{c}_1({\rm S}^{(1)}_d);{\bf \Xi}({\rm X})\big),{\bf g}^{\rm A'B'}({\rm X}){\bf Q}^{(2)}_{\rm A'B'}\!\big(\hat{c}_2({\rm S}^{(2)}_d); {\bf\Xi}({\rm X})\big), {\bf Q}^{(3)}_{\rm pert}\!\big(\hat{c}_3({\rm S}^{(3)}_d) ;{\bf \Xi}({\rm X}),\{{\bf g}_{\rm A'B'}({\rm X})\}\big)\Big) \nonumber\\
&&\mathcal V({\rm X})
= \sqrt{-\det~{\bf g}_{11,d}\!\big({\bf \Xi}({\rm X}),\, {\bf g}_{\rm A'B'}(X)\big)}, ~~~~~~(\mathcal Q_1({\rm X}), \mathcal Q_2({\rm X}, \mathcal Q_3({\rm X})) = \nonumber\\
&&
\Big({\bf Q}^{(1)}_{\rm pert}\!\big(c_1(\{{\rm S}^{(i)}_d\});{\bf \Xi}({\rm X})\big), {\bf g}^{\rm A'B'}({\rm X}){\bf Q}^{(2)}_{\rm A'B'}\!\big(c_2(\{{\rm S}^{(i)}_d\}); {\bf\Xi}({\rm X})\big),{\bf Q}^{(3)}_{\rm pert}\!\big(c_3(\{{\rm S}^{(i)}_d\}) ;{\bf \Xi}({\rm X}),\{{\bf g}_{\rm A'B'}({\rm X})\}\big)\Big),\nonumber
\nd}
where $\{{\rm S}^{(i)}_d\}$ denotes the (multi)index set being summed over, with weights $h_{\{{\rm S}^{(i)}_d\}}$. We have also hidden the information of $({\rm S}^{(i)}_d, {\bf \Xi}, {\bf g}_{\rm A'B'})$ when we use $\mathcal Q_i$ and $\hat{\mathcal Q}_i$ instead of ${\bf Q}_i$. But these nuances are easy to recover once the more complete variables are used. The advantage of using the definitions \eqref{Tking} is to rewrite the action \eqref{kidcascaman} in the following compact and manageable form:
\bg\label{eq:Sfull}
{\bf S}^{\rm sors}_{\rm tot} \;=\; {\rm M}_p^{11}\,\sum_{d}\;\sum_{\{{\rm S}^{(i)}_d\}=0}^{\infty} h_{\{{\rm S}^{(i)}_d\}}
\int d^{11}{\rm X}\;
\mathcal V({\rm X})\,\mathcal F({\rm X})\,
\bigotimes_{i=1}^{3}\mathcal E_i^{(d,\{{\rm S}_d\})}({\rm X})\,
\nd
where the information of the distribution of the fluctuation determinants, that we discussed in \eqref{pauorwg}, is hiding in the definition of $\mathcal F({\rm X})$, and the information about the volume measure is controlled by $\mathcal V({\rm X})$. The $\bigotimes$ symbol, instead of the usual product, takes care of the issue pointed out in \eqref{kidcascaman}.
The logarithmic variation of the 11D volume factor is:

{\footnotesize
\bg\label{eq:Kdef}
\mathcal K_{\rm C'D'}({\rm Z}) = 
 \frac{\partial \ln\mathcal V({\rm X})}{\partial {\bf g}^{\rm C'D'}({\rm Z})}
= -\tfrac12\,\big({\bf g}_{11,d}\big)_{\rm MN}({\rm X})\,
\frac{\partial \big({\bf g}_{11,d}\big)^{\rm MN}({\rm X})}{\partial {\bf g}^{\rm C'D'}({\rm Z})} = -{1\over 2} {\bf g}_{\rm C'D'}({\rm Z})~\delta^{11}({\rm X}-{\rm Z}),
\nd}
which is of course the standard answer. We can also ask how does this effect the integral of the integral function ${\rm I}_d^{(i)}({\rm X})$. This can be easily determined by keeping track of the nested integral carefully. The answer is the following:

{\scriptsize
\bg\label{eq:dId}
\frac{\delta {\rm I}^{(i)}_d({\rm X})}{\delta {\bf g}^{\rm C'D'}({\rm Z})}
={\Theta\!\big({\rm Y} - {\rm Y_Z}\big)\over {\rm M}_p^{11}}\;
\delta^{11-d}(x - x_{\rm Z})\;
\sqrt{|{\rm det}~{\bf g}_d({\rm Z})|}\,
\left[
\tfrac12\,({\bf g}_d)_{mn}({\rm Z})\,\frac{\partial ({\bf g}_d)^{mn}({\rm Z})}{\partial {\bf g}^{\rm C'D'}({\rm Z})}\;\big|\mathcal Q_i({\rm Z})\big|
+ \sgn\!\big(\mathcal Q_i({\rm Z})\big)\,\frac{\partial \mathcal Q_i({\rm Z})}{\partial {\bf g}^{\rm C'D'}({\rm Z})}
\right],\nonumber\\
\nd}
where $({\bf g}_d)_{mn}$ should be related to the pull-back metric on the world-volume of the instantons. Since we haven't yet gone to the level of the Glauber-Sudarshan states, we should consider all the allowed off-shell metric components here (and earlier), implying that the first term inside the square bracket in \eqref{eq:dId} is generically non-zero. The second term involve the following differentials: 

{\footnotesize
\bg\label{bessbritt}
&&\frac{\partial \mathcal Q_1({\rm Z})}{\partial {\bf g}^{\rm C'D'}({\rm Z})} = \frac{\partial{\bf Q}^{(1)}_{\rm pert}\!\big(c_1(\{{\rm S}_d\});{\bf \Xi}({\rm Z})\big)}{\partial {\bf g}^{\rm C'D'}({\rm Z})} = 0 \\
&& \frac{\partial \mathcal Q_3({\rm Z})}{\partial {\bf g}^{\rm C'D'}({\rm Z})} = \frac{\partial{\bf Q}^{(3)}_{\rm pert}\!\big(c_3(\{{\rm S}_d\}) ;{\bf \Xi}({\rm X}),\{{\bf g}_{\rm A'B'}({\rm X})\}\big)}{\partial {\bf g}^{\rm C'D'}({\rm Z})}\nonumber\\
&& \frac{\partial \mathcal Q_2({\rm Z})}{\partial {\bf g}^{\rm C'D'}({\rm Z})} = {\bf Q}^{(2)}_{\rm A'B'}\!\big(c_2(\{{\rm S}_d\}); {\bf\Xi}({\rm Z})\big) + {\bf g}^{\rm A'B'}({\rm Z})\,\frac{\partial{\bf Q}^{(2)}_{\rm C'D'}\!\big(c_2(\{{\rm S}_d\}); {\bf\Xi}({\rm Z})\big)}{\partial {\bf g}^{\rm C'D'}({\rm Z})} ={\bf Q}^{(2)}_{\rm C'D'}\!\big(c_2(\{{\rm S}_d\}); {\bf\Xi}({\rm Z})\big),\nonumber
\nd}
that shows the expected non-zero contributions from $\mathcal Q_2({\rm Z})$ and 
$\mathcal Q_3({\rm Z})$ and none from $\mathcal Q_1({\rm Z})$. (Similar decompositions are expected for $\hat{\mathcal Q}_i({\rm Z})$.) We can now combine the results from \eqref{eq:Sfull} to \eqref{bessbritt}, to express the variation of the action \eqref{kidcascaman} (or it's compact form \eqref{eq:Sfull}) with respect to the off-shell metric component as:
\bg\label{deltaS}
\frac{\delta {\bf S}^{\rm sors}_{\rm tot}({\bf \Xi, g}_{\rm C'D'}}{\delta {\bf g}^{\rm C'D'}({\rm Z})}
&= & 
\sum_{d}\;\sum_{\{S^{(i)}_d\}=0}^{\infty} h_{\{S^{(i)}_d\}}
\Bigg\{
\underbrace{\mathcal V({\rm Z})\,\mathcal F({\rm Z})\,
\bigotimes_{i=1}^{3}\mathcal E_i^{(d,\{{\rm S}_d\})}({\rm Z})}_{\text{evaluate at } {\rm X=Z}}
\Bigg[
-\tfrac12\,{\bf g}_{\rm C'D'}({\rm Z})\\
&+ & \;\frac{1}{\mathcal F({\rm Z})}\!
\Bigg( {\bf Q}^{(2)}_{\rm C'D'}\!\big(c_2(\{{\rm S}_d\}; {\bf\Xi}({\rm Z})\big) + \frac{\partial{\bf Q}^{(3)}_{\rm pert}\!\big(c_3(\{{\rm S}_d\}) ;{\bf \Xi}({\rm Z}),\{{\bf g}_{\rm A'B'}({\rm Z})\}\big)}{\partial {\bf g}^{\rm C'D'}({\rm Z})}\Bigg)\Bigg]\nonumber\\
&- & {\rm M}_p^{11}\,\int d^{11}{\rm X}\;
\mathcal V({\rm X})\,\mathcal F({\rm X})\,
\bigotimes_{i=1}^{3}\mathcal E_i^{(d,\{{\rm S}_d\})}({\rm X})\;
\sum_{j=1}^{3} {\rm S}^{(j)}_d\,{\rm M}_p^d\;
\frac{\delta {\rm I}^{(j)}_d({\rm X})}{\delta {\bf g}^{\rm C'D'}({\rm Z})}
\Bigg\}, \nonumber \nd
where in the third line, since \eqref{eq:dId} has a hidden delta function $\delta^{11-d}(x-x_{\rm Z})$, the eleven-dimensional integral reduces to a lower dimensional integral much like what we saw earlier for the BBS, KKLT and other instanton solutions. Moreover, if 
the internal $d$-dimensional metric ${\bf g}_d$ is independent of ${\bf g}^{\rm C'D'}$, then
$\tfrac12 ({\bf g}_d)_{mn}\,\frac{\partial ({\bf g}_d)^{mn}}{\partial {\bf g}^{\rm C'D'}}=0$ in \eqref{eq:dId}. Similarly, if 
$\mathcal Q_3$ is independent of some specific off-shell metric component(s), say the set $\{{\bf g}_{\rm C'D'}\}$, then the variation of this with respect to those metric components will be zero. 

Our aim now is to use \eqref{deltaS} within the path-integral \eqref{tgrlbahlu3} and then integrate out the off-shell metric components. Unfortunately this is easier said than done. Because of the complicated nature of \eqref{deltaS} we will have to first carefully extract out the terms that contain the off-shell pieces and then integrate them out. Since the off-shell terms are lodged deep inside the determinants; the $\mathcal F$ terms $\mathcal F({\rm Z}), \mathcal F({\rm X})$; the $\mathcal Q_i$ terms $\mathcal Q_i({\rm Z}), \mathcal Q_i({\rm X})$; and the $\mathcal I_d^{(i)}$ terms ${\rm I}_d^{(i)}({\rm Z}), {\rm I}_d^{(i)}({\rm X})$ both within and without the exponentials, this will require considerable care. We can then express \eqref{deltaS} in the following suggestive way:

{\footnotesize
\bg\label{deltaS2}
\frac{\delta {\bf S}^{\rm sors}_{\rm tot}({\bf \Xi, g}_{\rm C'D'})}{\delta {\bf g}^{\rm C'D'}({\rm Z})}
&= & -{1\over 2} 
\sum_{d}\;\sum_{\{S^{(i)}_d\}=0}^{\infty} h_{\{S^{(i)}_d\}}\sqrt{-{\bf g}_{11}({\rm Z})}\Big(1 + \mathcal F_{11}({\bf \Xi}({\rm Z}), \{{\bf g}_{\rm A'B'}({\rm Z})\})\Big)\\
&& \times \Big(\mathcal Q_1({\rm Z}) +  \mathcal Q_2({\rm Z}) + \mathcal Q_3({\rm Z})\Big)\bigotimes_{i = 1}^3 {\rm exp}\left[-{\rm S}^{(i)}_d {\rm M}_p^d {\rm I}_d^{(i)}({\rm Z})\right] ~{\bf g}_{\rm C'D'}({\rm Z}) \nonumber\\
&& +
\sum_{d}\;\sum_{\{S^{(i)}_d\}=0}^{\infty} h_{\{S^{(i)}_d\}}\sqrt{-{\bf g}_{11}({\rm Z})}\Big(1 + \mathcal F_{11}({\bf \Xi}({\rm Z}), \{{\bf g}_{\rm A'B'}({\rm Z})\})\Big) \nonumber\\
&&\times \left({\bf Q}^{(2)}_{\rm C'D'}\!\big({\rm Z}\big) + \frac{\partial{\bf Q}^{(3)}_{\rm pert}\!\big({\rm Z},\{{\bf g}_{\rm A'B'}({\rm Z})\}\big)}{\partial {\bf g}^{\rm C'D'}({\rm Z})}\right) \bigotimes_{i = 1}^3 {\rm exp}\left[-{\rm S}^{(i)}_d {\rm M}_p^d {\rm I}_d^{(i)}({\rm Z})\right]\nonumber\\ 
&& - {\rm M}_p^{11}\,
\sum_{d}\;\sum_{\{S^{(i)}_d\}=0}^{\infty} h_{\{S^{(i)}_d\}}\int d^{11}{\rm X}\sqrt{-{\bf g}_{11}({\rm X})}\Big(1 + \mathcal F_{11}({\bf \Xi}({\rm X}), \{{\bf g}_{\rm A'B'}({\rm X})\})\Big) \nonumber\\
&& \times \Big(\mathcal Q_1({\rm X}) +  \mathcal Q_2({\rm X}) + \mathcal Q_3({\rm X})\Big)\bigotimes_{i = 1}^3 {\rm exp}\left[-{\rm S}^{(i)}_d {\rm M}_p^d {\rm I}_d^{(i)}({\rm X})\right]\sum_{j=1}^{3} {\rm S}^{(j)}_d\,{\rm M}_p^d\;
\frac{\delta {\rm I}^{(j)}_d({\rm X})}{\delta {\bf g}^{\rm C'D'}({\rm Z})}\nonumber
\nd}
from where it is easy to infer that the $\mathcal F_{11}$ extension of the determinant will be related to the non-perturbative completion of the ${\bf S}^t_{\rm tot}$ factor from \eqref{tgrlbahlu} and \eqref{violetgrm3}\footnote{This may also be regarded as the {\it definition} of $\mathcal F_{11}({\bf \Xi}({\rm Z}), \{{\bf g}_{\rm A'B'}({\rm Z})\})$.}; and the $\mathcal Q_i$ factors will be related to the decomposition of the energy-momentum tensor in \eqref{stanfordtag} and \eqref{violetgrm3} $-$ albeit its non-perturbative completion $-$ once we insert \eqref{deltaS2} inside the path-integral. We can divide the result into two sets: one that has no off-shell degrees of freedom in the perturbative fluctuation sector, and the other that has at least one off-shell degree of freedom in that sector. We will also have to be careful about the off-shell pieces in $\det~{\bf g}_d$, {\it i.e.} in the determinant of the metric of the internal sub-manifold on which we have our wrapped instantons. Thus what we are aiming for is a path-integral structure of the following form:
\bg\label{lidiamey}
&& -{1\over 2} 
\sum_{d}\;\sum_{\{S^{(i)}_d\}=0}^{\infty} h_{\{S^{(i)}_d\}}\int \left[{\cal D}\hat{\bf \Xi}\right] e^{-{\bf S}^{\rm kin}_{\rm tot}(\hat{\bf \Xi}) }\sqrt{-{\bf g}_{11}({\rm Z})}~\mathcal Q_1({\rm Z}) ~\otimes
e^{-{\rm S}^{(1)}_d {\rm M}_p^d {\rm I}_d^{(1)}({\rm Z})} \nonumber\\
&& \times ~\mathbb{D}^\dagger(\sigma; {\bf \Xi})\mathbb{D}(\sigma; {\bf \Xi})  \int \Big[\{{\cal D}{\bf g}_{\rm A'B'}\}\Big] 
\Big(1 + {\cal F}_{11}({\bf \Xi}({\rm Z}), \{{\bf g}_{\rm A'B'}({\rm Z})\})\Big) {\bf g}_{\rm C'D'}({\rm Z})\nonumber\\
&& \times ~e^{-{\bf S}^{\rm kin}_{\rm tot}(\{{\bf g}_{\rm A'B'}\}) -{\bf S}^{\rm sors}_{\rm tot}({\bf \Xi}, \{{\bf g}_{\rm A'B'}\}) -
{\bf S}^{\rm ghosts}_{\rm tot}(\hat{\bf \Xi}, \{{\bf g}_{\rm A'B'}\})} \bigotimes_{i = 2}^3 e^{-{\rm S}^{(i)}_d {\rm M}_p^d {\rm I}_d^{(i)}({\rm Z})}
\nonumber\\
&& -{1\over 2} 
\sum_{d}\;\sum_{\{S^{(i)}_d\}=0}^{\infty} h_{\{S^{(i)}_d\}}\int \left[{\cal D}\hat{\bf \Xi}\right] e^{-{\bf S}^{\rm kin}_{\rm tot}(\hat{\bf \Xi}) }\sqrt{-{\bf g}_{11}({\rm Z})}~\otimes
e^{-{\rm S}^{(1)}_d {\rm M}_p^d {\rm I}_d^{(1)}({\rm Z})} \nonumber\\
&&  \times ~\mathbb{D}^\dagger(\sigma; {\bf \Xi})\mathbb{D}(\sigma; {\bf \Xi})\int \Big[\{{\cal D}{\bf g}_{\rm A'B'}\}\Big] 
\Big(1 + {\cal F}_{11}({\bf \Xi}({\rm Z}), \{{\bf g}_{\rm A'B'}({\rm Z})\})\Big) {\bf g}_{\rm C'D'}({\rm Z})\nonumber\\
&& \times ~e^{-{\bf S}^{\rm kin}_{\rm tot}(\{{\bf g}_{\rm A'B'}\}) -{\bf S}^{\rm sors}_{\rm tot}({\bf \Xi}, \{{\bf g}_{\rm A'B'}\}) -
{\bf S}^{\rm ghosts}_{\rm tot}(\hat{\bf \Xi}, \{{\bf g}_{\rm A'B'}\})} \bigotimes_{i = 2}^3 e^{-{\rm S}^{(i)}_d {\rm M}_p^d {\rm I}_d^{(i)}({\rm Z})}
\Big(\mathcal Q_2({\rm Z}) + \mathcal Q_3({\rm Z})\Big)\nonumber\\
&& -{1\over 2} 
\sum_{d}\;\sum_{\{S^{(i)}_d\}=0}^{\infty} h_{\{S^{(i)}_d\}}\int \left[{\cal D}\hat{\bf \Xi}\right] e^{-{\bf S}^{\rm kin}_{\rm tot}(\hat{\bf \Xi}) }\sqrt{-{\bf g}_{11}({\rm Z})}~\mathcal Q^{(2)}_{\rm C'D'}({\rm Z}) ~\otimes
e^{-{\rm S}^{(1)}_d {\rm M}_p^d {\rm I}_d^{(1)}({\rm Z})} \nonumber\\
&& \times~\mathbb{D}^\dagger(\sigma; {\bf \Xi})\mathbb{D}(\sigma; {\bf \Xi}) \int \Big[\{{\cal D}{\bf g}_{\rm A'B'}\}\Big] 
\Big(1 + {\cal F}_{11}({\bf \Xi}({\rm Z}), \{{\bf g}_{\rm A'B'}({\rm Z})\})\Big)\nonumber\\
&& \times ~e^{-{\bf S}^{\rm kin}_{\rm tot}(\{{\bf g}_{\rm A'B'}\}) -{\bf S}^{\rm sors}_{\rm tot}({\bf \Xi}, \{{\bf g}_{\rm A'B'}\}) -
{\bf S}^{\rm ghosts}_{\rm tot}(\hat{\bf \Xi}, \{{\bf g}_{\rm A'B'}\})} \bigotimes_{i = 2}^3 e^{-{\rm S}^{(i)}_d {\rm M}_p^d {\rm I}_d^{(i)}({\rm Z})}
\nonumber\\
&& -{1\over 2} 
\sum_{d}\;\sum_{\{S^{(i)}_d\}=0}^{\infty} h_{\{S^{(i)}_d\}}\int \left[{\cal D}\hat{\bf \Xi}\right] e^{-{\bf S}^{\rm kin}_{\rm tot}(\hat{\bf \Xi}) }\sqrt{-{\bf g}_{11}({\rm Z})}~\otimes
e^{-{\rm S}^{(1)}_d {\rm M}_p^d {\rm I}_d^{(1)}({\rm Z})}\times \mathbb{D}^\dagger(\sigma, {\bf \Xi}) \nonumber\\
&& \times~\mathbb{D}(\sigma, {\bf \Xi})\int \Big[\{{\cal D}{\bf g}_{\rm A'B'}\}\Big] 
\Big(1 + {\cal F}_{11}({\bf \Xi}({\rm Z}), \{{\bf g}_{\rm A'B'}({\rm Z})\})\Big) \left(\frac{\partial{\bf Q}^{(3)}_{\rm pert}\!\big({\rm Z},\{{\bf g}_{\rm A'B'}({\rm Z})\}\big)}{\partial {\bf g}^{\rm C'D'}({\rm Z})}\right)\nonumber\\
&& \times ~{\bf g}_{\rm C'D'}({\rm Z})~e^{-{\bf S}^{\rm kin}_{\rm tot}(\{{\bf g}_{\rm A'B'}\}) -{\bf S}^{\rm sors}_{\rm tot}({\bf \Xi}, \{{\bf g}_{\rm A'B'}\}) -
{\bf S}^{\rm ghosts}_{\rm tot}(\hat{\bf \Xi}, \{{\bf g}_{\rm A'B'}\})} \bigotimes_{i = 2}^3 e^{-{\rm S}^{(i)}_d {\rm M}_p^d {\rm I}_d^{(i)}({\rm Z})}
\nonumber\\
&& - ~
\sum_{d}\;{\rm M}_p^{d}\,\sum_{\{S^{(i)}_d\}=0}^{\infty} h_{\{S^{(i)}_d\}}\int \left[{\cal D}\hat{\bf \Xi}\right]\Big[\{{\cal D}{\bf g}_{\rm A'B'}\}\Big]\int d^{d}{\rm Y}\sqrt{-{\bf g}_{11}({\rm Y}, x_{\rm Z})}~e^{-{\bf S}'_{\rm tot}(\hat{\bf \Xi}, {\bf g}_{\rm A'B'})}\nonumber\\
&&\otimes ~ {\rm exp}\left[-{\rm S}^{(1)}_d {\rm M}_p^d {\rm I}_d^{(1)}({\rm Y}, x_{\rm Z})-{\rm S}^{(2)}_d {\rm M}_p^d {\rm I}_d^{(2)}({\rm Y}, x_{\rm Z})-{\rm S}^{(3)}_d {\rm M}_p^d {\rm I}_d^{(3)}({\rm Y}, x_{\rm Z})\right] \\
&& \times   ~\mathbb{D}^\dagger(\sigma; {\bf \Xi})\mathbb{D}(\sigma; {\bf \Xi}) \sum_{j = 1}^3\mathcal Q_j({\rm Y}, x_{\rm Z}) 
\Big(1 + \mathcal F_{11}({\bf \Xi}({\rm Y}, x_{\rm Z}), \{{\bf g}_{\rm A'B'}({\rm Y}, x_{\rm Z})\})\Big)
\nonumber\\
&& \otimes ~\Theta({\rm Y - Y_Z}) \sqrt{|{\rm det}~{\bf g}_d({\rm Z})|}\left[{1\over 2}{\bf g}_{\rm C'D'}({\rm Z}) \sum_{p = 1}^3 \vert \hat{\mathcal Q}_p({\rm Z})\vert + 
\sum_{l=2}^{3} {\rm S}^{(l)}_d\,{\rm sgn}(\hat{\mathcal Q}_l({\rm Z}))~
\frac{\delta \hat{\mathcal Q}_l({\rm Z})}{\delta {\bf g}^{\rm C'D'}({\rm Z})}\right], \nonumber
\nd
where ${\rm Z} \equiv ({\rm Y}_{\rm Z}, x_{\rm Z})$ and the integration variable is over ${\rm Y}$. The action ${{\bf S}'_{\rm tot}(\hat{\bf \Xi}, {\bf g}_{\rm A'B'})}$ is given in \eqref{erinburn}, and we expect 
$\sqrt{|{\rm det}~{\bf g}_d({\rm Z})|} = \sqrt{|{\bf g}_d({\rm Z})|}\Big(1 + {\cal F}_{d}({\bf \Xi}({\rm Z}), \{{\bf g}_{\rm A'B'}({\rm Z})\})\Big)$, with $\mathcal F_d$ being similar function as $\mathcal F_{11}$. It should also be clear from \eqref{erinburn} that $|{\bf g}_d({\rm Z})|$ is the determinant of the metric of the sub-manifold on which we have our wrapped instantons expressed in terms of the on-shell degrees of freedom ${\bf \Xi}({\rm Z})$. This distinguishes $|\det~{\bf g}_d({\rm Z})|$ from $|{\bf g}_d({\rm Z})|$. (The modulus sign is just a convention to keep the determinant positive in the Lorentzian signature.)
Note also two things: {\Su one}, the appearance of $\hat{\mathcal Q}_i({\rm Z})$ in the last line of \eqref{lidiamey} as expected from \eqref{Tking}, and {\Su two}, a careful distinction between $\otimes$ and $\times$, again as expected from \eqref{Tking}.

What is important for us is the last bit of \eqref{lidiamey} dealing with the non-perturbative effects at the Minkowski (or the solitonic) level. These are the last four lines of \eqref{lidiamey}, as the perturbative effects are red-herrings in the problem. We can rewrite this in the following suggestive way:

{\footnotesize
\bg\label{matilda70y}
&& \sum_{d}\;{\rm M}_p^{d}\,\sum_{\{S^{(i)}_d\}=0}^{\infty} h_{\{S^{(i)}_d\}}\int \left[{\cal D}\hat{\bf \Xi}\right]\Big[\{{\cal D}{\bf g}_{\rm A'B'}\}\Big]\int d^{d}{\rm Y}\sqrt{-{\bf g}_{11}({\rm Y}, x_{\rm Z})}~e^{-{\bf S}'_{\rm tot}(\hat{\bf \Xi}, {\bf g}_{\rm A'B'})}\nonumber\\
&& \times   ~\mathbb{D}^\dagger(\sigma; {\bf \Xi})\mathbb{D}(\sigma; {\bf \Xi}) \sum_{j = 1}^3\mathcal Q_j({\rm Y}, x_{\rm Z}) 
\Big(1 + \mathcal F_{11}({\bf \Xi}({\rm Y}, x_{\rm Z}), \{{\bf g}_{\rm A'B'}({\rm Y}, x_{\rm Z})\})\Big)
\nonumber\\
&&\otimes\Bigg\{ {\rm exp}\left[-{\rm S}^{(1)}_d {\rm M}_p^d {\rm I}_d^{(1)}({\rm Y}, x_{\rm Z})-{\rm S}^{(2)}_d {\rm M}_p^d {\rm I}_d^{(2)}({\rm Y}, x_{\rm Z})-{\rm S}^{(3)}_d {\rm M}_p^d {\rm I}_d^{(3)}({\rm Y}, x_{\rm Z})\right] \\
&& \times ~\Theta({\rm Y - Y_Z}) \sqrt{|{\rm det}~{\bf g}_d({\rm Z})|}\Bigg[{1\over 2}{\bf g}_{\rm C'D'}({\rm Z}) \sum_{p = 1}^3 \vert \hat{\mathcal Q}_p({\rm Z})\vert + 
\sum_{l=2}^{3} {\rm S}^{(l)}_d\,{\rm sgn}(\hat{\mathcal Q}_l({\rm Z}))~
\frac{\delta \hat{\mathcal Q}_l({\rm Z})}{\delta {\bf g}^{\rm C'D'}({\rm Z})}\Bigg]\Bigg\} \nonumber\\ 
&& = \sum_{d}\;{\rm M}_p^{d}\,\sum_{\{S^{(1, 2)}_d\}=0}^{\infty} h_{\{S^{(1, 2)}_d\}}\int \left[{\cal D}\hat{\bf \Xi}\right] \sqrt{-{\bf g}_{11}({\rm Y_Z}, x_{\rm Z})} ~e^{-{\bf S}^{\rm kin}_{\rm tot}(\hat{\bf \Xi})}~\mathbb{D}^\dagger(\sigma; {\bf \Xi})\mathbb{D}(\sigma; {\bf \Xi})\nonumber\\
&&\times~{\rm S}_d^{(2)} \int d^{d}{\rm Y}~\sqrt{{\bf g}_{11}({\rm Y}, x_{\rm Z}) \over{\bf g}_{11}({\rm Y_Z}, x_{\rm Z})} ~\Theta({\rm Y-Y_Z}) \mathcal Q_1({\rm Y}, x_{\rm Z}) \sqrt{|{\bf g}_d({\rm Y_Z}, x_{\rm Z})|} \otimes \Big[e^{-{\rm S}^{(1)}_d {\rm M}_p^d {\rm I}_d^{(1)}({\rm Y}, x_{\rm Z})}
\nonumber\\
&& \times ~ {\rm sgn}(\hat{\mathcal Q}_2({\rm Y_Z}, x_{\rm Z}))~
\frac{\delta \hat{\mathcal Q}_2({\rm Y_Z}, x_{\rm Z})}{\delta {\bf g}^{\rm C'D'}({\rm Y_Z}, x_{\rm Z})}\Big] \int \Big[\{{\cal D}{\bf g}_{\rm A'B'}\}\Big] e^{-{\bf S}^{\rm kin}_{\rm tot}({\bf g}_{\rm A'B'}) - 
{\bf S}^{\rm sors}_{\rm tot}({\bf \Xi}, {\bf g}_{\rm A'B'}) -
{\bf S}^{\rm ghosts}_{\rm tot}(\hat{\bf \Xi}, {\bf g}_{\rm A'B'})} \nonumber\\
&&~\otimes \Big[e^{-{\rm S}^{(2)}_d {\rm M}_p^d {\rm I}_d^{(2)}({\rm Y}, x_{\rm Z})}\Big] 
\times \Big(1 + \mathcal F_{11}({\bf \Xi}({\rm Y}, x_{\rm Z}), \{{\bf g}_{\rm A'B'}({\rm Y}, x_{\rm Z})\})\Big) \Big(1 + \mathcal F_{d}({\bf \Xi}({\rm Y_Z}, x_{\rm Z}), \{{\bf g}_{\rm A'B'}({\rm Y_Z}, x_{\rm Z})\})\Big) \nonumber\\
&& + ~...... \equiv \int \left[{\cal D}\hat{\bf \Xi}\right] \sqrt{-{\bf g}_{11}({\rm Y_Z}, x_{\rm Z})} ~e^{-\hat{\bf S}_{\rm tot}(\hat{\bf \Xi})}~\mathbb{D}^\dagger(\sigma; {\bf \Xi})\mathbb{D}(\sigma; {\bf \Xi})~{\bf T}^{\rm NP}_{\rm C'D'}({\bf \Xi}({\rm Y_Z}, x_{\rm Z})\nonumber
\nd}
where the dotted terms include ${\rm S}_d^{(3)} \ge 0$ and other terms. Let us make the following observations related to \eqref{matilda70y}. {\Su One}, we note that energy-momentum tensor is only defined for ${\rm S}_d^{(2)} \ge 1$, so for vanishing ${\rm S}_d^{(2)}$ one has to look at the ${\rm S}_d^{(3)} \ge 1$ sector. {\Su Two}, $\sqrt{-{\bf g}_{11}({\rm Z})}$ could be identified to parts of $\lambda_3$ in \eqref{butterflyeffect}. But now the analysis is more involved because of the presence of the internal determinants and the instanton saddles at the Minkowski level. {\Su Three}, since ${\rm S}_d^{(3)} > 0$, there are always non-local terms contributing to the energy-momentum tensor. And {\Su four}, to connect \eqref{erinburn} with \eqref{tomiraja} and therefore also to the splitting of the action in \eqref{matilda70y}, let us take the various terms in \eqref{erinburn} and split them further in the following way:

{\scriptsize
\bg\label{shonalifess}
&& {\bf S}_{\rm tot}^{\rm sors}({\bf \Xi}, {\bf g}_{\rm A'B'}) = 
{\bf S}_{\rm tot}^{\rm sors:1}({\bf \Xi}) + 
{\bf S}_{\rm tot}^{\rm sors:2}({\bf \Xi}, {\bf g}_{\rm A'B'}) = 
{\bf S}_{\rm NP}({\bf \Xi}) + {\bf S}_{\rm tot}^{\rm sors:2}({\bf \Xi}, {\bf g}_{\rm A'B'})\\
&& {\bf S}_{\rm tot}^{\rm kin}(\hat{\bf \Xi}, {\bf g}_{\rm A'B'}) = {\bf S}_{\rm tot}^{\rm kin}(\hat{\bf \Xi}) + 
{\bf S}_{\rm tot}^{\rm kin}({\bf g}_{\rm A'B'}) = {\bf S}_{\rm kin}({\bf \Xi}) + \hat{\bf S}_{\rm ghosts}^{(1)}({\bf \Upsilon}_g) +{\bf S}_{\rm tot}^{\rm kin}({\bf g}_{\rm A'B'})\nonumber\\
&& {\bf S}_{\rm tot}^{\rm ghosts}(\hat{\bf \Xi}, {\bf g}_{\rm A'B'}) = {\bf S}_{\rm tot}^{\rm ghosts:1}(\hat{\bf \Xi}) + 
{\bf S}_{\rm tot}^{\rm ghosts:2}(\hat{\bf \Xi}, {\bf g}_{\rm A'B'}), ~~\hat{\bf S}_{\rm ghosts}(\hat{\bf \Xi}) = \hat{\bf S}^{(1)}_{\rm ghosts}({\bf \Upsilon}_g) + {\bf S}_{\rm tot}^{\rm ghosts:1}(\hat{\bf \Xi}) + \hat{\bf S}_{\rm ghosts}^{\rm nloc}(\hat{\bf \Xi}), \nonumber
\nd}
which gives us a precise map to relate \eqref{erinburn} with ${\bf S}_{\rm kin}({\bf \Xi}), {\bf S}_{\rm NP}({\bf \Xi})$ and parts of $\hat{\bf S}_{\rm ghosts}(\hat{\bf \Xi})$ from \eqref{tomiraja}. The remaining non-local pieces from \eqref{tomiraja} can be derived from \eqref{shonalifess} by integrating away the off-shell degrees of freedom $\{{\bf g}_{\rm A'B'}\}$. Plugging \eqref{shonalifess} into \eqref{matilda70y} leads to the following expression for the energy-momentum tensor ${\bf T}_{\rm C'D'}^{\rm NP}$:

{\scriptsize
\bg\label{omakaludi}
{\bf T}_{\rm C'D'}^{\rm NP}({\rm Y_Z}, x_{\rm Z}) & = & 
\sum_{d}\;{\rm M}_p^{d}\,\sum_{\{S^{(1, 2)}_d\}=0}^{\infty} h_{\{S^{(1, 2)}_d\}}~ e^{{\bf S}_{\rm nloc}({\bf \Xi}) + \hat{\bf S}_{\rm ghosts}^{\rm nloc}(\hat{\bf \Xi}) }\int d^{d}{\rm Y}~\sqrt{{\bf g}_{11}({\rm Y}, x_{\rm Z}) \over{\bf g}_{11}({\rm Y_Z}, x_{\rm Z})}
\int \Big[\{{\cal D}{\bf g}_{\rm A'B'}\}\Big]\nonumber\\
&&\times~{\rm S}_d^{(2)}  ~\Theta({\rm Y-Y_Z}) 
~\otimes \Big[e^{-{\rm S}^{(2)}_d {\rm M}_p^d {\rm I}_d^{(2)}({\rm Y}, x_{\rm Z})}\Big] ~\times \mathcal Q_1({\rm Y}, x_{\rm Z}) \sqrt{|{\bf g}_d({\rm Y_Z}, x_{\rm Z})|} \otimes \Big[e^{-{\rm S}^{(1)}_d {\rm M}_p^d {\rm I}_d^{(1)}({\rm Y}, x_{\rm Z})}
\nonumber\\
&& \times ~ {\bf Q}^{(2)}_{\rm C'D'}\!\big(\hat{c}_2({\rm S}^{(2)}_d); {\bf\Xi}({\rm Y_Z}, x_{\rm Z})\big)\Big]\times
  e^{-{\bf S}^{\rm kin}_{\rm tot}({\bf g}_{\rm A'B'}) - 
{\bf S}^{\rm sors:2}_{\rm tot}({\bf \Xi}, {\bf g}_{\rm A'B'}) -
{\bf S}^{\rm ghosts:2}_{\rm tot}(\hat{\bf \Xi}, {\bf g}_{\rm A'B'})} \nonumber\\
&&
\times \Big(1 + \mathcal F_{11}({\bf \Xi}({\rm Y}, x_{\rm Z}), \{{\bf g}_{\rm A'B'}({\rm Y}, x_{\rm Z})\})\Big) \Big(1 + \mathcal F_{d}({\bf \Xi}({\rm Y_Z}, x_{\rm Z}), \{{\bf g}_{\rm A'B'}({\rm Y_Z}, x_{\rm Z})\})\Big) + .... , \nd}
where the dotted terms are the ones with ${\rm S}_d^{(3)} > 0$ including other related terms. We see that the action entering the path-integral over the off-shell states is ${\bf S}^{\rm kin}_{\rm tot}({\bf g}_{\rm A'B'}) +
{\bf S}^{\rm sors:2}_{\rm tot}({\bf \Xi}, {\bf g}_{\rm A'B'}) +
{\bf S}^{\rm ghosts:2}_{\rm tot}(\hat{\bf \Xi}, {\bf g}_{\rm A'B'})$ which, if were not due to the $\mathcal F_{11}, \mathcal F_d$ and ${\rm I}_d^{(2)}$ pieces, would have simply given us $-{\bf S}_{\rm nloc}({\bf \Xi}) - \hat{\bf S}_{\rm ghosts}^{\rm nloc}(\hat{\bf \Xi})$ thus canceling against similar terms in the first line of \eqref{omakaludi}. In fact, even if we ignore the $\mathcal F_{11}$ and the $\mathcal F_d$ pieces, the presence of ${\rm I}_d^{(2)}$ terms implies that the non-perturbative energy-momentum tensor from an expression like \eqref{omakaludi} is always non-local. On the other hand, the non-perturbative energy-momentum tensor for the on-shell degrees of freedom can have local pieces because we can always take ${\rm S}_d^{(1)} > 0$ keeping ${\rm S}_d^{(2)} = {\rm S}_d^{(3)} = 0$. The perturbative contributions however can have local pieces for both the on and the off-shell cases.


Let us now see whether we can quantify the non-locality from the path-integral representation in \eqref{omakaludi}. The nested path-integral that we are interested in can be expressed from \eqref{omakaludi} in the following way:

{\scriptsize
\bg\label{fesserstairs}
\mathbb{A}({\bf \Xi}({\rm Y, Y_Z}, x_{\rm Z}))~e^{-\mathbb{S}_{\rm nloc}(\hat{\bf \Xi})} &\equiv& \int \Big[\{{\cal D}{\bf g}_{\rm A'B'}\}\Big] ~e^{-{\rm S}^{(2)}_d {\rm M}_p^d {\rm I}_d^{(2)}({\rm Y}, x_{\rm Z})}
e^{-{\bf S}^{\rm kin}_{\rm tot}({\bf g}_{\rm A'B'}) - 
{\bf S}^{\rm sors:2}_{\rm tot}({\bf \Xi}, {\bf g}_{\rm A'B'}) -
{\bf S}^{\rm ghosts:2}_{\rm tot}(\hat{\bf \Xi}, {\bf g}_{\rm A'B'})} \nonumber\\
&&~\times \Big(1 + \mathcal F_{11}({\bf \Xi}({\rm Y}, x_{\rm Z}), \{{\bf g}_{\rm A'B'}({\rm Y}, x_{\rm Z})\})\Big) \Big(1 + \mathcal F_{d}({\bf \Xi}({\rm Y_Z}, x_{\rm Z}), \{{\bf g}_{\rm A'B'}({\rm Y_Z}, x_{\rm Z})\})\Big), \nonumber\\ \nd}
where we expect $\mathbb{S}_{\rm nloc}(\hat{\bf \Xi})$ to eventually cancel with ${\bf S}_{\rm nloc}({\bf \Xi}) + \hat{\bf S}_{\rm ghosts}^{\rm nloc}(\hat{\bf \Xi})$ from \eqref{omakaludi}. This is not too hard to justify because the form of the action appearing in \eqref{fesserstairs} is in some sense adjusted to reflect this. However it will be interesting to quantify $\mathbb{A}({\bf \Xi}({\rm Y, Y_Z}, x_{\rm Z}))$. For that we need to know not only ${\bf S}^{\rm sors:2}_{\rm tot}({\bf \Xi}, {\bf g}_{\rm A'B'})$ and  $
{\bf S}^{\rm ghosts:2}_{\rm tot}(\hat{\bf \Xi}, {\bf g}_{\rm A'B'})$, but also ${\rm I}_d^{(2)}({\rm Y}, x_{\rm Z})$ more precisely. This is a hard exercise because it will require a more careful study of the interactions at the Minkowski level that we do not undertake here. But we can study a toy model of the aforementioned set-up to quantify aspects of $\mathbb{A}({\bf \Xi}({\rm Y, Y_Z}, x_{\rm Z}))$. 

\subsubsection{A toy model for non-local interactions \label{nolock}}

A simple toy model that we can extract from \eqref{fesserstairs} is by (a) ignoring the ghosts as well as the determinant contributions $\mathcal F_{11}$ and $\mathcal F_d$, and (b) taking one off-shell component ${\bf g}_{\rm A'B'}$ by expressing it as a real scalar $\phi$; and one on-shell component ${\bf g}_{\rm AB}$ by expressing it as another real scalar $\psi$. The path-integral \eqref{fesserstairs} then reduces to the following:
\bg\label{oma2byrae1}
\mathbb{Z}[Q_1,Q_2] \;=\; \int \mathcal D\phi \;
\exp\!\left\{-\int d^4x \Big[(\partial_\mu\phi)^2 \;+\; f(\phi)\,Q_1(\psi)\Big]\right\} {\rm exp}\Big[- \phi(y)\,Q_2(\psi)\Big],\nonumber\\
\nd
where $f(\phi)$ is a polynomial interaction in terms of powers of $\phi$, and $Q_i(x)$ are external functions independent of $\phi$.
$Q_2$ may be identified to ${\bf Q}^{(2)}_{\rm C'D'}$ and $Q_1$ may be identified to the subset of on-shell interactions from ${\bf S}_{\rm tot}^{\rm sors:2}$ with the off-shell part collected in $f(\phi)$. We can generalize \eqref{oma2byrae1} by including other on-shell fields. An example would be the following:
\begin{equation}\label{jorjor}
\mathbb Z[Q_1,Q_2] \;=\; \int \mathcal D\phi \;
\exp\!\left\{-\int d^4x \Big[(\partial_\mu\phi)^2 \;+\; f(\phi(x))\,Q_1(\psi(x))\Big]\right\}
\,\exp\!\Big[-\phi(y)\,Q_2(\chi(y))\Big].
\end{equation}
with $\chi(y)$ being another on-shell field. 
For brevity we will set $Q_1(x) \equiv Q_1(\psi(x))$, $q_2(y) \equiv Q_2(\chi(y))$. Note that both the $y$-dependent insertions in \eqref{oma2byrae1} and \eqref{jorjor} appear from ${\rm exp}\left(-{\rm S}_d^{(2)} {\rm M}_p^d {\rm I}_d^{(2)}\right)$, but we have simplified the scenario by ignoring the integrals. Consider the following two cases.

\subsection*{Case 1. Solvable linear case $f(\phi)=g\,\phi$}

In this case the action is Gaussian with source ${\rm J}(x) = g\,Q_1(x)$.  
Let $G(x-x')$ be the regulated Green’s function of the massless scalar:
\begin{equation}
-\partial^2G(x-x')=\delta^{(4)}(x-x').
\end{equation}
which we can use to compute the 
mean $\mu(x)$ and the two-point function. Since the source is a linear function of $\phi(x)$, the classical background is simple $\mu(x) \equiv \langle\phi(x)\rangle_{\rm J}$. This immediately reproduces the familiar results:
\begin{equation}
\mu(x)=\int d^4x'\,G(x-x')\,{\rm J}(x'), \qquad
\langle\phi(x)\phi(x')\rangle_{c}=G(x-x'),
\end{equation}
where $c$ implies connected components.
At this point we can insert in the $y$ dependent operator that appears from removing the integral from ${\rm I}_d^{(2)}(y)$ in \eqref{Tking}. 
The insertion is $\exp[-q_2(y)\phi(y)]$, and for a Gaussian field one can take functional derivatives with respect to ${\rm J}$ to get the following expression:
\begin{equation}
\big\langle e^{-q_2(y)\phi(y)}\big\rangle_{\rm J}
=\exp\!\Big[-q_2(y)\,\mu(y)+\tfrac12\,q_2^2(y)\,G(0)\Big].
\end{equation}
where $G(0) = G(y - y)$ which appears from expressing $q(x)$ as $q(y)\delta^4(x-y)$. This way the source ${\rm J}(x)$ changes to ${\rm J}(x) - q(x)$. This means the computation of $\mathbb{Z}[Q_1, Q_2] \equiv \mathbb{Z}[Q_1, q_2]$ boils down to computing the generating function with a shifted source. This gives us:

{\footnotesize
\begin{equation}
\frac{\mathbb Z[Q_1,q_2]}{\mathbb Z[0,0]}
=\exp\!\left[
\frac12\!\int d^4x\,d^4x'\;{\rm J}(x)\,G(x-x')\,{\rm J}(x')
\;-\;q_2(y)\!\int d^4x\,G(y-x)\,{\rm J}(x)
\;+\;\frac12\,q_2^2(y)\,G(0)
\right], 
\end{equation}}
where $\mathbb{Z}[0, 0]$ is the standard Gaussian value of the path-integral which may be absorbed by choosing appropriately a normalized partition function. We can now plug in the functional form of the source term ${\rm J}(x)$ and the field $q_2(y)$. These provide the following 
explicit form:

{\footnotesize
\begin{equation}\label{skjor}
\frac{\mathbb Z[Q_1,q_2]}{\mathbb Z[0,0]}
=\exp\!\left[
\frac{g^2}{2}\!\int\! d^4x\,d^4x'\;Q_1(x)\,G(x-x')\,Q_1(x')
\;-\;g\,Q_2(y)\!\int\! d^4x\,G(y-x)\,Q_1(x)
\;+\;\frac12\,Q_2^2(y)\,G(0)
\right].
\end{equation}}
Let us now make the following observations on the final result \eqref{skjor}. First, the term $G(0)$ is a UV divergent piece but can be removed by the standard normal-ordering procedure. Secondly, we can easily infer the non-local behavior from the various terms above:
\vskip.2in
\begin{itemize}
  \item The term $\int d^4x\,d^4x'\,Q_1(\psi(x))\,G(x-x')\,Q_1(\psi(x'))$ is bilocal.
  \item The mixed term $\int d^4x\,Q_2(\chi(y))\,G(y-x)\,Q_1(\psi(x))$ couples the single point $y$ to all $x$.
  \item The contact term $\tfrac12 Q_2^2(\chi(y))\, G(0)$ is a local self-contraction but is harmless.
\end{itemize}
\vskip.1in

\noindent Therefore to conclude, we see that integrating out the massless field $\phi(x)$ does lead to the expected non-localities in the system. The final non-local action is expressed in terms of the on-shell fields $\psi(x)$ and $\chi(y)$.

\subsection*{Case 2. General polynomial form of $f(\phi)$}

Our above analysis took a simpler, but solvable, model to illustrate the underlying non-local structure for the corresponding energy-momentum tensor. Question is whether such a clean outcome survives if we take more complicated form for $f(\phi)$. To see this, let us consider the following polynomial form for $f(\phi)$:
\begin{equation}
f(\phi)=\sum_{n\ge 0}\frac{a_n}{n!}\,\phi^n,
\end{equation}
where $n!$ is not necessary for the present analysis but becomes helpful when we allow Borel resummation. Note that, compared to the previous exercise, we cannot use the trick of shifting the source to solve the system. Instead we take the following:
\begin{equation}
\frac{\mathbb Z[Q_1,q_2]}{\mathbb Z[0,0]}
=\left\langle
\exp\!\Big(-\!\int d^4x\,Q_1(x)\,f(\phi(x))\Big)\;
\exp\Big({-q_2(y)\phi(y)}\Big)
\right\rangle_0,
\end{equation}
where $0$ denotes the free vacuum. The functional form for $q_2(y)$ and $Q_1(x)$ remains the same as before. The only change is now the insertion, which is a more complicated function.
The linked-cluster expansion gives:
\begin{equation}
\log\frac{\mathbb Z[Q_1,q_2]}{\mathbb Z[0,0]}
=\sum_{m=0}^{\infty}\frac{(-1)^m}{m!}
\!\!\int\! d^4x_1\cdots d^4x_m \prod_{i=1}^m Q_1(x_i)\;
\Big\langle \prod_{i=1}^m f\!\big(\phi(x_i)\big)\; \exp\Big({-q_2(y)\phi(y)}\Big)\Big\rangle_{0,c},
\end{equation}
where $c$ denotes the connected components. Note that now
each connected Gaussian average produces sums of propagators $G$ that connect all insertion points $x_i$ and the marked point $y$. For example
at order $\mathcal O(Q_1)$ we have:
  \begin{equation}\label{duigarcia}
  -\sum_{n\ge 0}\frac{a_n}{n!}\,\int d^4x\,Q_1(x)\,\Big\langle \phi(x)^n \exp\Big({-q_2(y)\phi(y)}\Big)\Big\rangle_0,
  \end{equation}
  which connects all $n$-th order insertion points at $x$ to a given marked point $y$, where $n \in \mathbb{Z}_+$. For $n = 1$, it is easy to see that:
  \begin{equation}\label{torbmey}
  \big\langle \phi(x)\,\exp\left({-q_2(y)\phi(y)}\right)\big\rangle_0 = -\,q_2(y)\,G(x-y)\,\exp\left({\tfrac12 q_2^2(y) G(0)}\right),
  \end{equation}
  which may be inserted in \eqref{duigarcia} to make the non-locality more transparent. Similarly 
 at order $\mathcal O(Q_1^2)$ (with $a_1^2$ for illustration) the connected components give us:
  \begin{equation}\label{claudhostess}
  {a_1^2\over 2}\!\int\! d^4x\,d^4x'\,Q_1(x)Q_1(x')\,
  \Big[G(x-x')+q_2^2(y) G(x-y)G(x'-y)\Big],
  \end{equation}
with {\it no} overall $\exp\left({\tfrac12 q_2^2(y) G(0)}\right)$ multiplying it. On the other hand, 
if one does not take $\ln ~\mathbb{Z}$ $-$ {\it i.e.} work with ordinary (unconnected) expectations $-$ the factor $\exp\left({\tfrac12 q_2^2(y) G(0)}\right)$ appears multiplicatively from self-contractions at $y$ (as in \eqref{torbmey}). In practice, it is removed by \emph{normal ordering} the insertion:
\bg\label{natalclaud}
: \exp\left({-q_2(y)\phi(y)}\right):\;=\;\exp\left({-q_2(y)\phi(y)-\frac12 q_2^2(y) G(0)}\right),
\nd
or equivalently by composite-operator renormalization. After this subtraction, the $\mathcal O(Q_1^2)$ kernel coincides with the connected result in \eqref{claudhostess}. Therefore, to summarize: 
for the $\mathcal O(a_1^2)$ coefficient multiplying $Q_1(x)Q_1(x')$, the clean expression is \eqref{claudhostess}
while the factor $\exp\left({\tfrac12 q_2^2(y) G(0)}\right)$ is either relegated to the $Q_1^0$ vacuum piece in $\ln~ \mathbb{Z}$ or removed by normal ordering/renormalization of the local insertion at $y$. This further implies that integrating out $\phi$ produces an effective functional with explicitly non-local kernels built from the Green’s function $G(x)$. In {\bf figure \ref{feynmanNL}} we have shown some Feynman-like diagrams to illustrate the non-local interactions.


\begin{figure}[t]
\centering


\caption{Feynman-like diagrams depicting the non-local interactions.  
Panels (A)–(F) are labeled beneath each box with the corresponding interactions specified. Note: Squares = $Q(x)$ insertions, circles = $y$, lines = propagators $G_{xx'} = G(x-x')$. The insertion $y$ appears from  $\exp\left(q_2(y)\phi(y)\right)$. }
\label{feynmanNL}
\end{figure}

\subsubsection{EOMs for the cross-term components ${\bf R}_{m\rho}, {\bf R}_{im}$ and ${\bf R}_{i\rho}$ \label{pabondi}}

Our analysis in section \ref{nolock} gave us not only a way to quantify $-$ albeit for a simpler toy model $-$ the functional form for $\mathbb{A}({\bf \Xi}({\rm Y, Y_Z}, x_{\rm Z}))$ in \eqref{fesserstairs}, but also to quantify the non-locality parameter $\mathbb{F}(x-y)$ in terms of the Green's function $G(x-y)$. The latter identification is not generic because, although for the simpler models the non-localities are captured by $G(x-y)$ (see also {\bf figure \ref{feynmanNL}}), for generic analysis we expect the non-localities to be captured by a more complicated function which may not even depend on the relative distance $x-y$. Thus $\mathbb{F}(x-y) \to \mathbb{F}(x, y)$ and this may be used in \eqref{kimkarol}. Net result is that $\mathbb{A}({\bf \Xi}({\rm Y, Y_Z}, x_{\rm Z}))$ takes the form:

{\footnotesize
\bg\label{natgarcie}
\mathbb{A}({\bf \Xi}({\rm Y, Y_Z}, x_{\rm Z})) &= & {\cal G}\left[\{{\bf Q}_{\rm pert}^{(j)}(\hat{e}_j({\rm S}_d^{(j)}); {\bf \Xi}({\rm Y}, x_{\rm Z}))\}, \{{\bf Q}_{\rm pert}^{(j)}(\hat{d}_j({\rm S}_d^{(j)}); {\bf \Xi}({\rm Y_Z}, x_{\rm Z}))\}, \mathbb{F}({\rm Y, Y_Z})\right] \\
&\times & {\rm exp}\Big\{
{\cal H}\left[\{{\bf Q}_{\rm pert}^{(j)}(\hat{f}_j({\rm S}_d^{(j)}); {\bf \Xi}({\rm Y}, x_{\rm Z}))\}, \{{\bf Q}_{\rm pert}^{(j)}(\hat{g}_j({\rm S}_d^{(j)}); {\bf \Xi}({\rm Y_Z}, x_{\rm Z}))\}, \mathbb{F}({\rm Y, Y_Z})\right]\Big\}, \nonumber \nd}
which is expressed in terms of a polynomial functions $\mathcal G[{\bf Q}_{\rm pert}, \mathbb{F}]$ and $\mathcal H[{\bf Q}_{\rm pert}, \mathbb{F}]$ that are defined with ${\bf Q}_{\rm pert}^{(j)}({e}_j({\rm S}_d^{(j)}); {\bf \Xi})$ $-$ via parameters 
${e}_j({\rm S}_d^{(j)})$ and the on-shell field components ${\bf \Xi}$ $-$ and the non-locality function $\mathbb{F}({\rm Y, Y_Z})$. We will also take $j \ge 4$ so as not to interfere with the earlier trans-series form in \eqref{lidiamey}. A more elaborate trans-series form for the energy-momentum tensor appears from \eqref{natgarcie} after integrating out the off-shell states which is generically non-local. Combining everything together provides the following form for the energy-momentum tensor:

{\scriptsize
\bg\label{omalidia2}
{\bf T}_{\rm C'D'}^{\rm NP}({\rm Z}) & = & 
\sum_{d}\;{\rm M}_p^{d}\,\sum_{\{S^{(i)}_d\}=0}^{\infty}{\rm S}_d^{(2)} h_{\{S^{(i)}_d\}}~\int d^{d}{\rm Y}~\sqrt{{\bf g}_{11}({\rm Y}, x_{\rm Z}) \over{\bf g}_{11}({\rm Y_Z}, x_{\rm Z})}~
~\Theta({\rm Y-Y_Z})~ \mathcal Q_1({\rm Y}, x_{\rm Z}) \sqrt{|{\bf g}_d({\rm Z})|} \nonumber\\
&\otimes & \Big\{\exp\left({-{\rm S}^{(1)}_d {\rm M}_p^d {\rm I}_d^{(1)}({\rm Y}, x_{\rm Z})}\right)~ {\cal G}\left(\{{\bf Q}_{\rm pert}^{(j)}(\hat{e}_j; {\bf \Xi}({\rm Y}, x_{\rm Z}))\}, \{{\bf Q}_{\rm pert}^{(j)}(\hat{d}_j; {\bf \Xi}({\rm Z}))\}, \mathbb{F}({\rm Y, Y_Z})\right) \nonumber\\
&\times & {\rm exp}\Big[
{\cal H}\left(\{{\bf Q}_{\rm pert}^{(j)}(\hat{f}_j; {\bf \Xi}({\rm Y}, x_{\rm Z}))\}, \{{\bf Q}_{\rm pert}^{(j)}(\hat{g}_j; {\bf \Xi}({\rm Z}))\}, \mathbb{F}({\rm Y, Y_Z})\right)\Big] \mathcal Q_{\rm C'D'}^{(2)}(\hat{c}_2;{\bf \Xi}({\rm Z}))\Big\} + ...,
\nd}
where the dotted terms are additional contributions that we discussed earlier, and the other parameters are defined as: $(\hat{e}_j, \hat{d}_j, .., \hat{c}_2) = (\hat{e}_j({\rm S}_d^{(j)}),
\hat{d}_j({\rm S}_d^{(j)}), .., \hat{c}_2({\rm S}_d^{(2)}))$. Taking 
$\mathcal G(x) = \sum\limits_{n = 1}^\infty g_n x^n$ and $\mathcal H(x) = \sum\limits_{n = 1}^\infty h_n x^n$ tell us that they would only contribute as positive powers of $\bar{g}_s$. Negative powers of $\bar{g}_s$, if any, can only come from the $\sqrt{|{\bf g}_d({\rm Z})|}$
which appears from the sub-manifold on which we have our wrapped instantons. This is similar to what we encountered earlier when we studied the contributions from the non-perturbative states to the EOMs of the on-shell metric components. The question now is only in the scaling of $\mathcal Q_{\rm C'D'}^{(2)}(\hat{c}_2({\rm S}_d^{(2)});{\bf \Xi}({\rm Z}))$: should we consider the scalings proposed in {\bf Table \ref{omabiratag}}, or should we simply take the scalings at face value and not opt towards a Lorentz invariant formalism as in \eqref{omaolivia}? The answer lies in what physical observables that we want to compute. For us it is the energy-momentum tensors associated with the cross-term metric components whose expectation values over the Glauber-Sudarshan states vanish, and therefore it becomes necessary to know the scalings of the corresponding {\it perturbative} quantum series as in {\bf Table \ref{omabiratag}}. The EOMs take the form as in \eqref{nikclau}:
\bg\label{ninfelne}
\boxed{{\bf R}_{\rm C'D'}(\langle {\bf \Xi}\rangle_\sigma) =  {\bf T}^{\rm NP}_{\rm C'D'}(\langle {\bf \Xi}\rangle_\sigma)} \nd
where it is assumed that the perturbative contributions come from ${\rm S}_d^{(i)} = 0$ sectors. These may be extracted from \eqref{omalidia2} that we derived earlier.

Our aim in the following would be to first study the EOMs associated with the first and the third rows of {\bf Table \ref{niksmit008}} and see whether we can quantify the 18-tuple $-$ 17-tuple coming from \eqref{melisrain} and $x_{18}$ coming from the non-locality factor $-$ and specify the quantum terms from \eqref{brittbaba007}. After which we will go for the more intriguing case appearing in the second row of {\bf Table \ref{niksmit008}}. 

Let us start with the EOM associated with the off-shell metric components ${\bf g}_{im}$ with $x^i \in {\bf R}^2$ and $y^m \in {\cal M}_4$. It is clear that $\langle {\bf g}_{im}\rangle_\sigma = 0$, and therefore the EOM takes the form \eqref{ninfelne} with $({\rm C', D'}) = (i, m)$. The $\bar{g}_s$ scaling of the energy-momentum tensor ${\bf T}_{im}^{\rm NP}(\langle{\bf \Xi}\rangle_\sigma)$ in \eqref{omalidia2} can be inferred from the three series $\mathcal Q_1, \mathcal G({\bf Q}_{\rm pert})$ and $\mathcal Q^{(2)}_{im}$ and the determinant of the metric $\sqrt{|{\bf g}_d|}$. As before, we will only consider the BBS instantons and therefore ${\cal M}_d = {\cal M}_4 \times {\cal M}_2$. The $\bar{g}_s$ scalings of $\mathcal Q_1$  and $\mathcal G({\bf Q}_{\rm pert})$ can only be positive powers of $\bar{g}_s$ and the $\bar{g}_s$ scaling of $\mathcal Q^{(2)}_{im}$ can be inferred from the fifth row and second column of {\bf Table \ref{omabiratag}}. The determinant factor $\sqrt{|{\bf g}_d|}$, on the other hand scales as $-2 + 2\hat\sigma_e(t) + {\hat\alpha_e(t) + \hat\beta_e(t)\over 2}$. Now using the fact that ${\bf R}_{im}(\langle{\bf \Xi}\rangle_\sigma)$ from {\bf Table \ref{niksmit21}} scales as $0$, the contributions from the BBS instantons to the EOMs would scale as:

{\footnotesize
\bg\label{emynina}
\theta_{nl}  =  {11\over 3} -{\hat\zeta_e(t)\over 2} - {5\hat\sigma_e(t)\over 2} - {\hat\alpha_e(t) + \hat\beta_e(t)\over 2}
 =  \begin{cases}
(0,0,1,0,0,1,0,0,0,0,2,0,0,0,0,0,0,0)\\
(0,0,0,1,1,0,0,0,0,0,2,0,0,0,0,0,0,0)\\
(0,0,0,0,1,1,0,0,0,0,1,0,0,0,0,0,1,0)\end{cases} \nd}
providing three possible solutions for the 18-tuple \eqref{melisrain}. Note that $x_{18} = 0$, thus mirroring the story that we had with the on-shell metric components. The exponential pieces from the non-perturbative BBS instantons do not change the dominant scaling as we saw in the discussion from \eqref{daniniHA} and from {\bf figure \ref{llotus}}. To find out the precise quantum terms contributing here, one may compare \eqref{emynina} with \eqref{nahirmey}. For example, the 18-tuple with the choice $x_5 = x_6 = x_{11} = x_{17} = 1$ and all other $x_i = 0$ easily fits in with the choice from \eqref{nahirmey}. In fact all three choices in \eqref{emynina} show corrections to quartic order in curvatures. This is similar to what we saw for the on-shell cases.

Our next cross-term equation is for ${\bf R}_{i\rho}$, with $\rho \in \mathcal M_2$. There are two values we take for $\rho$, namely $\rho = \alpha$ and $\rho = \beta$. For either of these cases the Ricci tensor 
scales as $0$ as seen from {\bf Table \ref{niksmit23}}. This implies:

{\footnotesize
\bg\label{emynina2}
\theta_{nl}  =  {11\over 3} -{\hat\zeta_e(t)\over 2} - {2\hat\sigma_e(t)} - \hat\alpha_e(t) - {\hat\beta_e(t)\over 2}
 =  \begin{cases}
(0,0,1,0,1,1,0,0,0,0,1,0,0,0,0,0,0,0)\\
(0,0,0,1,0,0,0,0,0,0,2,1,0,0,0,0,0,0)\\
(0,0,0,1,2,0,0,0,0,0,1,0,0,0,0,0,0,0)\\
(0,0,0,0,0,1,0,0,0,0,1,1,0,0,0,0,1,0)\\
(0,0,0,0,2,1,0,0,0,0,0,0,0,0,0,0,1,0)\end{cases} \nd}
where we used {\bf Table \ref{omabiratag}} to fix some part of the scalings of the energy-momentum tensor. \eqref{emynina2}
may now be compared to \eqref{nahirmey} to fit in with either \eqref{botsuga} or \eqref{brittbaba007}. In a similar vein, when $\rho = \beta$, the quantum series contributing from the BBS instantons now scales as:

{\footnotesize
\bg\label{emynina3}
\theta_{nl}  =  {11\over 3} -{\hat\zeta_e(t)\over 2} - {2\hat\sigma_e(t)} - {\hat\alpha_e(t)\over 2} - {\hat\beta_e(t)}
 =  \begin{cases}
(0,0,1,0,0,2,0,0,0,0,1,0,0,0,0,0,0,0)\\
(0,0,1,0,0,0,0,0,0,0,2,0,1,0,0,0,0,0)\\
(0,0,0,1,1,1,0,0,0,0,1,0,0,0,0,0,0,0)\\
(0,0,0,0,1,2,0,0,0,0,0,0,0,0,0,0,1,0)\\
(0,0,0,0,1,0,0,0,0,0,1,0,1,0,0,0,1,0)\end{cases} \nd}
where the difference from \eqref{emynina2} is only in the choice of the 18-tuples. As before the precise connection to \eqref{botsuga} or \eqref{brittbaba007} again appears from comparing the values of the 18-tuples from \eqref{emynina3} with \eqref{nahirmey}. Needless to say, for both \eqref{emynina2} and \eqref{emynina3} we have $x_{18} = 0$, thus no temporal dependence of the non-locality functions and the quantum series are constructed from the quartic order curvature terms.

Our final cross-term equation is for ${\bf R}_{m\rho}$ with two possible set of components: ${\bf R}_{m\alpha}$ and ${\bf R}_{m\beta}$. Looking at {\bf Table \ref{niksmit9}} we see that they scale with respect to $\bar{g}_s$ as $0$. We can again fix some parts of the $\bar{g}_s$ scalings of the energy-momentum tensor from {\bf Table \ref{omabiratag}} as before. The result for the 18-tuples for the set of components ${\bf R}_{m\alpha}$ becomes:

{\footnotesize
\bg\label{emynina4}
\theta_{nl}  =  {8\over 3} - {5\hat\sigma_e(t)\over 2} - {\hat\alpha_e(t)} - {\hat\beta_e(t)\over 2}
 =  \begin{cases}
(0,0,0,0,2,1,0,0,0,0,1,0,0,0,0,0,0,0)\\
(0,0,0,0,0,1,0,0,0,0,2,1,0,0,0,0,0,0)\end{cases} \nd}
which when compared to \eqref{nahirmey} fixes the precise connection to 
\eqref{botsuga} or \eqref{brittbaba007}. As before this identification fixes the contributions from the exponential pieces of the BBS instantons with quartic order curvature corrections. In a similar vein the result for the 18-tuples for the set of components ${\bf R}_{m\beta}$ becomes:

{\footnotesize
\bg\label{emynina5}
\theta_{nl}  =  {8\over 3} - {5\hat\sigma_e(t)\over 2} - {\hat\alpha_e(t)\over 2} - {\hat\beta_e(t)}
 =  \begin{cases}
(0,0,0,0,1,0,0,0,0,0,2,0,1,0,0,0,0,0)\\
(0,0,0,0,1,2,0,0,0,0,1,0,0,0,0,0,0,0)\end{cases} \nd}
thus fixing the quartic order curvature contributions from the BBS instantons by comparing the values of the 18-tuples with \eqref{nahirmey}. We can also ask what happens at the zero instanton sector. For the curvature tensor ${\bf R}_{im}$, the $\bar{g}_s$ scaling for $\theta_{nl}$ leads to the following choice of the 18-tuple:

{\footnotesize
\bg\label{emynina6}
\theta_{nl}  =  {5\over 3} -{\hat\zeta_e(t)\over 2} - {\hat\sigma_e(t)\over 2} =
(0,0,0,0,0,0,0,0,0,0,0,0,0,0,0,0,1,0) \nd}
implying that there are no non-trivial quantum terms that can contribute to the zero instanton sector, and to get useful contributions we need to add in at least one BBS instanton. (We don't study the effects of the KKLT and other instantons as they do not have significant contributions to the on-shell EOMs as seen from section \ref{counting}.) For the curvature tensor ${\bf R}_{i\rho}$, the 18-tuples become:

{\footnotesize
\bg\label{emynina7}
\theta_{nl}  =  {5\over 3} -{\hat\zeta_e(t)\over 2}  - {1\over 2}( {\hat\alpha_e(t)},{\hat\beta_e(t)})
 =  \begin{cases}
(0,0,1,0,0,0,0,0,0,0,0,0,0,0,0,0,0,0)\\
(0,0,0,1,0,0,0,0,0,0,0,0,0,0,0,0,0,0)\end{cases} \nd}
where the two rows are related to $\rho = \alpha$ and $\rho = \beta$ respectively. One can easily see that there are no non-trivial quantum terms contributing to the zero instanton sector and we have to again resort to at least one BBS instanton sector to get any contribution that can solve \eqref{ninfelne}. Finally for the curvature components ${\bf R}_{m\rho}$, the zero instanton sector provides the following scalings:

{\footnotesize
\bg\label{emynina8}
\theta_{nl}  =  {2\over 3} -{\hat\sigma_e(t)\over 2}  - {1\over 2}( {\hat\alpha_e(t)},{\hat\beta_e(t)})
 =  \begin{cases}
(0,0,0,0,1,0,0,0,0,0,0,0,0,0,0,0,0,0)\\
(0,0,0,0,0,1,0,0,0,0,0,0,0,0,0,0,0,0)\end{cases} \nd}
where again the two rows are respectively related to $\rho = \alpha$ and $\rho = \beta$. The conclusion also remains the same: there are no non-trivial contributions from the zero instanton sector that can solve \eqref{ninfelne}, and one needs to introduce at least one BBS instanton. 
Once we include the contributions from the BBS instantons, comparing \eqref{emynina} till \eqref{emynina5}, suggests that the dominant quantum contributions come from quartic order in curvatures, exactly as we saw for the on-shell EOMs in section \ref{counting}.

\subsubsection{EOMs for the cross-term components ${\bf R}_{0n}, {\bf R}_{0\rho}$ and ${\bf R}_{0i}$ \label{aidapril}}

So far our exhaustive counting of the number of solutions for the on-shell cases in section \ref{counting} and part of the off-shell cases in section \ref{pabondi} did {\it not} produce any relationship between the parameters $(\hat\zeta_e(t), \hat\sigma_e(t), \hat\alpha_e(t), \hat\beta_s(t), \hat\eta_e(t))$. We also found multiple possibilities at the non-perturbative levels with quartic curvature contributions from the BBS instantons. While this spell consistencies in our overall program, the question of whether we can somehow find some equations relating these parameters is important. In the following we will analyze the EOMs for the cross-term components  ${\bf R}_{0n}, {\bf R}_{0\rho}$ and ${\bf R}_{0i}$ in two ways: {\Su one}, without imposing any relations between the aforementioned parameters and {\Su two}, by imposing a generic relation. Our aim is to see what happens in the latter case.

To make the analysis quantitatively, let us look at the $\bar{g}_s$ scalings of the curvature components ${\bf R}_{0n}, {\bf R}_{0i}$ and ${\bf R}_{0\alpha}$ from {\bf Tables \ref{niksmit11}}, {\bf \ref{niksmit18}} and {\bf \ref{niksmit24}}. They all scale as $-1 + {\hat\alpha_e(t) + \hat\beta_e(t)\over 4}$. On the other hand, the scalings at the zero instanton sector for the energy-momentum tensors are given in {\bf Table \ref{omabiratag}}. Our earlier experience has shown that unless we go to the non-zero instanton sectors, it is impossible to solve the Schwinger-Dyson equations. This continues to be the case here too. Thus the EOMs are the ones from \eqref{ninfelne}, which implies the inclusion of the non-perturbative effects (both local and non-local ones). Solving these equations for the various components suggest the following $\bar{g}_s$ scalings for the energy-momentum tensors:
\bg\label{lidiangtee}
&& {\bf T}^{\rm NP}_{0i}: ~\theta_{nl} \equiv {\rm L}_1 = {11\over 3} - \hat\zeta_e(t) - 2\hat\sigma_e(t) - {\hat\alpha_e(t) + \hat\beta_e(t)\over 4}\nonumber\\
&& {\bf T}^{\rm NP}_{0m}: ~\theta_{nl}\equiv {\rm L}_2 = {8\over 3} - {\hat\zeta_e(t)\over 2} - {5\hat\sigma_e(t)\over 2} - {\hat\alpha_e(t) + \hat\beta_e(t)\over 4}\nonumber\\
&& {\bf T}^{\rm NP}_{0\alpha}: ~\theta_{nl}\equiv {\rm L}_3 = {8\over 3} - {\hat\zeta_e(t)\over 2} - 2\hat\sigma_e(t) - {3\hat\alpha_e(t) + \hat\beta_e(t)\over 4}\nonumber\\
&& {\bf T}^{\rm NP}_{0\beta}: ~\theta_{nl} \equiv {\rm L}_4 = {8\over 3} - {\hat\zeta_e(t)\over 2} - 2\hat\sigma_e(t) - {\hat\alpha_e(t) + 3\hat\beta_e(t)\over 4}, \nd
where $\theta_{nl}$ for all the above cases should be compared to \eqref{brittbaba007}, much like the way we did earlier starting with \eqref{melisrain}. Comparing all the four scalings from \eqref{lidiangtee}, we can easily convince ourselves that there are {non-negative integer values} of $x_i$ that can solve:
\bg\label{streeber}
\sum_{i=1}^{17} x_i\,{\rm B}_i(\hat\sigma_e,\hat\zeta_e,\hat\alpha_e,\hat\beta_e)\;-\;\frac{2}{3}x_{18} = \begin{cases}
{11\over 3} - \hat\zeta_e(t) - 2\hat\sigma_e(t) - {\hat\alpha_e(t) + \hat\beta_e(t)\over 4}\\
{8\over 3} - {\hat\zeta_e(t)\over 2} - {5\hat\sigma_e(t)\over 2} - {\hat\alpha_e(t) + \hat\beta_e(t)\over 4}\\
{8\over 3} - {\hat\zeta_e(t)\over 2} - 2\hat\sigma_e(t) - {3\hat\alpha_e(t) + \hat\beta_e(t)\over 4}\\
{8\over 3} - {\hat\zeta_e(t)\over 2} - 2\hat\sigma_e(t) - {\hat\alpha_e(t) + 3\hat\beta_e(t)\over 4}\end{cases} \nd
{\it without} incorporating any relation between the parameters $(\hat\zeta_e(t), \hat\sigma_e(t), \hat\alpha_e(t), \hat\beta_s(t), \hat\eta_e(t))$, at least in the gravitational sector of \eqref{brittbaba007} or \eqref{botsuga}. In the flux sector from \eqref{brittbaba007}, we see that all the flux components have a dominant scaling of ${4\over 3}^\pm$ and therefore the only possible solutions could be:
\bg\label{deutscheL}
&& x_{l+17} = 2, ~~~~~~~~~~~~~~~~ {\rm all ~other}~~~x_{i+17} = 0 \nonumber\\
&& x_{i+17} = x_{j+17} = 1, ~~~~~ {\rm all ~ other} ~~~ x_{k+17} = 0, \nd
which would unfortunately not lead to any non-trivial interactions in the theory. (This is easy to justify from the ${8\over 3}^\pm$ factors 
appearing in \eqref{streeber}. For the ${11\over 3}^\pm$ factor, there are possibilities of non-trivial interactions solving \eqref{streeber}, but since the remaining three cases do not work, the whole procedure fails to provide solutions.) The ${\rm B}_i$ appearing above can be inferred from \eqref{brittbaba007}, which we also display in {\bf Table \ref{millerkhul}} for convenience.

\begin{table}[tb]  
 \begin{center}
\renewcommand{\arraystretch}{3.5}
\resizebox{\textwidth}{!}    
{\begin{tabular}{||c|c||c|c||c|c||c|c||c|c||c|c||}\hline
${\rm B}_1$ & $\tfrac{4}{3}-\tfrac{\hat\zeta_e(t)}{2}$ & $
{\rm B}_2 $ & $ \tfrac{5}{3}+\tfrac{\hat\alpha_e(t)}{4}+\tfrac{\hat\beta_e(t)}{4}-\hat\zeta_e(t)$ & 
${\rm B}_3 $ & $ \tfrac{5}{3}-\tfrac{\hat\alpha_e(t)}{2}-\tfrac{\hat\zeta_e(t)}{2}
$ & $
{\rm B}_4 $  & $ \tfrac{5}{3}-\tfrac{\hat\beta_e(t)}{2}-\tfrac{\hat\zeta_e(t)}{2}
$ &  $ {\rm B}_5 $ & $ \tfrac{2}{3}-\tfrac{\hat\sigma_e(t)}{2}-\tfrac{\hat\alpha_e(t)}{2}
$ & $
{\rm B}_6  $ & $  \tfrac{2}{3}-\tfrac{\hat\sigma_e(t)}{2}-\tfrac{\hat\beta_e(t)}{2}$ \\ \hline
$ {\rm B}_7 $ & $ \tfrac{2}{3}-\tfrac{\hat\alpha_e(t)}{2}+\tfrac{\hat\alpha_e(t)}{4}+\tfrac{\hat\beta_e(t)}{4}-\tfrac{\hat\zeta_e(t)}{2} $ & $
{\rm B}_8 $ & $ \tfrac{2}{3}-\tfrac{\hat\beta_e(t)}{2}+\tfrac{\hat\alpha_e(t)}{4}+\tfrac{\hat\beta_e(t)}{4}-\tfrac{\hat\zeta_e(t)}{2} $ & $
{\rm B}_9  $ &  $  \tfrac{2}{3}+\tfrac{\hat\alpha_e(t)}{4}+\tfrac{\hat\beta_e(t)}{4}-\tfrac{\hat\sigma_e(t)}{2}-\tfrac{\hat\zeta_e(t)}{2} $  & $
{\rm B}_{10}  $ & $ \tfrac{8}{3}-\hat\zeta_e(t) $ & $
{\rm B}_{11} $ &  $ \tfrac{2}{3}-\hat\sigma_e(t) $ & $
{\rm B}_{12} $ & $\tfrac{2}{3}-\hat\alpha_e(t)$ \\ \hline
${\rm B}_{13} $ & $ \tfrac{2}{3}-\hat\beta_e(t) $ & $
{\rm B}_{14} $ & $ \tfrac{2}{3}+\tfrac{\hat\alpha_e(t)}{2}+\tfrac{\hat\beta_e(t)}{2}-\hat\zeta_e(t)$
& $
{\rm B}_{15} $ & $  \tfrac{2}{3}-2\hat\alpha_e(t)+\hat\beta_e(t) $ & $
{\rm B}_{16}  $ & $  \tfrac{2}{3}-2\hat\beta_e(t)+\hat\alpha_e(t) $  & $
{\rm B}_{17} $ & $ \tfrac{5}{3}-\tfrac{\hat\sigma_e(t)}{2}-\tfrac{\hat\zeta_e(t)}{2} 
$ & ${\rm B}_{18} $ & $ -{2\over 3}$\\ \hline 
\end{tabular}}
\renewcommand{\arraystretch}{1}
\end{center}
 \caption[]{The values of the ${\rm B}_i(\hat\zeta_e, \hat\sigma_e, \hat\alpha_e, \hat\beta_e)$ appearing in \eqref{streeber}. They are taken from the gravitational parts of \eqref{brittbaba007}.}
 \label{millerkhul}
 \end{table}

The solutions for the four target values ${\rm L}_k$ can now be easily determined by comparing both the sides of \eqref{streeber}. For ${\rm L}_1$, this takes the form:

{\footnotesize
\bg\label{olwastoon}
{11\over 3} - \hat\zeta_e(t) - 2\hat\sigma_e(t) - {\hat\alpha_e(t) + \hat\beta_e(t)\over 4} = \begin{cases}
(0, 0, 0, 0, 1, 1, 0, 0, 1, 0, 0, 0, 0, 0, 0, 0, 1, 0)\\ 
(0, 0, 0, 0, 0, 1, 1, 0, 0, 0, 1, 0, 0, 0, 0, 0, 1, 0) \\
(0, 0, 1, 0, 0, 1, 0, 0, 1, 0, 1, 0, 0, 0, 0, 0, 0, 0)\\ 
(0, 0, 0, 0, 1, 0, 0, 1, 0, 0, 1, 0, 0, 0, 0, 0, 1, 0)\\ 
(0, 0, 0, 1, 1, 0, 0, 0, 1, 0, 1, 0, 0, 0, 0, 0, 0, 0)\\
 (0, 1, 0, 0, 1, 1, 0, 0, 0, 0, 1, 0, 0, 0, 0, 0, 0, 0)\\
 (0, 0, 0, 1, 0, 0, 1, 0, 0, 0, 2, 0, 0, 0, 0, 0, 0, 0)\\
 (0, 0, 1, 0, 0, 0, 0, 1, 0, 0, 2, 0, 0, 0, 0, 0, 0, 0)
    \end{cases} \nd}
which are eight 18-tuples that consistently solve the ${\bf R}_{0i}$ EOMs with quartic order curvature terms from the BBS instantons. For the ${\rm L}_2$ target from \eqref{lidiangtee}, \eqref{streeber} yields:

{\footnotesize
\bg\label{olwastoon2}
{8\over 3} - {\hat\zeta_e(t)\over 2} - {5\hat\sigma_e(t)\over 2} - {\hat\alpha_e(t) + \hat\beta_e(t)\over 4} = \begin{cases}
(0, 0, 0, 0, 1, 1, 0, 0, 1, 0, 1, 0, 0, 0, 0, 0, 0, 0)\\
(0, 0, 0, 0, 0, 1, 1, 0, 0, 0, 2, 0, 0, 0, 0, 0, 0, 0)\\ 
(0, 0, 0, 0, 1, 0, 0, 1, 0, 0, 2, 0, 0, 0, 0, 0, 0, 0) \end{cases} \nd}
giving rise to three 18-tuples that provide quartic curvature corrections from the BBS instantons required to solve the EOMs for 
${\bf R}_{0n}$ in \eqref{ninfelne}. In a similar vein, for the target ${\rm L}_3$ we get the following values:

{\footnotesize
\bg\label{olwastoon3}
{8\over 3} - {\hat\zeta_e(t)\over 2} - {2\hat\sigma_e(t)} - {3\hat\alpha_e(t) + \hat\beta_e(t)\over 4} = \begin{cases}
(0, 0, 0, 0, 2, 1, 0, 0, 1, 0, 0, 0, 0, 0, 0, 0, 0, 0)\\ 
(0, 0, 0, 0, 0, 1, 0, 0, 1, 0, 1, 1, 0, 0, 0, 0, 0, 0)\\ 
(0, 0, 0, 0, 1, 1, 1, 0, 0, 0, 1, 0, 0, 0, 0, 0, 0, 0)\\ 
(0, 0, 0, 0, 2, 0, 0, 1, 0, 0, 1, 0, 0, 0, 0, 0, 0, 0)\\
 (0, 0, 0, 0, 0, 0, 0, 1, 0, 0, 2, 1, 0, 0, 0, 0, 0, 0)\end{cases} \nd}
 which are five 18-tuples required to solve the EOMs for ${\bf R}_{0\alpha}$, again with quartic curvature terms from the BBS instantons. If we replace $\alpha$ by $\beta$ directions on $\mathcal M_4$, the 18-tuples for the ${\bf R}_{0\beta}$ EOMs become:

{\footnotesize
\bg\label{olwastoon4}
{8\over 3} - {\hat\zeta_e(t)\over 2} - {2\hat\sigma_e(t)} - {\hat\alpha_e(t) + 3\hat\beta_e(t)\over 4} = \begin{cases} 
(0, 0, 0, 0, 1, 2, 0, 0, 1, 0, 0, 0, 0, 0, 0, 0, 0, 0)\\ 
(0, 0, 0, 0, 0, 2, 1, 0, 0, 0, 1, 0, 0, 0, 0, 0, 0, 0)\\ 
(0, 0, 0, 0, 1, 0, 0, 0, 1, 0, 1, 0, 1, 0, 0, 0, 0, 0)\\ 
(0, 0, 0, 0, 1, 1, 0, 1, 0, 0, 1, 0, 0, 0, 0, 0, 0, 0)\\ 
(0, 0, 0, 0, 0, 0, 1, 0, 0, 0, 2, 0, 1, 0, 0, 0, 0, 0)\end{cases} \nd}
which are again five 18-tuples but differing slightly from \eqref{olwastoon3}. Comparing with \eqref{olwastoon}, \eqref{olwastoon2}, and \eqref{olwastoon3}, we see that the five parameters $(\hat\zeta_e(t), \hat\sigma_e(t), \hat\alpha_e(t), \hat\beta_e(t), \hat\eta_e(t))$ still {\it do not} get fixed.  On one hand, this is good because it shows that all the metric EOMs follow similar patterns, and they all require quartic order curvature terms coming from the non-perturbative BBS instantons. On the other hand, a slight concern may be that the whole analysis appears to work for all values of the five parameters ${p}_e(t) \equiv (\hat\zeta_e(t), \hat\sigma_e(t), \hat\alpha_e(t), \hat\beta_e(t), \hat\eta_e(t))$. However for the consistency of our construction $-$ shown for example in {\bf Table \ref{milleren4}} $-$ we need to keep all the parameters arbitrarily small, {\it i.e.} $|p_e(t)| \ll 1$. This, as we shall discuss in section \ref{dynamical}, is possible if at least $\hat\zeta_e(t)$ becomes a function of $ (\hat\sigma_e(t), \hat\alpha_e(t), \hat\beta_e(t))$.

Our conjecture then is the following. To maintain $|p_e(t)| \ll 1$, all the four parameters $(\hat\zeta_e(t), \hat\sigma_e(t), \hat\alpha_e(t), \hat\beta_e(t))$ in the gravitational sector of \eqref{brittbaba007}, cannot be independent of each other\footnote{How this fixes $\hat\eta_e(t)$ will be discussed in section \ref{dynamical}.}. We therefore propose the following polynomial relation between the various parameters:
\bg\label{irishdeka}
\hat\zeta_e(t) = \mathbb{P}[\hat\sigma_e(t), \hat\alpha_e(t), \hat\beta_e(t)] & = & \sum_{(n, m, p)\ge 0} c_{mnp} ~\hat\sigma^m_e(t)~ \hat\alpha^n_e(t) ~\hat\beta^p_e(t)\\
& = & a\hat\sigma_e(t) + {p\hat\alpha_e(t) + q\hat\beta_e(t)\over 2} + 
\sum_{n, m, p} c'_{mnp}~ \hat\sigma^m_e(t)~ \hat\alpha^n_e(t) ~\hat\beta^p_e(t), \nonumber \nd 
where $c_{100} = a, c_{010} = {p\over 2}$, $c_{001} = {q\over 2}$ and 
$c'_{mnp} = c_{mnp} - \{c_{100}, c_{010}, c_{001}\}$ thus capturing all the higher order terms in $(\hat\sigma_e(t), \hat\alpha_e(t), \hat\beta_e(t))$. However since $(\hat\sigma_e(t), \hat\alpha_e(t), \hat\beta_e(t)) << 1$, the linear term in \eqref{irishdeka} will dominate and therefore we propose that:
\bg\label{irishdeka2}
\boxed{\ \hat\zeta_e(t) \;=\; a\,\hat\sigma_e(t) \;+\; \frac{p\,\hat\alpha_e(t) + q\,\hat\beta_e(t)}{2}\ ,\qquad a,p,q\in\mathbb Z\ }
\nd
is sufficient to realize, for each of the four RHS choices, a nonnegative integer solution to the 18-tuple
$(x_1,\dots,x_{18})$ in \eqref{streeber}, with $\hat\alpha_e(t), \hat\beta_e(t), \hat\sigma_e(t), \hat\zeta_e(t) \neq 0$. At this stage we impose no relation between  $\hat\alpha_e(t), \hat\beta_e(t)$ and $ \hat\sigma_e(t)$. The question however is whether we can also fix $(a, p, q)$ from \eqref{streeber}.

To see how far we can go with a generic solution like \eqref{irishdeka2}, our strategy would be to substitute this in the four target values on the RHS of \eqref{streeber}, and see how they match-up with the LHS given by the ${\rm B}_i$ values from {\bf Table \ref{millerkhul}}. The substitution leads to the following relations:
\bg\label{irishdeka3}
&& {\rm L}_1=\frac{11}{3}-(a{+}2)\,\hat\sigma_e(t)-\Big(\frac{p}{2}{+}\frac14\Big)\hat\alpha_e(t)-\Big(\frac{q}{2}{+}\frac14\Big)\hat\beta_e(t) \nonumber\\
&& {\rm L}_2=\frac{8}{3}-\Big({a\over 2} + {5\over 2}\Big)\,\hat\sigma_e(t)-\Big(\frac{p}{4}{+}\frac14\Big)\hat\alpha_e(t)-\Big(\frac{q}{4}{+}\frac14\Big)\hat\beta_e(t)\nonumber\\
&& {\rm L}_3=\frac{8}{3}-\Big({a\over 2} + 2\Big)\,\hat\sigma_e(t)-\Big(\frac{p}{4}{+}\frac34\Big)\hat\alpha_e(t)-\Big(\frac{q}{4}{+}\frac14\Big)\hat\beta_e(t)\nonumber\\
&& {\rm L}_4 =\frac{8}{3}-\Big({a\over 2} + 2\Big)\,\hat\sigma_e(t)-\Big(\frac{p}{4}{+}\frac14\Big)\hat\alpha_e(t)-\Big(\frac{q}{4}{+}\frac34\Big)\hat\beta_e(t),
\nd
where we see that the substitution simply shifts the coefficients of $\hat\sigma_e(t), \hat\alpha_e(t)$ and $\hat\beta_e(t)$ by multiples of ${a\over 2}, {p\over 4}$ and ${q\over 4}$ respectively. In other words, 
in every case the coefficient triple lies in the lattice:
\bg\label{peribindu55}
\Lambda_{\hat\sigma_e,\hat\alpha_e,\hat\beta_e}\;=\; {\mathbb{Z}\over 2} ~\times ~ {\mathbb{Z}\over 4} ~ \times ~ {\mathbb{Z}\over 4}, \nd
and the constant term lies in ${\mathbb{Z}\over 3}$. The triples $(a, p, q)$ would simply denote a point on the $\Lambda_{\hat\sigma_e,\hat\alpha_e,\hat\beta_e}$ lattice.

For the LHS of \eqref{streeber}, with ${\rm B}_i$ defined in {\bf Table \ref{millerkhul}}, the story is similar but with some interesting subtleties. The LHS generators 
 span the same lattice \eqref{peribindu55} (ignoring constants) as may be seen from the $\hat\zeta_e(t)$ factors in {\bf Table \ref{millerkhul}}. The terms populating the $\hat\zeta_e(t)$ equation 
can be collected as:
\bg\label{514616leatag}
-x_2 - x_{10} - x_{14} -{x_1\over 2} - {x_3\over 2} - {x_4\over 2} -{x_7\over 2} - {x_8\over 2} -{x_9\over 2} -{x_{17} \over 2}, \nd
which provide the necessary shifts in the lattices for the $\hat\sigma_e(t), \hat\alpha_e(t)$ and $\hat\beta_e(t)$ coefficients. For example the coefficient for the $\hat\sigma_e(t)$ equation shifts in the following way:
\bg\label{pascsubs}
&& -{x_5\over 2} - {x_6\over 2} - {x_9\over 2} - x_{11} - {x_{17}\over 2} 
~\rightarrow ~ -{ax_1\over 2} - ax_2 - ax_3 - {ax_4\over 2} -{x_5\over 2} - {x_6\over 2} -{ax_7\over 2} - {ax_8\over 2} \nonumber\\
&& -\left({a\over 2} + {1\over 2}\right)x_9 - a x_{10} - x_{11} - ax_{14}
-\left({a\over 2} + {1\over 2}\right)x_{17} = \begin{cases}
    ~-(a+2)\\
    ~-\left({a \over 2} + {5\over 2}\right)\\
    ~-\left({a\over 2} + 2\right)
\end{cases} \nd
where on the RHS we have shown the three target values that may be read up from \eqref{irishdeka3}. In a similar vein, the coefficient of 
$\hat\alpha_e(t)$ equation shifts in the following way:
\bg\label{kill46}
&&{x_2\over 4} - {x_3\over 2} - {x_5\over 2} -{x_7\over 4} + {x_8\over 4} + {x_9\over 4} - x_{12} + {x_{14}\over 2} - 2x_{15} + x_{16} 
~\rightarrow ~  - {px_1\over 4} + \left({1\over 4} - {p\over 2}\right)x_2
\nonumber\\
&-&  \left({1\over 2} + {p\over 4}\right)x_3 -{px_4\over 4} -{x_5\over 2}  - \left({1\over 4} + {p\over 4}\right)x_7 +  \left({1\over 4} - {p\over 4}\right)x_8 + \left({1\over 4} - {p\over 4}\right)x_9 - {px_{10}\over 2} - x_{12}\nonumber\\
&+&  \left({1\over 2} - {p\over 2}\right)x_{14} - 2x_{15} + x_{16} -{px_{17}\over 4} = \begin{cases}
~ -\left({p\over 2} + {1\over 4}\right)\\
~ - \left({p\over 4} + {1\over 4}\right)\\
~ - \left({p\over 4} + {3\over 4}\right)
\end{cases}
\nd
where we can read up the three target values on the RHS from \eqref{irishdeka3}. As in \eqref{pascsubs}, we note the proliferation of the terms in \eqref{kill46} induced by \eqref{514616leatag}. Similar proliferation can be seen for the coefficients of the $\hat\beta_e(t)$ equation:
\bg\label{horizval}
&&{x_2\over 4} - {x_4\over 2} - {x_6\over 2} -{x_7\over 4} - {x_8\over 4} + {x_9\over 4} - x_{13} + {x_{14}\over 2} + x_{15} -2x_{16} 
~\rightarrow ~  - {qx_1\over 4} + \left({1\over 4} - {q\over 2}\right)x_2
\nonumber\\
&-&  {qx_3\over 4} +  \left({1\over 2} + {q\over 4}\right)x_4 -{x_6\over 2}  + \left({1\over 4} - {q\over 4}\right)x_7 -  \left({1\over 4} + {q\over 4}\right)x_8 + \left({1\over 4} - {q\over 4}\right)x_9 - {qx_{10}\over 2} - x_{13}\nonumber\\
&+&  \left({1\over 2} - {q\over 2}\right)x_{14} + x_{15} - 2x_{16} -{qx_{17}\over 4} = \begin{cases}
~ -\left({q\over 2} + {1\over 4}\right)\\
~ - \left({q\over 4} + {1\over 4}\right)\\
~ - \left({q\over 4} + {3\over 4}\right)
\end{cases}
\nd
with the three targets on the RHS appearing from \eqref{irishdeka3}. Our aim would be to solve for the 17-tuples (keeping $x_{18}= 0$ so as to be consistent with what we had earlier) with non-negative integer values. Three subtleties appear now: {\Su one}, for each of the four targets from \eqref{irishdeka3} we need to find the requisite 17-tuples {\it taking the same values for $(a, p, q)$.} {\Su Two}, the ordering of the targets appearing on the RHS of \eqref{pascsubs}, \eqref{kill46} and \eqref{horizval} have to be first aligned with the ones on the RHS of \eqref{irishdeka3}.  And {\Su three}, the constant terms on the LHS of \eqref{irishdeka3} {\it do not} get shifted by \eqref{514616leatag}, and therefore they impose the following {\it two} additional constraints:
\bg\label{harizztag}
4x_1 + 5\left(\sum_{i = 2}^4 x_i + x_{17}\right) + 2\left(\sum_{l = 5}^9 x_l + \sum_{j = 11}^{16} x_j - x_{18}\right) + 8x_{10} = \begin{cases}
    ~11\\
    ~~\\
    ~8
\end{cases} \nd
where the LHS resembles somewhat the LHS of \eqref{falooda} but the RHS is different. (See also {\bf Table \ref{millerkhul2}}.) The question now the following: without altering the chosen values for $(a, p, q)$ can we find non-negative integer values for the 17-tuples that also solve \eqref{harizztag}?


\begin{table}[tb]  
 \begin{center}
\renewcommand{\arraystretch}{1.3}
\resizebox{\textwidth}{!}    
{
}
\renewcommand{\arraystretch}{1}
\end{center}
 \caption[]{Table showing the values of the coefficients $(\mathcal C_i, \mathcal S_i, \mathcal A_i, \mathcal B_i)$ For each term on the LHS of \eqref{irishdeka3} of the form 
 $\big(\mathcal C_i + \mathcal S_i\,\hat\sigma_e(t) + \mathcal A_i\,\hat\alpha_e(t) + \mathcal B_i\,\hat\beta_e(t)\big)x_i$.}
 \label{millerkhul2}
 \end{table}

This appears to be a complicated problem at face value. We not only have to keep $(a, p, q)$ fixed to some chosen value, but also make sure that the 18-tuples have non-negative integer entries for all the four targets. The answer however turns out to be surprisingly simple! For the target ${\rm L}_1$ from \eqref{irishdeka3}, we have:

{\scriptsize
\bg\label{bathory1}
\dfrac{11}{3}-(a{+}2)\hat\sigma_e(t)-\left(\dfrac{p}{2}{+}\dfrac14\right)\hat\alpha_e(t)-\left(\dfrac{q}{2}{+}\dfrac14\right)\hat\beta_e(t) = \begin{cases}
(0, 0, 0, 0, 1, 1, 0, 0, 1, 0, 0, 0, 0, 0, 0, 0, 1, 0)\\ 
(0, 0, 0, 0, 0, 1, 1, 0, 0, 0, 1, 0, 0, 0, 0, 0, 1, 0) \\
(0, 0, 1, 0, 0, 1, 0, 0, 1, 0, 1, 0, 0, 0, 0, 0, 0, 0)\\ 
(0, 0, 0, 0, 1, 0, 0, 1, 0, 0, 1, 0, 0, 0, 0, 0, 1, 0)\\ 
(0, 0, 0, 1, 1, 0, 0, 0, 1, 0, 1, 0, 0, 0, 0, 0, 0, 0)\\
 (0, 1, 0, 0, 1, 1, 0, 0, 0, 0, 1, 0, 0, 0, 0, 0, 0, 0)\\
 (0, 0, 0, 1, 0, 0, 1, 0, 0, 0, 2, 0, 0, 0, 0, 0, 0, 0)\\
 (0, 0, 1, 0, 0, 0, 0, 1, 0, 0, 2, 0, 0, 0, 0, 0, 0, 0)
\end{cases} \nd}
which is exactly what we had in \eqref{olwastoon} and clearly shows the presence of quartic order in curvature from \eqref{brittbaba007}. Moreover all the conditions from \eqref{pascsubs}, \eqref{kill46} and \eqref{horizval} are satisfied, including the constraint from \eqref{harizztag} as shown for one specific 18-tuple from \eqref{bathory1} in the first row of {\bf Table \ref{millerkhul3}}. Similarly, for the next target ${\rm L}_2$ from 
\eqref{irishdeka3}, we have:

{\scriptsize
\bg\label{bathory2}
\dfrac{8}{3}- \left({a\over 2} + {5\over 2}\right)\hat\sigma_e(t)-\left(\dfrac{p}{4}{+}\dfrac14\right)\hat\alpha_e(t)-\left(\dfrac{q}{4}{+}\dfrac14\right)\hat\beta_e(t) = \begin{cases}
(0, 0, 0, 0, 1, 1, 0, 0, 1, 0, 1, 0, 0, 0, 0, 0, 0, 0)\\
(0, 0, 0, 0, 0, 1, 1, 0, 0, 0, 2, 0, 0, 0, 0, 0, 0, 0)\\ 
(0, 0, 0, 0, 1, 0, 0, 1, 0, 0, 2, 0, 0, 0, 0, 0, 0, 0) \end{cases} \nd}
which again unsurprisingly matches with \eqref{olwastoon2} and suggests the presence of quartic order in curvatures on the BBS instantons. As in \eqref{bathory1}, one may show consistency with all the four conditions from \eqref{pascsubs}, \eqref{kill46}, \eqref{horizval} and 
\eqref{harizztag}. This is depicted for one specific 18-tuple in the second row of {\bf Table \ref{millerkhul3}}. (One may easily work out for all other 18-tuples in \eqref{bathory1}.) For the next target ${\rm L}_3$, the details match with \eqref{olwastoon3}:

{\scriptsize
\bg\label{bathory3}
\dfrac{8}{3}-\left({a\over 2} + 2\right)\hat\sigma_e(t)-\left(\dfrac{p}{4}{+}\dfrac34\right)\hat\alpha_e(t)-\left(\dfrac{q}{4}{+}\dfrac14\right)\hat\beta_e(t) = \begin{cases}
(0, 0, 0, 0, 2, 1, 0, 0, 1, 0, 0, 0, 0, 0, 0, 0, 0, 0)\\ 
(0, 0, 0, 0, 0, 1, 0, 0, 1, 0, 1, 1, 0, 0, 0, 0, 0, 0)\\ 
(0, 0, 0, 0, 1, 1, 1, 0, 0, 0, 1, 0, 0, 0, 0, 0, 0, 0)\\ 
(0, 0, 0, 0, 2, 0, 0, 1, 0, 0, 1, 0, 0, 0, 0, 0, 0, 0)\\
 (0, 0, 0, 0, 0, 0, 0, 1, 0, 0, 2, 1, 0, 0, 0, 0, 0, 0)\end{cases} \nd}
along with the appearance of the correction to quartic order in curvatures. 
Consistencies with the four conditions from \eqref{pascsubs}, \eqref{kill46}, \eqref{horizval} and 
\eqref{harizztag} follow again as shown in the third row of {\bf Table \ref{millerkhul3}}. Finally, for the fourth target ${\rm L}_4$, we get:

{\scriptsize
\bg\label{bathory4}
\dfrac{8}{3}-\left({a\over 2} + 2\right)\hat\sigma_e(t)-\left(\dfrac{p}{4}{+}\dfrac14\right)\hat\alpha_e(t)-\left(\dfrac{q}{4}{+}\dfrac34\right)\hat\beta_e(t)\begin{cases} 
(0, 0, 0, 0, 1, 2, 0, 0, 1, 0, 0, 0, 0, 0, 0, 0, 0, 0)\\ 
(0, 0, 0, 0, 0, 2, 1, 0, 0, 0, 1, 0, 0, 0, 0, 0, 0, 0)\\ 
(0, 0, 0, 0, 1, 0, 0, 0, 1, 0, 1, 0, 1, 0, 0, 0, 0, 0)\\ 
(0, 0, 0, 0, 1, 1, 0, 1, 0, 0, 1, 0, 0, 0, 0, 0, 0, 0)\\ 
(0, 0, 0, 0, 0, 0, 1, 0, 0, 0, 2, 0, 1, 0, 0, 0, 0, 0)\end{cases} \nd}
with similar conclusion as in \eqref{olwastoon4} which may also be verified from the fourth row of {\bf Table \ref{millerkhul3}}. Collecting \eqref{bathory1}, \eqref{bathory2}, \eqref{bathory3} and \eqref{bathory4} provides the last set of keys to complete the consistency of the whole construction, including the justification that the acceleration of our universe may indeed be provided consistently by the BBS instantons.

\begin{table}[tb]  
 \begin{center}
\renewcommand{\arraystretch}{1.3}
\resizebox{\textwidth}{!}    
{\begin{tabular}{||c|c|c||}\hline
Target & Non-zero $x_i$ (each listed $x_i=1$) & Check of coefficients \\ \hline
${\rm L}_1$ &
$x_2,\;x_5,\;x_6,\;x_{11}$ &
$\begin{aligned}
\text{Const:}&\; {5/3}+{2/3} + {2/3} + {2/3} ={11/3}\\
\hat\sigma_e(t):&\; (-a) + (-\tfrac12)+(-\tfrac12)+(-1)=-(a+2)\\
\hat\alpha_e(t) :&\; (\tfrac14-\tfrac{p}{2})+(-\tfrac12)=-(\tfrac{p}{2}+\tfrac14)\\
\hat\beta_e(t):&\; (\tfrac14-\tfrac{q}{2})+(-\tfrac12)=-(\tfrac{q}{2}+\tfrac14)
\end{aligned}$\\ \hline
${\rm L}_2$ &
$x_5,\;x_6,\;x_9,\;x_{11}$ &
$\begin{aligned}
\text{Const:}&\; 2/3+2/3+2/3+2/3=8/3\\
\hat\sigma_e(t):&\; (-\tfrac12)+(-\tfrac12)+(-\tfrac{a+1}{2})+(-1)=-(\tfrac{a+5}{2})\\
\hat\alpha_e(t):&\; (-\tfrac12)+(\tfrac{1-p}{4})=-(\tfrac{p}{4}+\tfrac14)\\
\hat\beta_e(t):&\; (-\tfrac12)+(\tfrac{1-q}{4})=-(\tfrac{q}{4}+\tfrac14)
\end{aligned}$\\ \hline
${\rm L}_3$ &
$x_5,\;x_6,\;x_7,\;x_{11}$ &
$\begin{aligned}
\text{Const:}&\; 4\times(2/3)=8/3\\
\hat\sigma_e(t):&\; (-\tfrac12)+(-\tfrac12)+(-\tfrac{a}{2})+(-1)=-(\tfrac{a+4}{2})\\
\hat\alpha_e(t):&\; (-\tfrac12)+\big(-\tfrac{p+1}{4}\big)=-(\tfrac{p}{4}+\tfrac34)\\
\hat\beta_e(t):&\; (-\tfrac12)+\big(\tfrac{1-q}{4}\big)=-(\tfrac{q}{4}+\tfrac14)
\end{aligned}$\\ \hline
${\rm L}_4$ &
$x_5,\;x_6,\;x_8,\;x_{11}$ &
$\begin{aligned}
\text{Const:}&\; 4\times(2/3)=8/3\\
\hat\sigma_e(t):&\; (-\tfrac12)+(-\tfrac12)+(-\tfrac{a}{2})+(-1)=-(\tfrac{a+4}{2})\\
\hat\alpha_e(t):&\; (-\tfrac12)+\big(\tfrac{1-p}{4}\big)=-(\tfrac{p}{4}+\tfrac14)\\
\hat\beta_e(t):&\; (-\tfrac12)+\big(-\tfrac{q+1}{4}\big)=-(\tfrac{q}{4}+\tfrac34)
\end{aligned}$\\ \hline
\end{tabular}}
\renewcommand{\arraystretch}{1}
\end{center}
 \caption[]{Consistency check performed by choosing one specific 18-tuple from each of the four entries in \eqref{bathory1}, \eqref{bathory2}, \eqref{bathory3} and \eqref{bathory4} that solve ${\rm LHS}= {\rm L}_k$ for \emph{any} $(a,p,q) \in \mathbb{Z}$ in \eqref{irishdeka3}.}
 \label{millerkhul3}
 \end{table}

However a few questions still remain. {\Su One}: what justifies the {\it uniqueness} of the set of results in \eqref{bathory1}, \eqref{bathory2}, \eqref{bathory3} and \eqref{bathory4}? And {\Su two}: can we fix $(a, p, q)$ appearing in our ans\"atze \eqref{irishdeka2}? Let us start with the uniqueness problem: can we allow other solutions? The answer will turn out to be {\it no}. To justify this, we first note that after substituting \eqref{irishdeka2}, each LHS generators in \eqref{streeber} contributes:
\begin{equation}
x_i\Big(\mathcal C_i + \mathcal S_i\,\hat\sigma_e(t) + \mathcal A_i\,\hat\alpha_e(t) + \mathcal B_i\,\hat\beta_e(t)\Big),
\end{equation}
where the coefficients $(\mathcal C_i, \mathcal S_i, \mathcal A_i, \mathcal B_i)$ (for $i=1,\dots,17$) can be read up from {\bf Table \ref{millerkhul2}};
and we impose $x_{18}=0$ (so that the non-locality factor remains time-independent). After substitution, the four RHS targets are ${\rm L}_k$ as given in \eqref{streeber} with the associated 18-tuples depicted above (with simple checks performed in {\bf Table \ref{millerkhul3}}). In the following let us demonstrate the uniqueness for the set in \eqref{bathory1}.

\paragraph{Constant budget and structure.}
Since all the 18-tuples in \eqref{bathory1} used $\mathcal C_i>0$ and $x_{18}=0$, matching the constant term forces:
\begin{equation}
\frac{11}{3} \;=\; \frac{5}{3} \;+\; \frac{2}{3} \;+\; \frac{2}{3} \;+\; \frac{2}{3},
\end{equation}
so every solution uses exactly one $5/3$ term and three $2/3$ terms. No other combinations are either allowed or possible, thus fixing only a class of solutions.

\paragraph{Pair identities and restricting to a set.}
After the same substitution for $\hat\zeta_e(t)$, the following equalities of $4$-vectors hold:
\begin{equation}
v_2 + v_{11} \;=\; v_9 + v_{17},\qquad
v_2 + v_{5} \;=\; v_7 + v_{17},\qquad
v_2 + v_{6} \;=\; v_8 + v_{17},
\end{equation}
where $v_i=(\mathcal C_i, \mathcal S_i, \mathcal A_i, \mathcal B_i)$ denotes the per-term coefficient vector. These identities allow swapping pairs inside a solution without changing the total coefficients of $(\hat\sigma_e(t), \hat\alpha_e(t), \hat\beta_e(t))$ and the constant term. This allows a {\it set} of solutions instead of a single 18-tuple.

\paragraph{Why there are no more.}
Any alternative uses either a different constant multiset (e.g.\ ${5\over 3}+{4\over 3}+{2\over 3}$, which cannot match the $p,q$ coefficients for arbitrary $p,q$), or exceeds the constant budget (adding more than four terms), or fails to match the $a$-dependent $\hat\sigma_e(t)$ coefficient. Up to the pair identities above, the eight listed 18-tuples in \eqref{bathory1} are the only non-negative integer solutions with $x_{18} = 0$.

Our above analysis more or less completes our discussion of the cross-term EOMs and the consistency of the solutions in \eqref{bathory1} to \eqref{bathory4}. However the question of fixing $(a, p, q)$ in \eqref{irishdeka2} still remains. Can we determine them? In the following section we will see how far we can go in fixing the remaining ambiguities. In the process we will also be able to connect to the more recent developments in the {\it dynamical} nature of dark energy and the corresponding equation of state (EoS).


\subsection{Dynamical dark energy and the equation of state \label{dynamical}}

This brings us to the last stage of our work related to answering questions like: how does the dynamical nature of the dark energy get manifested through our choice of the warp-factors? What really causes the acceleration of our universe now viewed as an excited state? How does the axion decay constant behave in the presence of dynamical dark energy? What is the equation of state of our universe? And many related ones. In the following we will provide answers to some of the above questions, while leaving the more difficult ones for later research. We will start with the dynamical nature of the dark energy. 

\subsubsection{Dynamical dark energy and the axion decay constant \label{aidaprose}}

\begin{table}[tb]  
 \begin{center}
\renewcommand{\arraystretch}{2.6}
\resizebox{\textwidth}{!}{
}
\renewcommand{\arraystretch}{1}
\end{center}
 \caption[]{\Su Revisiting {\bf Table \ref{milleren2}} where we discussed the dynamical duality sequence that takes a M-theory configuration \eqref{giuraman1} to heterotic SO(32) theory with the gauge group broken to $\left({\rm SO(8)}\right)^4$. Now, following section \ref{sec4.2.1}, we introduce two more warp-factors ${\rm F}_4$ and ${\rm F}_5$ and study the consequence with the difference that we ignore the sub-dominant ${\rm M}_p$ and $g_s$ corrections to the duality sequence. Rest of the details are the same from {\bf Table \ref{milleren2}}. (We have taken ${\rm H}(y) = {\rm H}_o({\bf x}) \equiv 1$ to avoid clutter.)}
 \label{milleren222}
 \end{table}

In the previous sections we managed to show not only how we can solve all the Einstein's equations consistently, but also count the number of possible solutions. The only downside of the analysis appears to be that we could not fix the parameters $(a, p, q)$ in \eqref{irishdeka2}. Our aim in this section is to rectify this situation. In the process we will be able to quantify somewhat the behavior of dark energy as advocated in \eqref{marapaug}.

As a start, we will revisit the dynamical duality sequence given in 
{\bf Table \ref{milleren2}} in the light of what we did in section \ref{sec4.2.1}. The result appears in {\bf Table \ref{milleren222}}. The difference from section \ref{sec4.2.1} is that now we will not worry too much about the sub-dominant ${\rm M}_p$ and $g_s$ corrections to the duality sequence. These corrections are important, but probably not so for the kind of discussion we want to have here. 

The result shown in {\bf Table \ref{milleren222}} matches exactly with the metric configuration advocated in \eqref{gethmia}, modulo the ${\rm M}_p$ and $g_s$ corrections. Keeping the four-dimensional Newton's constant time-independent amounts to the condition ${\rm F}^3_4 = ({\rm F_1F_2})^4$, matching with \eqref{didigoth}. The heterotic coupling is similarly ${{\rm F_1}\over \sqrt{\rm F_4}}$, which also matches with the one in \eqref{CWsara}. Putting everything together implies that for the 
$(SO(8))^4$ heterotic theory we have:
\bg\label{sarkalupizz}
\bar{g}_s^{-2} {\rm F}_1 \sqrt{\rm F_4} = {1\over \Lambda(t)t^2}, \nd
as in \eqref{ccoleman}, with $\Lambda(t)$ providing the dynamical dark energy as defined in \eqref{marapaug}. From our earlier description for the generalized $SO(32)$ case we can express ${\rm F}_i(t)$ in terms of parameters appearing in say {\bf figure \ref{boxer}} as, ${\rm F_4}(t) = \bar{g}_s^{\zeta_e(t)}, {\rm F}_1(t) = \bar{g}_s^{\beta_e(t)}, {\rm F}_2(t) = \bar{g}_s^{\alpha_e(t)}$ and ${\rm F}_5(t) = \bar{g}_s^{\eta_e(t)}$, where $\bar{g}_s$ is given by the remarkable simplification advocated in \eqref{johnsonsteel}, {\it i.e.}
\bg\label{johnsonsteel2}
{g_s\over {\rm H}(y) {\rm H}_o({\bf x})} = \left({\Lambda} t^2\right)^{1\over 2 - \beta_e(t)}, \nd
where we used the bare cosmological constant term $\Lambda$ compared to the ``running" $\Lambda(t)$ from \eqref{marapaug}.
Combining \eqref{johnsonsteel2} and \eqref{sarkalupizz} with the volume preservation condition immediately gives us the following relations between the parameters:
\bg\label{sundarina}
&& \eta_e(t) = -\zeta_e(t), ~~~~ \zeta_e(t) = {4\over 3}\left(\alpha_e(t) + \beta_e(t)\right) \nonumber\\
&& {\alpha_e(t) + \beta_e(t)\over 2- \beta_e(t)} = {3\log\left(1 + {\check{\Lambda}(t) \over \Lambda}\right)\over 2\vert\log(\Lambda t^2)\vert} = {3\over 2\vert\log(\Lambda t^2)\vert} \sum_{n = 1}^\infty {(-1)^{n+1}\over n} \left[{\check{\Lambda}(t)\over \Lambda}\right]^n, \nd
where we see that, knowing $\beta_e(t)$ determines $\alpha_e(t)$ directly from the observation as to how the dark energy is changing with time from \eqref{marapaug}. In fact in \eqref{bdtran} we had proposed the forms for $(\alpha_e(t), \beta_e(t))$ in terms of $(\alpha(t), \beta(t))$ and a series in sub-dominant powers of $\bar{g}_s$. The parameters $(\alpha(t), \beta(t))$ on the other hand were fixed from the axionic constraints to take take trans-series forms as in \eqref{horseman}. If we had a {\it constant} dark energy, {\it i.e.} a cosmological constant $\Lambda$, $\alpha_e(t) = -\beta_e(t)$, which is more interesting than $\alpha(t) = -\beta(t)$ for the simplified $SO(32)$ case. Similarly $\zeta_e(t)$ from \eqref{sundarina} continues to remain sub-dominant as expected from section \ref{sec4.2.1}.

Of course we should remember that the results from \eqref{sundarina} are {\it not exact} because we have been consistently ignoring the sub-dominant ${\rm M}_p$ and $g_s$ corrections. These may be easily inserted in by following the precise duality chasing studied in section \ref{sec4.2.1}, but the inclusion of such is not going to change any of our conclusions. We will therefore leave this exercise for our diligent readers.

\begin{table}[tb]  
 \begin{center}
\renewcommand{\arraystretch}{3.6}
\resizebox{\textwidth}{!}    
{
}
\renewcommand{\arraystretch}{1}
\end{center}
 \caption[]{\Su A more detailed version of {\bf Table \ref{milleren4}} where we include additional warp-factors ${\rm F}_4$ and ${\rm F}_5$.
 As before, the dynamical duality sequence takes a M-theory configuration \eqref{giuraman2} to the unbroken ${\rm E}_8 \times {\rm E}_8$ heterotic theory. The coordinate choices have remained the same: $(\theta_1, \theta_2) = (y^4, y^5)$ with the four-manifold ${\cal M}_4$ spanned by the coordinates $(y^6, y^7, y^8, y^9)$. As in {\bf Table \ref{milleren4}}, we have not specified the precise numbers of ${\rm O}p$'s and ${\rm D}p$'s in various theories for global charge cancellations. The dashed lines in the last column denote the points where the branes, O-planes or other defects are located. See also {\bf figure \ref{tab4row4}} for the configuration in the fourth row. The presence of ${\rm F}_4$ and ${\rm F}_5$ warp-factors changes slightly the results from {\bf Table \ref{milleren4}}. (We have also taken ${\rm H}(y) = {\rm H}_o({\bf x}) \equiv 1$ to avoid clutter.)}
 \label{milleren444}
 \end{table}

The story gets more interesting for heterotic ${\rm E_8 \times E_8}$ case. The modified duality chasing from the original {\bf Table \ref{milleren4}} appears in {\bf Table \ref{milleren444}}. The war-factors ${\rm F}_i(t)$, for $i = 1, .., 3$ are defined in \eqref{khadsalim}, and the remaining ones take the form: ${\rm F}_4(t) = \bar{g}_s^{\hat\zeta_e(t)}$ and ${\rm F}_5(t) = \bar{g}_s^{\hat\eta_e(t)}$. Demanding that volume of the internal six-manifold, with local topology of $\hat{\bf S}^1_{\theta_1} \times \mathcal M_4 \times {\bf S}^1_{11}$, remains time-independent gives us ${\rm F}_4^2(t) = {\rm F}_1(t) {\rm F}^4_2(t) {\rm F}_3^3(t)$, leading us to:
\bg\label{rickrghor}
\hat\zeta_e(t) = 2\hat\sigma_e(t) + {\hat\alpha_e(t) + 3\hat\beta_e(t)\over 2},~~~~ \hat\eta_e(t) = -\hat\zeta_e(t), \nd
ignoring the sub-dominant ${\rm M}_p$ and $g_s$ corrections to the duality sequence in {\bf Table \ref{milleren444}}. In \eqref{rickrghor}, $\hat\alpha_e(t)$ and $\hat\beta_e(t)$ are taken from \eqref{bdtran2} which in turn are expressed in terms of $(\hat\alpha(t), \hat\beta(t))$ and sub-dominant series in powers of $\bar{g}_s$. The parameters $(\hat\alpha(t), \hat\beta(t))$ are fixed from the axionic constraints to take the trans-series form as in \eqref{horseman}. Comparing \eqref{rickrghor} with \eqref{irishdeka2} fixes uniquely the values for $(a, p, q)$ to the following:
\bg\label{laurphil}
a = 2, ~~~  p = 1, ~~~ q = 3, \nd
which became possible because of our choice of the warp-factors ${\rm F}_i(t)$, starting from section \ref{sec4.2.1} onwards. Recall that both the flux and the Einstein EOMs, {\it i.e.} the corresponding Schwinger-Dyson equations, do not impose additional constraints on the aforementioned parameters. New constraints come from the dynamical nature of the dark energy which modifies relations like \eqref{sarkalupizz} and \eqref{johnsonsteel2} to now take the following form:
\bg\label{gulabtagra}
\bar{g}_s^{-2} \sqrt{{\rm F}_1(t) {\rm F}_3(t) {\rm F}_4(t)} = {1\over \Lambda(t) t^2}, ~~~~ \bar{g}_s ={g_s\over {\rm H}(y) {\rm H}_o({\bf x})} = \left({\Lambda} t^2\right)^{2\over 4 - \hat\alpha_e(t) - \hat\beta_e(t)},\nd
with $\Lambda(t)$ as given by \eqref{marapaug}. The choice of the coupling $g_s$ in terms of the bare cosmological constant $\Lambda$ guarantees that the ``running" of dark energy is controlled dominantly by $\hat\zeta_e(t)$, similar to what we saw for the $SO(32)$ case. The details differ a bit from the second line in \eqref{sundarina}, and using \eqref{rickrghor} we get:
\bg\label{cuthictran}
{4\hat\sigma_e(t) + \hat\alpha_e(t) + 3\hat\beta_e(t)\over 4- \hat\alpha_e(t)- \hat\beta_e(t)} = {2\log\left(1 + {\check{\Lambda}(t) \over \Lambda}\right)\over \vert\log(\Lambda t^2)\vert} = {2\over \vert\log(\Lambda t^2)\vert} \sum_{n = 1}^\infty {(-1)^{n+1}\over n} \left[{\check{\Lambda}(t)\over \Lambda}\right]^n \nd
where $-{1\over \sqrt{\Lambda}} < t < 0$. The equation \eqref{cuthictran}
confirms the fact that if $\check{\Lambda}(t)$ vanishes $-$ so that the dark energy is given by a cosmological constant $-$ then $\hat\zeta_e(t)$ will equivalently vanish making ${\rm F}_4 = {\rm F}_5^{-1} = 1$ (at least if we ignore the sub-dominant ${\rm M}_p$ and $g_s$ corrections to the duality sequence in {\bf Table \ref{milleren444}}). In general however, even with a cosmological constant $\Lambda$ the warp-factors could remain 
non-trivial in M-theory, {\it i.e.} in the first row of {\bf Table \ref{milleren444}}, so long as they are being compensated by the sub-dominant but {\it non-trivial} 
${\rm M}_p$ and $g_s$ corrections to the duality chain. This aspect has been clearly demonstrated in section \ref{sec4.2.1}, which the readers may look up for details. (As a consistency check, one may recover the $SO(32)$ results by applying the flows from {\bf figure \ref{boxer}} to \eqref{cuthictran}.)

\begin{figure}[h]
\centering
\begin{tabular}{c}
\includegraphics[width=5in]{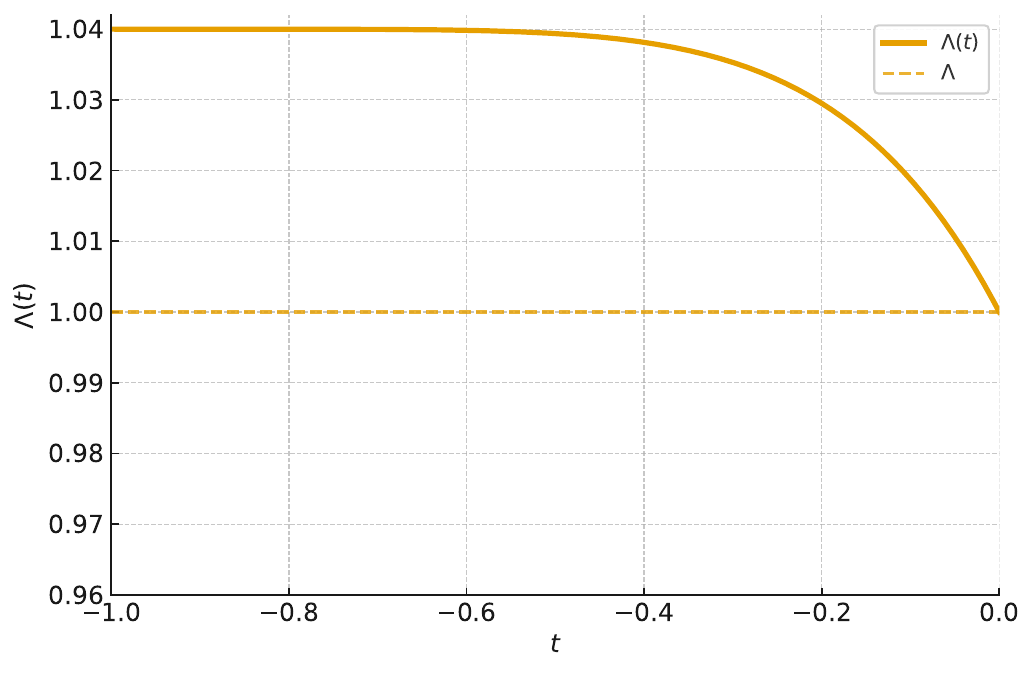}
\end{tabular}
\caption[]{A tentative plot of the slow decay of $\Lambda(t) = \Lambda + \check{\Lambda}(t)$ from \eqref{marapaug} in the temporal domain $-{1\over \sqrt{\Lambda}} < t< 0$,  where $\check{\Lambda}(t) \to 0$ as $ t \to 0$. We have taken $\Lambda = 1$ and assumed a $4\%$ decrease of the dark energy in the above temporal domain.}
\label{ccrunning}
\end{figure}

We can also make a prediction for the axion decay constant $f_a$, that we briefly discussed in \cite{axion} and in \eqref{shamitagra}, for our universe with a dynamical dark energy. The result for $f_a$ quoted in 
\eqref{shamitagra} changes slightly to accommodate the warp factor ${\rm F}_4(t)$ as $f_a \propto {{\rm F}_1(t) {\rm F}_3^3(t)\over {\rm F}_4(t)}$. (It also changes the heterotic coupling and the distance between the two Horava-Witten walls in similar ways.) Using \eqref{rickrghor} and \eqref{cuthictran}, the axion decay constant becomes:
\bg\label{mayroseboro}
f_a(t) \propto \Big(\Lambda t^2\Big)^{\hat\alpha_e(t)[2 + \mathcal F(t)] + \hat\beta_e(t)[6 + \mathcal F(t)] - 4\mathcal F(t)\over 4 -\hat\alpha_e(t) - \hat\beta_e(t)},  ~~-{1\over \sqrt{\Lambda}} < t < 0,   \nd 
where $\mathcal F(t) \equiv {2\log\left(1 + {\check{\Lambda}(t) \over \Lambda}\right)\over \vert\log(\Lambda t^2)\vert}$. We see that if $\check{\Lambda}(t) \to 0$ as $t \to 0$, so that at late time we recover $\Lambda$ as the value of the dark energy, \eqref{mayroseboro} goes back to the result we had in \eqref{TItostrat} and in \cite{axion}. A tentative running of the dark energy is shown in {\bf figure \ref{ccrunning}}. Thus the behavior of the axionic decay constant expectedly does not change very much as long as $\check{\Lambda}(t)$ shows a very small decay.  Incidentally the connection to the heterotic coupling $g_{\rm het}$ and the distance $\rho_{\rm HW}$ between the Horava-Witten walls, as advocated in \eqref{shamitagra}, continues to hold (albeit with some modifications) as:
\bg\label{meykadlo}
(g_{\rm het}, \rho_{\rm HW}) \propto \Big(\Lambda t^2\Big)^{\hat\alpha_e(t)[2 + \mathcal F(t)] + \hat\beta_e(t)[6 + \mathcal F(t)] - 4\mathcal F(t)\over 16 -4\hat\alpha_e(t) - 4\hat\beta_e(t)}, \nd
in the same temporal domain of $-{1\over \sqrt{\Lambda}} < t < 0$ using the flat-slicing. Due to the sub-dominant nature of $\hat\zeta_e(t)$, the theory continues to remain weakly coupled with a finite distance between the Horava-Witten walls at late time shown in {\bf figure \ref{axdecay}}.

A natural direction for future research is to develop the GS/Horava--Witten framework as an {alternative} to the non-standard ${\rm E_8\times E_8}$ embeddings proposed in~\cite{weigand} to decouple the axion--photon coupling from the axion--gluon coupling. 
In that construction, the decoupling is achieved through a \emph{static diagonal embedding} of $U(1)_{\rm Y}$ across both ${\rm E_8}$ factors, producing an axion that couples to photons but not to QCD. 
Such a mechanism relies on a time-independent compactification obeying the global topological constraint
\begin{equation}
c_2({\rm V}_{\text{vis}})+c_2({\rm V}_{\text{hid}})=c_2({\rm TX}),
\label{eq:bianchi_static}
\end{equation}
which severely restricts visible-sector bundles and typically forbids a vanishing $SU(3)$ instanton number $k_3=0$ within a single ${\rm E_8}$.

In contrast, the time-dependent GS/Horava--Witten setup satisfies anomaly cancellation \emph{dynamically} rather than topologically. 
Because both the warp factor ${\rm F}_i(t)$ and the heterotic coupling $g_{\rm het}$ evolve in time, the eleven-dimensional Bianchi identity:
\begin{equation}
d{\bf G}^{({\rm global})}_4 = \delta(y)\!\big(\mathrm{Tr}~{\bf F}_{\text{vis}}^2-\tfrac{1}{2}\mathrm{Tr}~{\bf R}^2\big)
+ \delta(y-\pi\rho)\!\big(\mathrm{Tr}~{\bf F}_{\text{hid}}^2-\tfrac{1}{2}\mathrm{Tr}~{\bf R}^2\big),
\label{eq:bianchi_dynamic}
\end{equation}
is balanced at every instant. In writing \eqref{eq:bianchi_dynamic}, we have split the dynamical five-brane sources $\hat{\mathbb{N}}_5(t)$ from \eqref{tiptip3} into two sets of small-instantons  controlled respectively by the gauge fluxes on the two Horava-Witten walls.
Equation \eqref{eq:bianchi_dynamic} remains globally satisfied, while the distribution of curvature and five--brane
charge between the two ${\rm E_8}$ boundaries can evolve dynamically through the motion or
recombination of small--instanton configurations.  This evolution allows visible--sector
bundles whose $SU(3)$ component has vanishing instanton number:
\begin{equation}
k_3 = 0, \qquad (k_2, k_{\rm Y})\neq 0,
\end{equation}
provided that the hidden sector and five--brane sources compensate to maintain the
global constraint \eqref{eq:bianchi_static}.  Upon dimensional reduction of the Green--Schwarz anomaly cancellation term
${\bf B}_2\wedge \mathbb{X}^{({\rm tot})}_8({\bf R}, {\bf F})$, the resulting four--dimensional
interaction takes the form\footnote{Defining $\mathbb{X}^{({\rm tot})}_8({\bf R}, {\bf F}) = \mathbb{X}_8({\bf R}) -{1\over 96} ~{\rm tr}~{\bf R}^2~{\rm tr}~{\bf F}^2 + {1\over 96}~{\rm tr}~{\bf F}^4 - {1\over 384}\left({\rm tr}~{\bf F}^2\right)^2$, with $\mathbb{X}_8({\bf R})$ as in \eqref{olivecostamey}, and dimensionally reducing ${\bf B}_2$ over the internal six-manifold $\mathcal M_6$ using a harmonic 2-form $\omega_a$, we can easily recover \eqref{zotka} with 
$k_i = {1\over 8\pi^2}\int_{\mathcal M_6} w_a \wedge {\rm tr}\big({\bf F}_i \wedge {\bf F}_i\big)$ where ${\bf F}_1 \equiv {\bf F}_{\rm Y}$, ${\bf F}_2 \equiv {\bf W}$, and $k_1 = k_{\rm Y}$.}:
\begin{equation}\label{zotka}
\mathcal{L}_{\mathrm{eff}}
 \supset \frac{a(t)}{f_a(t)}
  \big(k_{\rm Y}\,{\bf F}_{\rm Y}\widetilde {\bf F}_Y + k_2\, {\bf W}\widetilde {\bf W}\big),
  \qquad g_{aG}(t)=\frac{\alpha_s}{2\pi f_a(t)}\,k_3=0,
\end{equation}
where $f_a(t)$ evolves according to the quasi--de Sitter warp hierarchy as in \eqref{mayroseboro}.
Hence, rather than requiring a static diagonal embedding to geometrically cancel the
QCD anomaly, the GS/HW framework achieves the same ``axion--photon but no axion--QCD''
pattern through an allowed redistribution of topological charge between the two
${\rm E_8}$ walls and bulk five--brane sources, while the time dependence of $f_a(t)$
governs the cosmological evolution of the effective axion couplings.

\subsubsection{Accelerating universe and the equation of state \label{kalasona}}

Let us start by asking the following question: what causes our universe, now viewed as an excited state over a supersymmetric Minkowski background, to {\it accelerate}? We will provide an answer to this using the flat-slicing that we have been using so far, but it is instructive to discuss briefly the story using a global slicing from 
\eqref{peribindu00}.

Looking at the global slicing we see that as time increases the space increases, {\it i.e.} we are creating new spaces using the dark energy as a source. In string theory this is clearly not possible as a vacuum configuration \cite{joydeep}, so a realization of such a process is through the excited state over a supersymmetric Minkowski vacuum. The ``creation" of new spaces then becomes possible because the excited state now spans more spaces over the Minkowski solution. This immediately leads to the question of the source of such an expansion: what objects in M-theory help us to {\it inflate} the excited state? The answer as we have seen from section \ref{eomgbeta} and specifically from {\bf Table \ref{niksmit0100}} is that the BBS instantons \cite{bbs} are the main sources. The reason is that the typical contribution of the BBS instantons to the emergent energy-momentum tensor is of the form:
\bg\label{aida2bone}
\mathbb{T}_{\mu\nu}^{({\rm BBS}; 1)} \propto {\bar{g}_s^{\theta_{nl} + \theta'_{nl} - {8\over 3} + \hat\zeta_e}\over \bar{g}_s^{2-2\hat\sigma_e -{\hat\alpha_e +\hat\beta_e\over 2}}}~ {\rm exp}\left(-{\bar{g}_s^{\theta'_{nl}}\over \bar{g}_s^{2-2\hat\sigma_e -{\hat\alpha_e + \hat\beta_e\over 2}}}\right), \nd
where for $\theta_{nl} = {8\over 3}^\pm$ $-$ which is the value of the quantum series from \eqref{botsuga} and \eqref{brittbaba007} that consistently fit in over all the flux and the metric EOMs $-$ scales as 
$\bar{g}_s^{-2 + 2\hat\sigma_e(t) + {\hat\alpha_e(t) + \hat\beta_e(t)\over 2}}$. As $\bar{g}_s \to 0$, which is the late time in the flat-slicing, the energy-momentum tensor {\it increases} exactly in the same proportion as the Einstein tensor (compare with the first row in {\bf Table \ref{niksmit0100}} and \eqref{holudlili}). As we showed in sections \ref{eomgbeta} and \ref{counting}, \eqref{aida2bone} forms the dominant contribution (along with the non-local BBS instantons). Other effects are sub-dominant.

The story does not end here. Finding the source of the acceleration suggests that we should also verify the equation of state (EoS). In cosmology, the EoS typically comes from the pressure and the energy density of the system once we dimensionally reduce to four spacetime dimensions. Our Schwinger-Dyson equations lead to:
\bg\label{aidatara}
{\bf G}_{\mu\nu}(\langle {\bf \Xi}\rangle_\sigma) = \mathbb{T}_{\mu\nu}^{\rm tot}(\langle{\bf \Xi}\rangle_\sigma), \nd
as seen from \eqref{coffiemarieso}, where $\mathbb{T}_{\mu\nu}^{\rm tot}(\langle{\bf \Xi}\rangle_\sigma)$ is the {\it total} energy-momentum tensor that includes the BBS instanton contributions from \eqref{aida2bone}, plus other sub-dominant non-perturbative contributions.

Let us now look at the metric configuration for the generalized $SO(32)$ case from {\bf Table \ref{milleren222}}. For simplicity we will mostly concentrate on this particular case here as the story for the generalized ${\rm E_8 \times E_8}$ case will be very similar. The metric configuration that appears from the last row in {\bf Table \ref{milleren222}} is:

{\footnotesize
\bg\label{engbldcafmey}
ds^2 &= & {\bar{g}_s^{-2}\over {\rm H}^2(y)} \left({\rm F}_1(t) \sqrt{{\rm F}_4(t)} + \mathcal O(\bar{g}_s, {\rm M}_p)\right) \left(-dt^2 + \eta_{ij} dx^i dx^j\right)\\
&+& {\rm H}^2(y)\left[\left({{\rm F}_1(t) {\rm F}_2(t)\over \sqrt{{\rm F}_4(t)}} + \mathcal O(\bar{g}_s, {\rm M}_p)\right) {\bf g}_{mn}(y) dy^m dy^n +\left({1 \over \sqrt{{\rm F}_4(t)}} + \mathcal O(\bar{g}_s, {\rm M}_p)\right) {\bf g}_{\alpha\beta}(y) dy^\alpha dy^\beta \right], \nonumber \nd}
where we inserted the $\mathcal O(\bar{g}_s, {\rm M}_p)$ corrections entering the duality sequence that we discussed in detail in section \ref{sec4.2.1}, and we have kept ${\rm H}_o({\bf x}) \equiv 1$ for simplicity. In addition to the $\mathcal O(\bar{g}_s, {\rm M}_p)$ corrections, the other important ingredient is ${\rm F}_4(t) \equiv \bar{g}_s^{\hat\zeta_e(t)}$. This becomes identity (or alternatively $\hat\zeta_e(t) = 0$) when $\check{\Lambda}(t) = 0$ and is therefore indicative of the presence of dynamical dark energy as evident from 
\eqref{sundarina}. With only a cosmological constant $\Lambda$, the metric \eqref{engbldcafmey} is replaced by the one appearing from the second-last row in {\bf Table \ref{milleren2}}. The metric configuration for the generalized ${\rm E_8 \times E_8}$ takes the following form:

{\footnotesize
\bg\label{engbldcafmey2}
ds^2 &= & {\bar{g}_s^{-2}\over {\rm H}^2(y)} \left(\sqrt{{\rm F}_1(t){\rm F}_3(t){\rm F}_4(t)} + \mathcal O(\bar{g}_s, {\rm M}_p)\right) \left(-dt^2 + \eta_{ij} dx^i dx^j\right) \nonumber\\
& + & {\rm H}^2(y)\left({{\rm F}_2(t) \sqrt{{\rm F}_1(t){\rm F}_3(t)\over {{\rm F}_4(t)}}} + \mathcal O(\bar{g}_s, {\rm M}_p)\right) {\bf g}_{mn}(y) dy^m dy^n \\
&+ & {\rm H}^2(y)\left[\left({\sqrt{ {\rm F}_3(t) \over {\rm F}_1(t){\rm F}_4(t)}} + \mathcal O(\bar{g}_s, {\rm M}_p)\right) {\bf g}_{\theta_1\theta_1}(y) d\theta_1^2 +\left({\sqrt{ {\rm F}_3(t) {\rm F}_4(t)\over {\rm F}_1(t)}} + \mathcal O(\bar{g}_s, {\rm M}_p)\right) {\bf g}_{\theta_2\theta_2}(y) d\theta_2^2 \right], \nonumber \nd}
which can be inferred from the second-last row of {\bf Table \ref{milleren444}}. The above metric almost matches with the one from \eqref{engbldcafmey} in the limit ${\rm F}_1(t) = {\rm F}_3(t)$ but differs slightly in the placement of the ${\rm F}_4(t)$ operator along the $\theta_2$ direction (we called ${\bf S}^1_{11}$ in {\bf Table \ref{milleren444}}). Comparing \eqref{rickrghor} and \eqref{cuthictran}, we see that the presence of ${\rm F}_4(t)$ (or equivalently $\hat\zeta_e(t)$) signifies the presence of a dynamical dark energy $\Lambda(t)$. As emphasized above, the $\mathcal O(\bar{g}_s, {\rm M}_p)$ corrections to the warp-factors are crucial. But we should clarify what we mean by these corrections because $\mathcal F_i(t)$ warp-factors are only functions of $t$ (or alternatively, are functions of $\bar{g}_s$ as $t \equiv t(\bar{g}_s)$). If we denote the collection of the $\mathcal F_i(t)$ warp-factors appearing on the various terms in \eqref{engbldcafmey} and \eqref{engbldcafmey2} symbolically as $\mathcal G_f(t)$, then consistency will require us to replace the corrections in \eqref{engbldcafmey} and \eqref{engbldcafmey2} more appropriately as:
\bg\label{engtagmey}
\mathcal G_f(t) + \mathcal O(\bar{g}_s, {\rm M}_p) \to  \left(\mathcal G_f(t) + \mathcal O(\bar{g}_s)\right) \times \mathcal O({\rm M}_p), \nd
implying a splitting where we expect the $\mathcal O({\rm M}_p)$ corrections to be absorbed by the $y$-dependent functions: the warp-factor ${\rm H}(y)$ and the internal metric components. Our study of the EOMs have revealed that splitting the metric into a product of a $\bar{g}_s$ dependent piece and a $y$ dependent function is possible, so a simple picture like \eqref{engtagmey} is self-consistent.

Coming back to the generalized $SO(32)$ case, some redefinitions of the warp-factors appearing in \eqref{engbldcafmey} can significantly reduce the effort to compute the EoS. For example using \eqref{engtagmey}, redefinitions like:
\bg\label{garnerju}
&& a^2(t) \equiv  \bar{g}_s^{-2} \left({\rm F}_1(t) \sqrt{{\rm F}_4(t)} + \mathcal O(\bar{g}_s)\right) = {1\over \Lambda(t) t^2} \nonumber\\
&& \mathcal F_1(t) \equiv {1 \over \sqrt{{\rm F}_4(t)}} + \mathcal O(\bar{g}_s), ~~~~ \mathcal F_2(t) \equiv {{\rm F}_1(t) {\rm F}_2(t)\over \sqrt{{\rm F}_4(t)}} + \mathcal O(\bar{g}_s), 
\nd
can help us to export directly the results of ${\bf G}_{ij}$ and ${\bf G}_{00}$ from \eqref{mameybike} and \eqref{mameybike2} by simply making the replacements ${\rm F}_1(t) \to \mathcal F_1(t)$ and ${\rm F}_2(t) \to \mathcal F_2(t)$ there. We should also rename $t \to \eta$ to be more consistent with the literature. Switching to conformal time $\eta$  gives us:
\bg\label{indigpaap1}
\frac{\dot a^{2}}{a^{4}}=\frac{\Hc^{2}}{a^{2}}, 
\qquad
\frac{\ddot a}{a^{3}}=\frac{\Hc'+\Hc^{2}}{a^{2}},
\nd
where $\mathcal H$ is the Hubble constant (and should not be confused with the warp-factor ${\rm H}(y)$). We have defined $\Hc={a'\over a}$ and prime $' = {d\over d\eta}$ in the usual way. Using these definitions, we can rewrite \eqref{mameybike} and \eqref{mameybike2} as:
\bg\label{ladigi}
&& {{\bf G}_{00}}(y, \eta) = {a^{2}(\eta)} 
\left[{\cal C}(y, \eta)- {\cal L}(y, \eta) + {3\Hc^{2}(\eta)\over a^2(\eta)}+\frac{{\cal T}_{00}(\eta)}{a^{2}(\eta)}\right] \nonumber\\
&& {\bf G}_{ij}(y, \eta) = {a^{2}(\eta)\delta_{ij}} \left[-{\cal C}(y, \eta)+{\cal L}(y, \eta)-{2\Hc'(\eta)+\Hc^{2}(\eta)\over a^2(\eta)}+\frac{{\cal T}(\eta)}{a^{2}(\eta)}\right],
\nd
where ${\cal C}(y, \eta)$, ${\cal L}(y, \eta)$ collect the internal curvature pieces, and ${\cal T}_{ij}(\eta) \equiv {\cal T}\delta_{ij}$, ${\cal T}_{00}(\eta)$ are the time-derivative brackets involving $\mathcal F_{1}(\eta)$ and $\mathcal F_{2}(\eta)$. They are defined from \eqref{mameybike} and \eqref{mameybike2} in the following way:

{\footnotesize
\bg\label{indigoladigi}
&& {\cal C}(y, \eta) = 
\frac{{\bf R}(y,\eta)}{2{\rm H}^{4}(y)} 
+ \frac{4\,{\bf g}^{\alpha\beta}\partial_{\alpha}{\rm H}(y)\,\partial_{\beta}{\rm H}(y)}{{\rm H}^{6}(y)\mathcal F_{1}(\eta)} 
+ \frac{4\,{\bf g}^{mn}\partial_{m}{\rm H}(y)\,\partial_{n}{\rm H}(y)}{{\rm H}^{6}(y)\mathcal F_{2}(\eta)}\nonumber\\
&& {\cal L}(y, \eta) = 
\frac{\Box_{\alpha} {\rm H}^{4}(y)}{2{\rm H}^{8}(y)\mathcal F_{1}(\eta)} 
+ \frac{\Box_{m} {\rm H}^{4}(y)}{2{\rm H}^{8}(y)\mathcal F_{2}(\eta)}\\
&& {\cal T}_{00}(\eta) = 
\frac{\mathcal F_{1}'^{2}(\eta)}{4\mathcal F_{1}^{2}(\eta)} 
+ 3{\cal H}(\eta)\frac{\mathcal F_{1}'(\eta)}{\mathcal F_{1}(\eta)} 
+ \frac{3\mathcal F_{2}'^{2}(\eta)}{2\mathcal F_{2}^{2}(\eta)} 
+ 6{\cal H}(\eta)\frac{\mathcal F_{2}'(\eta)}{\mathcal F_{2}(\eta)} 
+ \frac{2\mathcal F_{1}'(\eta)\mathcal F_{2}'(\eta)}{\mathcal F_{1}(\eta)\mathcal F_{2}(\eta)} \nonumber\\
&& {\cal T}(\eta) = 
\frac{\mathcal F_{1}'^{2}(\eta)}{4\mathcal F_{1}^{2}(\eta)} 
- {\cal H}(\eta)\frac{\mathcal F_{1}'(\eta)}{\mathcal F_{1}(\eta)} 
- \frac{\mathcal F_{1}''(\eta)}{\mathcal F_{1}(\eta)} 
- \frac{\mathcal F_{2}'^{2}(\eta)}{2\mathcal F_{2}^{2}(\eta)} 
- 2{\cal H}(\eta)\frac{\mathcal F_{2}'(\eta)}{\mathcal F_{2}(\eta)} 
- \frac{2\mathcal F_{2}''(\eta)}{\mathcal F_{2}(\eta)} 
- \frac{2\mathcal F_{1}'(\eta)\mathcal F_{2}'(\eta)}{\mathcal F_{1}(\eta)\mathcal F_{2}(\eta)}, \nonumber \nd}
where the warp-factors are defined in \eqref{garnerju}. Note that 
$\mathcal T_{ij}(\eta) \equiv {\cal T}\delta_{ij}$ and $\mathcal T_{00}(\eta)$ are by definition independent of the complications of the internal geometry and as such are functions of the conformal time $\eta$. Interestingly, if we demand that $\mathcal T_{ij}(\eta) = \mathcal T_{00}(\eta)$, then \eqref{indigoladigi} leads to the following constraint equation:
\bg\label{ladigi}
\mathcal A'(\eta) + 2\mathcal B'(\eta) + \mathcal A^{2}(\eta) + 4\mathcal B^{2}(\eta) + 4\mathcal A(\eta) \mathcal B(\eta) + 4\Hc(\eta)\mathcal A(\eta) + 8\Hc(\eta) \mathcal B(\eta) = 0, \nonumber\\ \nd
where 
$\mathcal A(\eta) \equiv {\mathcal F_{1}'(\eta)\over \mathcal F_{1}(\eta)}$, and $\mathcal B(\eta) \equiv {\mathcal F_{2}'(\eta)\over F_{2}(\eta)}$. This is the generic differential relation linking $\mathcal F_{1}(\eta)$ and $\mathcal F_{2}(\eta)$ that guarantees isotropy of the external $(3+1)$ spacetime. Defining $\mathbb{T}_{\mu\nu}^{\rm tot}(\langle{\bf \Xi}\rangle_\sigma) = 8\pi {\rm G}_{\rm N} \mathbb{T}_{\mu\nu}^{\rm eff}(\langle{\bf \Xi}\rangle_\sigma)$ in \eqref{aidatara}, we can express the energy density $\rho$ and the pressure $p$ in $3+1$ dimensions as:
\bg\label{milleranondo}
&&\rho = {{\bf G}_{00}\over 8\pi {\rm G}_{\rm N} a^2(\eta)} = {1\over8\pi {\rm G}_{\rm N}}\left[
{\cal C}(y, \eta)- {\cal L}(y, \eta) + {3\Hc^{2}(\eta)\over a^2(\eta)}+\frac{{\cal T}_{00}(\eta)}{a^{2}(\eta)}\right]\\
&& p = {1\over 3} \sum_i{{\bf G}_{ii}\over 8\pi {\rm G}_{\rm N} a^2(\eta)} = {1\over8\pi {\rm G}_{\rm N}}\left[ -{\cal C}(y, \eta)+{\cal L}(y, \eta)-{2\Hc'(\eta)+\Hc^{2}(\eta)\over a^2(\eta)}+\frac{{\cal T}(\eta)}{a^{2}(\eta)}\right], \nonumber \nd
where our choice of $\zeta_e(t) = {4\over 3}(\alpha_e(t) + \beta_e(t))$ in \eqref{sundarina} guarantees that the four-dimensional Newton constant ${\rm G_N}$ remains time-independent. With this we are ready to express the EoS in terms of the parameters of the theory. Using \eqref{milleranondo}, the EoS becomes:
\bg\label{millerfasa}
w(\eta) = {p\over \rho} &= & {a^2(\eta)\left[-{\cal C}(y, \eta)+{\cal L}(y, \eta)\right]-{2\Hc'(\eta)-\Hc^{2}(\eta)}+ {{\cal T}(\eta)} \over a^2(\eta)\left[{\cal C}(y, \eta)- {\cal L}(y, \eta)\right] + {3\Hc^{2}(\eta)}+ {{\cal T}_{00}(\eta)}} \nonumber\\
& = & -{1\over 3} - \frac{2}{3}\frac{\Hc'(\eta)}{\Hc^{2}(\eta)} + 
\delta w(\mathcal C, \mathcal L, \mathcal F_i), \nd
where $\delta w$ represent the corrections coming from the  back-reaction of the internal geometry on the external equation of state. These corrections are not important from the four-dimensional point of view, and $w(\eta)$ is solely captured by $\mathcal H(\eta) = {a'(\eta)\over a(\eta)}$ with $a(\eta)$ given by \eqref{garnerju}. It is also easy to check that if $a^2(\eta) = {1\over \Lambda \eta^2}$, then 
${\mathcal H'(\eta)\over \mathcal H^2(\eta)} = 1$ and $w(\eta) = -1$ as expected (ignoring the $\delta w$ corrections). On the other hand, if we take the scale factor in conformal time $\eta$ to be 
$a^{2}(\eta) = \frac{1}{\Lambda(\eta)\,\eta^{2}}$, then the EoS takes the following form:
\bg\label{ennataraida}
\boxed{w(\eta)
= -\frac{1}{3} - \frac{2}{3}\frac{\mathcal{H}'(\eta)}{\mathcal{H}^{2}(\eta)}
= -\frac{1}{3} - \frac{4}{3}\left[{2- \eta^2 \mathbb{X}'(\eta)\over (2 + \eta \mathbb{X}(\eta))^2}\right]}
\nd
where $\mathbb{X}(\eta) = {\Lambda'(\eta)\over \Lambda}$. 
Unless $\mathbb{X}(\eta) = 0$ (i.e.\ $\Lambda = \text{const}$) or $\Lambda(\eta)$ is tuned such that ${\mathcal{H}'(\eta)\over \mathcal{H}^{2}(\eta)}=1$, 
the equation of state is \emph{time-dependent}, corresponding to a \emph{dynamical dark energy} component. For small departures from de Sitter, assume $\mathbb{X}(\eta)$ and $\mathbb{X}'(\eta)$ are small, we can expand \eqref{ennataraida} to express it as a series:
\begin{equation}
w(\eta) \approx -1 + \frac{2}{3}\,\eta\,\mathbb{X}(\eta) + \frac{1}{3}\,\eta^{2} \mathbb{X}'(\eta) + \mathcal O(\mathbb{X}^2),
\end{equation}
showing that the deviation from $w(\eta) = -1$ is controlled at least by the first and second derivatives of $\ln~ \Lambda(\eta)$. As a check, when $\Lambda$ is constant, $\mathbb{X} = 0$ which implies that $w(\eta) = -1$, thus recovering the pure de Sitter.

The reader might recognize that the procedure described above, used to generate a dynamical equation of state for an accelerating universe, can be generalized to the case of an arbitrary FLRW cosmology. In this case, instead of choosing the exterior metric of the form ${1\over \Lambda(\eta) \eta^2}$ with $ \Lambda(\eta) = \Lambda + \check{\Lambda}(\eta)$ as in \eqref{marapaug}, one could just choose the scale factor $a(\eta)$ as the arbitrary function of (conformal) time $\eta$. The main subtlety would arise in how long can such a solution be trusted within string theory, i.e., what is the temporal domain of such a theory? In the case of an accelerating universe, we showed that choosing an ans\"atze of the form \eqref{johnsonsteel2} for the $SO(32)$ and \eqref{gulabtagra} for the ${\rm E_8 \times E_8}$ theories allows us to recover the standard domain for the flat-slicing of de Sitter. For other solutions that are of cosmological interest, such as matter-dominated $a(\eta)\propto \eta^2$ and radiation-dominated $a(\eta)\propto \eta$ eras, one needs to plug in those specific functions for the scale factor and check if such a simple temporal domain can be derived. The key reason of how our prescription would work for such more general cosmologies is that we would keep an extra warp factor (as in ${\rm F}_4(t)$) for the external spacetime which would require the existence of similar time-dependent warp factors for the internal metric (as shown in {\bf Tables \ref{milleren222}} and {\bf \ref{milleren444}}). The time-dependence of these warp factors would adjust themselves in a way that the $\bar{g}_s$ would still have a relatively straightforward  relationship with conformal time, analogous to \eqref{johnsonsteel2} and \eqref{gulabtagra}, for such cosmological solutions. As an example, looking at the second-last row in {\bf Table \ref{milleren444}}, we can define for the heterotic ${\rm E_8 \times E_8}$ frame:
\bg\label{millerchotu}
\bar{g}_s^{-2} \left(\sqrt{{\rm F}_1(\eta) {\rm F}_3(\eta) {\rm F}_4(\eta)} + \mathcal O(\bar{g}_s)\right) \equiv a^2(\eta), \nd 
keeping $\bar{g}_s$ as in \eqref{gulabtagra}. This way any generic $a^2(\eta)$ could be entertained without spoiling the trans-Planckian bound.
Of course, all of this needs to be explored in detail in future work, and is beyond the scope of this paper. However, let us emphasize that, in principle, this would allow us to describe (3+1)-d cosmological solutions consistently in string theory in our framework.

Before ending this section, let us perform a few more consistency checks.
Consider the $SO(32)$ case where ${\rm F}_4(t) = 1$, {\it i.e.} $\zeta_e(t) = 0$. This should in principle give us a de Sitter solution with only a cosmological constant $\Lambda$ ({\it i.e.} $\check{\Lambda}(t) = 0$ in \eqref{marapaug}). Going back to \eqref{ladigi}, we now require $\mathcal T_{ij} = - \mathcal T_{00}$. 
Imposing this on \eqref{indigoladigi} gives us:
\bg\label{chokhumedorod}
\mathcal A'(\eta) + 2\mathcal B'(\eta) + \frac{1}{2}\mathcal A^{2}(\eta) + \mathcal B^{2}(\eta) - 2\mathcal{H}(\eta)\mathcal A(\eta) - 4\mathcal{H}(\eta)\mathcal B(\eta) = 0.
\nd
where 
$\mathcal A(\eta) \equiv {\mathcal F_{1}'(\eta)\over \mathcal F_{1}(\eta)}$, and $\mathcal B(\eta) \equiv {\mathcal F_{2}'(\eta)\over F_{2}(\eta)}$ as in \eqref{ladigi}, but the difference from \eqref{ladigi} is the sign. The sign is important because for $a^2(\eta) = {1\over \Lambda \eta^2}$, we expect:
\bg\label{ananorok}
3\mathcal H^2(\eta) = 2\mathcal H'(\eta) + \mathcal H^2(\eta), \nd
as shown in \eqref{2indig2metro}. Plugging this in \eqref{millerfasa} immediately gives us $w(\eta) = -1$, implying that 
\eqref{chokhumedorod} is the generic relation between the time–dependent warp factors $\mathcal F_{1}(\eta)$ and $\mathcal F_{2}(\eta)$ required for the time–derivative sector to behave as a vacuum–energy component (not to be confused with vacuum solution). 

However there is something off about \eqref{chokhumedorod}. \eqref{chokhumedorod} is basically a Riccati-forced differential equation
whose solution depends on the presence of a source. This can be made clear by rewriting \eqref{chokhumedorod} in a slightly more convenient way by
introducing an auxiliary function $\phi(\eta)$ via $\mathcal{A}(\eta)=-{2\phi'(\eta)\over \phi(\eta)}$. This 
transforms \eqref{chokhumedorod} into a \emph{linear} equation for $\phi(\eta)$:
\bg\label{tolouch1}
\phi''(\eta) - 2\mathcal{H}(\eta)\phi'(\eta) + 
\frac{1}{2}\Big(2\mathcal{B}'(\eta) + \mathcal{B}^{2}(\eta) - 4\mathcal{H}(\eta)\mathcal{B}(\eta)\Big)\phi(\eta) = 0,
\nd
which suggests the presence of a source term from $\mathcal B(\eta)$. This means 
once the functional form for $\mathcal{B}=(\ln \mathcal{F}_2)'$ is provided, the solution for $\mathcal A(\eta)$ (or alternatively $\mathcal F_1(\eta)$) may be easily extracted from \eqref{tolouch1} as:
\bg\label{tolouch2}
\mathcal{F}_1(\eta) = c_1\,\phi(\eta)^{-2}, \nd
where $\phi(\eta)$ solves \eqref{tolouch1}.
This expresses $\mathcal{F}_1(\eta)$ uniquely (up to a constant $c_1$) once $\mathcal{F}_2(\eta)$ is chosen.  
Conversely, one can swap the roles of $(\mathcal{F}_1(\eta),\mathcal{A}(\eta))$ and 
$(\mathcal{F}_2(\eta),\mathcal{B}(\eta))$ to solve for $\mathcal{F}_2(\eta)$ given $\mathcal{F}_1(\eta)$. 

\begin{figure}[h]
\centering
\begin{tabular}{c}
\includegraphics[width=4in]{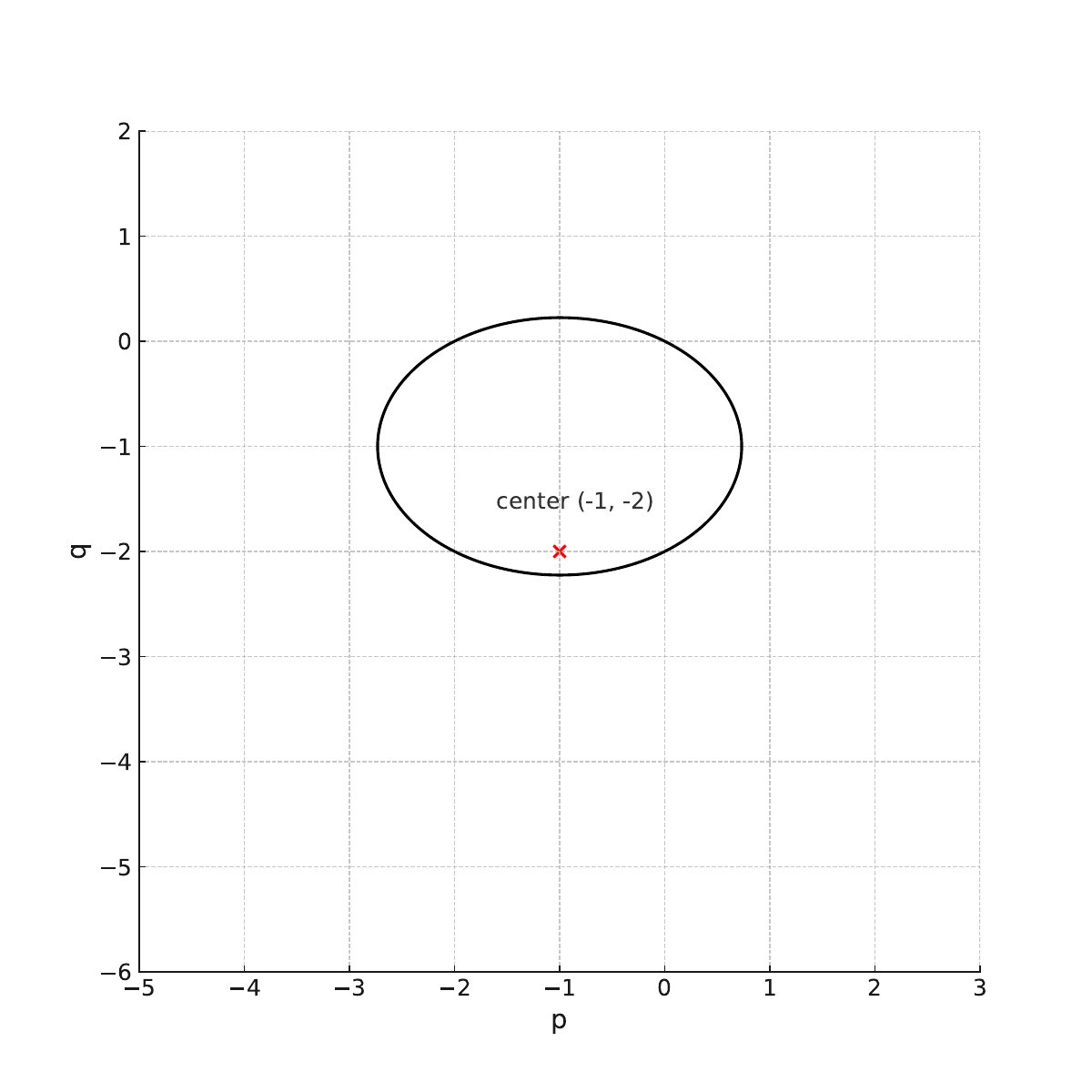}
\end{tabular}
\caption[]{Plot of the curve given by \eqref{toulouch6}. This can be expressed as an ellipse ${(p+1)^2\over 3^2} + {(q+2)^2\over (3/\sqrt{2})^2} = 1$, so it’s centered at $(-1, -2)$ with semi-axes
3 (along $p$) and ${3\over \sqrt{2}}$
 (along $q$). The plot shows the zero-level contour 
$f(p,q)=0$ with that center marked.}
\label{ellipseplot}
\end{figure}

Question now is what values of $\mathcal F_2(\eta)$ we should consider so that it can appear as a source term in \eqref{tolouch1}. However before dwelling into that, let us see how far we can go with \eqref{tolouch1} without specifying $\mathcal F_2(\eta)$. To start let us consider the following interesting 
monotonic combination:
\bg\label{tolouch3}
\mathcal S(\eta) \equiv \mathcal{A}(\eta) + 2\mathcal{B}(\eta) = \left[\ln(\mathcal{F}_1(\eta)\mathcal{F}_2^{2}(\eta))\right]'\,,
\nd
which vanishes when $\mathcal{F}_1(\eta)\mathcal{F}_2^{2}(\eta) = 1$. Assuming first that $\mathcal S(\eta) \ne 0$, then plugging this in 
the constraint \eqref{chokhumedorod}, we can easily infer the following two conditions:

{\footnotesize
\bg\label{touloch3}
\mathcal S'(\eta) - 2\mathcal{H}(\eta)\mathcal S(\eta) = -\Big(\tfrac{1}{2}\mathcal{A}^{2}(\eta) + \mathcal{B}^{2}(\eta)\Big) ~~ \implies ~~
\frac{d}{d\eta}\!\left[\frac{S(\eta)}{a^{2}(\eta)}\right]
= -\left[{\mathcal{A}^{2}(\eta) + 2\mathcal{B}^{2}(\eta) \over 2a^2(\eta)}\right] \le 0,
\nd}
where we have taken $\mathcal H(\eta) = {a'(\eta)\over a(\eta)}$. The second condition in \eqref{touloch3} is interesting because $a^2(\eta)$ is a monotonically increasing function in the temporal domain $-{1\over \sqrt{\Lambda}} < t < 0$. However this doesn't necessarily imply that $\mathcal S(\eta)$ is also monotonically decreasing. In other words, all we can say about the 
the weighted combination:
\begin{equation}\label{anashonal}
\frac{d}{d\eta}\!\left[
\frac{(\ln(\mathcal{F}_1(\eta)\mathcal{F}_2^{2}(\eta)))'}{a^{2}(\eta)}
\right] \le 0
\end{equation}
is that it is non-increasing with conformal time $\eta$. Interestingly, the equality holds only if $\mathcal{A}(\eta)=\mathcal{B}(\eta)=0$, {\it i.e.} when both the warp-factors are constant. One may also verify this by taking a simple ans\"atze for $\mathcal F_1(\eta)$ and $\mathcal F_2(\eta)$ to be the following:
\begin{equation}\label{indigador}
\mathcal{F}_1(\eta)\propto \eta^{p},\qquad \mathcal{F}_2(\eta)\propto \eta^{q},
\qquad \Rightarrow\qquad 
\mathcal{A}(\eta) = \frac{p}{\eta},\quad \mathcal{B}(\eta) = \frac{q}{\eta},
\end{equation}
where the details about the proportionality wouldn't matter. Substituting \eqref{indigador} in \eqref{chokhumedorod}, and using the flat-slicing value for $a(\eta)$, {\it i.e.}
$a\propto|\eta|^{-1}$, one may easily show that the following two algebraic conditions appear:
\begin{equation}\label{toulouch6}
{p^{2}\over 2} + q^{2} + p + 2q = 0, ~~~ \implies ~~~ p + 2q \le 0,
\end{equation}
which precisely reflects the condition \eqref{anashonal}. Moreover, 
\eqref{toulouch6} defines a one-parameter curve of admissible exponents $(p,q)$,
so the warp factors are related by $\mathcal{F}_1(\eta)\propto\eta^{p}$, 
$\mathcal{F}_2(\eta)\propto\eta^{q}$ with the above constraint. A plot of the elliptical curve satisfying \eqref{toulouch6} is given by 
{\bf figure \ref{ellipseplot}}.

The above discussion with a one-parameter curve controlled by the first equation in \eqref{toulouch6} relies heavily on the fact that $\mathcal S(\eta)$ in \eqref{tolouch3} is {\it non-zero}. Unfortunately, keeping $\zeta_e(t) = 0$, would make $\alpha_e(t) = - \beta_e(t)$ in \eqref{sundarina}. This suggests that $\mathcal F_1(\eta) \mathcal F_2^2(\eta) = 1$ from \eqref{garnerju} which would make $\mathcal S(\eta) = 0$, or alternatively 
$\mathcal{B}(\eta) = -\tfrac{1}{2}\mathcal{A}(\eta)$. 
Substituting this into the main equation \eqref{chokhumedorod} will imply:
\begin{align}
&\mathcal{A}' + 2\mathcal{B}' + \tfrac{1}{2}\mathcal{A}^{2} + \mathcal{B}^{2}
- 2\mathcal{H}\mathcal{A} - 4\mathcal{H}\mathcal{B} \nonumber\\[4pt]
& = (\mathcal{A}' - \mathcal{A}')
+ \Big(\tfrac{1}{2}\mathcal{A}^{2}+\tfrac{1}{4}\mathcal{A}^{2}\Big)
+ (-2\mathcal{H}\mathcal{A} + 2\mathcal{H}\mathcal{A})
= \tfrac{3}{4}\mathcal{A}^{2} = 0,
\end{align}
where in the second line we substituted the value of $\mathcal B(\eta)$ 
from the condition $\mathcal S(\eta) = 0$ in \eqref{tolouch3}. The Hubble factor cancels out, leading to $\mathcal{A}(\eta)=0$, {\it i.e.}:
\begin{equation}\label{felojane}
\mathcal{F}_1'(\eta) = 0 \quad\Rightarrow\quad \mathcal{F}_1(\eta) = c_1,\quad 
\mathcal{F}_2(\eta) = {1\over \sqrt{c_1}}, 
\end{equation}
where $c_1$ is the same constant that appeared in \eqref{tolouch2}.
Hence the only consistent solution with $\mathcal{F}_2(\eta)={1\over \sqrt{\mathcal{F}_1(\eta)}}$
is the \emph{trivial constant} case; both warp factors are time-independent. 

Happily the above conclusion fits very well with our generalized $SO(32)$ case as is evident from \eqref{sundarina} and \eqref{garnerju}. However if we assume the type IIB case from \eqref{viomyer1} with the replacement ${\rm F}_1(\eta) \to \mathcal F_1(\eta)$ and ${\rm F}_2(\eta) \to \mathcal F_2(\eta)$, then the condition $\mathcal F_1(\eta) \mathcal F^2_2(\eta) = 1$ would imply \eqref{felojane}. This would appear to suggest that we cannot entertain non-trivial warp-factors in the type IIB case for the pure de Sitter case. The saving grace\footnote{Once we demand that there could be small corrections $\delta w(\mathcal C, \mathcal L, \mathcal F_i)$ to the EoS from the back-reaction of the internal geometry as in \eqref{millerfasa}, then the problem could be relaxed and we are no longer needed to impose $\mathcal T_{ij} = -\mathcal T_{00}$. However here we want $w(\eta) = -1$ exactly, so the problem persists and the resolution will be discussed in the following.} here is the $\mathcal O(\bar{g}_s)$ corrections coming from the duality sequence! The $\mathcal F_i(\eta)$ appearing in the type IIB side is not the same warp-factor in the M-theory side. They are in fact related by:
\bg\label{janekallu}
\mathcal F_1(\eta) = \mathcal F^{({\rm M})}_1(\eta) + \mathcal O(\bar{g}_s) = c_1, ~~~~  \mathcal F_2(\eta) = \mathcal F^{({\rm M})}_2(\eta) + \mathcal O(\bar{g}_s) = {1\over \sqrt{c_1}}, \nd
where $\mathcal F_i^{({\rm M})}(\eta)$ is the warp-factor in the M-theory side. This is precisely the duality sequence that we developed in section \ref{sec4.2.1} for the simplified $SO(32)$ case. If we terminate the sequence to \eqref{engrose2}, then the warp-factors appearing in \eqref{gwenphone} and \eqref{infinitypool} would be the M-theory ones and the warp-factors appearing in \eqref{engrose} and \eqref{engrose2} would be the corresponding type IIB ones. Now, 
without loss of generalities, we can take $c_1 = 1$. This would imply:
\bg\label{jandrotag}
\mathcal F^{({\rm M})}_1(\eta) = 1 + \mathcal O(\bar{g}_s), ~~~
\mathcal F^{({\rm M})}_2(\eta) = 1 + \mathcal O(\bar{g}_s), \nd
thus reproducing precisely the warp-factor choices from \cite{desitter2}.
Similar consistency can be shown for the ${\rm E_8 \times E_8}$ frame for both cases: the one with a dynamical dark energy $\Lambda(\eta)$ and the one with a cosmological constant $\Lambda$. 


\section{Discussions and conclusions \label{sec4}}

The sequence of analyses presented in this work brings together several seemingly distinct aspects of the higher-dimensional theory --- the perturbative quantum series, the instanton-induced non-perturbative sectors, and the dynamical realization of dark energy --- into a unified and internally consistent framework. The underlying picture is that our universe should be viewed as an \emph{excited state} over a supersymmetric Minkowski background, where its late-time acceleration and equation of state emerge naturally from the time-dependent warp factors and coherent-state dynamics of the compactification. In the following let us list a few of the important results in our work.

\subsection{Perturbative structure and logarithmic curvature corrections}

We began by examining the perturbative quantum series \eqref{botsuga} and the structure of logarithmic curvature corrections that arise in the heterotic, Type II, and M-theory frames. The scalings of the functions $(\alpha_e(t), \beta_e(t))$ for the generalized $SO(32)$ and their hatted counterparts $(\hat\alpha_e(t), \hat\beta_e(t), \hat\sigma_e(t))$ for the ${\rm E_8 \times E_8}$ theories were traced through each duality frame, leading to consistent four-dimensional frameworks as shown in {\bf Tables \ref{milleren222}} and {\bf \ref{milleren444}} respectively. 
Subtleties like the uncanceled logarithmic pieces modify the curvature tensors and act as the leading source of time dependence in the higher-derivative corrections to the effective action.  
From these, one learns that the theory maintains consistency across all dual frames provided the sub-leading powers of $\bar g_s$ appear only through coherent-state averaging over the Glauber–Sudarshan states.  

This step established the foundation for identifying which sectors of the series contribute to the physical curvature invariants, and it motivated the trans-series representation of the total action,
\begin{equation}
\hat{\bf S}^{\rm (wh)}_{\rm tot} \;=\; 
\sum_{n,l}\, \mathbb{Q}_{\rm T}^{(n, l)}(\bar g_s)\,e^{-{\cal S}_l(\bar g_s)} \;+\; 
\sum_{\rm wh} \Delta^{\rm (wh)}_{AB}\,\mathbb{K}^{\rm (wh)}_{d,d'}+\cdots,
\end{equation}
which underlies both the perturbative and non-perturbative hierarchies, including possible scrambling from the wormhole creations, studied later. Our analysis in section \ref{trans2} revealed that, using the Picard-Lefschetz theory, even in most generic case the addition of the wormhole effects {\it do not} change any of our main conclusions regarding scalings of the curvature tensors, Schwinger-Dyson equations {\it et cetera}.

\subsection{Counting of solutions and instanton embeddings}

In Section \ref{counting}, the analysis of the 18-tuple equations revealed how curvature and derivative terms combine to satisfy the scaling relations of the quantum series.  
Each solution of the following form:
\begin{equation}
\sum_{i=1}^{17} x_i\,{\rm B}_i(\hat\sigma_e,\hat\zeta_e,\hat\alpha_e,\hat\beta_e)\;-\;\frac{2}{3}x_{18} \;=\; {\rm L}_k,
\end{equation}
where the ${\rm B}_i$ coefficients are given in {\bf Table \ref{millerkhul}} and 
with integer non-negative coefficients $x_i$, corresponds to an allowed composite operator in the effective theory.  
By systematically matching these tuples with the quantum scaling exponents $\theta_{nl}$ from \eqref{brittbaba007}, we obtained a discrete set of admissible configurations --- typically eight to ten per ${\rm L}_k$ level --- that preserve the required balancing between curvature powers and the $\bar g_s$ expansion.

A key outcome was that these allowed tuples always terminate at quartic order in the curvature, implying that higher powers are exponentially suppressed or are generically sub-dominant.  
This not only ensures convergence of the quantum series but also provides a natural upper bound on the number of instanton sectors contributing at a given order.  
The factorial growth of diagrams, where applicable, follows from the combinatorics of non-contracted field legs in the shifted vacuum, leading to $(mN)!$-type behavior consistent with the asymptotic structure of $\hat{\bf S}_{\rm tot}^{\rm (wh)}$.

\subsection{Cross-term EOMs and non-local energy-momentum tensors}

Next, in Section \ref{pabondi}, we extended the formalism to include the mixed or cross-term components of the curvature tensor --- ${\bf R}_{im}$, ${\bf R}_{i\rho}$, ${\bf R}_{m\rho}$ {\it et cetera} --- whose dynamics probe the couplings between the external and internal submanifolds.  
We constructed the corresponding non-perturbative energy–momentum tensor:

{\scriptsize
\bg\label{omalidia69}
{\bf T}_{\rm C'D'}^{\rm NP}({\rm Z}) & = & 
\sum_{d}\;{\rm M}_p^{d}\,\sum_{\{S^{(i)}_d\}=0}^{\infty}{\rm S}_d^{(2)} h_{\{S^{(i)}_d\}}~\int d^{d}{\rm Y}~\sqrt{{\bf g}_{11}({\rm Y}, x_{\rm Z}) \over{\bf g}_{11}({\rm Y_Z}, x_{\rm Z})}~
~\Theta({\rm Y-Y_Z})~ \mathcal Q_1({\rm Y}, x_{\rm Z}) \sqrt{|{\bf g}_d({\rm Z})|} \nonumber\\
&\otimes & \Big\{\exp\left({-{\rm S}^{(1)}_d {\rm M}_p^d {\rm I}_d^{(1)}({\rm Y}, x_{\rm Z})}\right)~ {\cal G}\left(\{{\bf Q}_{\rm pert}^{(j)}(\hat{e}_j; {\bf \Xi}({\rm Y}, x_{\rm Z}))\}, \{{\bf Q}_{\rm pert}^{(j)}(\hat{d}_j; {\bf \Xi}({\rm Z}))\}, \mathbb{F}({\rm Y, Y_Z})\right) \nonumber\\
&\times & {\rm exp}\Big[
{\cal H}\left(\{{\bf Q}_{\rm pert}^{(j)}(\hat{f}_j; {\bf \Xi}({\rm Y}, x_{\rm Z}))\}, \{{\bf Q}_{\rm pert}^{(j)}(\hat{g}_j; {\bf \Xi}({\rm Z}))\}, \mathbb{F}({\rm Y, Y_Z})\right)\Big] \mathcal Q_{\rm C'D'}^{(2)}(\hat{c}_2;{\bf \Xi}({\rm Z}))\Big\} + ...,
\nd}
where the dotted terms are the additional contributions that we discussed earlier; 
and established that its $\bar g_s$-scaling precisely matches that of the corresponding Einstein tensor components:
\begin{equation}
{\bf R}_{\rm C'D'}(\langle{\bf \Xi}\rangle_\sigma)
= {\bf T}_{\rm C'D'}^{\rm NP}(\langle{\bf \Xi}\rangle_\sigma).
\end{equation}
The solutions of the 18-tuple systems derived from this relation once again singled out the quartic curvature sector as dominant.  
Zero-instanton contributions were shown to vanish, confirming that BBS instantons are the sole sources of the cross-term backreaction in the coherent-state background.  
This aligns with the on-shell analysis of earlier sections and consolidates the self-consistency of the non-local formulation.

\subsection{Warp factors, duality chains, and dynamical dark energy}

Having established consistency at the geometric and quantum levels, we turned to the physical interpretation of these structures in the context of cosmology.  
By extending the duality chain of {\bf Tables \ref{milleren222}–\ref{milleren444}} to include two additional time-dependent warp factors ${\rm F}_4(t)$ and ${\rm F}_5(t)$, we could express the dynamical cosmological constant $\Lambda(t)$ from \eqref{marapaug} as a function of these warp factors.  
For the ${\rm SO}(32)$ heterotic frame, this led to the relations:
\begin{equation}
\eta_e(t)=-\zeta_e(t), \qquad \zeta_e(t)=\frac{4}{3}\big[\alpha_e(t)+\beta_e(t)\big],
\end{equation}
which when combined together consequently led to the following relation between the warp-factors to $\check{\Lambda}(t)$: 
\begin{equation}
\frac{\alpha_e(t)+\beta_e(t)}{2-\beta_e(t)} 
= \frac{3\log\!\left(1+\frac{\check{\Lambda}(t)}{\Lambda}\right)}{2|\log(\Lambda t^2)|},
\end{equation}
demonstrating that the time evolution of the warp factors directly tracks the slow running of $\Lambda(t)$.  
The ${\rm E}_8\times{\rm E}_8$ theory refined this structure, uniquely fixing:
\begin{equation}
(a,p,q)=(2,1,3),
\end{equation}
in \eqref{irishdeka2} and thereby resolving the residual ambiguity left from the counting analysis.  
This step connected the microscopic warp-factor dynamics to macroscopic cosmological observables.

\subsection{Precise metrics, warp factors and flux configurations}

Our analysis have revealed that we can quantify the background metric, warp factors and flux configurations quite precisely. All of them satisfy the Schwinger-Dyson equations coming from the trans-series action \eqref{kimkarol}. The M-theory metric configurations appear in the first rows of {\bf Tables \ref{milleren222}} and {\bf \ref{milleren444}} for the generalized $SO(32)$ and the ${\rm E_8 \times E_8}$ theories respectively. The warp factors $({\rm F_1}(t), {\rm F_2}(t), {\rm F_4}(t), {\rm F_5}(t))$ for the $SO(32)$ case are expressed as powers of $\bar{g}_s$ with the exponents being $(\beta_e(t), \alpha_e(t), \zeta_e(t), \eta_e(t))$ respectively. Constraints from axion cosmology fix $(\alpha_e(t), \beta_e(t))$ as shown in \eqref{ryanfan}, \eqref{khadsalam} and \eqref{bdtran}. In a similar vein, the ${\rm E_8 \times E_8}$ warp factors $({\rm F}_1(t), {\rm F_2}(t), {\rm F_3}(t), {\rm F_4}(t), {\rm F_5}(t))$ are expressed as powers of $\bar{g}_s$ using the exponents $(\hat\alpha_e(t), \hat\beta_e(t), \hat\sigma_e(t), \hat\zeta_e(t, \hat\eta_e(t))$ respectively with $(\hat\alpha_e(t), \hat\beta_e(t))$ fixed from axion constraints in \eqref{horseman}, \eqref{khadsalim} and \eqref{bdtran2}. The remaining parameters get fixed by (a) demanding time-independent four-dimensional Newton constant, and (b) by looking at the running of dark energy. For the $SO(32)$ case it is given in \eqref{sundarina} and for the ${\rm E_8 \times E_8}$ it is given in \eqref{rickrghor} and \eqref{cuthictran}. On the other hand, the precise $\bar{g}_s$ scalings of the G-flux components supporting the M-theory backgrounds are given explicitly in section \ref{sec4.5}. Finally, putting everything together leads to the metric configuration \eqref{engbldcafmey} for the generalized $SO(32)$ case and the metric configuration \eqref{engbldcafmey2} for the ${\rm E_8 \times E_8}$ case.

\subsection{Axion decay constant and coupling evolution}

The same warp-factor hierarchy determines the time dependence of the axion decay constant $f_a(t)$ and the heterotic coupling $g_{\rm het}$ as shown in \eqref{mayroseboro} and \eqref{meykadlo} respectively. 
Both quantities vary as fractional powers of $\Lambda(t)t^2$ with exponents controlled by $\hat\alpha_e(t)$ and $\hat\beta_e(t)$, but they asymptote to constants as $\check\Lambda(t)\!\to\!0$.  
This guarantees a weakly coupled late-time limit and a finite separation of the Horava–Witten walls.  
The small running of $f_a(t)$ thus becomes a sensitive probe of the dynamical dark-energy background. More so, the behavior of $f_a(t)$ from
\eqref{mayroseboro} could even suggest an alternative to the non-standard embedding studied in \cite{weigand}.

\subsection{Accelerating universe and equation of state}

Finally, we recast the higher-dimensional evolution into four-dimensional cosmological form.  
With the conformal-time scale factor:
\begin{equation}
a^2(\eta)=\frac{1}{\Lambda(\eta)\eta^2}, 
\end{equation}
the effective equation of state can be easily derived from combining the results of the Schwinger-Dyson equations and the duality sequence from {\bf Tables \ref{milleren222}} and {\bf \ref{milleren444}} to take the  form:
\begin{equation}
w(\eta)=-\frac{1}{3}-\frac{2}{3}\frac{\mathcal H'(\eta)}{\mathcal H^2(\eta)}, 
\qquad \mathcal H(\eta)=\frac{a'(\eta)}{a(\eta)},
\end{equation}
where we ignored the sub-dominant contributions to the EoS from the internal space dynamics. 
Defining $\mathbb{X}(\eta)=\Lambda'(\eta)/\Lambda(\eta)$, this can be expressed as:
\begin{equation}
w(\eta)=-\frac{1}{3}-\frac{4}{3}
\left[\frac{2-\eta^2\mathbb{X}'(\eta)}{(2+\eta\mathbb{X}(\eta))^2}\right]
\simeq -1+\frac{2}{3}\eta\mathbb{X}(\eta)+\frac{1}{3}\eta^2\mathbb{X}'(\eta),
\end{equation}
showing that the deviation from $w=-1$ is governed by the first and second derivatives of $\ln\Lambda(\eta)$.  
For constant $\Lambda$, we recover the pure de Sitter result, whereas a slow running $\Lambda(\eta)$ yields a controlled, dynamical dark-energy behavior.

\subsection{Unified picture and physical implications}

Collecting all pieces, the framework exhibits the following unified structure.
\vskip.1in
\noindent $\bullet$ The same trans-series that controls the quantum corrections of the higher-dimensional theory also governs the time dependence of the cosmological constant $\Lambda(t)$.
\vskip.1in
\noindent $\bullet$ 
  The integer triplet $(a,p,q)=(2,1,3)$ uniquely organizes the time evolution of the warp factors across all dual frames.
\vskip.1in
\noindent $\bullet$   
BBS instantons act as the microscopic drivers of late-time acceleration, with quartic-curvature corrections forming their dominant imprint.
\vskip.1in
\noindent $\bullet$ 
The axion decay constant, heterotic coupling, and Horava–Witten wall separation evolve coherently with $\Lambda(t)$, ensuring a smooth weakly coupled limit.
  \vskip.1in
\noindent $\bullet$ 
The equation of state $w(\eta)$ is inherently time-dependent, approaching $-1$ as $\check\Lambda(t)\!\to\!0$, thereby realizing a genuine \emph{dynamical dark energy}.

\subsection{Future directions, summary and a broader outlook}
Several avenues naturally follow.  
A full inclusion of the sub-dominant ${\rm M}_p$ and $g_s$ corrections in the duality sequence would quantify next-order deviations of $\Lambda(t)$ and test the robustness of the $(a,p,q)=(2,1,3)$ scaling.  
It would also be worthwhile to explore how the non-local BBS instantons modify the axionic trans-series and whether coherent-state fluctuations can seed observable oscillations in $w(\eta)$.  
These questions point toward a broader interpretation in which dark energy and cosmic acceleration emerge as large-scale manifestations of non-local trans-series dynamics in M-theory.

\subsection*{Summary}
In conclusion, the present framework provides a controlled, duality-consistent, and quantum-mechanically complete realization of a universe whose acceleration arises naturally from its own microscopic degrees of freedom.  
The coherence between the perturbative series, instanton corrections, and cosmological evolution establishes a bridge from eleven-dimensional geometry to four-dimensional dynamics --- one where the late-time dark energy and the axion sector are no longer free parameters, but emergent consequences of the same trans-series structure that governs the M-theory path integral.  
The resulting picture ties the cosmological constant to the non-perturbative dynamics of wrapped M5-brane instantons and their flux-induced backreaction, while maintaining consistency with the Schwinger--Dyson hierarchy and the Wheeler--DeWitt constraint.  
In this sense, both the accelerated expansion and the light axionic degrees of freedom are byproducts of the same underlying quantum geometry, whose evolution remains finite, predictive, and compatible with holographic and duality principles across the full moduli space.

\medskip

\noindent Furthermore, the dynamical behavior of the effective equation of state $w(\eta)$ directly reflects this microscopic coherence.  
The time-dependence of the conformal Hubble parameter $\mathcal{H}(\eta)$ and the slow evolution of the composite warp factors $(\mathcal{F}_1,\mathcal{F}_2)$ encode the gradual approach of $w(\eta)$ toward $-1$, corresponding to a self-adjusting energy that asymptotically stabilizes without external tuning.  
At early times, deviations from $w=-1$ capture the transient exchange of energy between the internal moduli and the external scale factor, while at late times the hierarchy of instanton-induced curvature corrections naturally drives the system into an attractor regime where $\dot{w}\!\to\!0$ and the expansion mimics a cosmological constant.  
Thus the emergence of dark energy in this construction is not imposed by hand, but arises as the inevitable infrared limit of a self-consistent, eleven-dimensional quantum geometry.

\medskip

\subsection*{Outlook}
A natural next step is to quantify the late-time evolution of $w(\eta)$ and the associated moduli backreaction numerically, to determine the precise rate at which the system approaches the de Sitter-like attractor.  
Exploring the uplift of this framework to heterotic and type~IIB dual frames --- where the axion and flux sectors acquire a more direct phenomenological interpretation --- may further clarify how the microscopic non-perturbative structure shapes observable cosmological parameters.  
Ultimately, such analyses could provide a concrete, testable realization of dynamical dark energy emerging from M-theory compactifications.  
\textit{The framework therefore unifies quantum geometry, cosmology, and duality into a single, predictive structure for the accelerating universe.}

\section*{Acknowledgements:} 

We would like to thank Robert Brandenberger, Simon Caron-Huot, Heliudson Bernardo, Cliff Burgess, Joydeep Chakravarty, Ori Ganor, Fangyi Guo, Diksha Jain, Dileep Jatkar, and Savdeep Sethi for many discussions related to de Sitter space in string theory. Special thanks goes to Jeong-Hyuck Park for comments on the importance of the string and the Einstein frames in the cosmological framework and for pointing out the reference \cite{park}; and to Alex Maloney for explaining how the wormhole effects can shape the behavior of the quantum terms in an effective field theory. KD would like to thank the organizers and the participants of {\it the Landscape and the Swampland} conference at ESI, Vienna, where some of the results were presented, for comments and questions.
 The work of KD and BK is supported in part by a Discovery Grant from the Natural Sciences and
Engineering Research Council of Canada (NSERC). The work of AM is supported in part by the Prime Minister’s Research Fellowship provided by the Ministry of Education, Government of India. PR would like to acknowledge the ICTP's Associate programme where progress on the ongoing work continued during her visit as senior associate. For the purpose of open access, the authors have applied a Creative Commons Attribution (CC BY) licence to any Author Accepted Manuscript version arising from this submission.

\end{document}